\documentclass[twocolumn]{aastex631}

\usepackage{threeparttable}
\usepackage{amsmath}
\usepackage{multirow}

\usepackage{soul}
\newcommand{\add}[1]{\textbf{\textcolor{blue}{#1}}}

\begin{document}


\title{Bright bursts with sub-millisecond structures of FRB\,20230607A in a highly magnetized environment}

\correspondingauthor{J.L. Han; Bing Zhang}
\email{hjl@nao.cas.cn; bing.zhang@unlv.edu}

\author[0000-0002-6423-6106]{DeJiang Zhou}
\affiliation{National Astronomical Observatories, Chinese Academy of Sciences, Beijing 100101, China}

\author[0000-0002-9274-3092]{J. L. Han} 
\affiliation{National Astronomical Observatories, Chinese Academy of Sciences, Beijing 100101, China}
\affiliation{Key Laboratory of Radio Astronomy and Technology, Chinese Academy of Sciences, Beijing 100101, China}

\author[0000-0002-9725-2524]{Bing Zhang} 
\affiliation{Nevada Center for Astrophysics, University of Nevada, Las Vegas, NV 89154-4002, USA}
\affiliation{Department of Physics and Astronomy, University of Nevada, Las Vegas, NV 89154-4002, USA}

\author[0000-0001-5105-4058]{WeiWei Zhu} 
\affil{National Astronomical Observatories, Chinese Academy of Sciences, Beijing 100101, China}
\affil{Institute for Frontiers in Astronomy and Astrophysics, Beijing Normal University, Beijing 102206, China}

\author[0000-0001-9036-8543]{Wei-yang Wang} 
\affil{School of Astronomy and Space Science, University of Chinese Academy of Sciences, Beijing 100049, China}
\affil{Department of Astronomy, School of Physics, Peking University, Beijing 100871, China}

\author[0000-0001-6374-8313]{Yuan-Pei Yang} 
\affil{South-Western Institute for Astronomy Research, Yunnan University, Kunming, Yunnan 650504, China}
\affil{Purple Mountain Observatory, Chinese Academy of Sciences, Nanjing 210023, China} 

\author[0000-0003-4721-4869]{Yuanhong Qu}  
\affil{Nevada Center for Astrophysics, University of Nevada, Las Vegas, NV 89154-4002, USA}
\affil{Department of Physics and Astronomy, University of Nevada, Las Vegas, NV 89154-4002, USA}

\author[0000-0002-8744-3546]{Yong-Kun Zhang} 
\affil{National Astronomical Observatories, Chinese Academy of Sciences, Beijing 100101, China}
\affil{School of Astronomy and Space Science, University of Chinese Academy of Sciences, Beijing 100049, China}

\author[0009-0008-1612-9948]{Yi Yan} 
\affil{National Astronomical Observatories, Chinese Academy of Sciences, Beijing 100101, China}
\affil{School of Astronomy and Space Science, University of Chinese Academy of Sciences, Beijing 100049, China}

\author[0000-0002-1056-5895]{Wei-Cong Jing} 
\affil{National Astronomical Observatories, Chinese Academy of Sciences, Beijing 100101, China}
\affil{School of Astronomy and Space Science, University of Chinese Academy of Sciences, Beijing 100049, China}

\author{Shuo Cao} 
\affil{Yunnan Observatories, Chinese Academy of Sciences, Kunming 650216, People's Republic of China}
\affil{School of Astronomy and Space Science, University of Chinese Academy of Sciences, Beijing 100049, China}

\author[0000-0001-5649-2591]{Jintao Xie} 
\affil{Research Center for Astronomical Computing, Zhejiang Laboratory, Hangzhou 311100, China}
\affil{School of Computer Science and Engineering, Sichuan University of Science and Engineering, Yibin 644000, China}

\author[0009-0009-3517-6640]{Xuan Yang} 
\affil{Purple Mountain Observatory, Chinese Academy of Sciences, Nanjing 210008, People's Republic of China}
\affil{School of Astronomy and Space Sciences, University of Science and Technology of China, \\ Hefei 230026, People's Republic of China}

\author{Shiyan Tian} 
\affil{Department of Astronomy, School of Physics, Huazhong University of Science and Technology,
Wuhan 430074, China}

\author[0000-0001-5931-2381]{Ye Li} 
\affil{Purple Mountain Observatory, Chinese Academy of Sciences, Nanjing 210008, People's Republic of China}

\author[0000-0001-7931-0607]{Dongzi Li} 
\affil{Cahill Center for Astronomy and Astrophysics, California Institute of Technology, Pasadena, CA, USA}

\author[0000-0001-8065-4191]{Jia-Rui Niu} 
\affil{National Astronomical Observatories, Chinese Academy of Sciences, Beijing 100101, China}
\affil{School of Astronomy and Space Science, University of Chinese Academy of Sciences, Beijing 100049, China}

\author[0000-0002-1381-7859]{Zi-Wei Wu} 
\affil{National Astronomical Observatories, Chinese Academy of Sciences, Beijing 100101, China}

\author[0000-0001-6021-5933]{Qin Wu} 
\affil{School of Astronomy and Space Science, Nanjing University, Nanjing 210093, China}

\author[0000-0002-0475-7479]{Yi Feng} 
\affiliation{Research Center for Astronomical Computing, Zhejiang Laboratory, Hangzhou 311100, China}
\affil{Institute for Astronomy, School of Physics, Zhejiang University, Hangzhou 310027, China}

\author[0000-0003-4157-7714]{Fayin Wang} 
\affil{School of Astronomy and Space Science, Nanjing University, Nanjing 210093, China}
\affil{Key Laboratory of Modern Astronomy and Astrophysics, Nanjing University, Nanjing 210093, China}

\author[0000-0002-3386-7159]{Pei Wang}
\affiliation{National Astronomical Observatories, Chinese Academy of Sciences, Beijing 100101, China}









\begin{abstract}
We report the observations of a repeating FRB 20230607A for 15.6 hours spanning 16 months using the Five-hundred-meter Aperture Spherical Radio Telescope (FAST) with the detection of 565 bursts. We present three bright bursts with detailed temporal/spectral structures. We also report that one burst carries a narrow component with a width of only 0.3 ms, which is surrounded by broader components. This suggests that repeaters can make both narrow and broad components in one burst. With the narrow spike, we precisely measure the dispersion measure (DM) of $362.85 \pm 0.15 \;{\rm pc\,cm^{-3}}$ and the Faraday rotation measures (RMs) of and $-12249.0\pm 1.5 \; {\rm rad\,m^{-2}}$. We also analyze the statistical distribution of the burst parameters,  including waiting times, temporal widths, central frequencies and frequency widths, fluences and energies, all showing typical distributions of known active repeaters. In particular, most bursts show narrow spectra with $\Delta\nu/\nu_0 = 0.125\pm 0.001$. This fact, together with the narrow 0.3 ms spike, strongly suggests a magnetospheric origin of the FRB emission. Based on a predicted correlation between RM and the luminosity of a persistent radio source (PRS) by Yang et al., we predict that PRS should have a specific luminosity of the order of $10^{29} \ {\rm erg \ s^{-1} \ Hz^{-1}}$ and encourage a search for such a PRS. 
\end{abstract}



\section{Introduction}\label{sec:intro}

Fast radio bursts \citep[FRBs,][]{Lorimer2007Sci...318..777L} are coherent radio flashes with a duration of milliseconds. Their dispersion measures (DMs) exceed the maximum DM affordable by the Milky Way \citep{2013Sci...341...53T}. Localizations of some FRBs have firmly placed them at cosmological distances \citep{2020Natur.581..391M}. Observationally there are two apparently different types of FRBs: repeating bursts and one-off bursts \citep{CHIME2019ApJ...885L..24C}. The physical origin and emission mechanism of FRBs have remained enigmatic \citep{ZhangB2023RvMP...95c5005Z}. 

Sensitive observations and comprehensive investigations of the details of FRB emission provide keys to understand the nature of FRBs. For instance, polarization measurements can elucidate whether they originate from a source with a magnetosphere \citep{LuoR2020Natur.586..693L,XuH2022Natur.609..685X,JiangJC2022RAA....22l4003J,NiuJR2024ApJ...972L..20N,JiangJC2024arXiv240803313J}.
A statistical study shows that the bursts from repeaters exhibit more complex burst morphology and larger temporal widths compared to one-off bursts, and that some bursts have very narrow spectra \citep{Pleunis2021ApJ...923....1P}.
%
Some repeating bursts have very long durations with connected pulses \citep{CHIME2021ApJS..257...59C, Zhou2022RAA....22l4001Z}. High-time resolution observations also revealed the microstructure of some bursts, with a time scale from nanoseconds to microseconds \citep{Farah2018MNRAS.478.1209F,Nimmo2021NatAs...5..594N,Nimmo2022NatAs...6..393N,Hewitt2023MNRAS.526.2039H,Snelders2023NatAs.tmp....5S}. These narrow spikes can help to very precisely measure the DM of some FRB bursts. 

When a large sample of repeated bursts are detected, the energy distributions of bursts can be analyzed \citep{LiD2021Natur.598..267L, XuH2022Natur.609..685X, ZhangYK2022RAA....22l4002Z, ZhangYK2023ApJ...955..142Z}. 
Statistics of the FRB energy distribution may be related to source energy release processes. For example, based on the observational data of FRB\,20121102A from the Arecibo telescope, a high-energy tail was identified beside the main peak in the energy distribution \citep{Jahns2023MNRAS.519..666J}. Such high-energy bursts have been also detected for FRB\,20201124A \citep{Kirsten2024},  FRB\,20220912A \citep{KonijnDC2024arXiv240710155K}, and FRB\,20200120E \citep{ZhangSB2024NatCo..15.7454Z}.


\begin{figure}[th]%
\centering
\includegraphics[width=0.98\columnwidth]{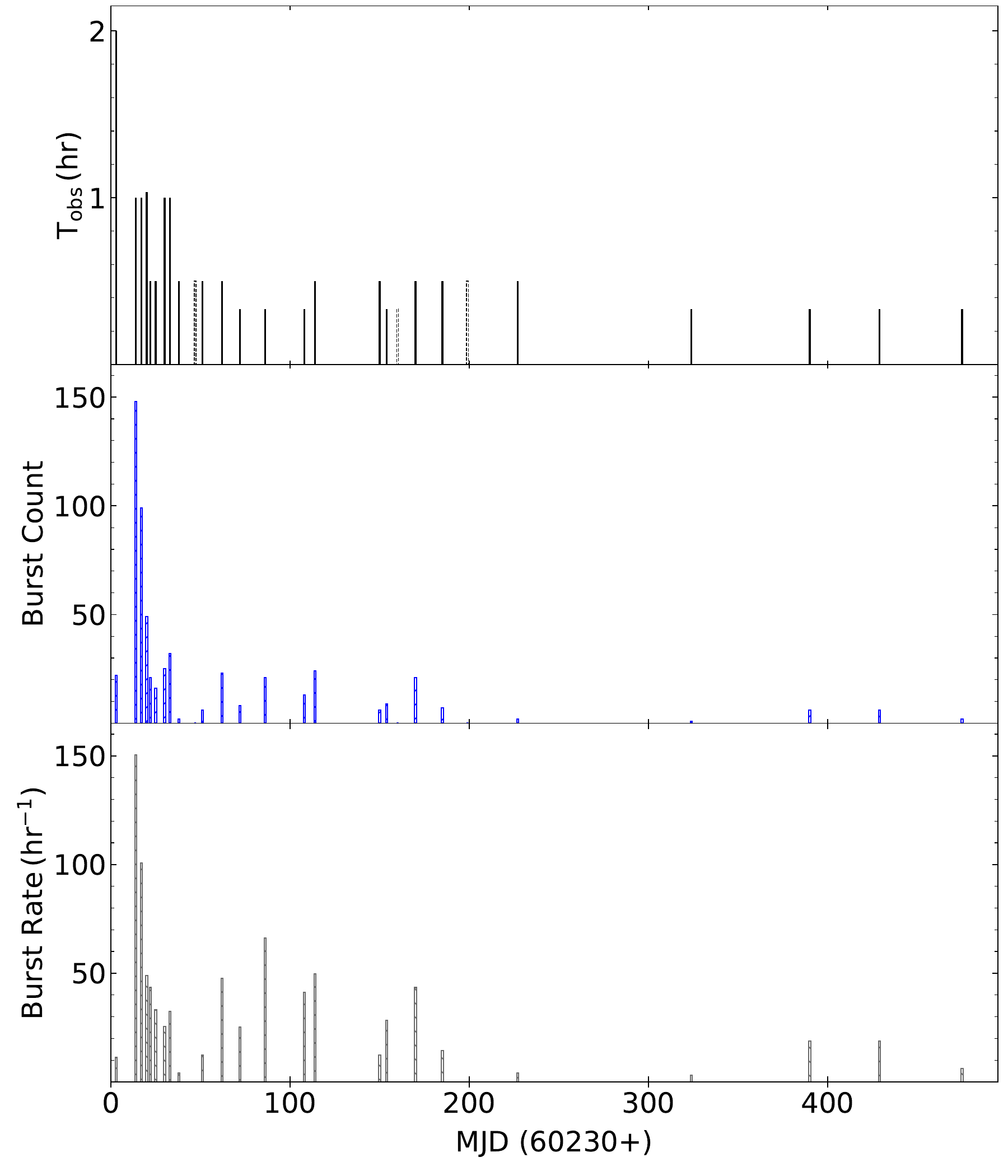}
\caption{The observation duration of the FAST, detected 
burst number and burst rate of each observation session for FRB\,20230607A. The relevant parameters are listed in Table \ref{tab1}.
}\label{figObs}
\end{figure}

Polarization measurements of FRBs provide special constraints of the FRB sources and environments. The extremely large Faraday rotation measures ($|\rm RMs|$) of the two well-known repeating FRBs, FRB\,20121102A \citep{Chatterjee2017Natur.541...58C,Michilli2018Natur.553..182M} and FRB\,20190520B \citep{NiuCH2022Natur.606..873N,AnnaT2023Sci...380..599A} have been detected. Incidentally, these two sources are also each associated with a persistent radio source (PRS) \citep{Chatterjee2017Natur.541...58C,Michilli2018Natur.553..182M,MarcoteB2017ApJ...834L...8M,NiuCH2022Natur.606..873N,AnnaT2023Sci...380..599A}, implying a strong synchrotron nebula in the vicinity of the FRB sources. 
%
\citet{Yang20,YangYP2022ApJ...928L..16Y} suggested that the luminosity of the PRS should be related to the RM if the PRS is located around the FRB emission region and produces the high RM. Such a predicted correlation has been confirmed by the new observation results of FRB\,20201124A, FRB\,20181030A, FRB\,20190407A, and FRB\,202401124A \citep{BhusareY2024ATel16820....1B,Bruni2024Natur.632.1014B,Bruni2024arXiv241201478B,Ibik2024arXiv240911533I,ZhangX2024ATel16695....1Z,ZhangX2025arXiv250114247Z}\footnote{Note that in the case of FRB\,202401124A, a flaring radio source was discovered \citep{ZhangX2024ATel16695....1Z,ZhangX2025arXiv250114247Z}. Long-term monitoring of the source is needed to confirm whether the source will settle down as a persistent radio source. }. 
In addition, significant variation of RM has been discussed in many repeaters, including FRB\,20201124A \citep{XuH2022Natur.609..685X,RaviV2022MNRAS.513..982R}, FRB\,20190520B \citep{AnnaT2023Sci...380..599A}, FRB\,20180916B \citep{Mckinven2023ApJ...950...12M} and a few others, suggesting a dynamically evolving magnetized environment around these sources. 

\begin{table*}[t]
\caption{FAST observations of FRB 20230607A. The columns are observation date with MJD, observation mode, effective observation time in minutes, sampling time, number of bursts detected, burst rate in number per hour, and sub-burst number. 
}\label{tab1}%
\small
\begin{tabular*}{\textwidth}{@{\extracolsep\fill}ccrrrrr}
\hline
Obs.Date/MJD        & Obs.Mode  & T$_{\rm obs}$     & t$_{\rm samp}$    & No. of Burst   & Burst Rate  & No. of Sub-burst\\
            &           & (min)               & ($\mu$s)     &           & ($\rm hr^{-1}$)            &\\
(1)         &   (2)     &  (3)              & (4)           & (5)       & (6)                       & (7)\\
\hline
20231016/60233    & snapshot  & 118             & 196.608       & 21 (M07)  & 10.7 & 26  \\
20231027/60244    & tracking  & 59              & 98.304        & 147 (M05) & 149.5& 193 \\
20231030/60247    & tracking  & 59              & 49.152        & 99        & 100.7& 138 \\
20231102/60250    & tracking  & 61              & 49.152        & 47        & 46.2 & 56  \\
20231104/60252    & tracking  & 29              & 196.608       & 21        & 43.4 & 25  \\
20231107/60255    & tracking  & 29              & 49.152        & 16        & 33.1 & 19  \\
20211112/60260    & tracking  & 59              & 49.152        & 25        & 25.4 & 41  \\
20211115/60263    & tracking  & 59              & 49.152        & 32        & 32.5 & 32  \\
20231120/60268    & tracking  & 29              & 49.152        & 2         & 4.1  & 2   \\
20231129/60277    & tracking  & 29              & 49.152        & 0         & 0.0  & 0   \\
20231203/60281    & tracking  & 29              & 49.152        & 6         & 12.4 & 7   \\
20231214/60292    & tracking  & 29              & 49.152        & 23        & 47.6 & 31  \\
20231224/60302    & tracking  & 19              & 49.152        & 8         & 25.2 & 10  \\
20240107/60316    & tracking  & 19              & 49.152        & 21        & 66.3 & 26  \\
20240129/60338    & tracking  & 19              & 49.152        & 13        & 41.1 & 14  \\
20240204/60344    & tracking  & 29              & 49.152        & 24        & 49.7 & 27  \\
20240311/60380    & tracking  & 29              & 196.608       & 6         & 12.4 & 6   \\
20240315/60384    & tracking  & 19              & 49.152        & 9         & 28.4 & 11  \\
20240321/60390    & tracking  & 19              & 98.304        & 0         & 0.0  & 0   \\
20240331/60400    & tracking  & 29              & 49.152        & 21        & 43.4 & 26  \\
20240415/60415    & tracking  & 29              & 49.152        & 7         & 14.5 & 8   \\
20240429/60429    & tracking  & 29              & 49.152        & 0         & 0.0  & 0   \\
20240527/60457    & tracking  & 29              & 49.152        & 2         & 4.1  & 2   \\
20240901/60554    & swiftcalibration  & 19      & 49.152        & 1         & 3.2  & 1   \\
20241106/60620 & swiftcalibration & 19 & 49.152  & 6  & 18.9 & 10  \\
20241215/60659 & swiftcalibration & 19 & 49.152  & 6  & 18.9 & 7   \\
20250130/60705 & swiftcalibration & 19 & 49.152  & 2  & 6.3  & 2   \\
\hline
Total             &           & 934 &               & 565       &      & 720 \\
\hline
\end{tabular*}
\end{table*}

Recently, a new repeating FRB, FRB\,20230607A, was discovered by CHIME and reported by CHIME/FRB Virtual Observatory Event (VOEvent) Service\footnote{\url{https://www.chime-frb.ca/voevents}}. 
%
Sensitive observations can reveal the properties of FRBs.
The Five-hundred-meter Aperture Spherical radio Telescope \citep[FAST,][]{Nan2006ScChG}, with the L-band 19-beam receiver, possesses an exceptionally high sensitivity, making it an ideal instrument for detection of numerous bursts from several FRBs \citep{LiD2021Natur.598..267L, JiangJC2022RAA....22l4003J, NiuCH2022Natur.606..873N, NiuJR2022RAA....22l4004N, XuH2022Natur.609..685X, ZhangYK2022RAA....22l4002Z, Zhou2022RAA....22l4001Z, ZhangYK2023ApJ...955..142Z, WuZW2024ApJ...969L..23W, WuZW2024SCPMA..6719512W}.
Following the report by the CHIME/VOEvent service on October 1st, 2023, we scheduled the FAST to observe and monitor it for a few months. The observation details are shown in Table~\ref{tab1}. In total, we observed the source for 15.6 hours and detected 565 bursts, including three bright ones. In this paper, we present the observational results of FRB\,20230607A and reveal its properties. A concise overview of the observations is given in Section \ref{sect2:obs}, and the results and discussions are presented in Section \ref{sect3:result}. The conclusions are given in Section \ref{sect5:con}.

\section{FAST observations and burst detection}
\label{sect2:obs}

FAST has an L-band 19-beam receiver, working in the frequency range from 1.0 GHz to 1.5 GHz \citep{JiangP2020RAA....20...64J}. The system temperature at this band is about 22~K. 
In general, the observation band of 500~MHz is channelized to 4096 frequency channels. Due to the low gain at two sides of the band, only data for an effective bandwidth of 437.5 MHz are analyzed after removing 31.25 MHz both at the high and low band edge frequencies.

\begin{figure*}[!thp]%
\centering
\includegraphics[width=0.24\textwidth]{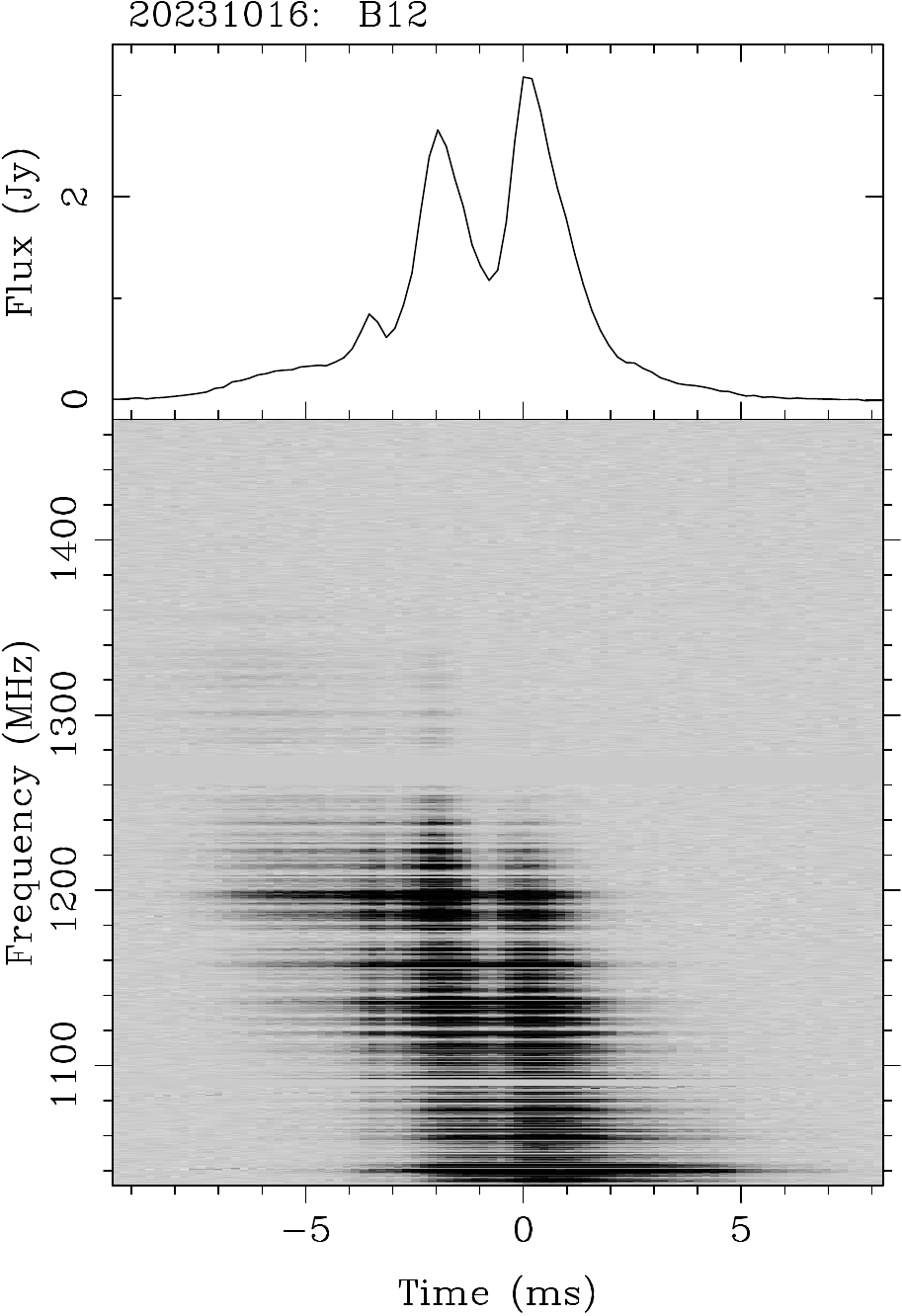}
\includegraphics[width=0.24\textwidth]{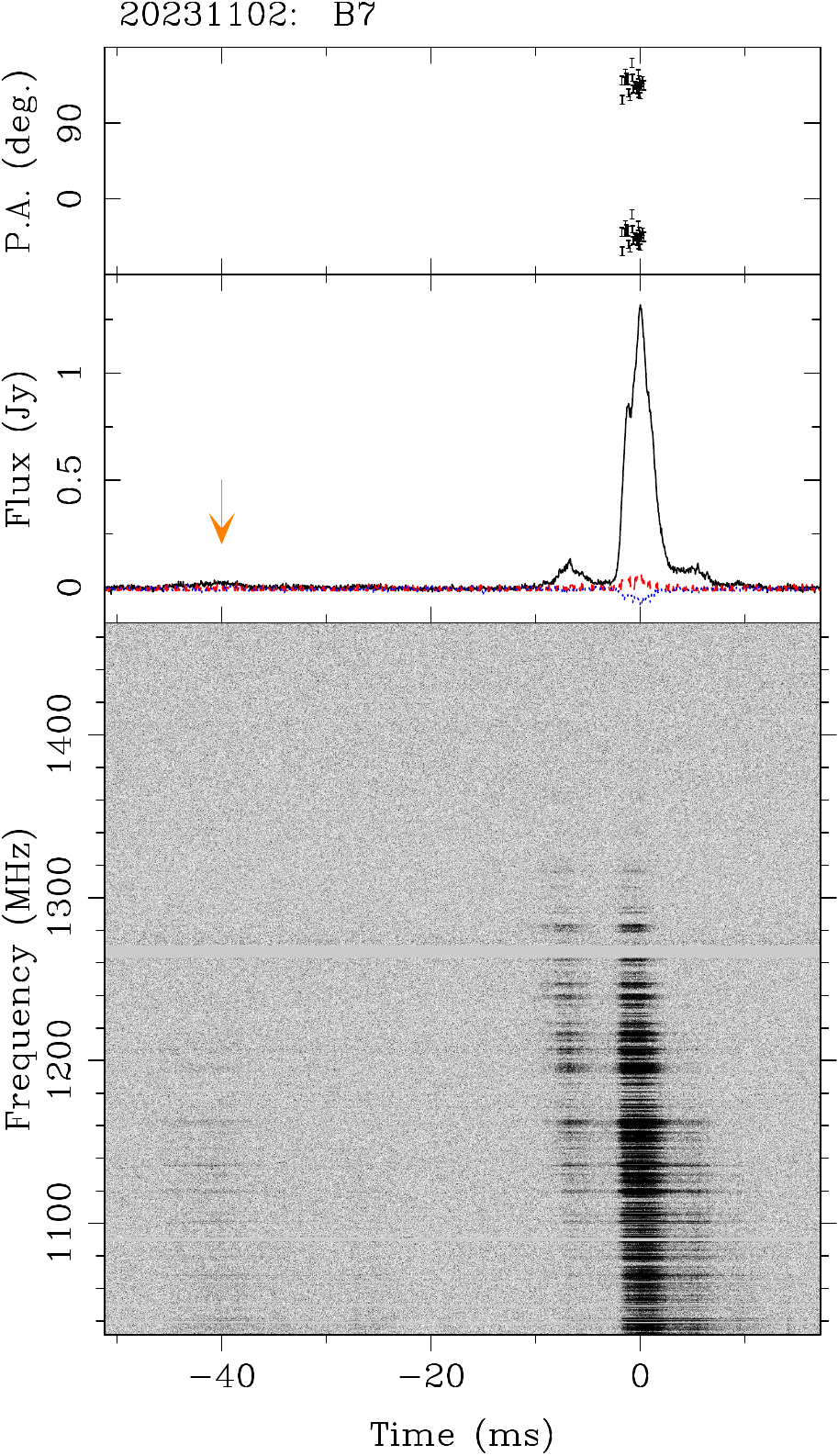}
\includegraphics[width=0.49\textwidth]{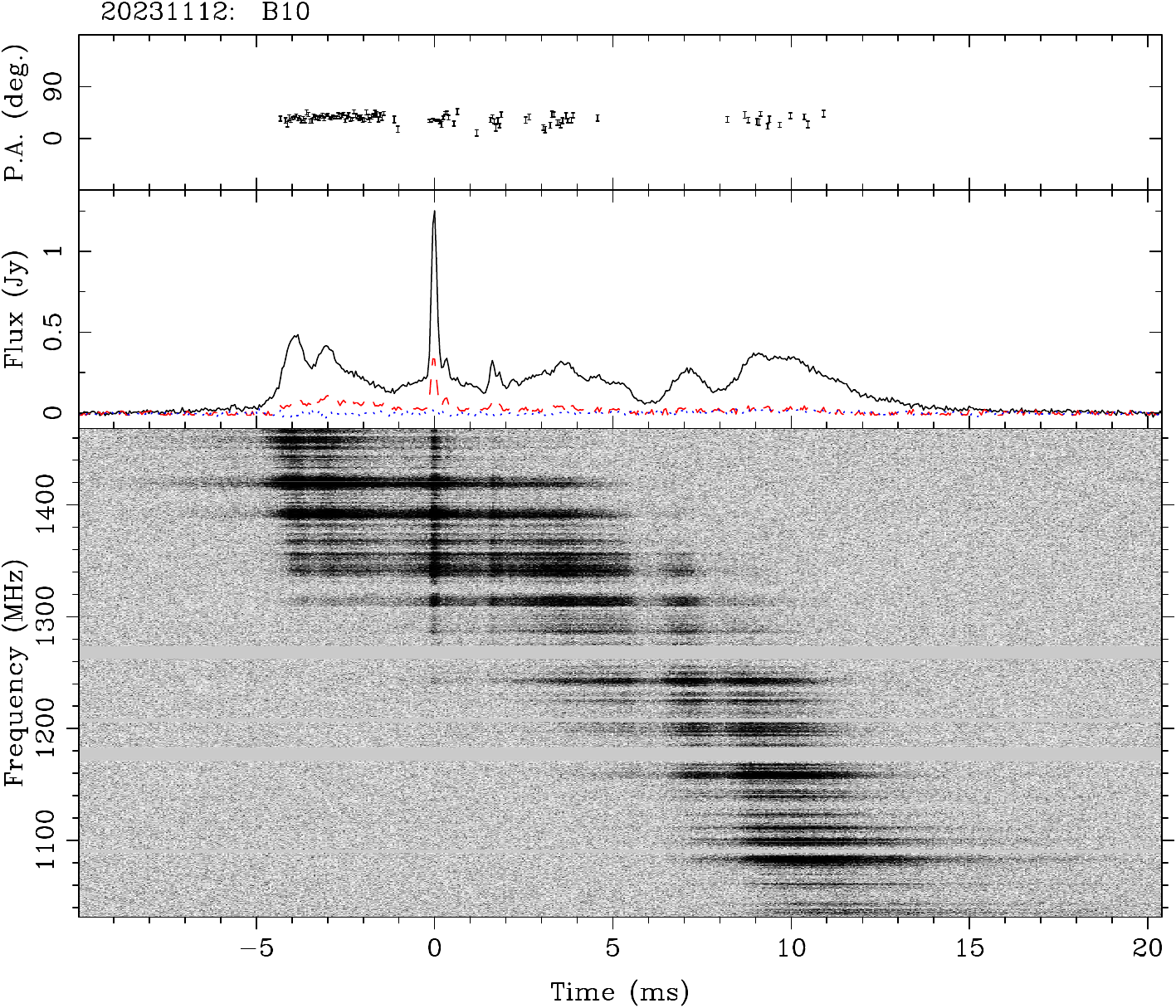}
\caption{Three bightest bursts of FRB\,20230607A. 
The bottom sub-panels show the 2 dimensional dynamic spectrum over frequency and time. The middle sub-panels show the polarization profiles, solid line for total intensity $I$, dashed line for linear polarization $L$ and dotted line for circular polarization $V$, and the top sub-panels show the polarized position angles. The observation date and burst number are marked on the top of panels. The arrow at around -40 ms of the burst B7 on 20231102 indicates a weak emission component.
}\label{bfig}
\end{figure*}

\subsection{FAST Observations}
After receiving the low-latency alerts of two repetitive bursts from FRB\,20230607A from the CHIME/FRB VOEvent Service \citep{CHIME/FRB2018ApJ...863...48C} on October 1st, 2023, we promptly proposed and then scheduled a two-hour monitor observation session on October 16th 2023 by FAST. Because of the large uncertainty of source position, we use the \texttt{snapshot mode} \citep{Han2021RAA....21..107H} for observations, which cover a sky area of 0.157 square-degree around the suggested position by using 4 adjacent pointings of the 19 beams, 30 minutes each. In each pointing all 19 beams track at given coordinates. Data of all $4\times19 = 76$ beams are recorded and then searched for single pulses in a DM range of 3-1000 $\rm pc\,cm^{-3}$ with a step of 1.0 $\rm pc\,cm^{-3}$ by using the artificial intelligence techniques to identify burst signatures on DM-Time images \citep{Zhou2023RAA....23j4001Z}. 

From the first observation session, 21 bursts were detected, including one particularly bright burst visible in several beams and even saturated in the FAST snapshot beam M07-P4. The coordinates of the burst source are then determined by using the signal-to-noise ratios (S/N) and the beam center positions assuming they all have the ideal responses as done in \cite{Han2021RAA....21..107H} for bright pulsars. The new coordinates of FRB\,20230607A are set to be RA = $\rm 21^h53^m35^s.9$, Dec = +11$^\circ$17$^\prime$04.2$^{\prime\prime}$ with uncertainties of $\sigma_{\rm RA} = \rm 3^s.0$ and  $\sigma_{\rm Dec}$ = 1$^\prime$06.8$^{\prime\prime}$ (68$\%$ confidence level). 

The subsequent FAST monitoring observations (see Table~\ref{tab1}), lasted almost for one year, targeting the new coordinate by using the central beam of the FAST L-band 19-beam receiver in the \texttt{tracking or swiftcalibration} mode.

The polarization data ($XX$, $X^*Y$, $XY^*$ and $YY$) are recorded for each frequency channel in all sessions except the first snapshot observations. The periodical calibration signals with a noise-diode-signal of 1~K in amplitude and a period of 0.2~s are injected into the feed during the first minute of each observation session. Based on these calibration signals, one can derive the polarization profiles of detected bursts \citep{WangPF2023RAA....23j4002W}, see details below.

\subsection{Burst detection and parameter estimation}
\label{Sec:BPar}

From the recorded data with a sample time of generally 49.152~$\mu$s  FRB\,20230607A, we employ the structure maximizing method \citep{Hessels2019ApJ...876L..23H} to find the best DM based on the average profile of the bright multi-component bursts, that is 362.85$\pm$0.15 ${\rm pc\,cm^{-3}}$ from a narrow-component burst on November 12, 2023, which has the smallest uncertainty (see Fig.\ref{bfig}). We then take this DM to detect other bursts. The burst profile is obtained after the data are de-dispersion with this DM.

For bursts with $S/N>7.0$, we adopt the iterative fitting methodology  of \citet{Zhou2022RAA....22l4001Z}  for the burst morphology: (1) Manually set the frequency window of FAST observations, (2) Perform a multi-Gaussian function fitting on the de-dispersed burst profile to extract the burst components, (3) For each burst component, fit the energy distribution along the frequency channels with a Gaussian function, to get the emission frequency peak ($\nu_0$) and the observed emission bandwidth $BW_{\rm e}$ defined by the Full Width at Tenth Maximum (FWTM), (4) Construct the refined profiles using the energy from extracted frequency channels, and (5) Execute a final multi-Gaussian decomposition to establish the time-of-arrivals (TOAs), the burst temporal width ($W_{\rm sb}$), and specific fluence ($F_{\nu}$) for both the whole burst and the components (or say sub-bursts).
Here, we take the system parameters of FAST provided by \citet{JiangP2020RAA....20...64J} to estimate the calibrated peak flux density ($S_{\rm peak}$) and burst fluence, converting data from each bin to mJy units, with an uncertainty of approximately 20\% due to system fluctuations and measurement errors, see the estimated values in Table~\ref{appTable1}. 

\begin{figure}[th]
\centering
\includegraphics[width=0.9\columnwidth]{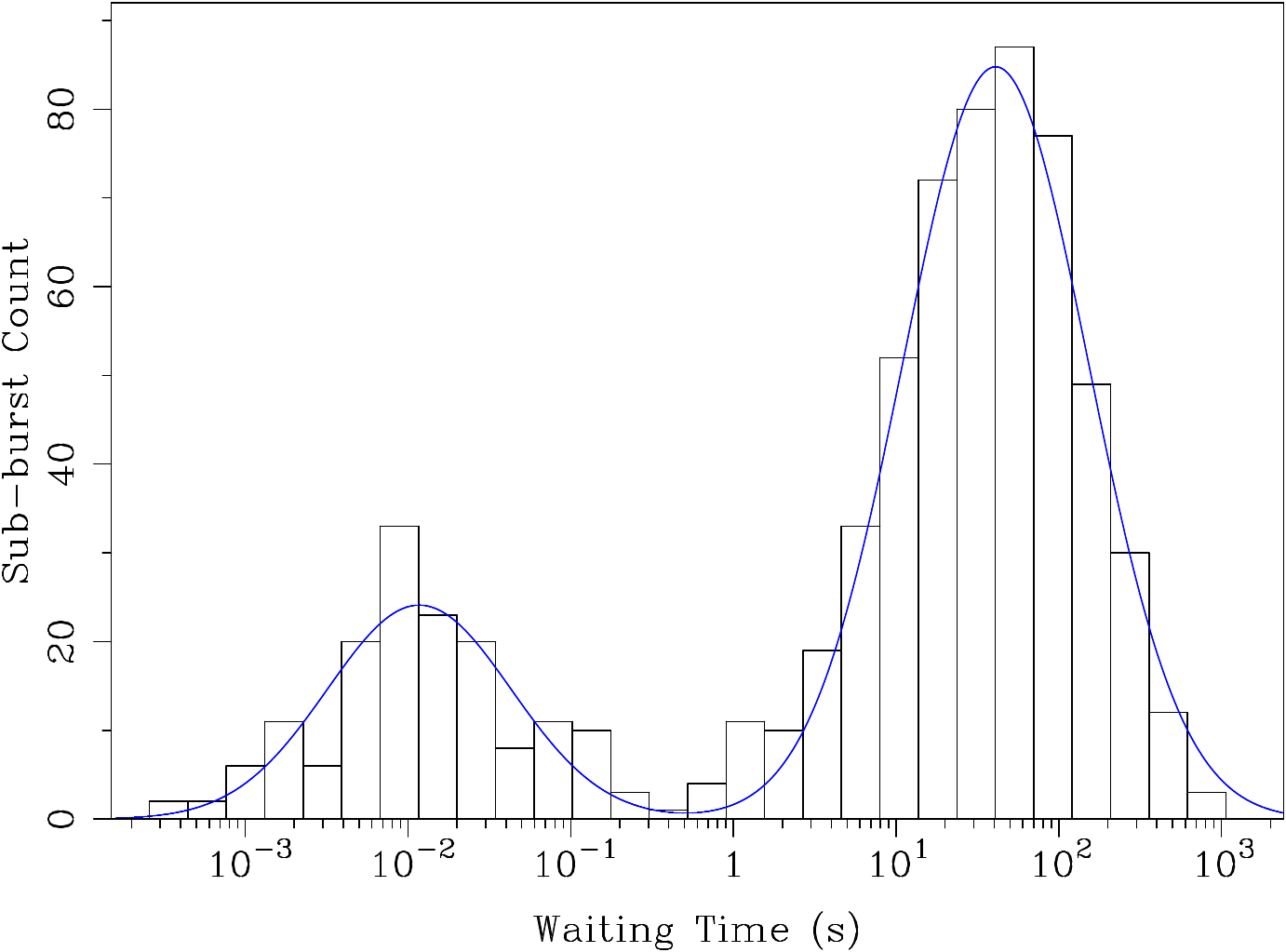}
\caption{The distribution of waiting time of FAST detected sub-bursts from FRB\,20230607A. The blue solid line represents the best-fit model with two log-normal functions with fitted peaks at 11.7\,ms and 40.9\,s. The minimum of the bridge is 507.6 ms.
}\label{fig:waitt}
\end{figure}

\begin{table*}[th]
\begin{threeparttable}
\caption{Properties of three bright bursts of FRB\,20230607A}\label{tab2}%
\small
\begin{tabular*}{\textwidth}{@{\extracolsep\fill}lcrr}
\hline
Date                                & 20231016              & 20231102              & 20231112          \\
Burst NO. 
& B12                   & B7                    & B10               \\ \hline
TOA \tnote{a} (MJD)                 & 60233.6029055482      & 60250.5562919966      & 60260.5286762679  \\
DM \tnote{b} ($\rm pc\,cm^{-3}$)    & 363.16$\pm$0.21       & 363.16$\pm$0.28       & 362.85$\pm$0.15   \\
Width (ms)                          & 12.5                  & 36.7                  & 36.0              \\
Peak flux density $S_{\rm peak}$ (Jy)& $>3280.5\pm$3.9       & 1315.4$\pm$7.7        & 1312.3$\pm$7.8    \\
Specific Fluence $F_{\nu}$ (mJy ms) & $>34851\pm$17          & 16076$\pm$30          & 11123$\pm$64    \\
Energy $\sum E_{\nu}$ ($\rm\times10^{39} erg$) & $>8.028\pm$0.004       & 4.129$\pm$0.008       & 4.223$\pm$0.002   \\ 
$\rm RM_{\rm obs}$ ($\rm rad\,m^{-2}$)  &                       & -12300.1$\pm$2.9         & -12249.0$\pm$1.5  \\
$\langle B_\parallel \rangle$ (${\mu}G$)&                   & -41.73$\pm$0.04       & -41.59$\pm$0.02   \\
$L/I$ ($\%$)                          &                       & 2.72$\pm$0.20         & 13.81$\pm$0.24      \\
$L_{\rm peak}/I_{\rm peak}$ ($\%$)&                       & 4.52$\pm$0.64         & 26.81$\pm$0.79    \\
$V/I$ ($\%$)                          &                       & -5.12$\pm$0.17        & 1.76$\pm$0.18     \\
$\rm \left |V \right |$/I ($\%$)    &                       & 4.45$\pm$0.17         & 2.52$\pm$0.18     \\
$f_{\rm depol}$                     &                       & 3.18$\times$10$^{-3}$ & 3.15$\times$10$^{-3}$ \\
\hline
\end{tabular*}
\begin{tablenotes}
\item[a] {TOA of each burst is referenced to the infinite frequency at the Solar System barycentre by the ephemeris DE440.}
\item[b] {The DM derived from each burst. }
\end{tablenotes}
\end{threeparttable}
\end{table*}

\begin{figure}[th]
\centering
\includegraphics[width=0.9\columnwidth]{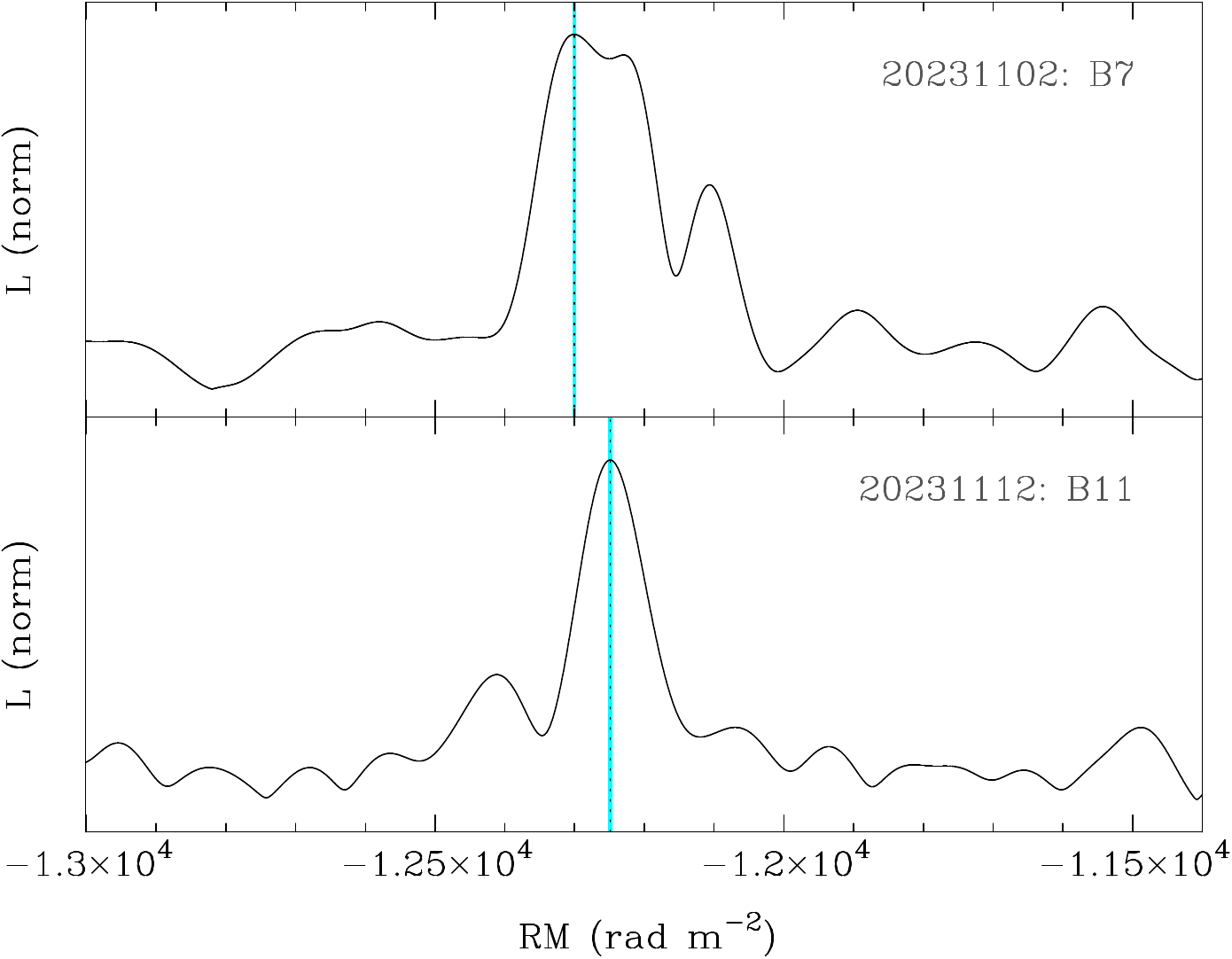}
\caption{The RMCLEAN results of the two bursts for the RM determinations. The vertical dotted lines indicated RM values at -12300.1 $\rm rad\,m^2$ (up) and -12249.0 $\rm rad\,m^2$ (bottom), respectively. 
}\label{fig:rmsynthesis}
\end{figure}

The waiting time distribution of sub-bursts from FRB\,20230607A is then obtained and shown in Figure \ref{fig:waitt}, which shows two peaks. The first peak typically corresponds to intervals between different components of a burst, while the second peak represents intervals between distinct bursts. There is a valley between the two peaks at 0.5\,s, separating different bursts. Such a distribution is very similar to what we have detected for other repeating FRBs by FAST  \citep{LiD2021Natur.598..267L, XuH2022Natur.609..685X, ZhangYK2023ApJ...955..142Z}. This leads us to define a \textit{burst} with these emission substructures within a time-window of 0.5~s, and \textit{sub-bursts} are these fine components.

\begin{figure*}[th]%
\centering
\includegraphics[width=0.9\linewidth]{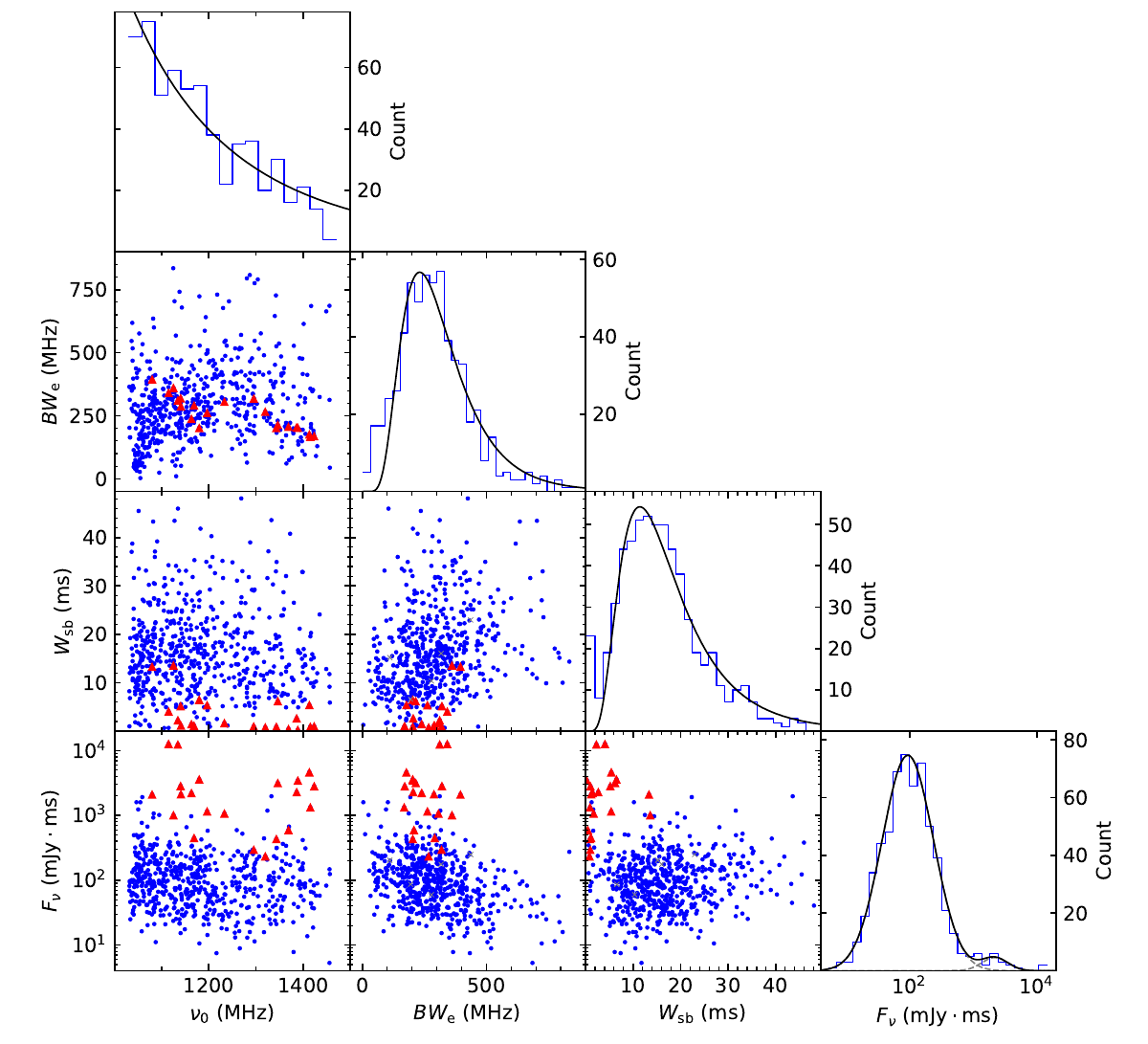}
\caption{Parameter distribution of sub-bursts of FRB\,20230607A for the emission peak frequency ($\nu_0$ from 1031.25 MHz to 1468.75 MHz in the FAST effective observation bandwidth), derived emission bandwidth ($BW_{\rm e}$), sub-burst temporal width ($W_{\rm sb}$) and the fluence $F_{\nu}$. The triangles represent the parameters from the three bright bursts, the other points correspond to those from weaker bursts. The distribution of $\nu_0$ can be fitted with a power-law function with an index of  $-4.79\pm0.32$, as shown on the top subpanel. A log-normal function is used to fit the distribution of $BW_{\rm e}$ and $W_{\rm sb}$ which have the fitted peak of $231_{-99}^{+171}$ MHz and $11.4_{-5.5}^{+10.6}$ ms, respectively. Two Gaussian functions are used to fit the distribution of log$F_{\nu}$ with peaks at log$F_{\nu} = 1.97\pm0.40$ and log$F_{\nu} = 3.34\pm0.21$.
}\label{fig:stat}
\end{figure*}

\section{Results and Discussions}
\label{sect3:result}

We observed FRB\,20230607A for  15.6 hours, and detected  565 bursts. The observational parameters and the number of detected bursts as well as the burst rates are listed in Table \ref{tab1} and plotted in Figure \ref{figObs}. Most bursts are faint but three of them are very bright. 

\subsection{Bright Bursts}

Three bright bursts were detected from FRB\,20230607A on 20231016, 20231102 and 20231112, all exhibiting multi-component structures (see Figure \ref{bfig}). The key parameters of FRB\,20230607A can be easily derived from these bright bursts, as listed in Table \ref{tab2}.

First is the accurate DM, which is the integration of electrons from the source to us along the path, ${\rm DM} = \int_0^d n_{\rm e} {\rm d}l$, where the $n_{\rm e}$ is the electron density, 
$l$ is the unit length, and $d$ is the distance to the source.
We employed the frequency-averaged profile structure maximization method \citep{Hessels2019ApJ...876L..23H} to get the DM values from the
three bursts. Despite different complex temporal-spectrum structures, DM values from these three bright bursts from three days spanning over more than one month are consistent with each other, suggesting that the DM of this FRB does not vary significantly. From the narrowest burst detected on November 12, 2023, we get a DM with the highest precision, with DM = 362.85$\pm$0.15 $\rm pc\,cm^{-3}$.  
%

We calibrate the polarization data of the two bright bursts detected in the tracking observations on 20231102 and 20231112. The polarization angle $PA$ is computed using the formula of $\psi = 1/2 \sqrt{U_{\rm i}/Q_{\rm i}}$. The linear polarization $L_{\rm i}$ for each data point is computed as $L_{\rm i} = \sqrt{Q_{\rm i}^2 + U_{\rm i}^2}$. To estimate the baseline bias $L_{\rm i}$, $|V_{\rm i}|$, and the errors associated with these polarization parameters, we adopt the method described by \citet{WangPF2023RAA....23j4002W}, and take $L_{\rm i} = \sqrt{|L_{\rm i}^2 - (\sigma_{\rm Q}^2 + \sigma_{\rm U}^2)|}$ if $L_{\rm i}^2 \geq (\sigma_{\rm Q}^2 + \sigma_{\rm U}^2)$ or $L_{\rm i} = -\sqrt{|L_{\rm i}^2 - (\sigma_{\rm Q}^2 + \sigma_{\rm U}^2)|}$ for otherwise, and $|V_{\rm i}| = \sqrt{|V^2_{\rm i} - \sigma_{\rm V}^2|}$ if $V^2_{\rm i} \geq \sigma_{\rm V}^2$ or $|V_{\rm i}| = -\sqrt{|V^2_{\rm i} - \sigma_{\rm V}^2|}$ for otherwise. We then search Faraday rotation measures (RMs) by using the \texttt{RMSYNTH1D} and \texttt{RMCLEAN1D} of \texttt{RM-Tools} package \citep{Purcell2020ascl.soft05003P} in the range from -20000 to 20000 $\rm rad\,m^{-2}$ with a step of 1.0 $\rm rad\,m^{-2}$. We measured ${\rm RM}_{\rm obs} =  -12300.1\pm2.9$ rad~m$^{-2}$ and $-12249.0\pm1.5$ rad~m$^{-2}$ from the two bursts, as shown in Figure \ref{fig:rmsynthesis}. The Faraday rotation measure is ${\rm RM}_{\rm obs}$ = 0.81 $\int_d^0 B_\parallel(l)n_{\rm e}(l) {\rm d}l$, where $B_\parallel(l)$  represents the average line-of-sight magnetic field strength, measured in $\mu$G. Combining the DM and RM values, the intervening medium has magnetic fields with an average field strength of $B_{||} = 1.232\,\rm{RM/DM} \simeq 42 \mu$G, with an average direction going away from us. This FRB source is located at a 
high Galactic latitude (GL=68.62$^{\circ}$, GB=$-32.26^{\circ}$), and there is no such a foreground Galactic cloud or intergalactic cloud of ionized gas to contribute such a DM and RM. Such a dense cloud is likely located in the host galaxy in the immediate front of the FRB source. The local magnetic field should be much stronger than $42 \mu$G we estimated, because the local DM is much smaller than the total DM we measured. 
 
After de-Faraday rotation, we get the linear polarization of the two bursts, which has the degree of approximately $\Pi_{\rm L} = {L/I} = \sum L_{\rm i}/\sum I_{\rm i} = (2.72 \pm 0.20) \%$ and $(13.81\pm0.24)\%$, respectively. We also measure the circular polarization degree as $\Pi_{\rm V} = {V/I} = \sum V_{\rm i}/\sum I_{\rm i} = (-5.12 \pm 0.17) \%$ and $(1.76 \pm 0.18) \%$, respectively. Considering the negligible depolarization of only 0.32$\%$ caused by the RM inside the frequency channels, such low polarization degrees in the frequency range of 1.0 -- 1.5 GHz need to attribute to other effects. Intrinsic FRB emission mechanisms typically produce higher polarization degrees than these observations \citep{QuYH2023MNRAS.522.2448Q}, which are consistent with the polarization observations of other active repeaters suggest \citep[e.g.][]{JiangJC2022RAA....22l4003J}. Even though it is not impossible that the intrinsic radiation mechanism of this source might produce lower polarization degrees, some external mechanisms, such as de-polarization effect due to multi-path RM variations, might be at play. The effect is more significant for larger RM FRBs \citep{FengY2022Sci...375.1266F}. FRB\,20230607A has the third highest RM known, so likely suffers from such a $\sigma_{\rm RM}$ effect.  


One important feature is in the bright burst (B10) observed on November 12, 2023 (Figure \ref{bfig}). Among many broad pulses, there is a very narrow spike. The time delay $\Delta{t_{\rm chan}}$ caused by DM smearing inside a channel:
\begin{equation}
    \Delta{t_{\rm chan}} = 8.3 \mu{s}\cdot {\rm DM} \cdot \frac{\Delta{f}}{\rm MHz} \cdot \left( \frac{f}{\rm GHz} \right) ^{-3}, 
\end{equation} 
where the channel bandwidth for FAST observations is $\Delta{f} = $0.1220703125\,MHz, and the $f$ is the central frequency of the channel. The width of this narrow spike is fitted as $W_{\rm obs} \sim 0.3\,ms$ (see the method in Section \ref{Sec:BPar}), after ignoring the scatter-broadening effect and correcting for the DM smearing of about  $\Delta{t_{\rm chan}}  = 188\,{\mu}s$ at 1250 MHz with the function of $W_{\rm obs}  = \sqrt{W_{\rm int}^2 + t_{\rm samp}^2 + \Delta{t_{\rm chan}^2}}$, hear the sampling time is $t_{\rm samp} = 49.152\,{\mu}s$, the intrinsic FWTM of this spike is only $W_{\rm int} \sim 229 \,{\mu}s$. This timescale corresponds to a light travel distance of $\sim$69\,km, representing an upper limit on the spatial extent of the coherent emission region for a rotating central engine, suggesting a magnetospheric origin of the emission. 

Such a curious feature sheds light onto our understanding of FRB emission mechanism. First, CHIME observations have suggested that repeaters and non-repeaters seem to have statistically different emission properties. While non-repeaters tend to have short widths, bursts of repeaters tend to be broad and often shows a frequency down drift with time \citep{Pleunis2021ApJ...923....1P}. The B10 burst indeed has many pulses with the frequency down drift. However, even though many bursts have broad widths, the brightest one has a very narrow spike. 
Our observation indicates that repeaters can make both broad and narrow pulses. 

Second, the distinct width difference between the narrow spike and the surrounding broader pulses constrain the FRB emission mechanism. First, the existence of a narrow spike in long-duration burst such as B10 essentially rules out any far away models invoking relativistic shocks and synchrotron maser, because a narrow spike needs a very small clump in the upstream which gives a very low radiative efficiency, inconsistent with the high flux of the spike \citep{Lu2022MNRAS.510.1867L,ZhangB2023RvMP...95c5005Z}. This gives yet another strong support to the magnetospheric origin of FRBs. Within the magnetospheric models, different sub-pulses correspond to emitters from different bundle of magnetic field lines. The sharp spike raises a challenge to such a model. One possibility is that the sharp spike and the other broader pulses originate from different emission regions, with the sharp spike coming from lower altitudes than the broader ones. Another possibility is that the narrow spike might be amplified by the caustic effect. Detailed modeling is needed to unveil the radiation geometry of this peculiar burst.


\begin{table*}
\centering
\caption{Burst numbers for different classifications for FRB 20230607A. }  
\label{tab:ClassProp}
%
\begin{tabular}{llllll}
\hline 
\multirow{2}{*}{Drifting mode} & Component &\multicolumn{4}{c}{Burst emerging in the FAST L-band}  \\
\cline{3-6}
    &  Count & Whole-band   & High part   & Middle part   & Lower part    \\
\hline 
\multirow{3}{*}{{Downward: 405}} 
    & one: 343  & D1-W: 60 & D1-H: 41 & D1-M: 129 & D1-L: 113 \\
\cline{2-6}
    & two: 58   & D2-W: 19 & D2-H: 3  & D2-M: 19  & D2-L: 17 \\
\cline{2-6}
    & multiple: 15 & Dm-W: 9  & Dm-H: 0  & Dm-M: 0   & Dm-L: 6  \\
\hline
Upward: 9 & 
          &    &     &  U2-M: 6  & U2-L: 3\\
\hline
Complex: 15  &  & \multicolumn{4}{c}{C: 15}   \\
\hline 
No Drifting: 18 & & \multicolumn{4}{c}{ND: 18} \\
\hline
No evidence: 107 & &    & NE-H: 12  &    & NE-L: 95\\ 
\hline 
\end{tabular} 
\end{table*}

\subsection{Burst classification}

There are 720 sub-bursts in all 565 bursts detected from 20230607A. Most bursts are weak but exhibit a diverse morphology: some bursts emerge in narrowband at some random frequencies; some bursts show frequency drifting between or within burst components. 

Following the classification scheme proposed by \citet{Zhou2022RAA....22l4001Z}, we classify the detected bursts of FRB 20230607A based on the number of components, their occurrence frequencies, and their drift states. We use `U' and `D' to indicate upward- and downward-frequency drifting; we use numbers `1', `2', and `m' to indicate the number of burst component for one, two, or multiple components, respectively; we use `W', `H', `M', and `L' to mark the observed burst energy detected in whole (wide) FAST L-band, or only in higher, middle, or lower part of the frequency band, respectively. For example, the class `D1-W' means a burst with one component only detected in the whole FAST L-band with a downward-drifting feature. Some bursts have a complex morphology, which we class them as ``C". Some bursts are detected in the water-fall plot of the frequency-time image and the pulses do not drift and we class them as `ND'. If the data are not good enough to judge if there is a frequency drift, we class them as 
`NE' for no evidence of drifting. 

We checked all detected bursts (see plots and details in the appendix in Table \ref{appTable1} and Fig.~\ref{appfig1}), and counted the burst numbers for each category, as listed in Table \ref{tab:ClassProp}. Among these 565 bursts, the majority exhibit downward frequency drifting, with only 9 bursts showing upward frequency drifting. 


\begin{figure}
\centering
\includegraphics[width=0.98\columnwidth]{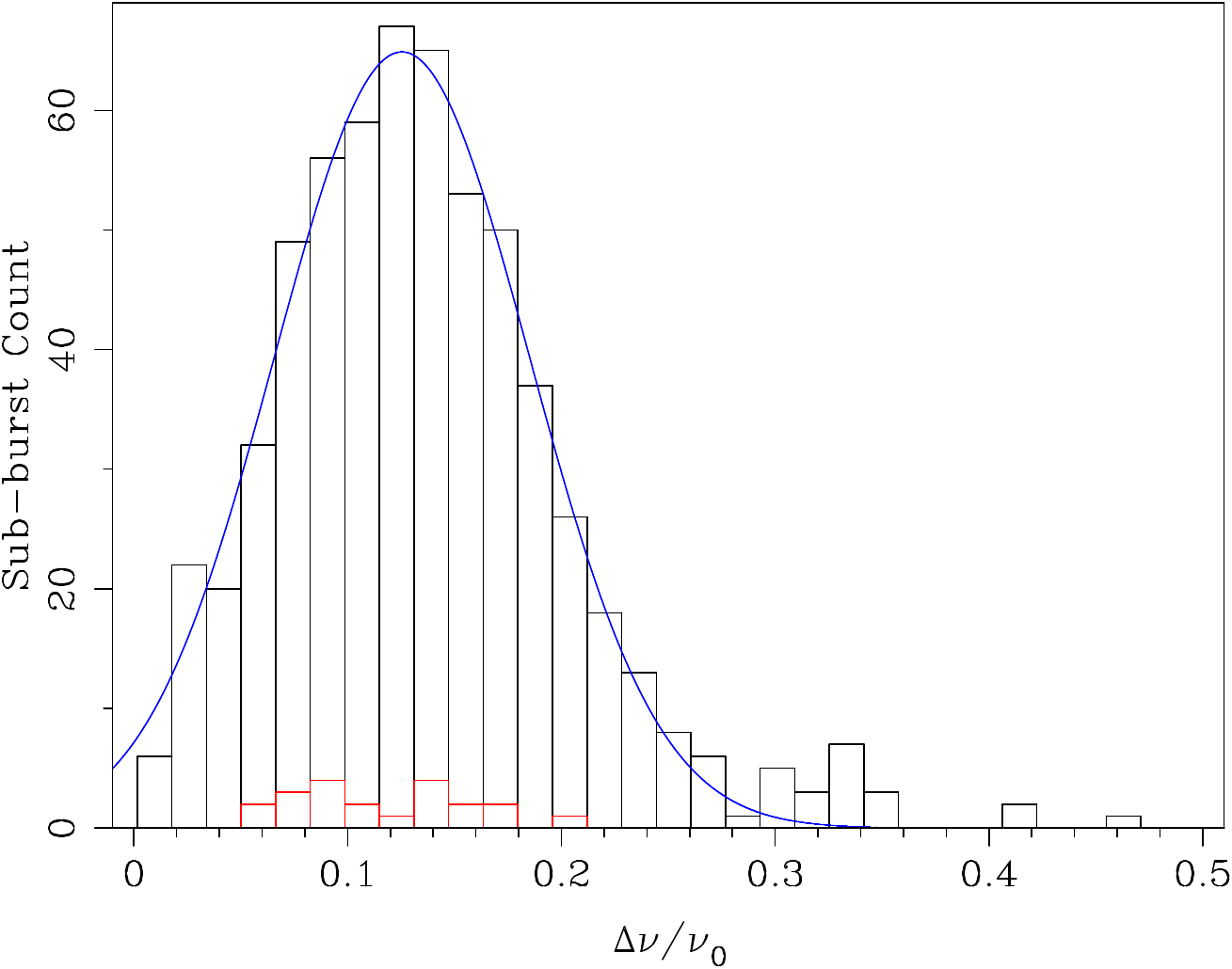}
\caption{The fraction distribution of emission bandwidth $\Delta\nu$ of the emission frequency peak $\nu_{0}$ for all the sub-bursts (black) and those in the three bright bursts (red) of FRB\,20230607A. The solid line represents the best-fit model with a Gaussian function with a peak at $\Delta\nu/\nu_{0} = 0.125 \pm 0.001 $ and a fitted width of $\sigma = 0.060 \pm 0.001$.
}\label{fig:deltanunu}
\end{figure}

\subsection{Distributions of sub-burst parameters}

We have analyzed all the bursts of FRB\,20230607A in our sample and obtained all the parameters of bursts and sub-bursts. In Figure \ref{fig:stat}, the distributions of these parameters are presented. To ensure good statistics, only those sub-bursts with an emission frequency $\nu_0$ peaked within the FAST band of 1031.25-1468.75MHz are considered in this subsection. 

As shown in Figure \ref{fig:stat}, the number of detected sub-bursts obviously decreases with the peak emission frequency $\nu_0$, which differs from the bimodal distribution reported for FRB\,20201124A detected by FAST \citep{Zhou2022RAA....22l4001Z} and FRBs observed by CHIME  \citep{LanmanAE2022ApJ...927...59L}. The number distribution can be fitted by a power-law function with an index of $-4.79\pm0.32$. The number distributions of $BW_{\rm e}$ and $W_{\rm sb}$ exhibit log-normal shapes, and they can be fitted with log-normal functions, with the peak values at $231_{-99}^{+171}$ MHz and $11.4_{-5.5}^{+10.6}$ ms, respectively. The distribution of fluence $F_{\rm \nu}$ (in units of mJy\,ms) has one main peak and a tail component for high values. We use two log-normal functions to fit the distribution, with two peaks at log$F_{\rm \nu} = 1.97\pm0.40$ and log$F_{\nu} = 3.34\pm0.21$, respectively.

For the sub-plot of the frequency vs. emission bandwidth in Figure \ref{fig:stat}, the fitted emission bandwidth narrows toward the edges of the observing band that may arise from scintillation-induced distortions in burst morphology. While our calibration and fitting procedures are robust (see Section \ref{Sec:BPar}), scintillation introduces nonphysical trends in these edge regions. Its impact is confined to a small subset of bursts and does not alter our conclusions on spectral properties and burst energetics. Critically, the narrowband bursts detected near the center of the observing band—unaffected by edge artifacts—remain statistically representative, and their inclusion ensures the validity of the overall trends.

Notice that most bursts from FRB\,20230607A have the emission limited in a narrow range inside the FAST observation band, similar to other FRBs \citep{Pleunis2021ApJ...923....1P, Zhou2022RAA....22l4001Z, ZhangYK2023ApJ...955..142Z}. The distribution of $\Delta\nu/\nu_0$, here $\Delta\nu$ is the Full Width at Half Maximum (FWHM, not FWTM for the Full Width at Tenth Maximum), is shown in Figure \ref{fig:deltanunu}. 
It is almost in a Gaussian distribution with a mean of  $\Delta\nu/\nu_{0} = 0.125 \pm 0.001 $ and a standard deviation of $\sigma = 0.060 \pm 0.001$.
The narrow band emission of FRBs, with a mean $\Delta\nu/\nu_{0}$ of 0.125 for FRB\,20230607A, 0.13 for FRB\,20220912A \citep{ZhangYK2023ApJ...955..142Z}, and 0.12 for FRB\,20240114A \citep{Kumar2024arXiv240612804K}, is now a common feature of active repeating FRB sources. Even though some propagation effects such as scintillation and filamentation instability may produce narrow spectra in some bursts \citep{Kumar2024ApJ...974..160K}, the ubiquitousness of such narrow spectra among all detected bursts suggest that they are most likely related to the intrinsic observational mechanisms. Since synchrotron maser models invoking relativistic shocks predict a minimum $\Delta \nu/\nu_0 \sim 0.58$ because of the high-latitude effect, our observations rule out these models and lend another strong support of the magnetospheric origin of FRB emission \citep{Kumar2024ApJ...974..160K}. One attractive mechanism to produce narrow spectra in a magnetar magnetosphere invokes inverse Compton scattering (ICS) of low frequency fast magnetosonic waves by relativistic particle bunches \citep{Zhang2022,Qu&Zhang2024}. The traditional coherent curvature radiation by bunches may also achieve a narrow spectrum if the bunches are properly spaced spatially \citep{Yang2023ApJ...956...67Y,WangWY2024A&A...685A..87W}.  


%

\begin{figure*}[th]%
\centering
\includegraphics[width=0.8\columnwidth]{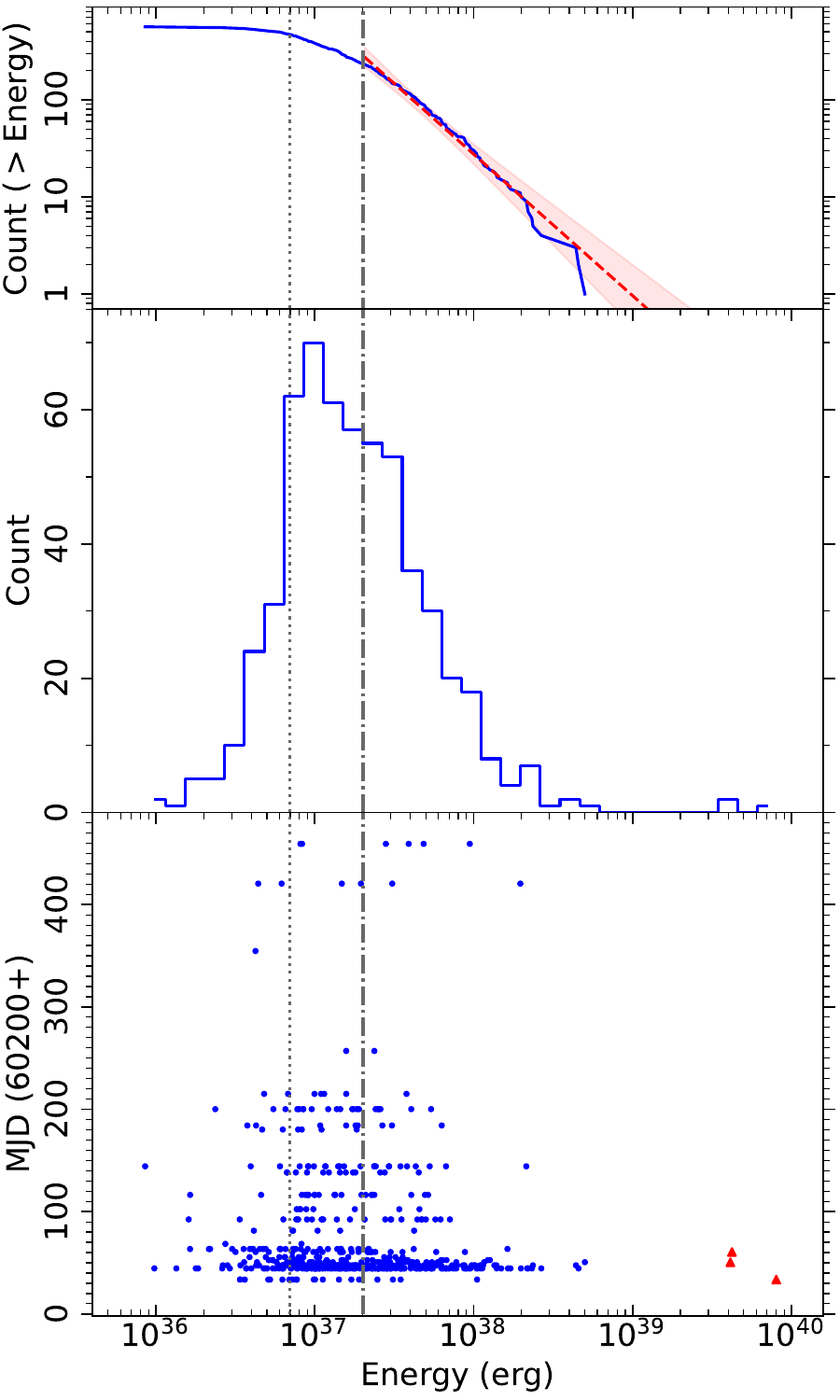}
\includegraphics[width=0.8\columnwidth]{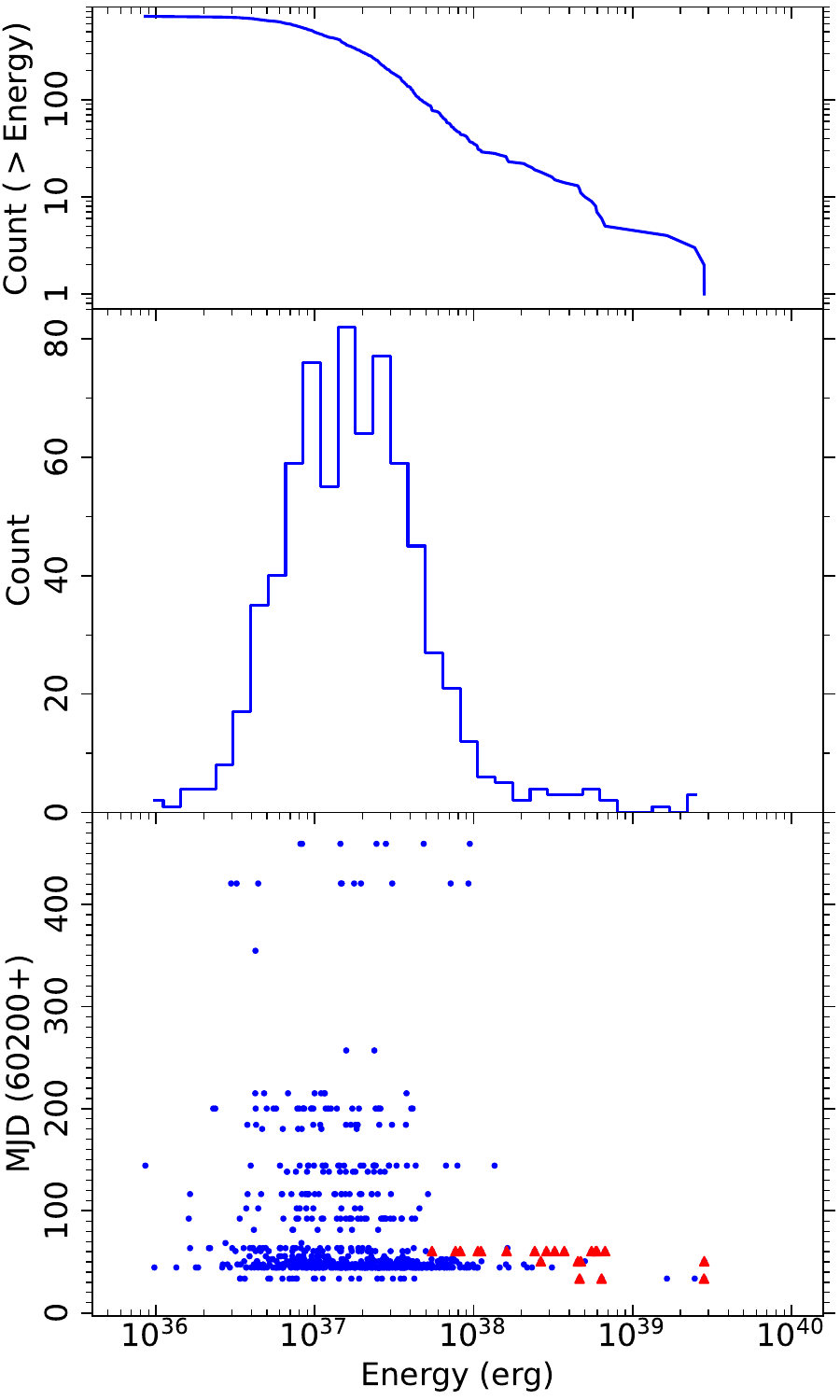}
\caption{The energy distribution of bursts (left) and sub-bursts (right) detected from FRB\,20230607A by FAST.
The bottom sub-panels show the time-dependent energy distribution of bursts and sub-bursts of FRB\,20230607A. The triangles represent the bright bursts, while the other points correspond to weaker bursts. The gray dotted and dash–dotted vertical lines indicate the detection threshold and completeness threshold of $\rm 7.0\times10^{36}\,erg$ and $\rm 2.0\times10^{37}\,erg$ (95$\%$ detection probability) , respectively, for bursts with the 1~ms temporal width or 0.2~ms sampling time for the S/N = 7. 
The mid-sub-panels show the number distribution of the sub-burst or burst of FRB\,20230607A against the energy distribution of bursts. 
The top sub-panels present the cumulative distribution of the energy function towards weak bursts. The red dashed line in the left panel for the bursts represents the best-fitted power-law function with an index of $\alpha=-2.46\pm0.05$ for the high-energy part with the 95\% confidence interval shown in the shaded region. 
}\label{fig:E}
\end{figure*}


\subsection{The Energy Distribution}

Except for the bursts detected on the first day with the snapshot mode, when the source was offset from the beam center and the scintillation effect complicated the flux analysis, the flux density of other bursts/sub-bursts can be reliably estimated from data through comparing with the system noise temperature. The fluence distribution for sub-bursts of FRB 20230607A has a prominent low-energy peak and a less conspicuous high-fluence tail (see Figure~\ref{fig:stat}).

Because the host galaxy of FRB\,20230607A is not known yet, before analyzing the energy of FRB events, we have to estimate the source distance from the intergalactic DM contribution. The DM contribution from the Milky Way can be estimated by using the YMW16 model \citep{YMW2016} with $\rm DM_{\rm MW}$ = 37.8$\rm\,pc\,cm^{-3}$. 
The excessive DM of ${\rm DM_{\rm extra} = 275.05\rm\;pc\,cm^{-3}}$ is attributed to contributions from the local medium and the intergalactic medium. The DM contribution from the host is unknown.
Following a generic statistical constraint by \cite{2020MNRAS.496L..28L}, the DM contribution from the host in the local galaxy frame is approximated as $\rm DM_{\rm host,LG} = 107_{-45}^{+24}\,{\rm pc\,cm^{-3} }$.
By deducting $\rm DM_{\rm MW}$, we derive the extra-galactic ${\rm DM_{\rm extra} =  DM_{\rm IGM}+DM_{\rm host,LG}}/(1+z)$ and use it to estimate the redshift $z$, where the DM contributed by the IGM is \citep{Deng2014ApJ}: 
\begin{equation}
{\rm DM_{\rm IGM}} = \frac{3cH_0\Omega_{\rm b}f_{\rm IGM}}{8\pi Gm_{\rm p}}\int^z_0\frac{\chi(z)(1+z)dz}{[\Omega_{\rm m}(1+z)^3+\Omega_{\Lambda}]^{\frac{1}{2}}},
\label{eq:DMIGM}
\end{equation} 
where $\chi(z) = 7/8$ is the free electron number per baryon in the universe, $f_{\rm IGM}\sim0.83$ the fraction of baryons in the intergalactic medium, where the $\Lambda$CDM cosmological parameters
are taken as $\Omega_m = 0.315 \pm 0.007$, $\Omega_b = 0.02237 \pm 0.00015$,
and $H_0 = 67.36 \pm 0.54\,\rm km\,s^{-1}\,Mpc^{-1}$ \citep{Planck2020AA}. We estimate $z \sim 0.22$, which gives a luminosity distance of $D_L\sim0.98~{\rm Gpc}$. 
This distance is subsequently used for the analysis of the burst energy and estimating the flux of a potential PRS. The isotropic-equivalent energy of a sub-burst can be estimated as
\begin{equation}
    E = \frac{4\pi D_L^2}{(1+z)} F_{\nu} BW_{\rm e}. 
\end{equation}
%

The temporal energy distribution of bursts and sub-bursts for FRB\,20230607A is displayed in Figure \ref{fig:E}. The three bright bursts with substantial energies are marked in red. The detection threshold $S/N \textgreater 7$ (corresponding energy is $\textgreater 7.0\times10^{36}$\,erg) was calculated via the radiometer equation, while the completeness threshold ($\sim2.0\times10^{37}$\,erg, corresponding to 95$\%$ detection probability) was determined through mock burst injection simulations following \citet{XuH2022Natur.609..685X}. The number distribution of the burst energy at the high-energy part can be described by a power-law function with an index of $\alpha=-2.46\pm0.05$, which is lower than the spectral index typically observed for pulsars \citep{BatesSD2013MNRAS.431.1352B, JankowskiF2018MNRAS.473.4436J}.

\begin{figure}[tp]%
\centering
\includegraphics[width=0.98\columnwidth]{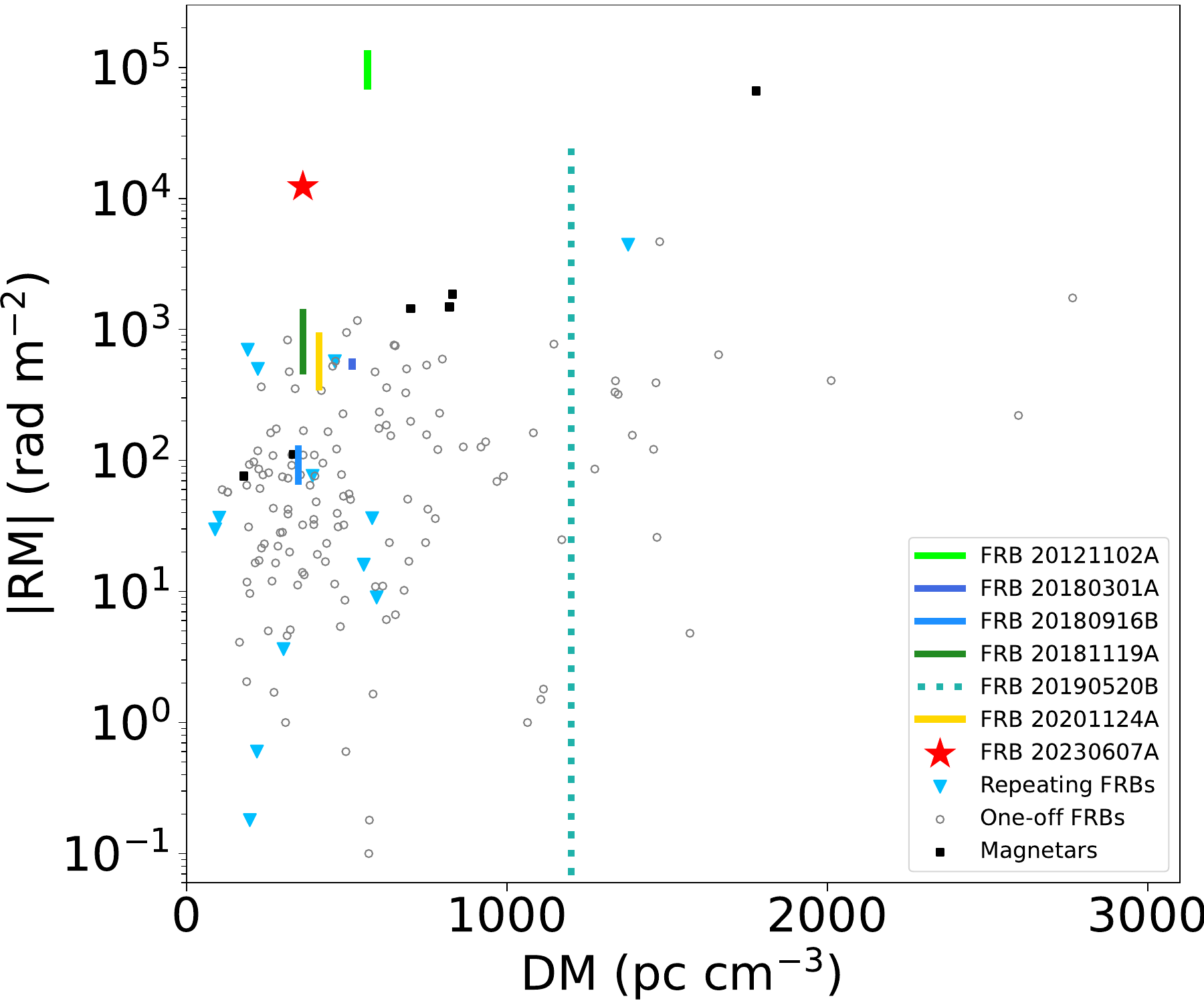}
\caption{The observed $\rm |RM|$ versus DM of FRBs, including FRB\,20230607A from this paper, other repeating FRBs and many one-off FRBs \citep{Petroff2016PASA...33...45P, Oslowski2019MNRAS.488..868O, LuoR2020Natur.586..693L, CHIME2021ApJS..257...59C, FengY2022Sci...375.1266F, Plavin2022MNRAS.511.6033P, XuH2022Natur.609..685X, AnnaT2023Sci...380..599A, Faber2023arXiv231214133F, Mckinven2023ApJ...950...12M, MckinvenR2023ApJ...951...82M, ShermanMB2023arXiv230806813S, Zhou2023RAA....23j4001Z, PandhiA2024arXiv240117378P}, and magnetars \citep{HobbsG2004IAUS..218..139H,ZhuWW2023SciA....9F6198Z}.
The large absolute RM of FRB\,20230607A implies a highly magnetized environment around the FRB source.
}\label{fig:dmrm}
\end{figure}

\begin{figure}[tp]%
\centering
\includegraphics[width=0.98\columnwidth]{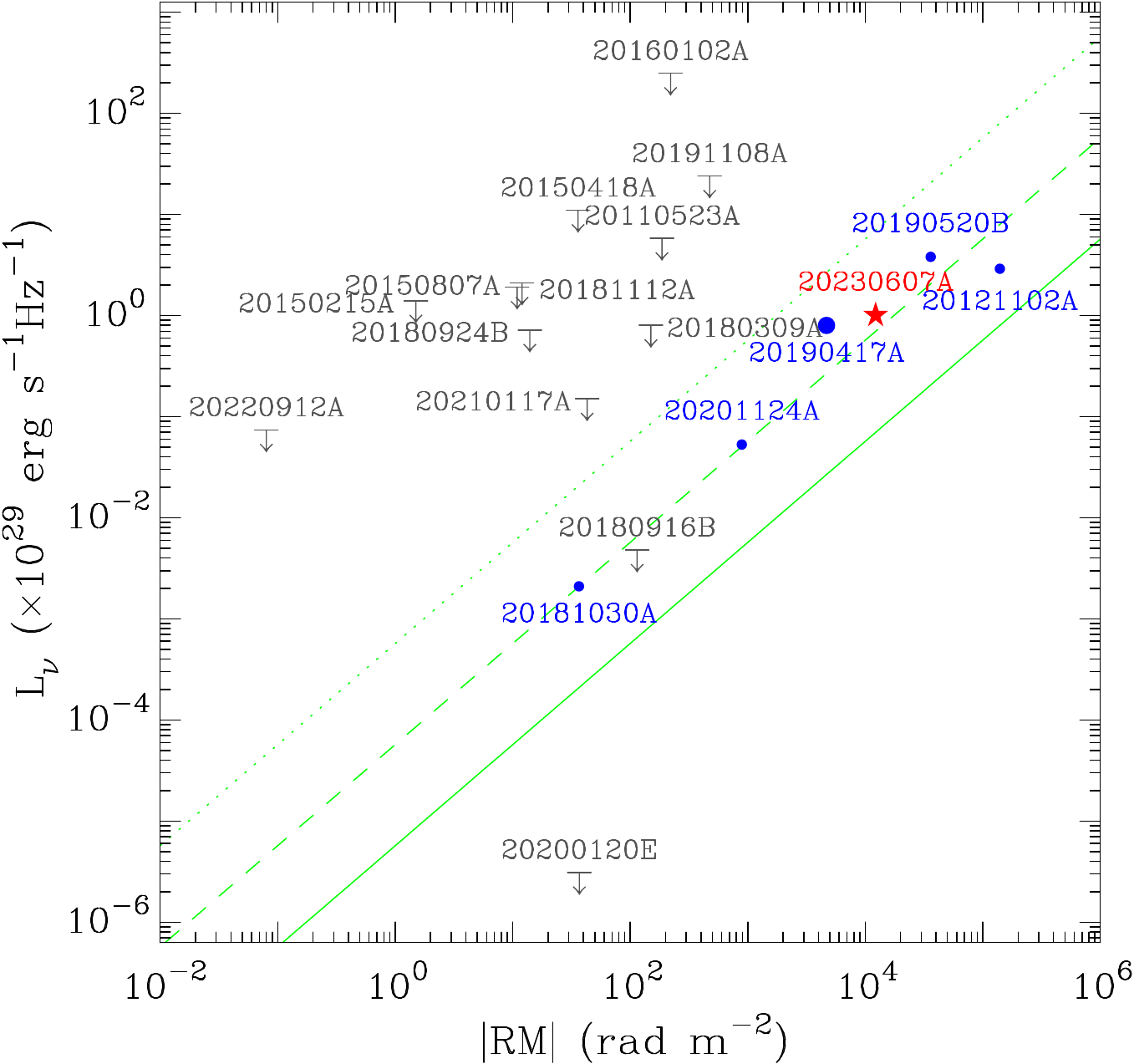}
\caption{The ${\rm |RM|}-L_\nu$ of the observed FRBs along with the theoretical prediction of 
\cite{YangYP2022ApJ...928L..16Y}. The dotted, dashed, and solid lines indicate the possible relation of $\zeta_e\gamma_{\rm th}^2(R/0.01\rm\,pc)^2$ =10, 1, 0.1, respectively. 
The data points are listed in Table \ref{tab:APPtab2}. The persistent radio source of FRB\,20230607A has not yet been detected. The red star indicates its predicted $L_\nu$ based on its measured ${\rm |RM|}$.}
\label{fig:RMLv}
\end{figure}

\subsection{Predicated persistent radio source from FRB\,20230607A}

With the well measured DM and RM, we plot FRB\,20230607A in the DM vs. ${\rm |RM|}$ space along with other repeating and one-off FRBs in Fig.~\ref{fig:dmrm}. One can see that it ranks the third in the measured ${\rm |RM|}$ values of FRBs, just behind FRB 20201122 \citep{Michilli2018Natur.553..182M} and FRB 20190520B \citep{NiuCH2022Natur.606..873N}. 
The combination of RM and DM gives 
an average parallel magnetic field 
of $-42\;\mu$G, 
much larger than the typical interstellar magnetic field of a few $\mu$G in the Milky Way, and that of the intergalactic medium of a few nG  \citep{xuhan14, xuhan22, car22}.
Such a high magnetic field strength suggests that FRB\,20230607A is located in an extremely magnetized local environment, similar to 
FRB\,20121102A  \citep{Michilli2018Natur.553..182M} and FRB\,20190520B \citep{AnnaT2023Sci...380..599A}, each 
associated with a persistent radio source (PRS) \citep{Chatterjee2017Natur.541...58C, NiuCH2022Natur.606..873N}. 

Physically, there could be a connection between RM and the luminosity of the PRS. This is because the PRS is likely powered by synchrotron radiation \citep{YangYP2016ApJ...819L..12Y,Murase16,Metzger17,Yang20}, the brightness of which also depends on the electron number density\footnote{Strictly speaking, it depends on the number density of relativistic electrons, which can be reasonably assumed to be proportional to the cold electron number density \citep{Yang20,YangYP2022ApJ...928L..16Y}.} and the strength of local magnetic fields. It has been suggested that ${\rm |RM|}$ could be a good proxy of the PRS luminosity, with a predicted scaling law of 
$L_{\nu}\propto(\zeta_e\gamma_{\rm th}^2R^2)|\rm RM|$ \citep{Yang20, YangYP2022ApJ...928L..16Y}, where $R$ is the radius of the region that contributes to the PRS and the RM, $\zeta_e$ is the electron fraction that generates synchrotron emission in the GHz band, $\gamma_{\rm th}$ the Lorentz factor defined by $\gamma_{\rm th}^2\equiv\int n_e(\gamma)d\gamma/\int[n_e(\gamma)/\gamma^2]d\gamma$, where $n_e(\gamma)$ is the differential distribution of electrons. In particular, for an electron distribution with a thermal and a non-thermal component, $\gamma_{\rm th}$ would be the typical Lorentz factor separating the two components. Such a predicted scaling has been recently verified by the detection of several more PRSs \citep{Bruni2024Natur.632.1014B,Ibik2024arXiv240911533I}.

Based on the large absolute RM value of FRB\,20230607A, 
the predicted specific luminosity of the PRS would be about $L_\nu\sim10^{29}~{\rm erg~s^{-1}Hz^{-1}}$, as indicated in Figure \ref{fig:RMLv}. 
At the distance $d\sim0.98~{\rm Gpc}$ estimated by the DM$_{\rm IGM}$, the maximum flux density of the PRS is around $F_{\rm \nu}\sim87~{\rm \mu Jy}$. 
We encourage observers to conduct a deep search of a possible PRS associated with FRB\,20230607A.

\section{Conclusion and discussion}
\label{sect5:con}

This paper presents the observational results of FRB 20230607A by FAST. In 15.6 hours spanning 16 months, we detected 565 bursts from the source. This allowed us to perform the statistical properties of the bursts, including 
waiting times, temporal widths, central frequencies and
frequency widths, fluences and energies. The distributions are consistent with other well-studied repeaters observed by FAST, suggesting that active repeaters share many properties in common.

In any case, several observational properties stand out from this source. First, one of the three brightest bursts, B10, showed a unique lightcurve, with a bright, narrow spike lasting 0.3 ms sticking out from the surrounding broader pulses, which in general show a frequency down-drifting. It has been known that repeating FRBs tend to show broader pulses than non-repeating ones \citep{Pleunis2021ApJ...923....1P}. Our observed feature suggests that repeaters can make both broad and narrow pulses.  Our observation suggests that at least some apparently non-repeating FRBs with narrow spikes could be the tip of the iceberg of an underlying repeater with fainter broader pulses. 

The 0.3\,ms spike also strongly suggests a magnetospheric origin of the FRB emission \citep{Lu2022MNRAS.510.1867L,ZhangB2023RvMP...95c5005Z}. Another strong evidence in support of such a conclusion is the narrow spectra widely observed in this source, with $\Delta\nu/\nu_0=0.125$, much smaller than the lower limit 0.58 predicted in the far-away models invoking relativistic shocks \citep{Kumar2024arXiv240612804K}.

We notice that 
the number of detected sub-bursts depends on observing frequency with a power-law index of $-4.79\pm0.32$, which means that many more bursts could be at even lower frequencies outside the FAST band.
The emission frequency bandwidth of sub-bursts is typically at $231_{-99}^{+171}$ MHz, similar to $277_{-84}^{+122}$ MHz for FRB\,20201124A \citep{Zhou2022RAA....22l4001Z}, suggesting a possible common narrowness of the FRB bursts. 

From the FAST detected bursts of FRB\,20230607A, the maximum burst rate was about 149.5 $\rm h^{-1}$. At other observing sessions, the burst rate was lower or even there was no detection.
There is no evidence for 
an activity periodicity like some other repeating FRBs \citep{Chime2020Natur.582..351C,Rajwade2020MNRAS.495.3551R}. It is also different from some other sources that show non-stopping long-term activities \citep{NiuCH2022Natur.606..873N}. 
This suggests that repeating FRBs have diverse long-term behaviors in terms of burst rate.


Finally, FRB\,20230607A has the third largest 
$\rm \left|RM_{\rm obs}\right|$,   
after FRB\,20121102A \citep{Michilli2018Natur.553..182M} and FRB 20190520B \citep{AnnaT2023Sci...380..599A}. Its $\rm \left|RM_{\rm obs}\right|$ is also close to that of the magnetar PSR J1745-2900 near the Galactic Center \citep{SchnitzelerDH2016MNRAS.459.3005S}.  
Based on the simple relationship between the luminosity of PRS and $\rm \left|RM_{\rm obs}\right|$ \citep{YangYP2022ApJ...928L..16Y}, we predict that there should be a PRS with a specific luminosity of $L_\nu \sim 10^{29} \ {\rm erg \ s^{-1} \ Hz^{-1}}$, which could be detectable. 
%

\section*{Acknowledgments}

This work made use of the data from FAST (Five-hundred-meter Aperture Spherical radio Telescope)(https://cstr.cn/31116.02.FAST).  FAST is a Chinese national mega-science facility, operated by National Astronomical Observatories, Chinese Academy of Sciences. This research is triggered by the CHIME/FRB VOEvent service. 
D.J.Z., J.L.H. and their NAOC group are supported by the National Natural Science Foundation of China (NSFC Nos. 11988101 and 12403059) and the Key Research Program of the Chinese Academy of Sciences (No. QYZDJ-SSW-SLH021).
Y.P.Y. is supported by the NSFC No. 12473047 and the National SKA Program of China (2022SKA0130100).
F.Y. is supported by the NSFC No. 12203045. 
We thank an anonymous referee for helpful comments.

\bibliography{sample631}{}
\bibliographystyle{aasjournal}

\appendix

\section{A Complete List of Detected Bursts and Sub-bursts of FRB\,20230607A}

All parameters of each burst and sub-burst analysis use the method of \citet{Zhou2022RAA....22l4001Z} and list in the Table \ref{appTable1}, which includes the TOA expressed in MJD, the signal-to-noise ratio (S/N), the burst or sub-burst width (W or W$_{\rm sb}$, in ms), the observed emission peak flux density (S$_{\rm peak}$, in mJy), the fluence (F$_{\nu}$, in mJy ms), the frequency of emission peak ($\rm \nu_{0}$, in MHz), the bandwidth (BW$_{\rm e}$, in MHz), the detected emit low and high frequency ($\rm \nu_{low}$, $\rm \nu_{high}$ respectively, in MHz), and the group for each burst. The fitted $\rm \nu_{0}$ is not given if it is out of the effective observation bandwidth in the range of 1031.25 MHz to 1468.75 MHz. The dynamic spectrum and burst profile for each burst are present in Figure \ref{appfig1}, together with the energy distribution and the Gaussian fitting over the observational frequency for each sub-burst.

\startlongtable

\begin{deluxetable*}{lcrrrrrrrrr}
\tablecaption{Burst and sub-burst parameters of FRB\,20230607A. \label{appTable1}
}
\tablewidth{0pt}
\tablehead{
\colhead{Burst} & \colhead{TOA} & S/N & \colhead{W} &  S$_{\rm peak}$  & F$_{\nu}$  & \colhead{$\rm \nu_{0}$}  & \colhead{BW$_{\rm e}$}  & \colhead{$\rm \nu_{low}$} & \colhead{$\rm \nu_{high}$} & \colhead{Group}\\
\colhead{No.} & \colhead{(MJD)} & \nocolhead{}  & \colhead{(ms)}  & \colhead{(mJy)} & \colhead{(mJy\,ms)} & \colhead{(MHz)} & \colhead{(MHz)} &
\colhead{(MHz)} & \colhead{(MHz)} & 
}
\decimalcolnumbers
\startdata
\multicolumn2l{\bf Date: 20231016:}\\
      1  & 60233.558168477 &       5.0 &      13.0 &        4.3 &       16.0 &        1420.6 &        352.6 &  1244.3 & 1468.8 & NE-H \\
      2  & 60233.570489883 &       9.4 &      10.5 &       15.1 &      107.3 &        1046.5 &         97.7 &  1031.3 & 1095.3 & NE-L \\
      3  & 60233.571091100 &       7.9 &      11.6 &        5.7 &       24.9 &        1103.7 &        355.5 &  1031.3 & 1281.4 & NE-L \\
      4  & 60233.581817599 &      10.1 &      16.7 &        5.8 &       42.7 &        1130.3 &        462.9 &  1031.3 & 1361.7 & D1-M \\
      5  & 60233.584250617 &      16.6 &      40.8 &        8.2 &      148.1 &        1372.0 &        222.3 &  1260.9 & 1468.8 & D1-H \\
      6  & 60233.585737573 &      16.9 &      39.3 &        7.9 &      144.8 &               &        169.9 &  1031.4 & 1201.2 & D1-L \\
      7  & 60233.589840754 &       7.0 &      34.1 &        3.6 &       21.8 &        1130.6 &        277.9 &  1031.3 & 1269.6 & D1-L \\
\enddata
\end{deluxetable*}

\begin{figure*}
\flushleft
\includegraphics[height=0.29\linewidth]{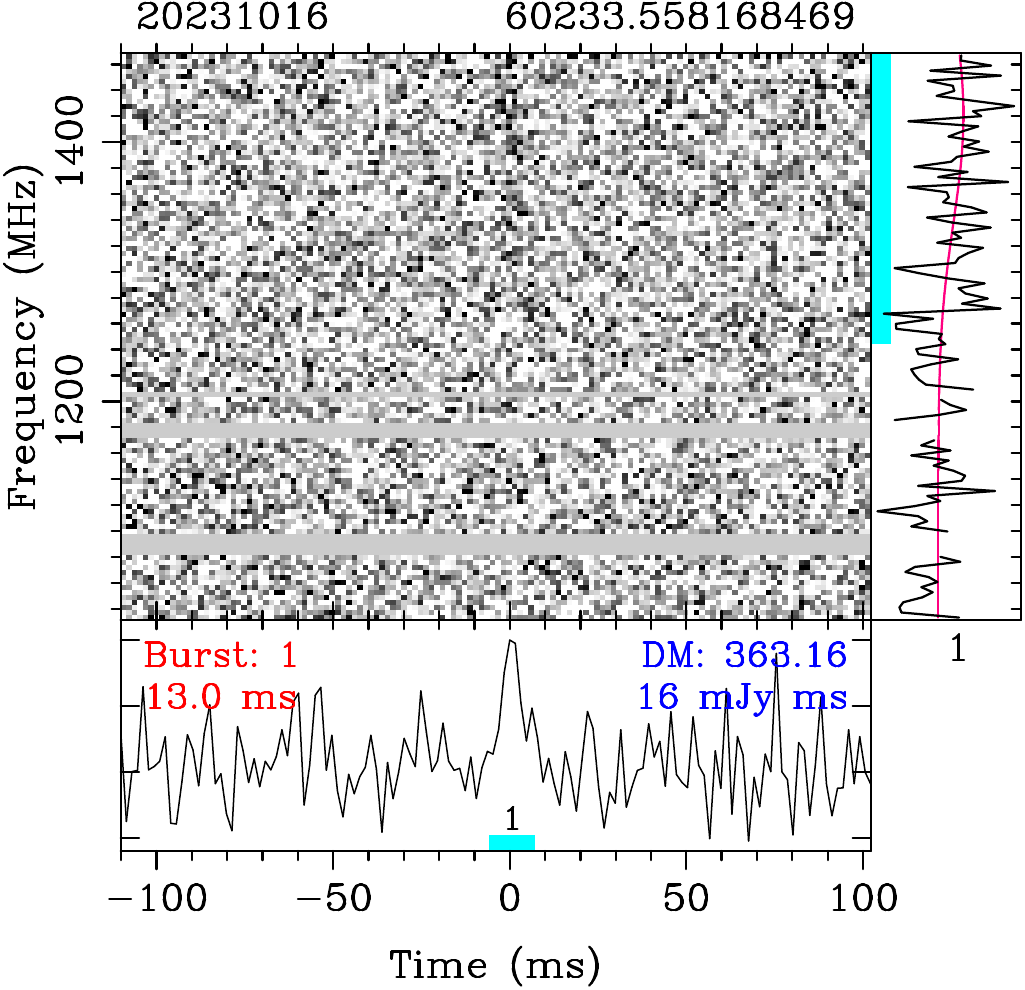}
\includegraphics[height=0.29\linewidth]{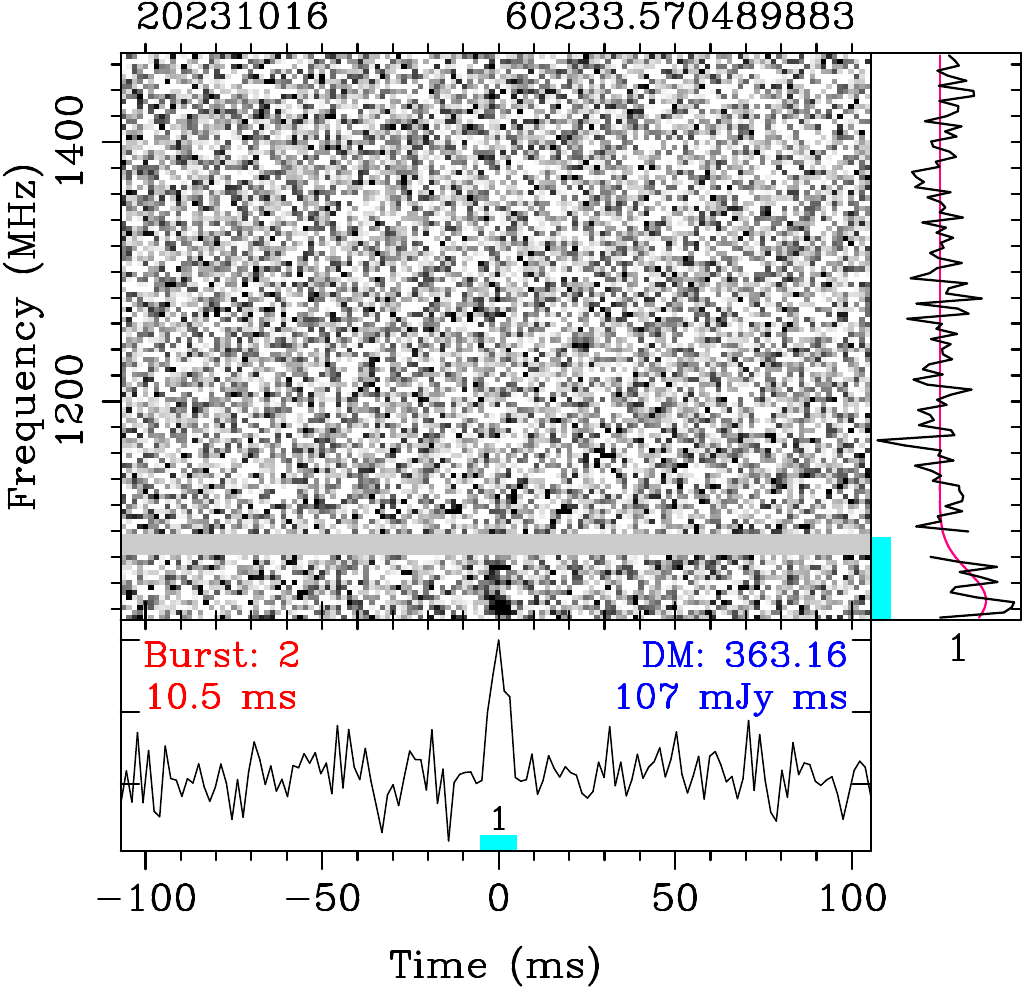}
\includegraphics[height=0.29\linewidth]{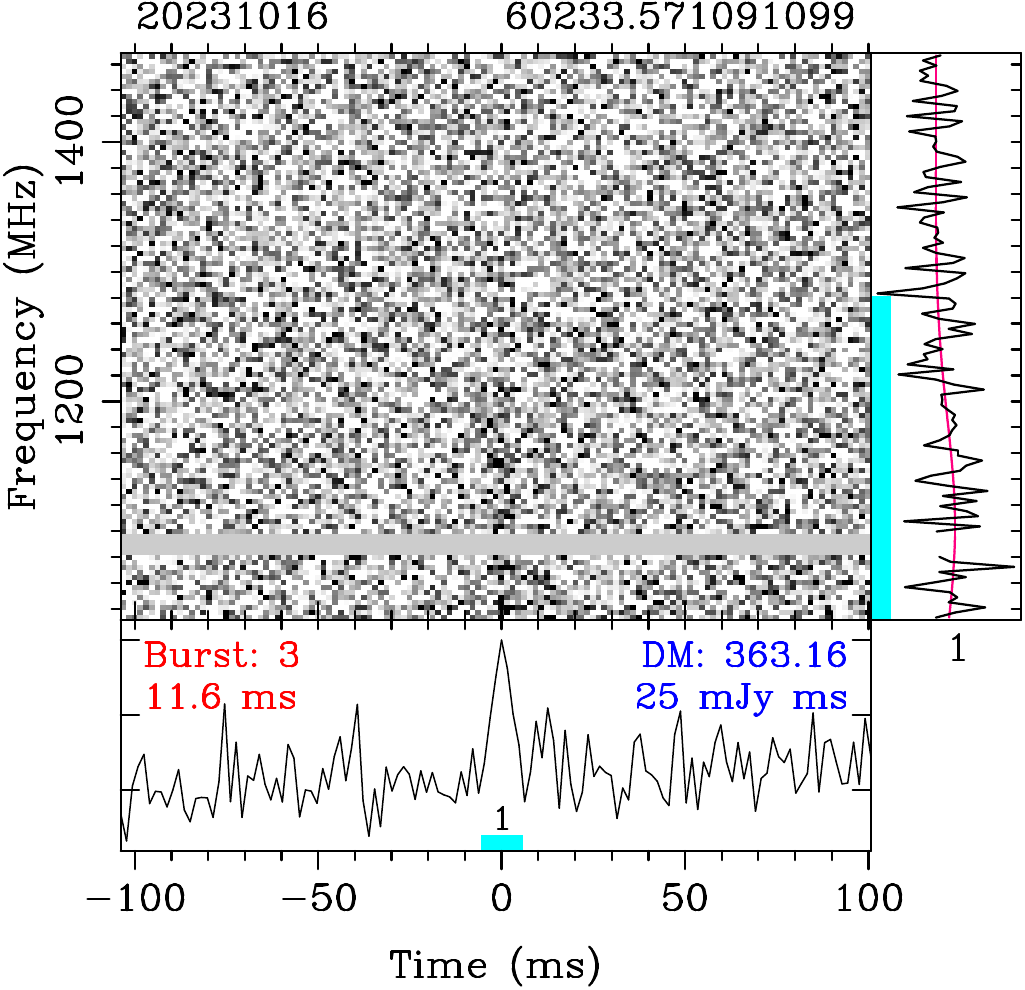}
\includegraphics[height=0.29\linewidth]{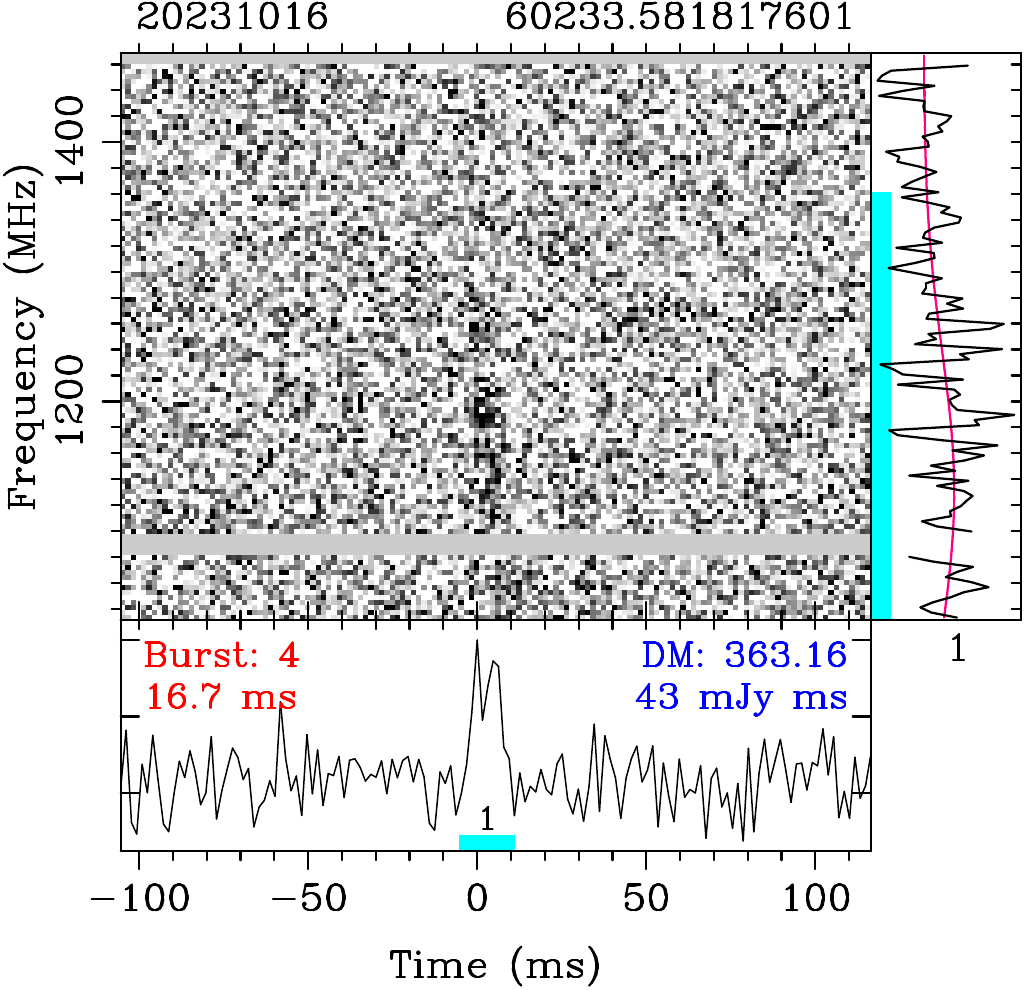}
\includegraphics[height=0.29\linewidth]{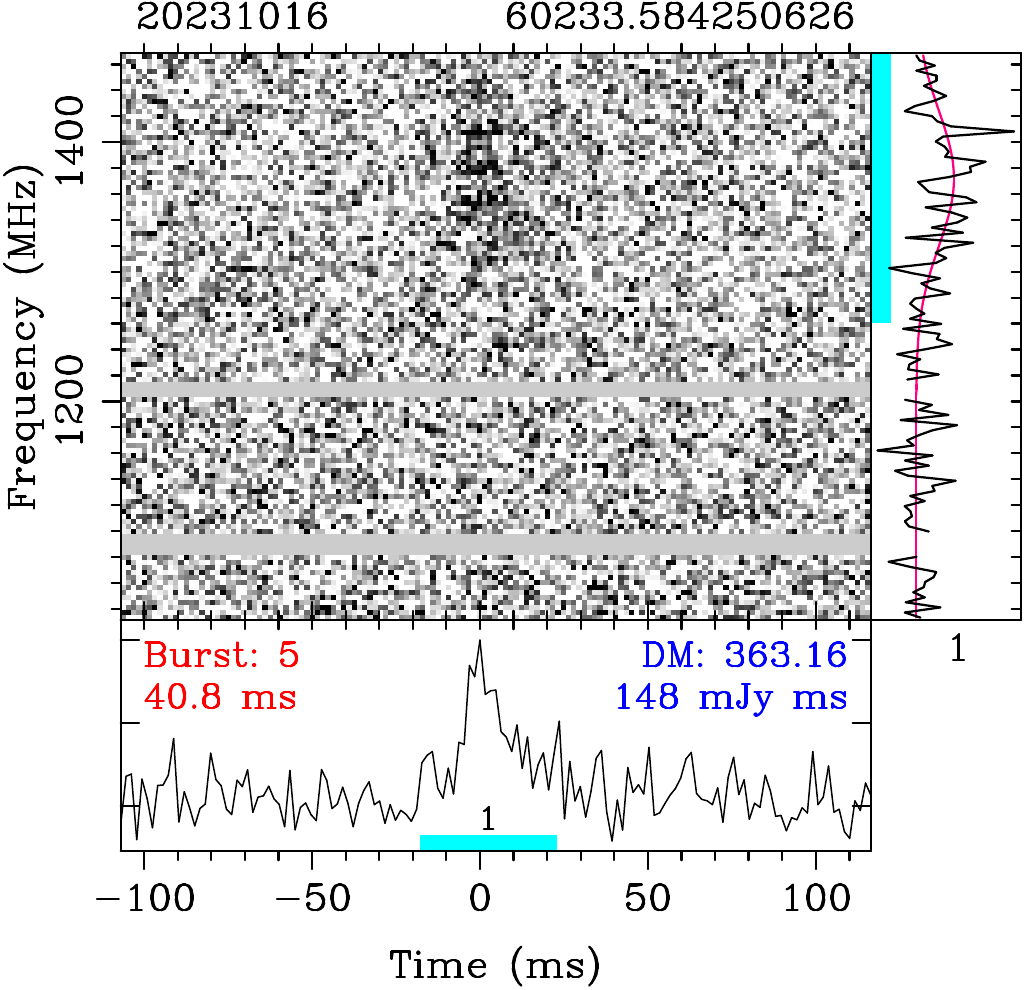}
\includegraphics[height=0.29\linewidth]{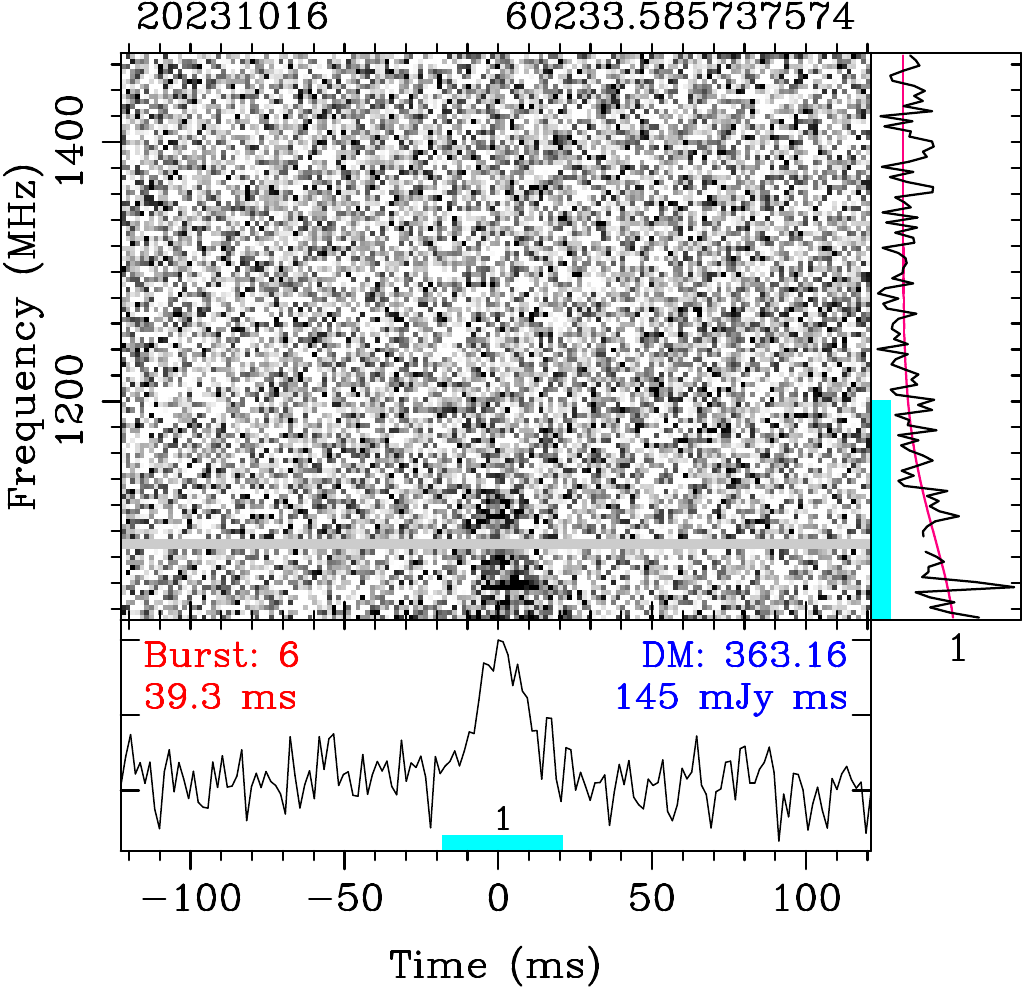}
\includegraphics[height=0.29\linewidth]{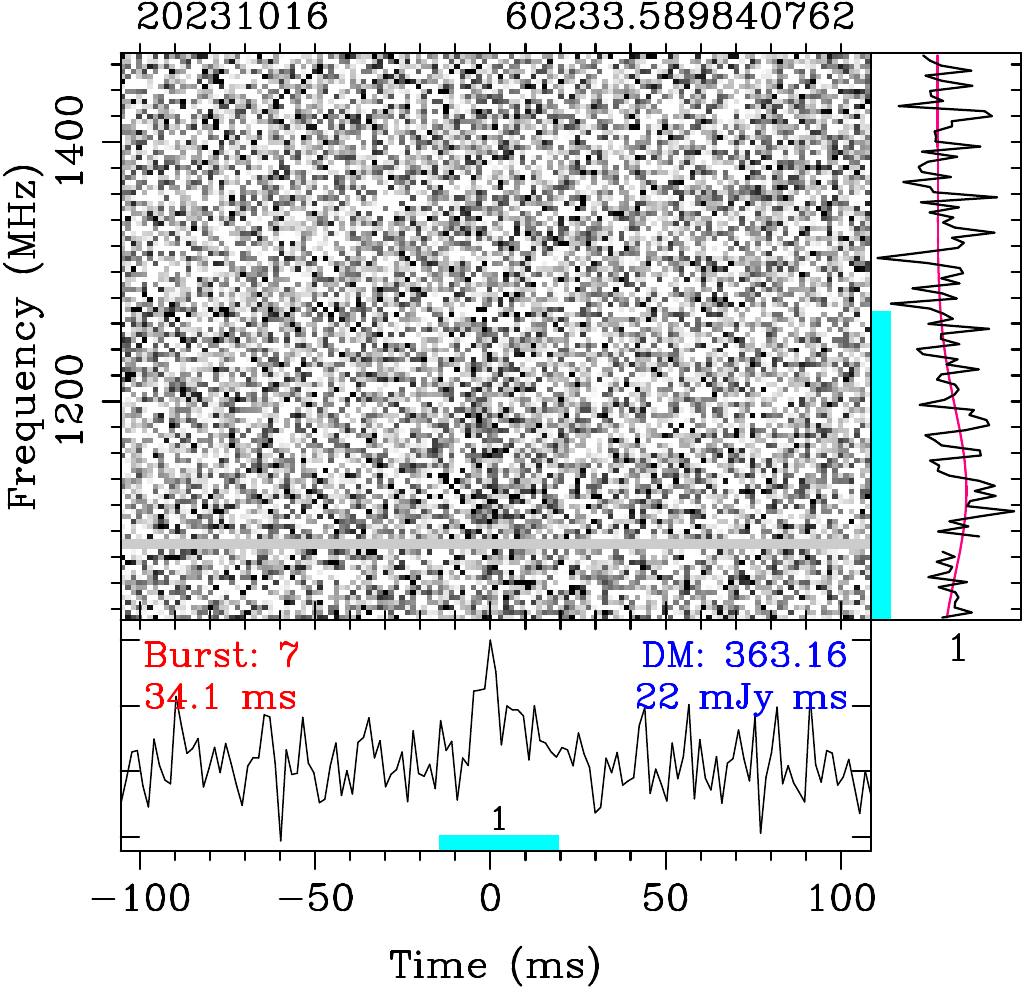}
\includegraphics[height=0.29\linewidth]{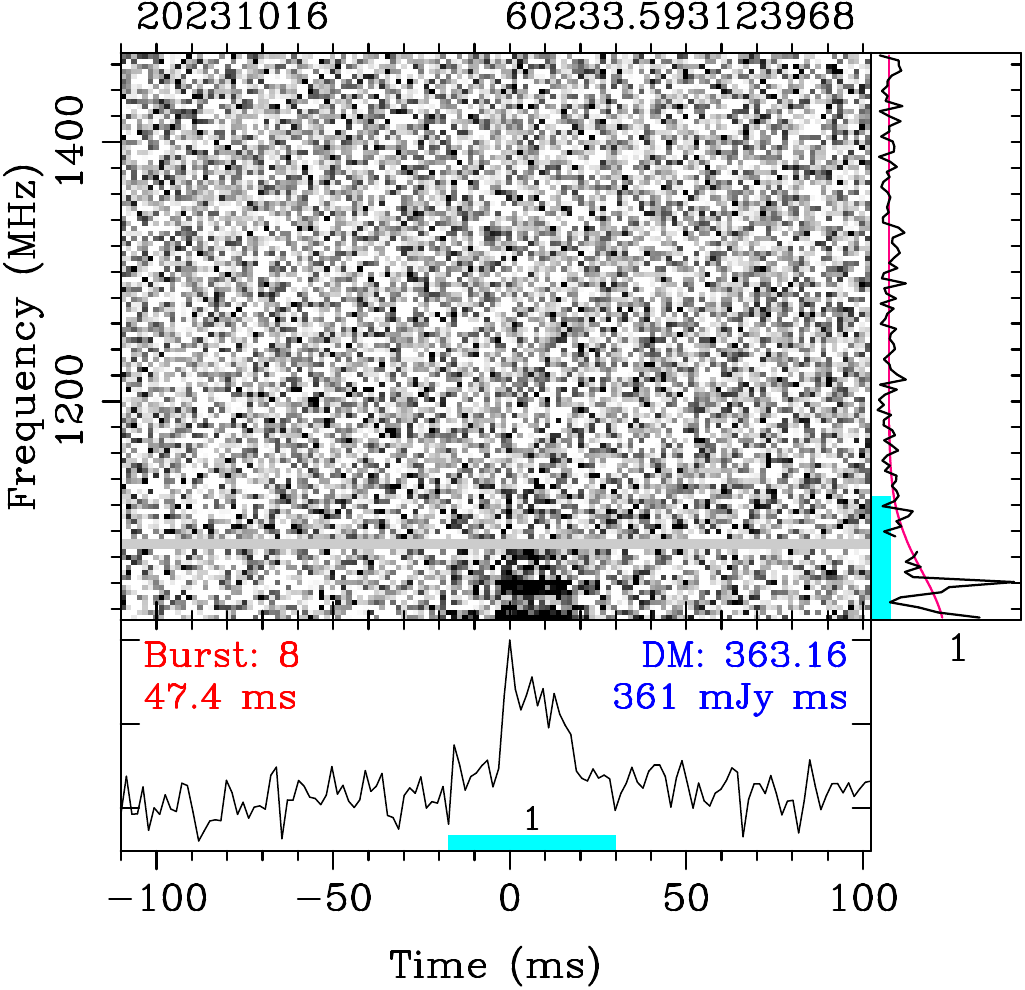}
\includegraphics[height=0.29\linewidth]{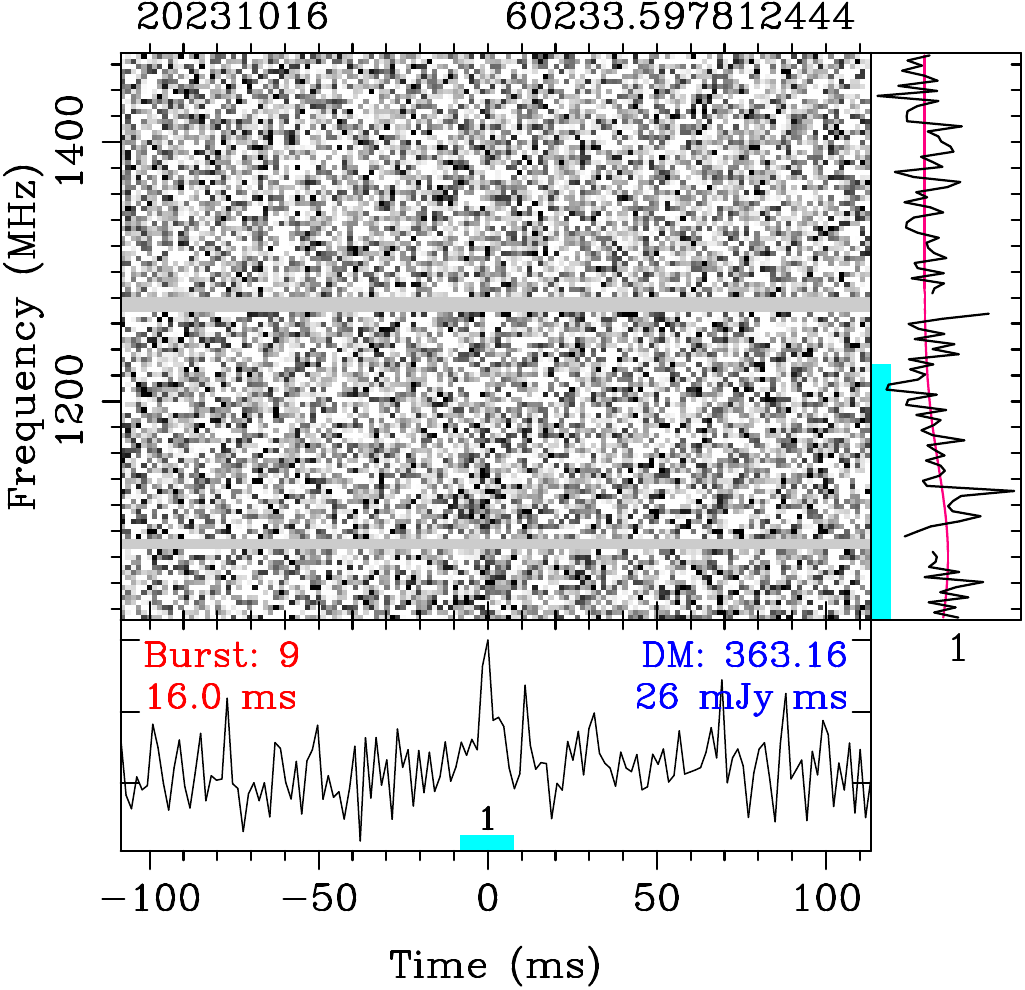}
\includegraphics[height=0.29\linewidth]{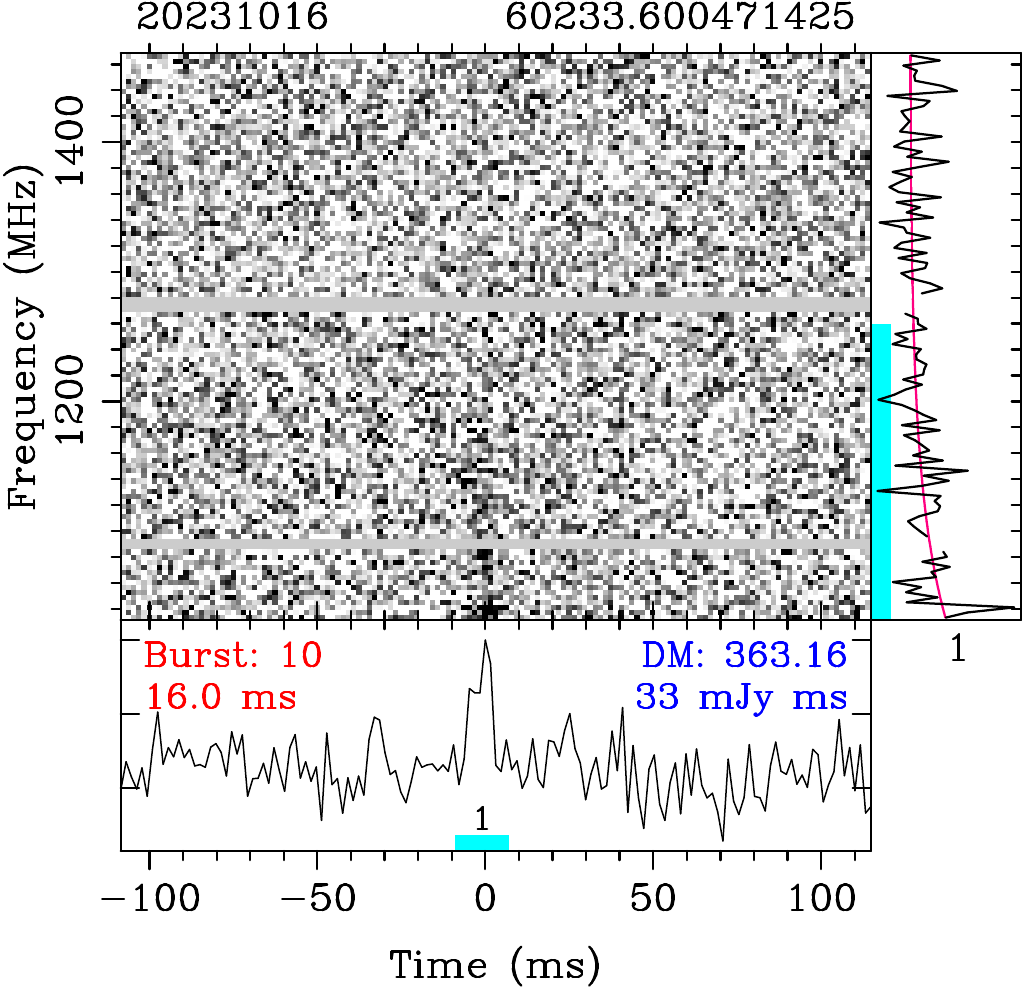}
\includegraphics[height=0.29\linewidth]{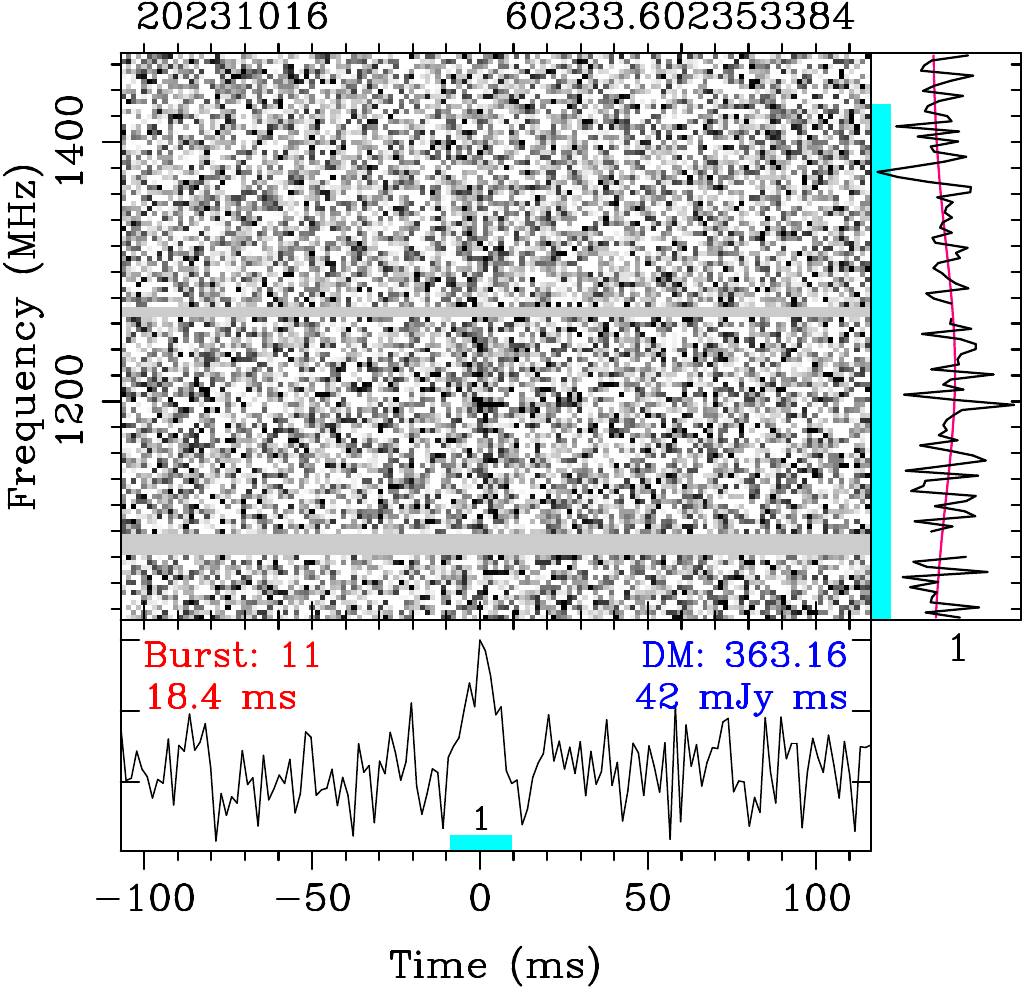}
\caption{The dynamic spectra of each burst for FRB\,20230607A. \add{The complete figure set (565 images of all the bursts) is available in the online journal.}}
\label{appfig1}
\end{figure*}
\addtocounter{figure}{-1}
\begin{figure*}
\flushleft
\includegraphics[height=0.29\linewidth]{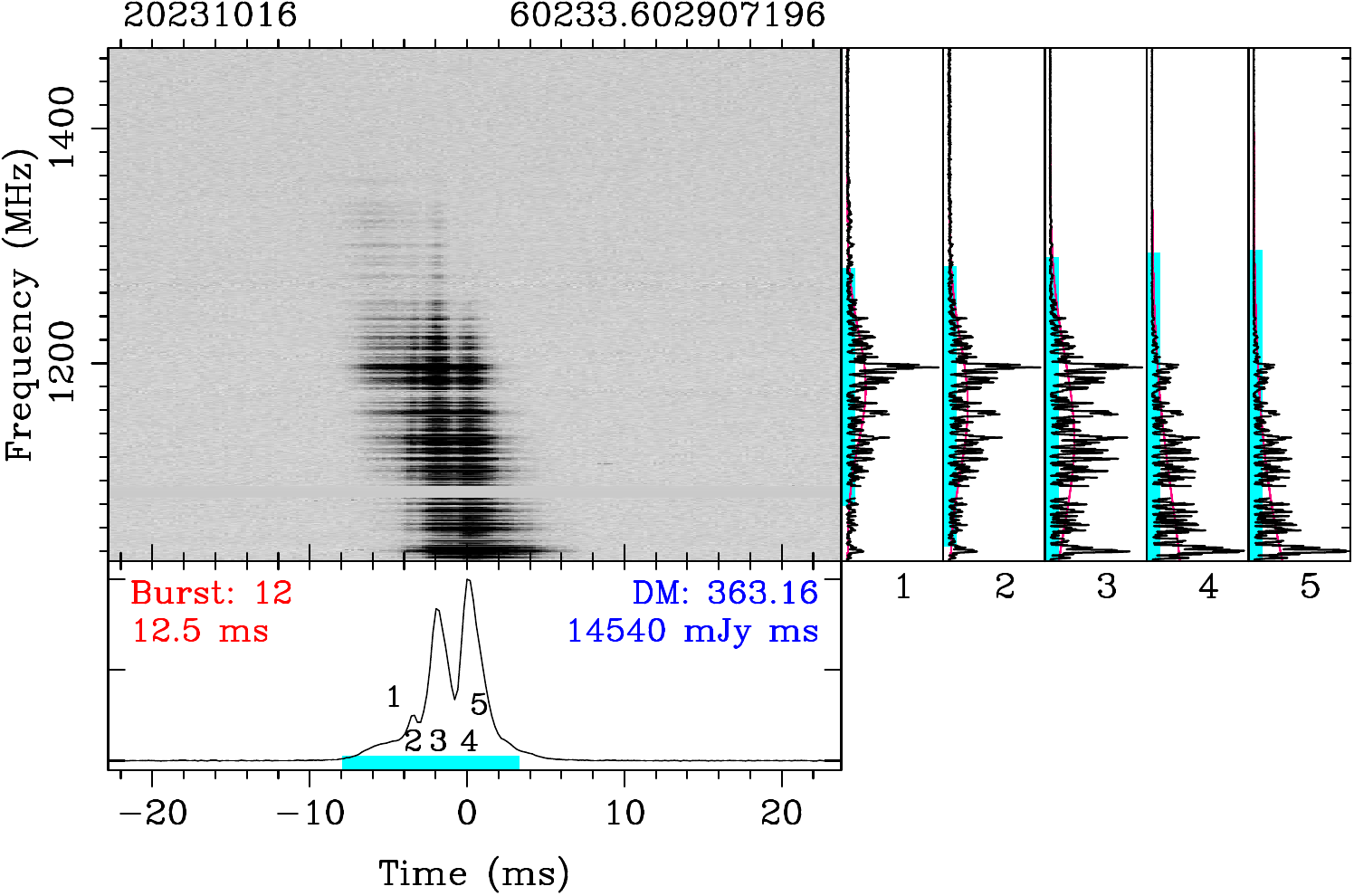}
\includegraphics[height=0.29\linewidth]{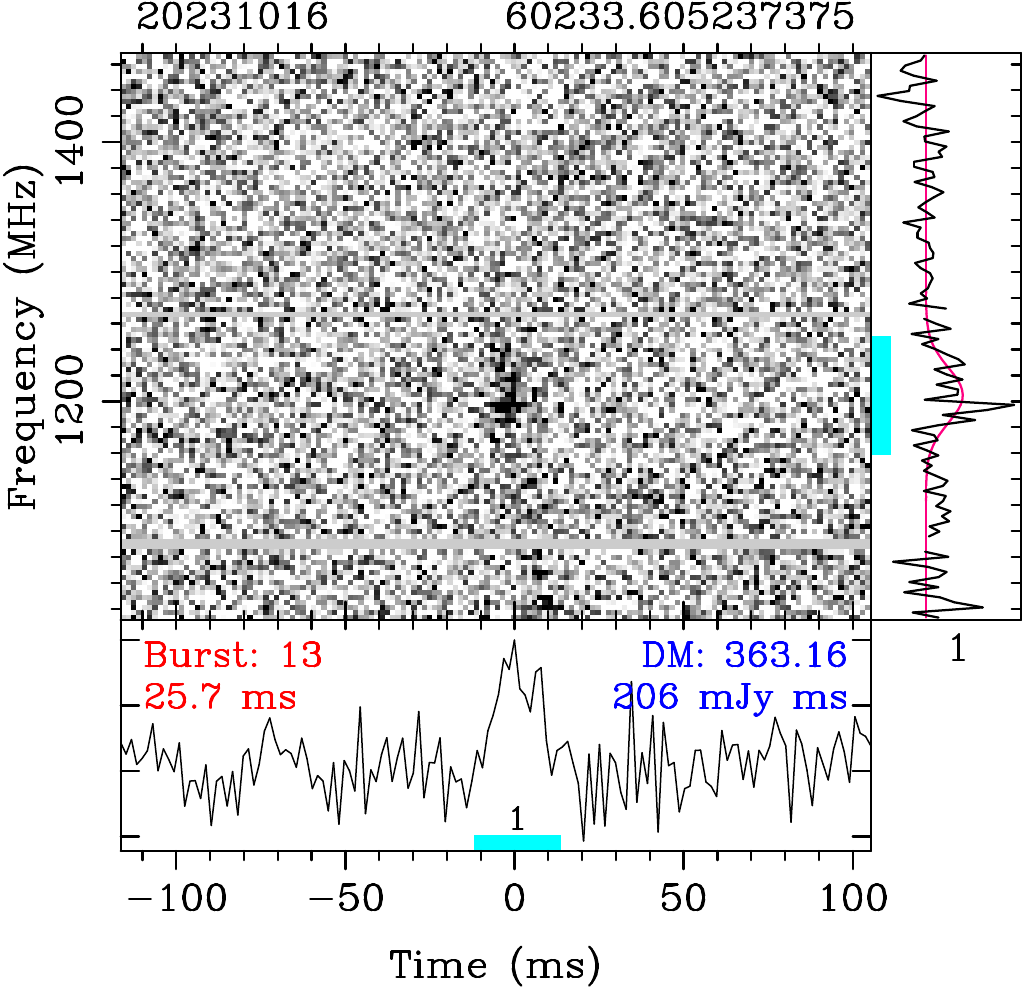}
\includegraphics[height=0.29\linewidth]{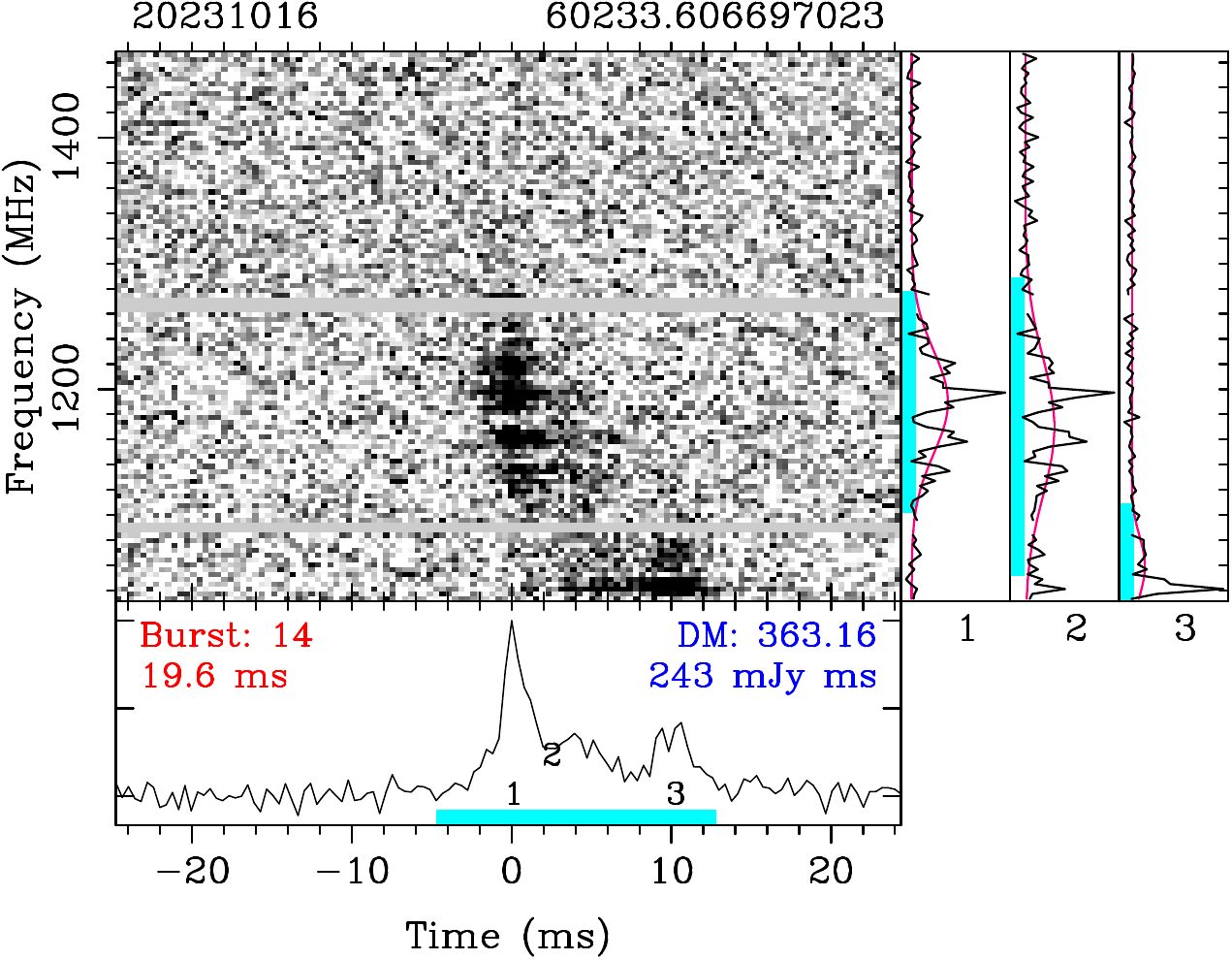}
\includegraphics[height=0.29\linewidth]{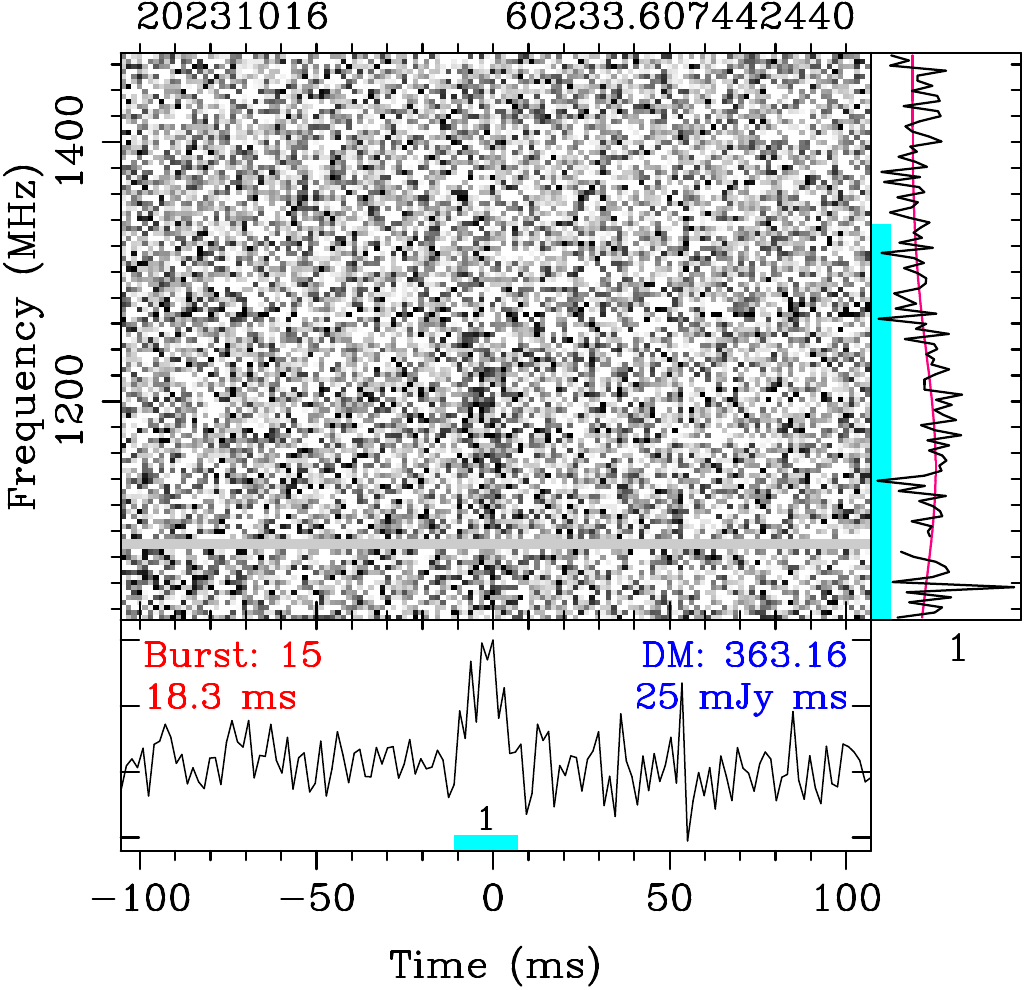}
\includegraphics[height=0.29\linewidth]{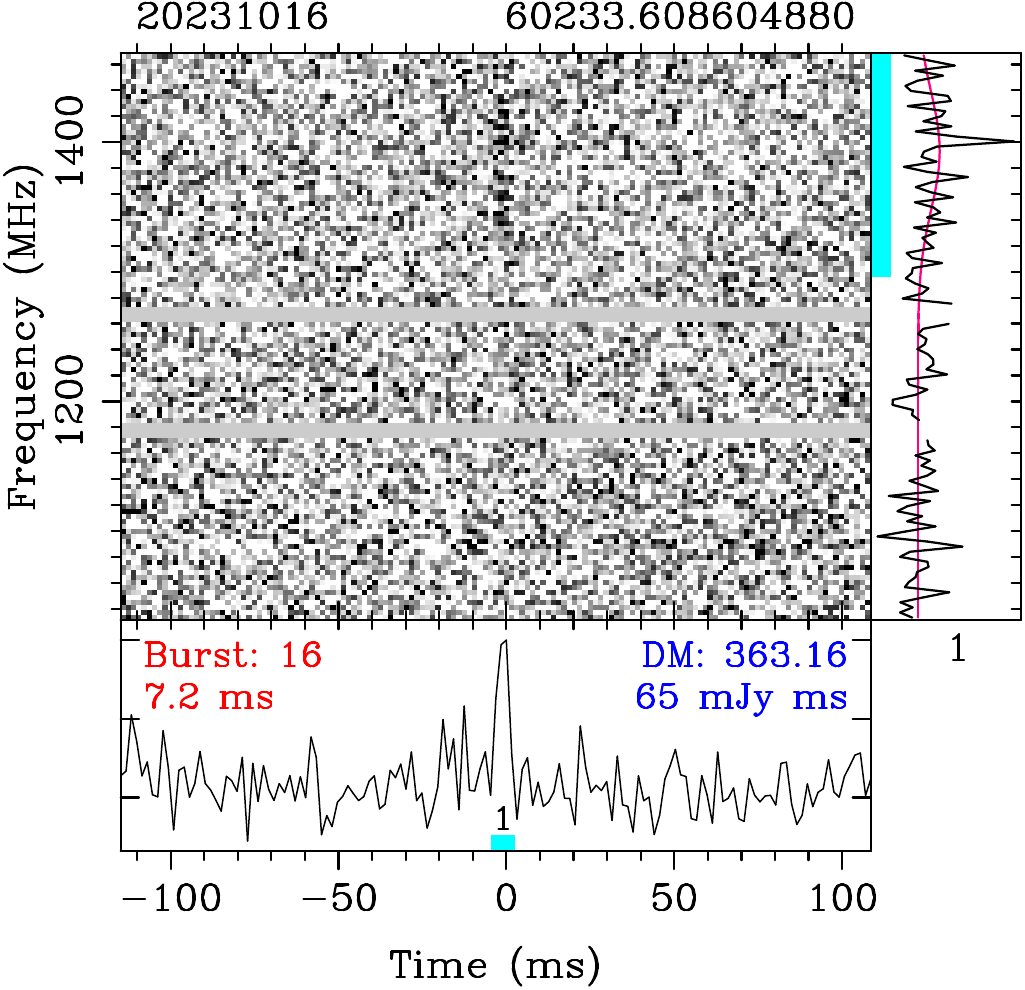}
\includegraphics[height=0.29\linewidth]{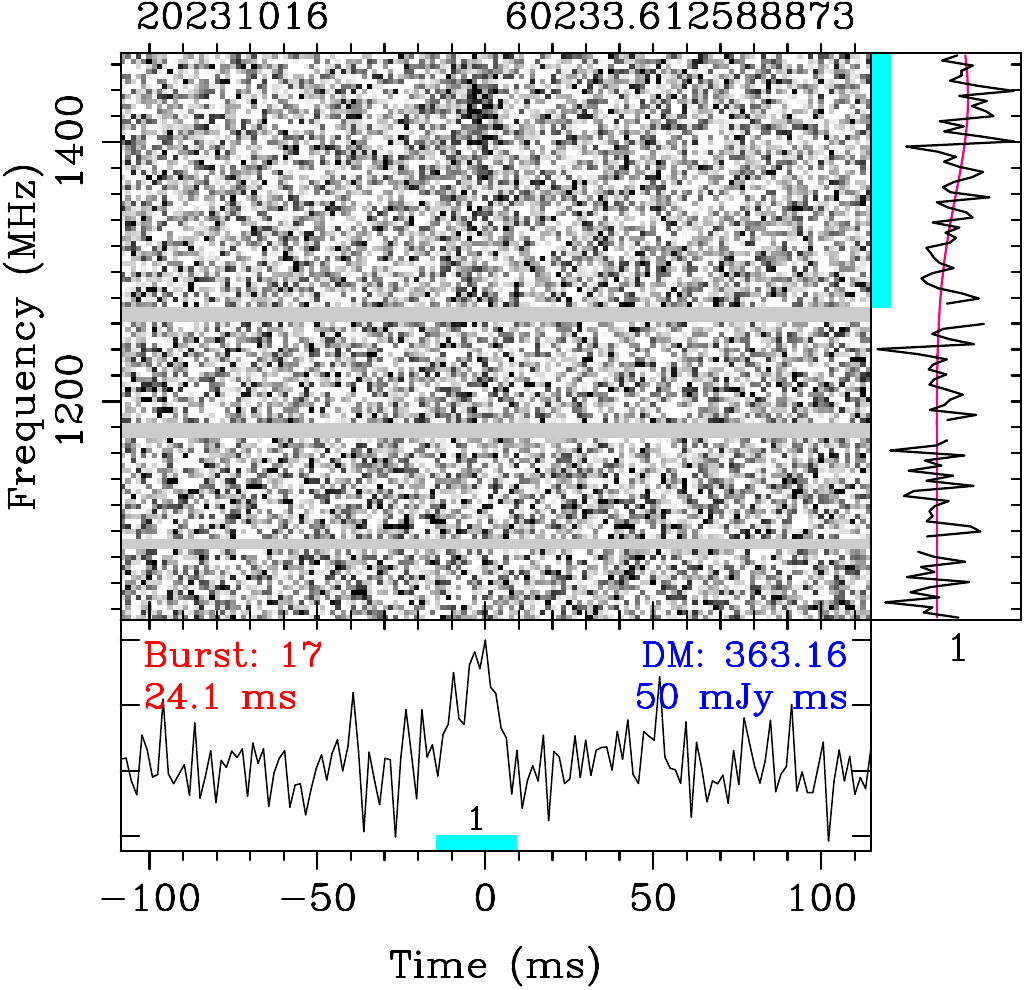}
\includegraphics[height=0.29\linewidth]{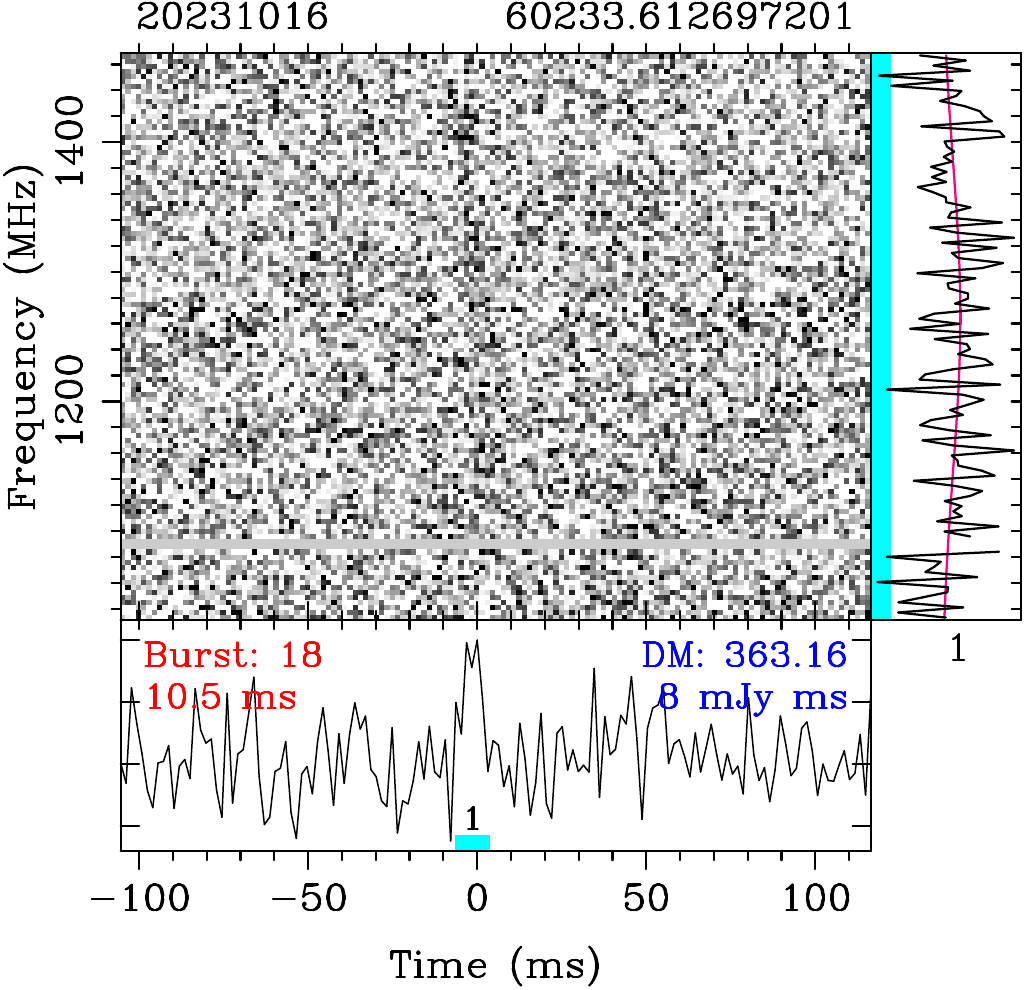}
\includegraphics[height=0.29\linewidth]{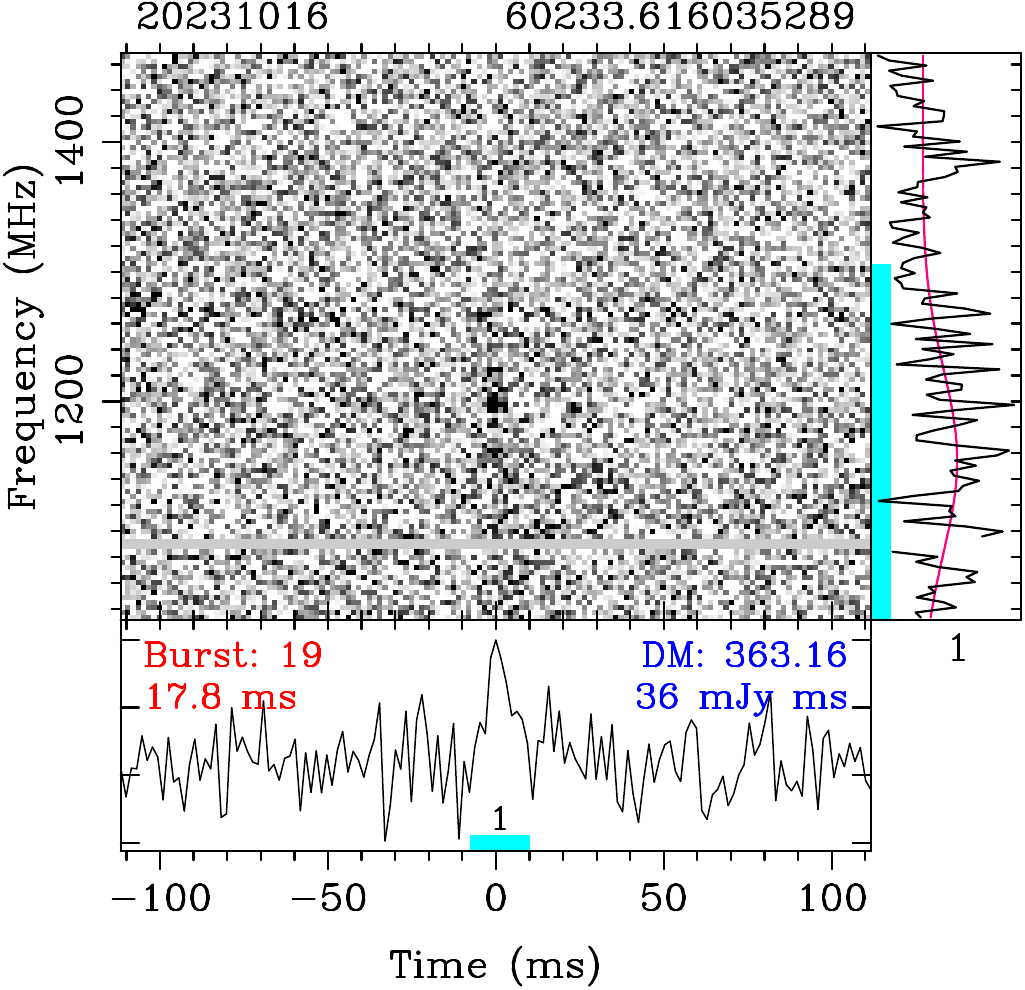}
\includegraphics[height=0.29\linewidth]{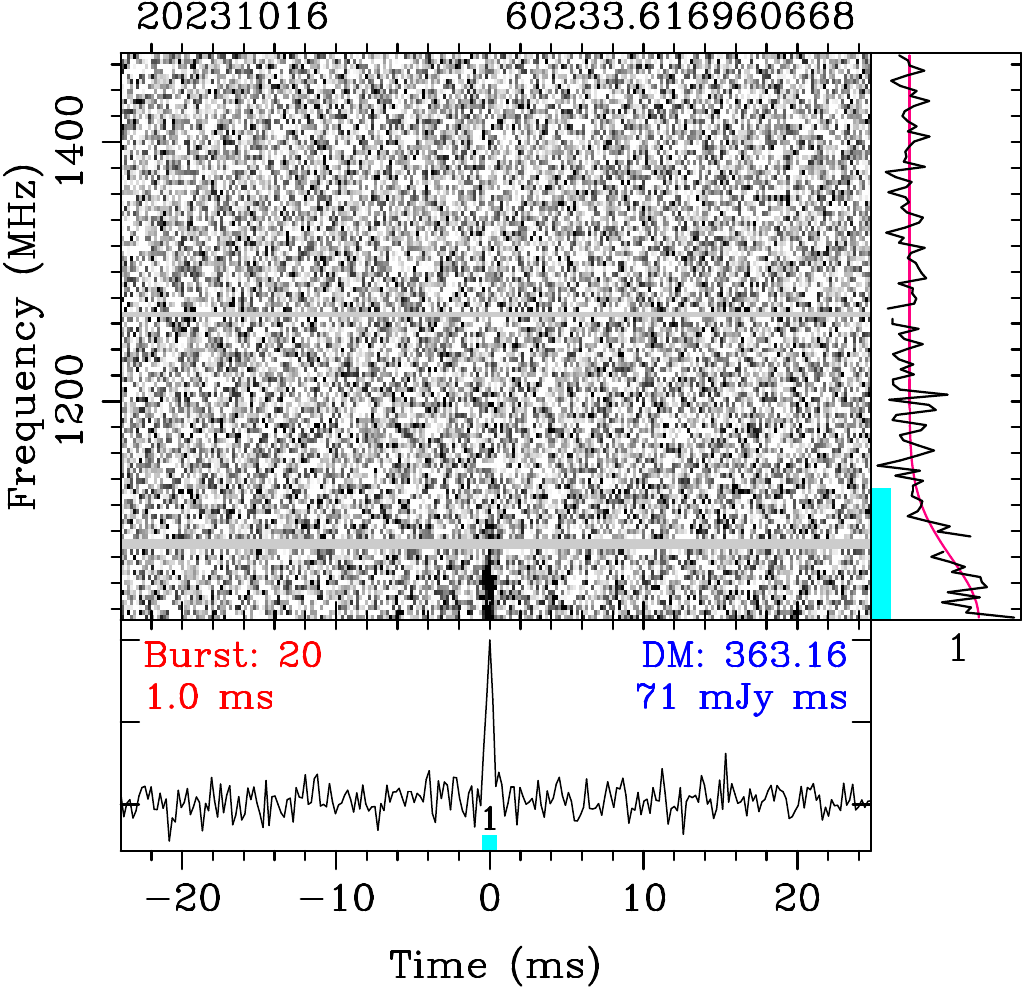}
\includegraphics[height=0.29\linewidth]{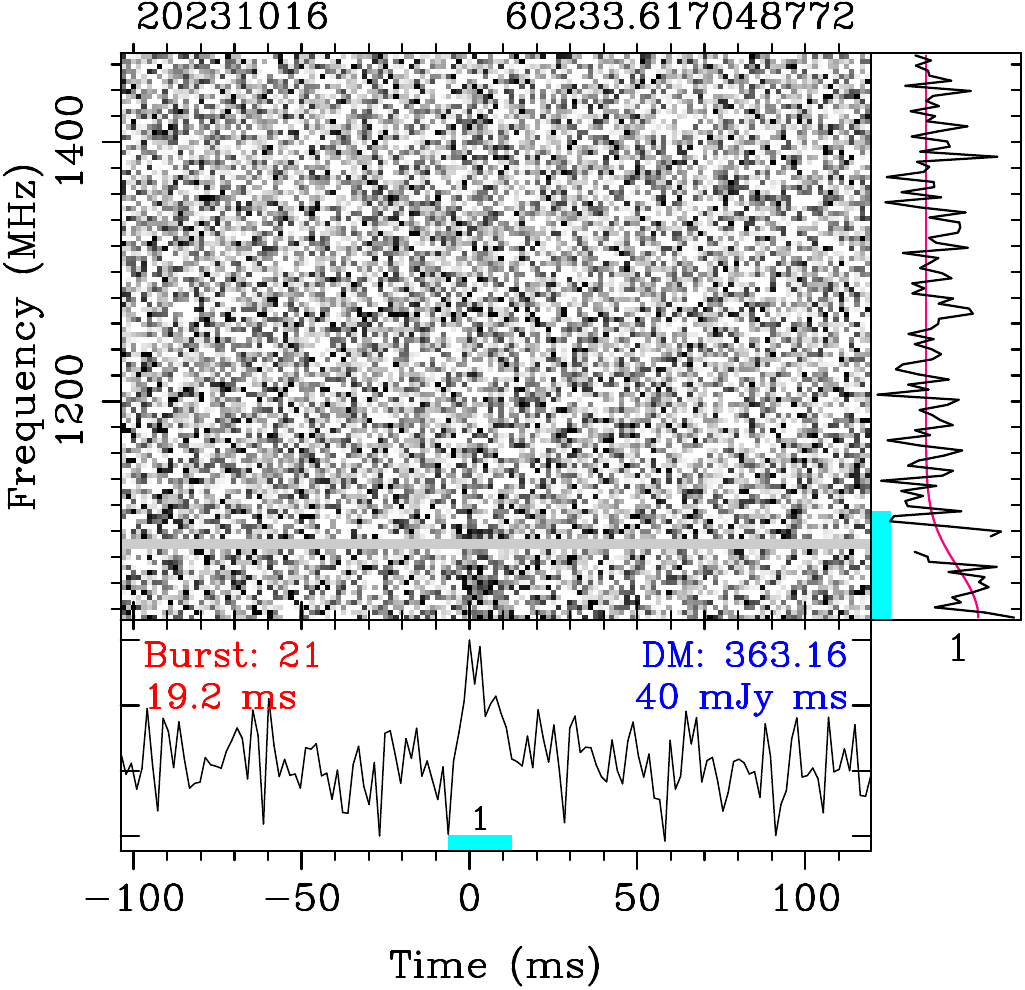}
\includegraphics[height=0.29\linewidth]{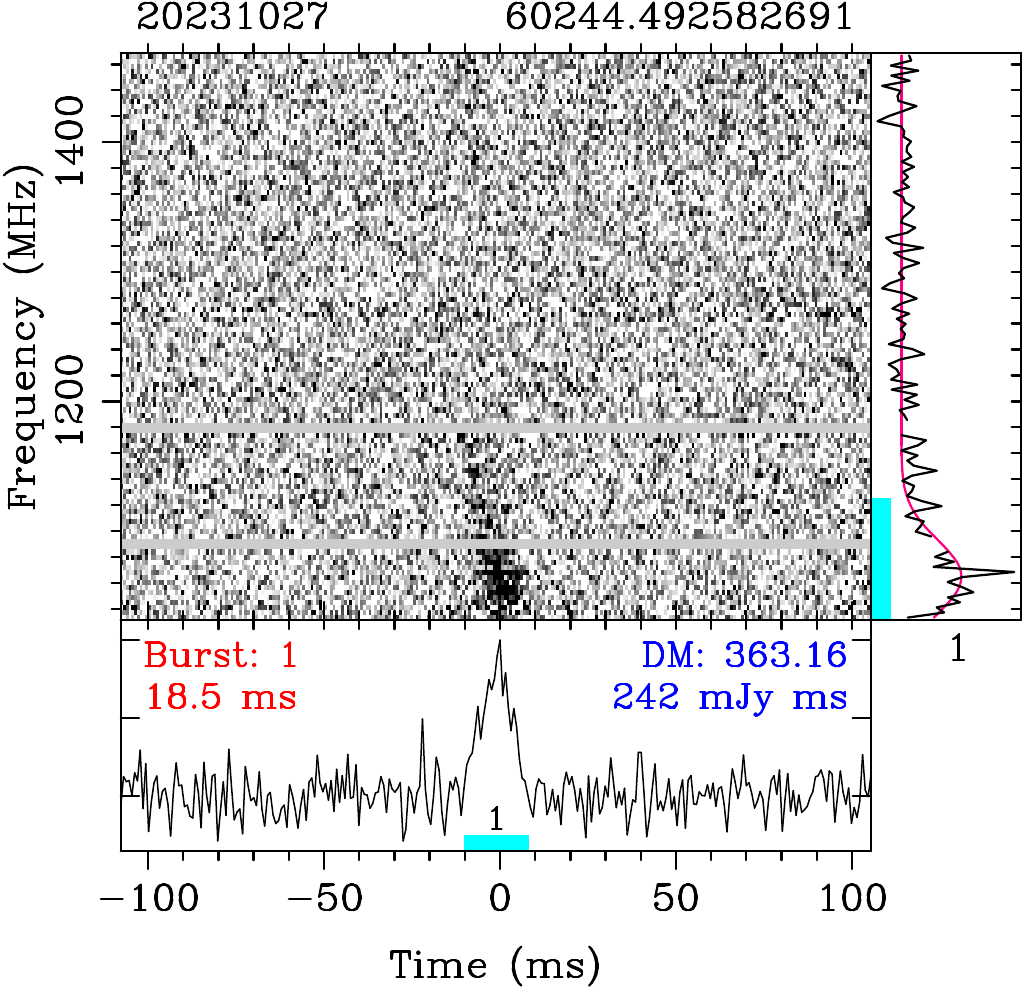}
\caption{({\textit{continued}})}
\end{figure*}
\addtocounter{figure}{-1}
\begin{figure*}
\flushleft
\includegraphics[height=0.29\linewidth]{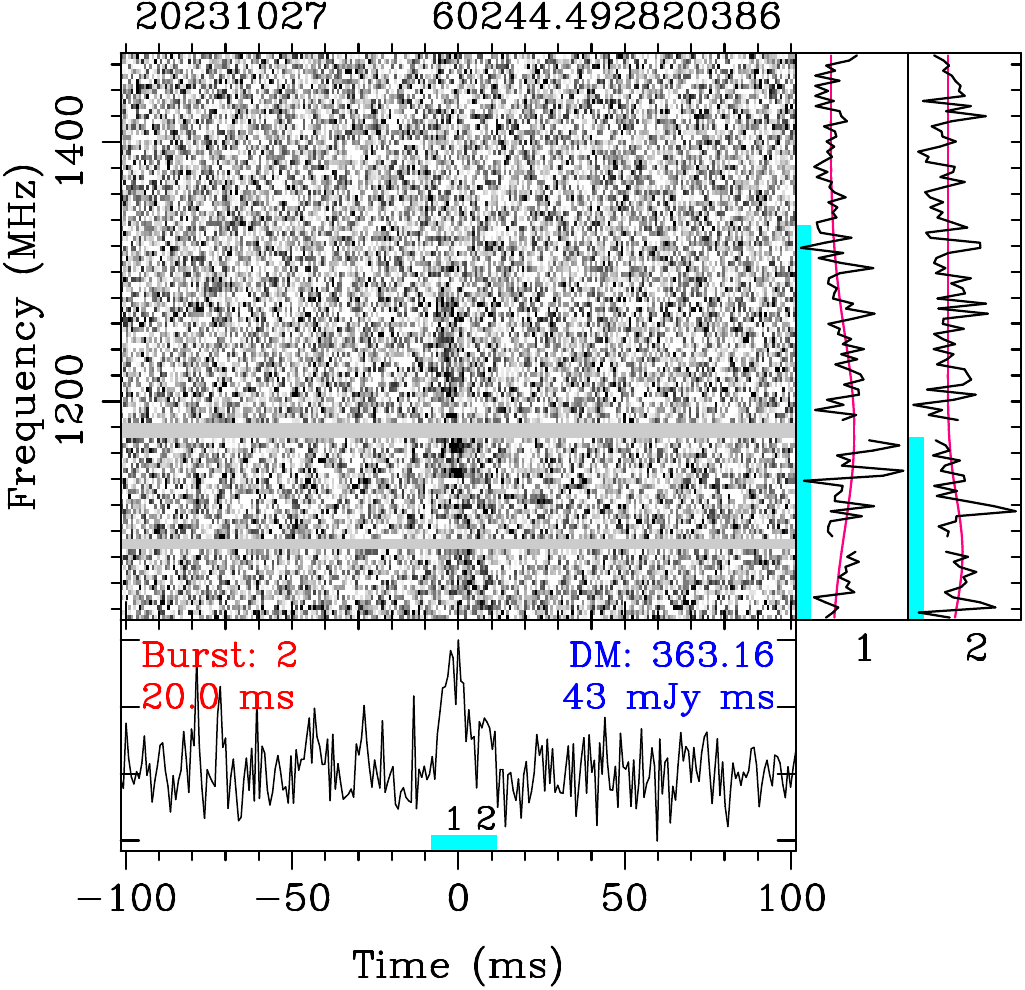}
\includegraphics[height=0.29\linewidth]{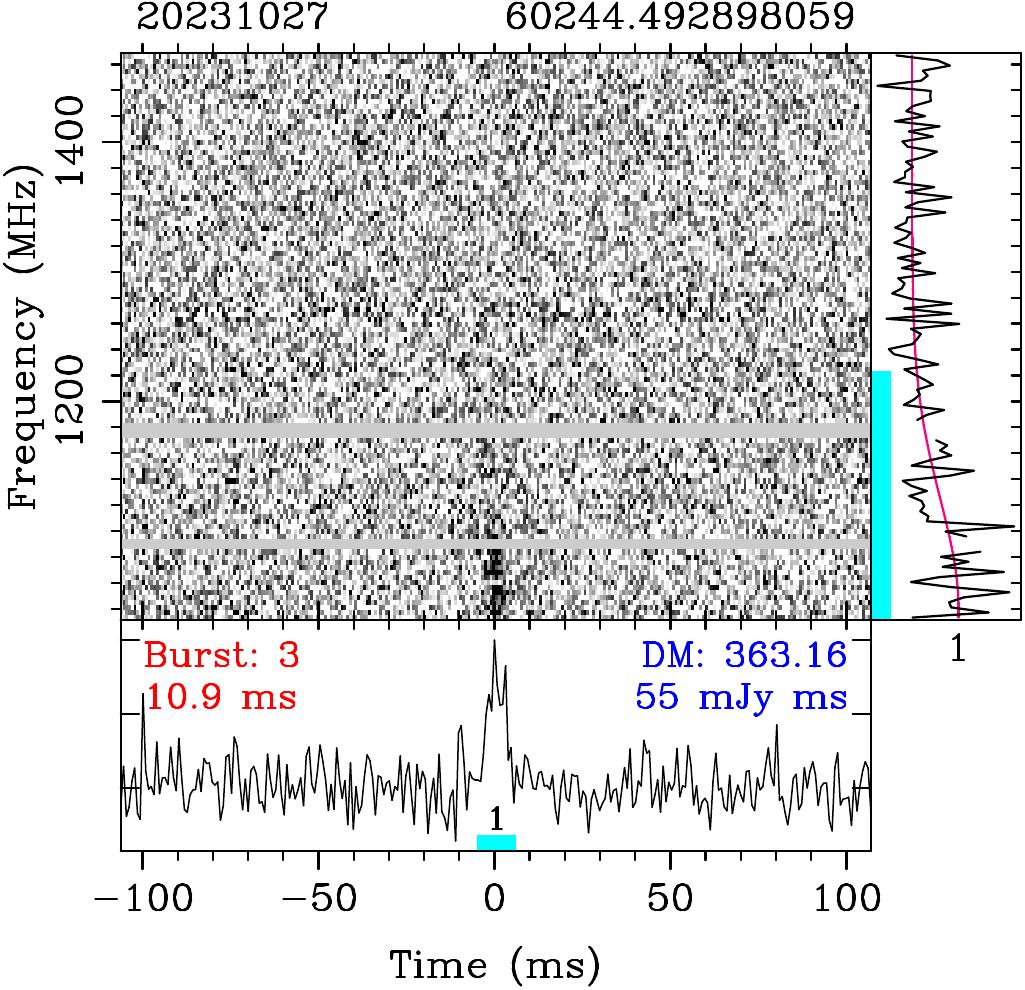}
\includegraphics[height=0.29\linewidth]{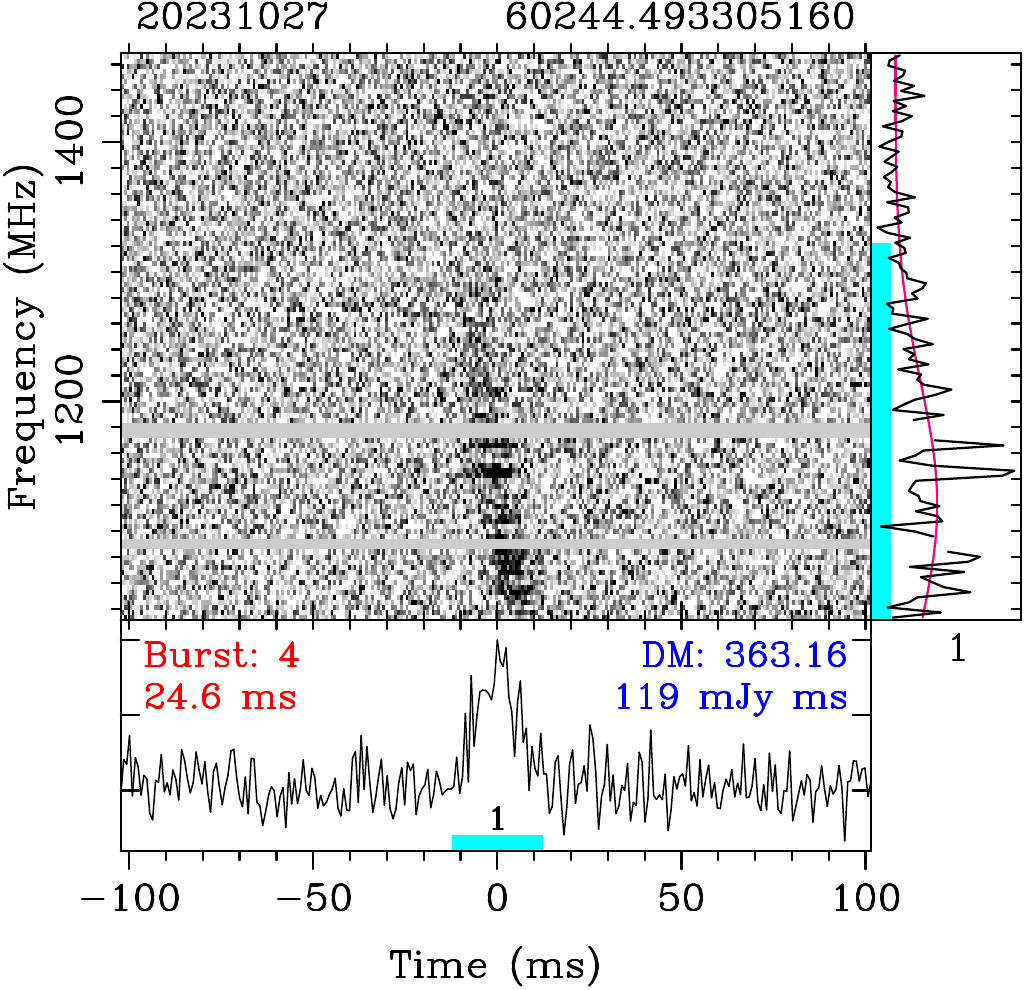}
\includegraphics[height=0.29\linewidth]{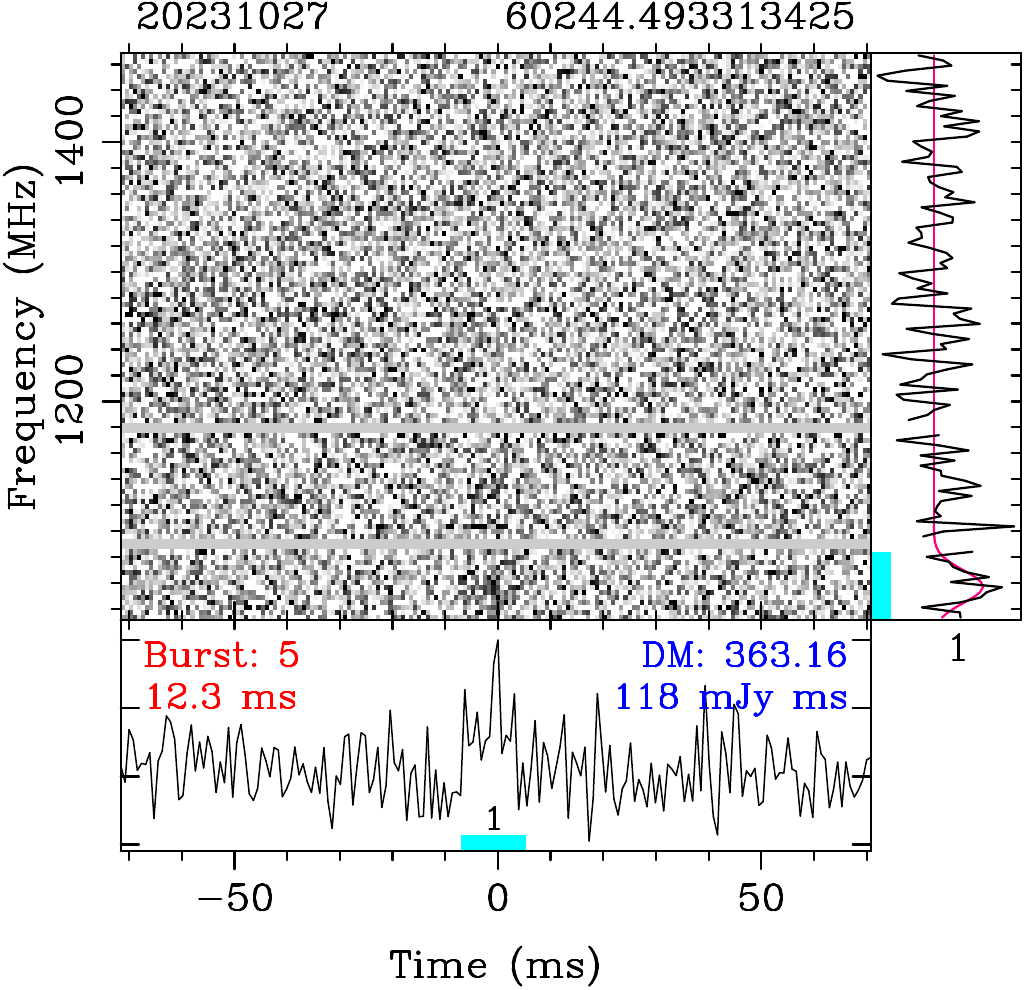}
\includegraphics[height=0.29\linewidth]{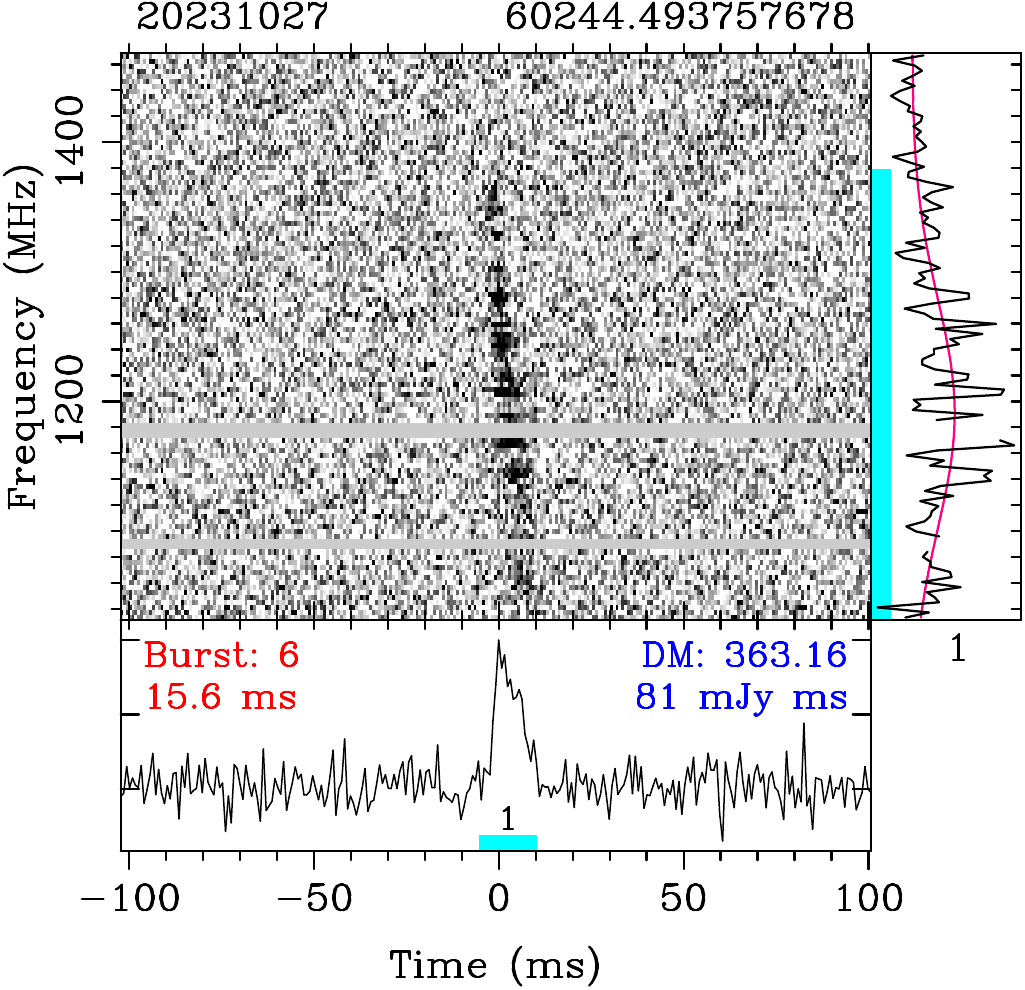}
\includegraphics[height=0.29\linewidth]{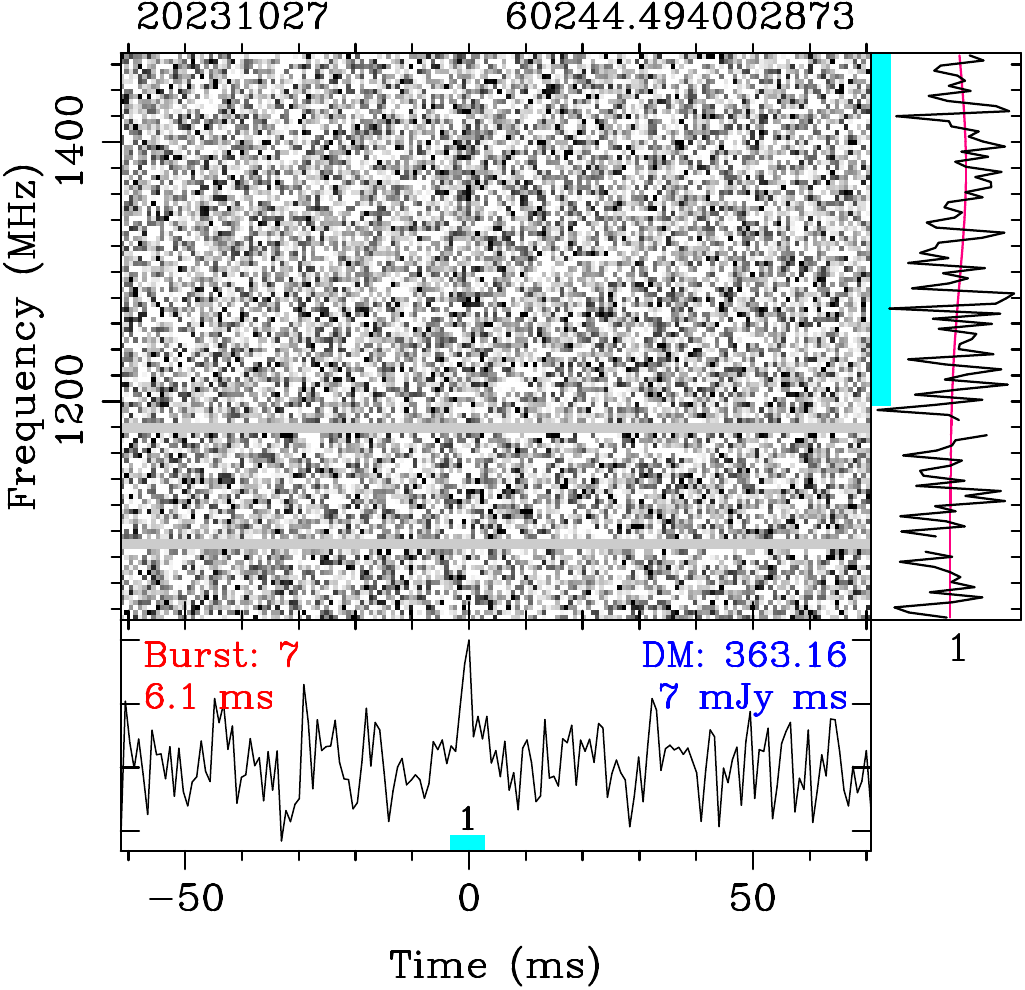}
\includegraphics[height=0.29\linewidth]{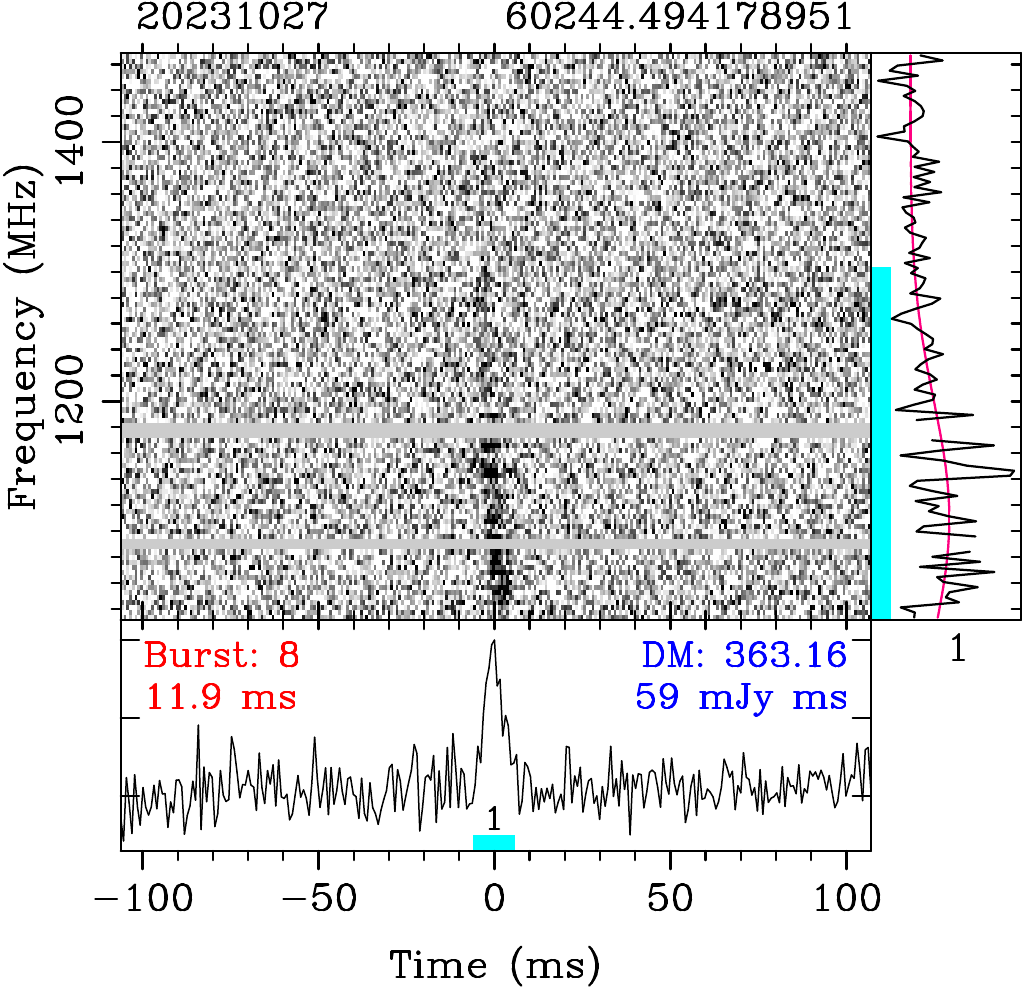}
\includegraphics[height=0.29\linewidth]{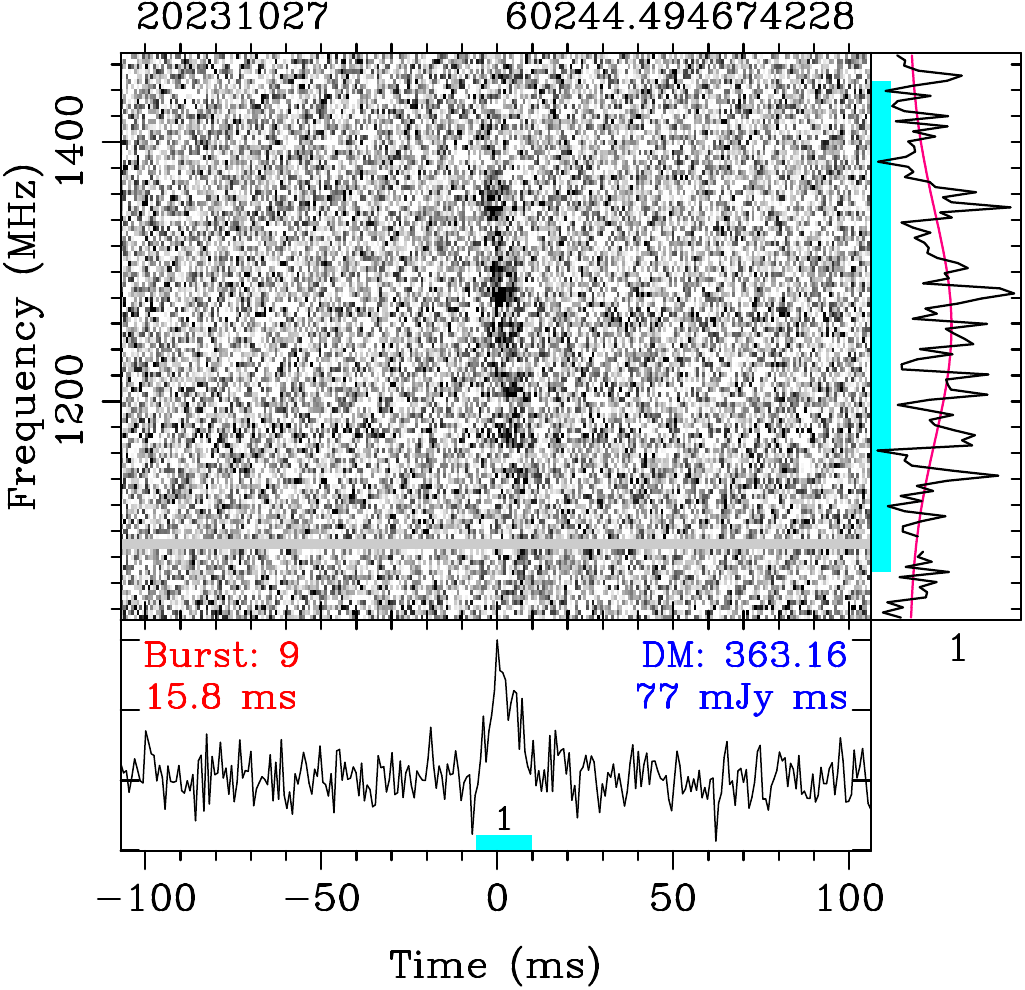}
\includegraphics[height=0.29\linewidth]{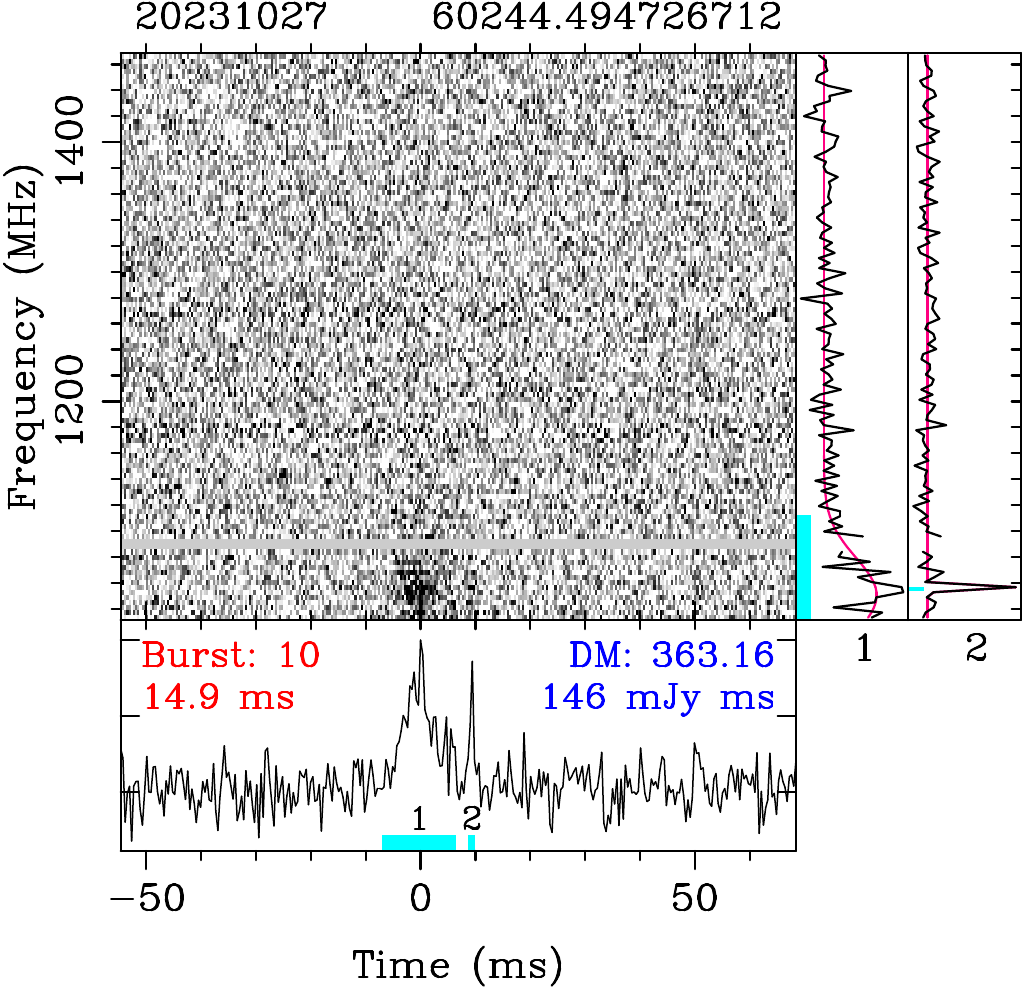}
\includegraphics[height=0.29\linewidth]{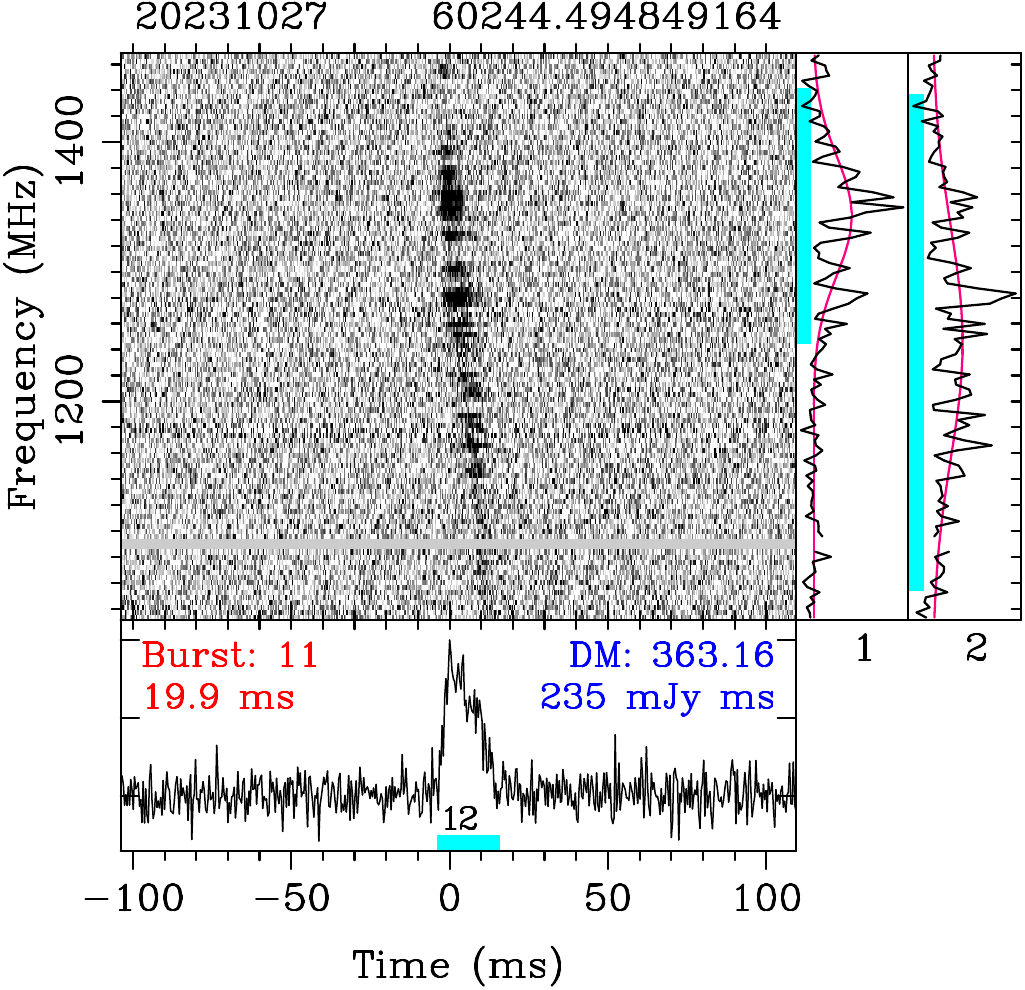}
\includegraphics[height=0.29\linewidth]{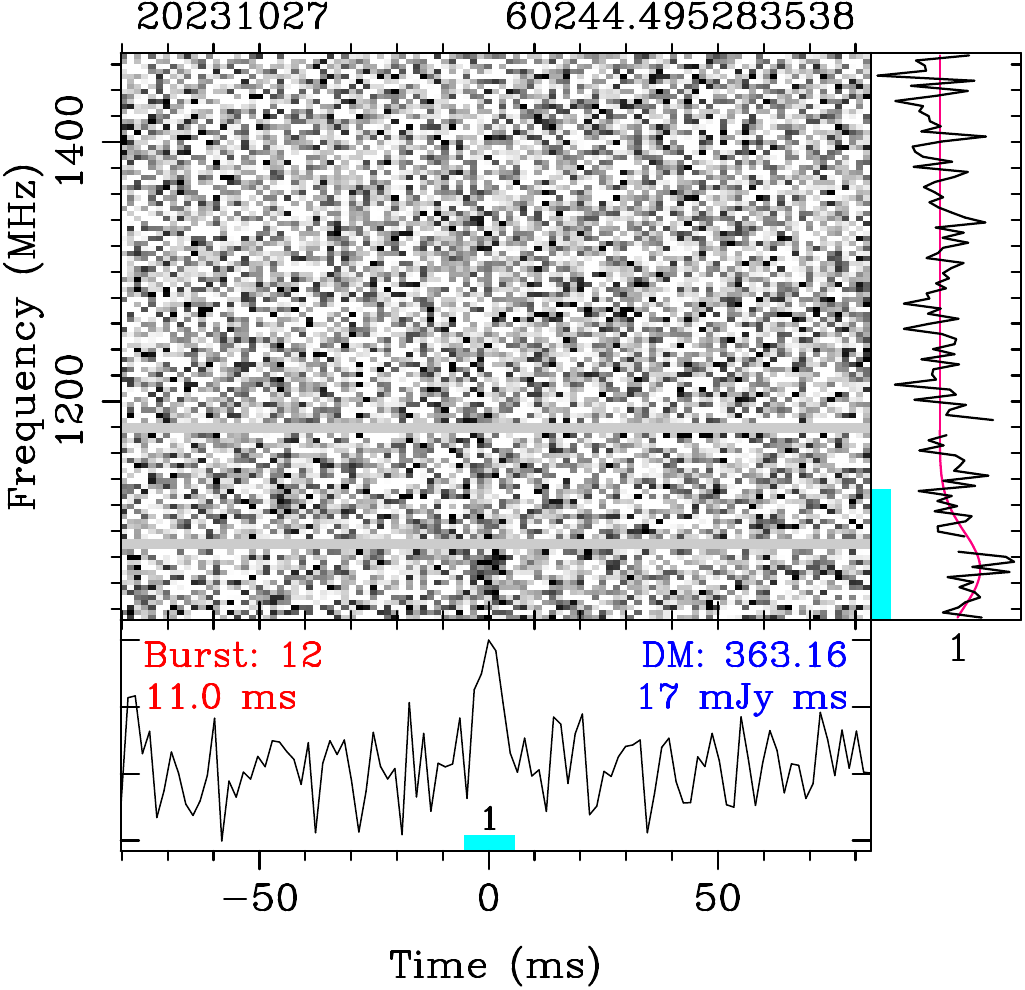}
\includegraphics[height=0.29\linewidth]{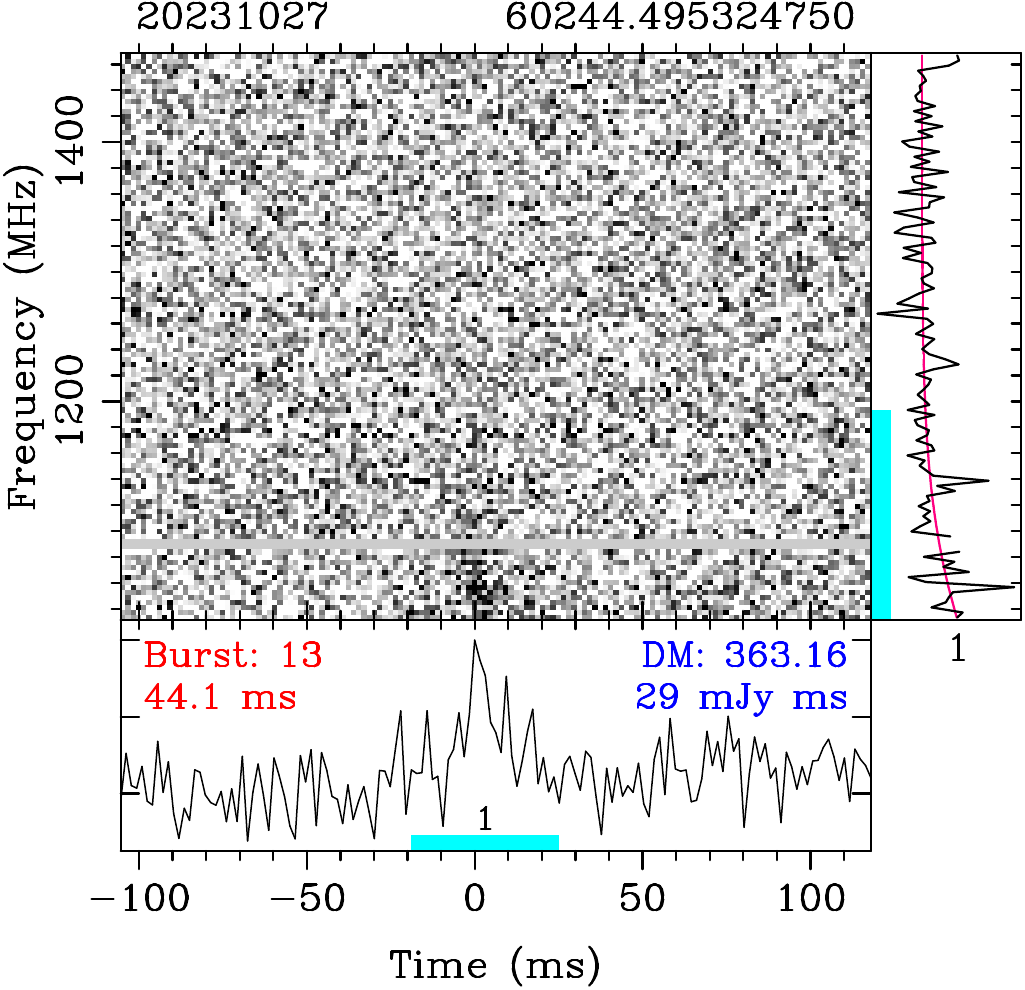}
\caption{({\textit{continued}})}
\end{figure*}
\addtocounter{figure}{-1}
\begin{figure*}
\flushleft
\includegraphics[height=0.29\linewidth]{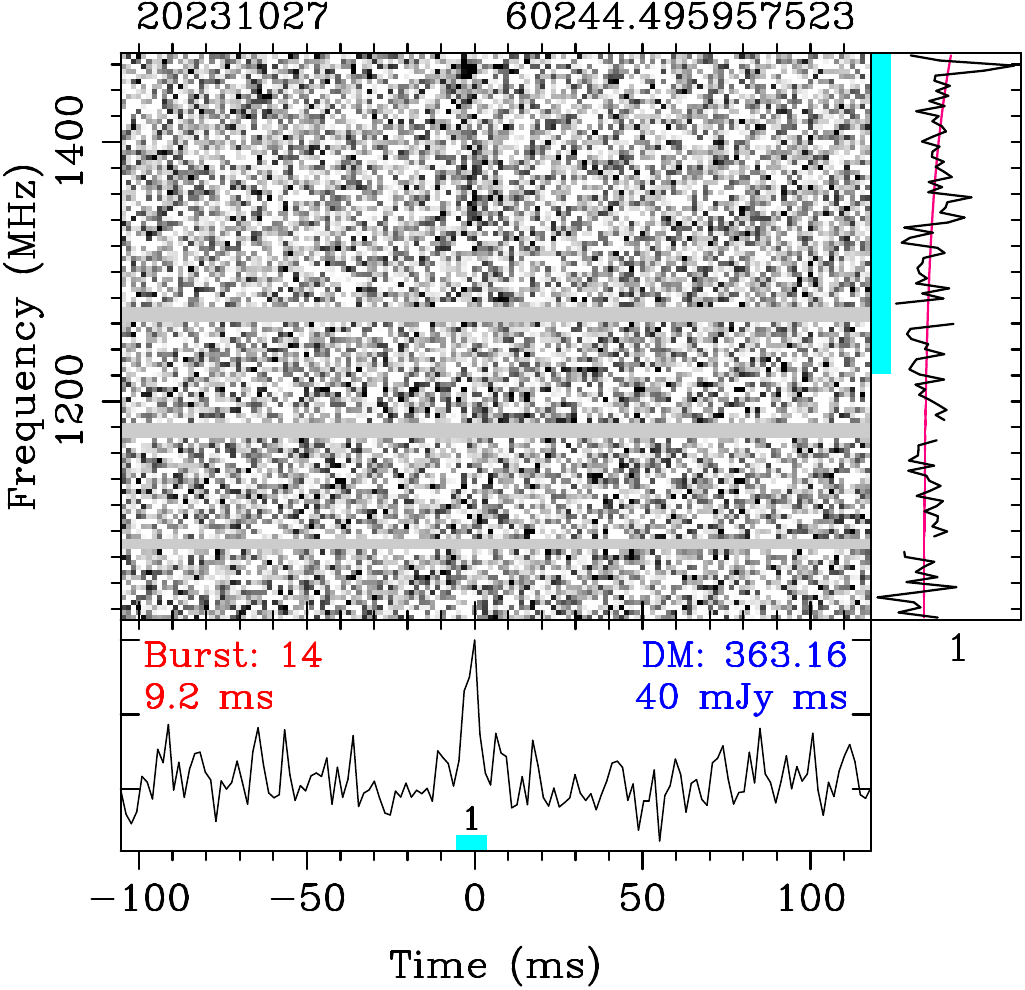}
\includegraphics[height=0.29\linewidth]{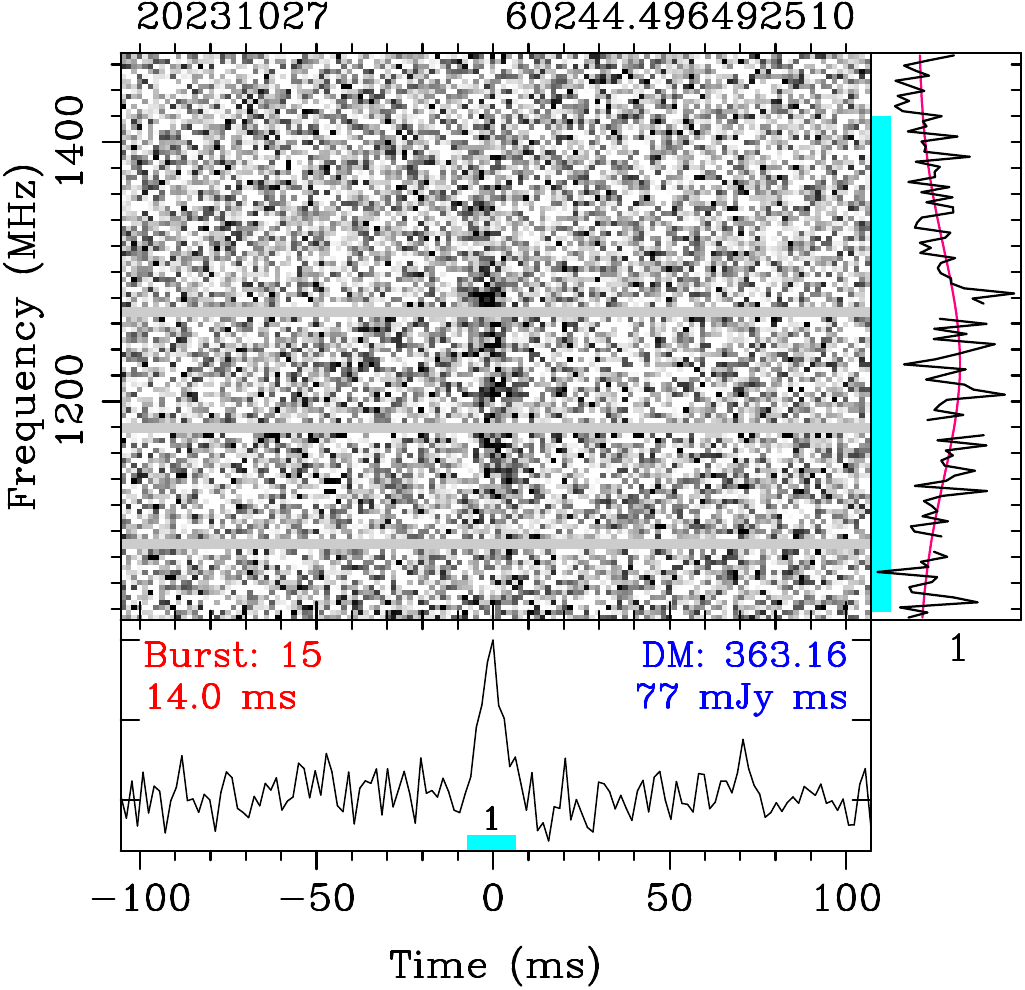}
\includegraphics[height=0.29\linewidth]{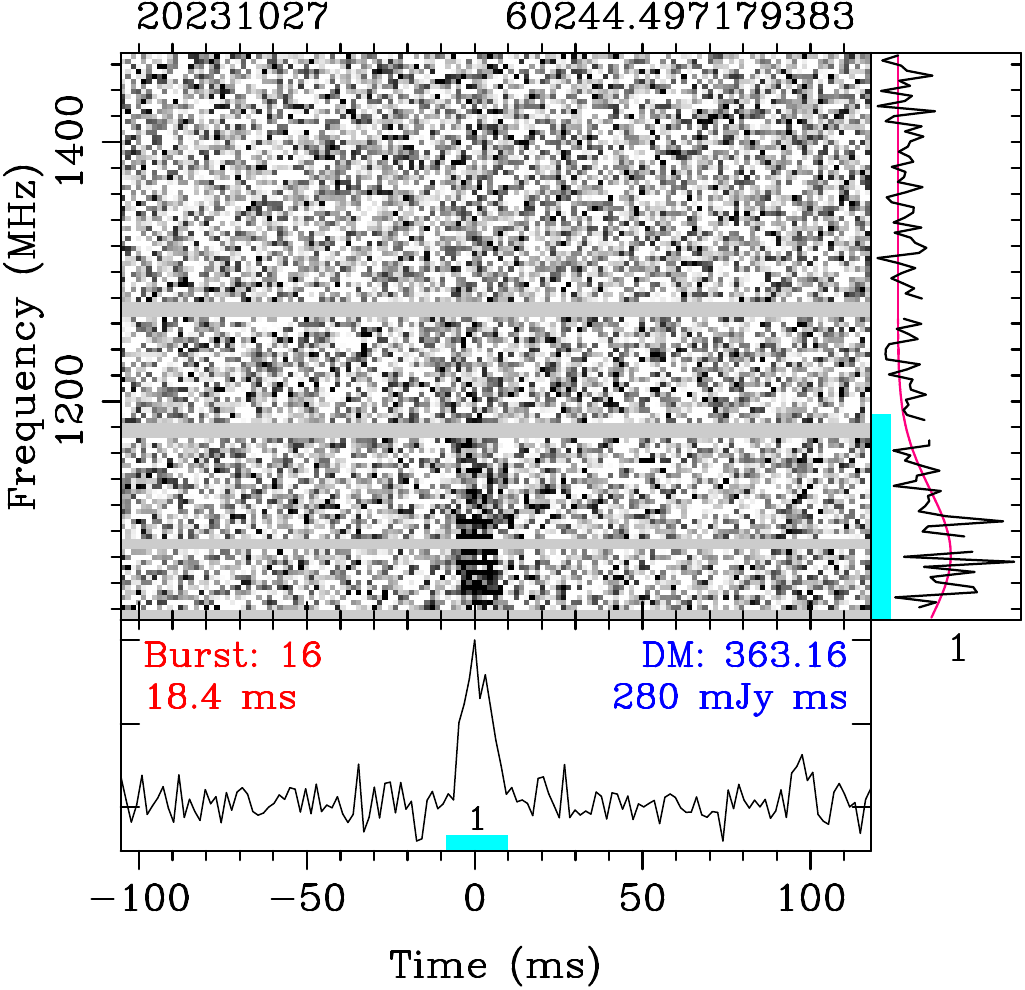}
\includegraphics[height=0.29\linewidth]{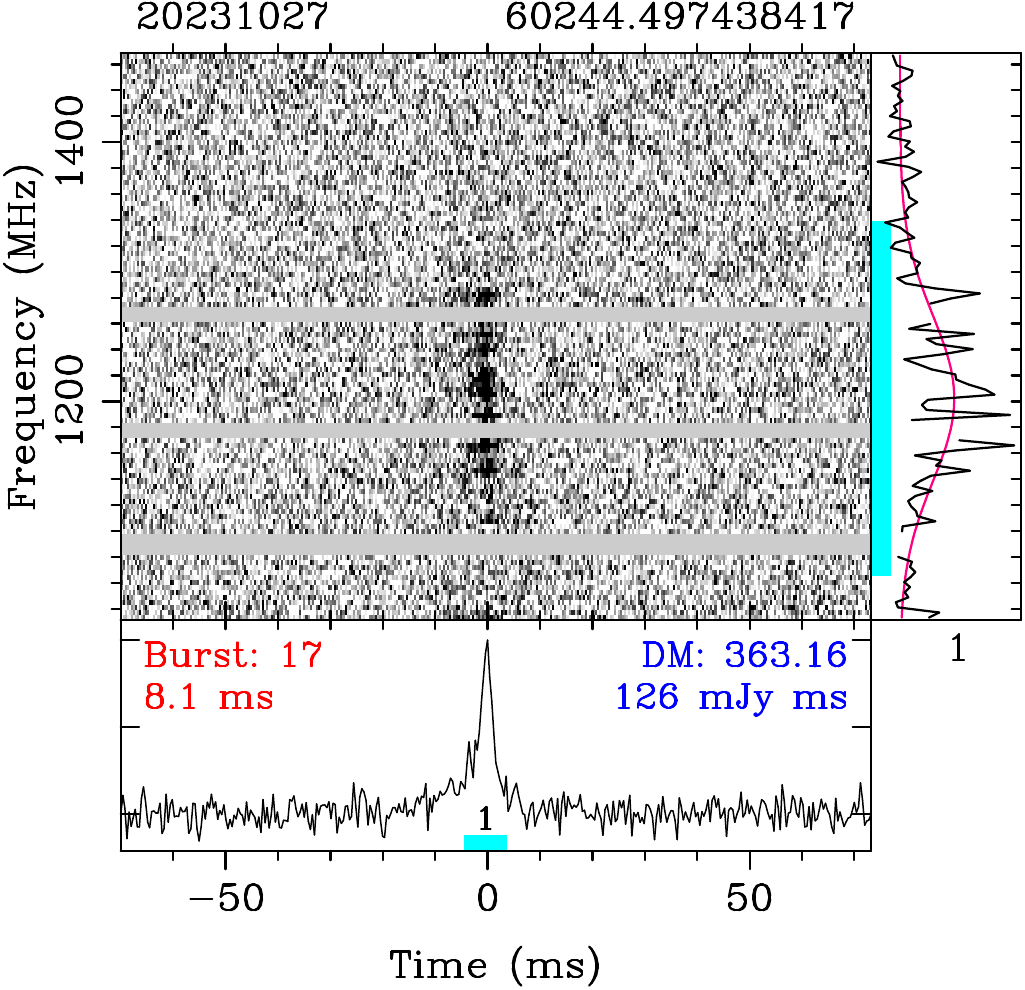}
\includegraphics[height=0.29\linewidth]{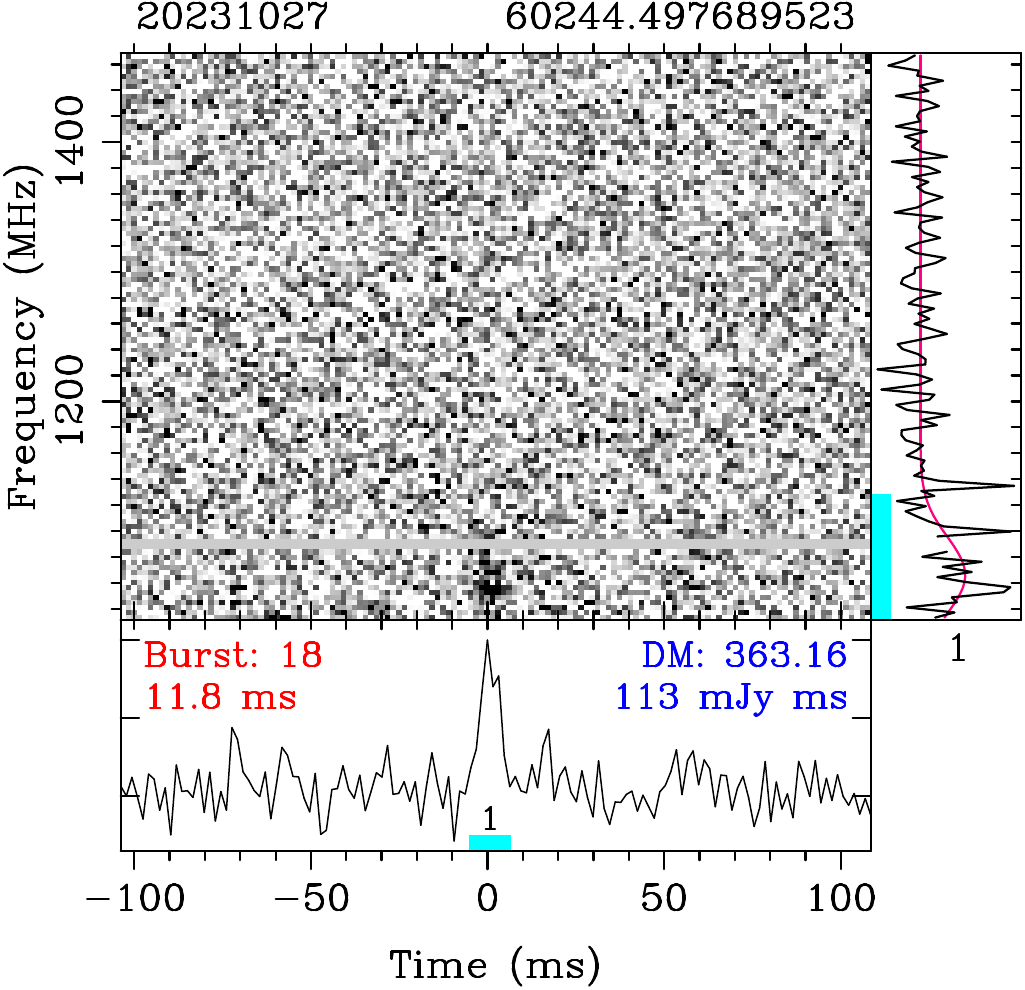}
\includegraphics[height=0.29\linewidth]{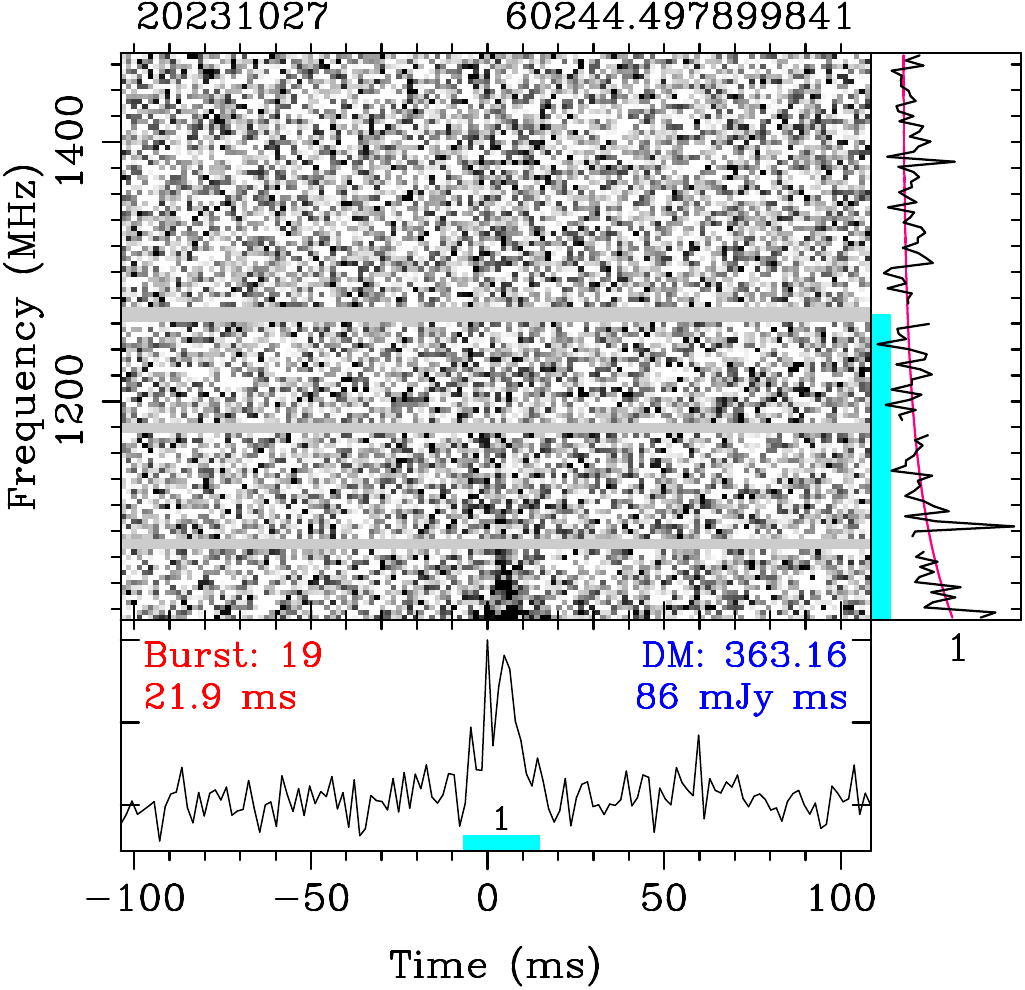}
\includegraphics[height=0.29\linewidth]{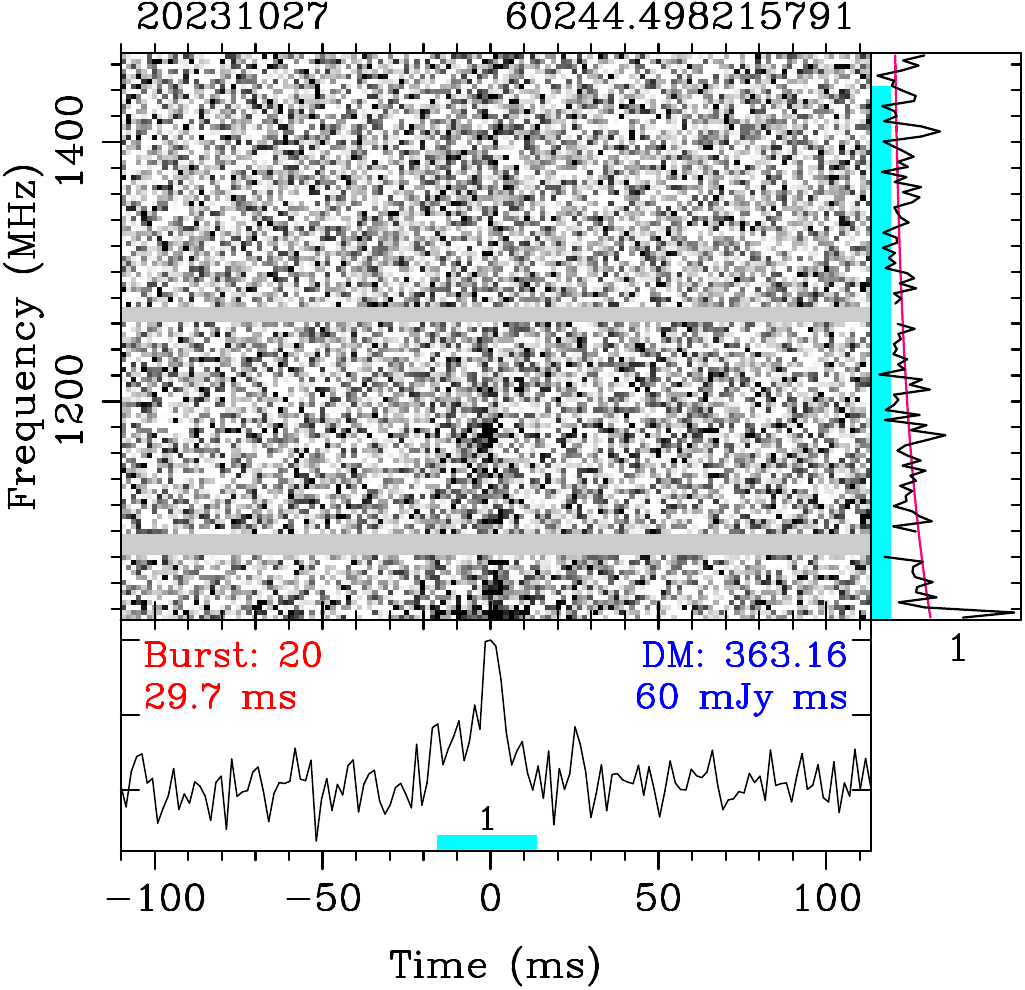}
\includegraphics[height=0.29\linewidth]{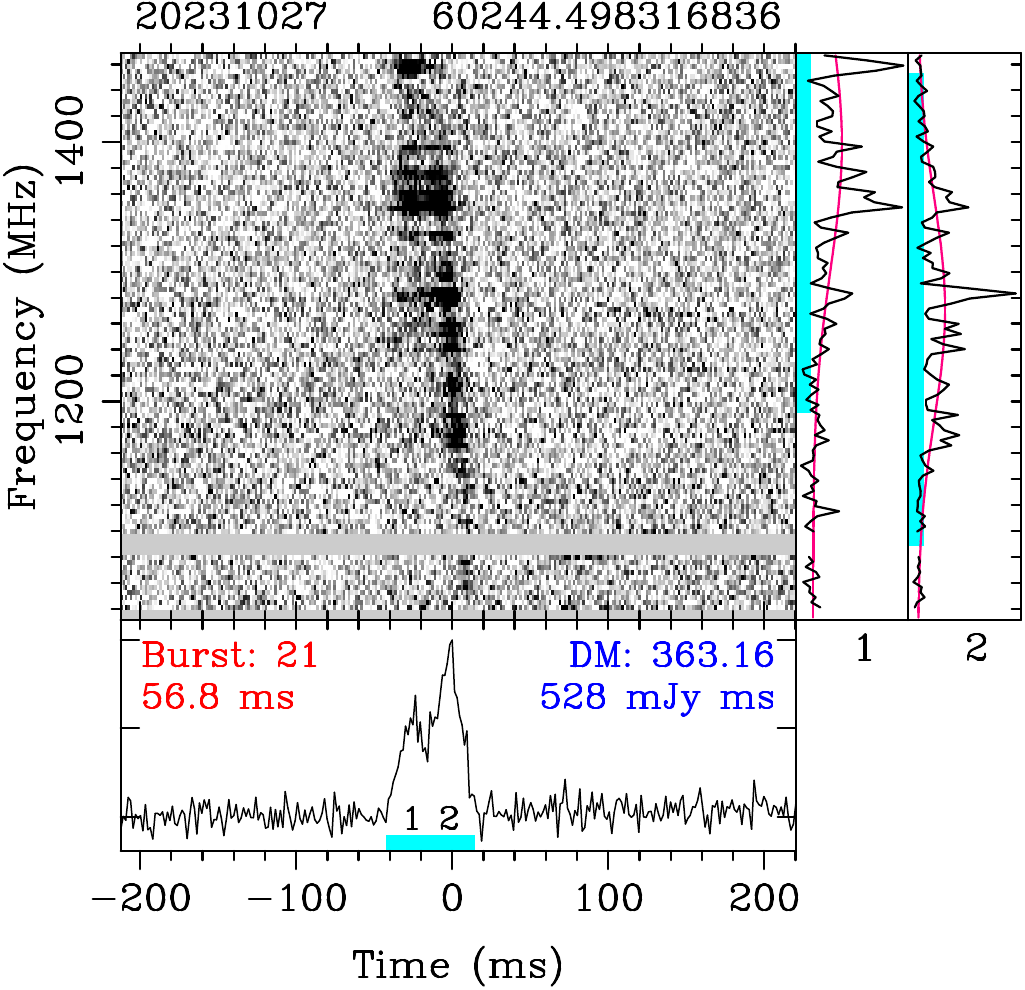}
\includegraphics[height=0.29\linewidth]{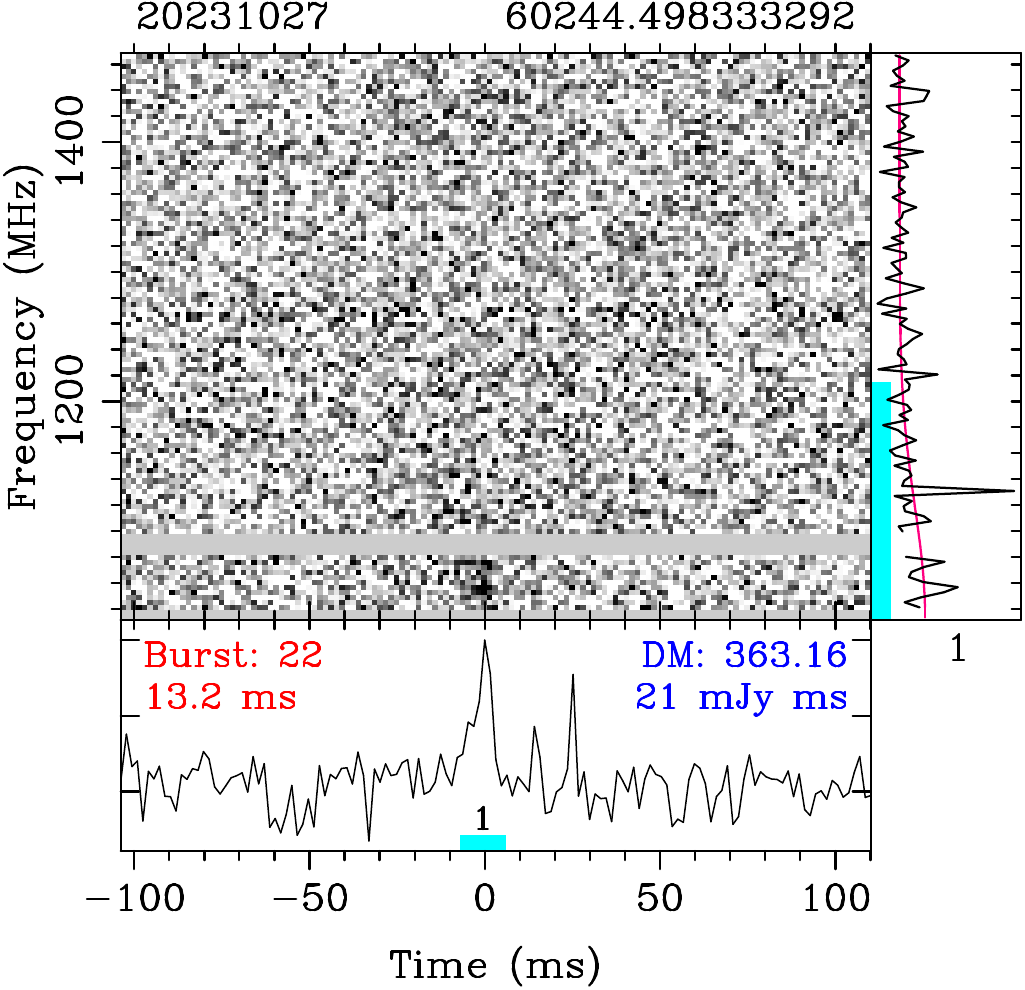}
\includegraphics[height=0.29\linewidth]{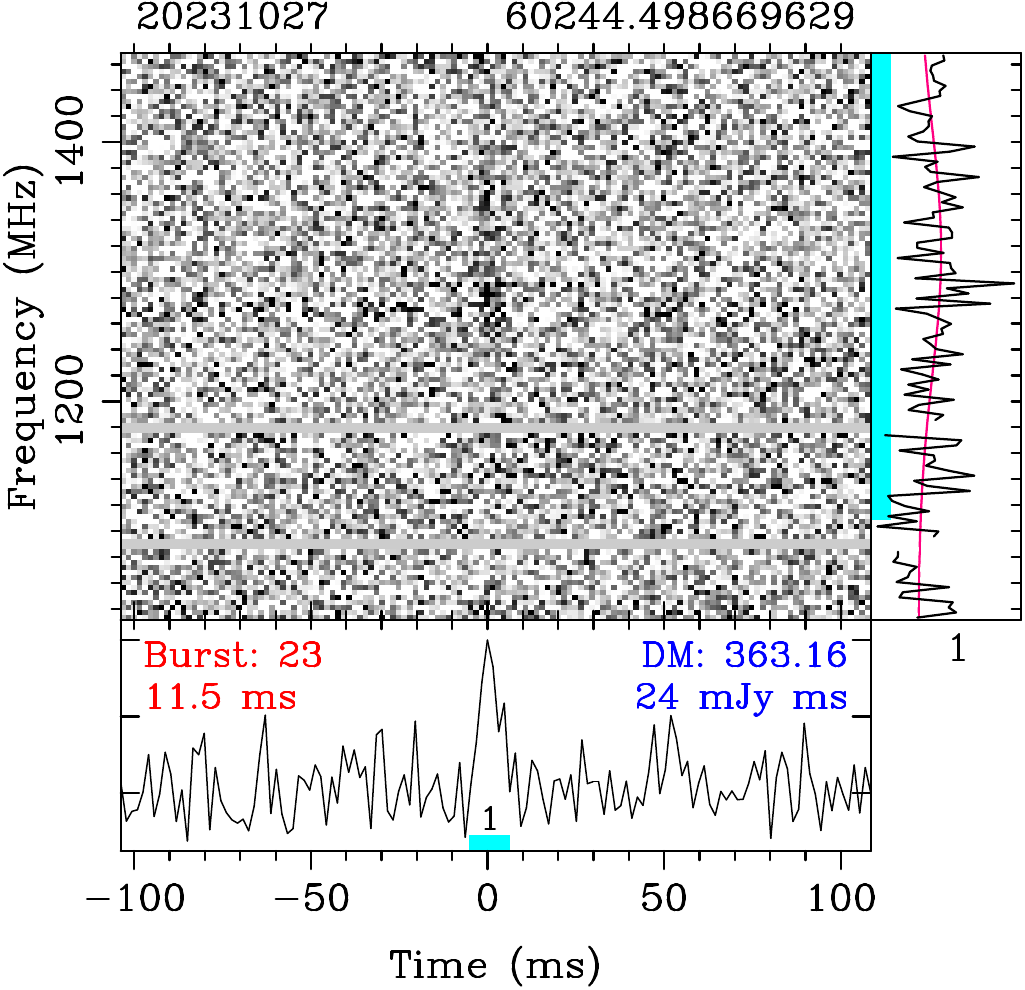}
\includegraphics[height=0.29\linewidth]{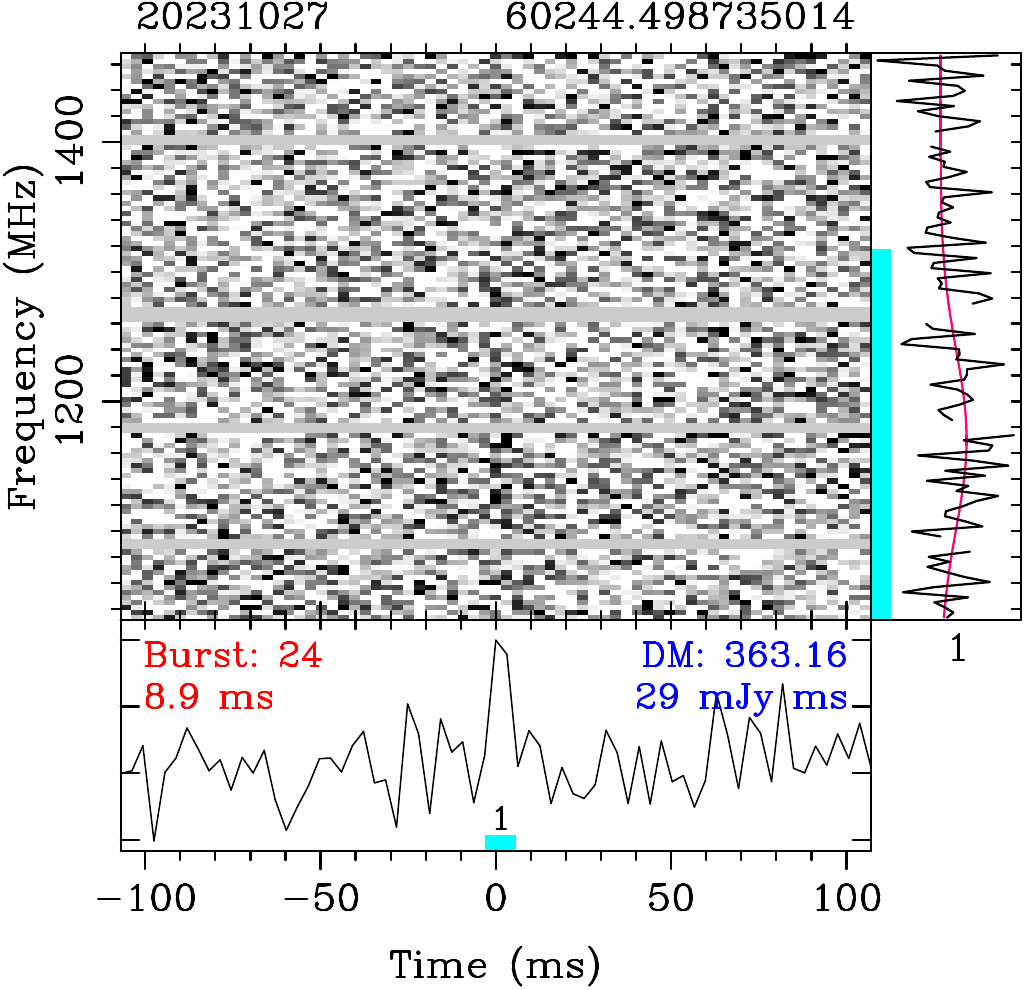}
\includegraphics[height=0.29\linewidth]{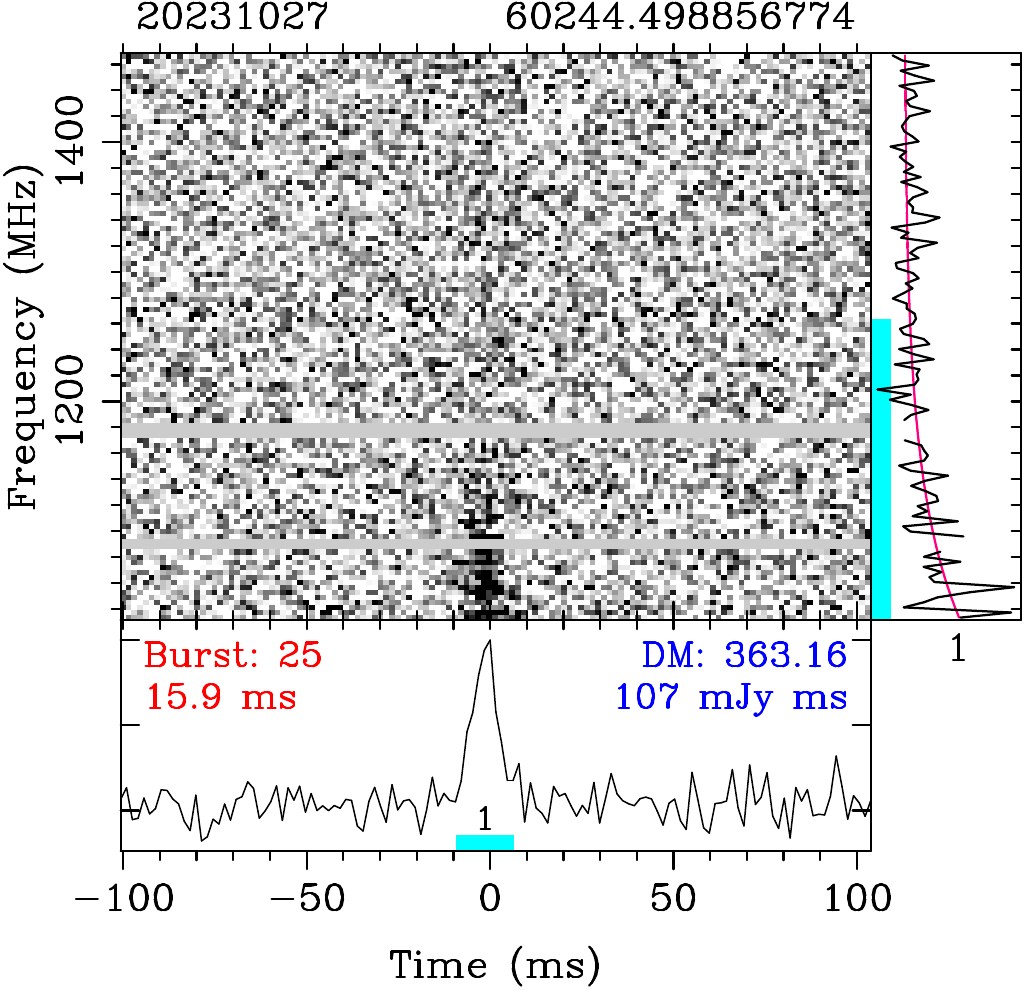}
\caption{({\textit{continued}})}
\end{figure*}
\addtocounter{figure}{-1}
\begin{figure*}
\flushleft
\includegraphics[height=0.29\linewidth]{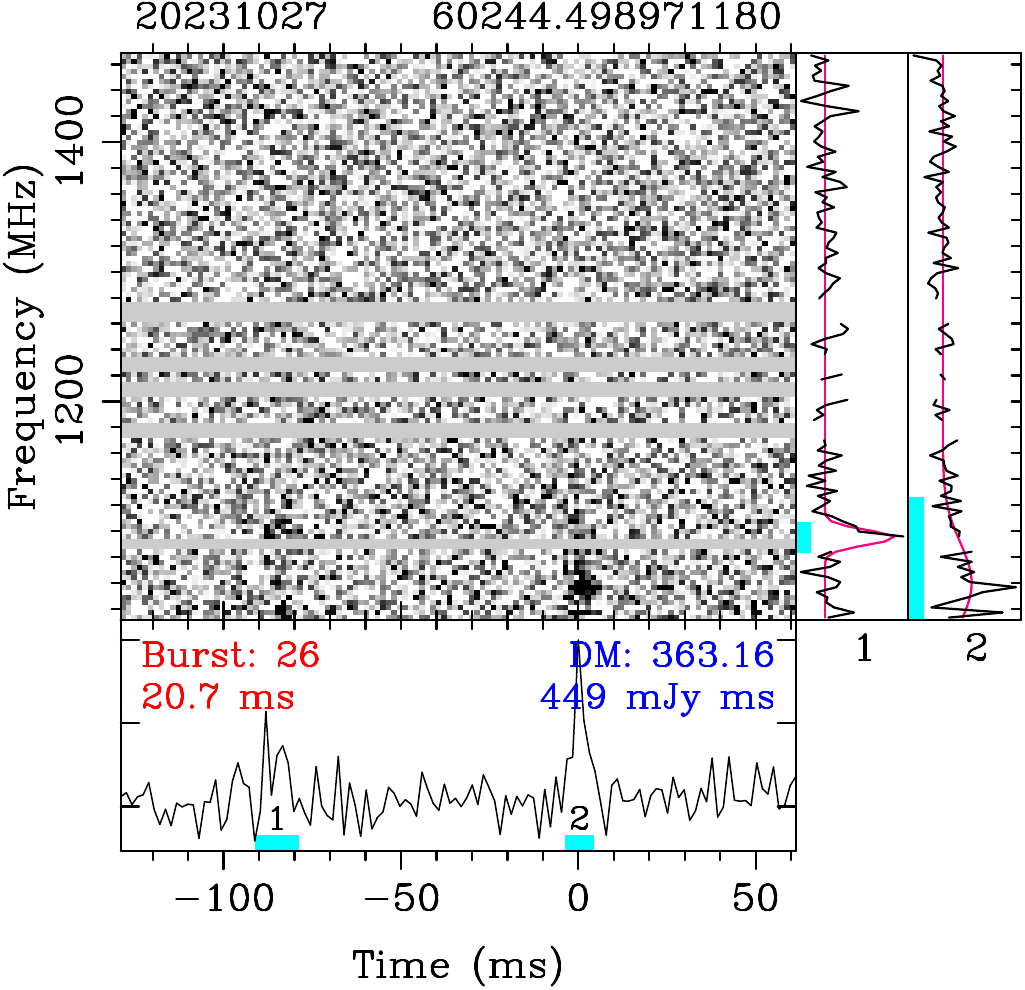}
\includegraphics[height=0.29\linewidth]{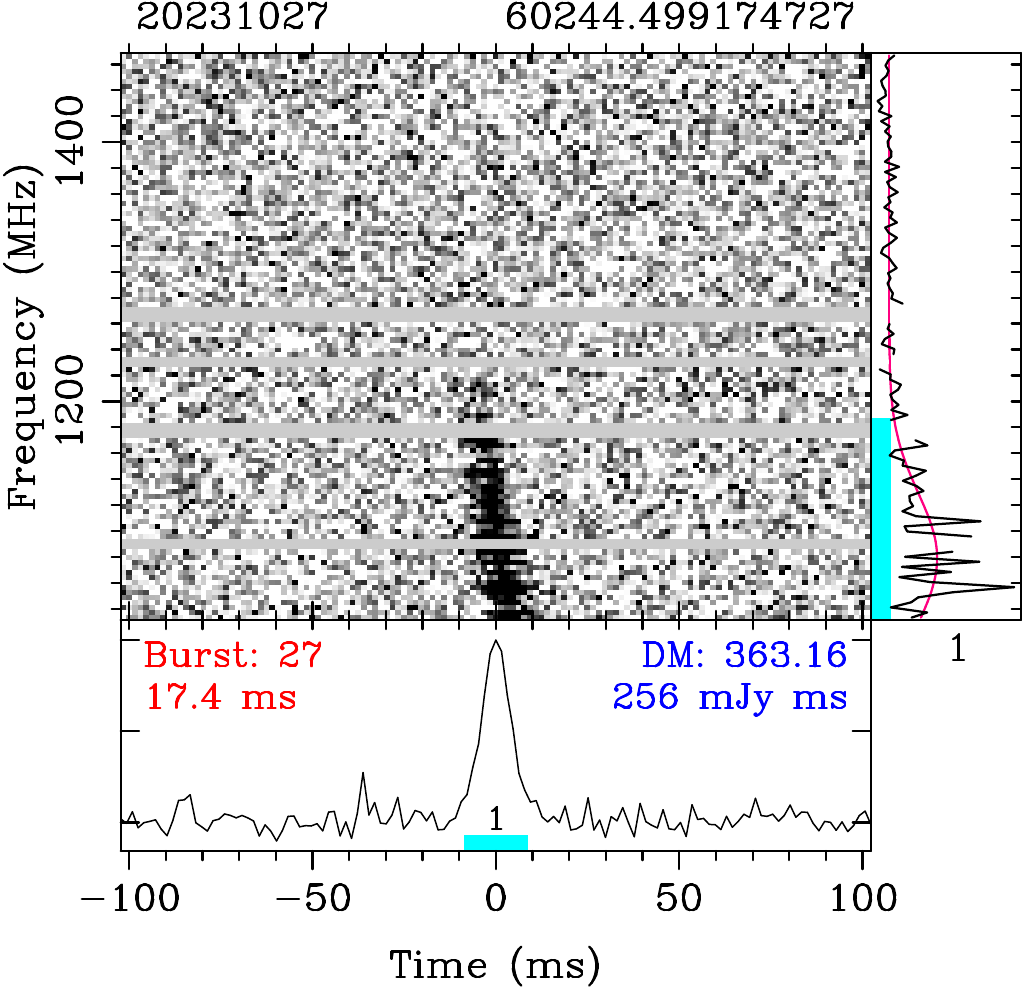}
\includegraphics[height=0.29\linewidth]{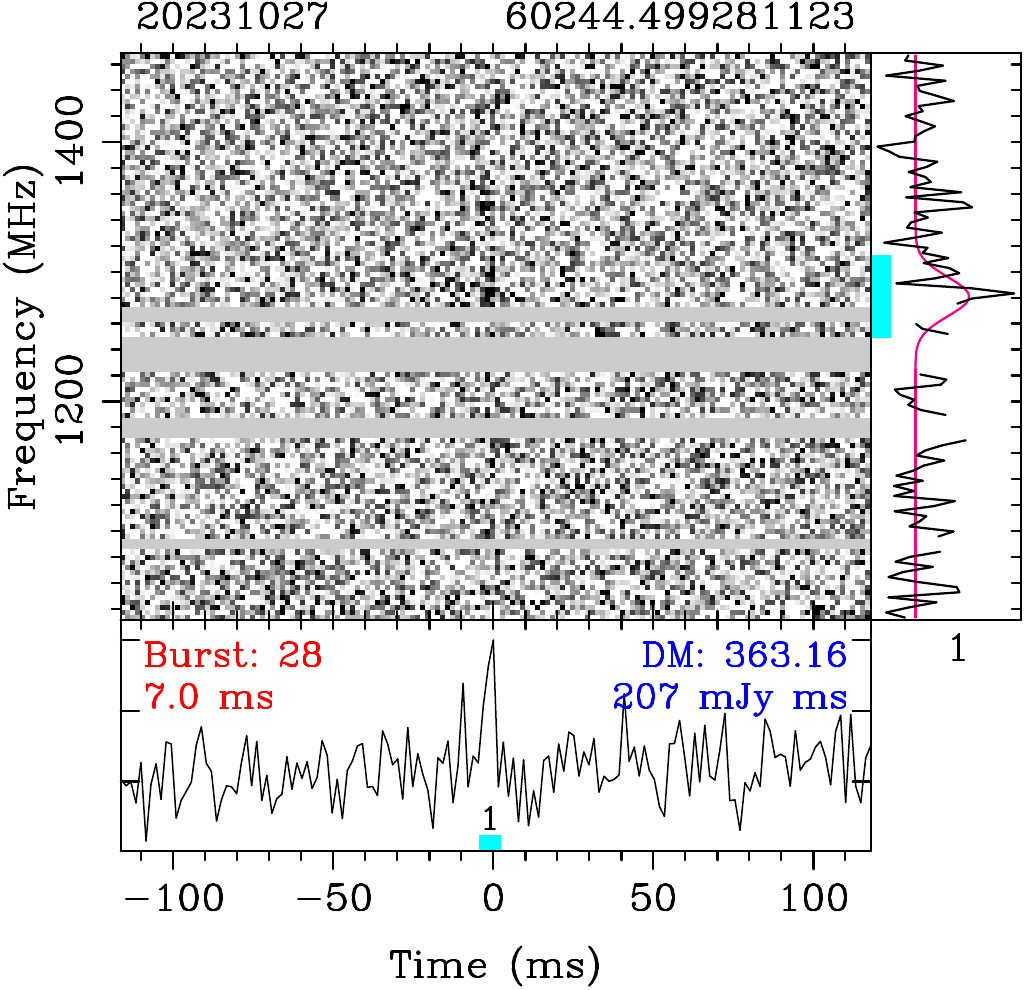}
\includegraphics[height=0.29\linewidth]{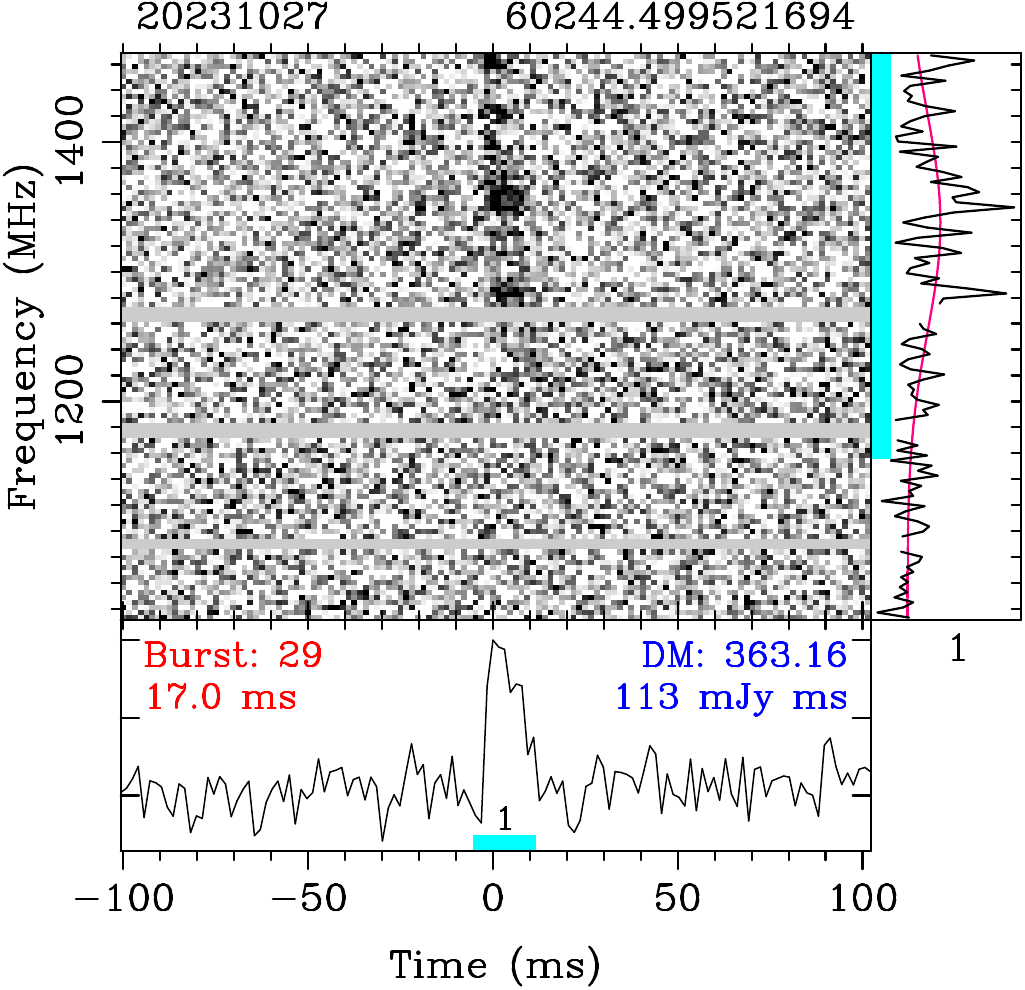}
\includegraphics[height=0.29\linewidth]{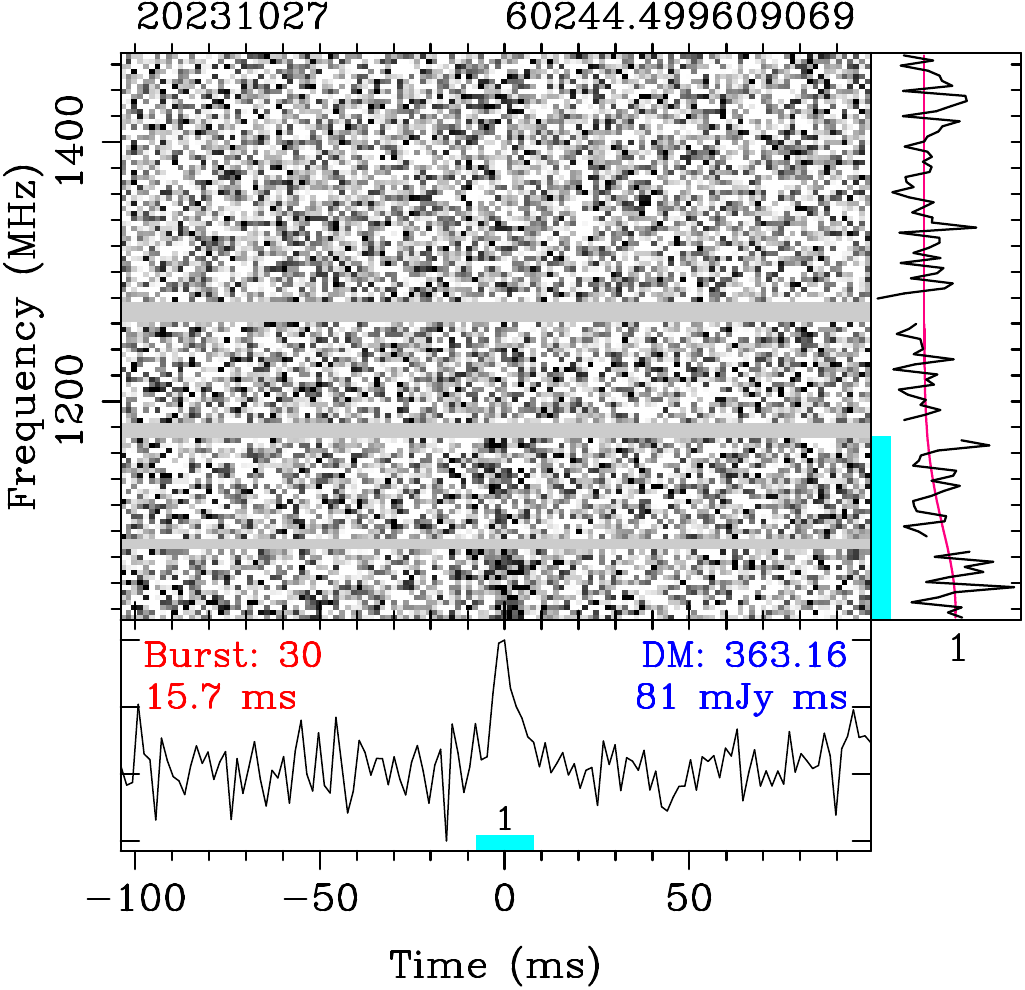}
\includegraphics[height=0.29\linewidth]{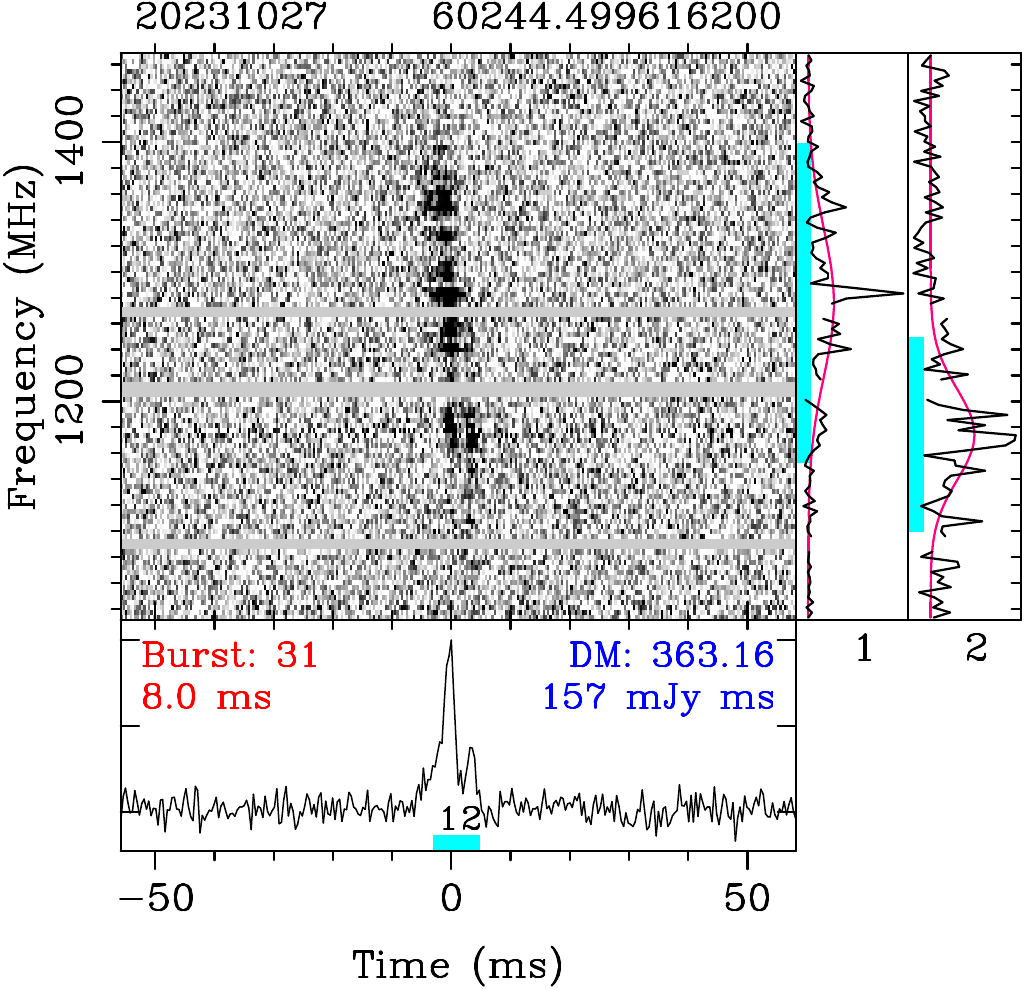}
\includegraphics[height=0.29\linewidth]{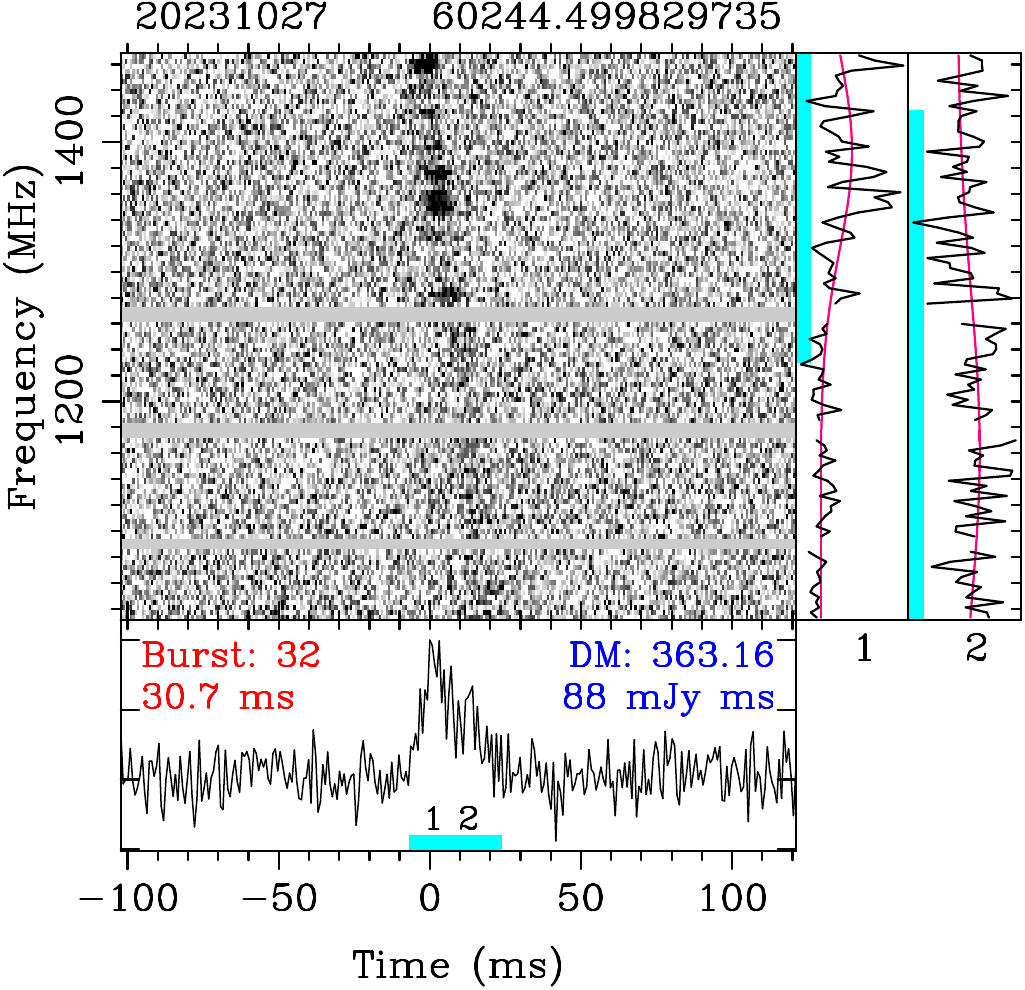}
\includegraphics[height=0.29\linewidth]{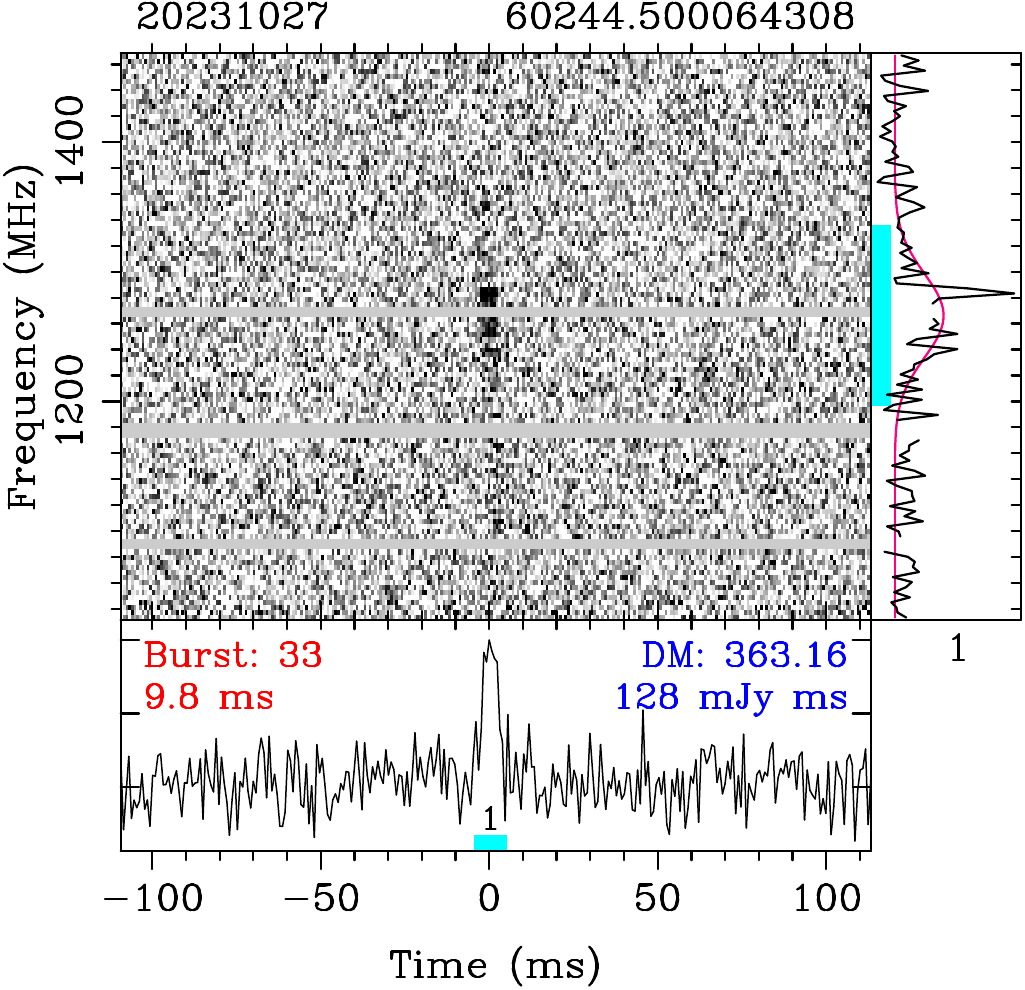}
\includegraphics[height=0.29\linewidth]{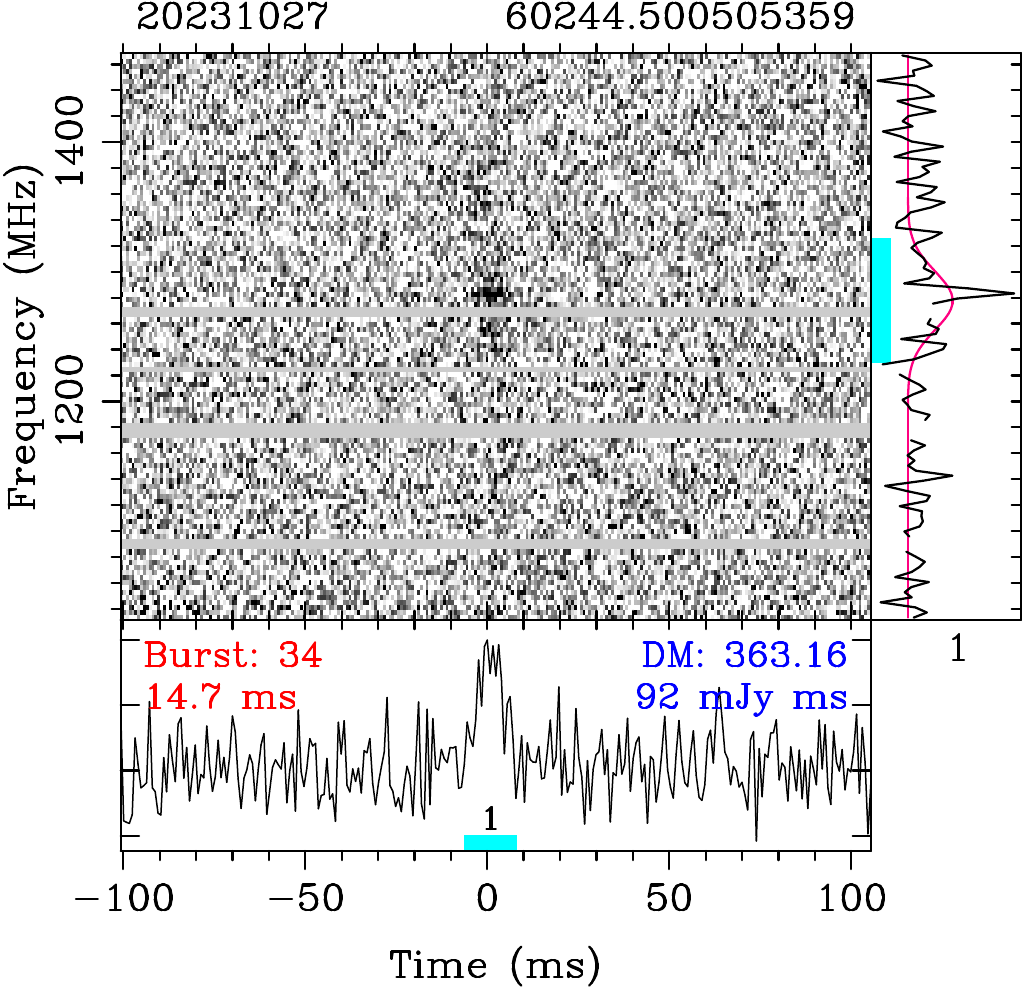}
\includegraphics[height=0.29\linewidth]{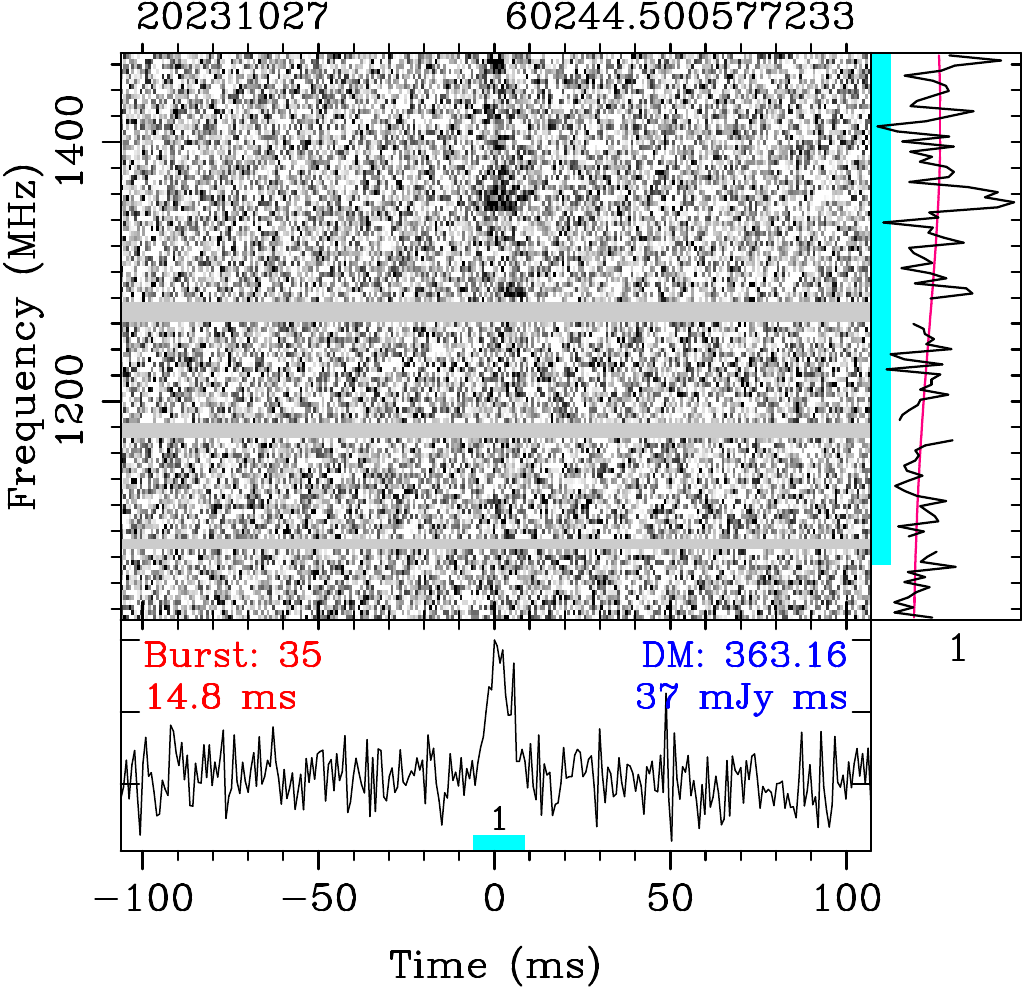}
\includegraphics[height=0.29\linewidth]{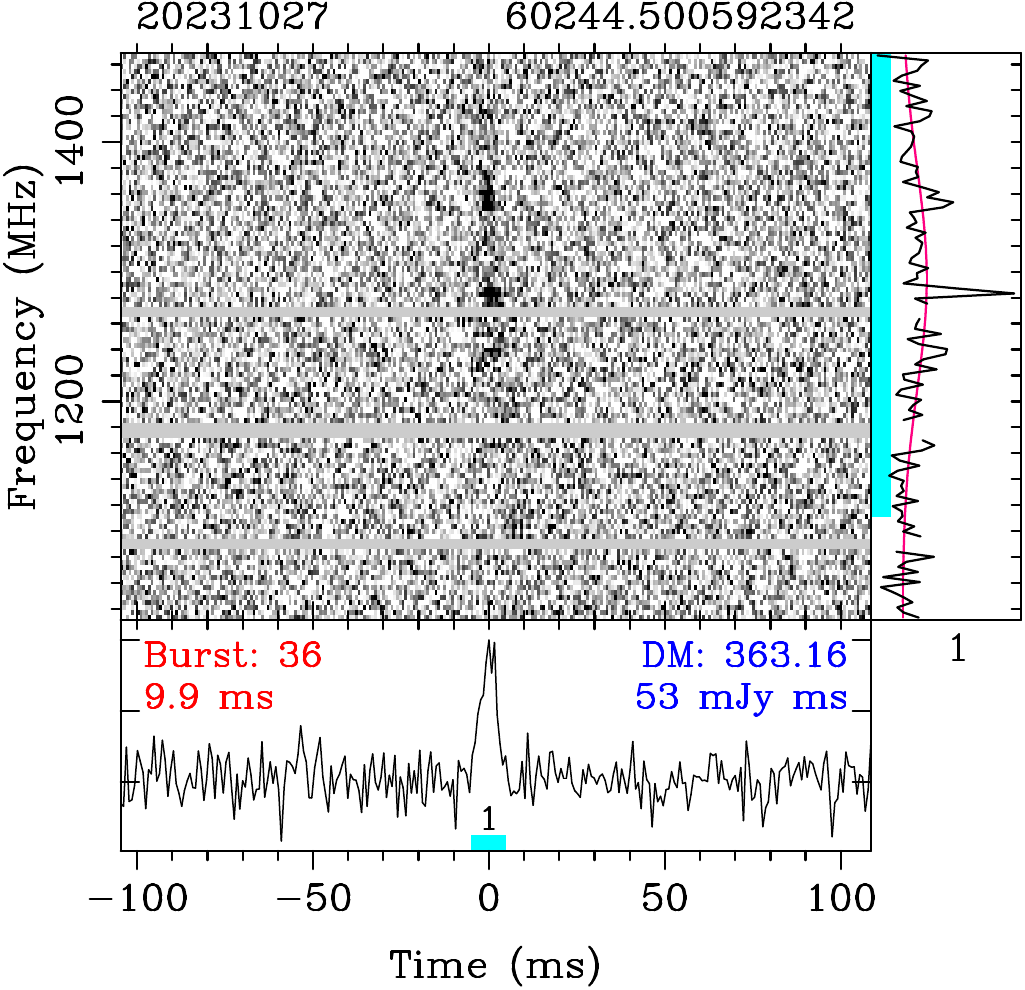}
\includegraphics[height=0.29\linewidth]{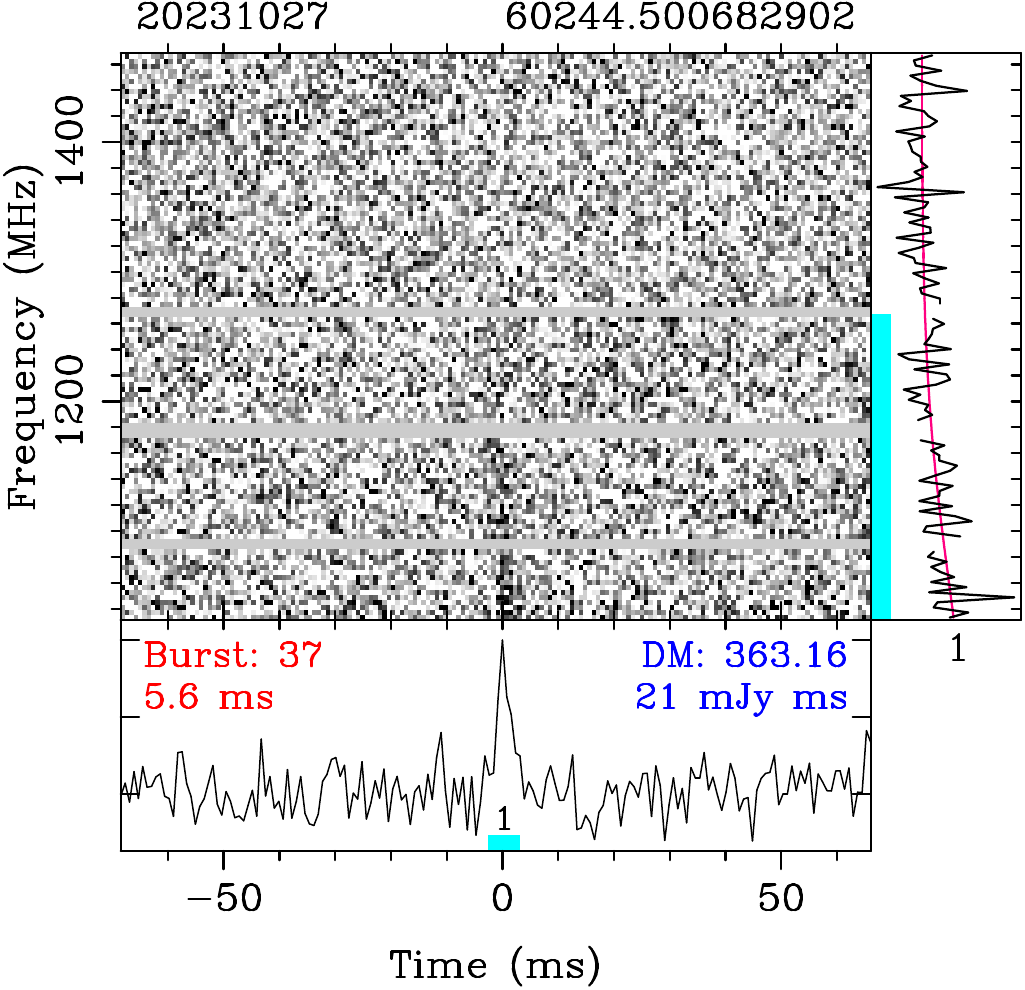}
\caption{({\textit{continued}})}
\end{figure*}
\addtocounter{figure}{-1}
\begin{figure*}
\flushleft
\includegraphics[height=0.29\linewidth]{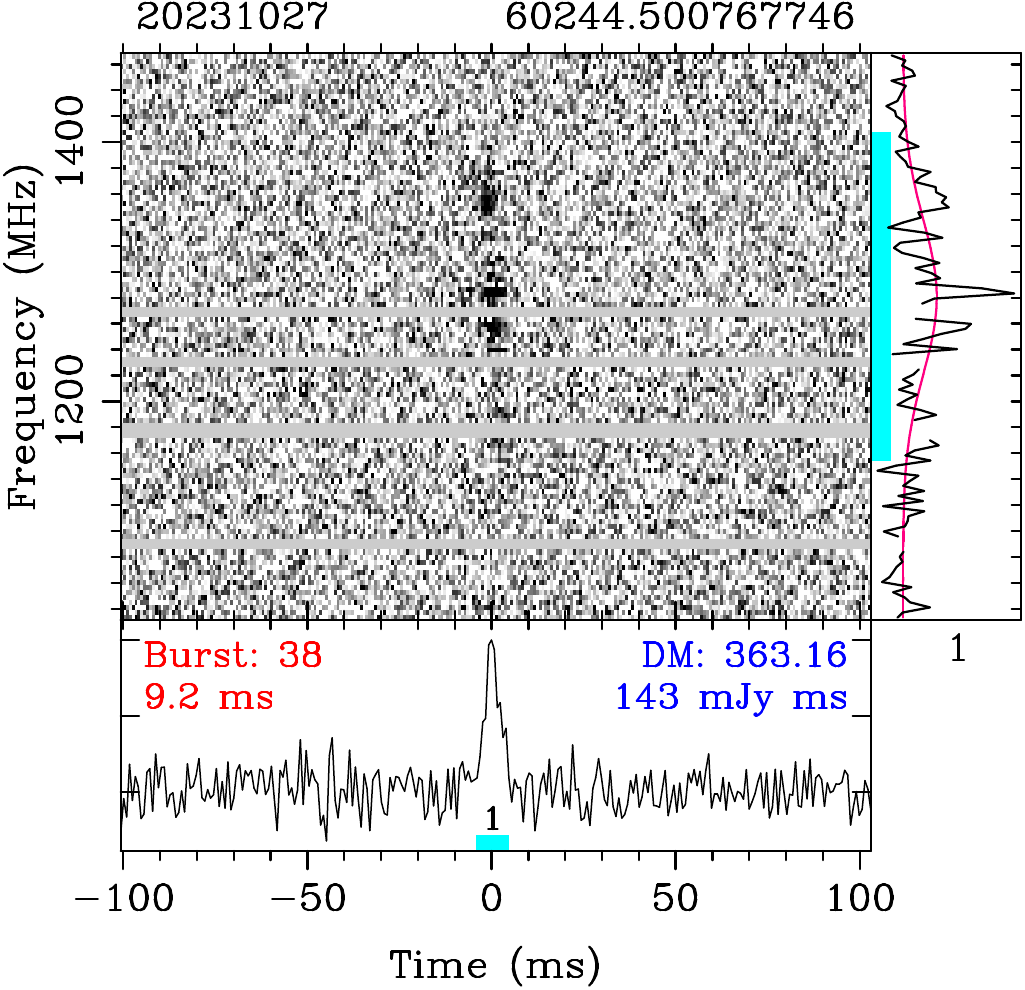}
\includegraphics[height=0.29\linewidth]{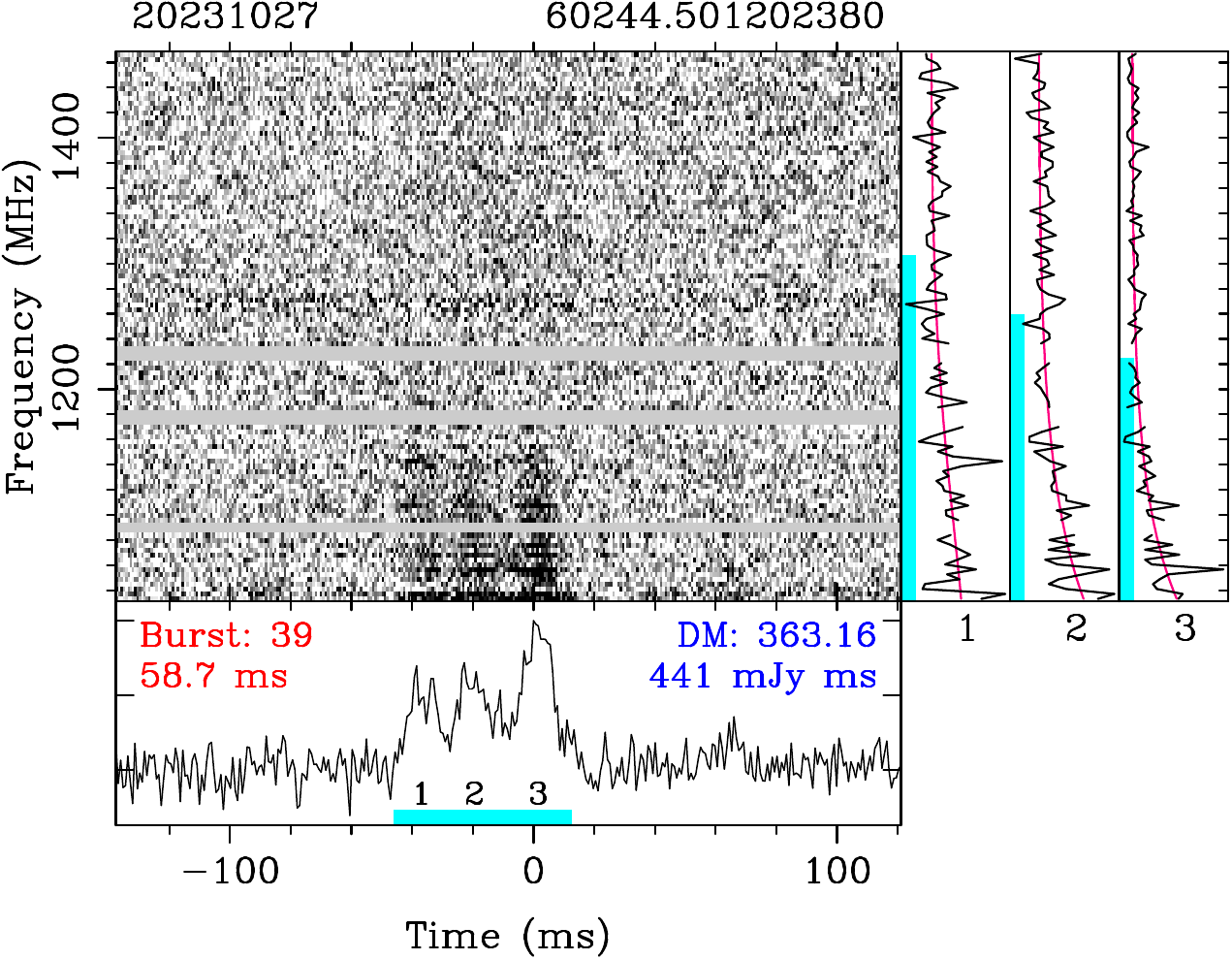}
\includegraphics[height=0.29\linewidth]{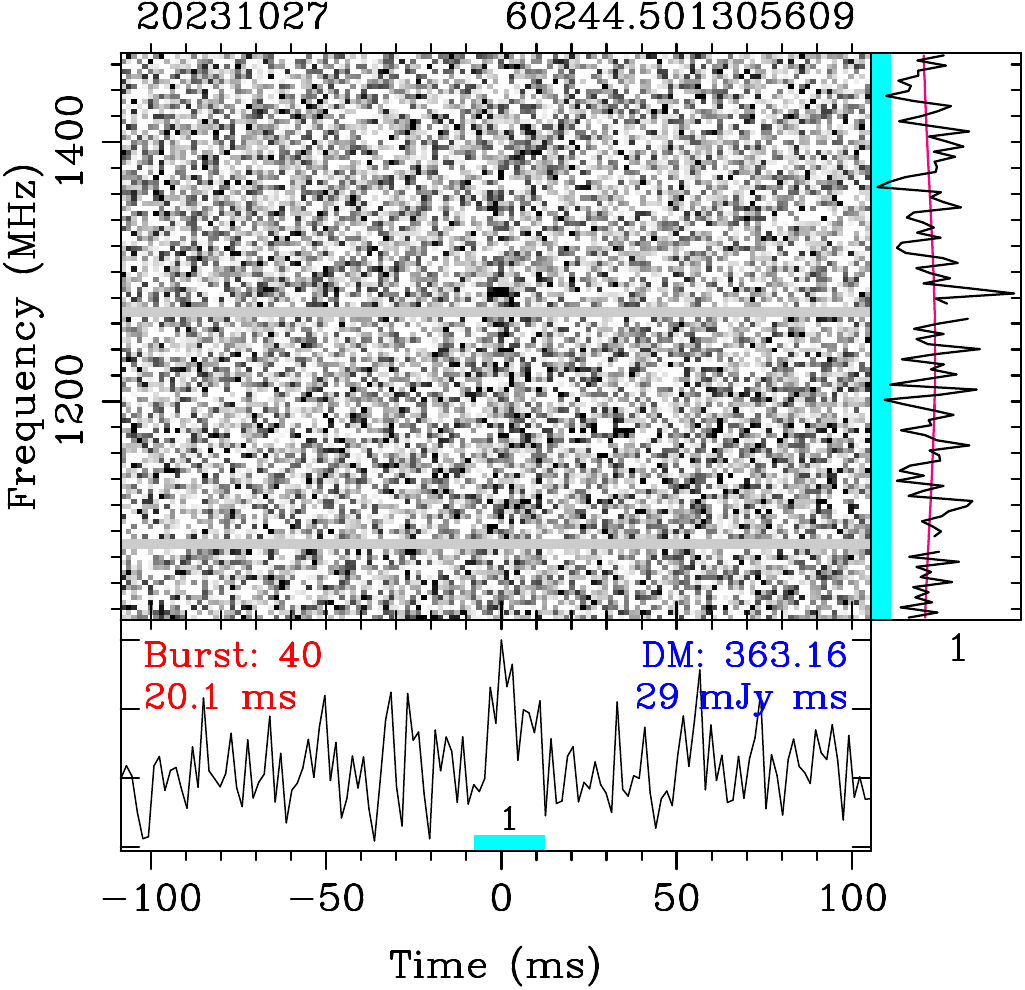}
\includegraphics[height=0.29\linewidth]{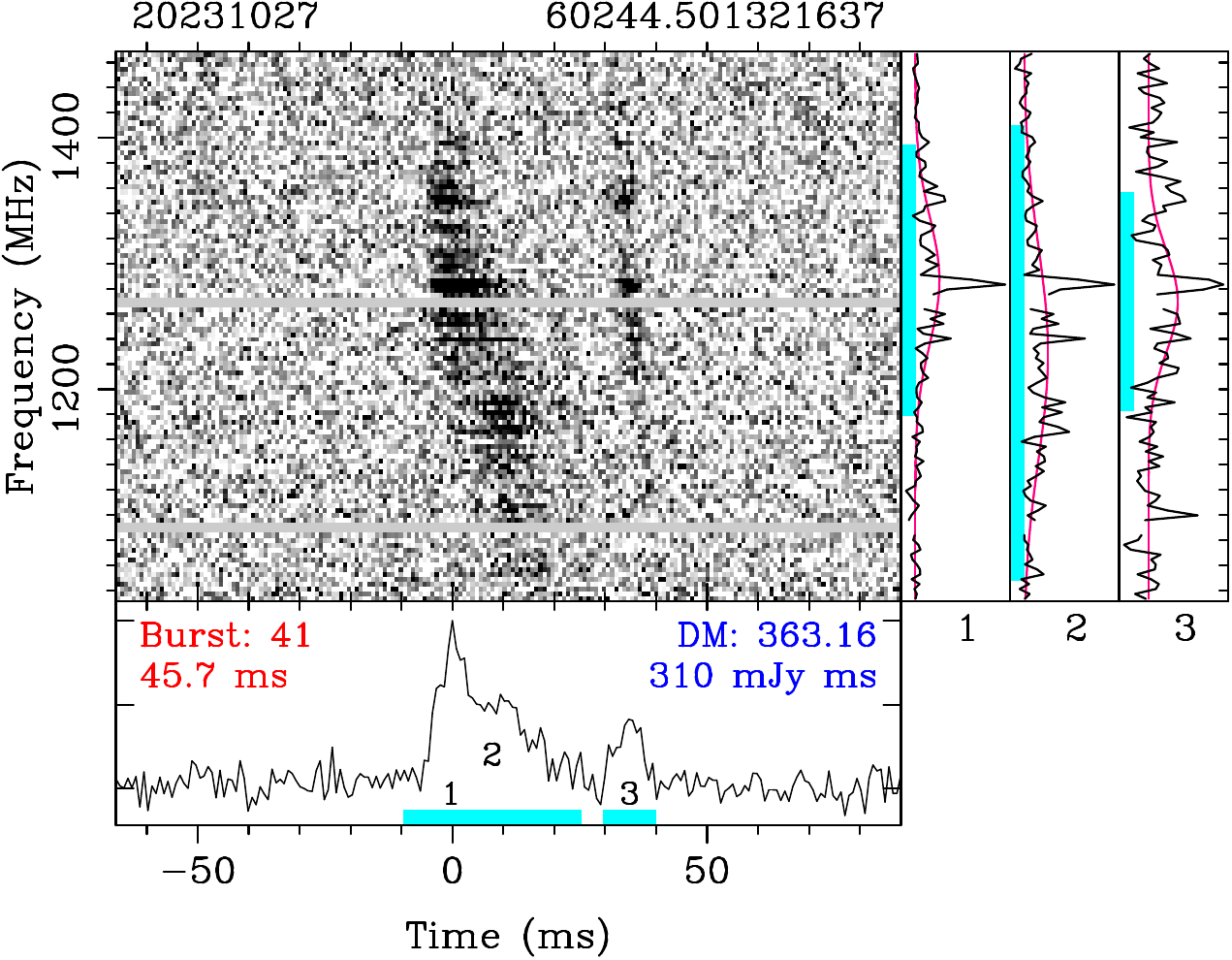}
\includegraphics[height=0.29\linewidth]{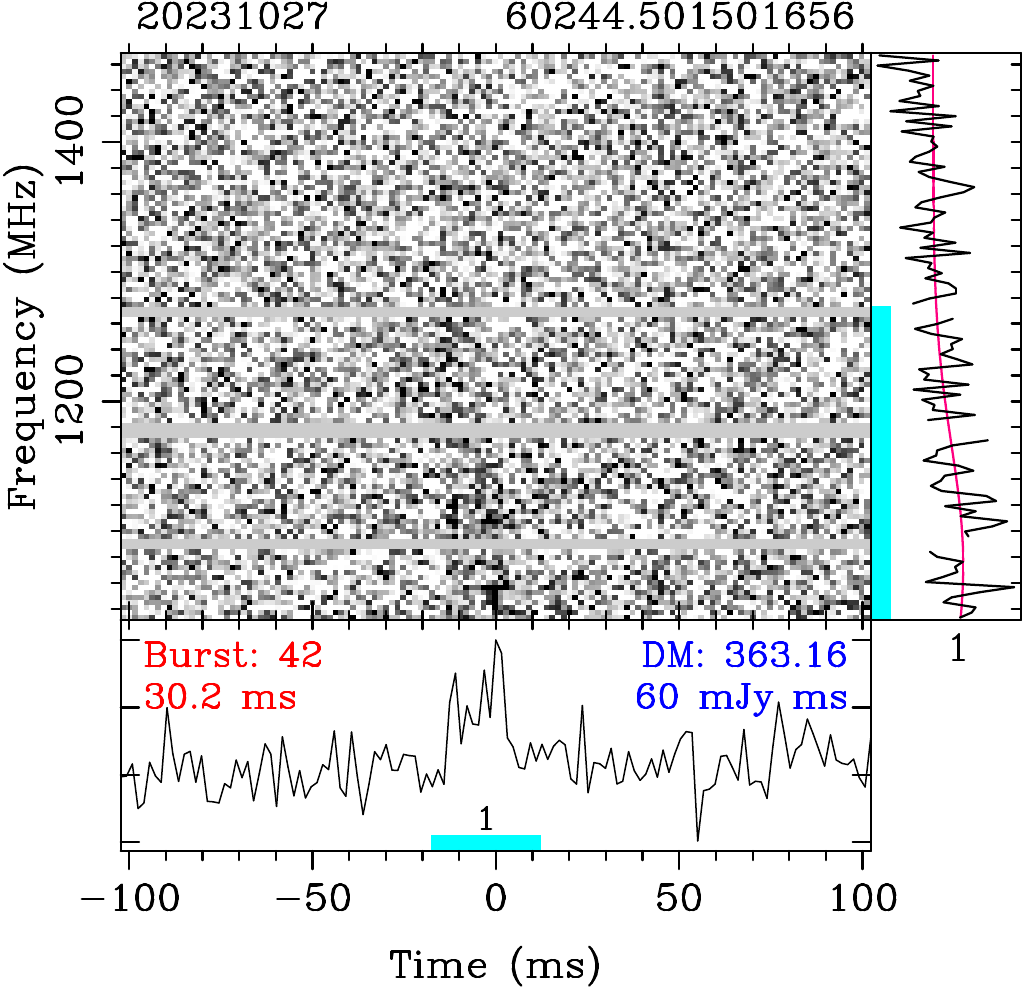}
\includegraphics[height=0.29\linewidth]{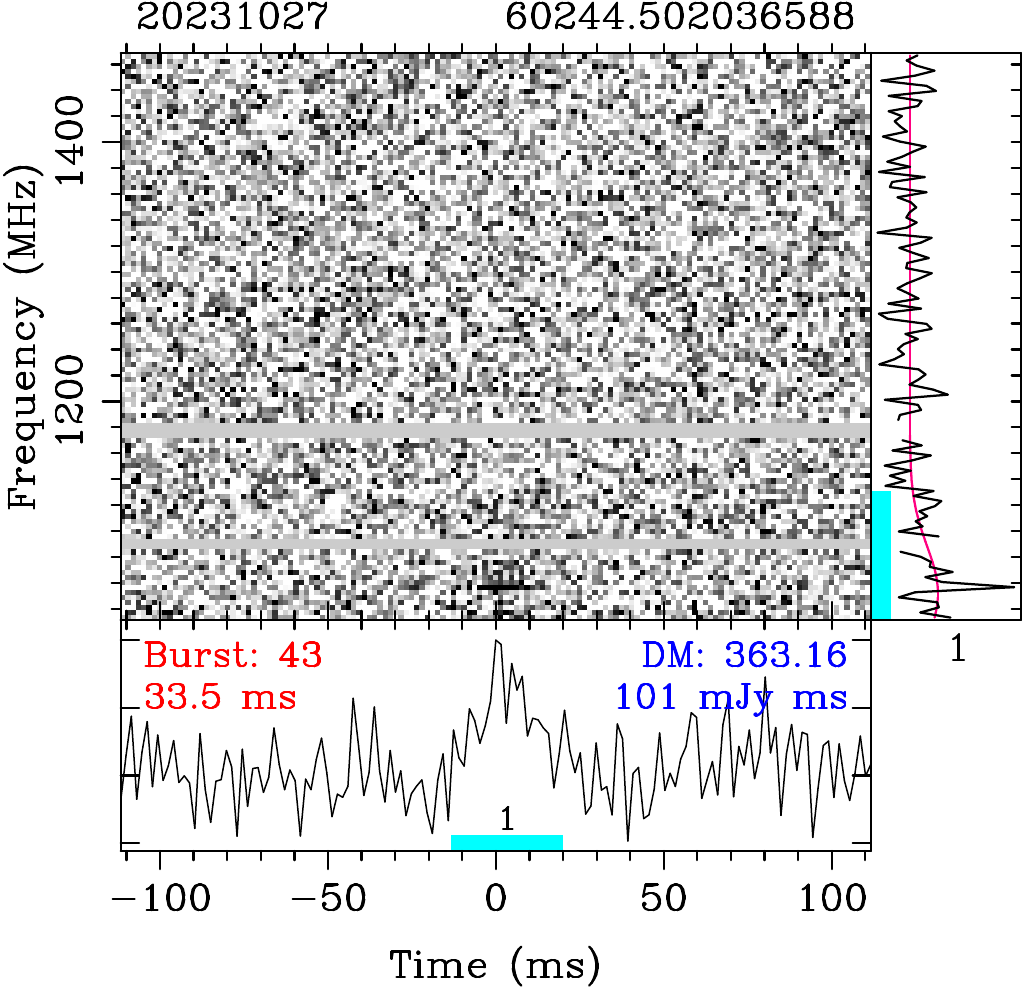}
\includegraphics[height=0.29\linewidth]{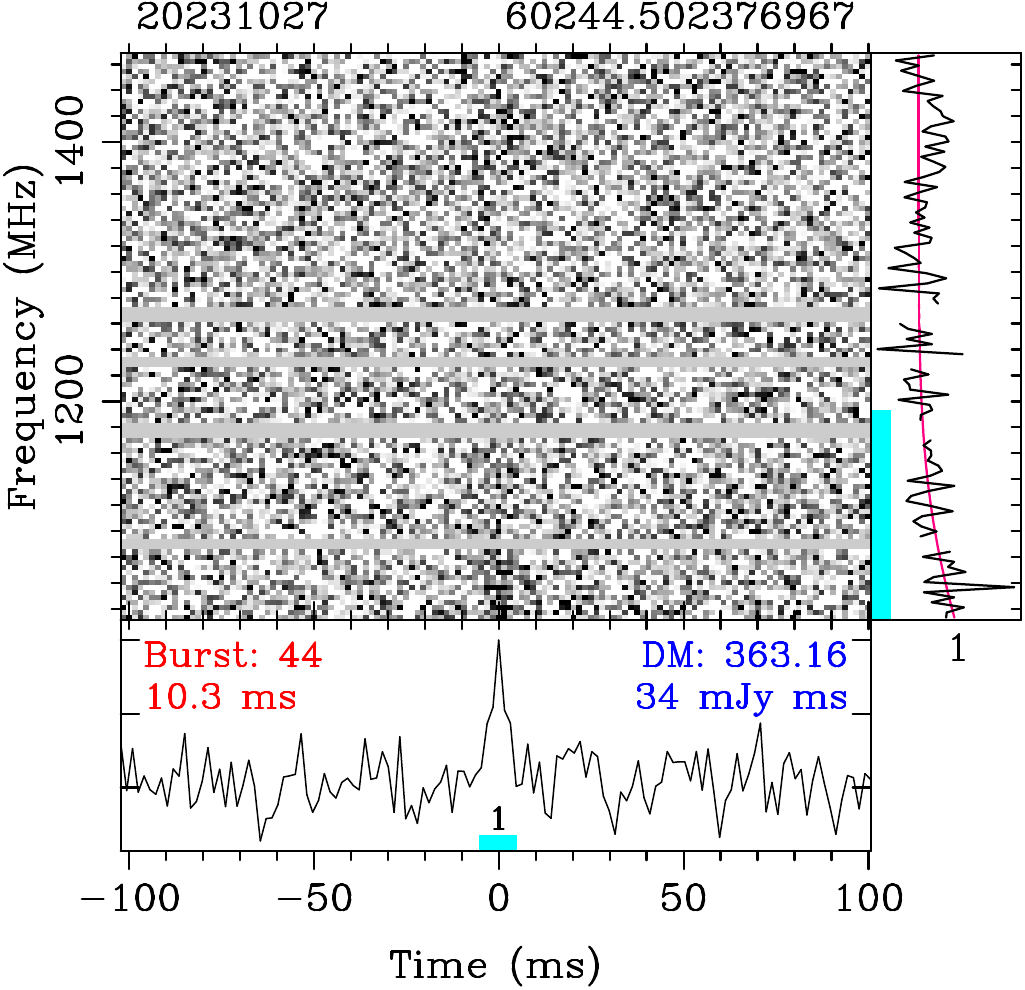}
\includegraphics[height=0.29\linewidth]{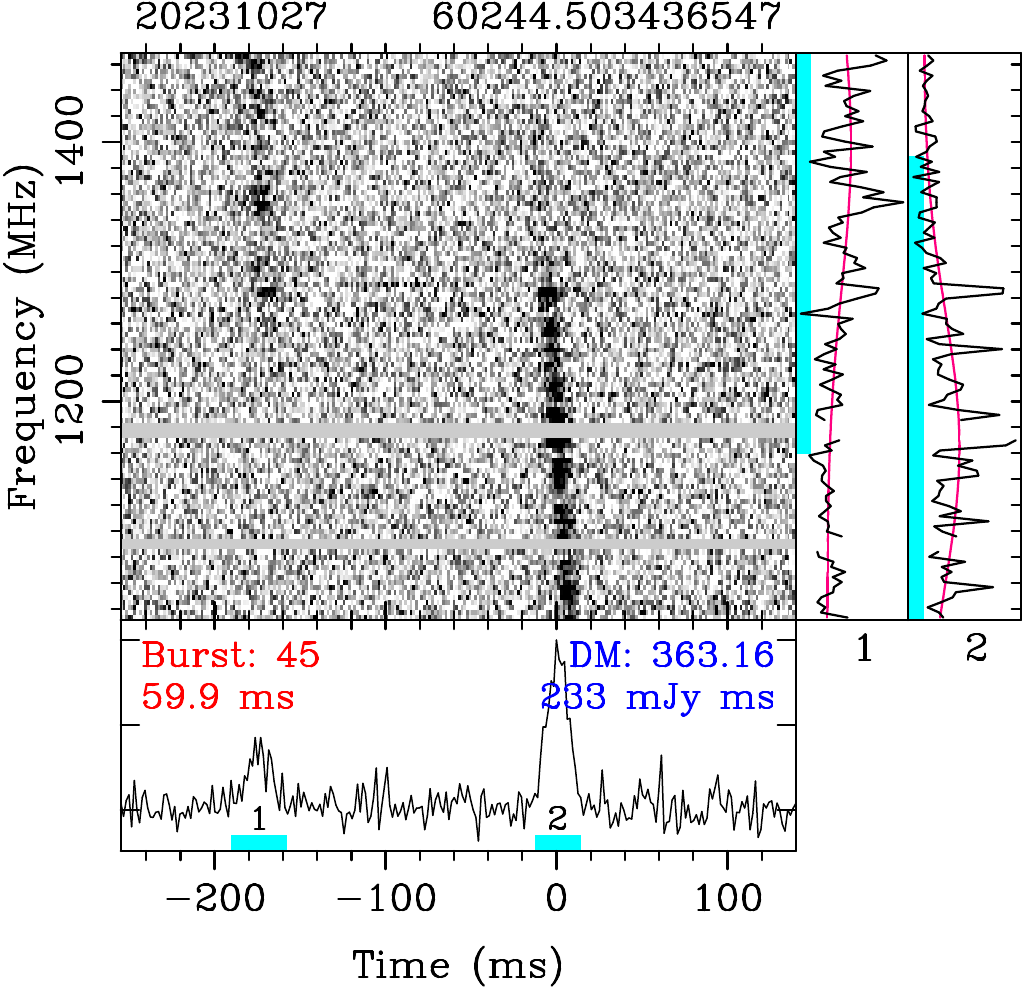}
\includegraphics[height=0.29\linewidth]{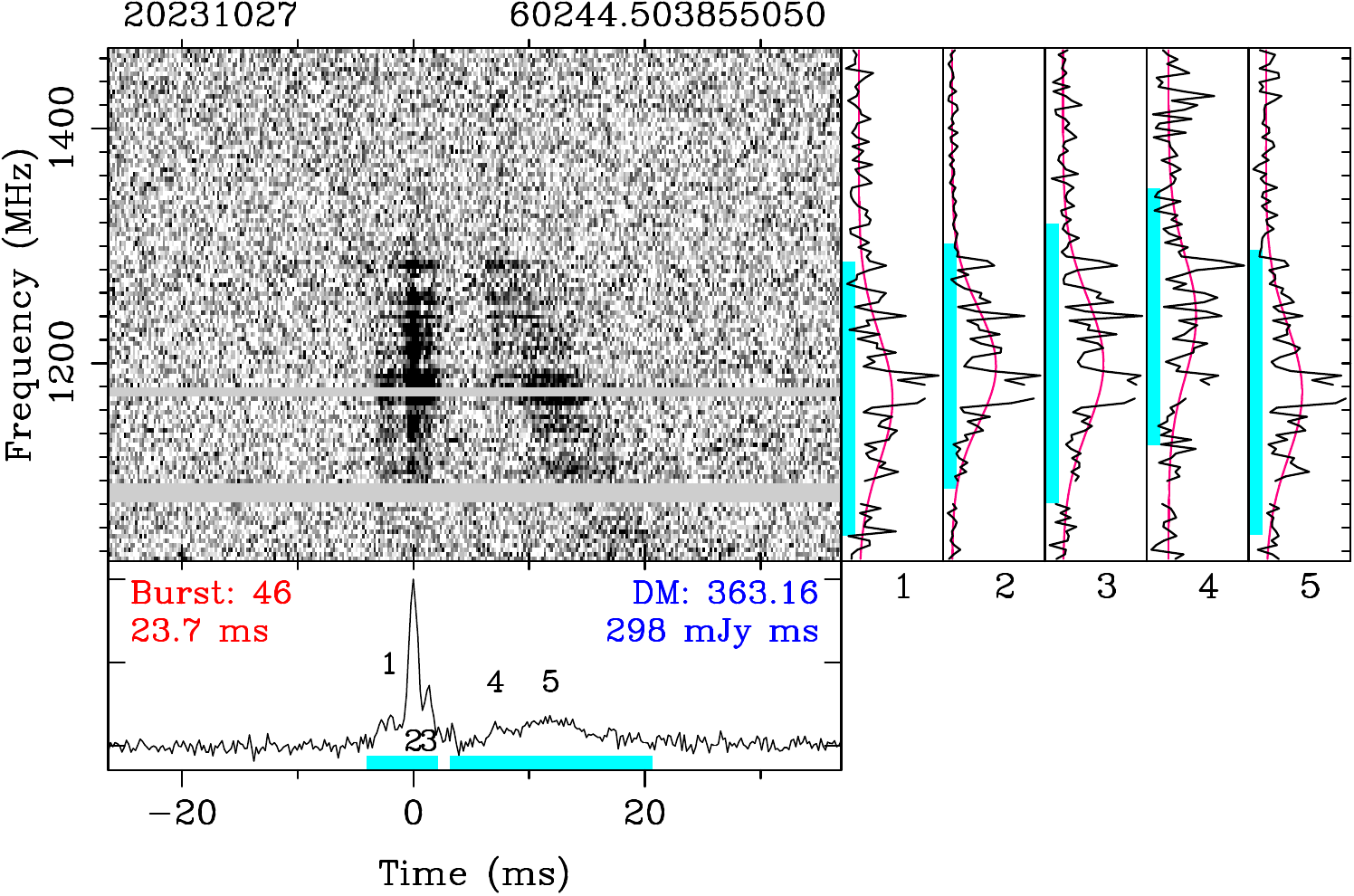}
\includegraphics[height=0.29\linewidth]{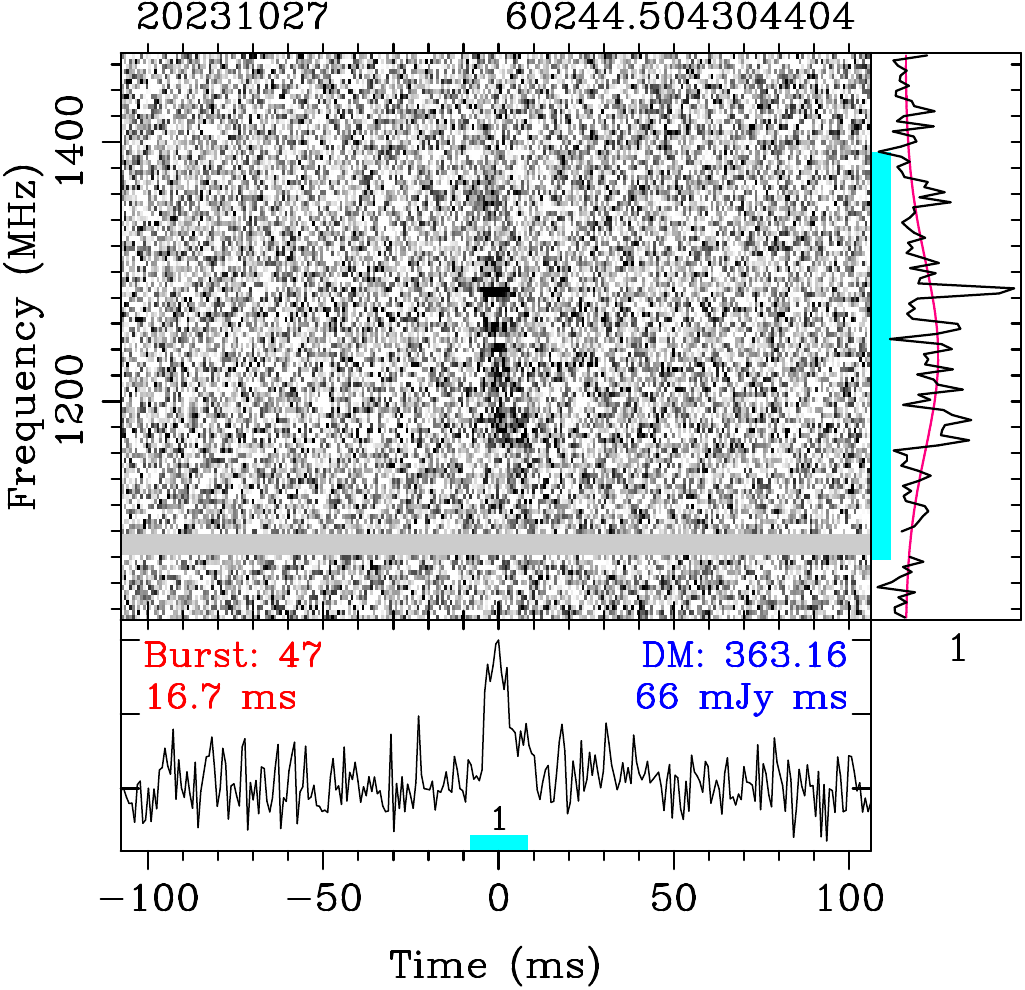}
\caption{({\textit{continued}})}
\end{figure*}
\addtocounter{figure}{-1}
\begin{figure*}
\flushleft
\includegraphics[height=0.29\linewidth]{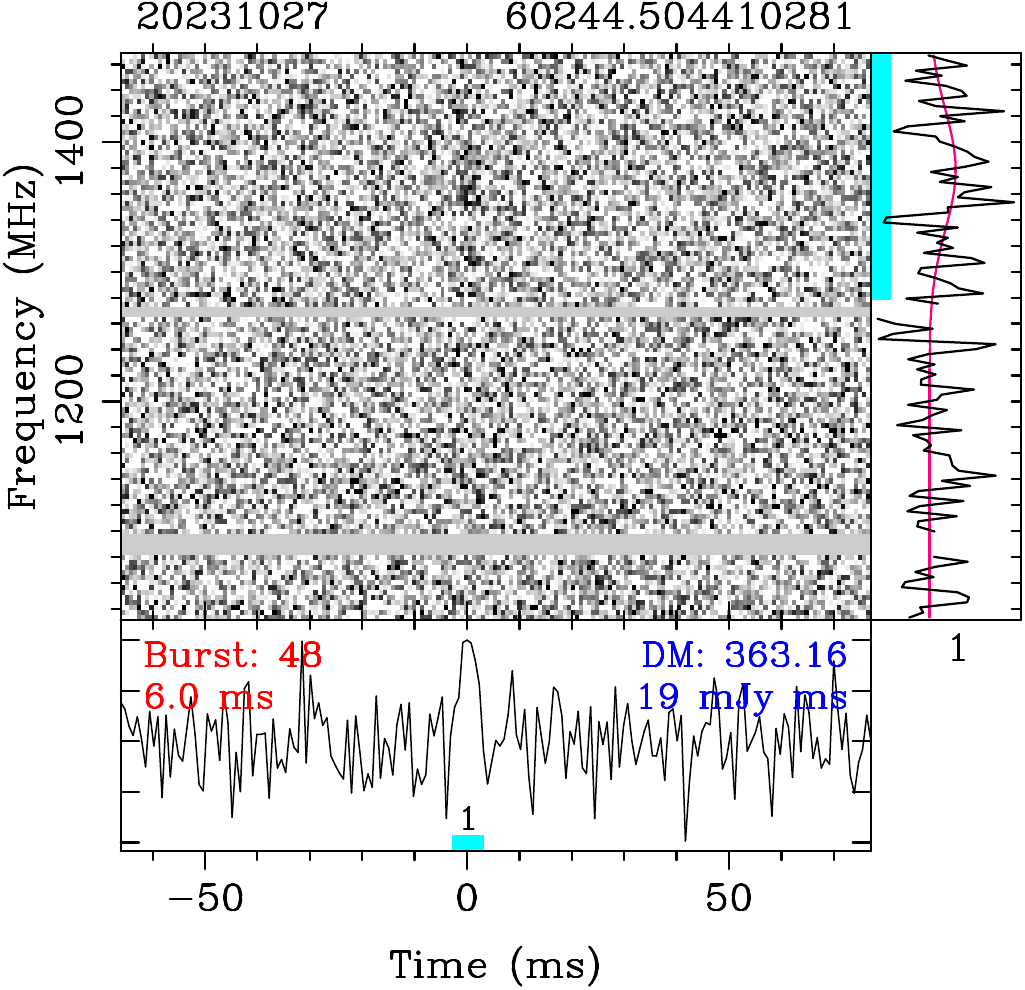}
\includegraphics[height=0.29\linewidth]{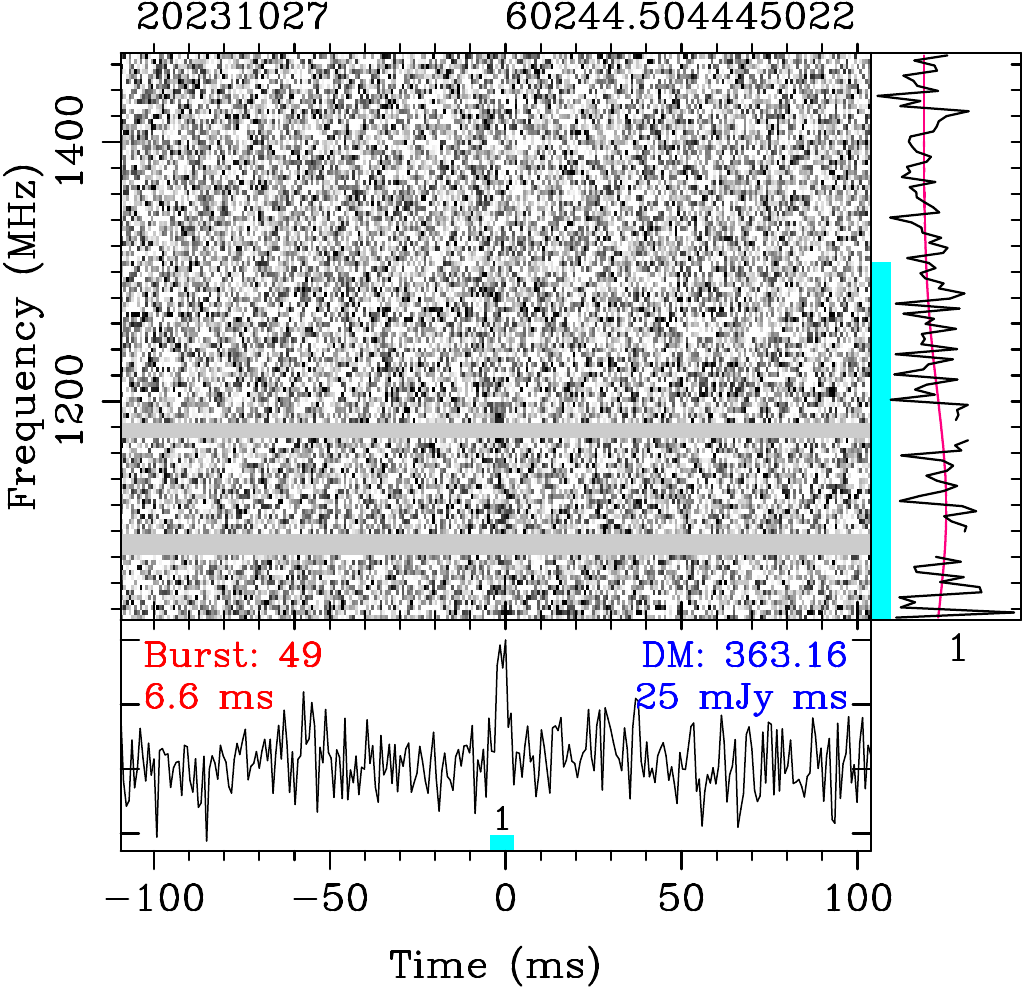}
\includegraphics[height=0.29\linewidth]{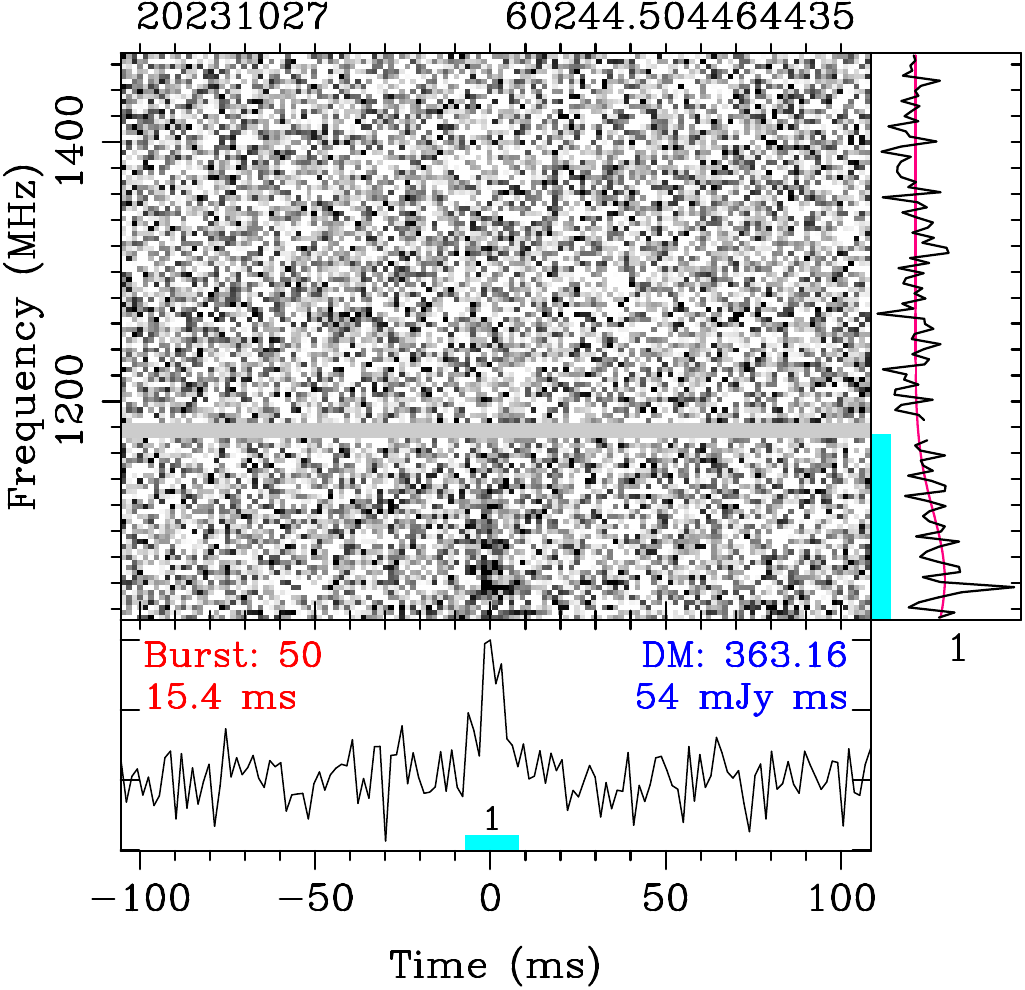}
\includegraphics[height=0.29\linewidth]{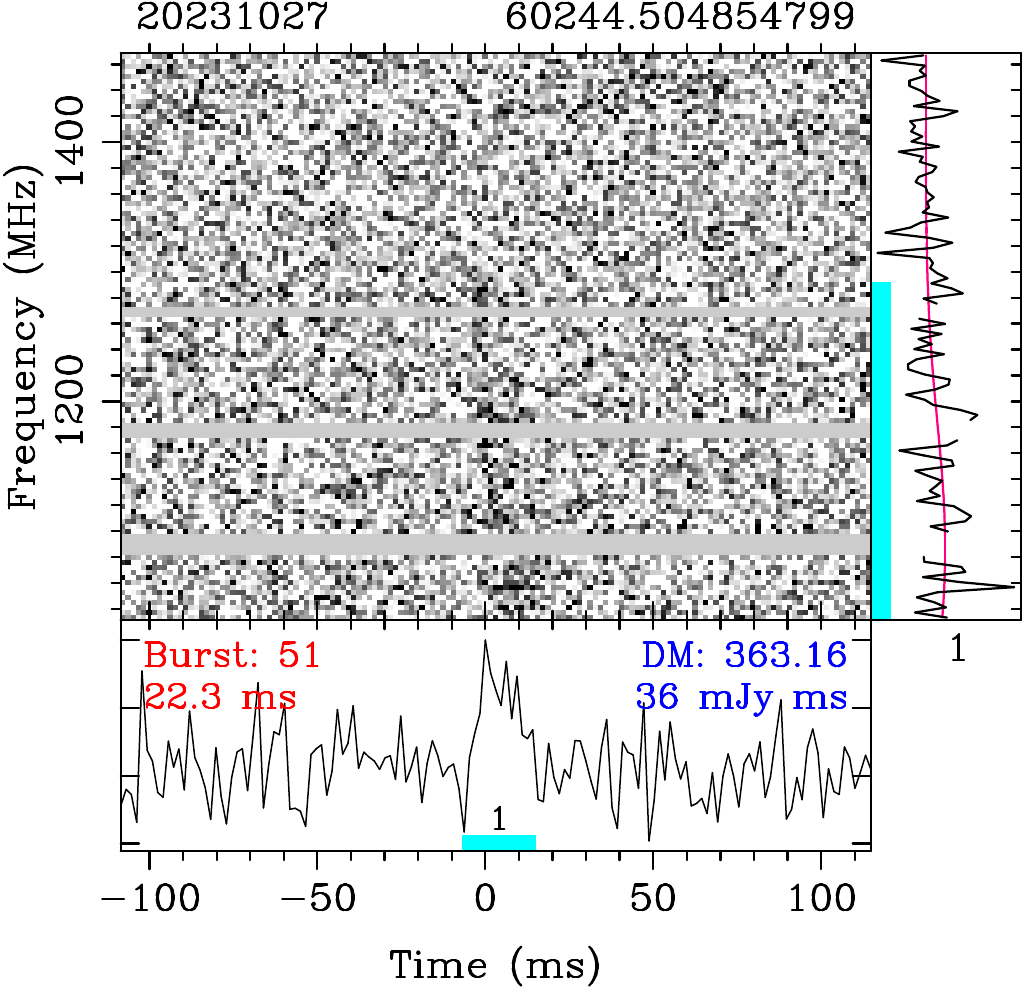}
\includegraphics[height=0.29\linewidth]{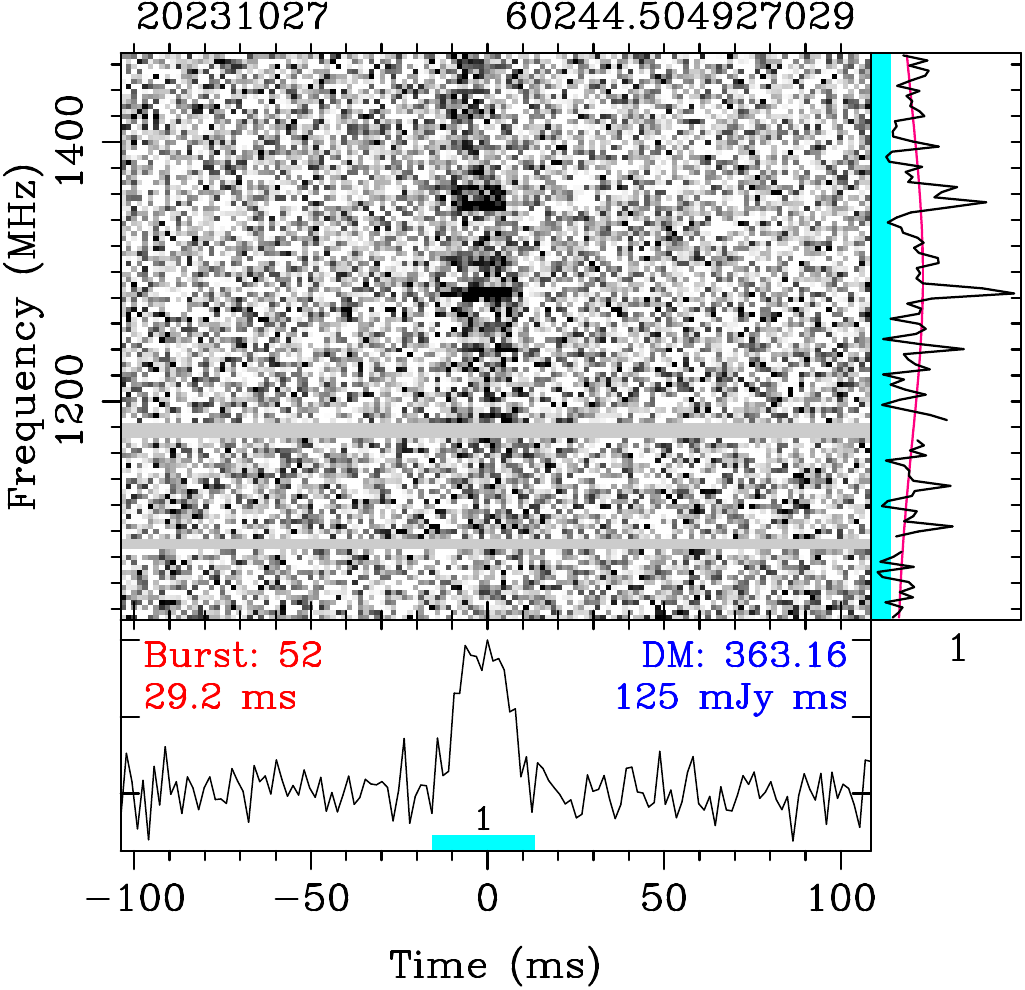}
\includegraphics[height=0.29\linewidth]{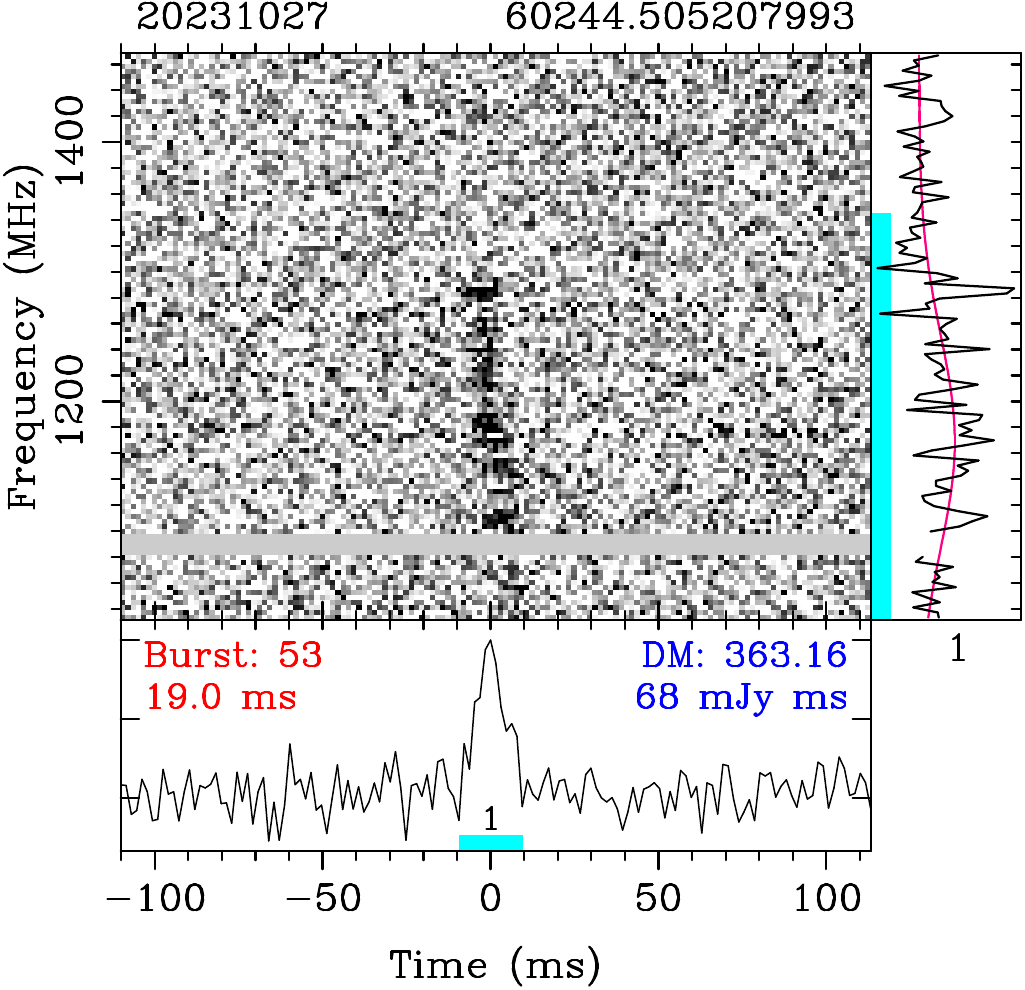}
\includegraphics[height=0.29\linewidth]{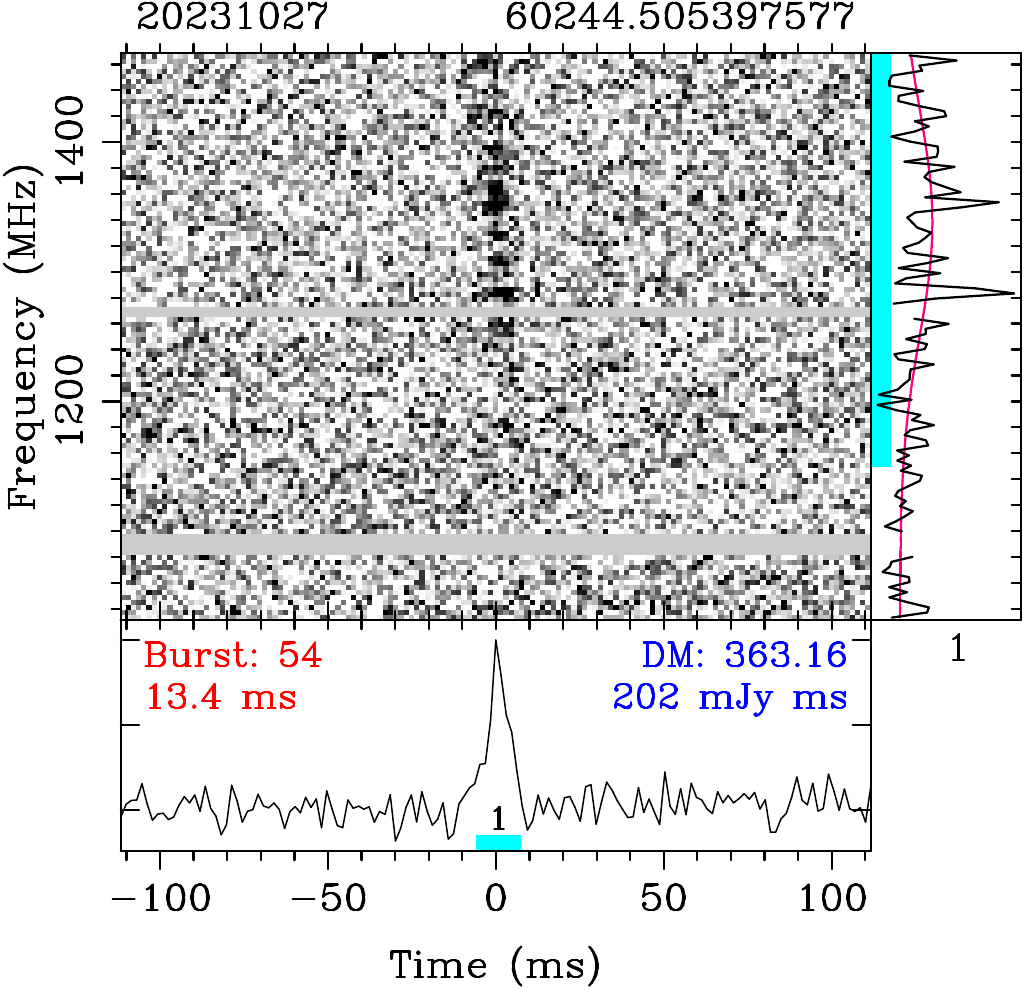}
\includegraphics[height=0.29\linewidth]{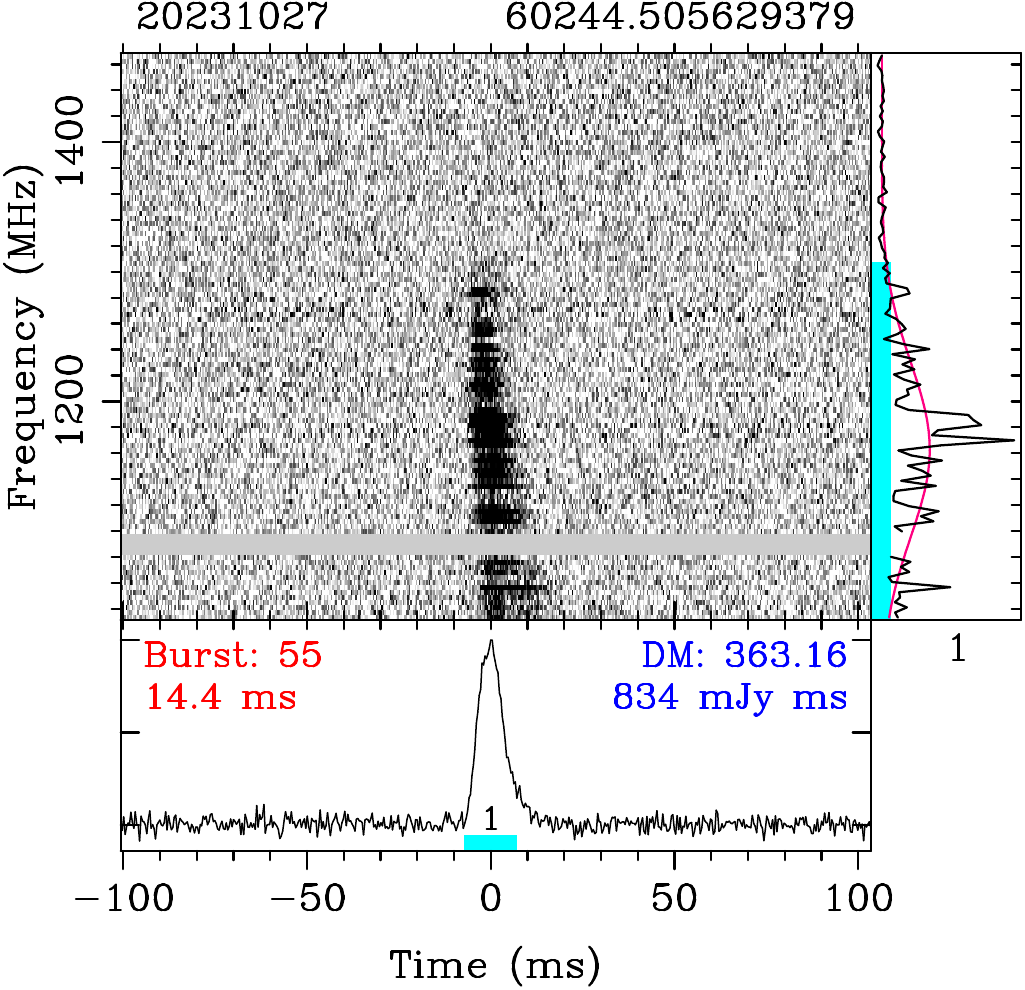}
\includegraphics[height=0.29\linewidth]{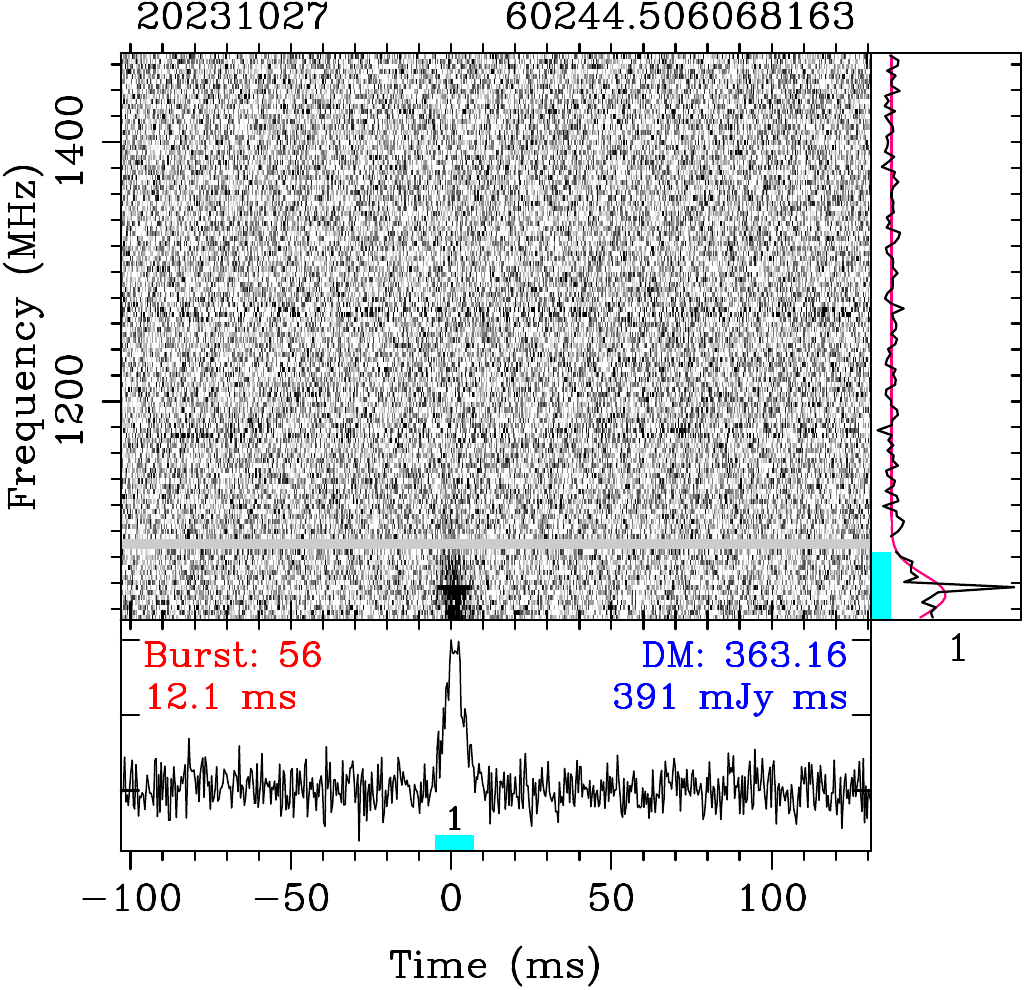}
\includegraphics[height=0.29\linewidth]{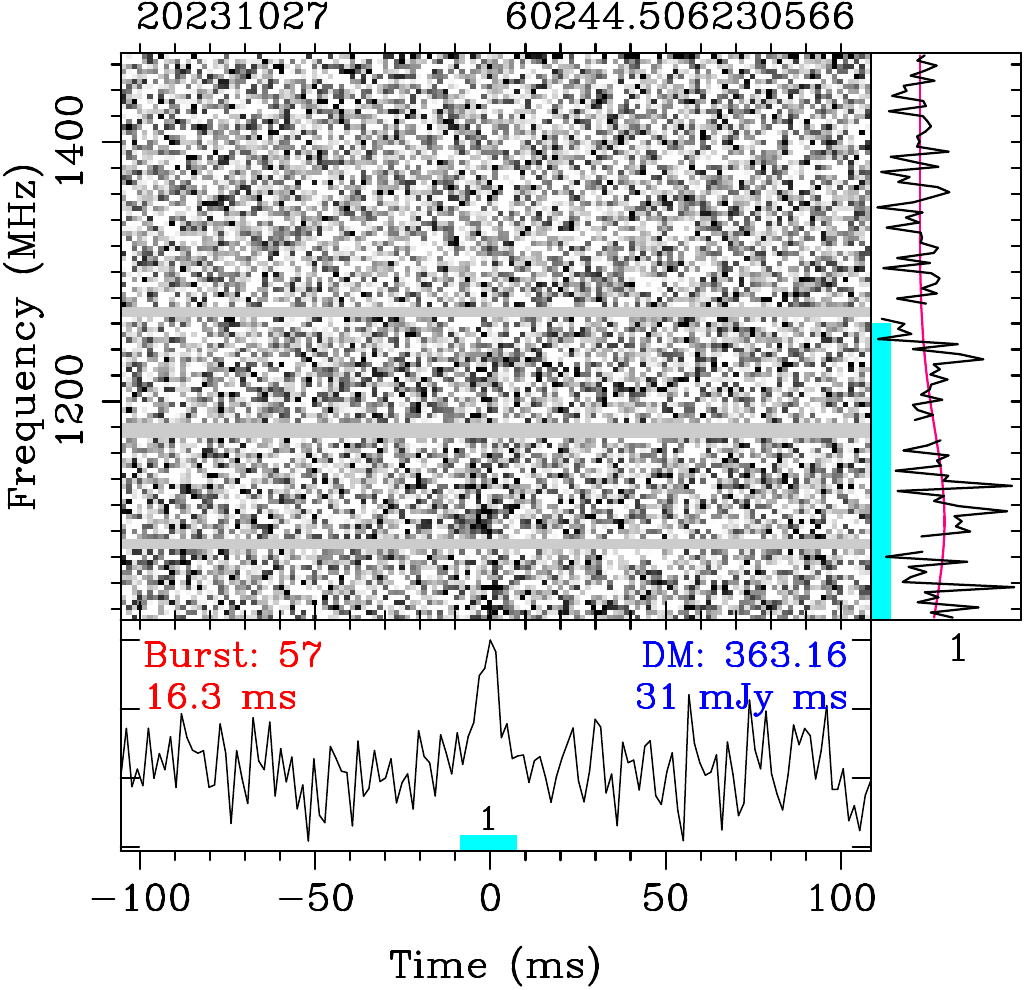}
\includegraphics[height=0.29\linewidth]{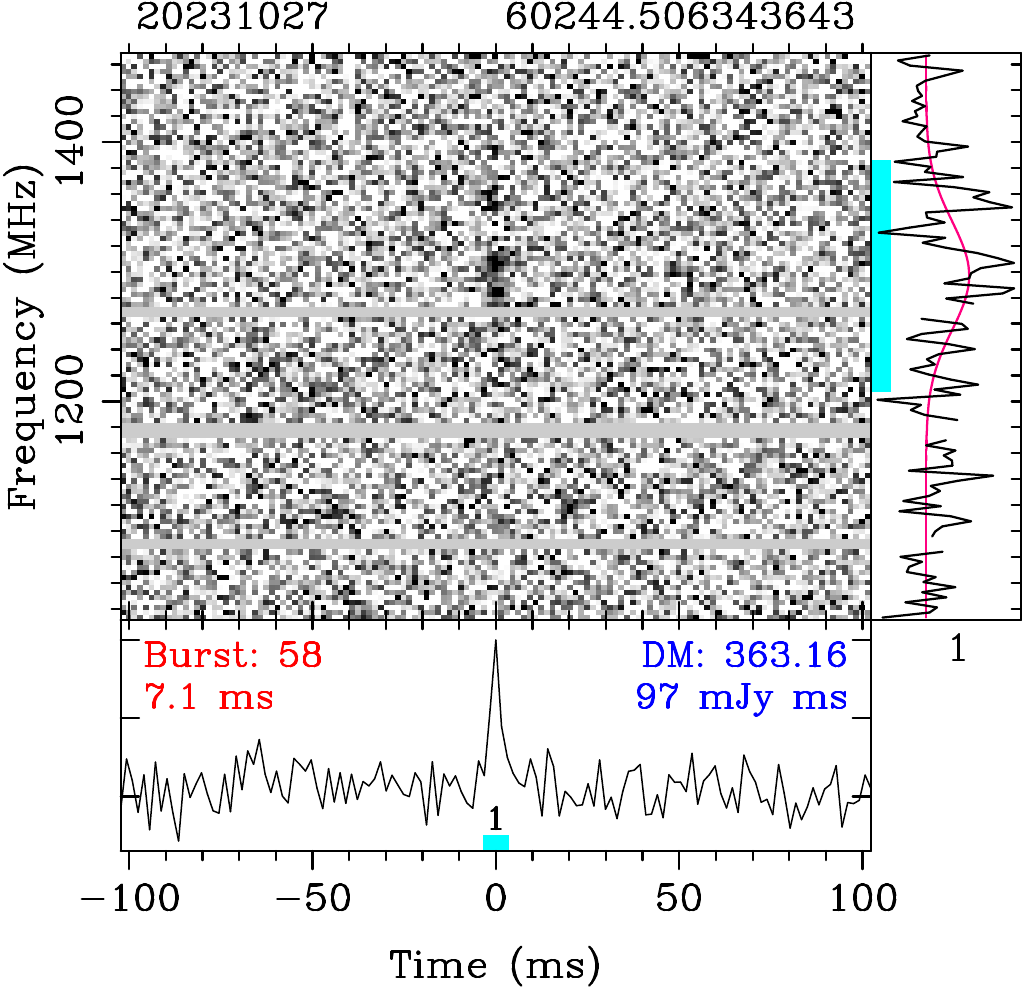}
\includegraphics[height=0.29\linewidth]{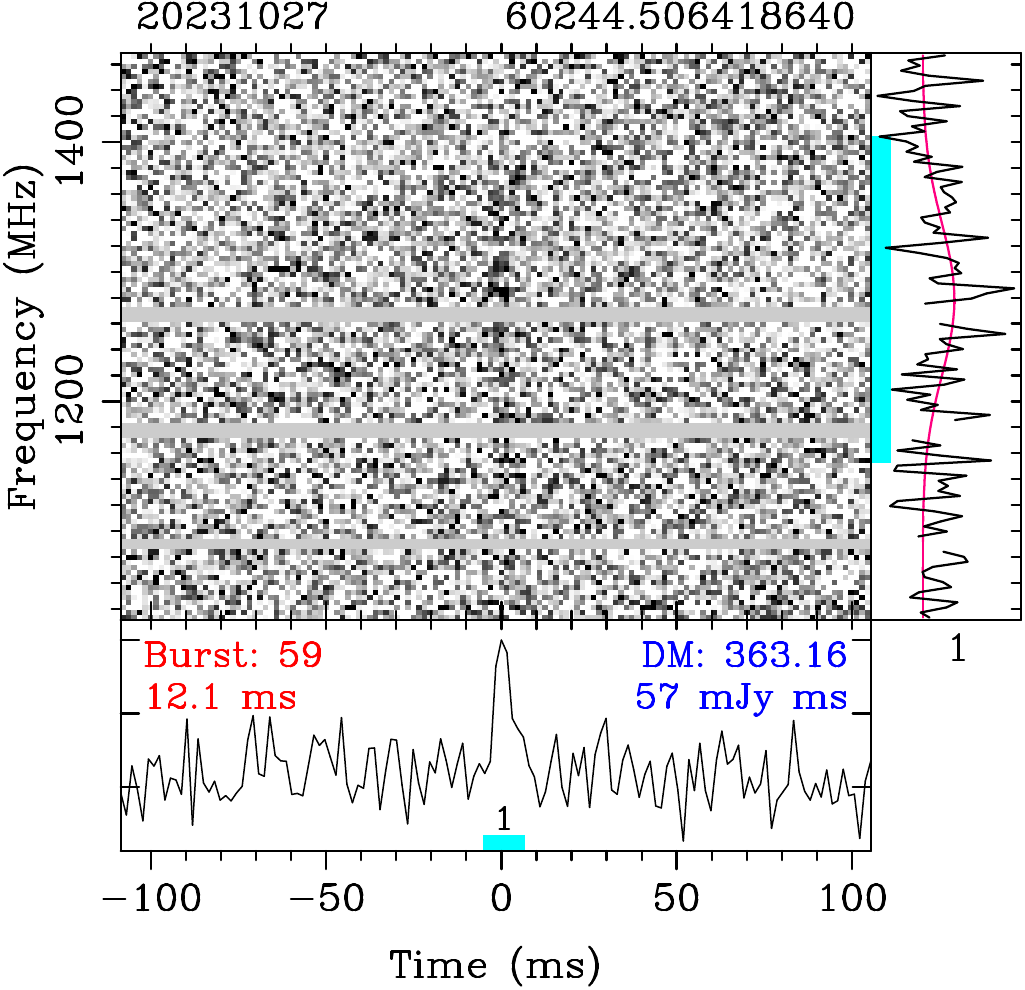}
\caption{({\textit{continued}})}
\end{figure*}
\addtocounter{figure}{-1}
\begin{figure*}
\flushleft
\includegraphics[height=0.29\linewidth]{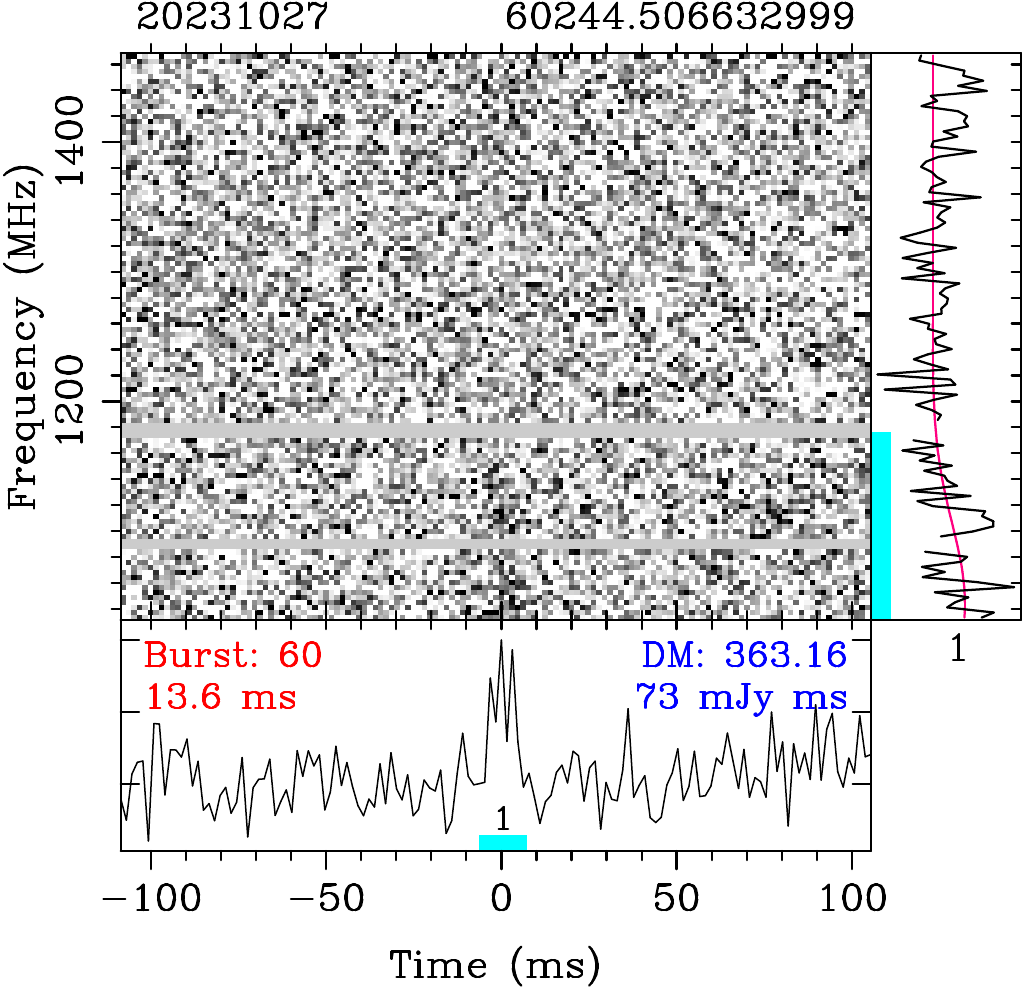}
\includegraphics[height=0.29\linewidth]{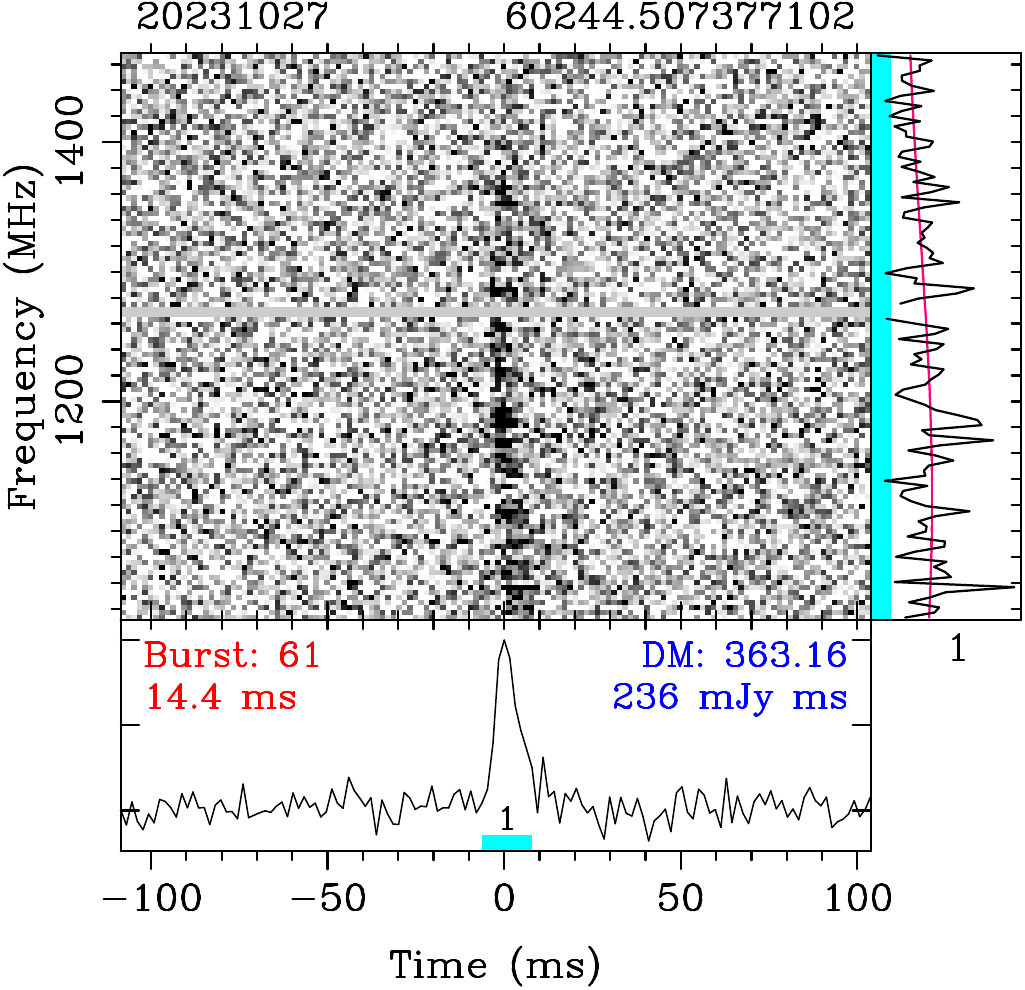}
\includegraphics[height=0.29\linewidth]{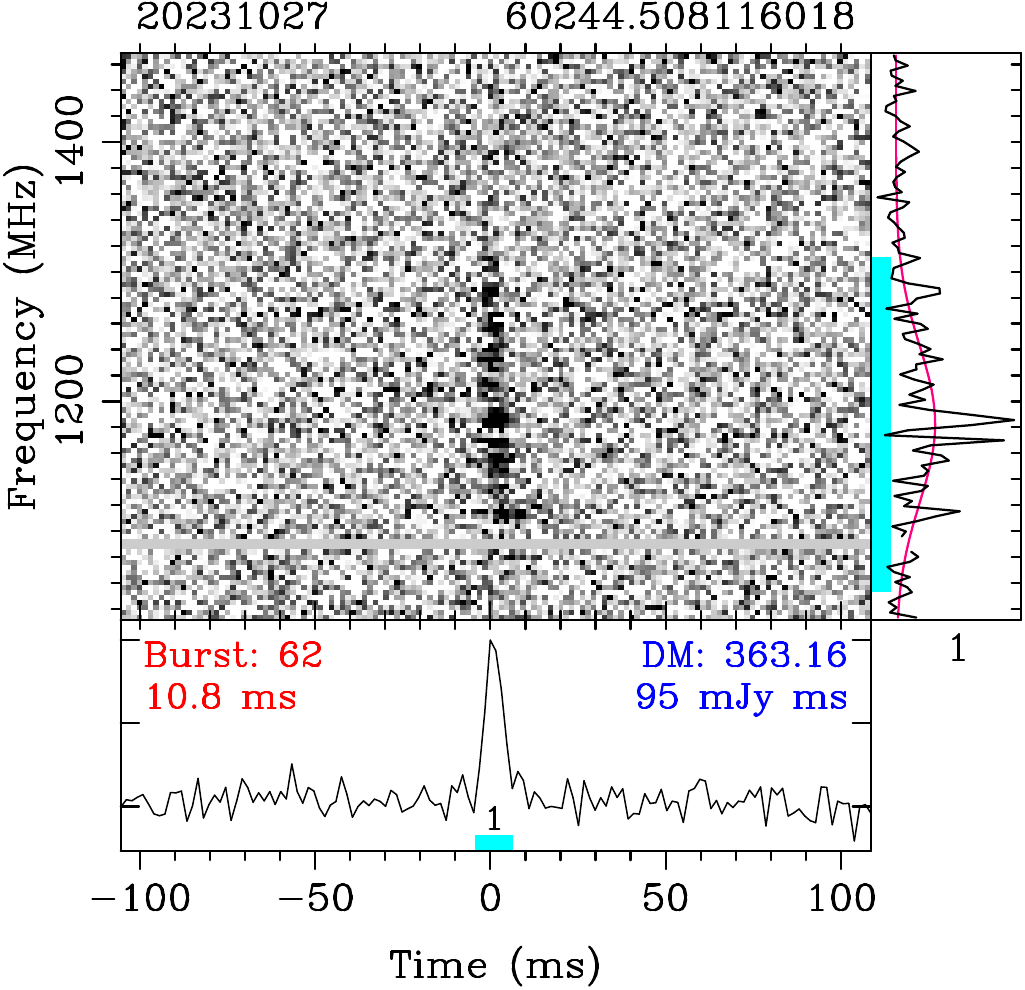}
\includegraphics[height=0.29\linewidth]{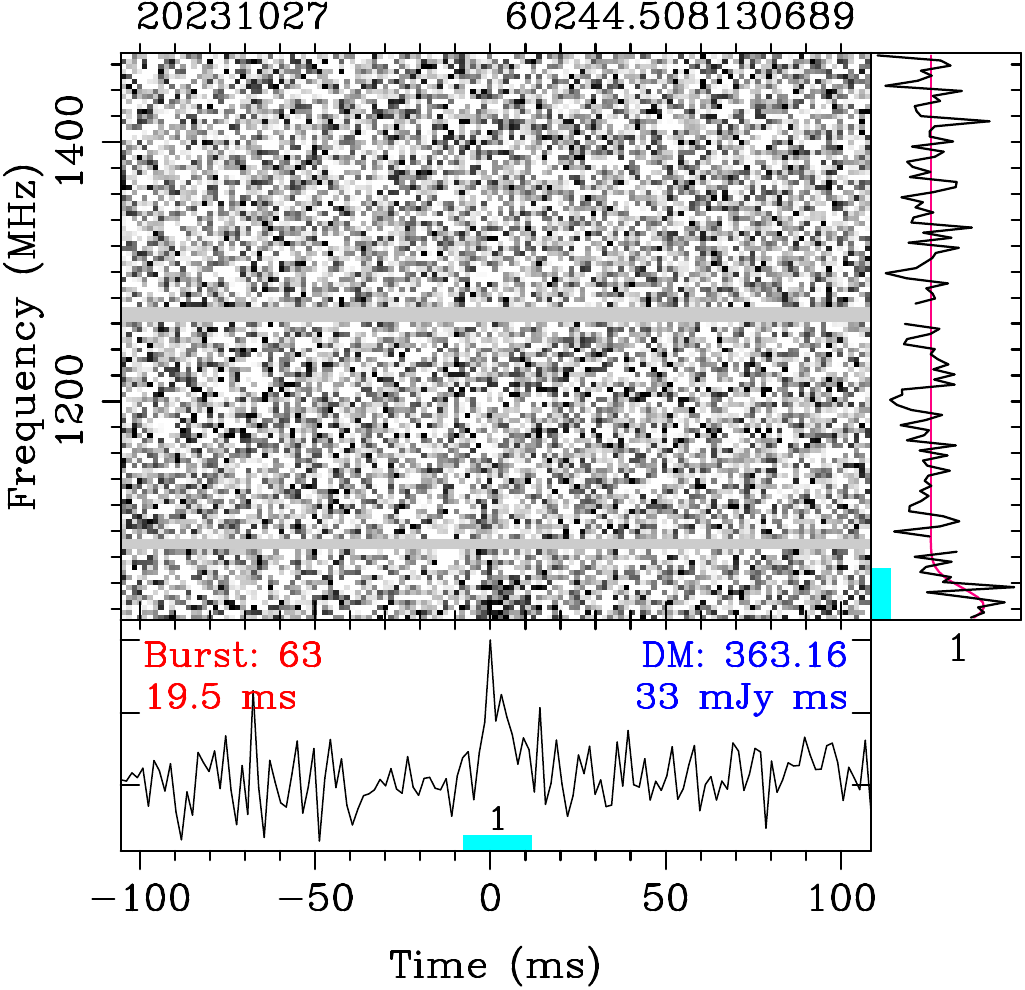}
\includegraphics[height=0.29\linewidth]{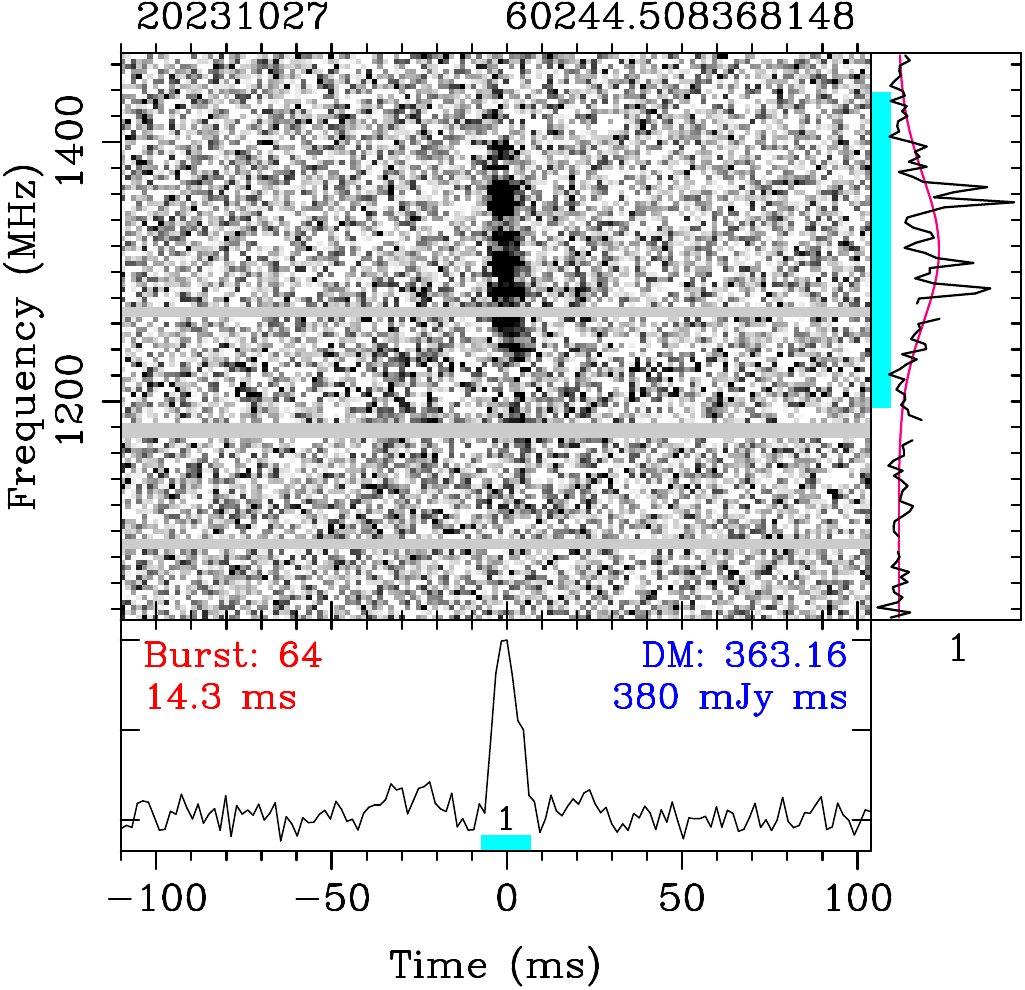}
\includegraphics[height=0.29\linewidth]{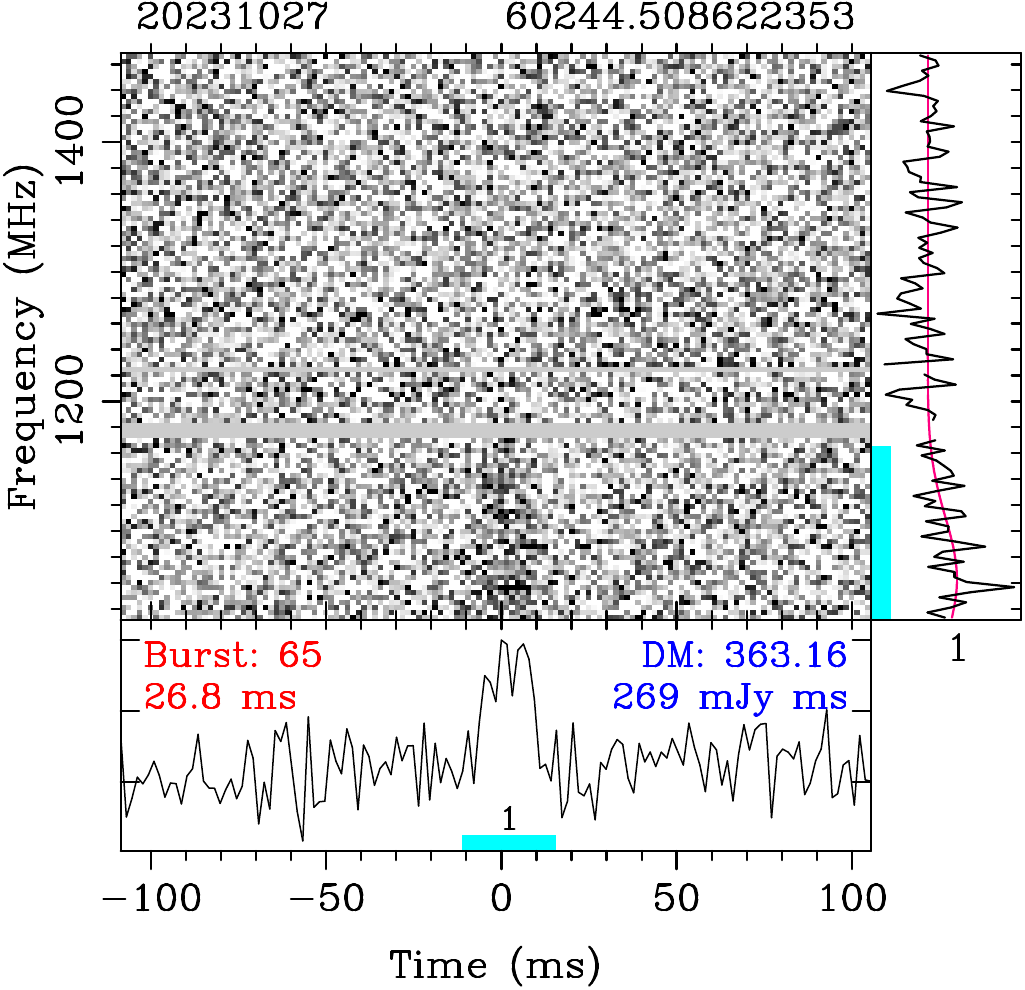}
\includegraphics[height=0.29\linewidth]{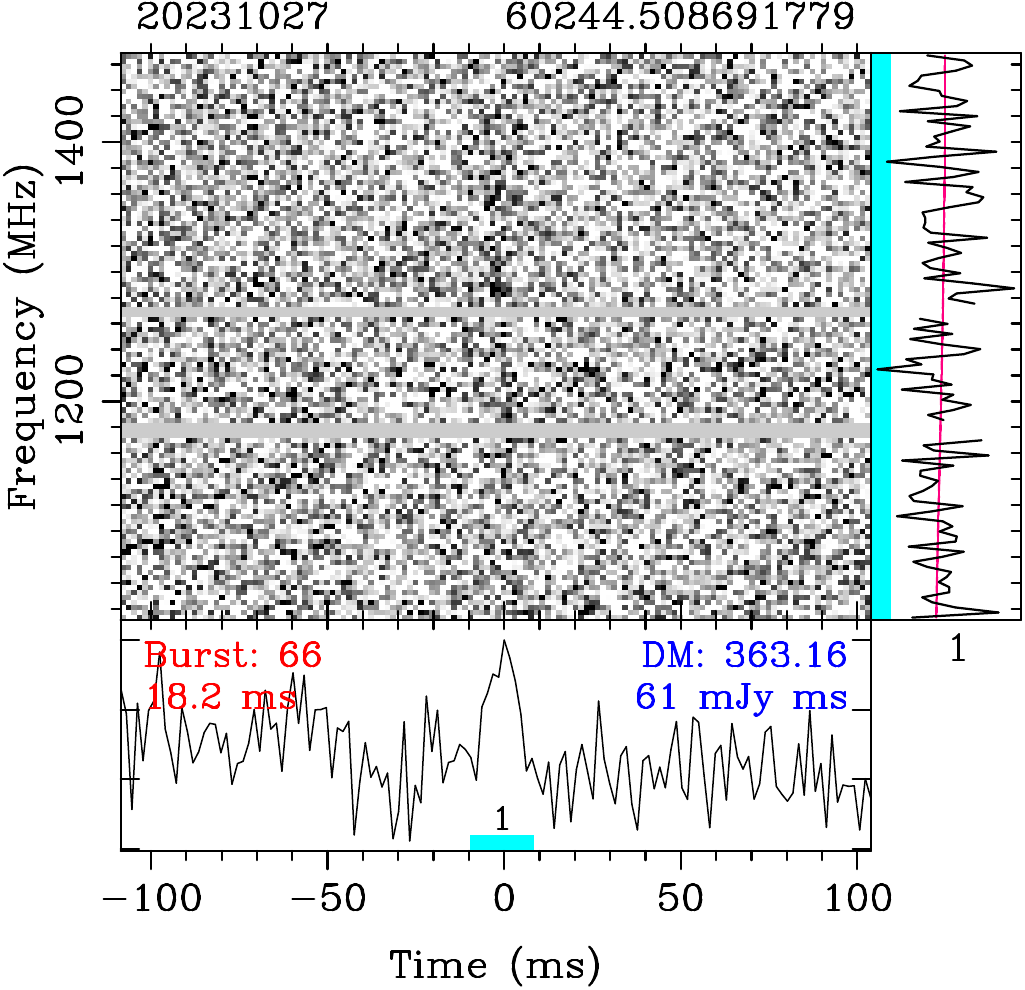}
\includegraphics[height=0.29\linewidth]{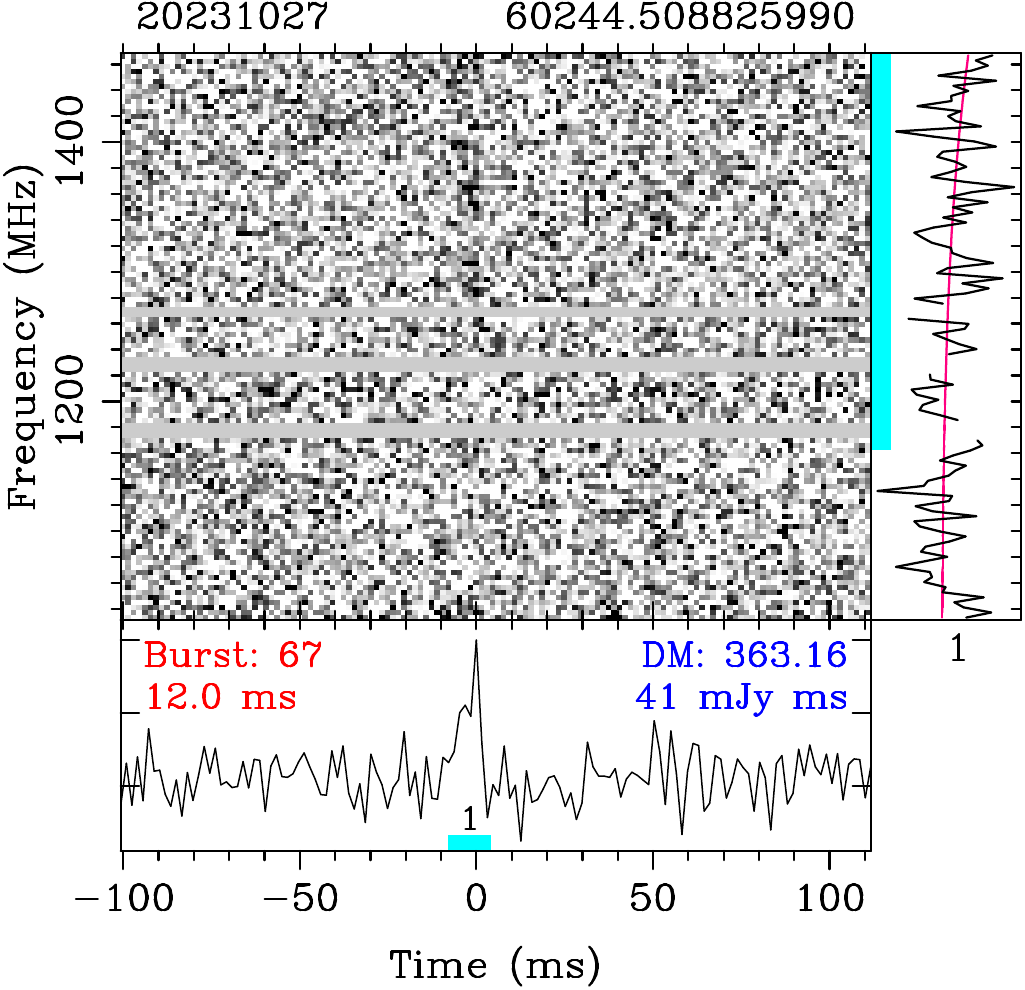}
\includegraphics[height=0.29\linewidth]{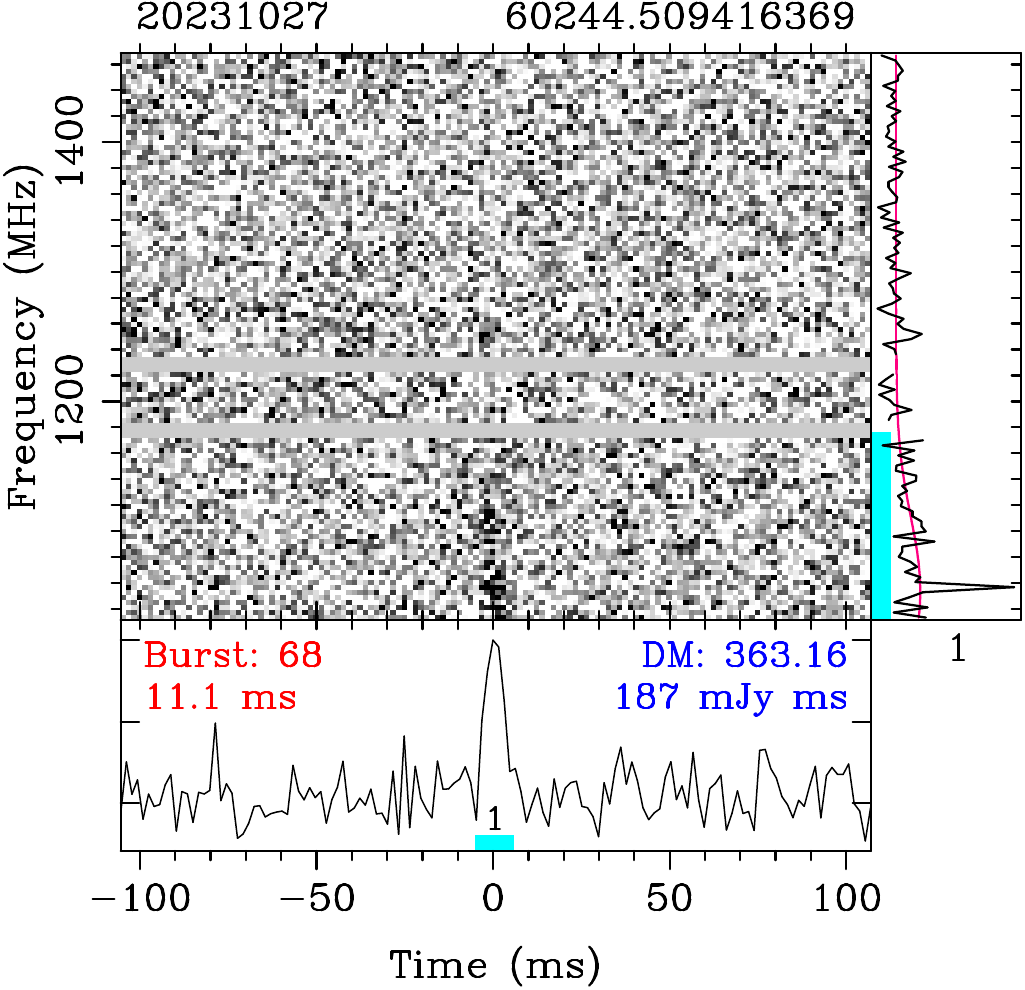}
\includegraphics[height=0.29\linewidth]{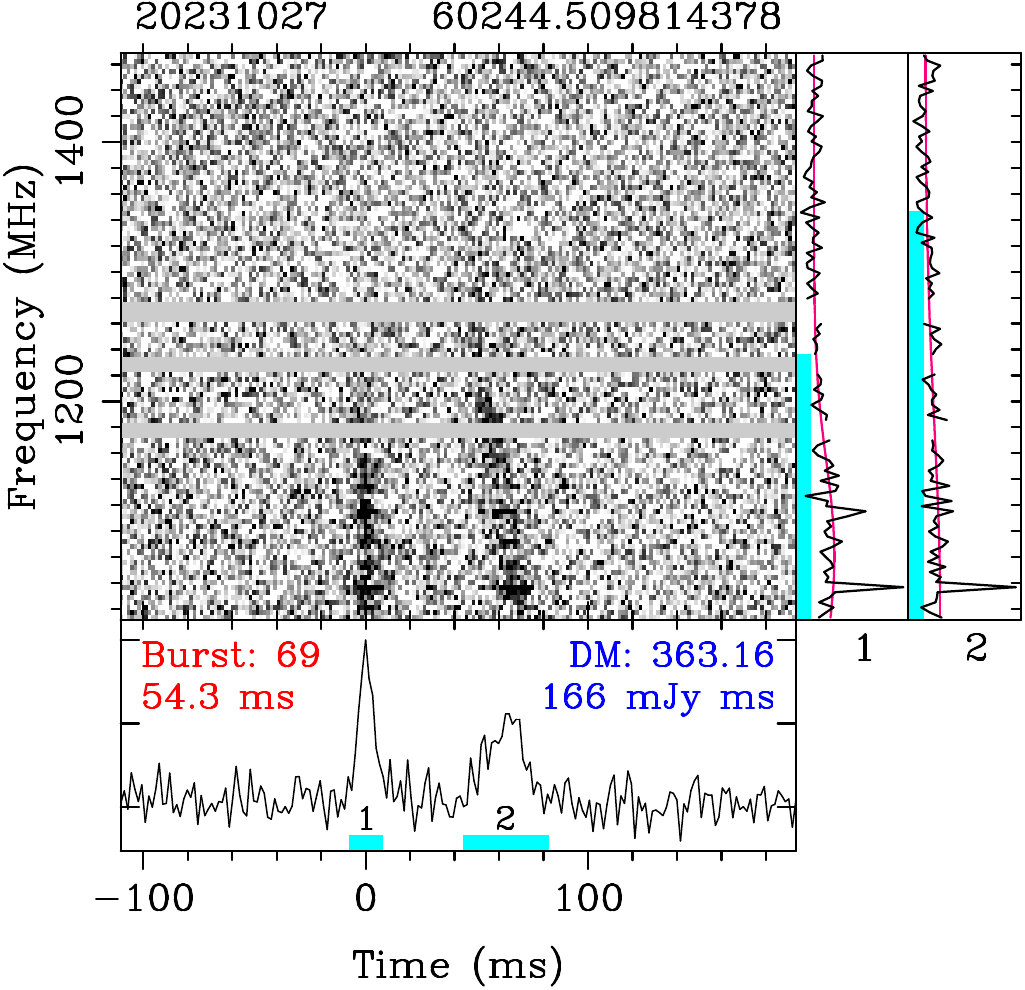}
\includegraphics[height=0.29\linewidth]{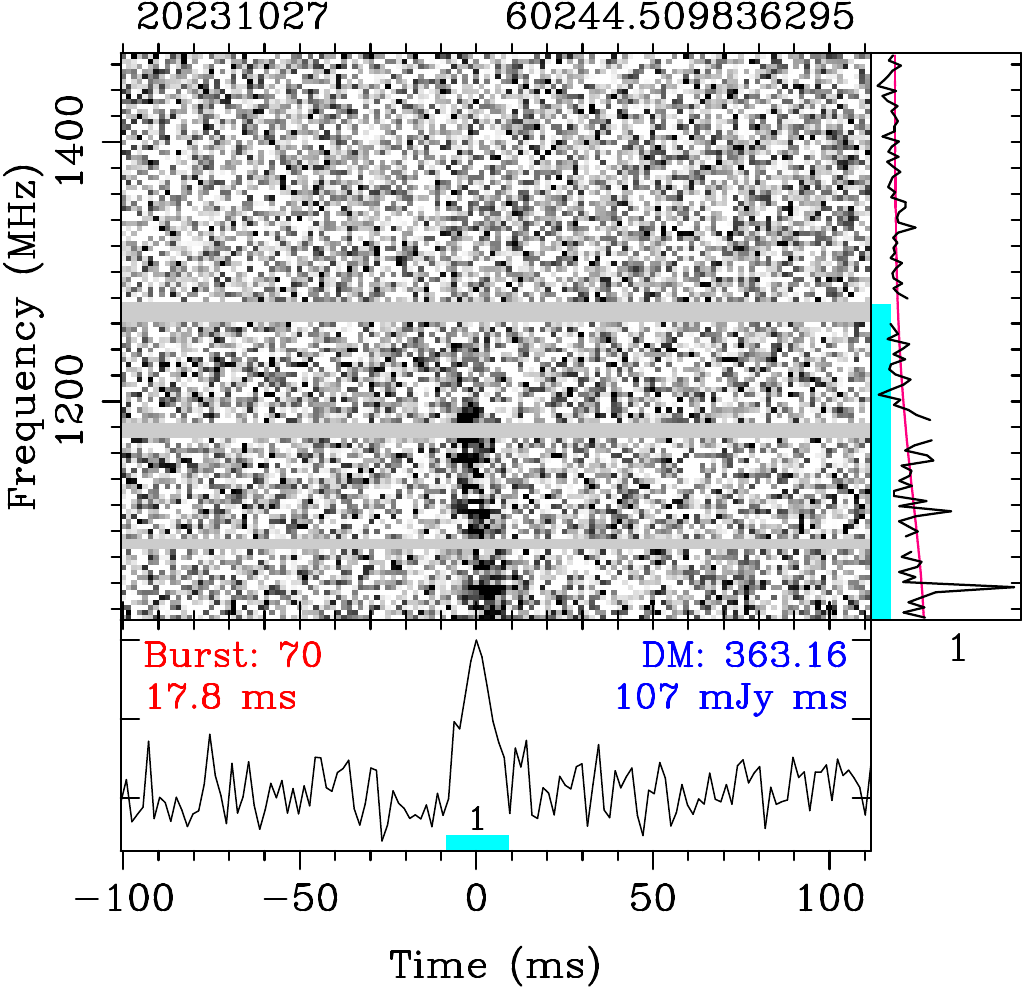}
\includegraphics[height=0.29\linewidth]{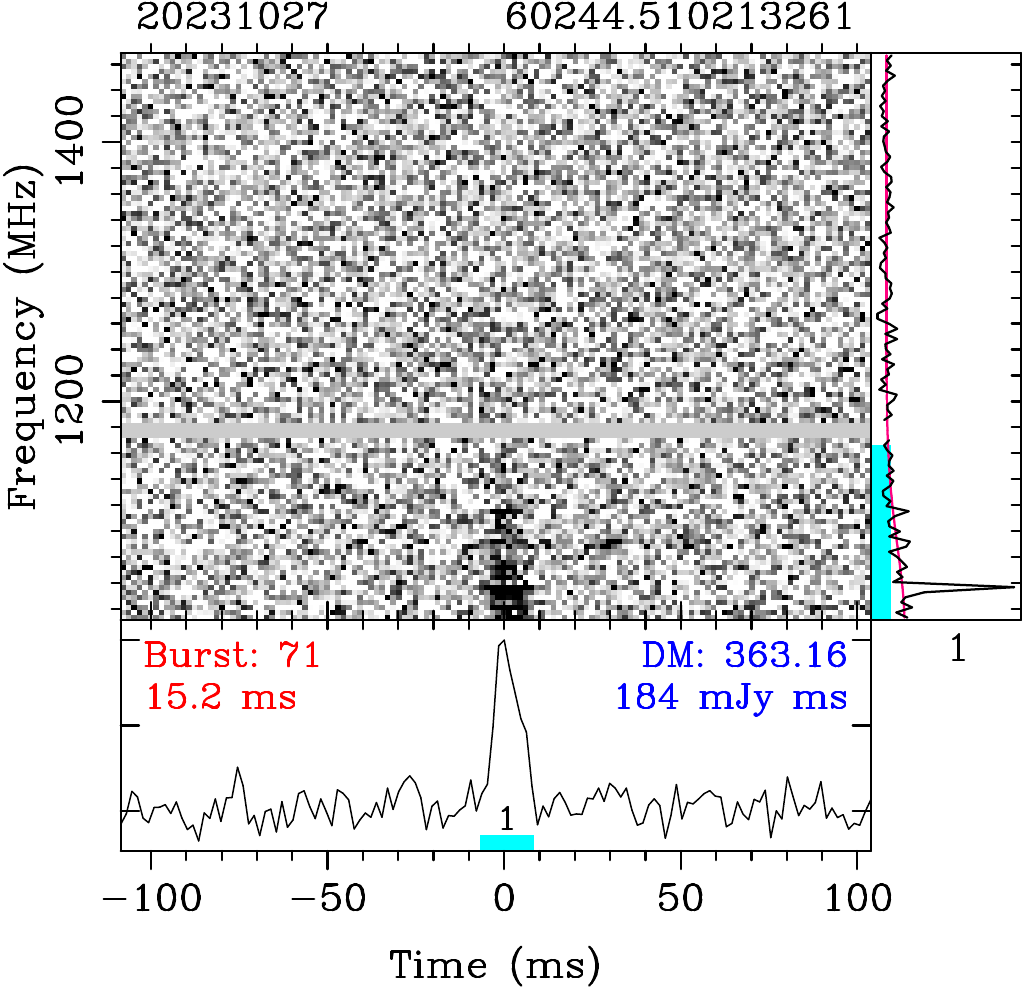}
\caption{({\textit{continued}})}
\end{figure*}
\addtocounter{figure}{-1}
\begin{figure*}
\flushleft
\includegraphics[height=0.29\linewidth]{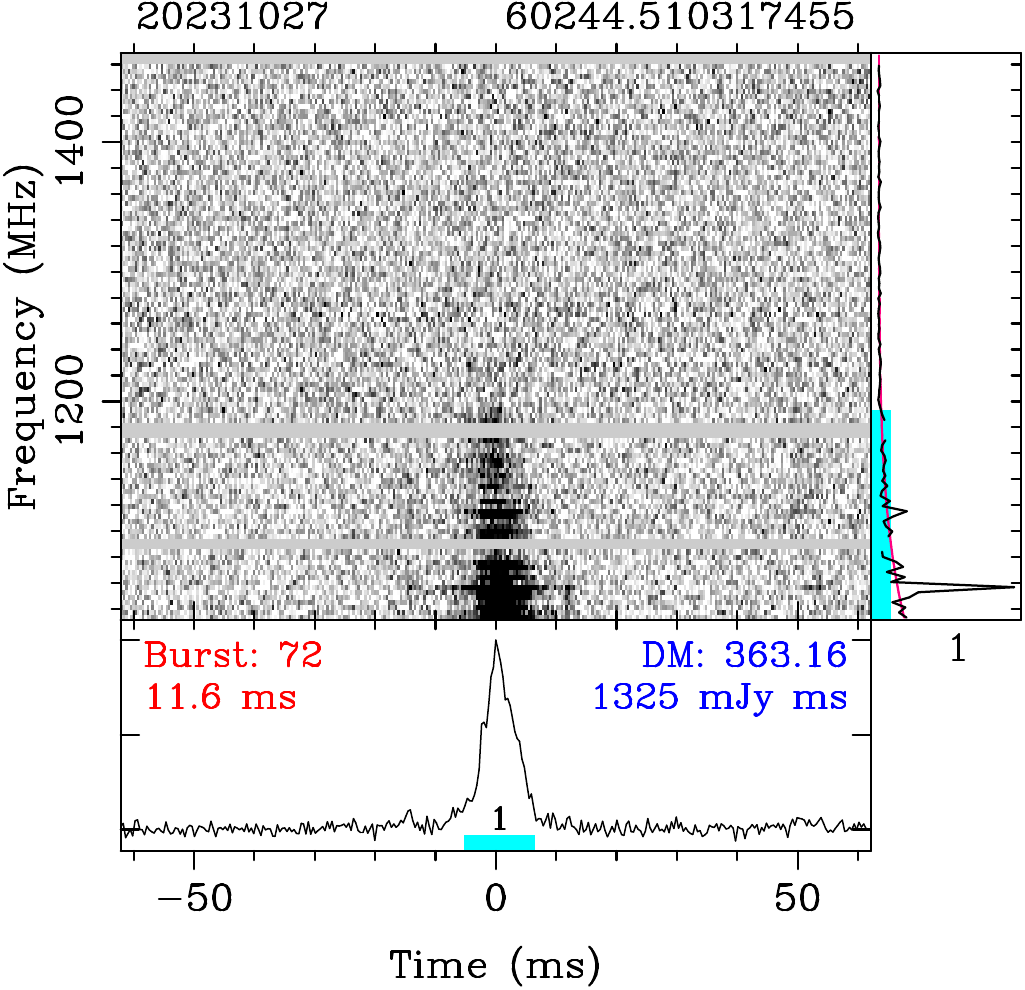}
\includegraphics[height=0.29\linewidth]{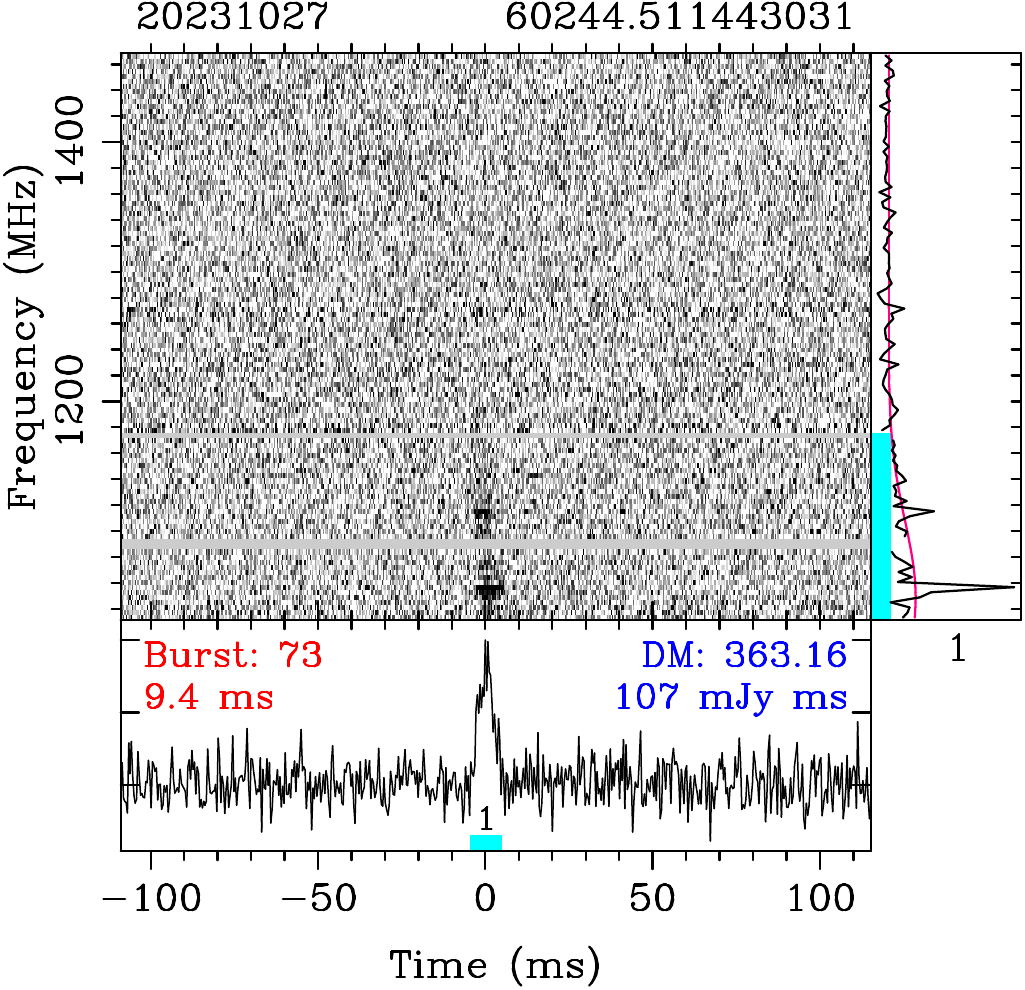}
\includegraphics[height=0.29\linewidth]{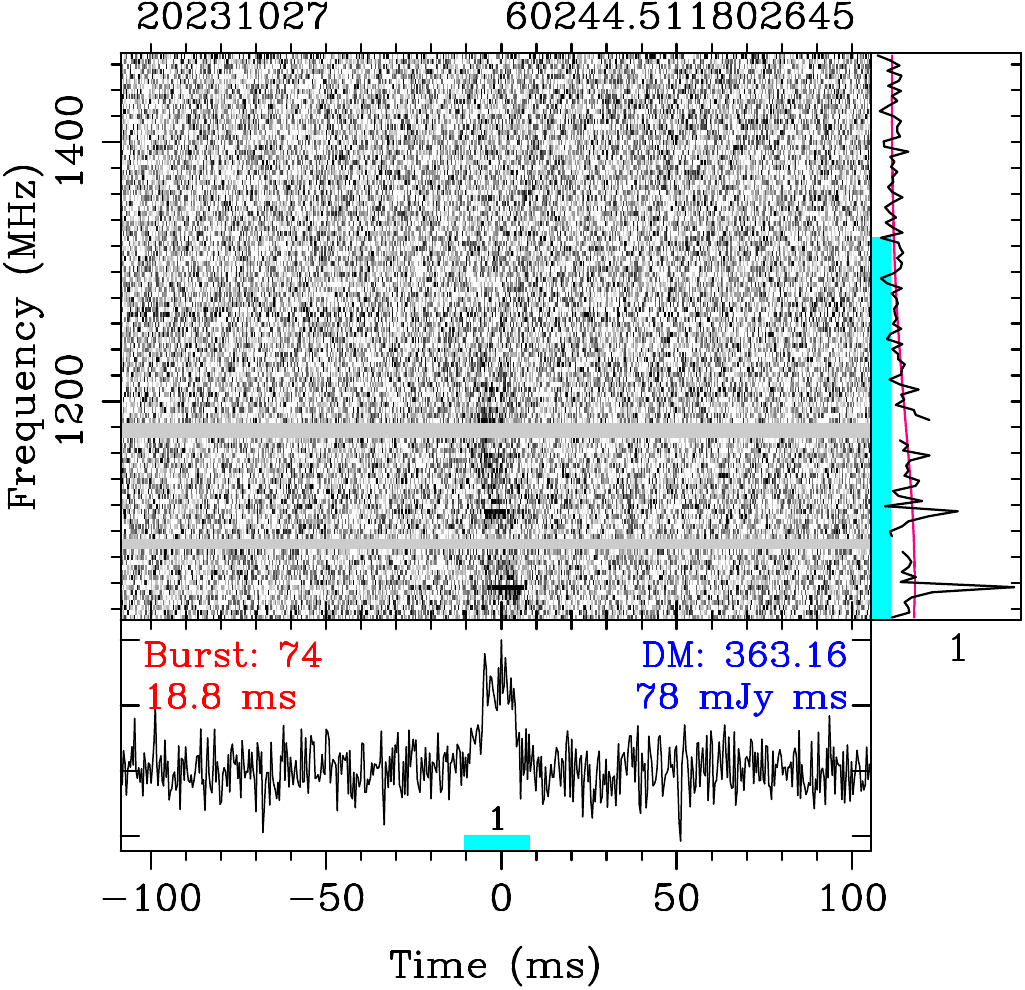}
\includegraphics[height=0.29\linewidth]{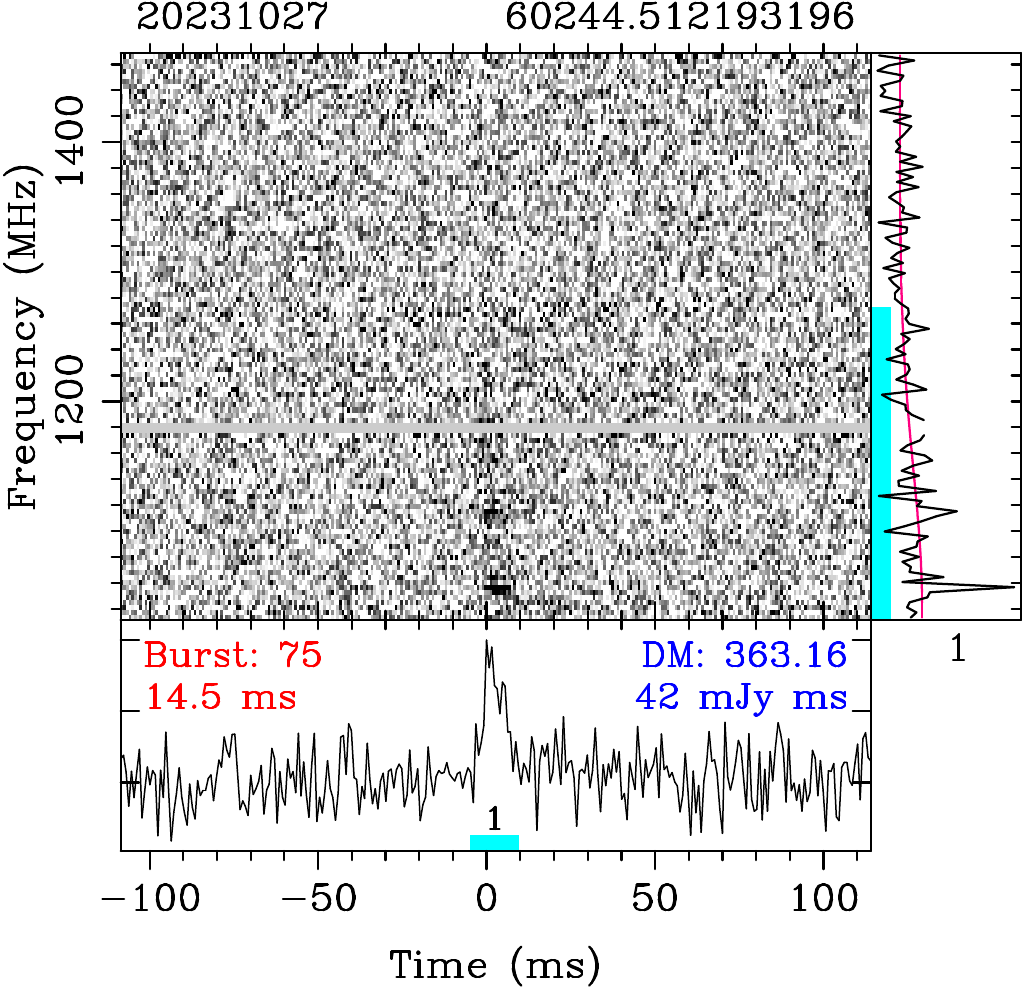}
\includegraphics[height=0.29\linewidth]{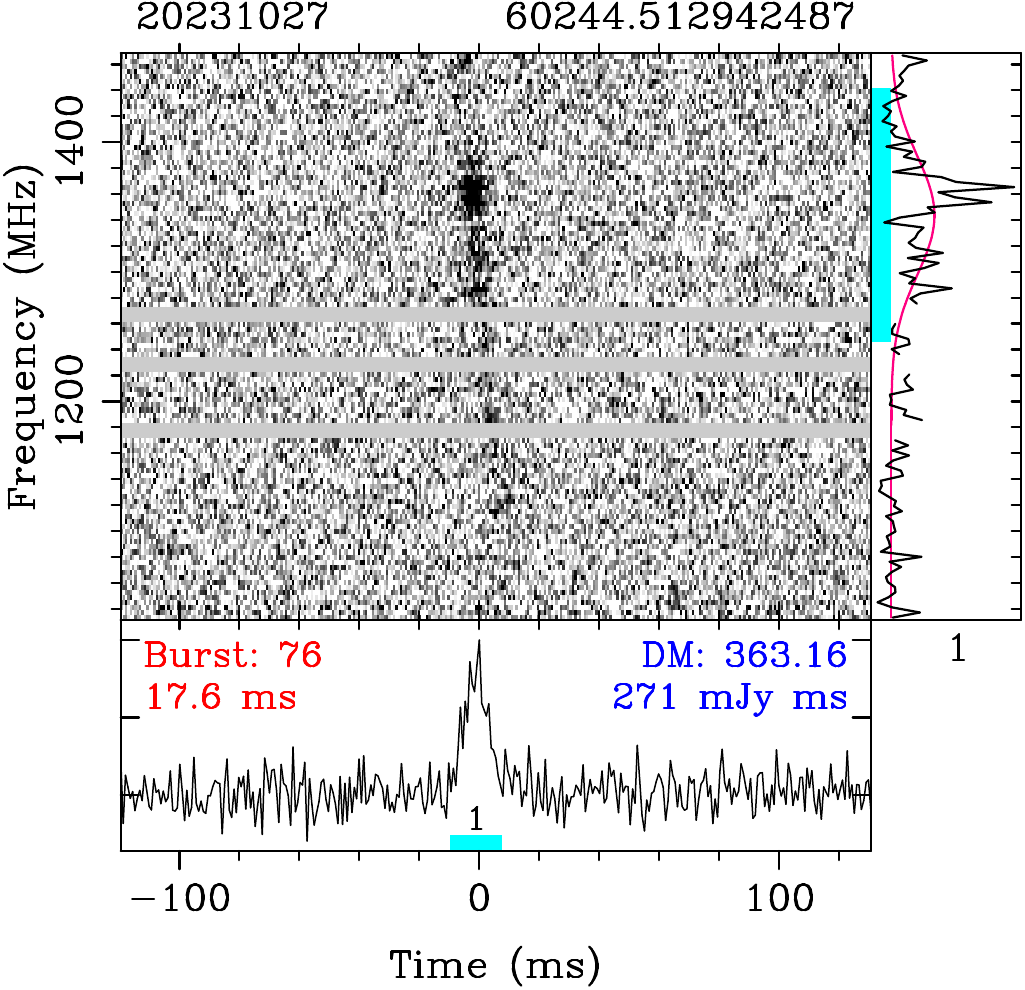}
\includegraphics[height=0.29\linewidth]{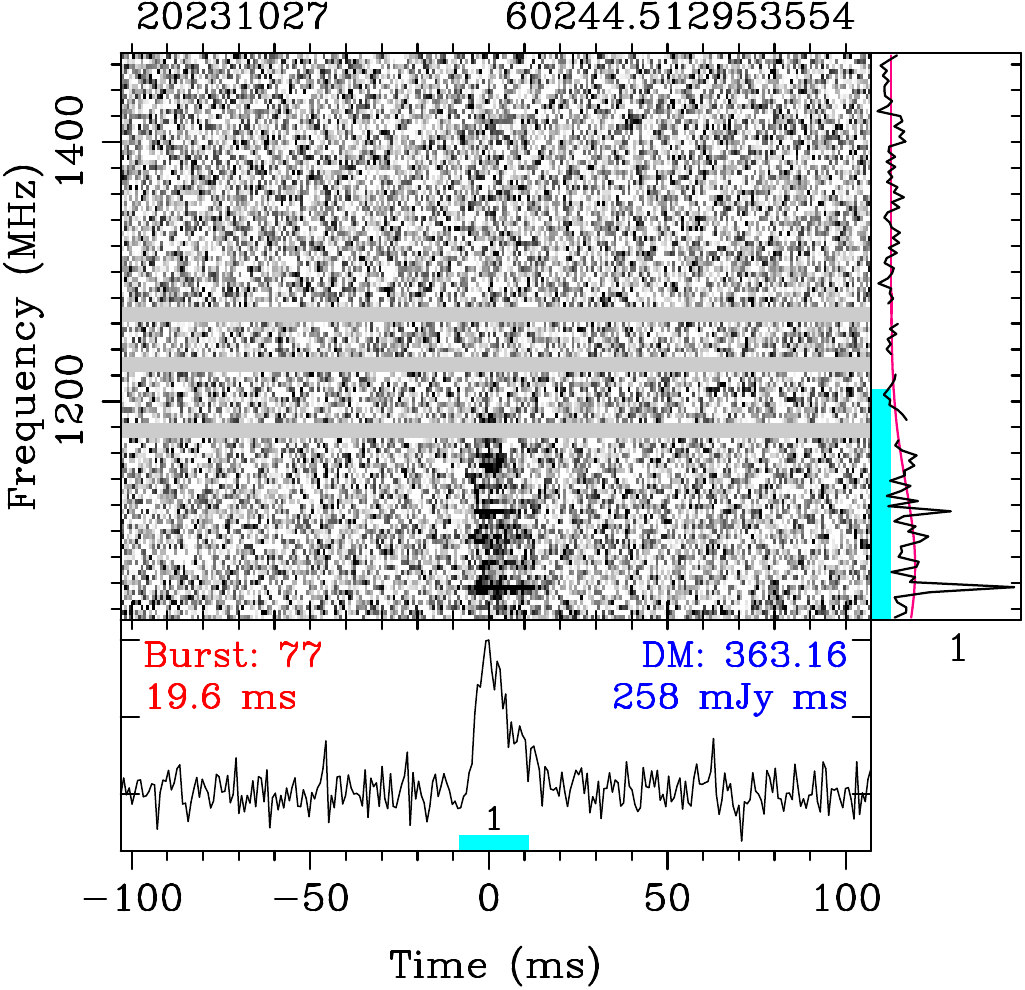}
\includegraphics[height=0.29\linewidth]{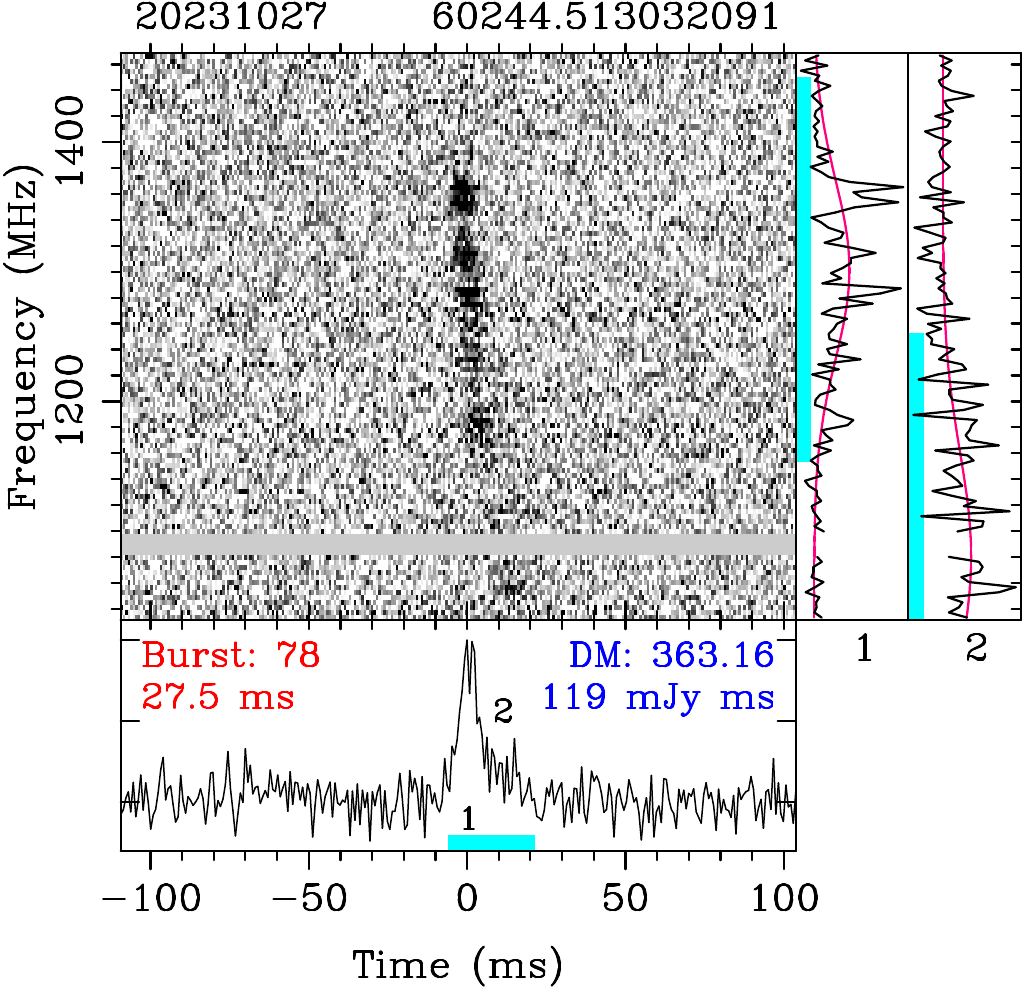}
\includegraphics[height=0.29\linewidth]{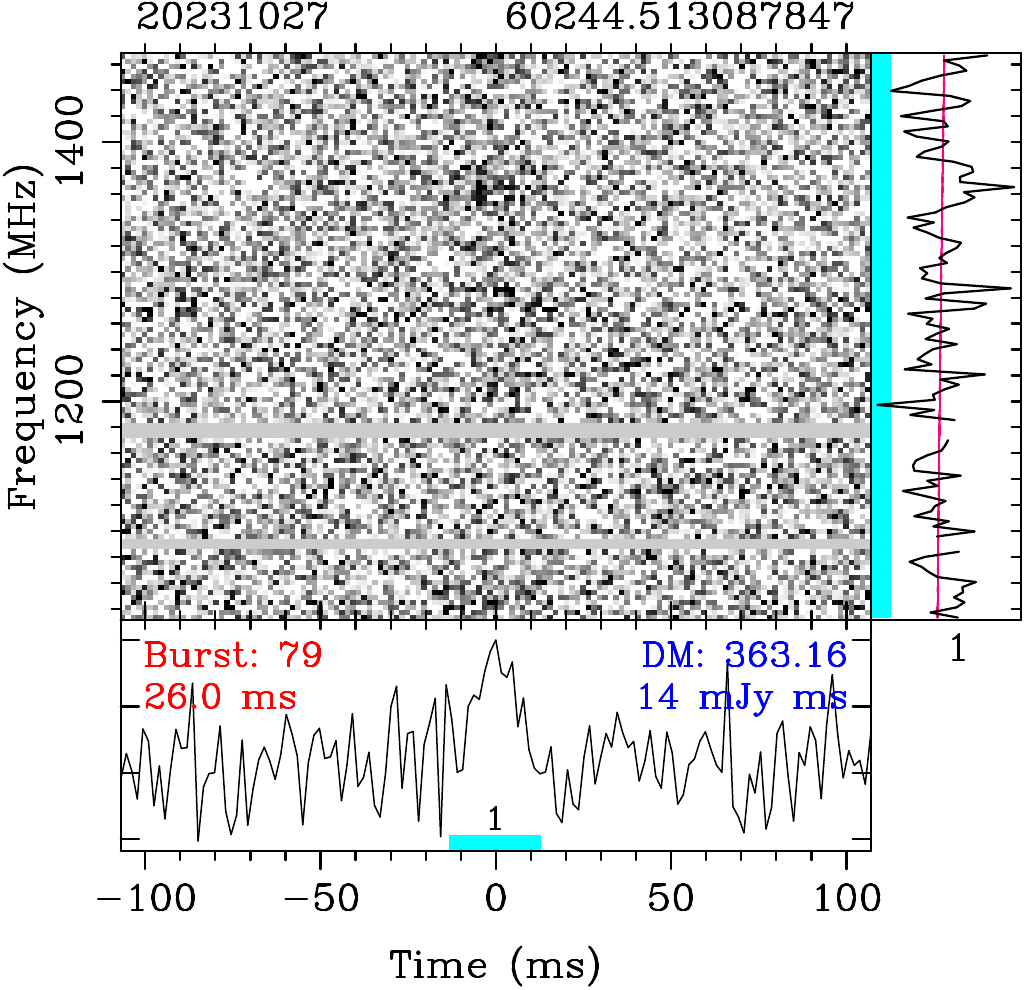}
\includegraphics[height=0.29\linewidth]{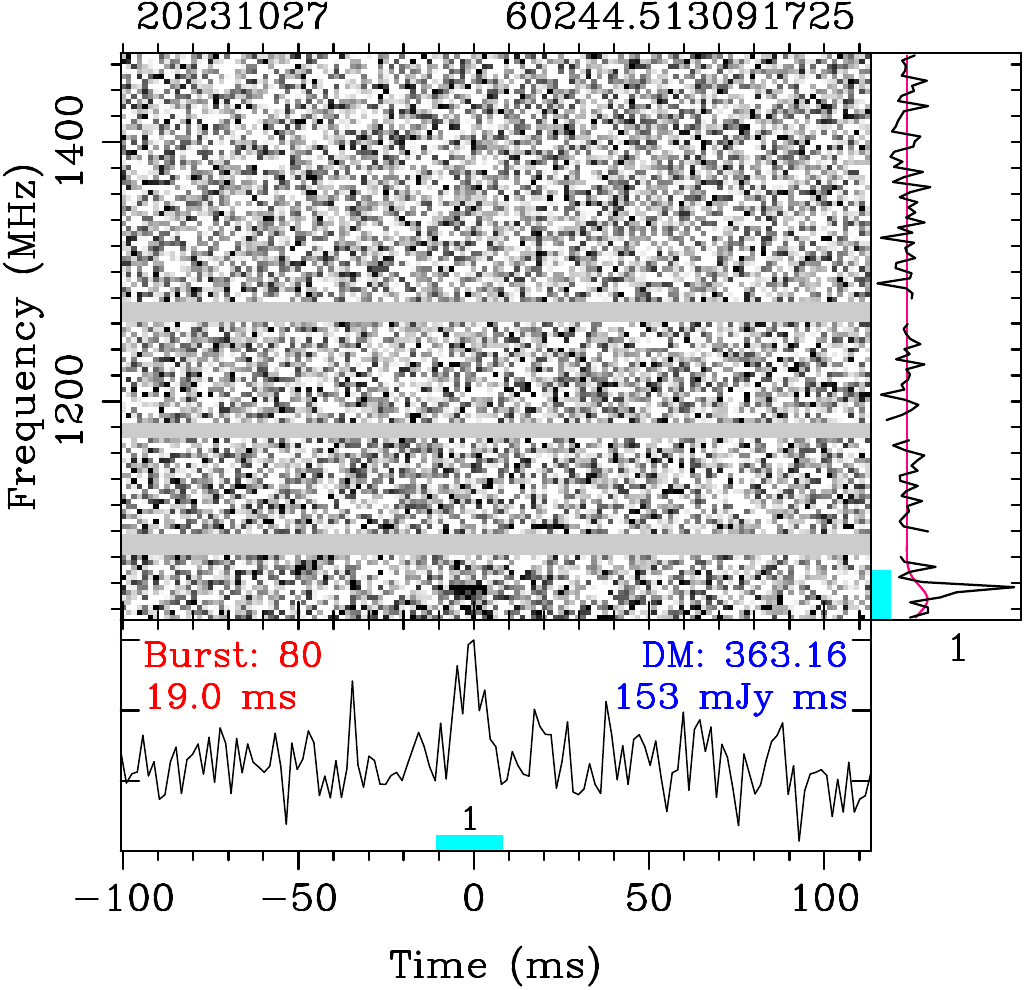}
\includegraphics[height=0.29\linewidth]{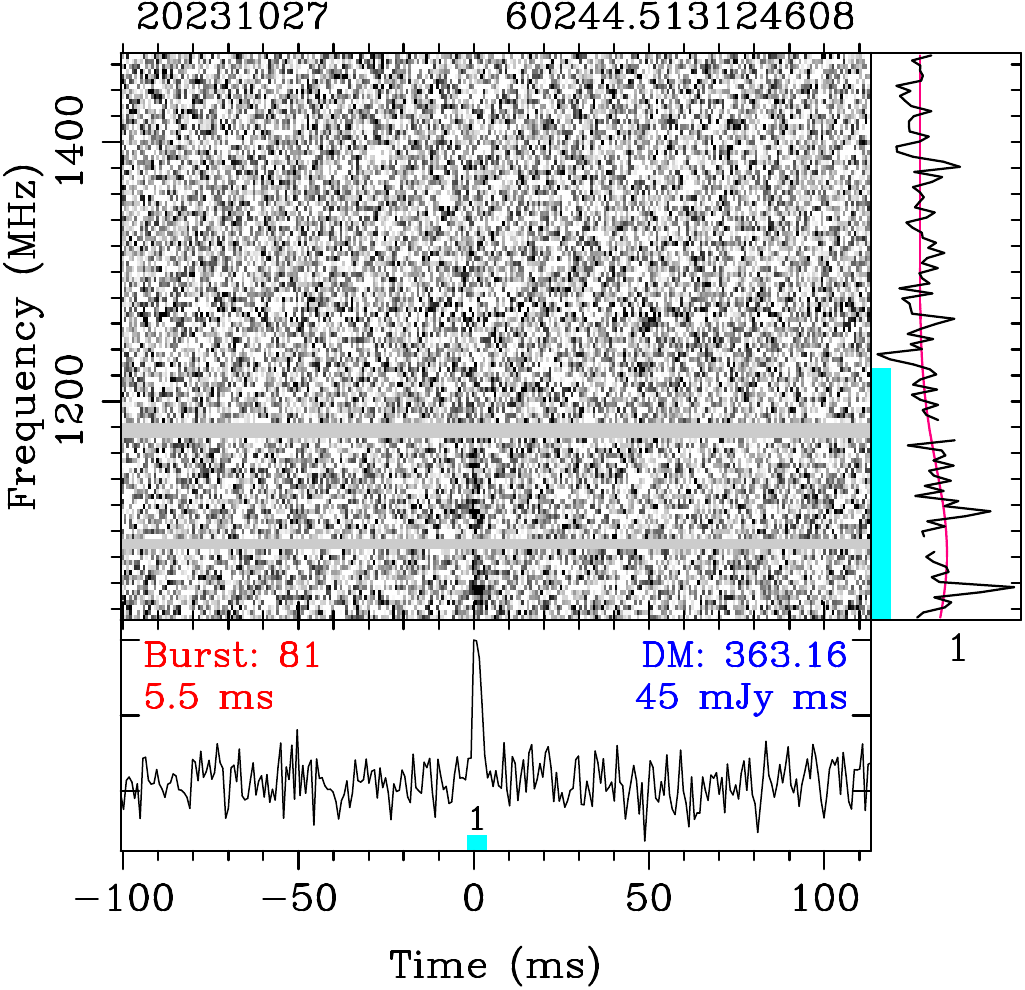}
\includegraphics[height=0.29\linewidth]{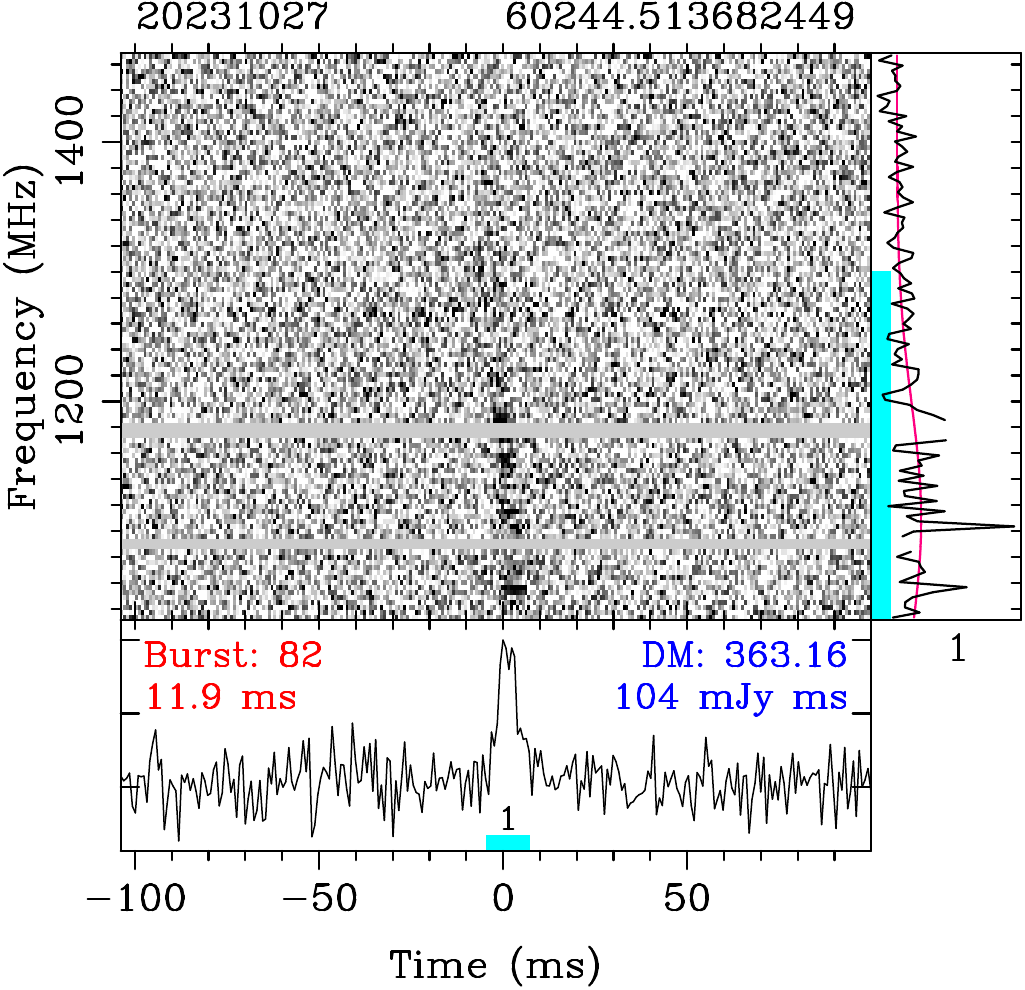}
\includegraphics[height=0.29\linewidth]{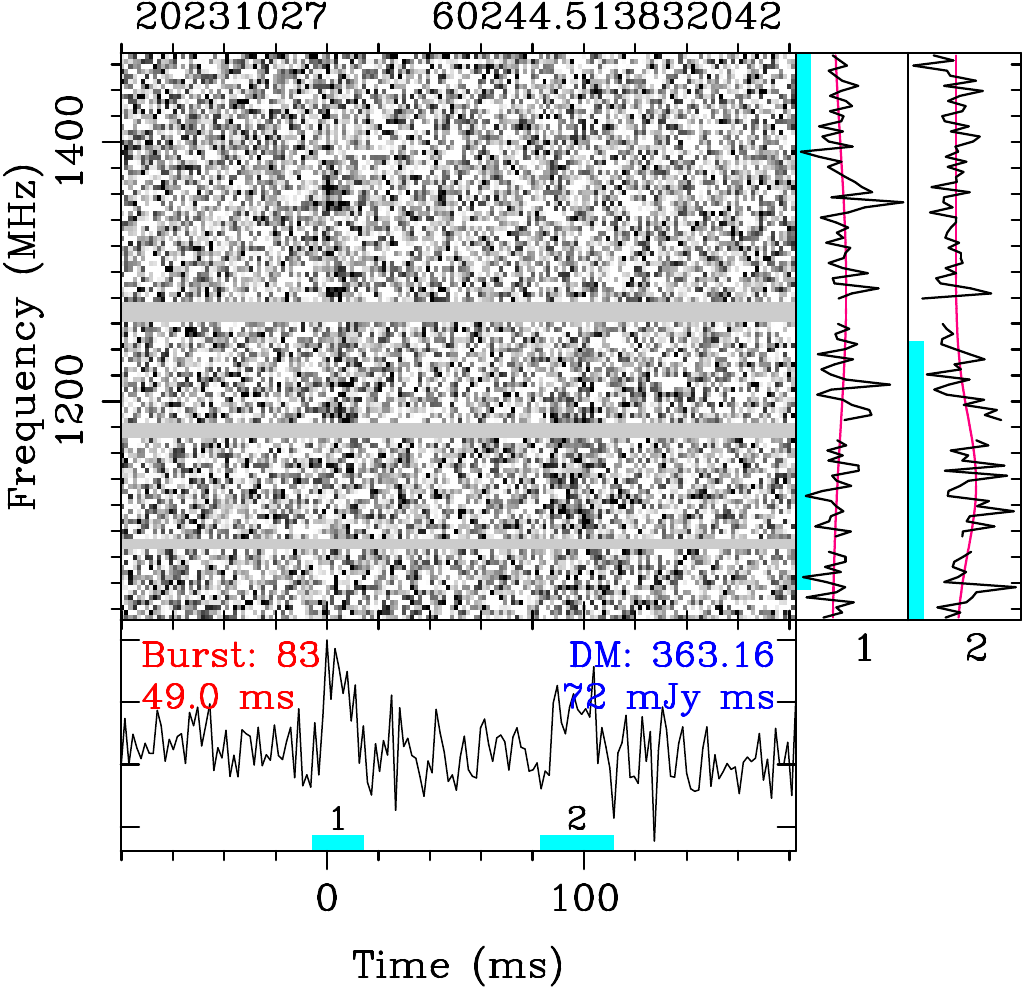}
\caption{({\textit{continued}})}
\end{figure*}
\addtocounter{figure}{-1}
\begin{figure*}
\flushleft
\includegraphics[height=0.29\linewidth]{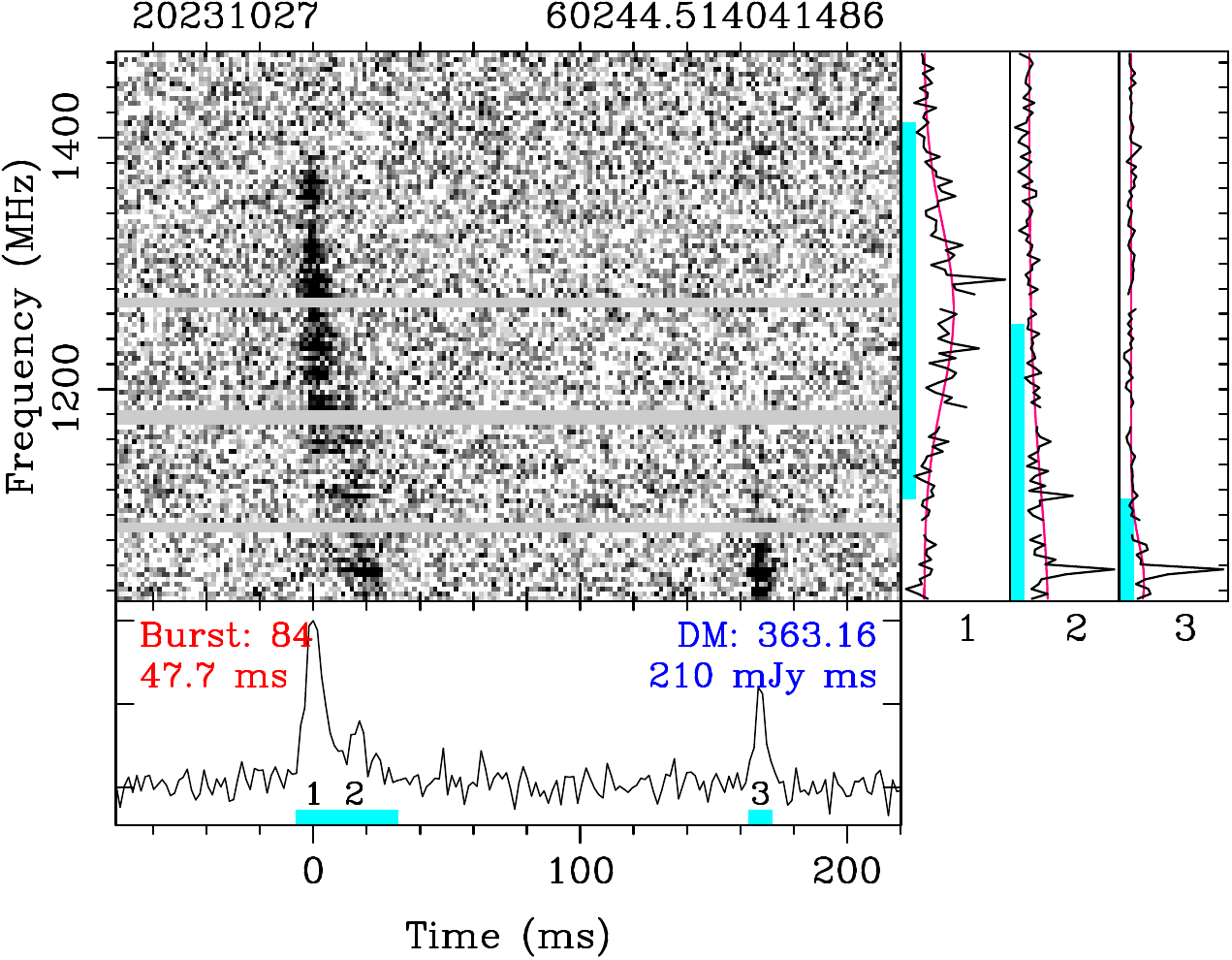}
\includegraphics[height=0.29\linewidth]{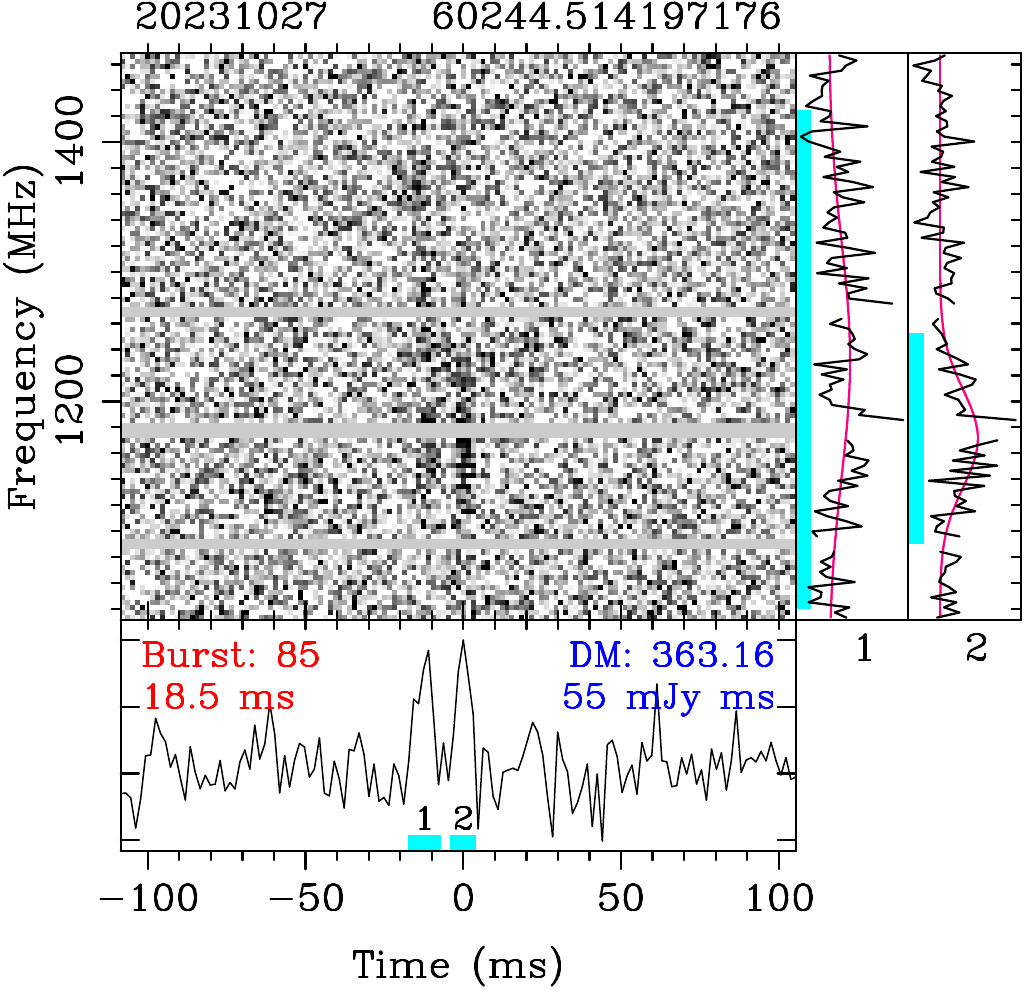}
\includegraphics[height=0.29\linewidth]{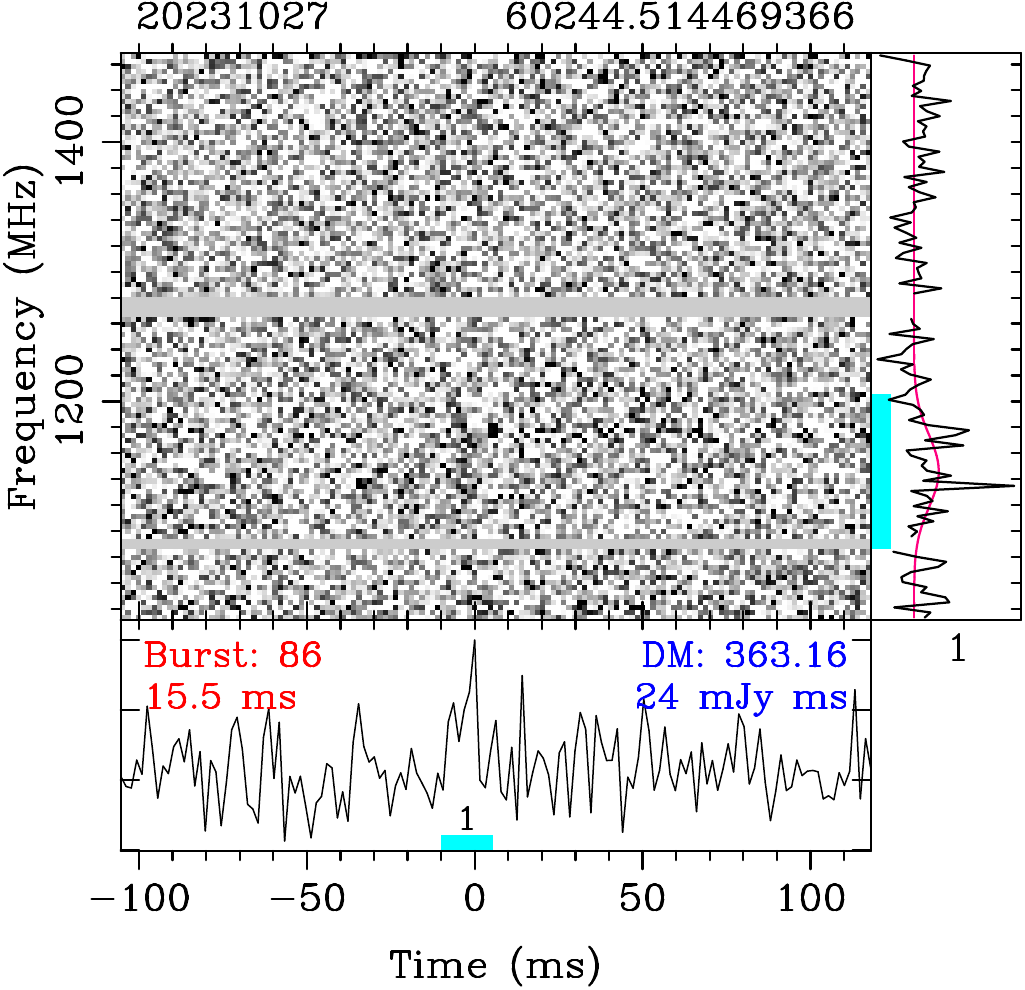}
\includegraphics[height=0.29\linewidth]{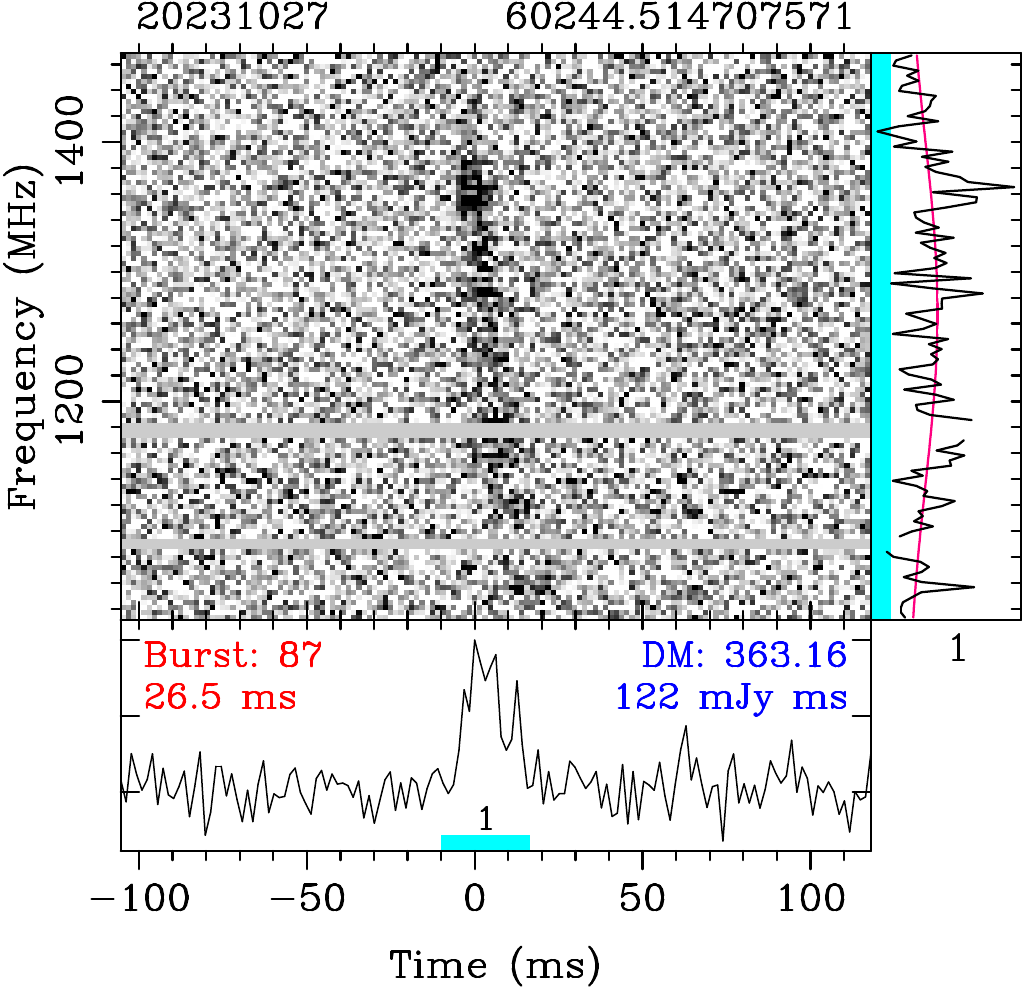}
\includegraphics[height=0.29\linewidth]{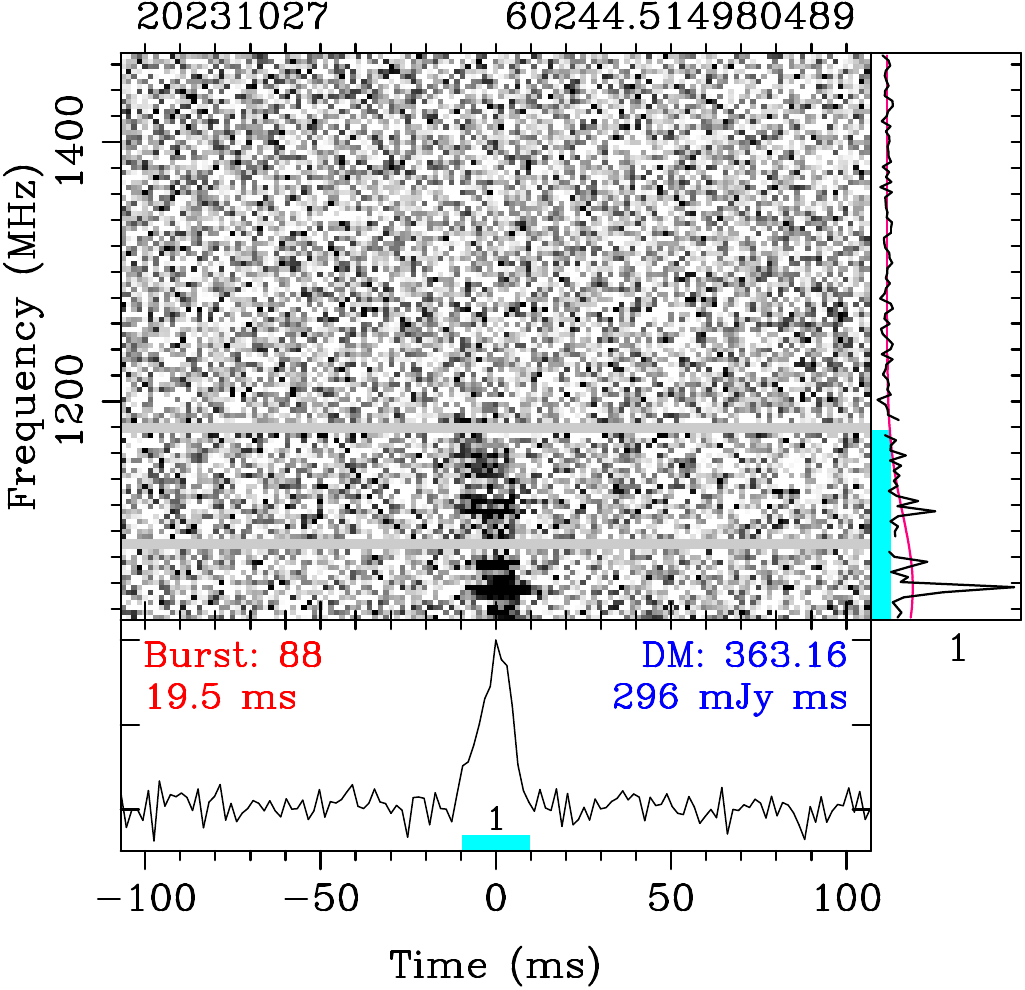}
\includegraphics[height=0.29\linewidth]{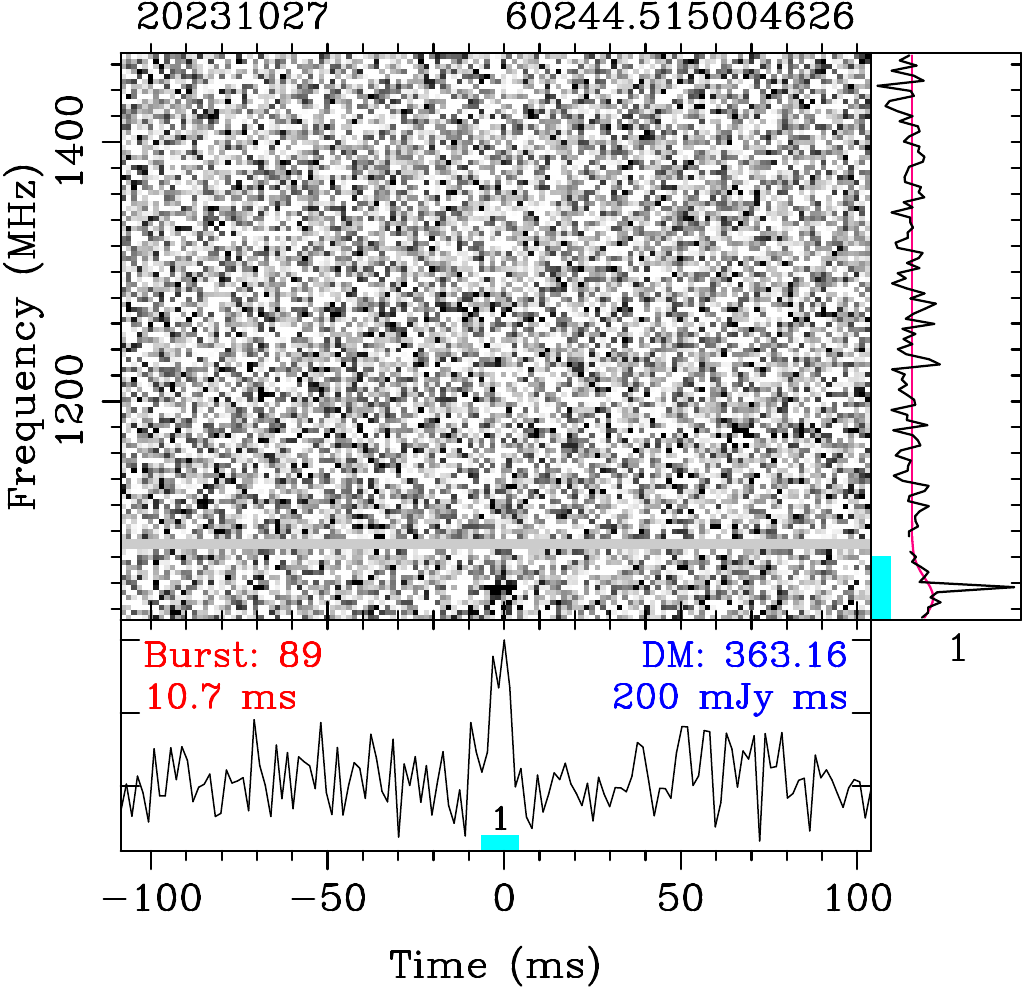}
\includegraphics[height=0.29\linewidth]{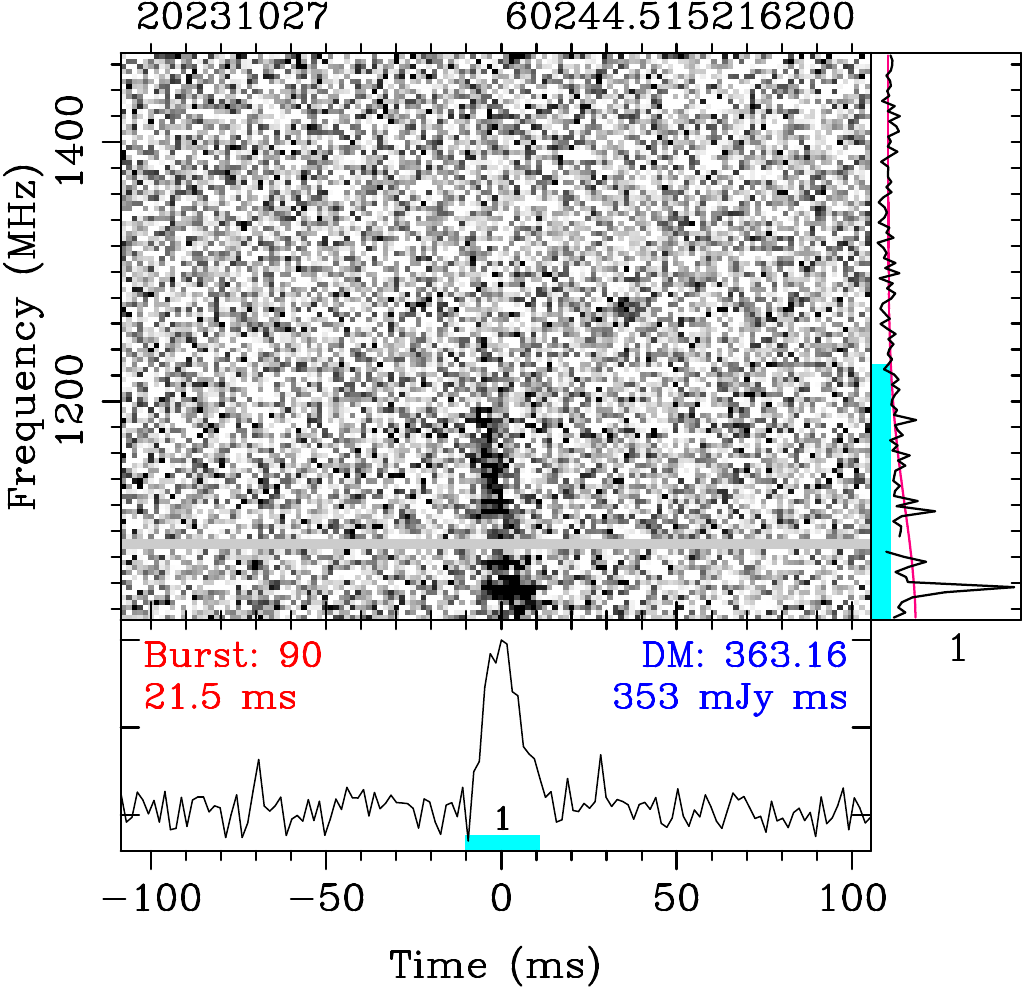}
\includegraphics[height=0.29\linewidth]{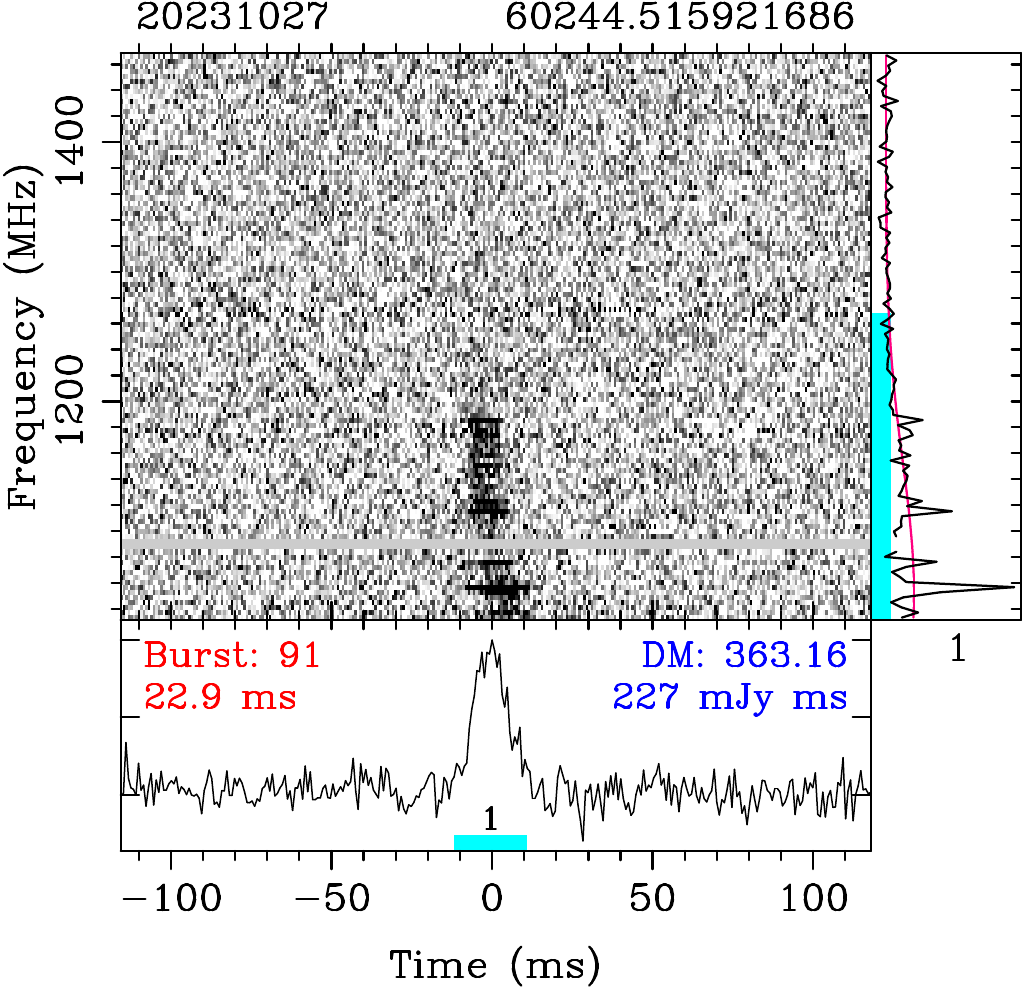}
\includegraphics[height=0.29\linewidth]{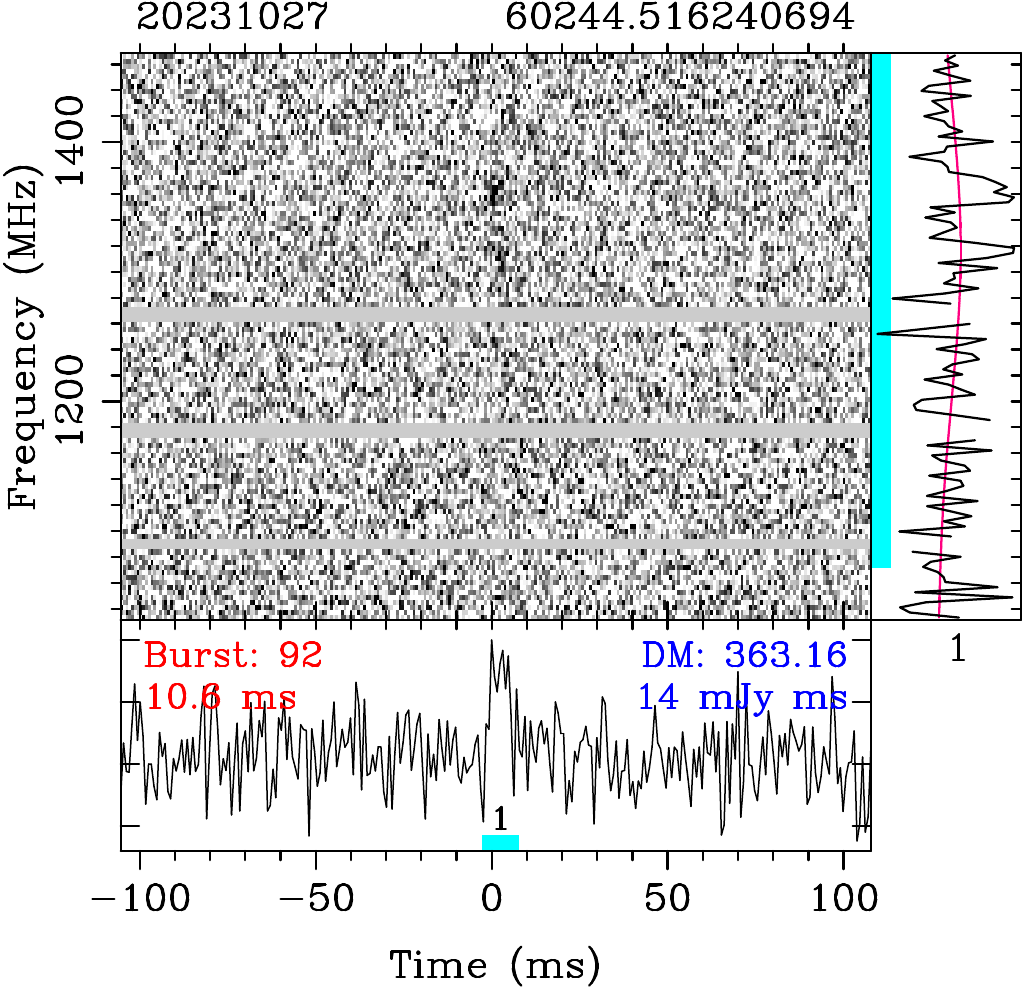}
\includegraphics[height=0.29\linewidth]{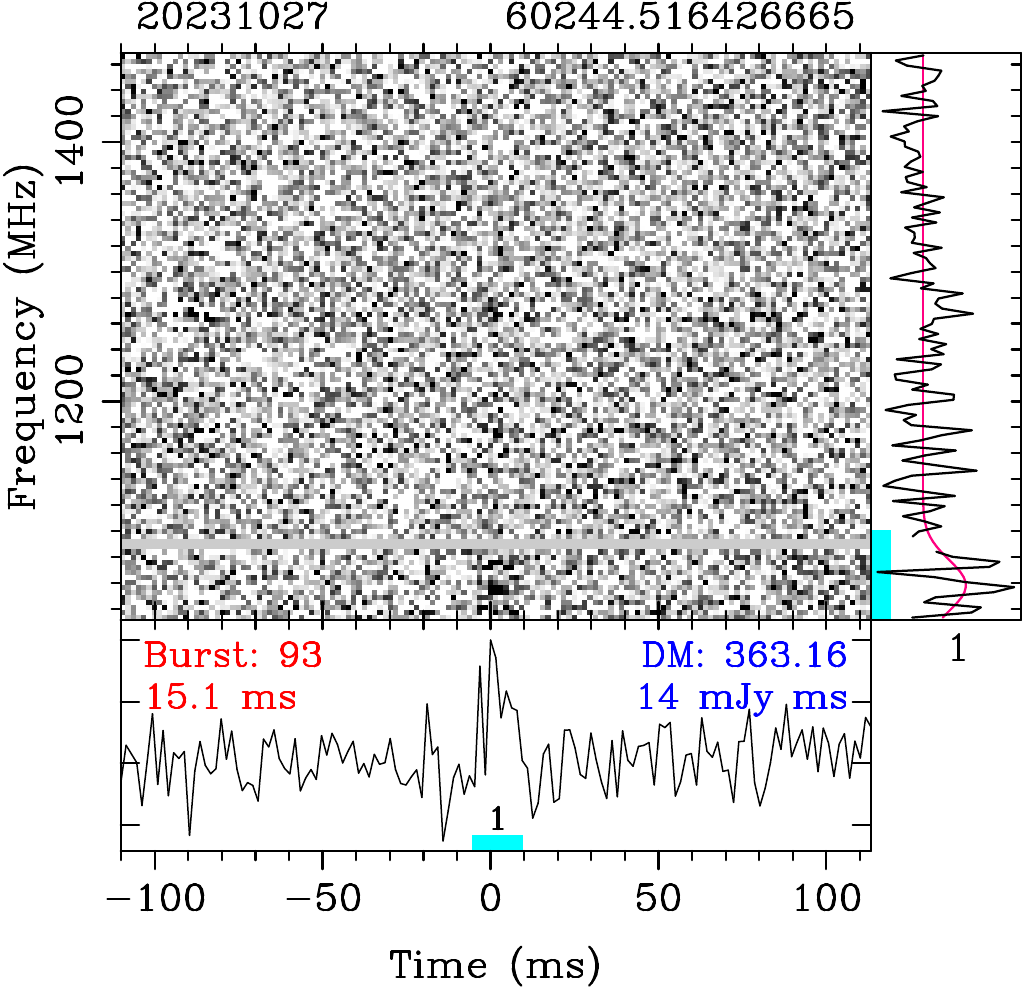}
\includegraphics[height=0.29\linewidth]{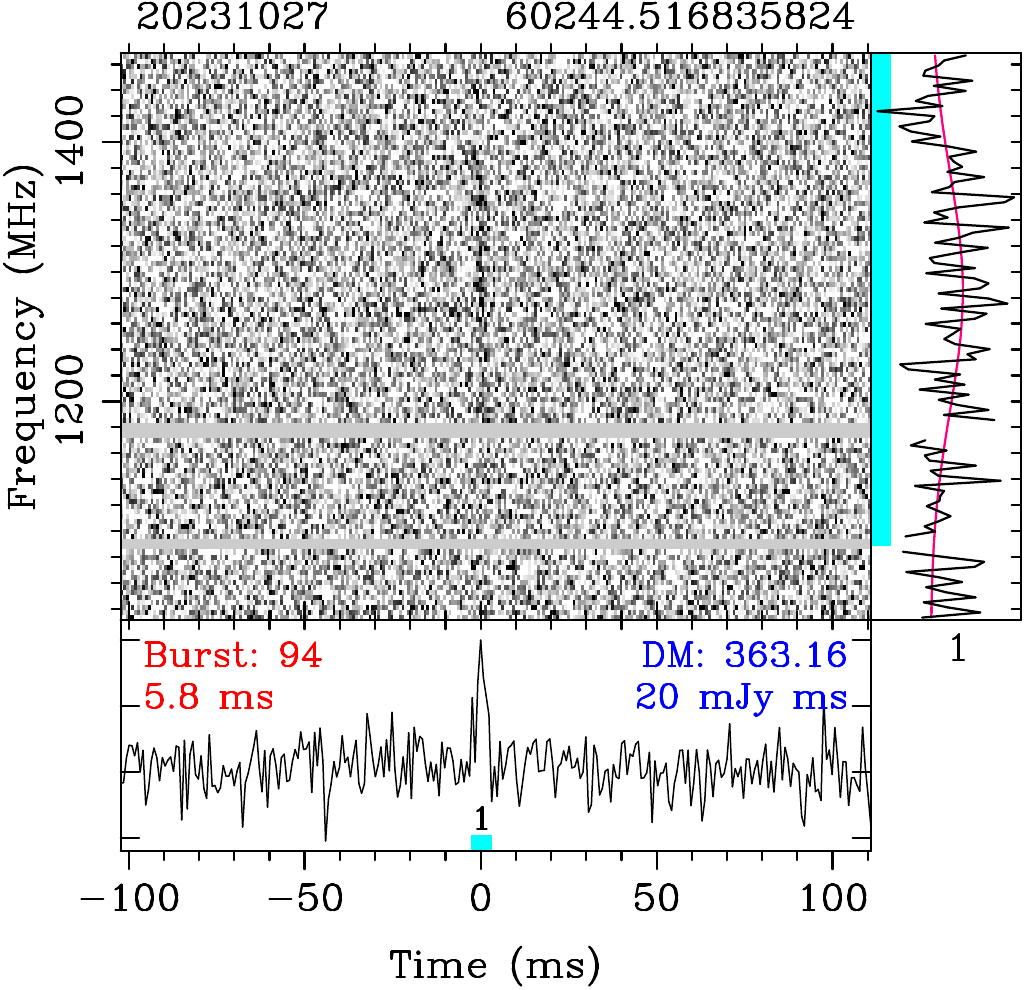}
\includegraphics[height=0.29\linewidth]{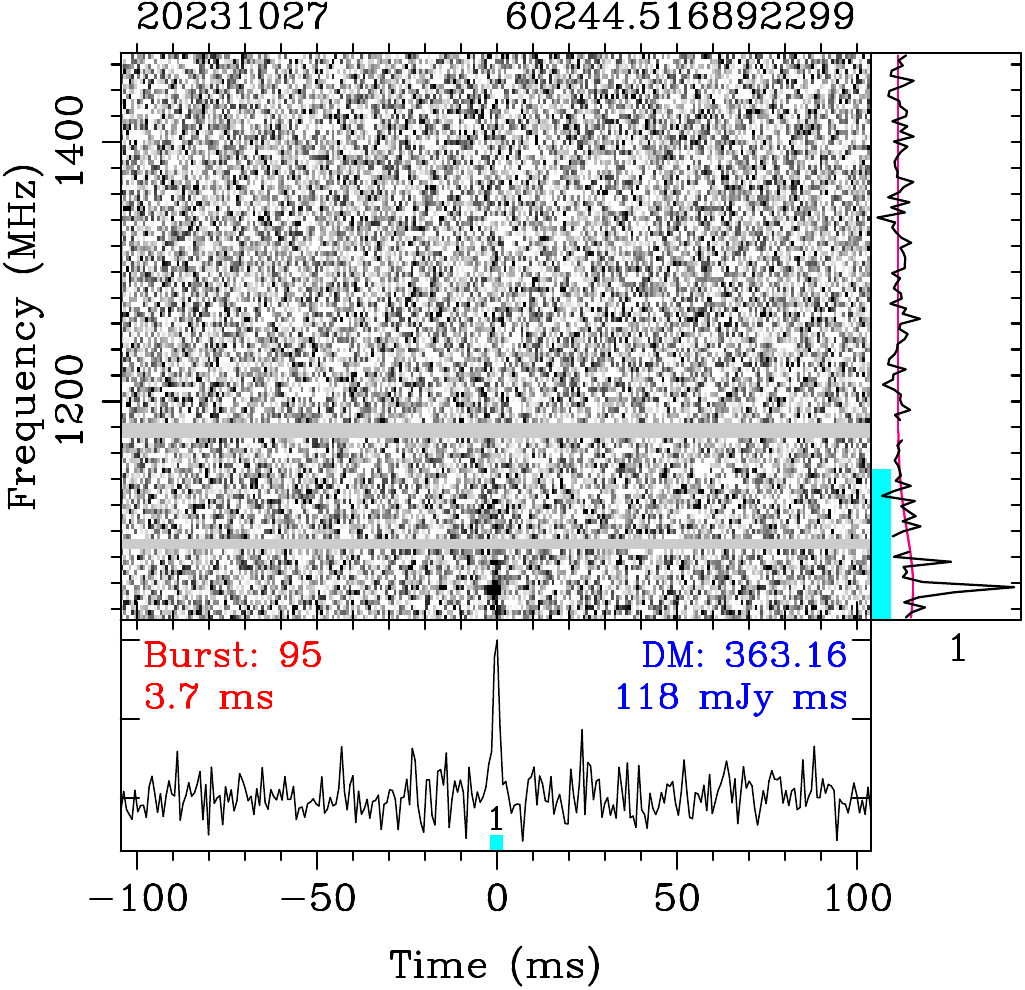}
\caption{({\textit{continued}})}
\end{figure*}
\addtocounter{figure}{-1}
\begin{figure*}
\flushleft
\includegraphics[height=0.29\linewidth]{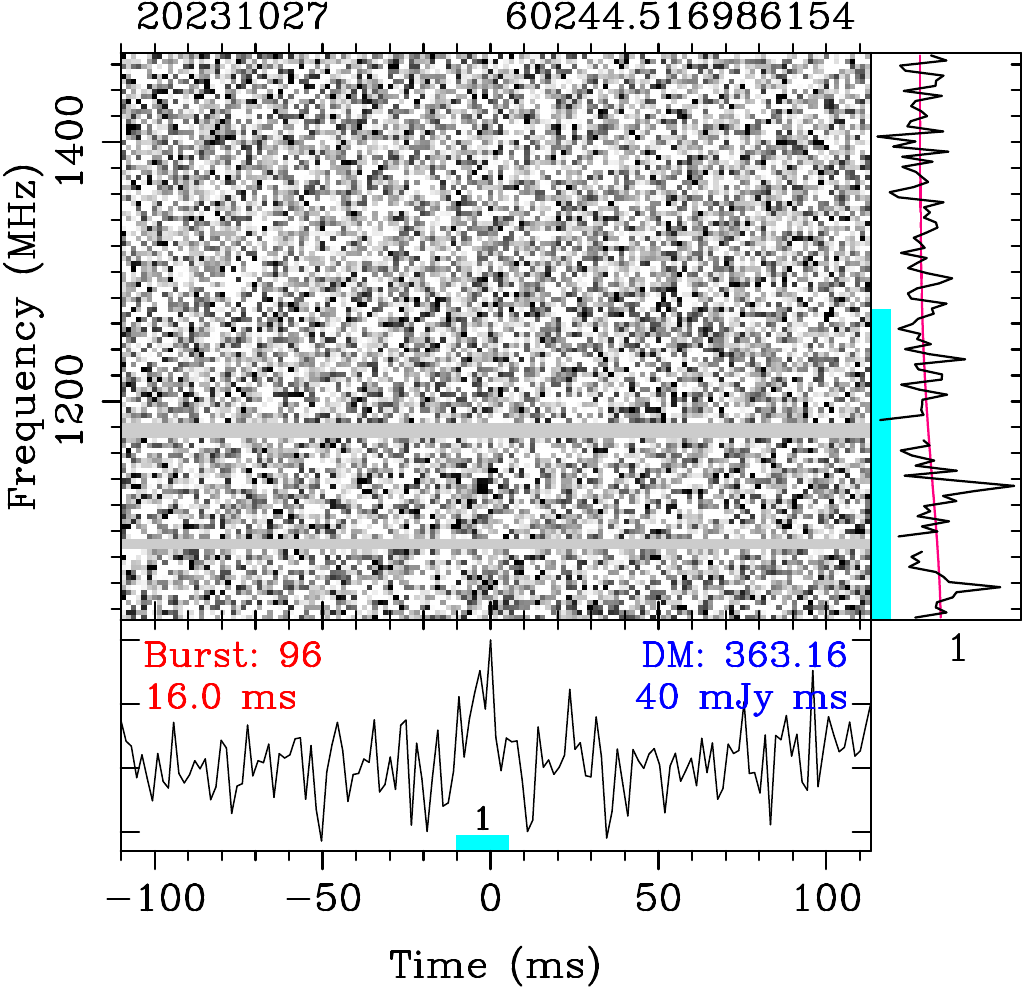}
\includegraphics[height=0.29\linewidth]{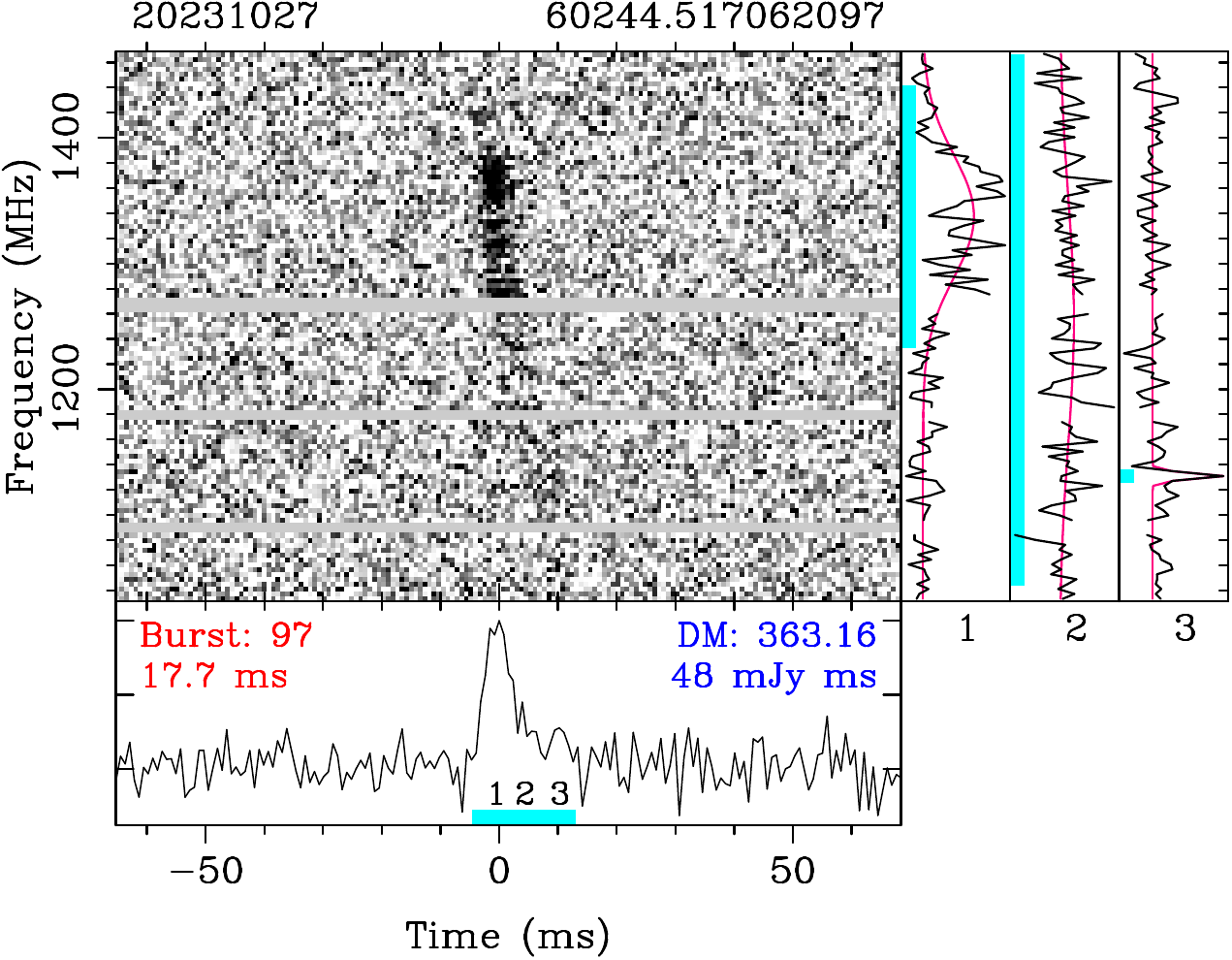}
\includegraphics[height=0.29\linewidth]{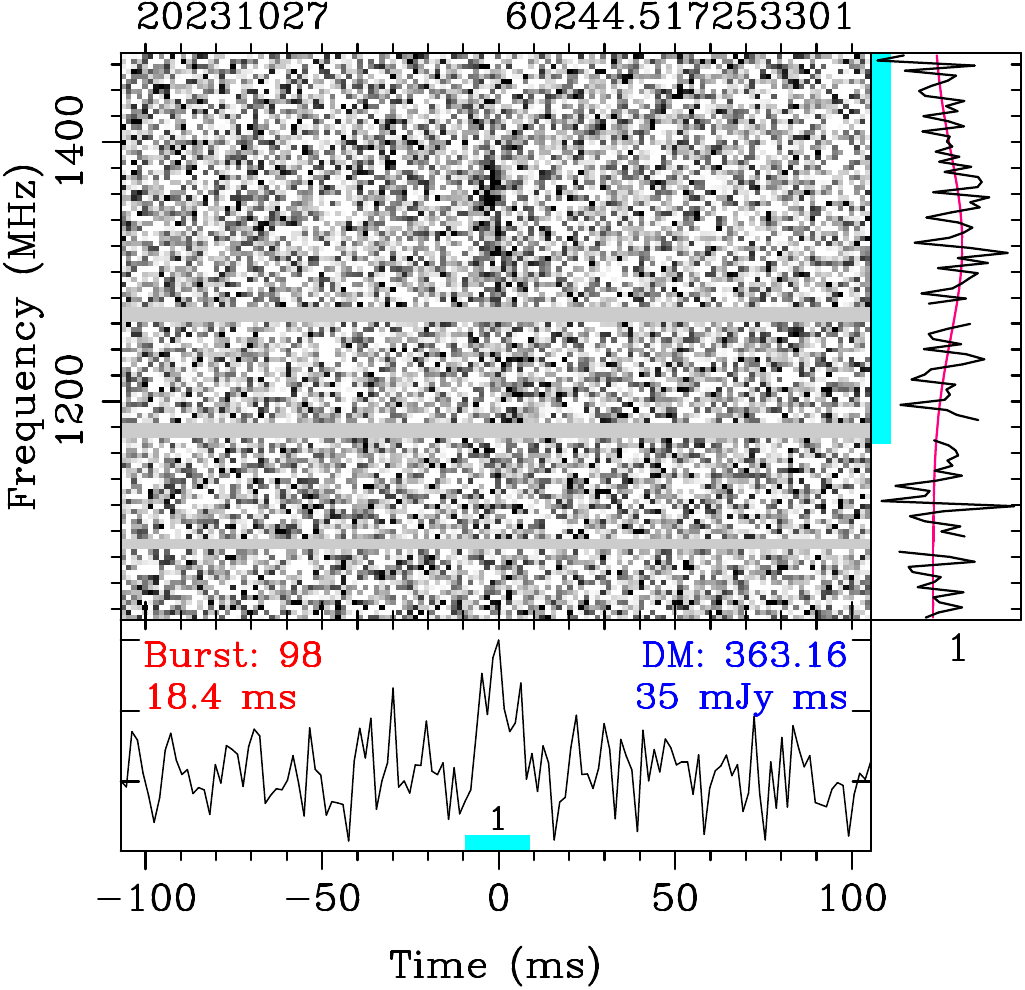}
\includegraphics[height=0.29\linewidth]{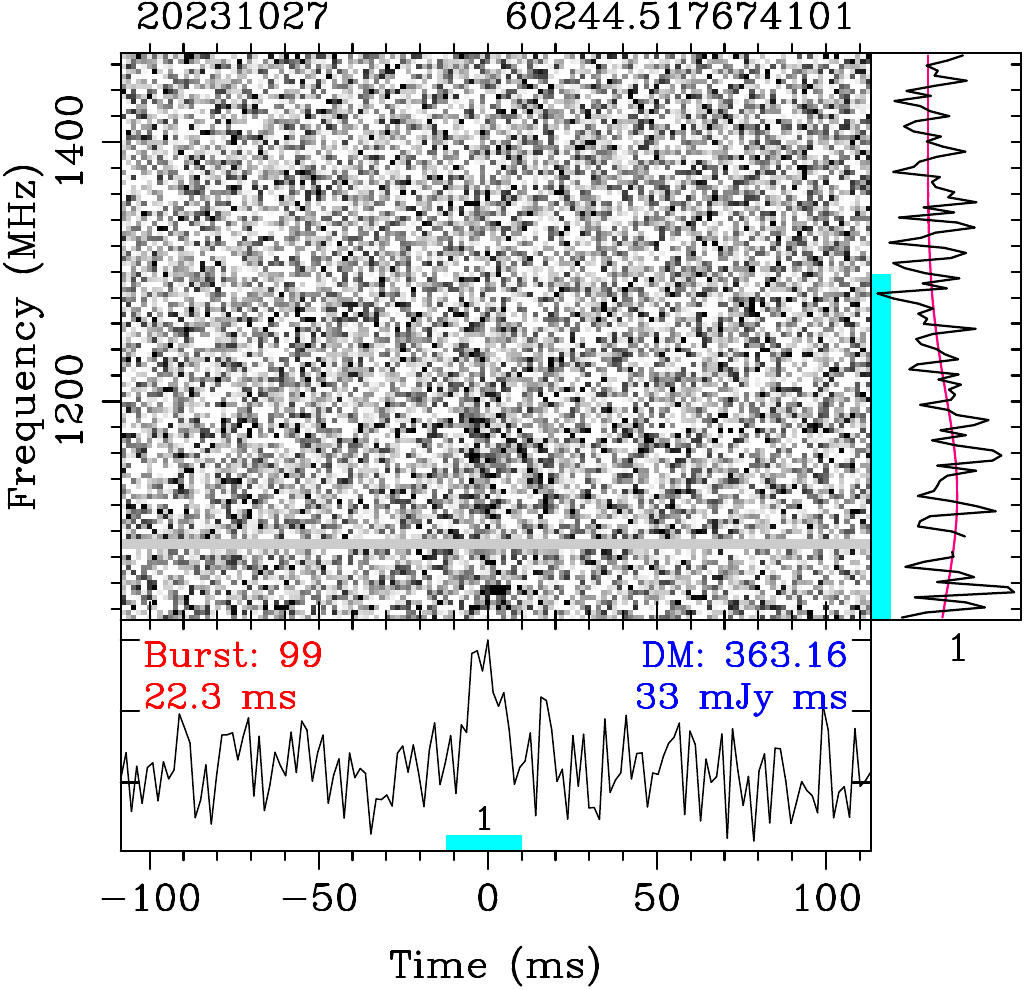}
\includegraphics[height=0.29\linewidth]{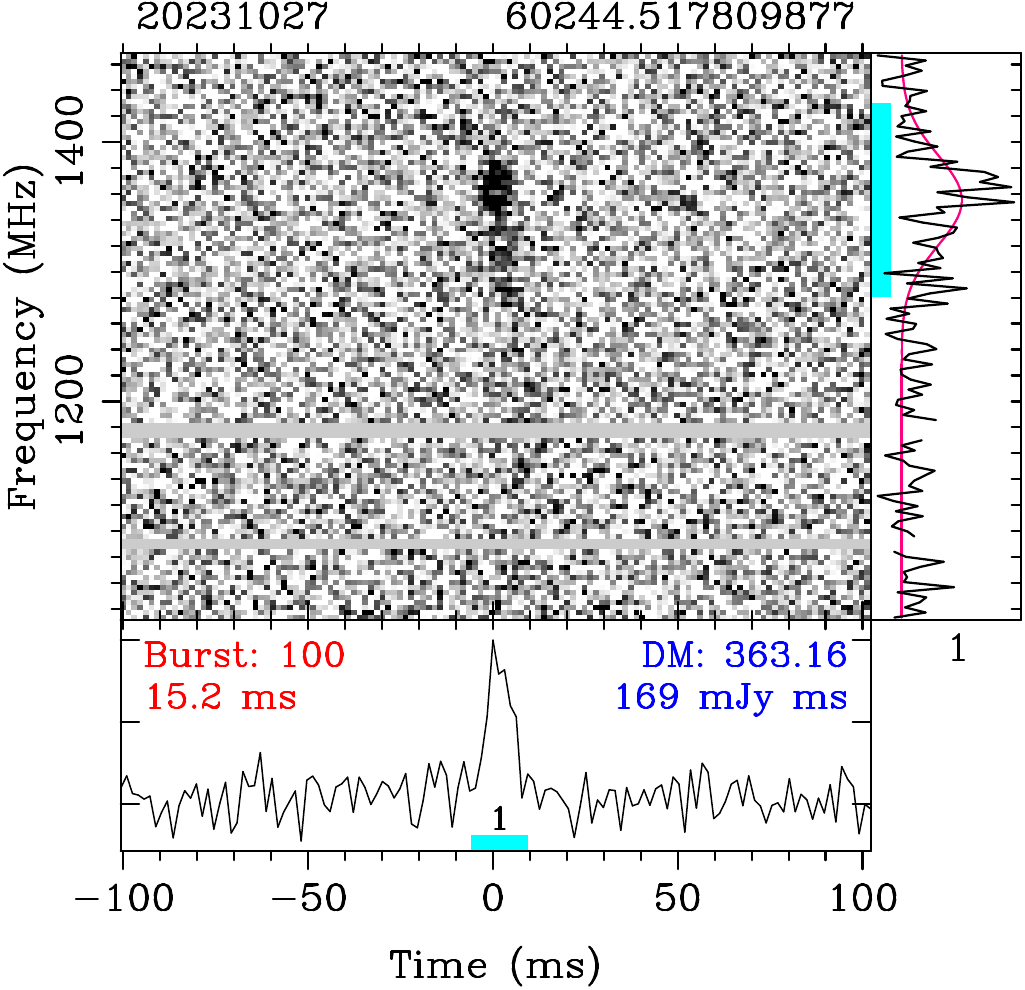}
\includegraphics[height=0.29\linewidth]{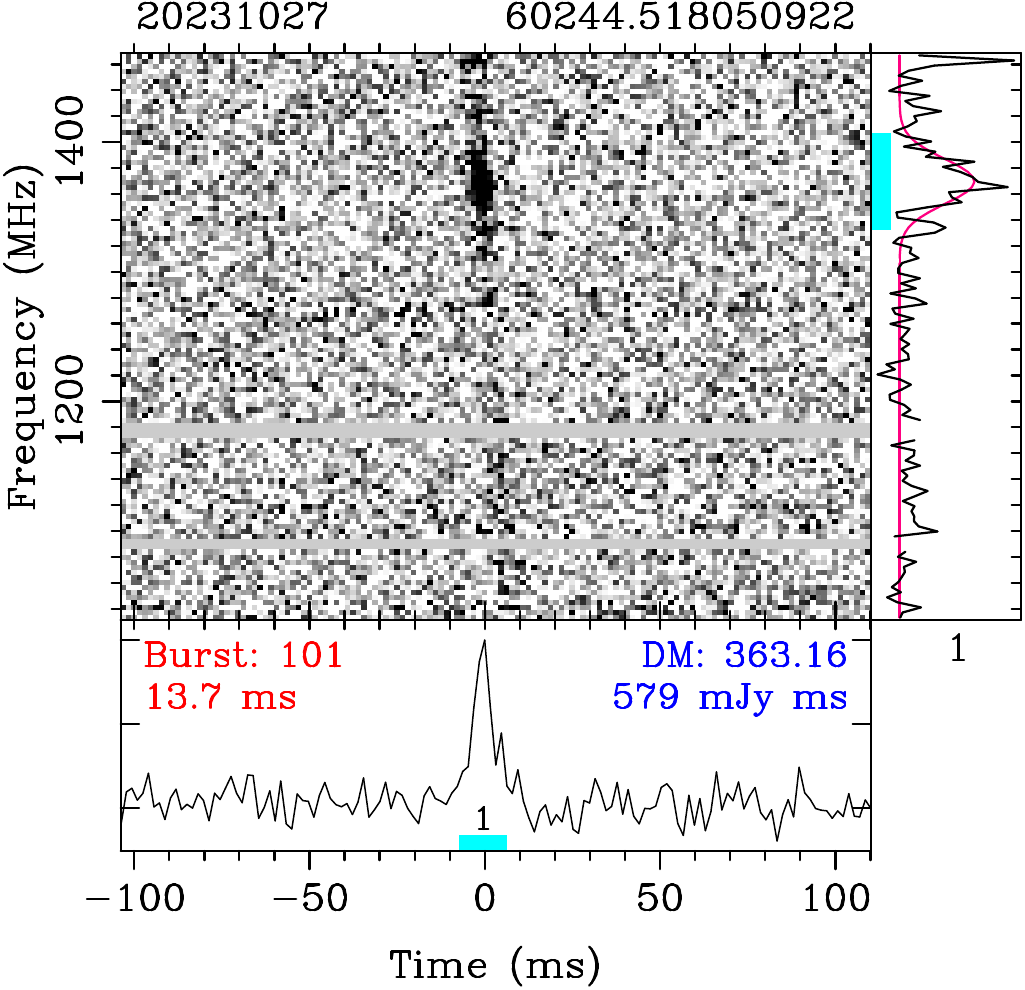}
\includegraphics[height=0.29\linewidth]{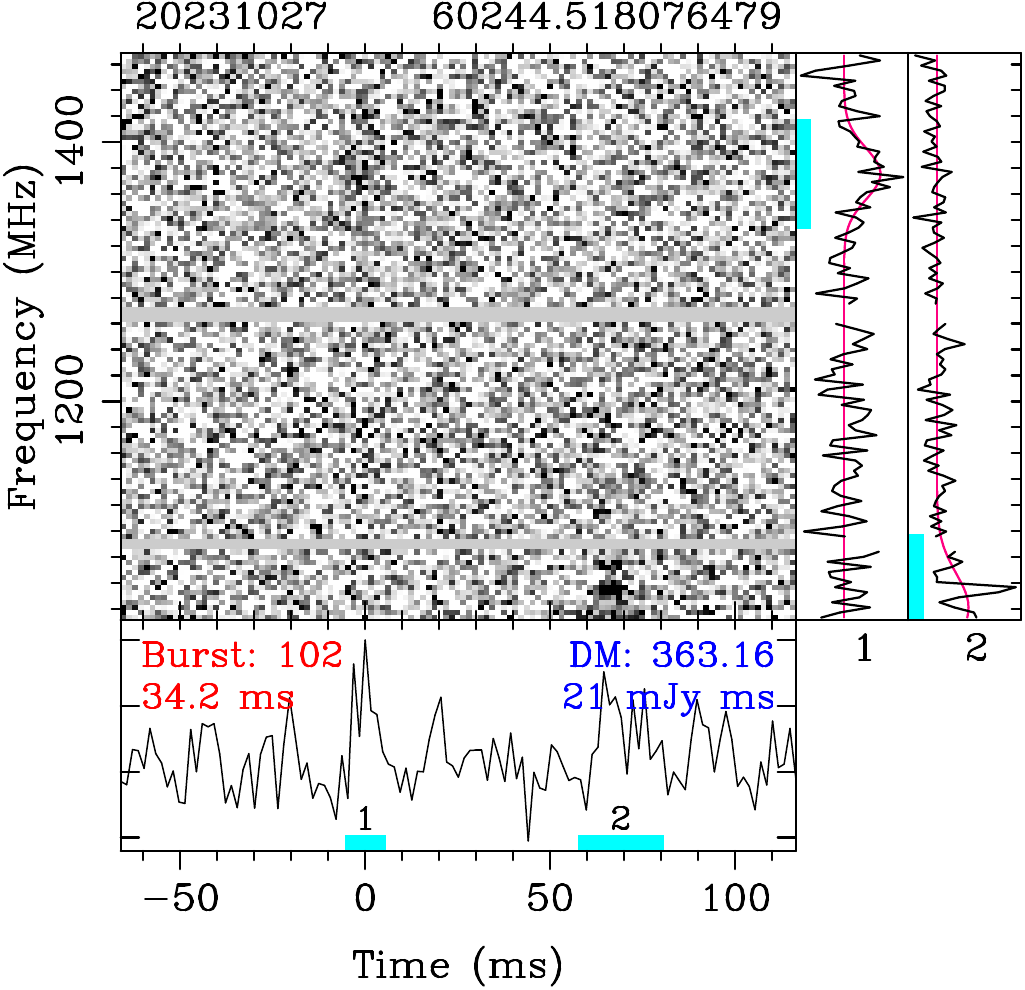}
\includegraphics[height=0.29\linewidth]{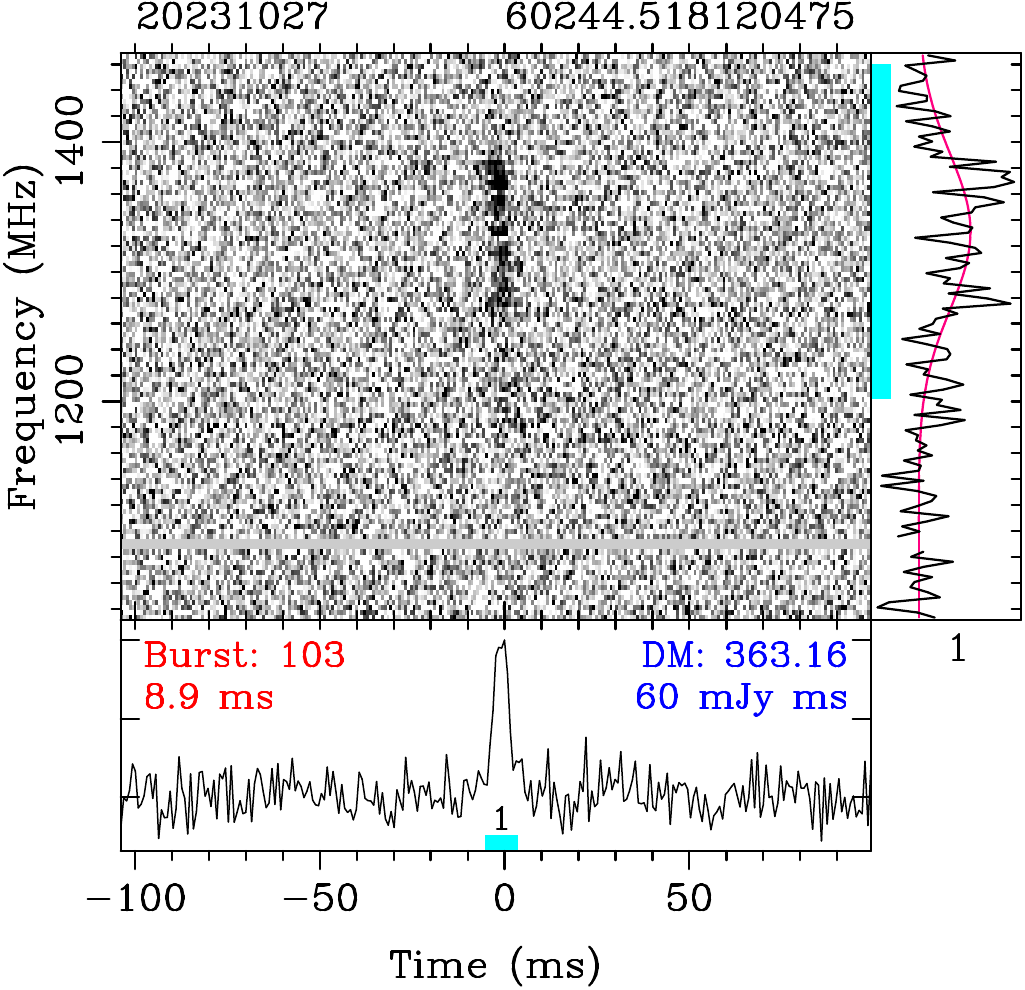}
\includegraphics[height=0.29\linewidth]{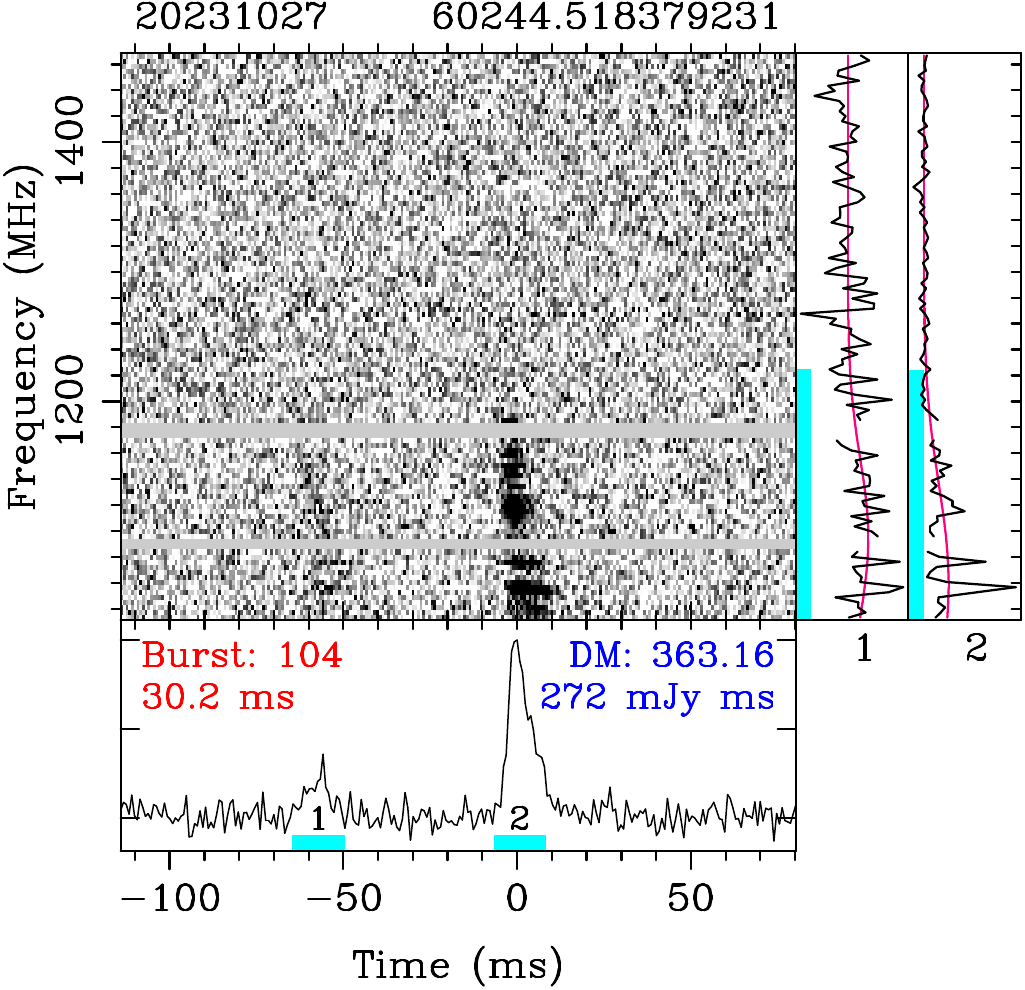}
\includegraphics[height=0.29\linewidth]{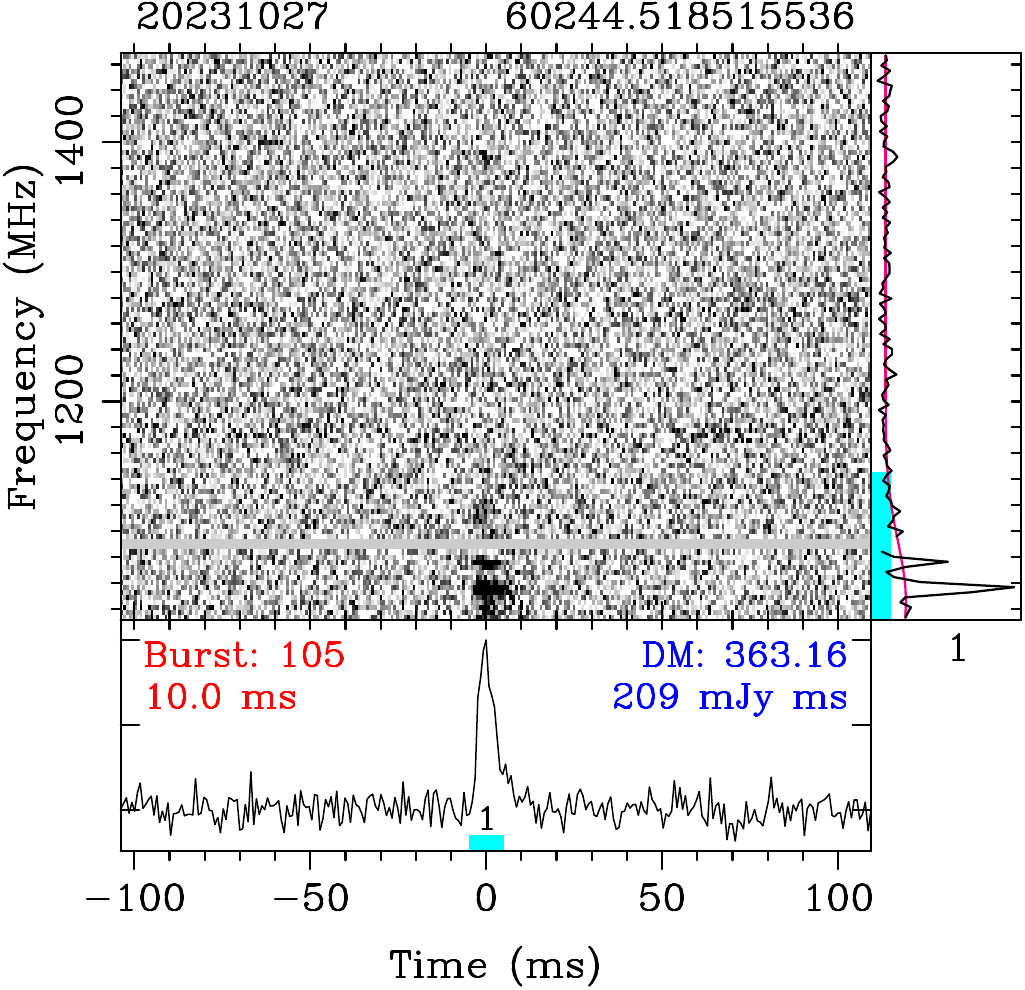}
\includegraphics[height=0.29\linewidth]{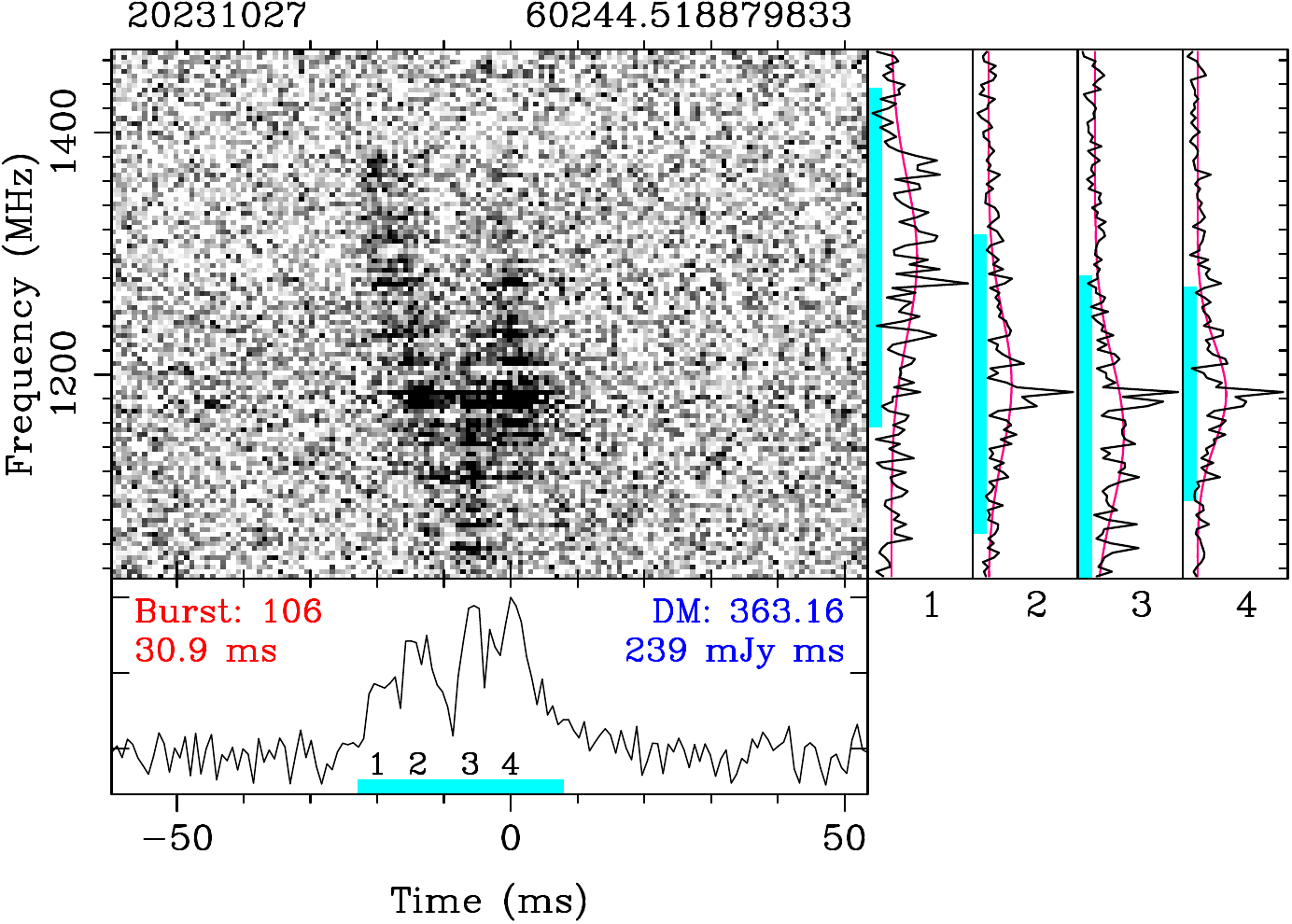}
\caption{({\textit{continued}})}
\end{figure*}
\addtocounter{figure}{-1}
\begin{figure*}
\flushleft
\includegraphics[height=0.29\linewidth]{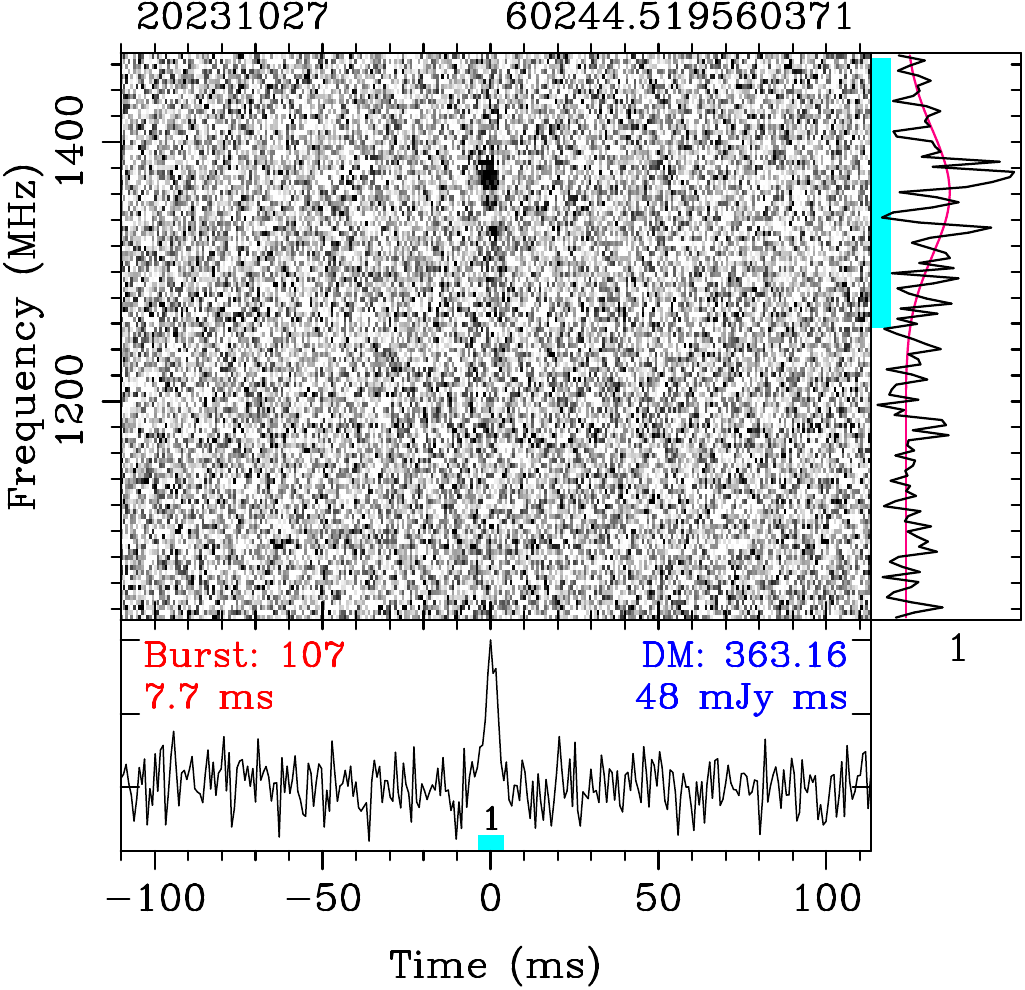}
\includegraphics[height=0.29\linewidth]{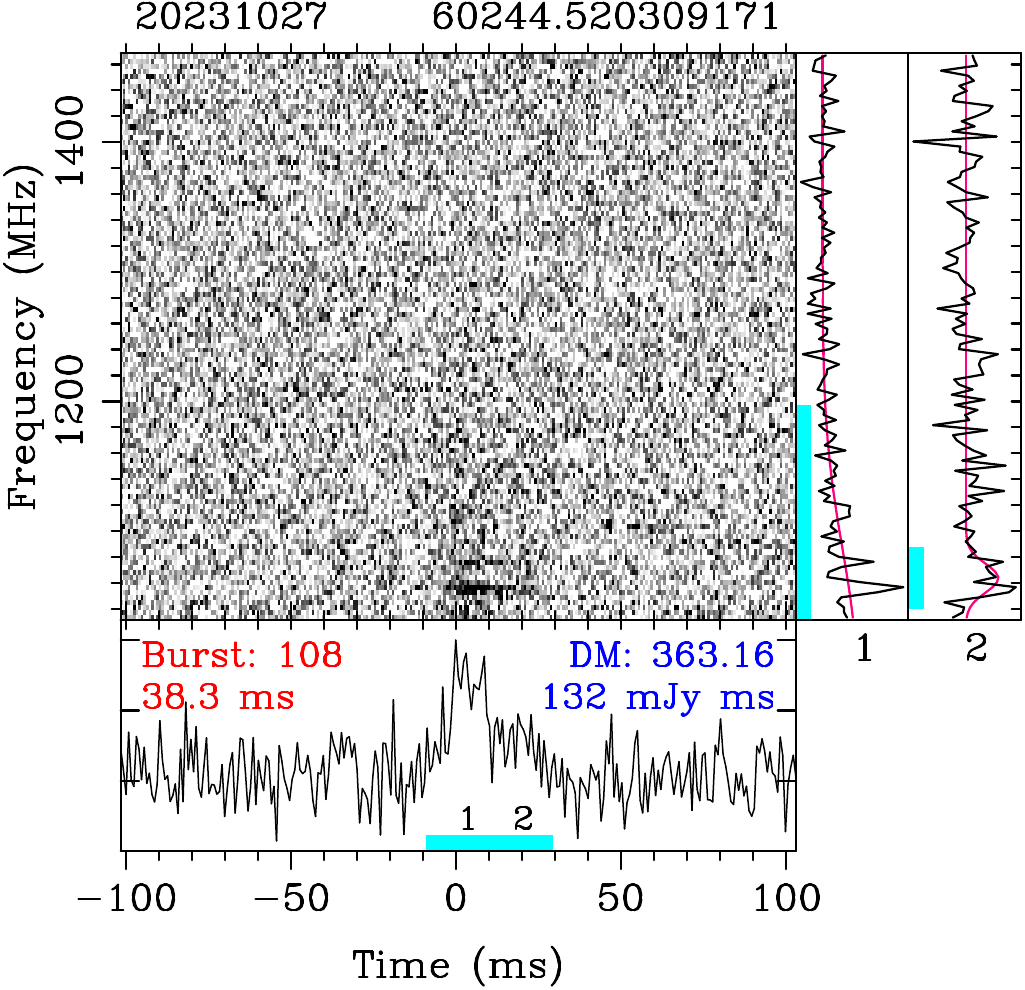}
\includegraphics[height=0.29\linewidth]{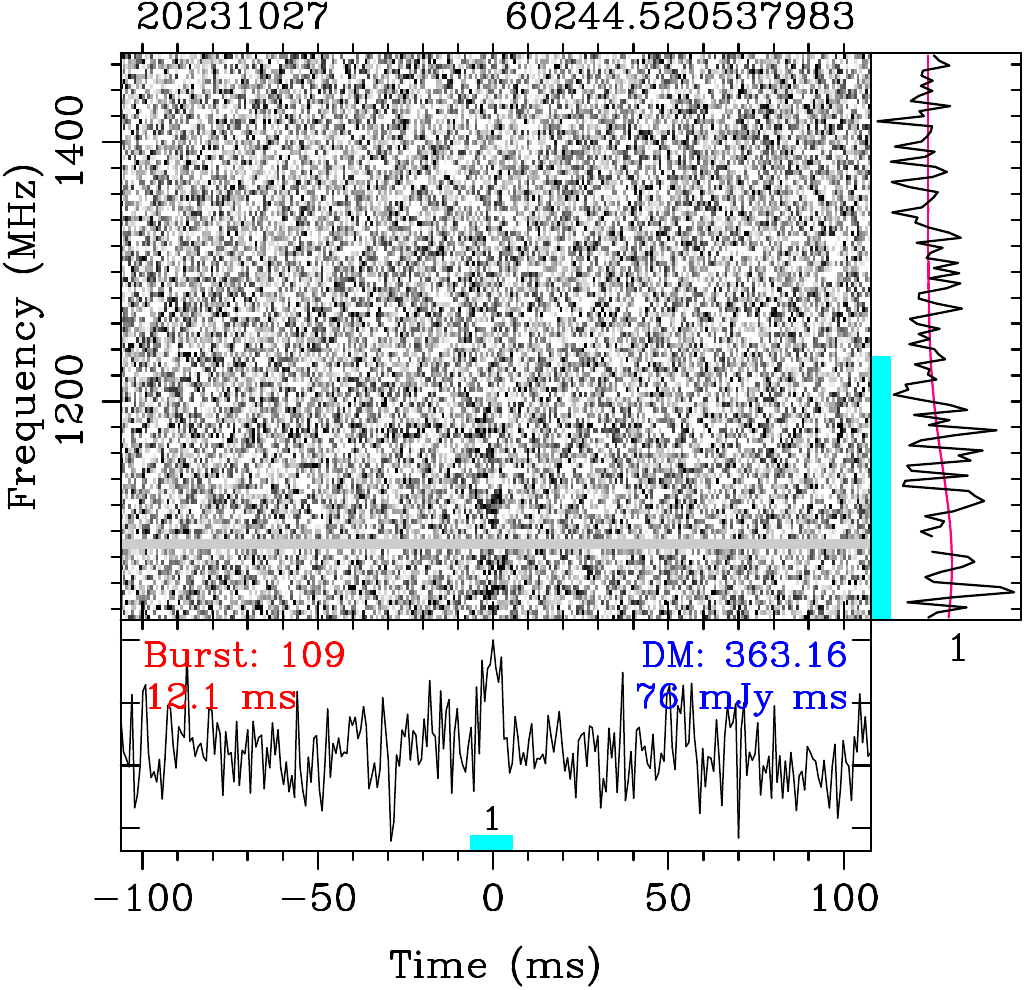}
\includegraphics[height=0.29\linewidth]{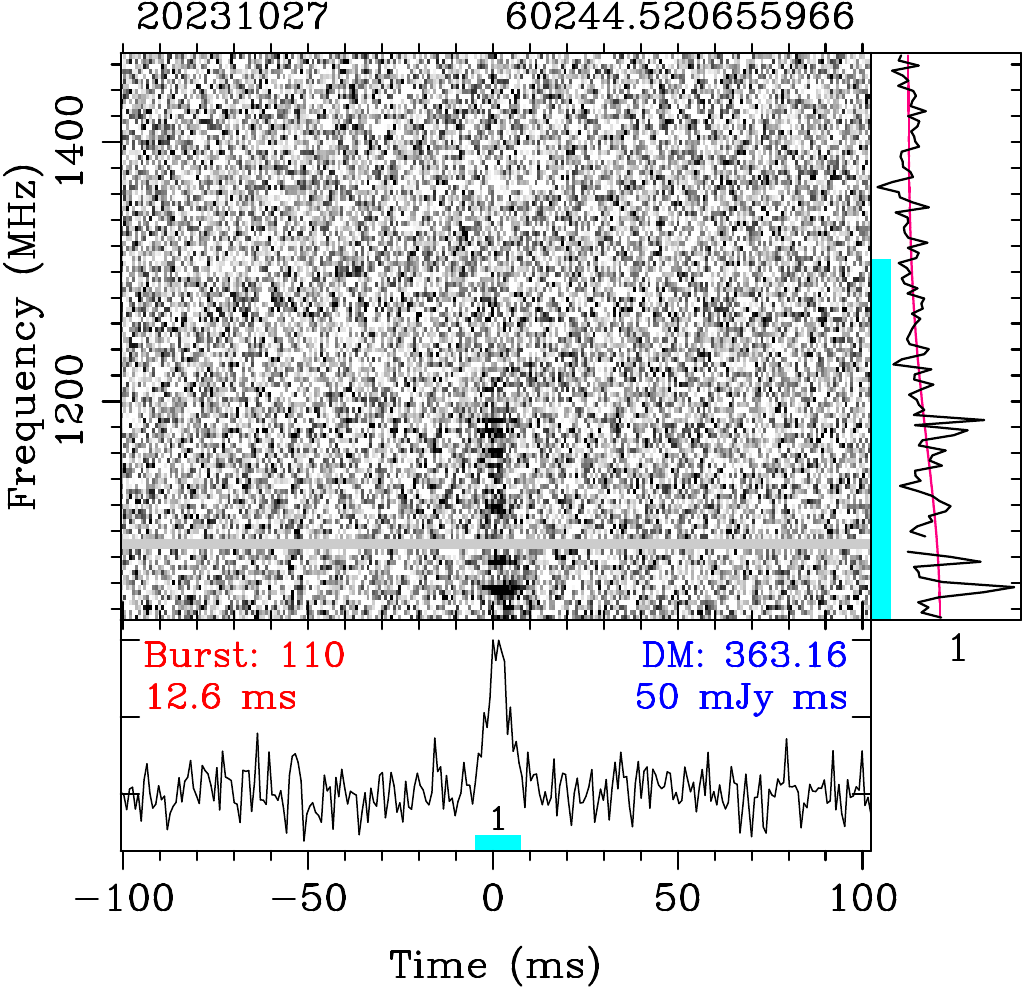}
\includegraphics[height=0.29\linewidth]{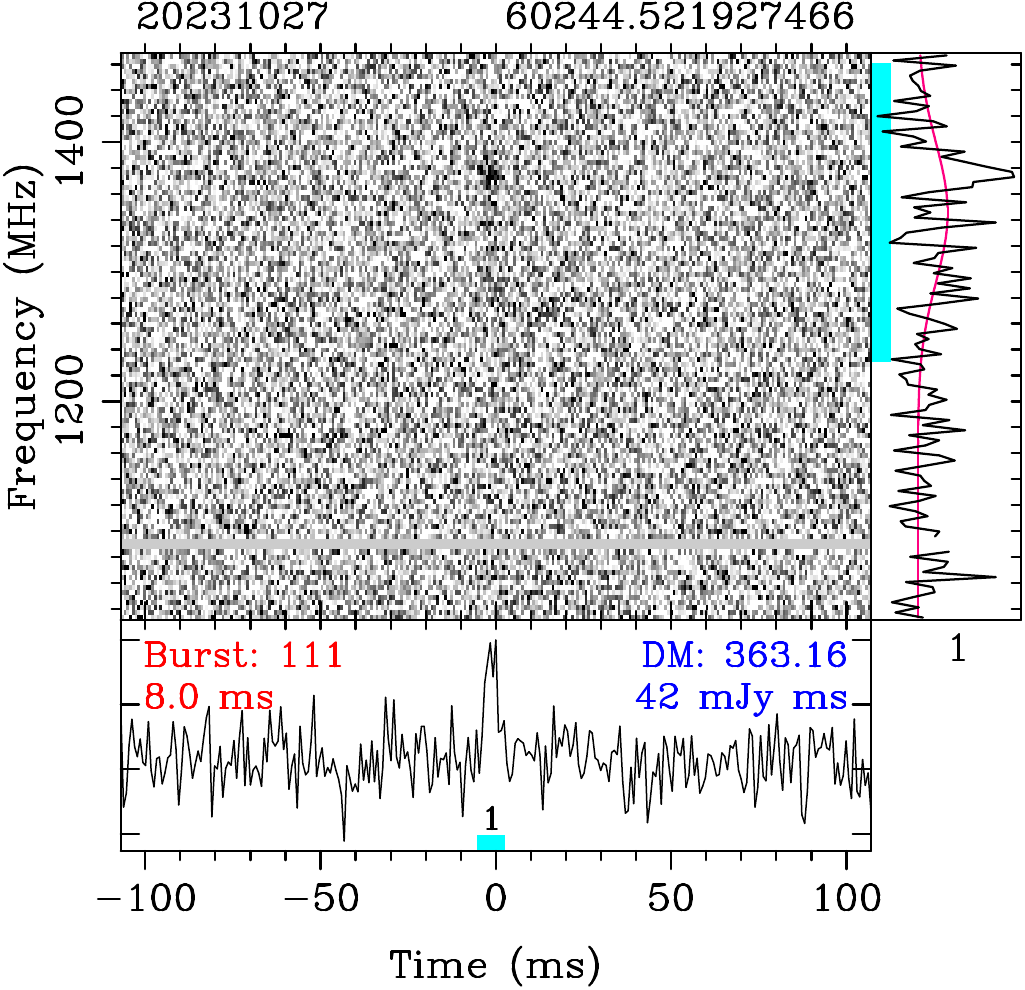}
\includegraphics[height=0.29\linewidth]{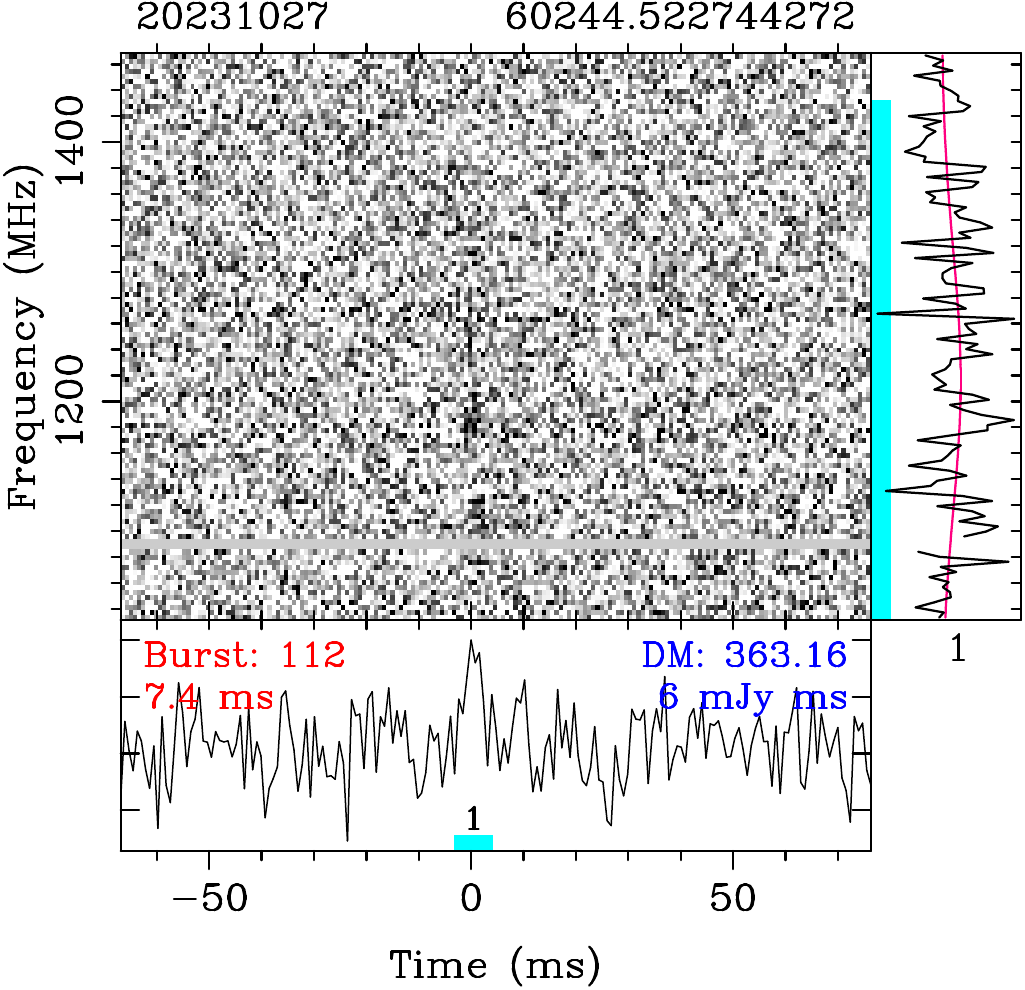}
\includegraphics[height=0.29\linewidth]{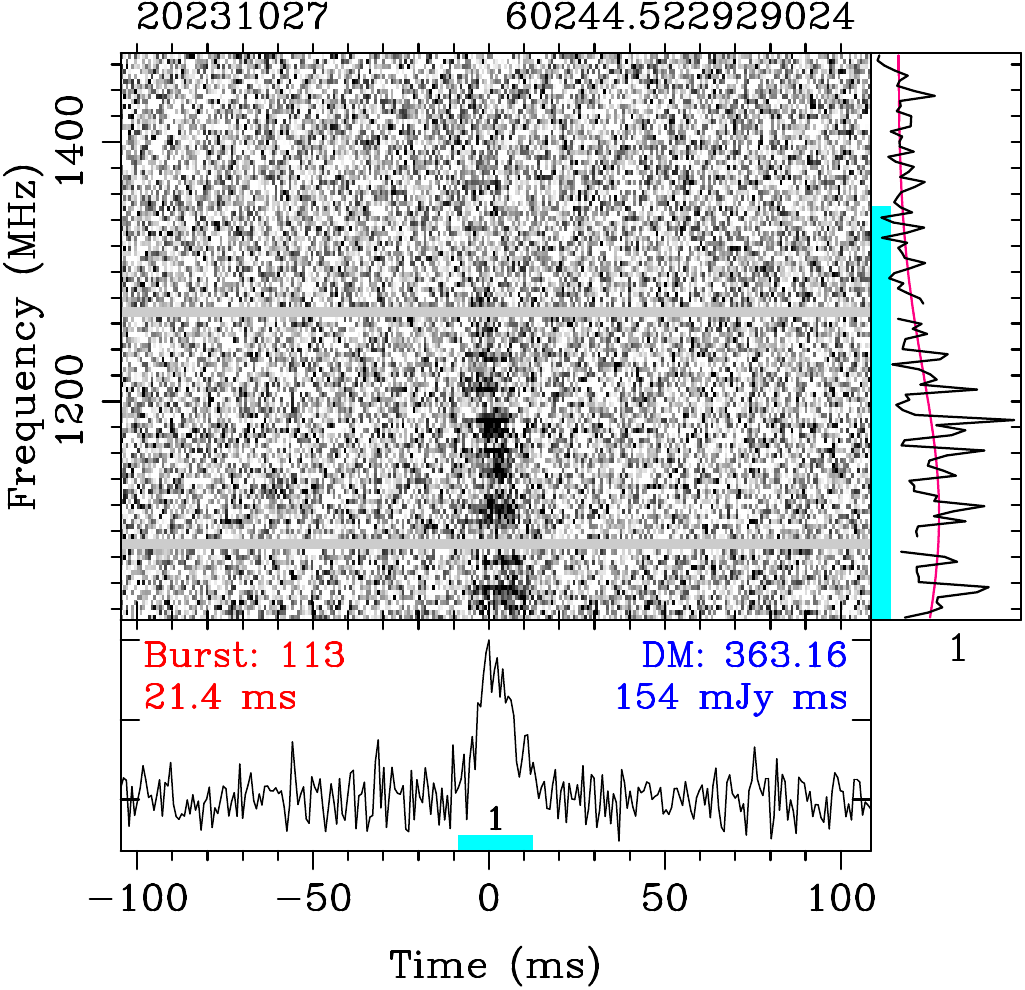}
\includegraphics[height=0.29\linewidth]{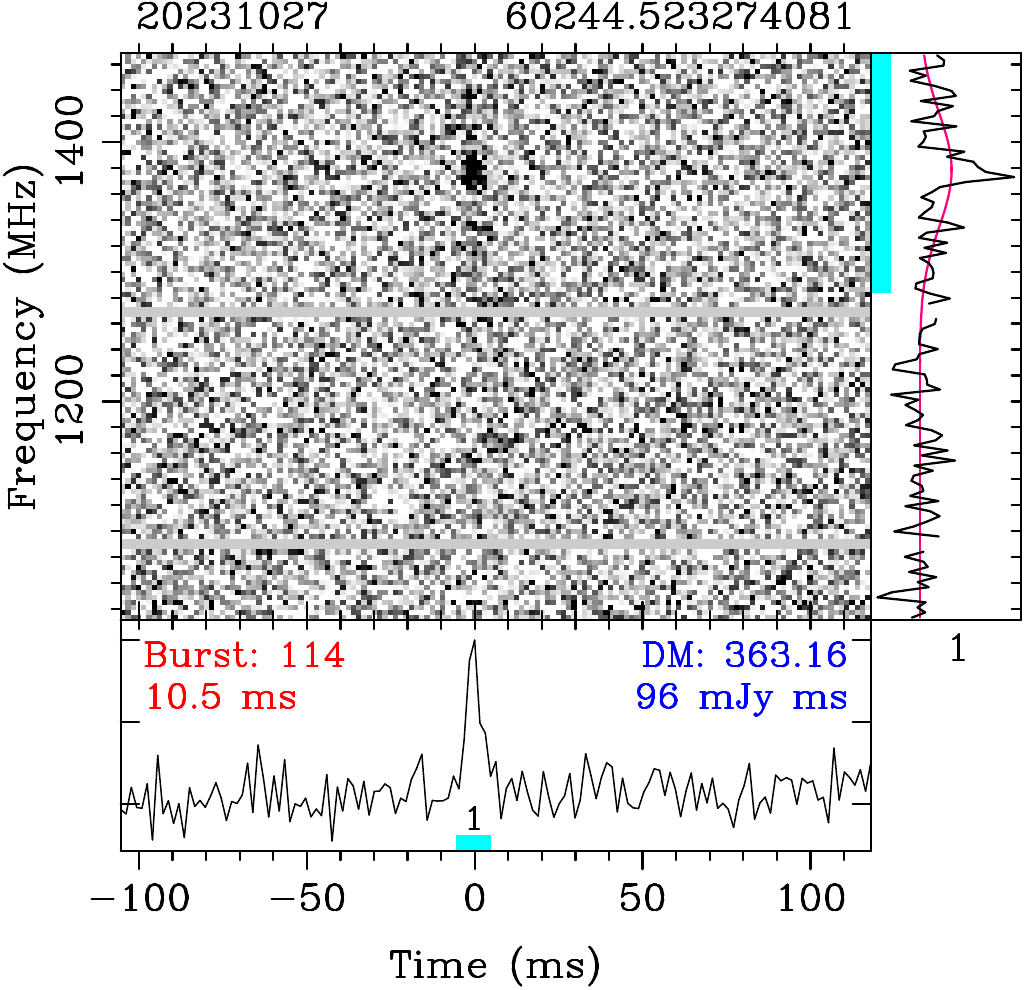}
\includegraphics[height=0.29\linewidth]{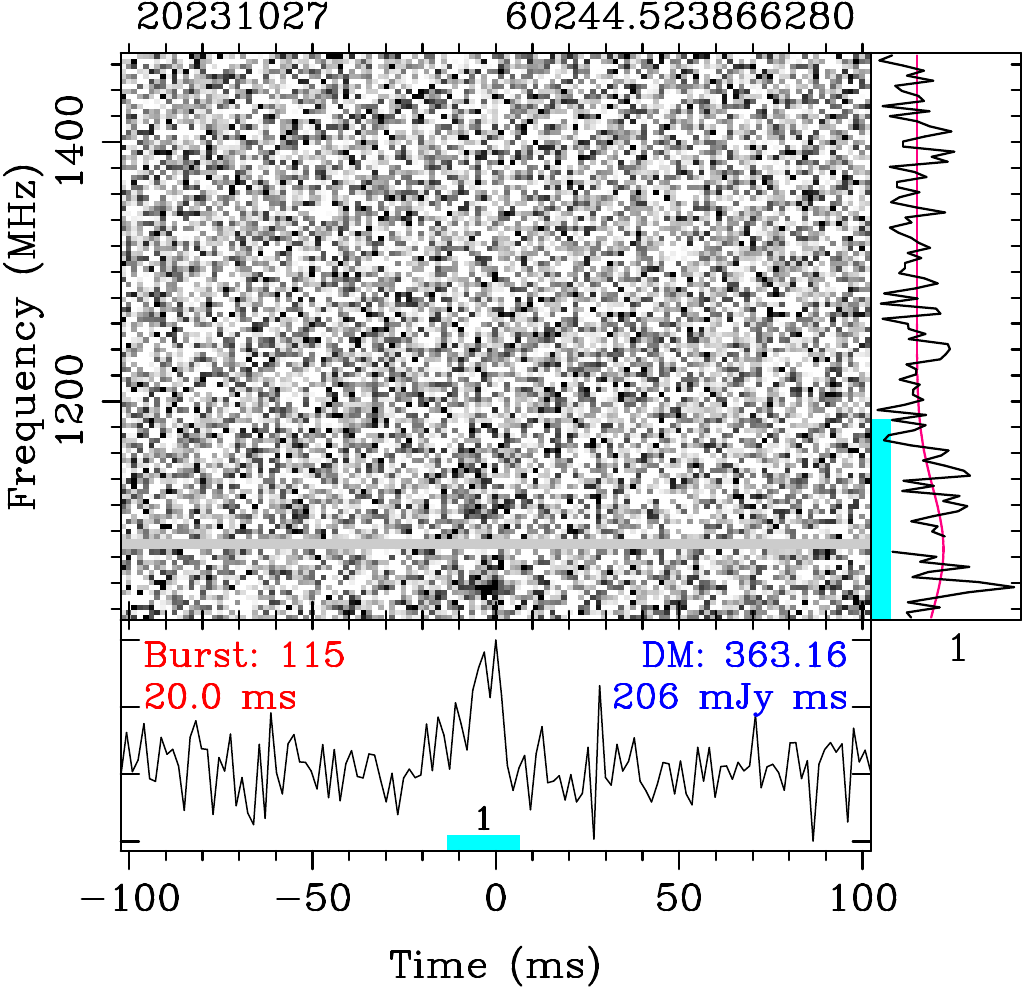}
\includegraphics[height=0.29\linewidth]{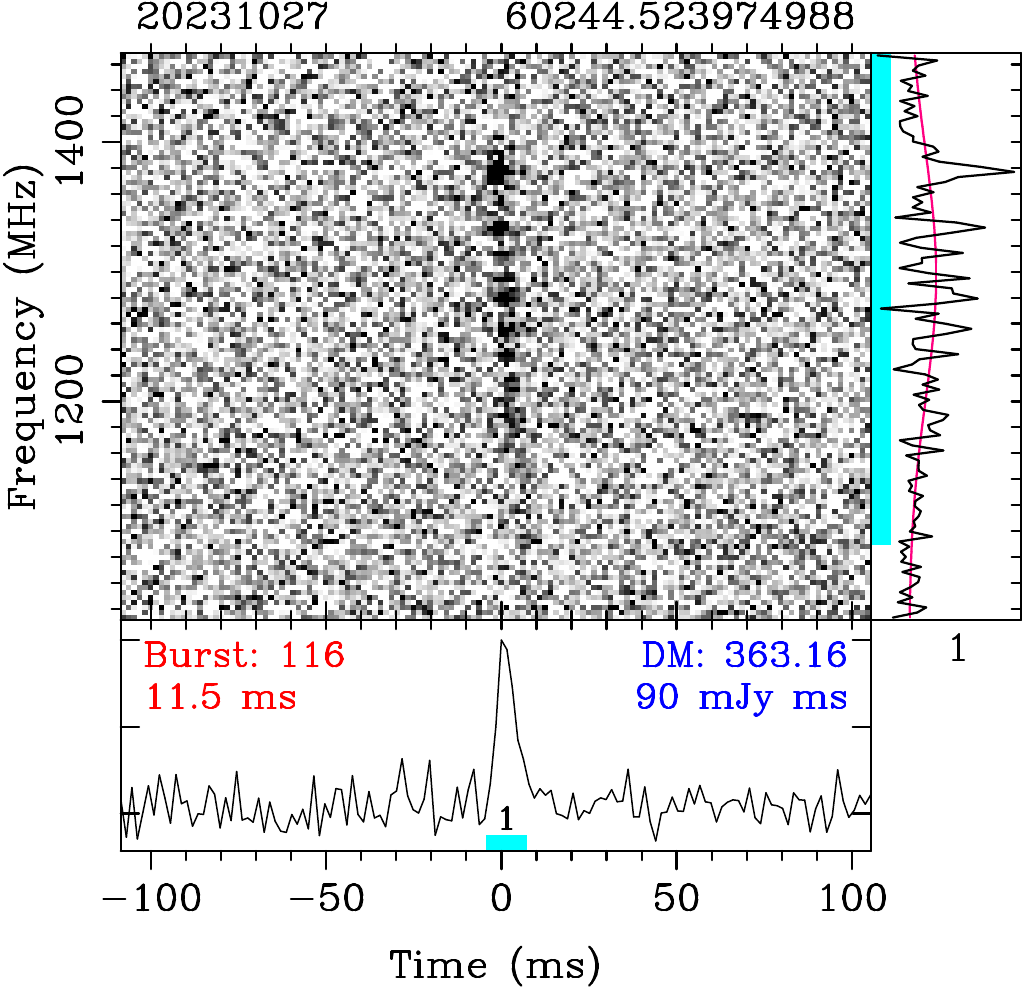}
\includegraphics[height=0.29\linewidth]{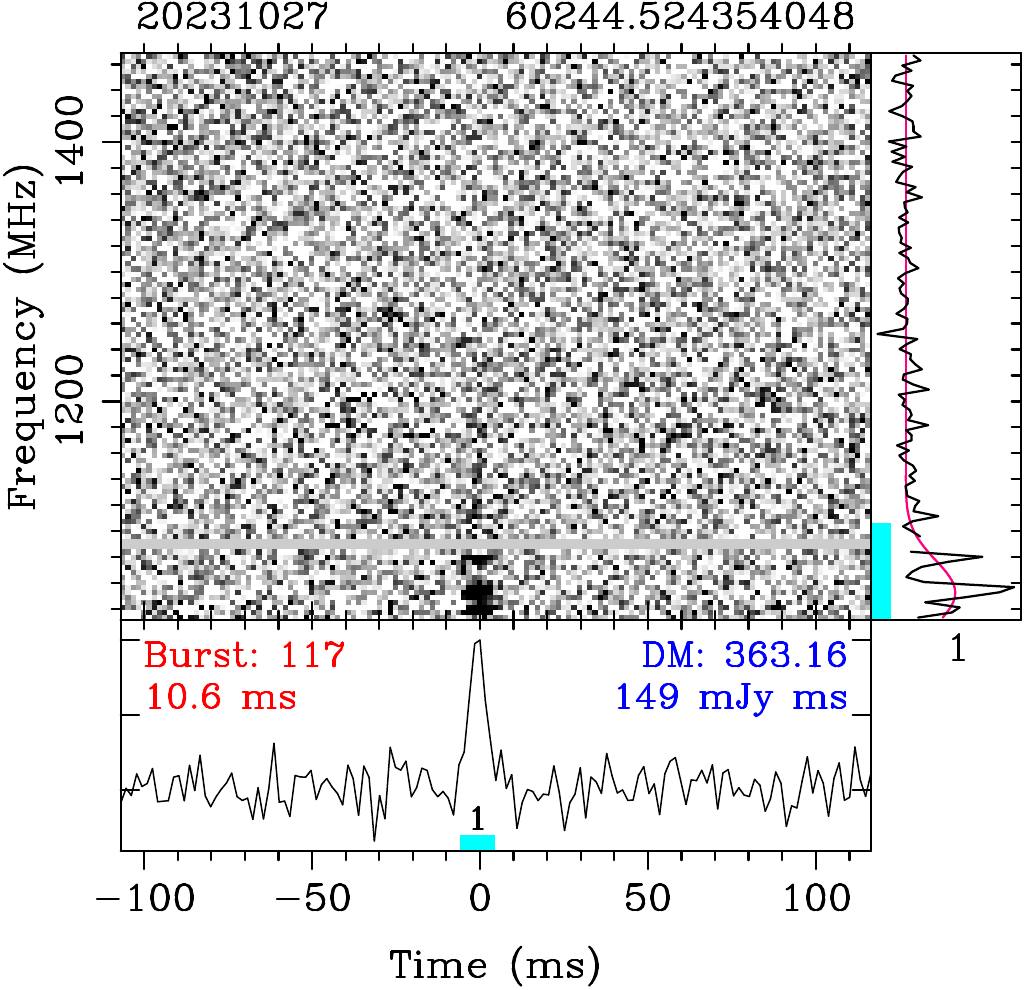}
\includegraphics[height=0.29\linewidth]{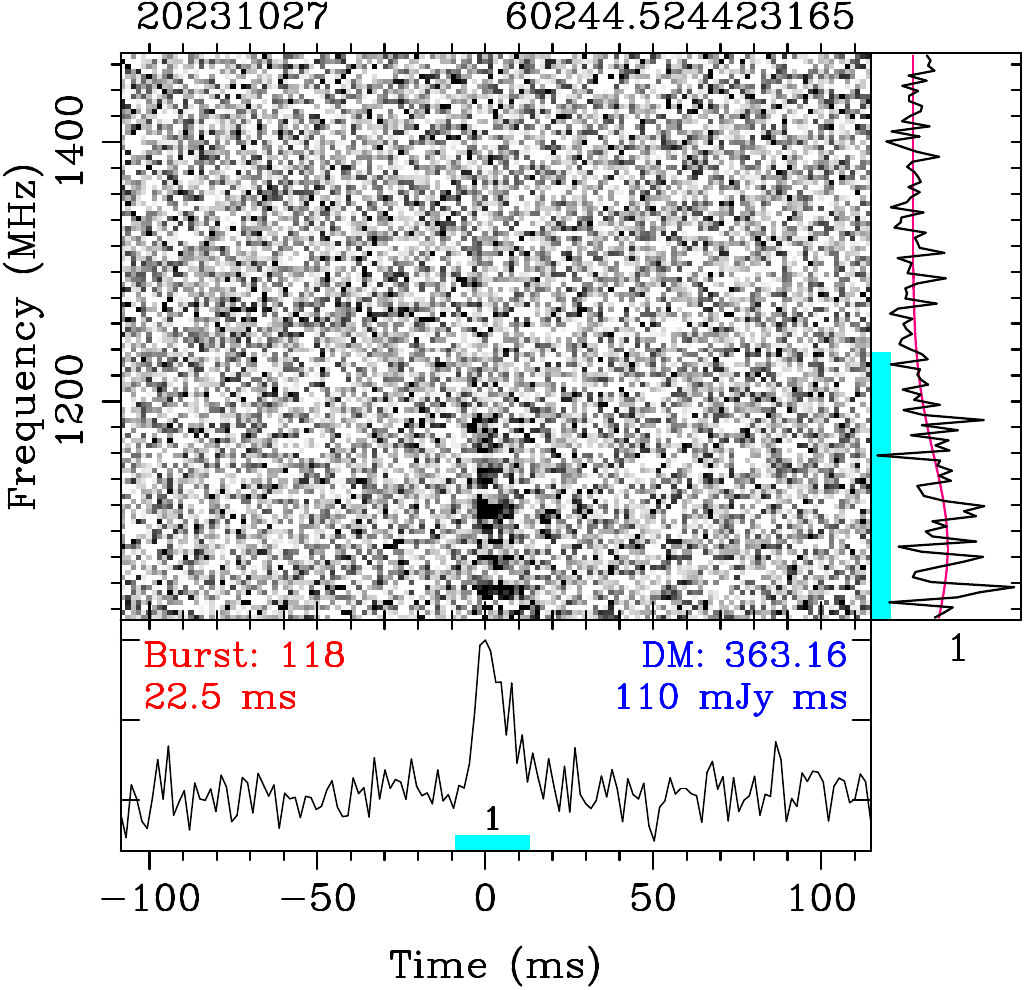}
\caption{({\textit{continued}})}
\end{figure*}
\addtocounter{figure}{-1}
\begin{figure*}
\flushleft
\includegraphics[height=0.29\linewidth]{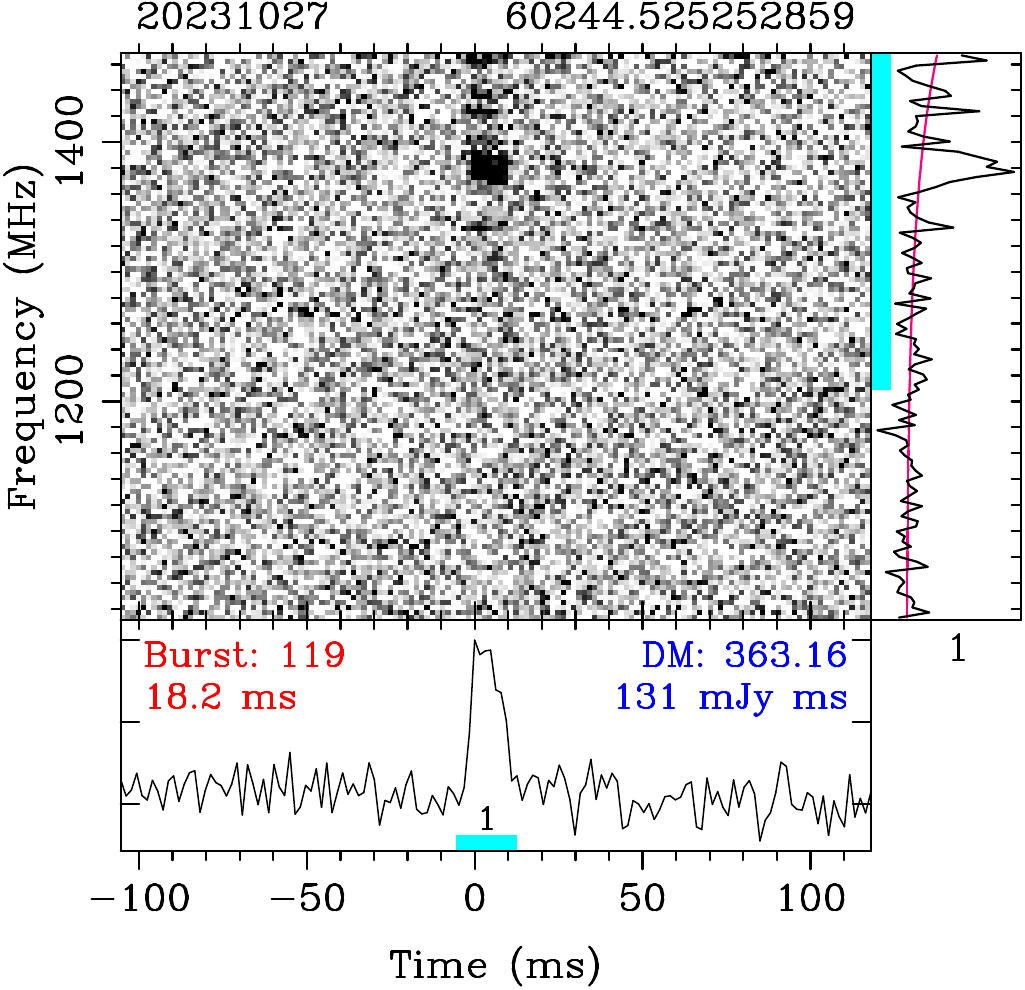}
\includegraphics[height=0.29\linewidth]{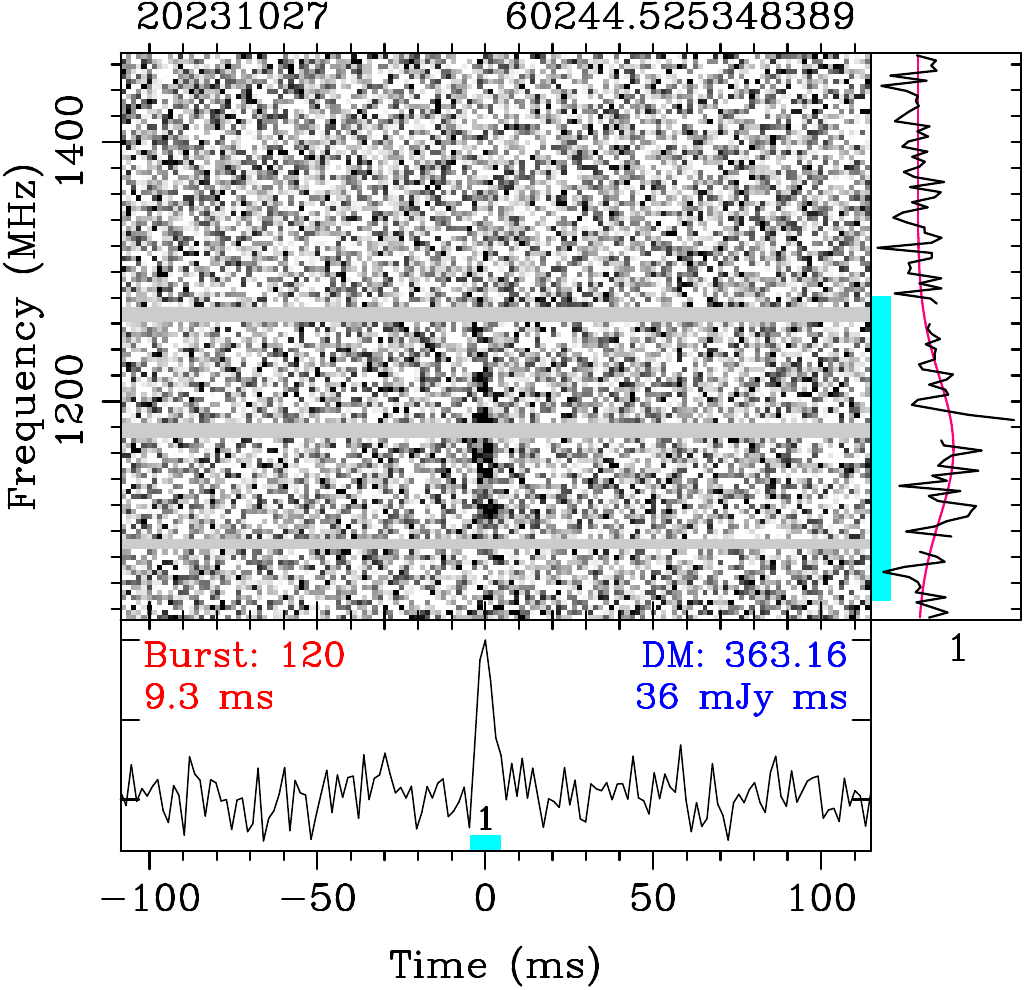}
\includegraphics[height=0.29\linewidth]{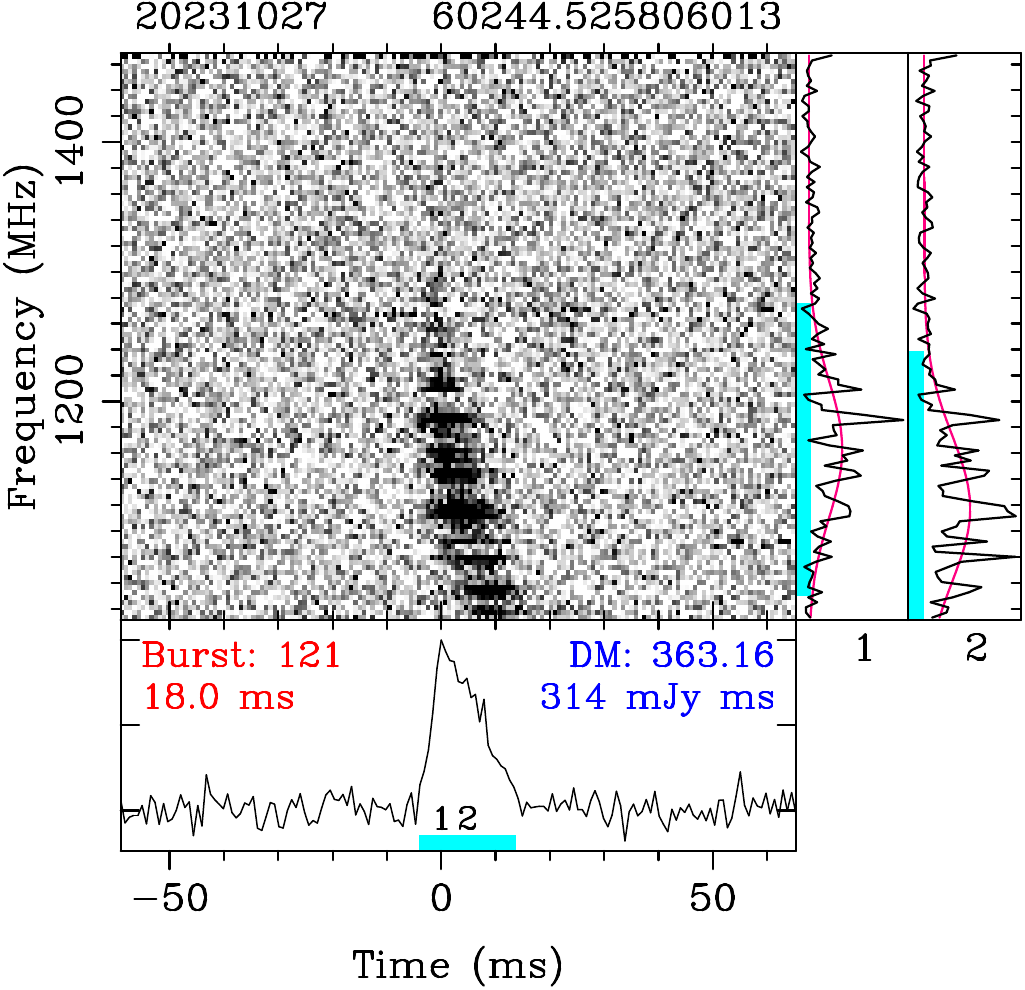}
\includegraphics[height=0.29\linewidth]{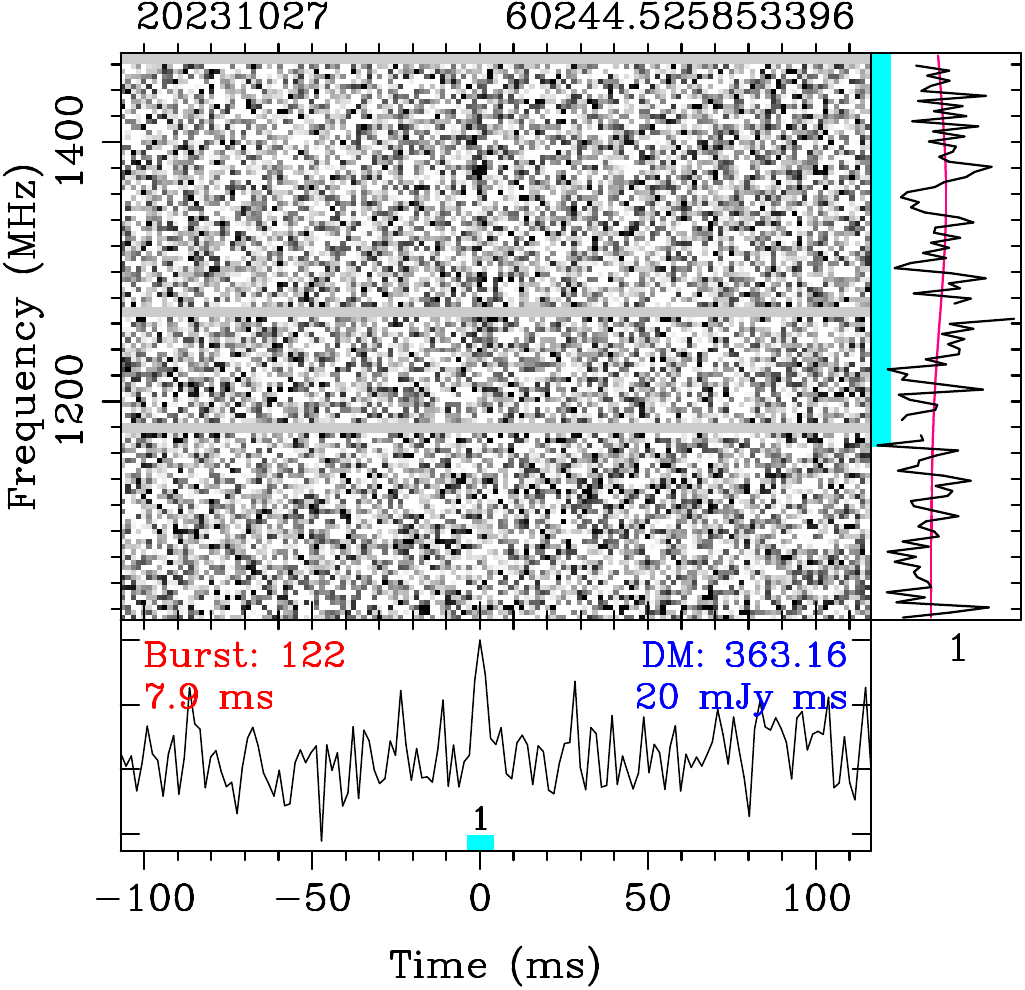}
\includegraphics[height=0.29\linewidth]{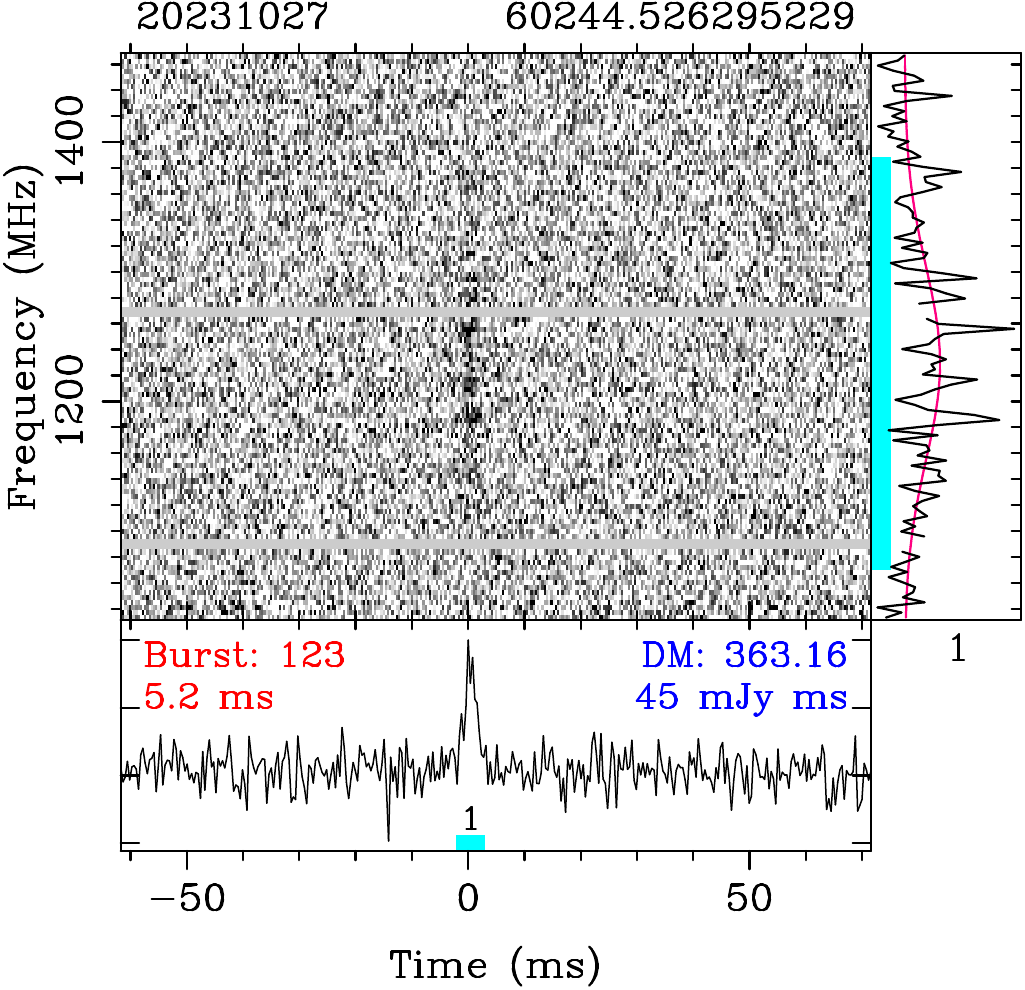}
\includegraphics[height=0.29\linewidth]{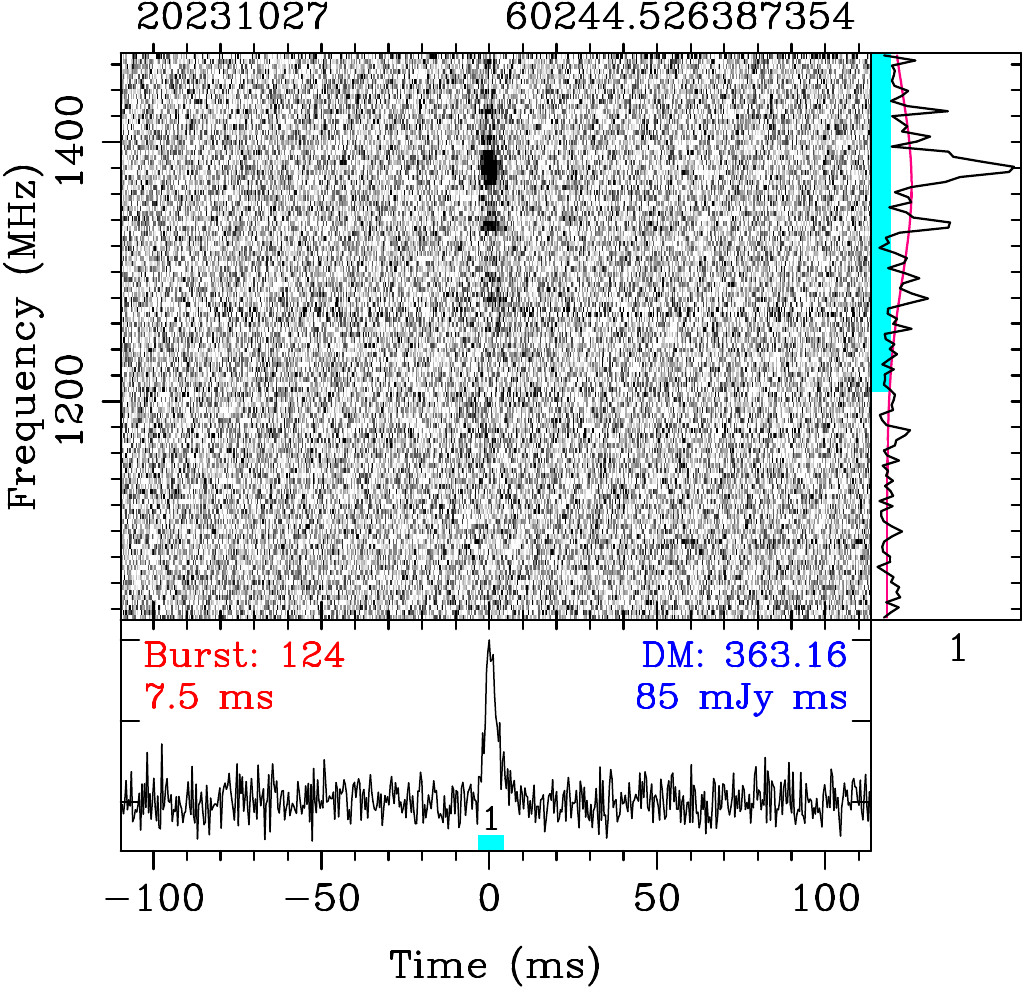}
\includegraphics[height=0.29\linewidth]{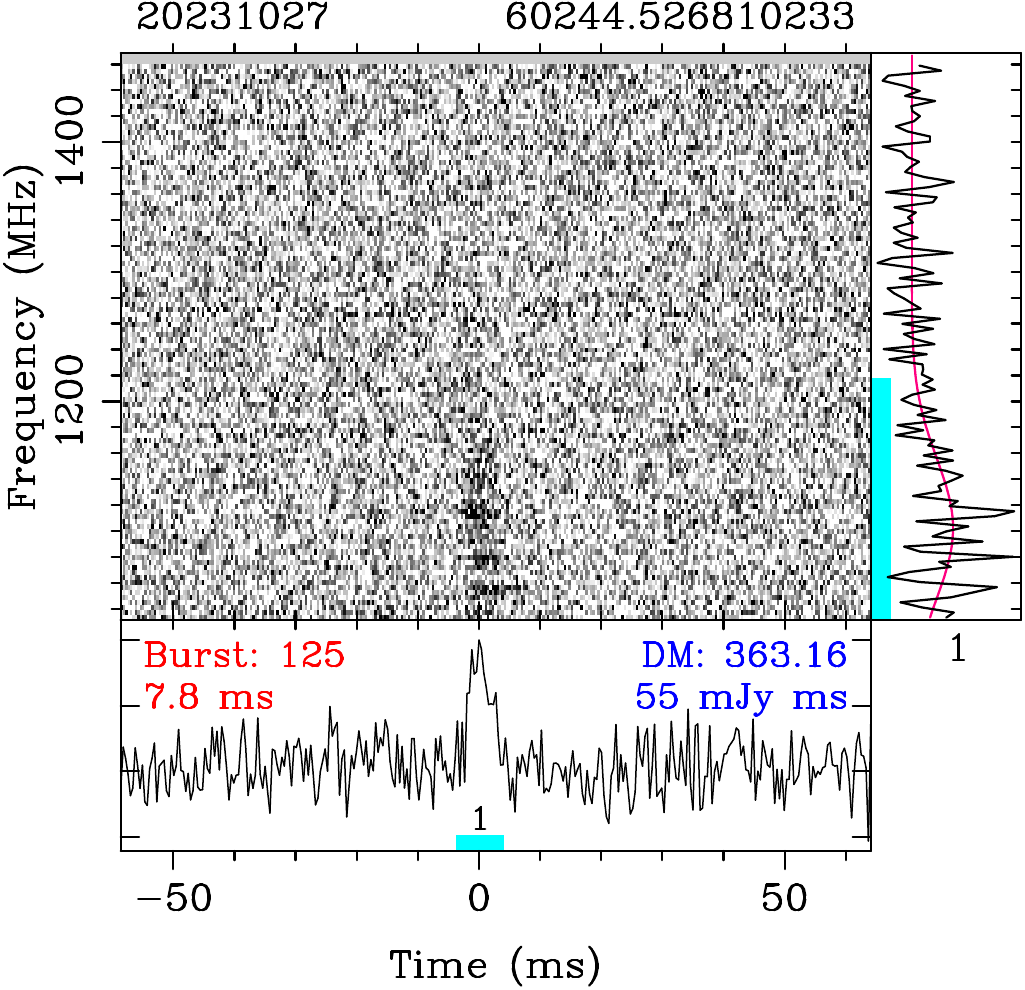}
\includegraphics[height=0.29\linewidth]{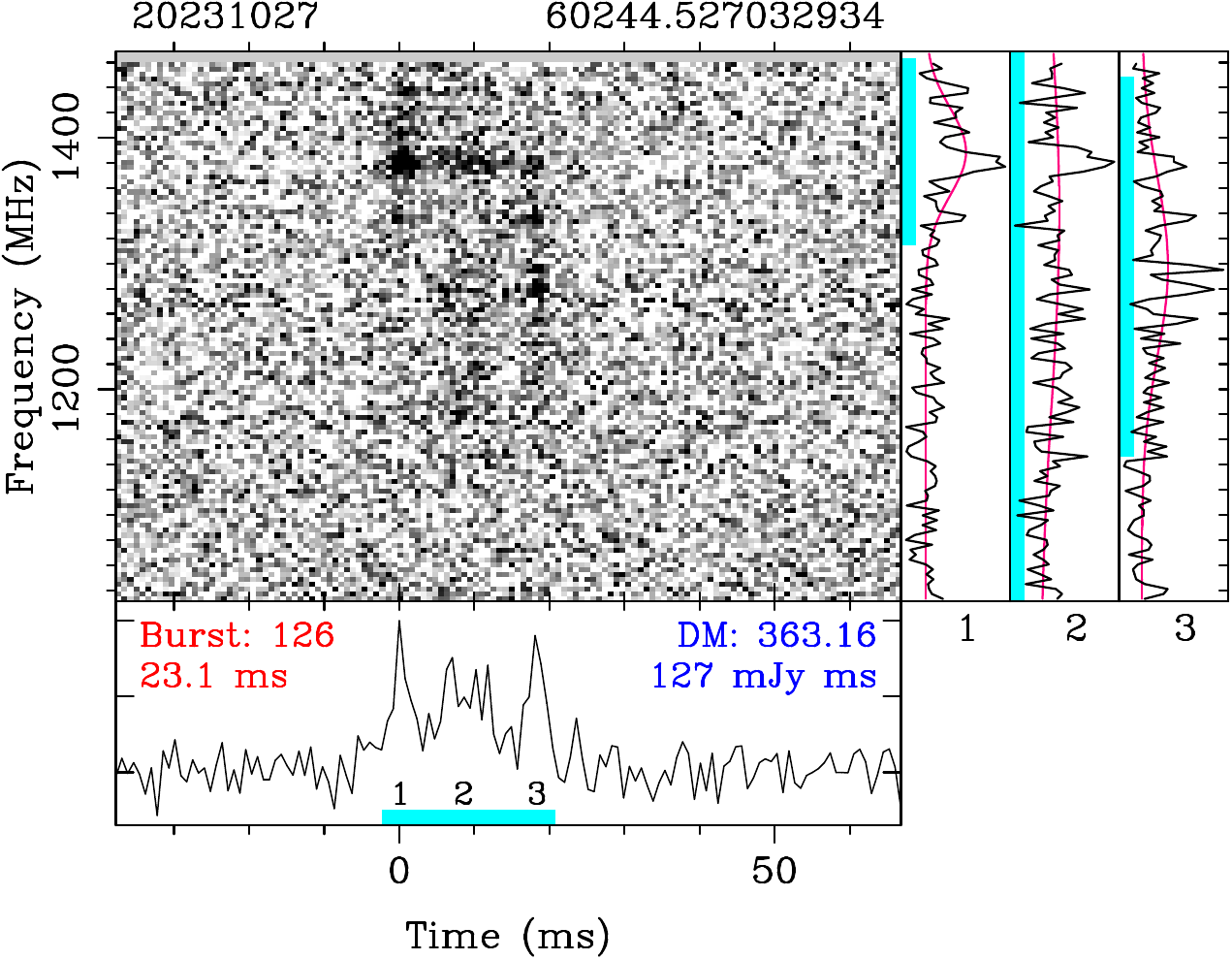}
\includegraphics[height=0.29\linewidth]{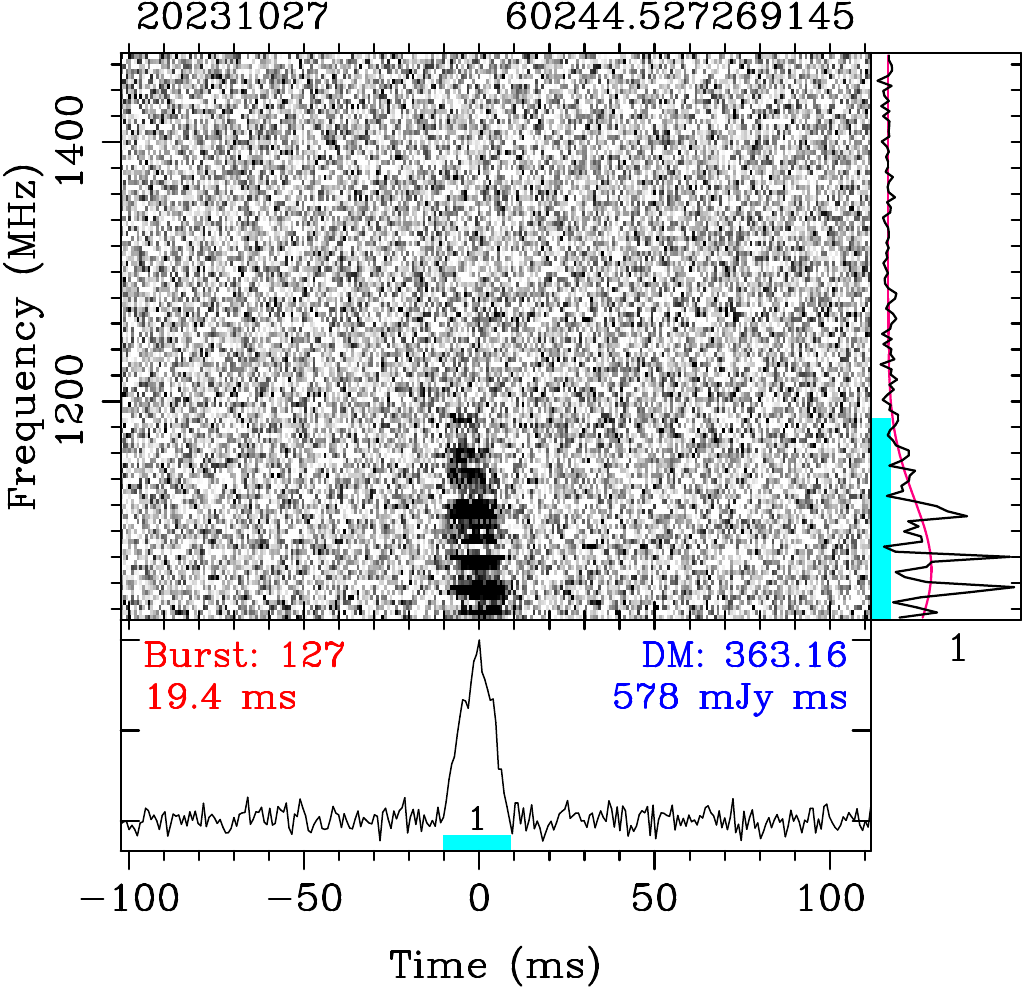}
\includegraphics[height=0.29\linewidth]{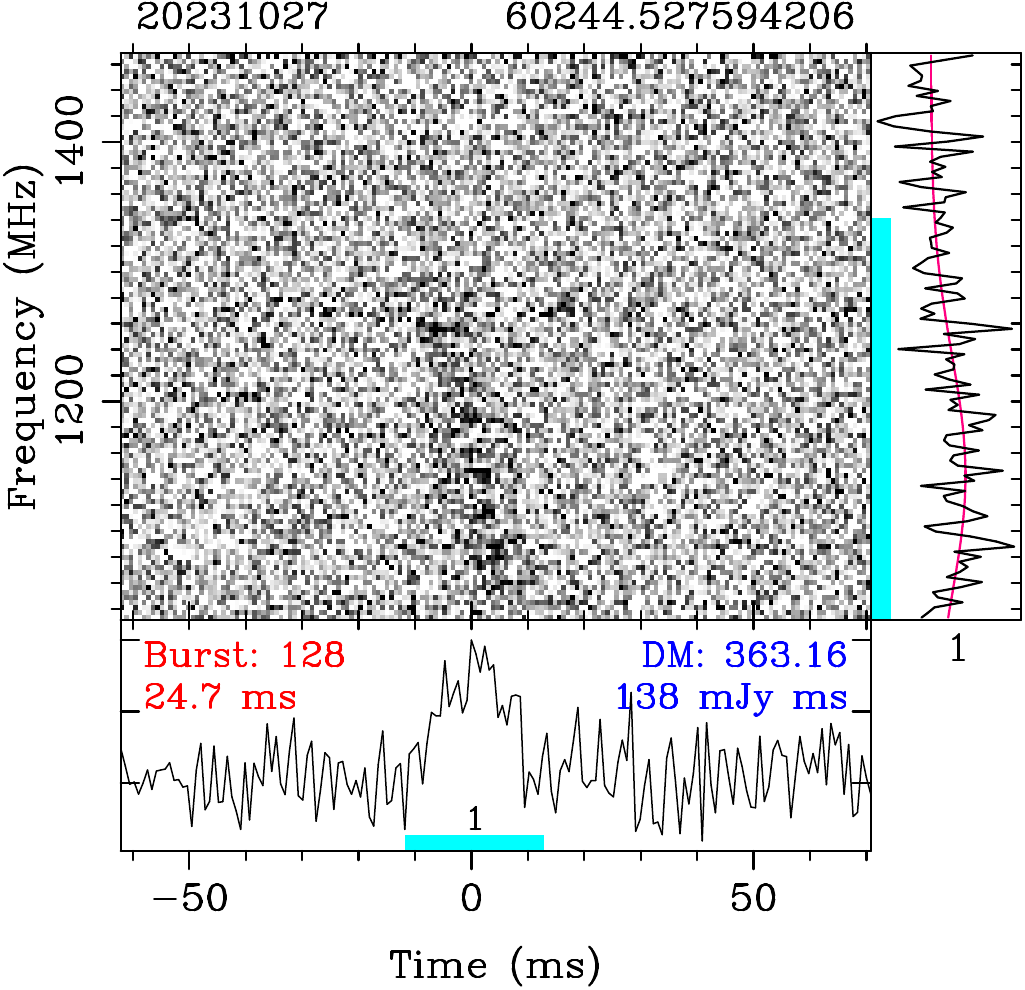}
\includegraphics[height=0.29\linewidth]{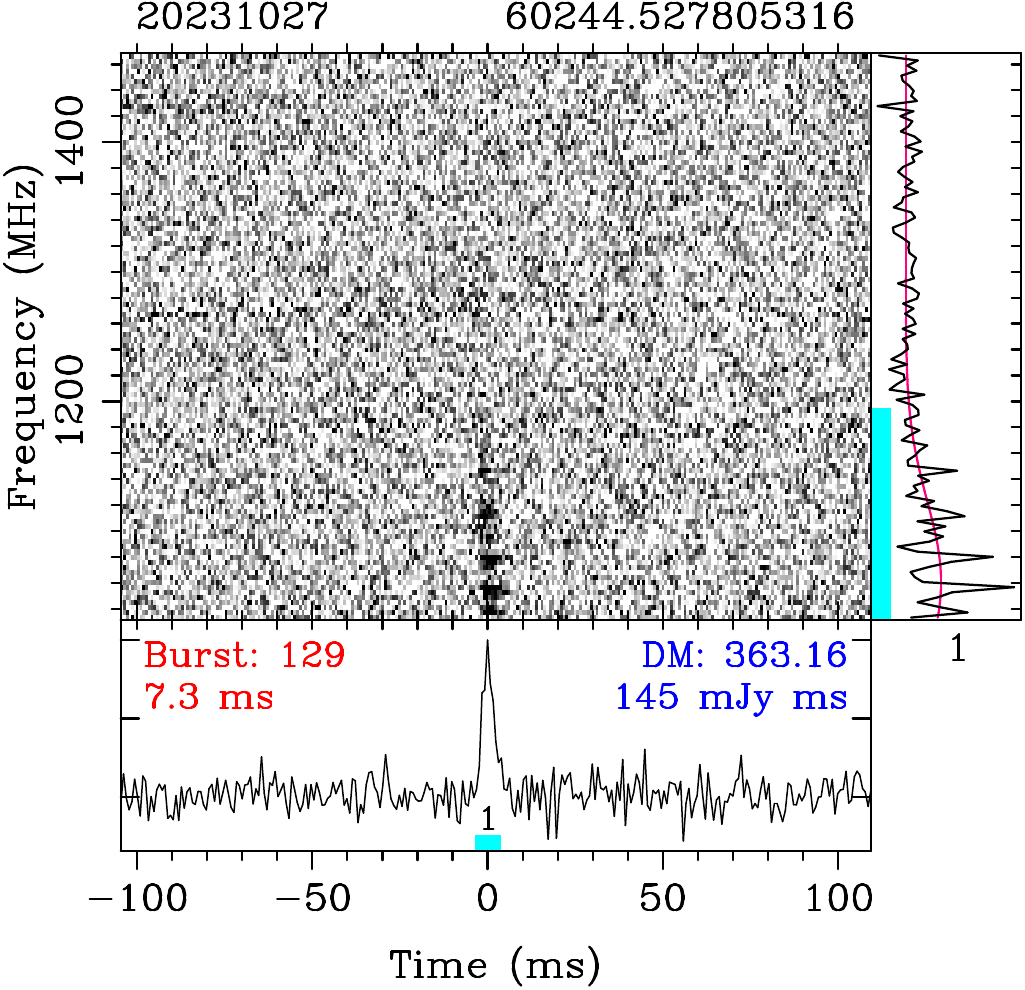}
\includegraphics[height=0.29\linewidth]{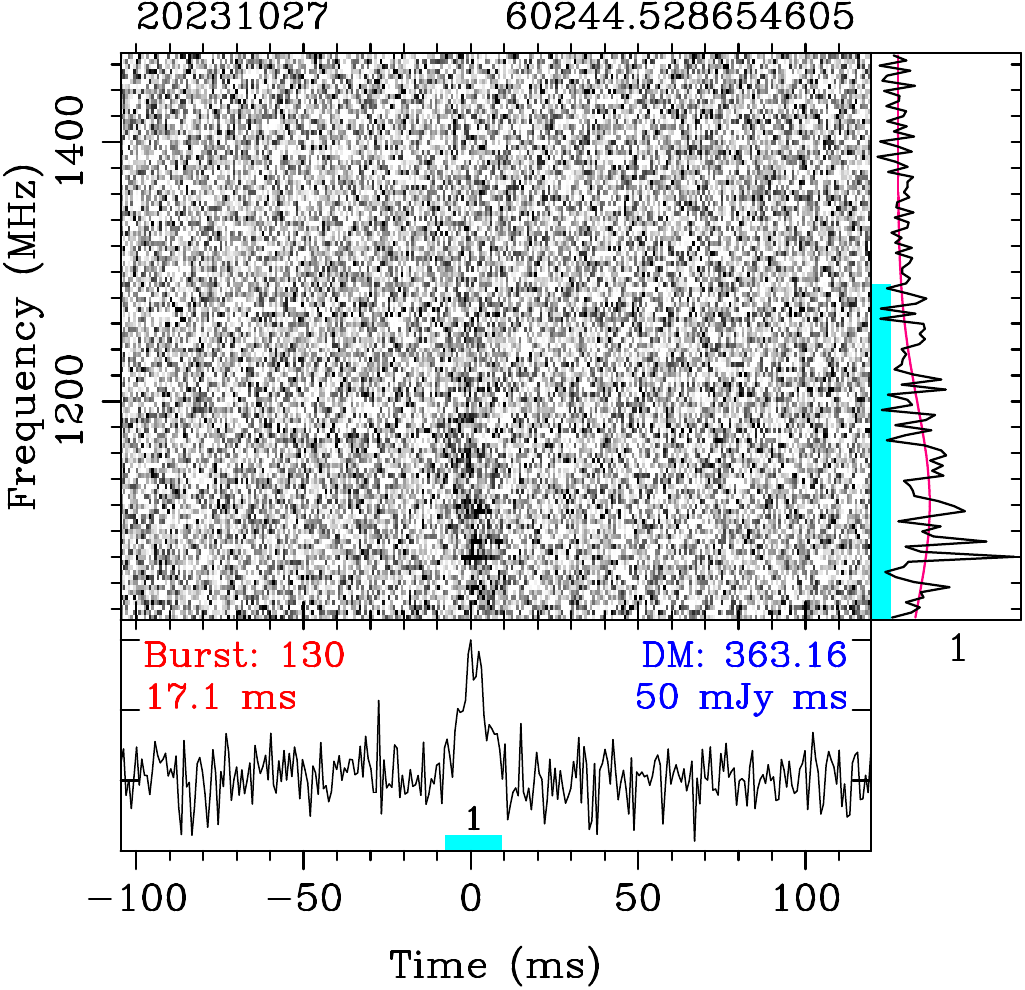}
\caption{({\textit{continued}})}
\end{figure*}
\addtocounter{figure}{-1}
\begin{figure*}
\flushleft
\includegraphics[height=0.29\linewidth]{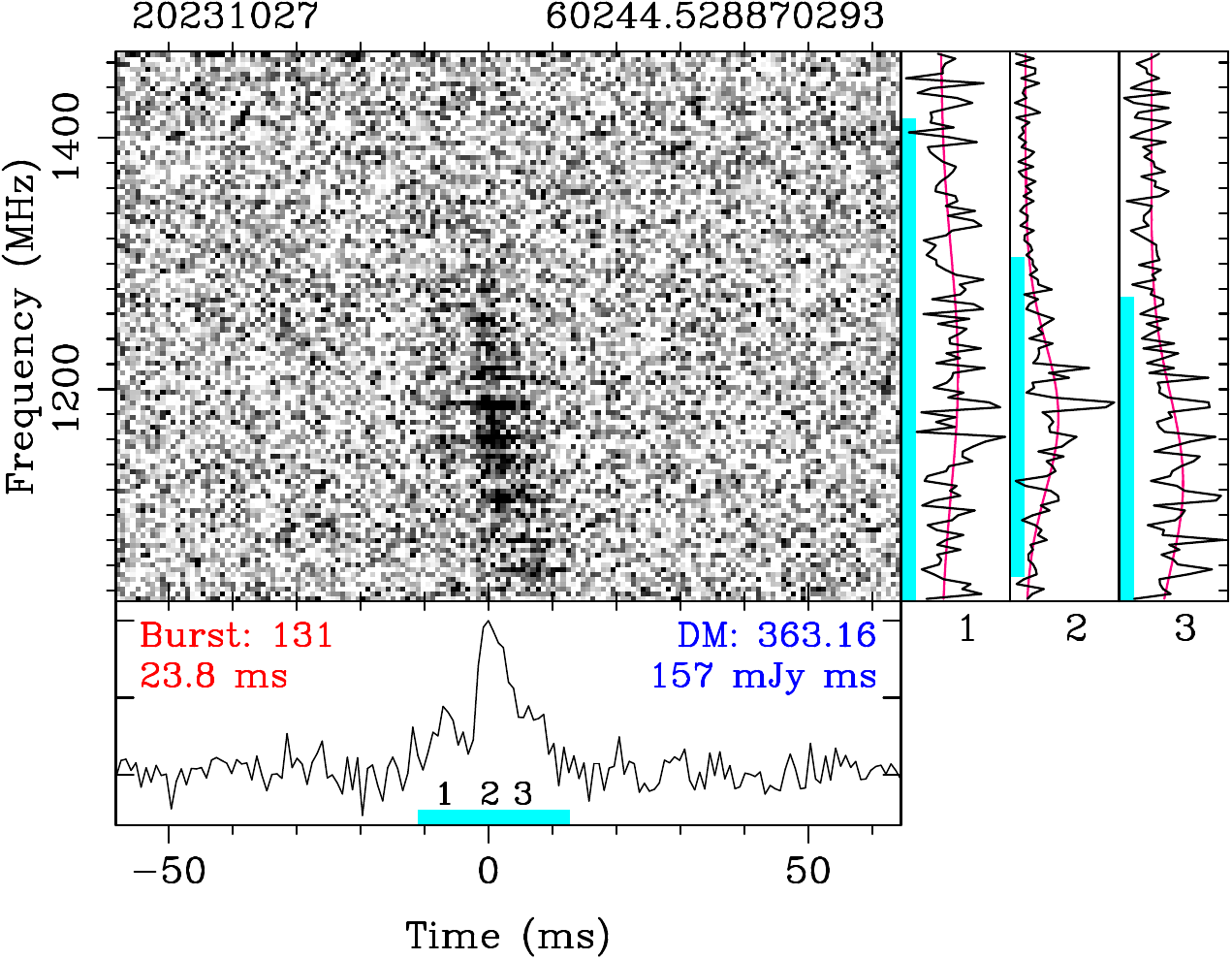}
\includegraphics[height=0.29\linewidth]{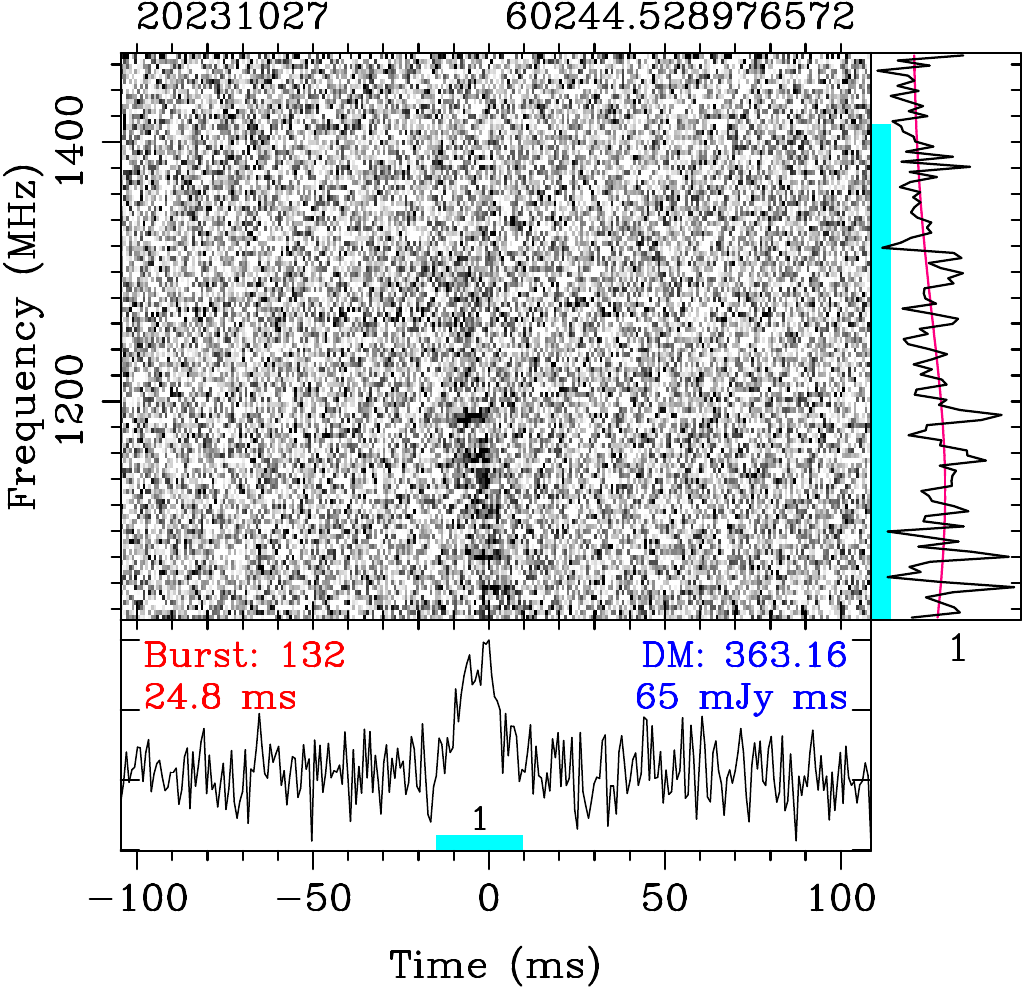}
\includegraphics[height=0.29\linewidth]{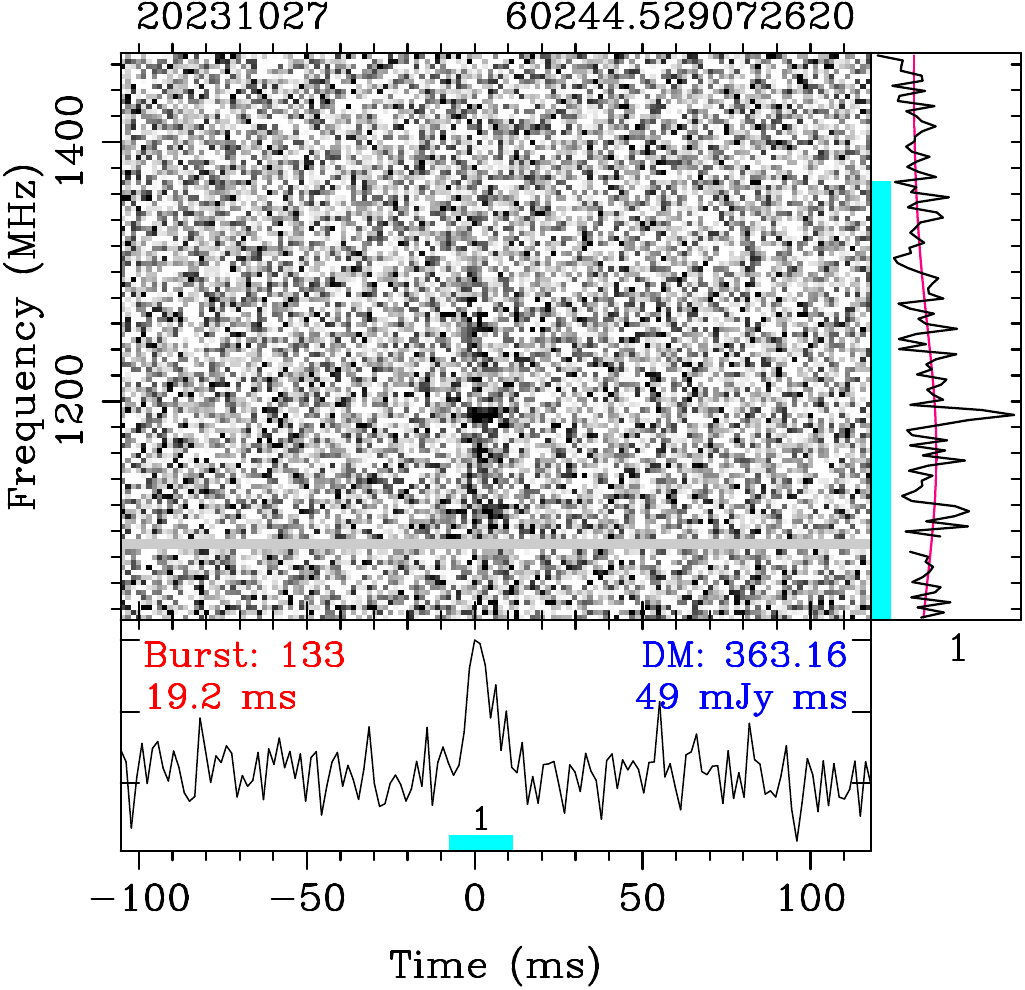}
\includegraphics[height=0.29\linewidth]{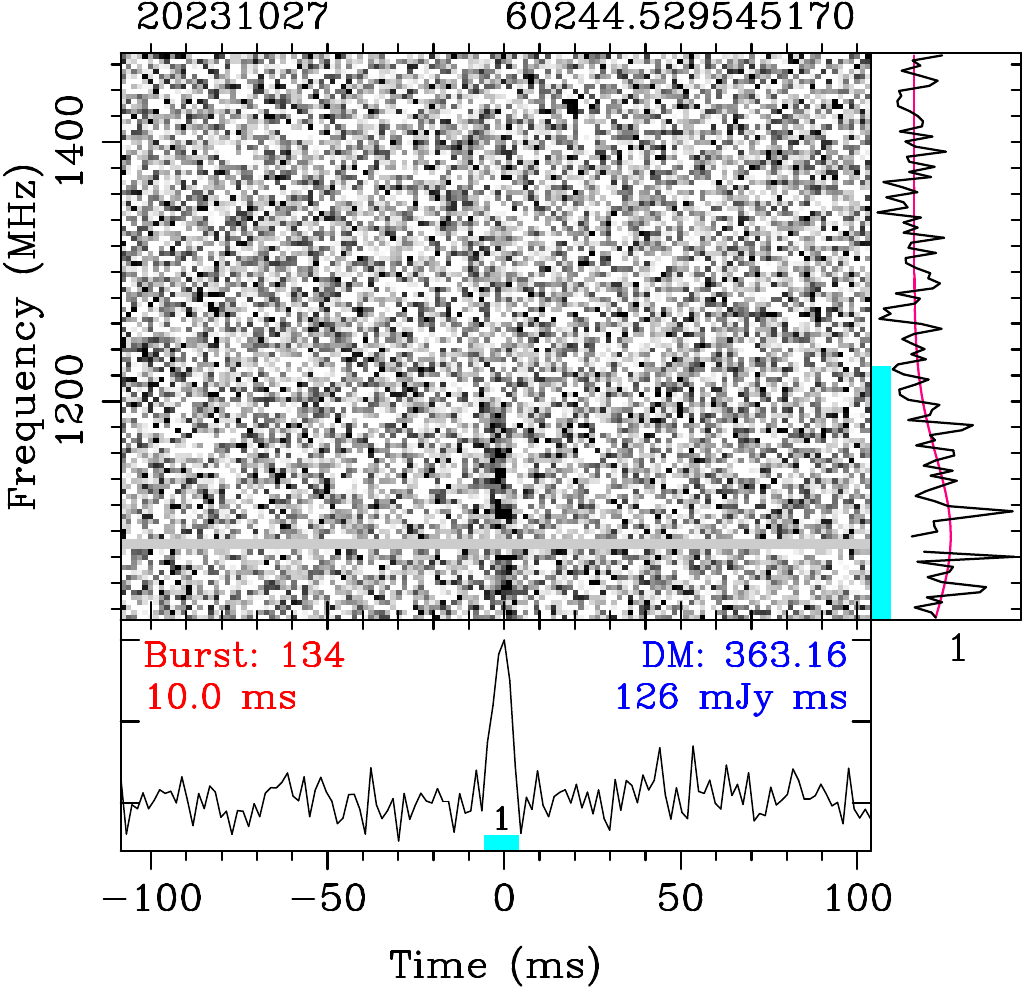}
\includegraphics[height=0.29\linewidth]{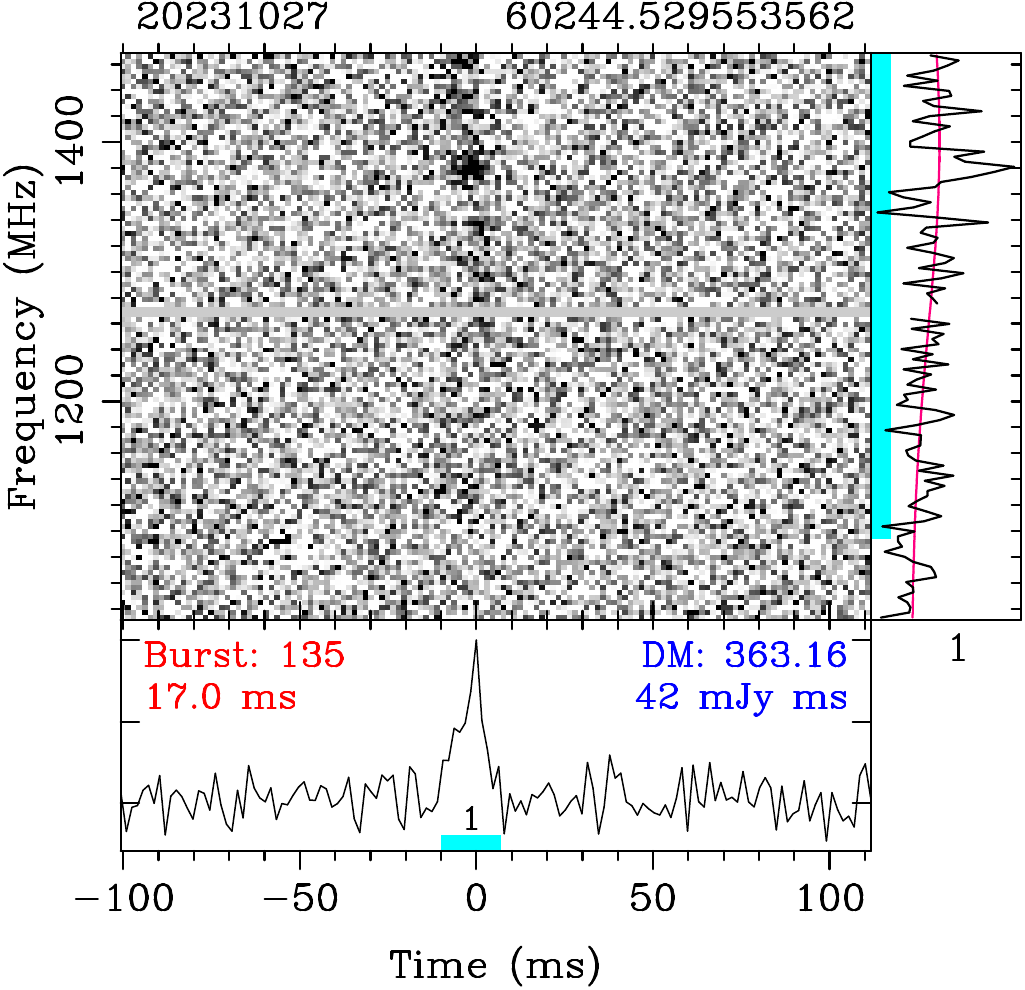}
\includegraphics[height=0.29\linewidth]{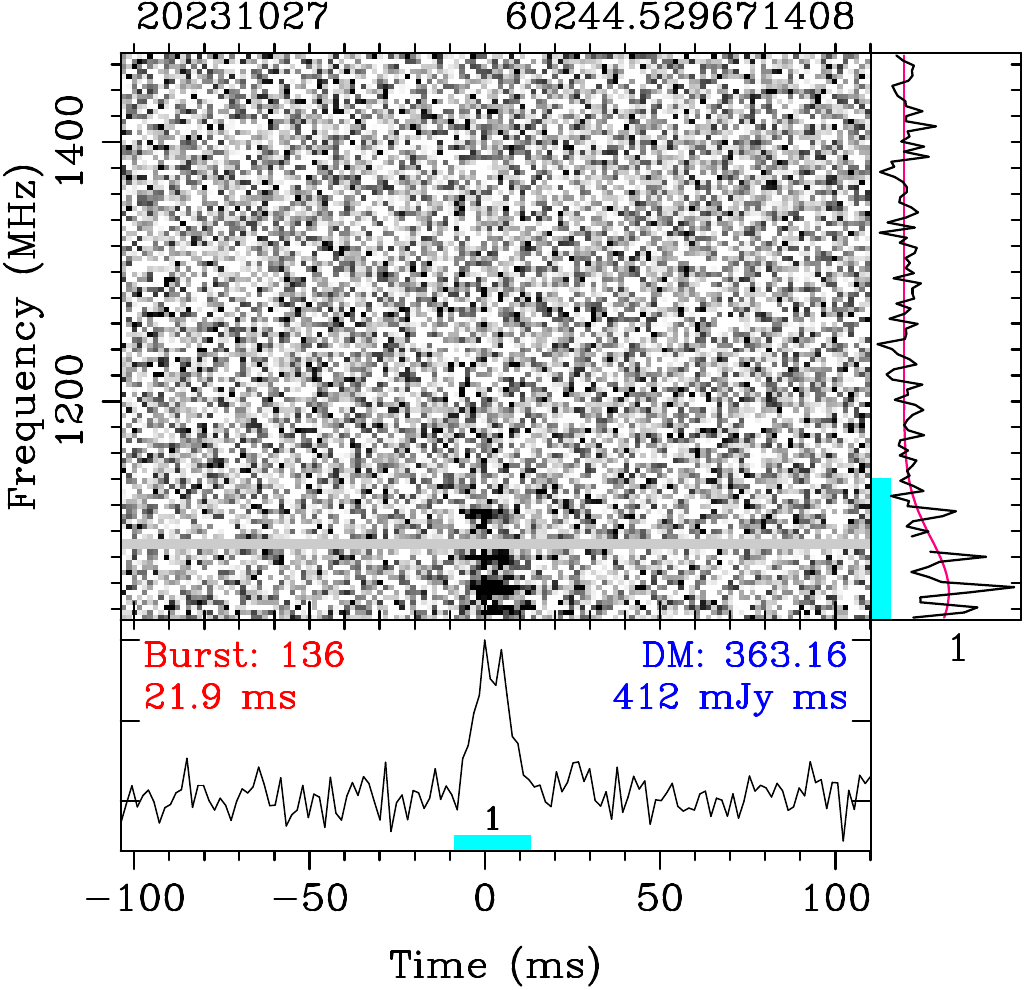}
\includegraphics[height=0.29\linewidth]{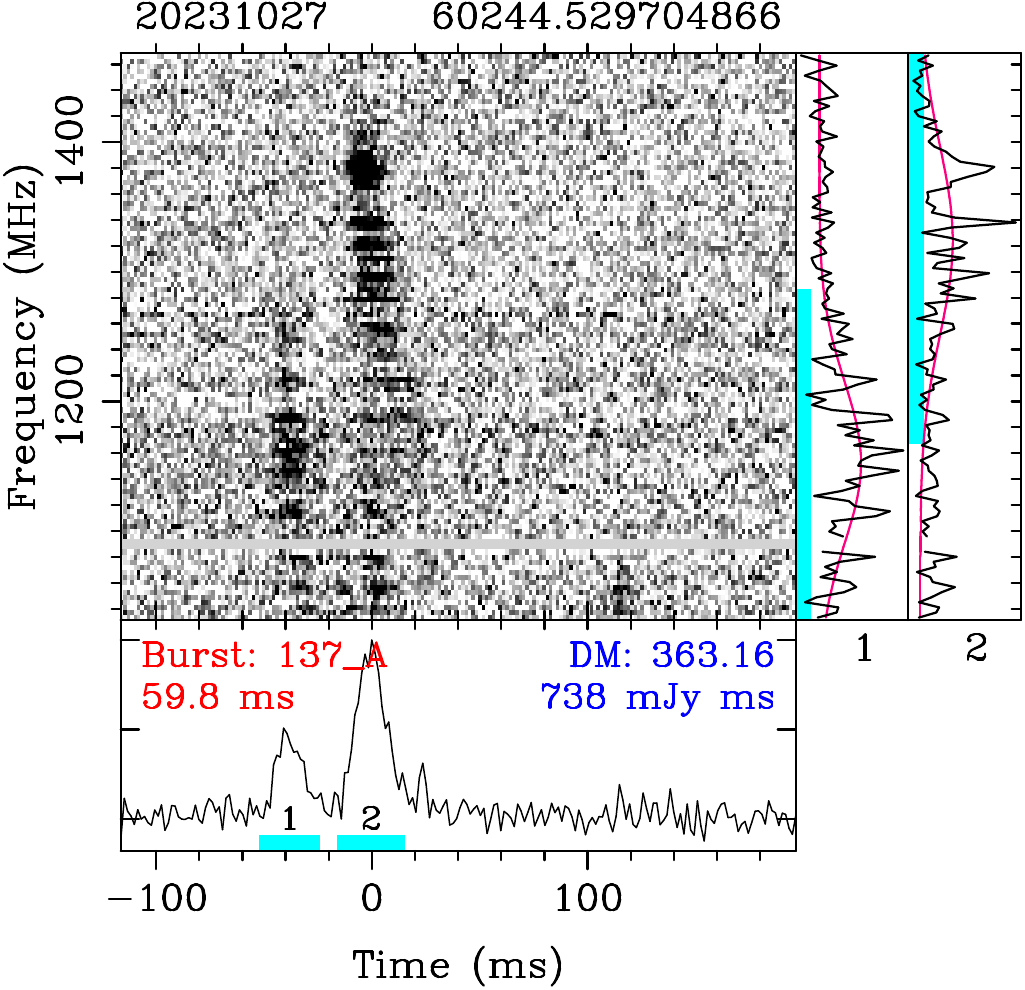}
\includegraphics[height=0.29\linewidth]{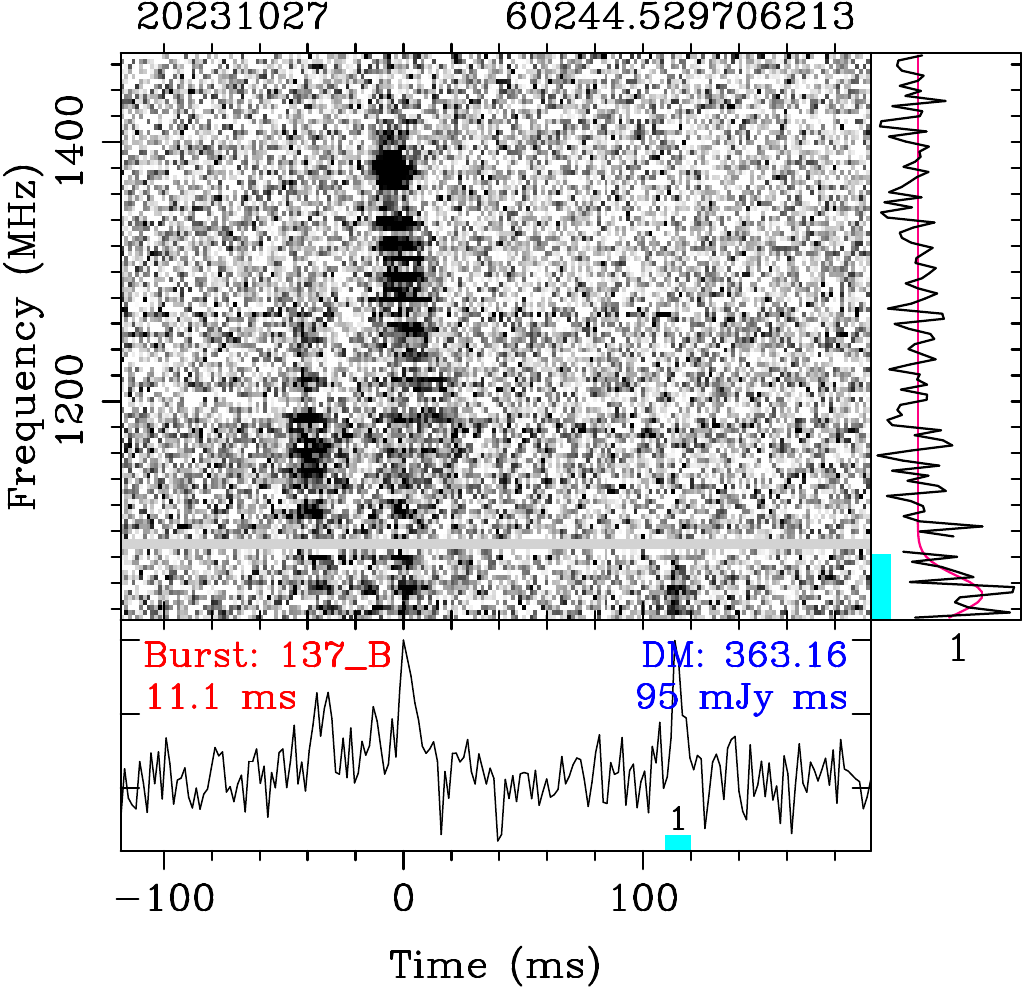}
\includegraphics[height=0.29\linewidth]{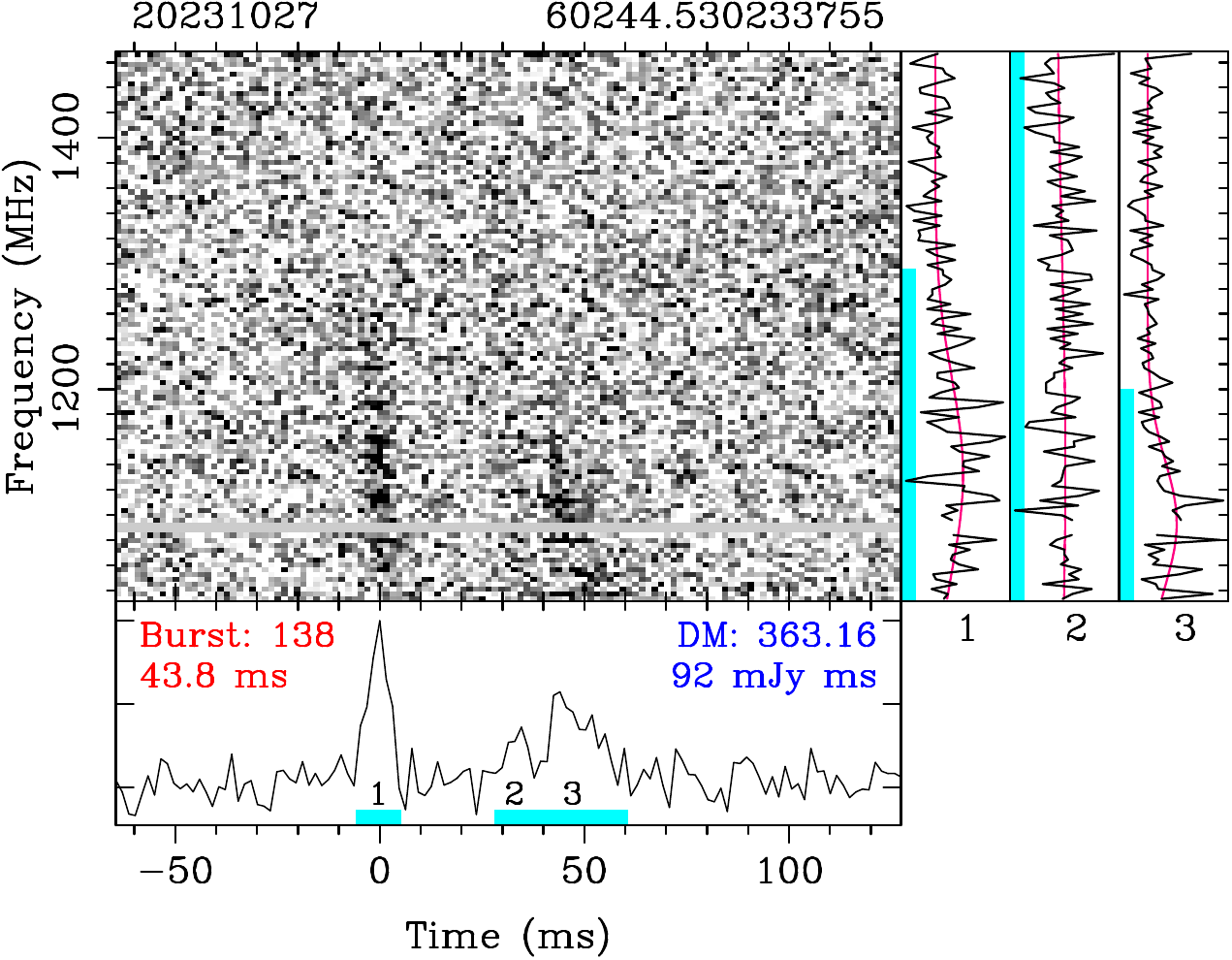}
\includegraphics[height=0.29\linewidth]{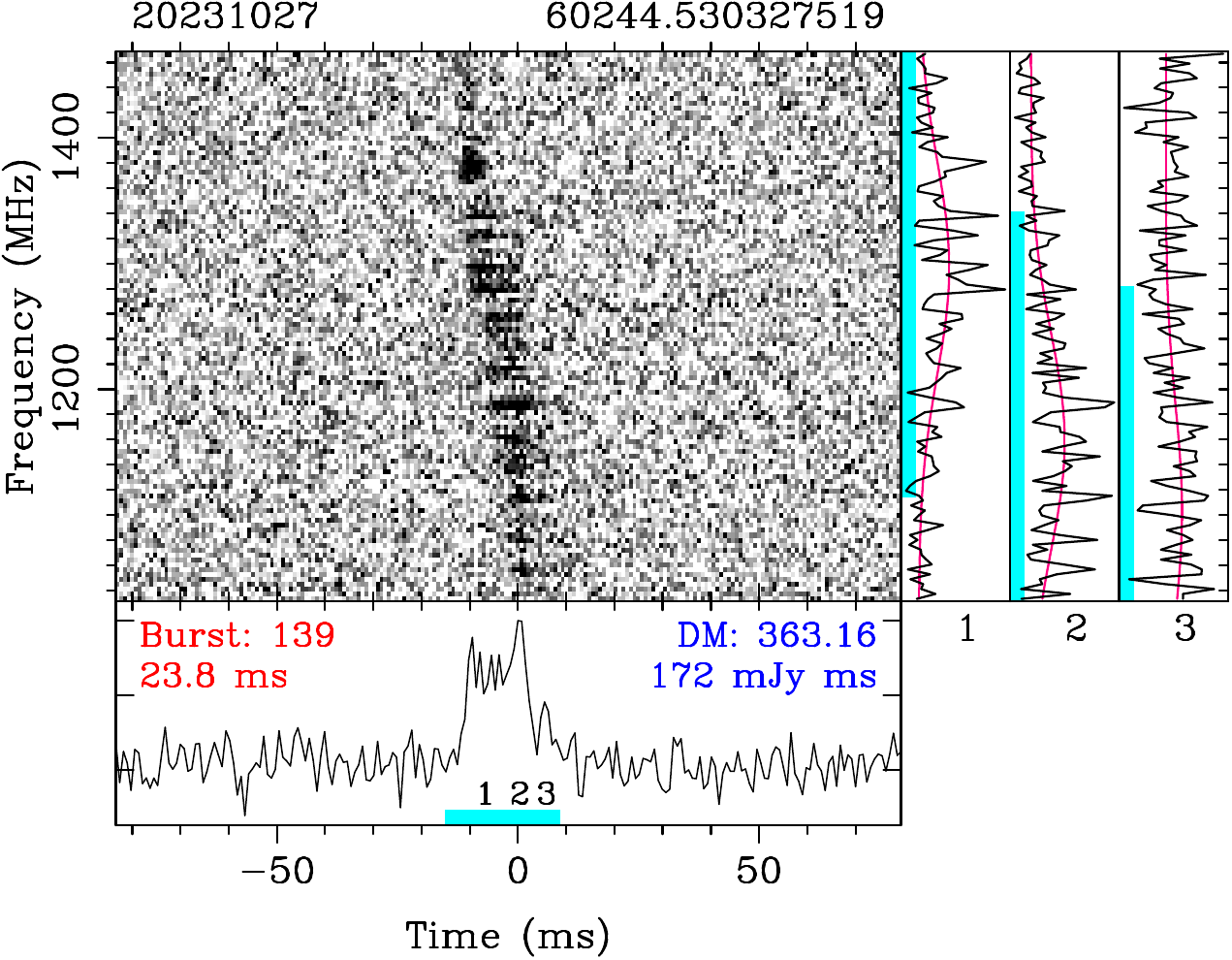}
\includegraphics[height=0.29\linewidth]{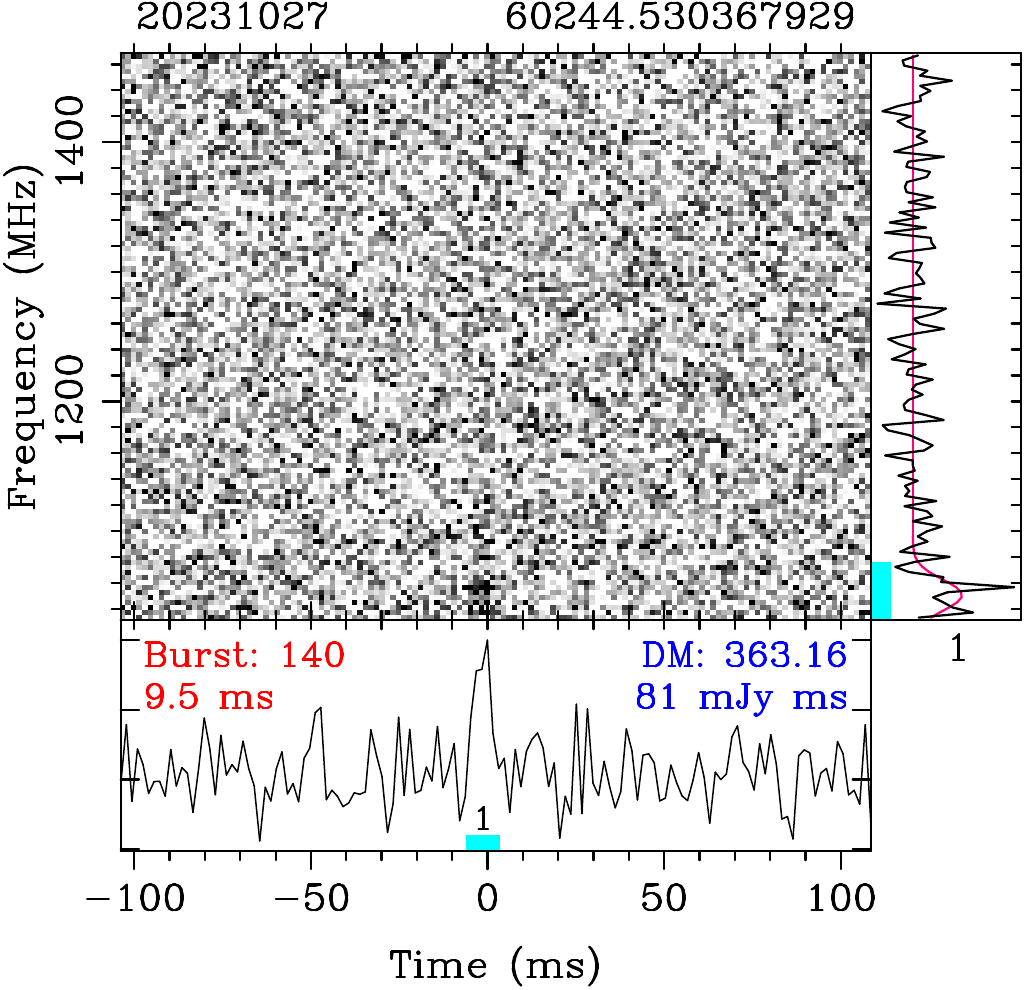}
\includegraphics[height=0.29\linewidth]{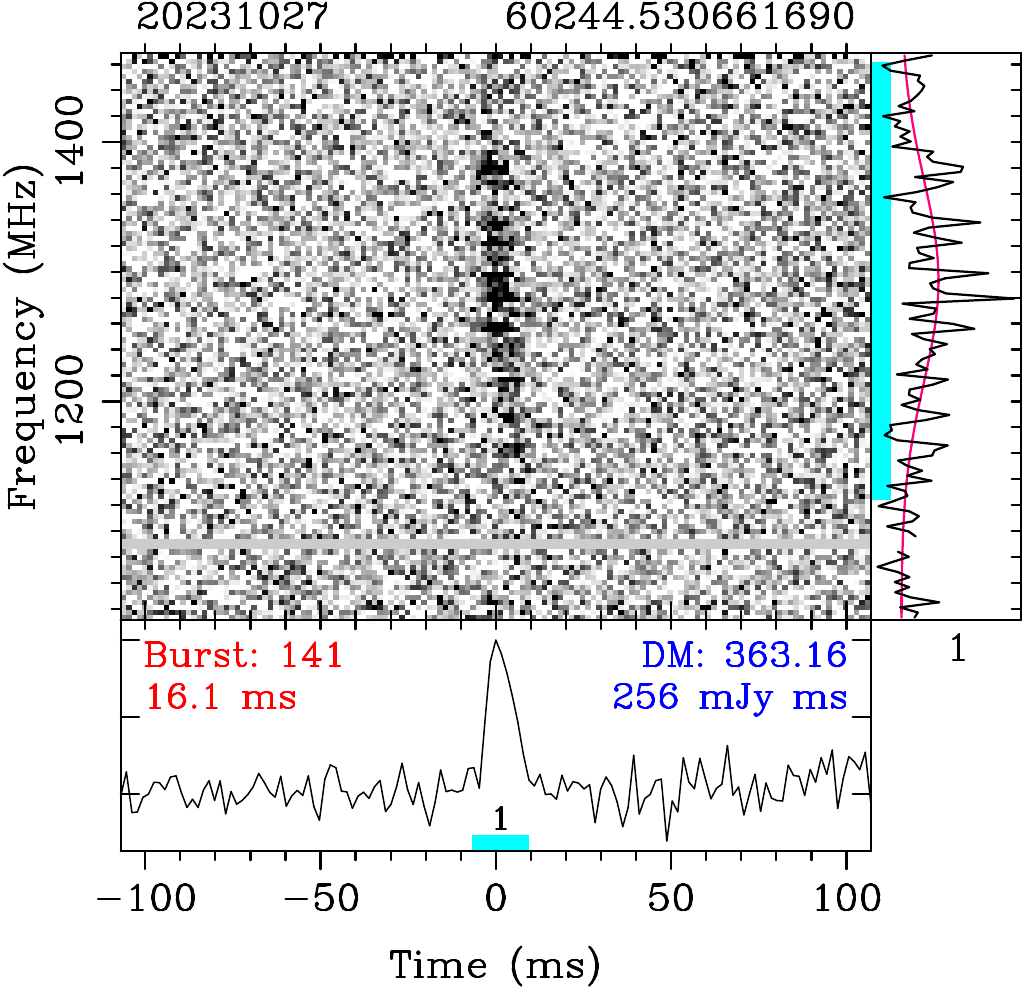}
\caption{({\textit{continued}})}
\end{figure*}
\addtocounter{figure}{-1}
\begin{figure*}
\flushleft
\includegraphics[height=0.29\linewidth]{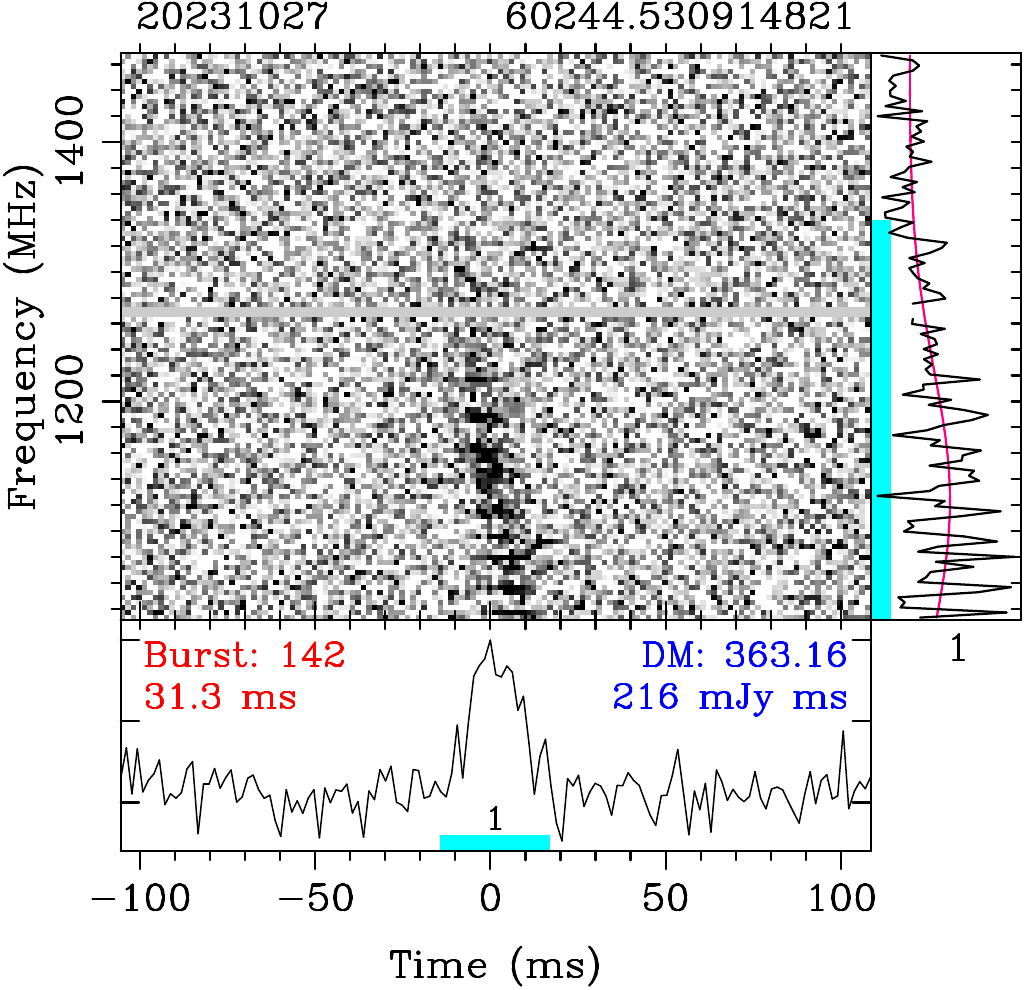}
\includegraphics[height=0.29\linewidth]{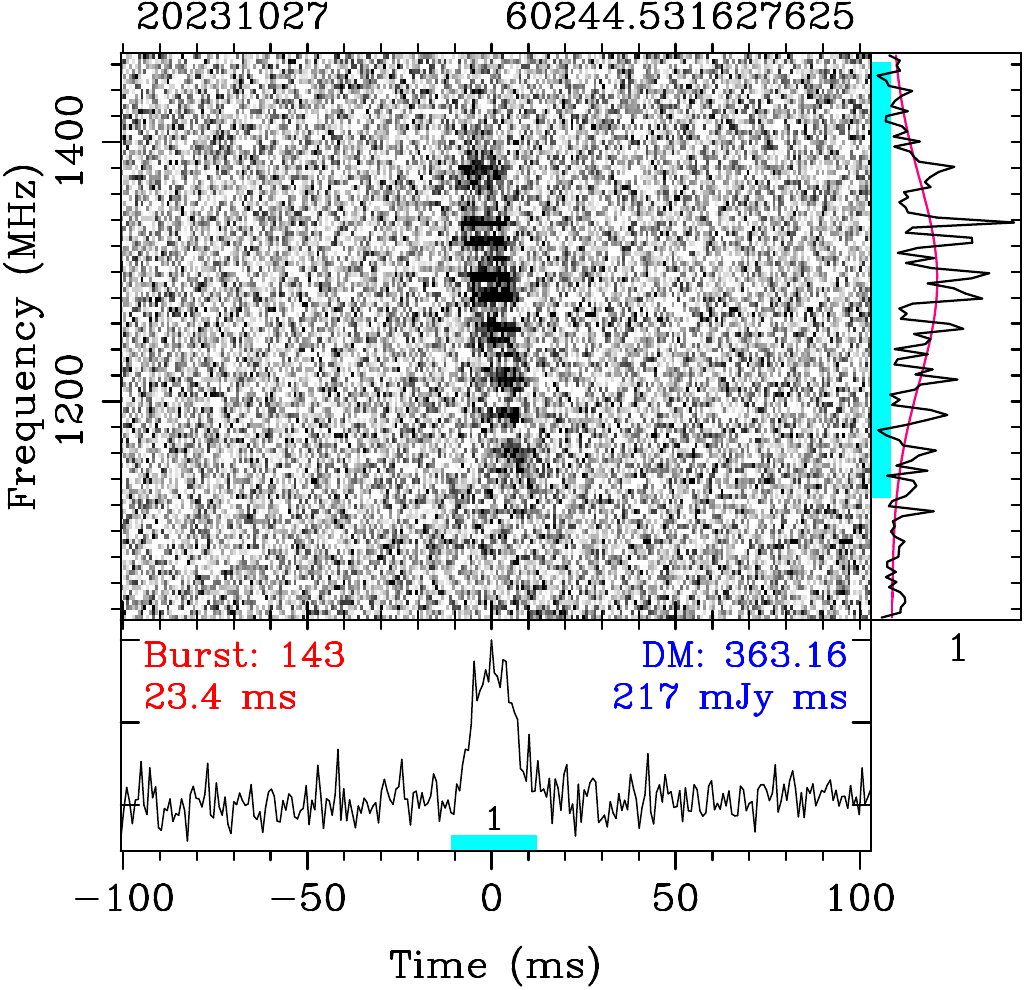}
\includegraphics[height=0.29\linewidth]{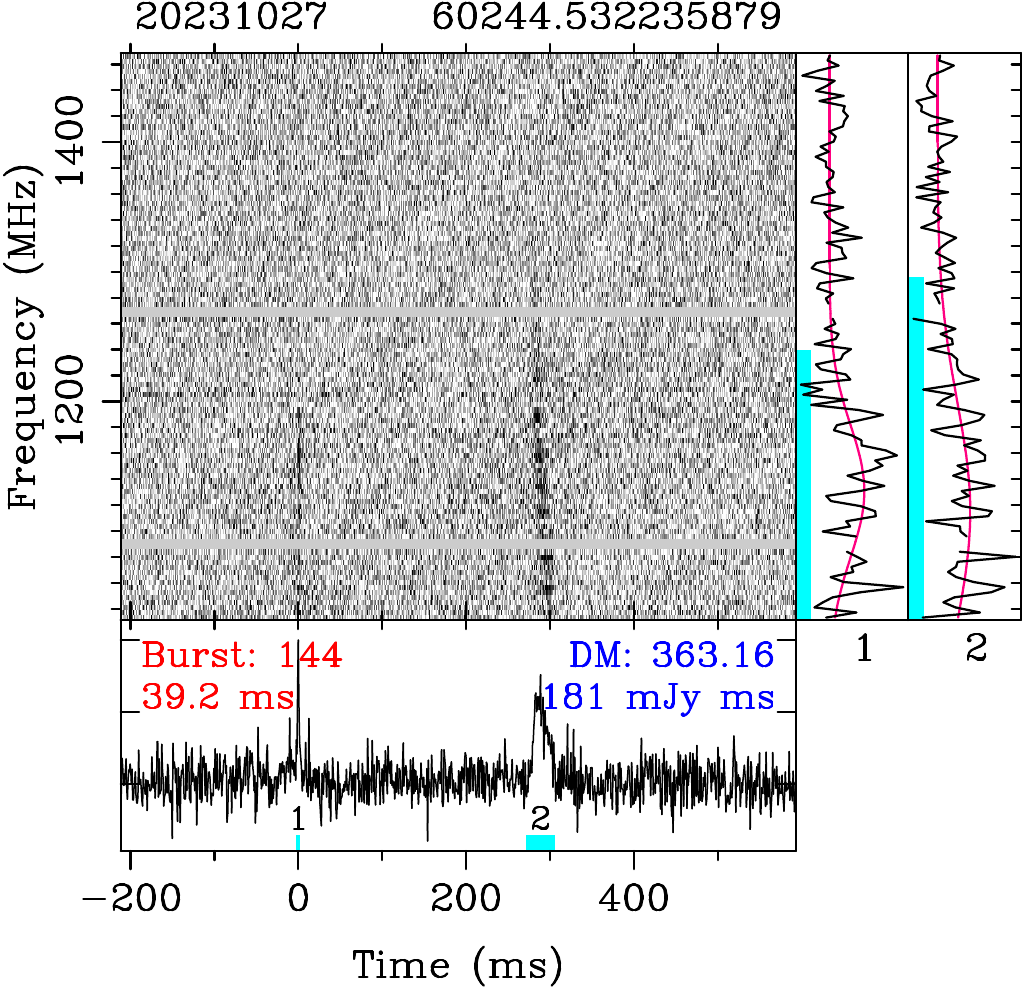}
\includegraphics[height=0.29\linewidth]{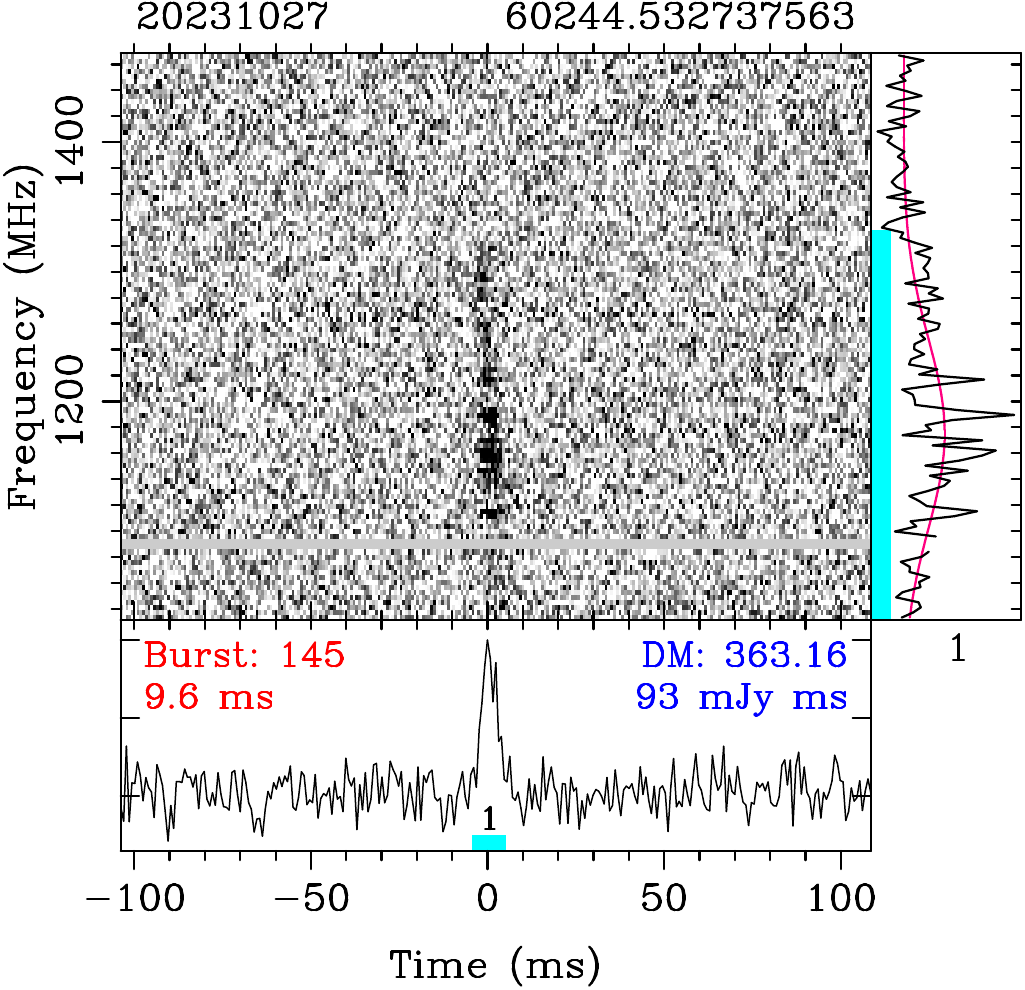}
\includegraphics[height=0.29\linewidth]{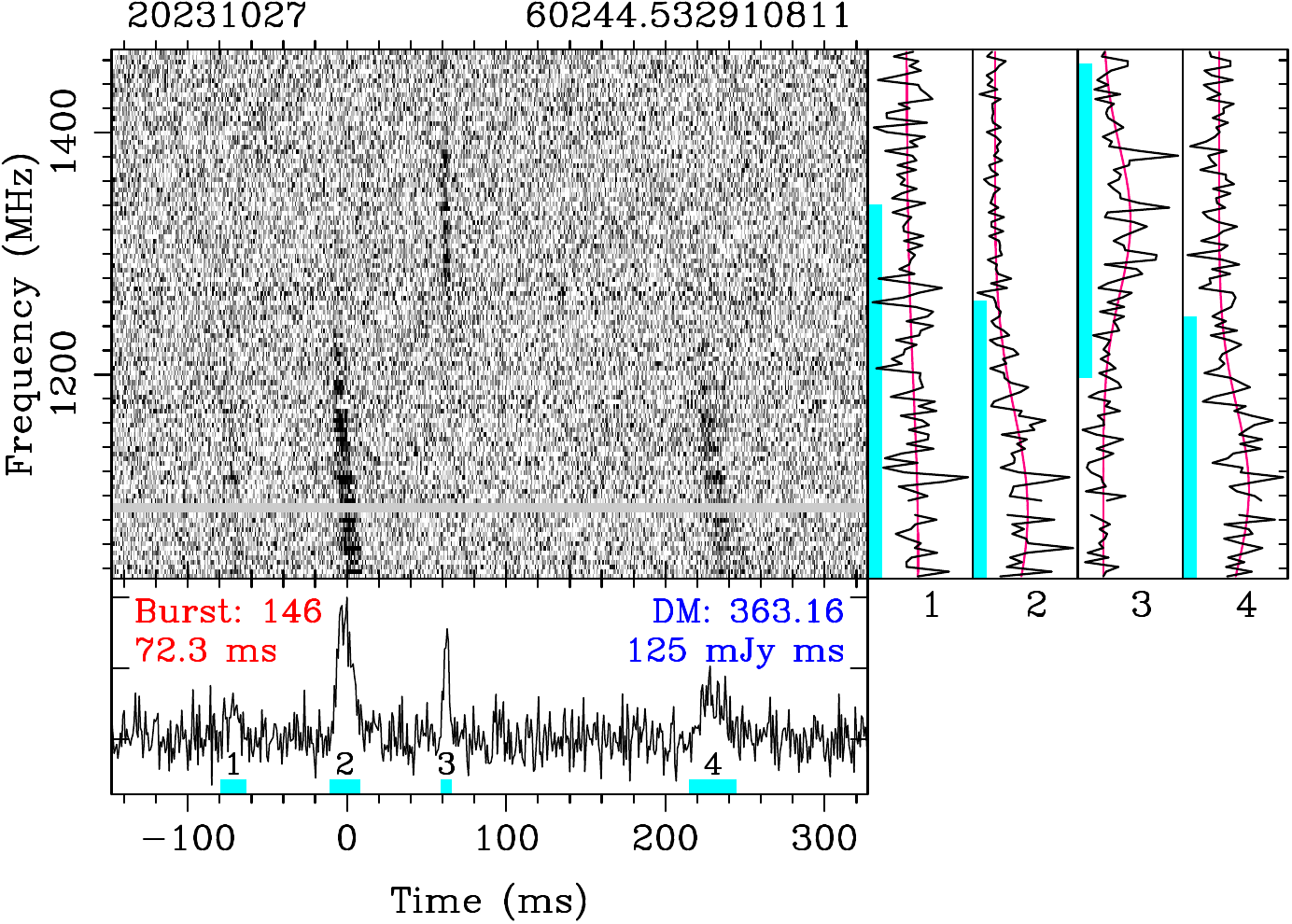}
\includegraphics[height=0.29\linewidth]{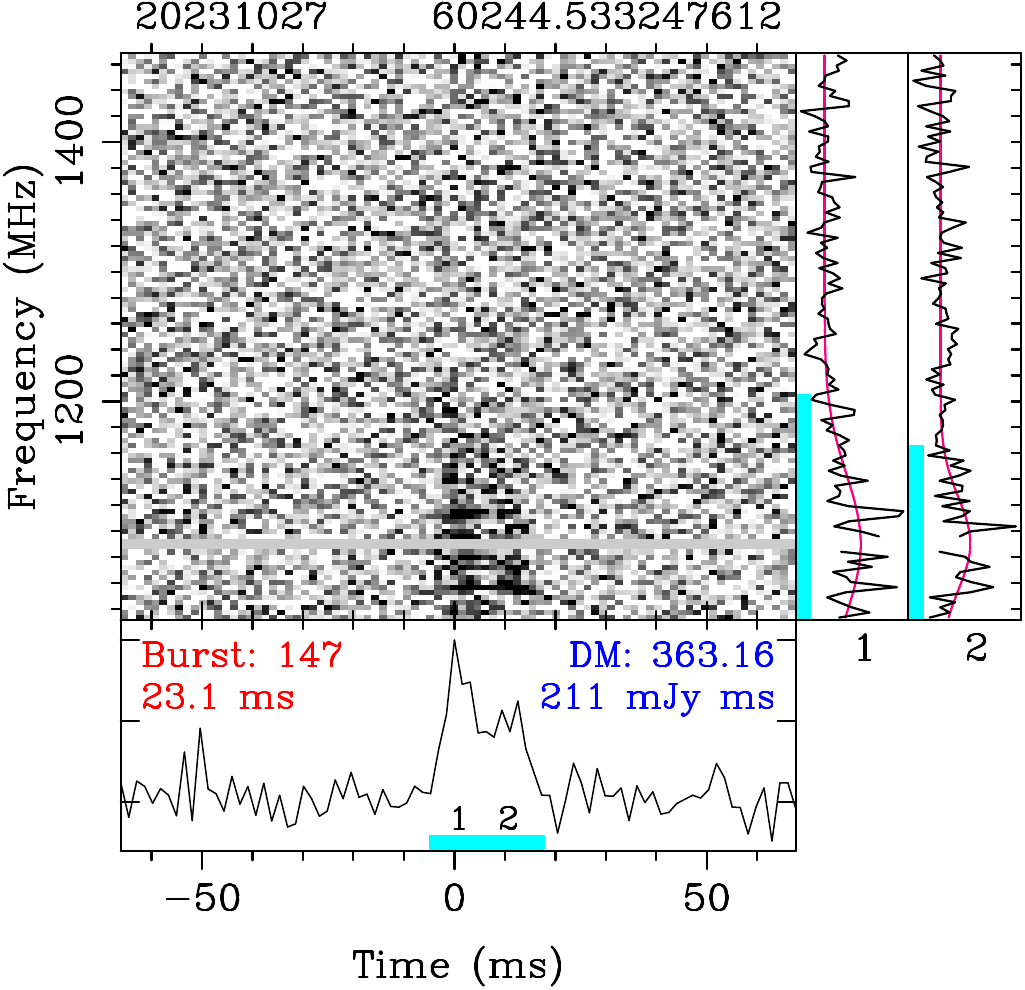}
\includegraphics[height=0.29\linewidth]{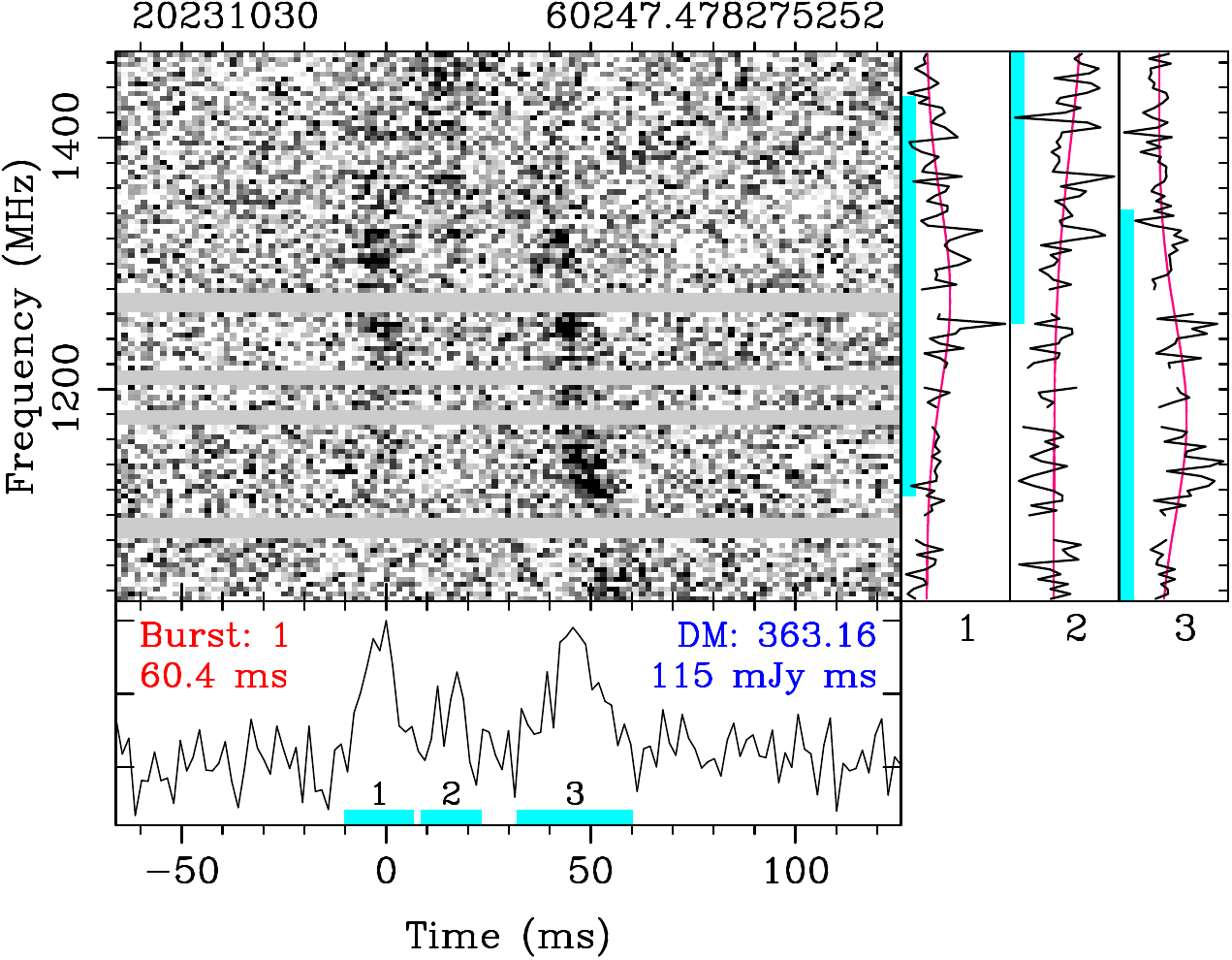}
\includegraphics[height=0.29\linewidth]{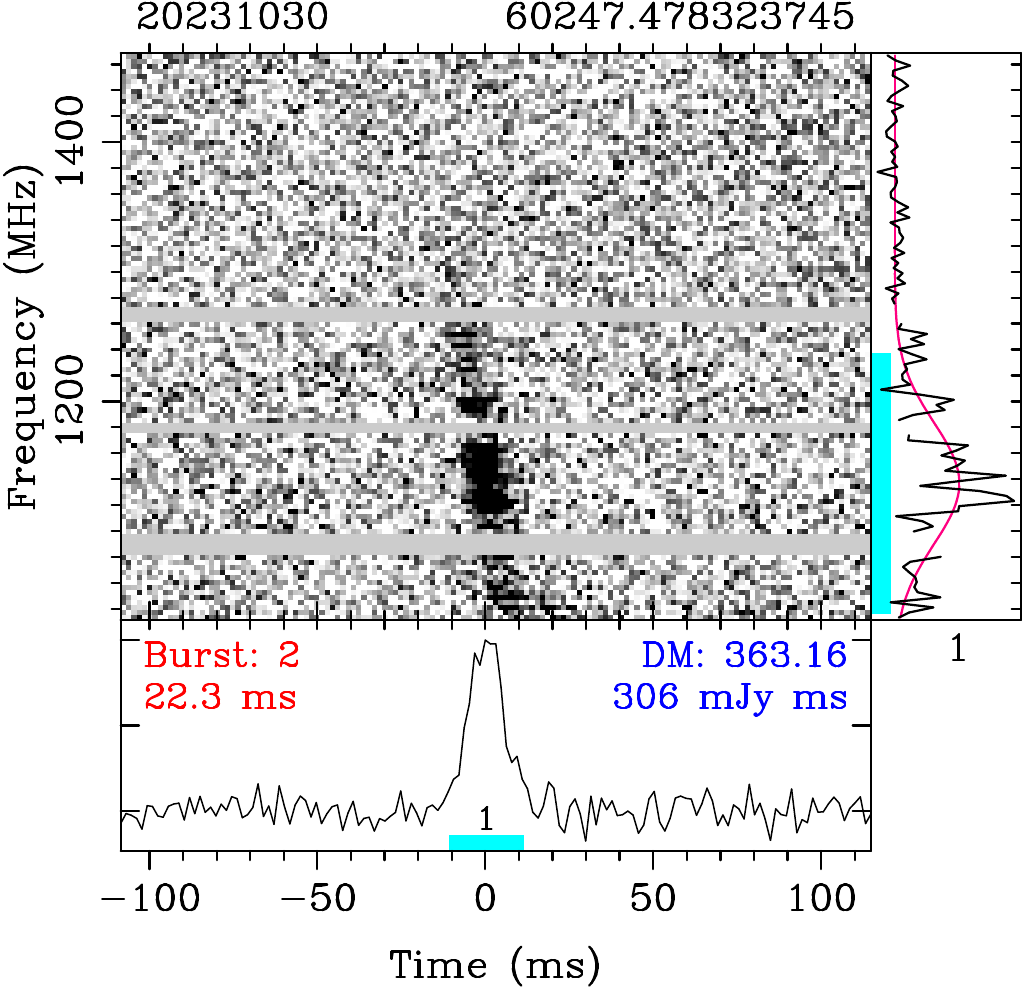}
\includegraphics[height=0.29\linewidth]{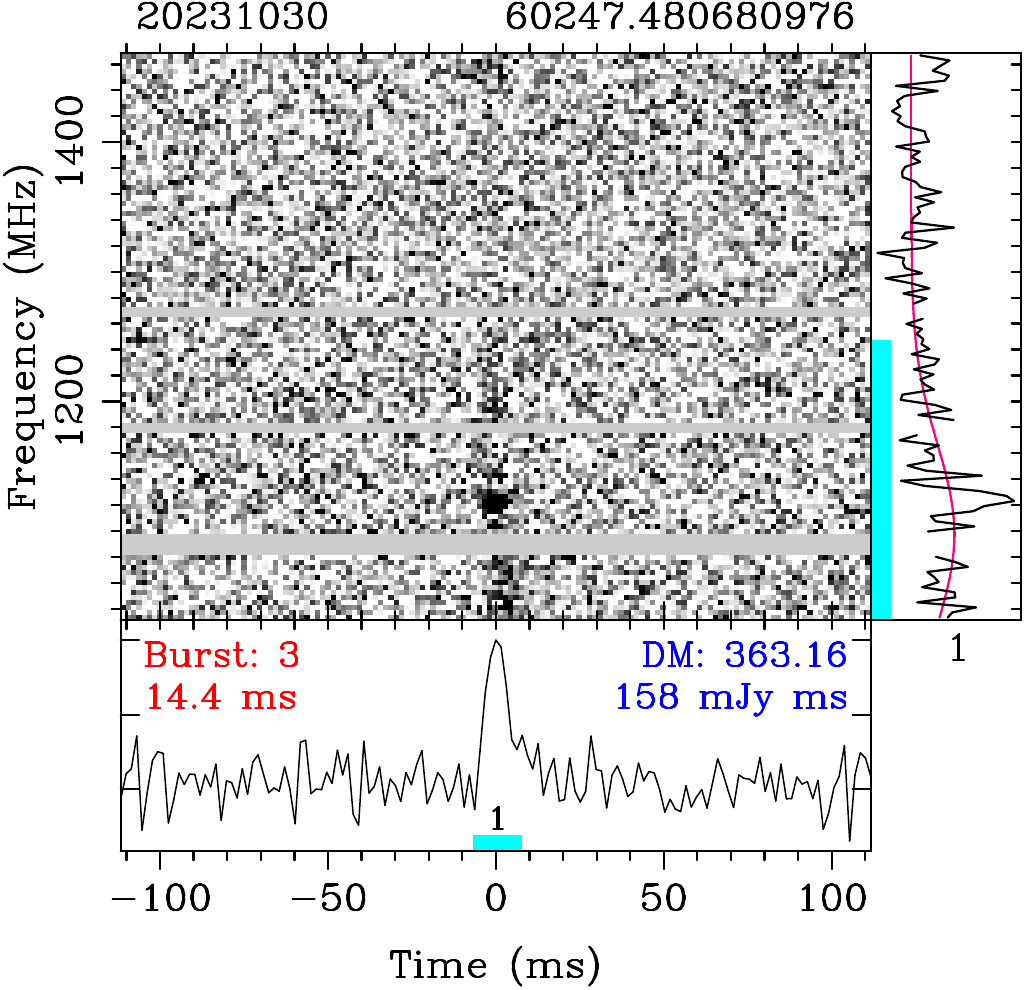}
\includegraphics[height=0.29\linewidth]{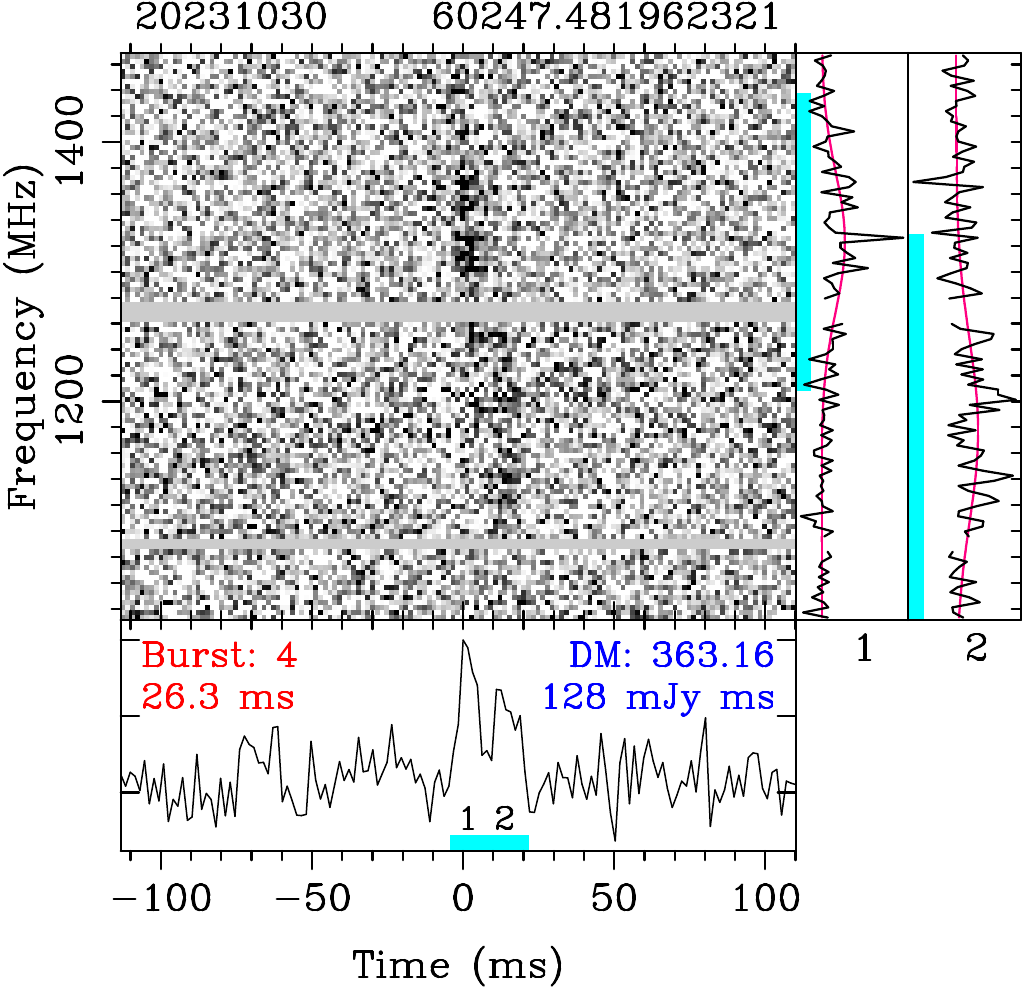}
\includegraphics[height=0.29\linewidth]{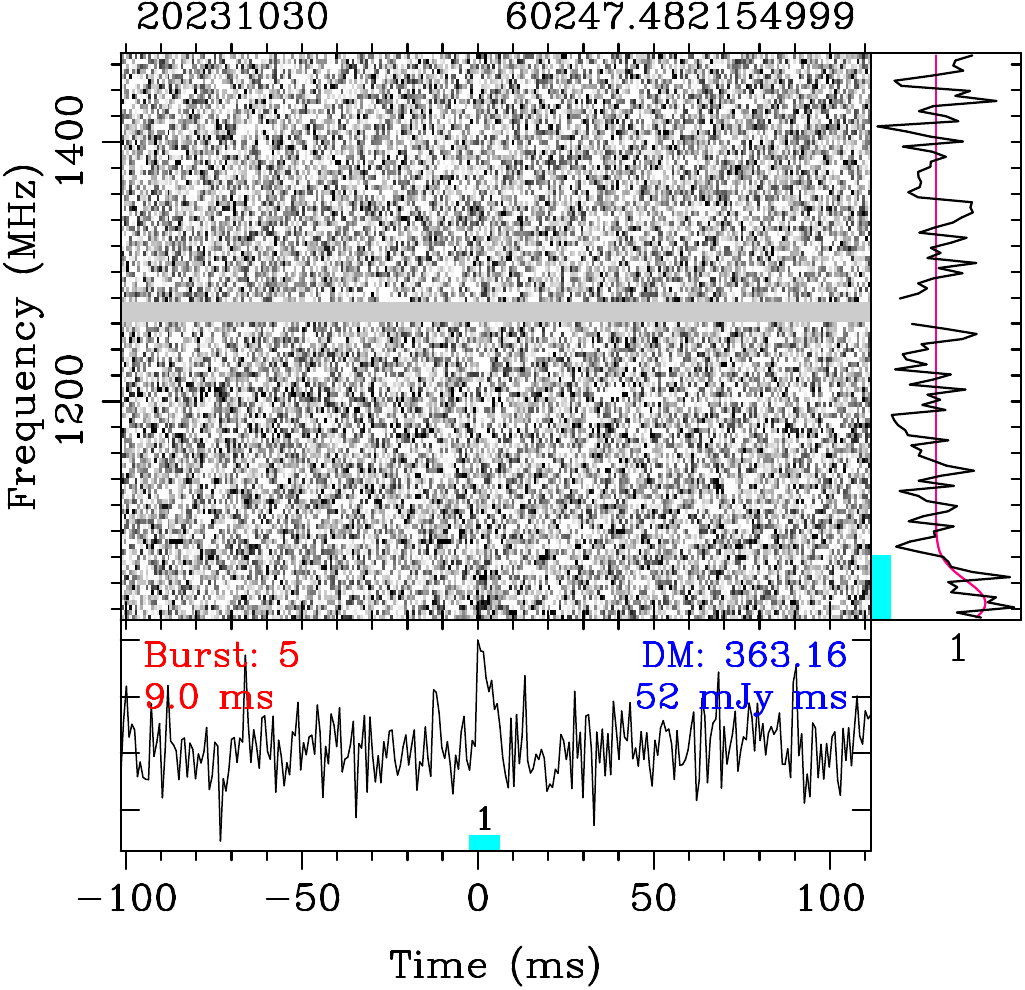}
\caption{({\textit{continued}})}
\end{figure*}
\addtocounter{figure}{-1}
\begin{figure*}
\flushleft
\includegraphics[height=0.29\linewidth]{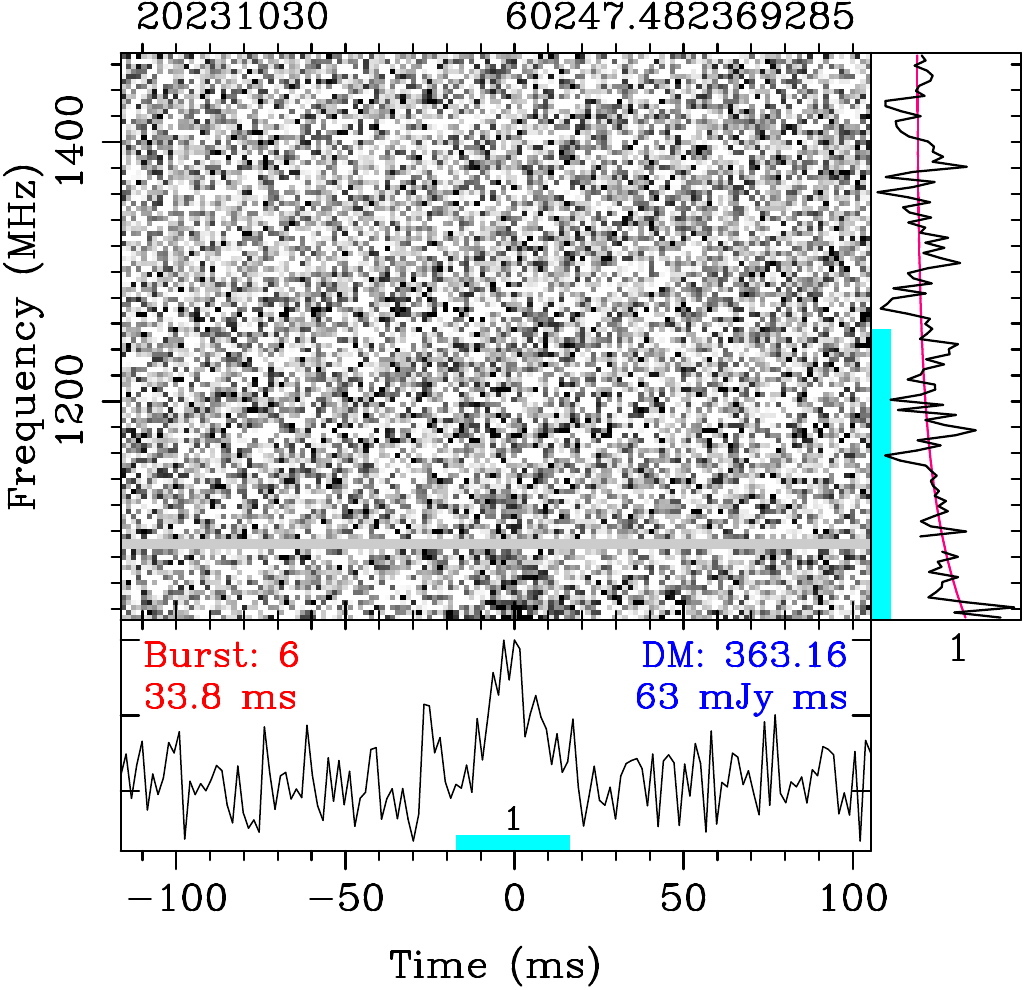}
\includegraphics[height=0.29\linewidth]{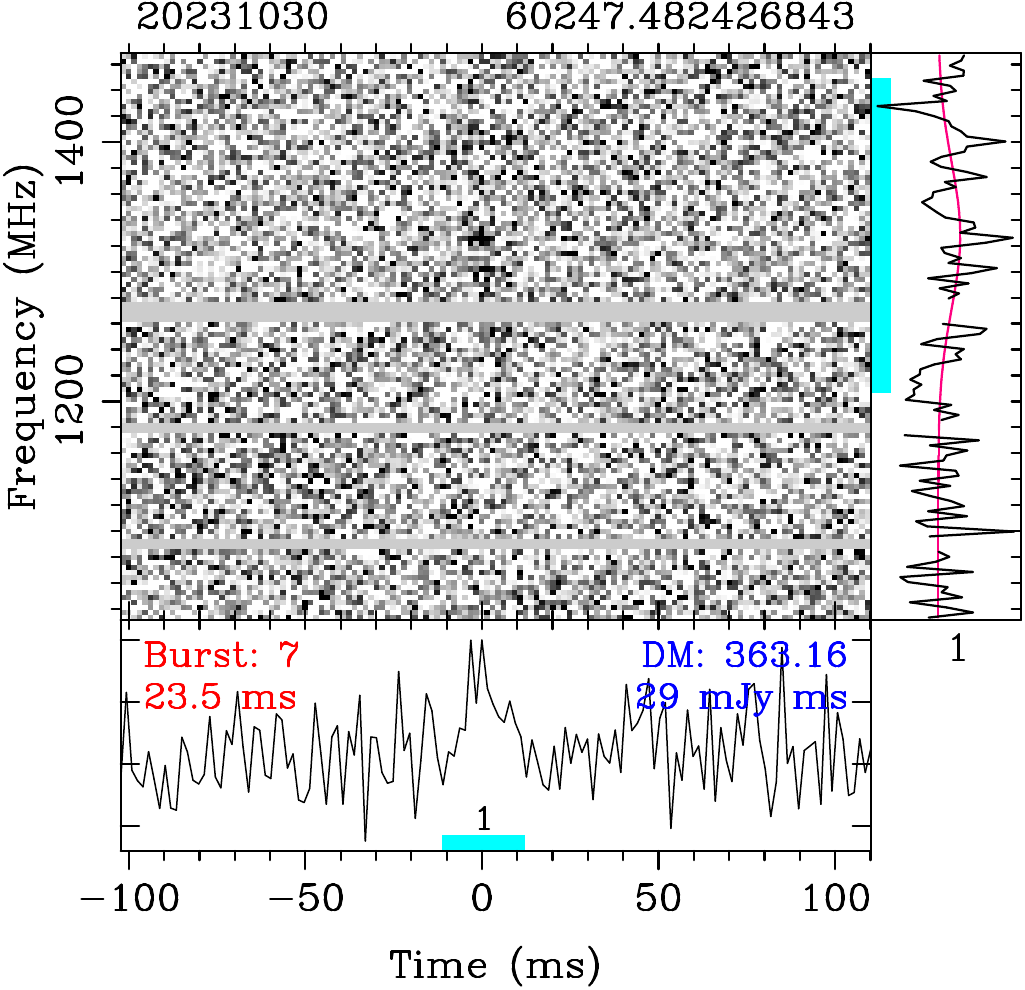}
\includegraphics[height=0.29\linewidth]{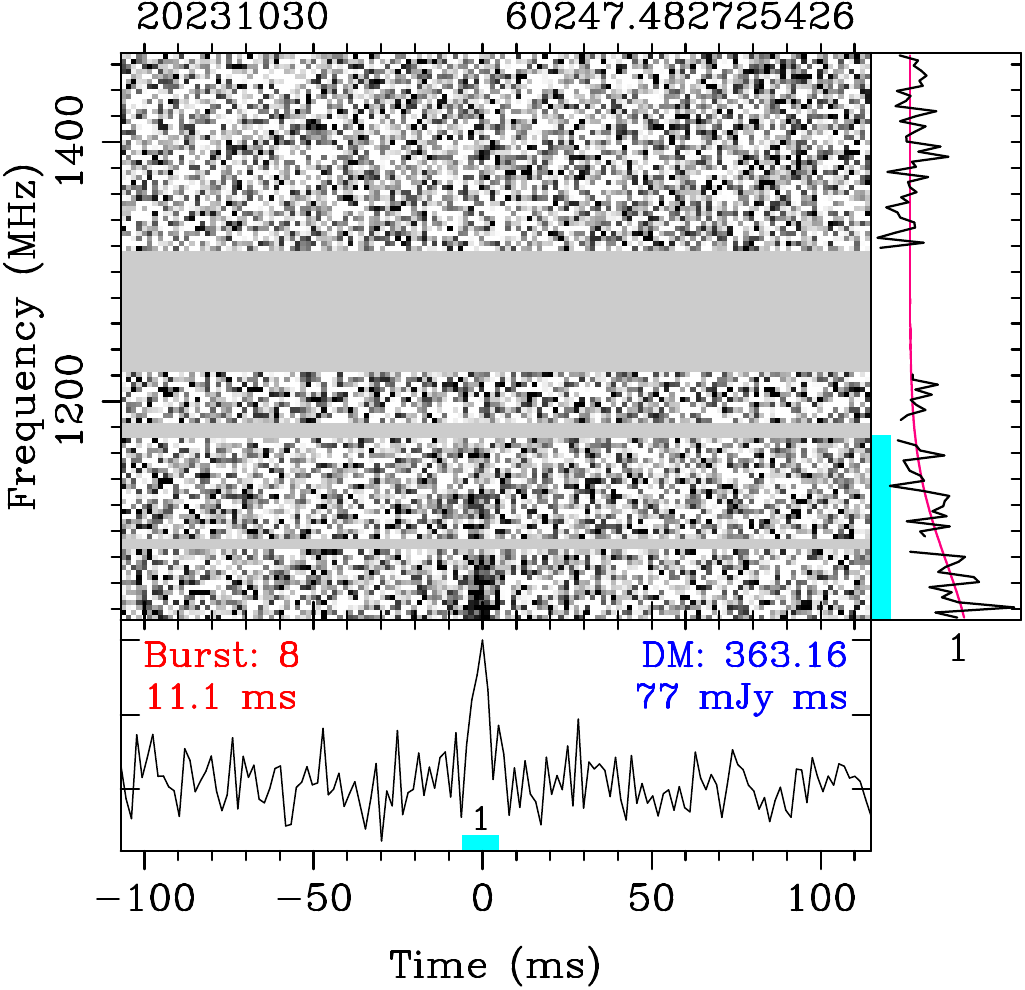}
\includegraphics[height=0.29\linewidth]{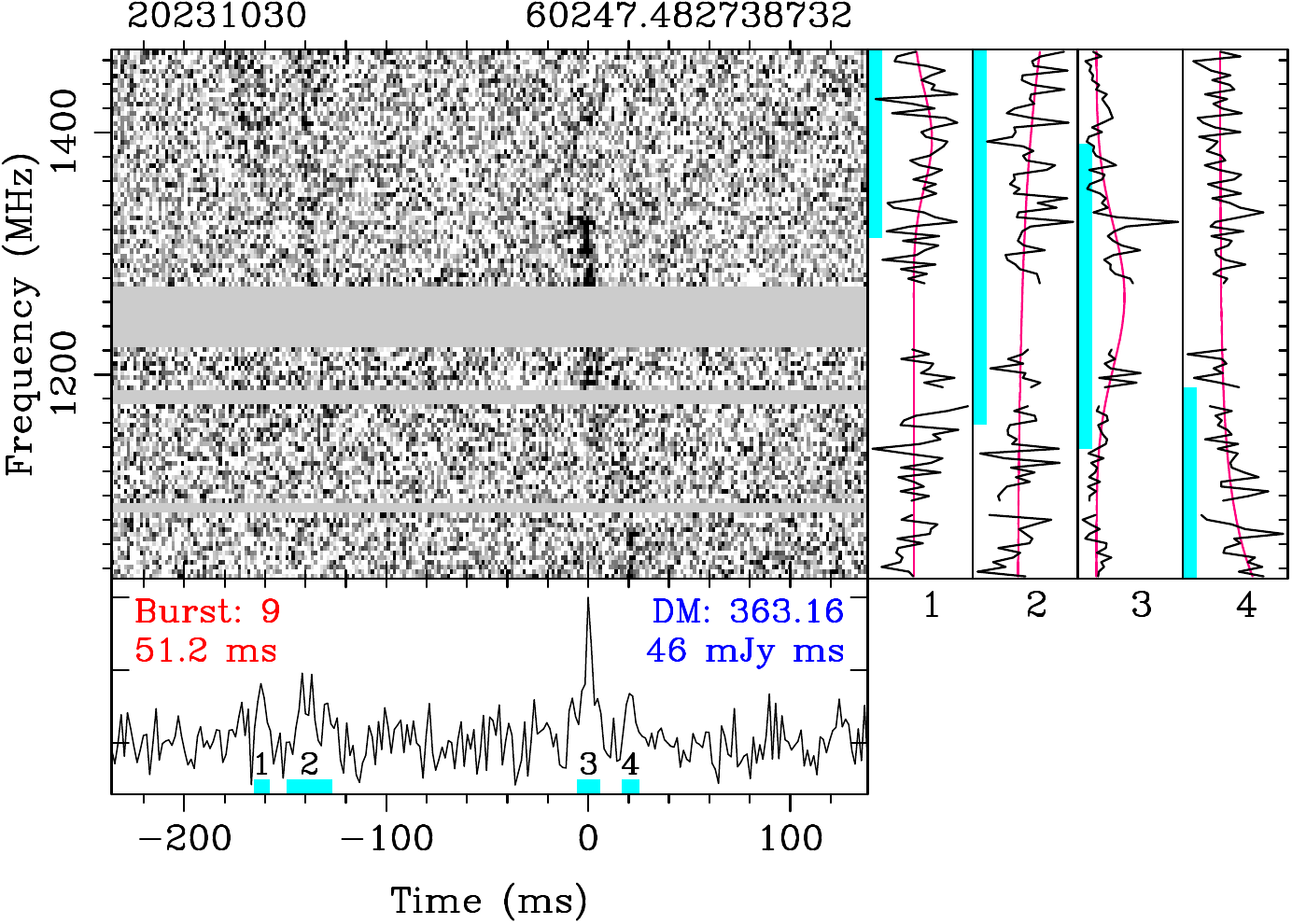}
\includegraphics[height=0.29\linewidth]{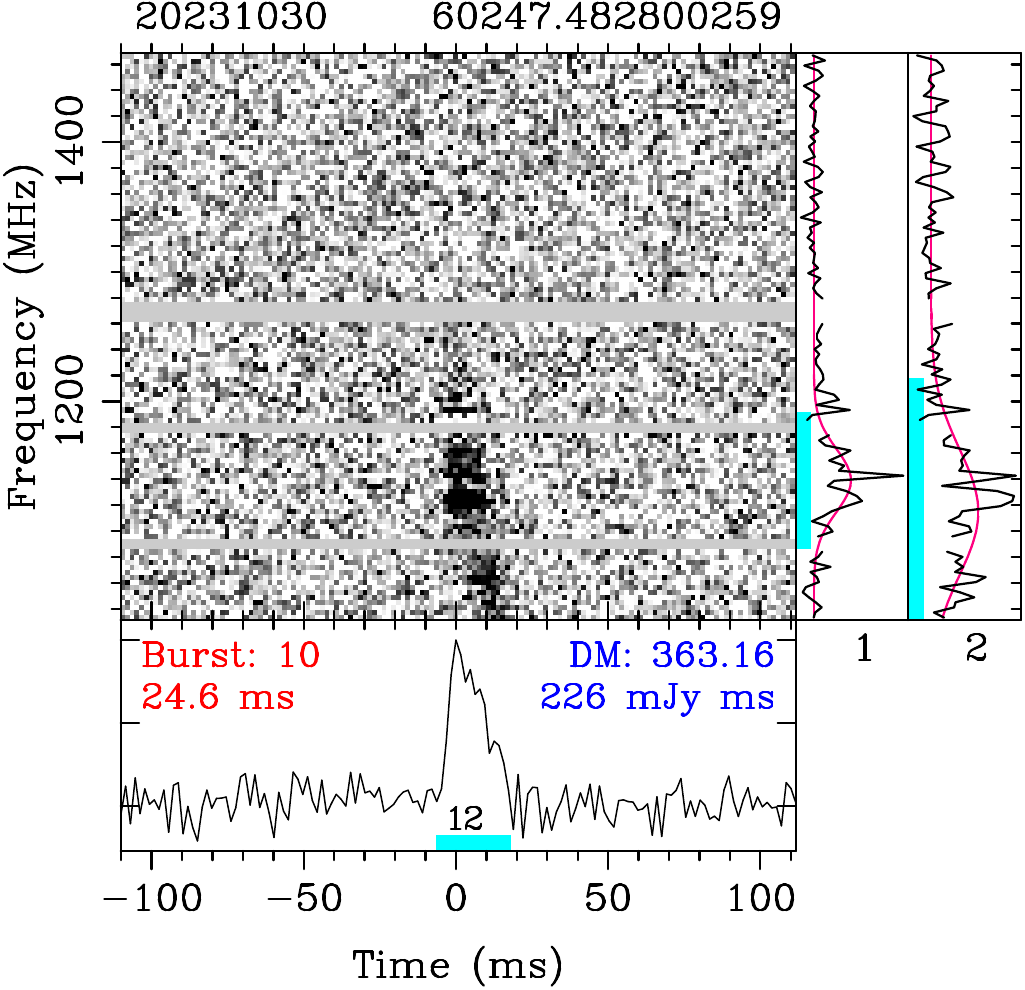}
\includegraphics[height=0.29\linewidth]{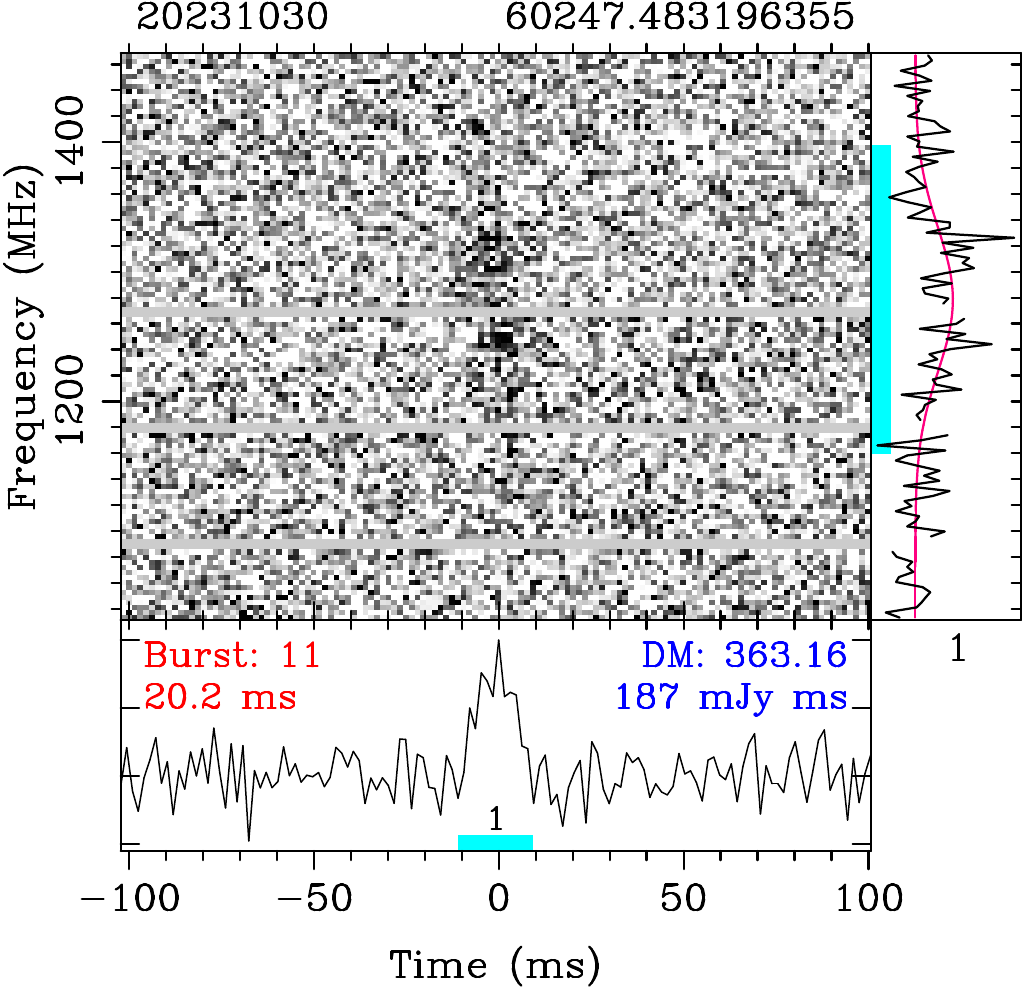}
\includegraphics[height=0.29\linewidth]{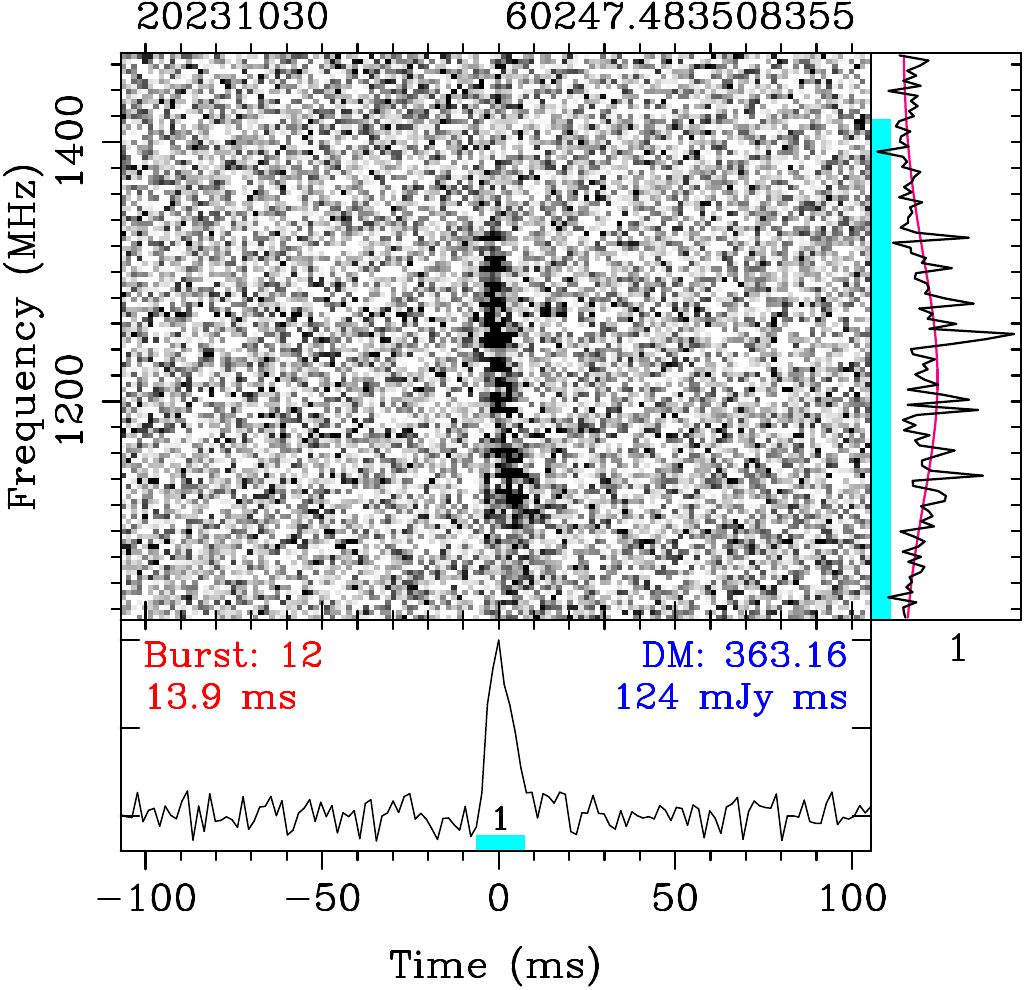}
\includegraphics[height=0.29\linewidth]{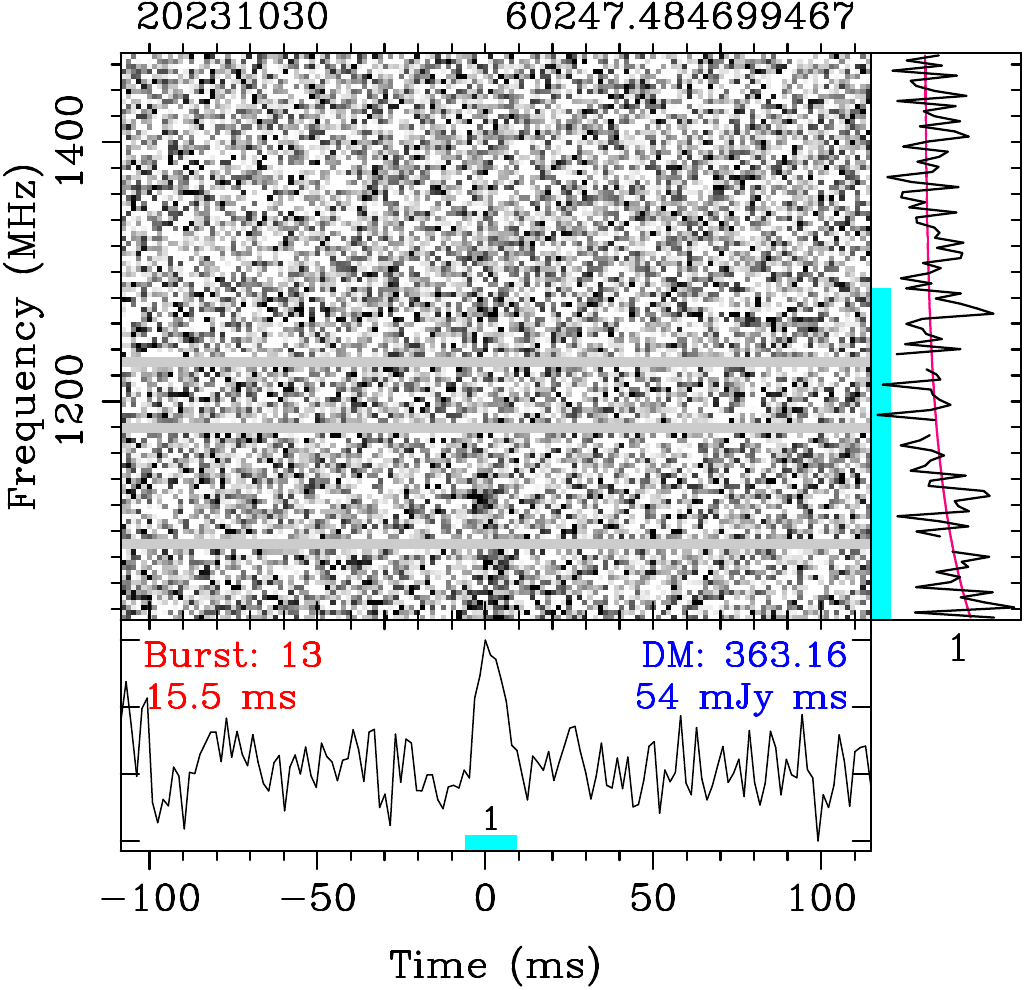}
\includegraphics[height=0.29\linewidth]{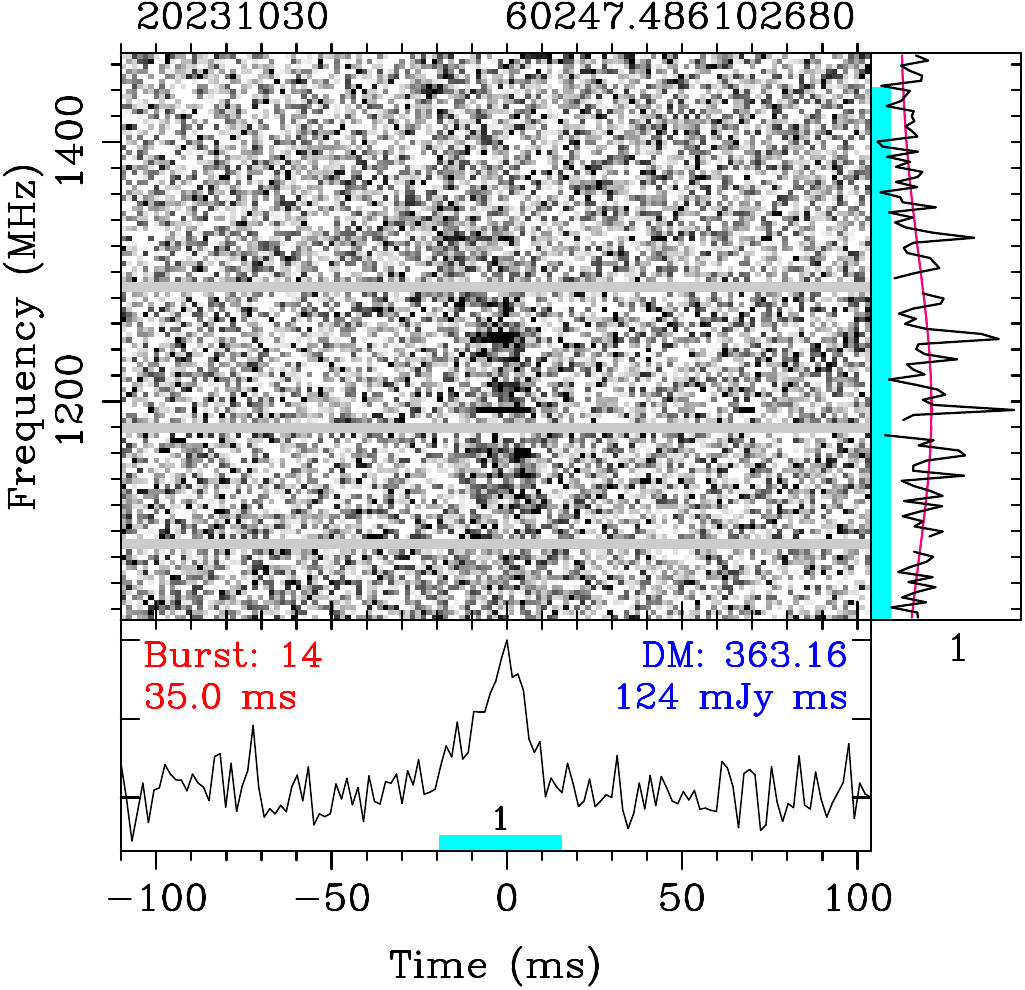}
\includegraphics[height=0.29\linewidth]{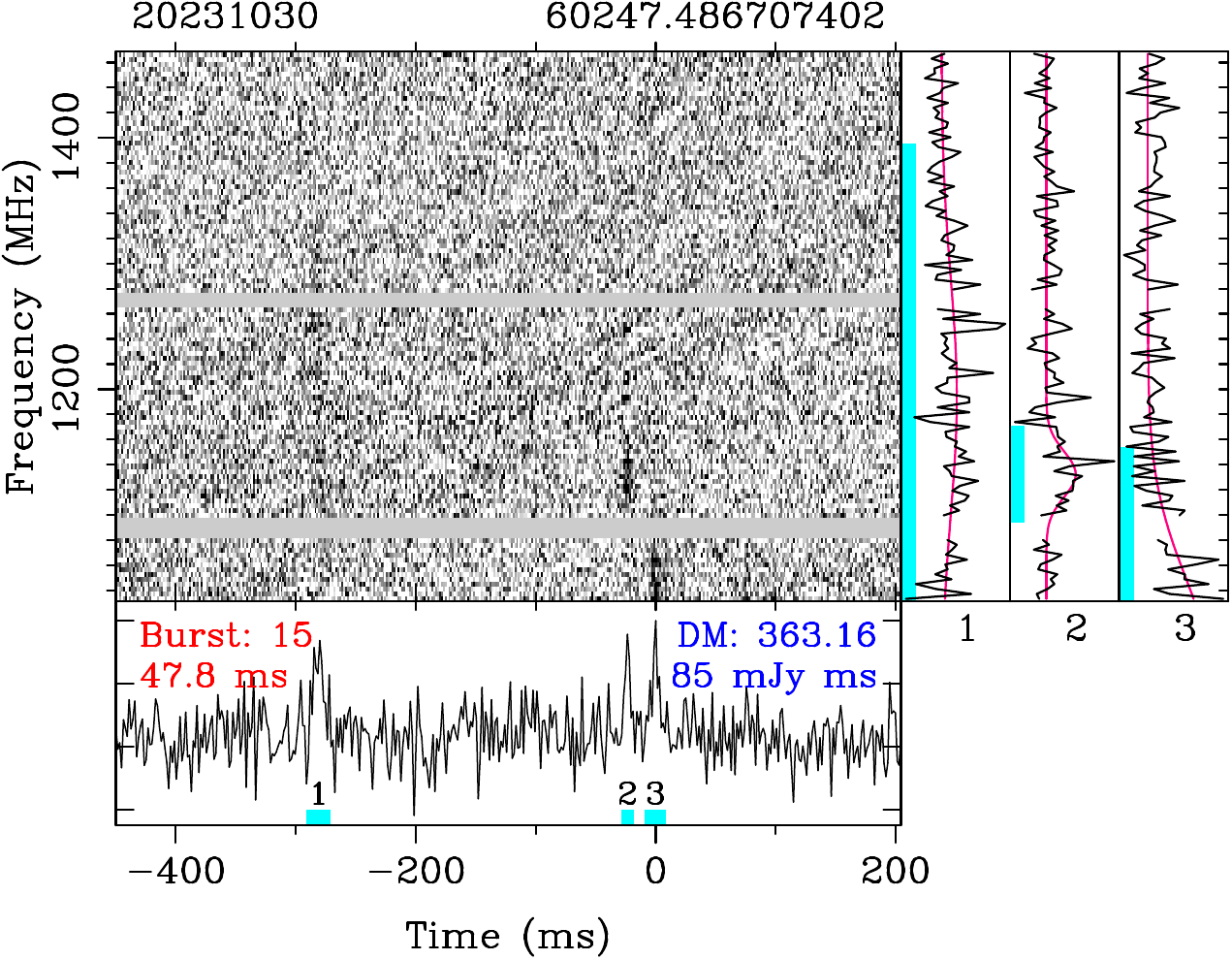}
\includegraphics[height=0.29\linewidth]{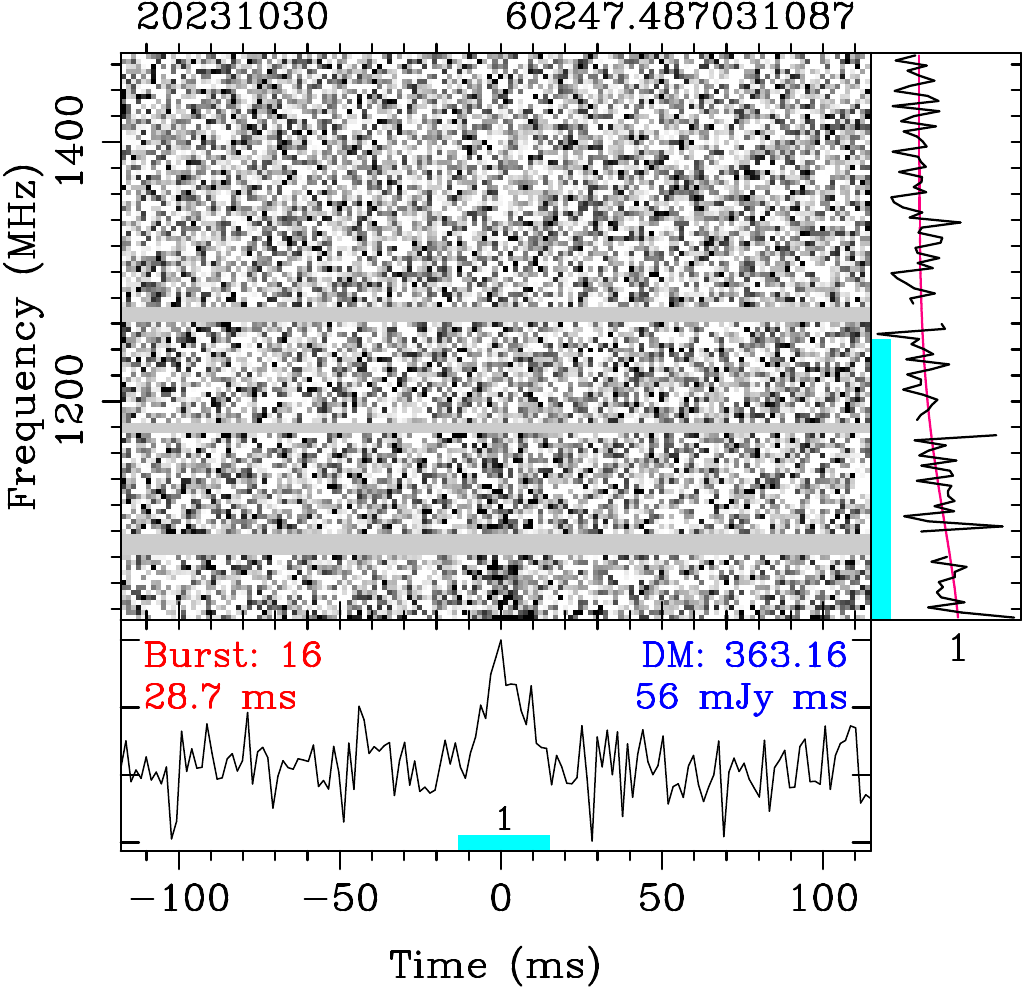}
\caption{({\textit{continued}})}
\end{figure*}
\addtocounter{figure}{-1}
\begin{figure*}
\flushleft
\includegraphics[height=0.29\linewidth]{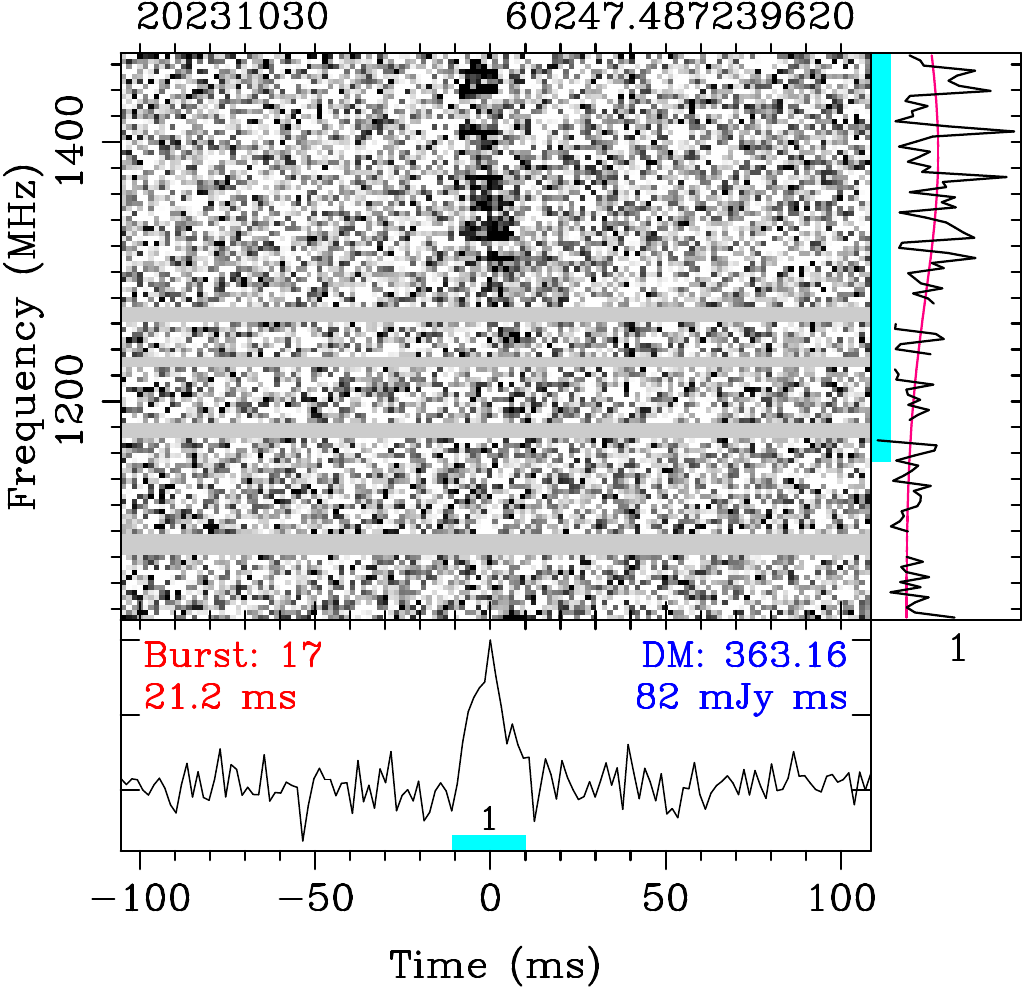}
\includegraphics[height=0.29\linewidth]{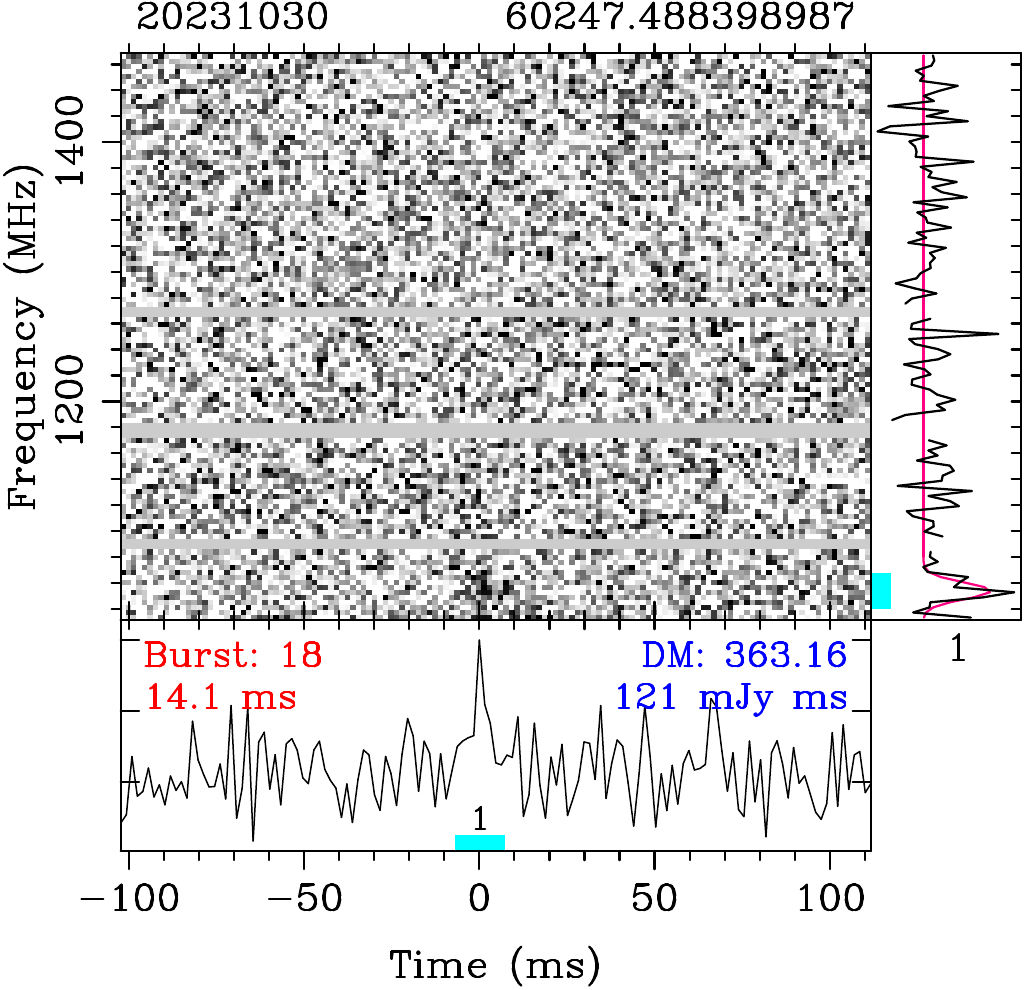}
\includegraphics[height=0.29\linewidth]{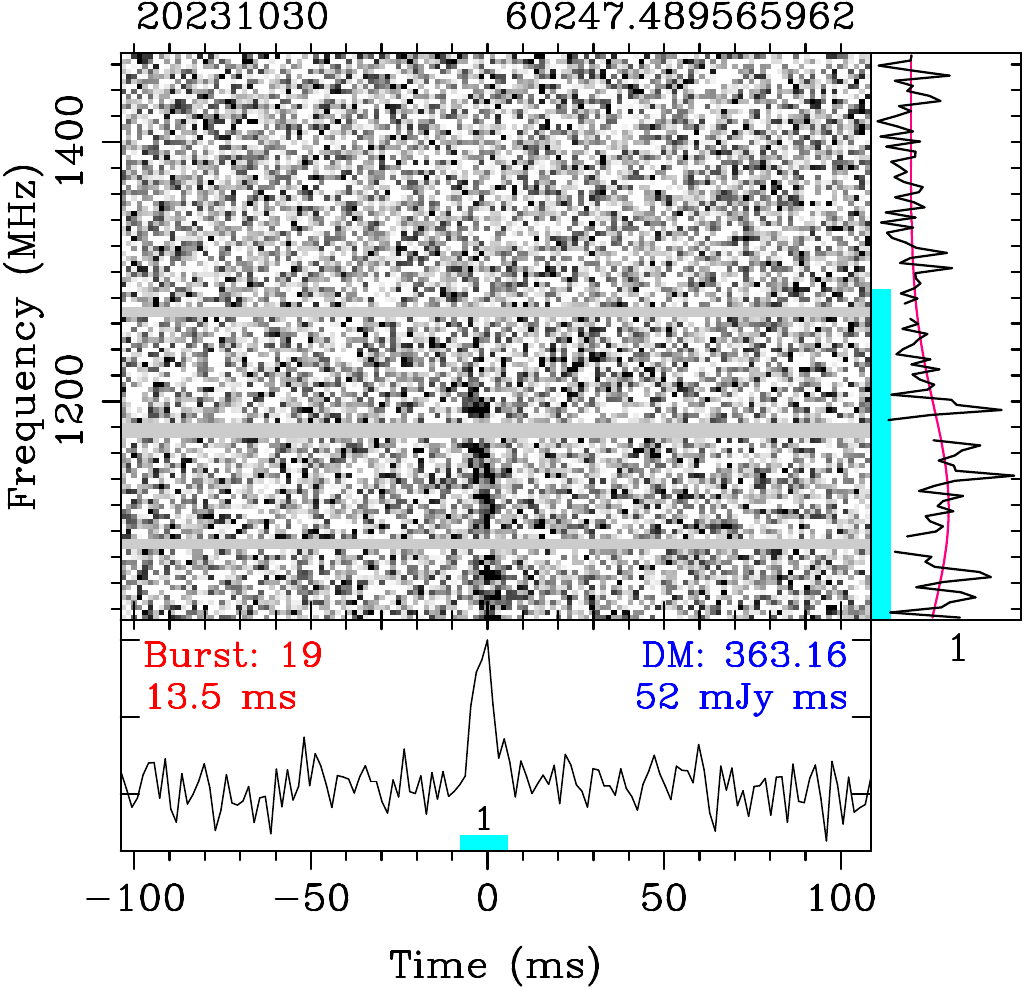}
\includegraphics[height=0.29\linewidth]{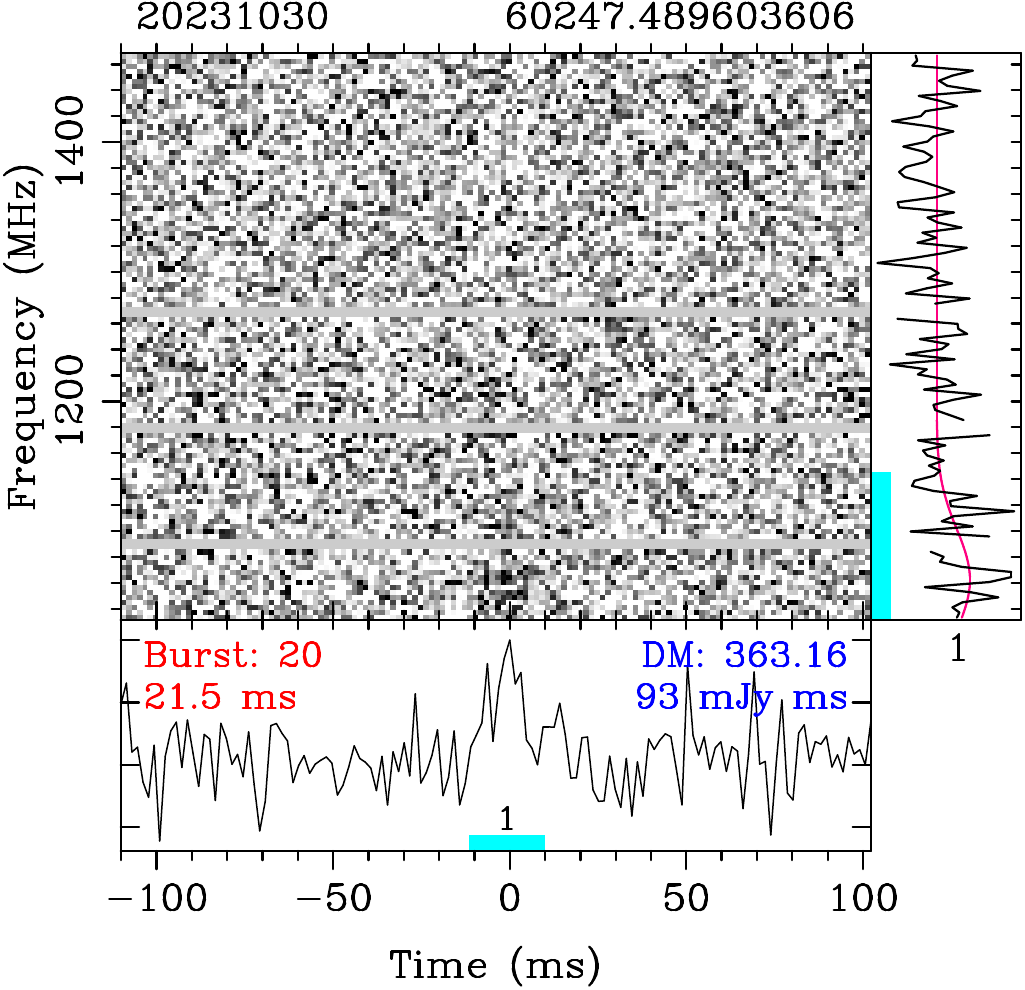}
\includegraphics[height=0.29\linewidth]{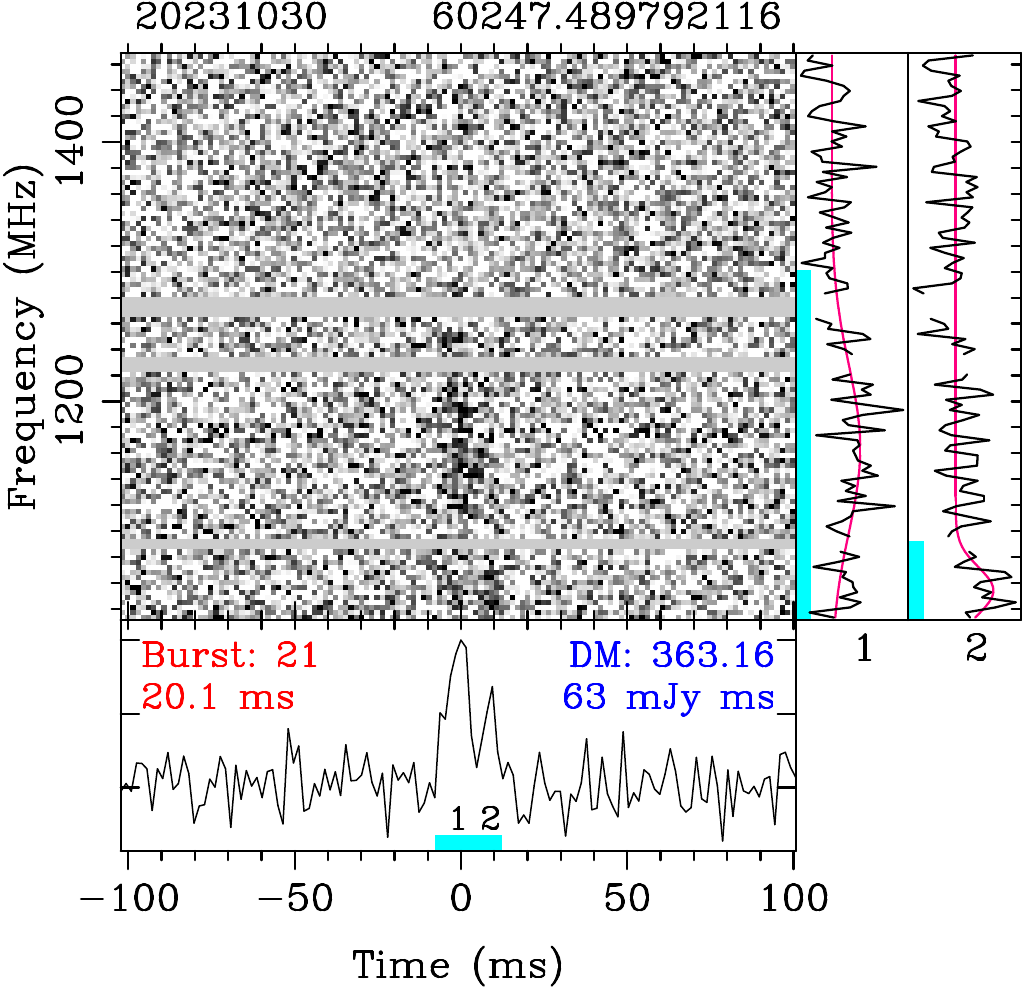}
\includegraphics[height=0.29\linewidth]{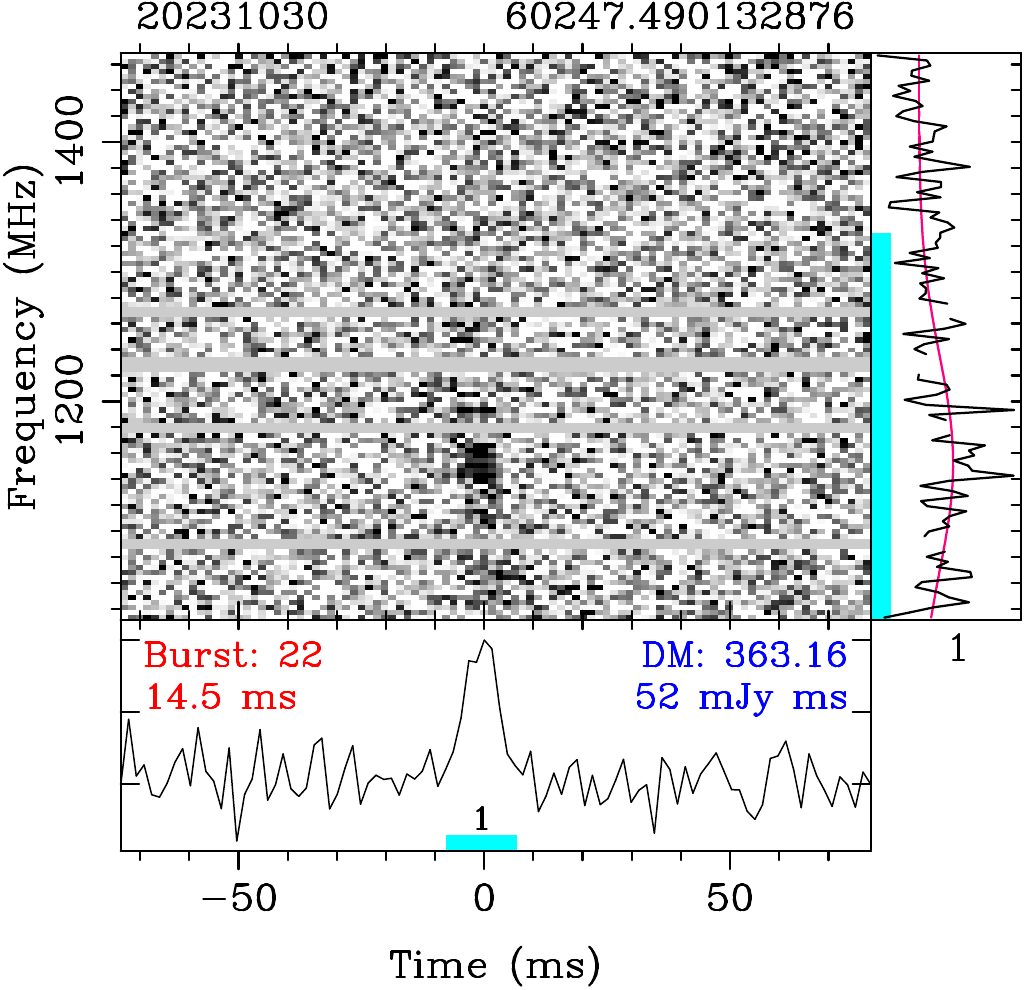}
\includegraphics[height=0.29\linewidth]{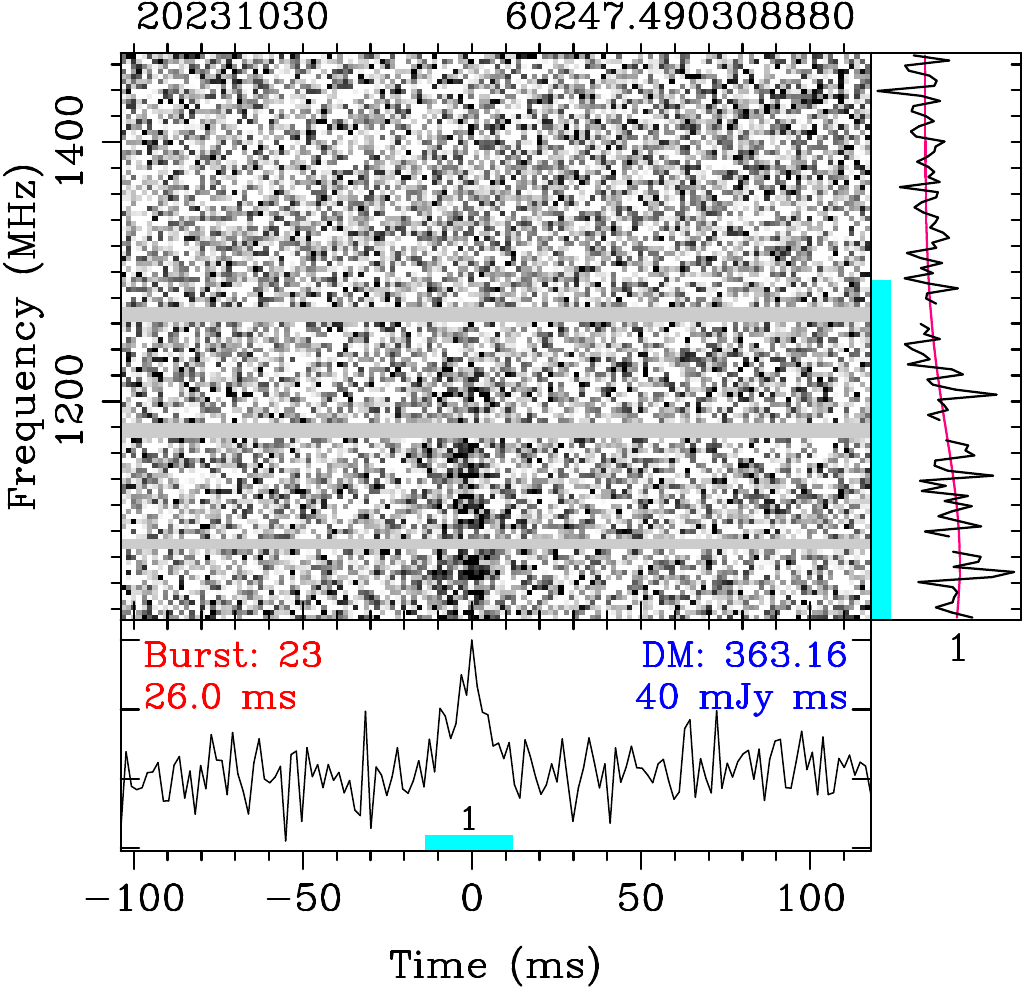}
\includegraphics[height=0.29\linewidth]{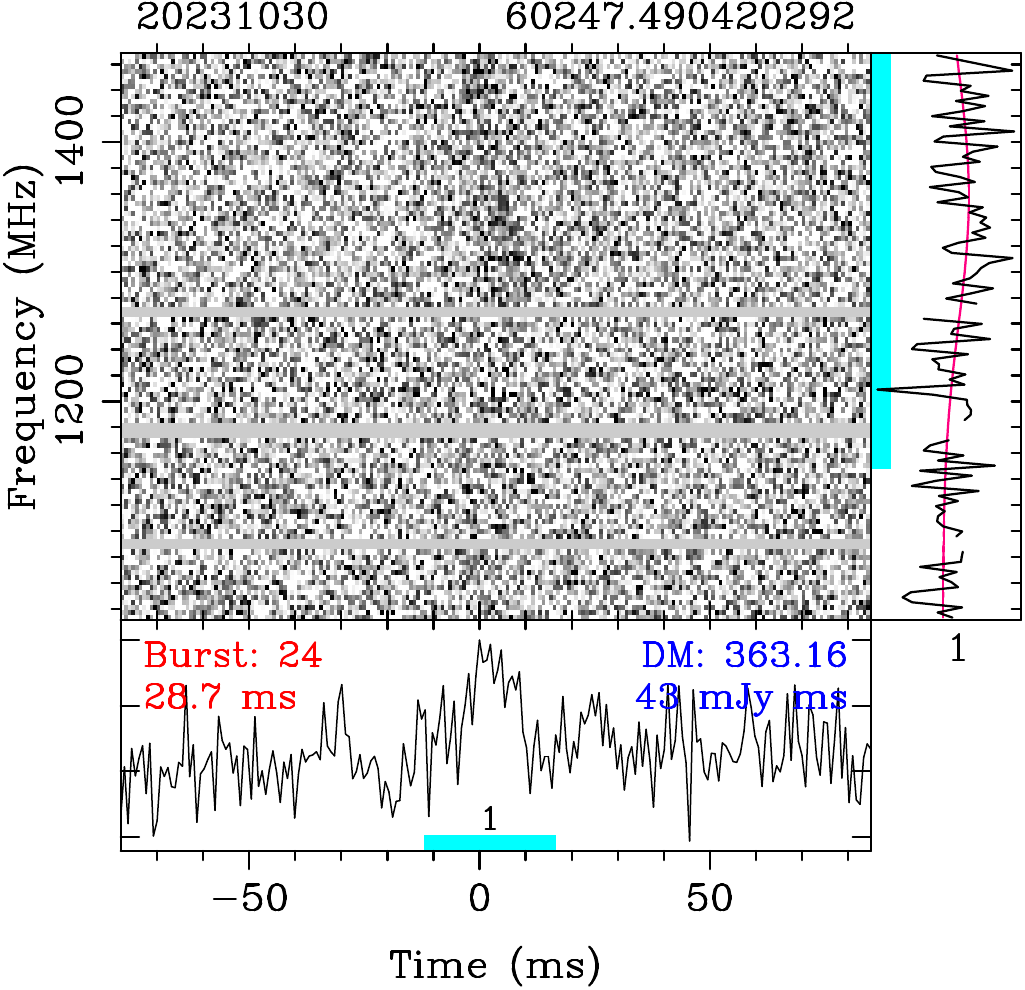}
\includegraphics[height=0.29\linewidth]{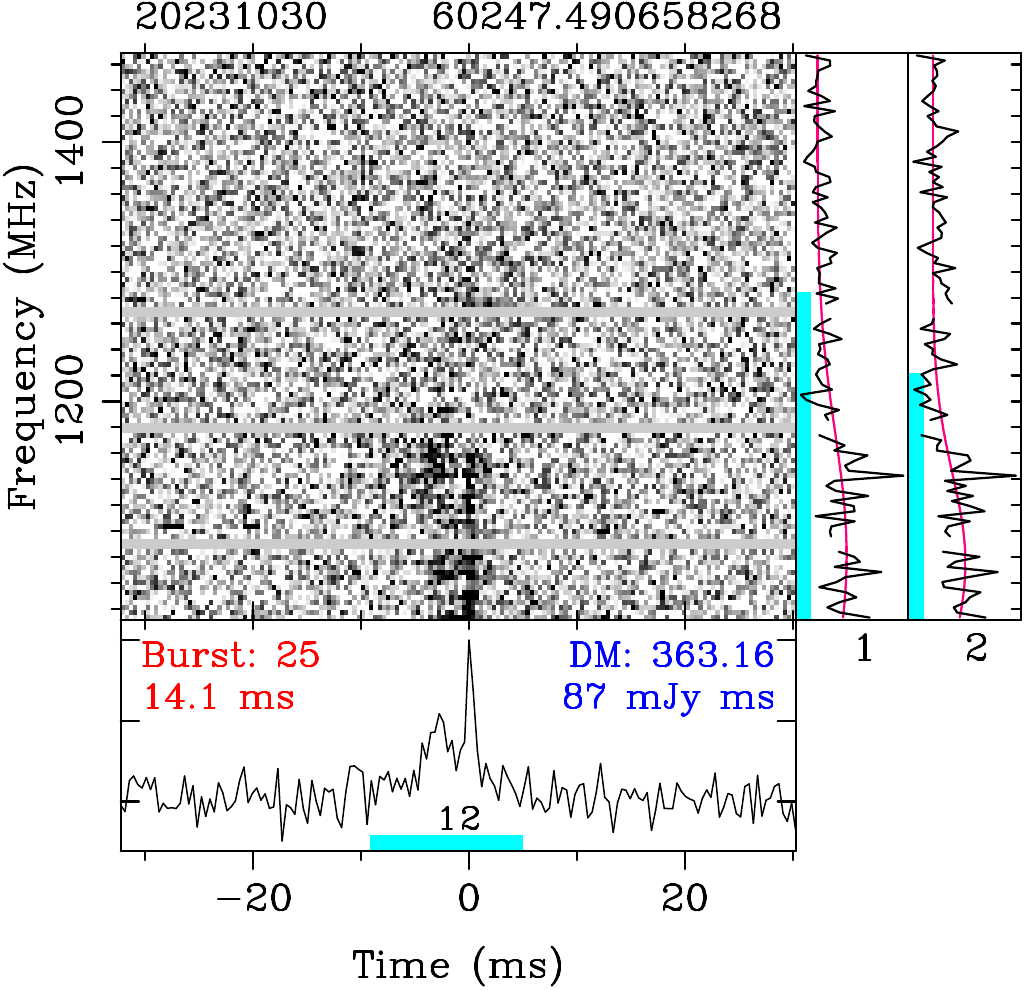}
\includegraphics[height=0.29\linewidth]{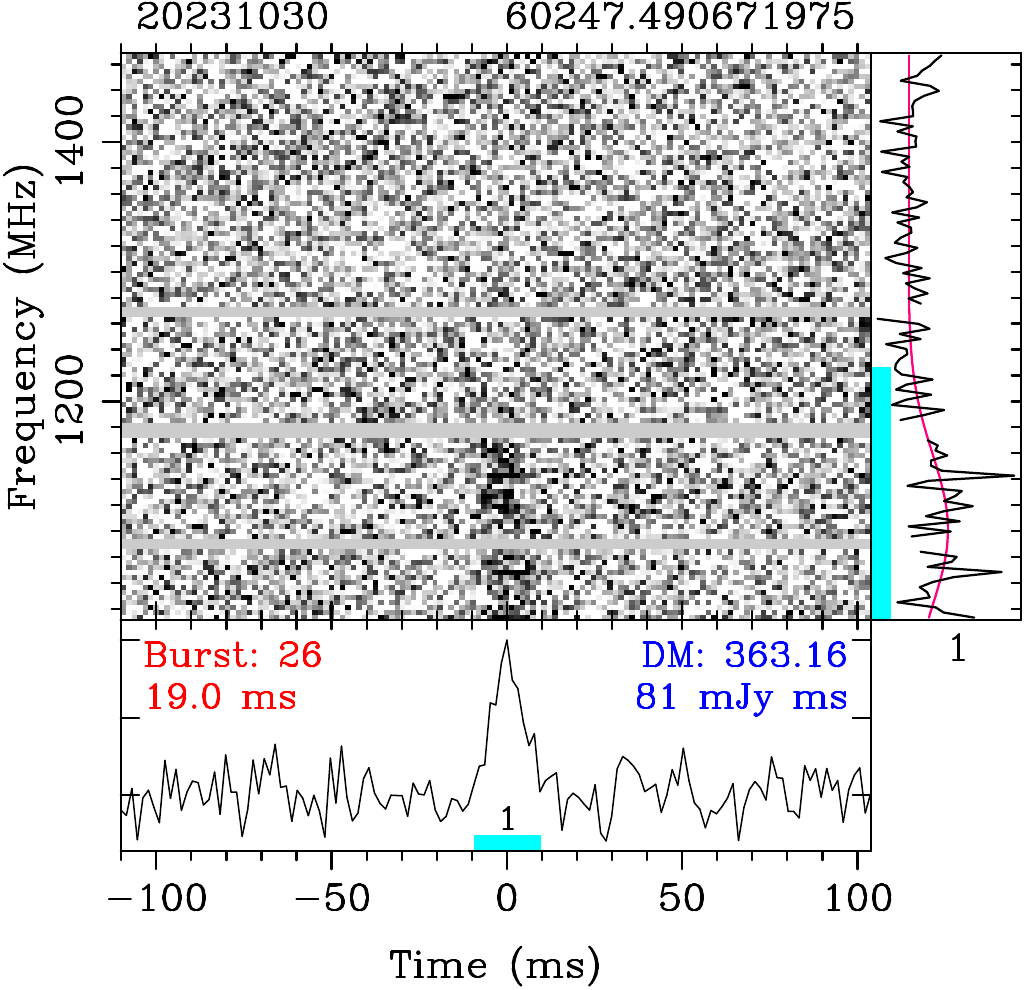}
\includegraphics[height=0.29\linewidth]{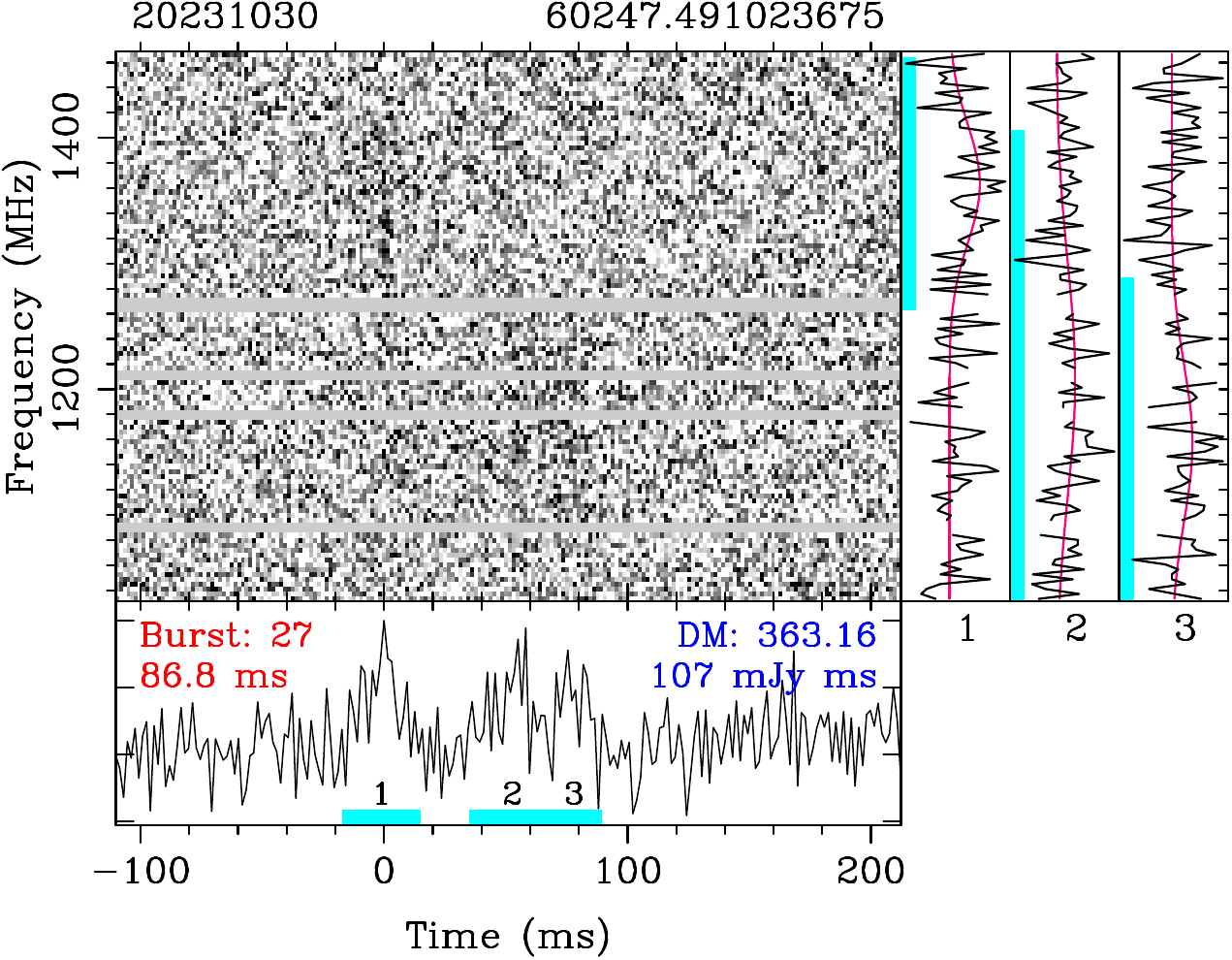}
\includegraphics[height=0.29\linewidth]{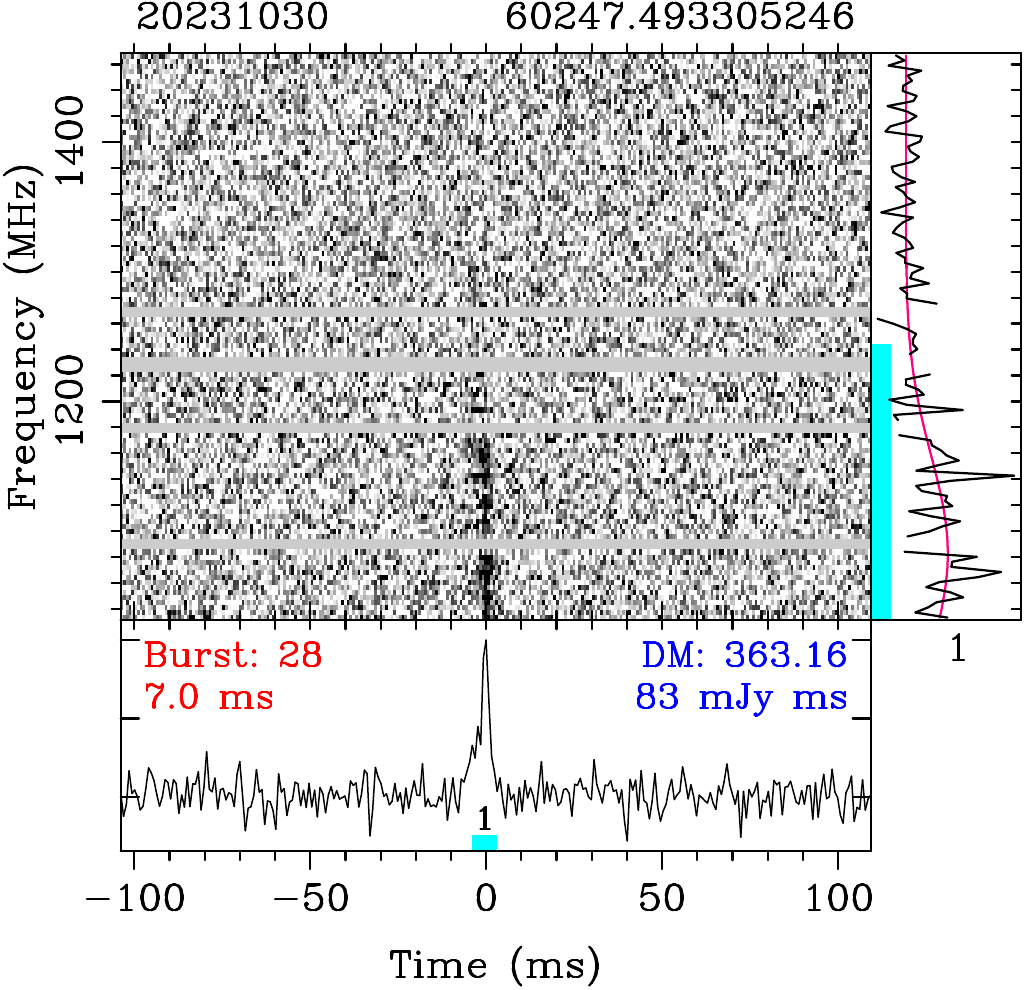}
\caption{({\textit{continued}})}
\end{figure*}
\addtocounter{figure}{-1}
\begin{figure*}
\flushleft
\includegraphics[height=0.29\linewidth]{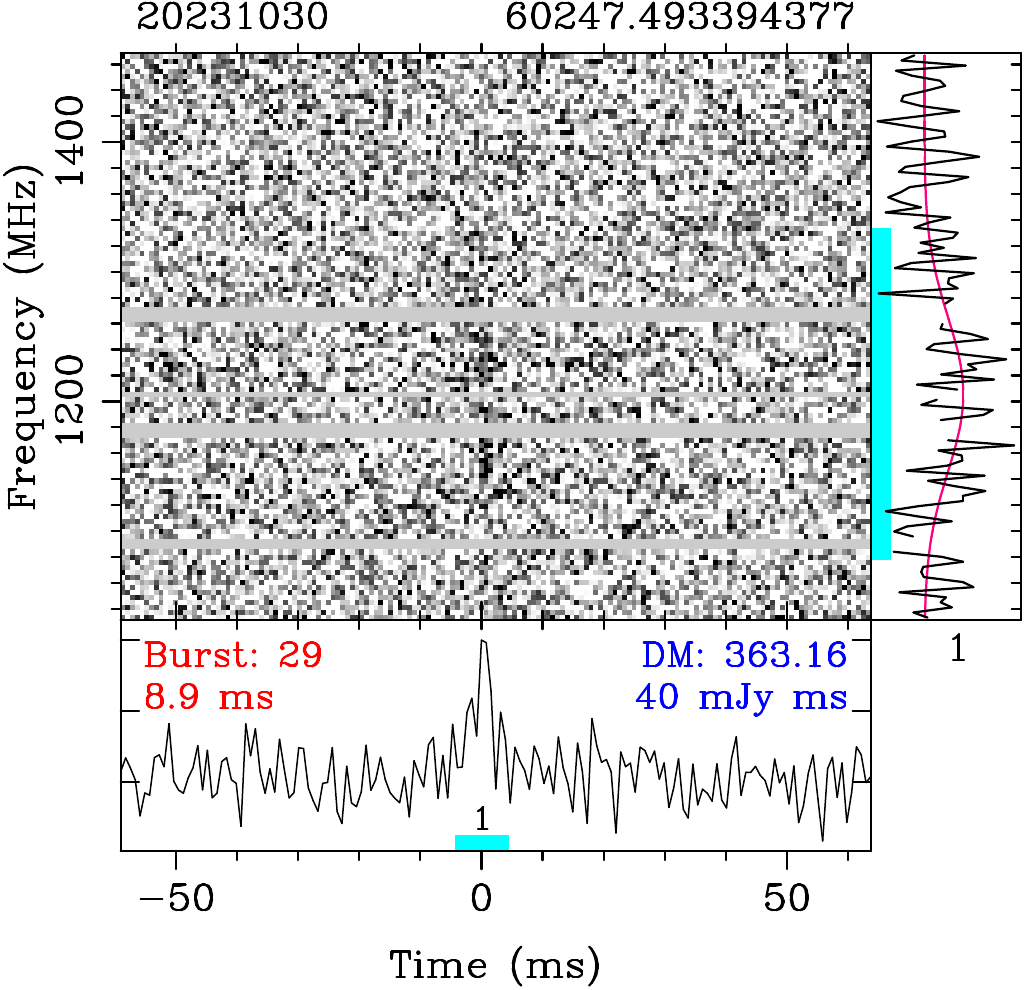}
\includegraphics[height=0.29\linewidth]{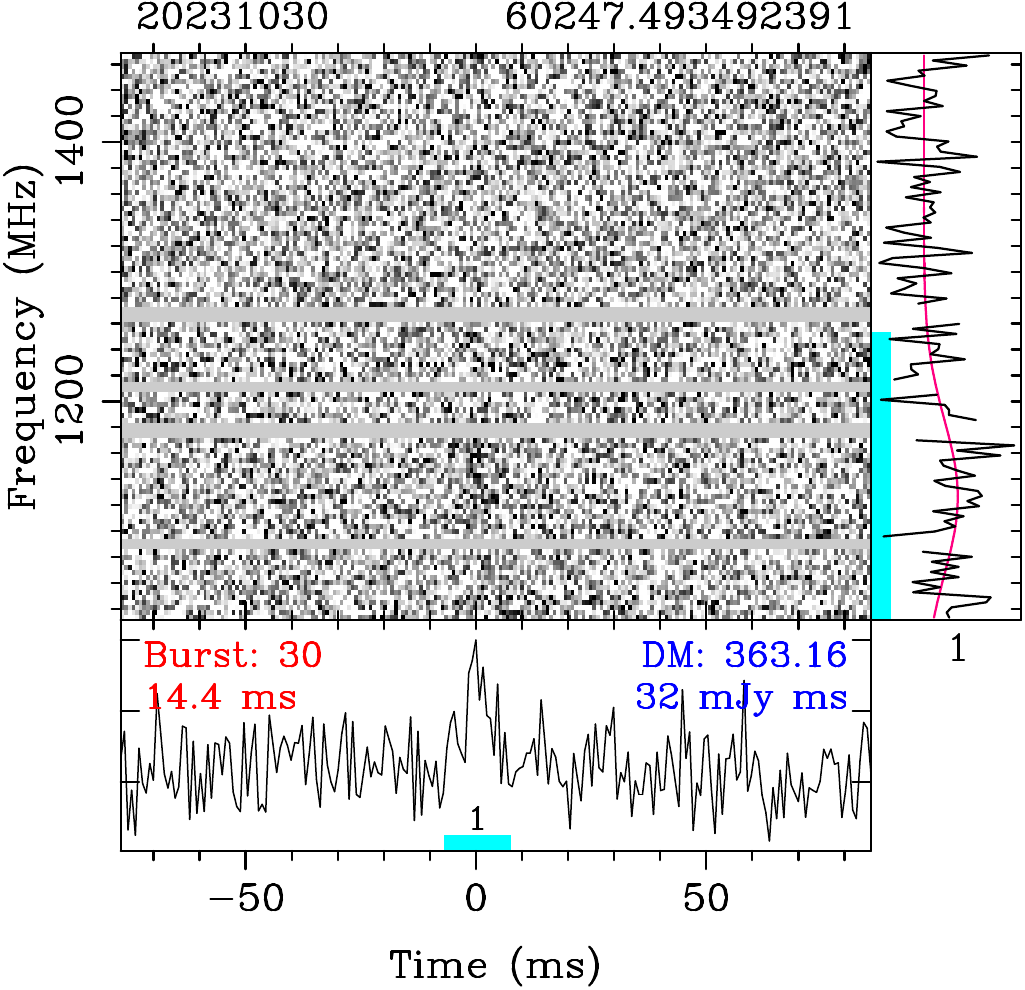}
\includegraphics[height=0.29\linewidth]{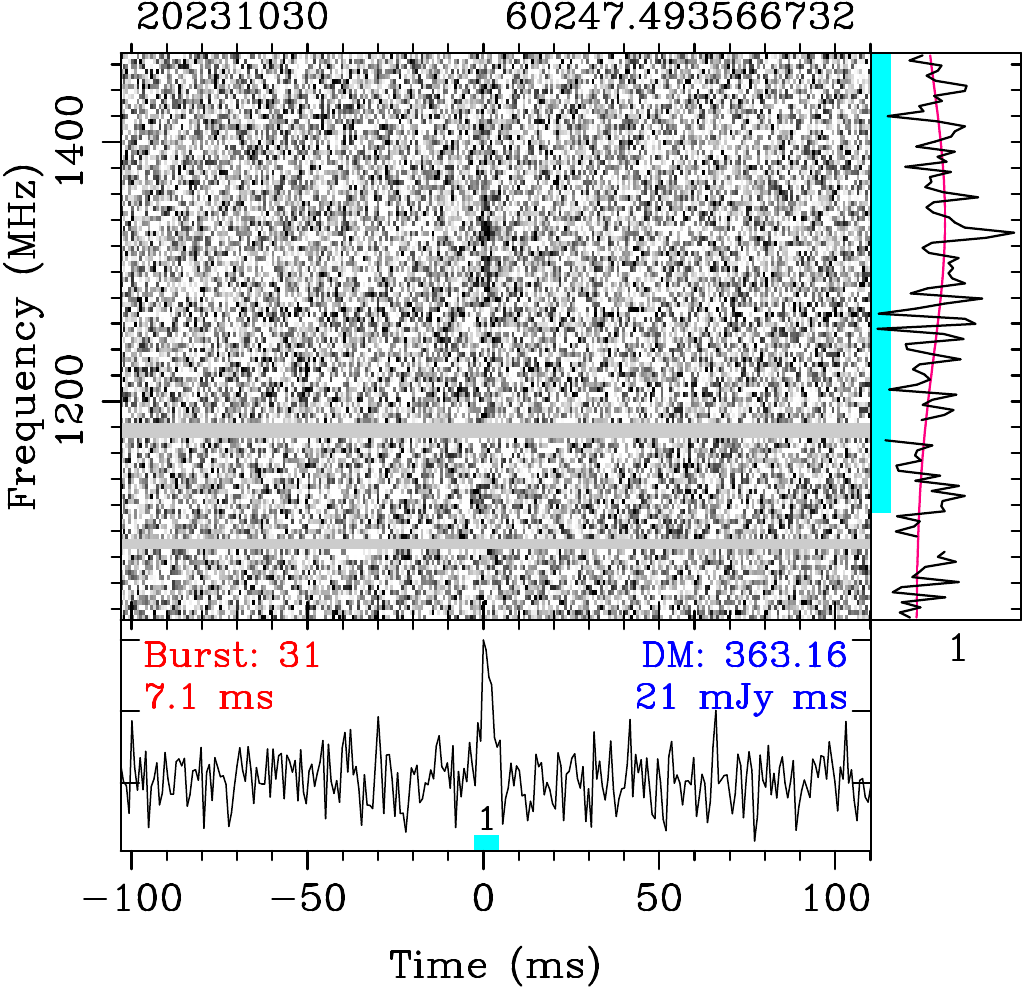}
\includegraphics[height=0.29\linewidth]{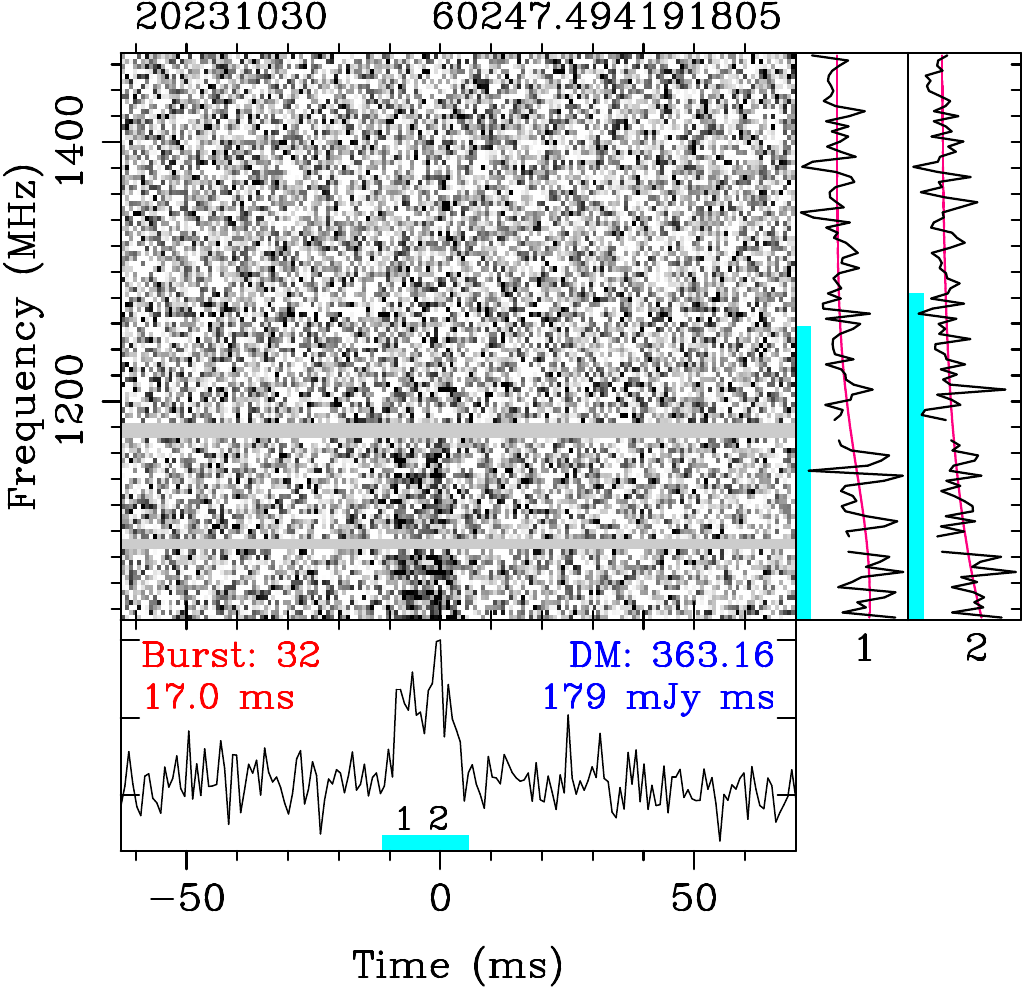}
\includegraphics[height=0.29\linewidth]{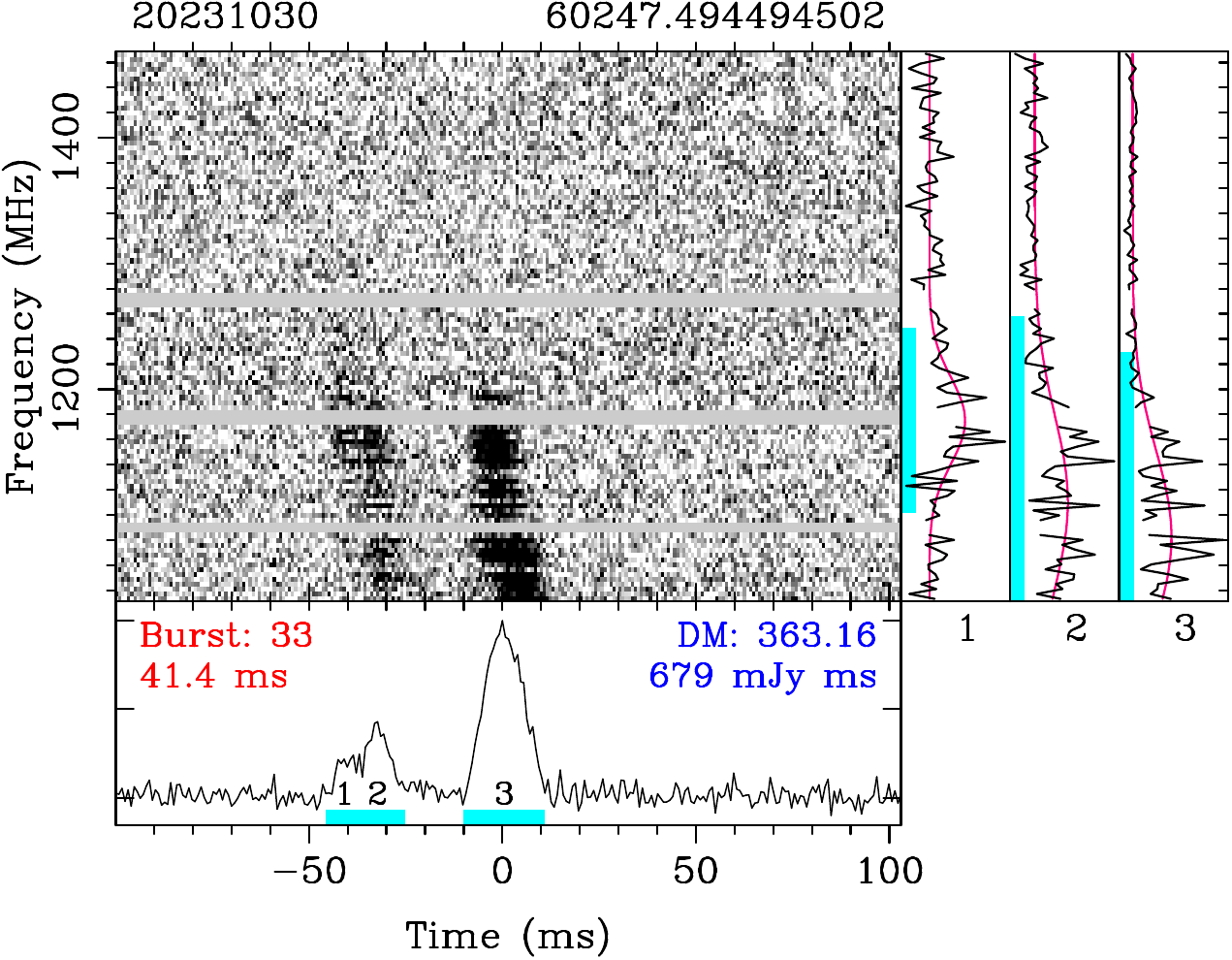}
\includegraphics[height=0.29\linewidth]{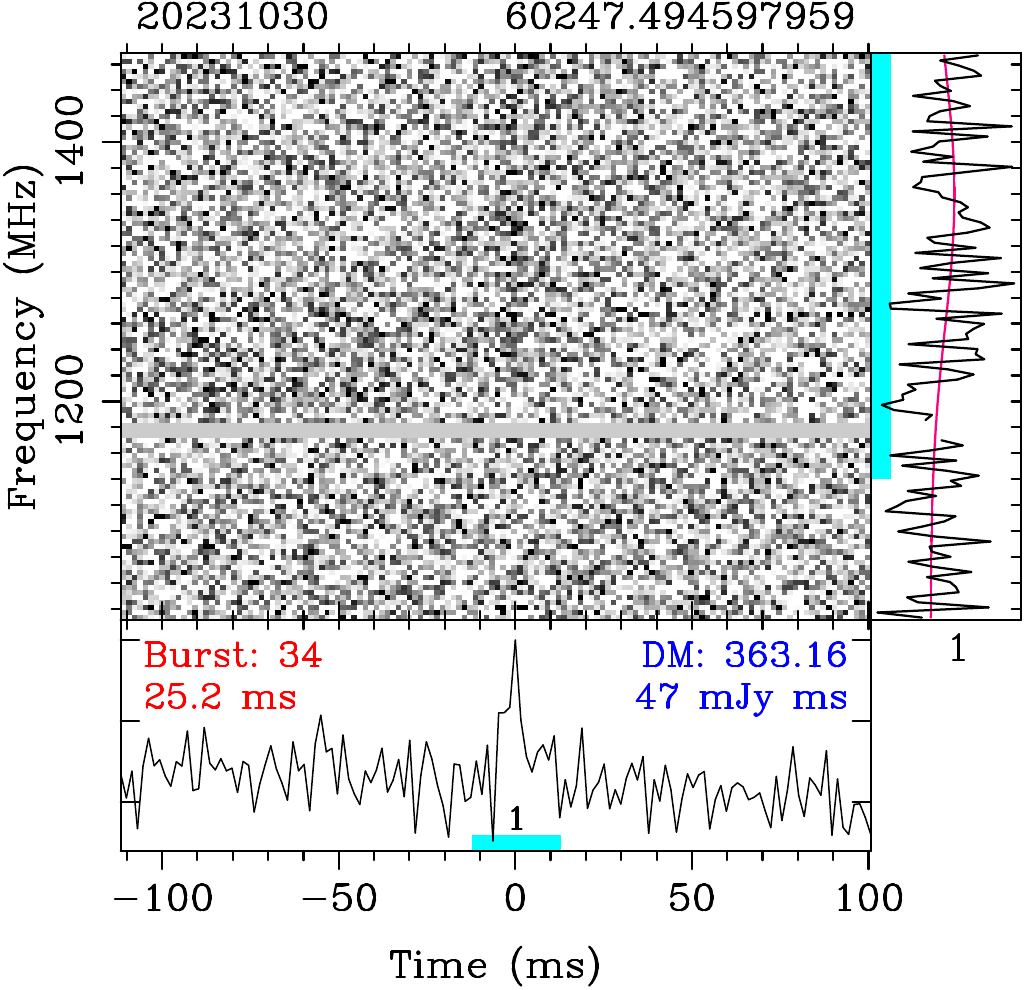}
\includegraphics[height=0.29\linewidth]{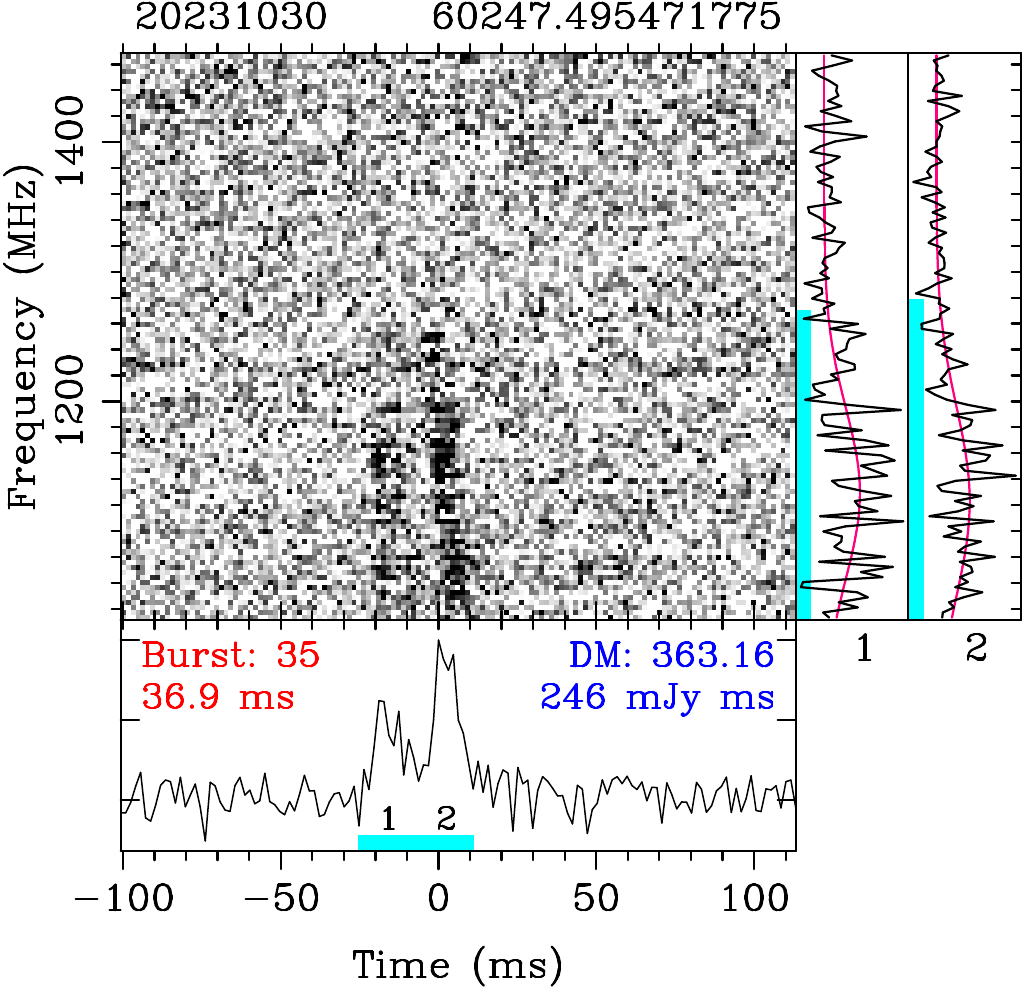}
\includegraphics[height=0.29\linewidth]{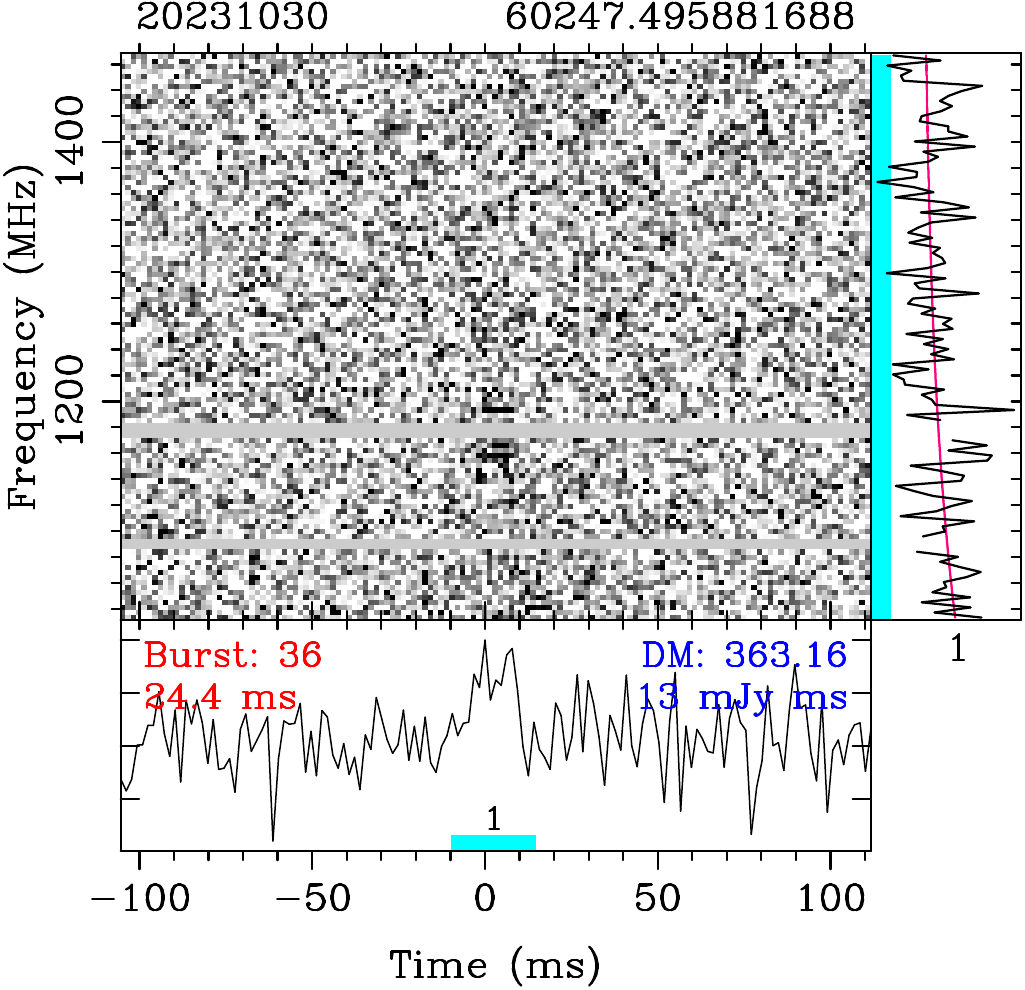}
\includegraphics[height=0.29\linewidth]{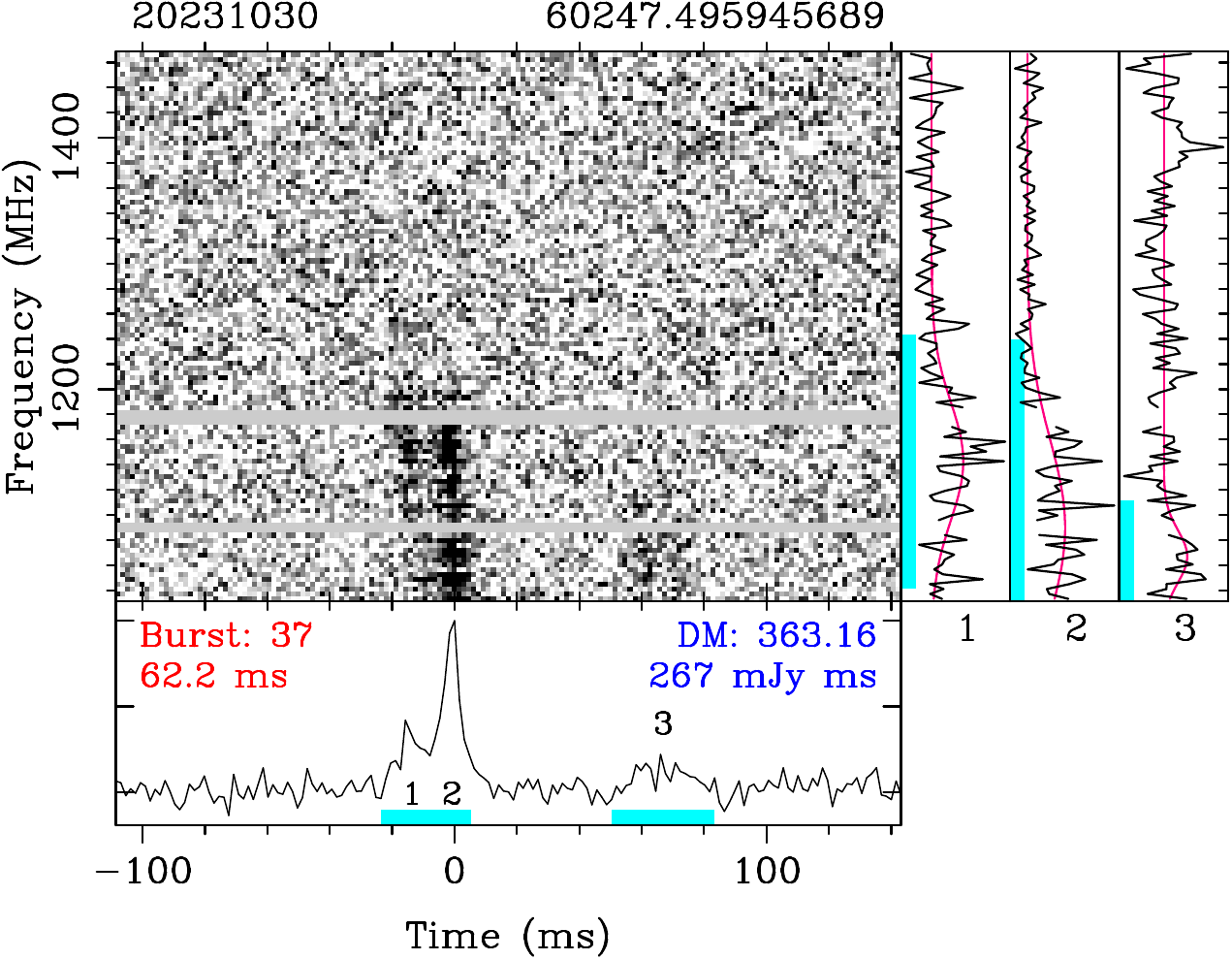}
\includegraphics[height=0.29\linewidth]{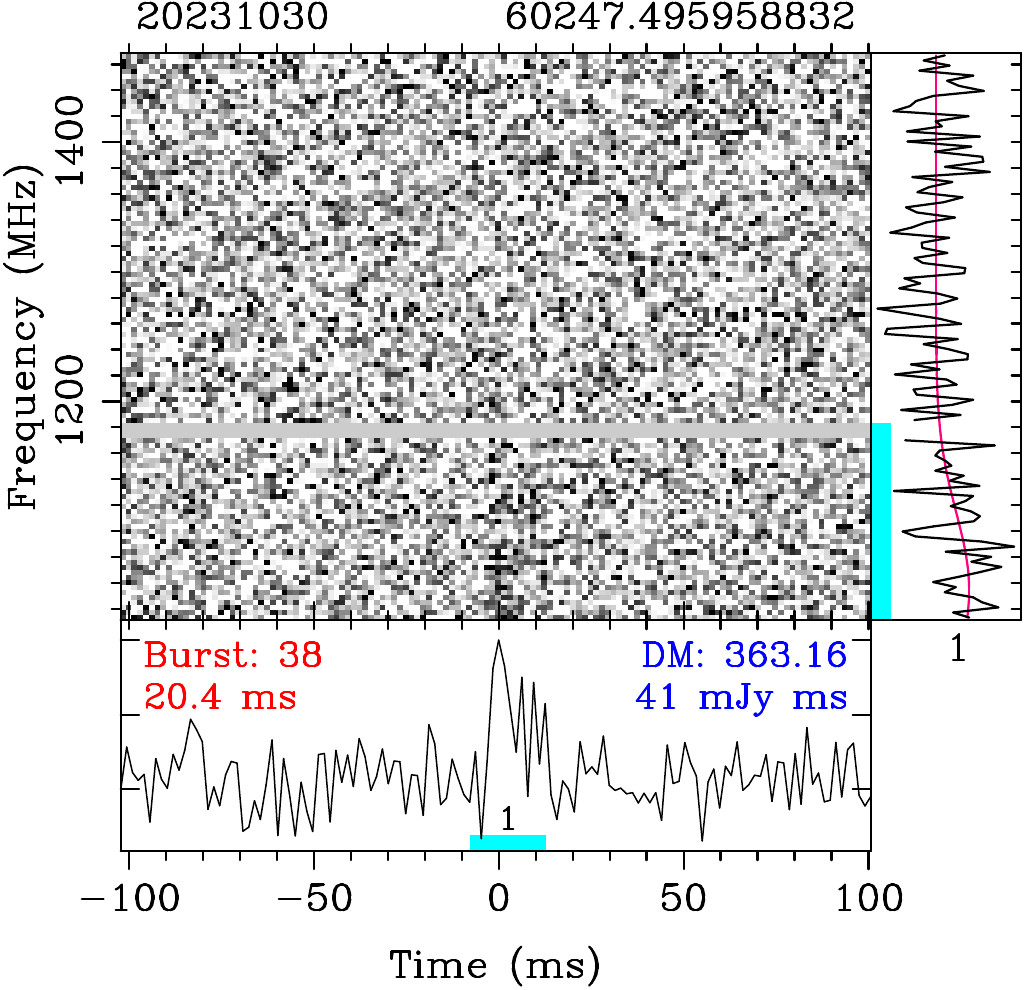}
\includegraphics[height=0.29\linewidth]{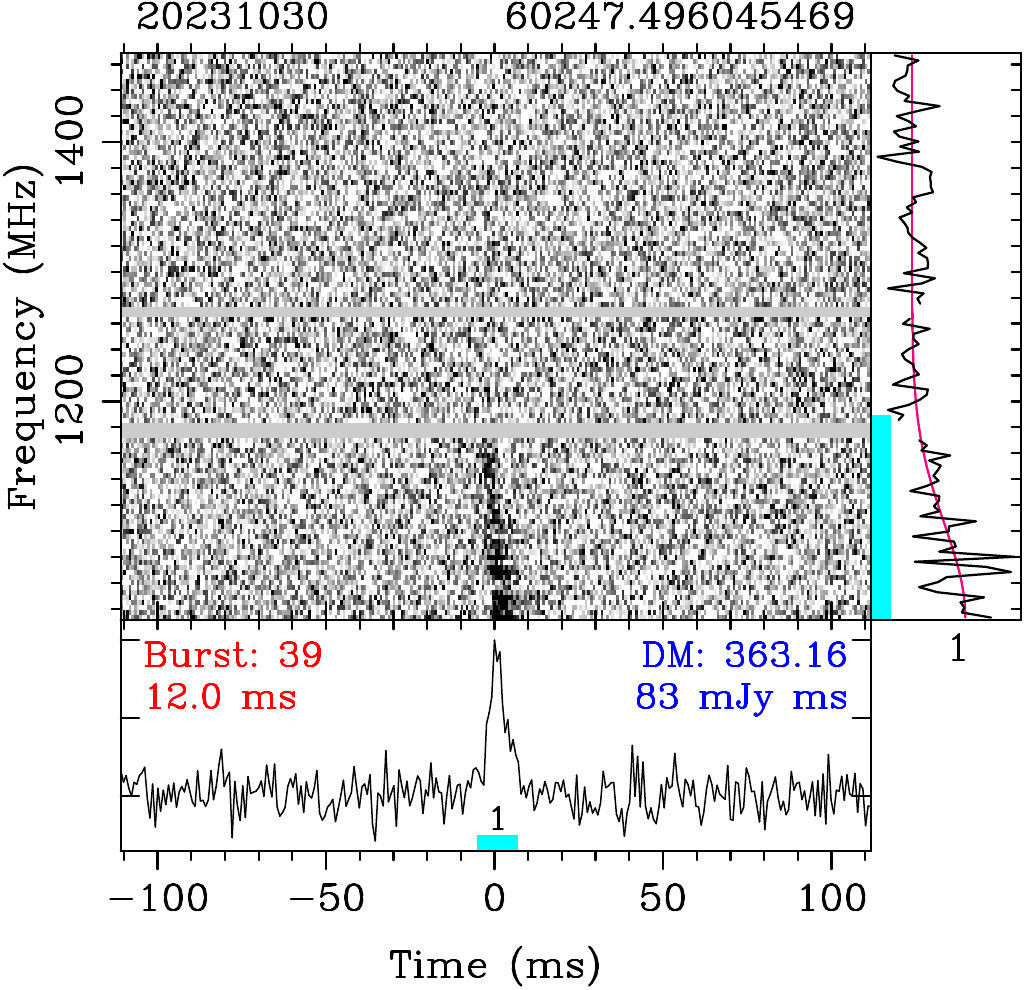}
\includegraphics[height=0.29\linewidth]{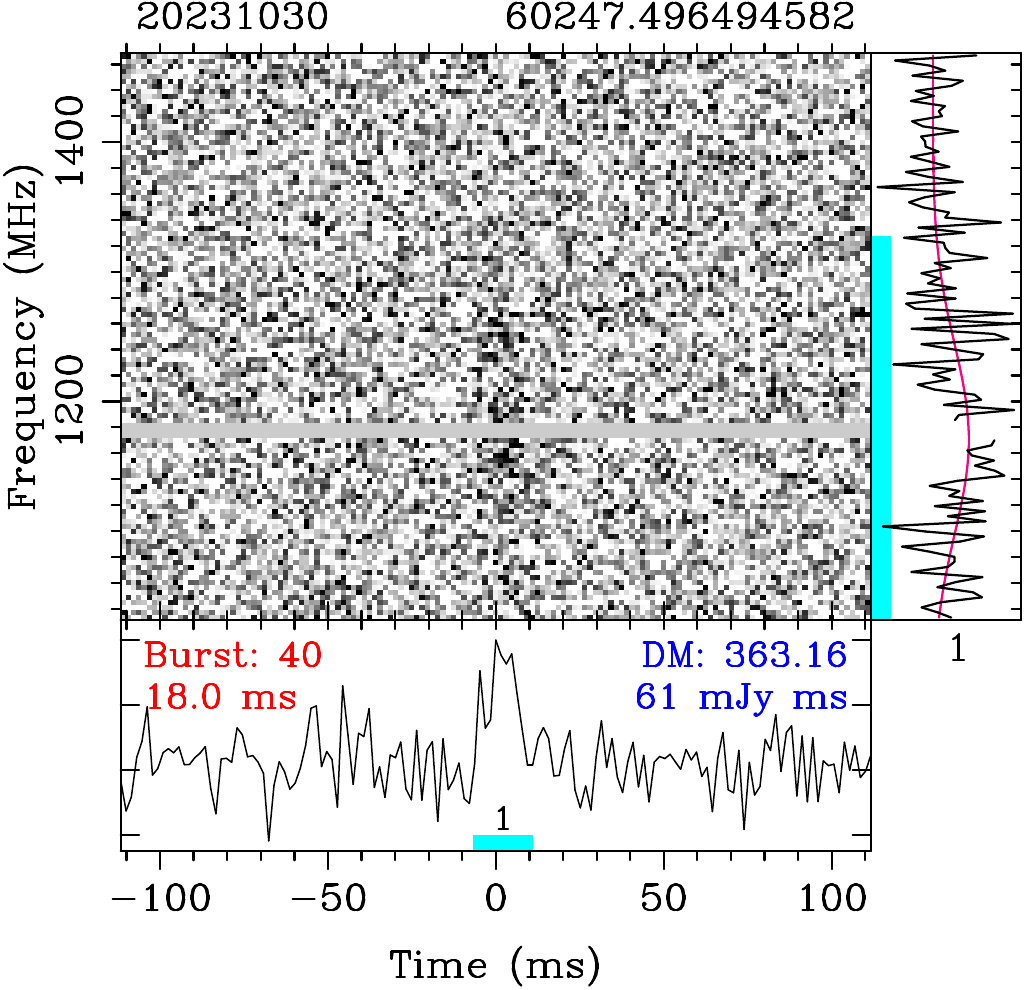}
\caption{({\textit{continued}})}
\end{figure*}
\addtocounter{figure}{-1}
\begin{figure*}
\flushleft
\includegraphics[height=0.29\linewidth]{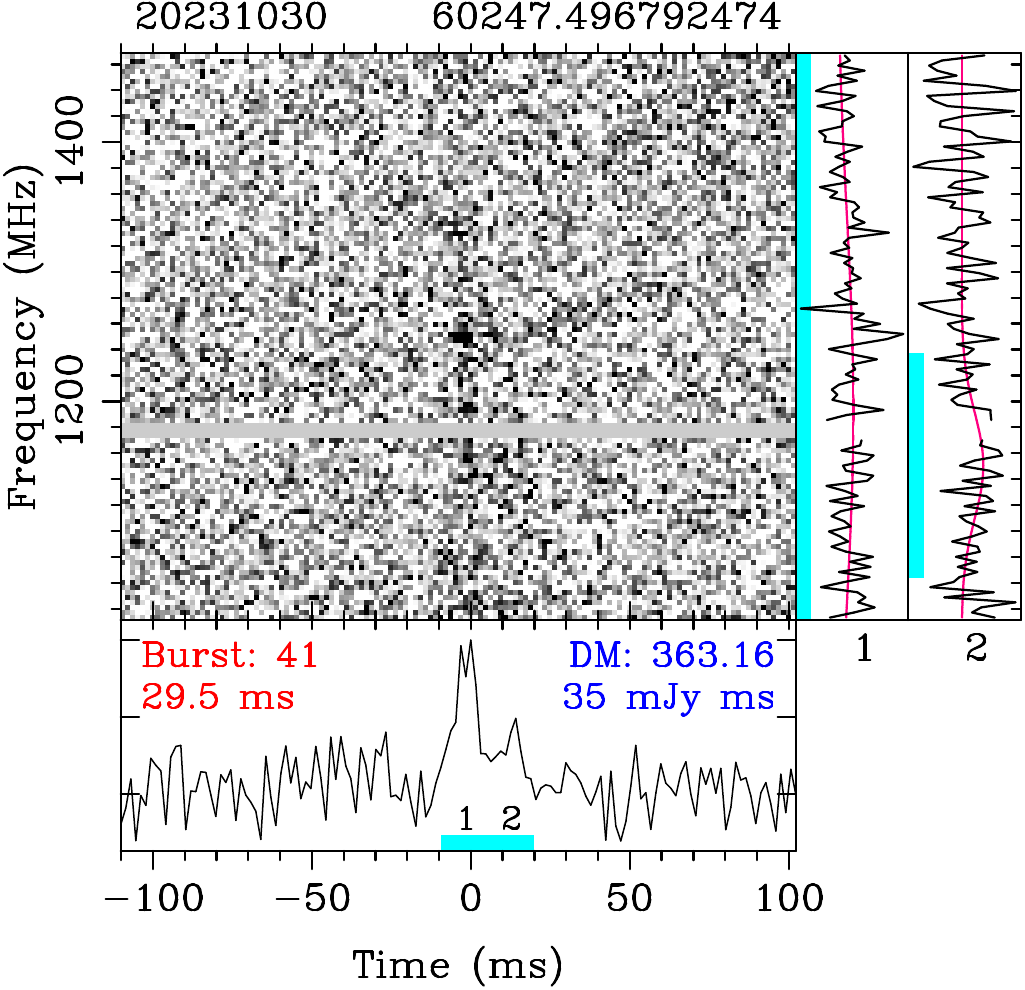}
\includegraphics[height=0.29\linewidth]{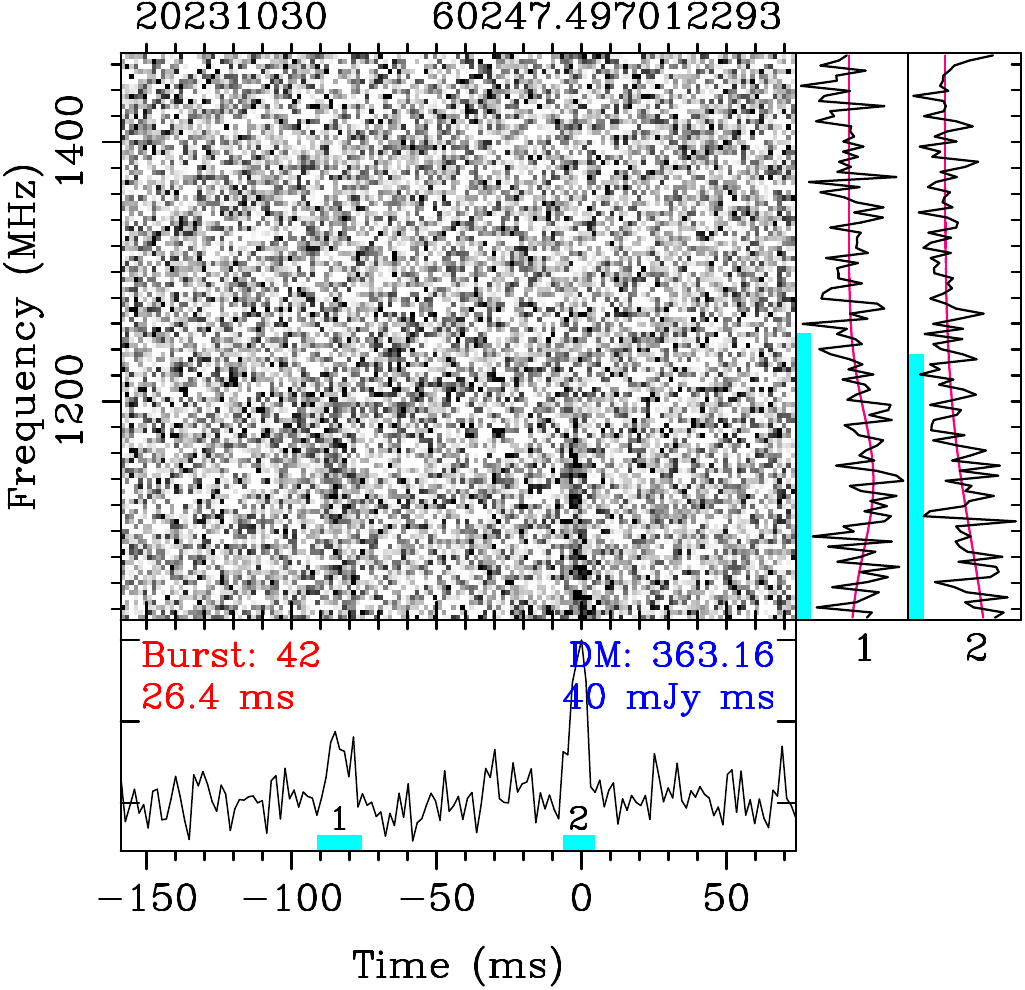}
\includegraphics[height=0.29\linewidth]{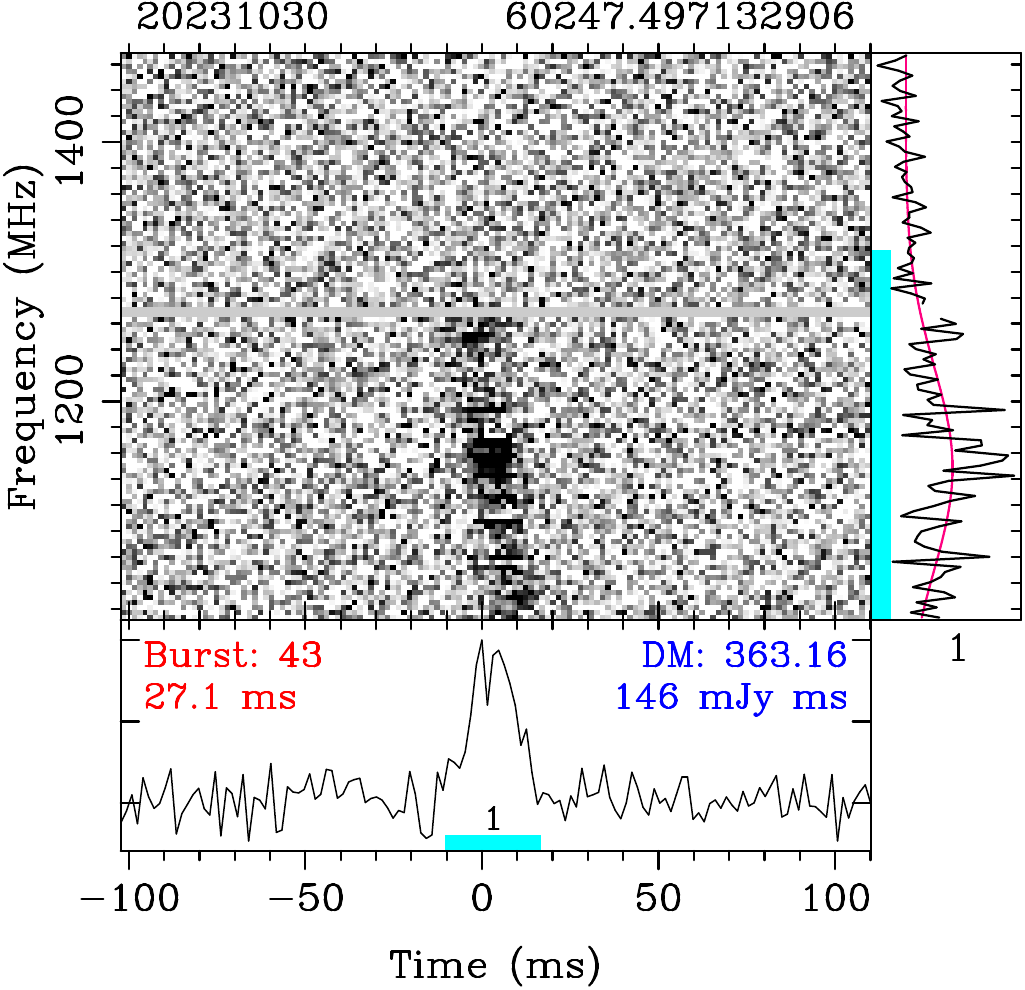}
\includegraphics[height=0.29\linewidth]{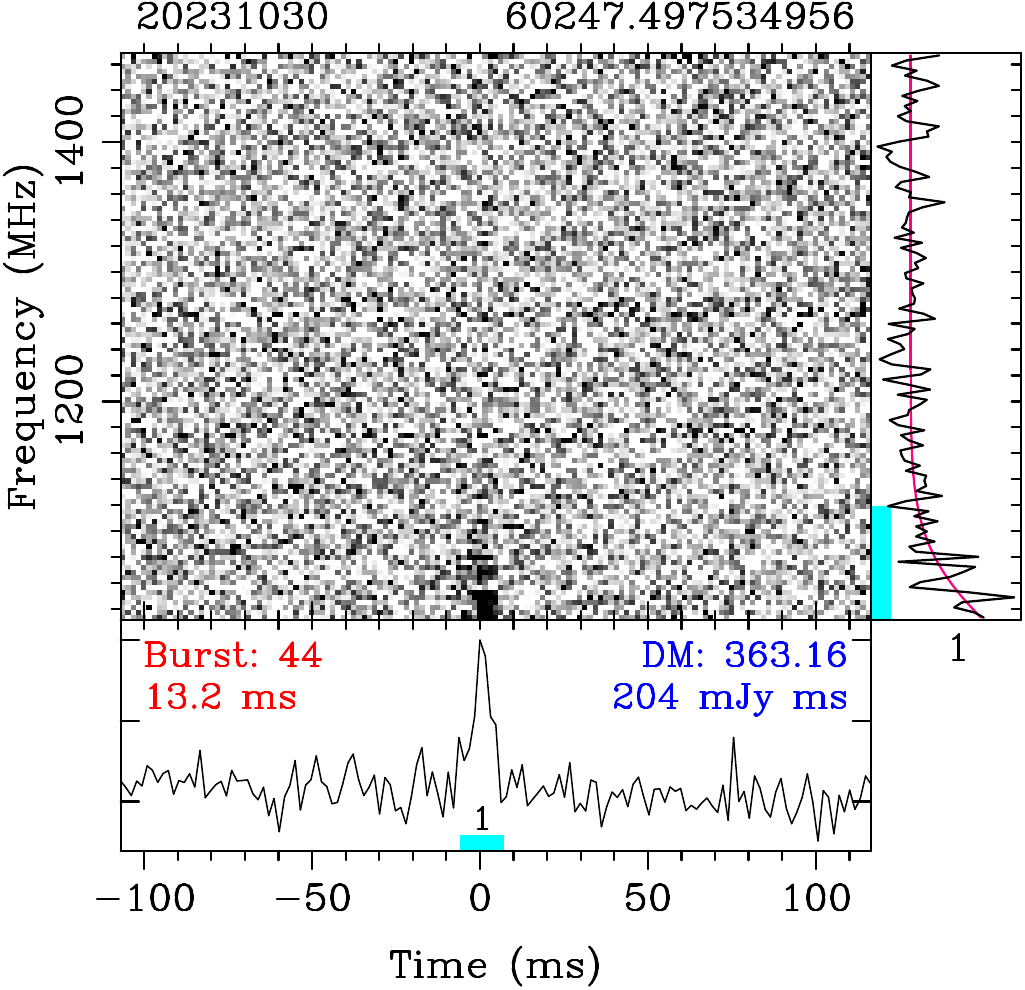}
\includegraphics[height=0.29\linewidth]{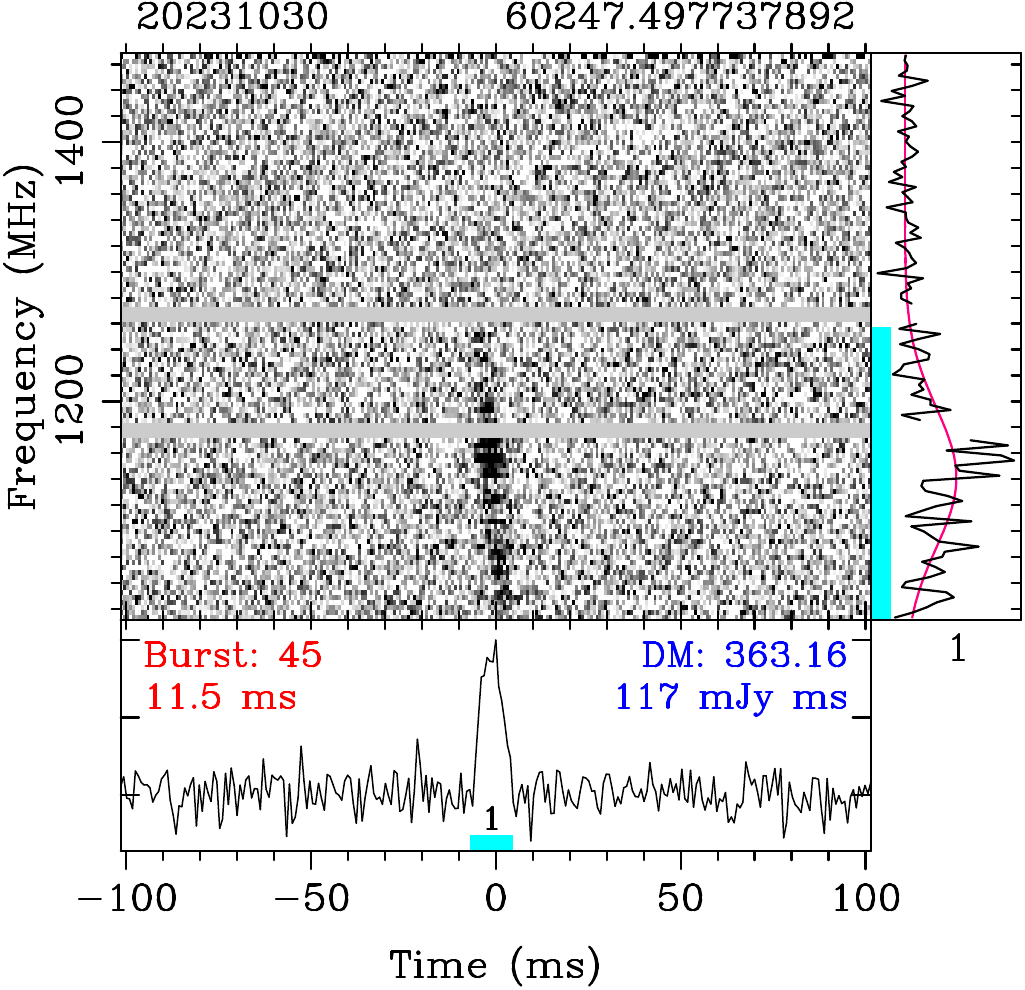}
\includegraphics[height=0.29\linewidth]{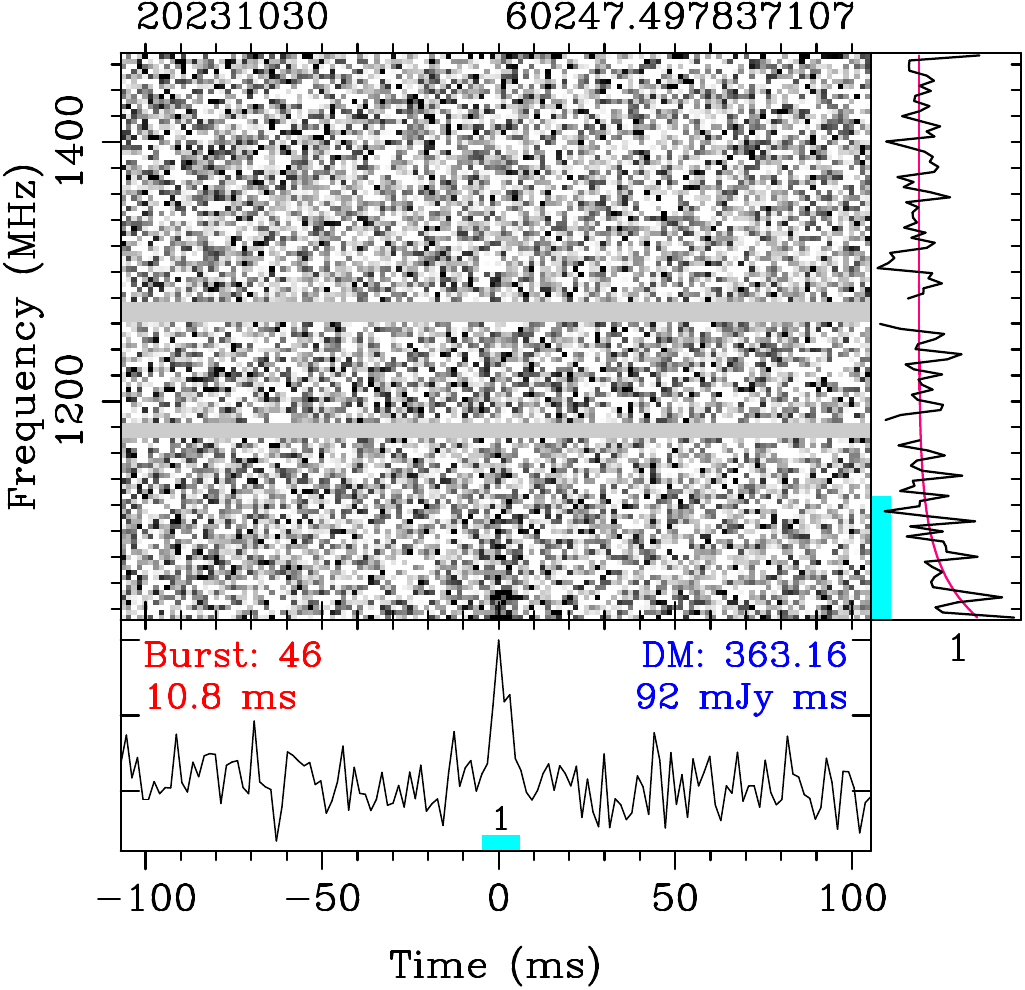}
\includegraphics[height=0.29\linewidth]{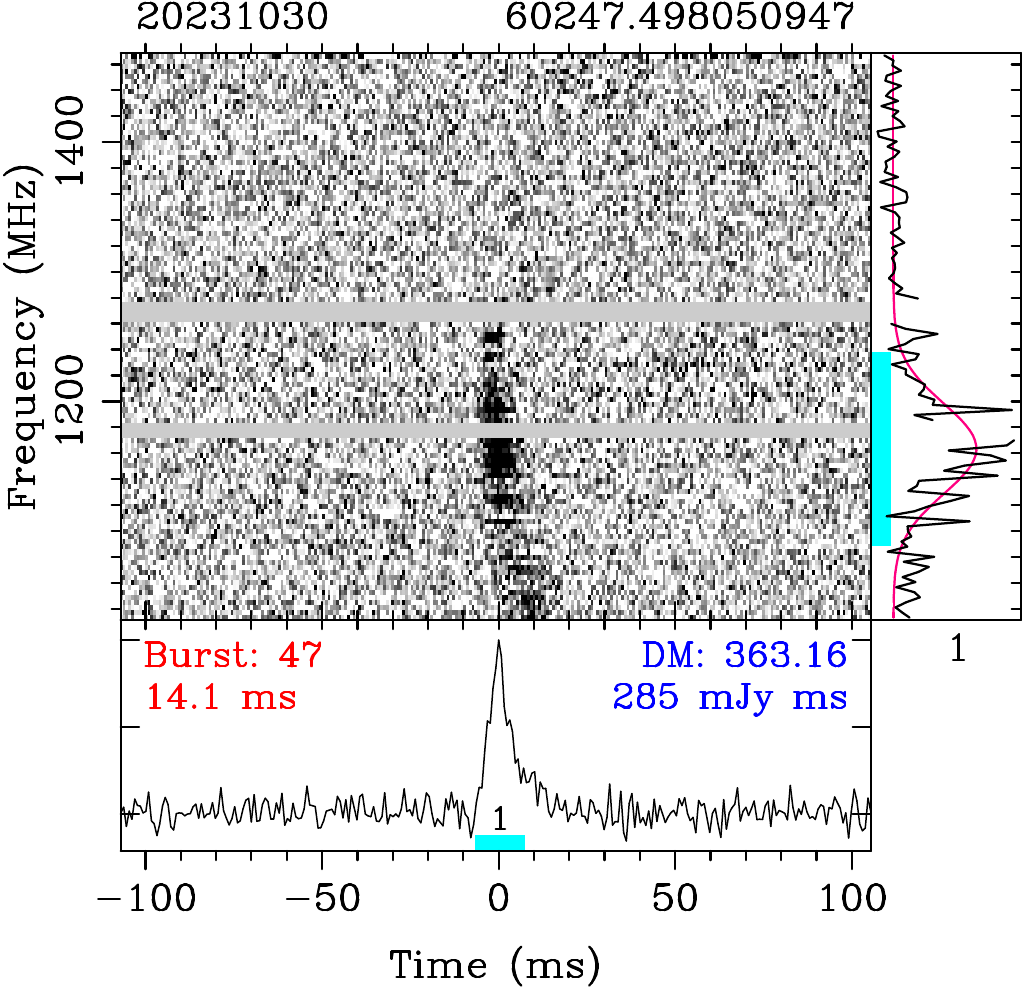}
\includegraphics[height=0.29\linewidth]{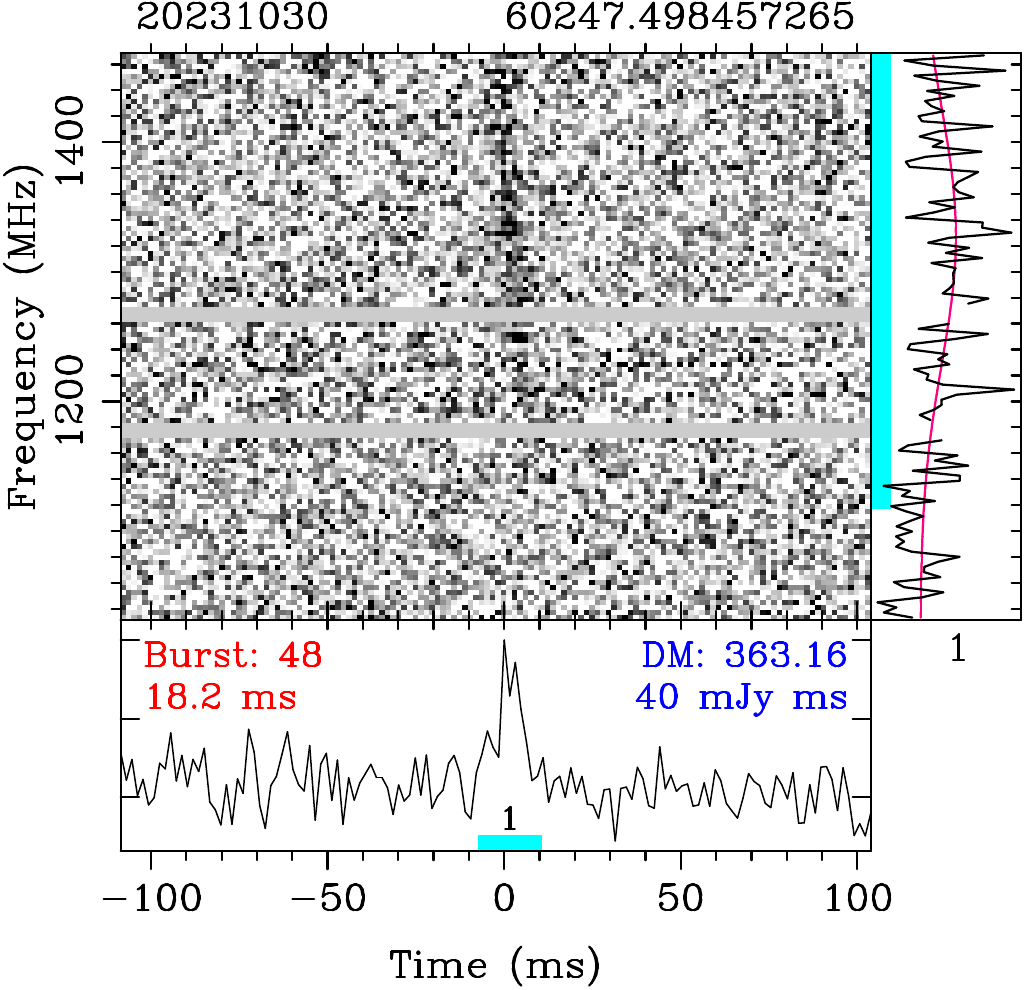}
\includegraphics[height=0.29\linewidth]{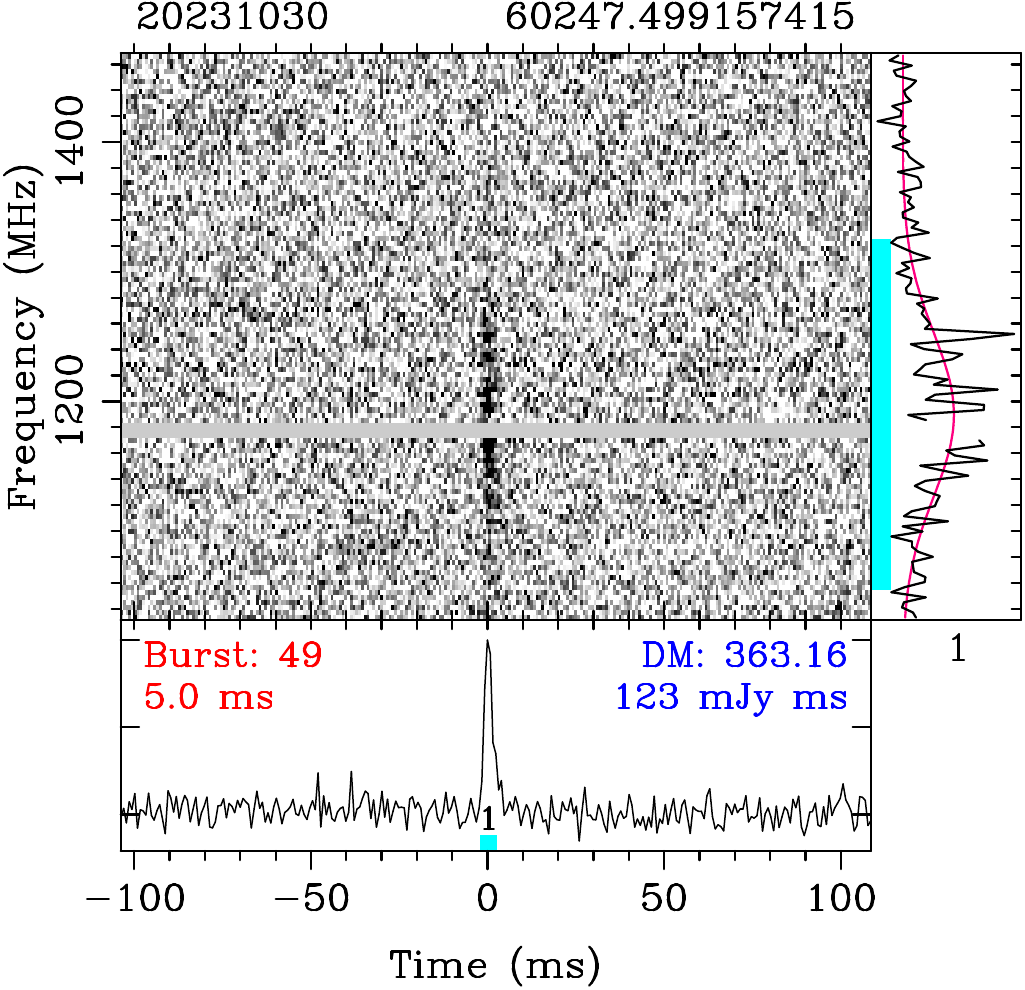}
\includegraphics[height=0.29\linewidth]{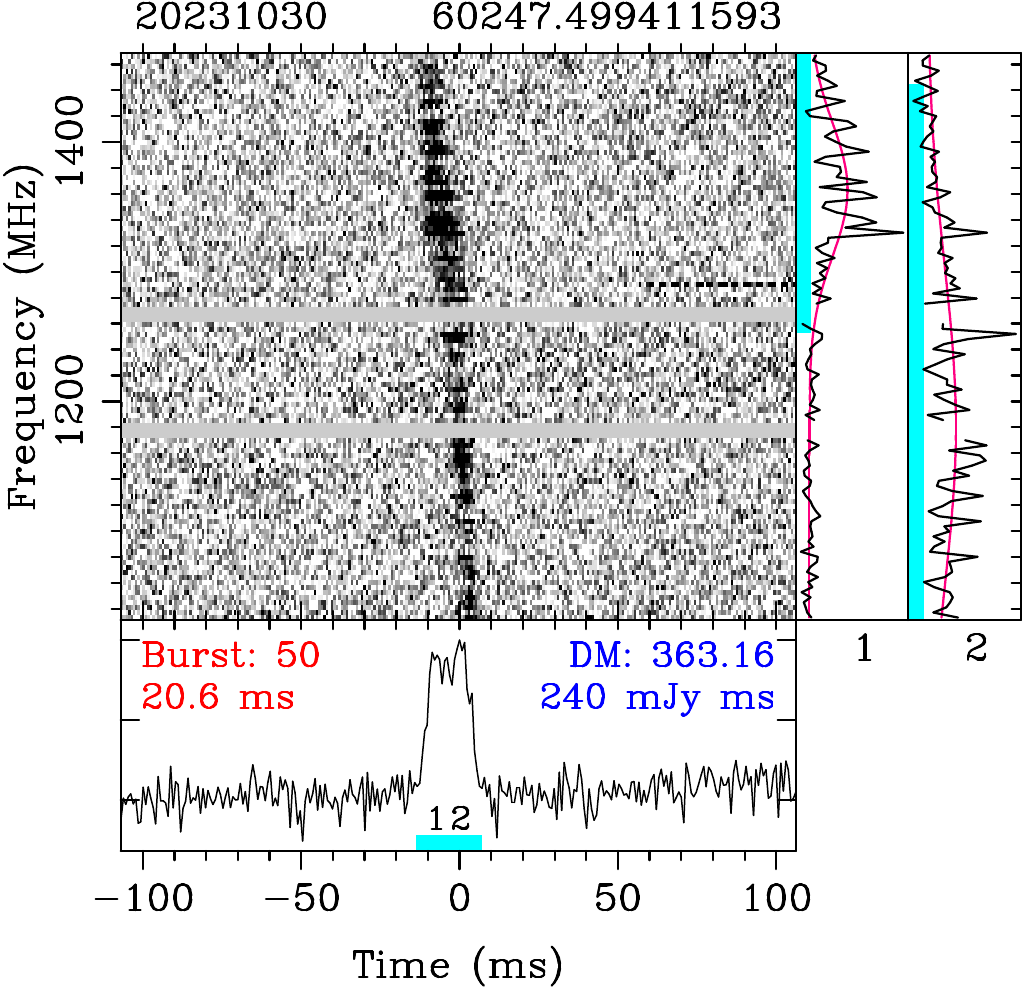}
\includegraphics[height=0.29\linewidth]{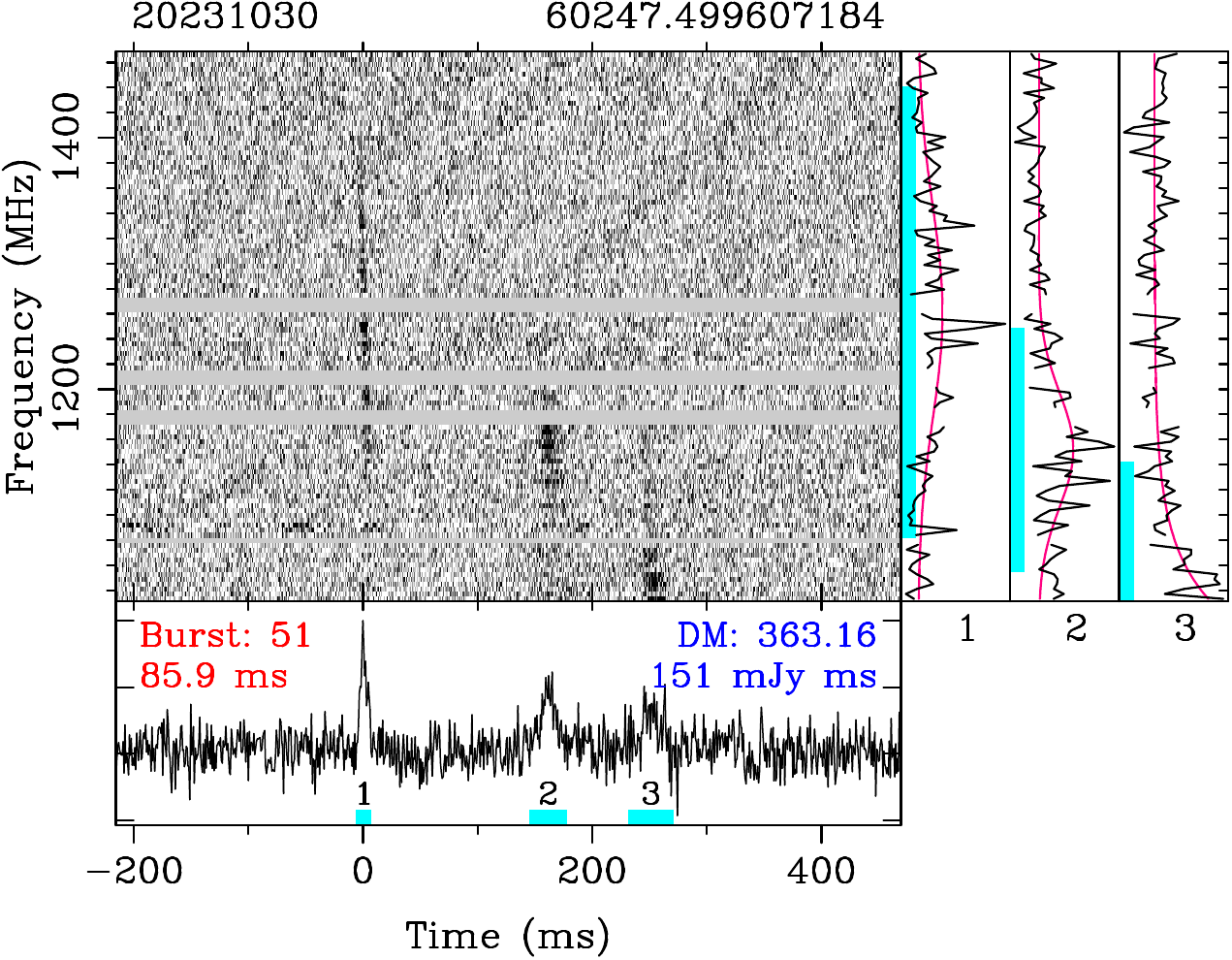}
\includegraphics[height=0.29\linewidth]{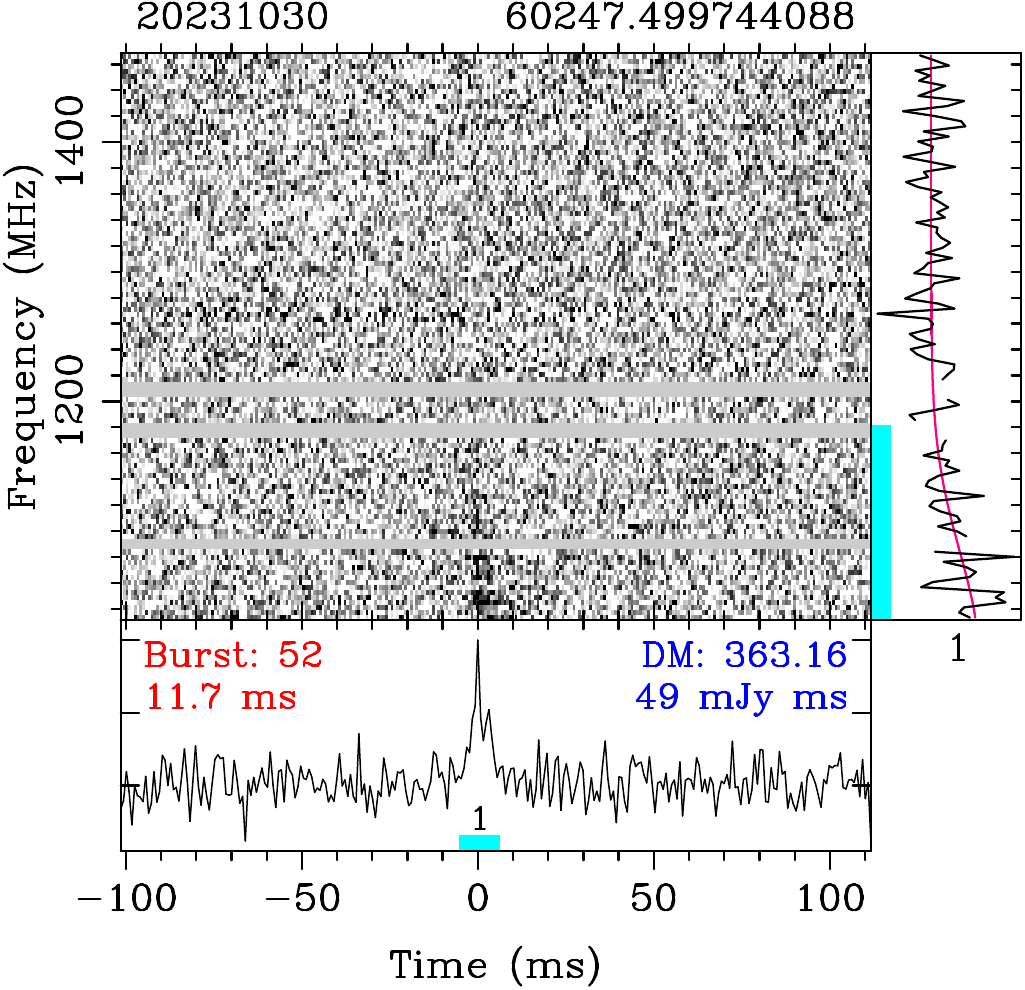}
\caption{({\textit{continued}})}
\end{figure*}
\addtocounter{figure}{-1}
\begin{figure*}
\flushleft
\includegraphics[height=0.29\linewidth]{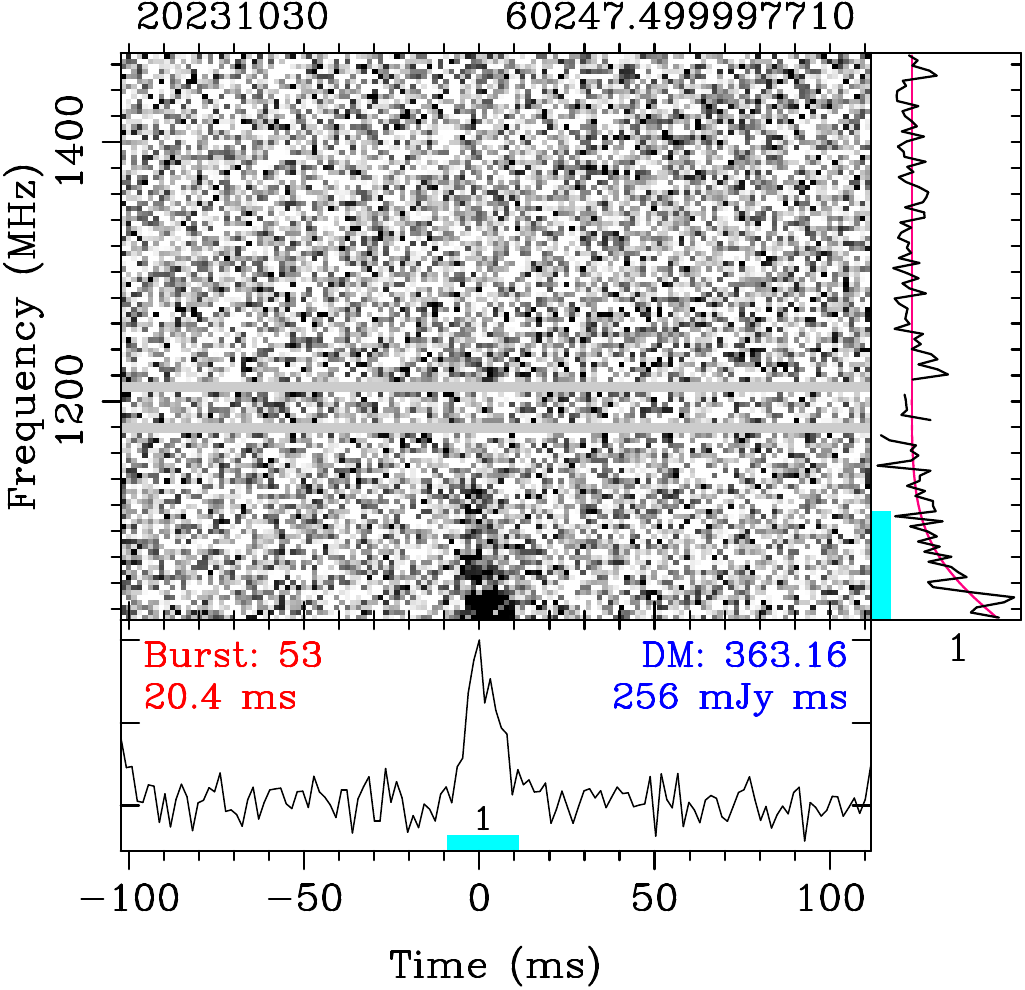}
\includegraphics[height=0.29\linewidth]{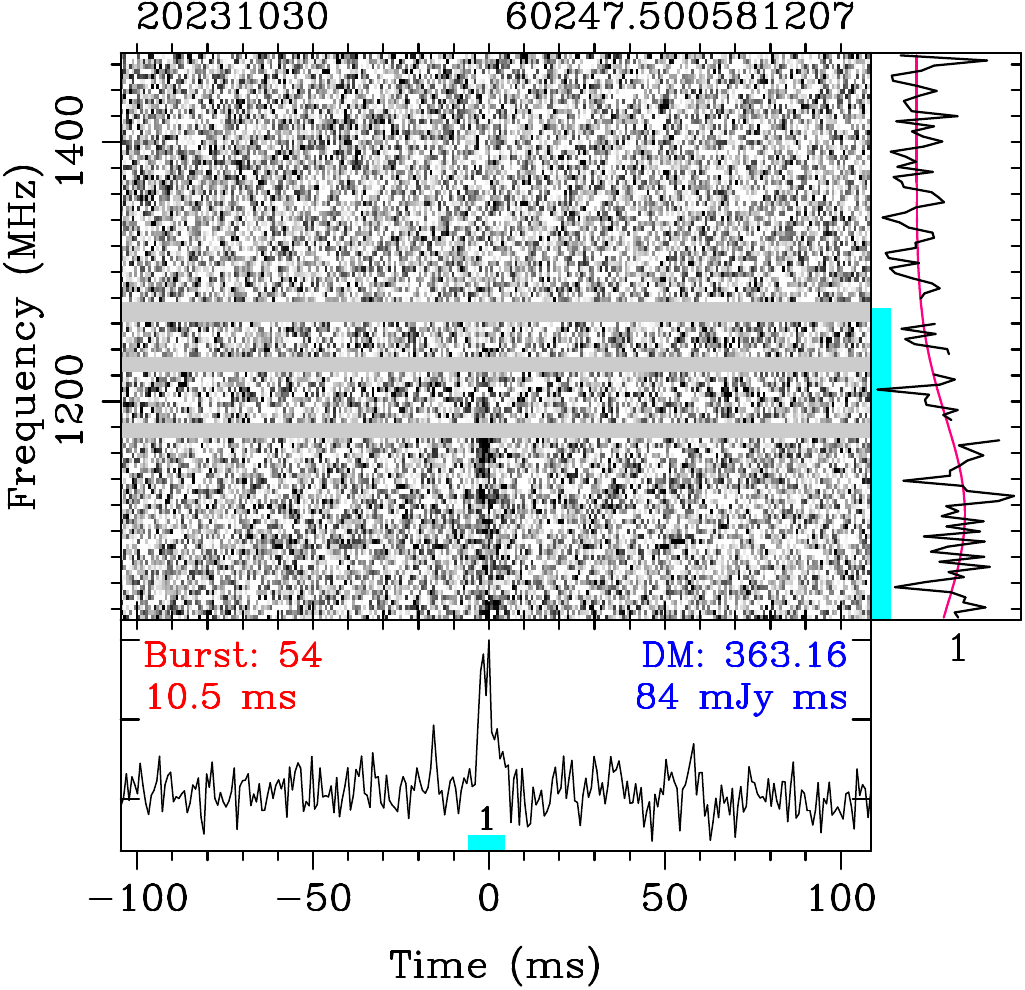}
\includegraphics[height=0.29\linewidth]{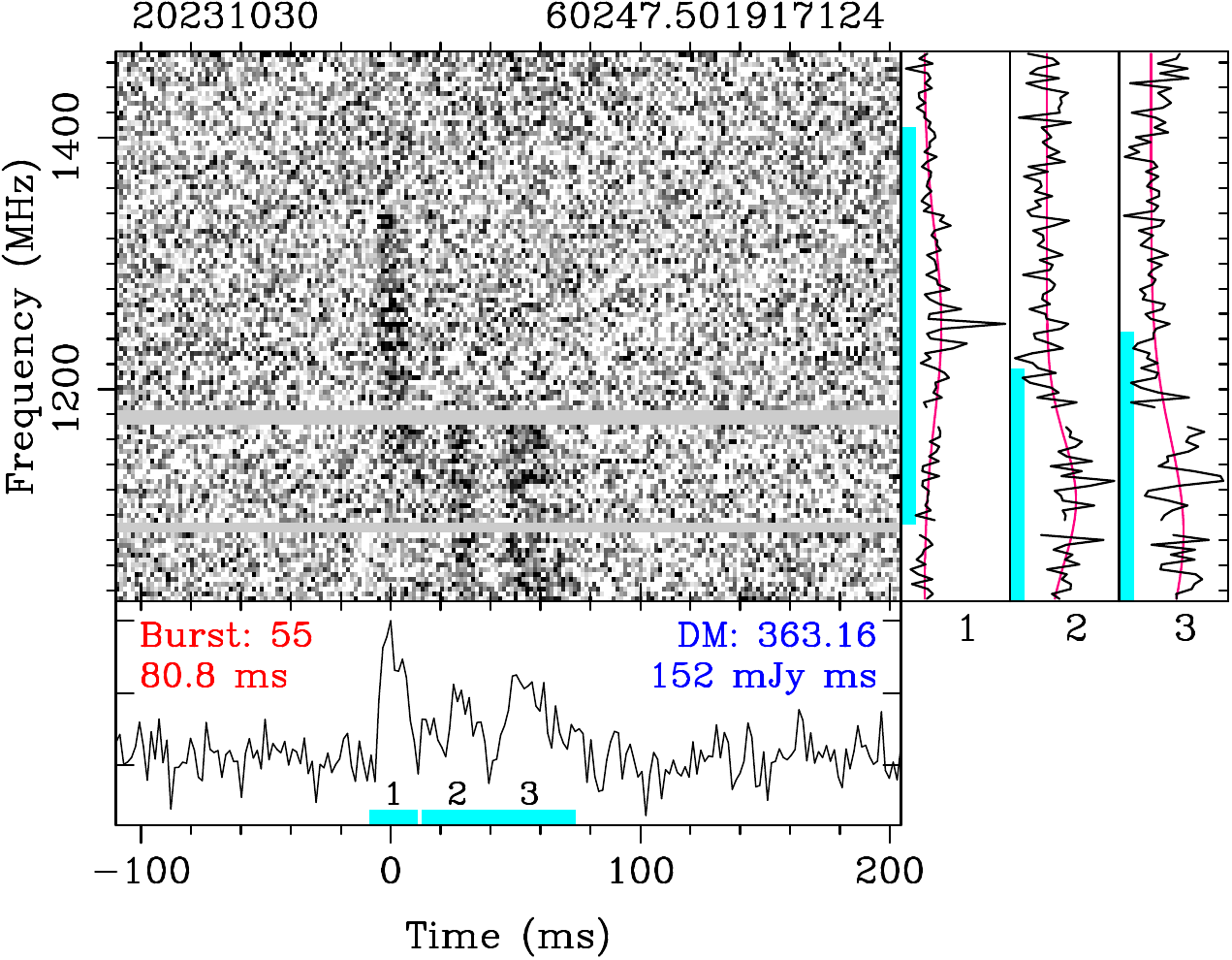}
\includegraphics[height=0.29\linewidth]{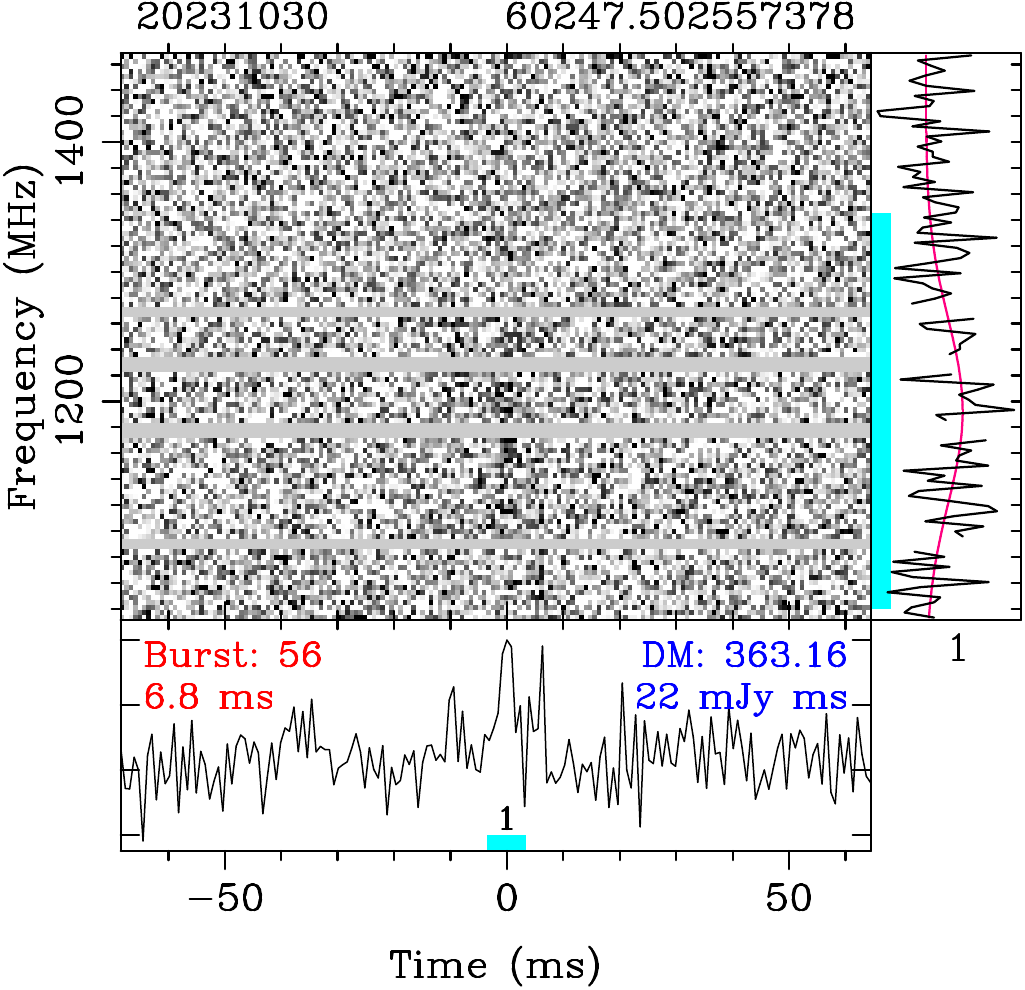}
\includegraphics[height=0.29\linewidth]{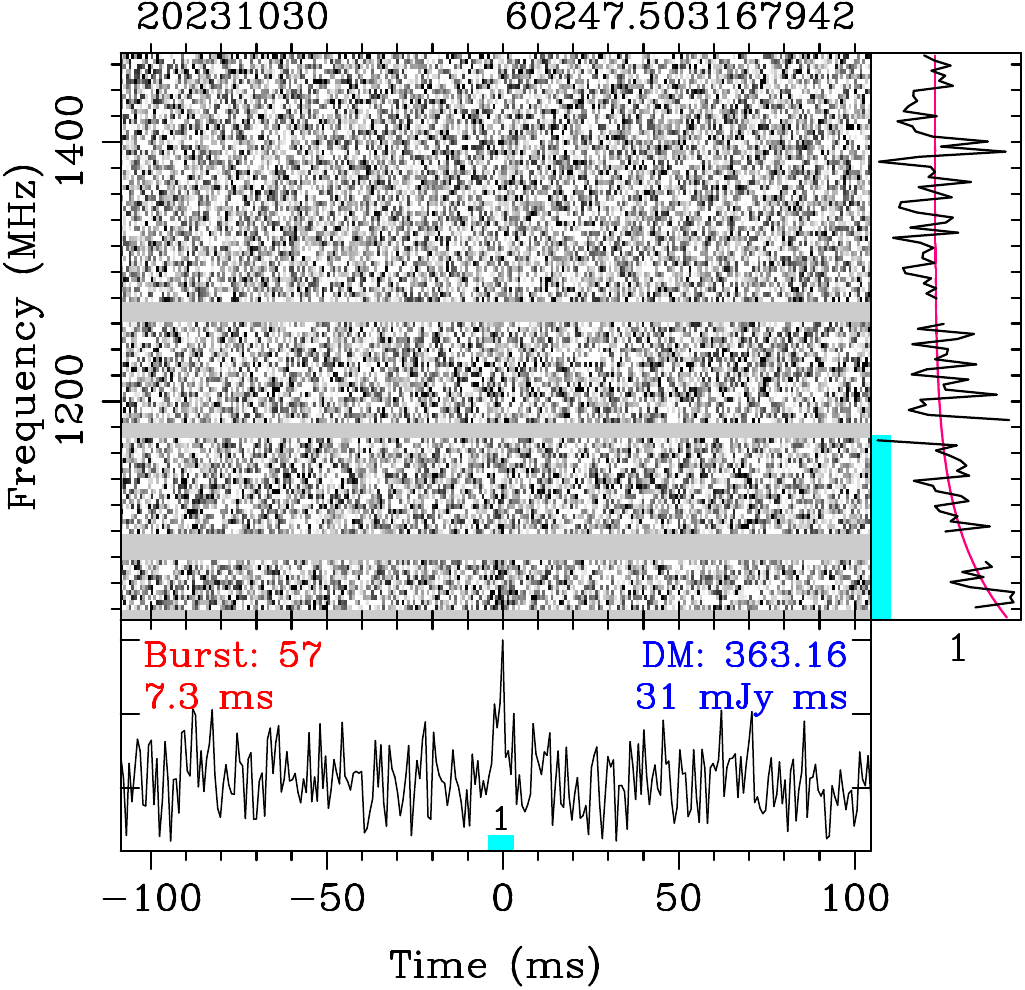}
\includegraphics[height=0.29\linewidth]{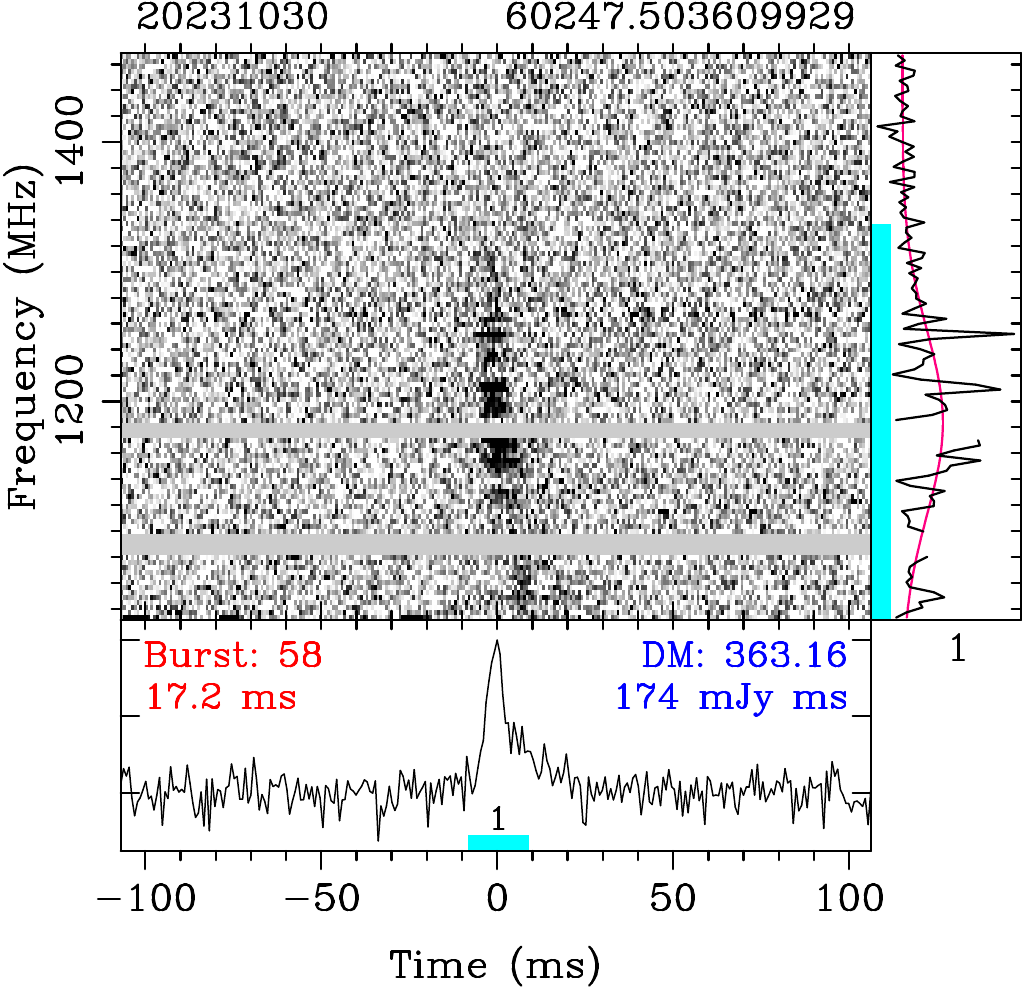}
\includegraphics[height=0.29\linewidth]{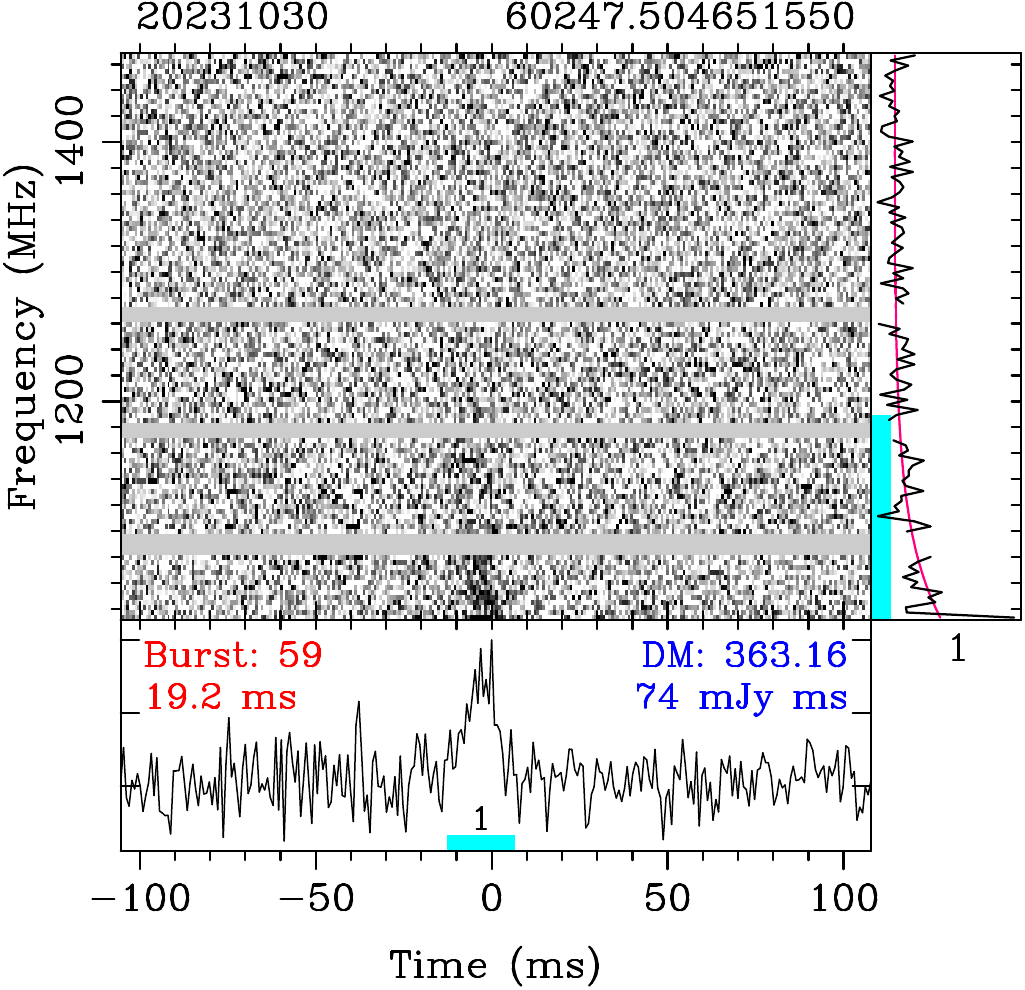}
\includegraphics[height=0.29\linewidth]{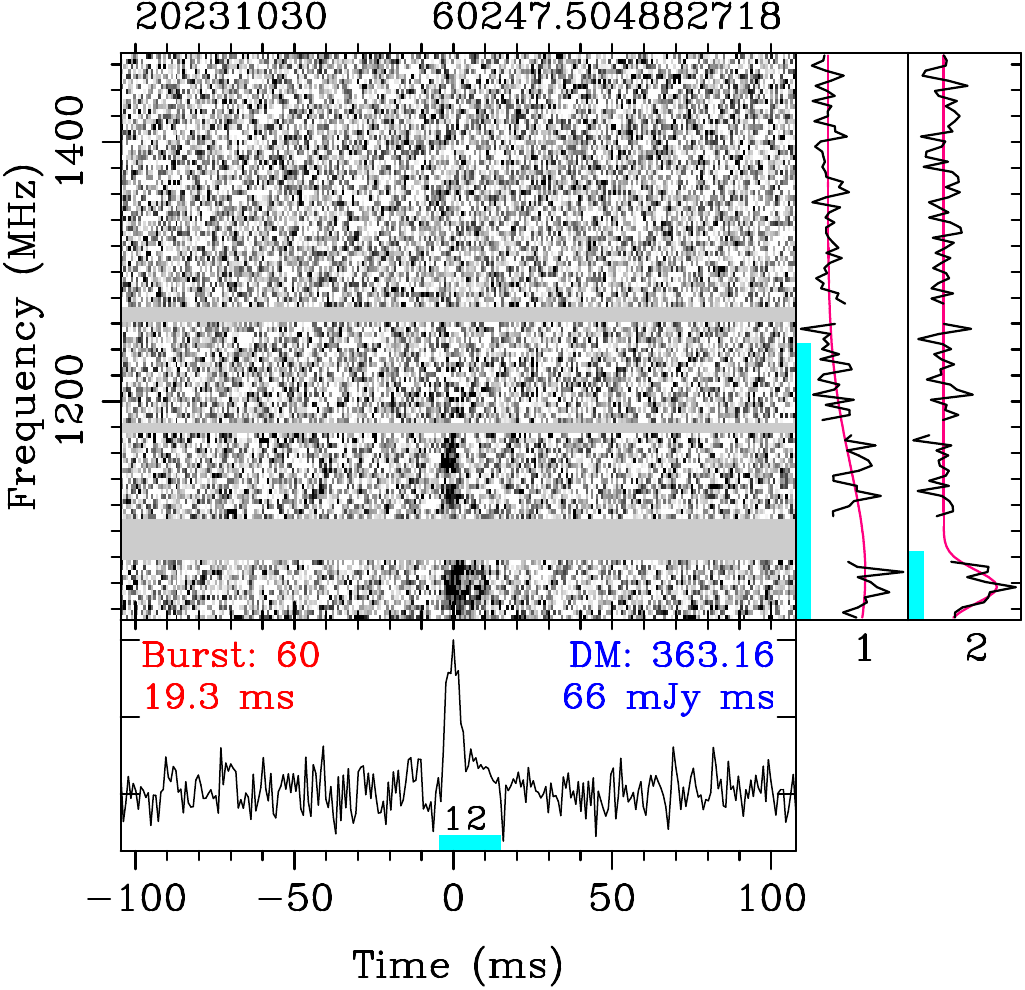}
\includegraphics[height=0.29\linewidth]{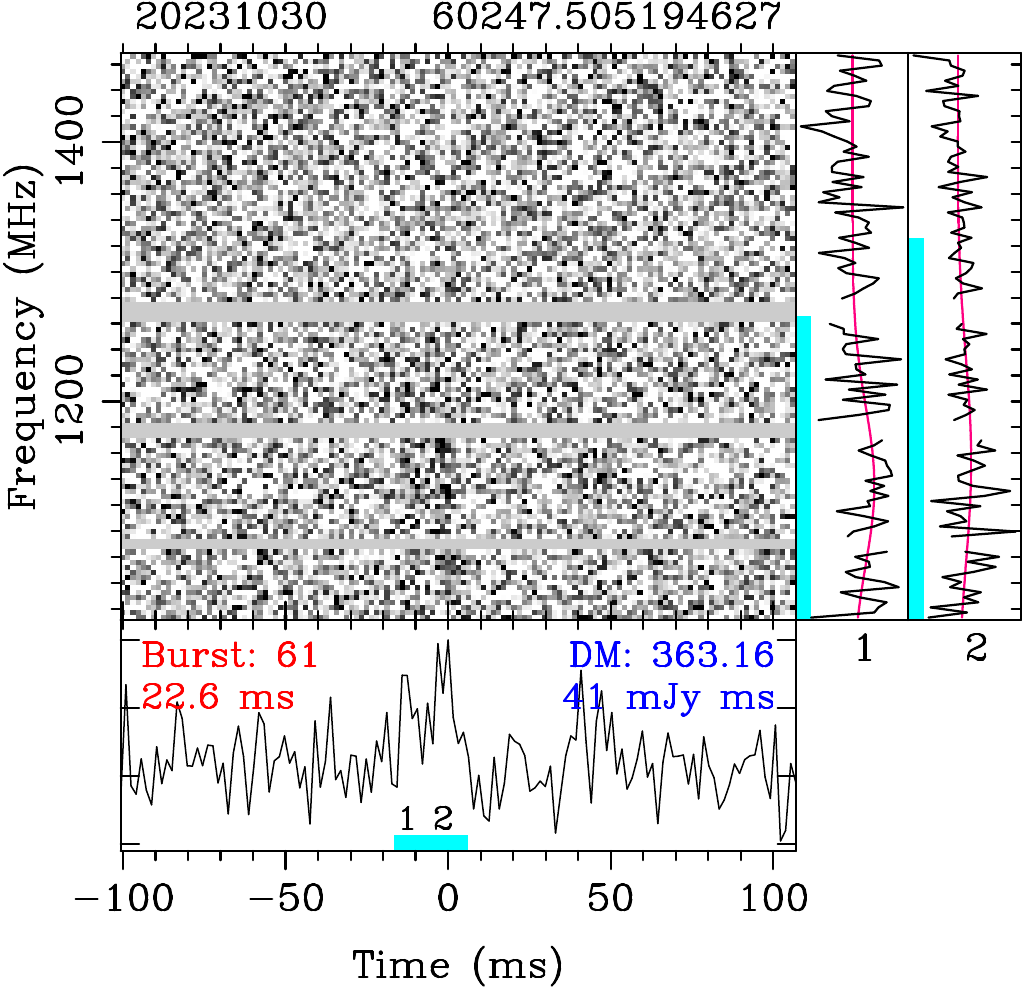}
\includegraphics[height=0.29\linewidth]{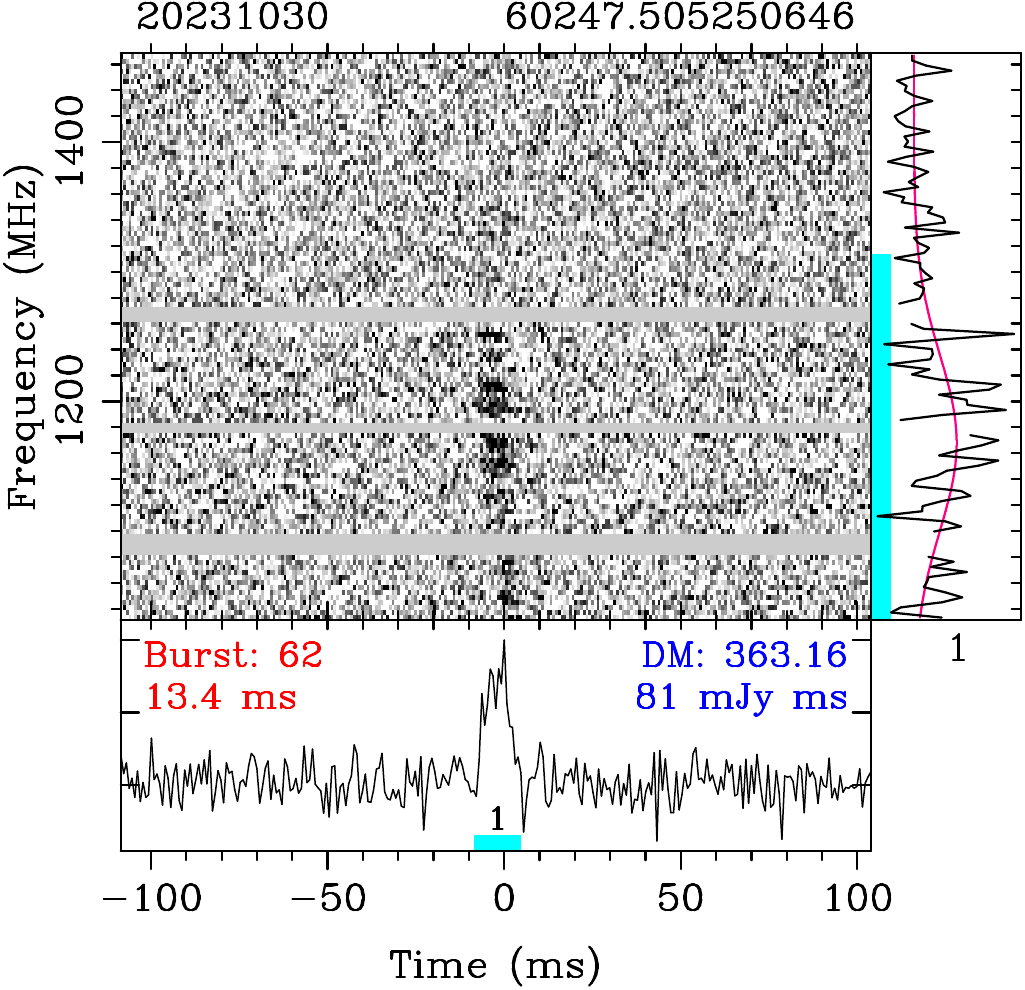}
\includegraphics[height=0.29\linewidth]{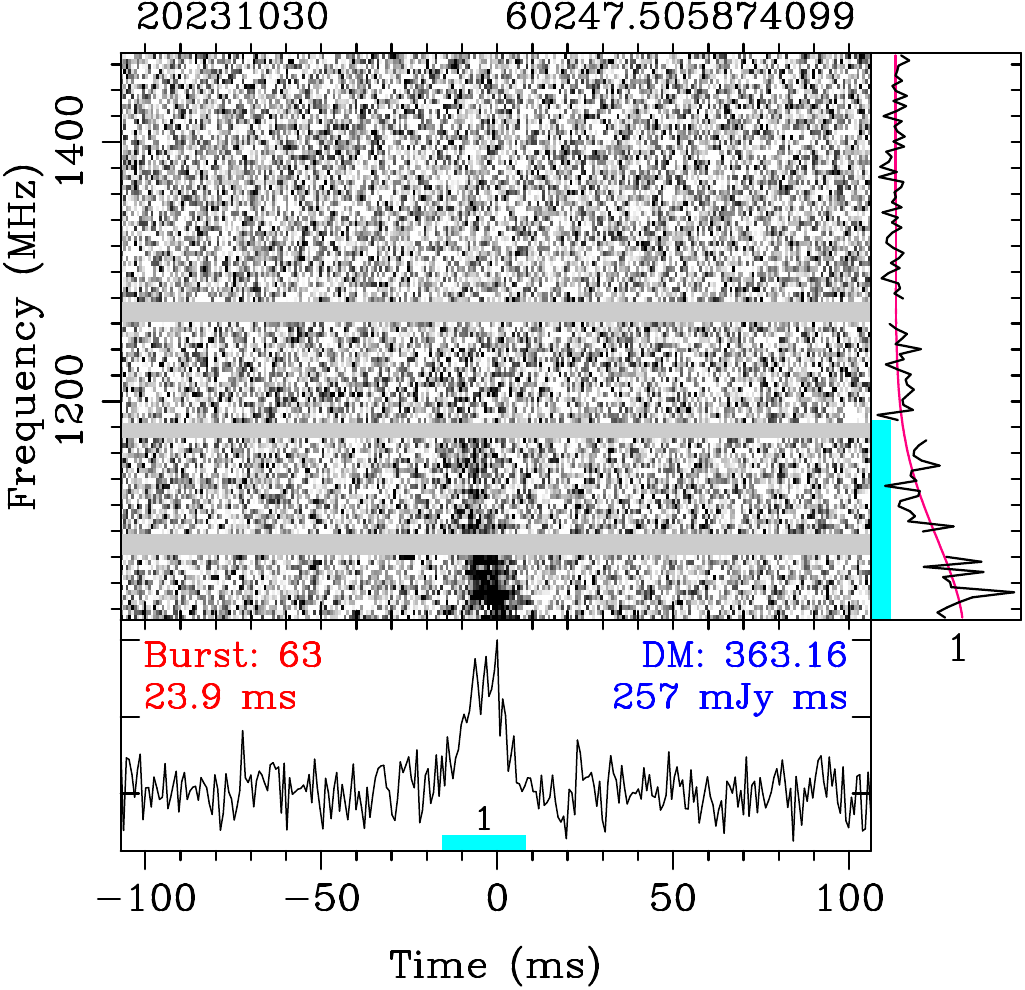}
\includegraphics[height=0.29\linewidth]{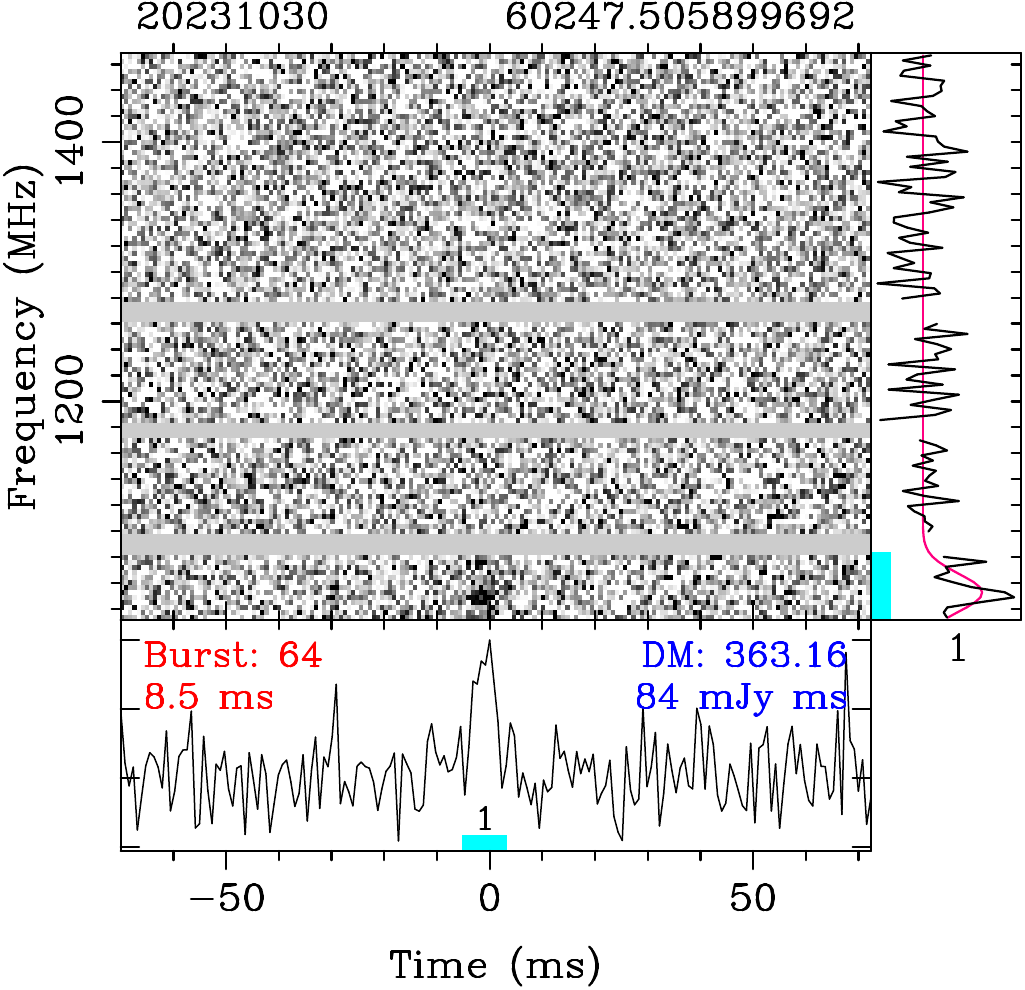}
\caption{({\textit{continued}})}
\end{figure*}
\addtocounter{figure}{-1}
\begin{figure*}
\flushleft
\includegraphics[height=0.29\linewidth]{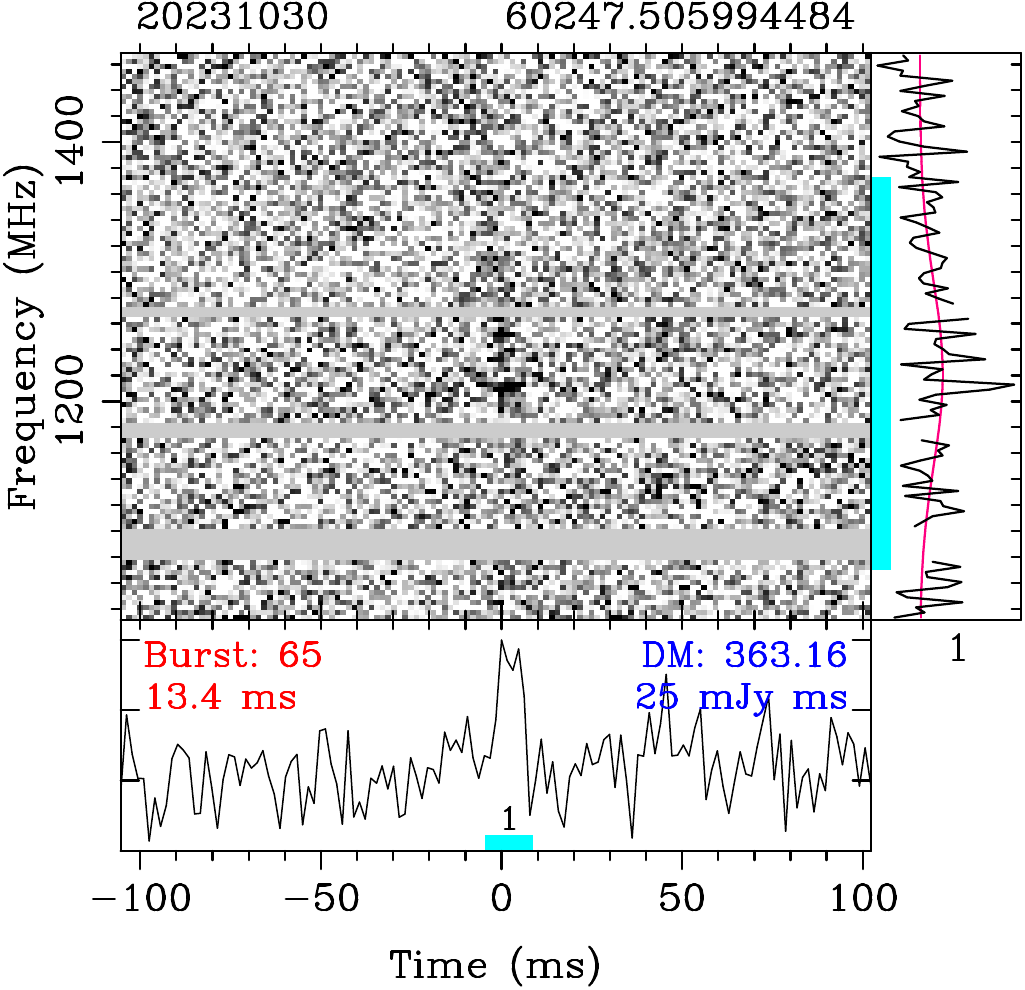}
\includegraphics[height=0.29\linewidth]{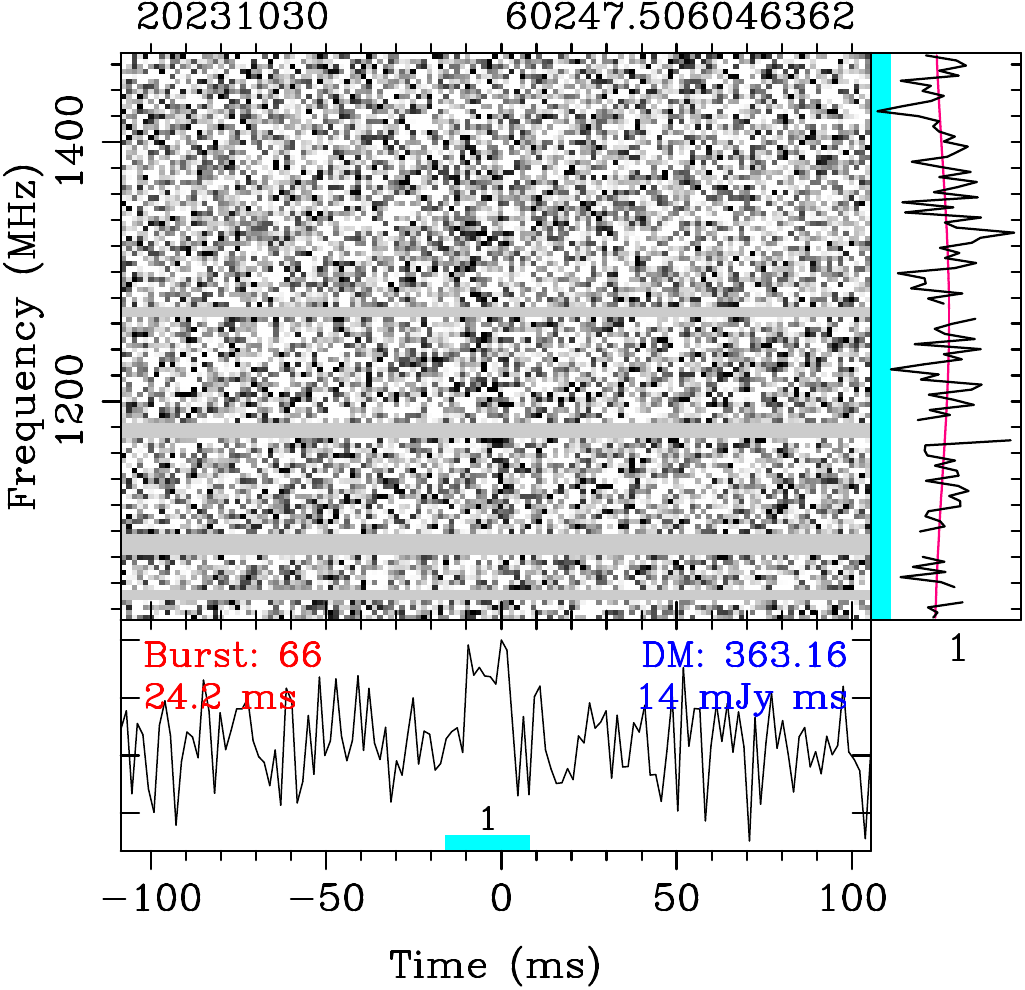}
\includegraphics[height=0.29\linewidth]{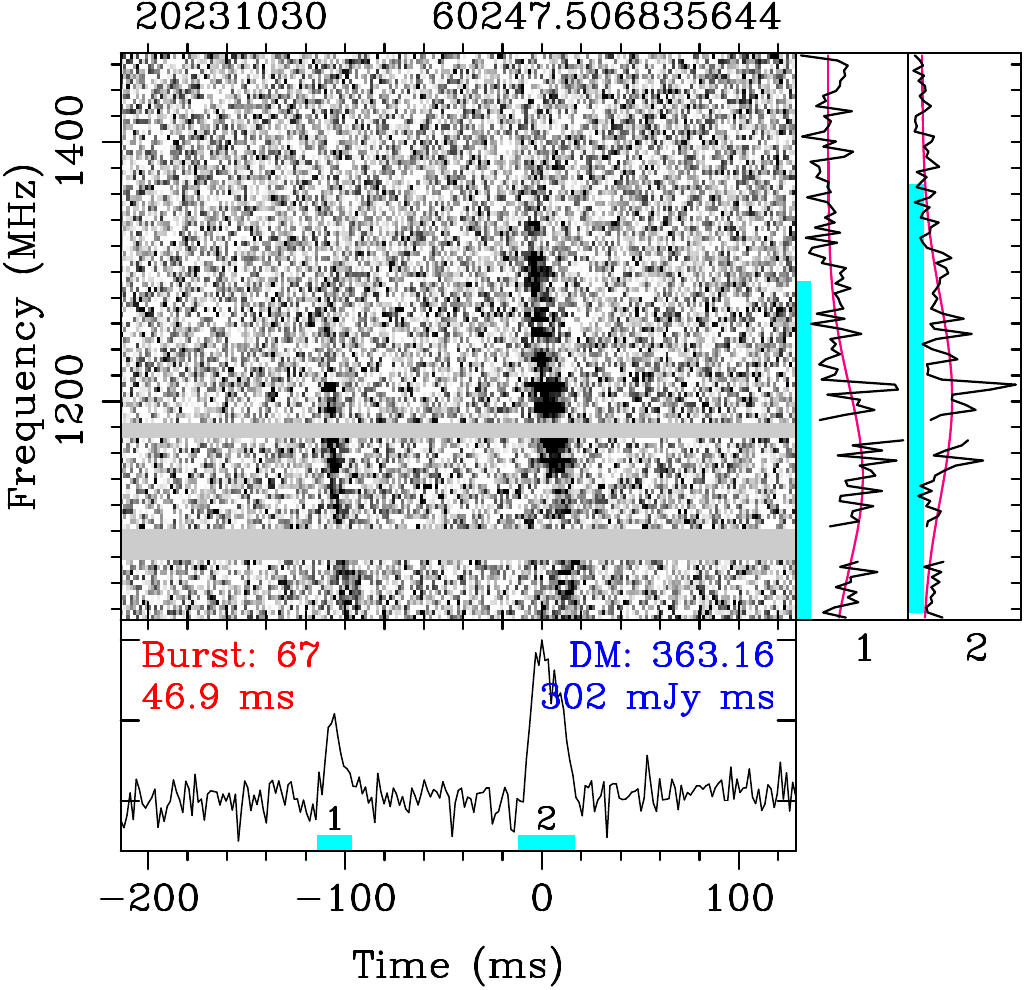}
\includegraphics[height=0.29\linewidth]{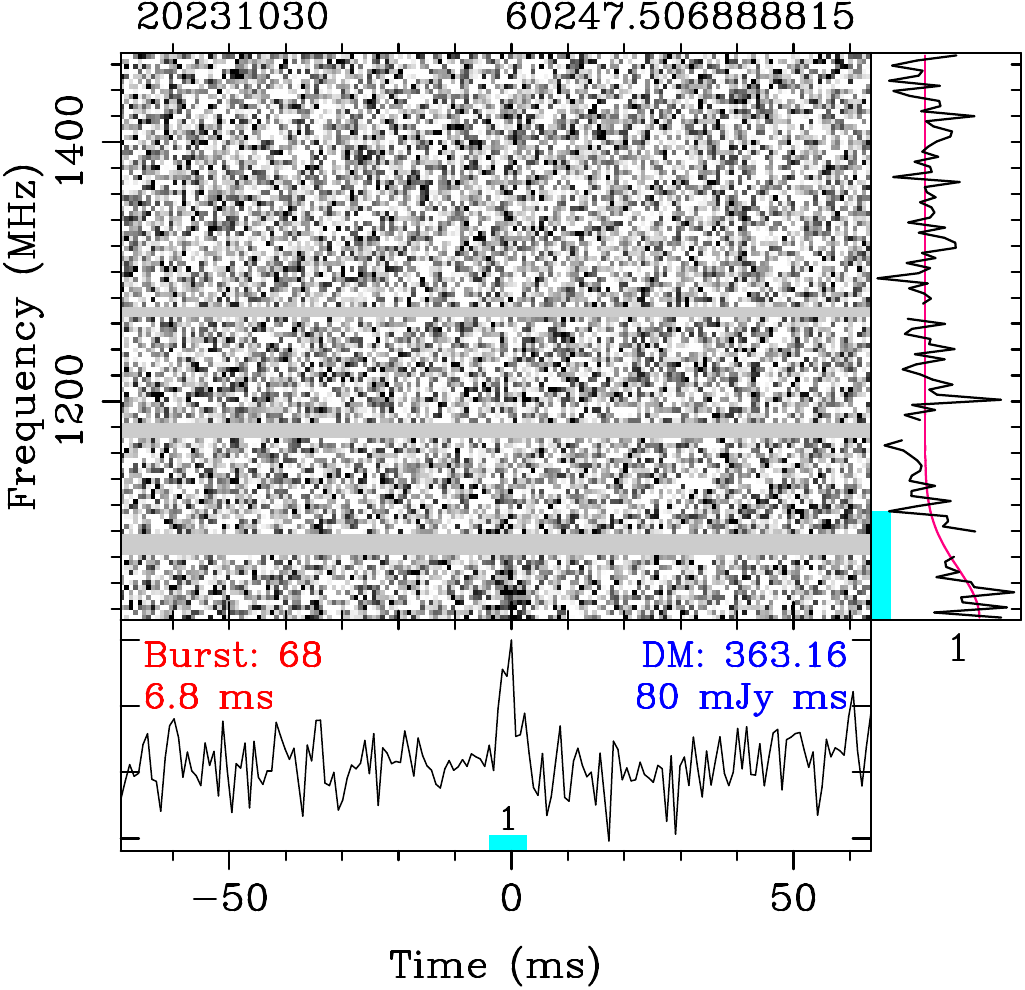}
\includegraphics[height=0.29\linewidth]{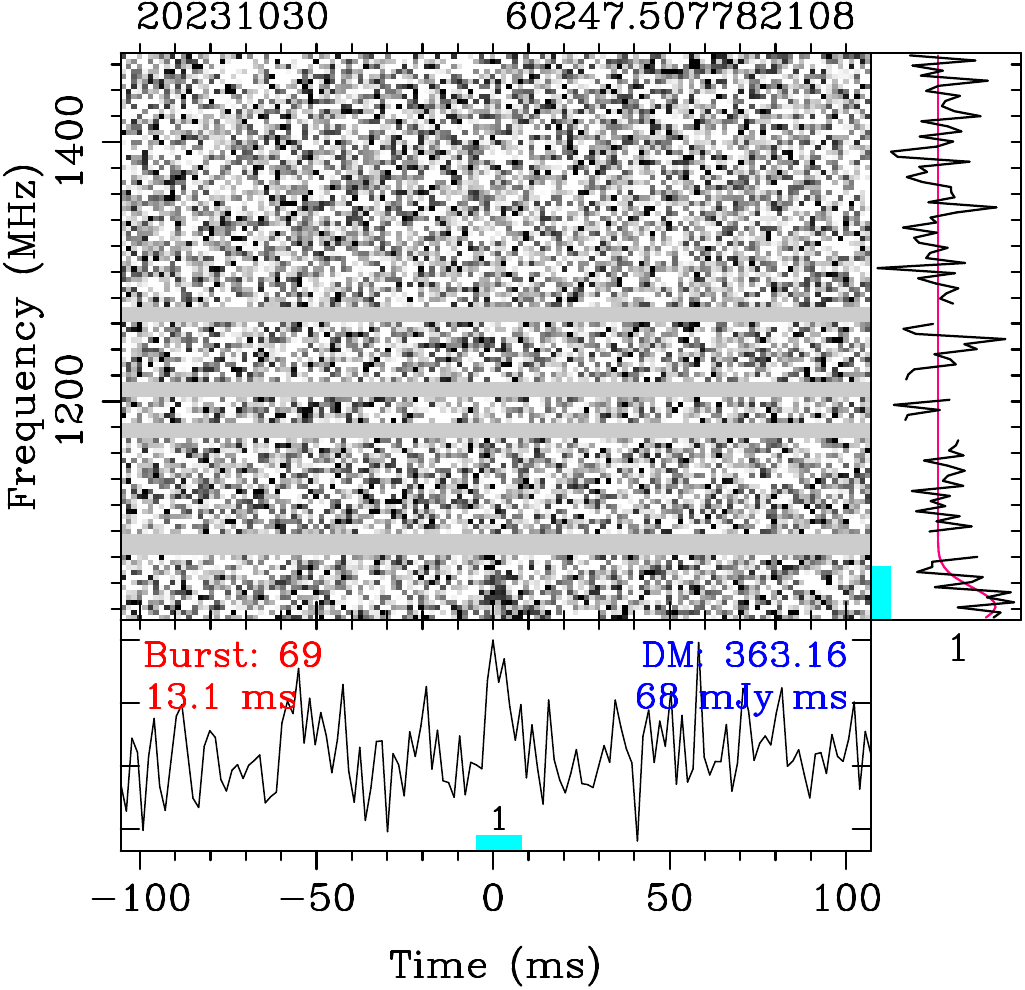}
\includegraphics[height=0.29\linewidth]{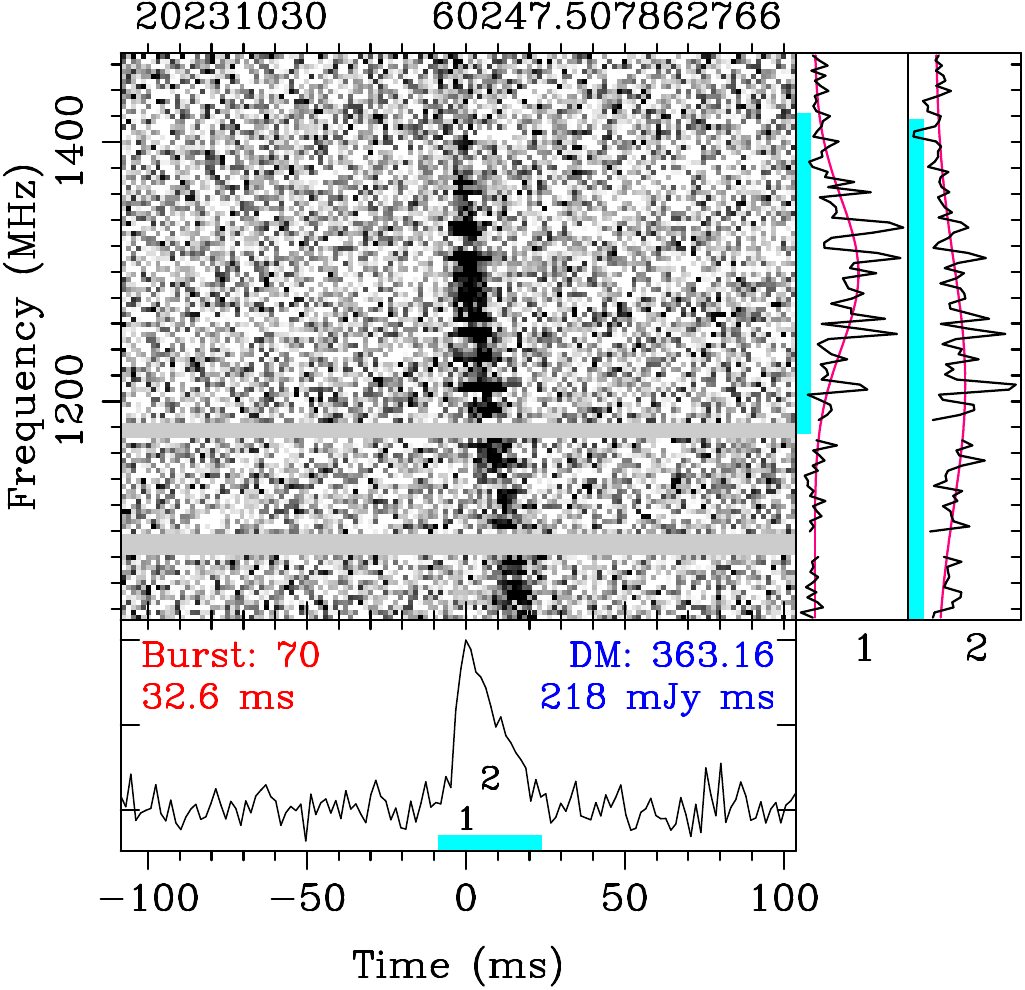}
\includegraphics[height=0.29\linewidth]{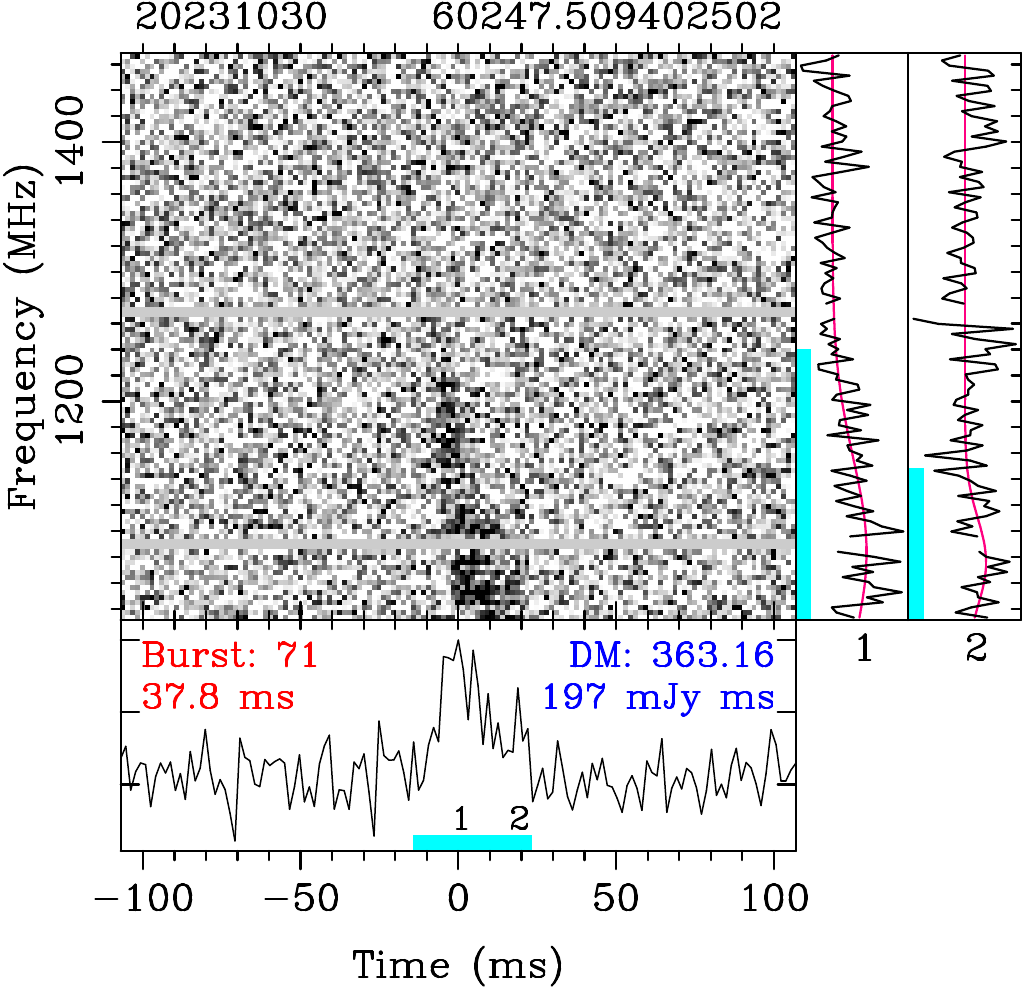}
\includegraphics[height=0.29\linewidth]{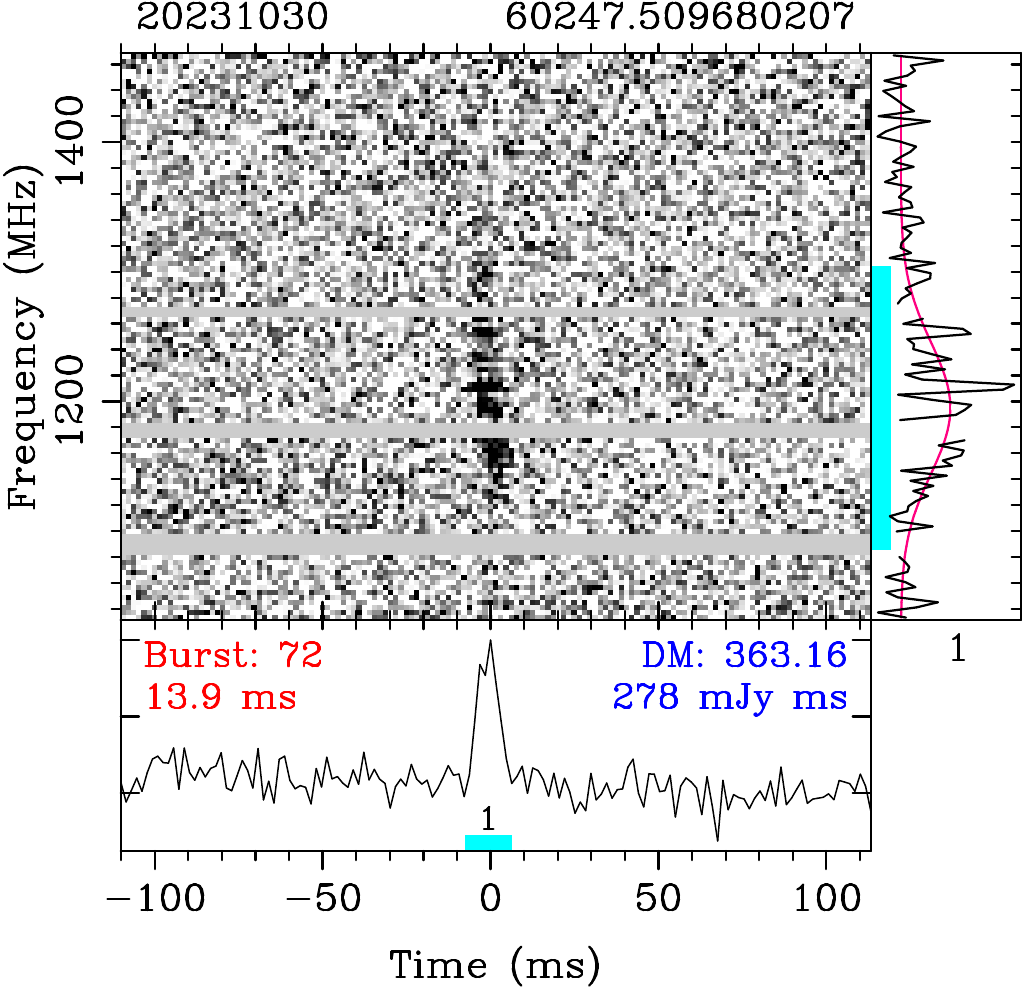}
\includegraphics[height=0.29\linewidth]{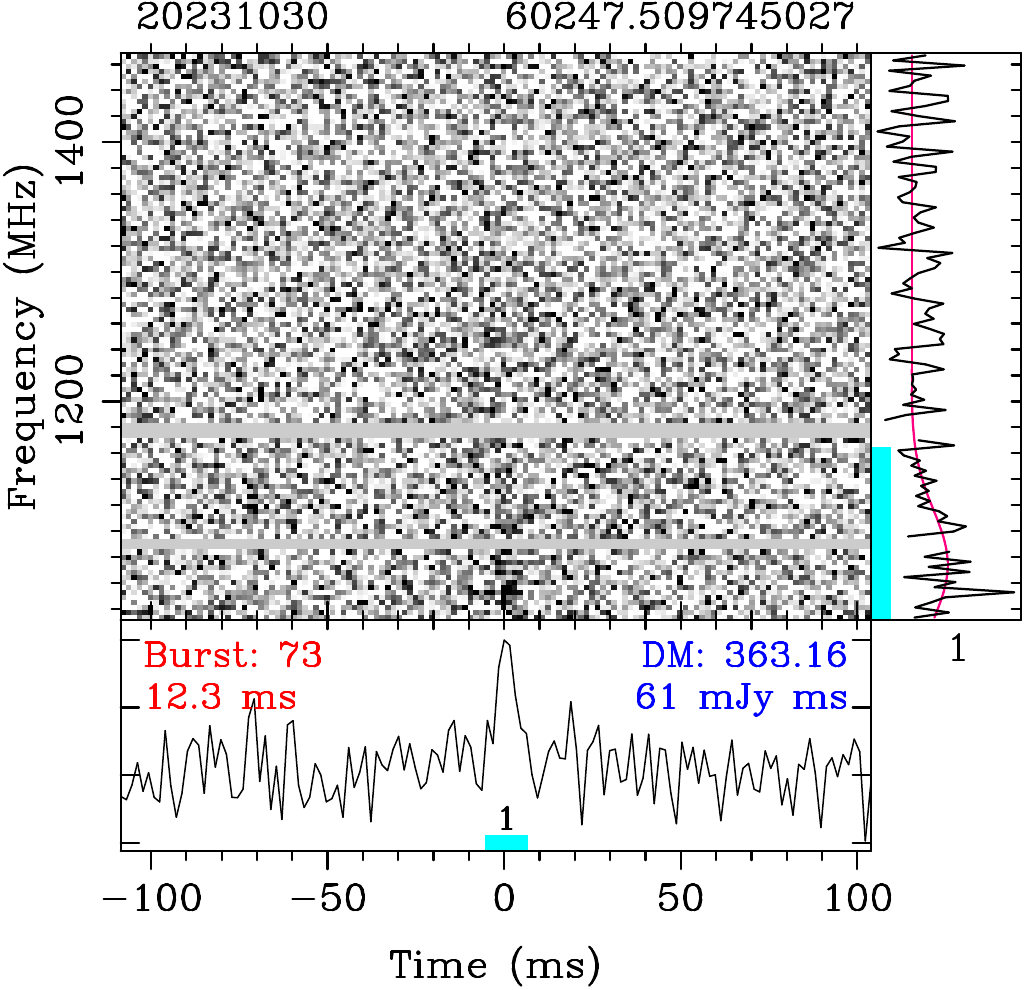}
\includegraphics[height=0.29\linewidth]{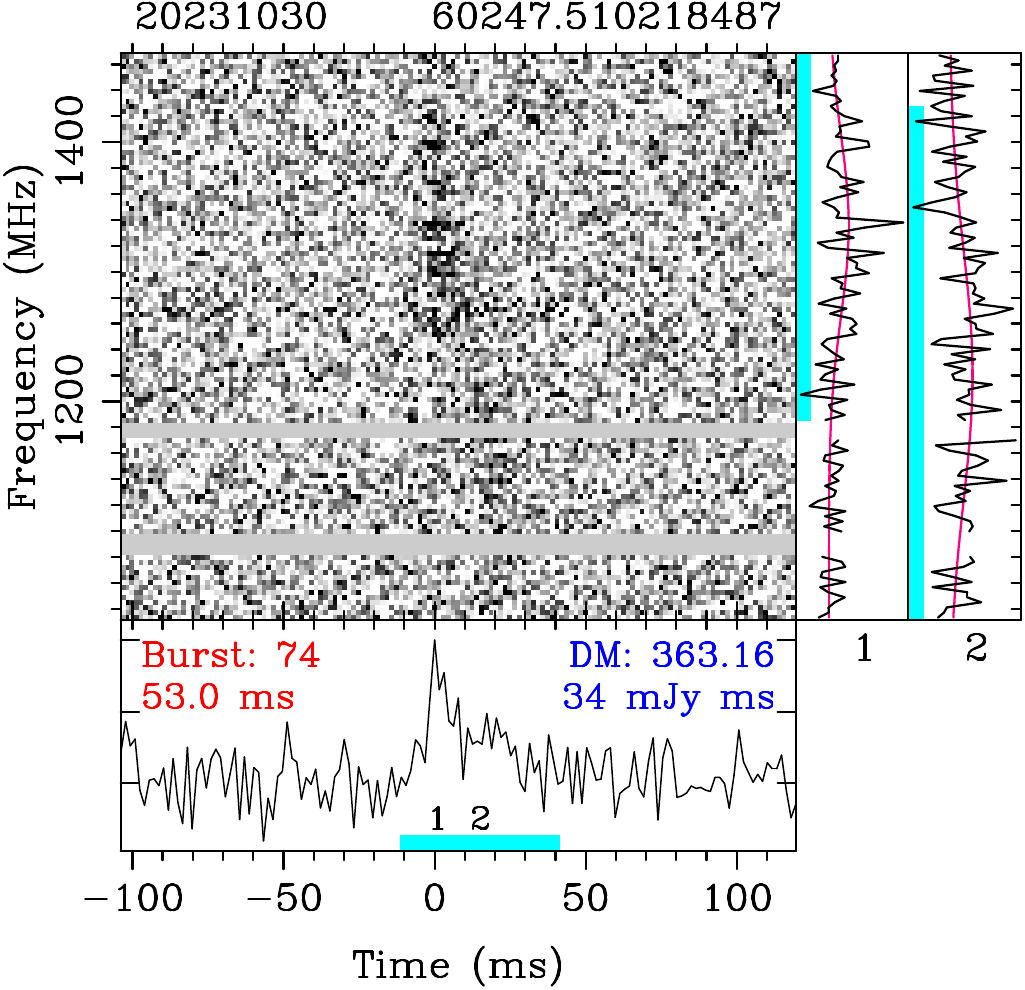}
\includegraphics[height=0.29\linewidth]{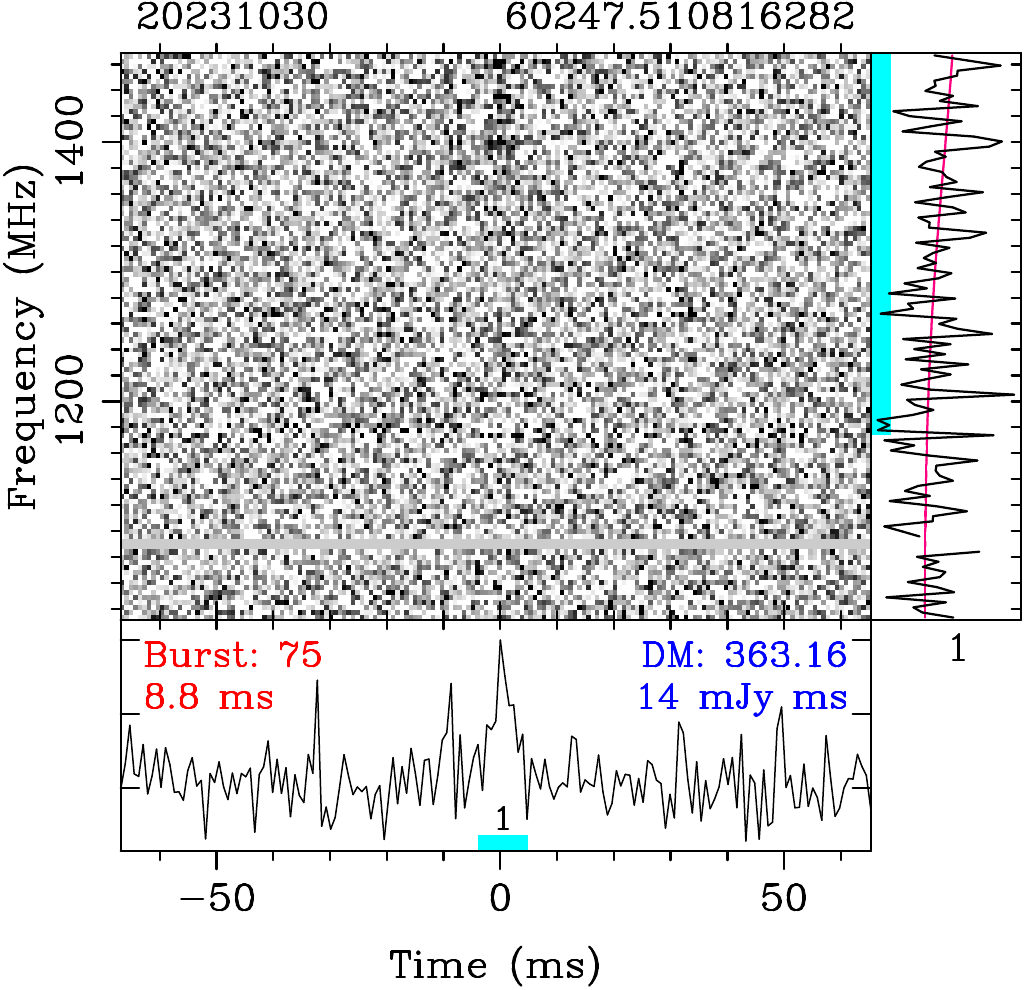}
\includegraphics[height=0.29\linewidth]{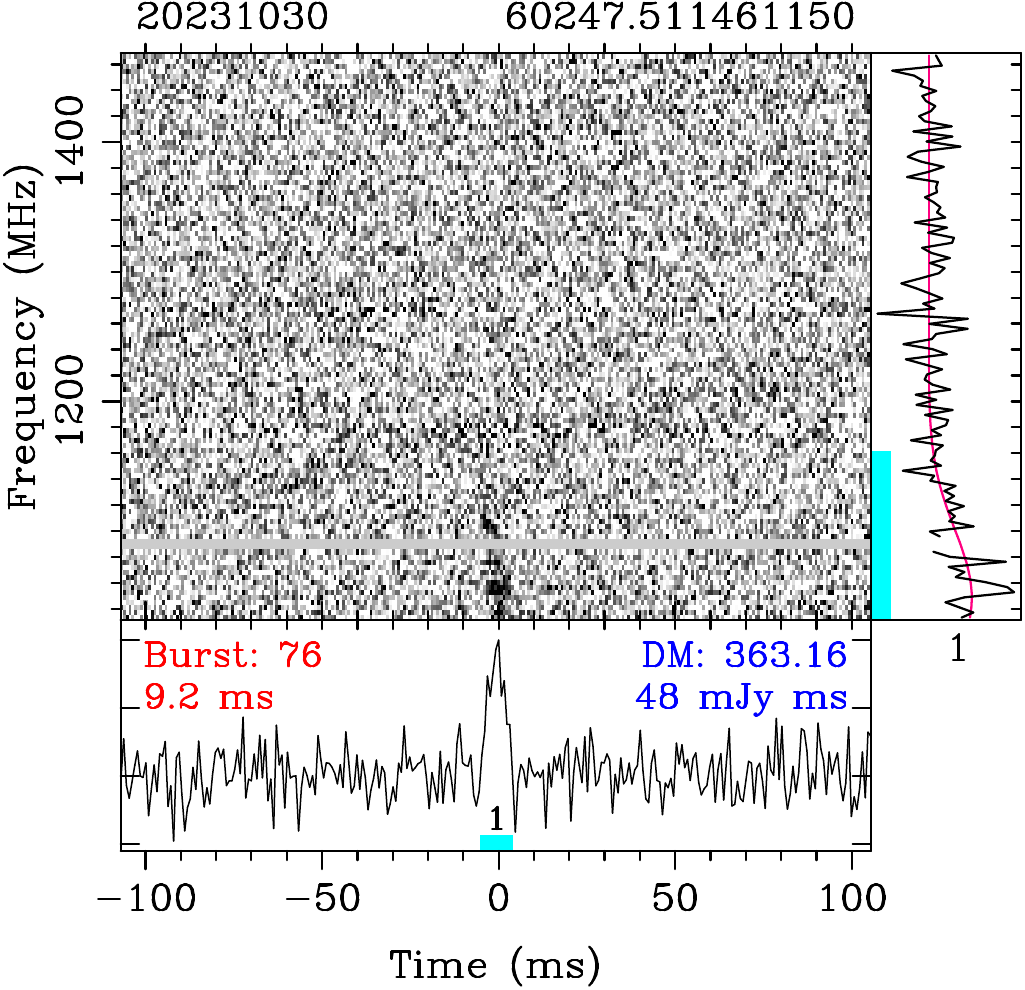}
\caption{({\textit{continued}})}
\end{figure*}
\addtocounter{figure}{-1}
\begin{figure*}
\flushleft
\includegraphics[height=0.29\linewidth]{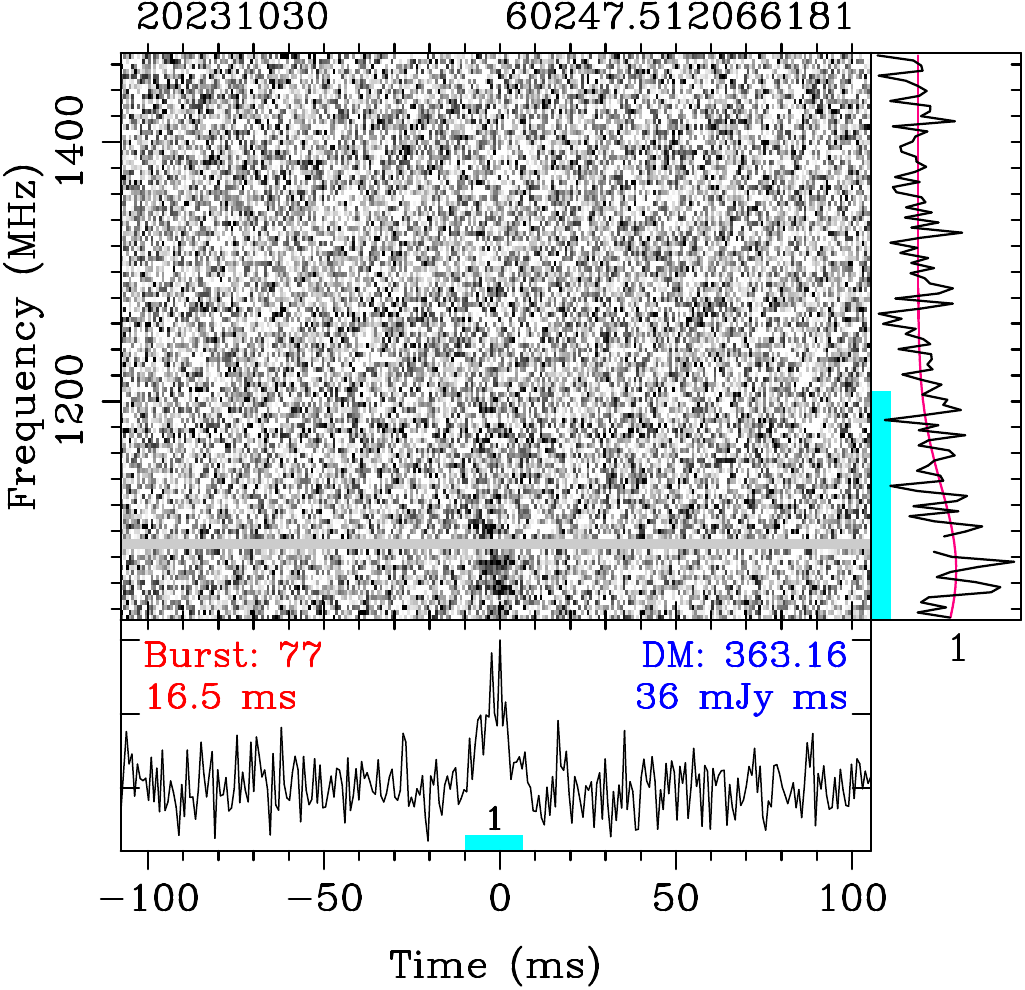}
\includegraphics[height=0.29\linewidth]{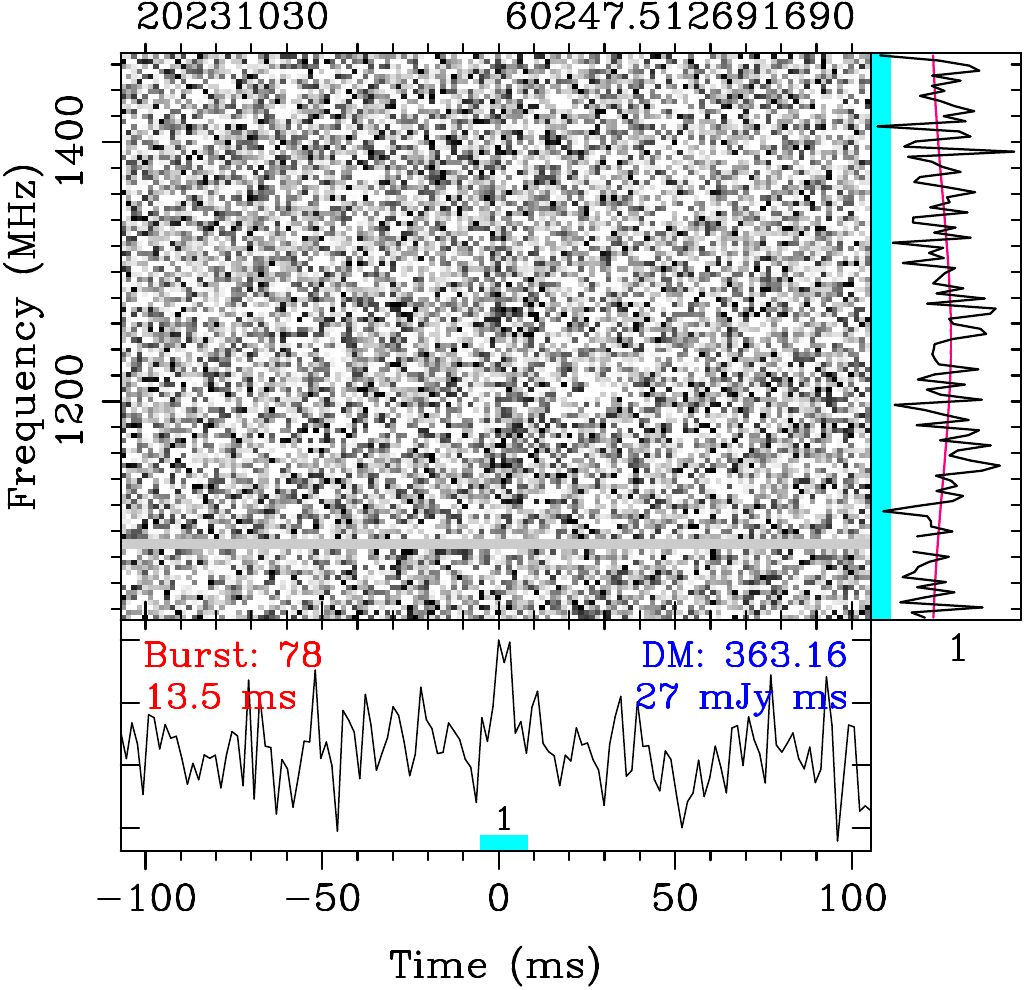}
\includegraphics[height=0.29\linewidth]{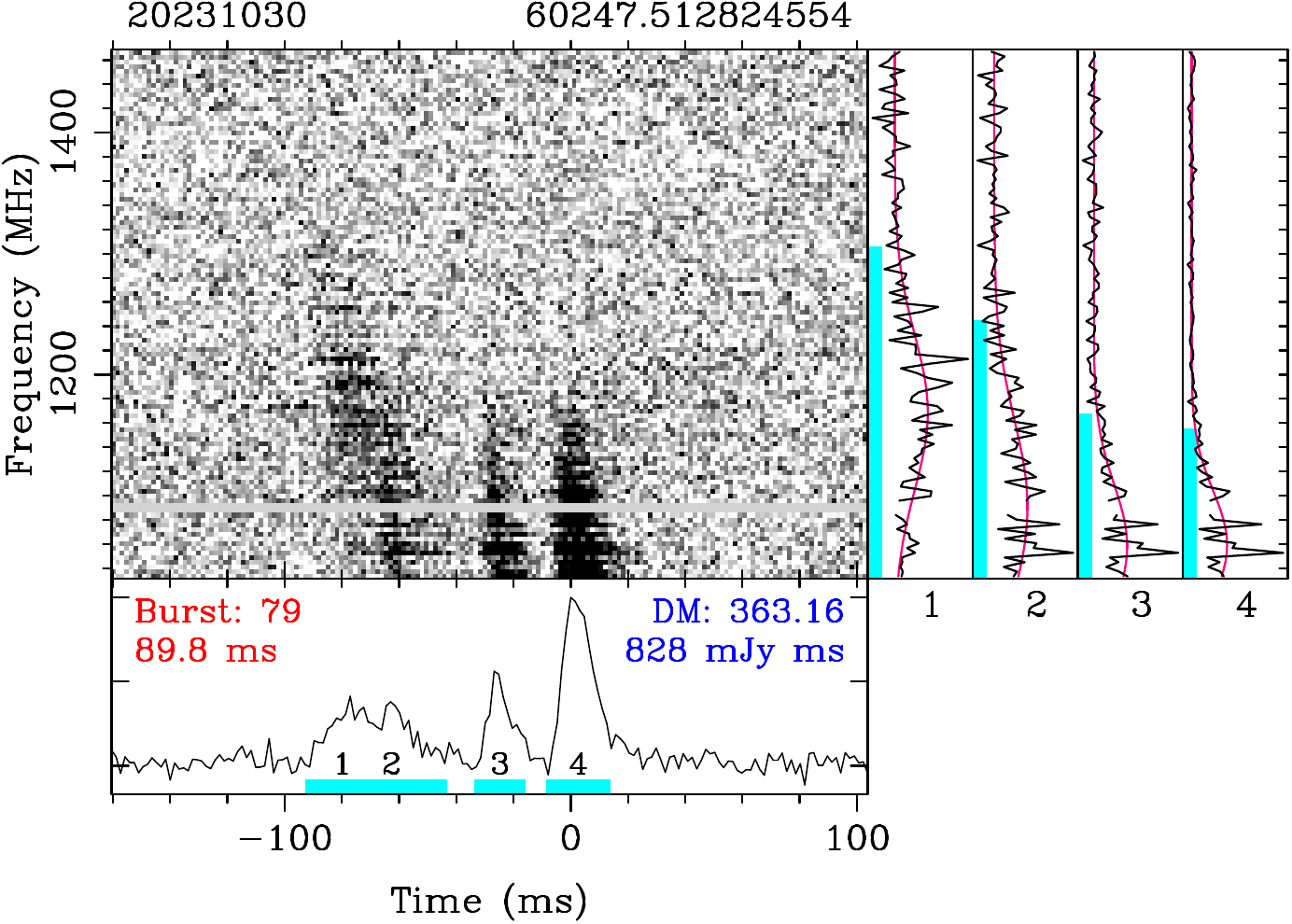}
\includegraphics[height=0.29\linewidth]{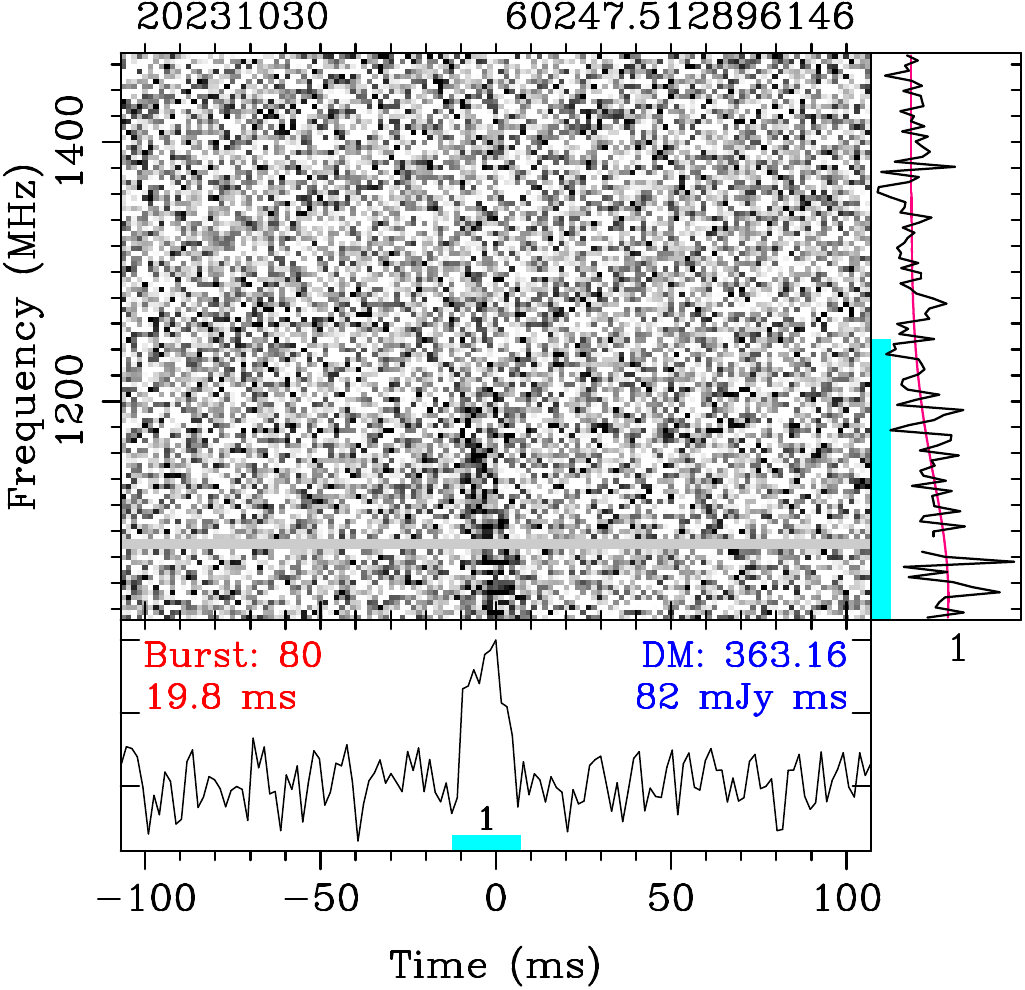}
\includegraphics[height=0.29\linewidth]{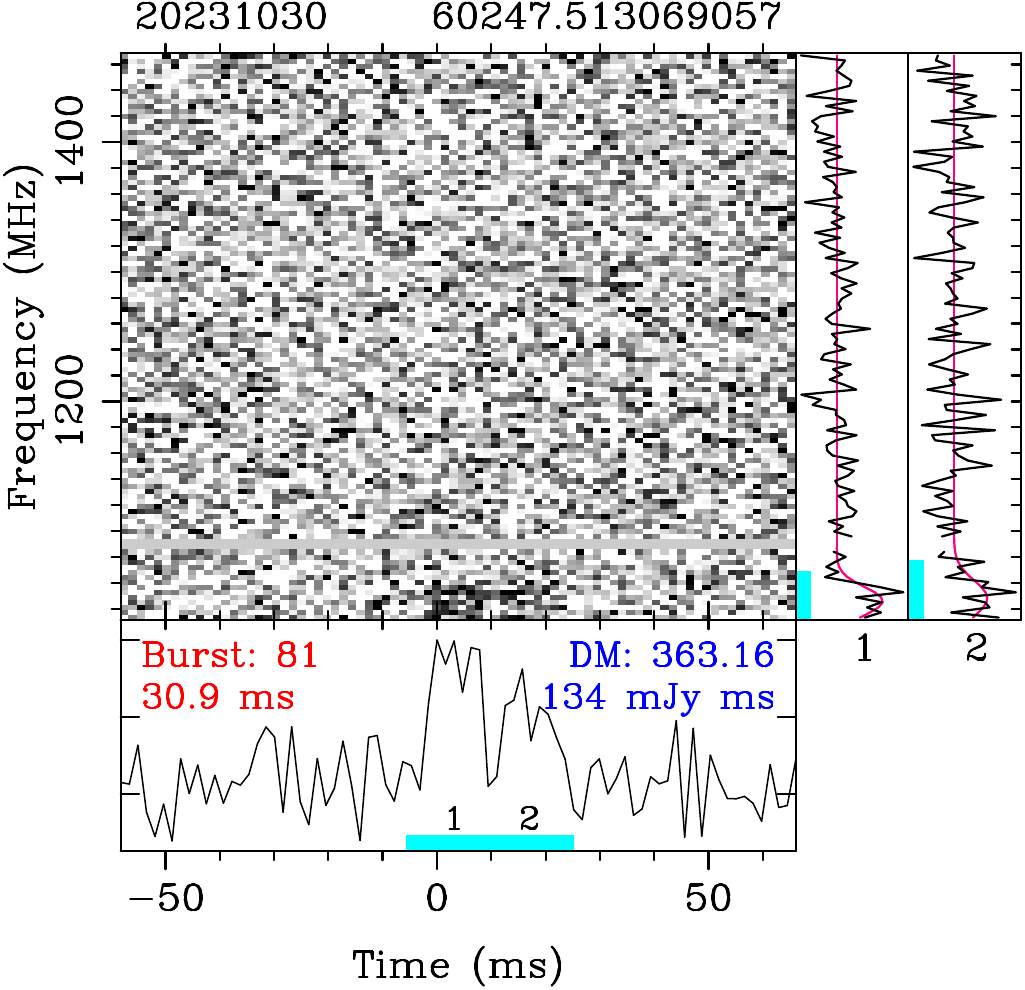}
\includegraphics[height=0.29\linewidth]{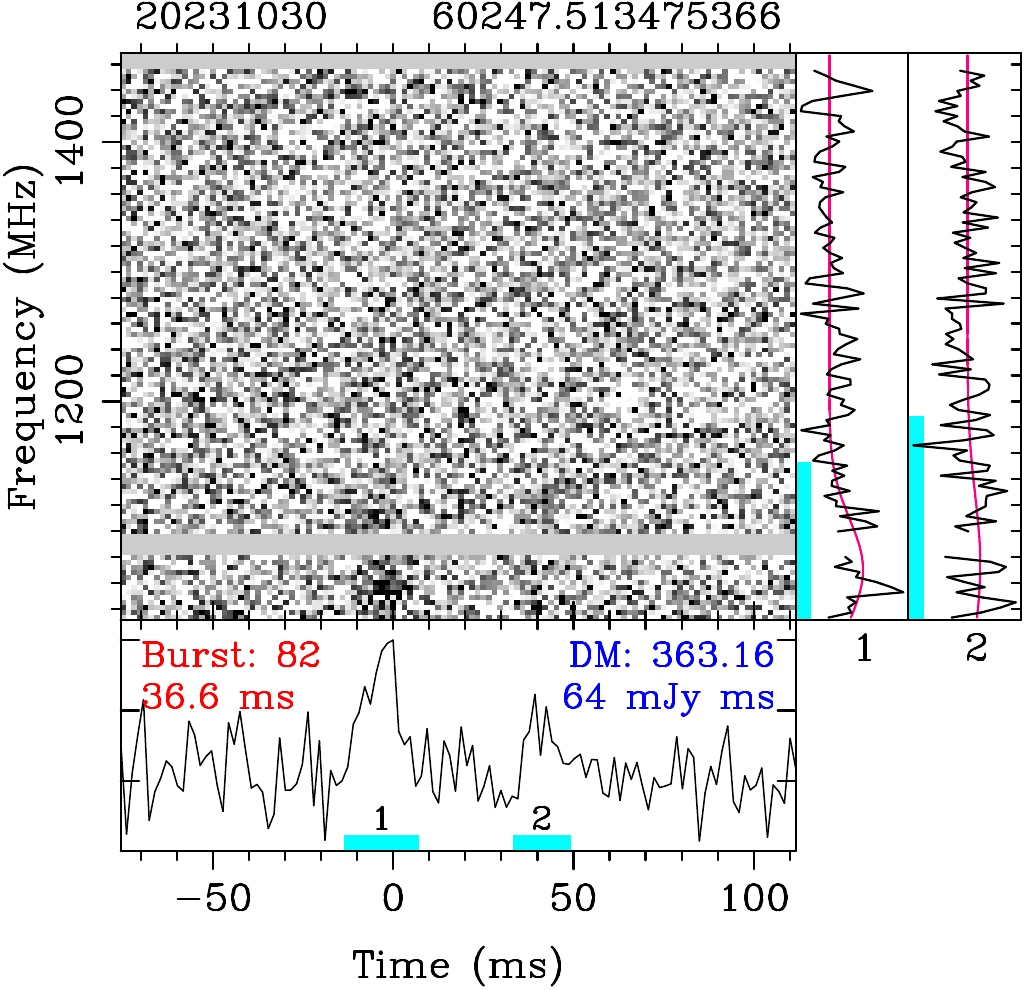}
\includegraphics[height=0.29\linewidth]{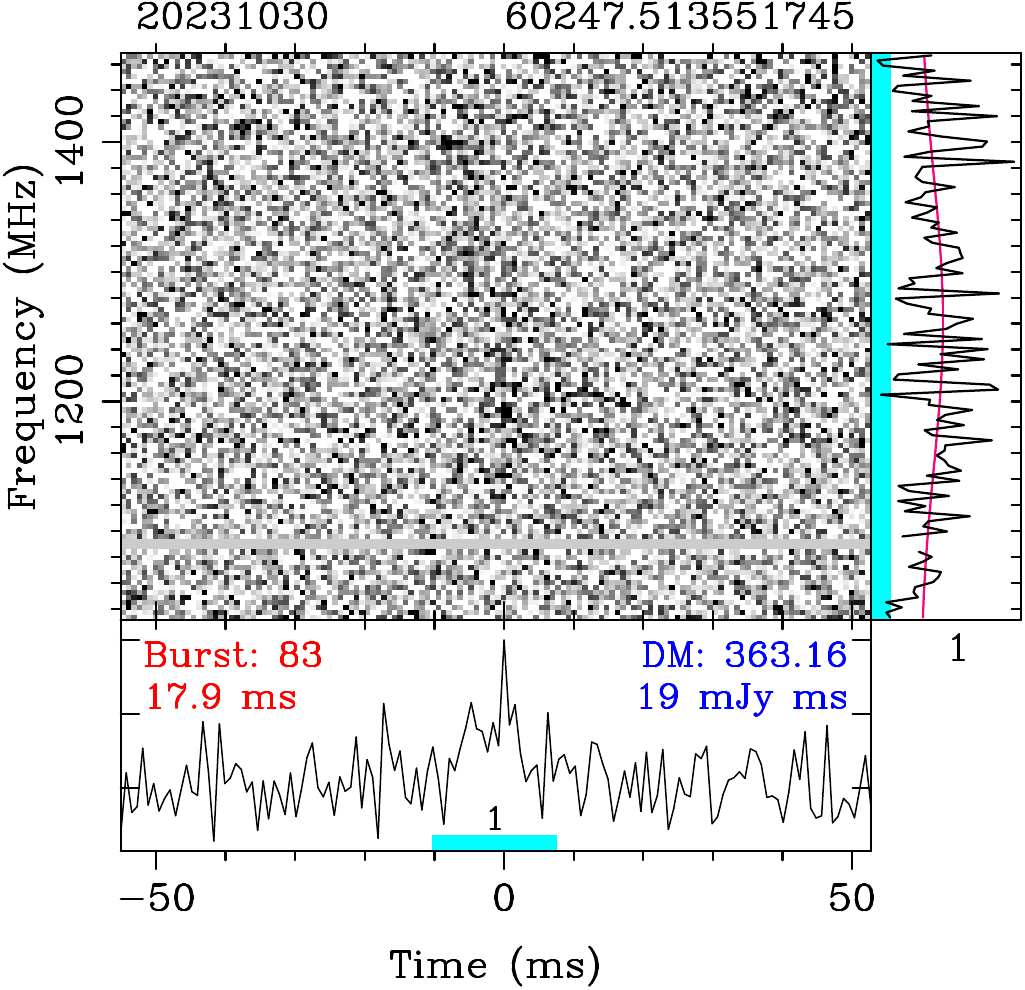}
\includegraphics[height=0.29\linewidth]{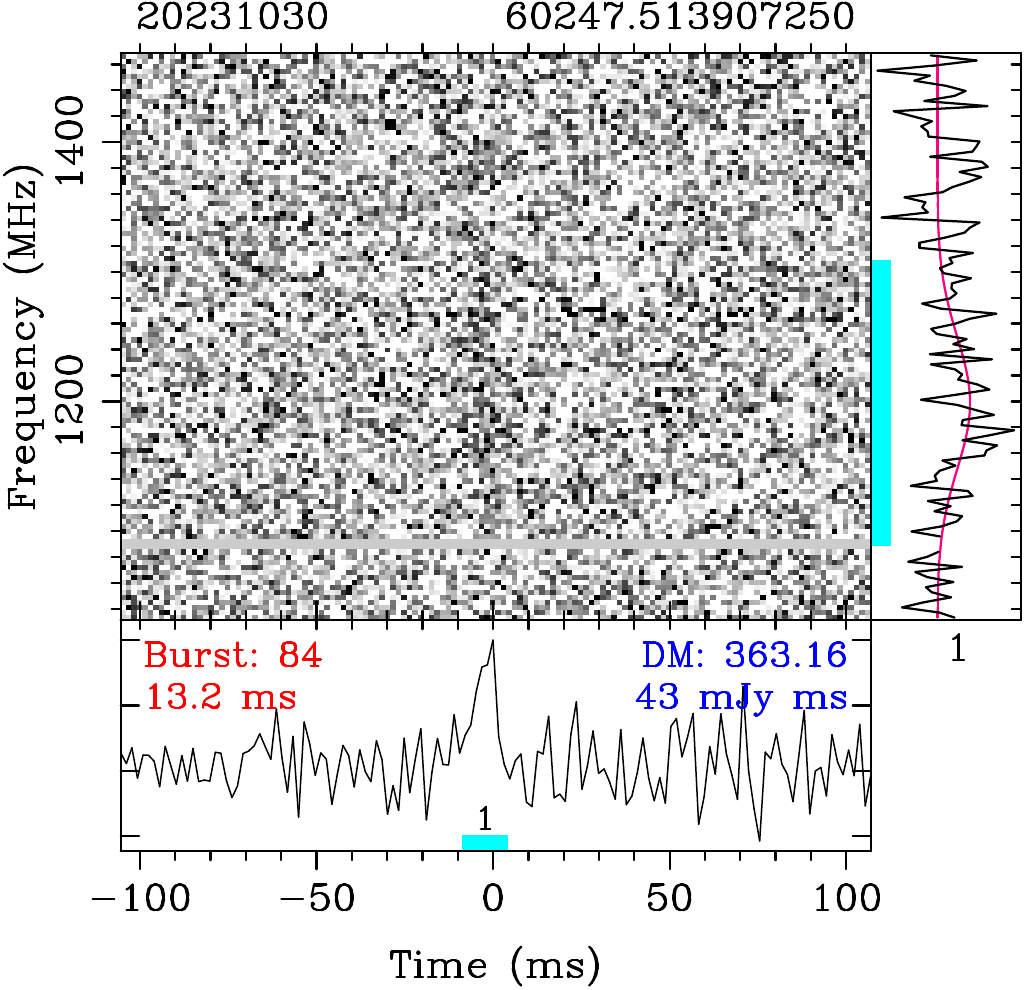}
\includegraphics[height=0.29\linewidth]{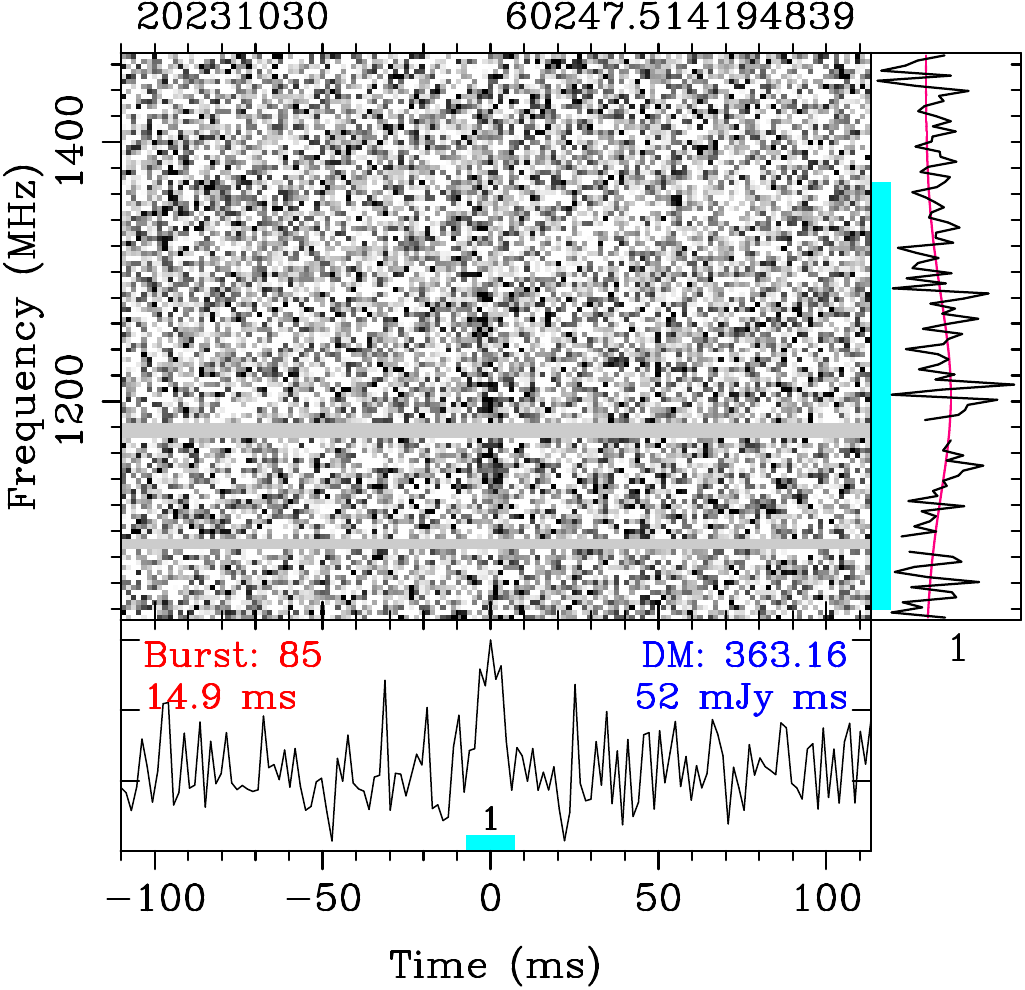}
\includegraphics[height=0.29\linewidth]{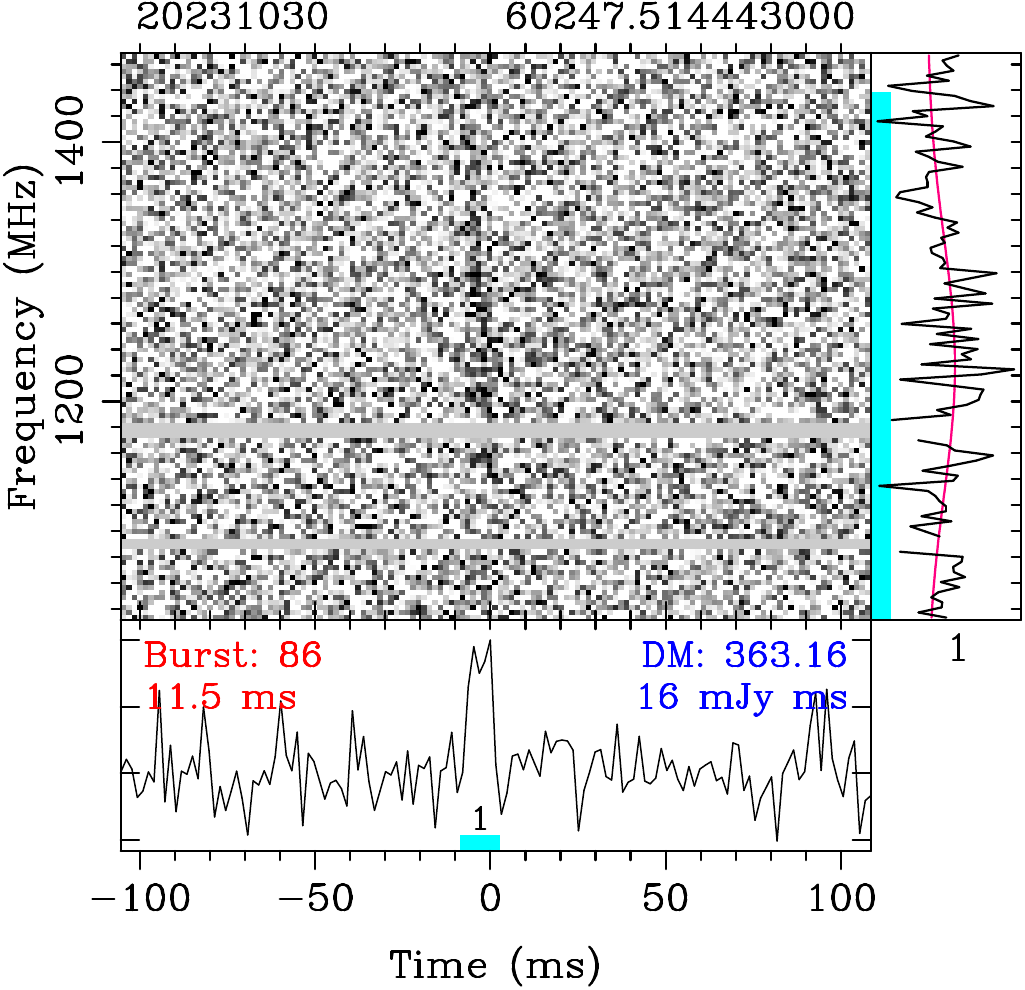}
\caption{({\textit{continued}})}
\end{figure*}
\addtocounter{figure}{-1}
\begin{figure*}
\flushleft
\includegraphics[height=0.29\linewidth]{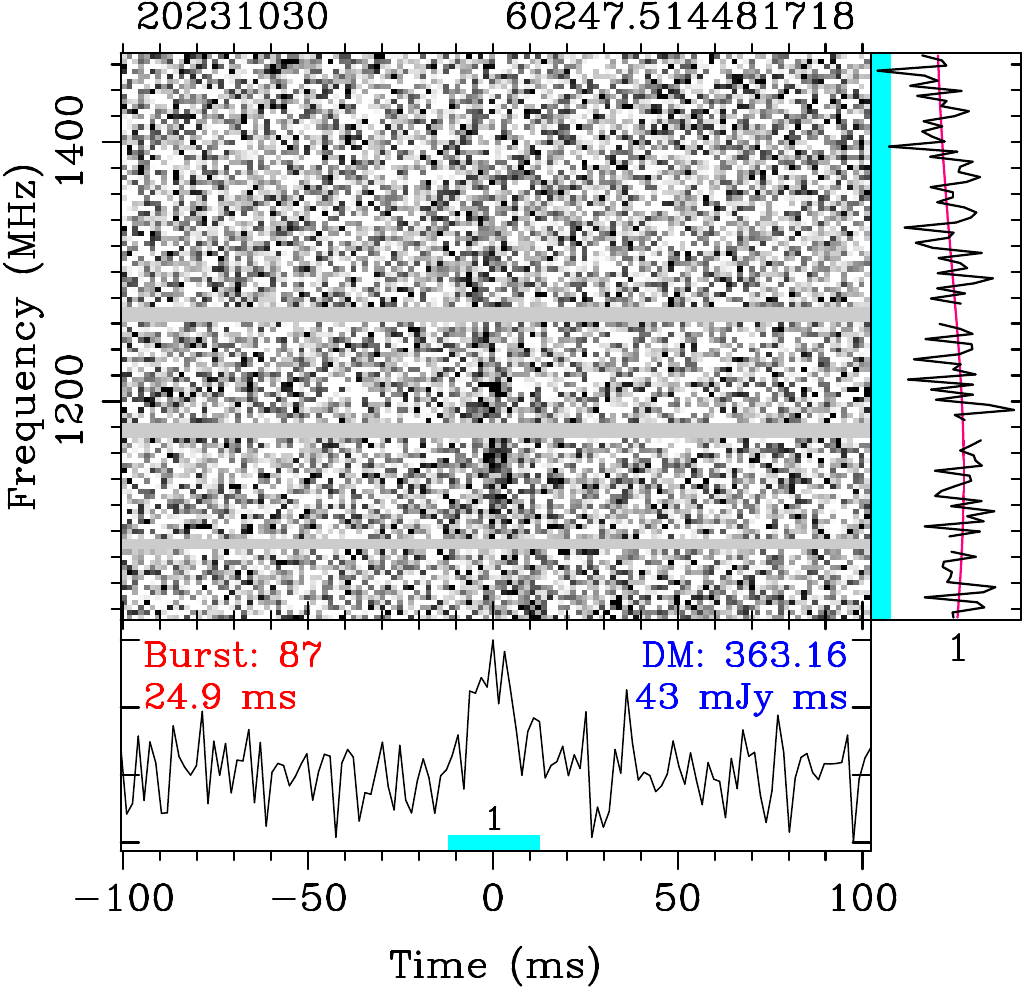}
\includegraphics[height=0.29\linewidth]{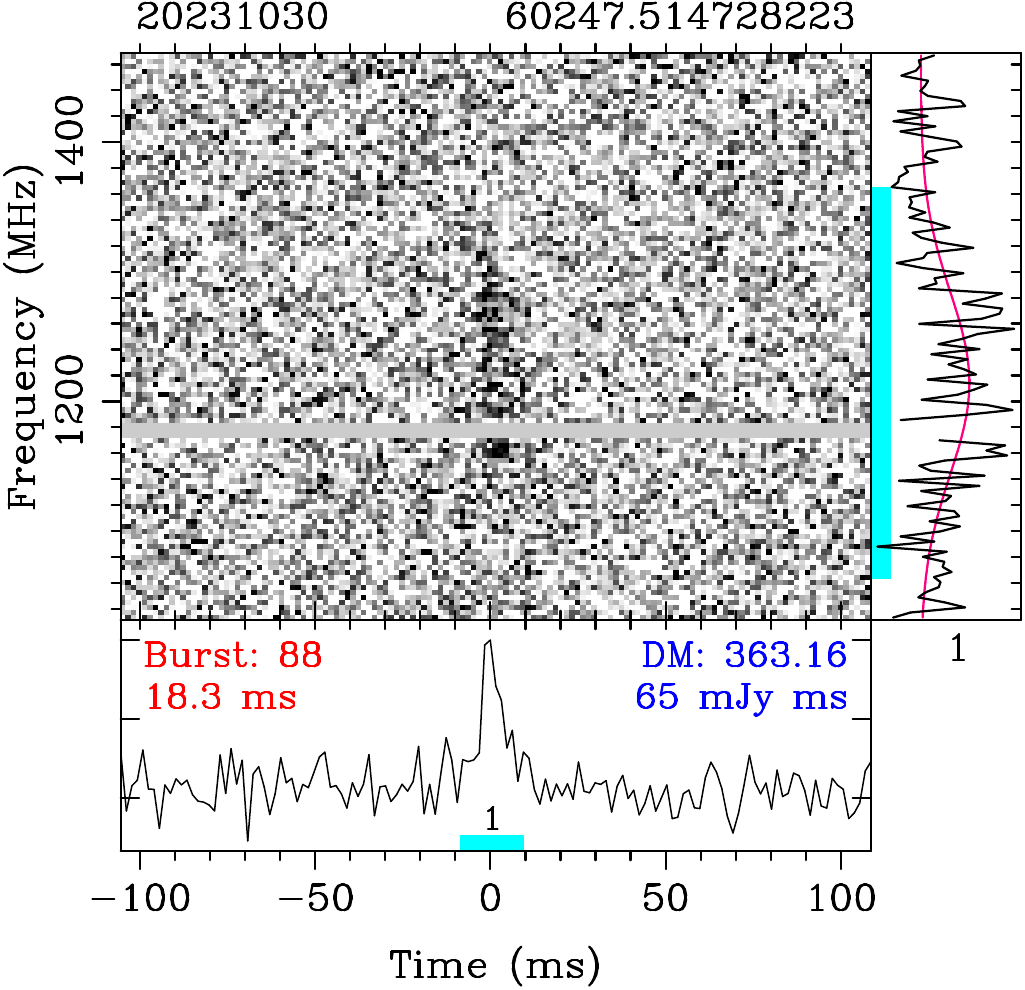}
\includegraphics[height=0.29\linewidth]{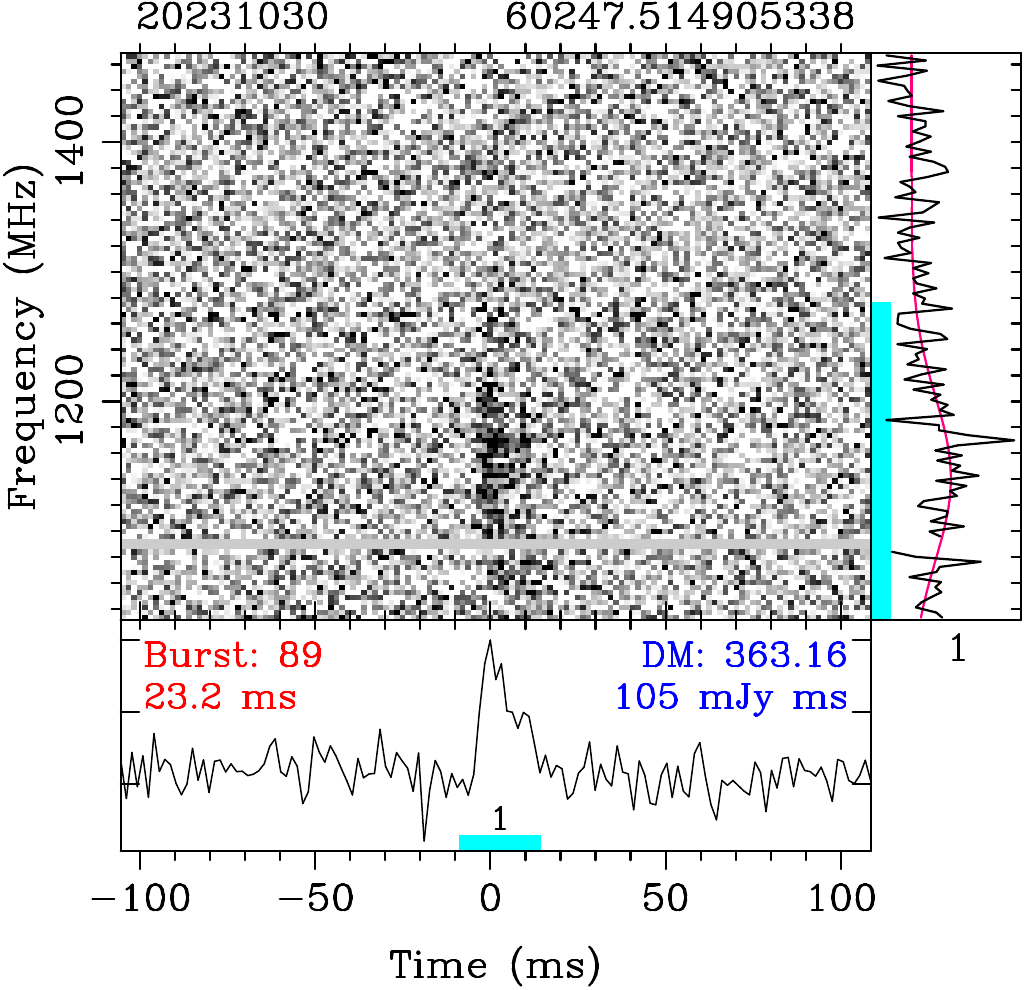}
\includegraphics[height=0.29\linewidth]{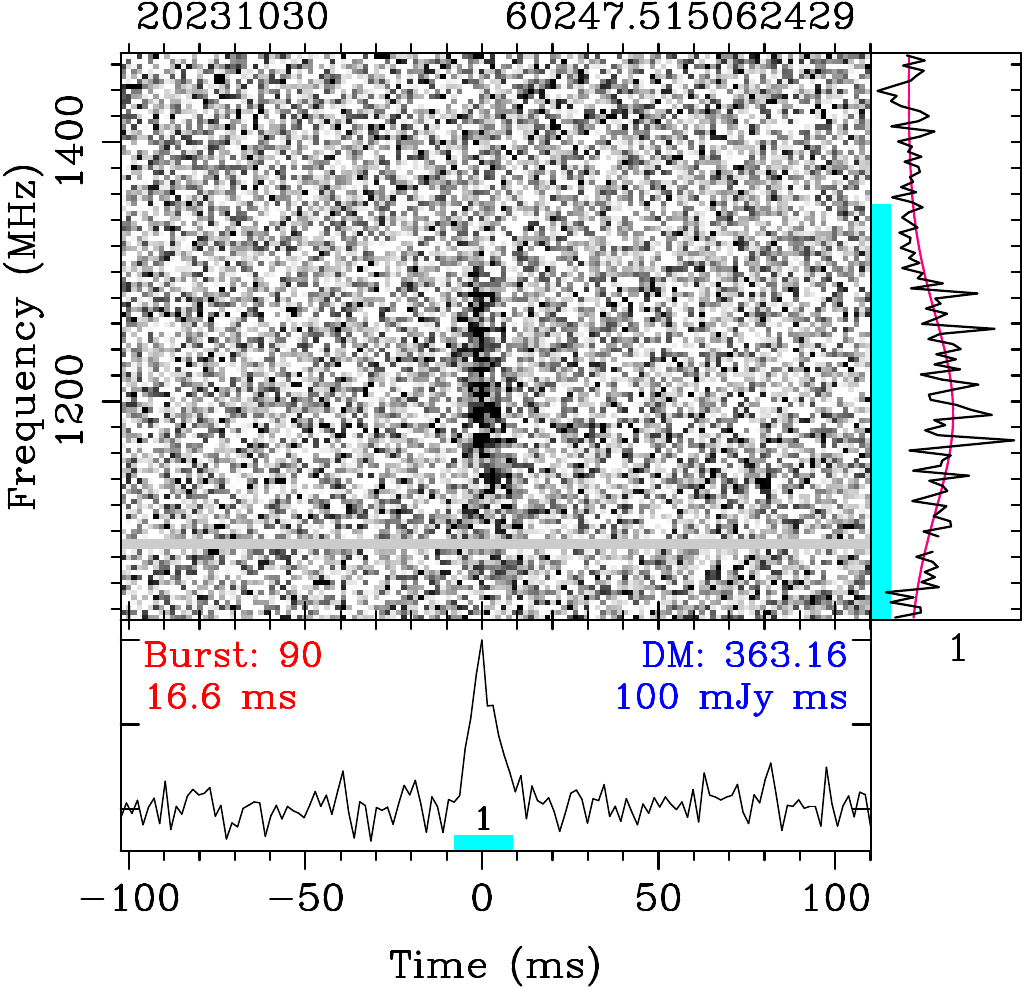}
\includegraphics[height=0.29\linewidth]{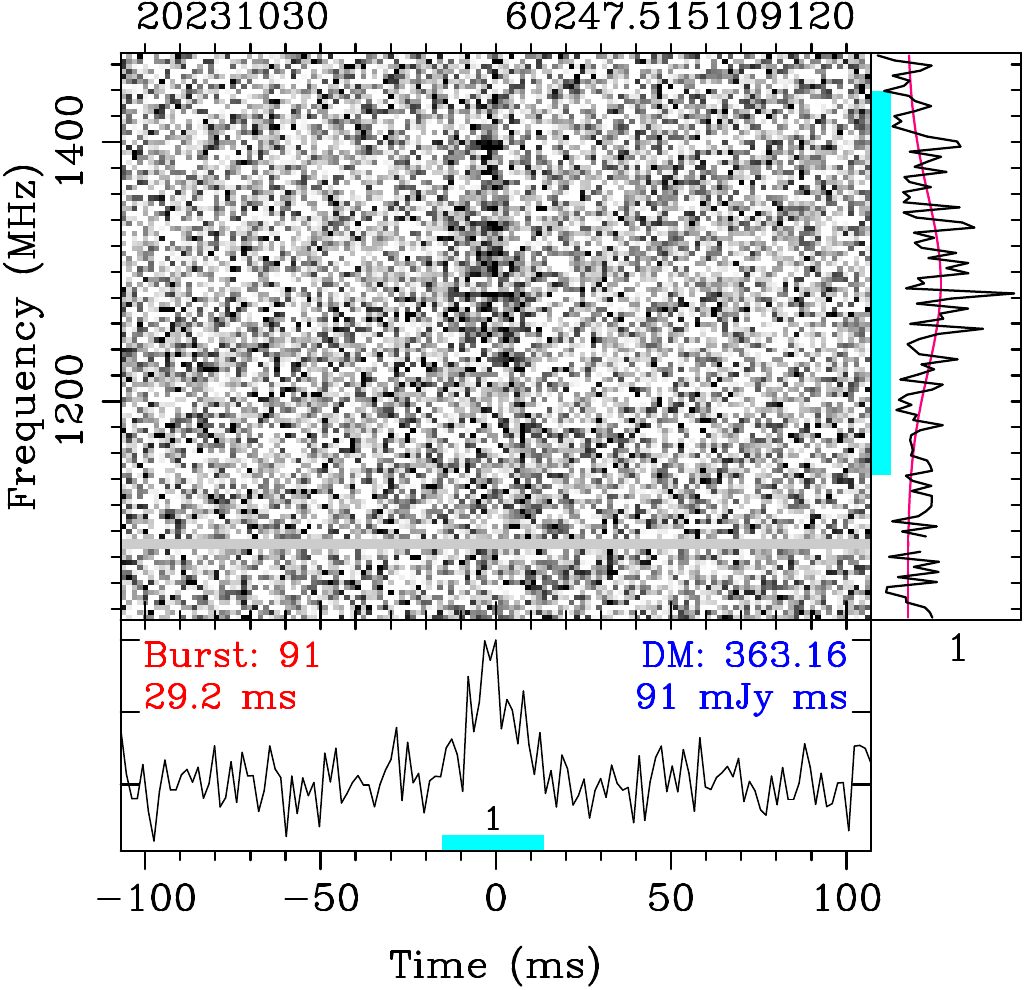}
\includegraphics[height=0.29\linewidth]{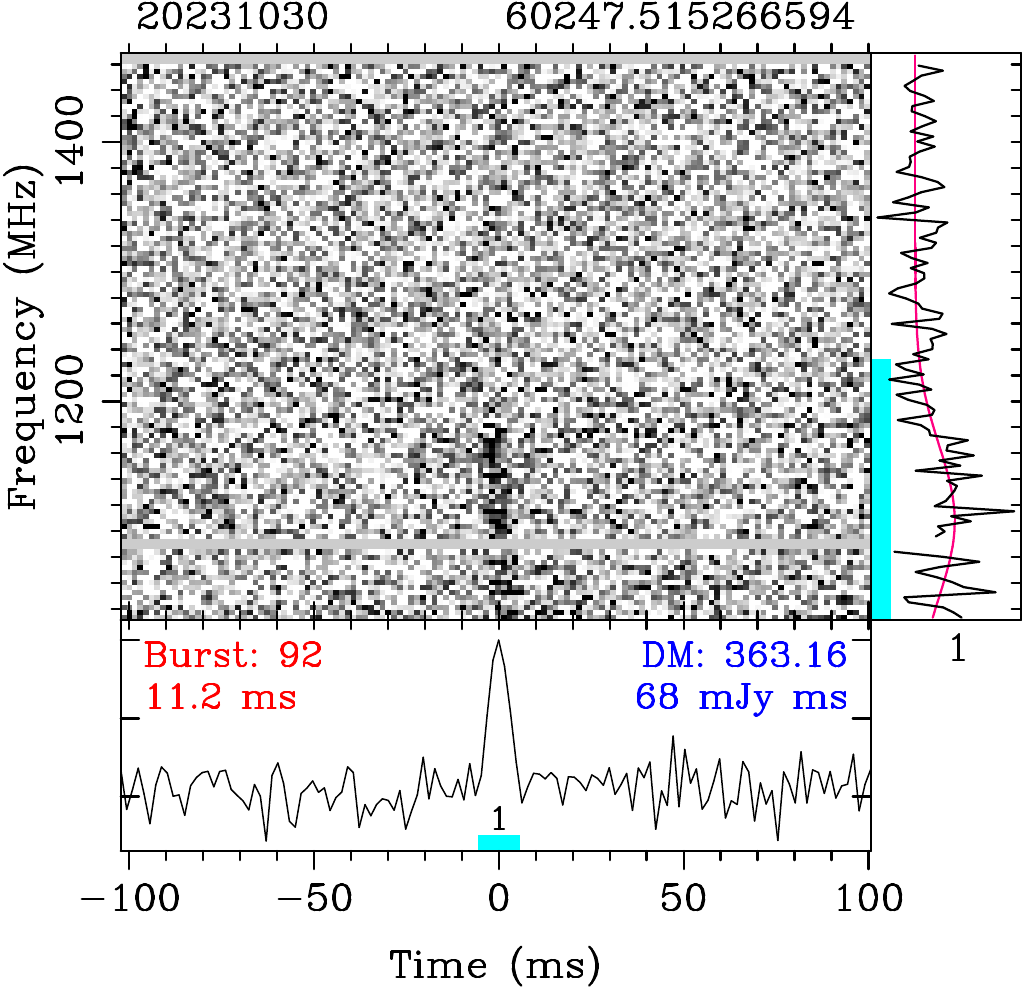}
\includegraphics[height=0.29\linewidth]{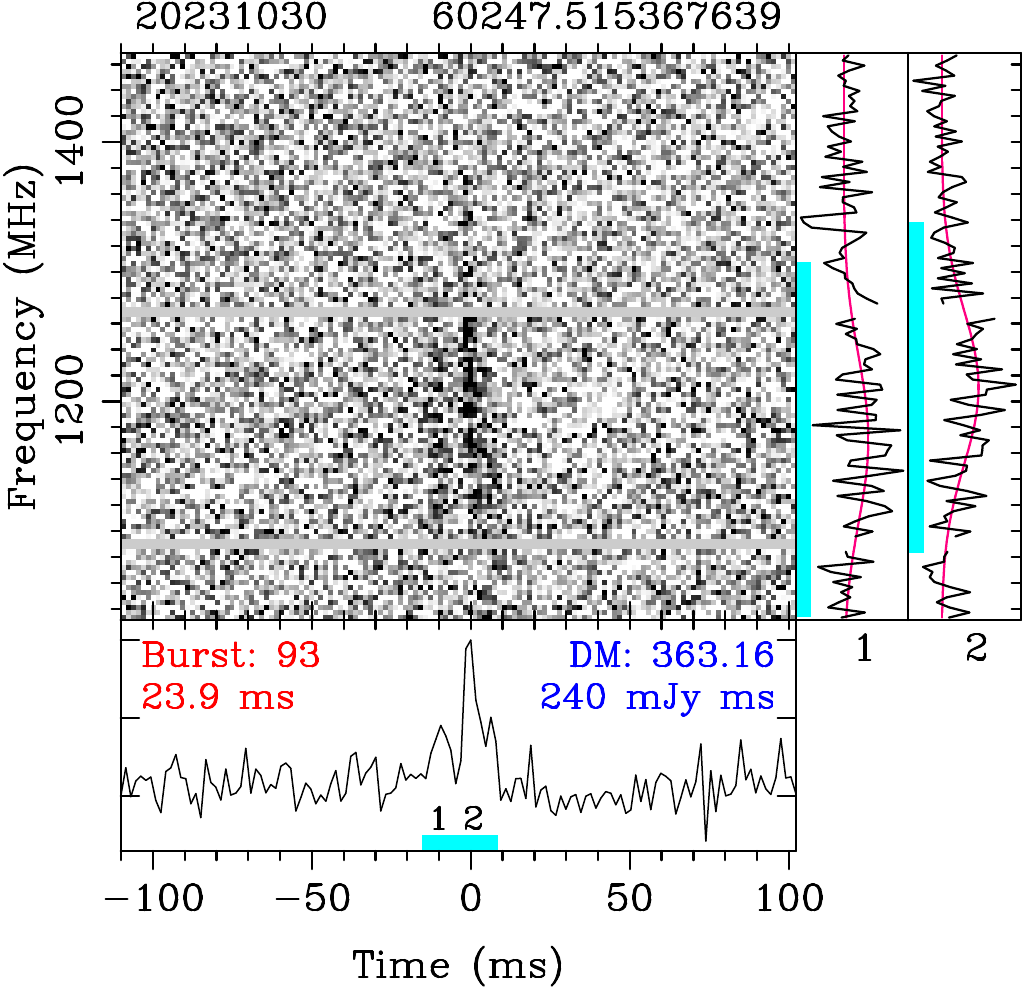}
\includegraphics[height=0.29\linewidth]{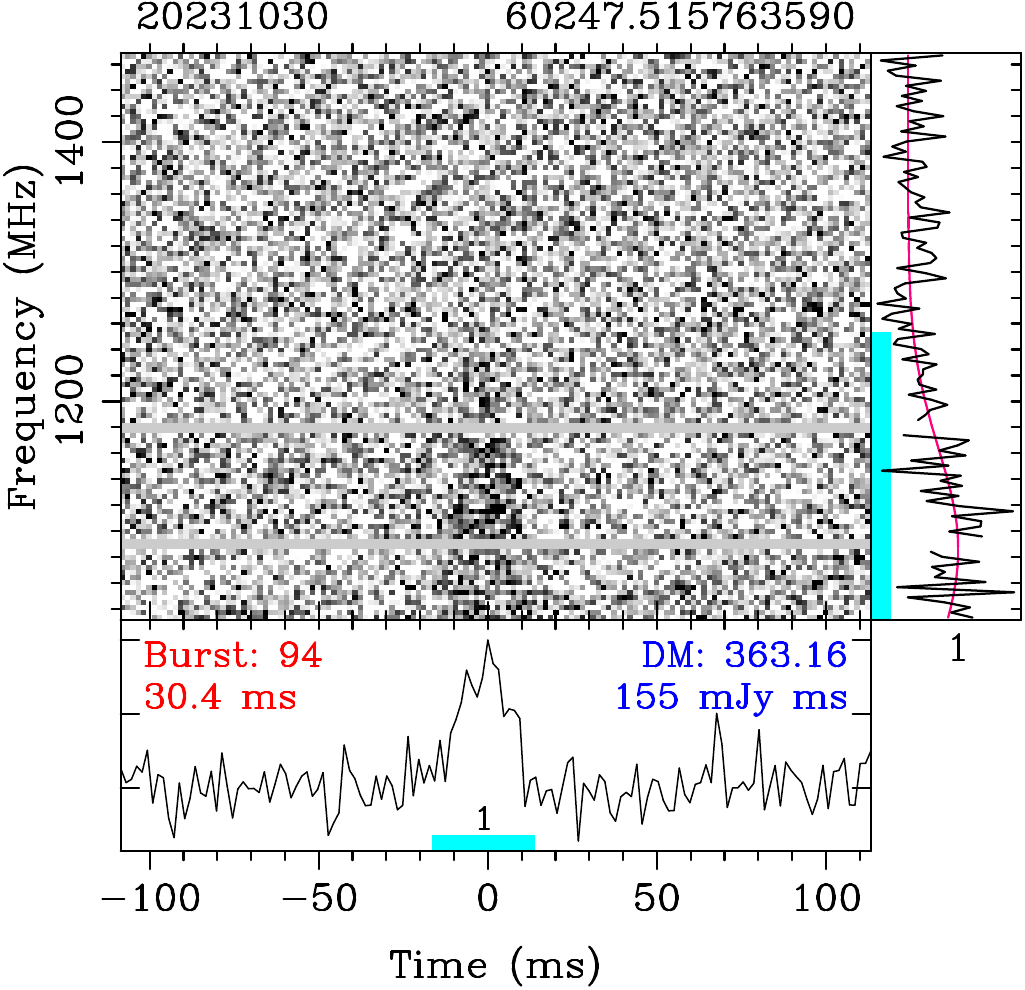}
\includegraphics[height=0.29\linewidth]{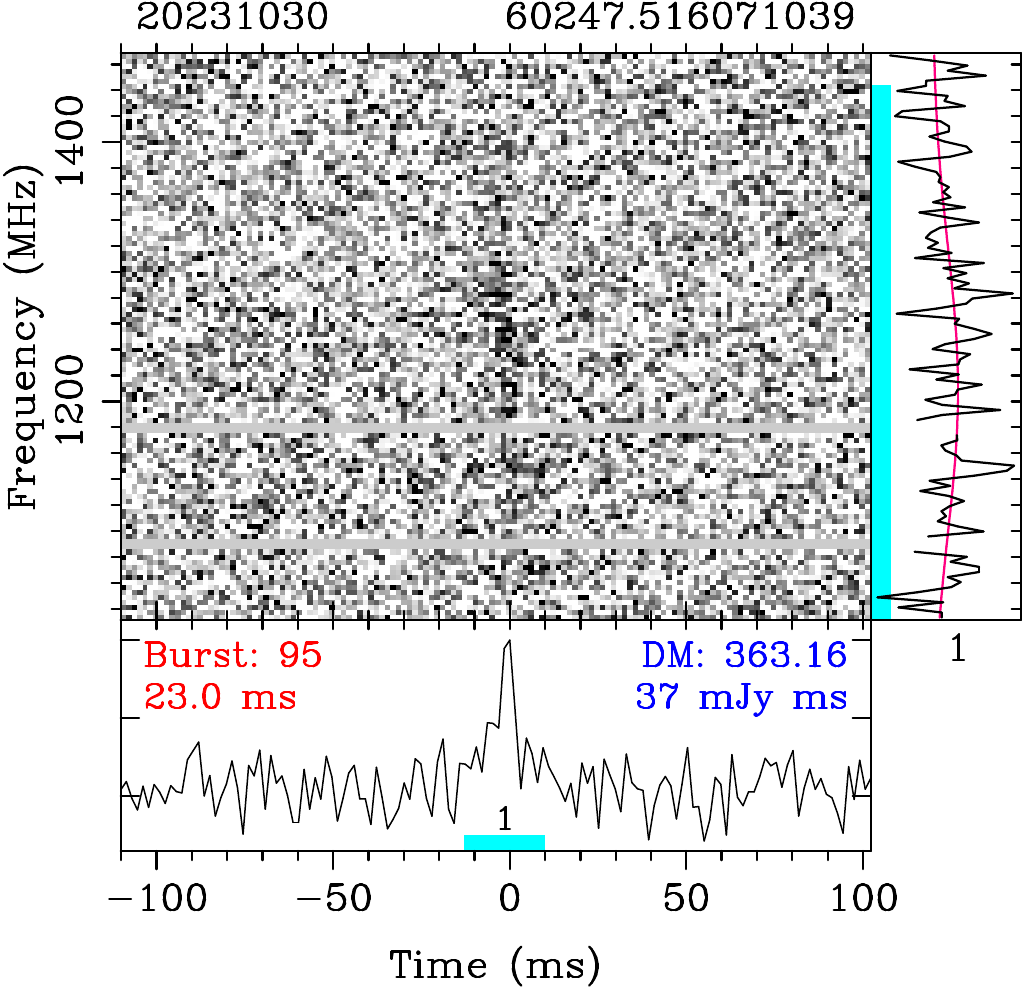}
\includegraphics[height=0.29\linewidth]{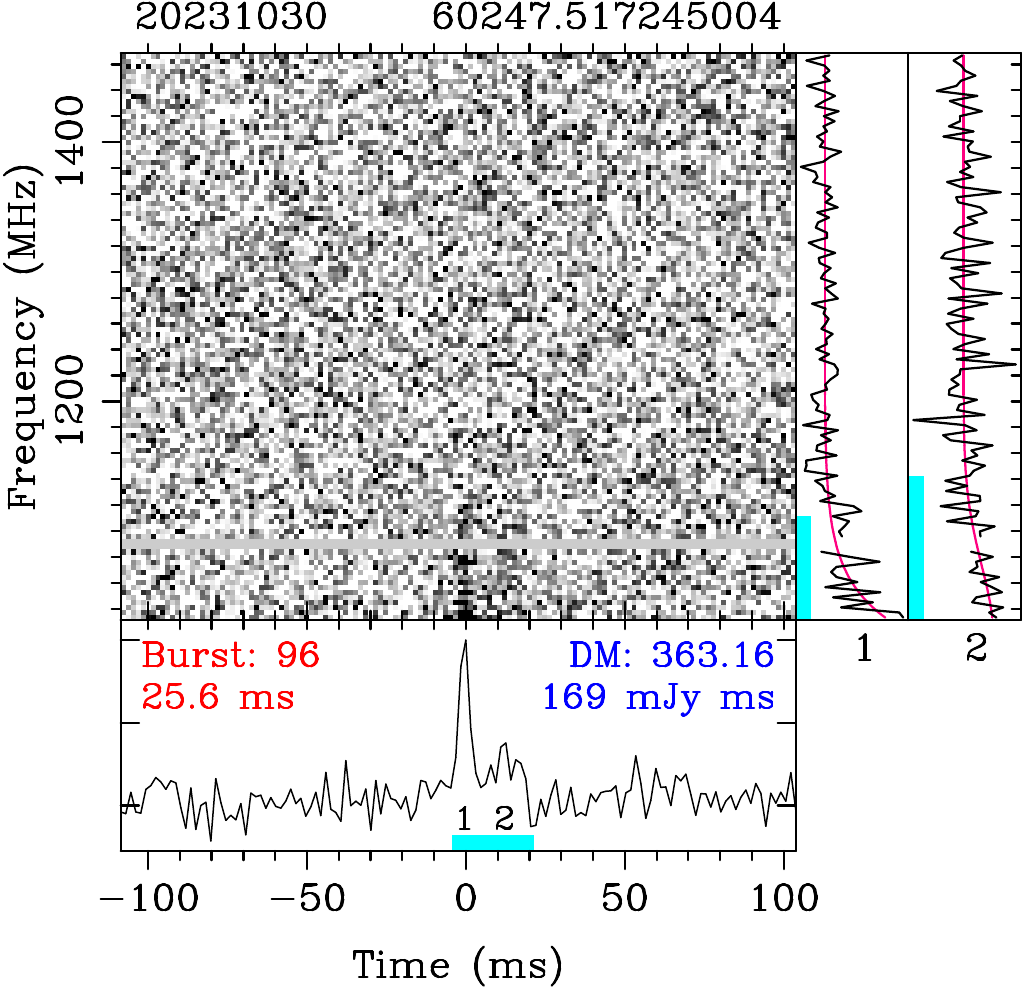}
\includegraphics[height=0.29\linewidth]{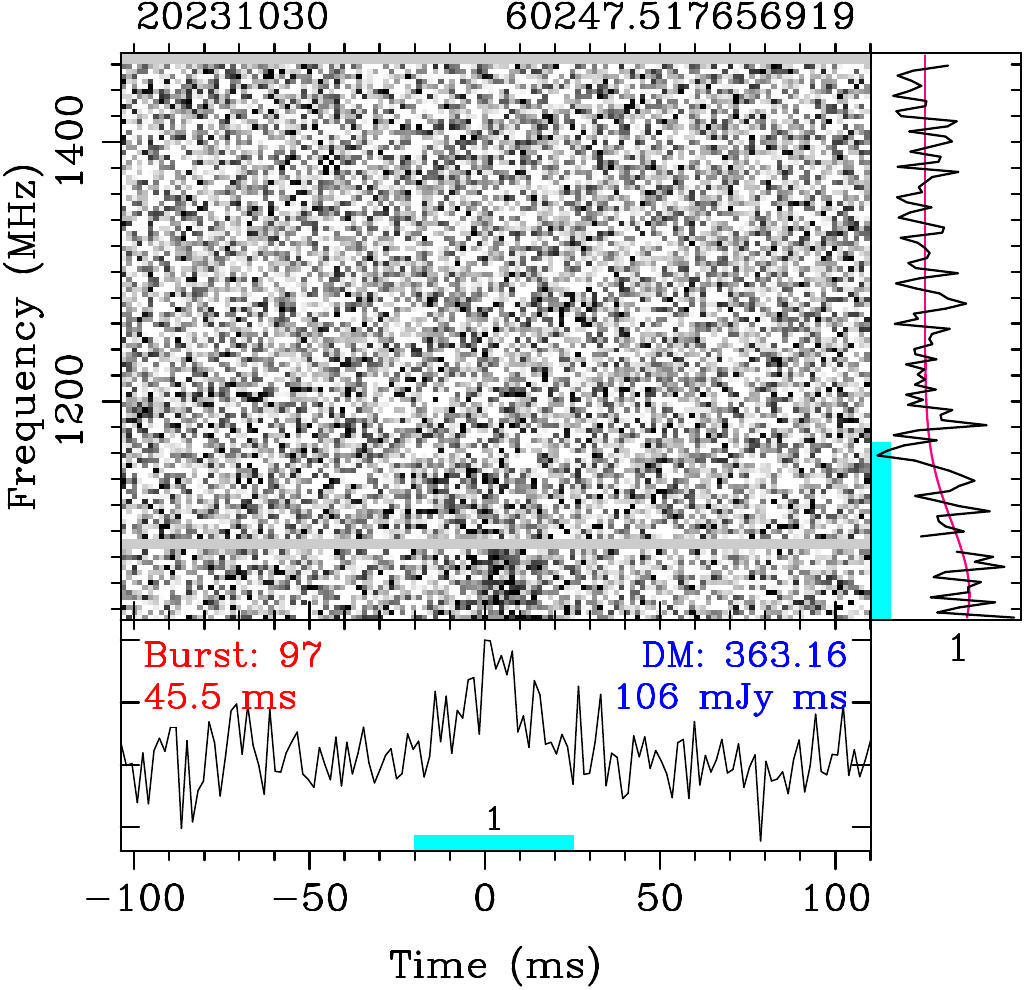}
\includegraphics[height=0.29\linewidth]{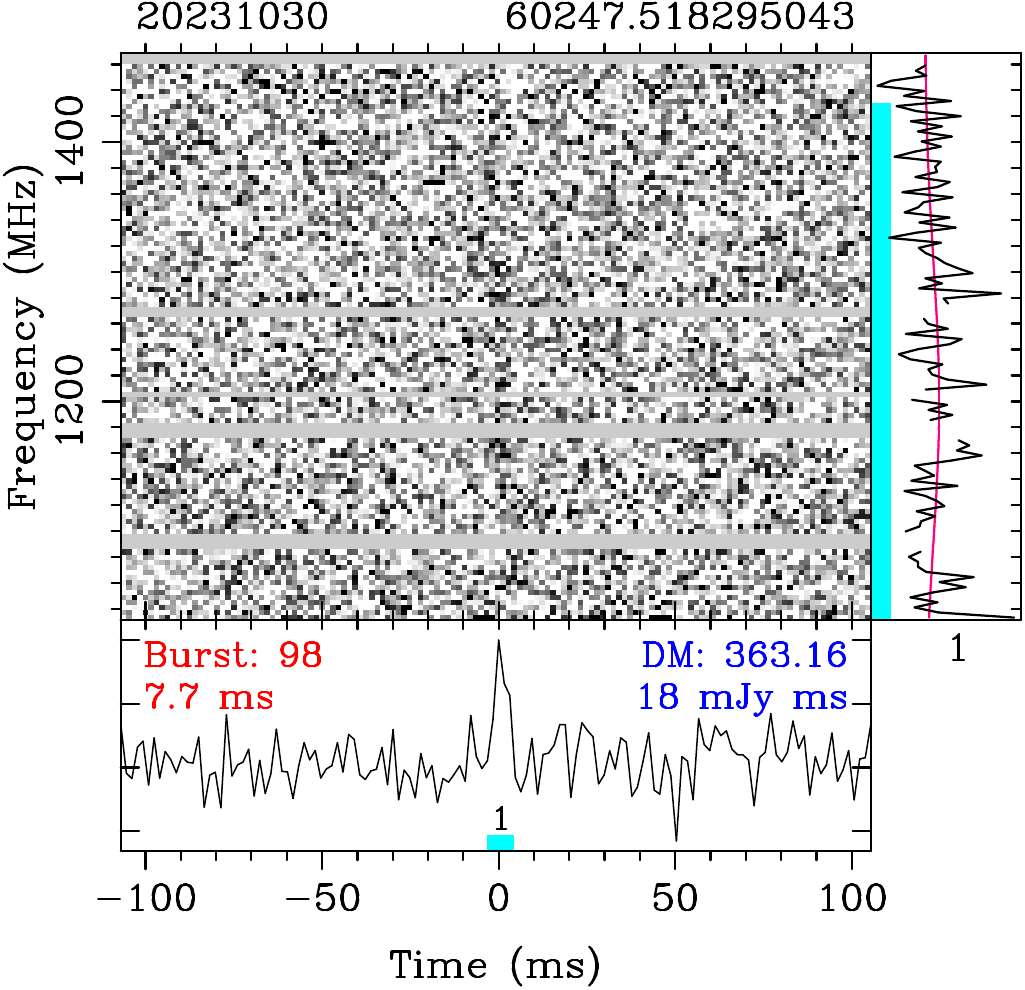}
\caption{({\textit{continued}})}
\end{figure*}
\addtocounter{figure}{-1}
\begin{figure*}
\flushleft
\includegraphics[height=0.29\linewidth]{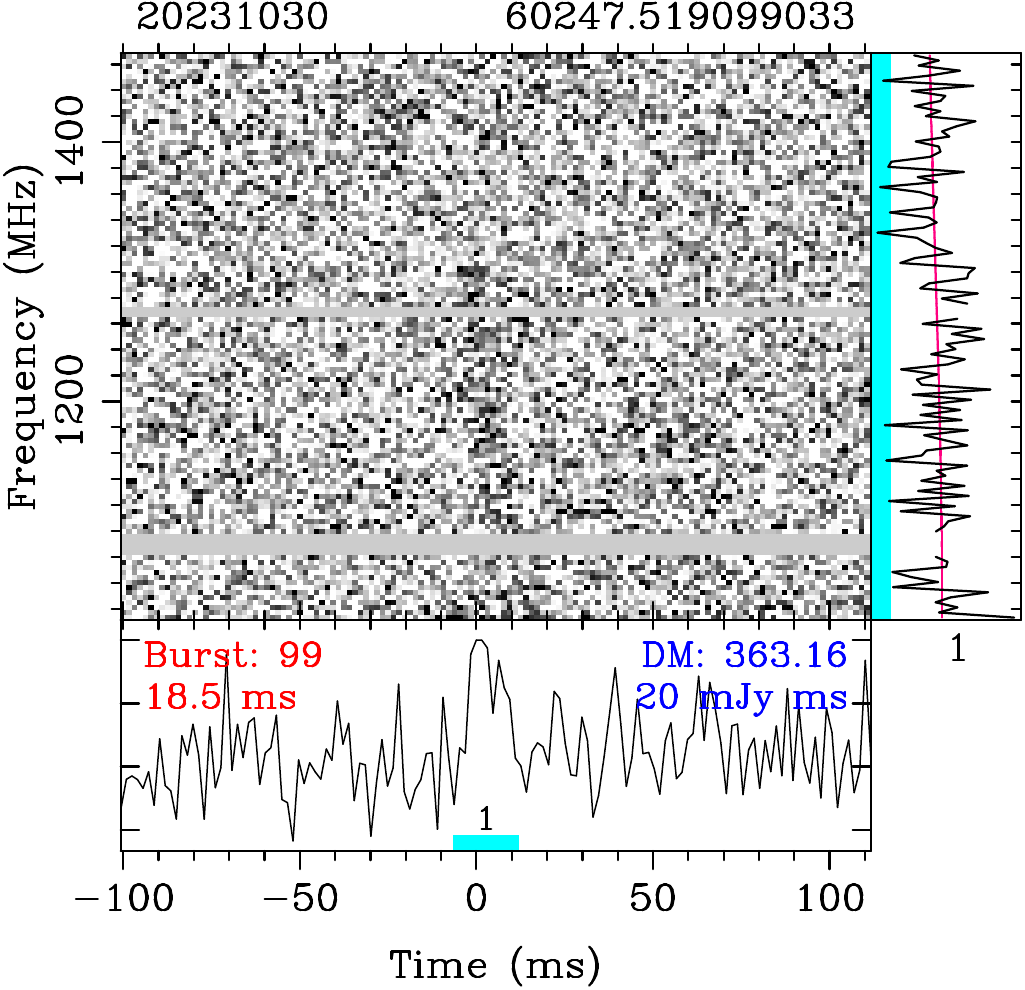}
\includegraphics[height=0.29\linewidth]{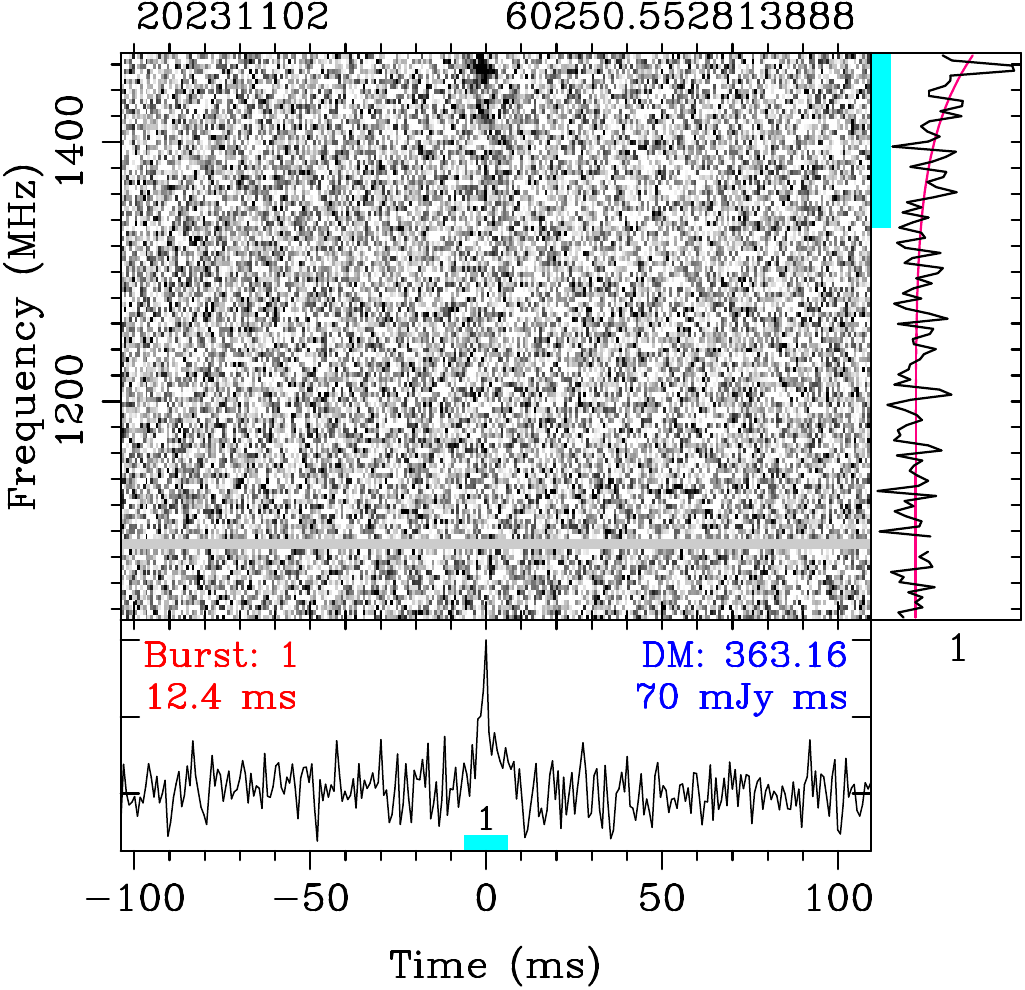}
\includegraphics[height=0.29\linewidth]{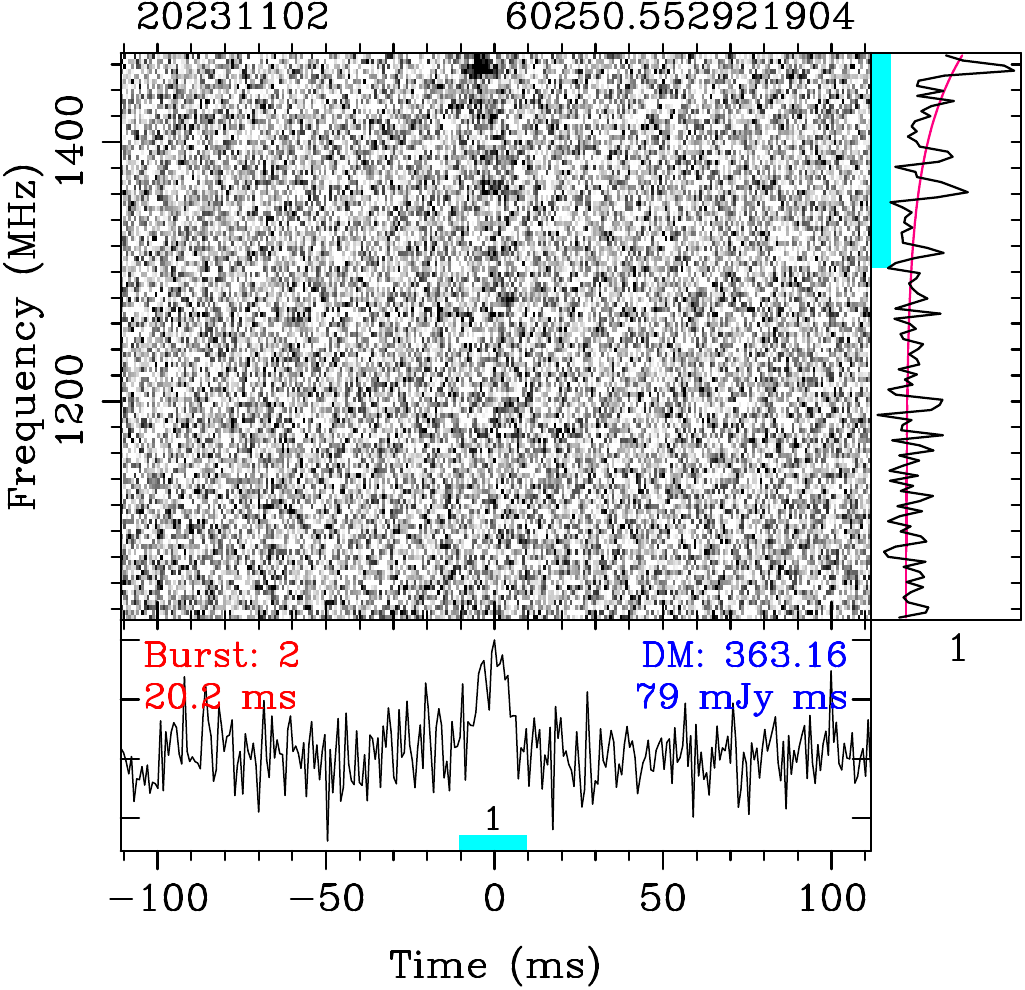}
\includegraphics[height=0.29\linewidth]{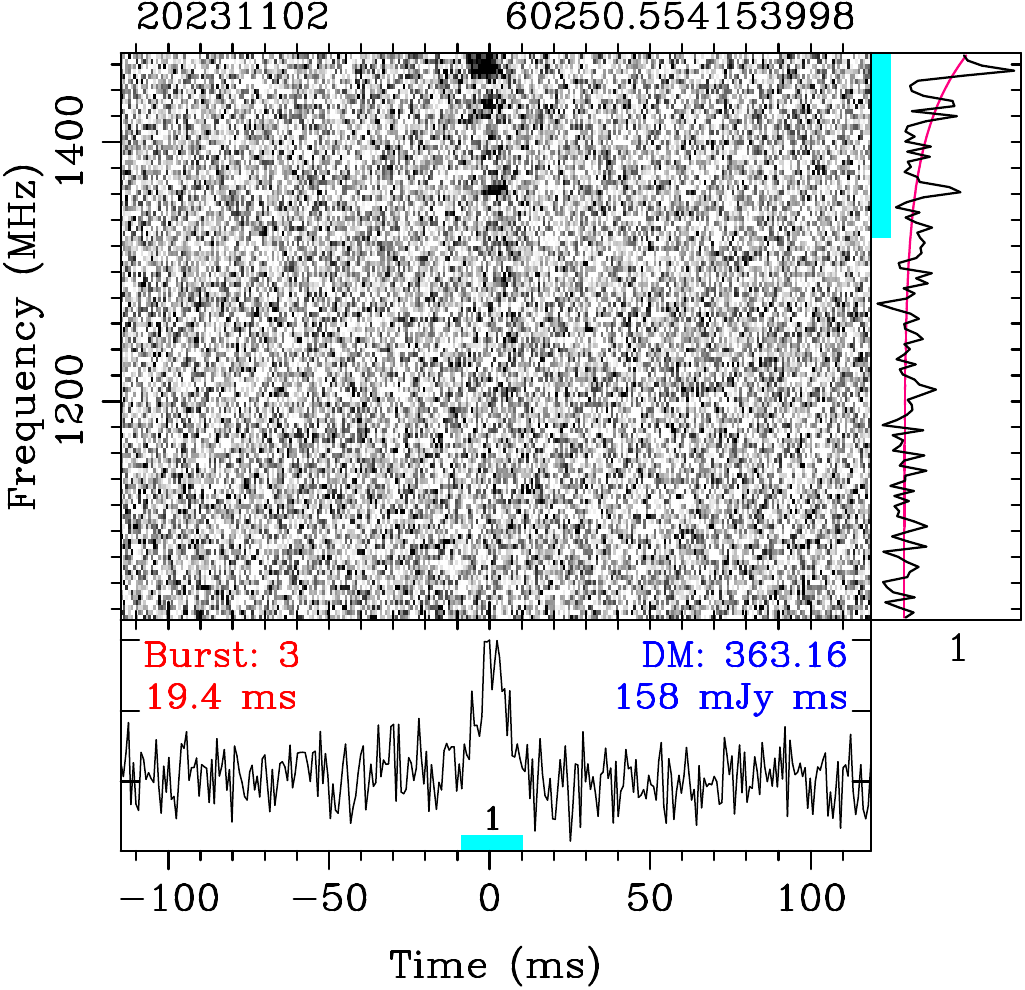}
\includegraphics[height=0.29\linewidth]{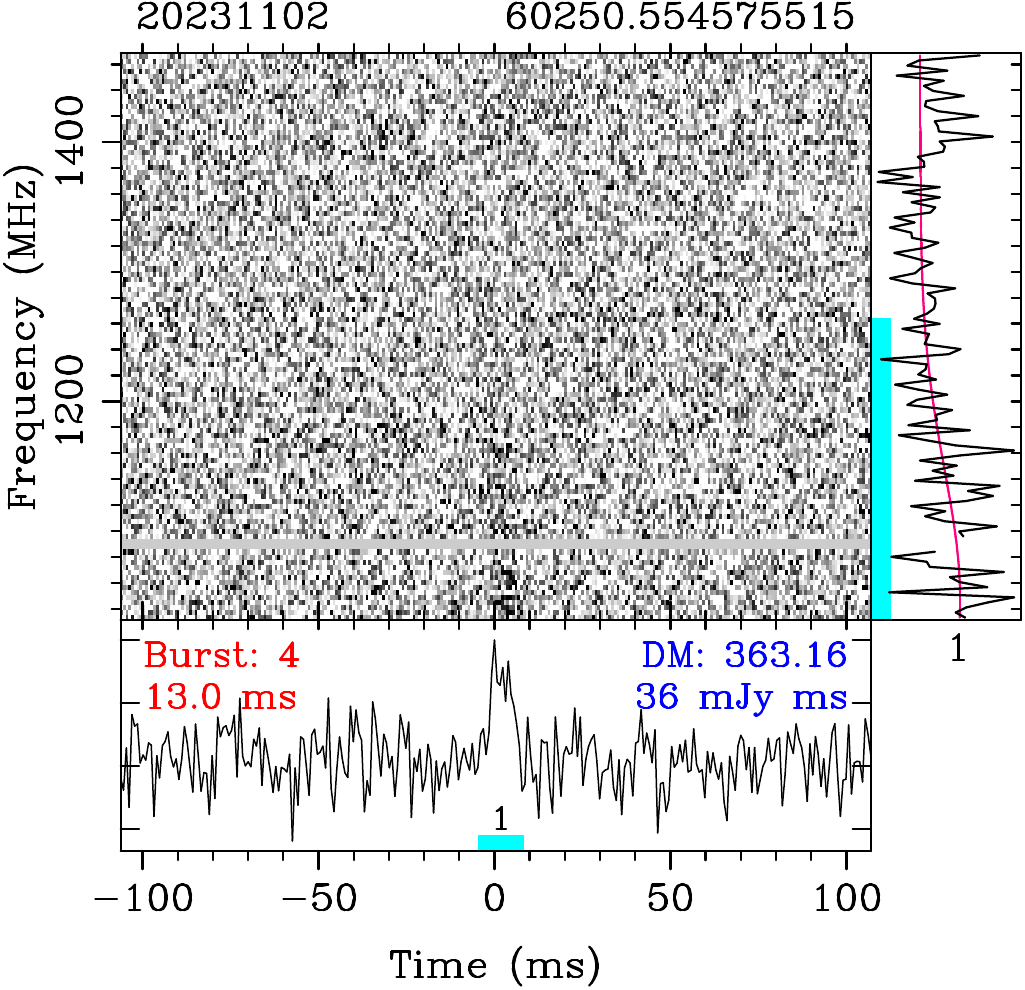}
\includegraphics[height=0.29\linewidth]{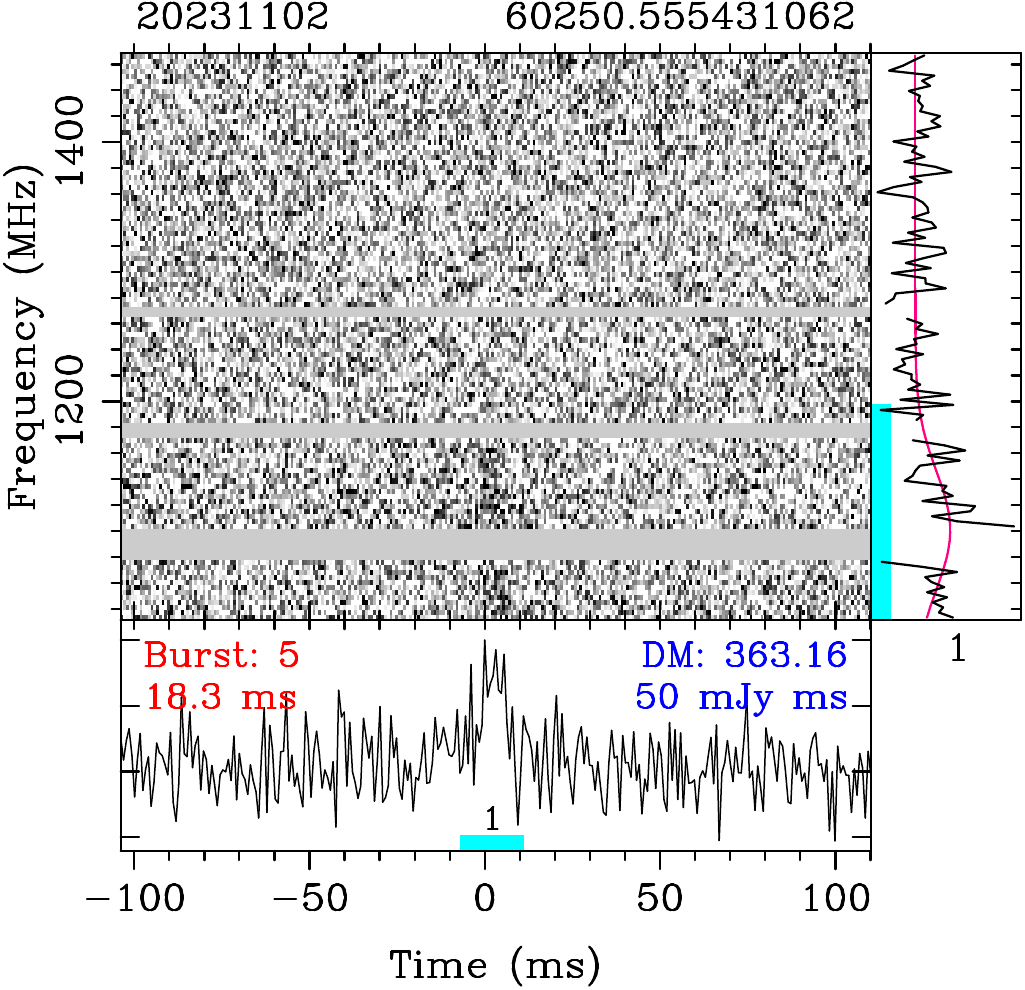}
\includegraphics[height=0.29\linewidth]{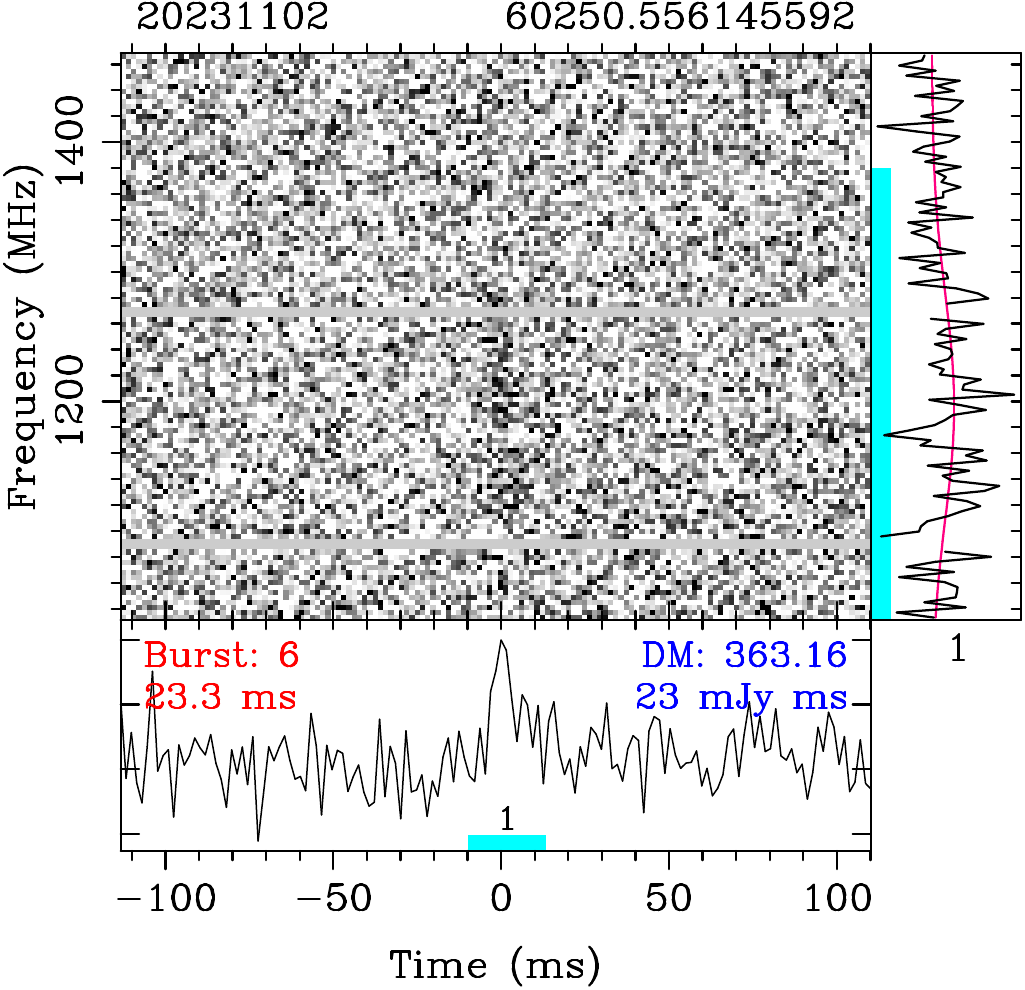}
\includegraphics[height=0.29\linewidth]{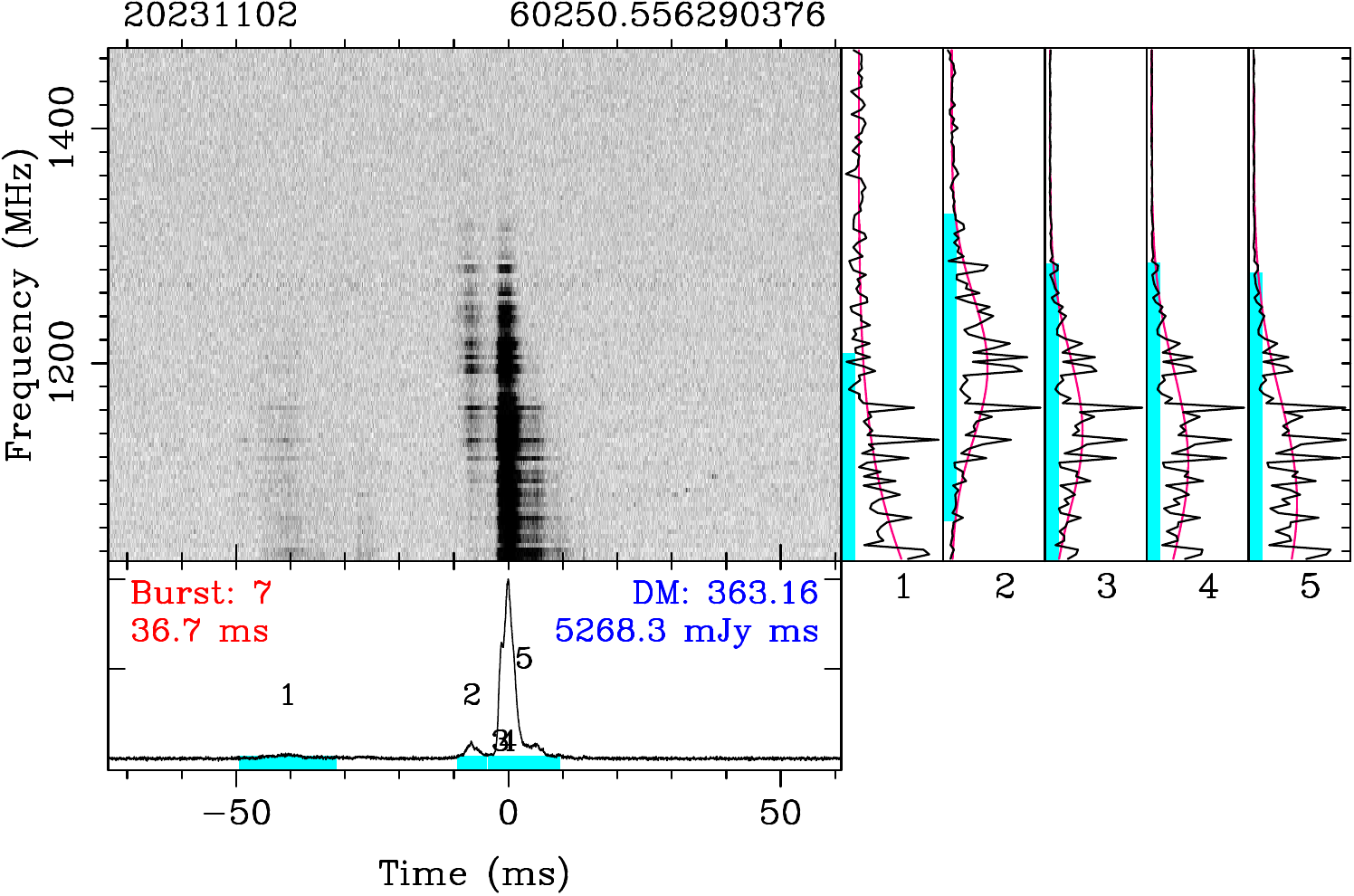}
\includegraphics[height=0.29\linewidth]{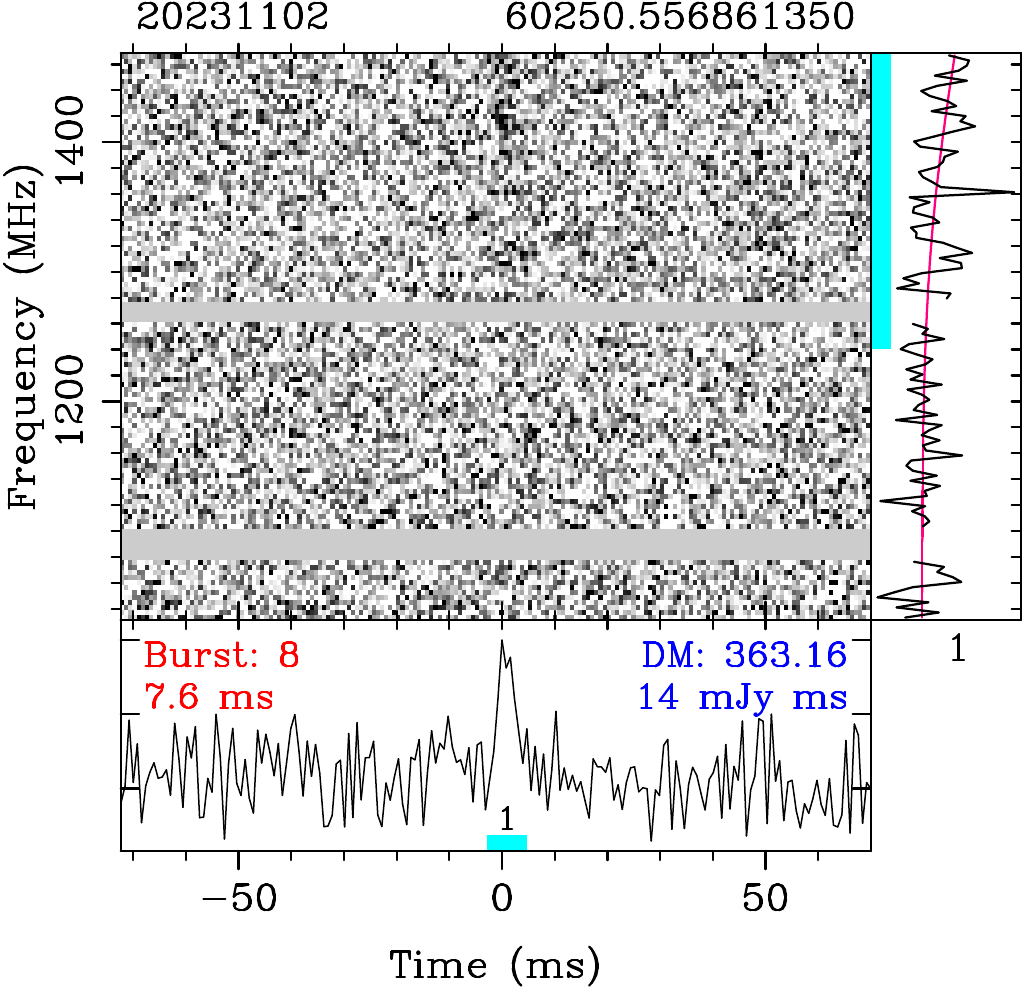}
\includegraphics[height=0.29\linewidth]{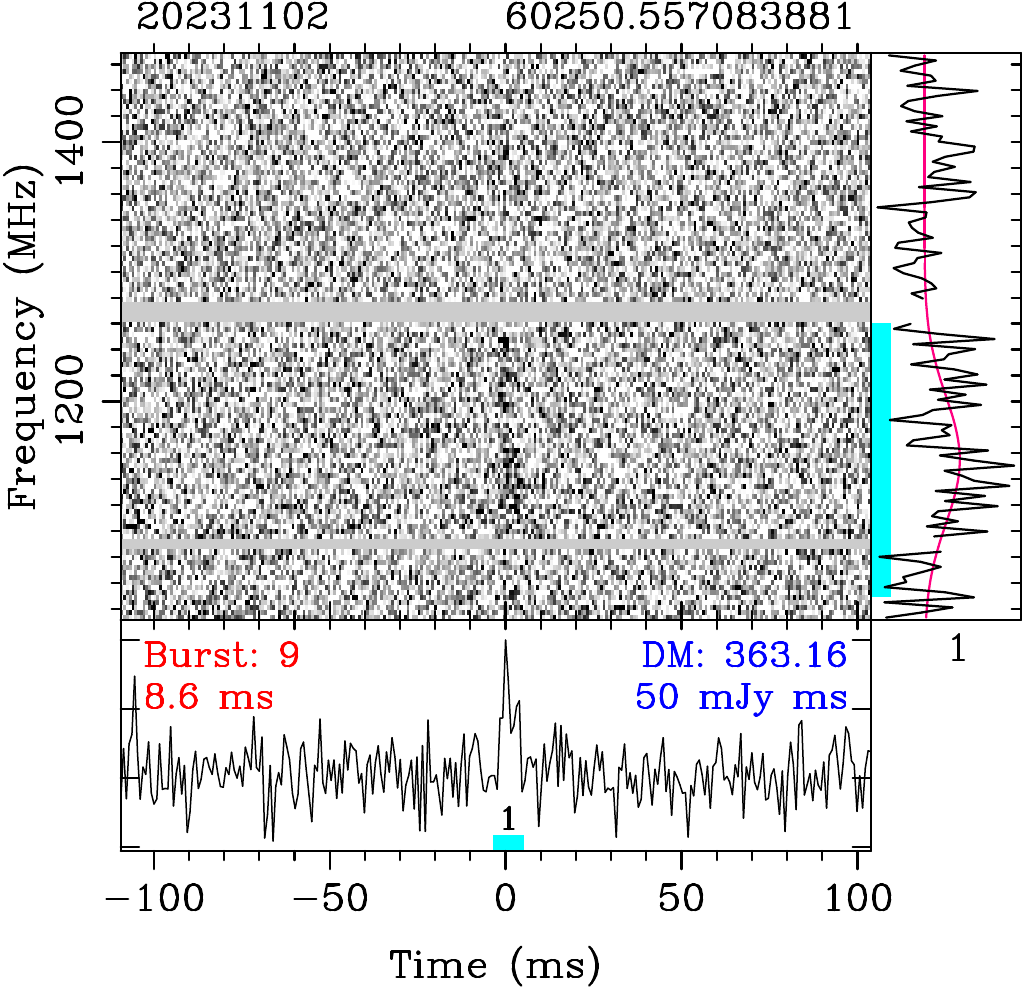}
\includegraphics[height=0.29\linewidth]{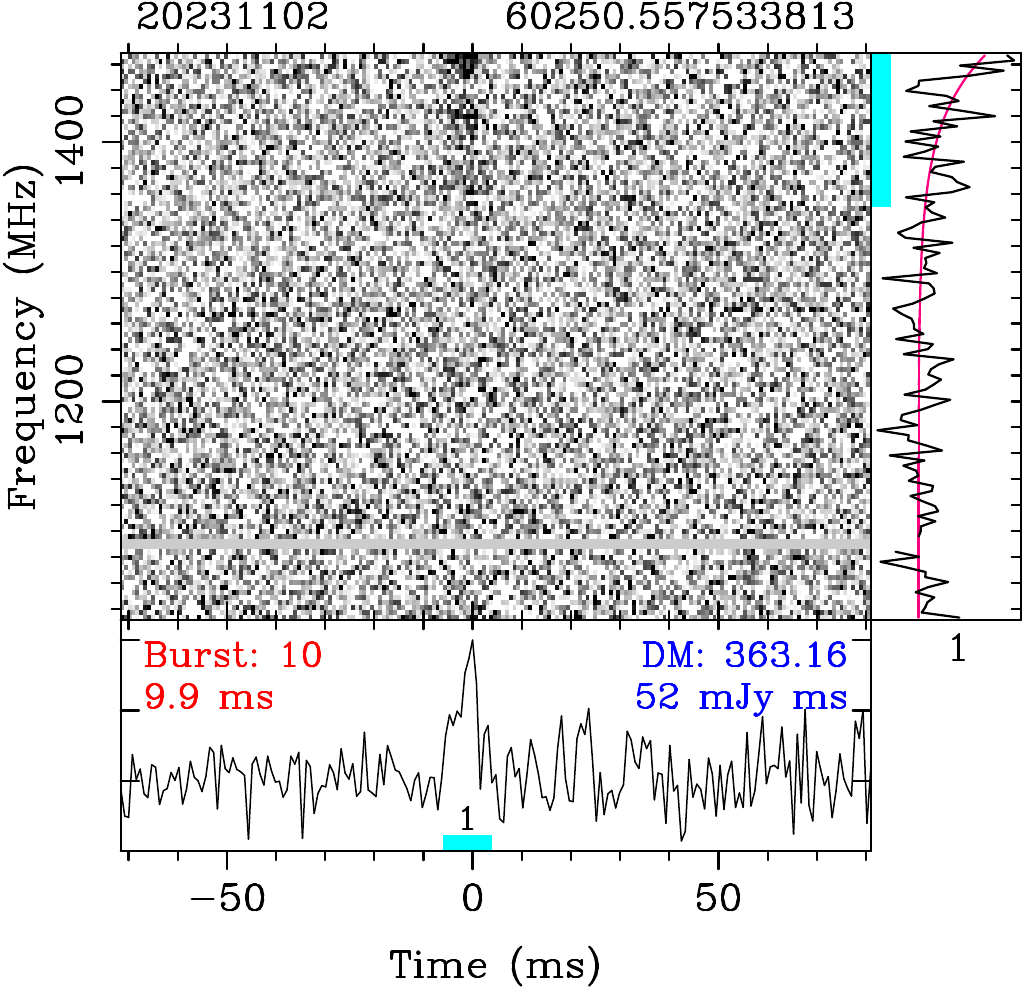}
\caption{({\textit{continued}})}
\end{figure*}
\addtocounter{figure}{-1}
\begin{figure*}
\flushleft
\includegraphics[height=0.29\linewidth]{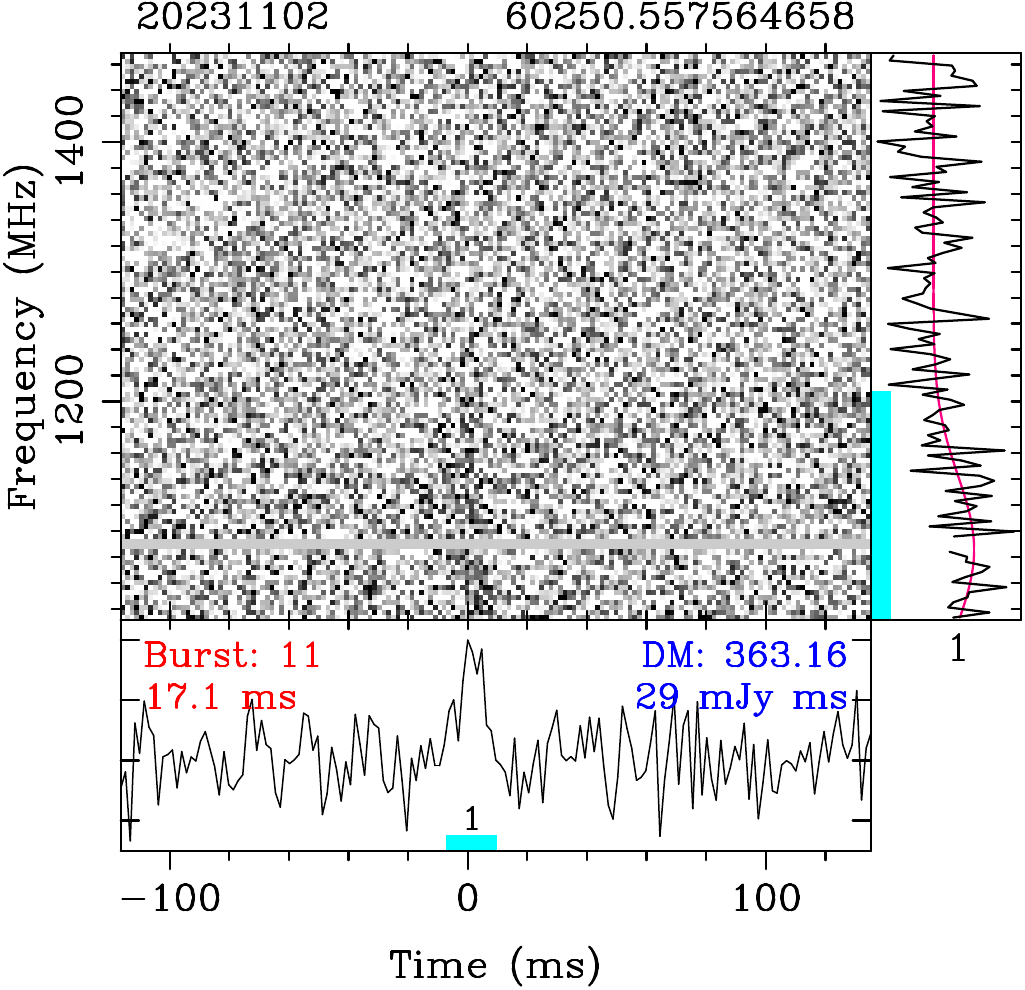}
\includegraphics[height=0.29\linewidth]{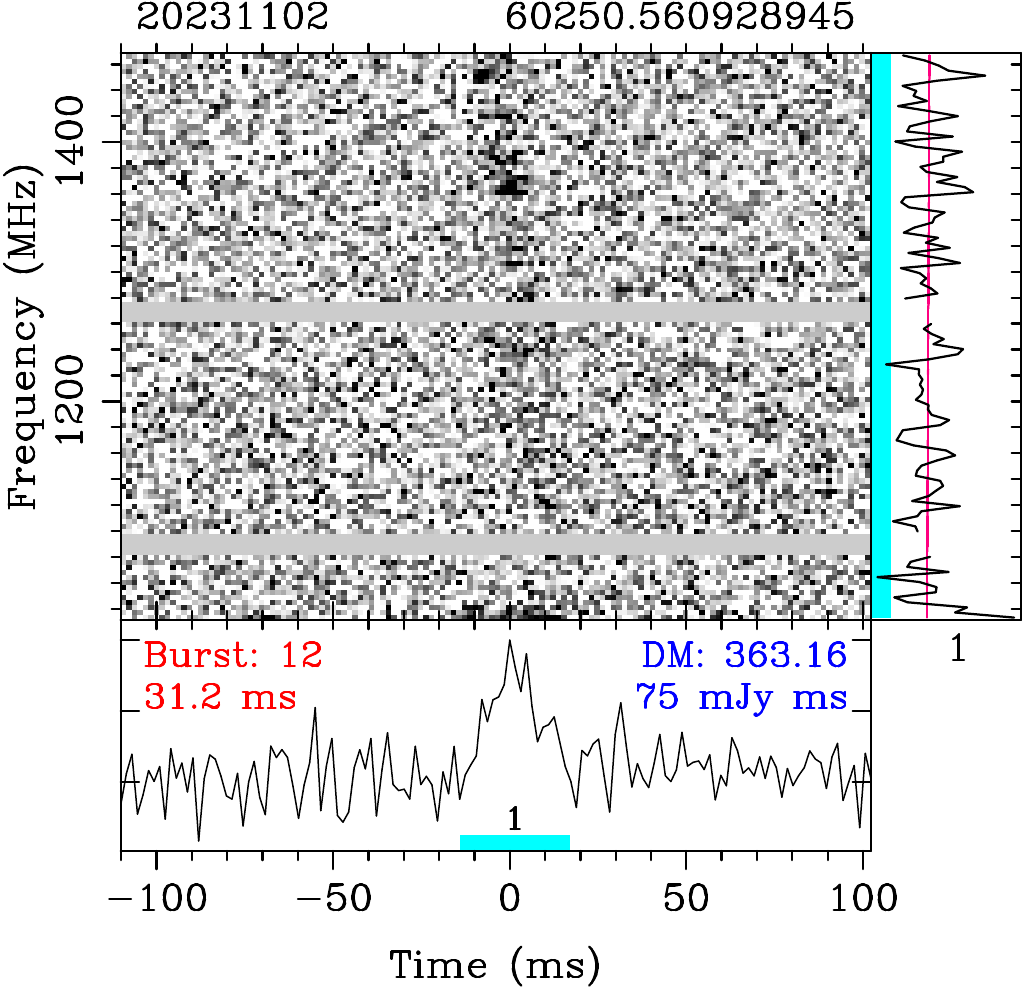}
\includegraphics[height=0.29\linewidth]{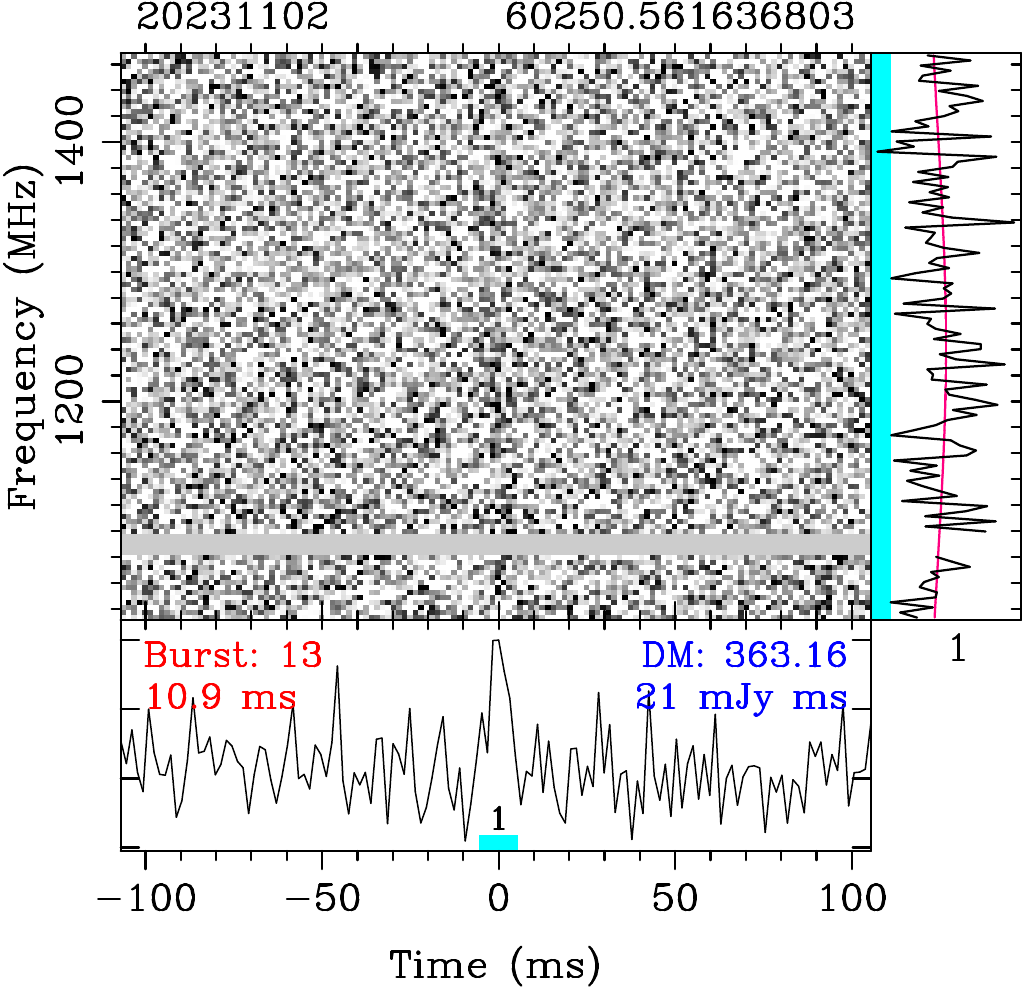}
\includegraphics[height=0.29\linewidth]{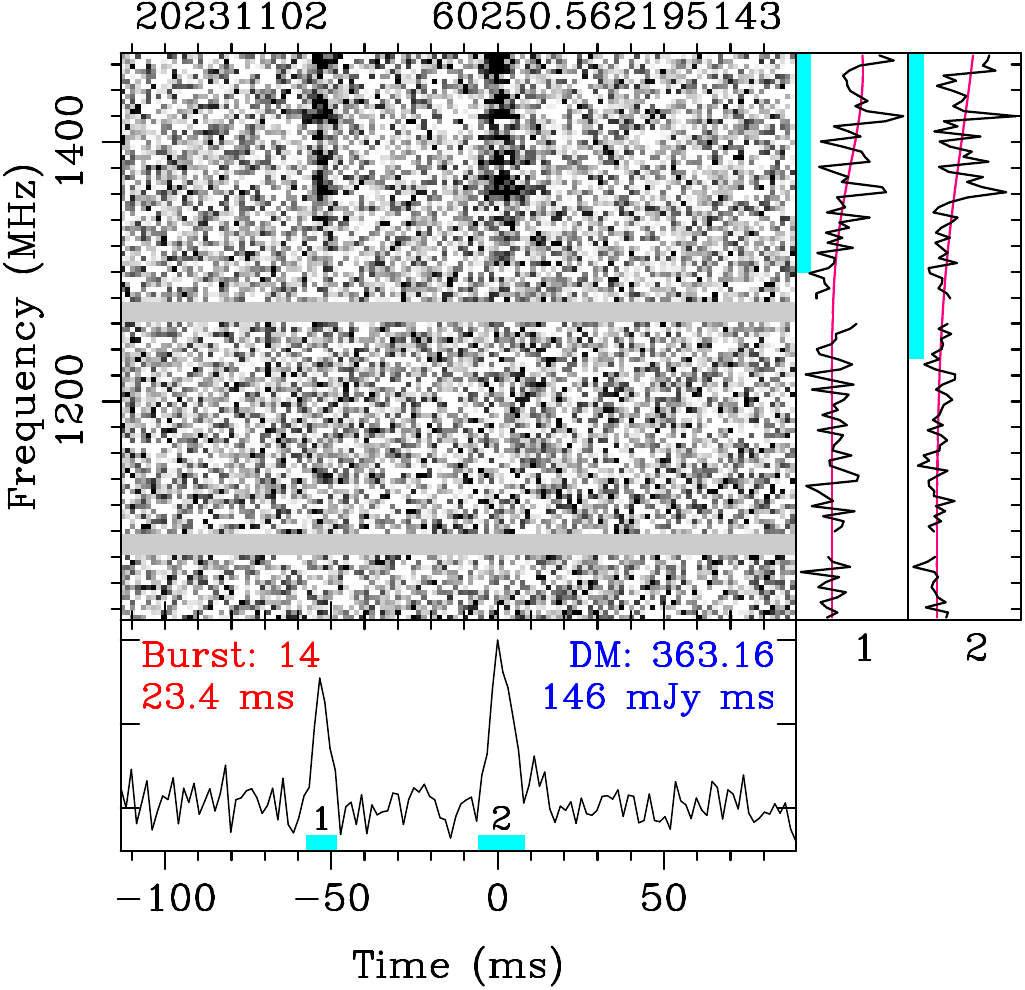}
\includegraphics[height=0.29\linewidth]{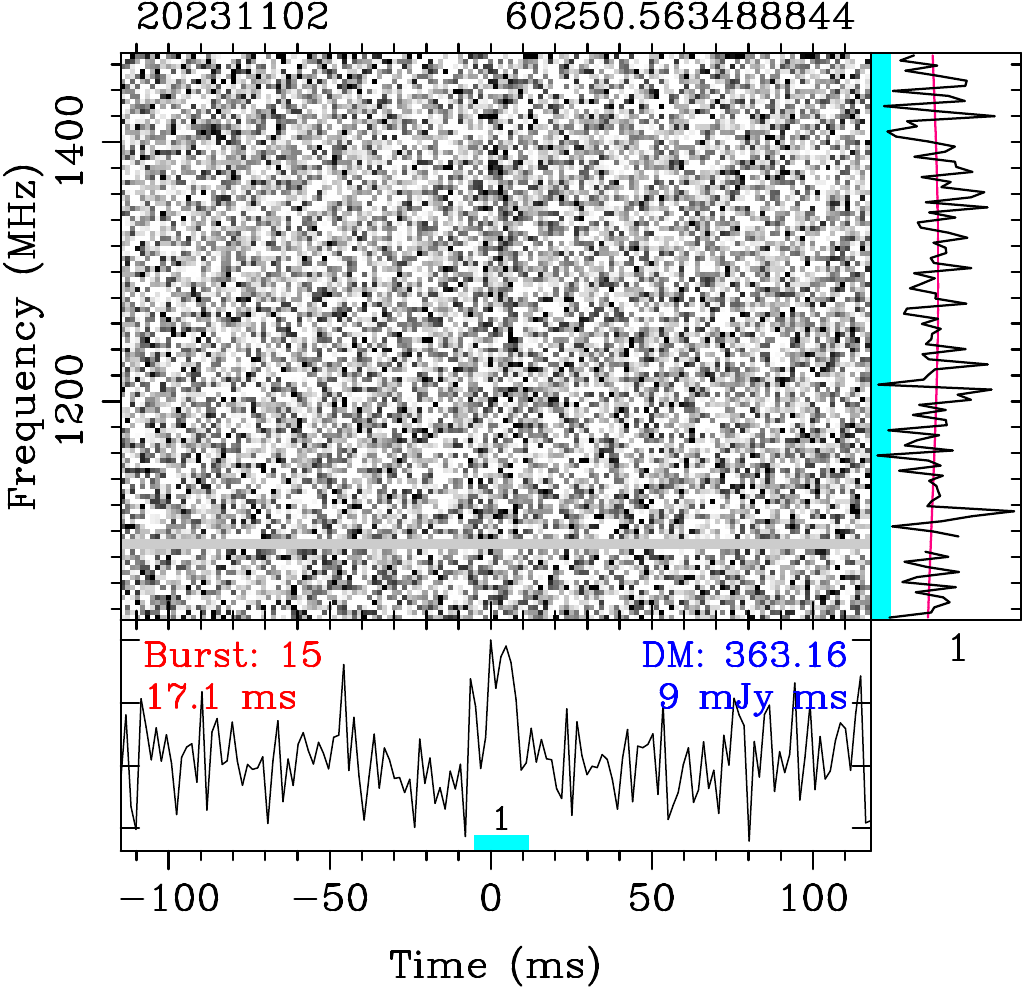}
\includegraphics[height=0.29\linewidth]{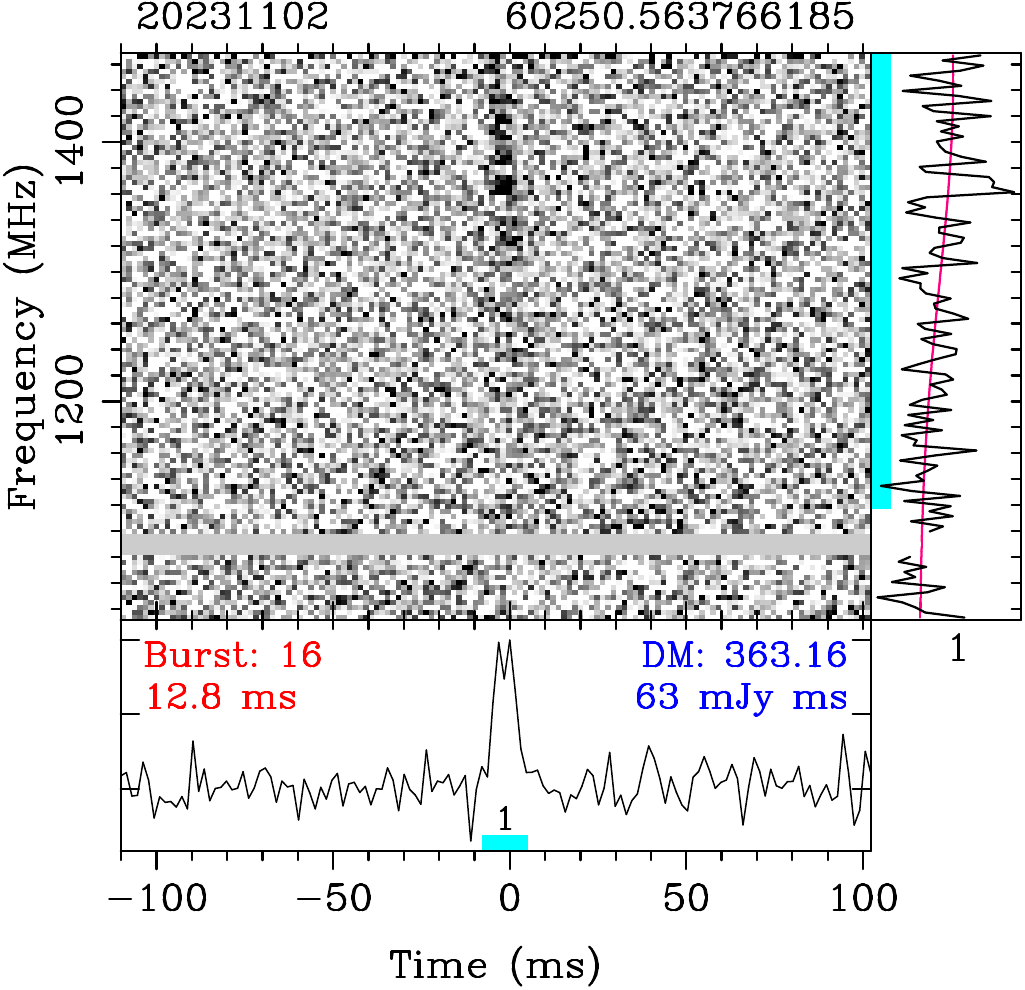}
\includegraphics[height=0.29\linewidth]{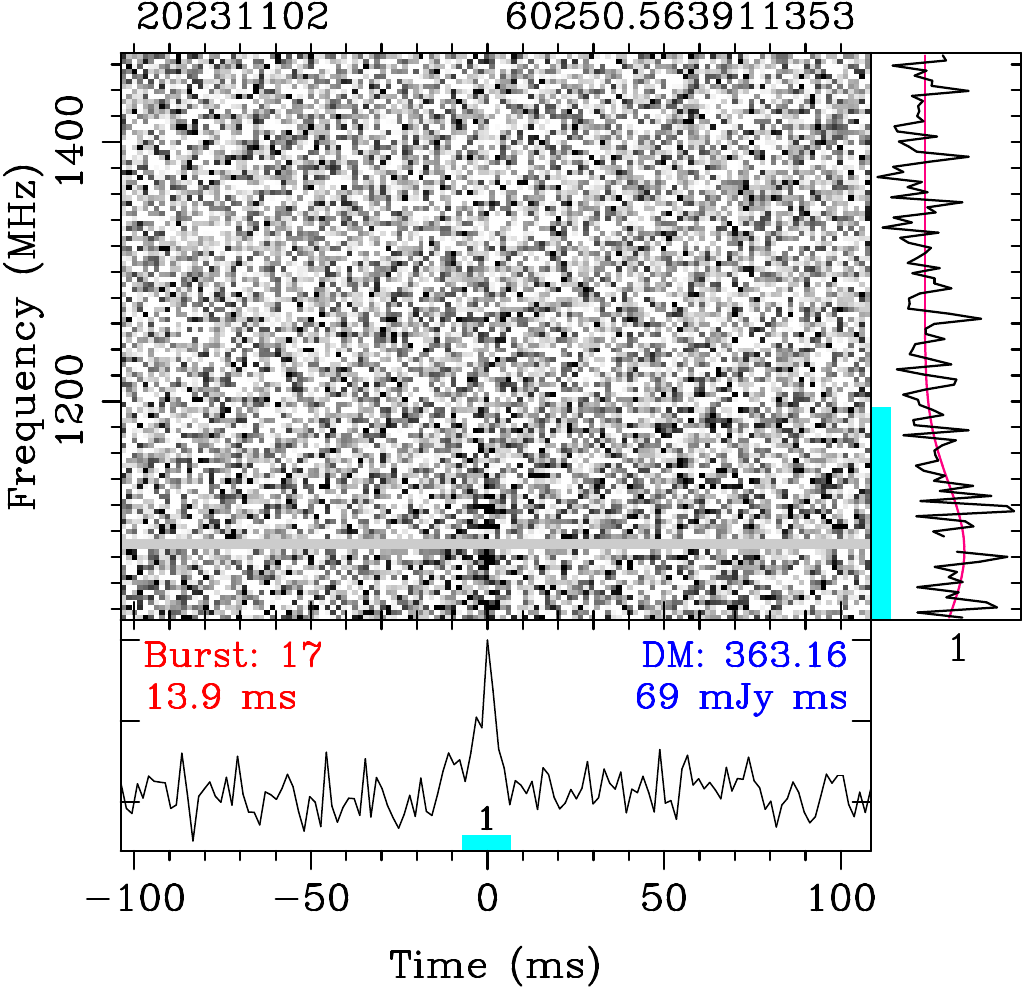}
\includegraphics[height=0.29\linewidth]{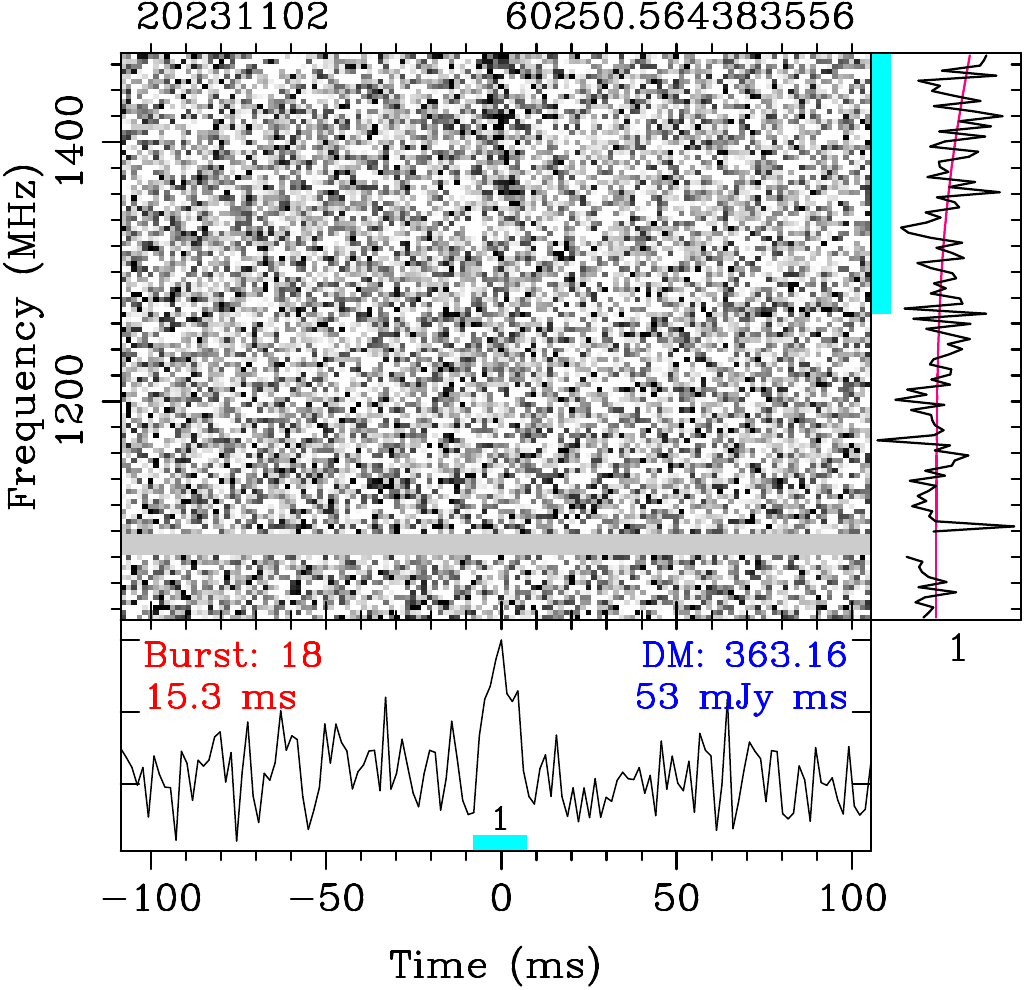}
\includegraphics[height=0.29\linewidth]{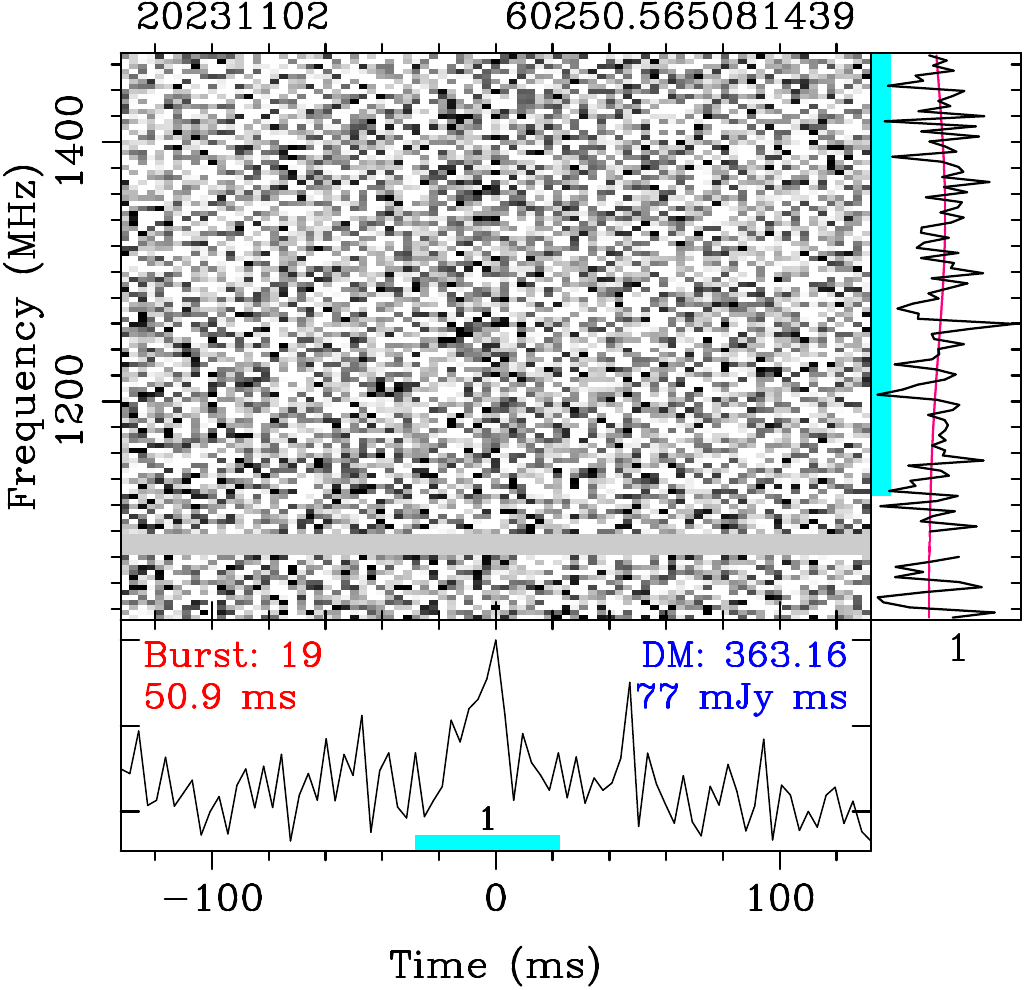}
\includegraphics[height=0.29\linewidth]{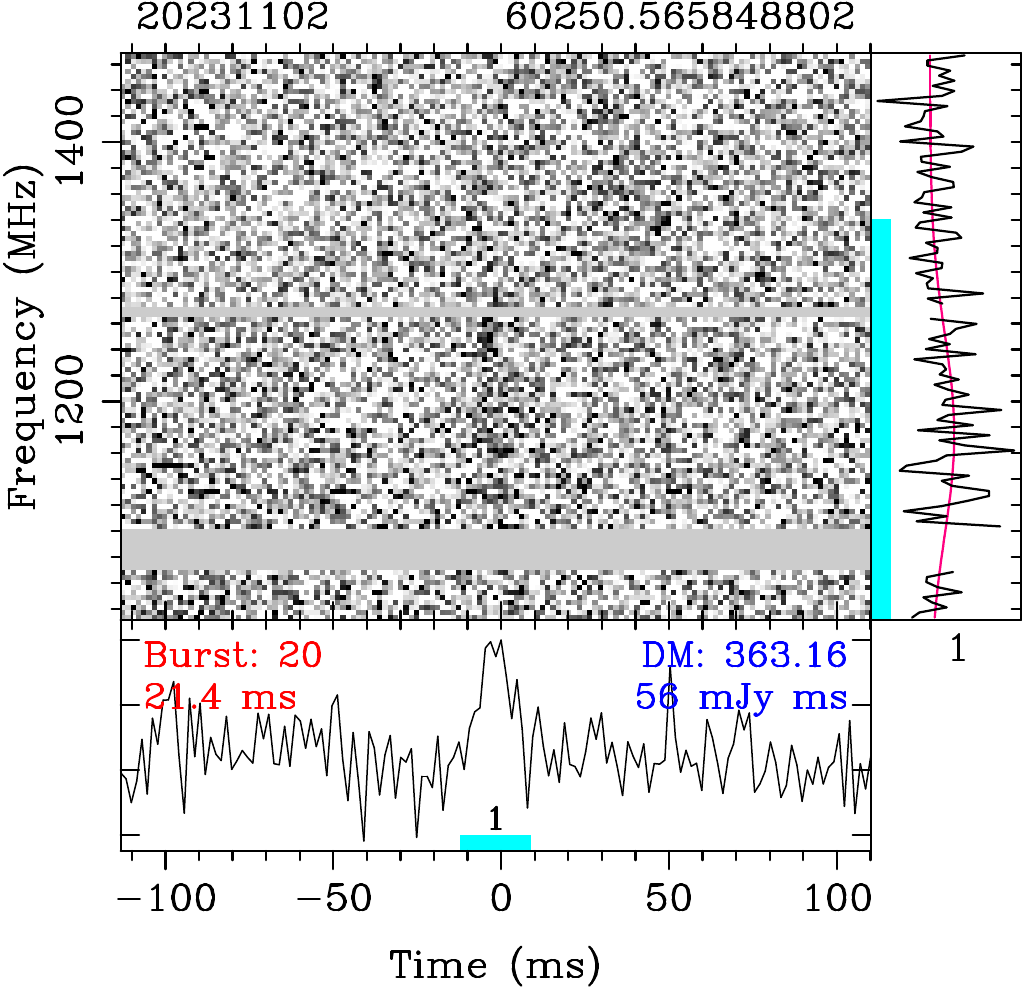}
\includegraphics[height=0.29\linewidth]{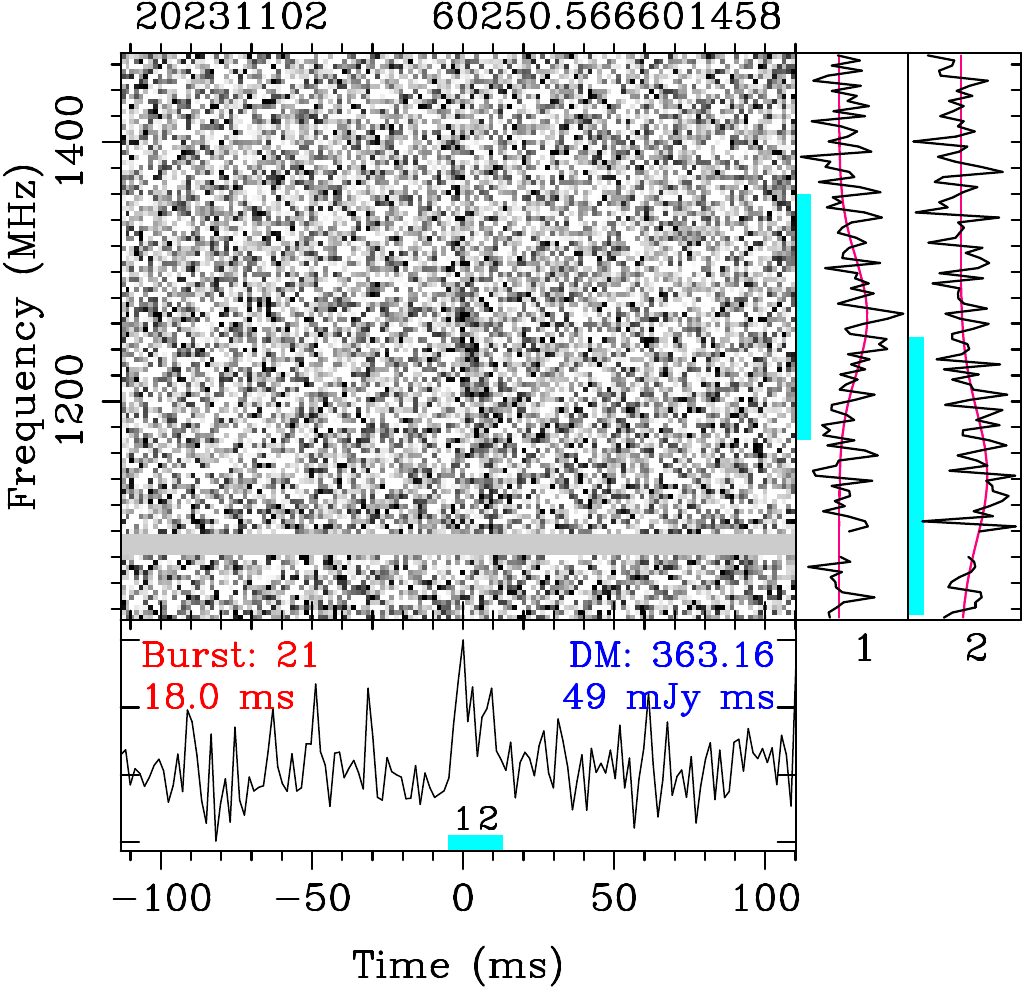}
\includegraphics[height=0.29\linewidth]{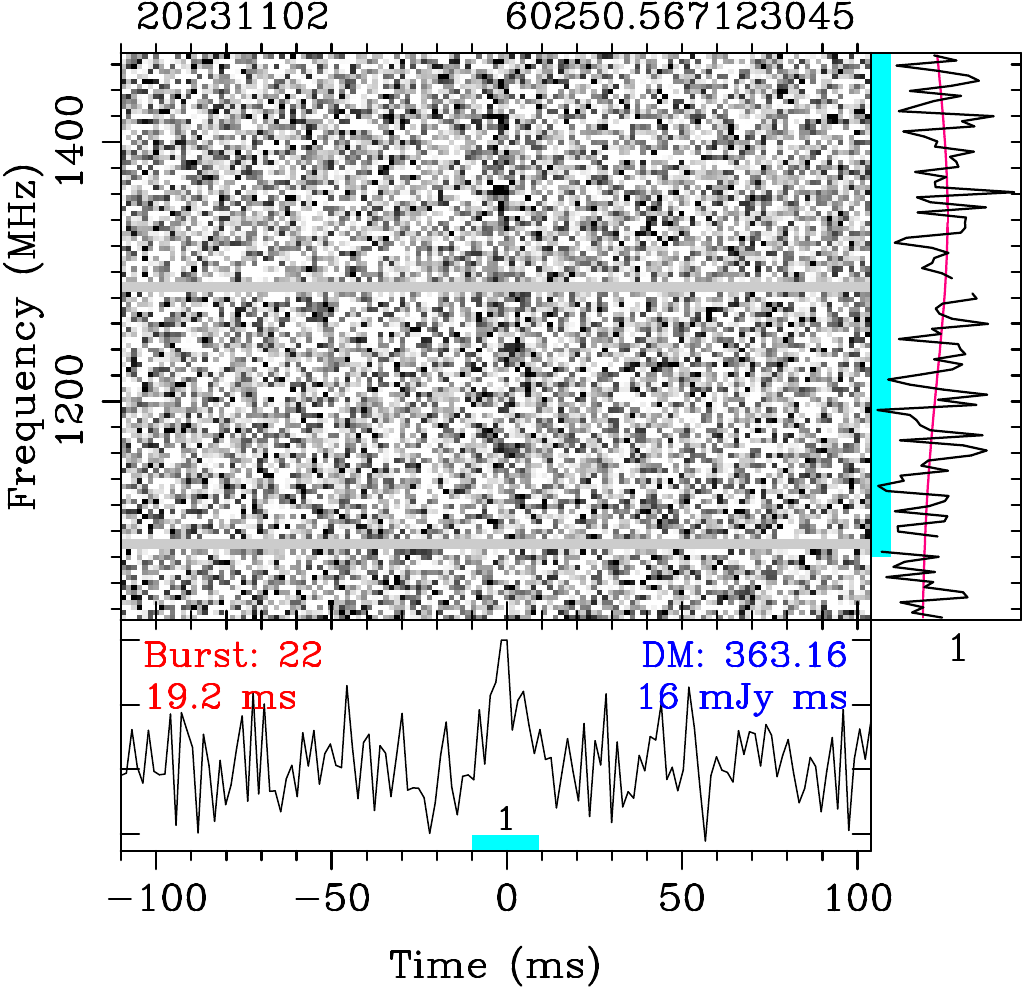}
\caption{({\textit{continued}})}
\end{figure*}
\addtocounter{figure}{-1}
\begin{figure*}
\flushleft
\includegraphics[height=0.29\linewidth]{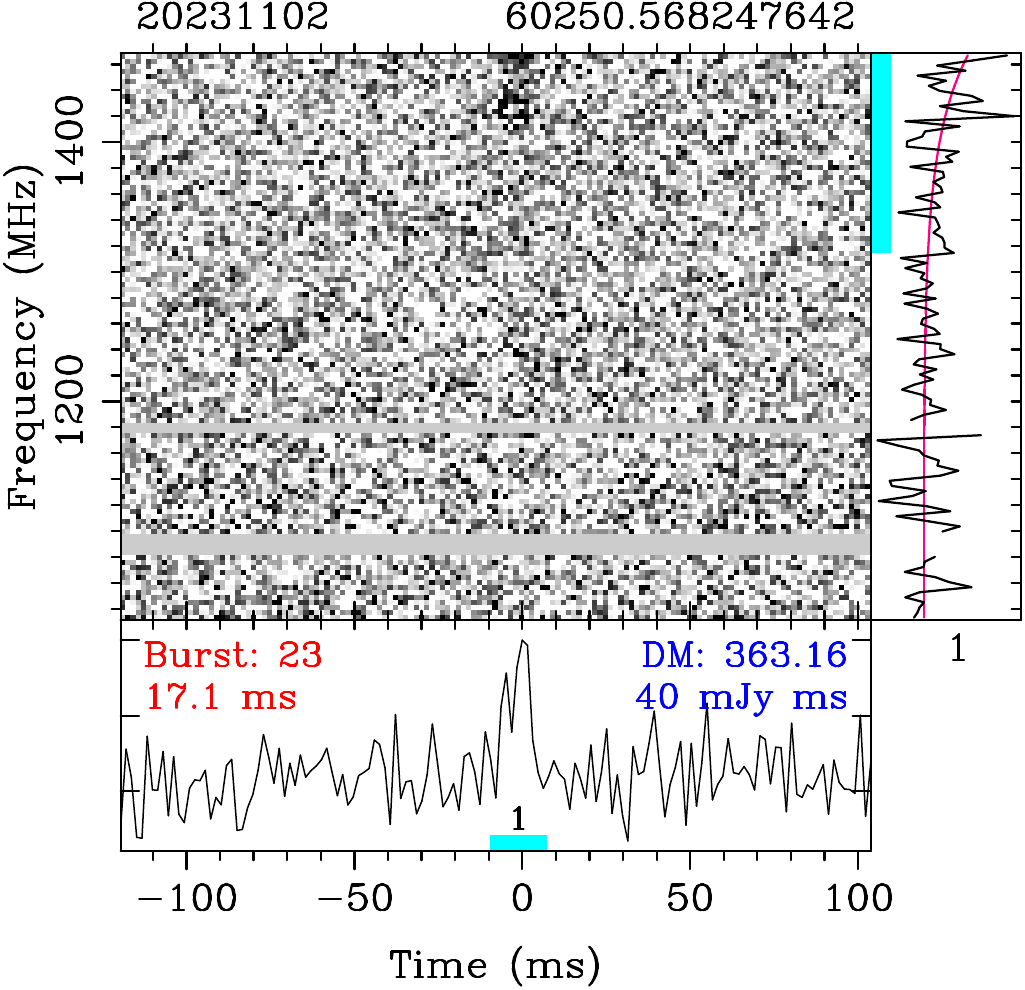}
\includegraphics[height=0.29\linewidth]{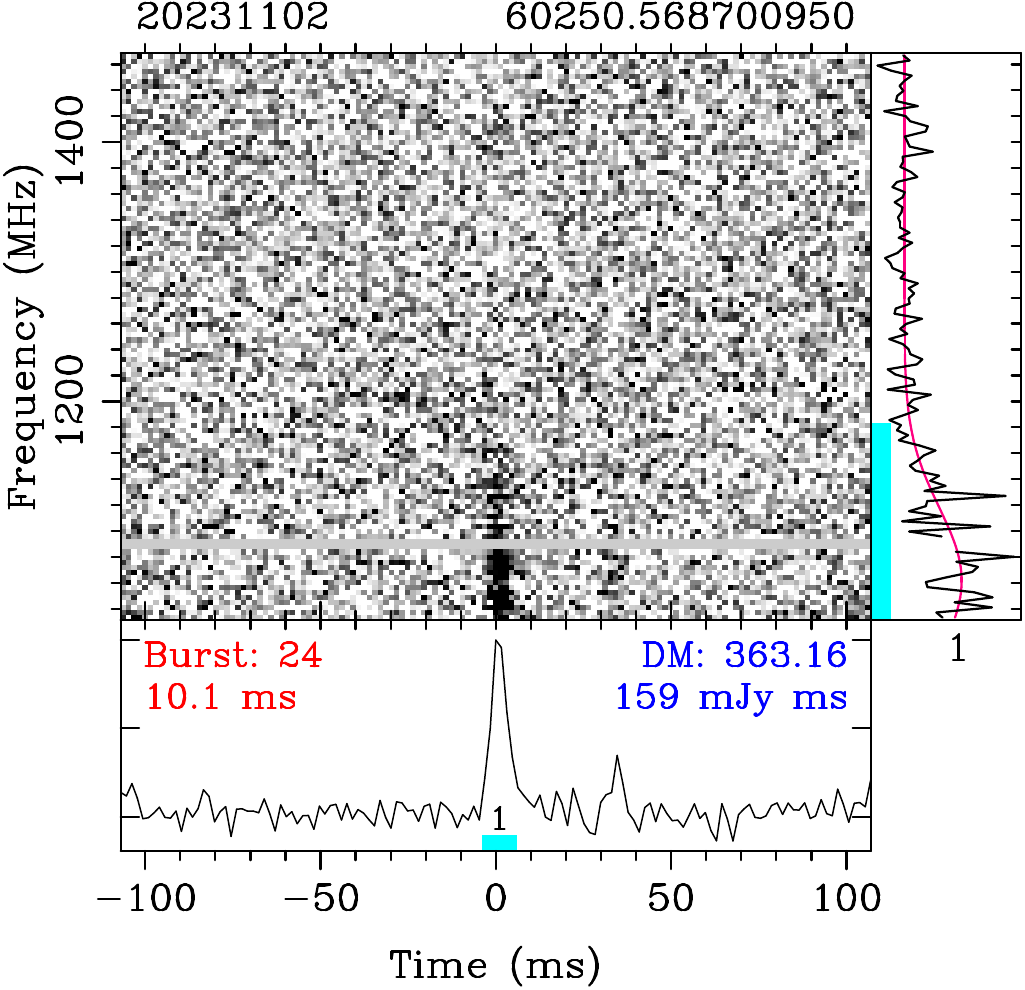}
\includegraphics[height=0.29\linewidth]{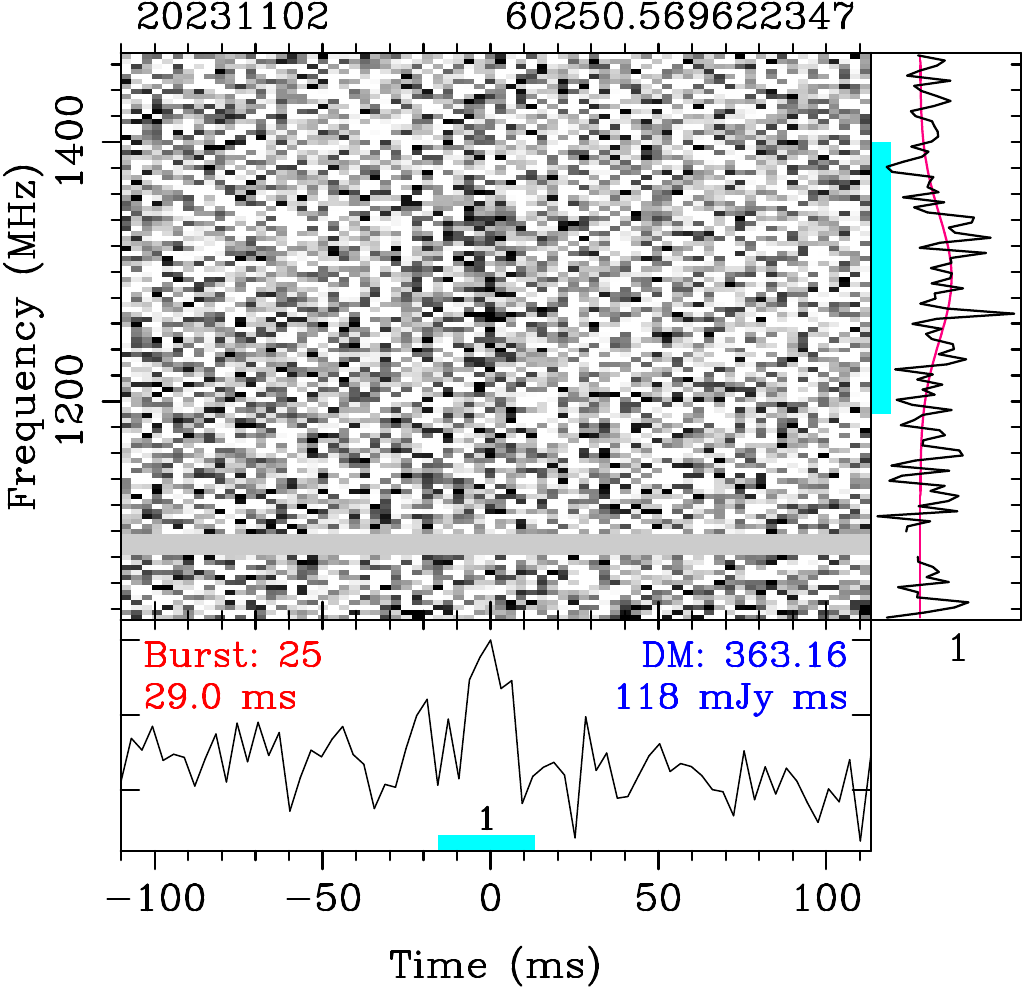}
\includegraphics[height=0.29\linewidth]{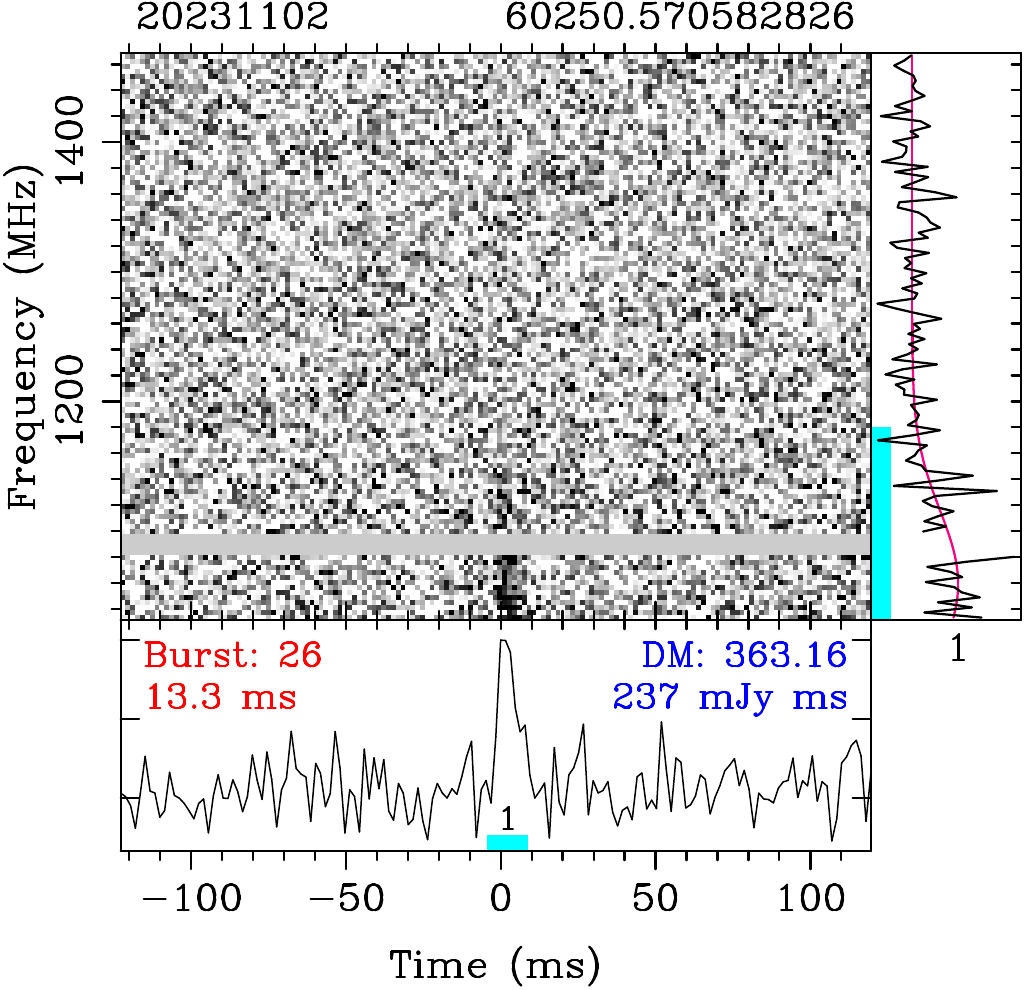}
\includegraphics[height=0.29\linewidth]{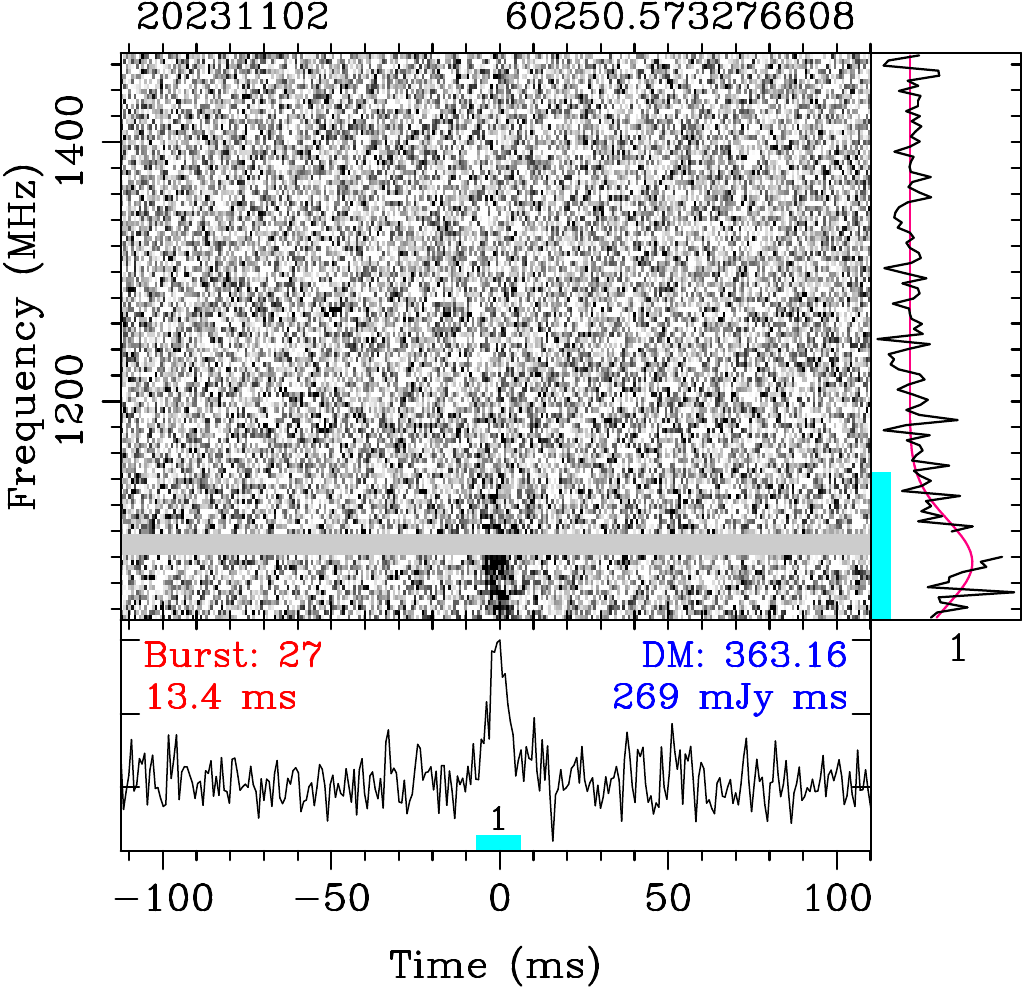}
\includegraphics[height=0.29\linewidth]{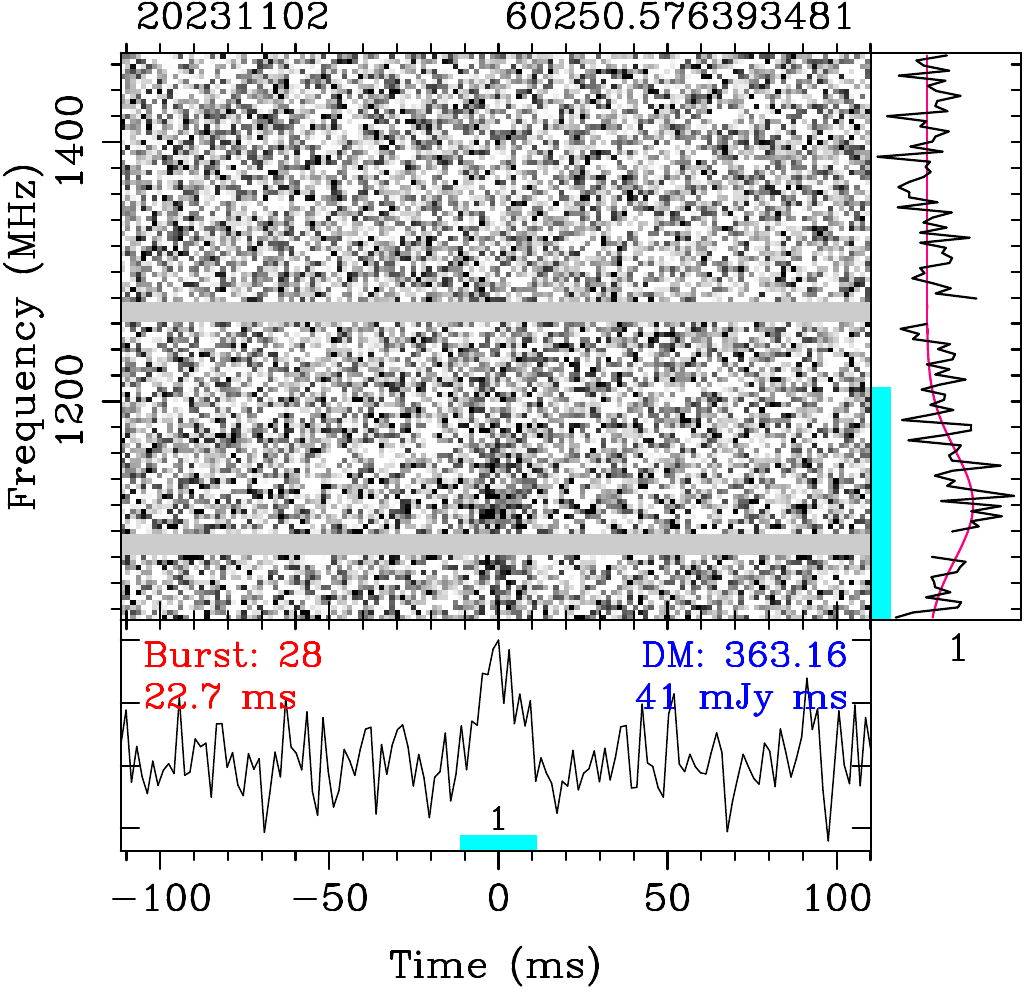}
\includegraphics[height=0.29\linewidth]{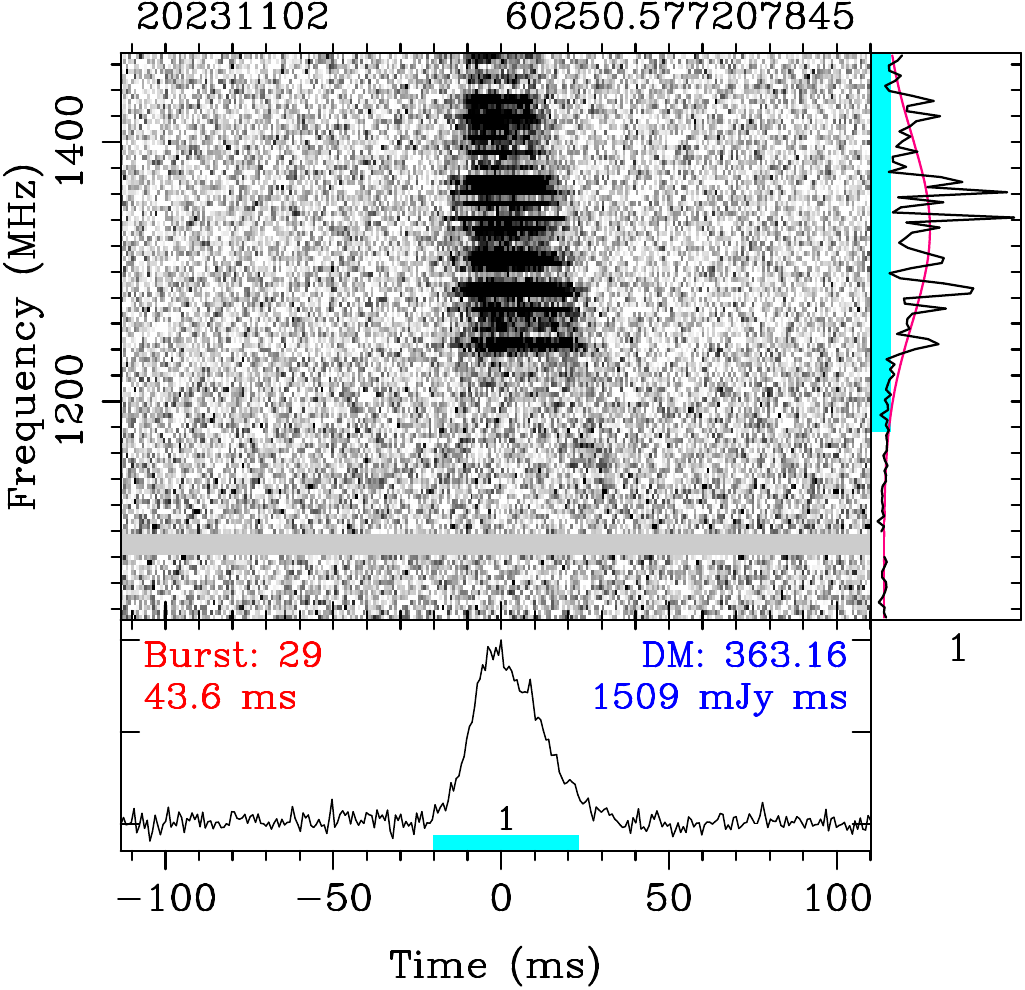}
\includegraphics[height=0.29\linewidth]{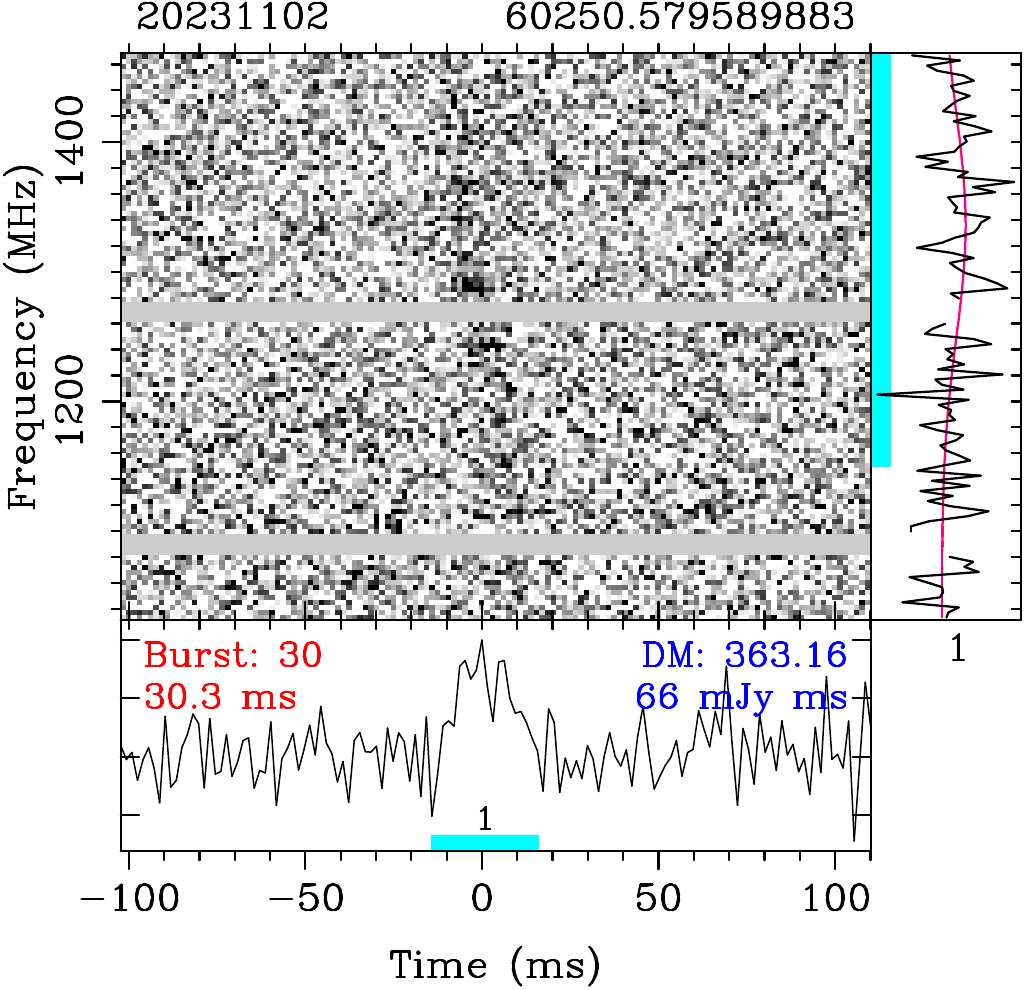}
\includegraphics[height=0.29\linewidth]{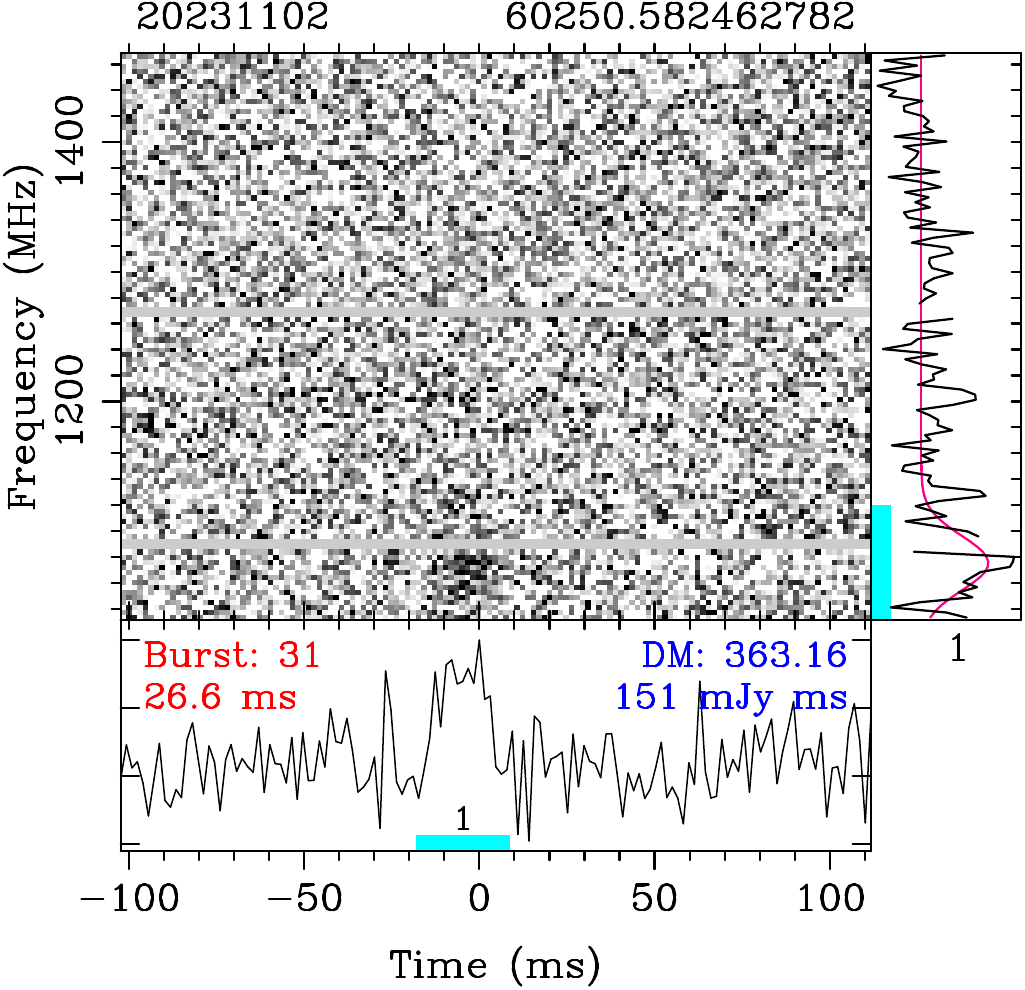}
\includegraphics[height=0.29\linewidth]{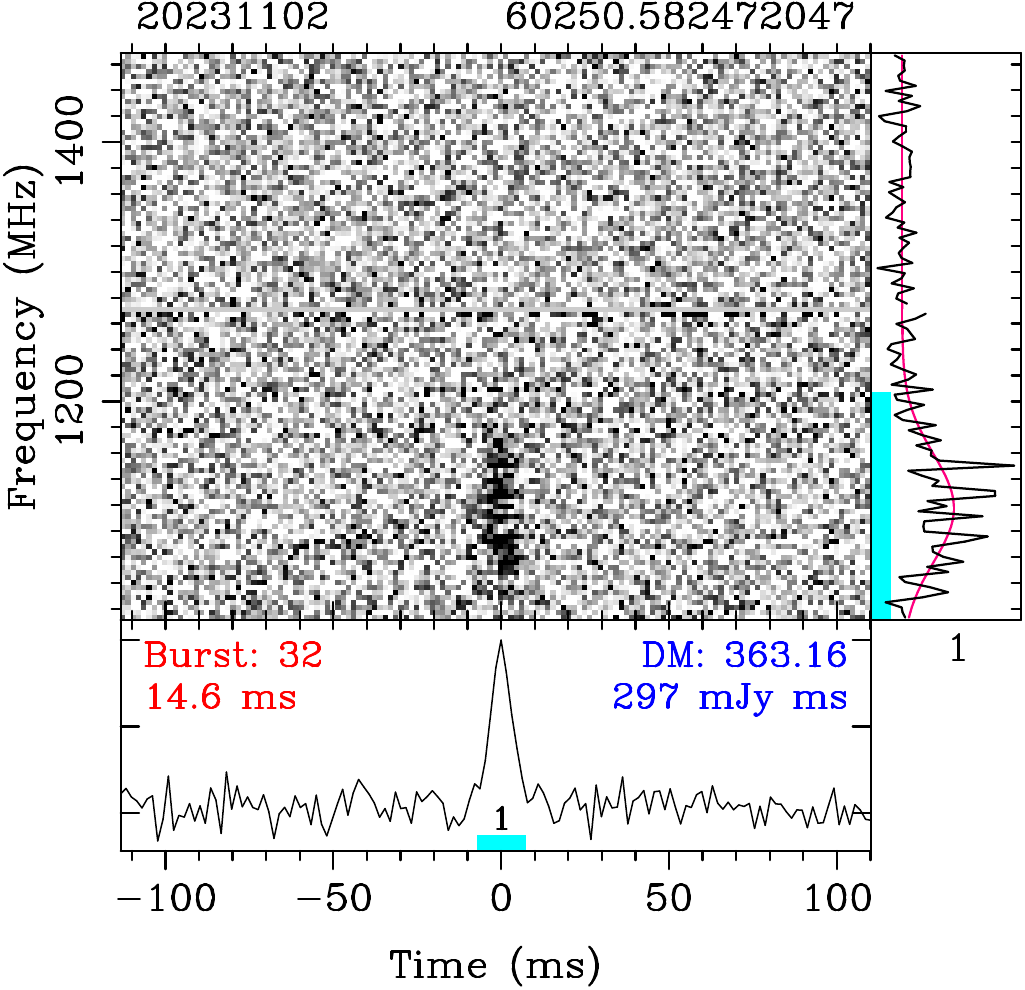}
\includegraphics[height=0.29\linewidth]{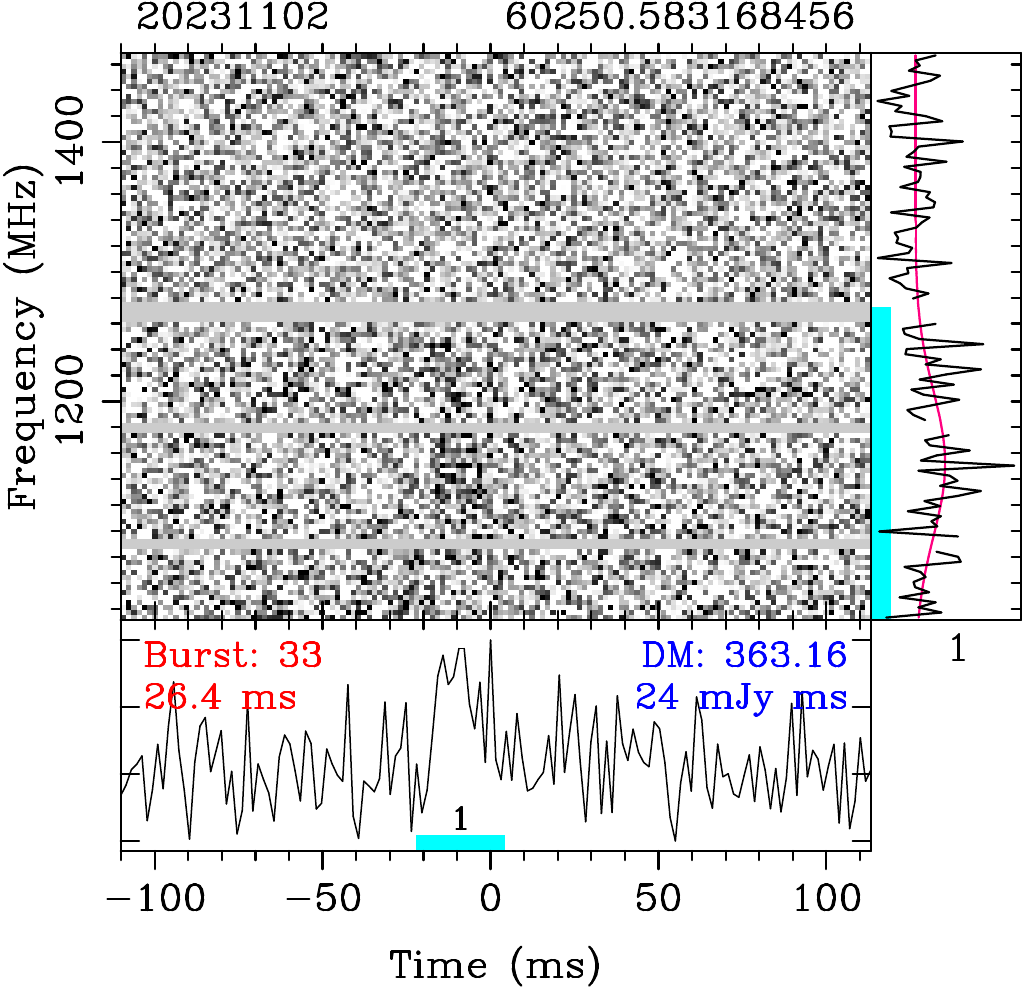}
\includegraphics[height=0.29\linewidth]{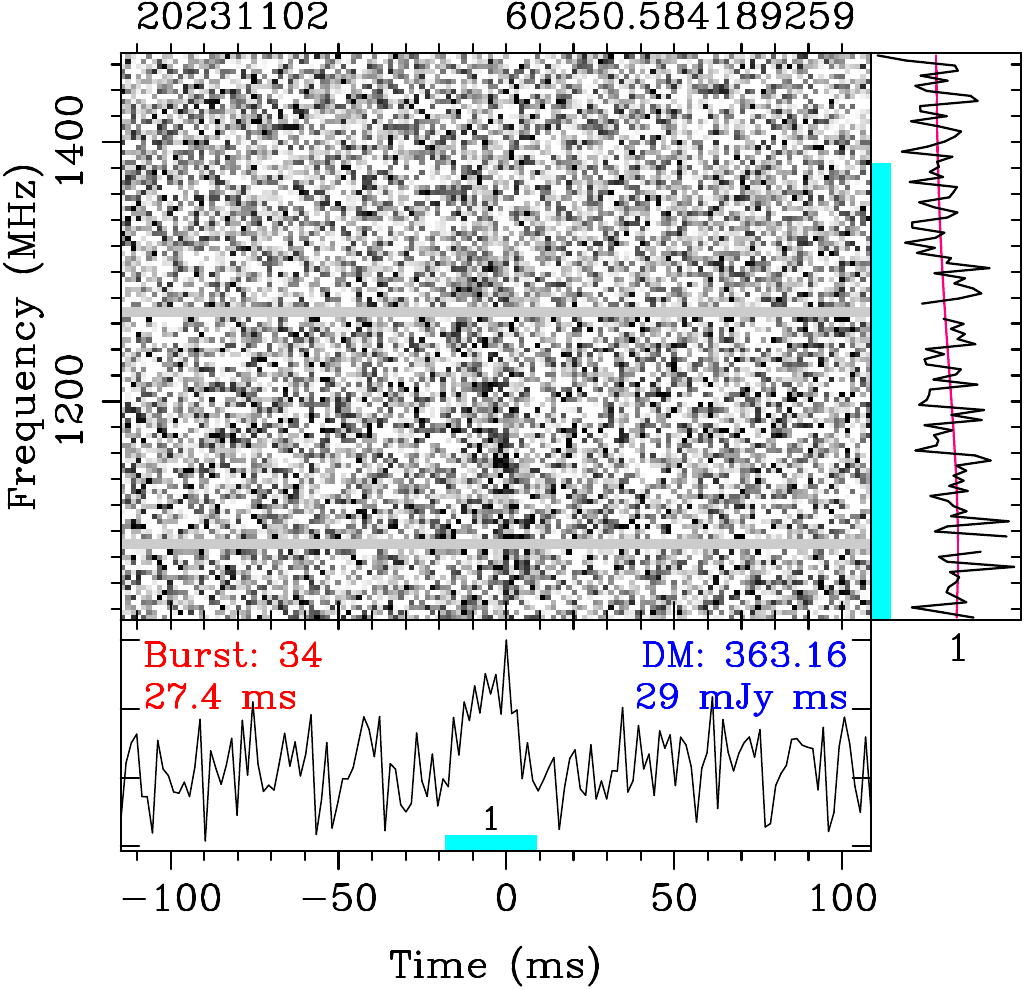}
\caption{({\textit{continued}})}
\end{figure*}
\addtocounter{figure}{-1}
\begin{figure*}
\flushleft
\includegraphics[height=0.29\linewidth]{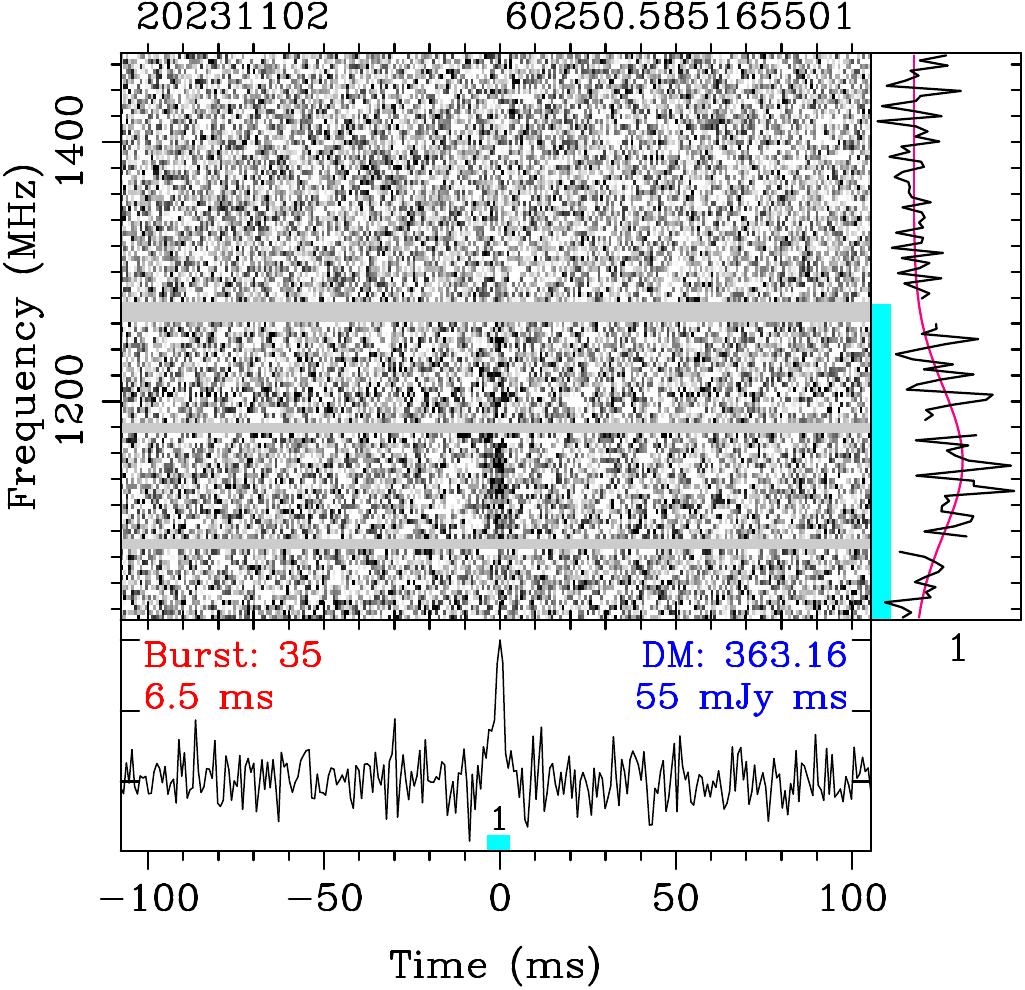}
\includegraphics[height=0.29\linewidth]{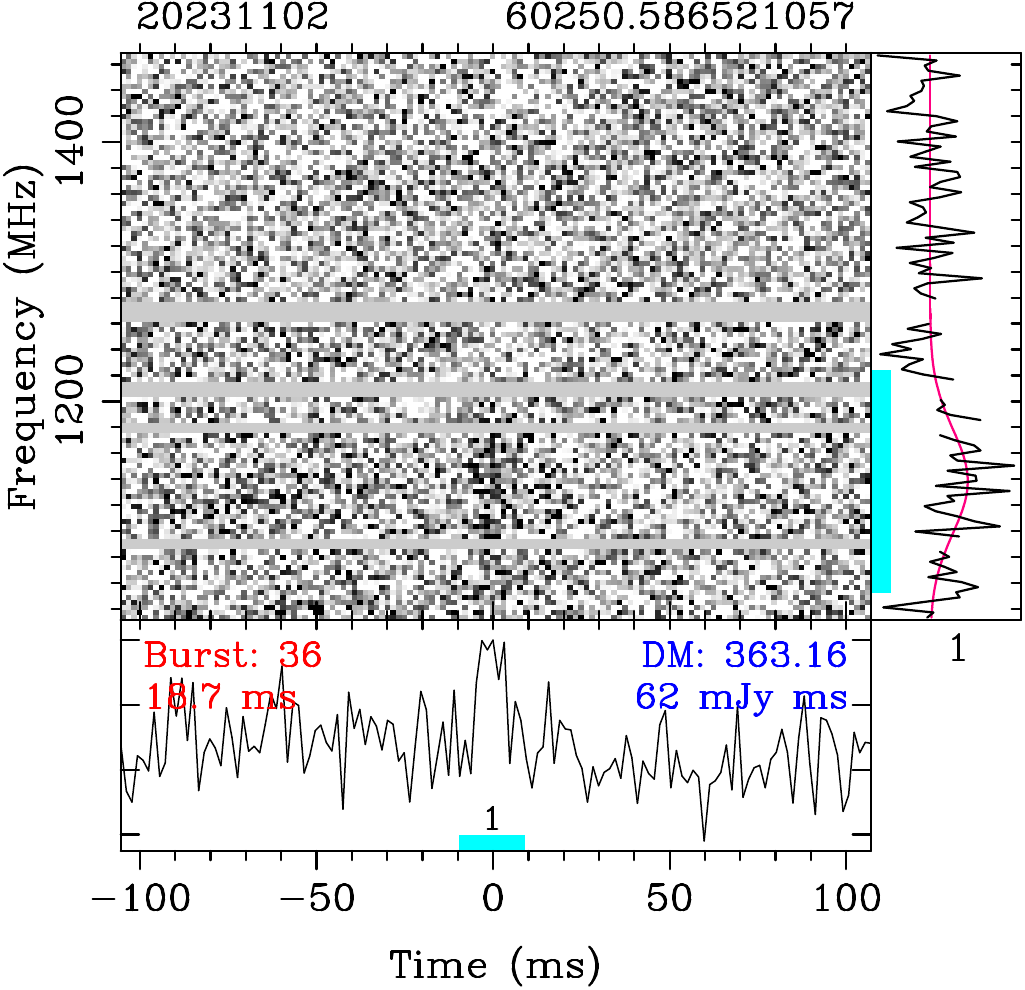}
\includegraphics[height=0.29\linewidth]{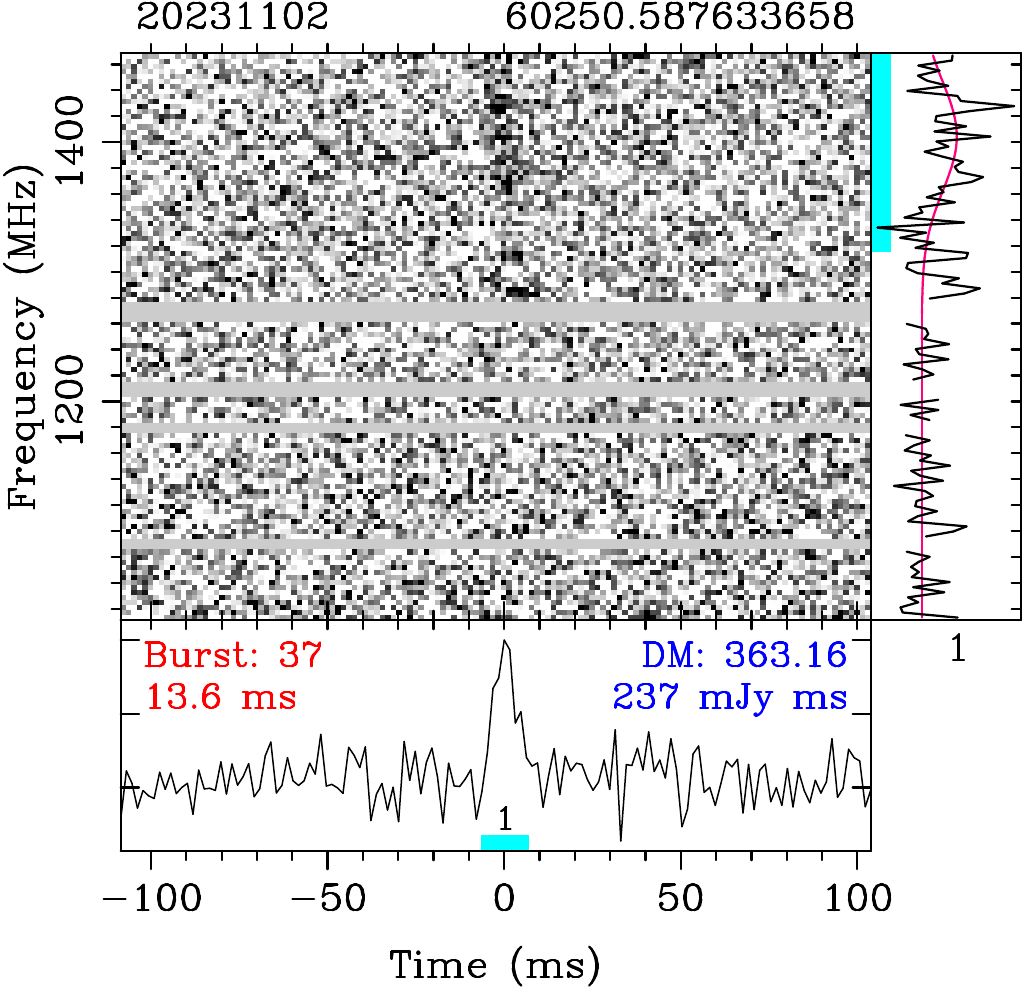}
\includegraphics[height=0.29\linewidth]{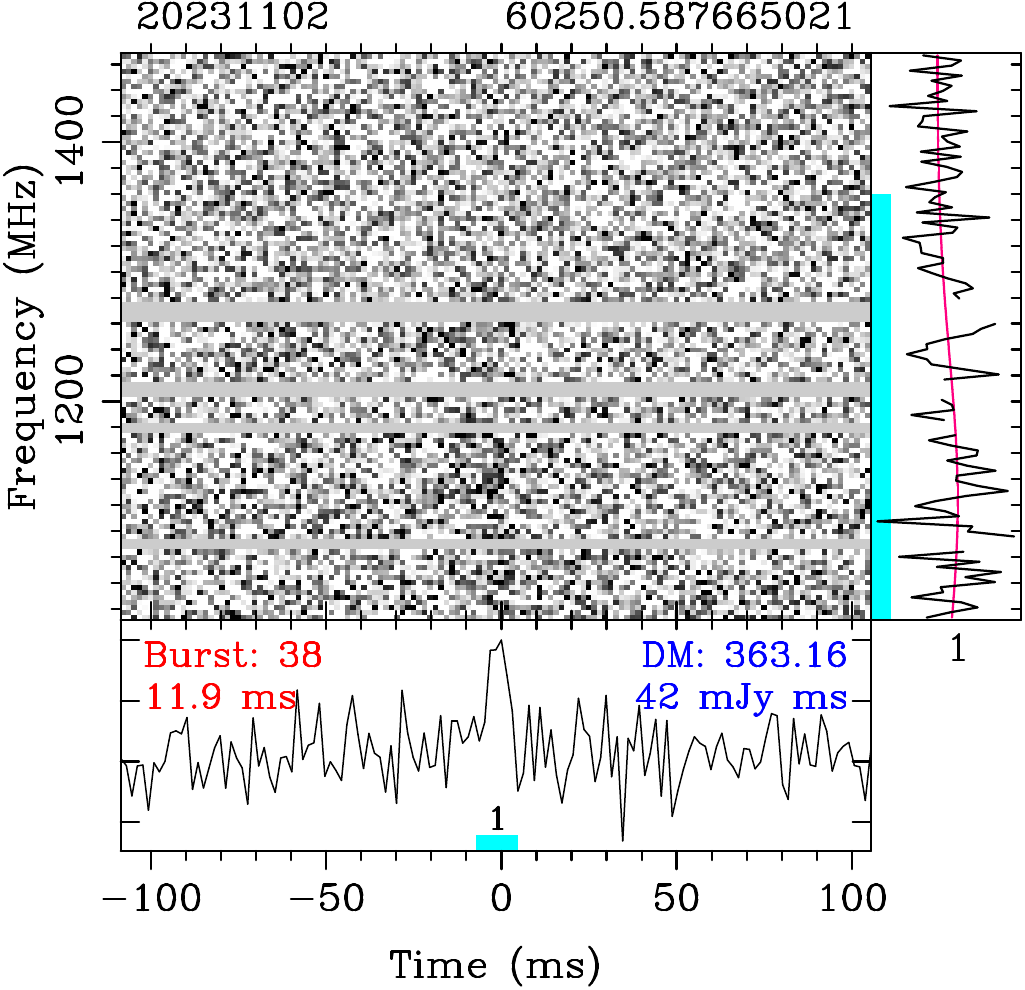}
\includegraphics[height=0.29\linewidth]{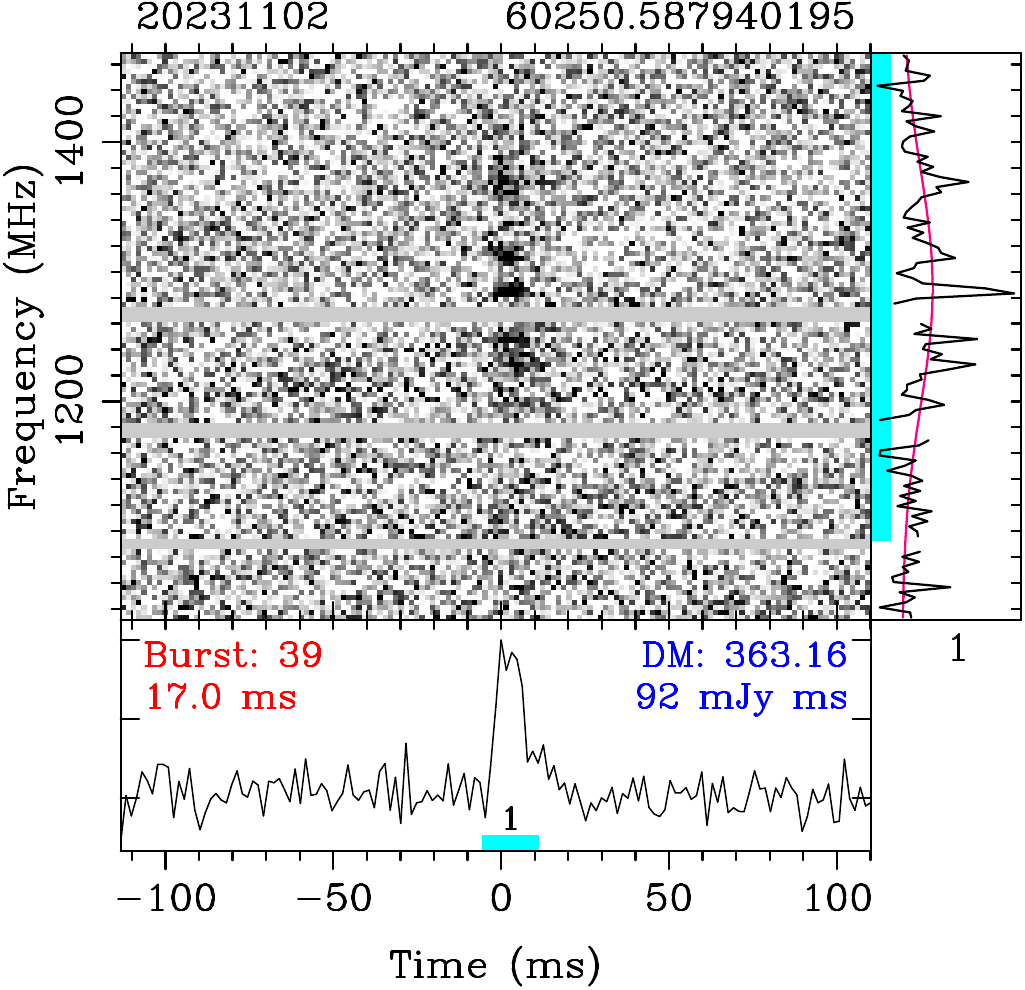}
\includegraphics[height=0.29\linewidth]{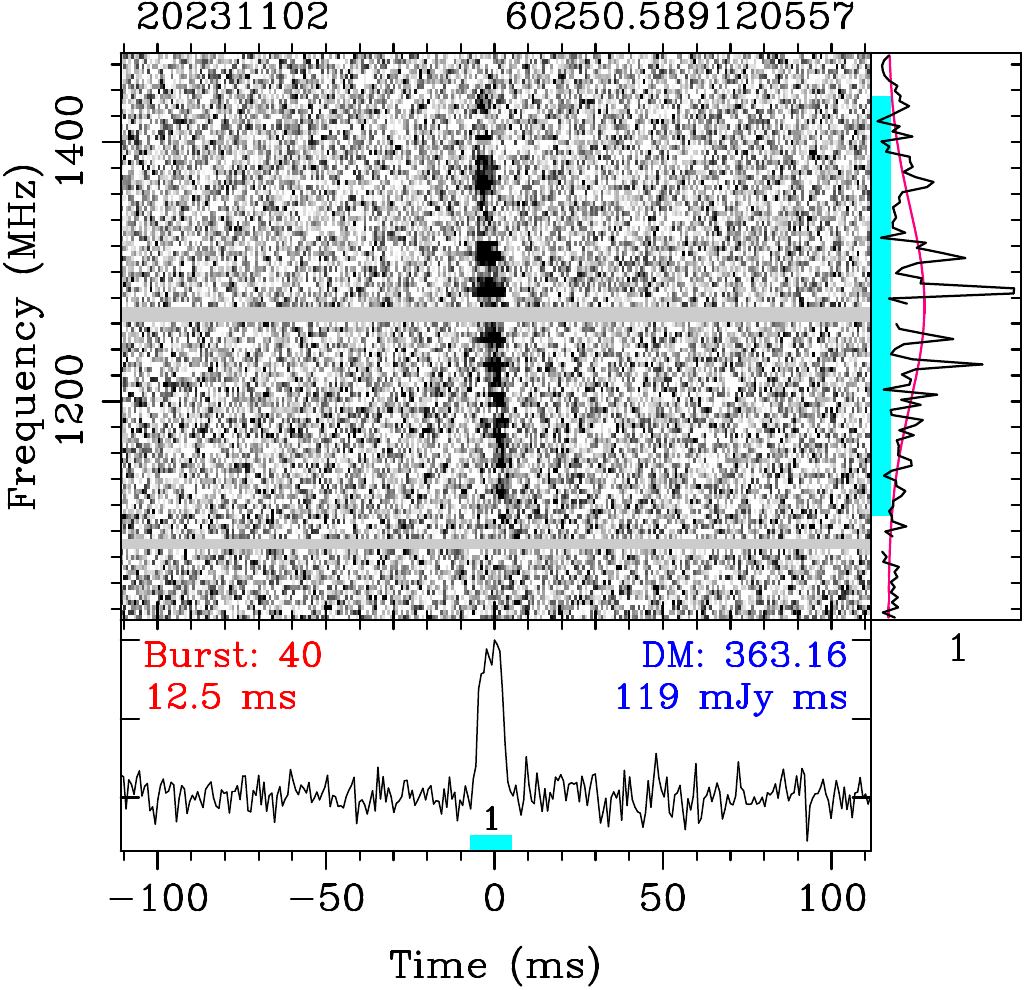}
\includegraphics[height=0.29\linewidth]{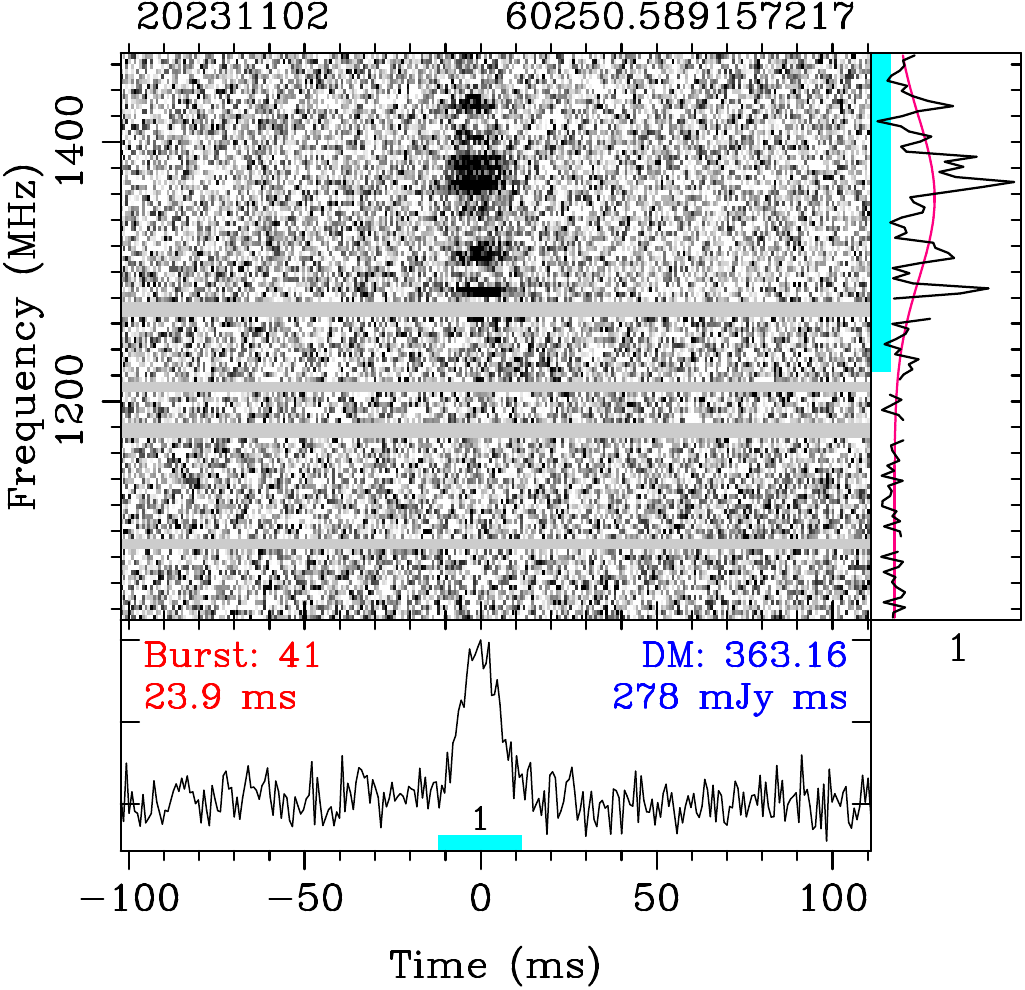}
\includegraphics[height=0.29\linewidth]{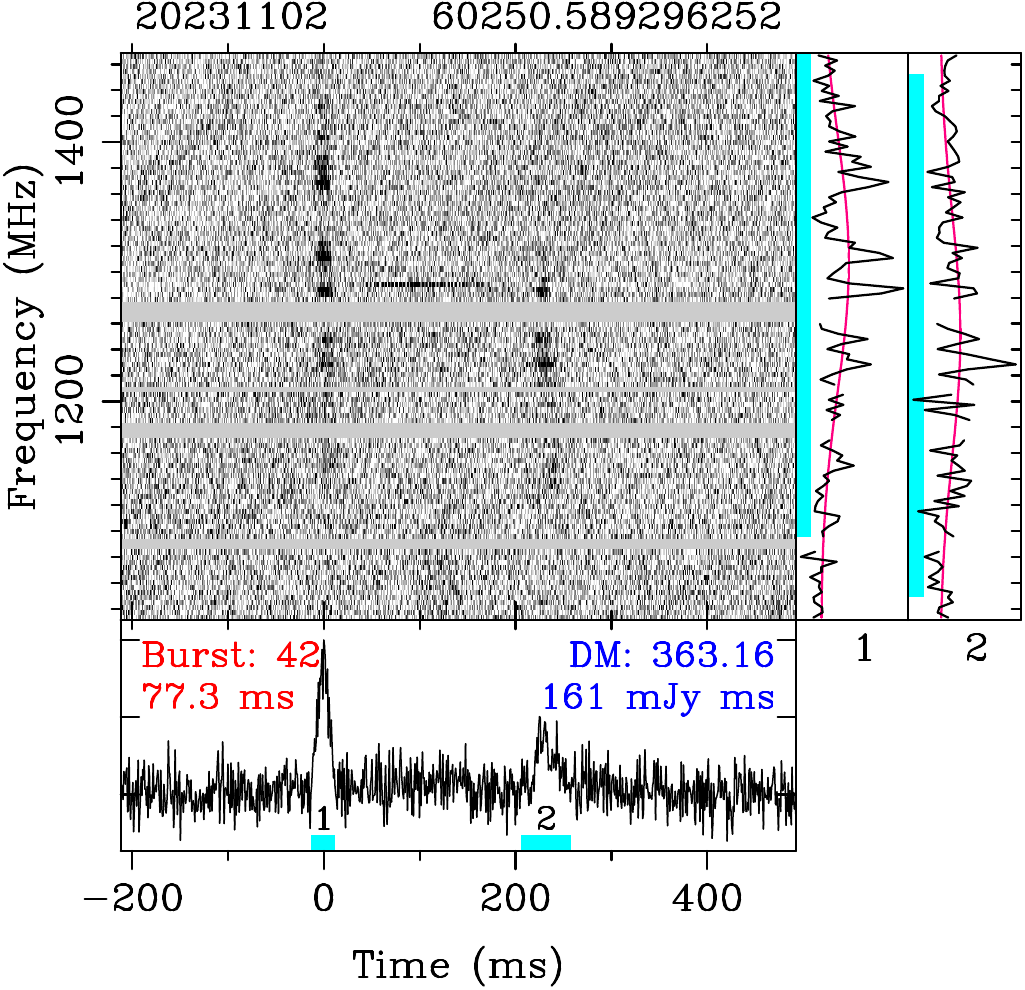}
\includegraphics[height=0.29\linewidth]{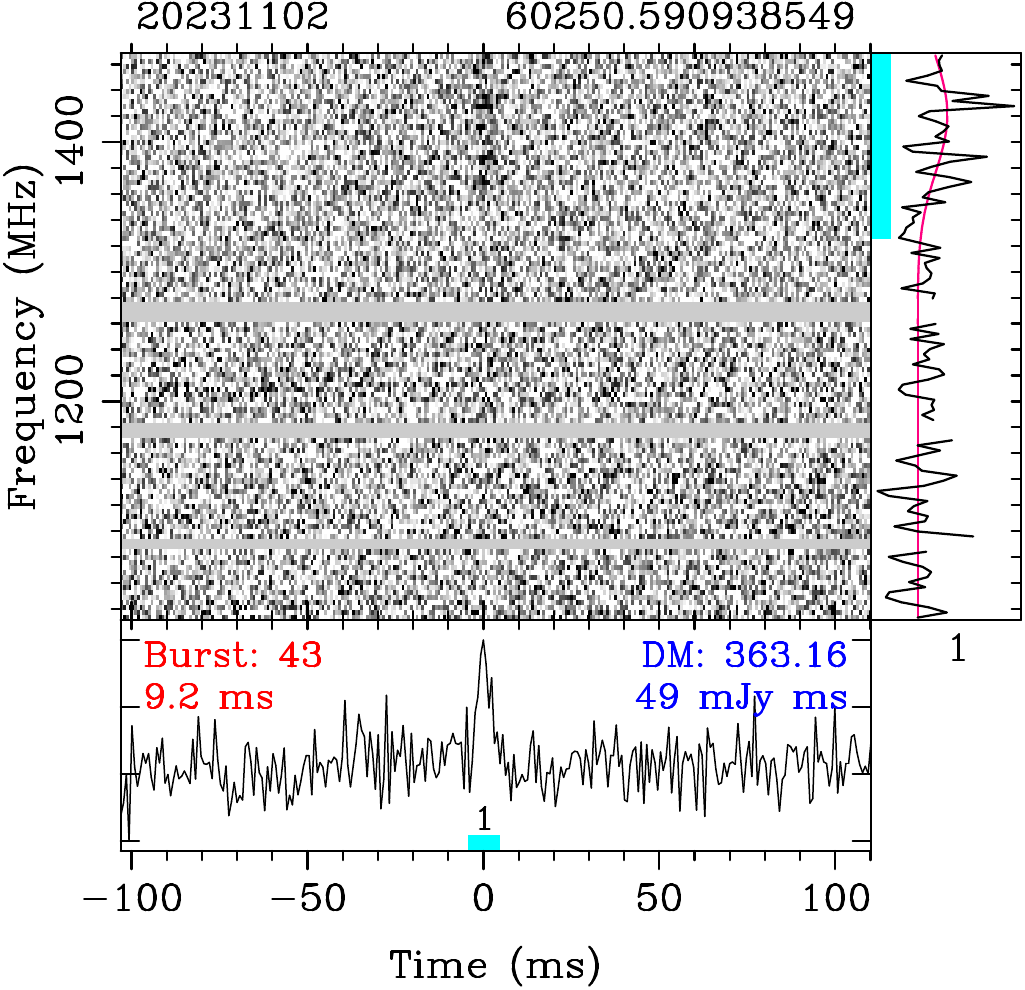}
\includegraphics[height=0.29\linewidth]{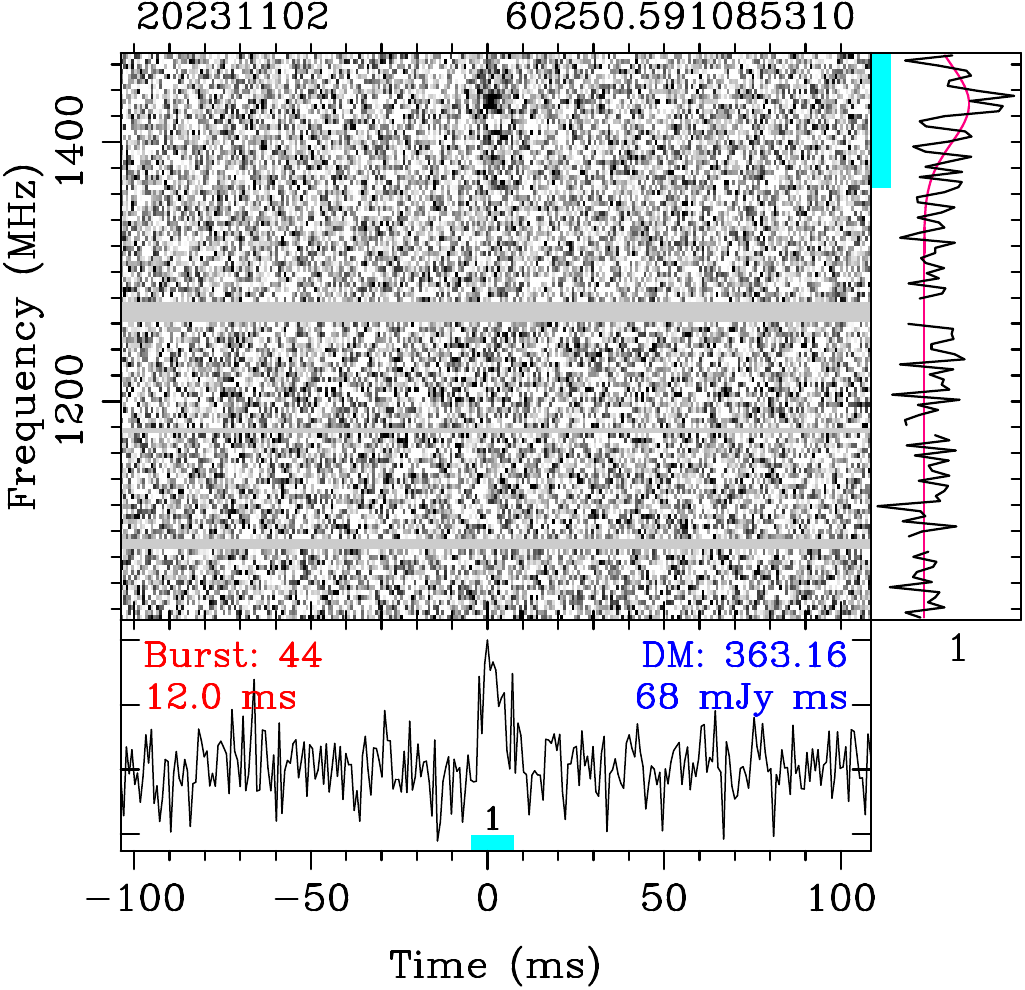}
\includegraphics[height=0.29\linewidth]{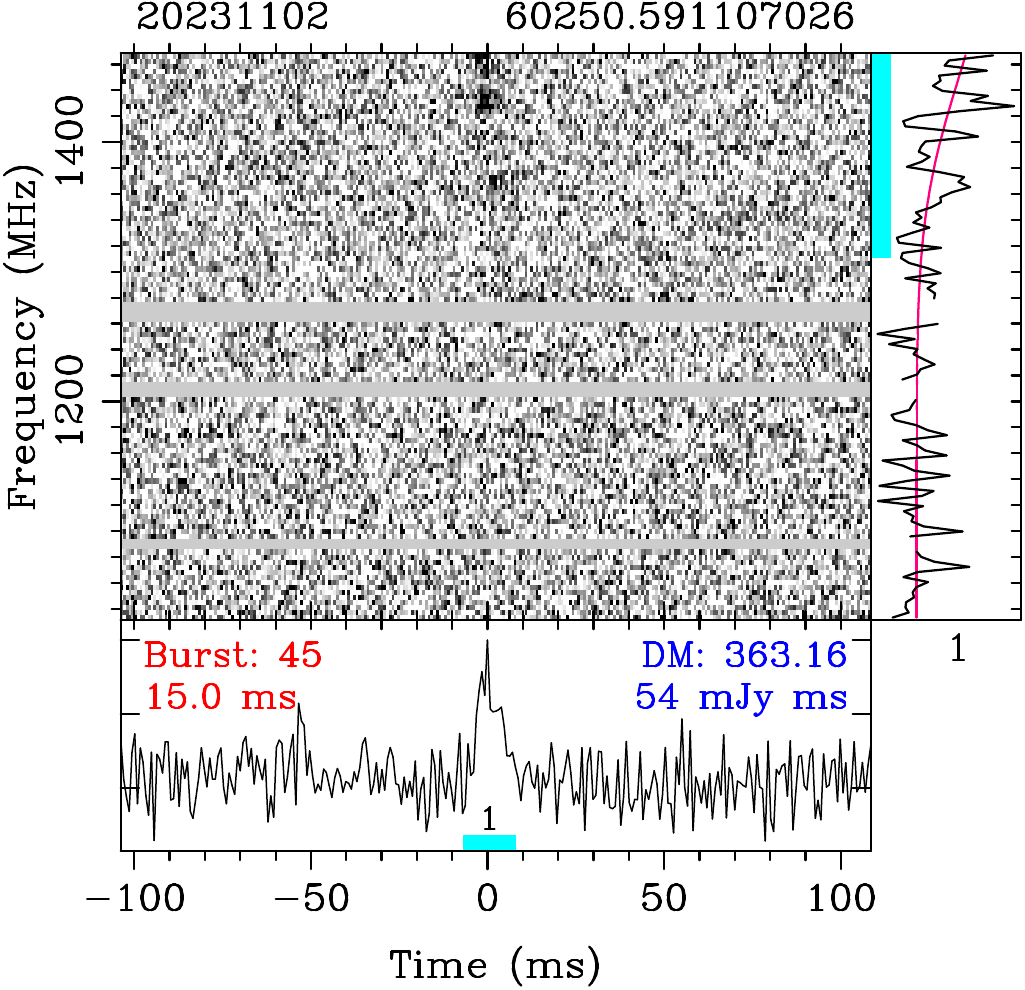}
\includegraphics[height=0.29\linewidth]{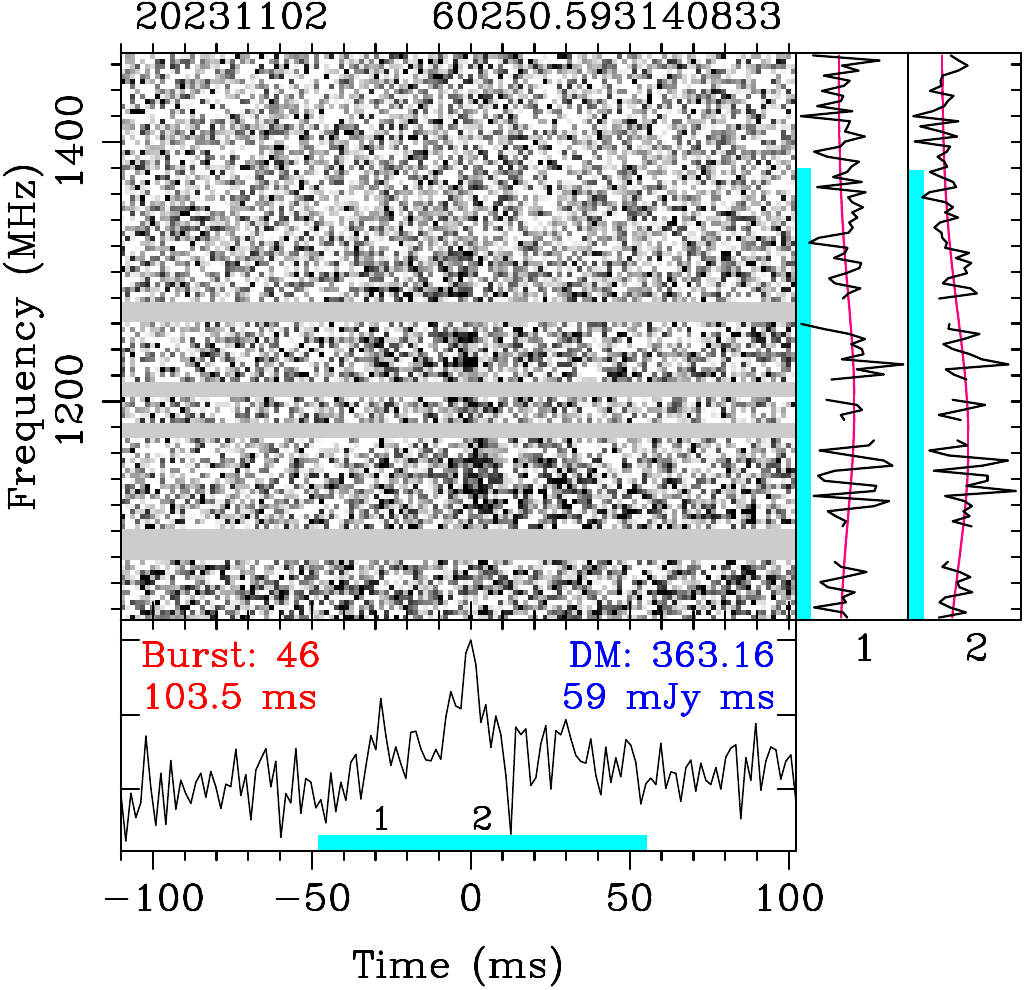}
\caption{({\textit{continued}})}
\end{figure*}
\addtocounter{figure}{-1}
\begin{figure*}
\flushleft
\includegraphics[height=0.29\linewidth]{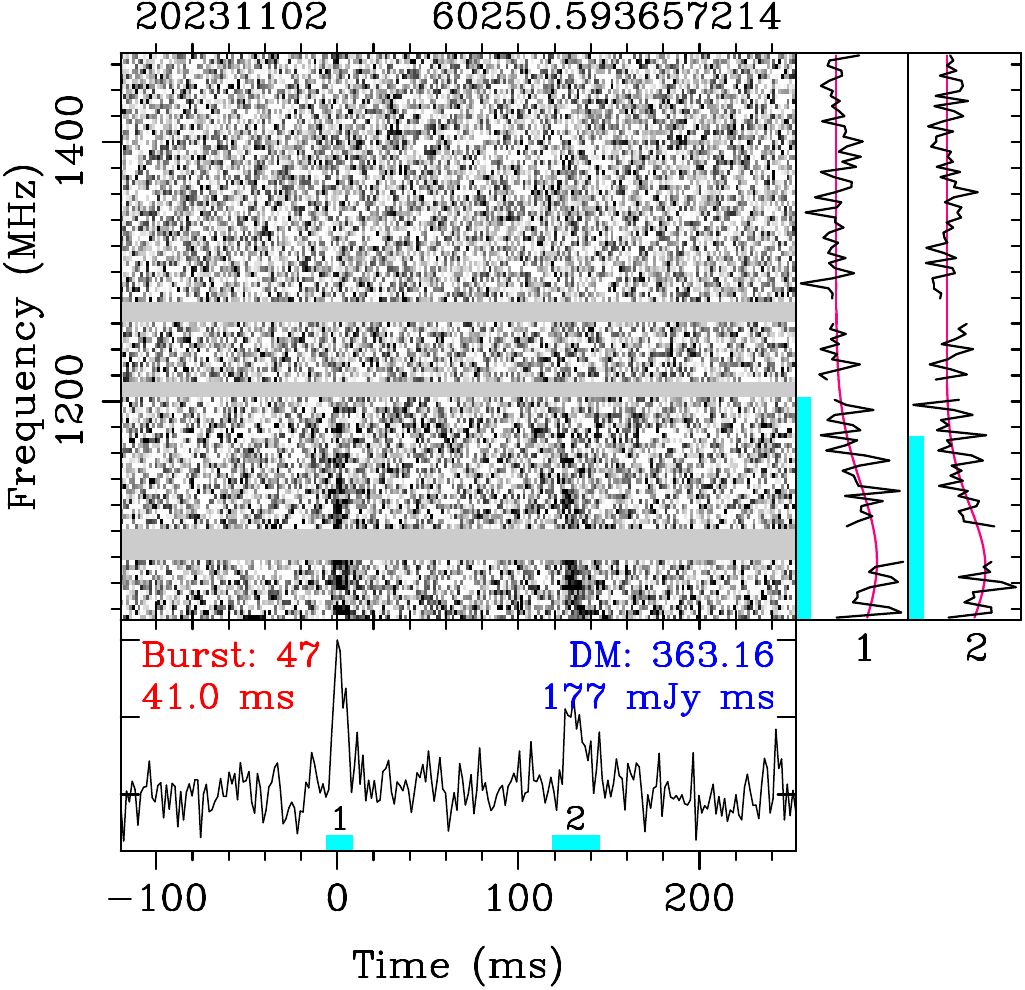}
\includegraphics[height=0.29\linewidth]{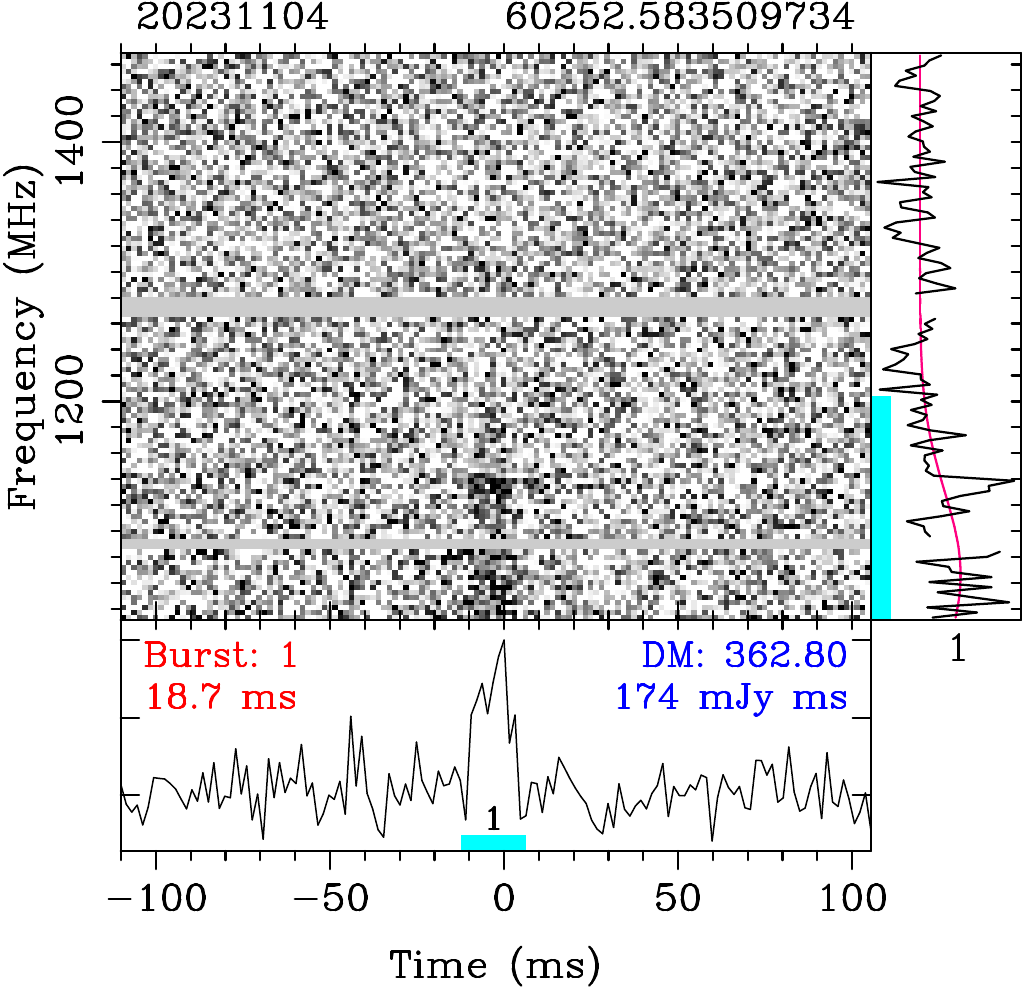}
\includegraphics[height=0.29\linewidth]{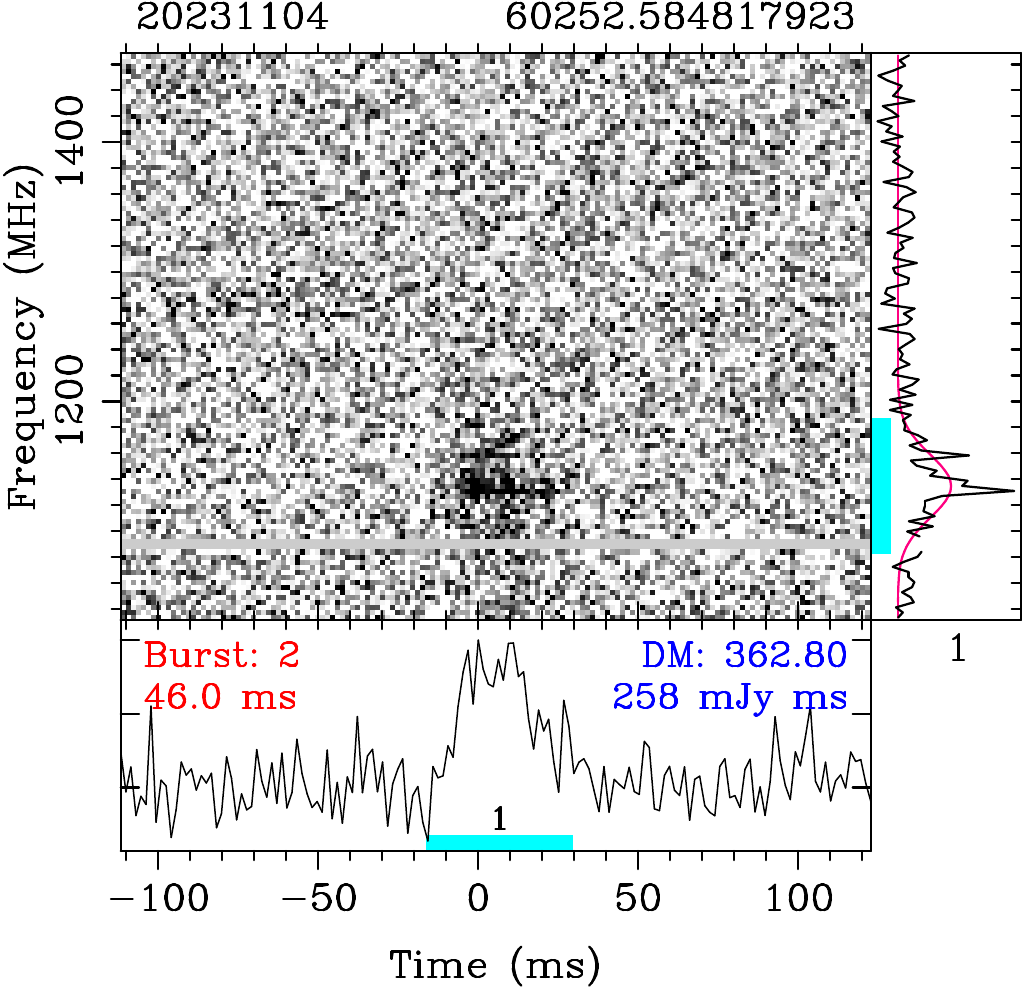}
\includegraphics[height=0.29\linewidth]{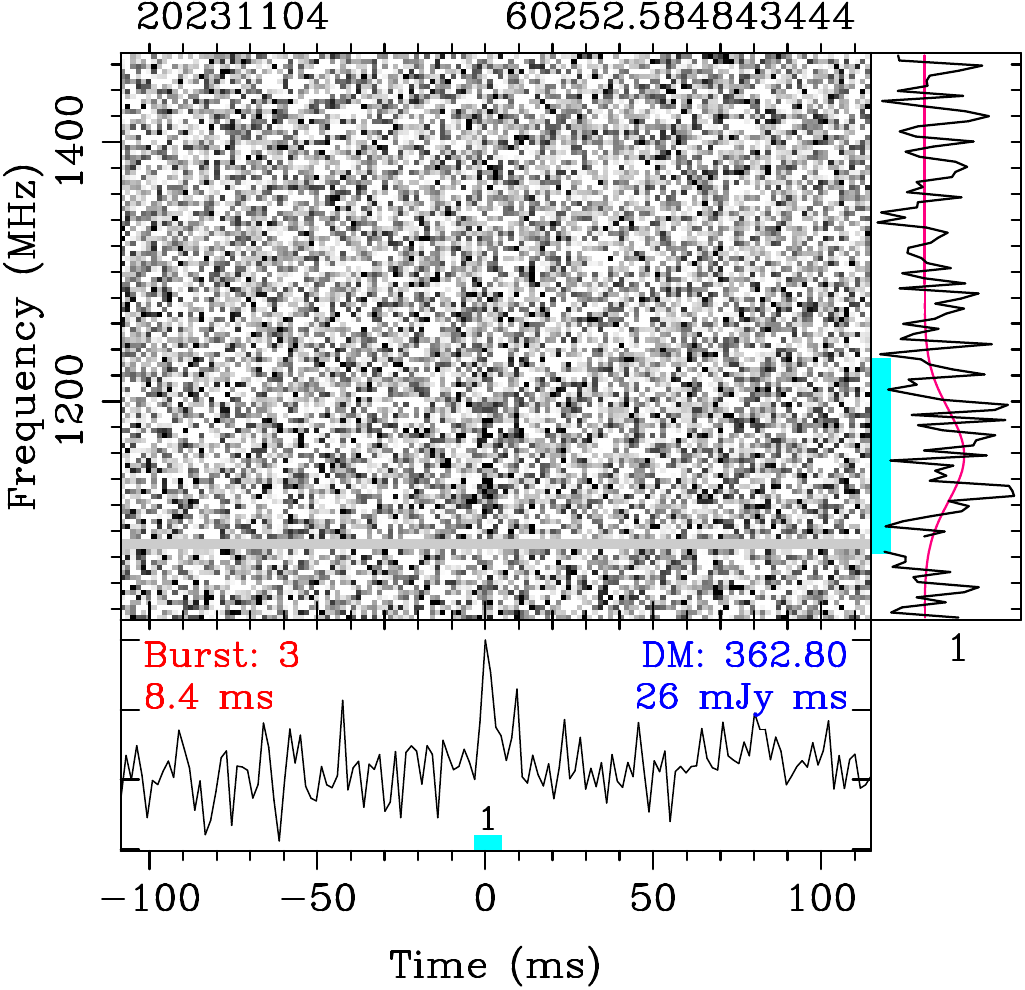}
\includegraphics[height=0.29\linewidth]{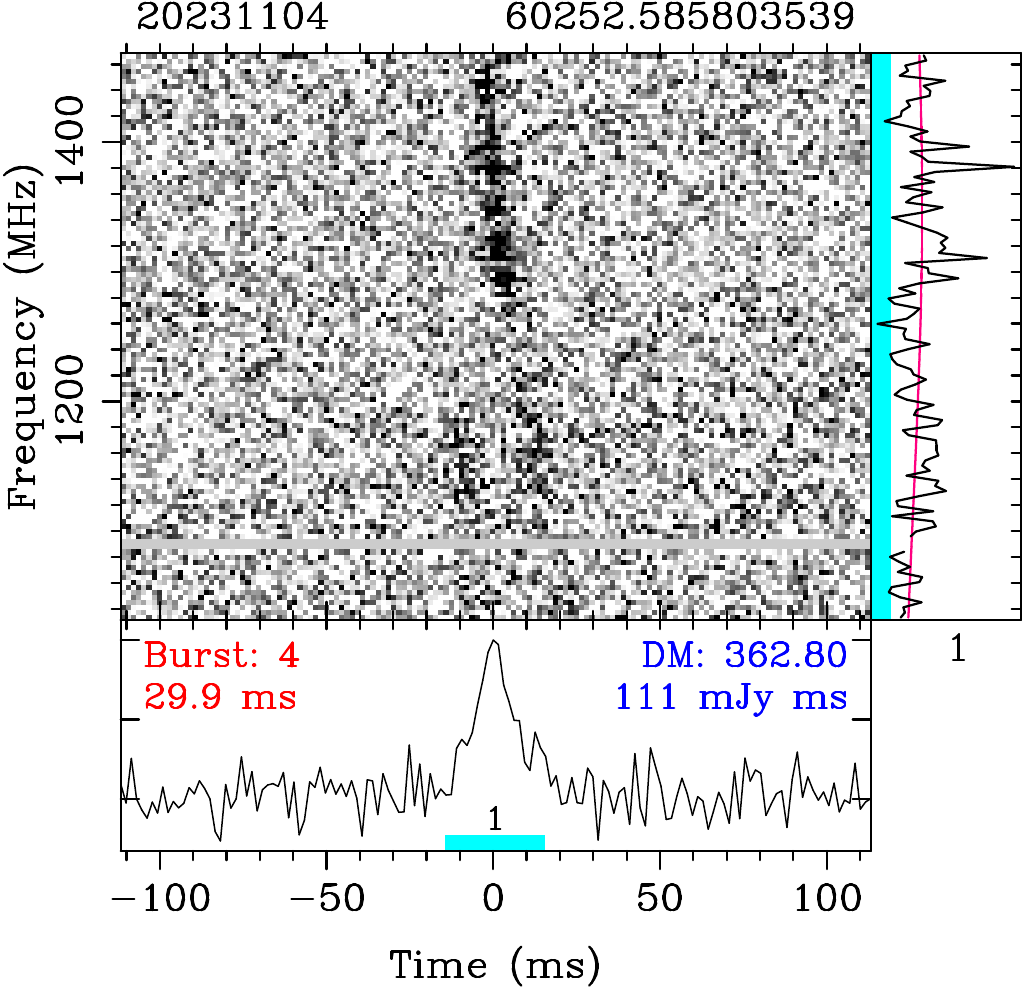}
\includegraphics[height=0.29\linewidth]{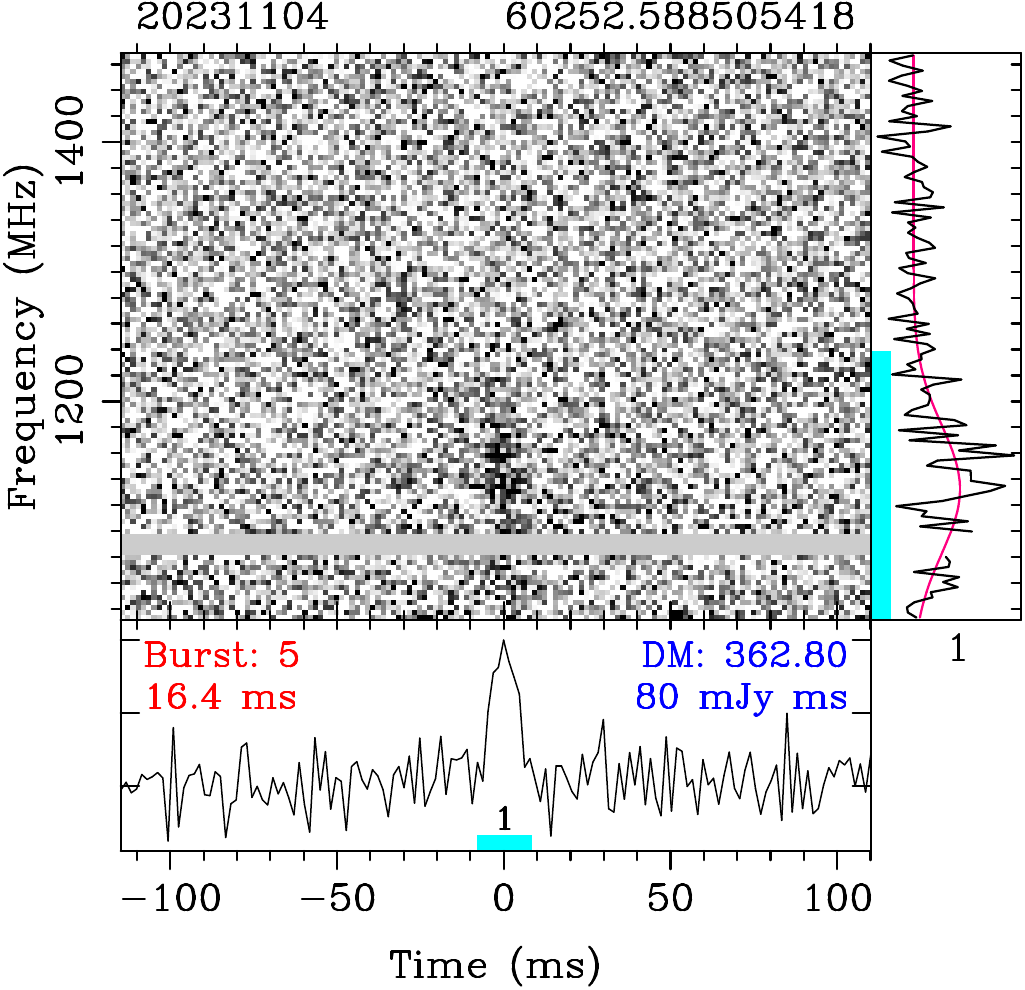}
\includegraphics[height=0.29\linewidth]{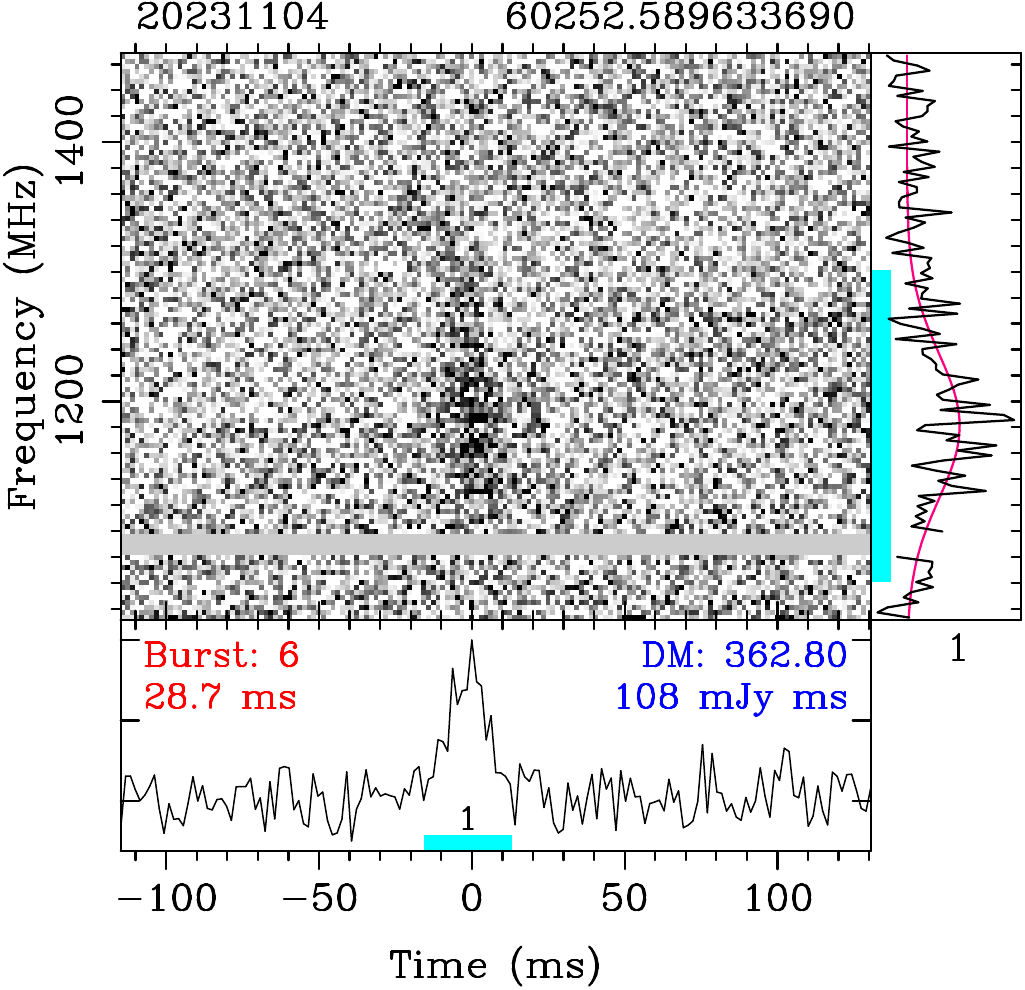}
\includegraphics[height=0.29\linewidth]{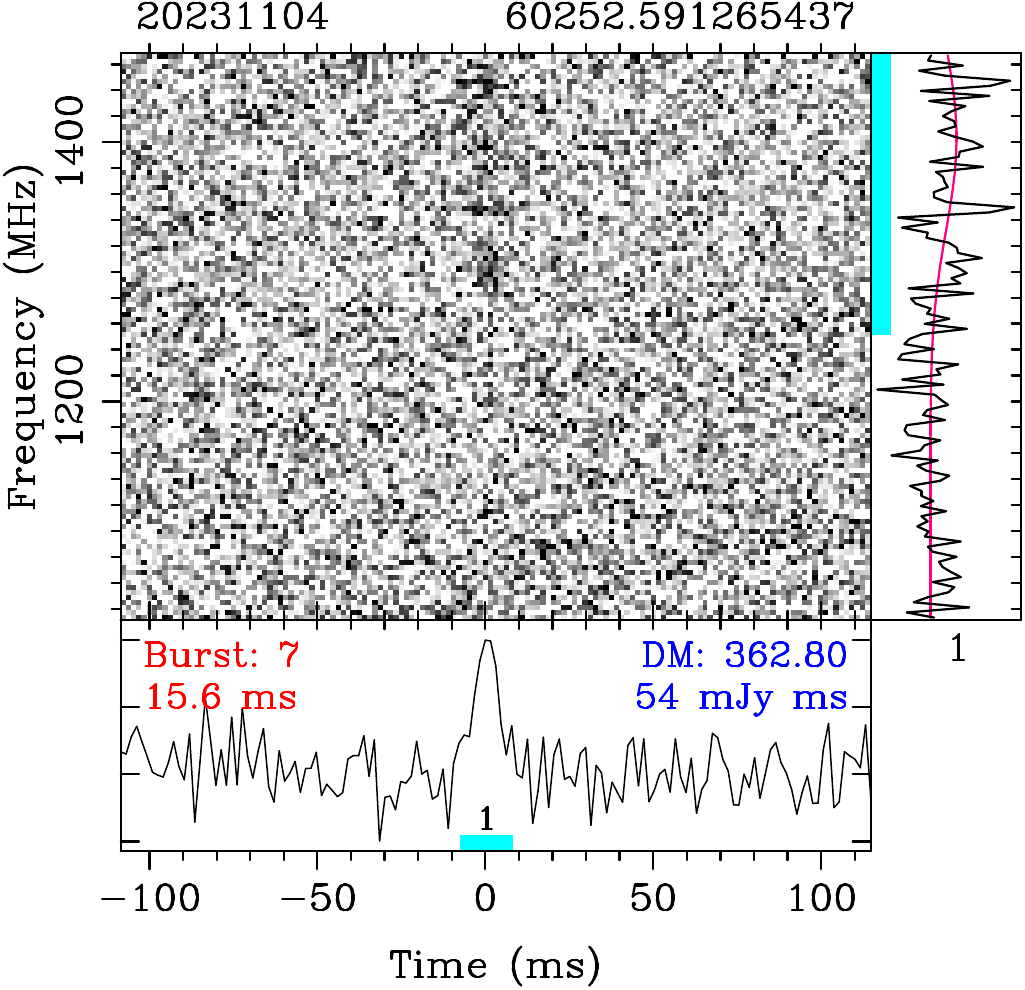}
\includegraphics[height=0.29\linewidth]{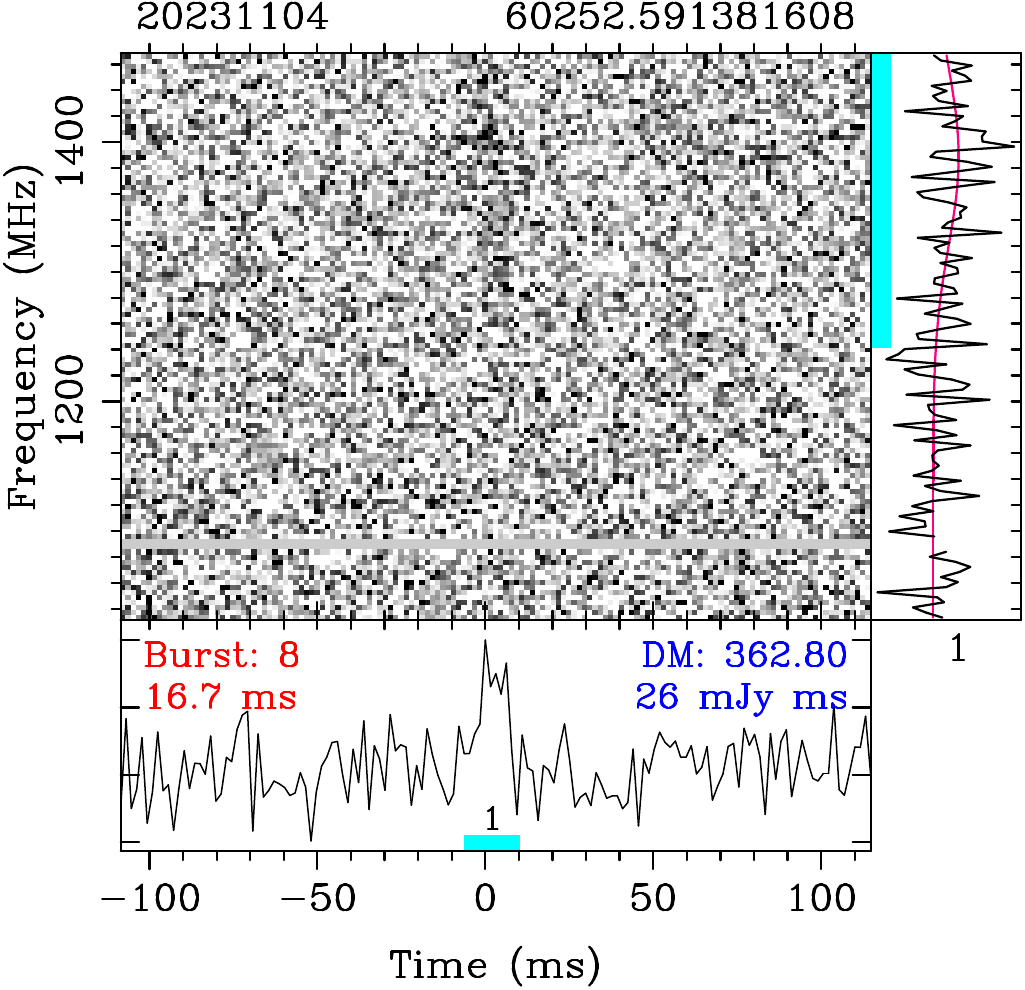}
\includegraphics[height=0.29\linewidth]{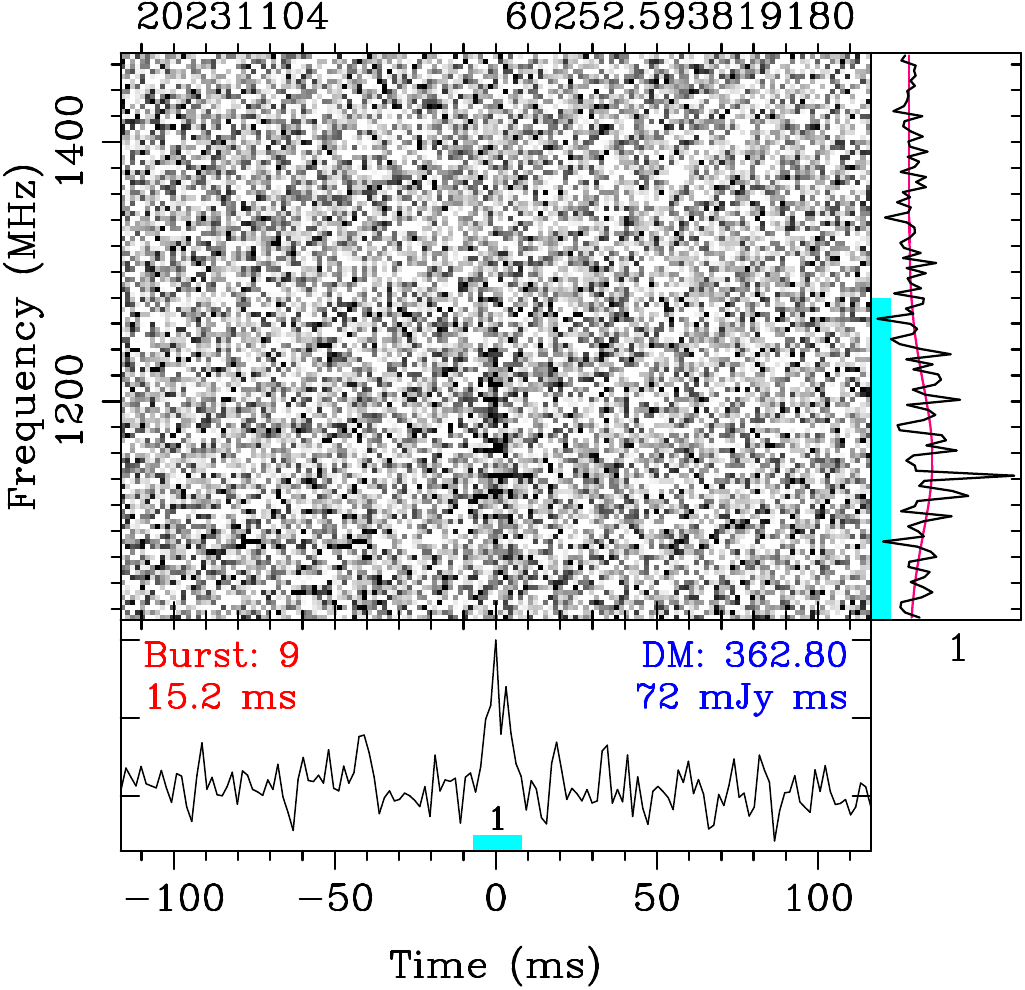}
\includegraphics[height=0.29\linewidth]{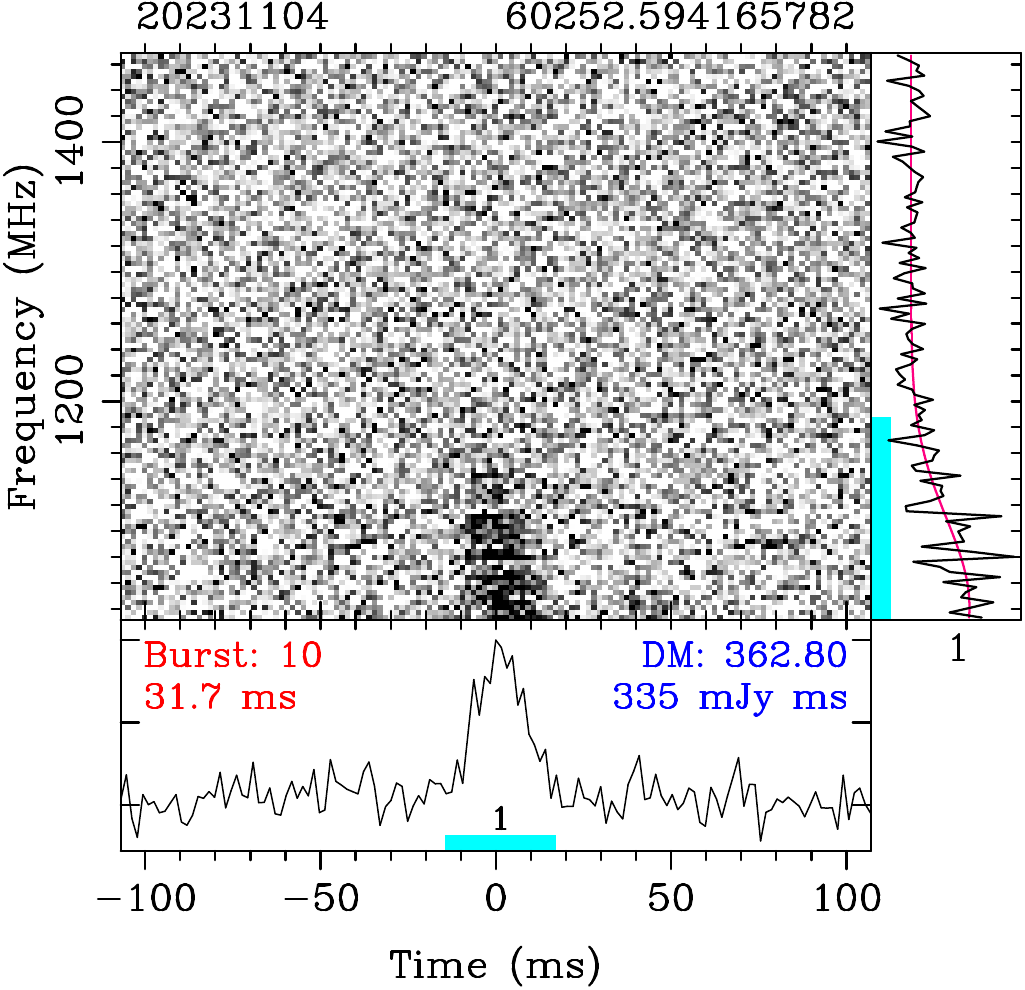}
\includegraphics[height=0.29\linewidth]{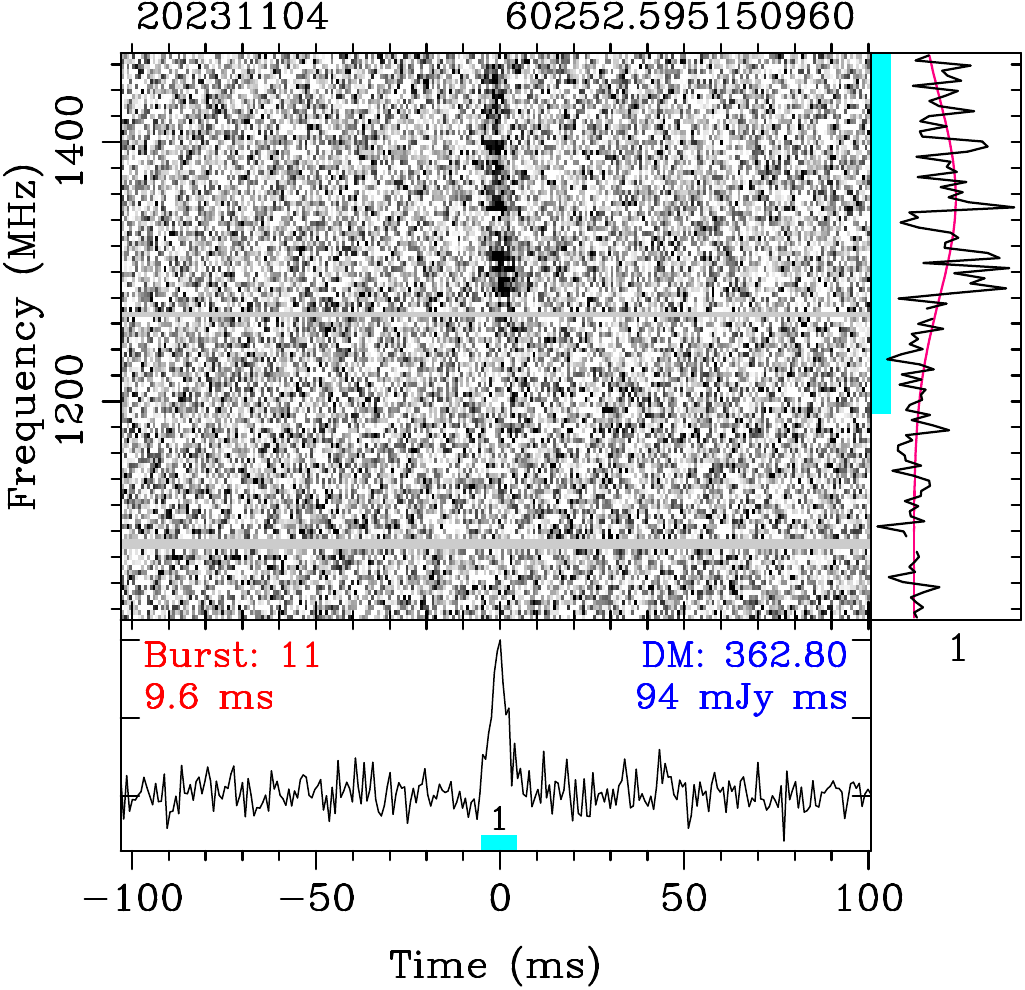}
\caption{({\textit{continued}})}
\end{figure*}
\addtocounter{figure}{-1}
\begin{figure*}
\flushleft
\includegraphics[height=0.29\linewidth]{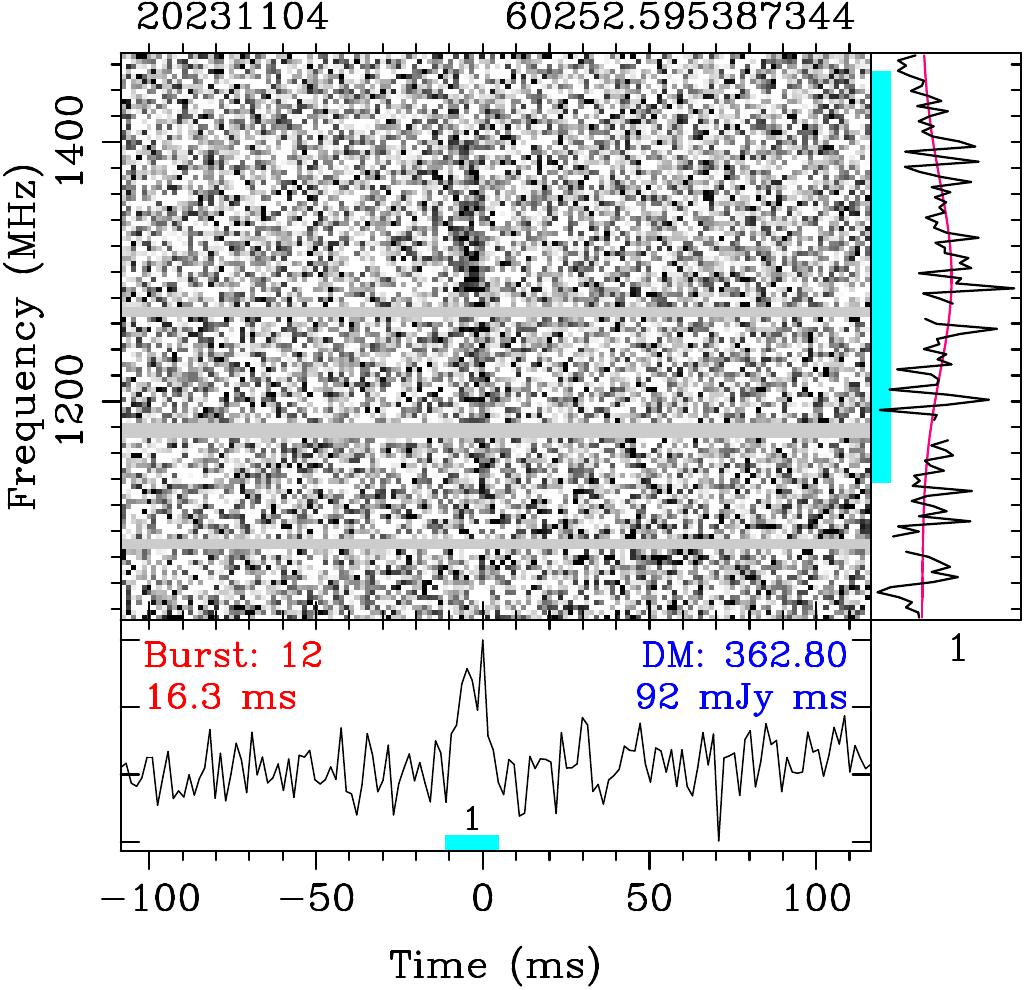}
\includegraphics[height=0.29\linewidth]{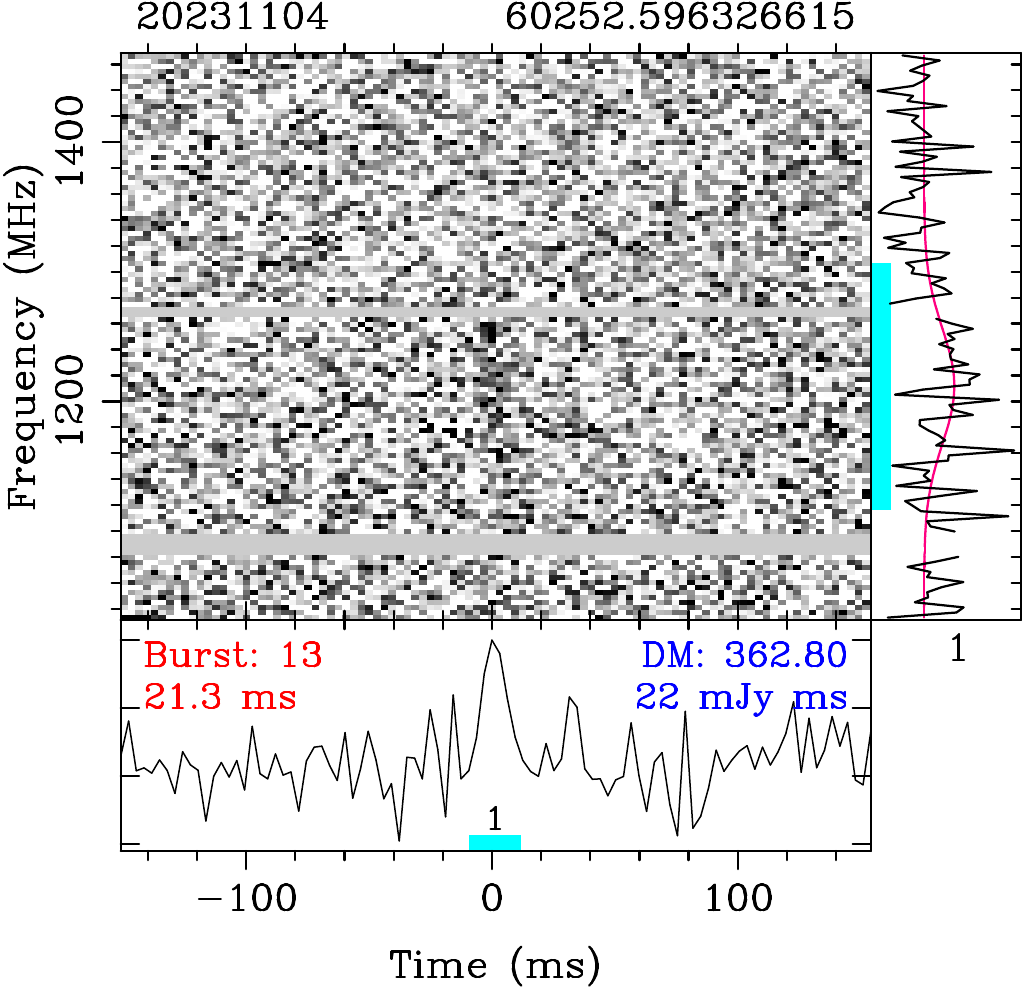}
\includegraphics[height=0.29\linewidth]{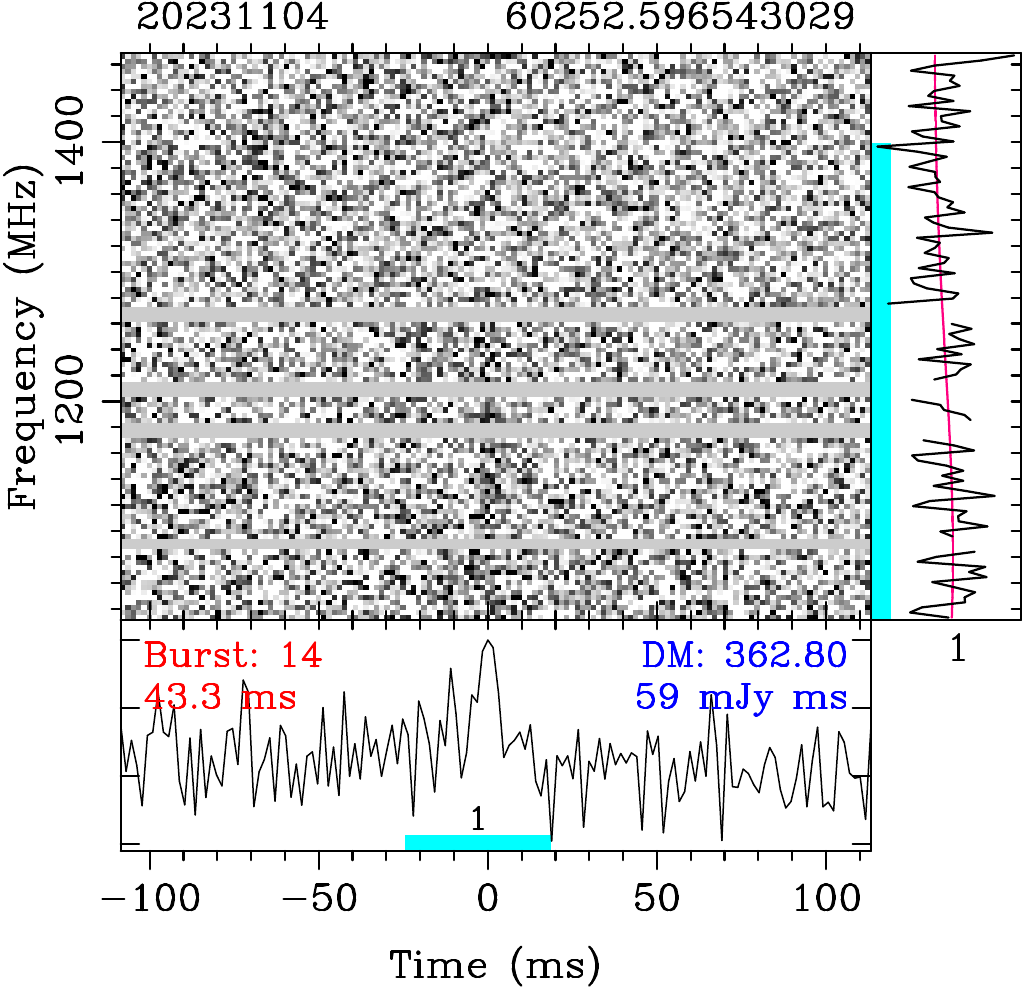}
\includegraphics[height=0.29\linewidth]{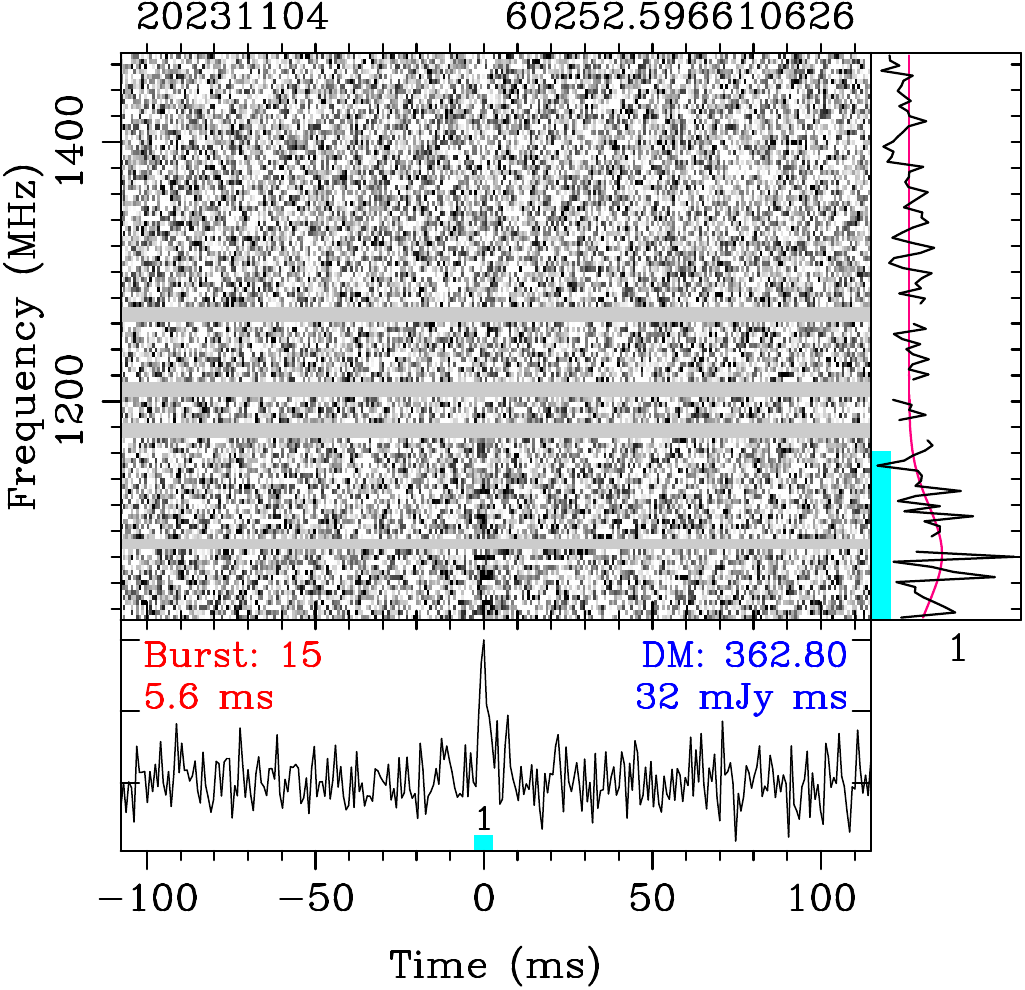}
\includegraphics[height=0.29\linewidth]{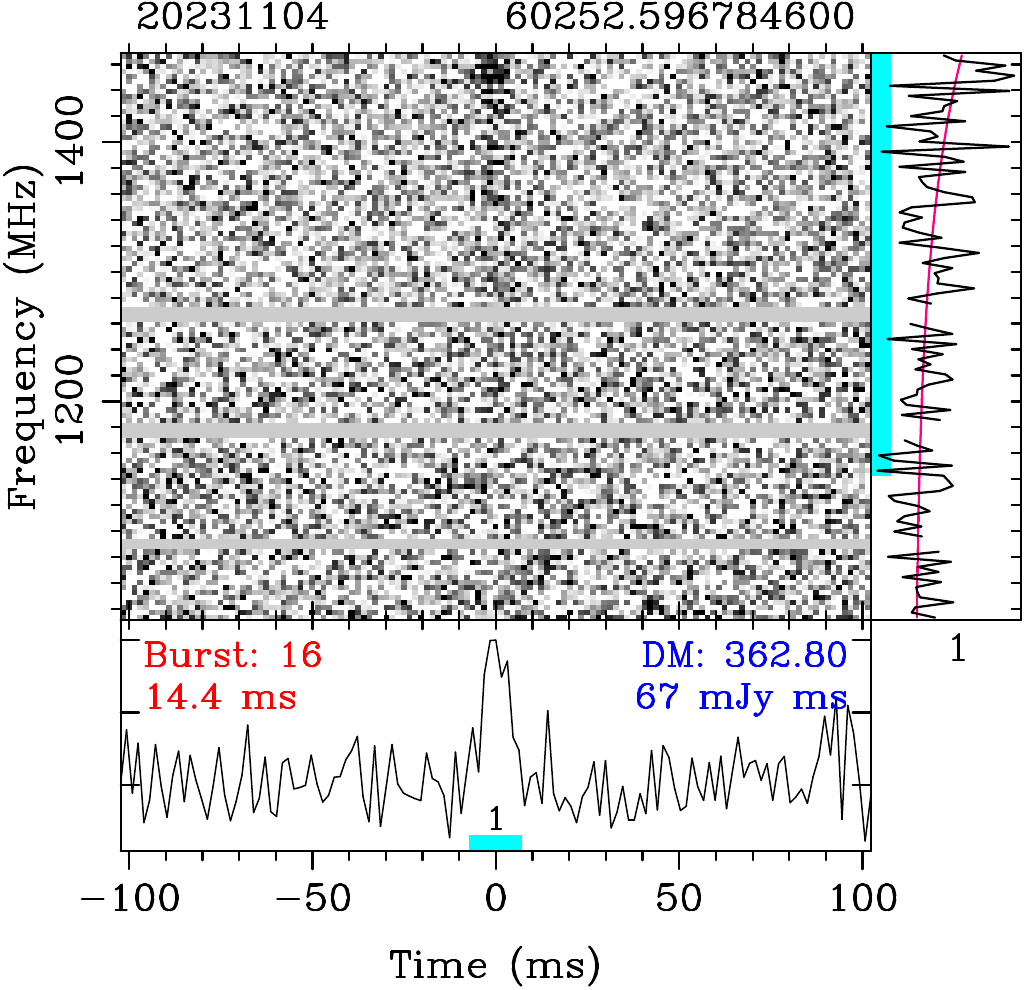}
\includegraphics[height=0.29\linewidth]{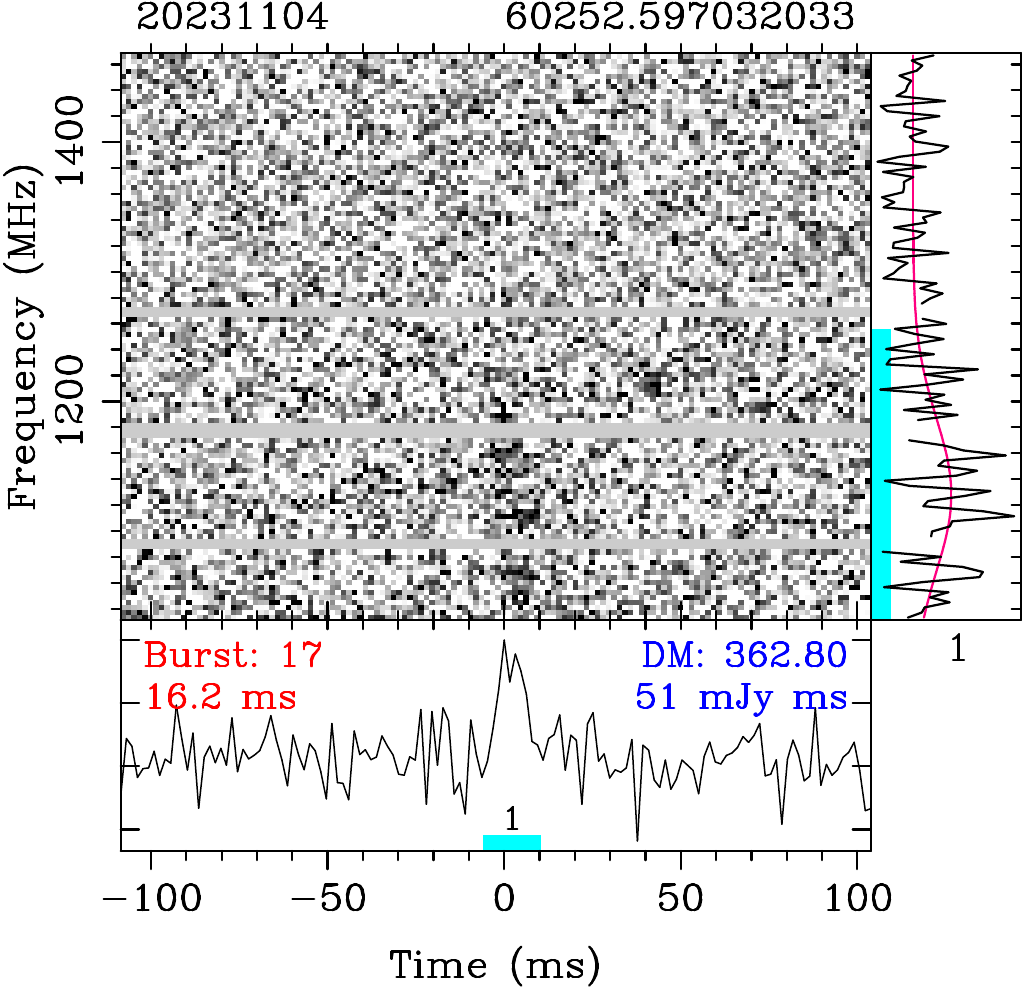}
\includegraphics[height=0.29\linewidth]{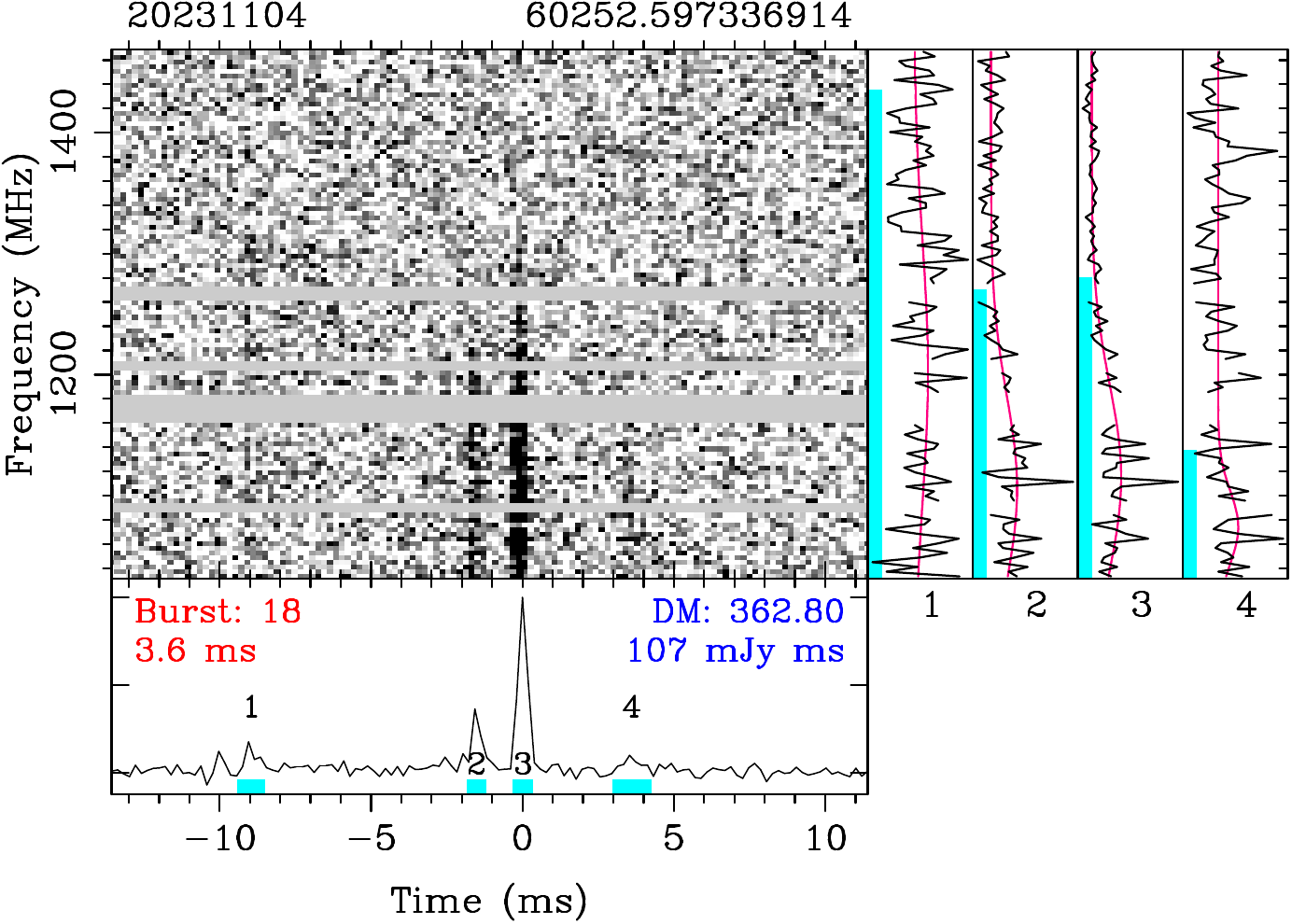}
\includegraphics[height=0.29\linewidth]{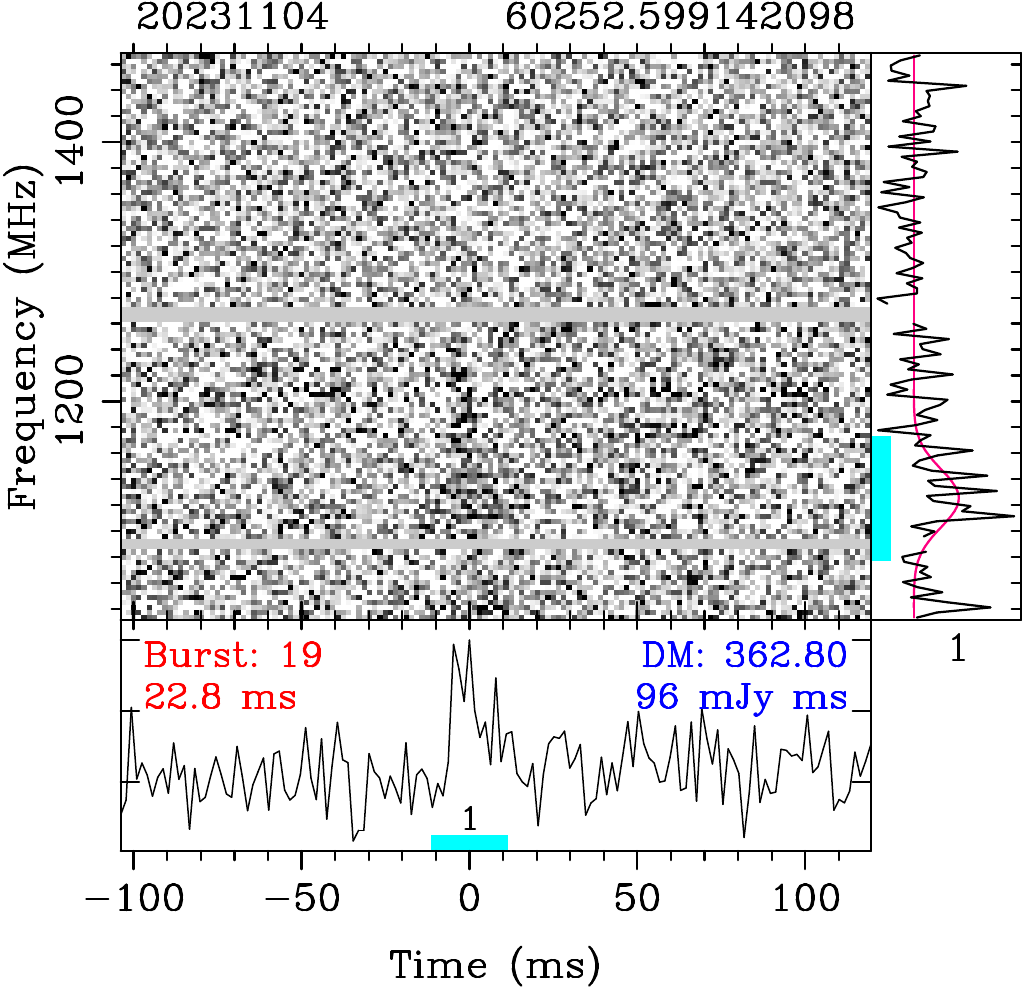}
\includegraphics[height=0.29\linewidth]{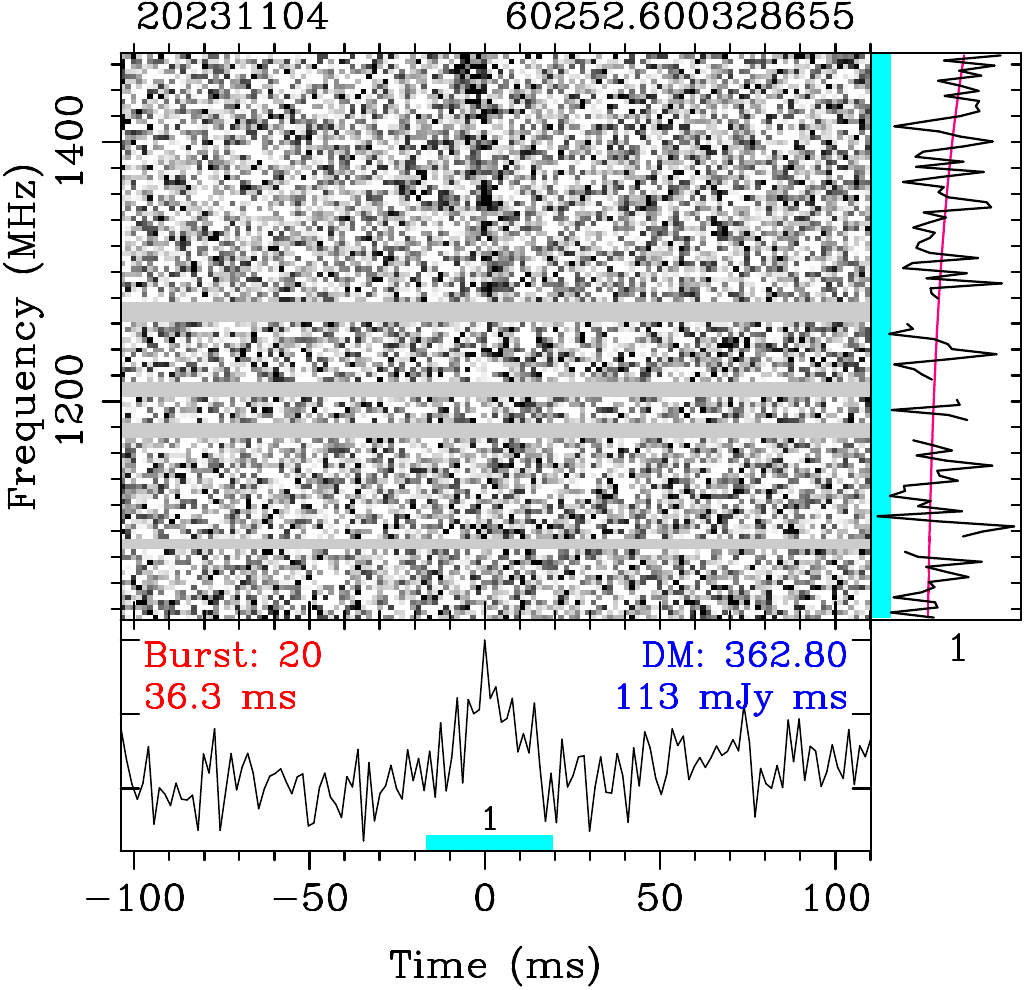}
\includegraphics[height=0.29\linewidth]{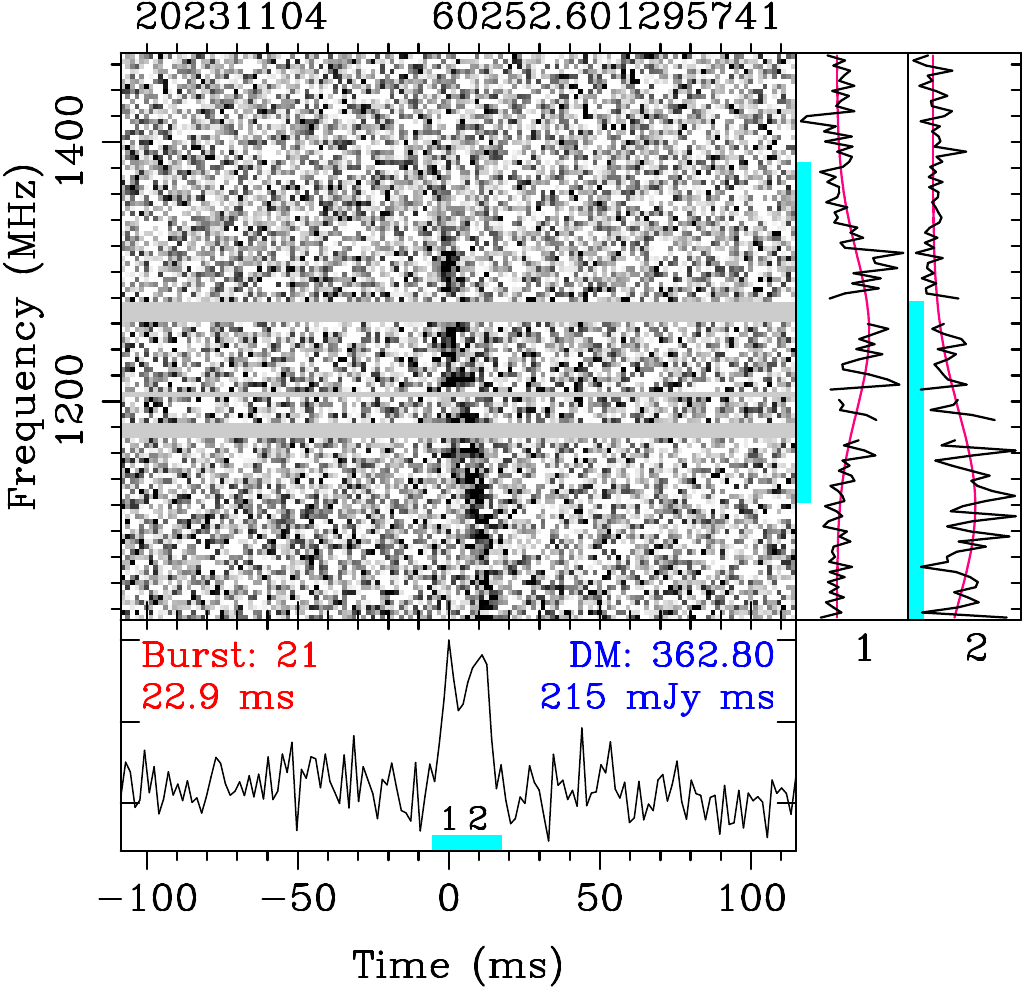}
\includegraphics[height=0.29\linewidth]{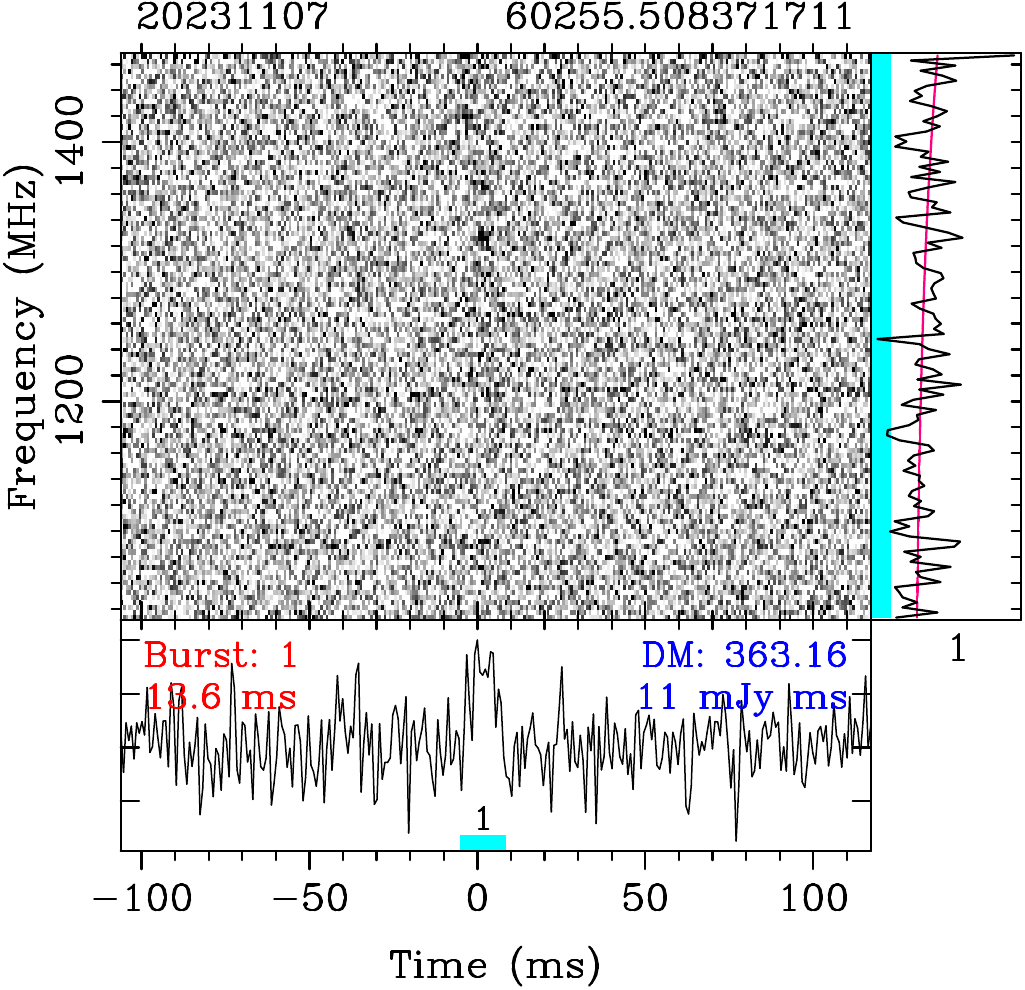}
\caption{({\textit{continued}})}
\end{figure*}
\addtocounter{figure}{-1}
\begin{figure*}
\flushleft
\includegraphics[height=0.29\linewidth]{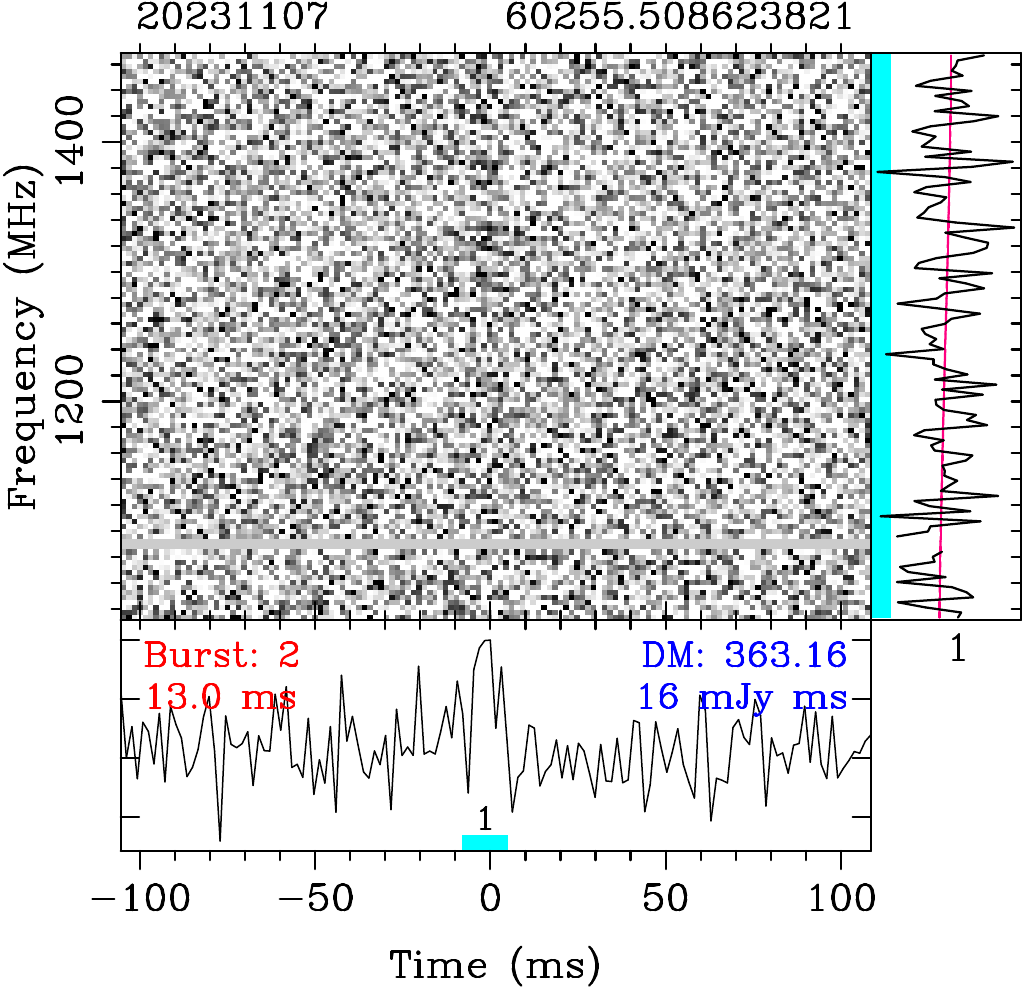}
\includegraphics[height=0.29\linewidth]{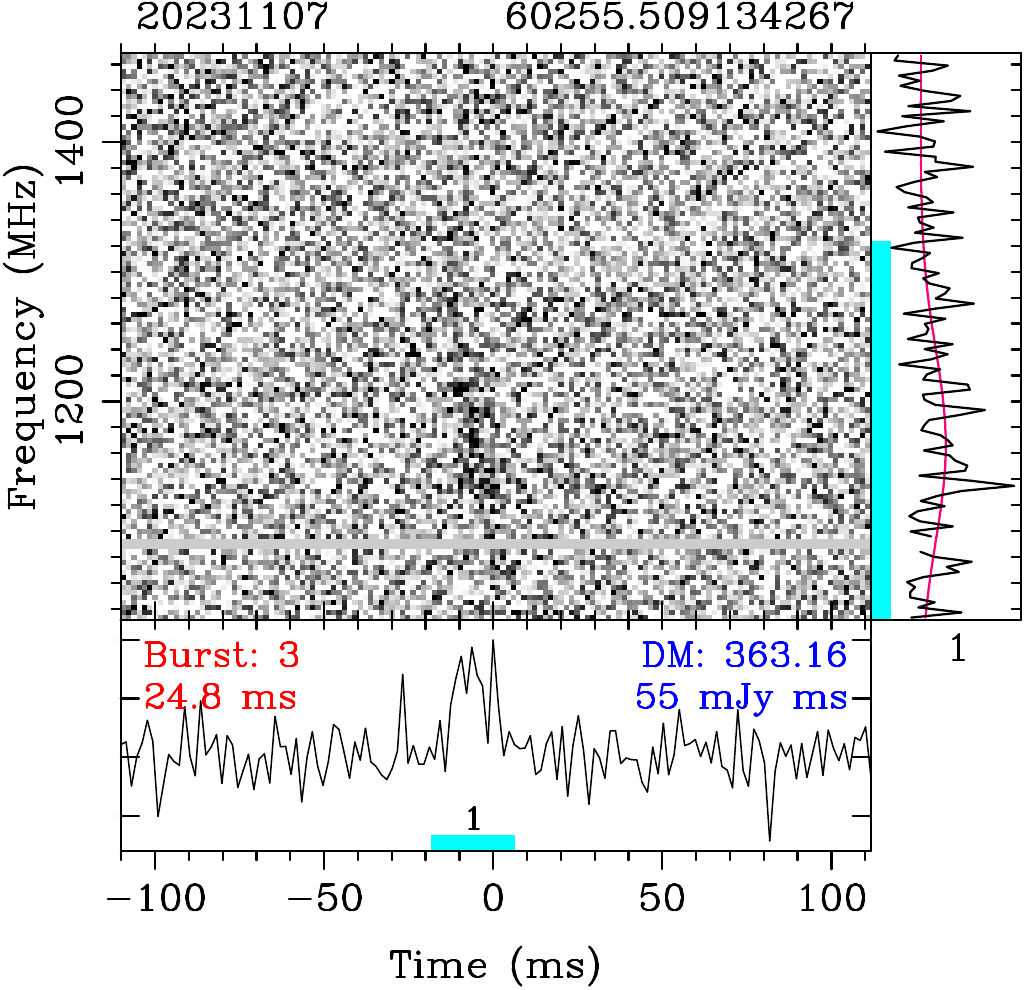}
\includegraphics[height=0.29\linewidth]{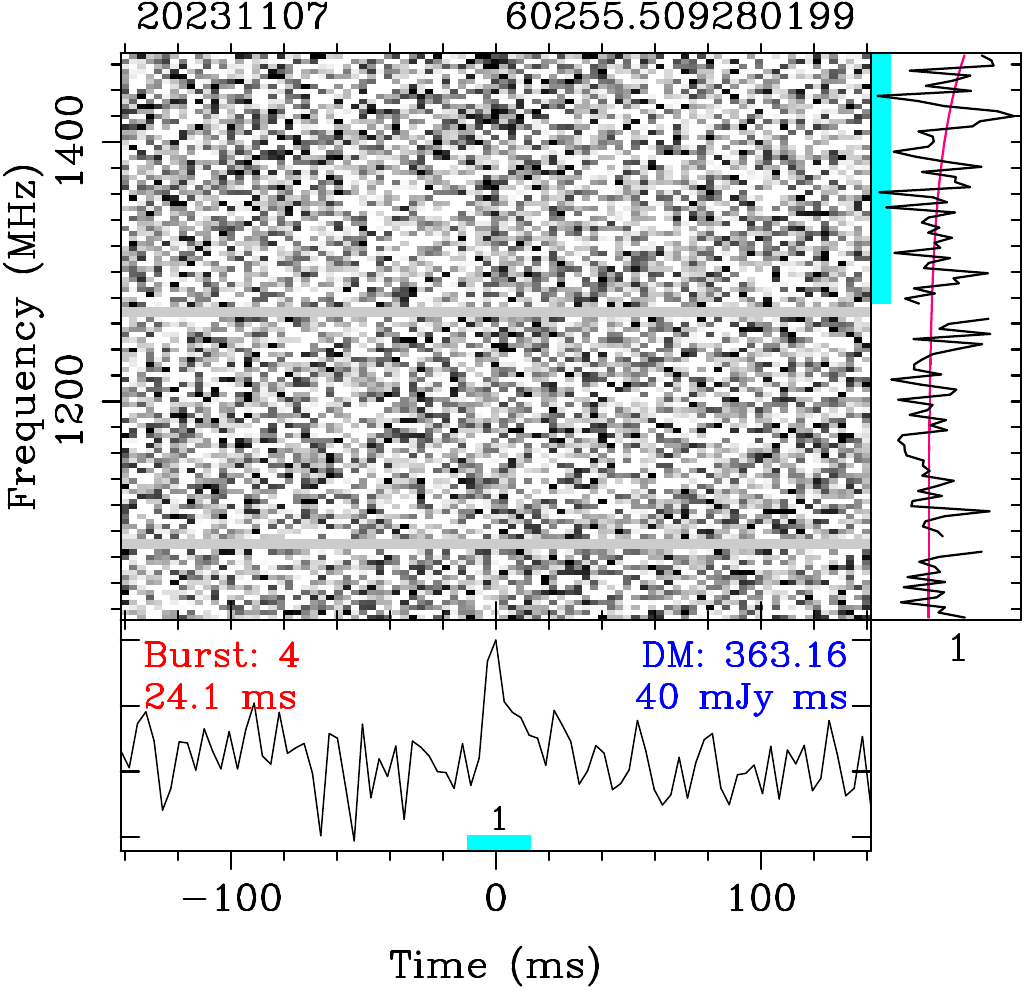}
\includegraphics[height=0.29\linewidth]{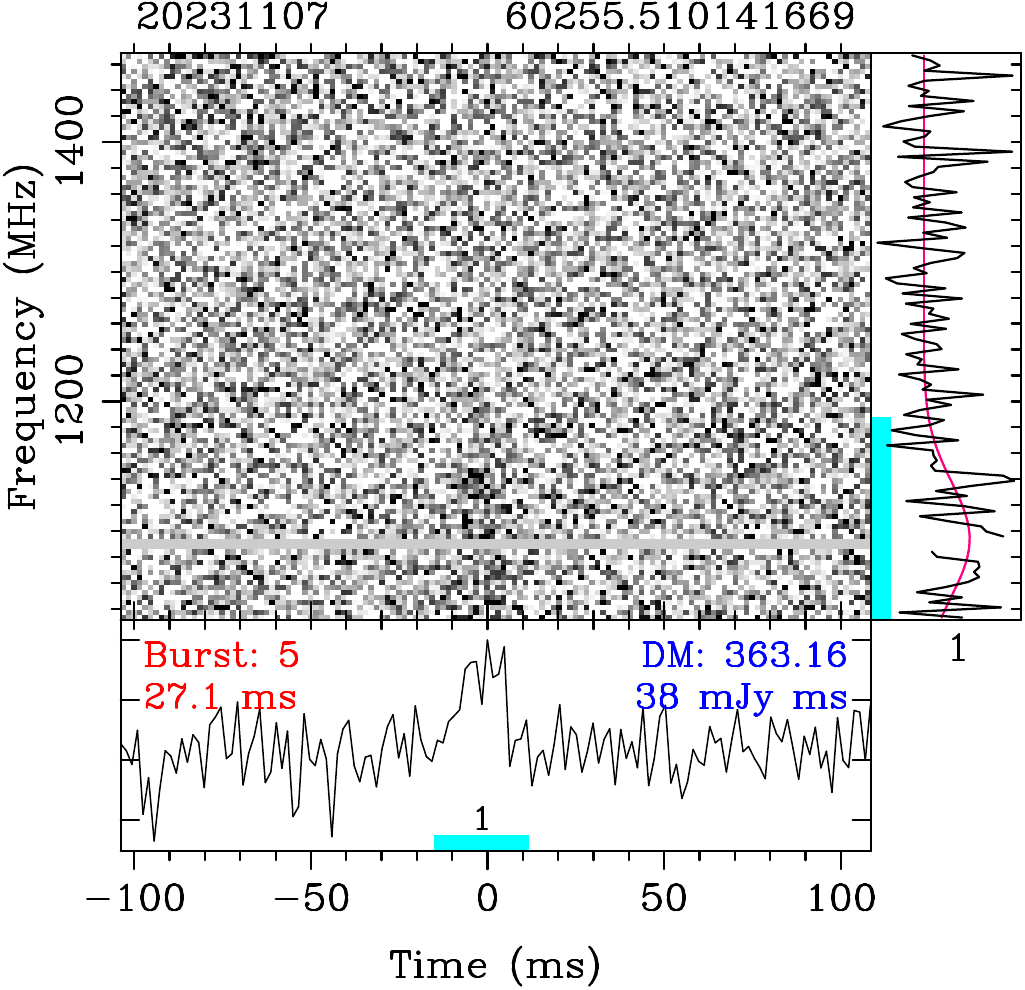}
\includegraphics[height=0.29\linewidth]{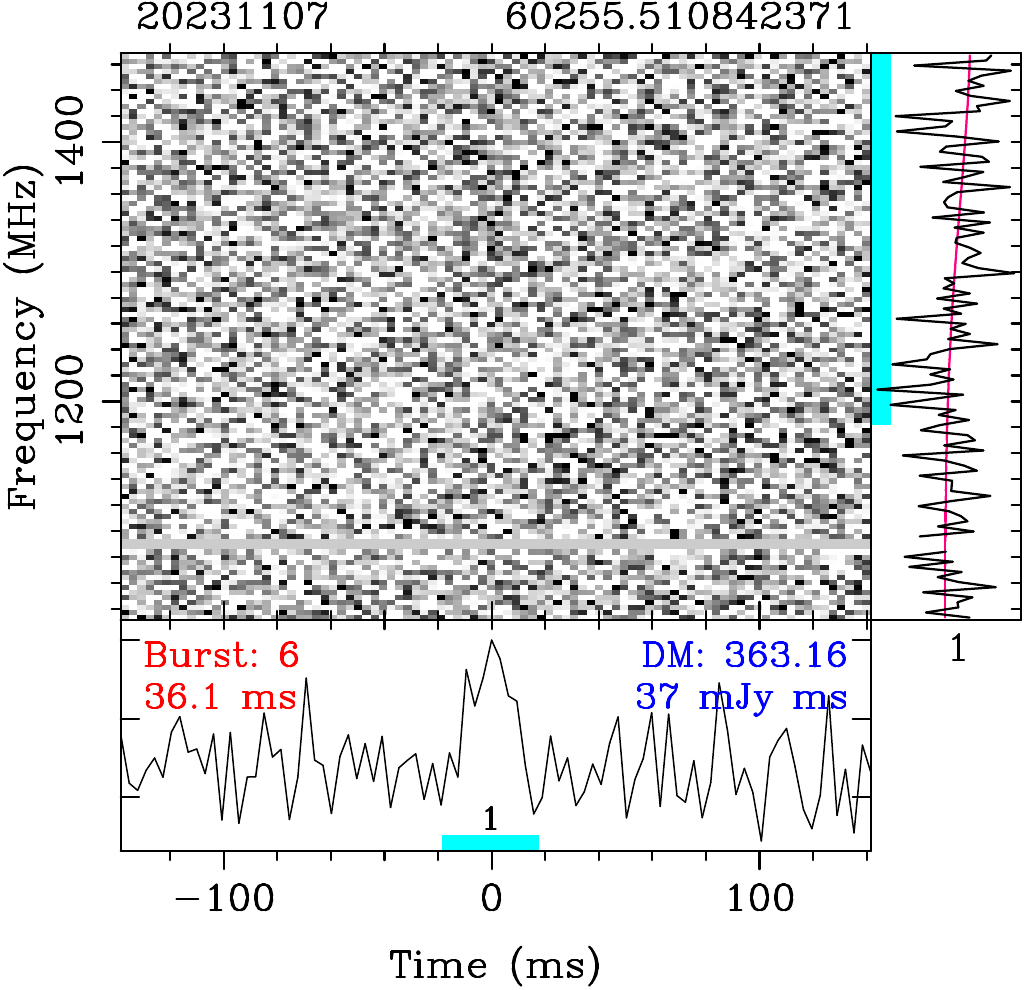}
\includegraphics[height=0.29\linewidth]{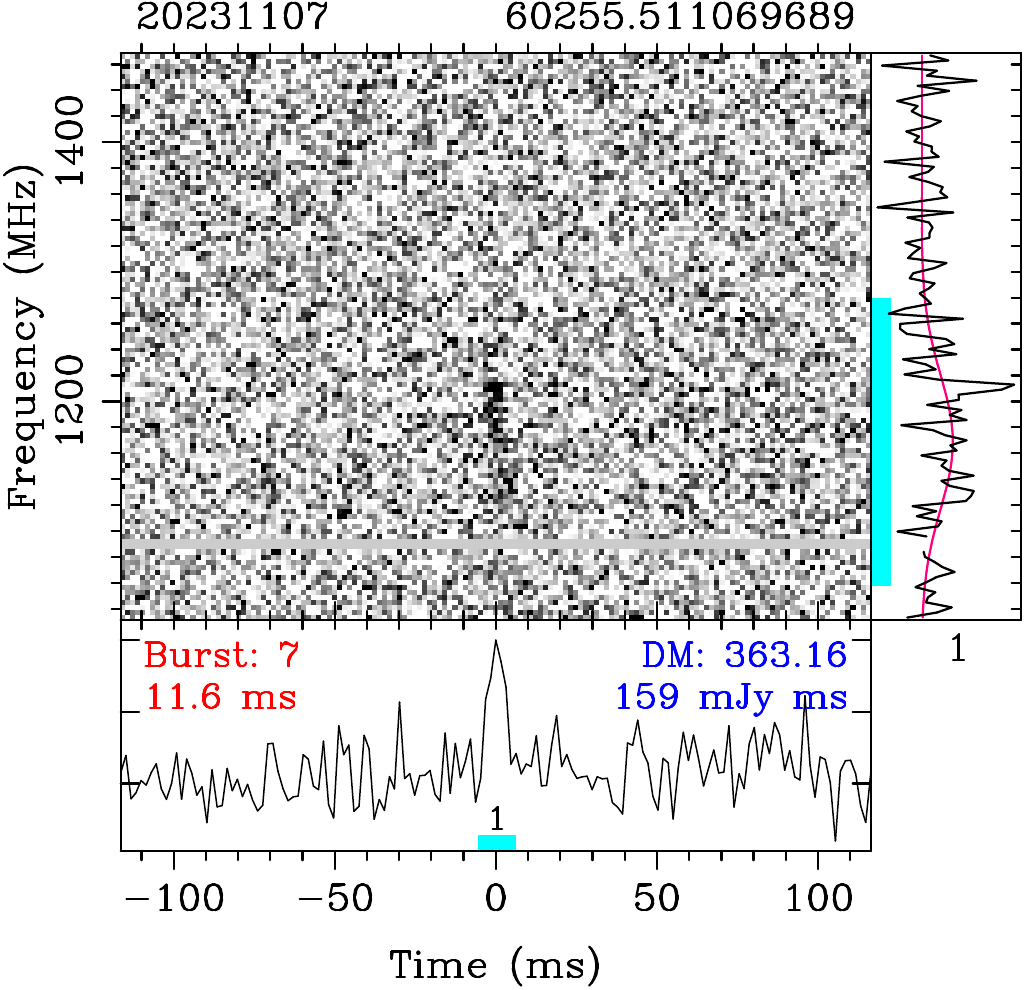}
\includegraphics[height=0.29\linewidth]{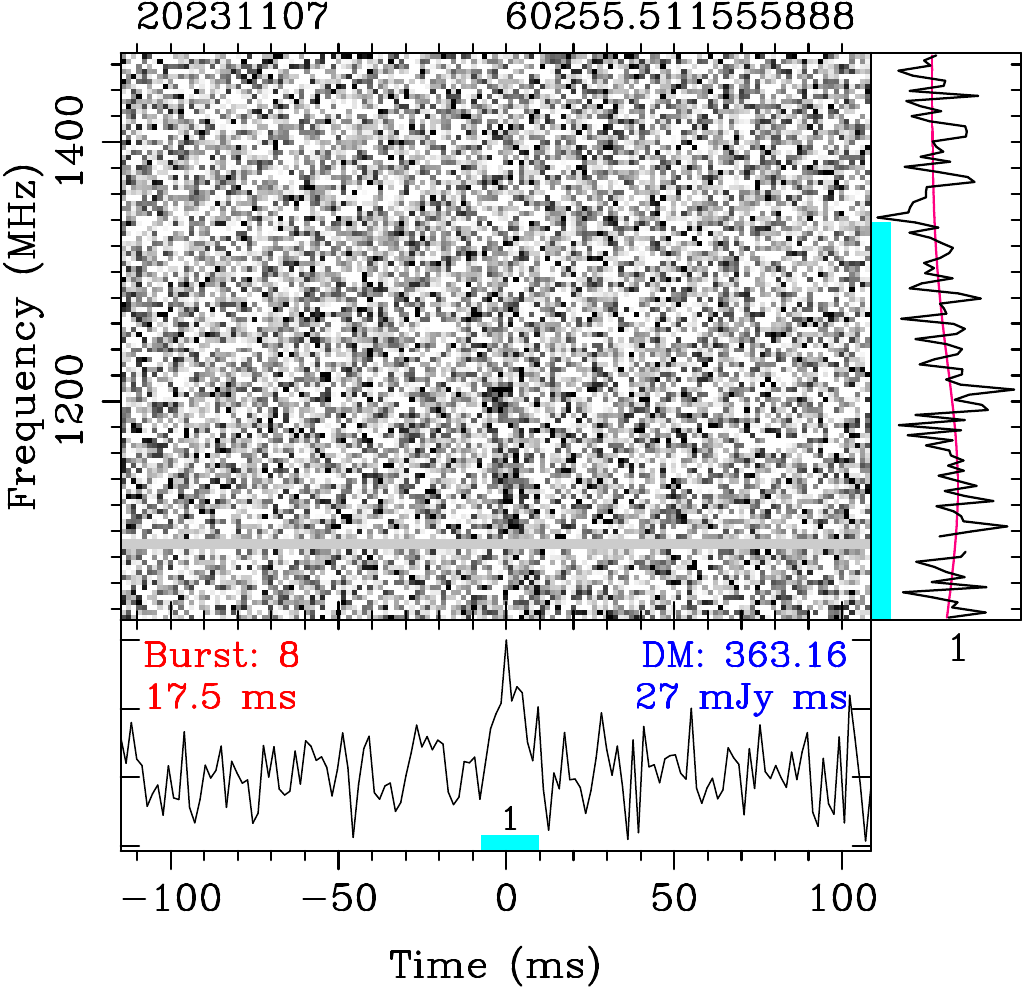}
\includegraphics[height=0.29\linewidth]{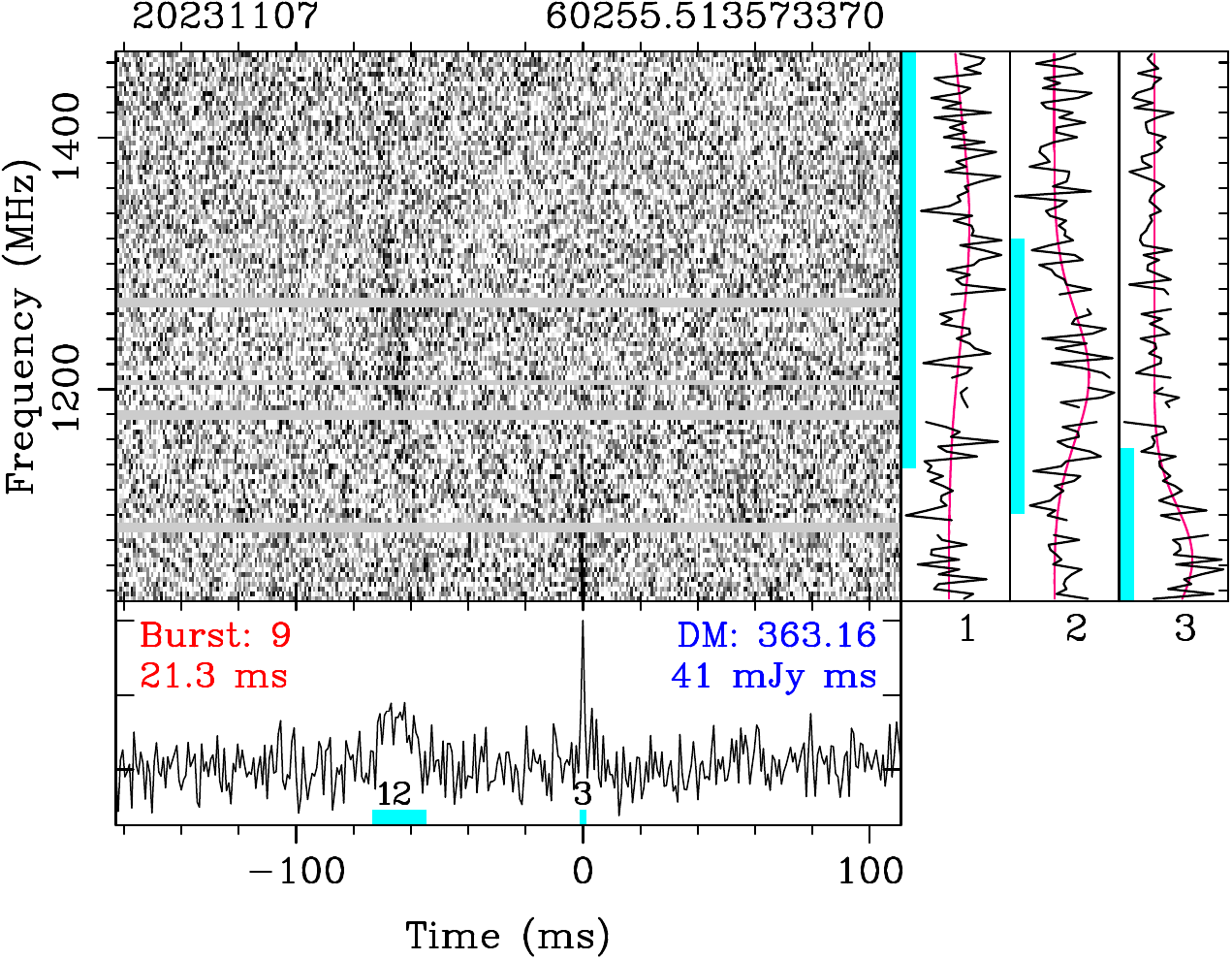}
\includegraphics[height=0.29\linewidth]{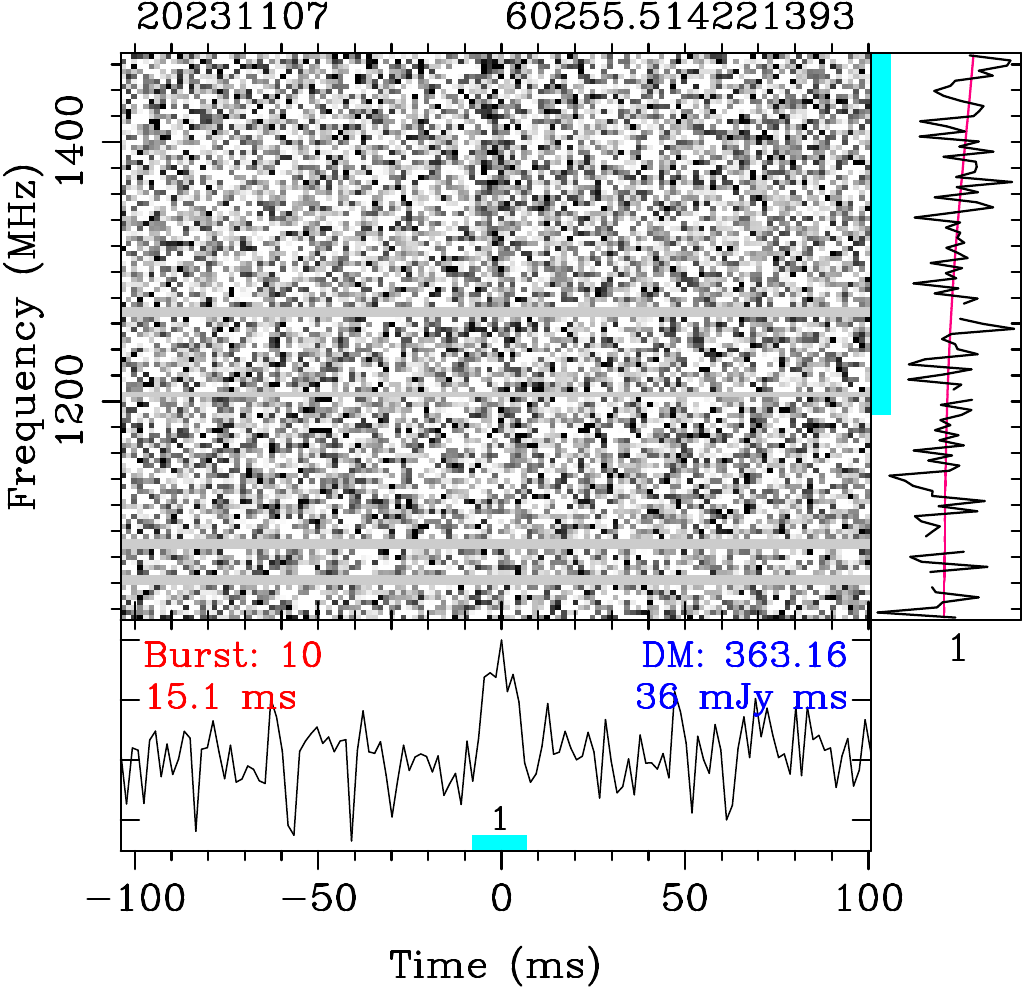}
\includegraphics[height=0.29\linewidth]{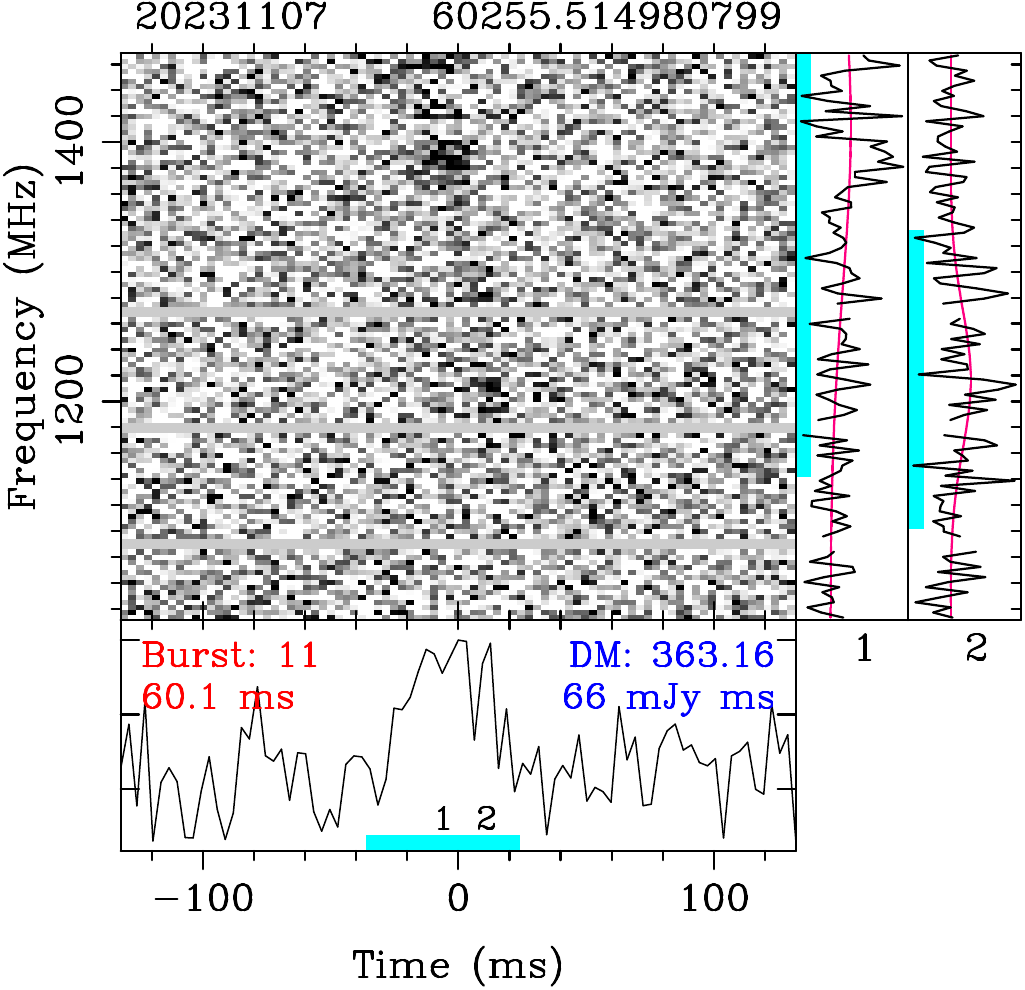}
\includegraphics[height=0.29\linewidth]{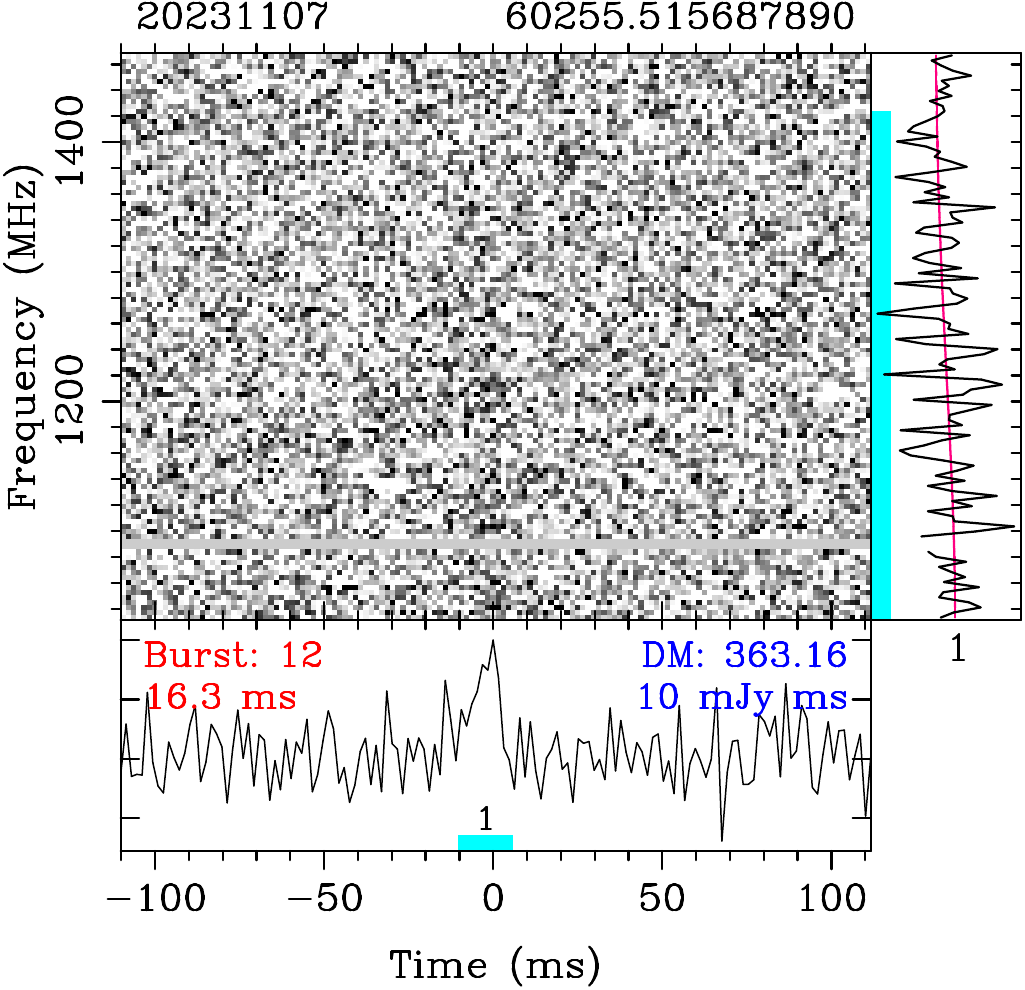}
\includegraphics[height=0.29\linewidth]{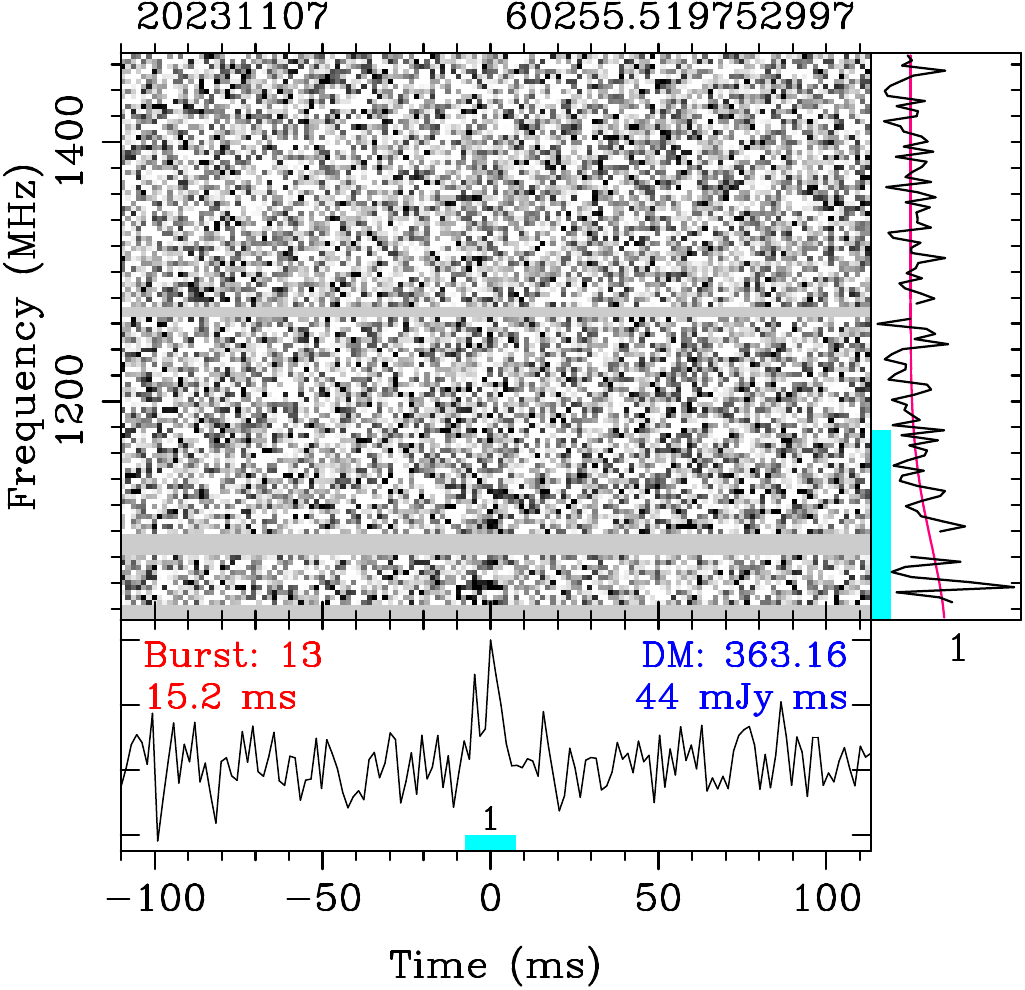}
\caption{({\textit{continued}})}
\end{figure*}
\addtocounter{figure}{-1}
\begin{figure*}
\flushleft
\includegraphics[height=0.29\linewidth]{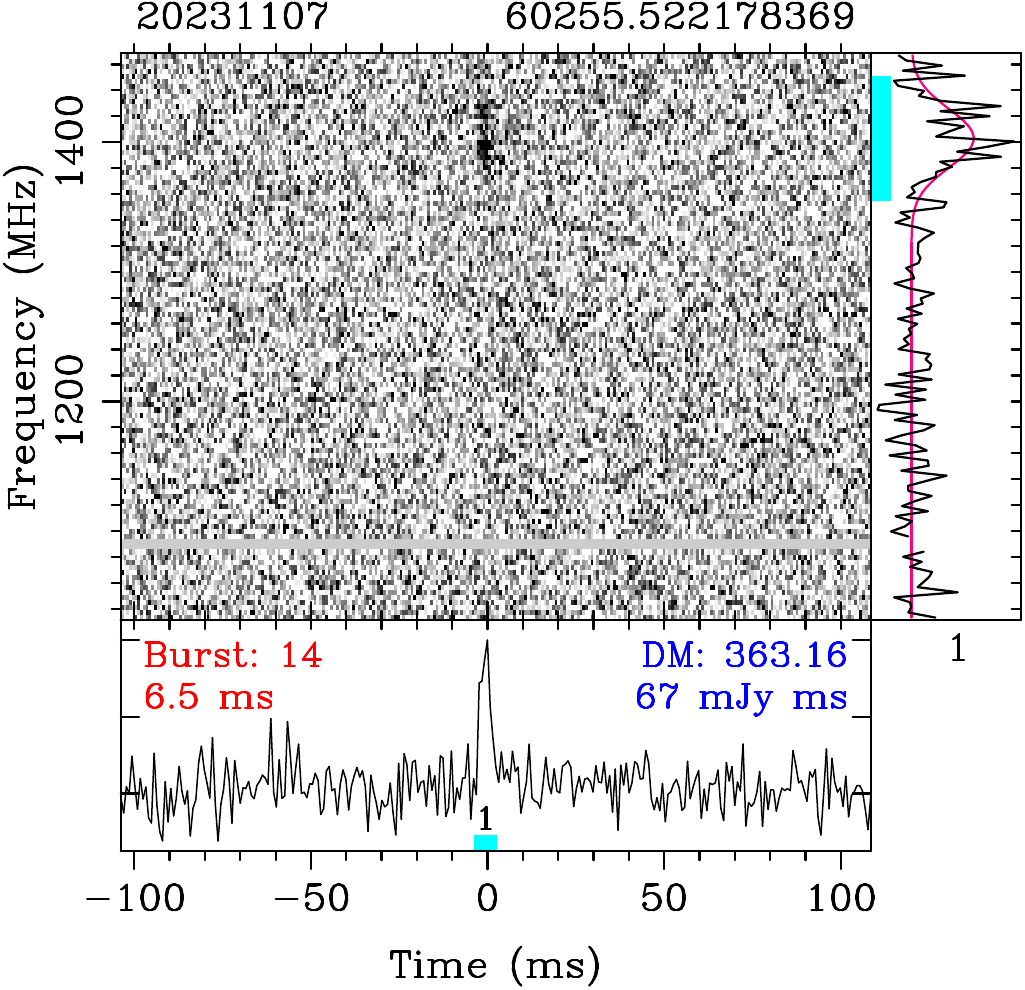}
\includegraphics[height=0.29\linewidth]{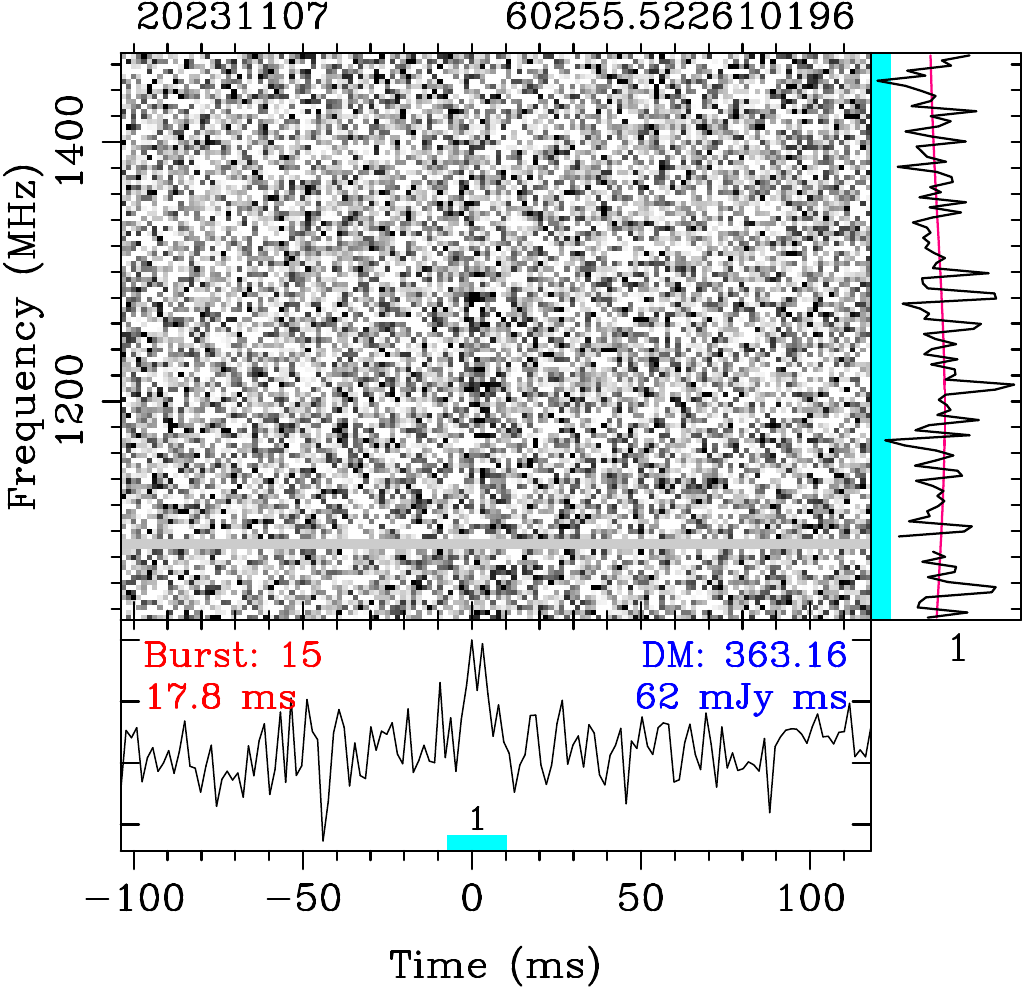}
\includegraphics[height=0.29\linewidth]{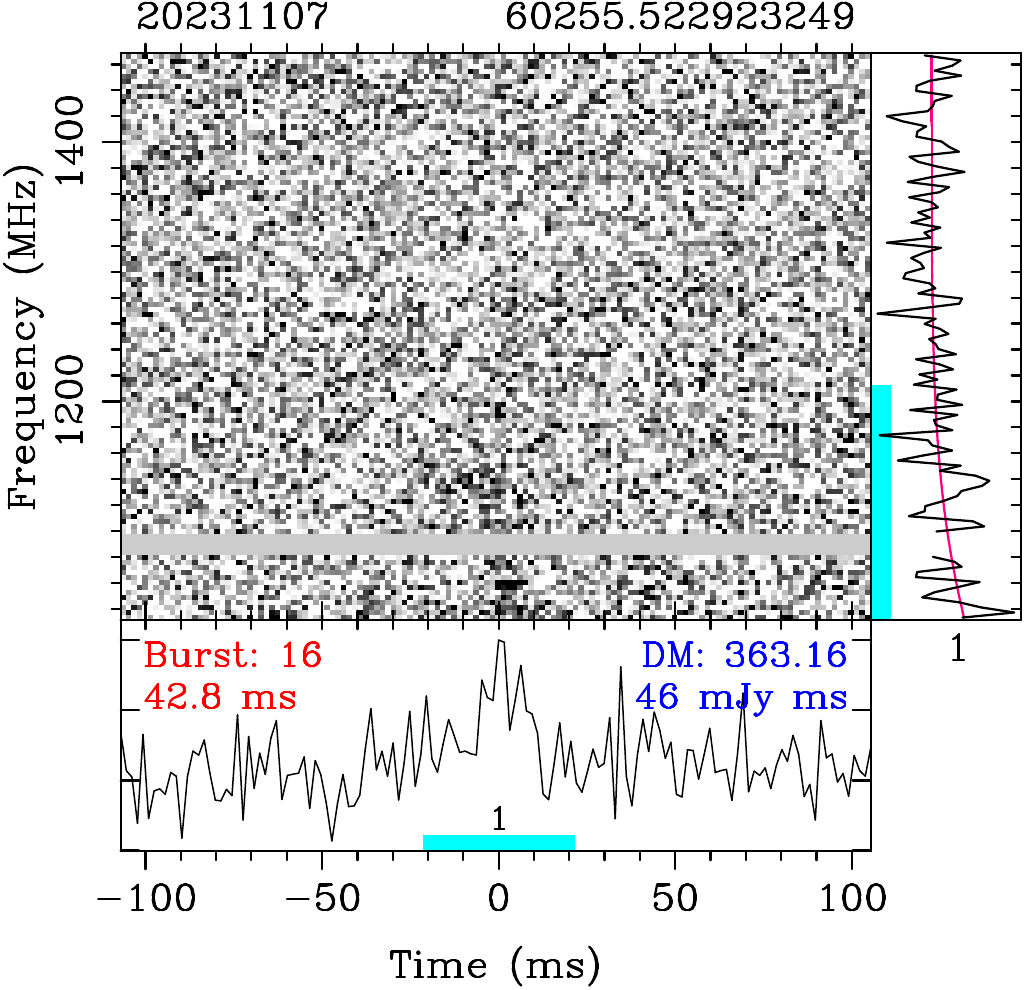}
\includegraphics[height=0.29\linewidth]{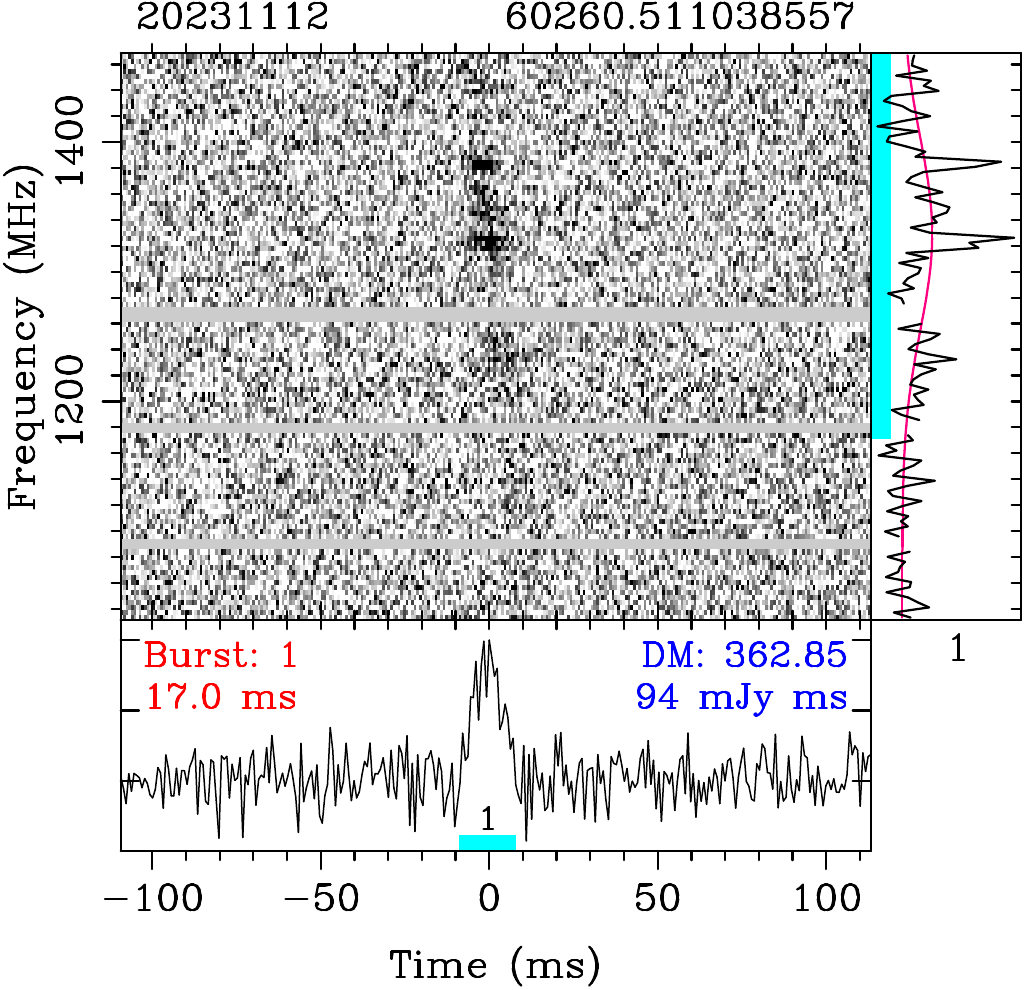}
\includegraphics[height=0.29\linewidth]{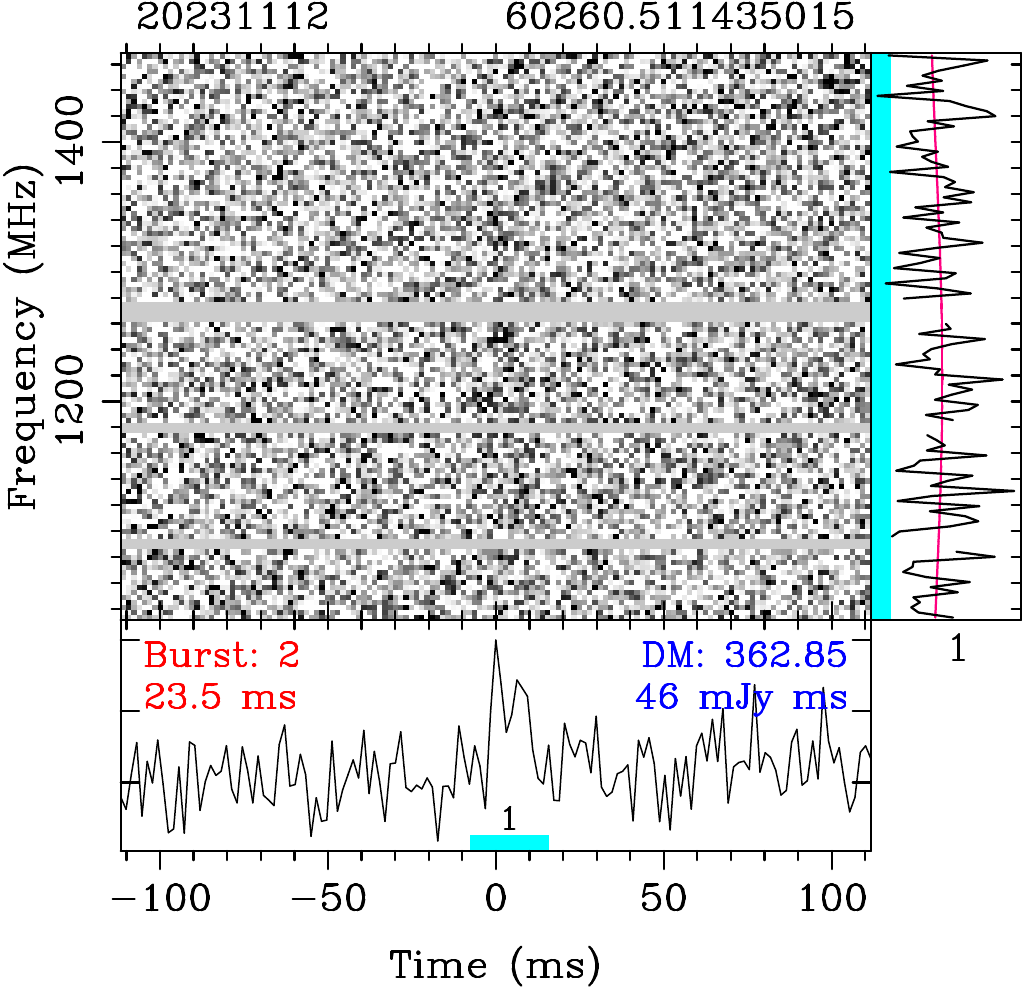}
\includegraphics[height=0.29\linewidth]{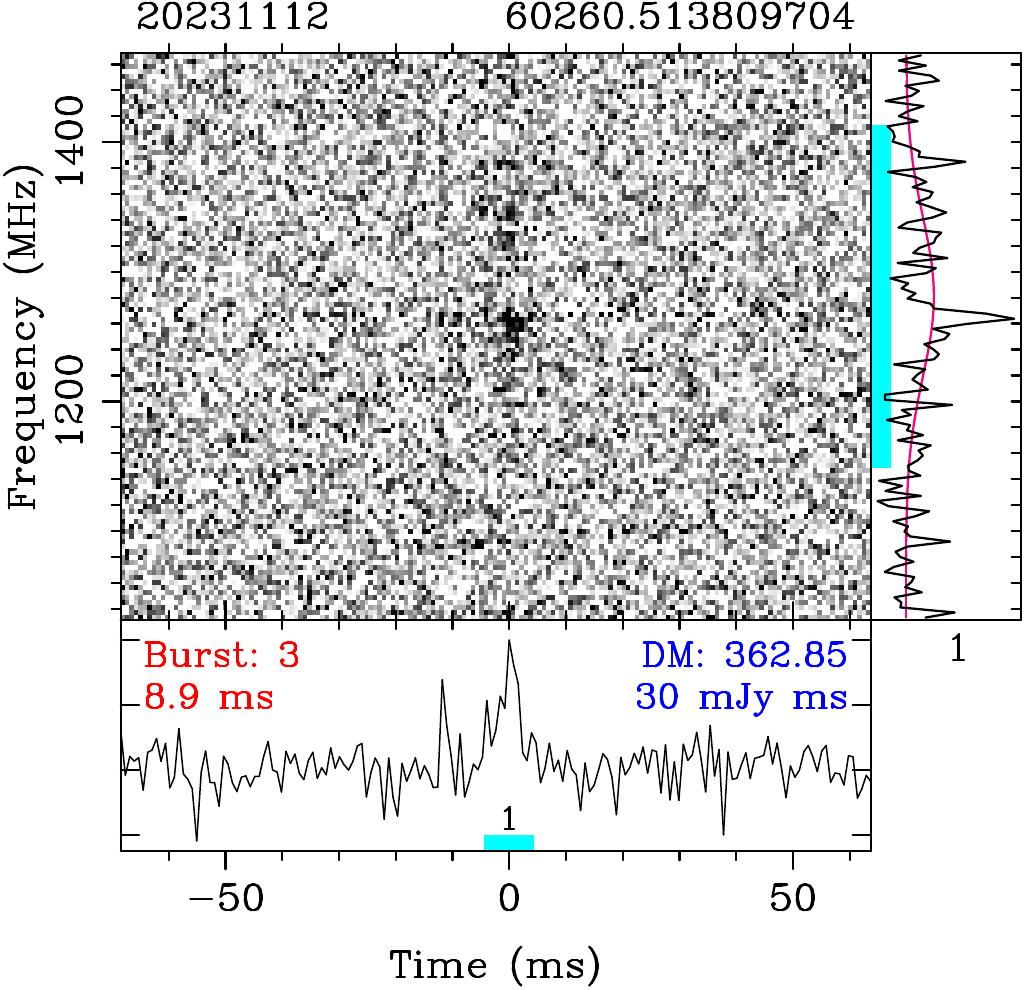}
\includegraphics[height=0.29\linewidth]{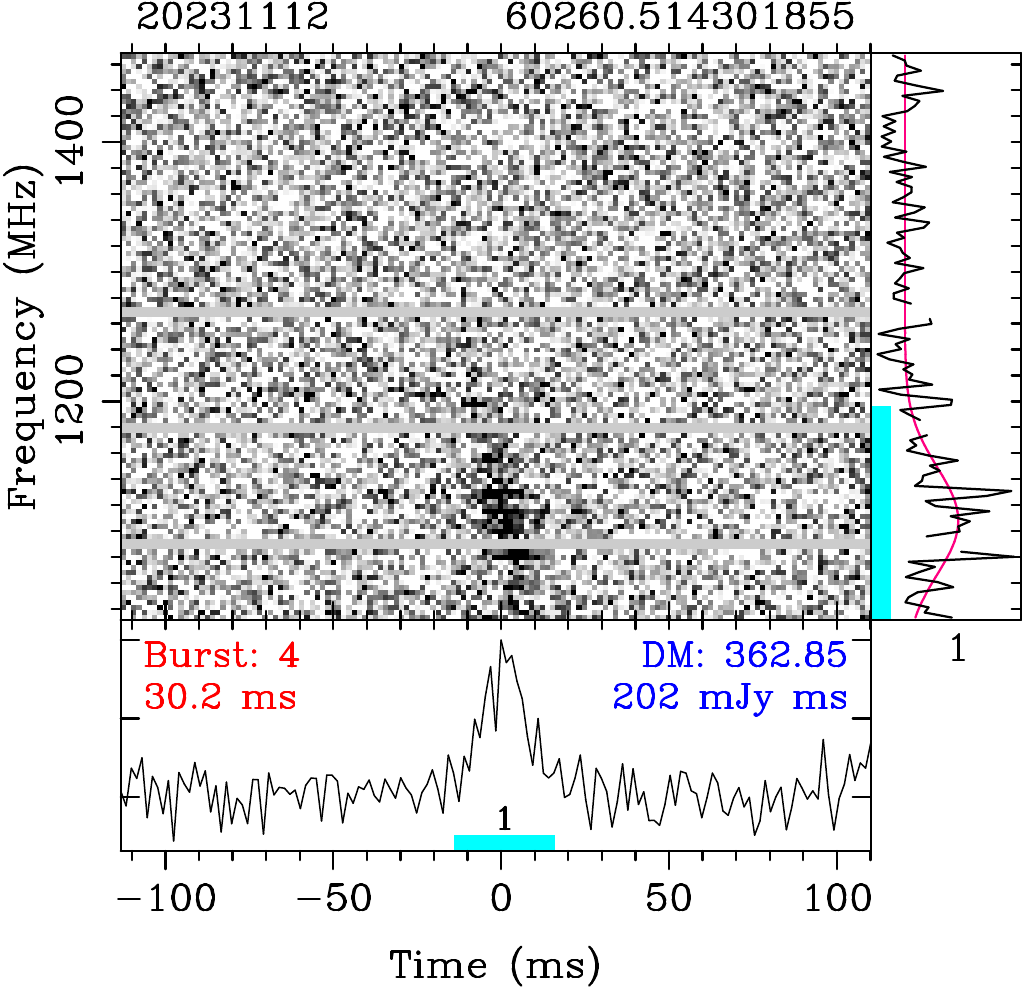}
\includegraphics[height=0.29\linewidth]{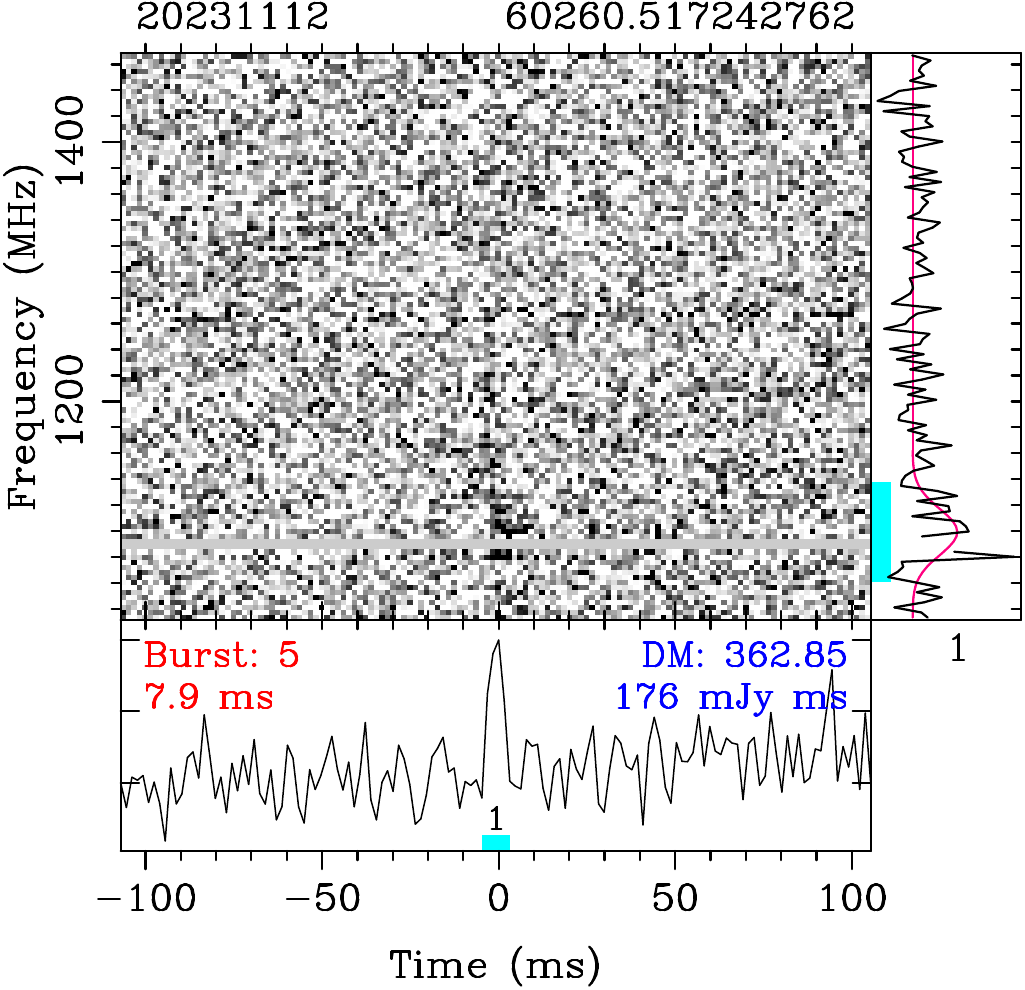}
\includegraphics[height=0.29\linewidth]{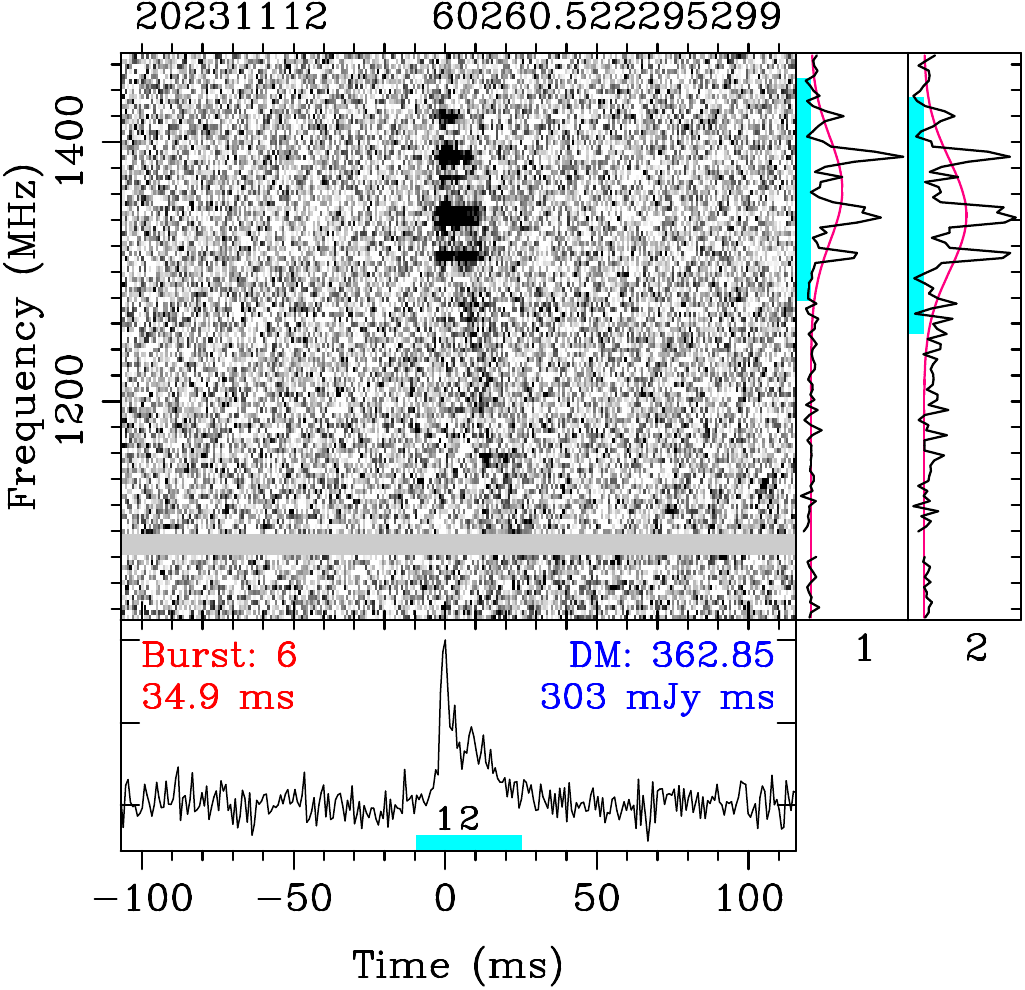}
\includegraphics[height=0.29\linewidth]{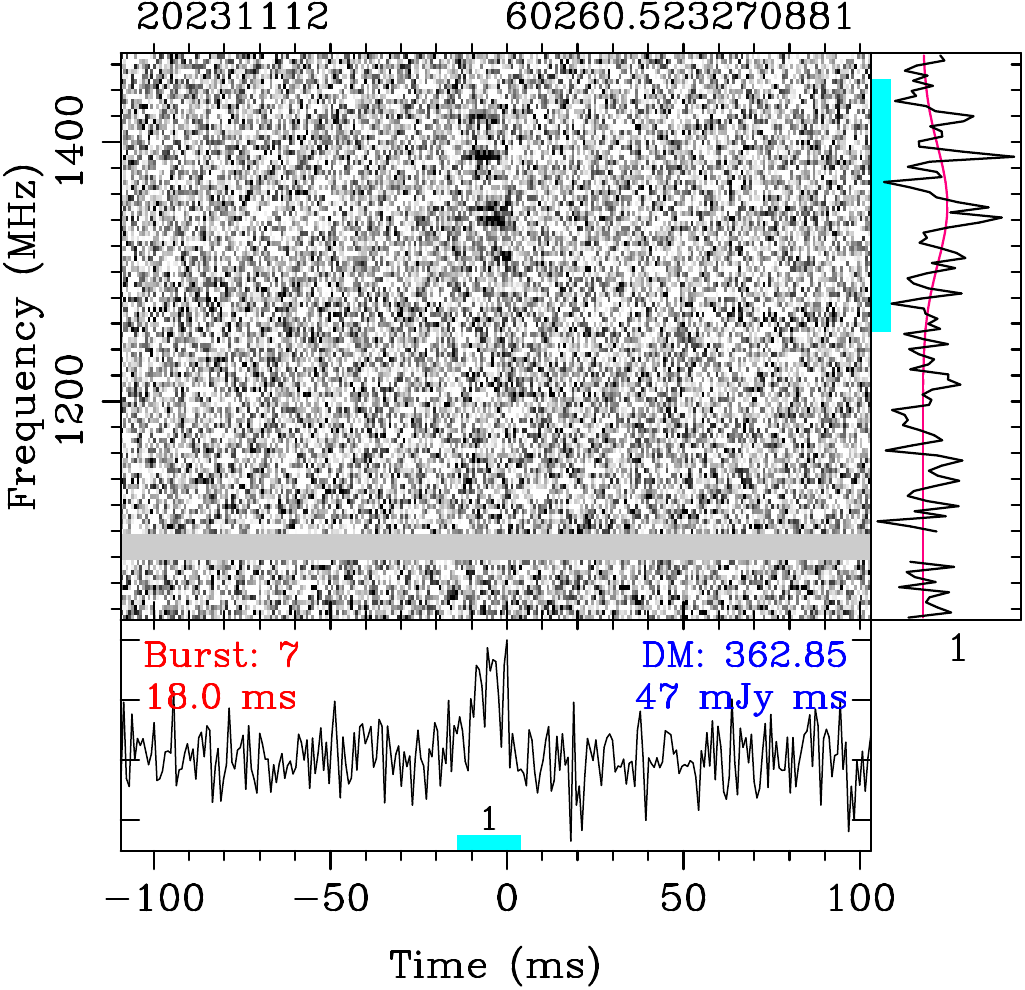}
\includegraphics[height=0.29\linewidth]{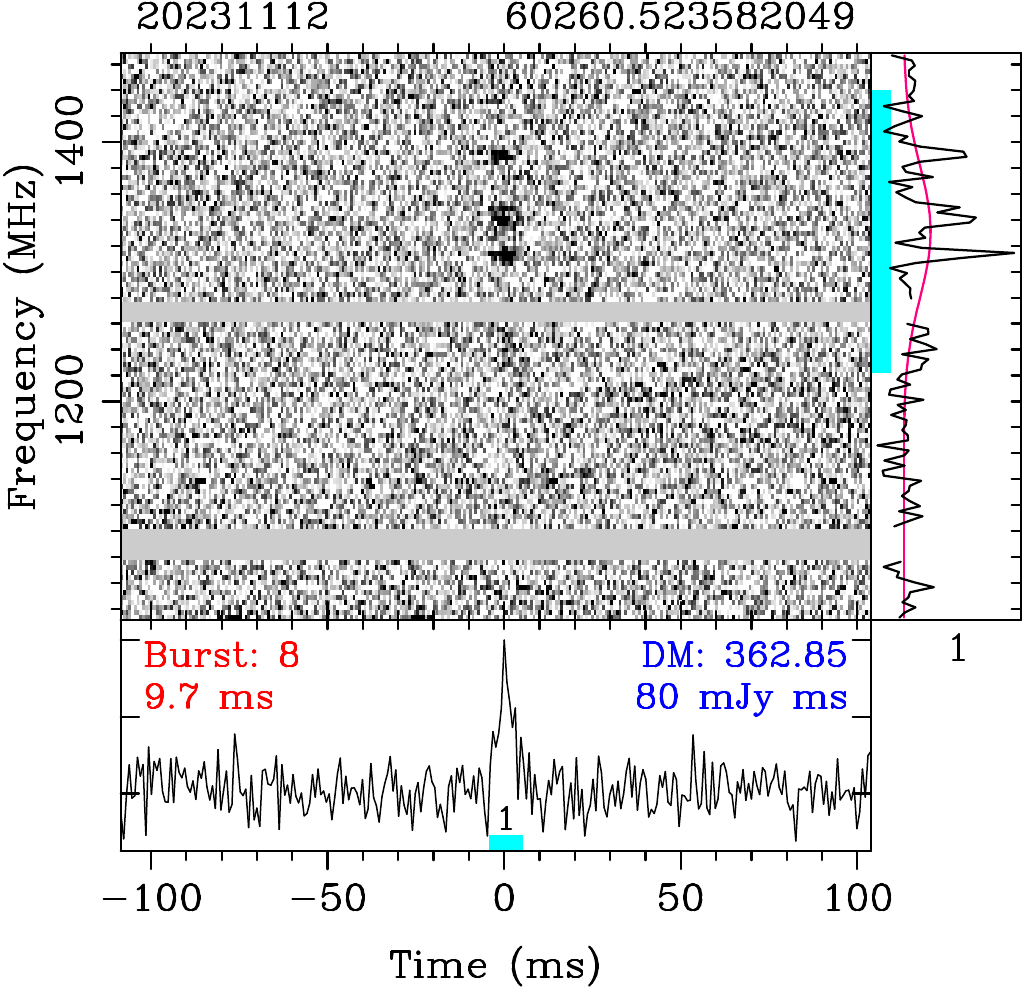}
\includegraphics[height=0.29\linewidth]{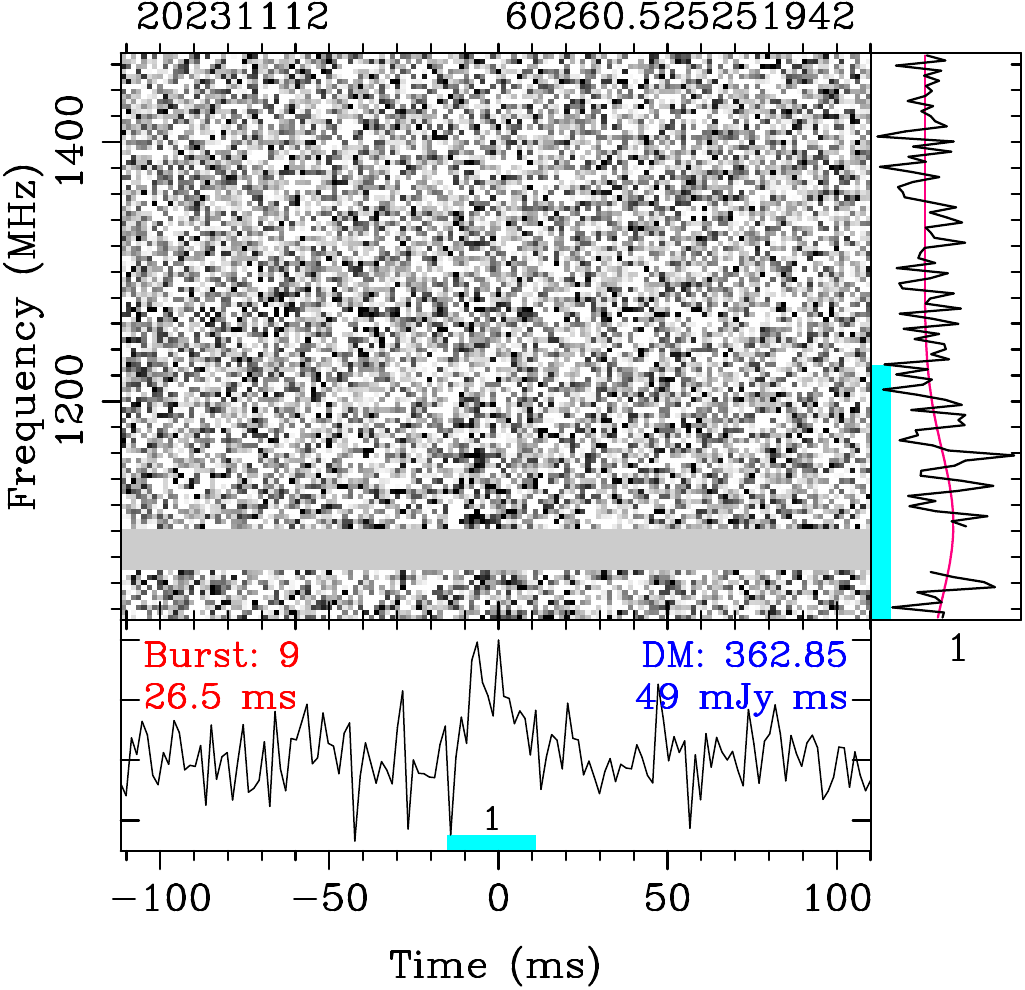}
\caption{({\textit{continued}})}
\end{figure*}
\addtocounter{figure}{-1}
\begin{figure*}
\flushleft
\includegraphics[height=0.29\linewidth]{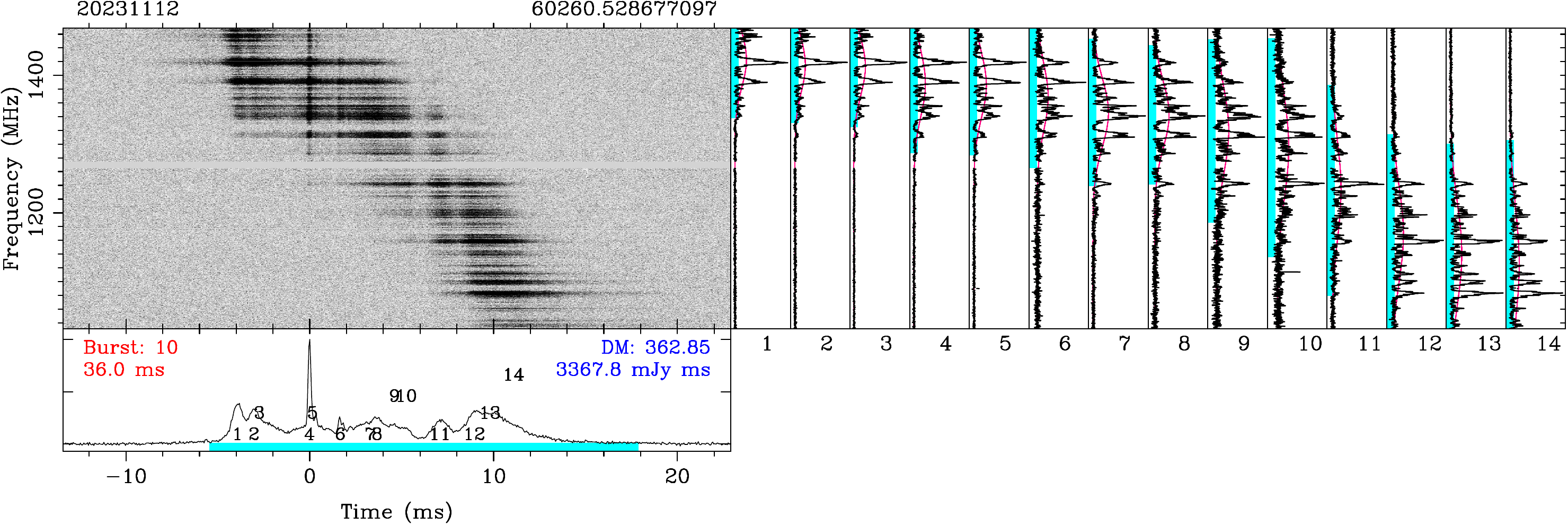}
\includegraphics[height=0.29\linewidth]{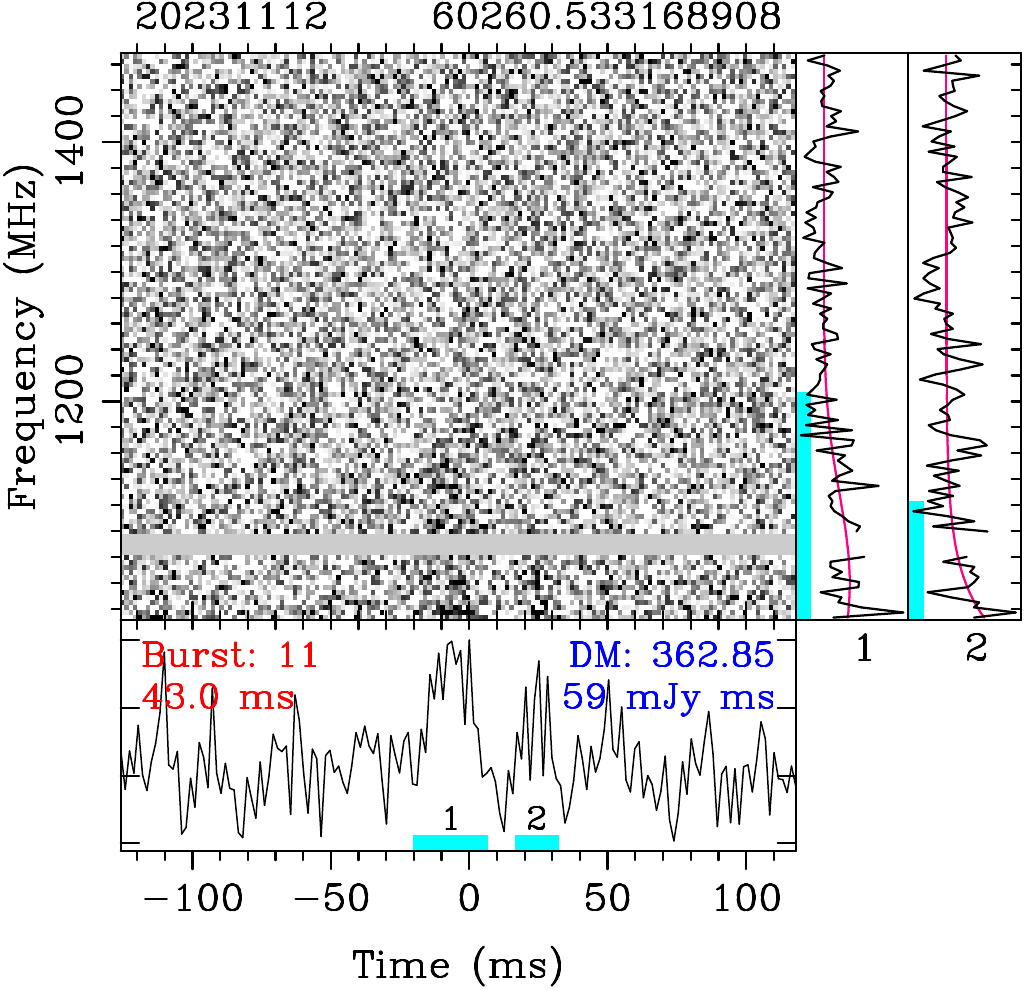}
\includegraphics[height=0.29\linewidth]{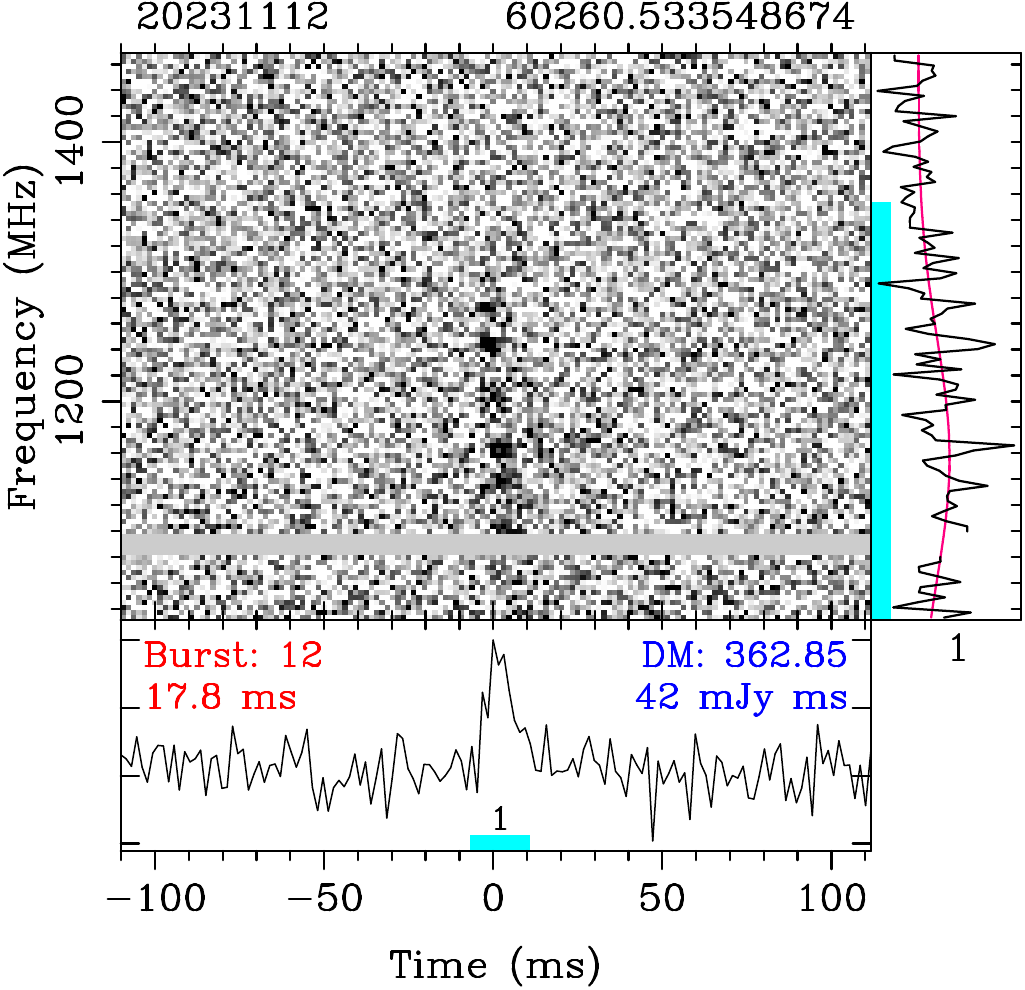}
\includegraphics[height=0.29\linewidth]{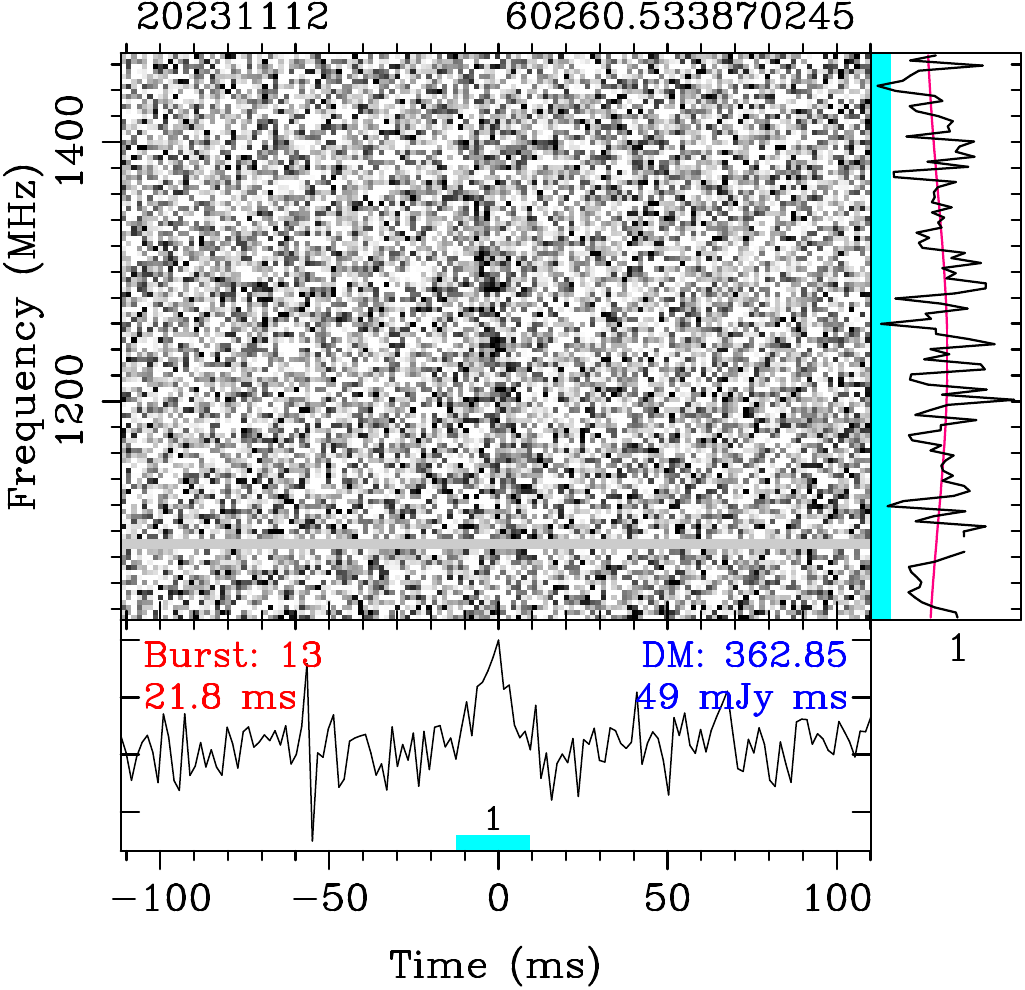}
\includegraphics[height=0.29\linewidth]{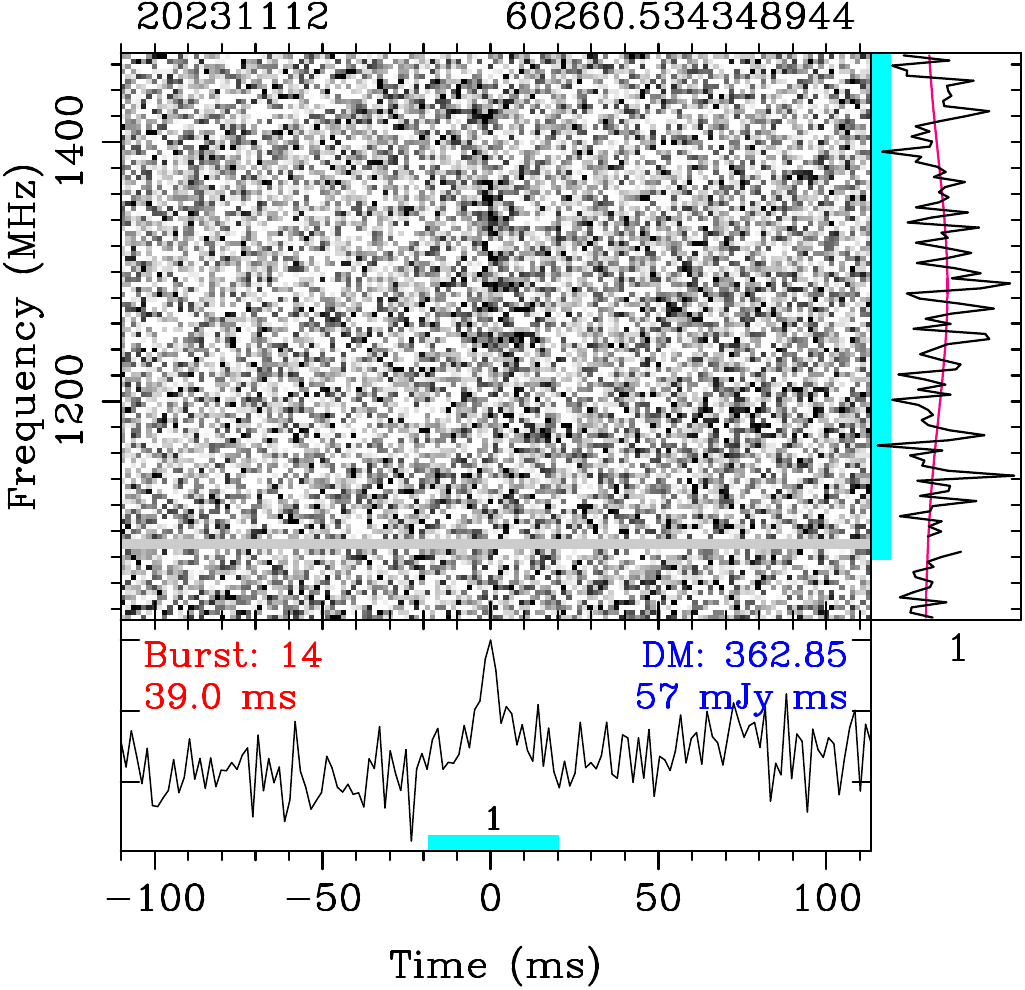}
\includegraphics[height=0.29\linewidth]{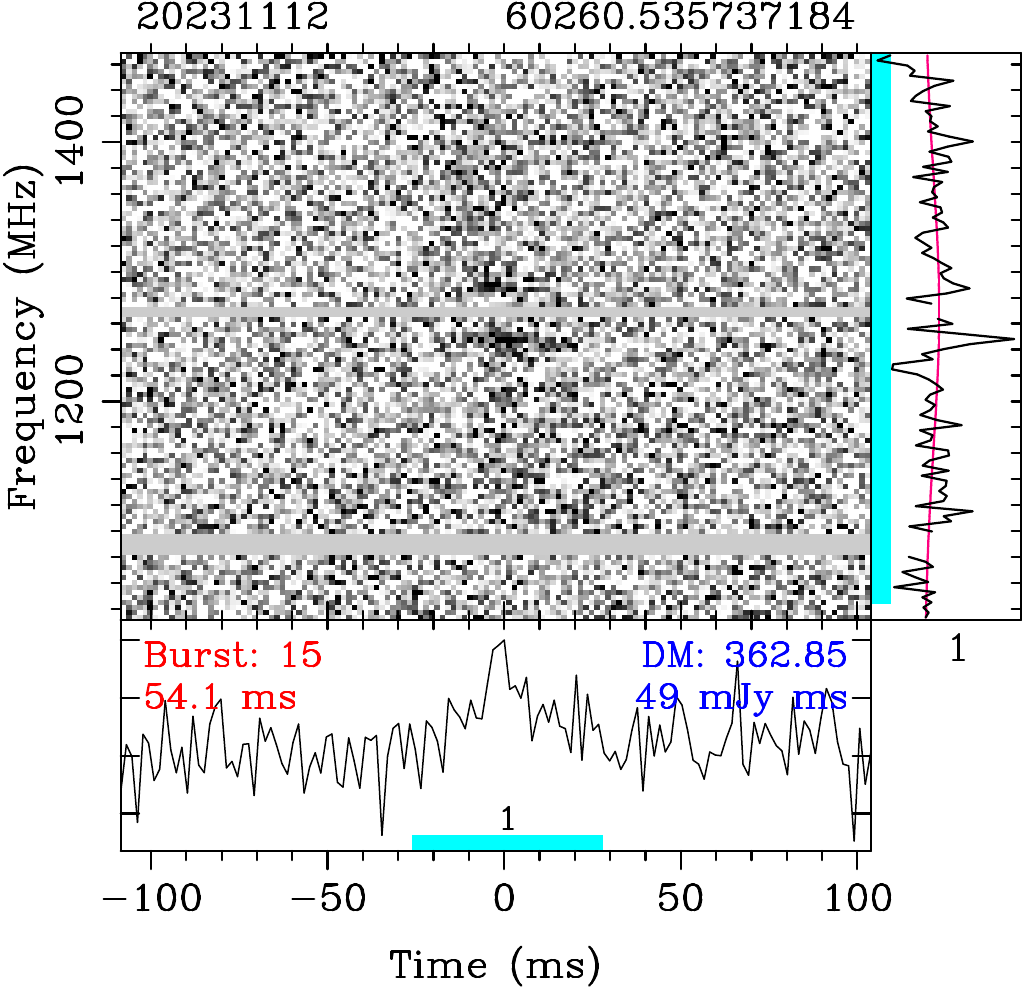}
\includegraphics[height=0.29\linewidth]{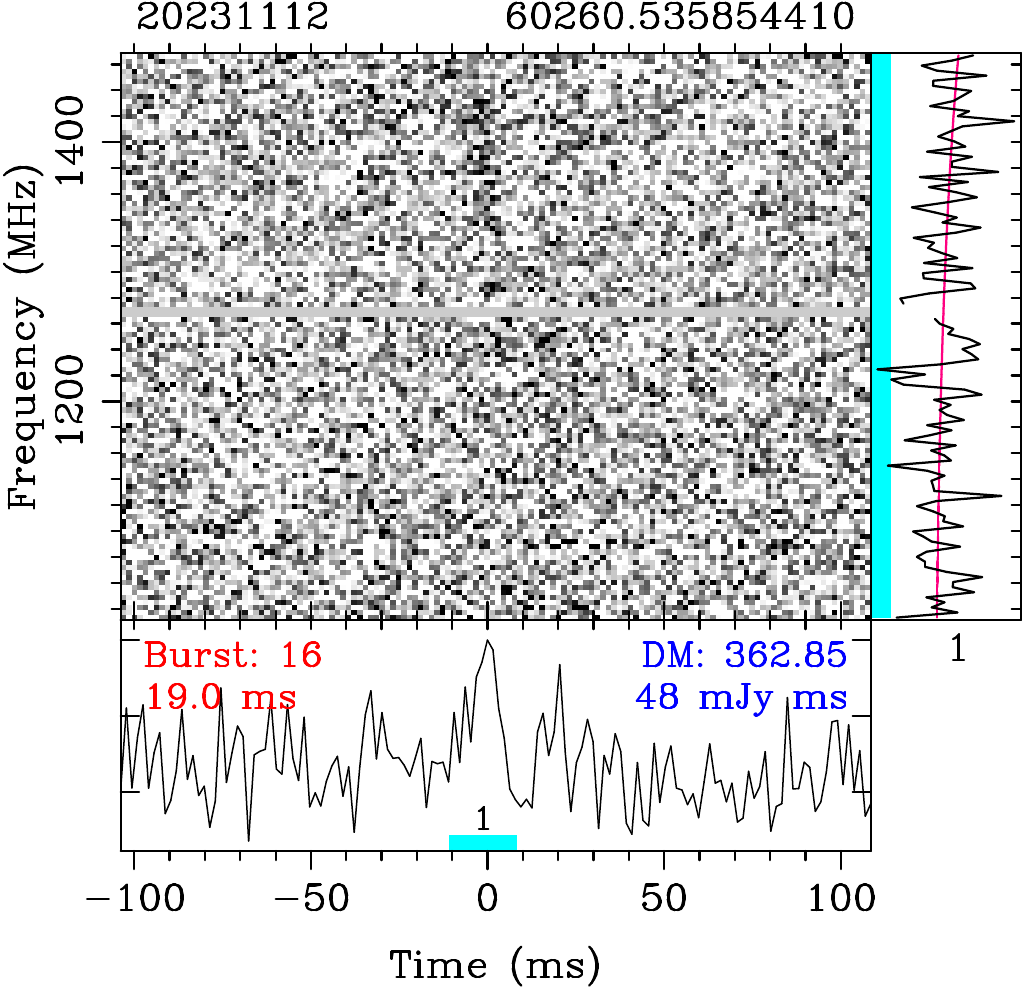}
\includegraphics[height=0.29\linewidth]{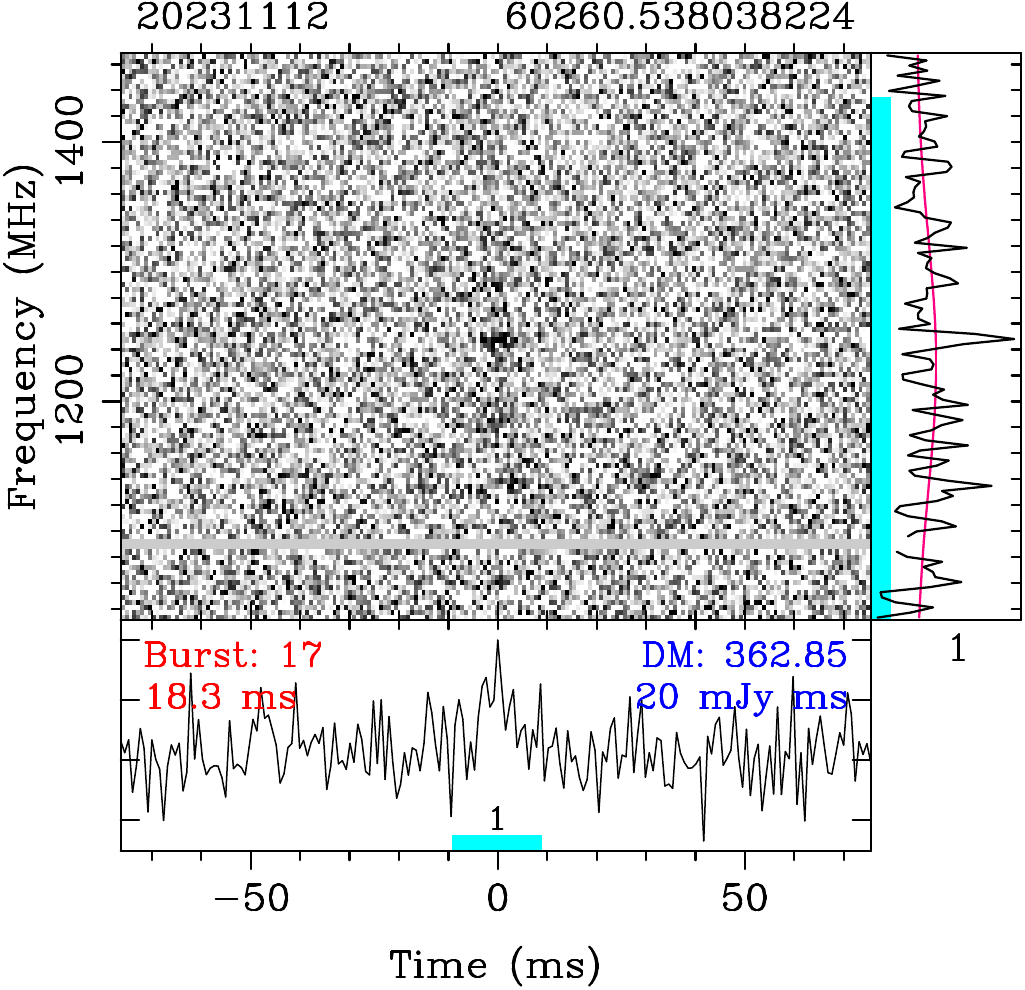}
\includegraphics[height=0.29\linewidth]{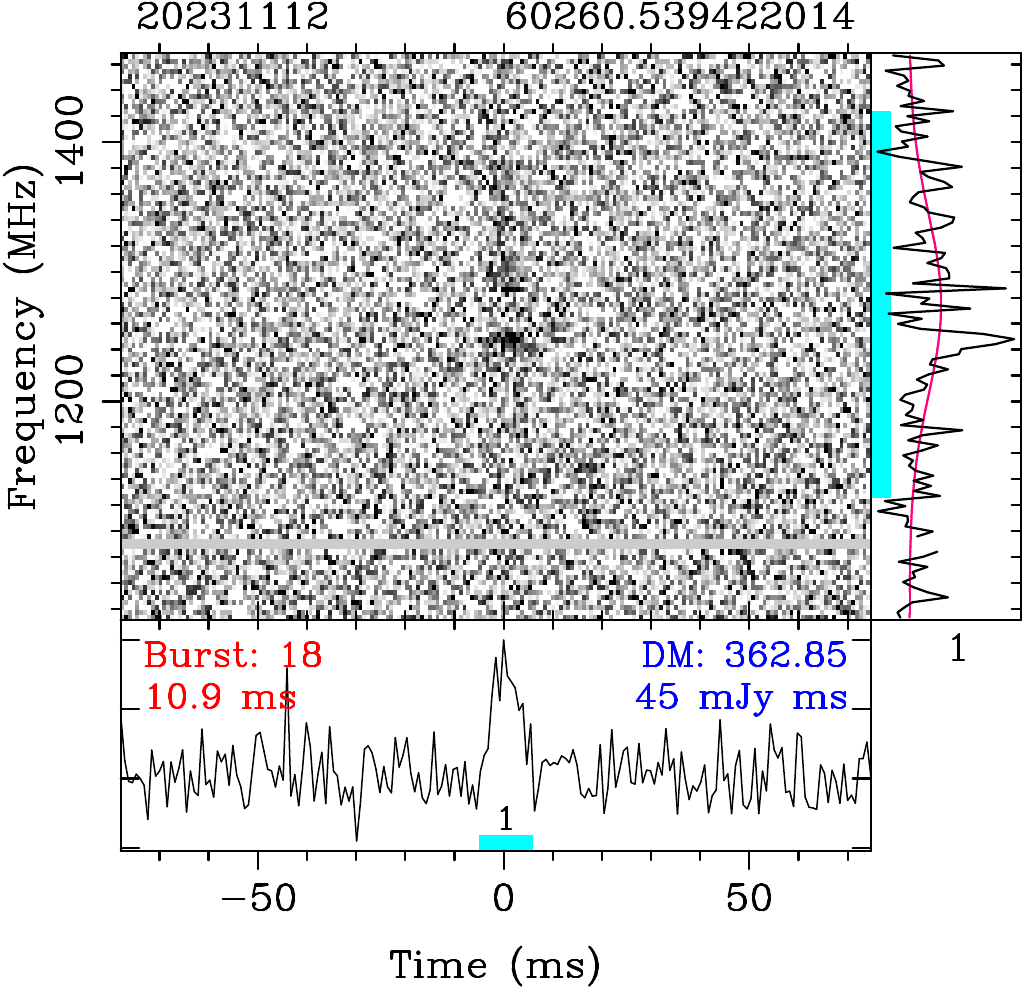}
\includegraphics[height=0.29\linewidth]{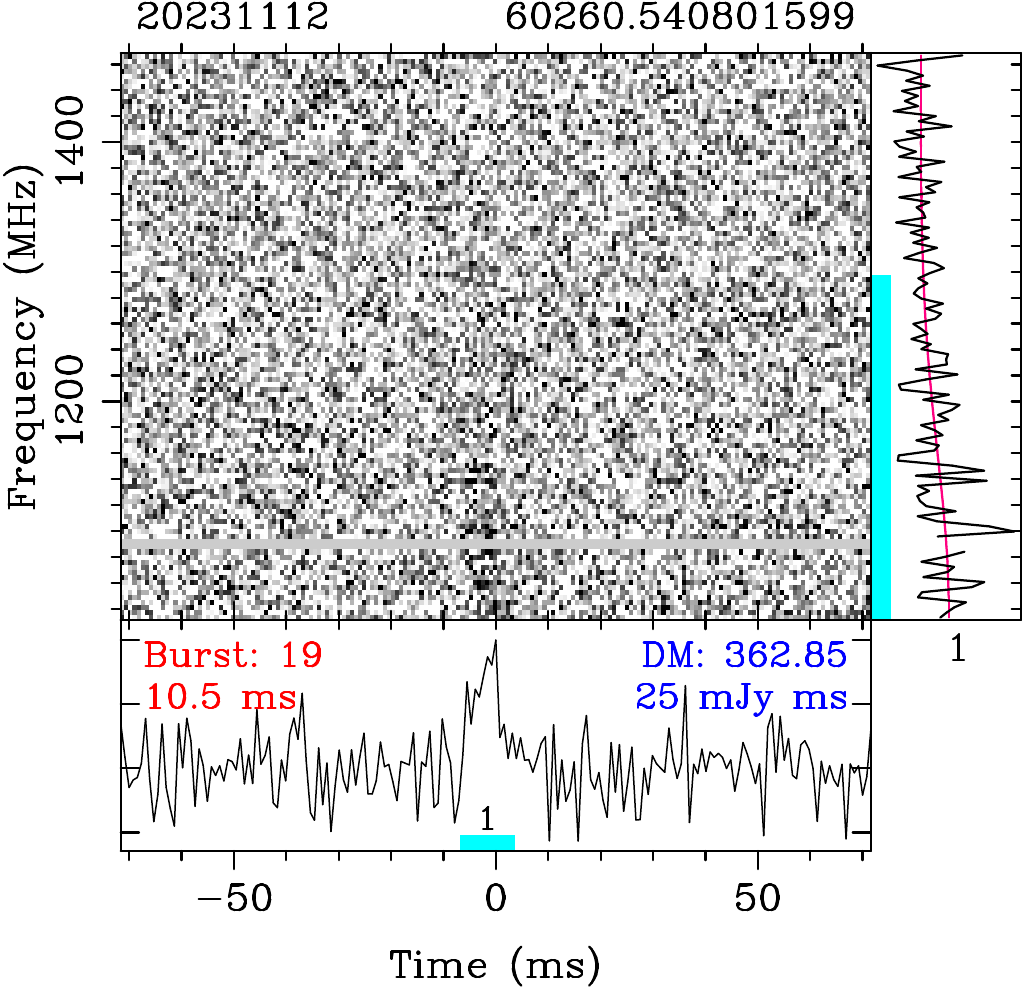}
\caption{({\textit{continued}})}
\end{figure*}
\addtocounter{figure}{-1}
\begin{figure*}
\flushleft
\includegraphics[height=0.29\linewidth]{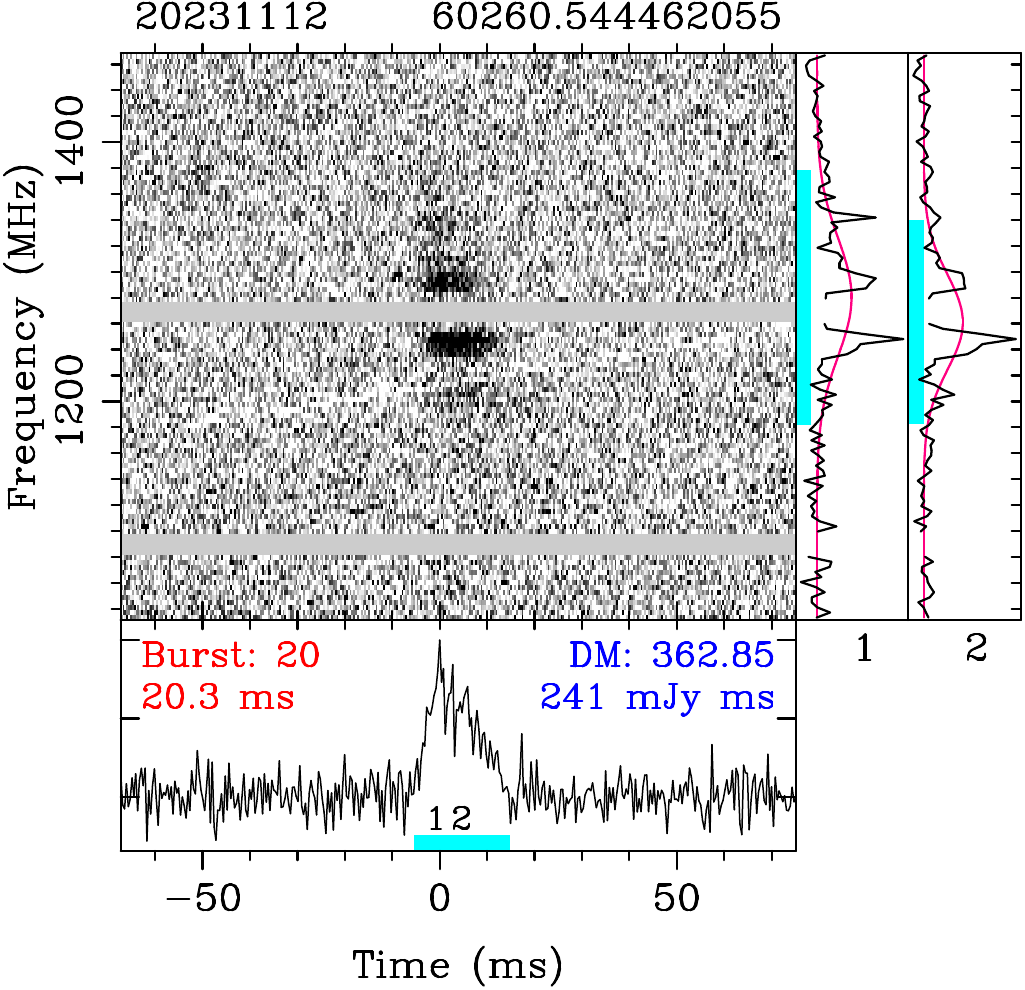}
\includegraphics[height=0.29\linewidth]{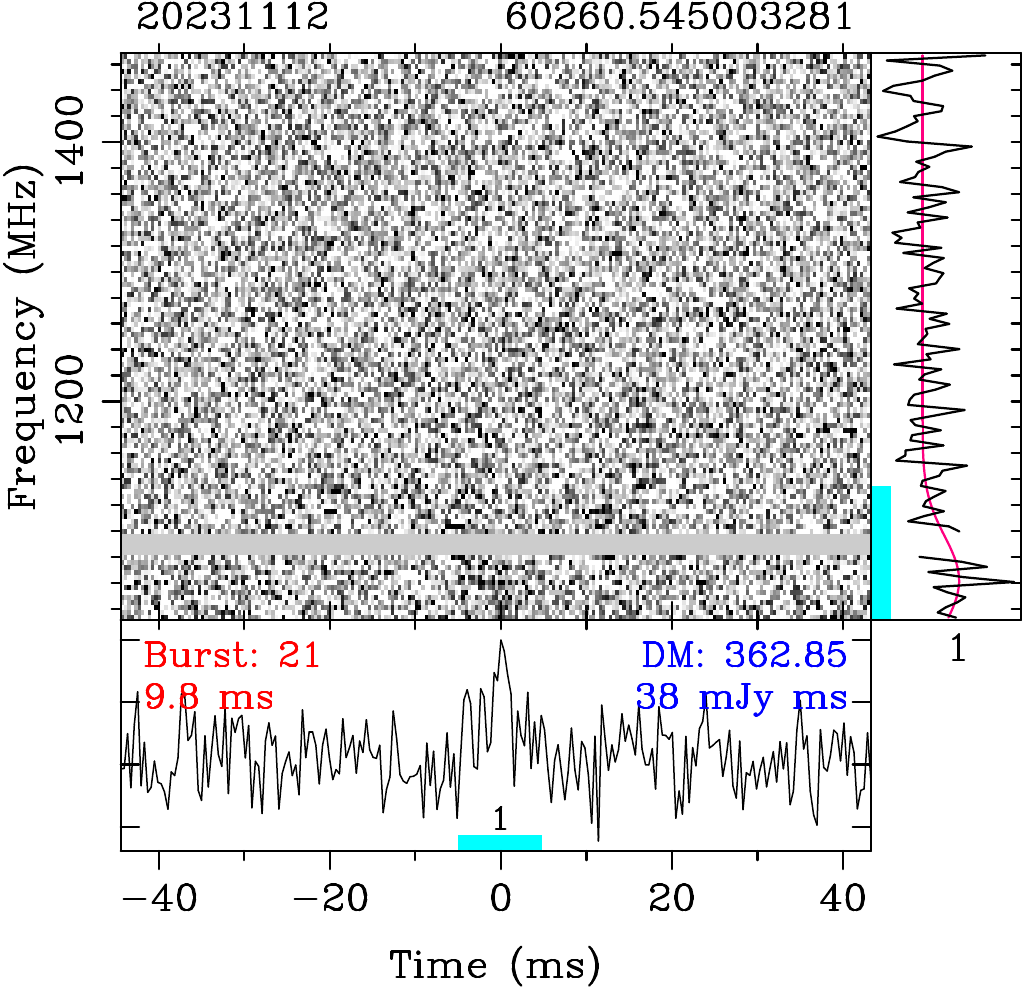}
\includegraphics[height=0.29\linewidth]{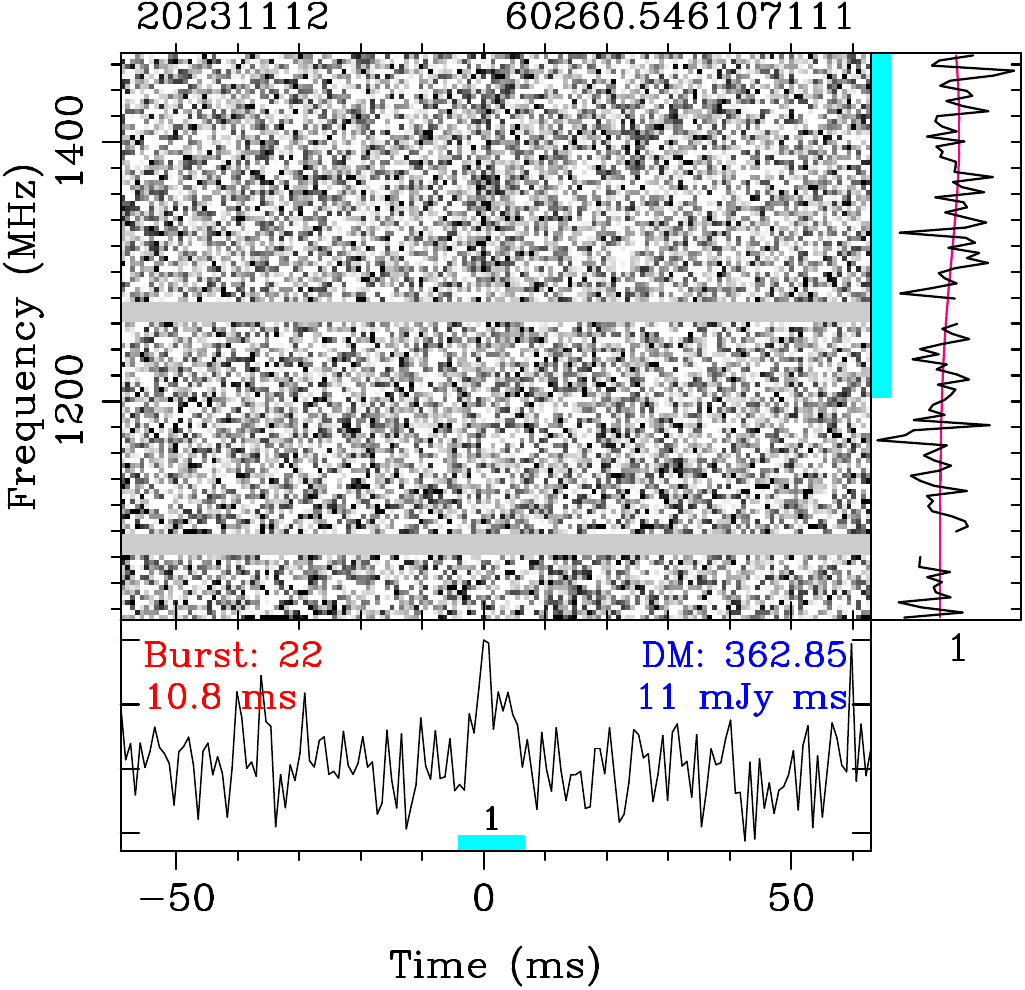}
\includegraphics[height=0.29\linewidth]{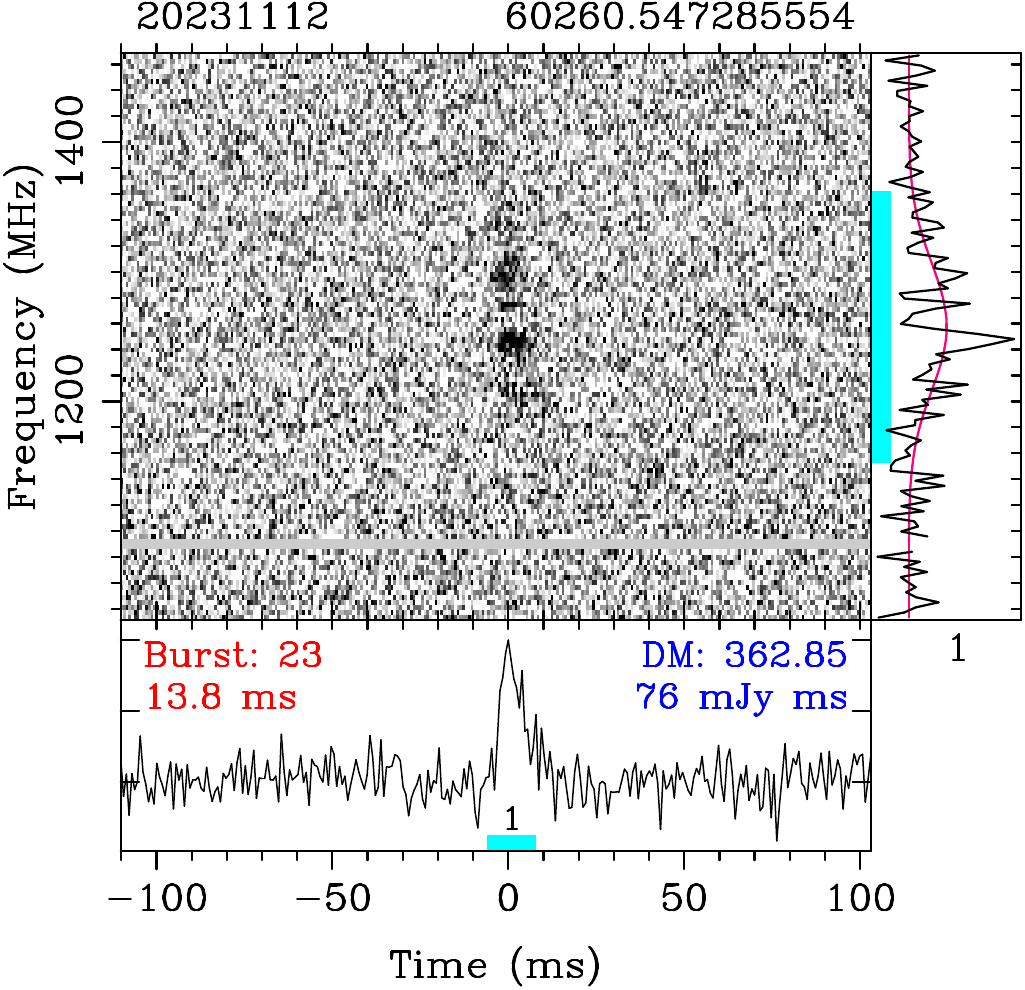}
\includegraphics[height=0.29\linewidth]{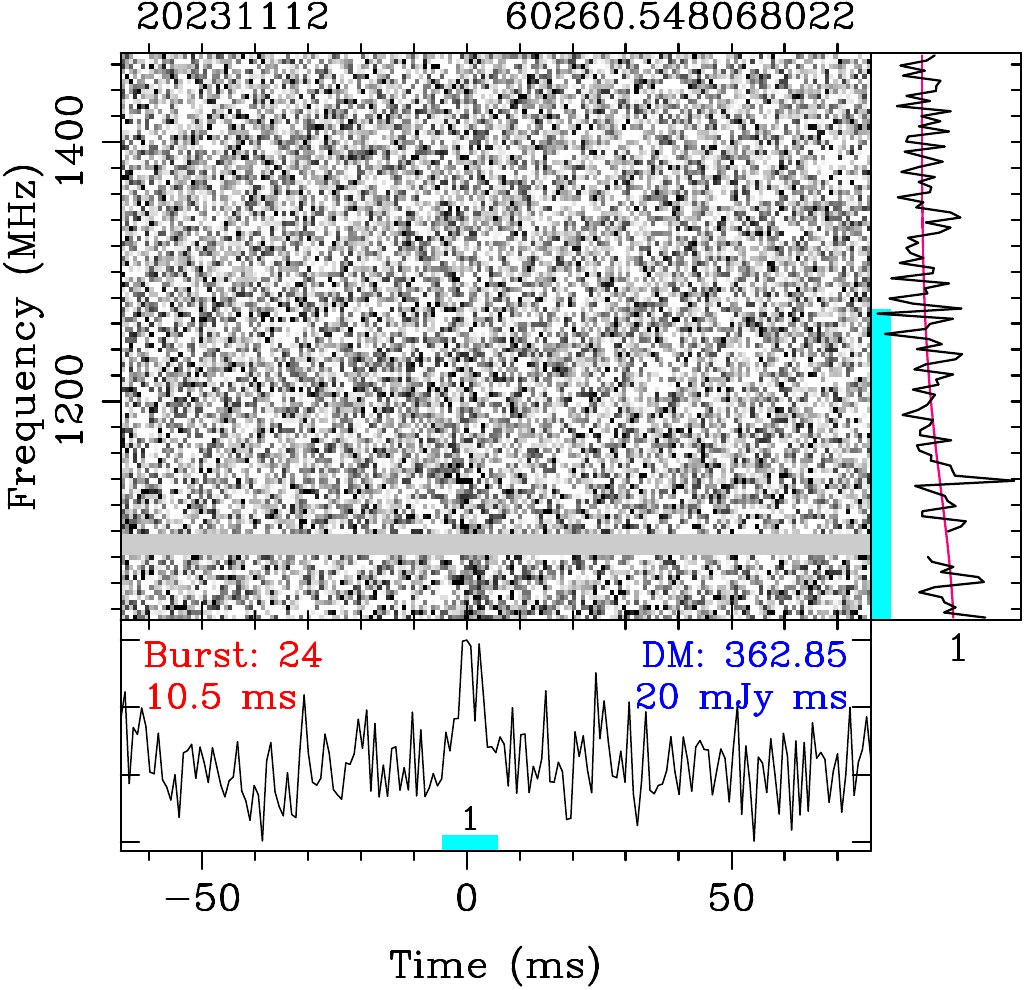}
\includegraphics[height=0.29\linewidth]{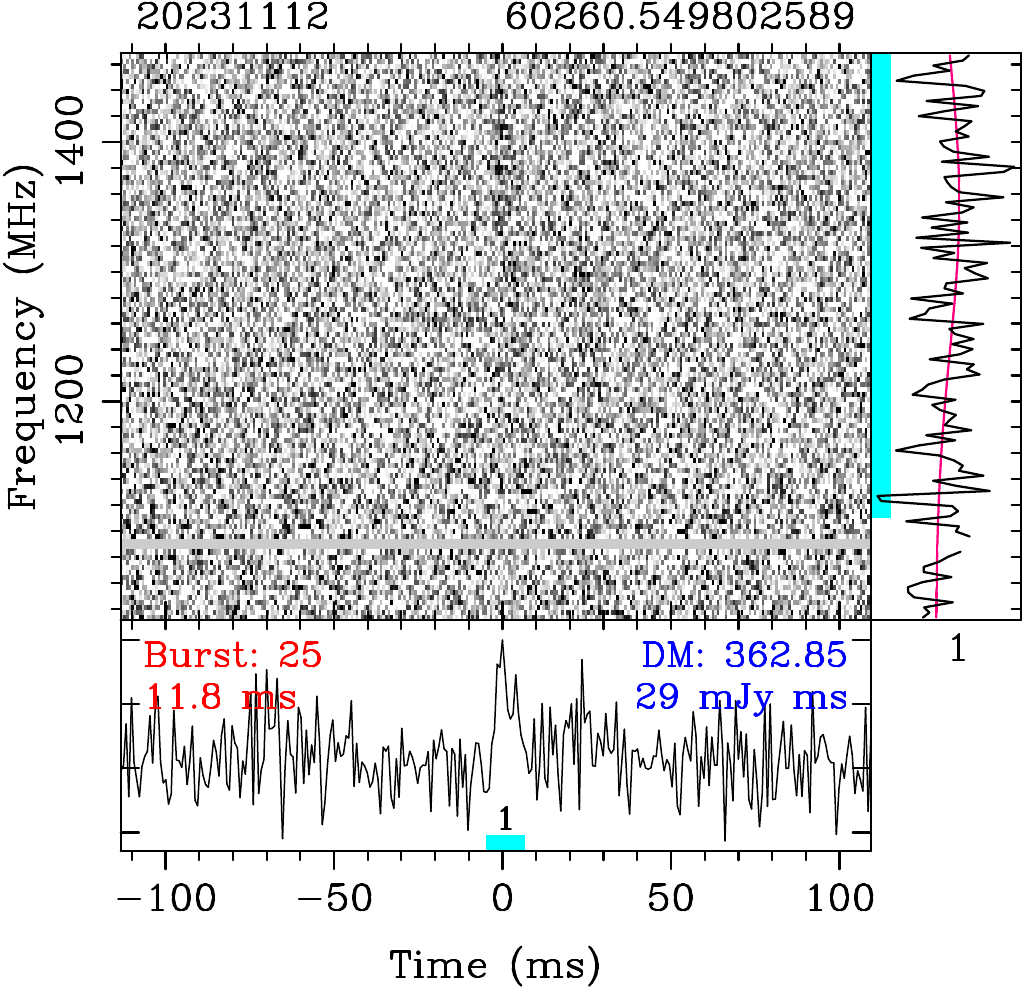}
\includegraphics[height=0.29\linewidth]{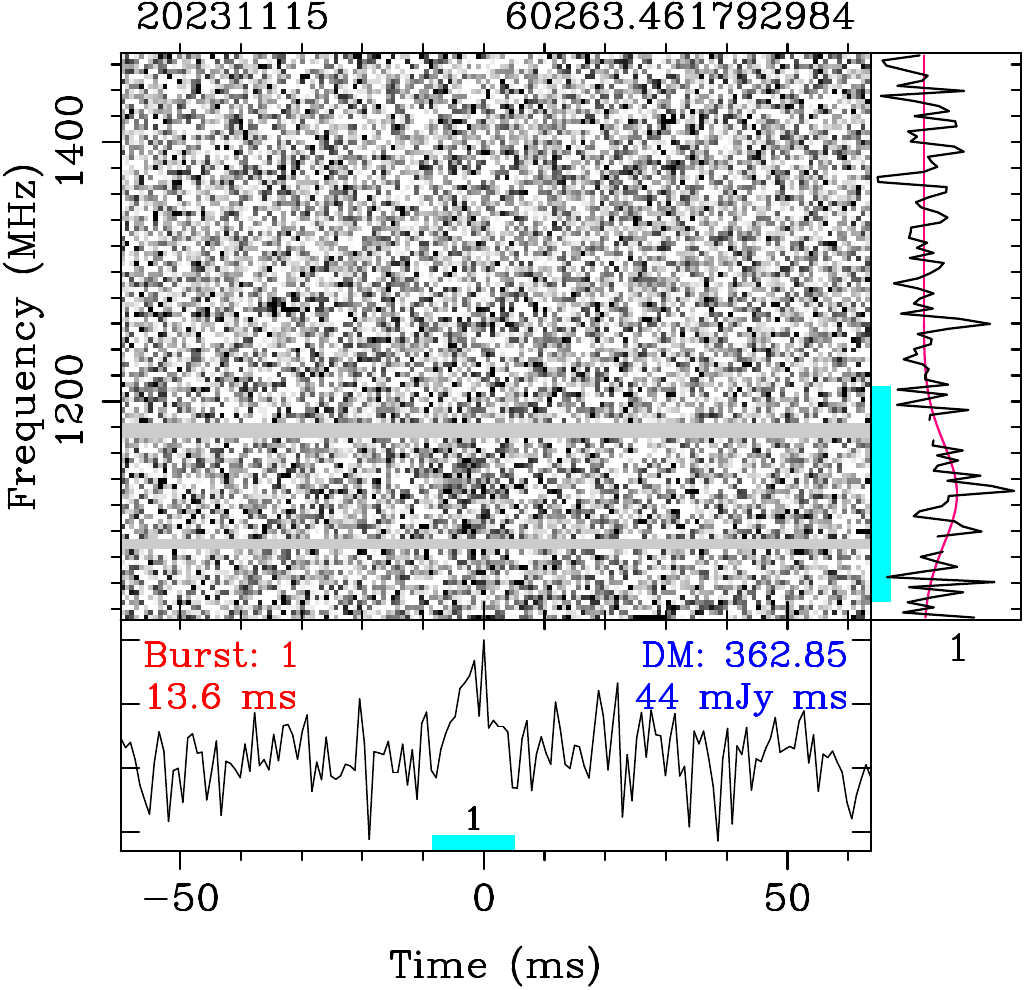}
\includegraphics[height=0.29\linewidth]{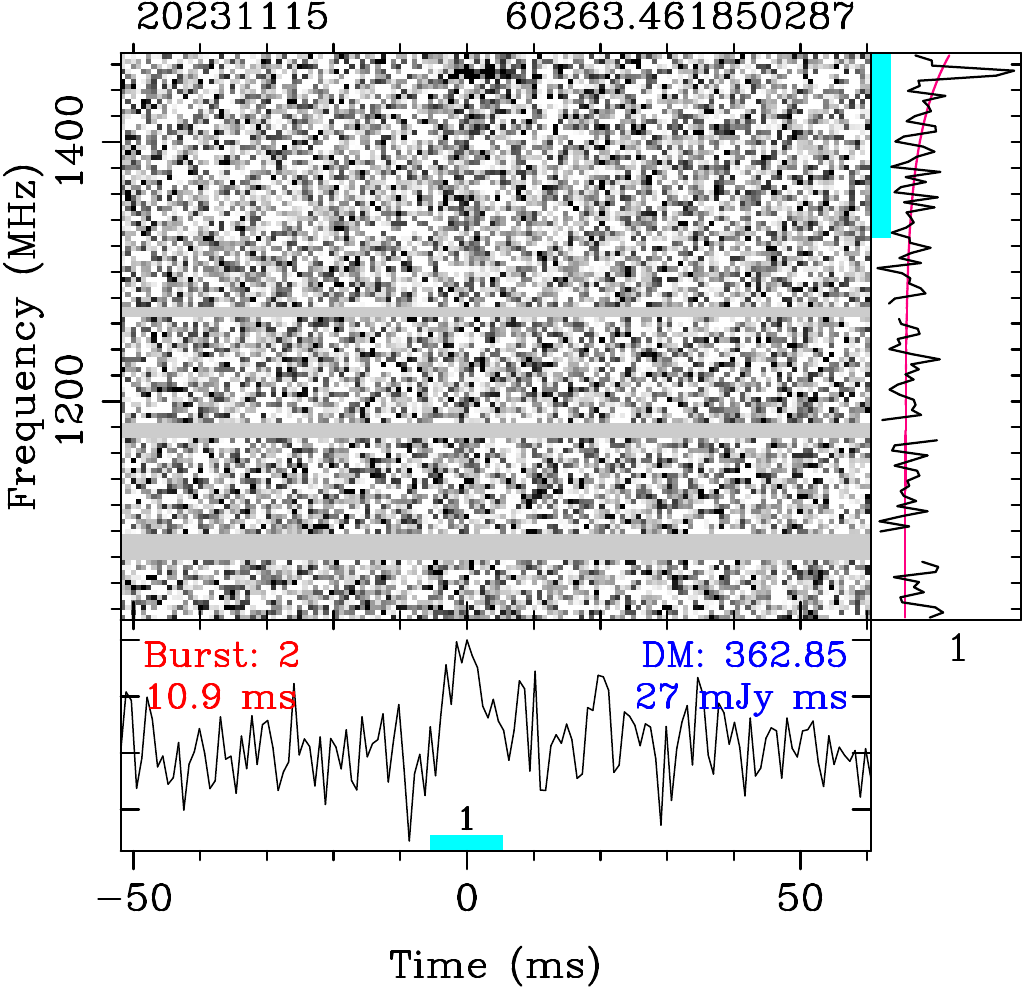}
\includegraphics[height=0.29\linewidth]{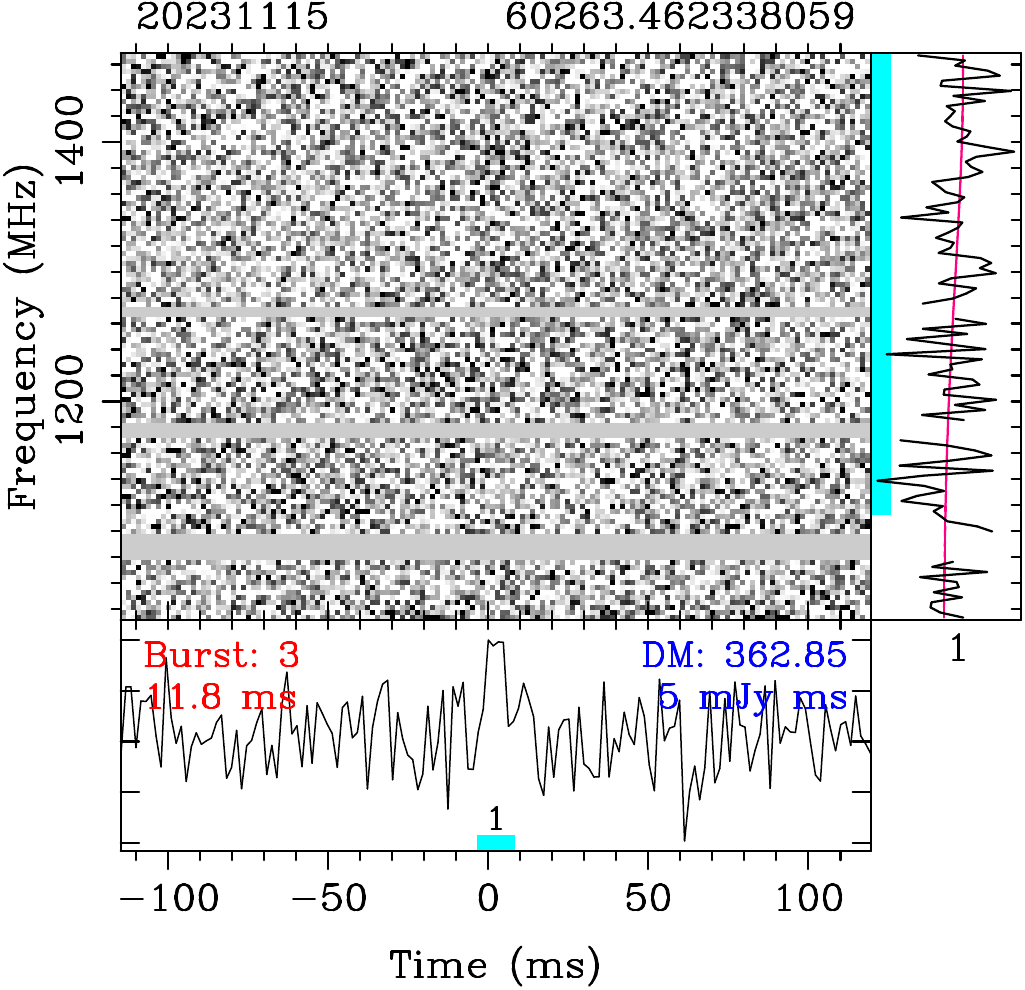}
\includegraphics[height=0.29\linewidth]{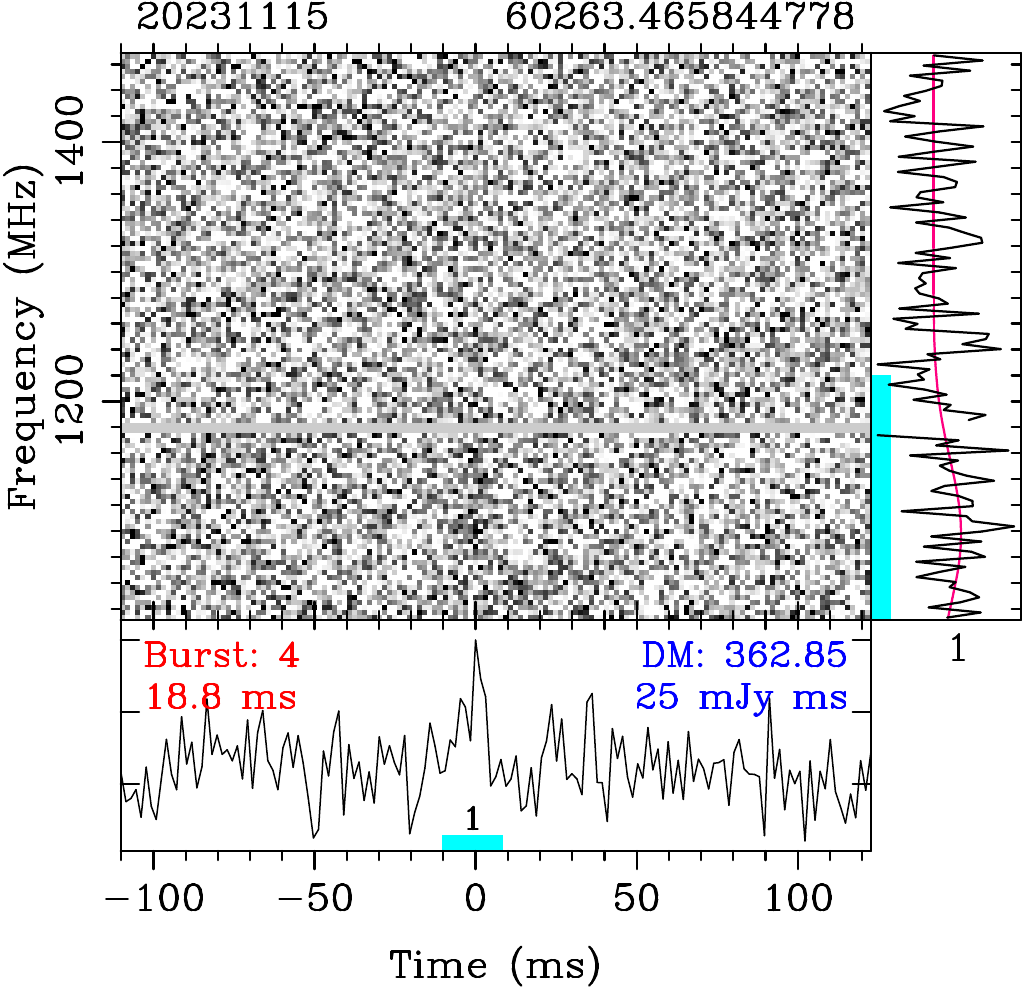}
\includegraphics[height=0.29\linewidth]{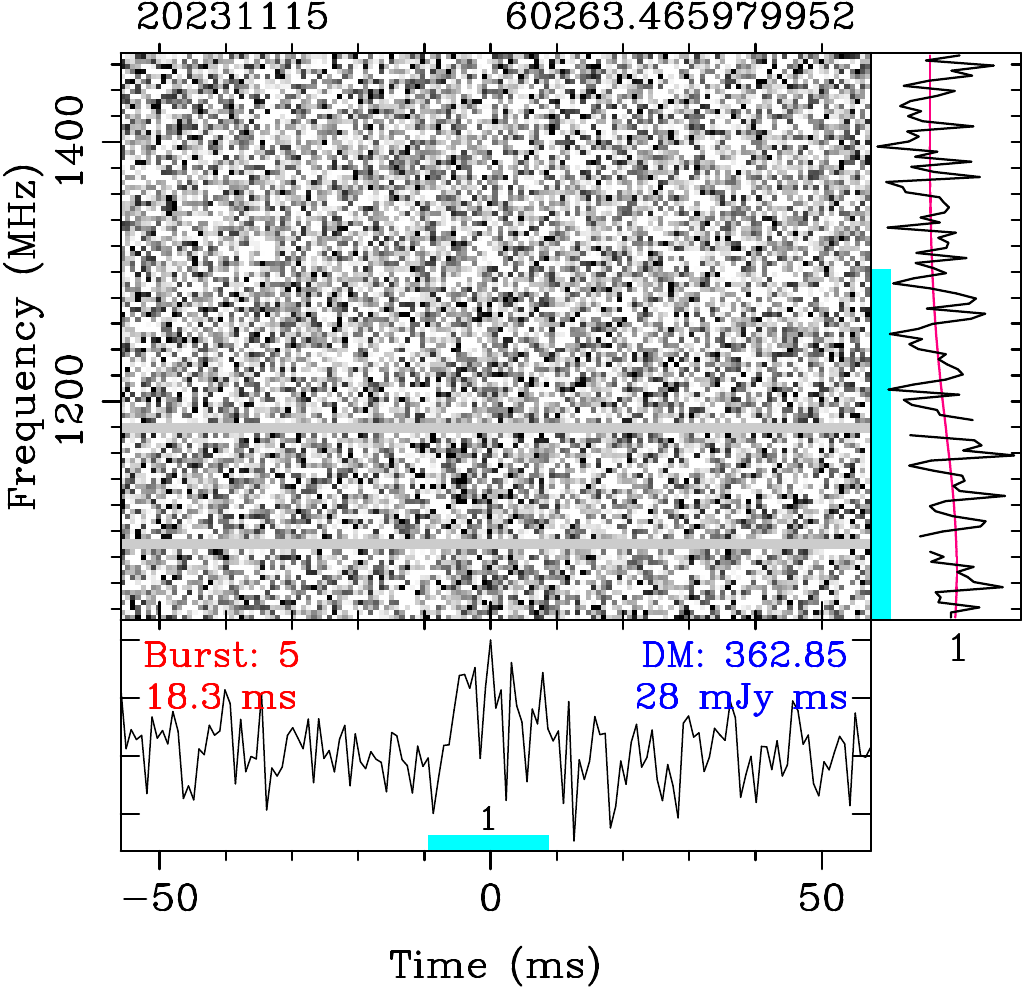}
\includegraphics[height=0.29\linewidth]{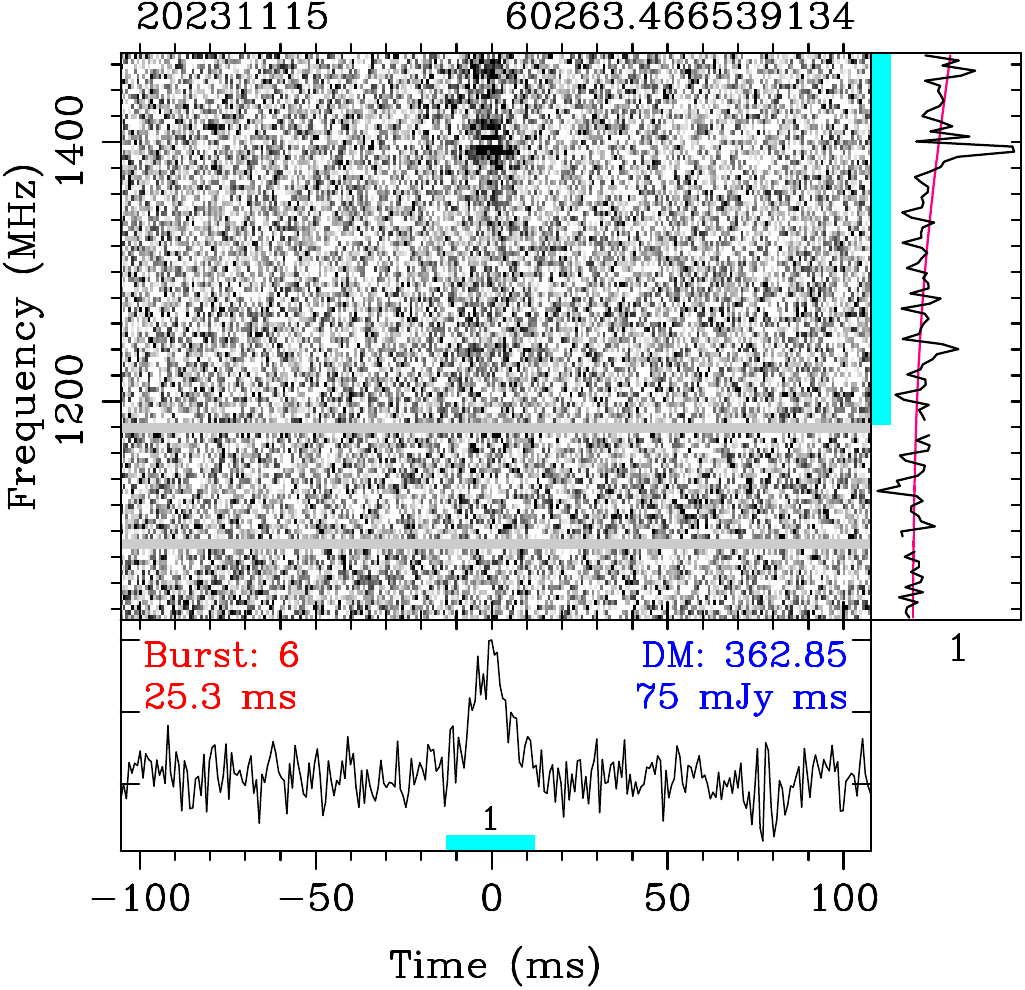}
\caption{({\textit{continued}})}
\end{figure*}
\addtocounter{figure}{-1}
\begin{figure*}
\flushleft
\includegraphics[height=0.29\linewidth]{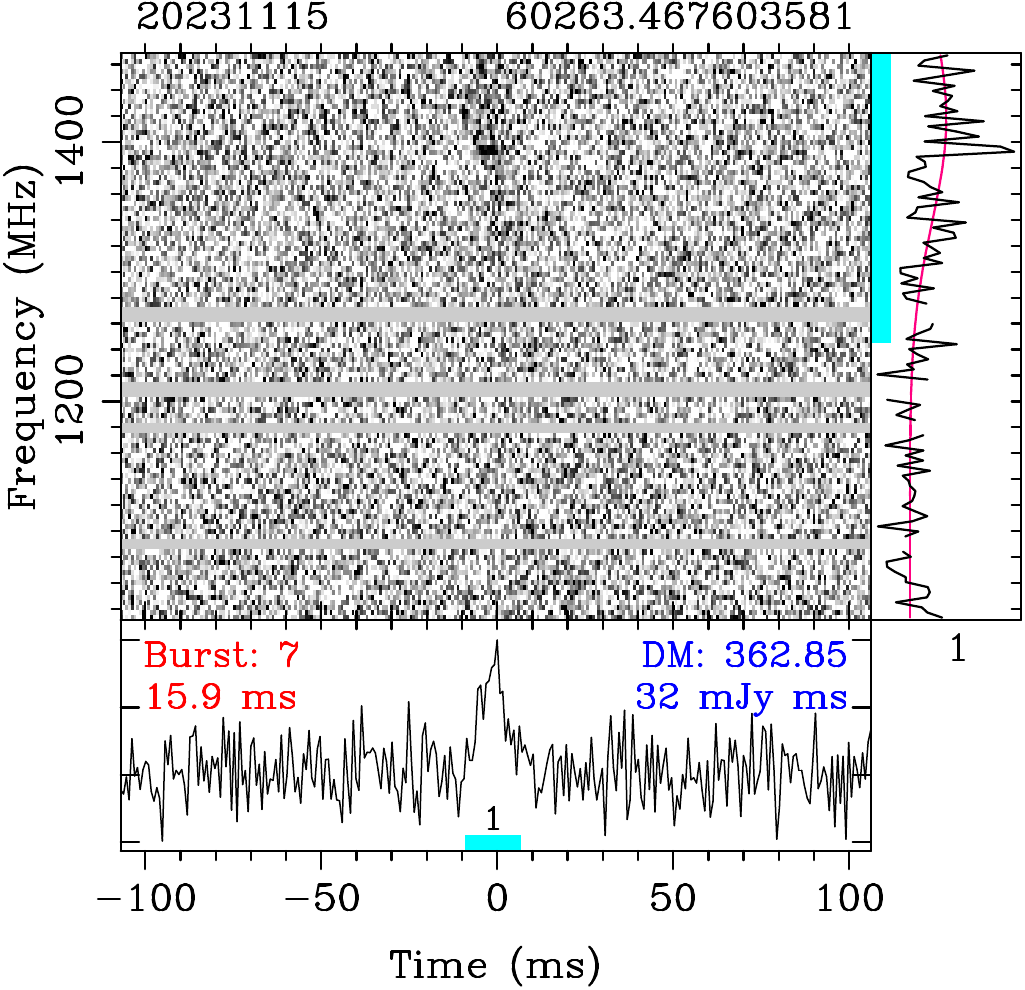}
\includegraphics[height=0.29\linewidth]{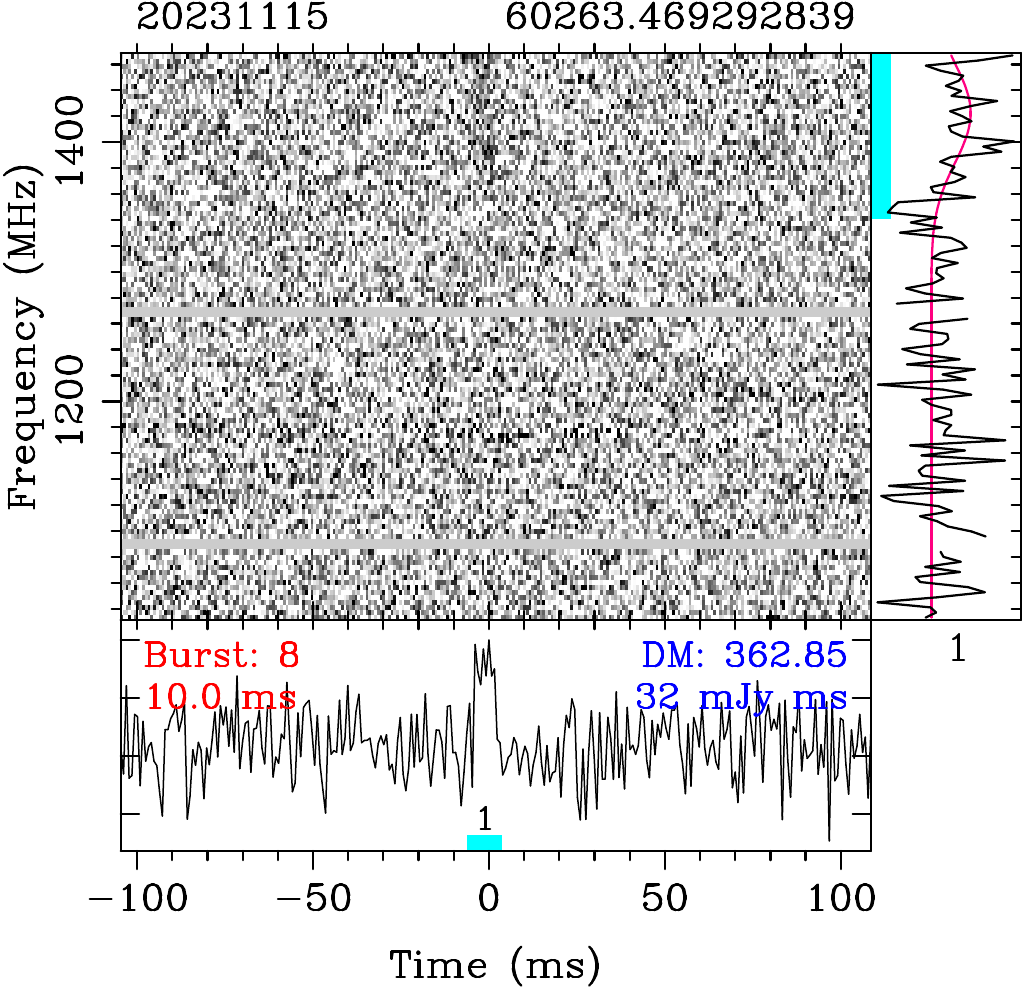}
\includegraphics[height=0.29\linewidth]{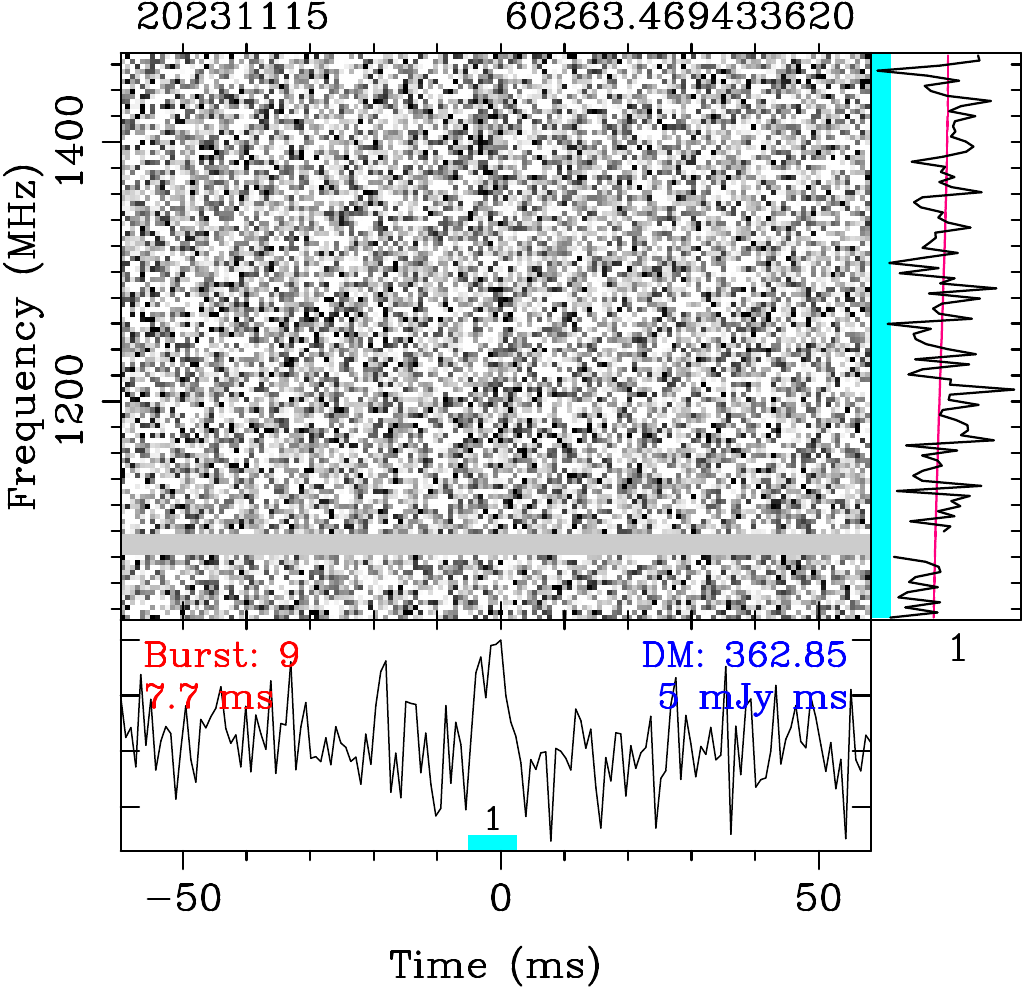}
\includegraphics[height=0.29\linewidth]{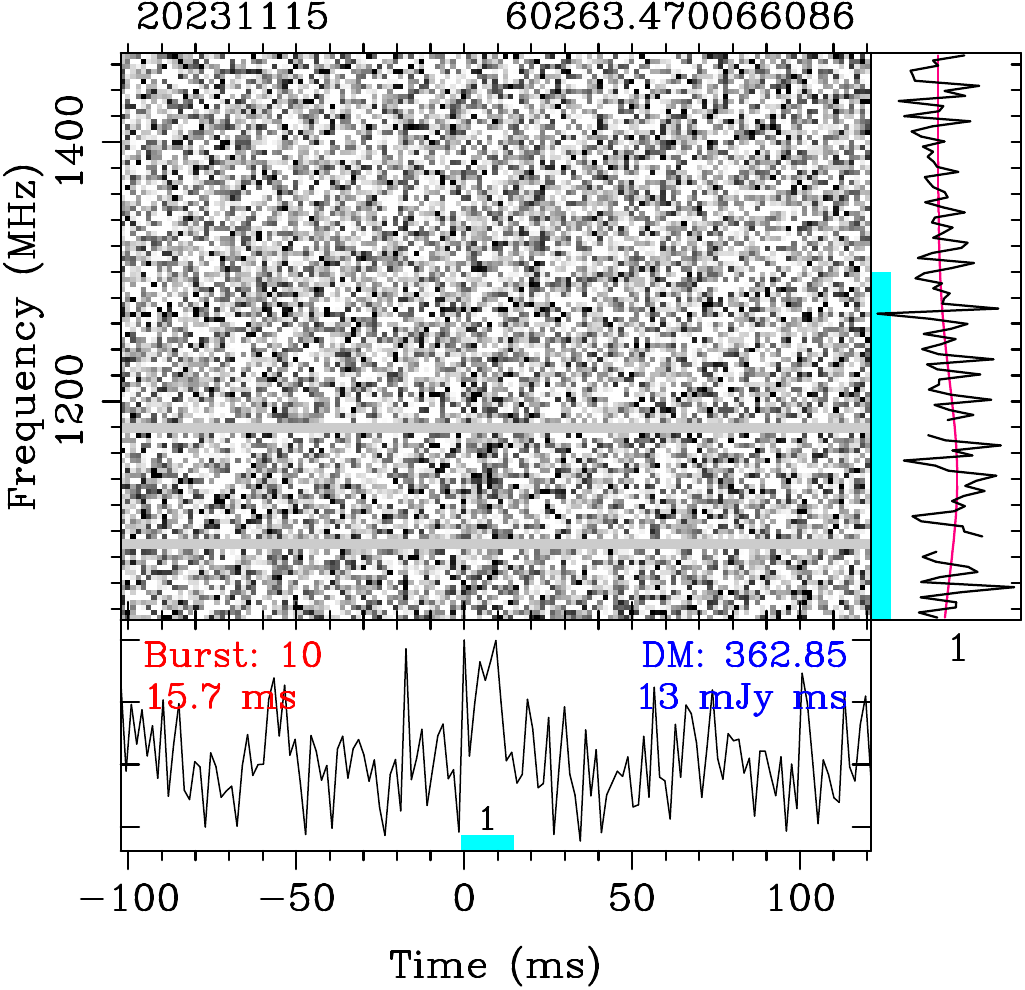}
\includegraphics[height=0.29\linewidth]{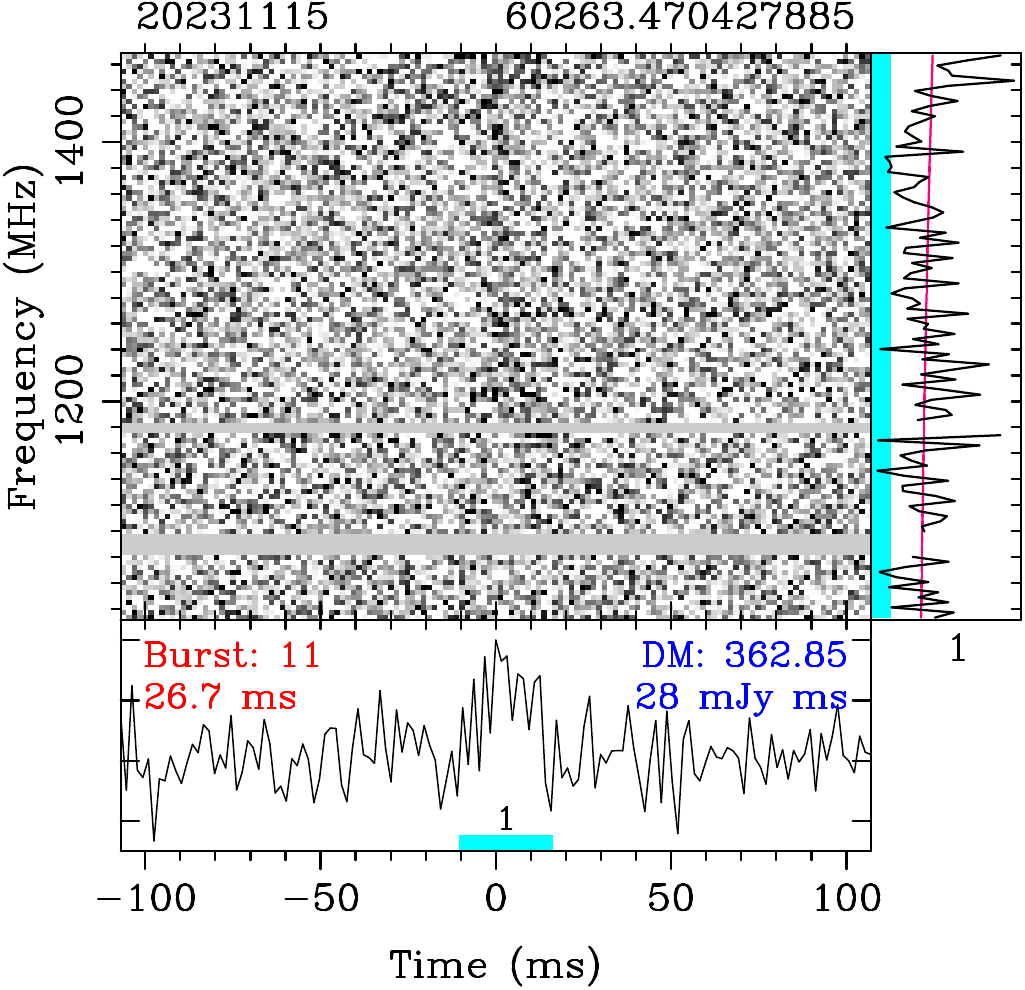}
\includegraphics[height=0.29\linewidth]{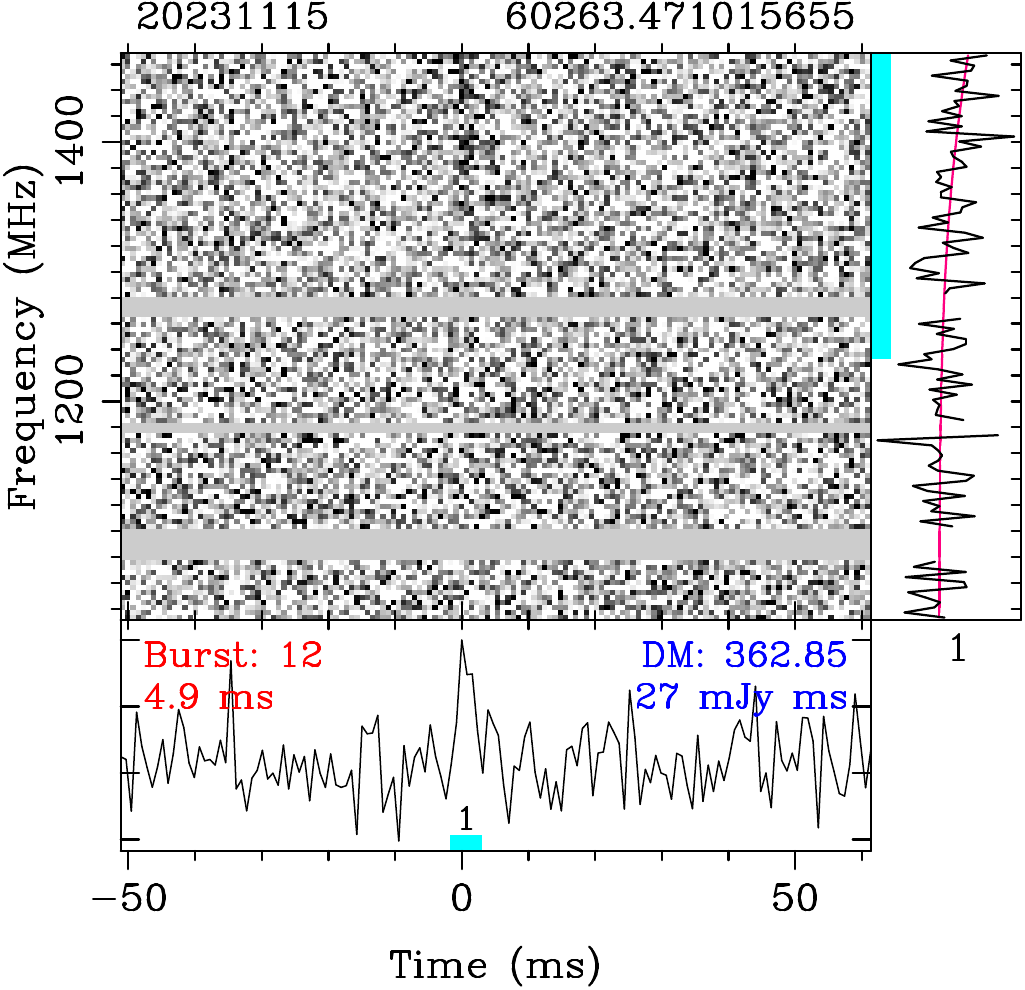}
\includegraphics[height=0.29\linewidth]{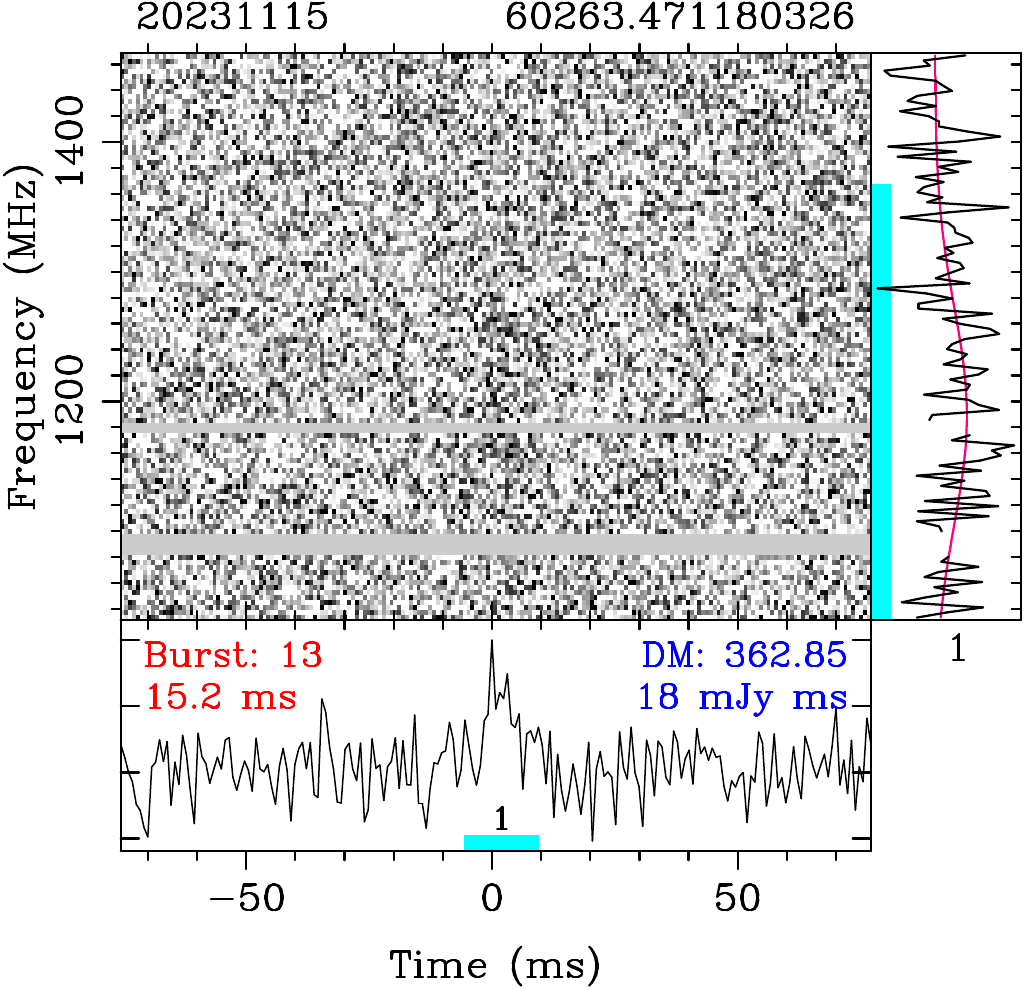}
\includegraphics[height=0.29\linewidth]{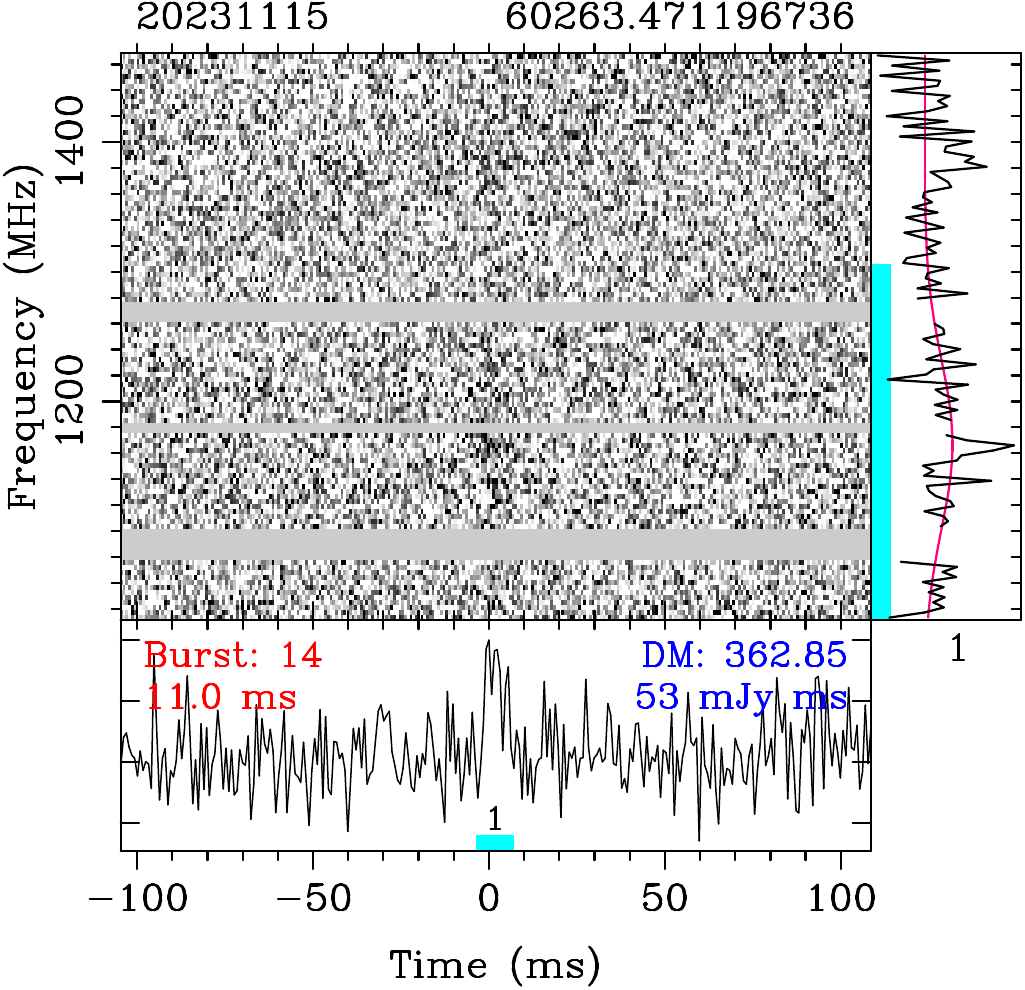}
\includegraphics[height=0.29\linewidth]{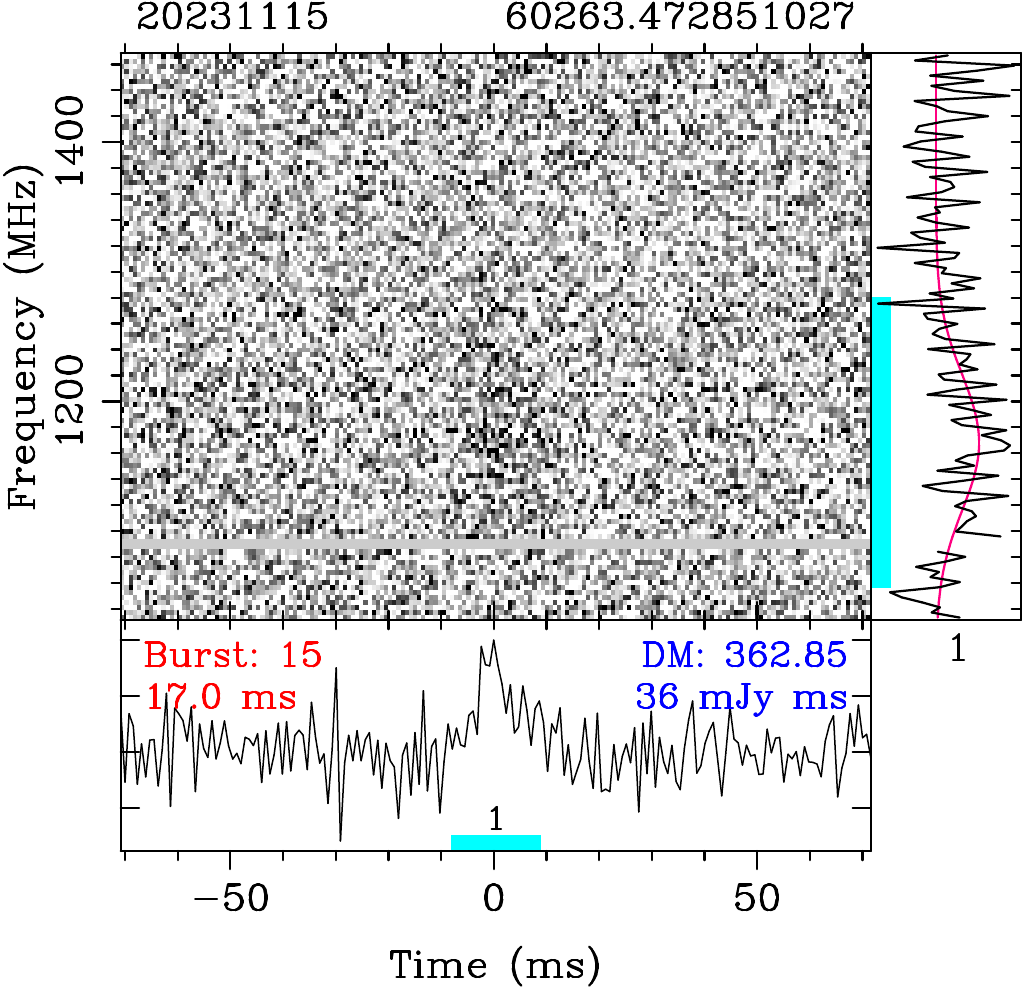}
\includegraphics[height=0.29\linewidth]{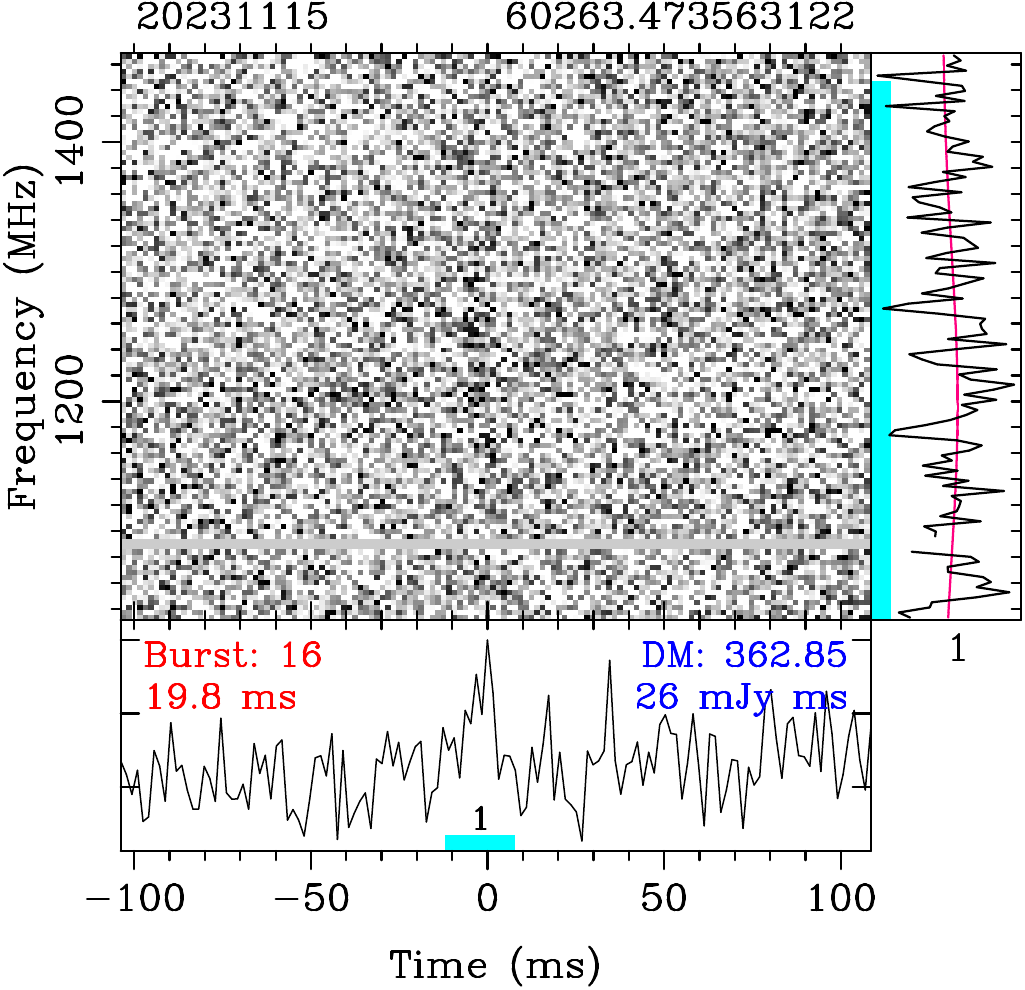}
\includegraphics[height=0.29\linewidth]{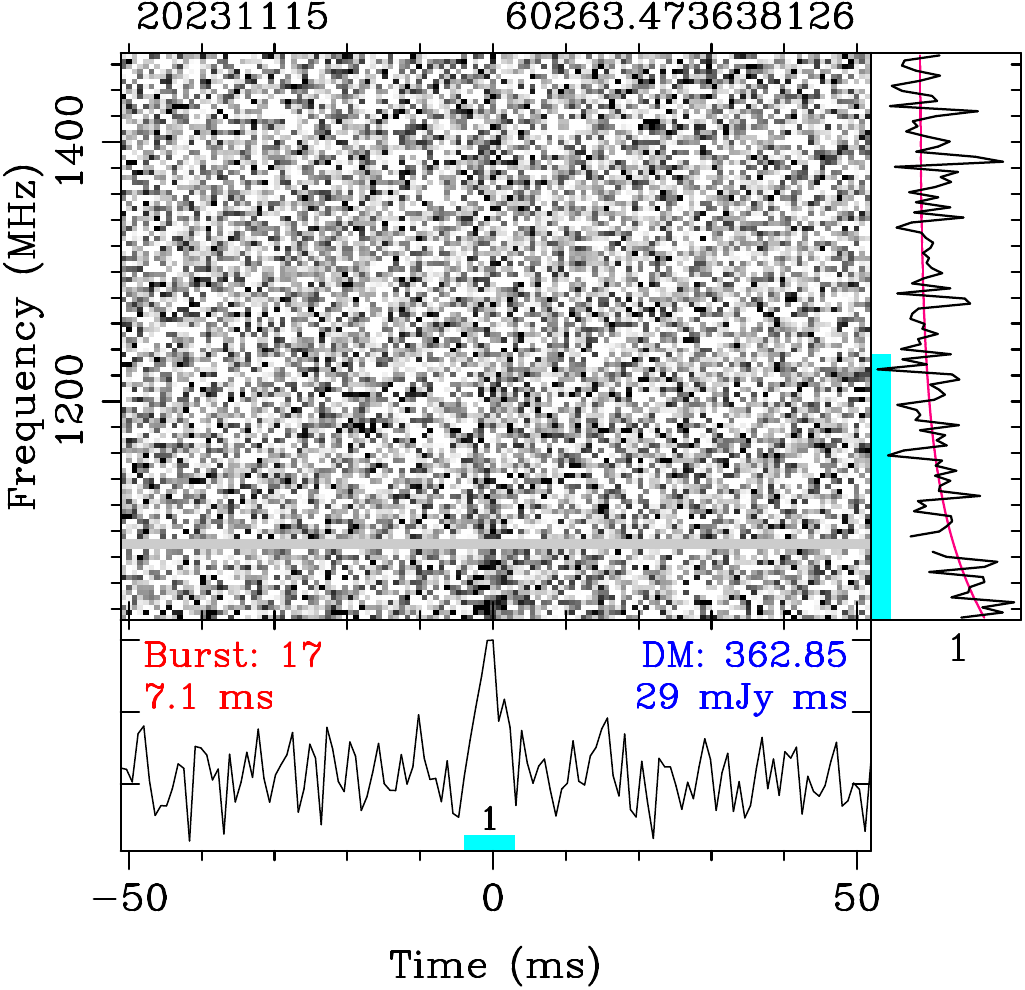}
\includegraphics[height=0.29\linewidth]{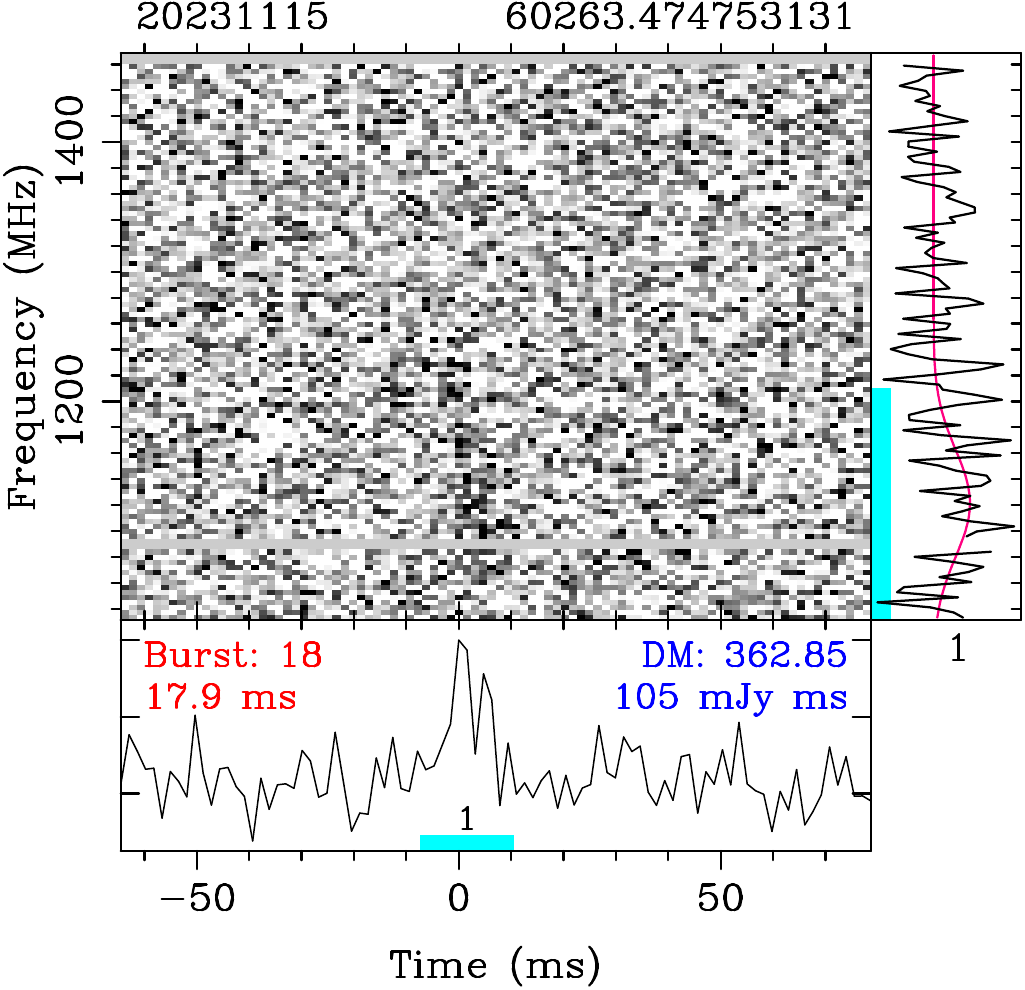}
\caption{({\textit{continued}})}
\end{figure*}
\addtocounter{figure}{-1}
\begin{figure*}
\flushleft
\includegraphics[height=0.29\linewidth]{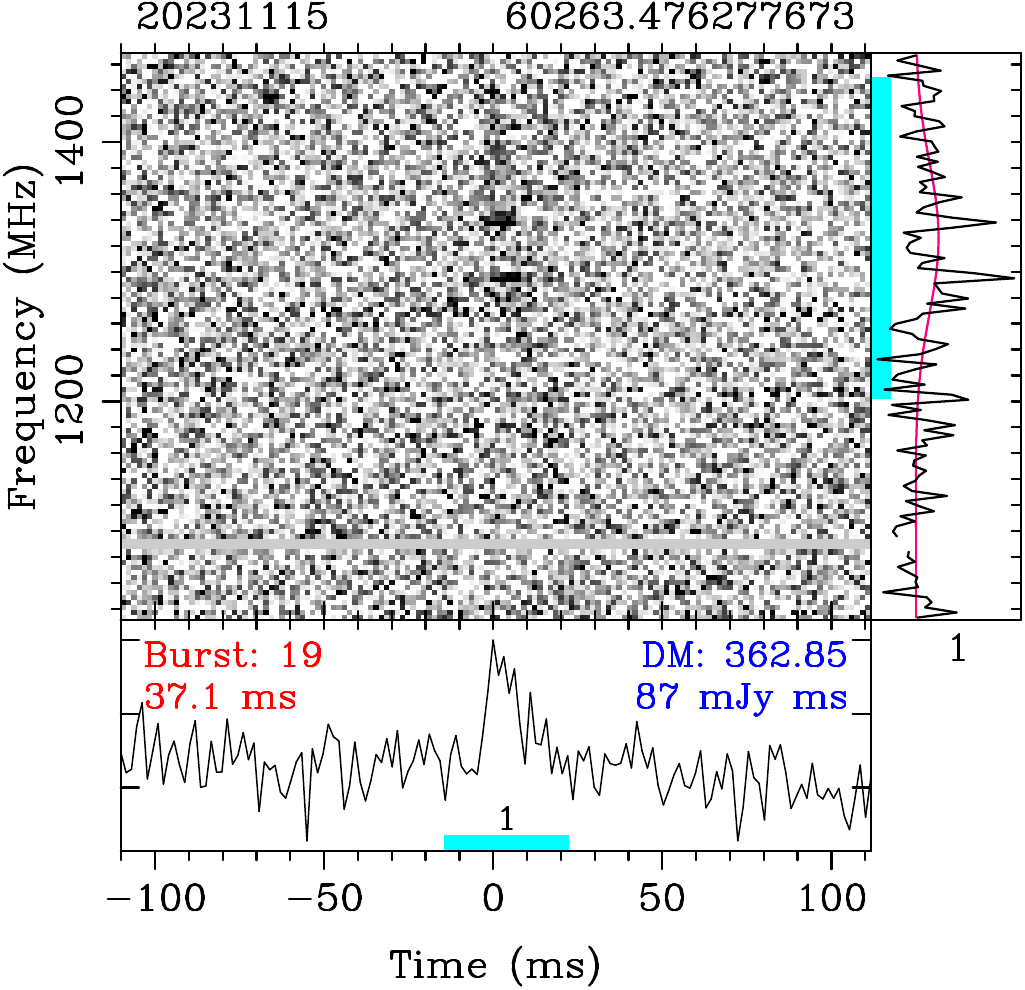}
\includegraphics[height=0.29\linewidth]{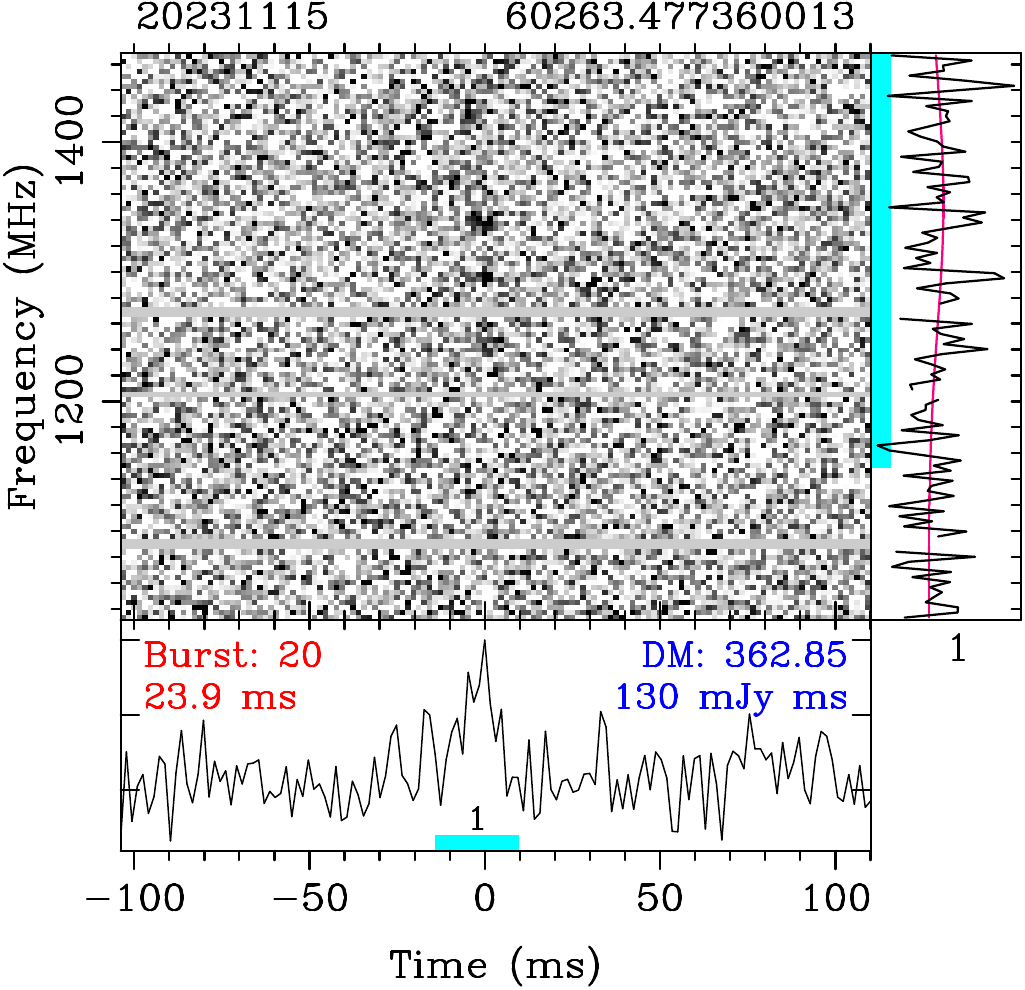}
\includegraphics[height=0.29\linewidth]{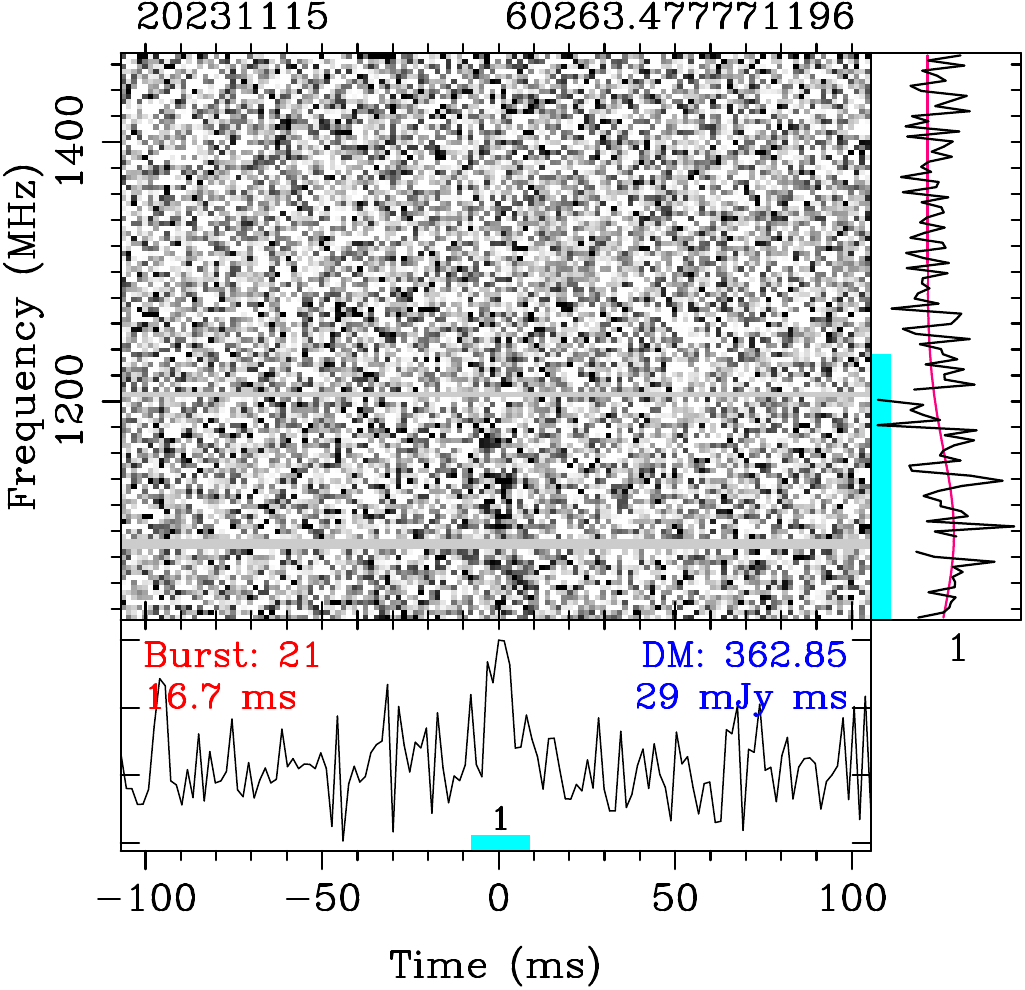}
\includegraphics[height=0.29\linewidth]{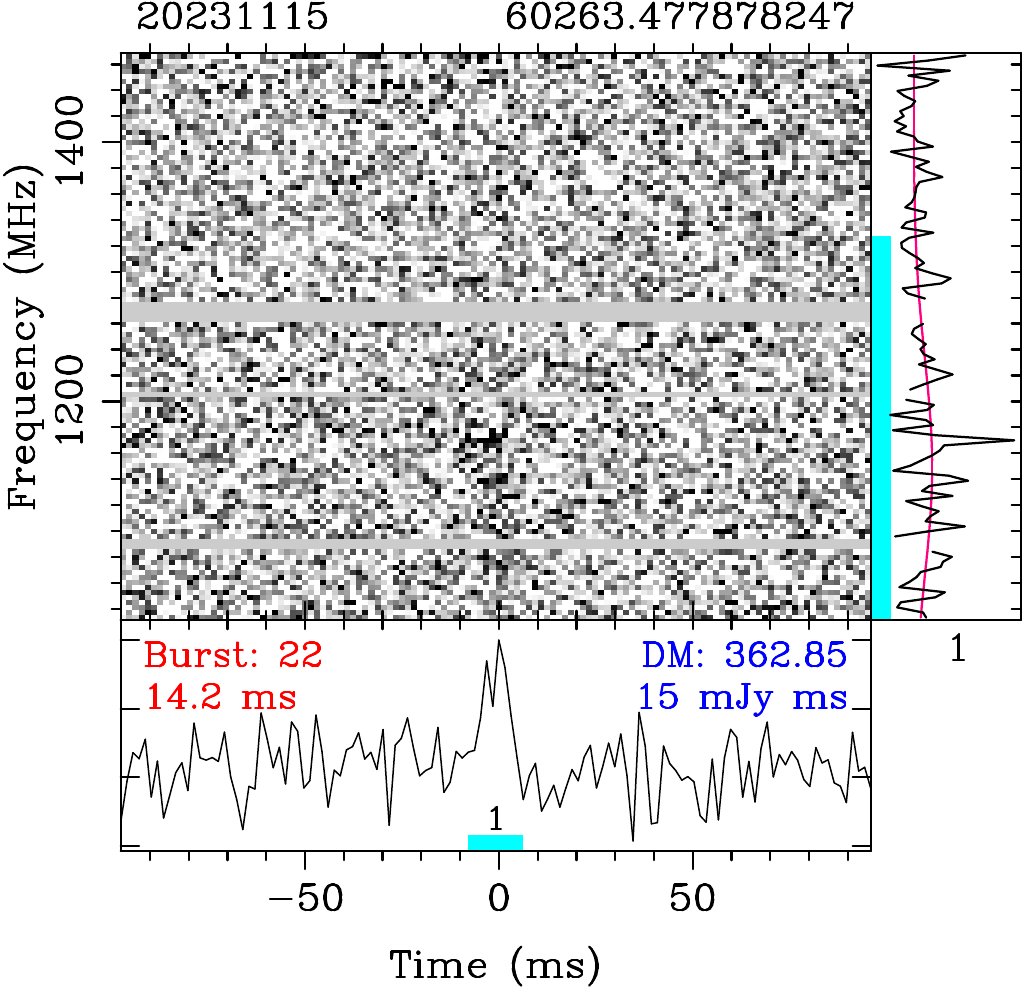}
\includegraphics[height=0.29\linewidth]{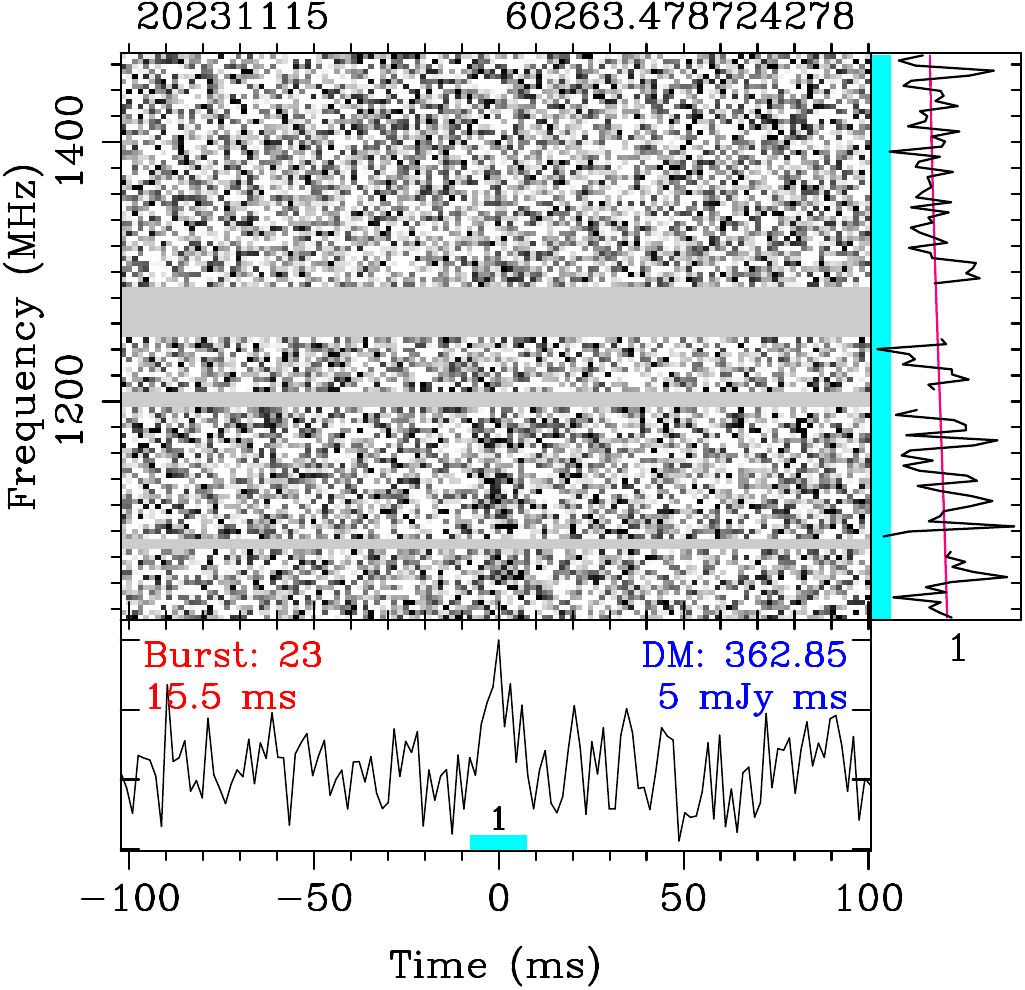}
\includegraphics[height=0.29\linewidth]{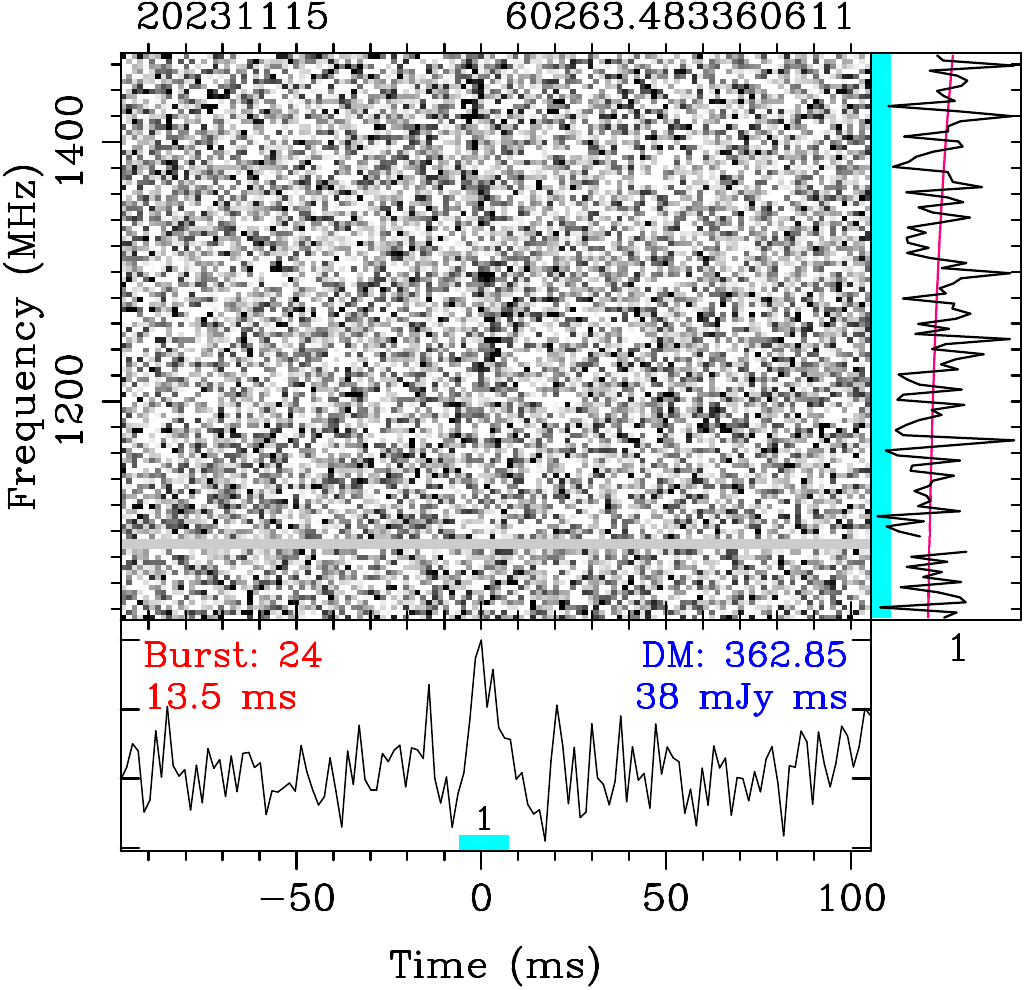}
\includegraphics[height=0.29\linewidth]{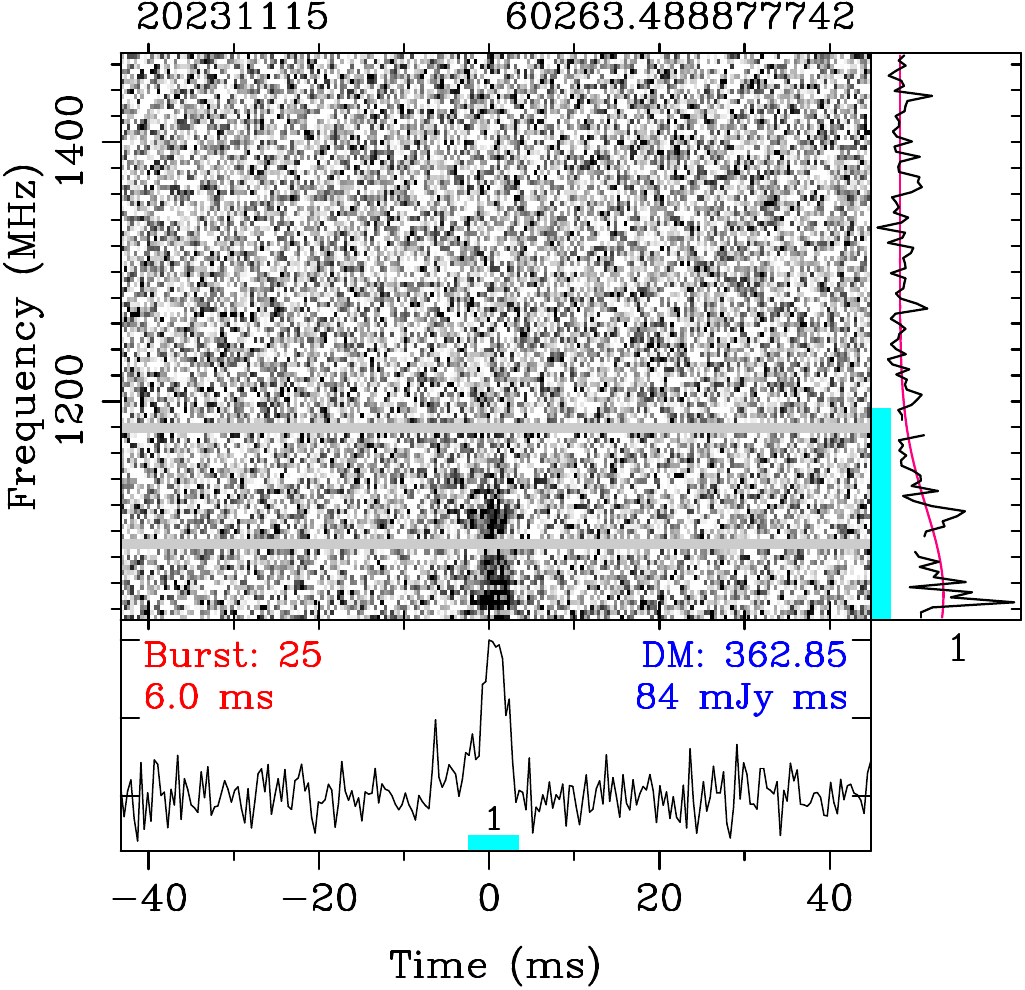}
\includegraphics[height=0.29\linewidth]{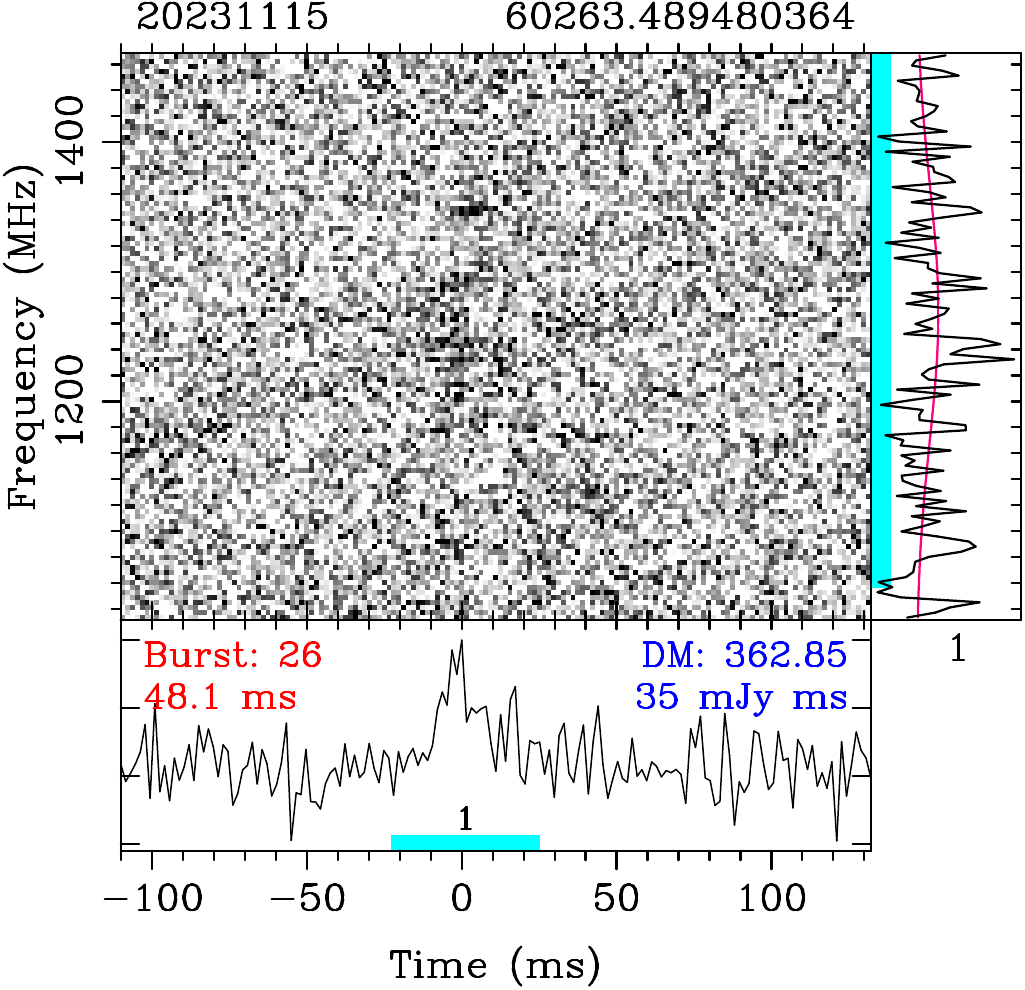}
\includegraphics[height=0.29\linewidth]{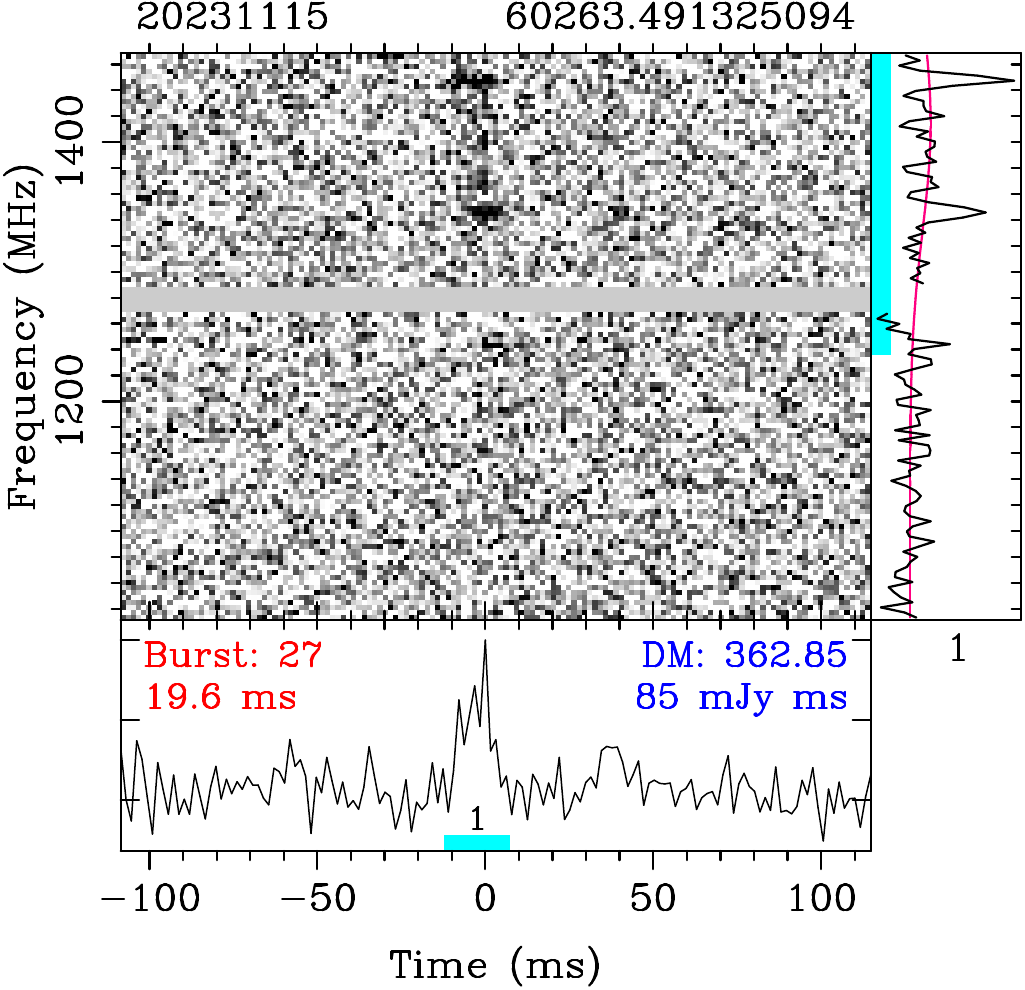}
\includegraphics[height=0.29\linewidth]{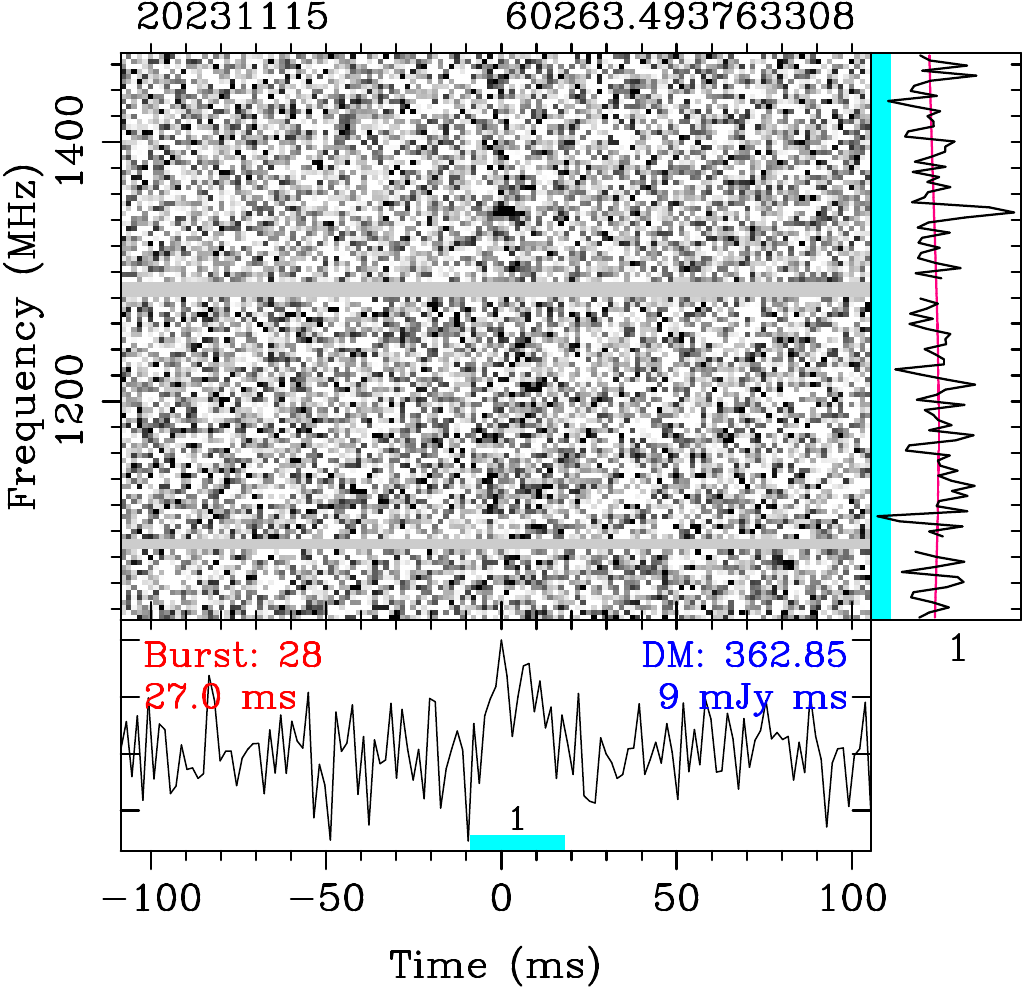}
\includegraphics[height=0.29\linewidth]{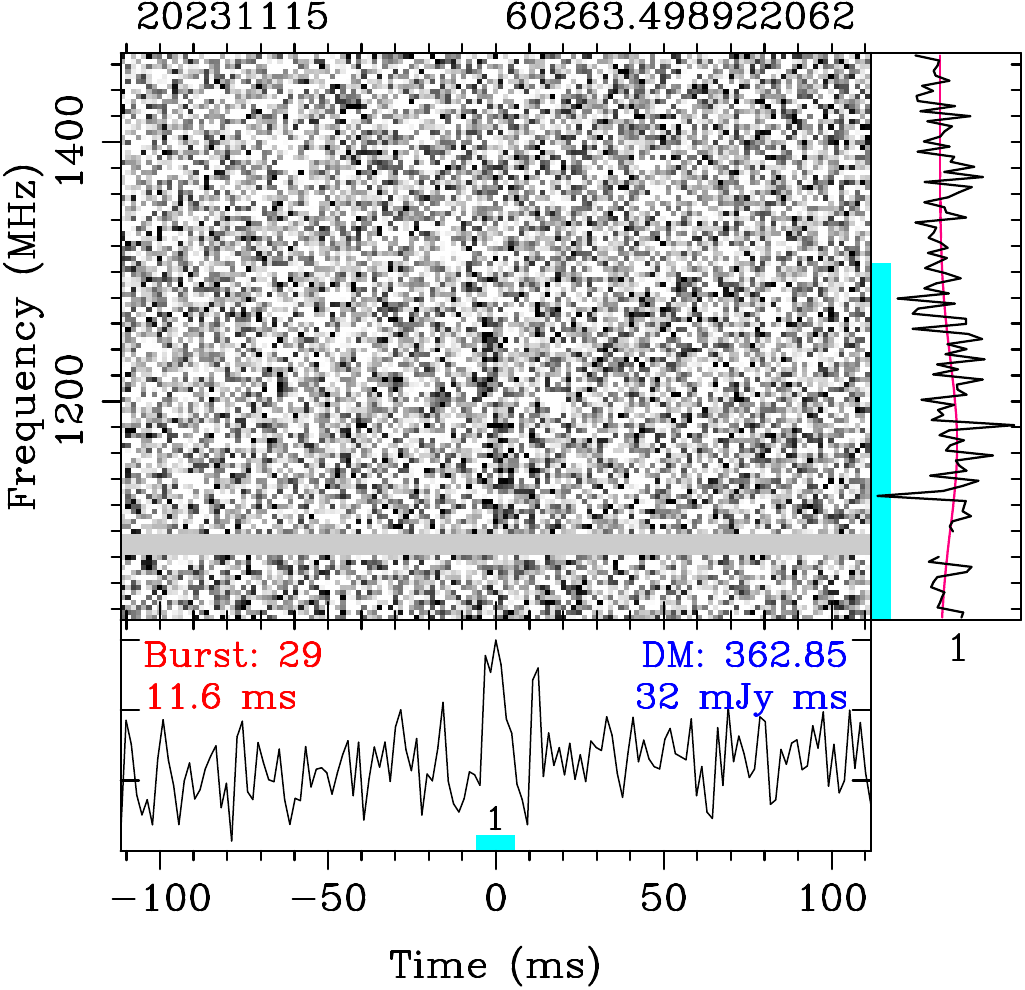}
\includegraphics[height=0.29\linewidth]{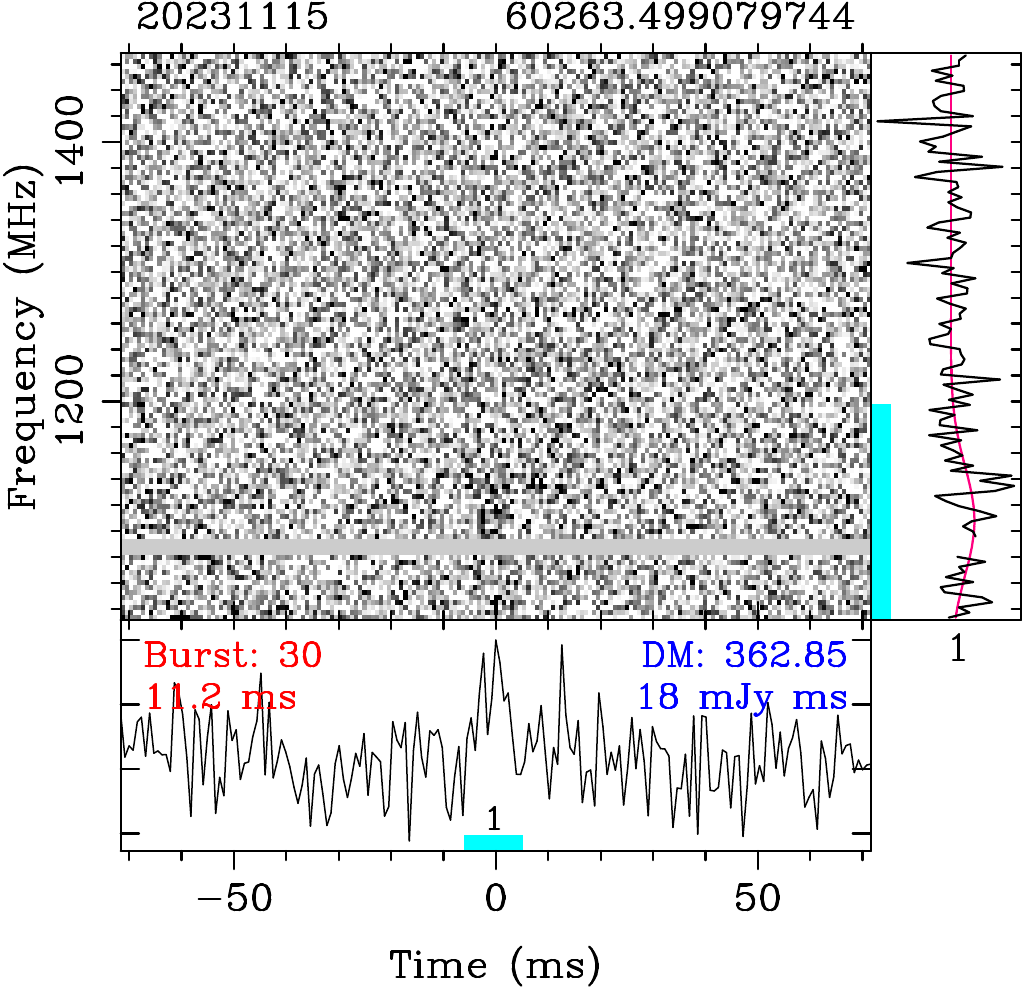}
\caption{({\textit{continued}})}
\end{figure*}
\addtocounter{figure}{-1}
\begin{figure*}
\flushleft
\includegraphics[height=0.29\linewidth]{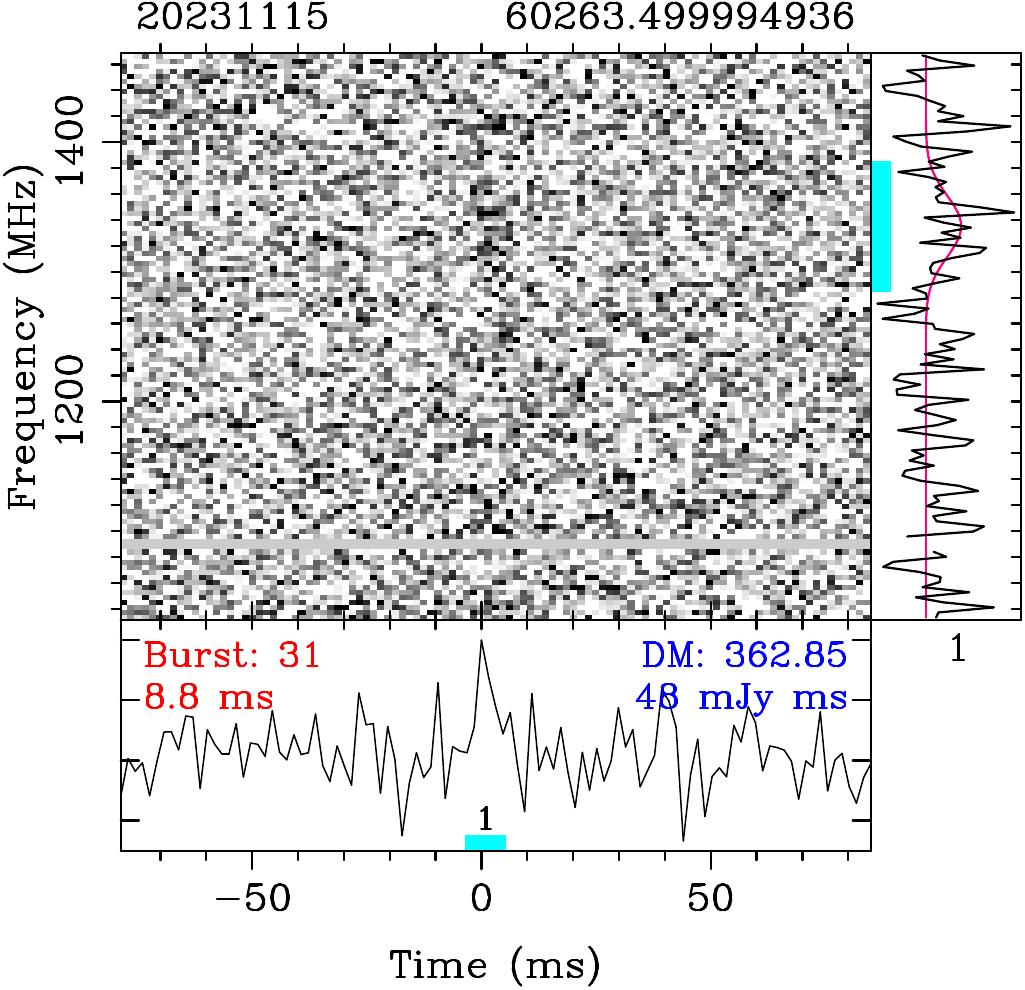}
\includegraphics[height=0.29\linewidth]{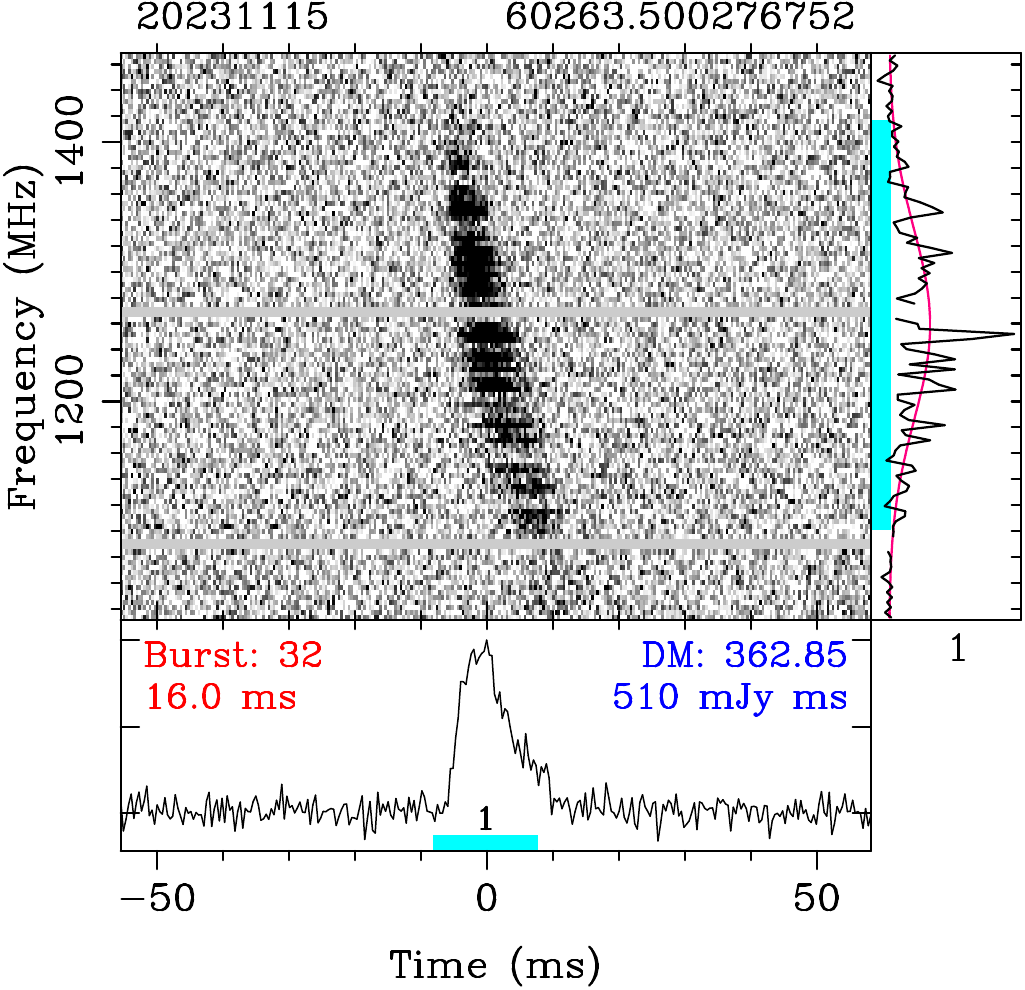}
\includegraphics[height=0.29\linewidth]{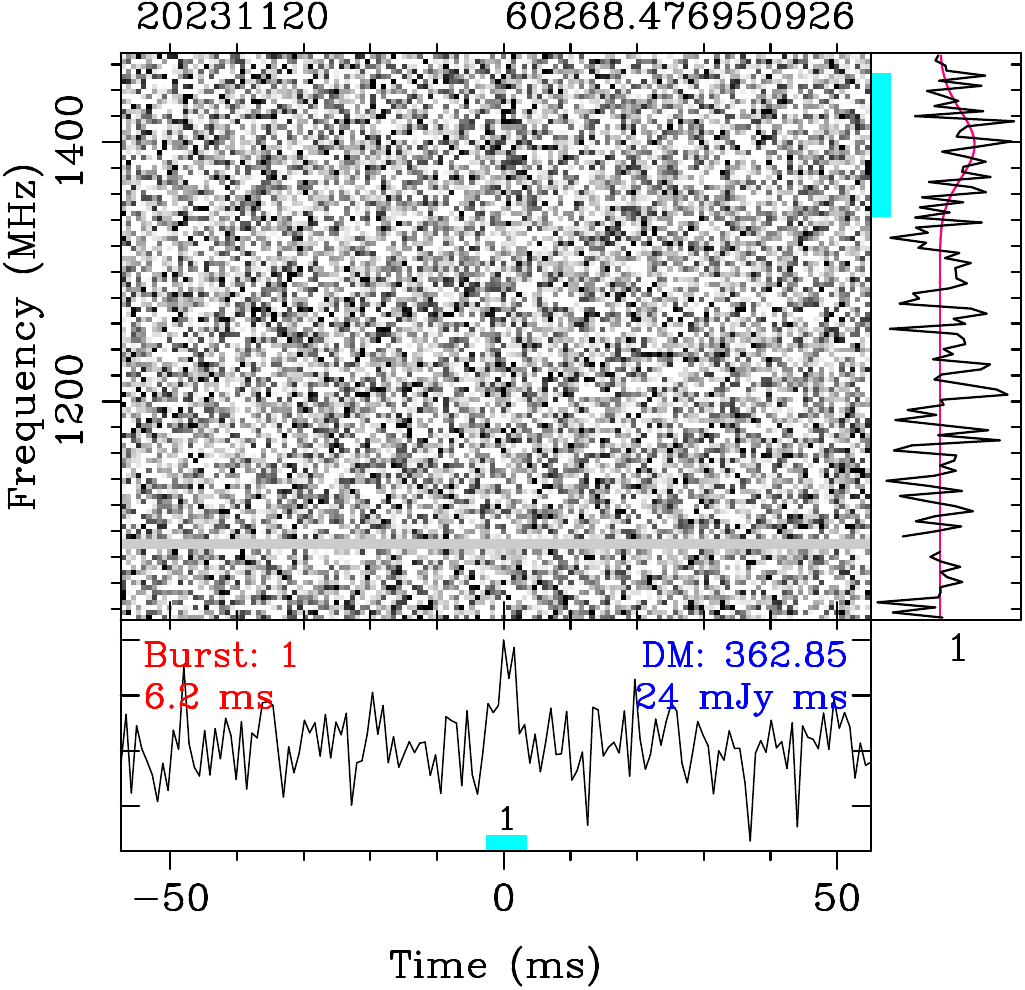}
\includegraphics[height=0.29\linewidth]{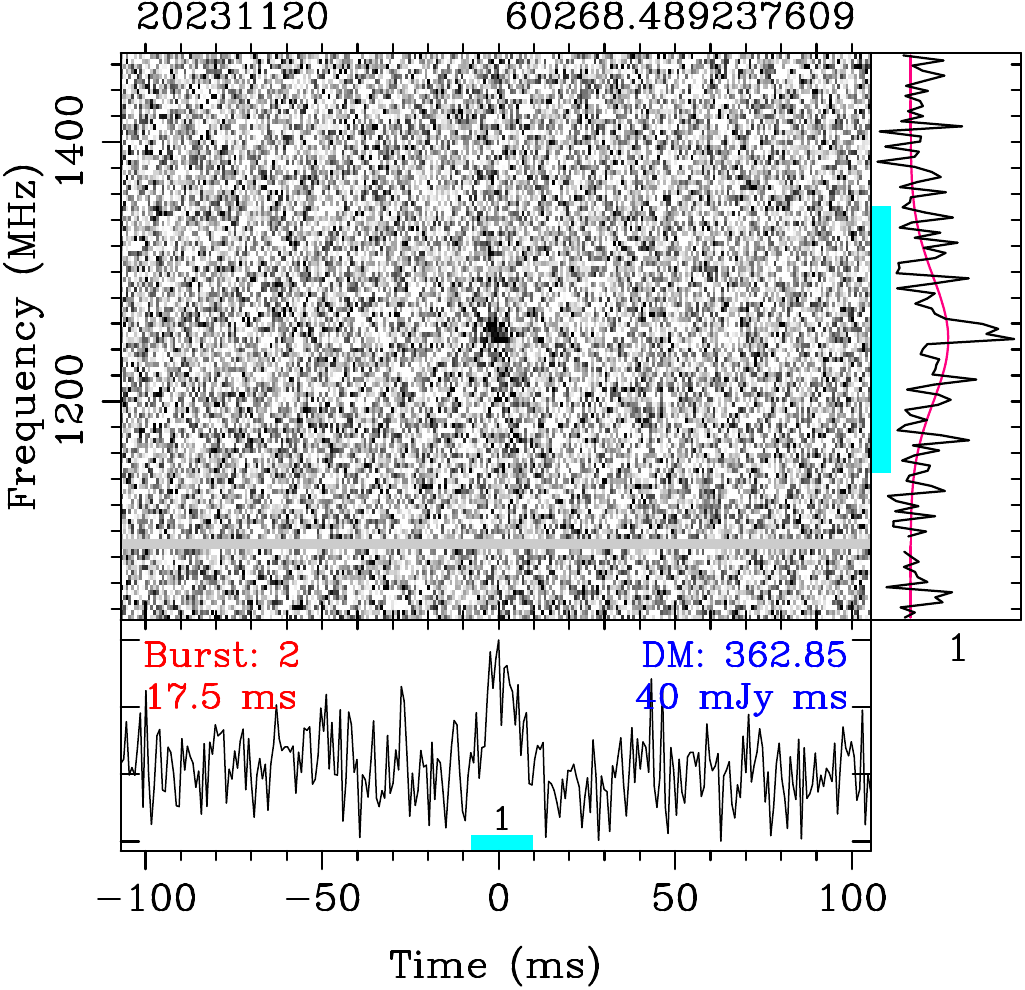}
\includegraphics[height=0.29\linewidth]{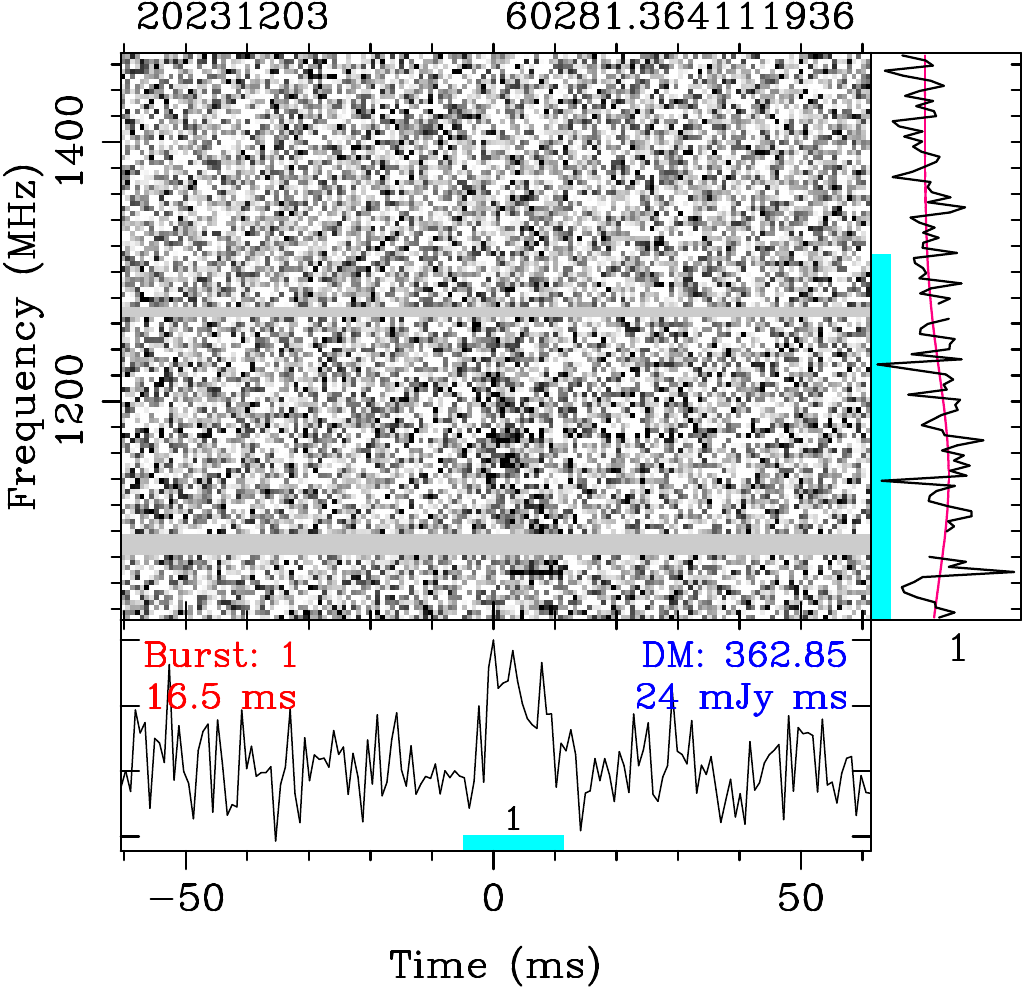}
\includegraphics[height=0.29\linewidth]{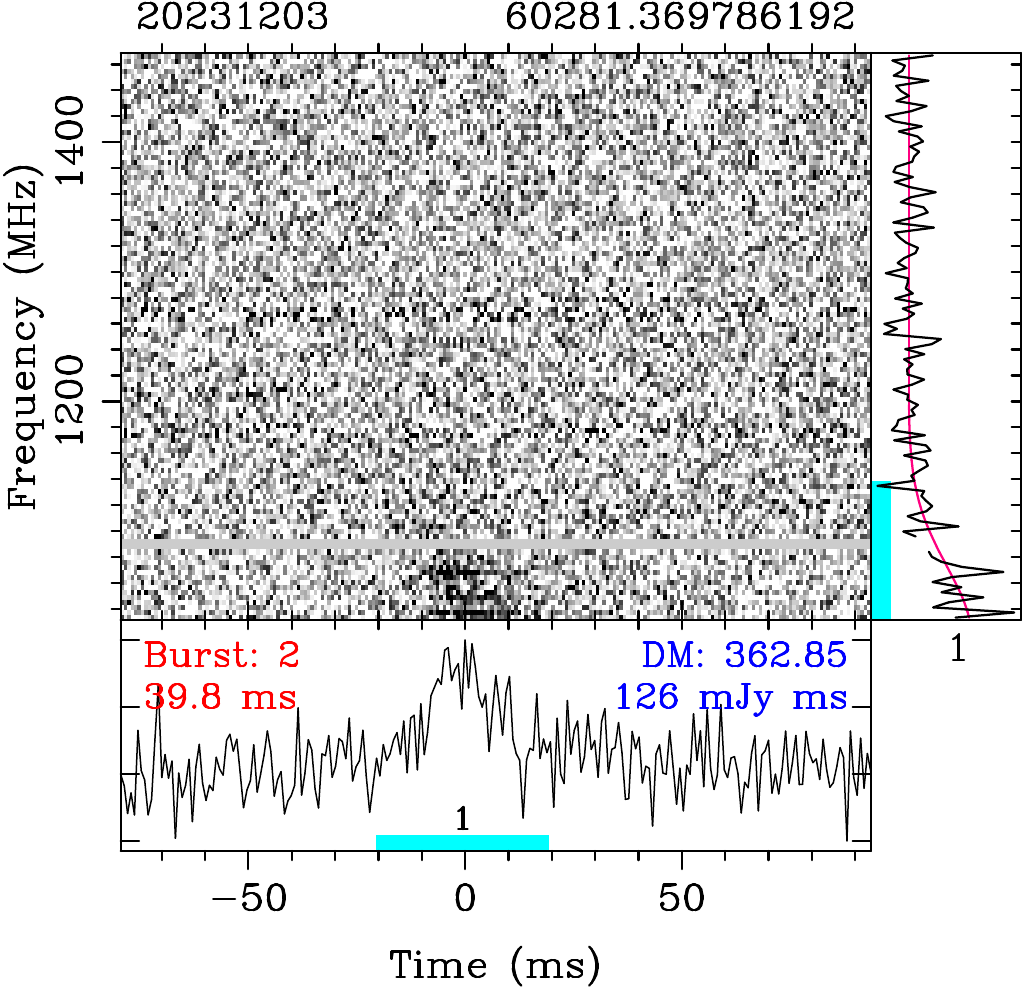}
\includegraphics[height=0.29\linewidth]{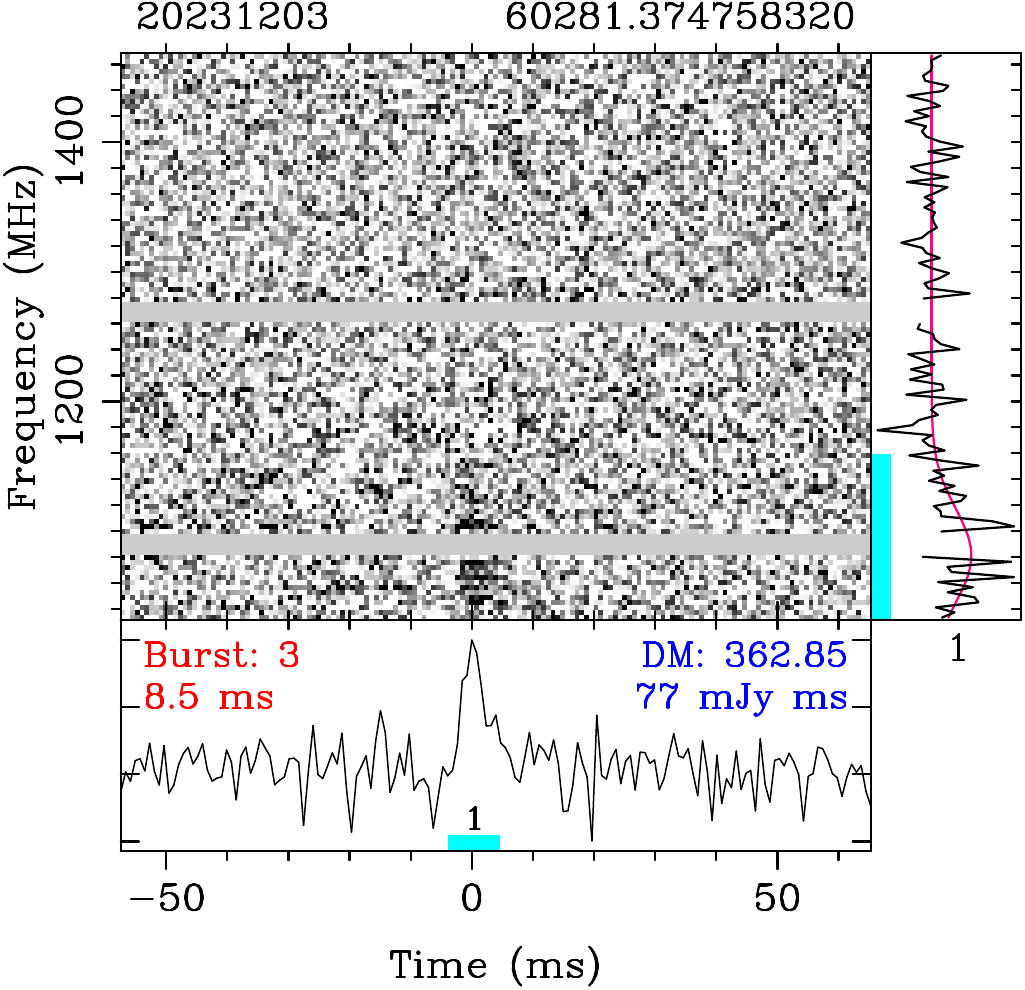}
\includegraphics[height=0.29\linewidth]{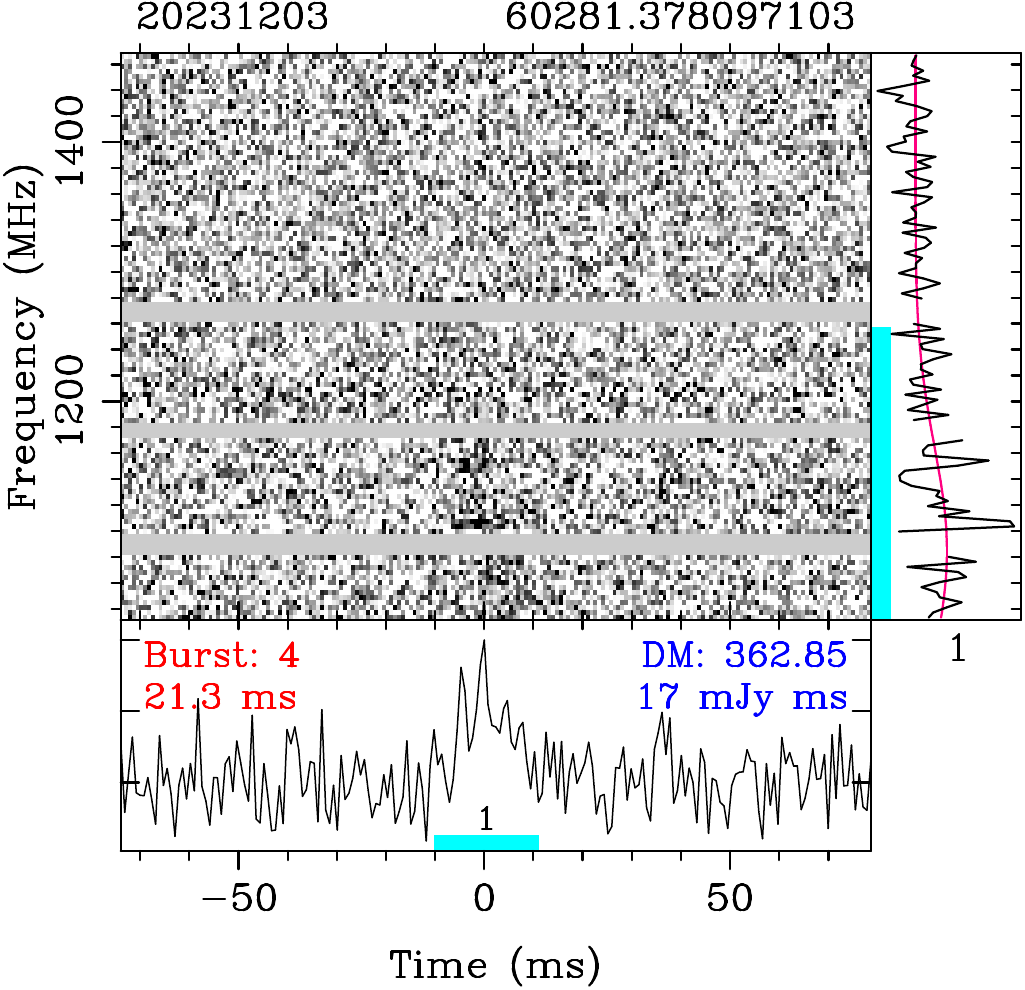}
\includegraphics[height=0.29\linewidth]{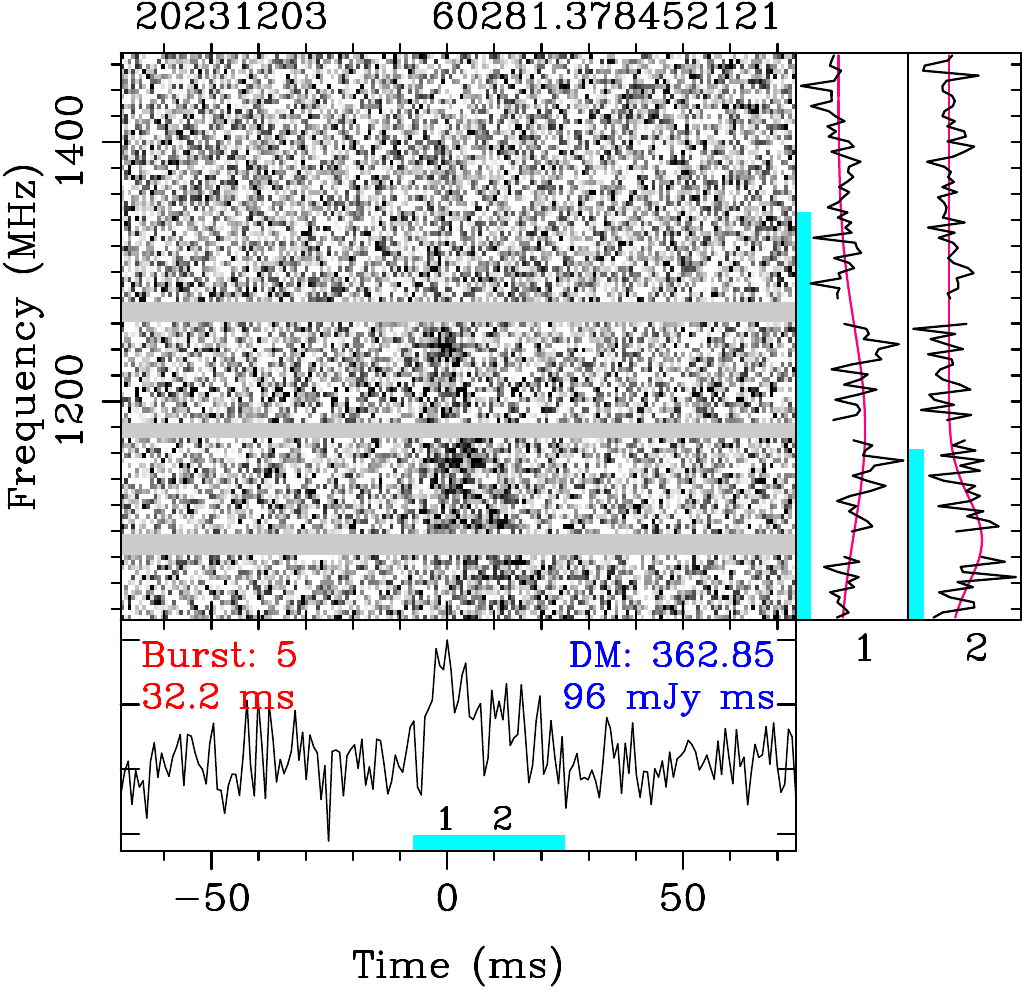}
\includegraphics[height=0.29\linewidth]{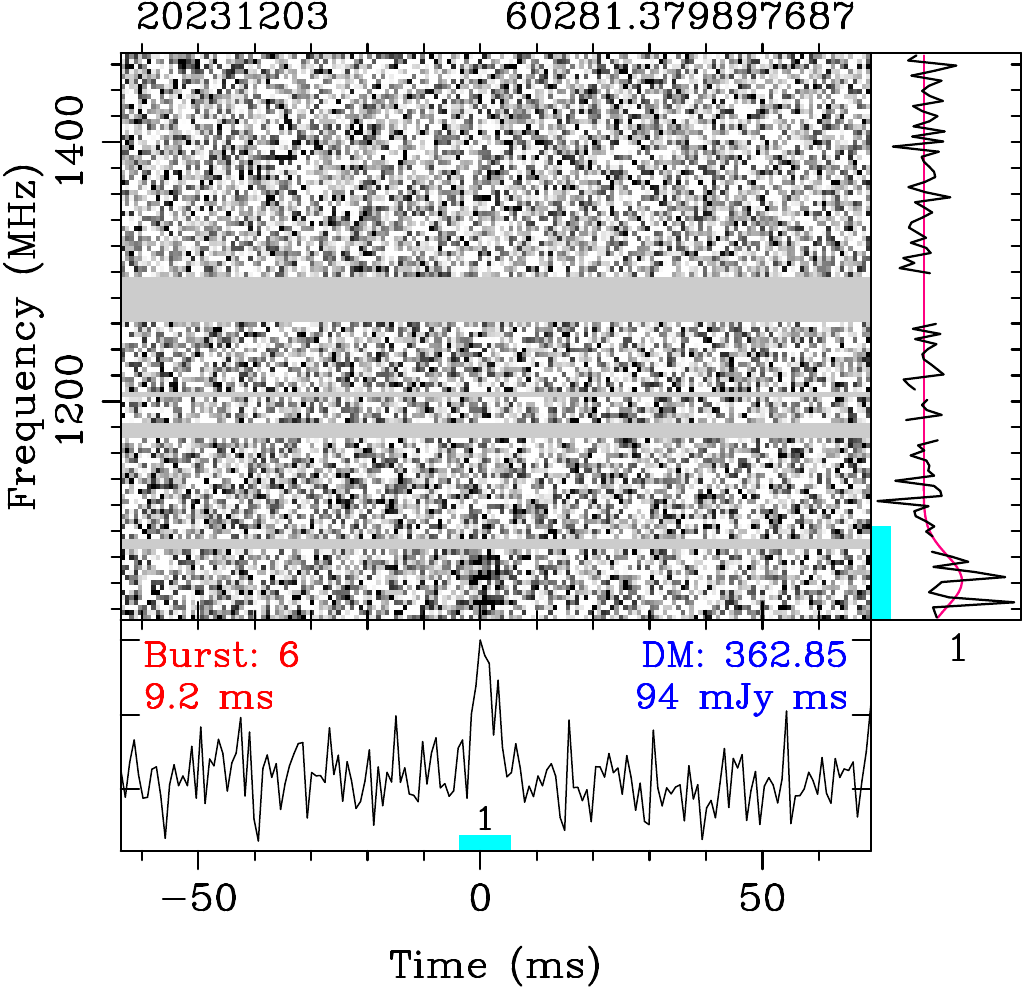}
\includegraphics[height=0.29\linewidth]{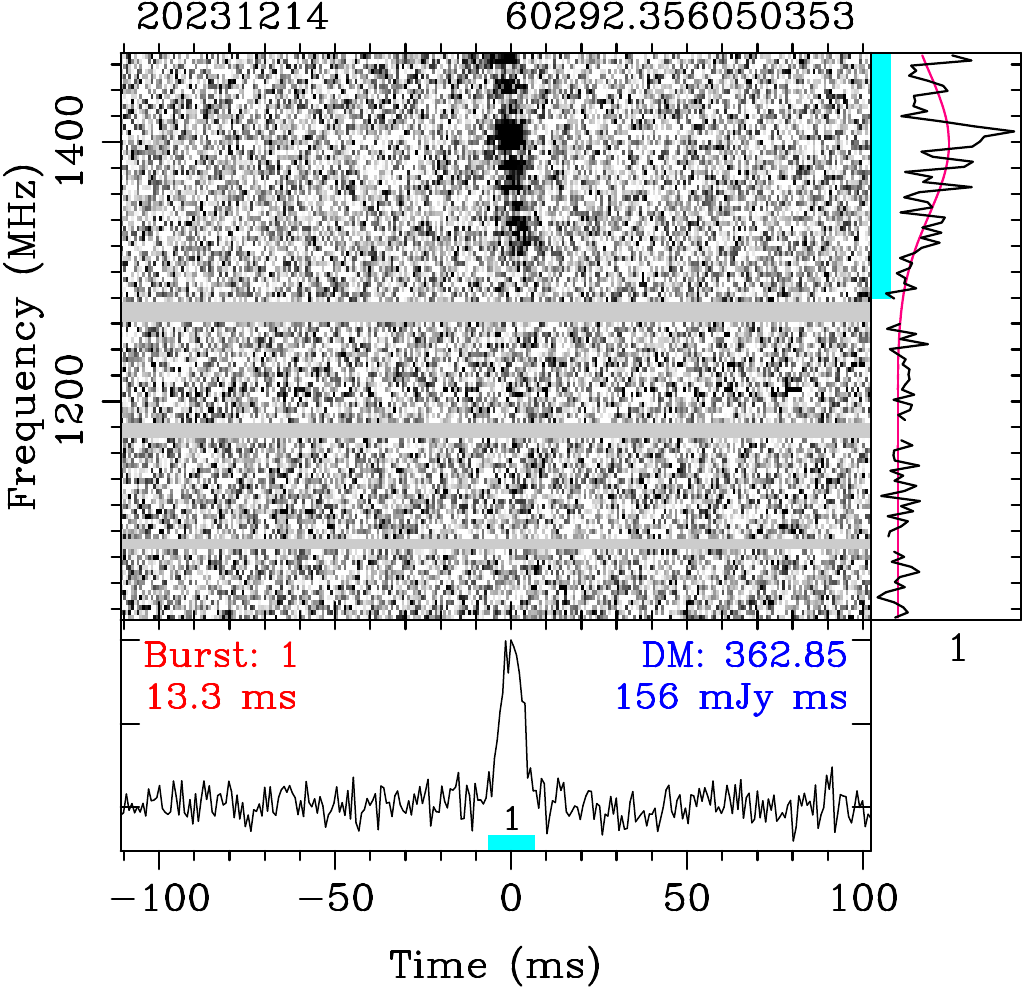}
\includegraphics[height=0.29\linewidth]{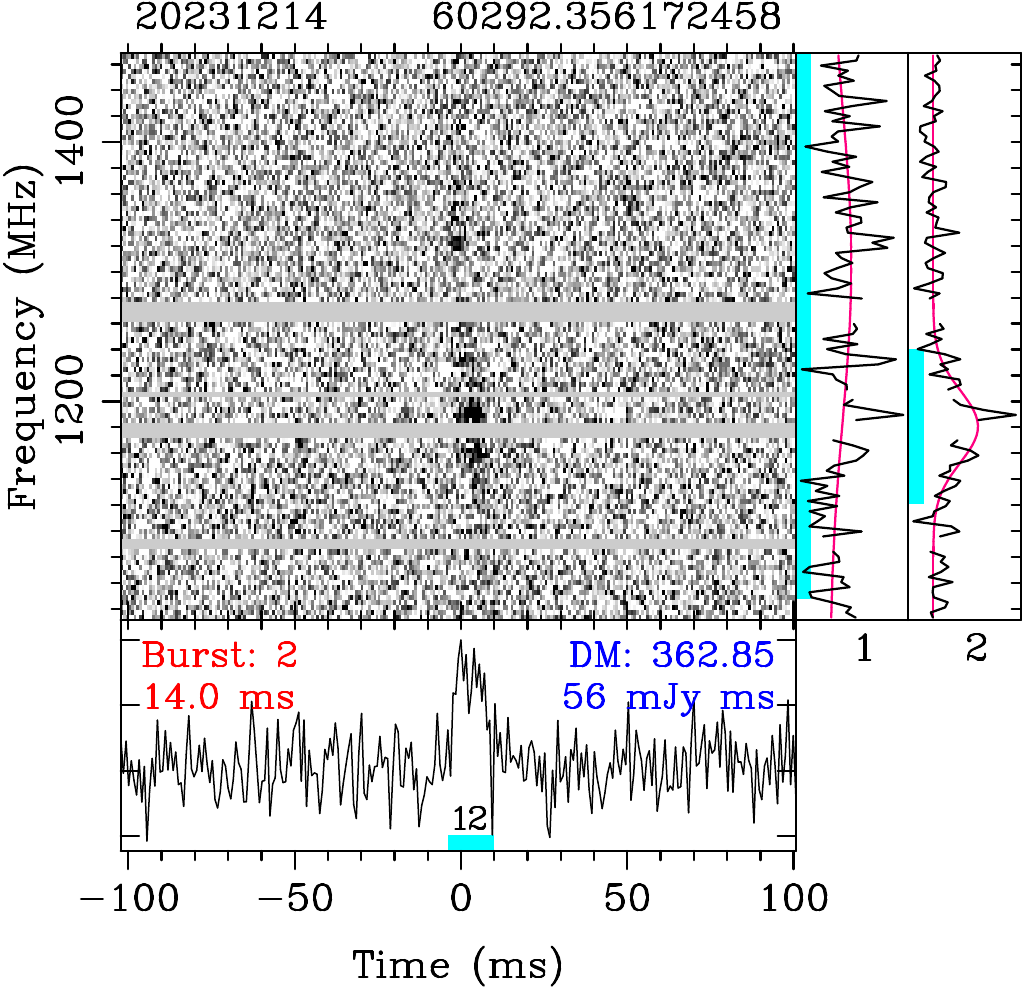}
\caption{({\textit{continued}})}
\end{figure*}
\addtocounter{figure}{-1}
\begin{figure*}
\flushleft
\includegraphics[height=0.29\linewidth]{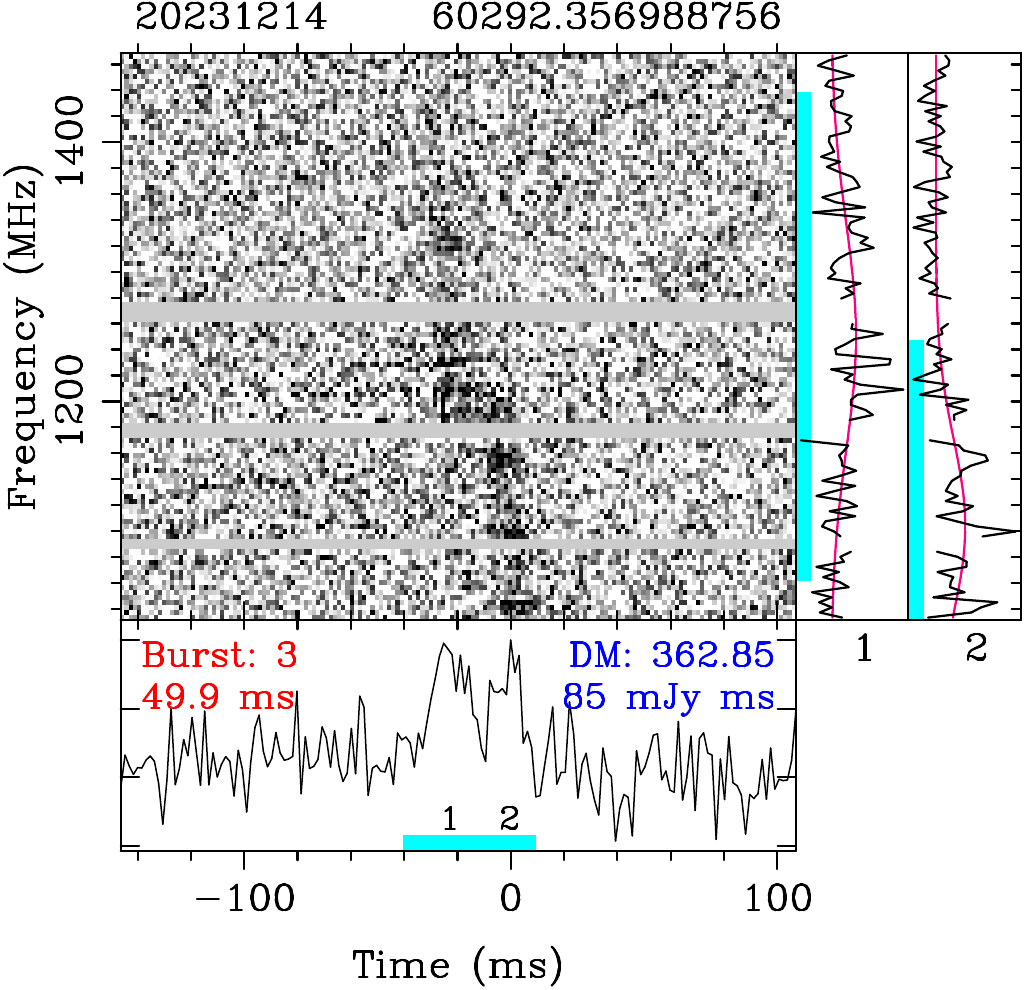}
\includegraphics[height=0.29\linewidth]{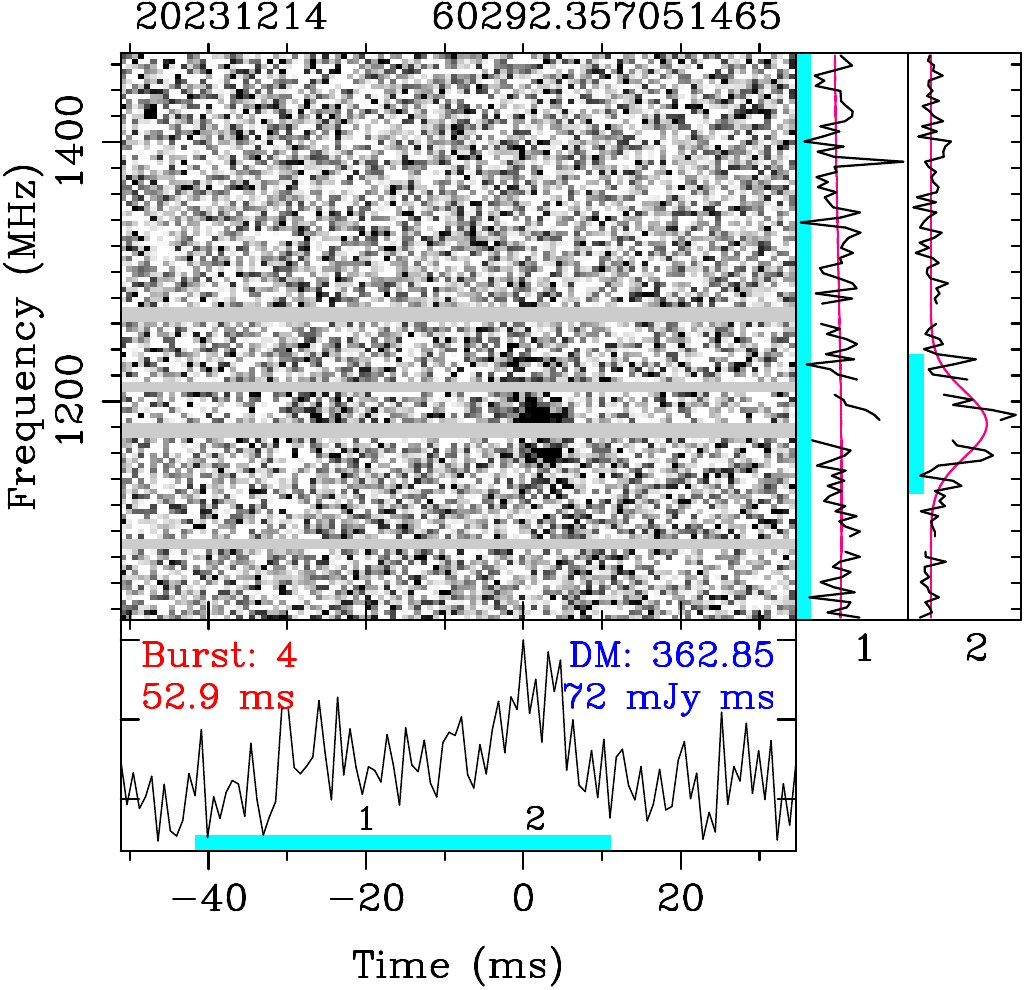}
\includegraphics[height=0.29\linewidth]{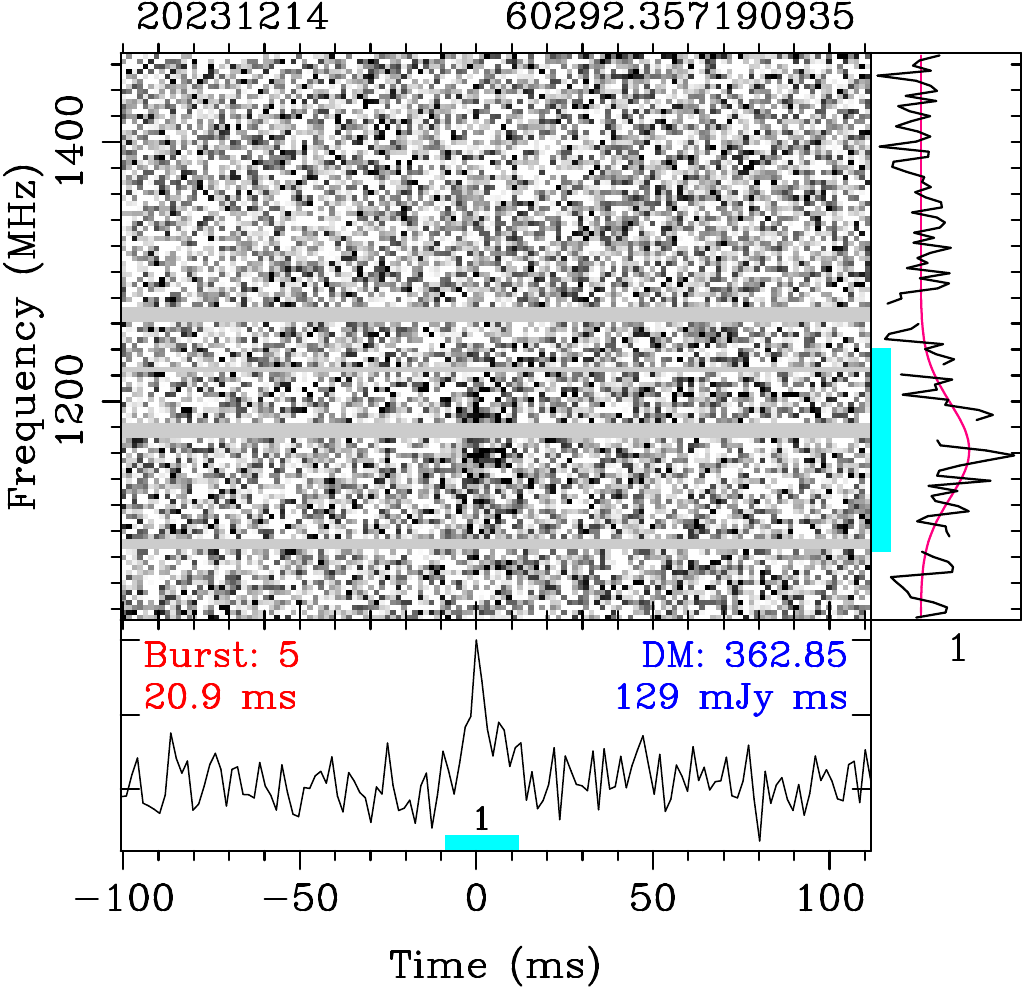}
\includegraphics[height=0.29\linewidth]{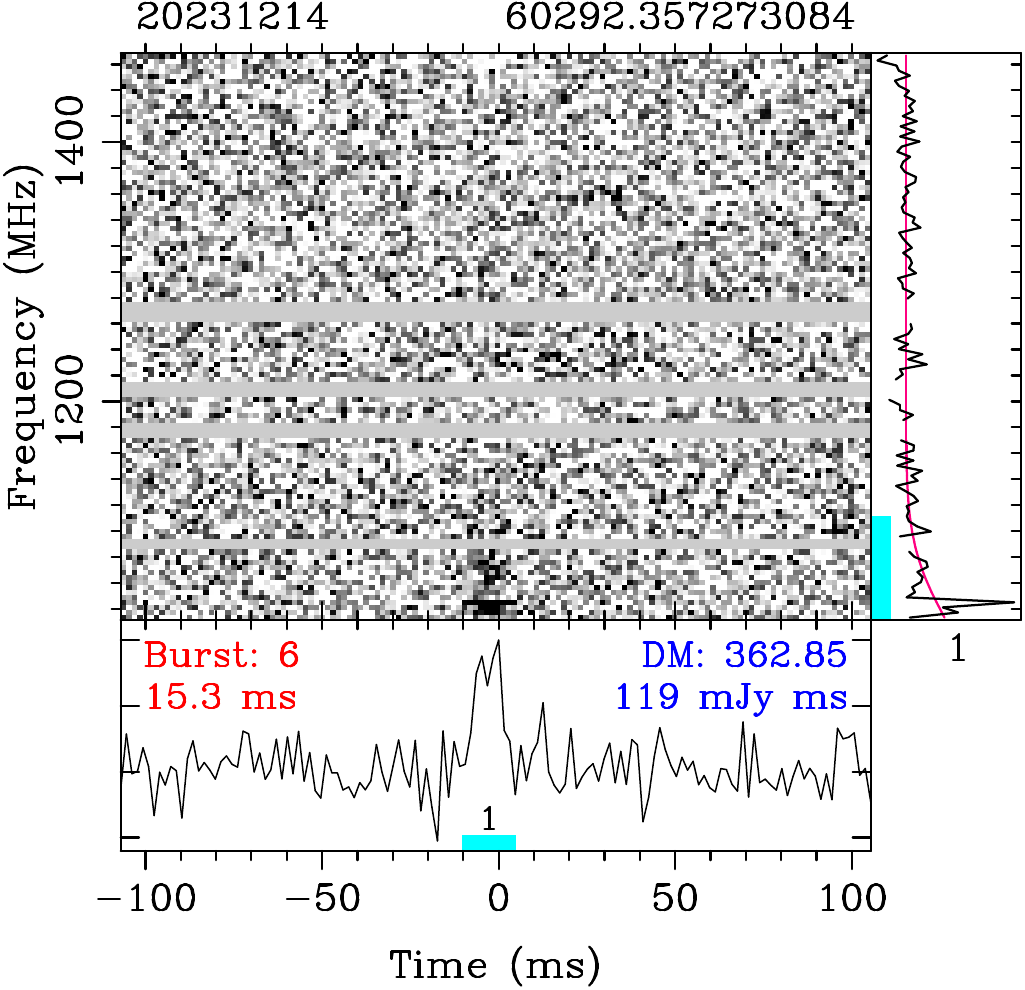}
\includegraphics[height=0.29\linewidth]{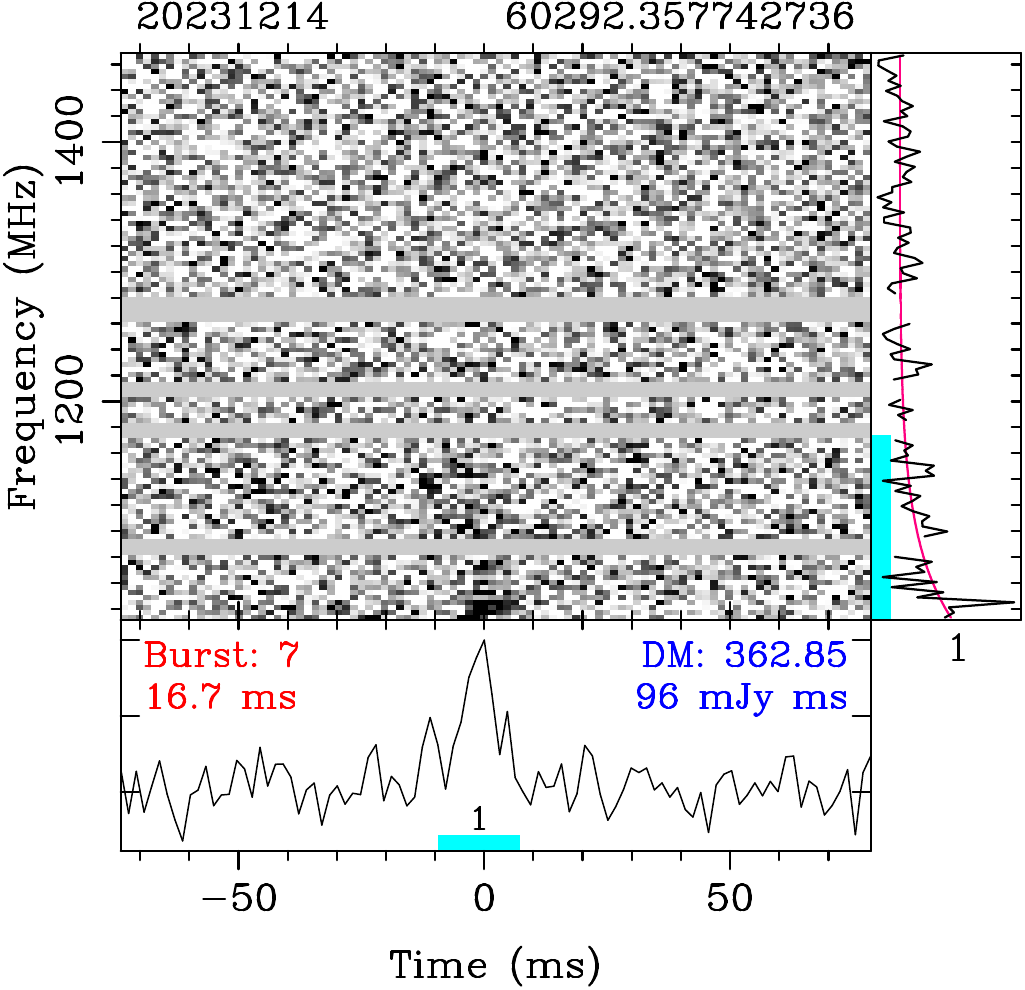}
\includegraphics[height=0.29\linewidth]{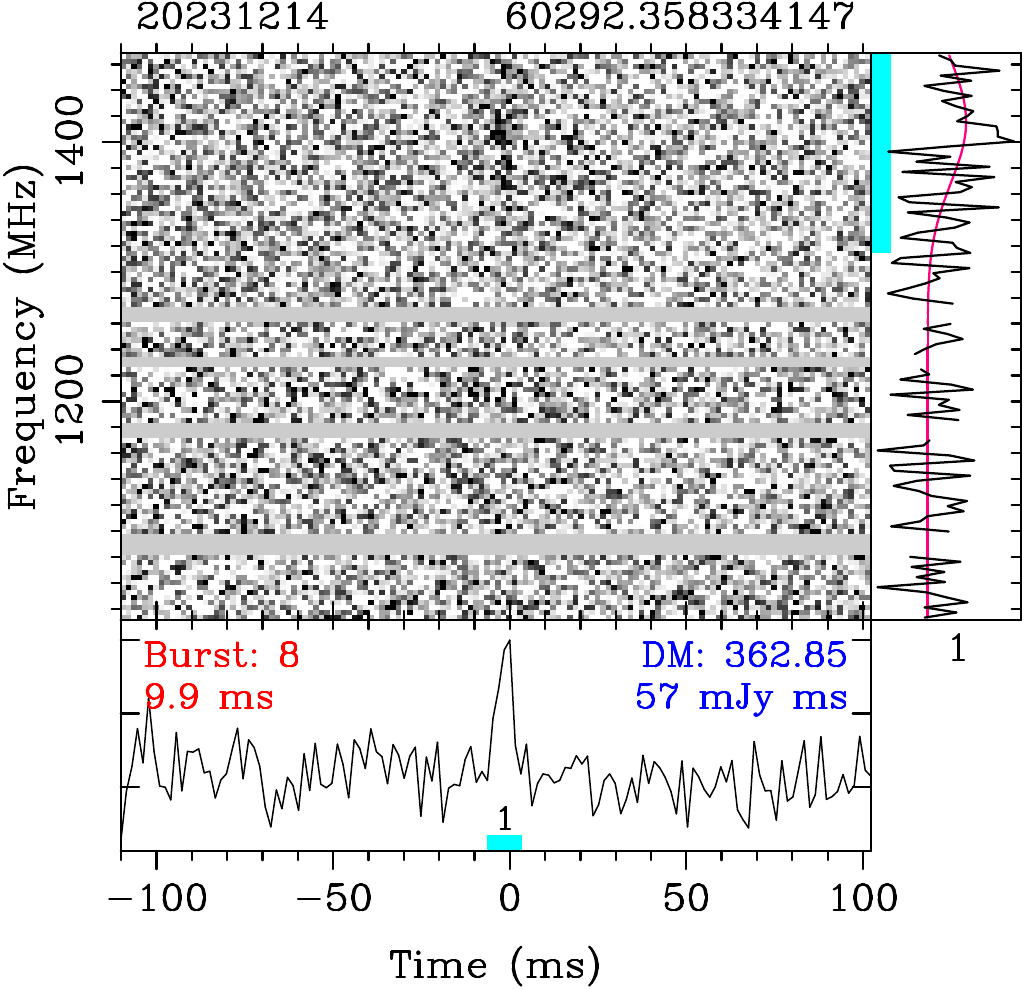}
\includegraphics[height=0.29\linewidth]{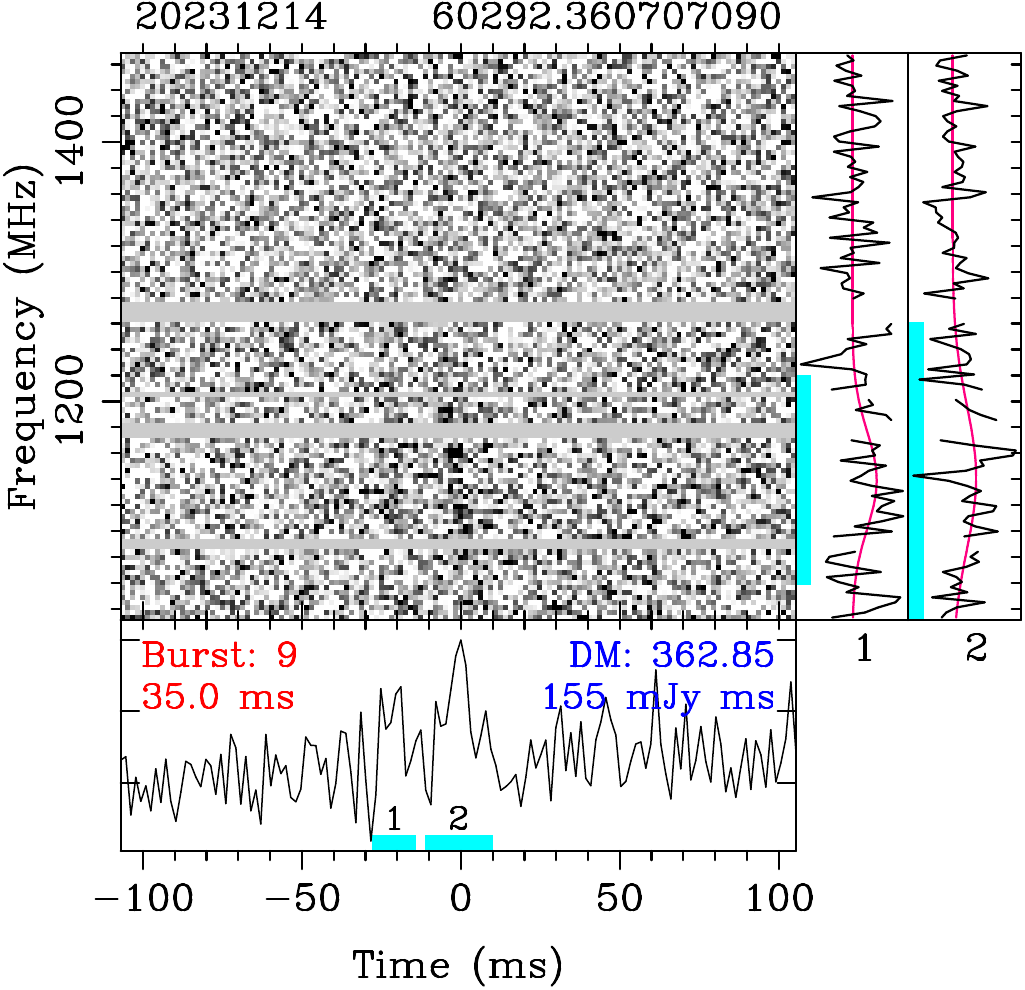}
\includegraphics[height=0.29\linewidth]{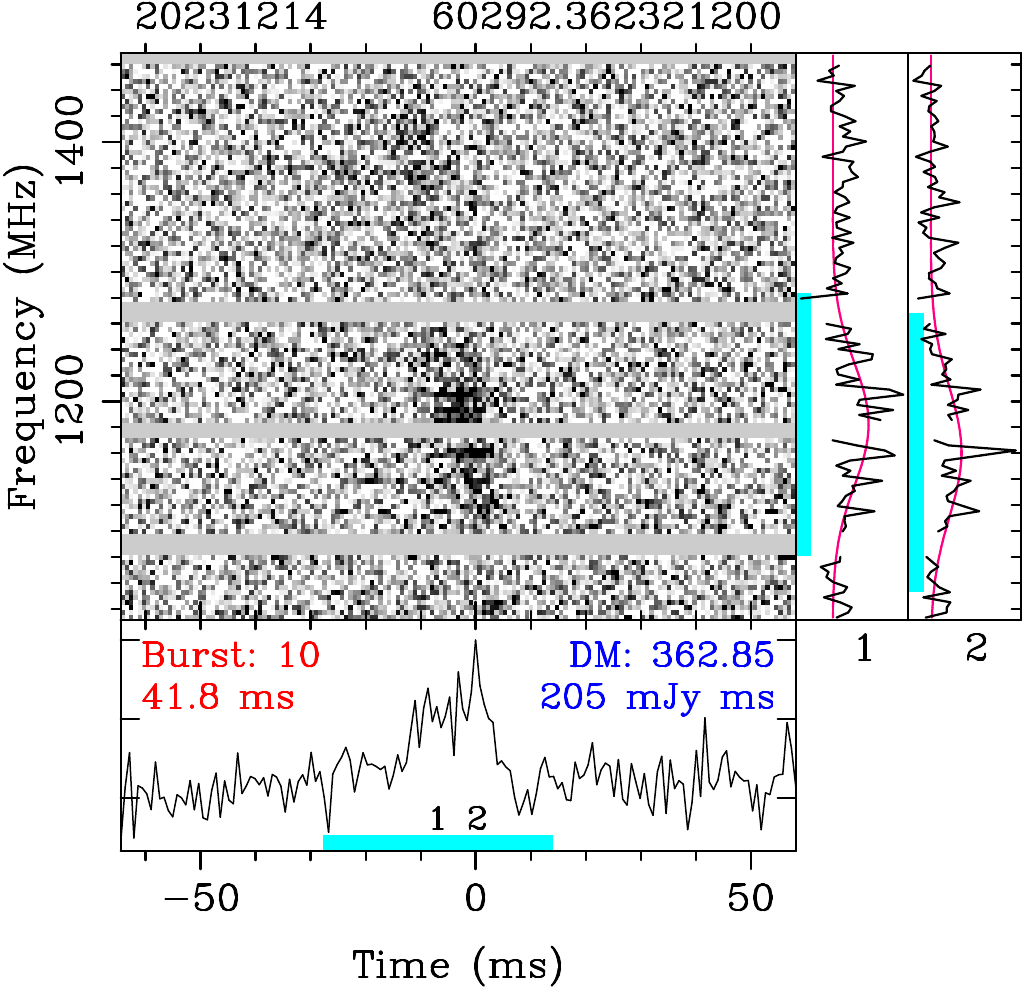}
\includegraphics[height=0.29\linewidth]{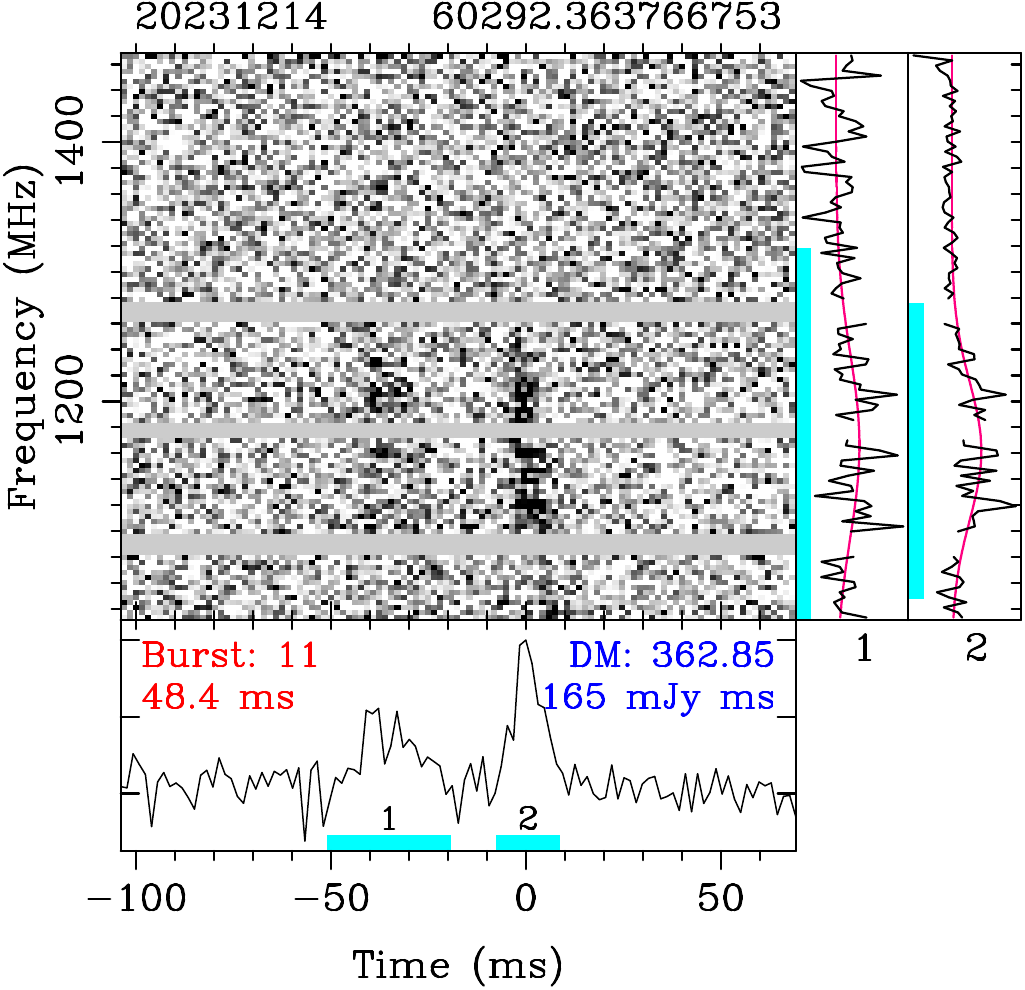}
\includegraphics[height=0.29\linewidth]{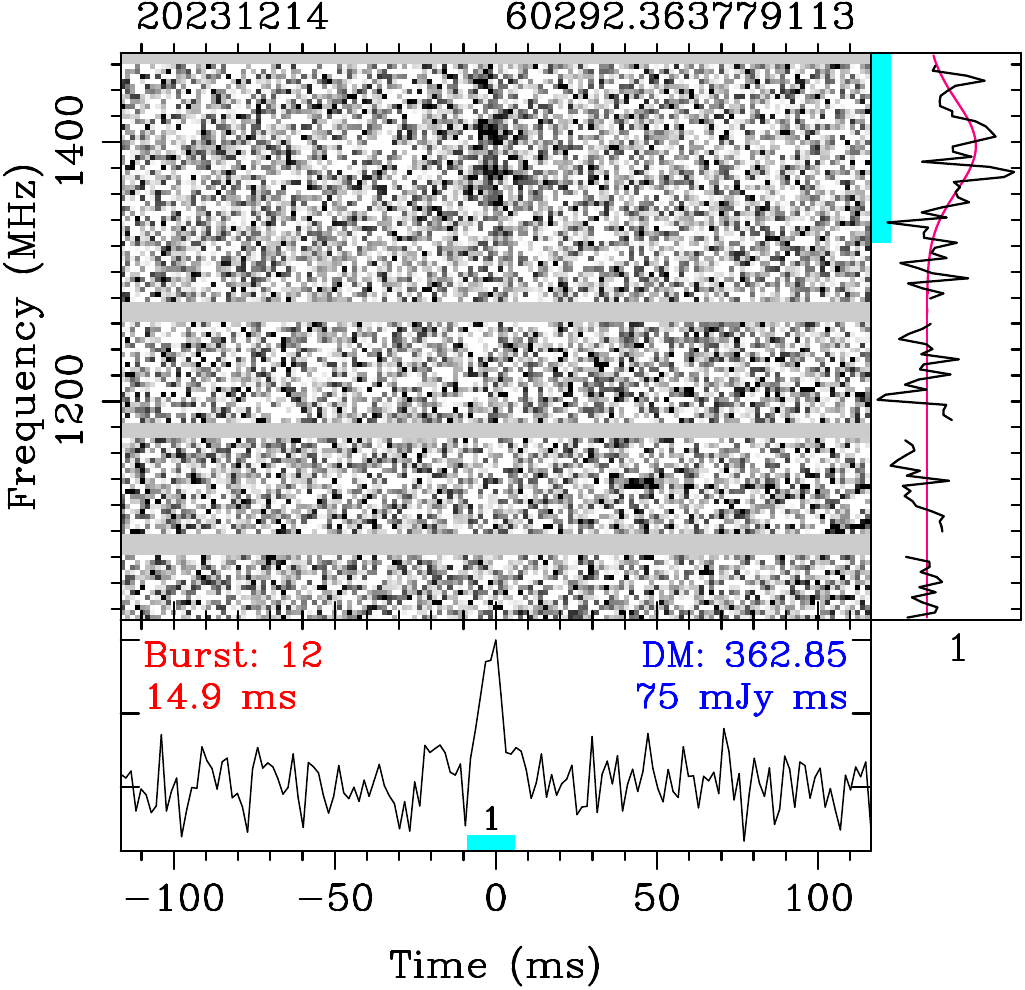}
\includegraphics[height=0.29\linewidth]{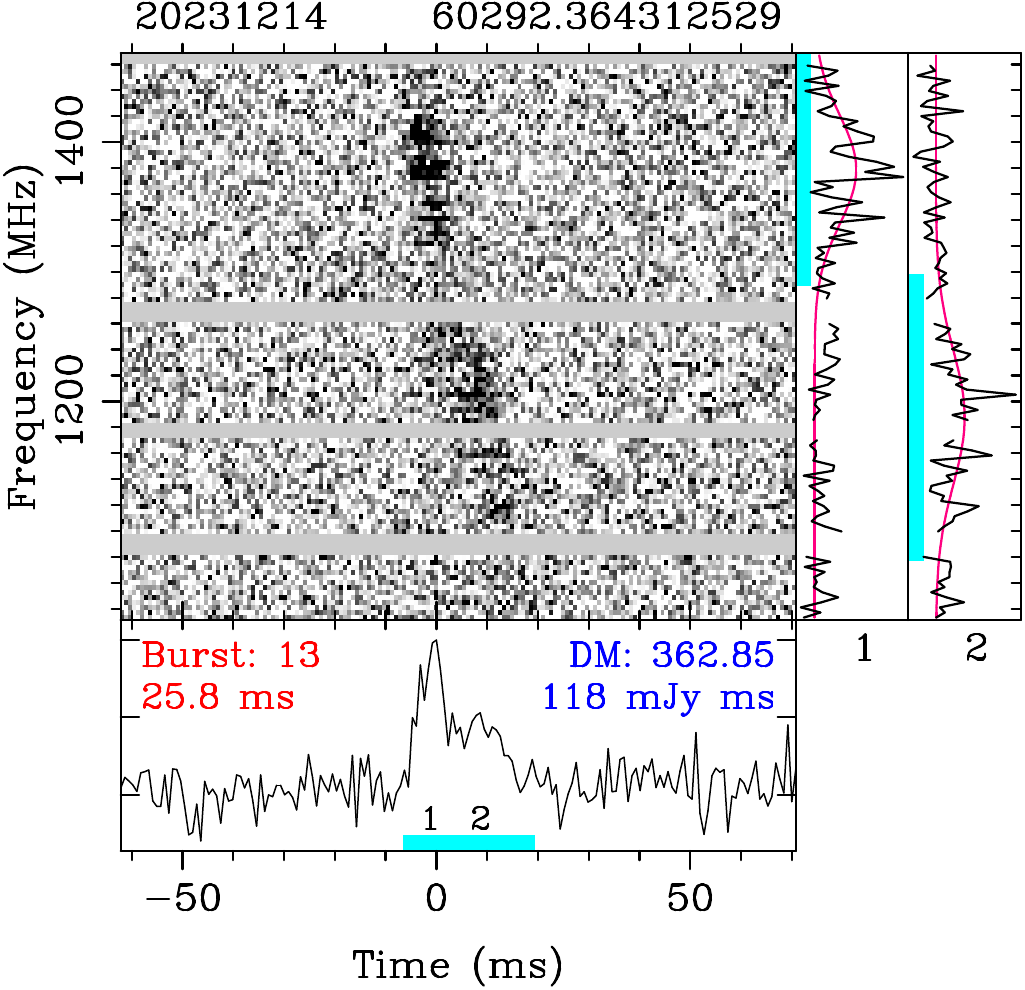}
\includegraphics[height=0.29\linewidth]{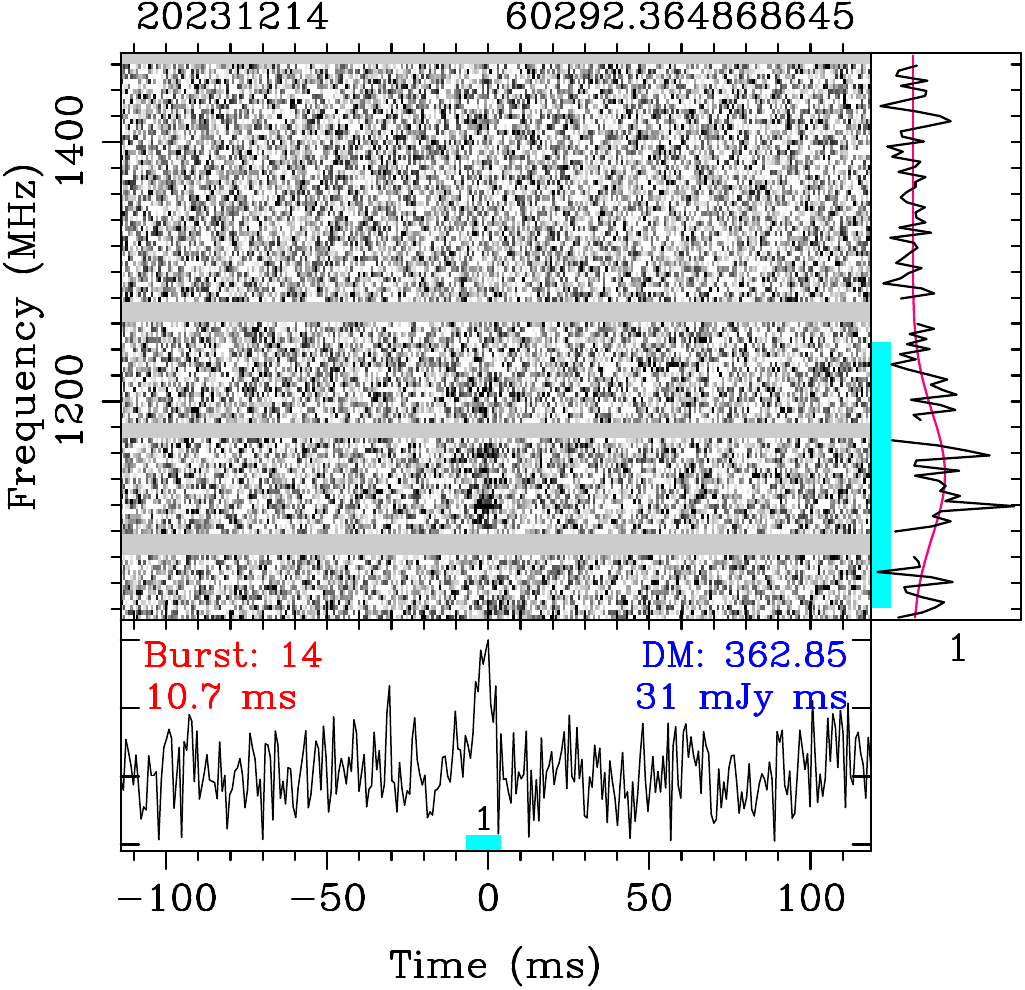}
\caption{({\textit{continued}})}
\end{figure*}
\addtocounter{figure}{-1}
\begin{figure*}
\flushleft
\includegraphics[height=0.29\linewidth]{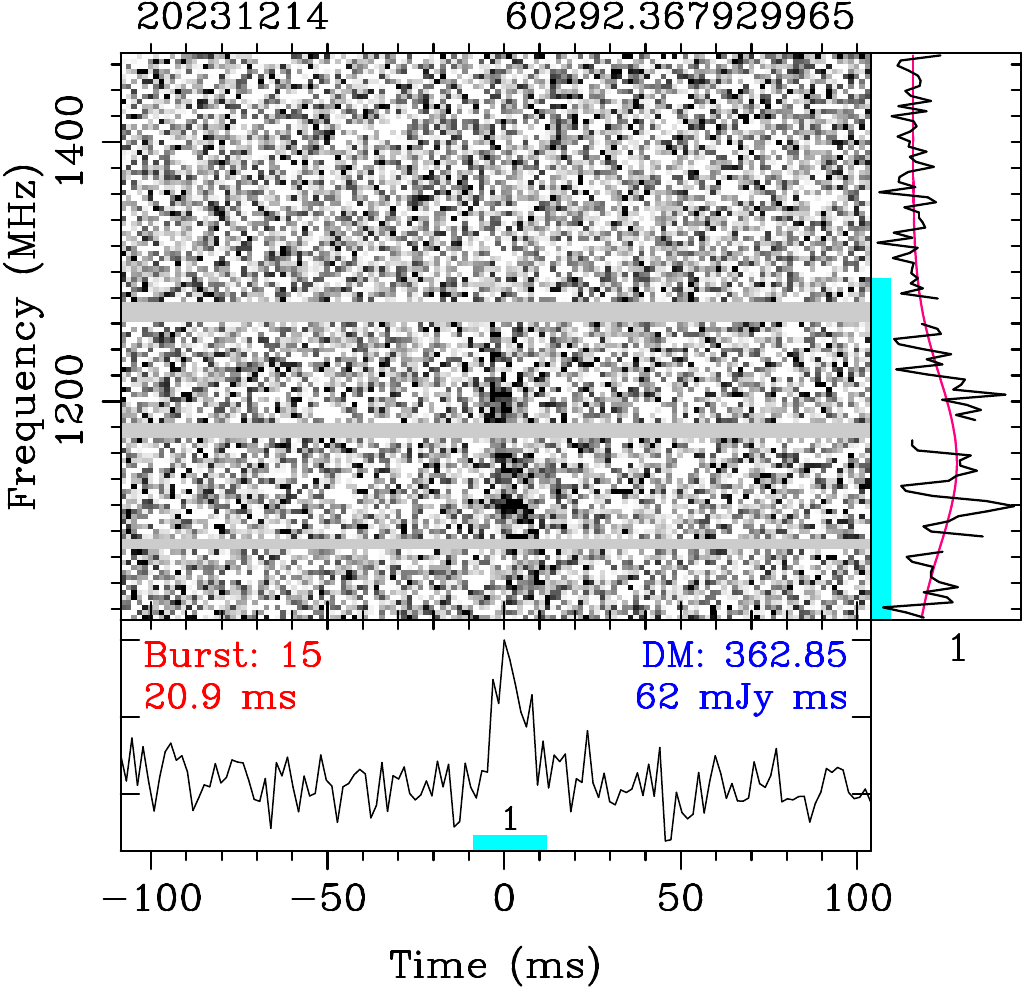}
\includegraphics[height=0.29\linewidth]{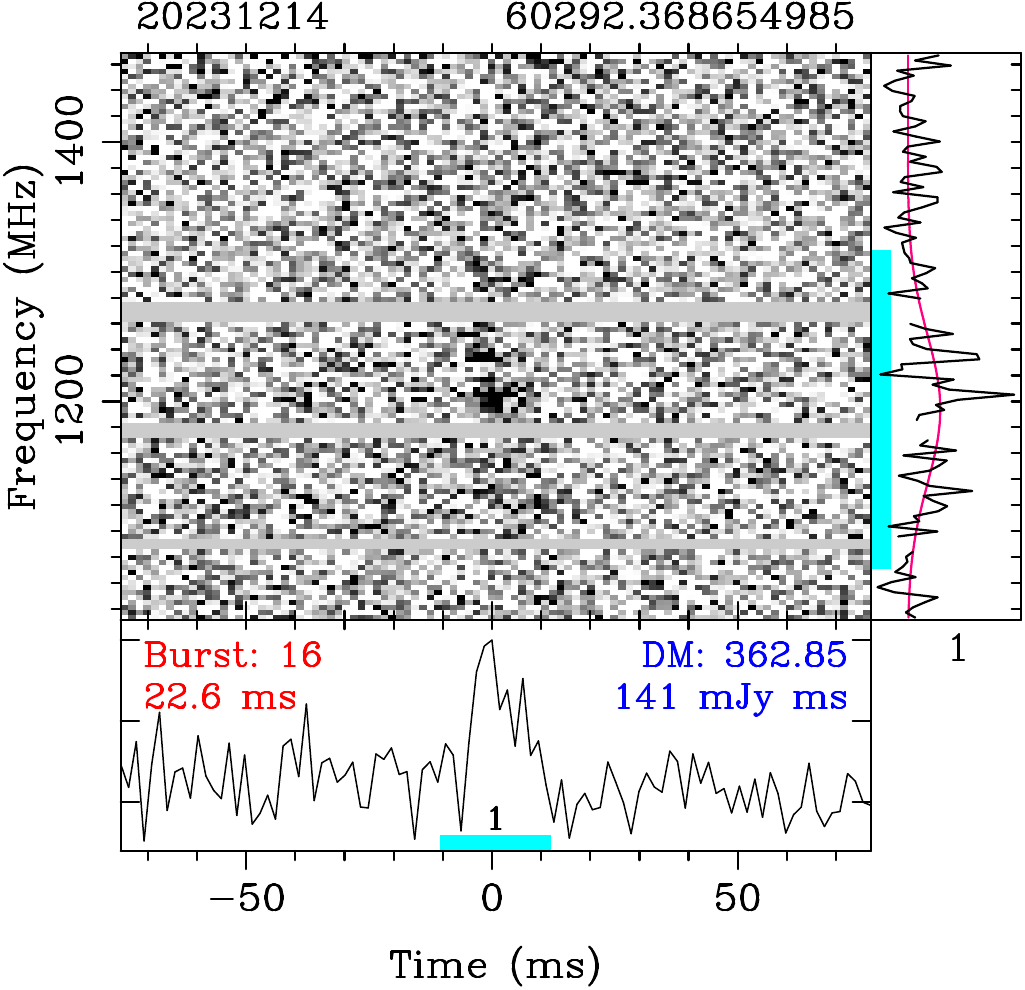}
\includegraphics[height=0.29\linewidth]{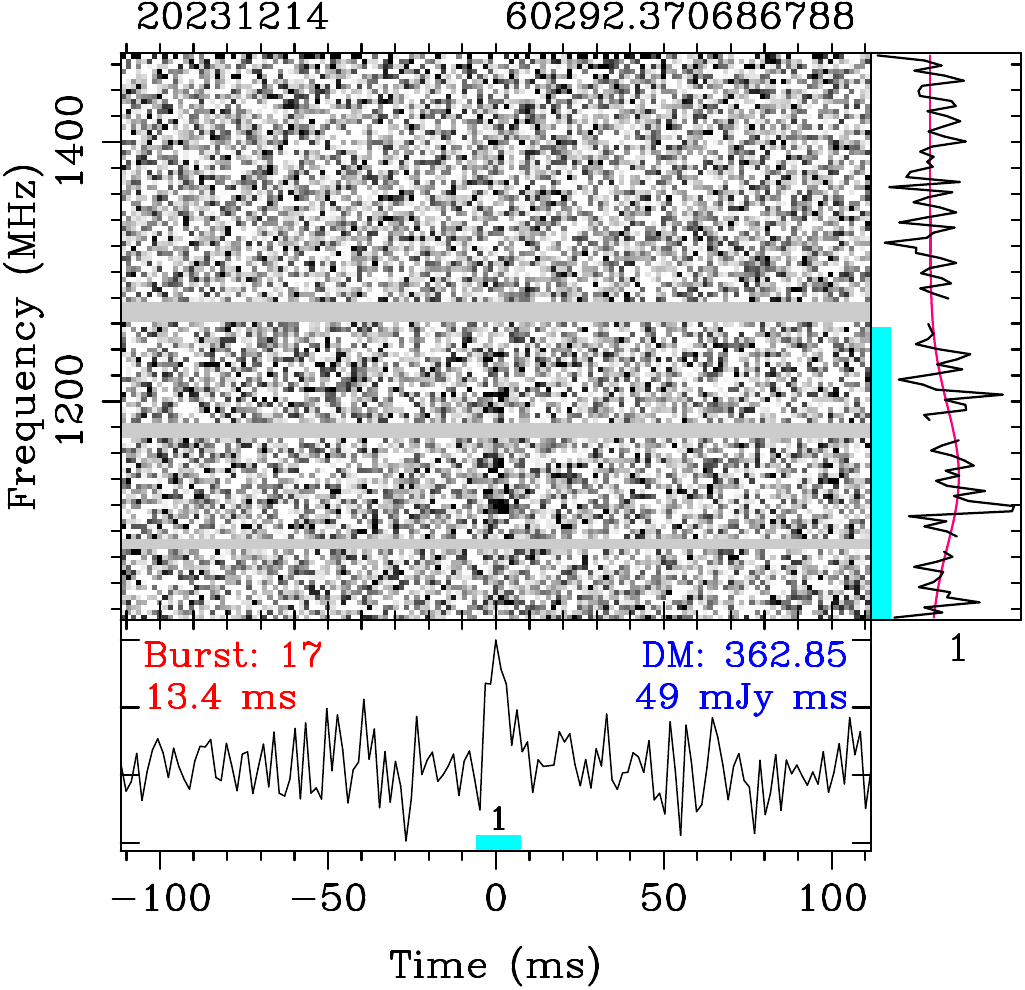}
\includegraphics[height=0.29\linewidth]{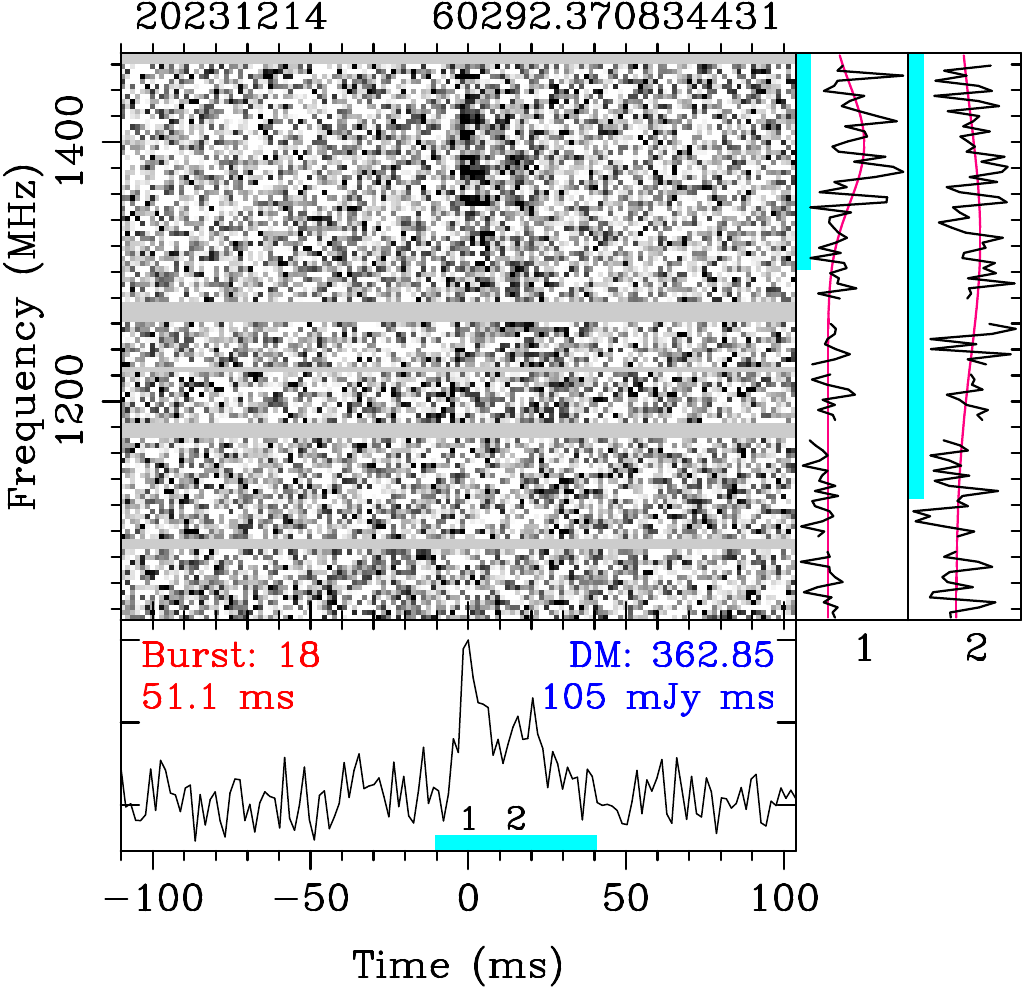}
\includegraphics[height=0.29\linewidth]{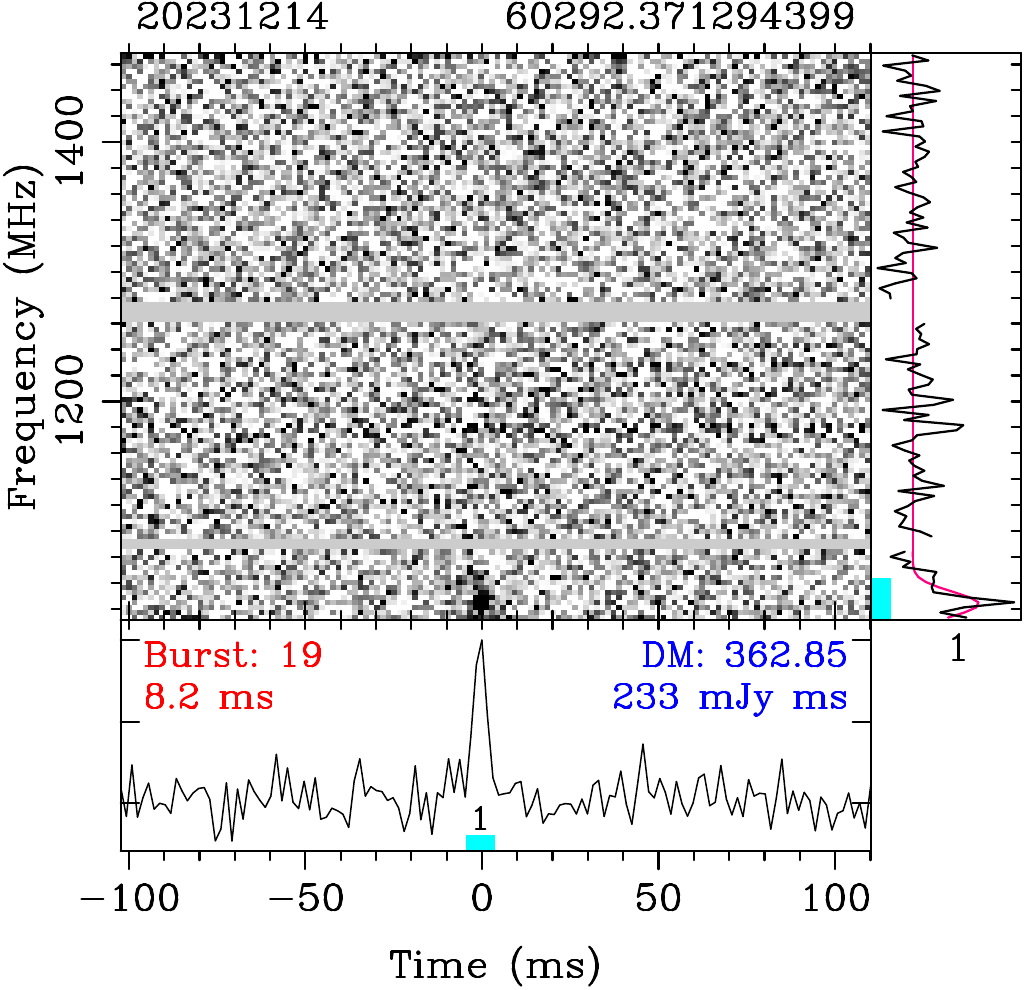}
\includegraphics[height=0.29\linewidth]{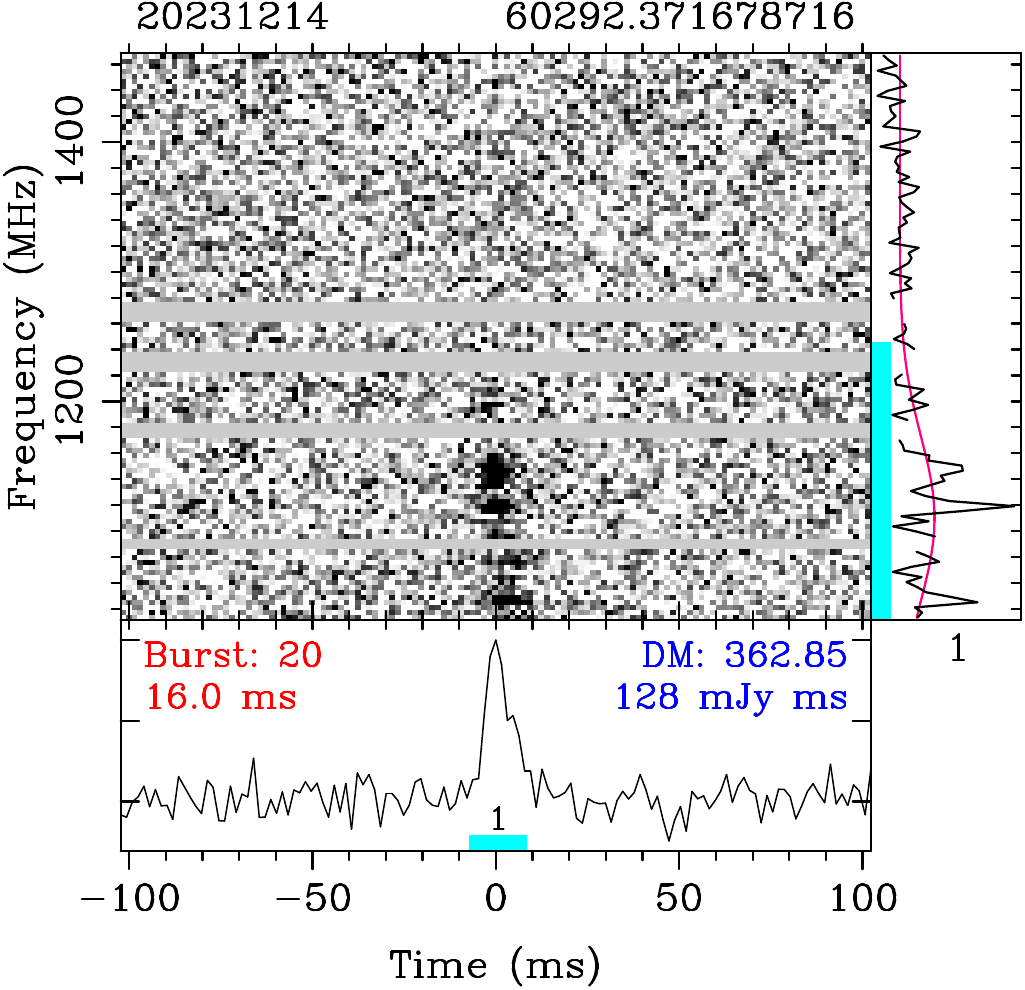}
\includegraphics[height=0.29\linewidth]{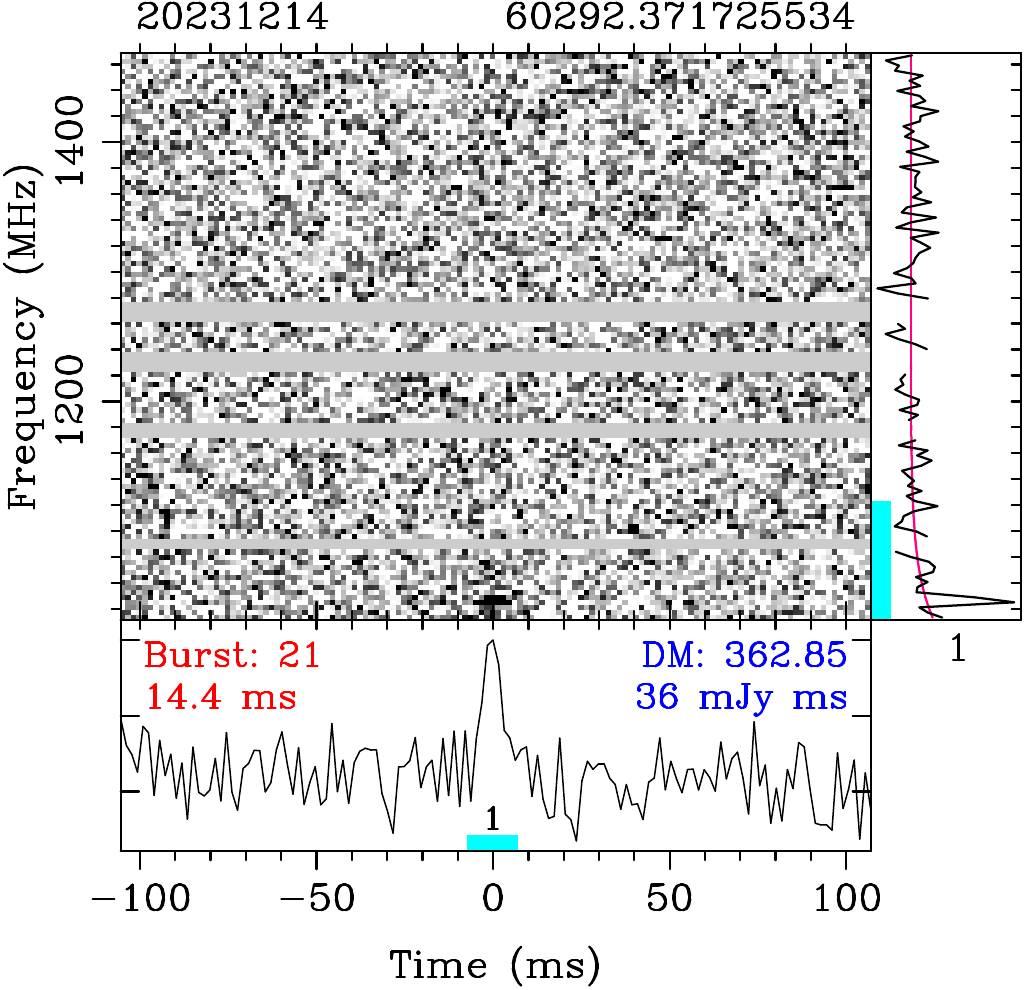}
\includegraphics[height=0.29\linewidth]{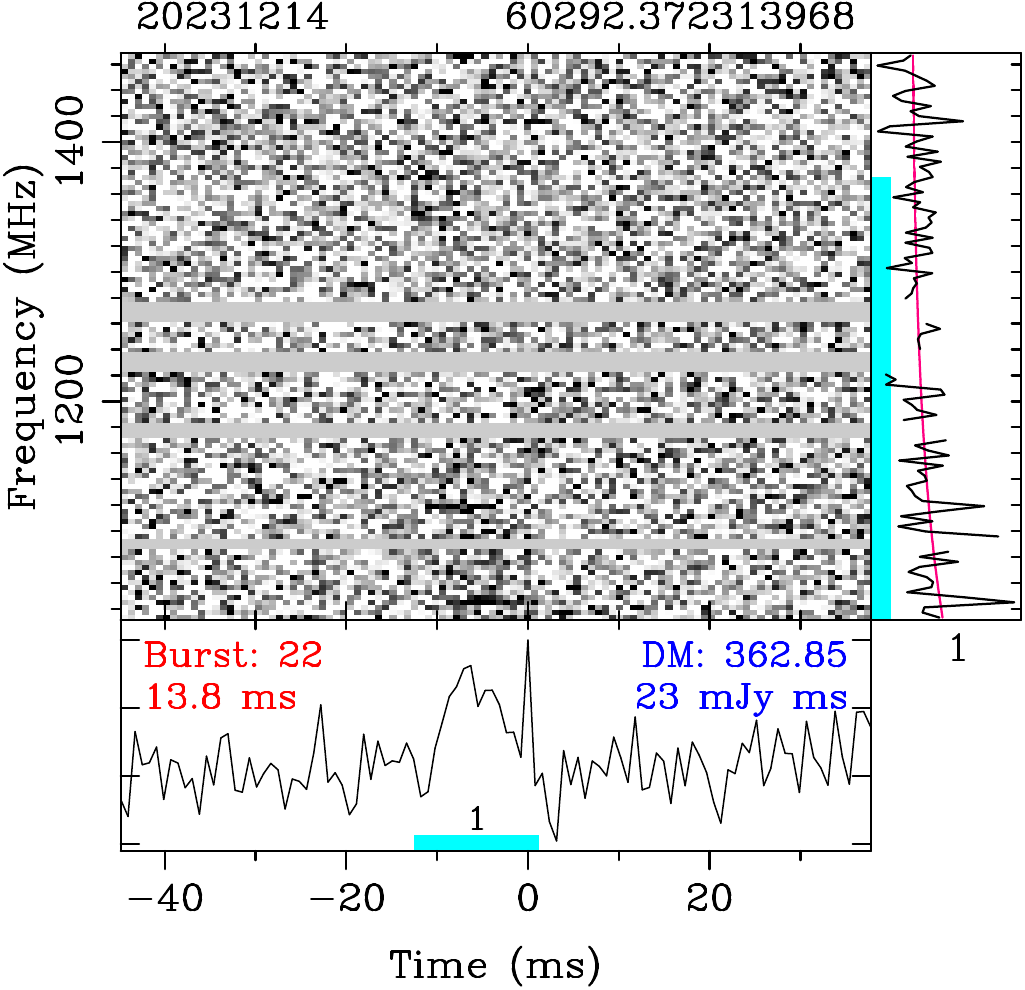}
\includegraphics[height=0.29\linewidth]{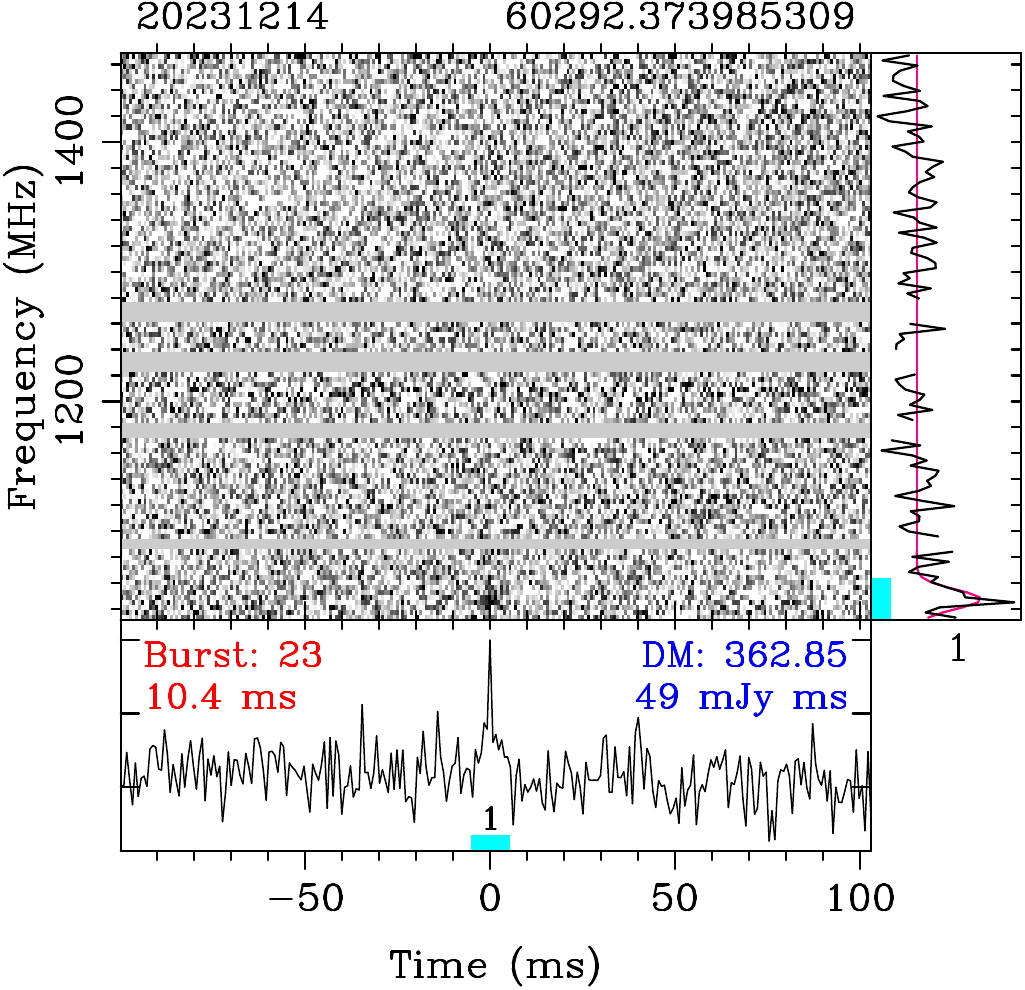}
\includegraphics[height=0.29\linewidth]{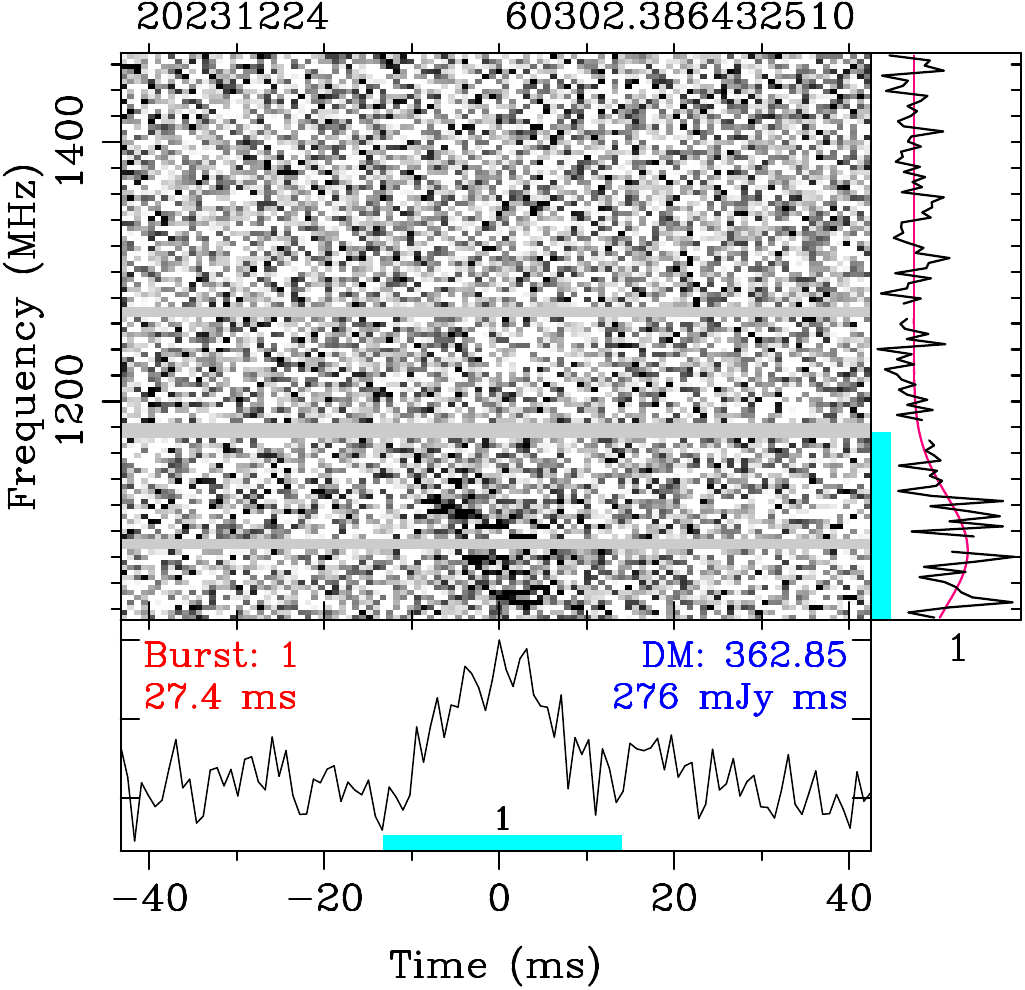}
\includegraphics[height=0.29\linewidth]{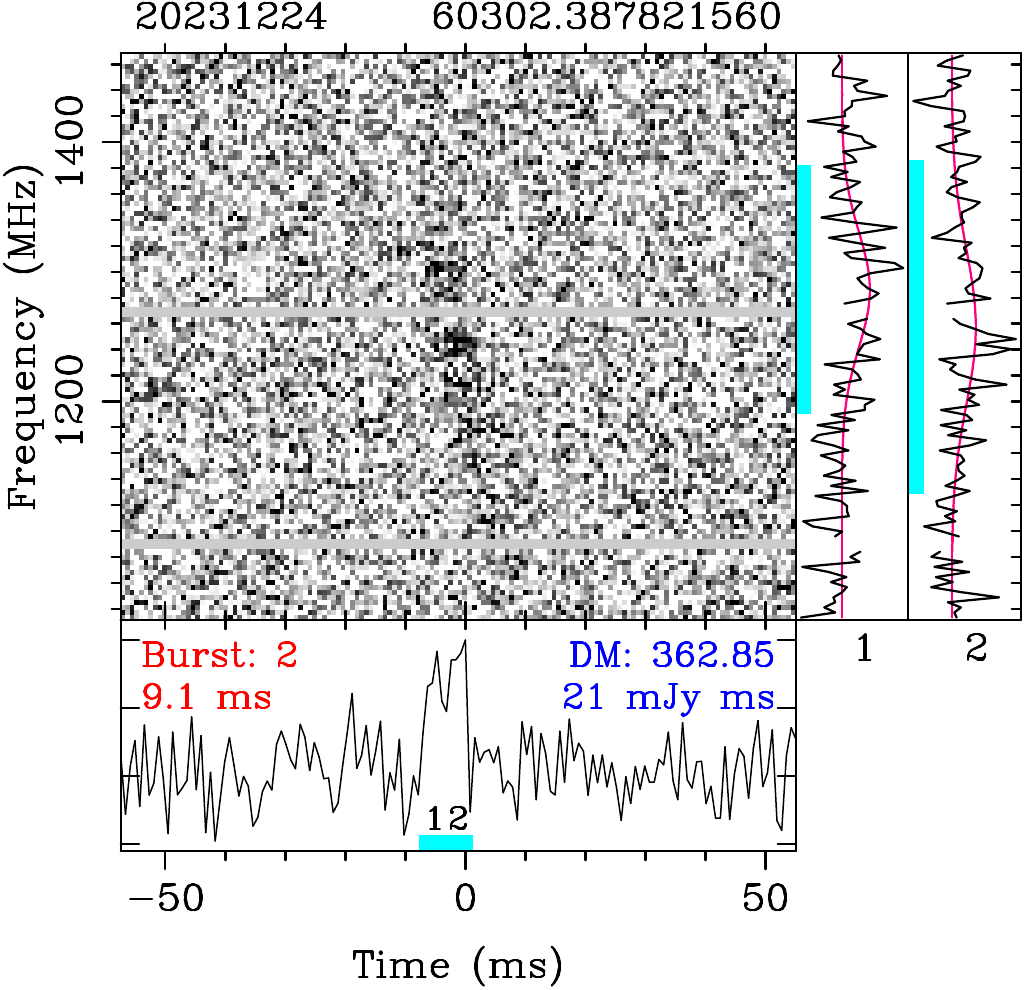}
\includegraphics[height=0.29\linewidth]{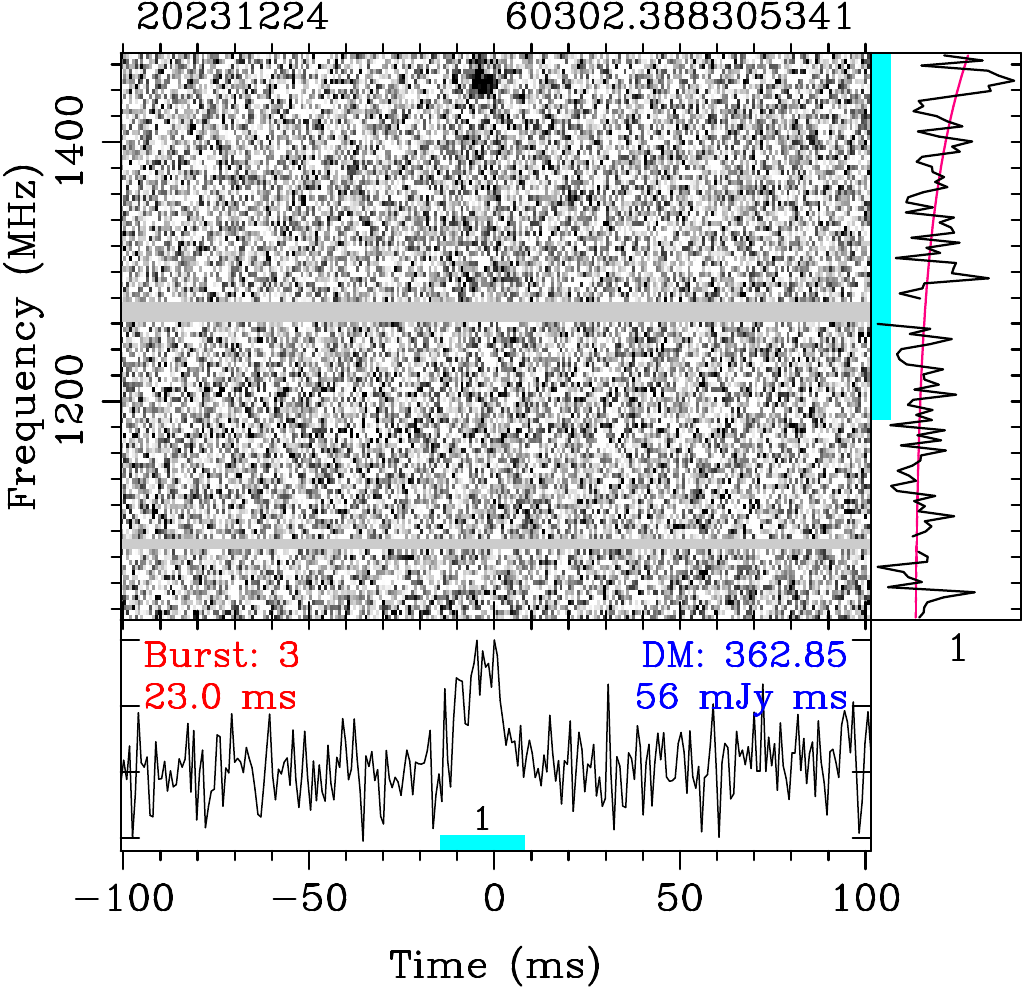}
\caption{({\textit{continued}})}
\end{figure*}
\addtocounter{figure}{-1}
\begin{figure*}
\flushleft
\includegraphics[height=0.29\linewidth]{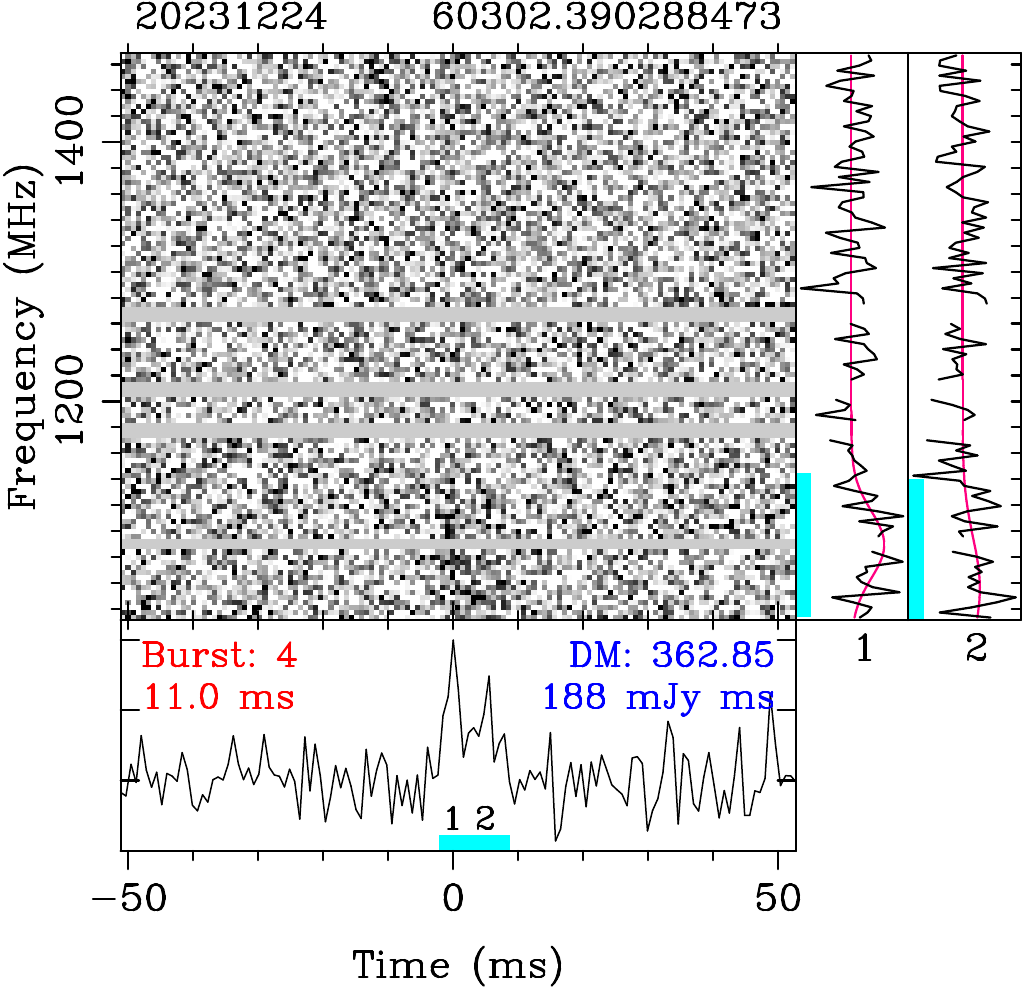}
\includegraphics[height=0.29\linewidth]{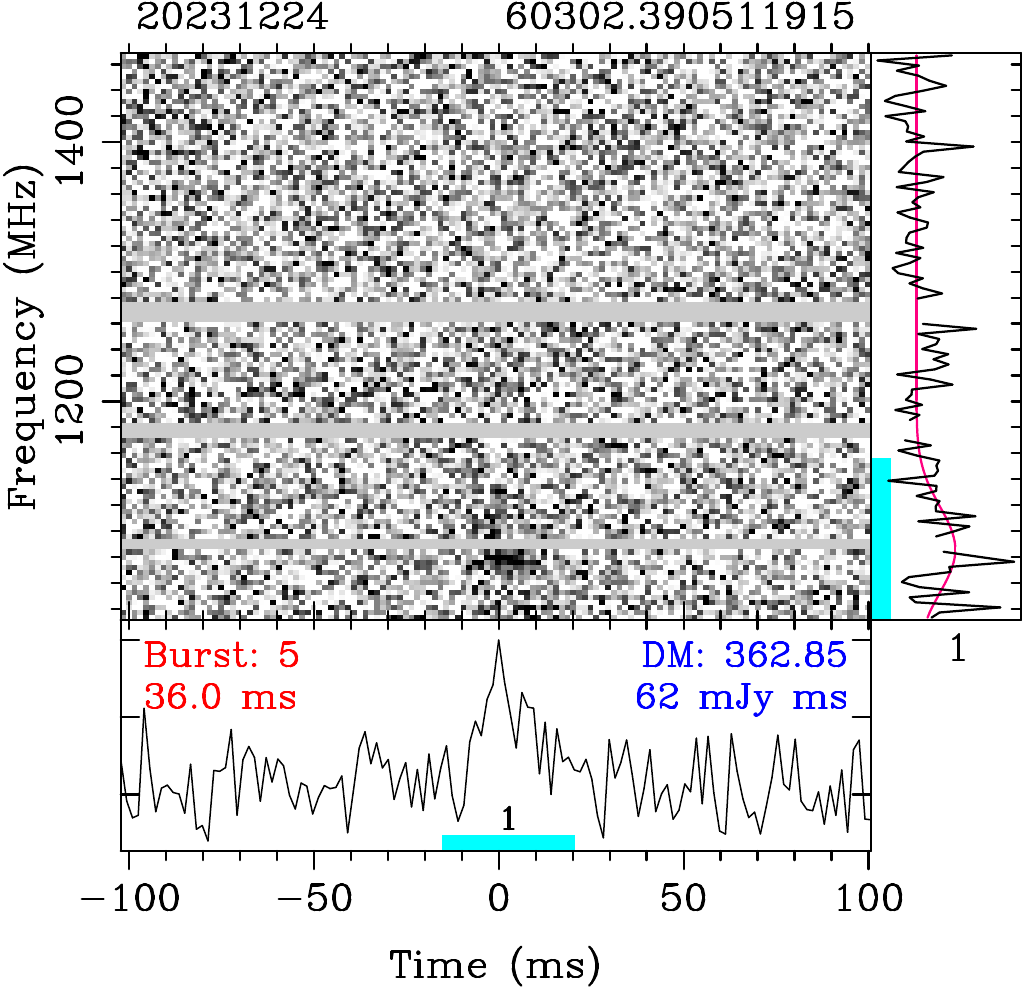}
\includegraphics[height=0.29\linewidth]{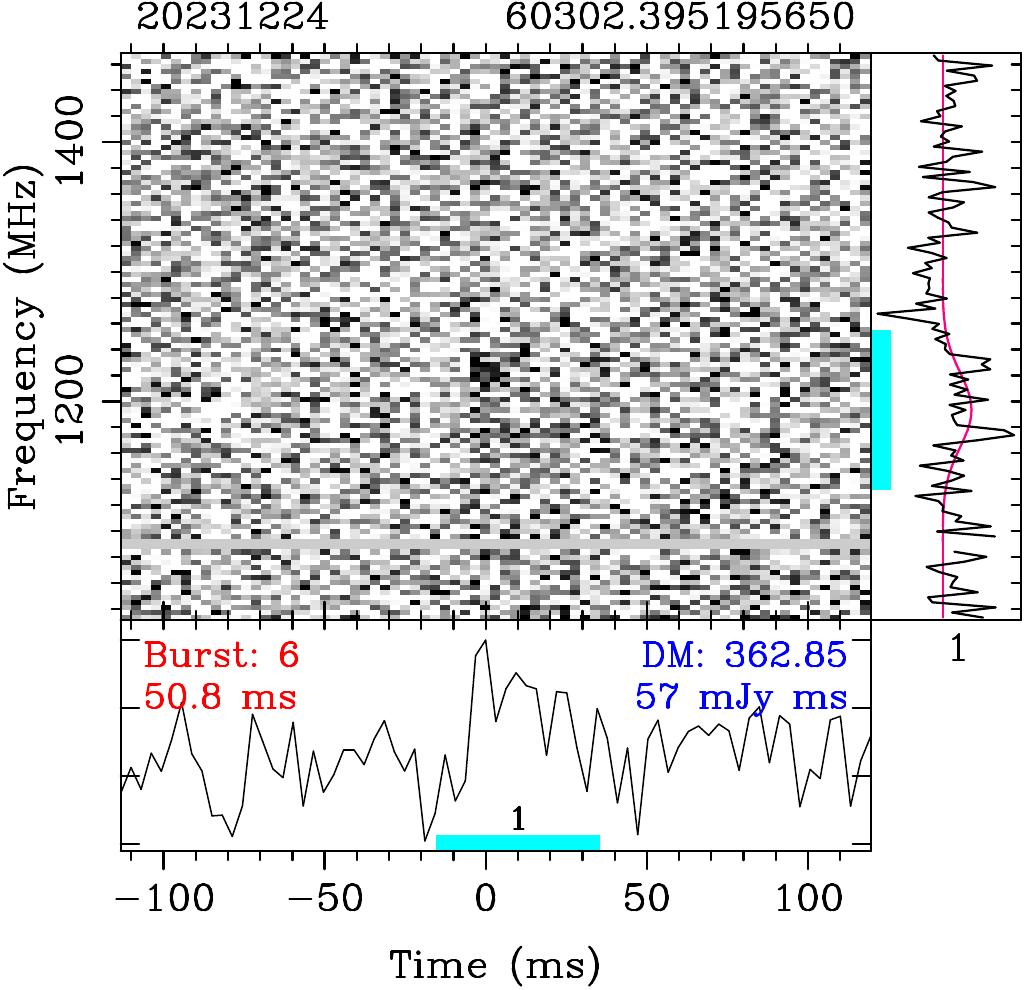}
\includegraphics[height=0.29\linewidth]{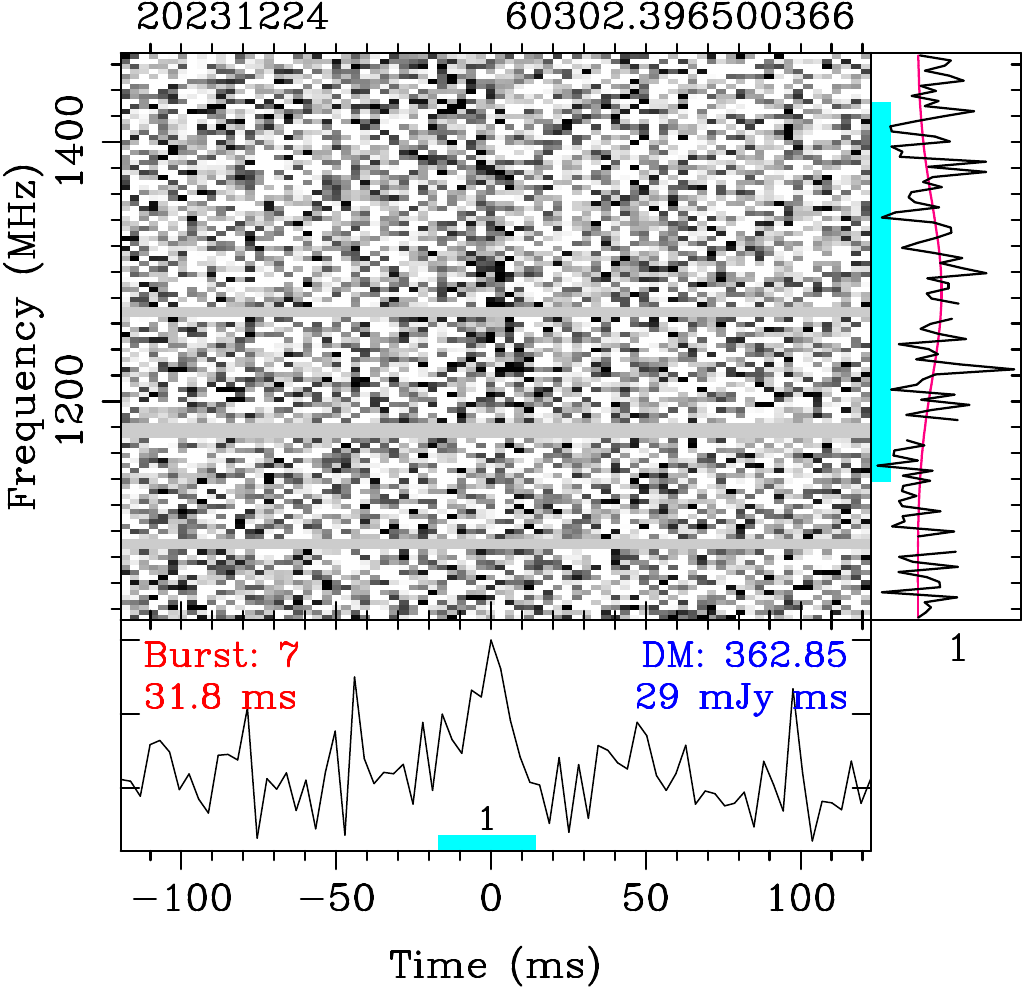}
\includegraphics[height=0.29\linewidth]{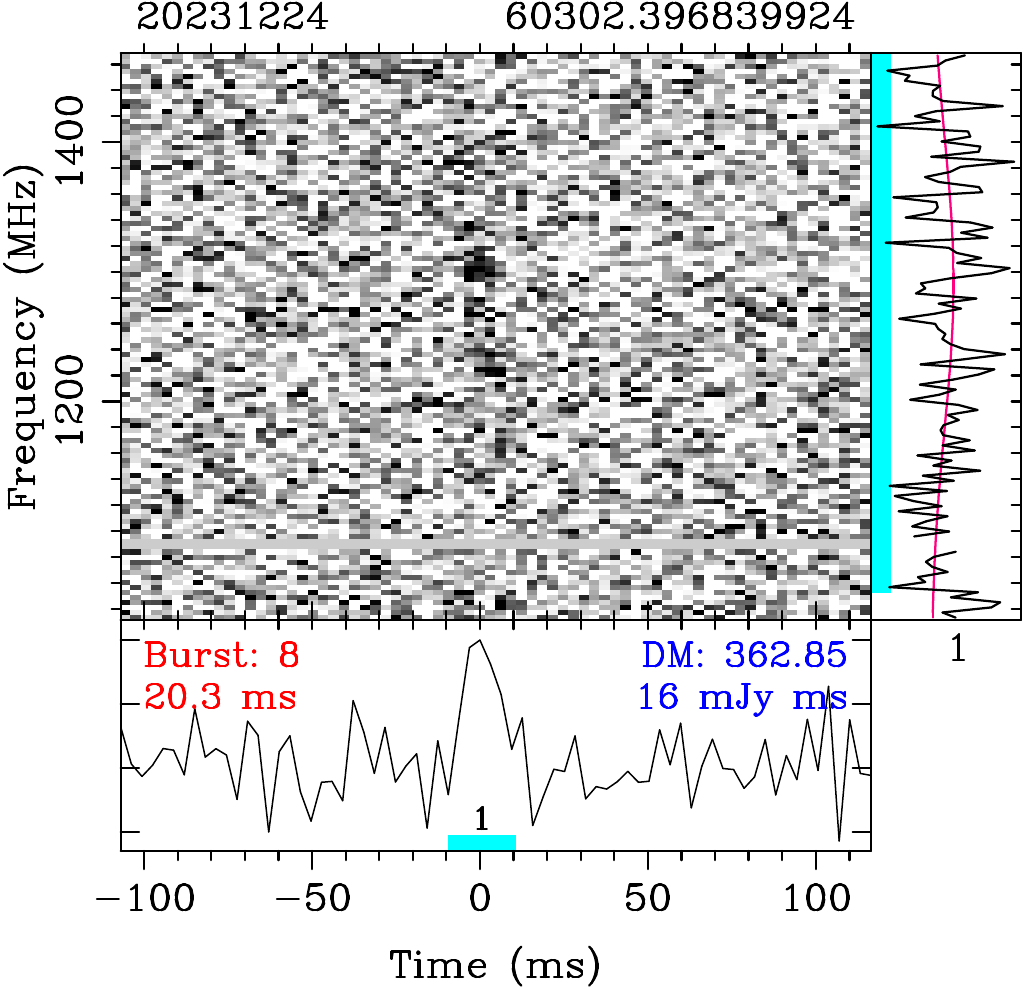}
\includegraphics[height=0.29\linewidth]{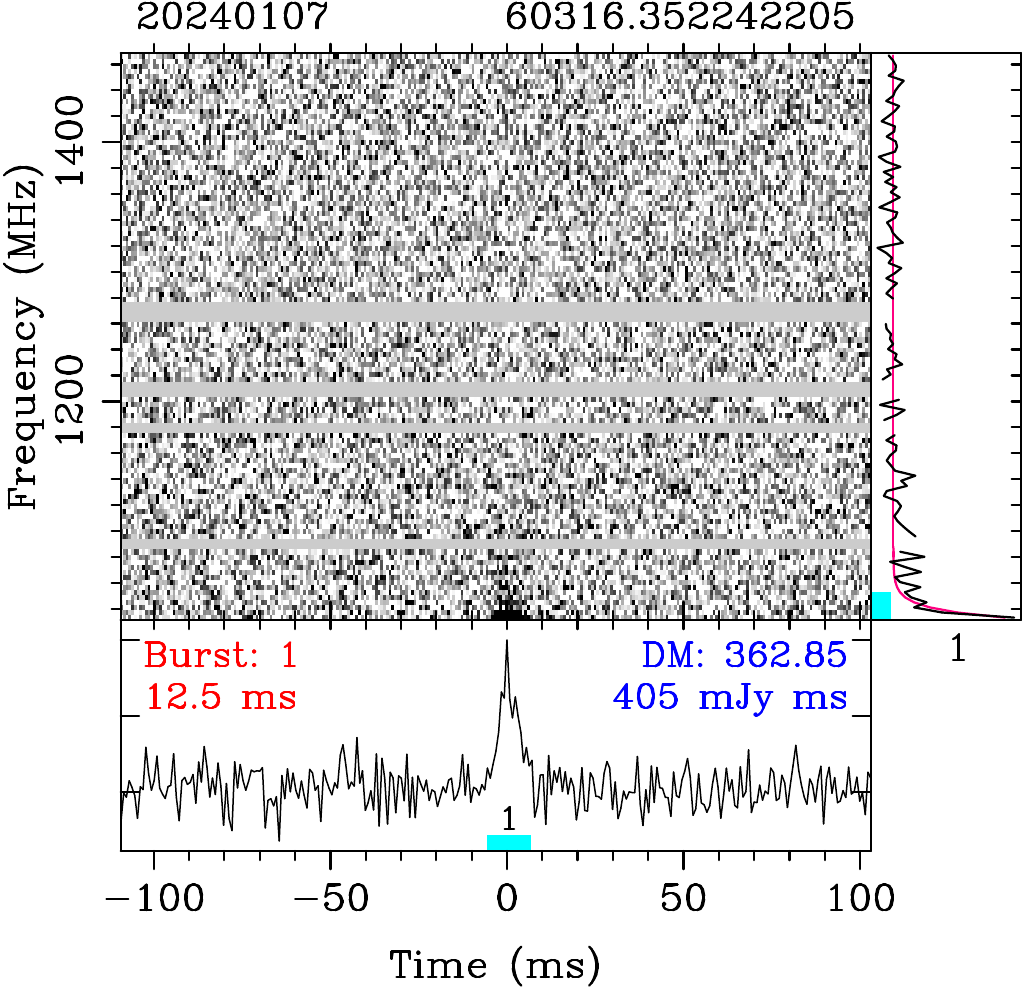}
\includegraphics[height=0.29\linewidth]{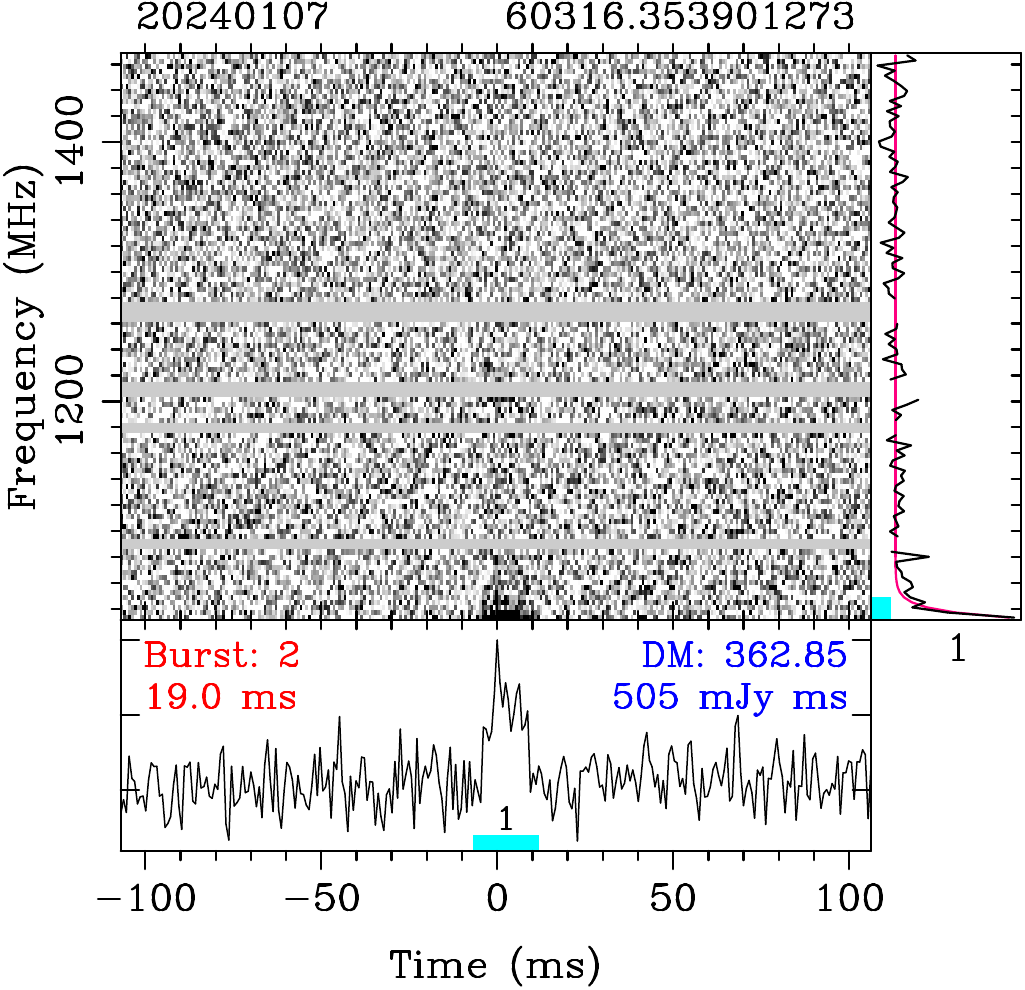}
\includegraphics[height=0.29\linewidth]{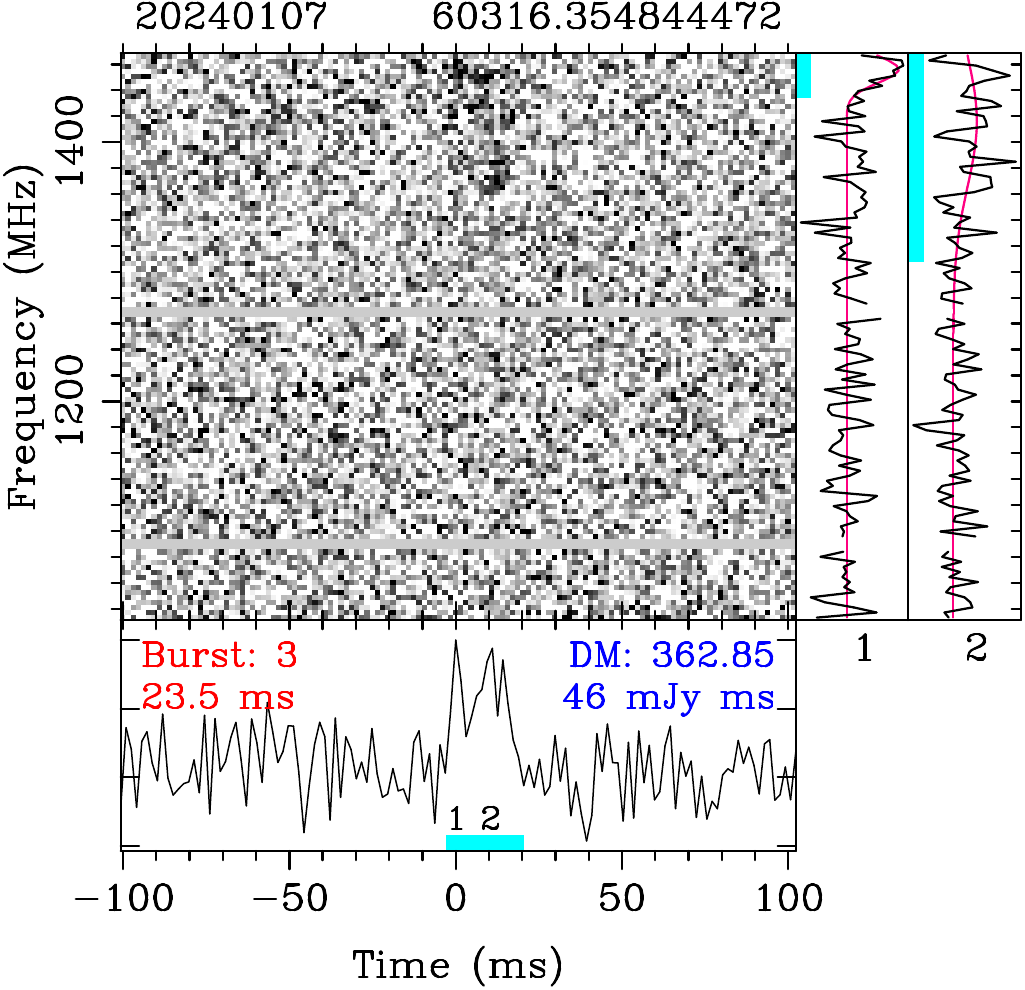}
\includegraphics[height=0.29\linewidth]{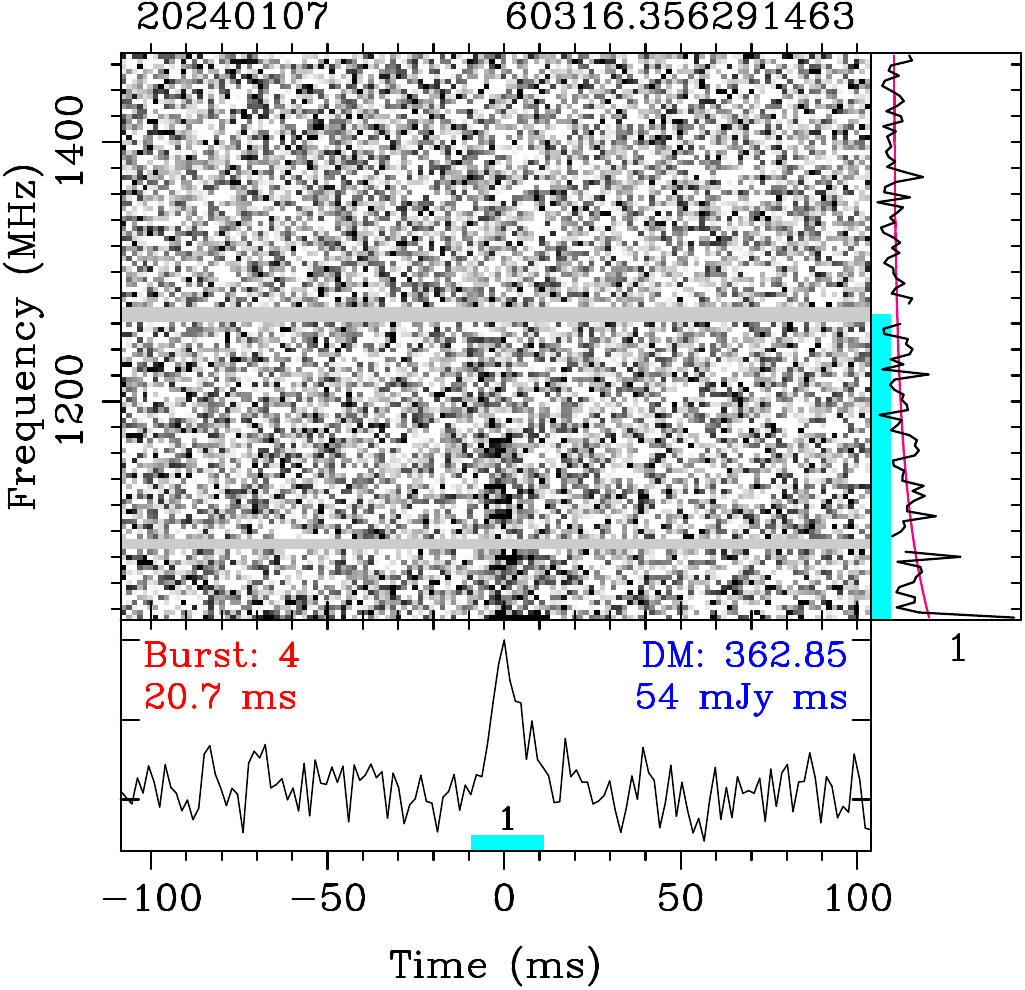}
\includegraphics[height=0.29\linewidth]{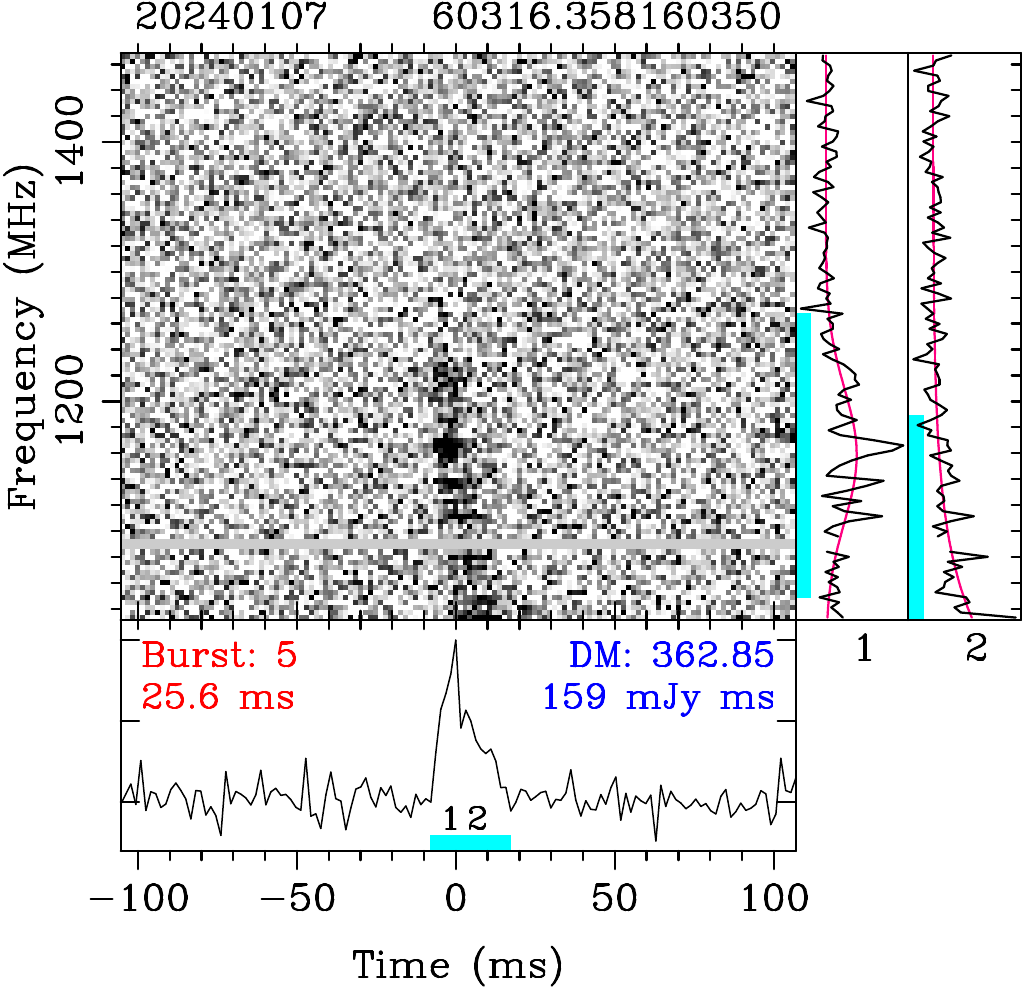}
\includegraphics[height=0.29\linewidth]{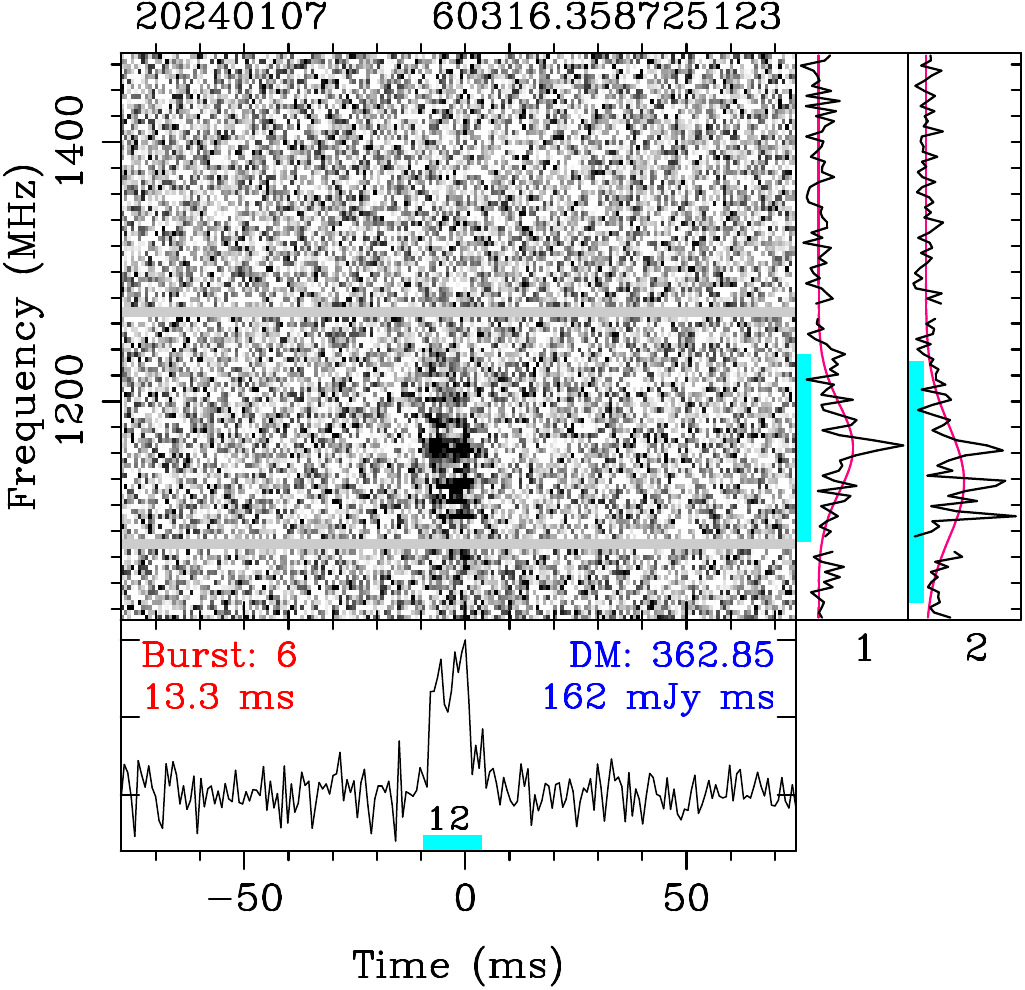}
\includegraphics[height=0.29\linewidth]{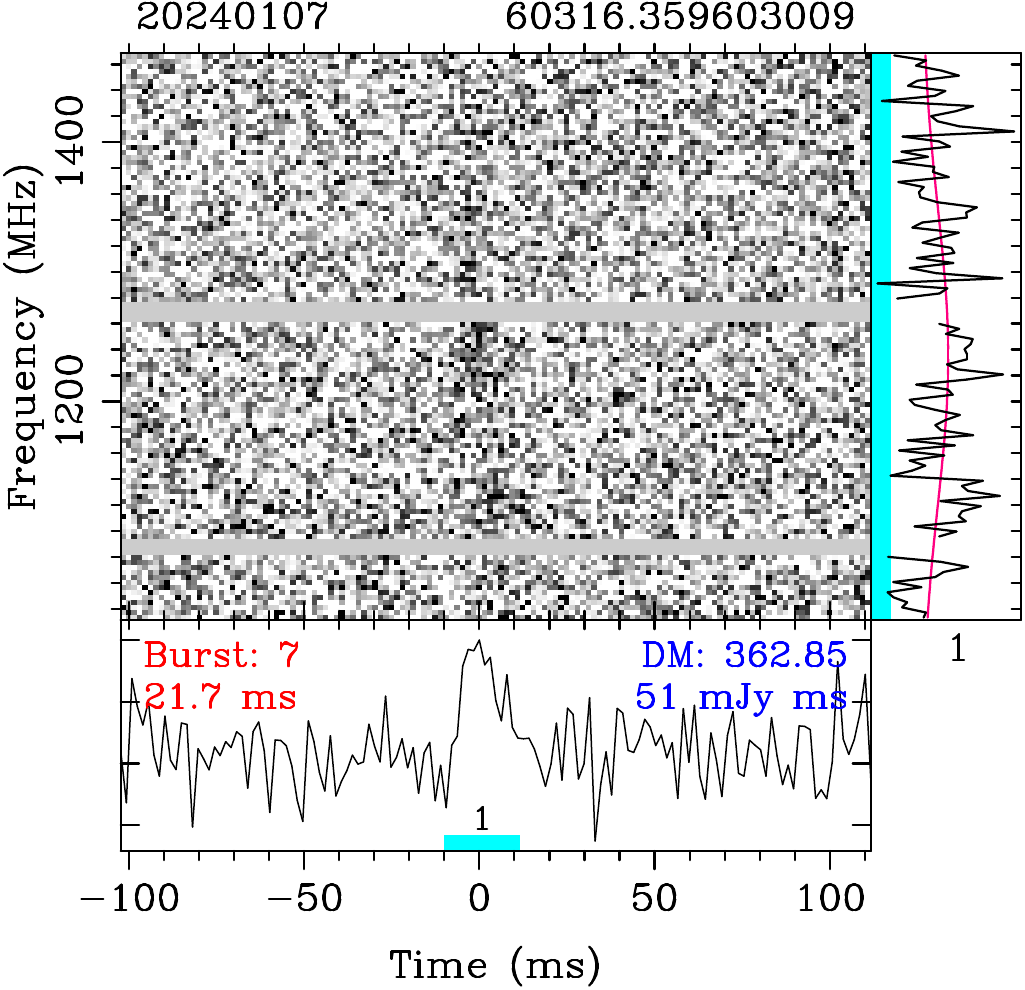}
\caption{({\textit{continued}})}
\end{figure*}
\addtocounter{figure}{-1}
\begin{figure*}
\flushleft
\includegraphics[height=0.29\linewidth]{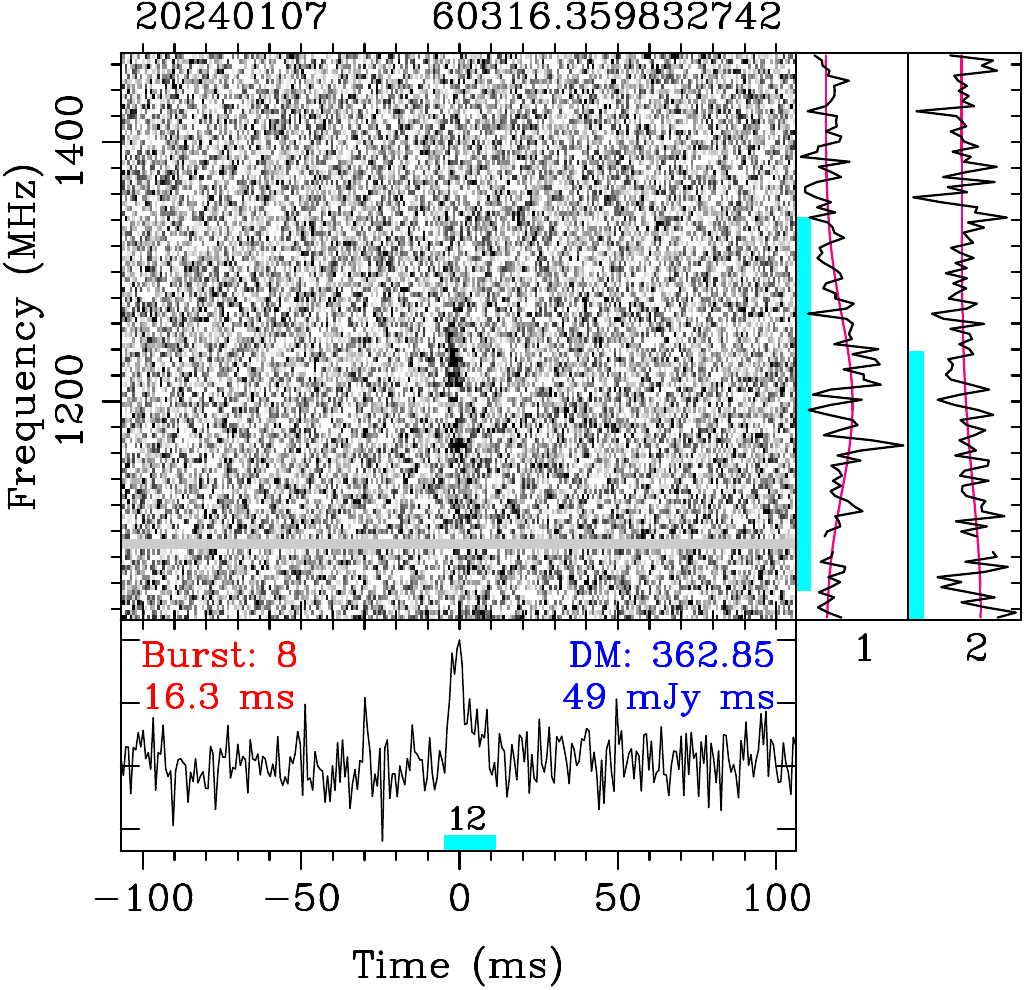}
\includegraphics[height=0.29\linewidth]{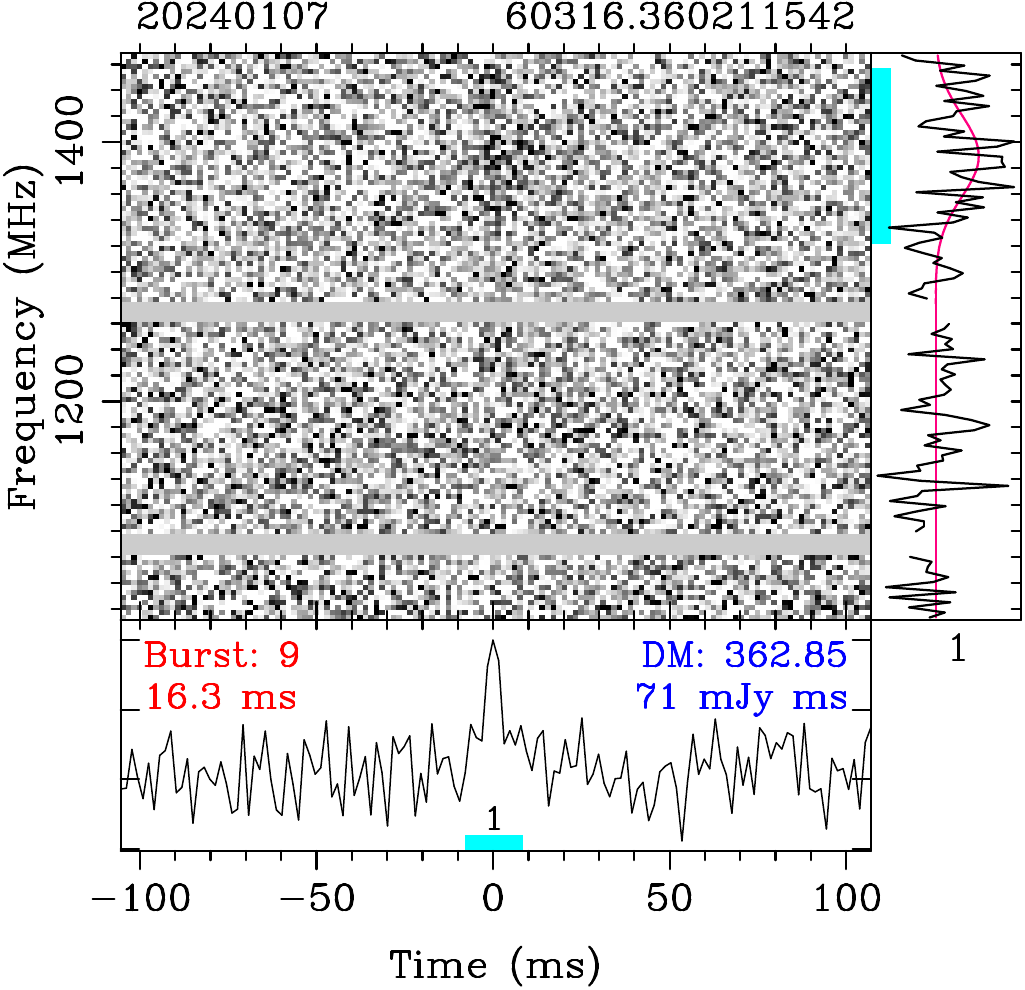}
\includegraphics[height=0.29\linewidth]{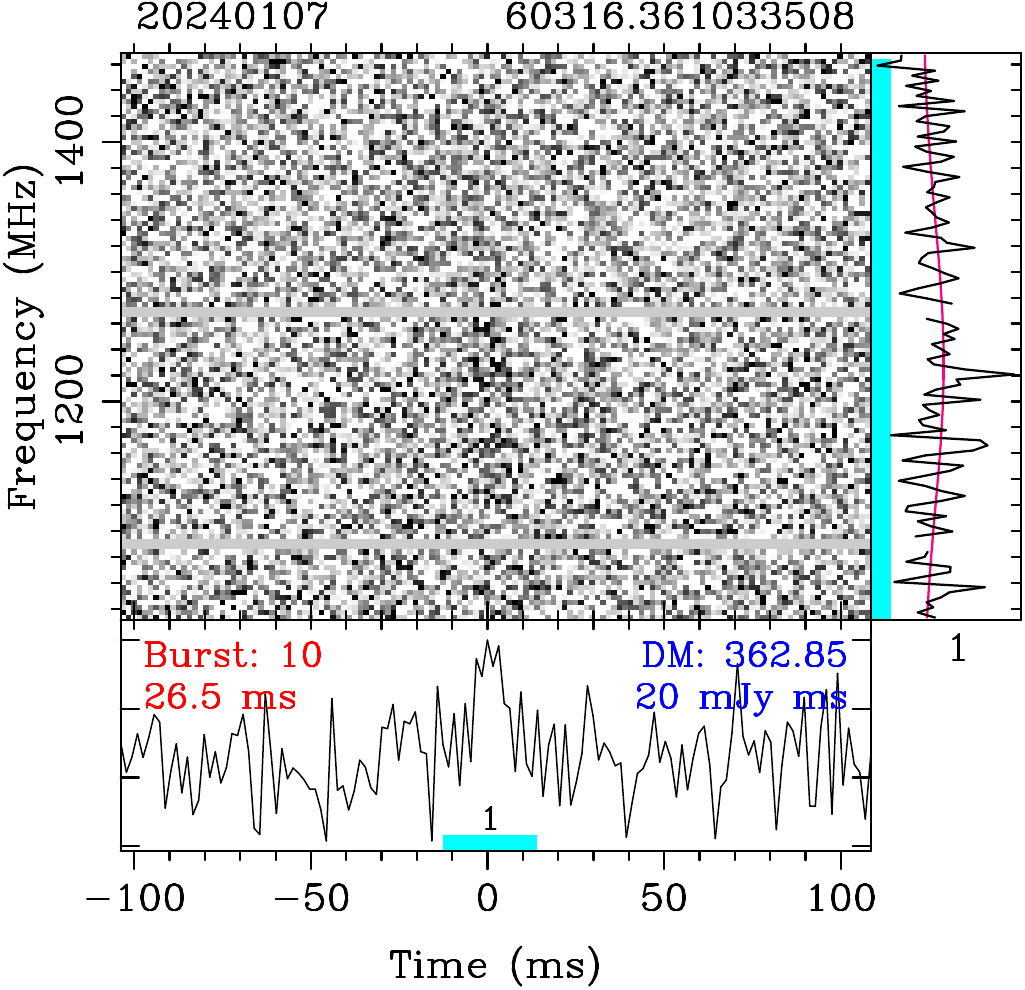}
\includegraphics[height=0.29\linewidth]{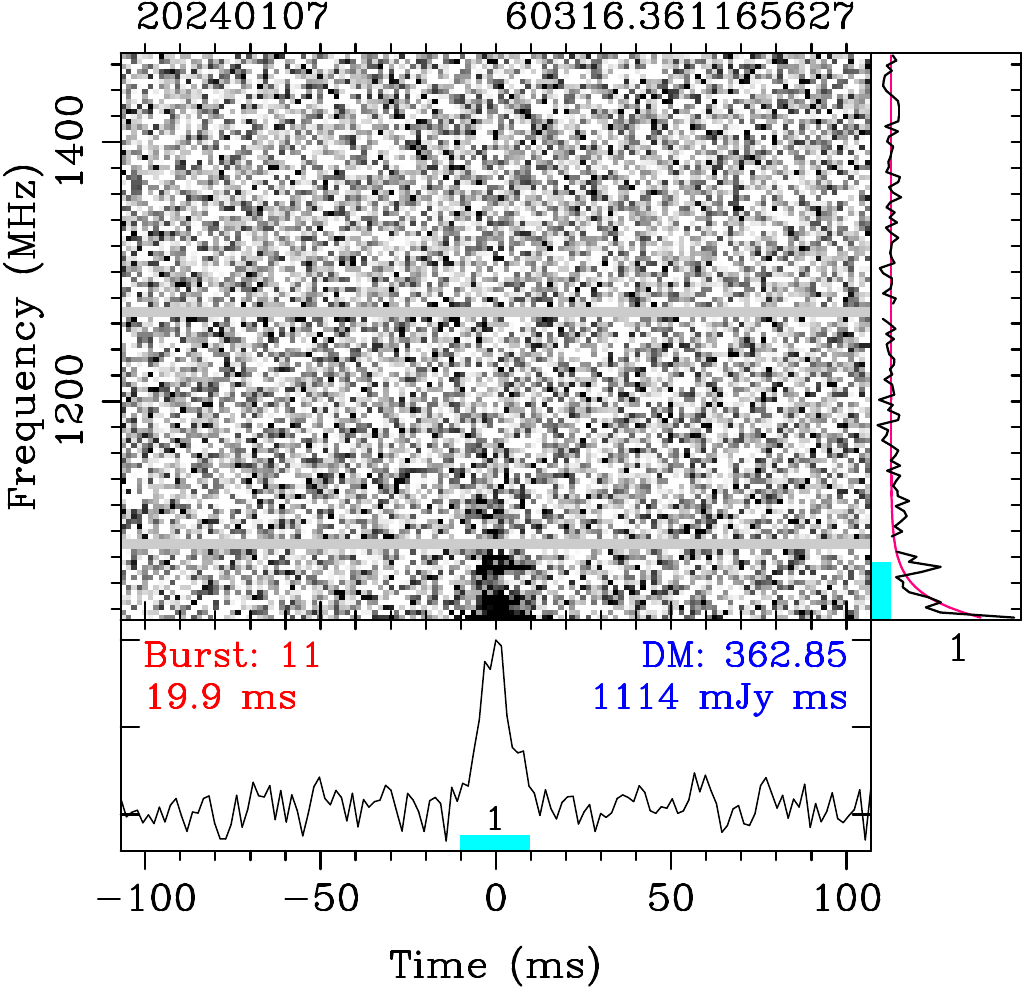}
\includegraphics[height=0.29\linewidth]{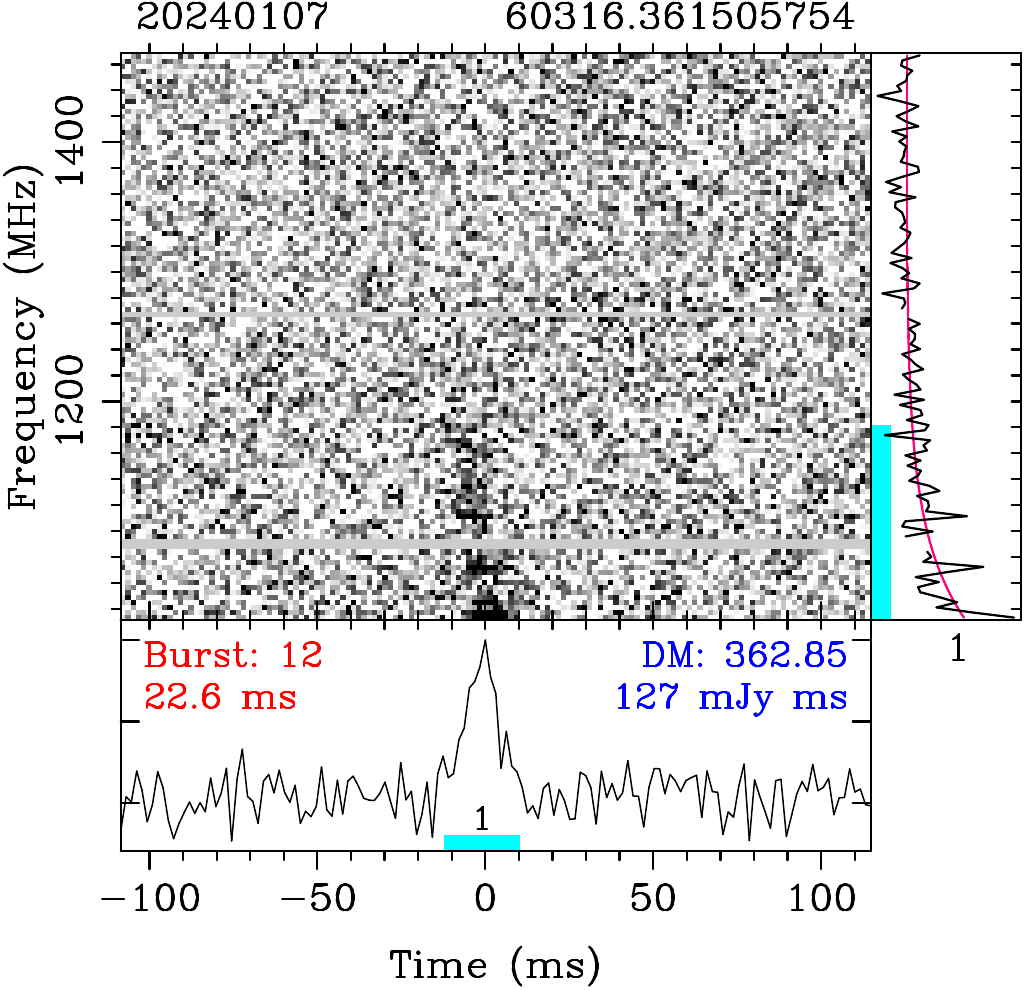}
\includegraphics[height=0.29\linewidth]{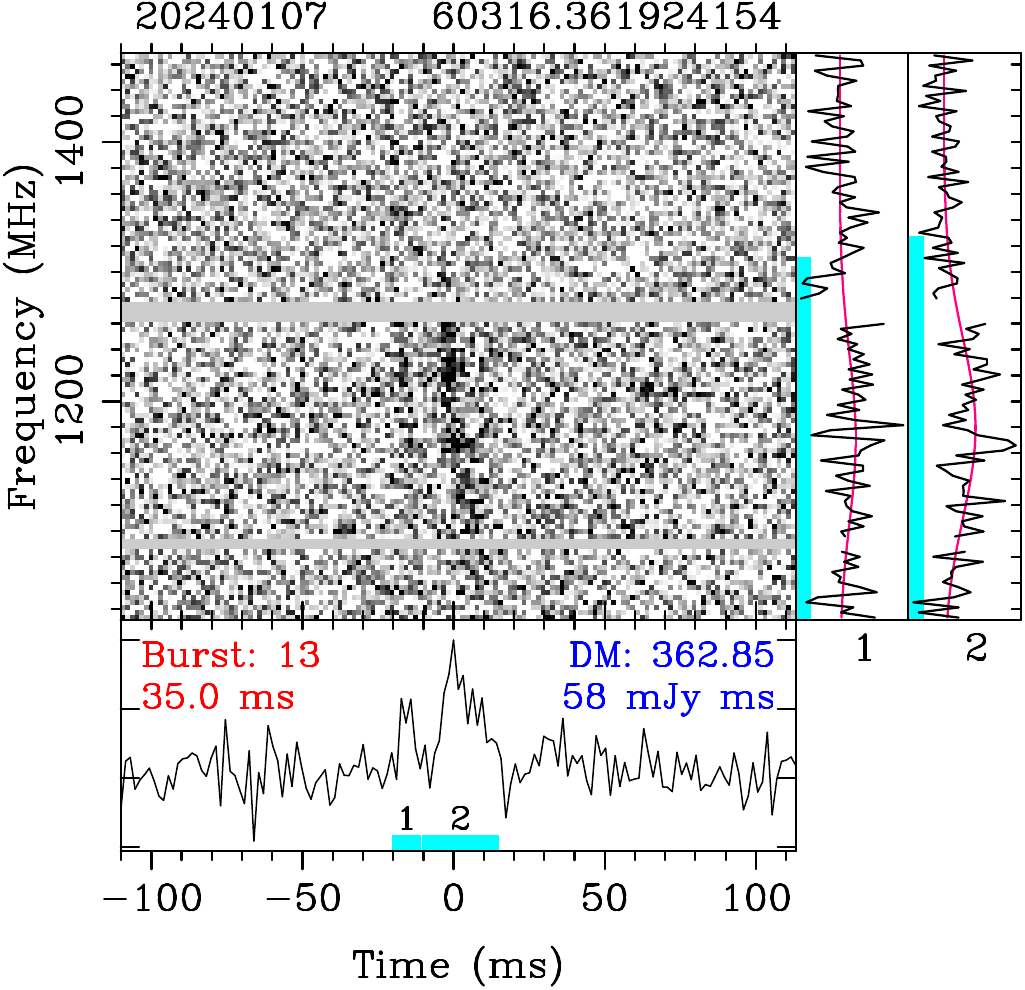}
\includegraphics[height=0.29\linewidth]{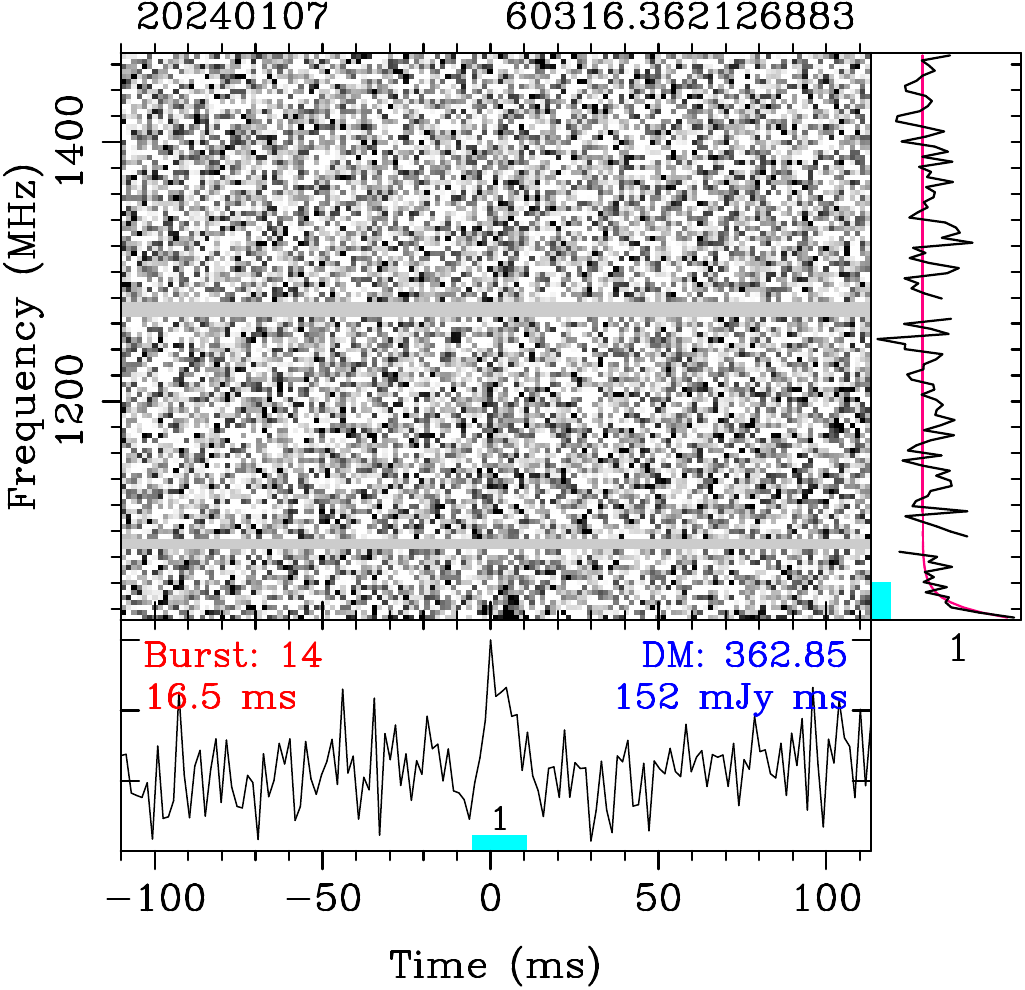}
\includegraphics[height=0.29\linewidth]{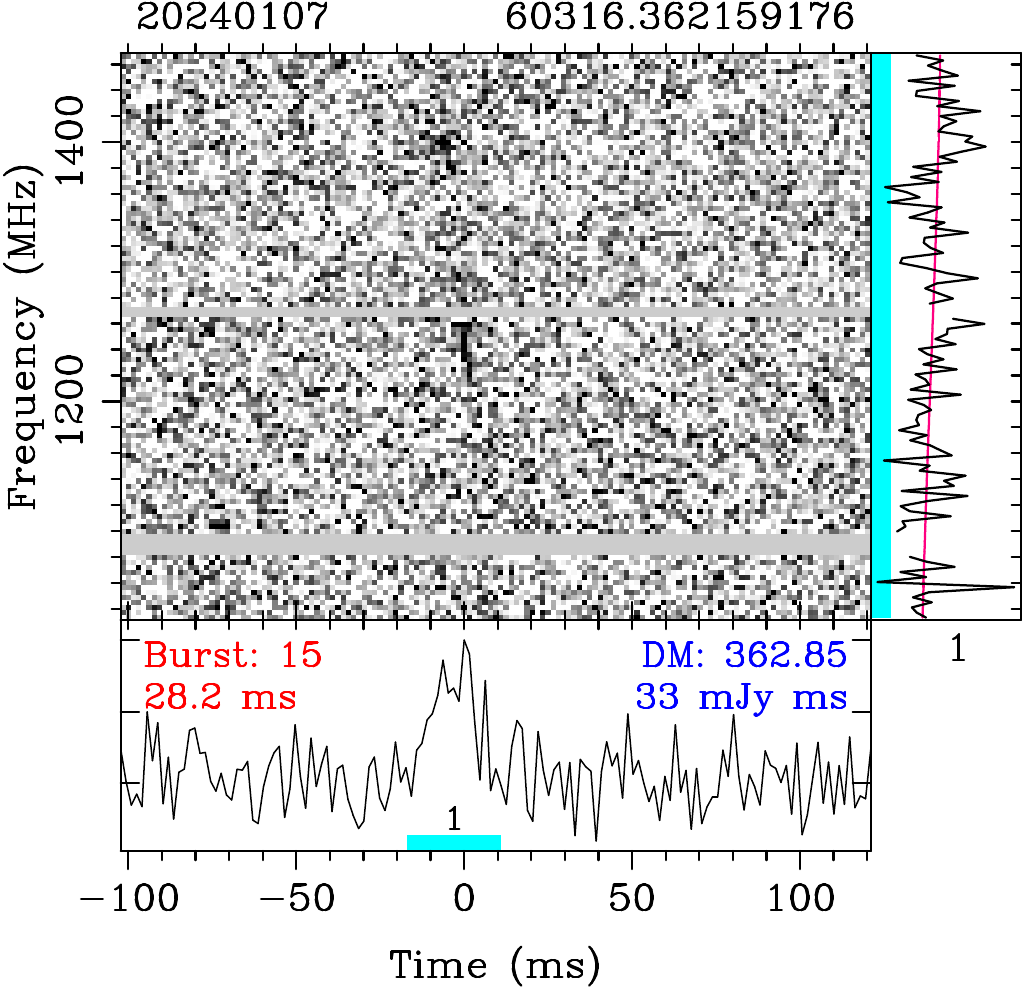}
\includegraphics[height=0.29\linewidth]{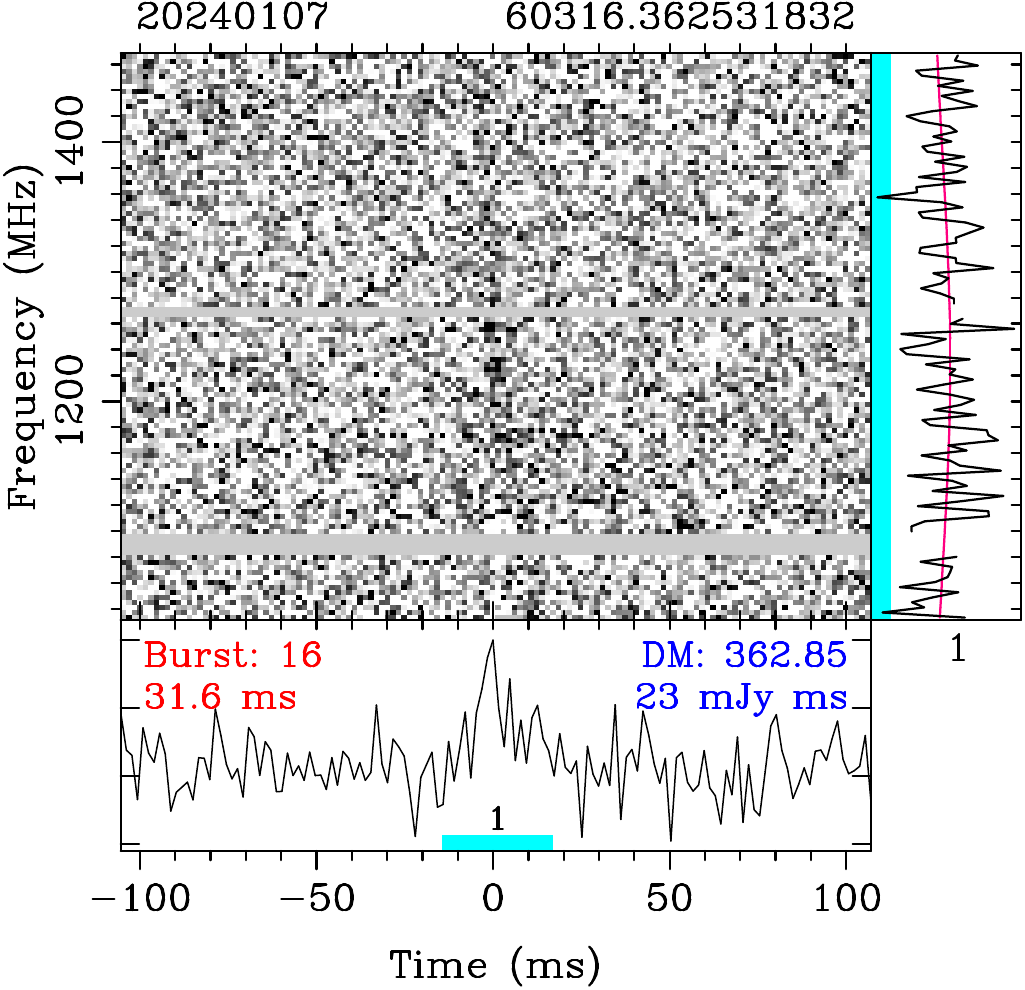}
\includegraphics[height=0.29\linewidth]{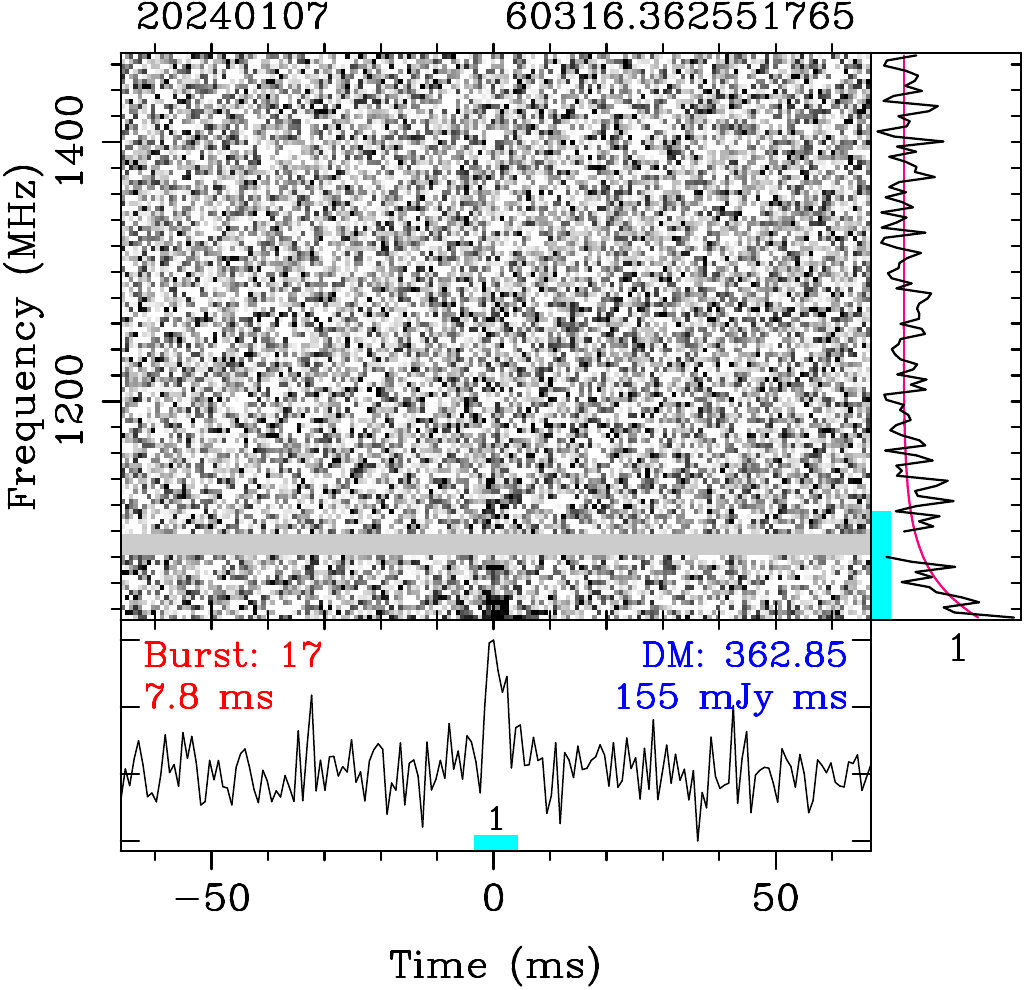}
\includegraphics[height=0.29\linewidth]{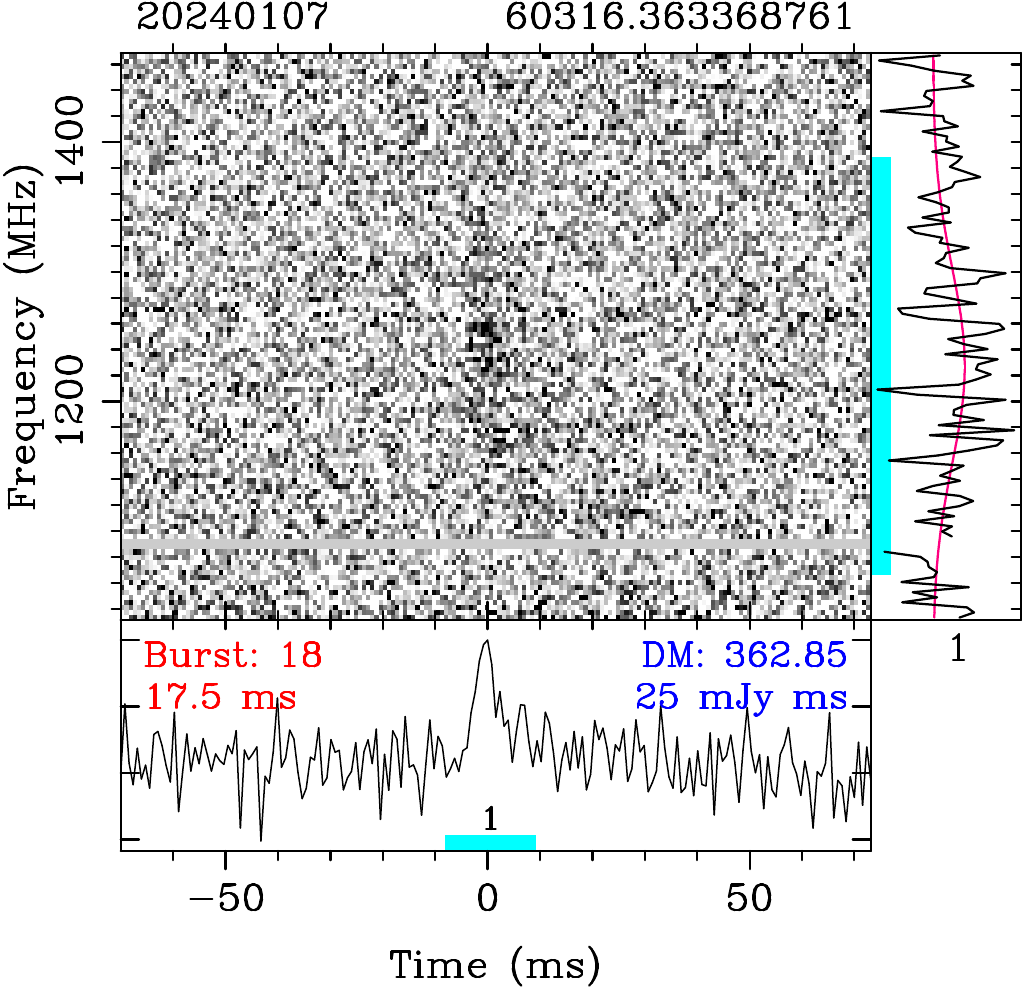}
\includegraphics[height=0.29\linewidth]{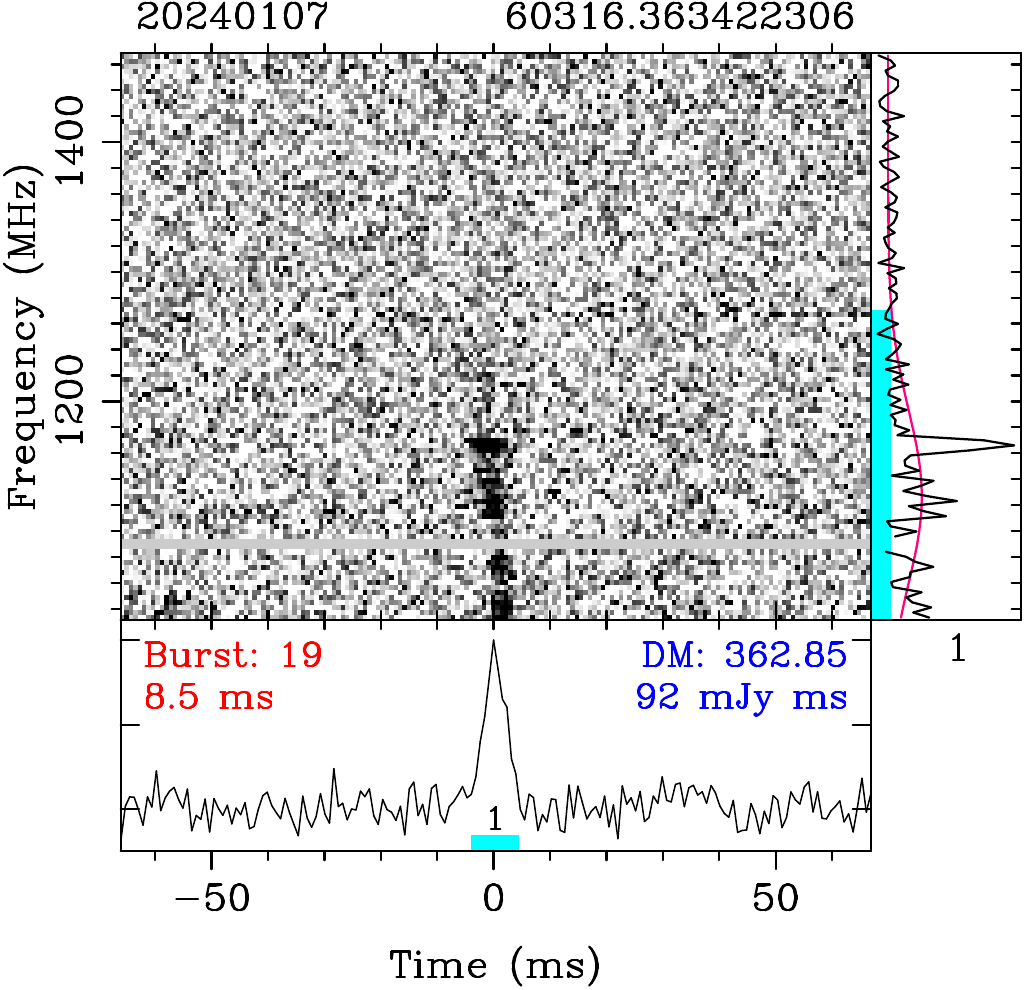}
\caption{({\textit{continued}})}
\end{figure*}
\addtocounter{figure}{-1}
\begin{figure*}
\flushleft
\includegraphics[height=0.29\linewidth]{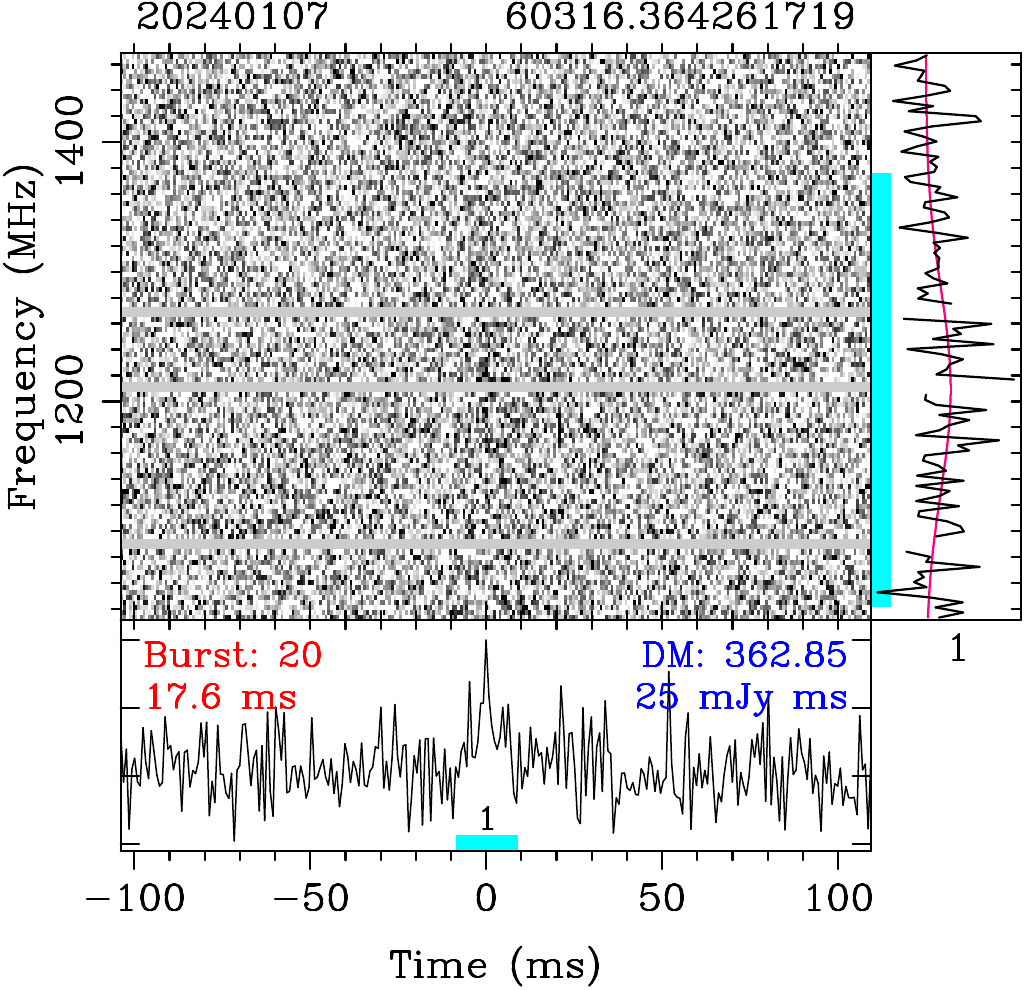}
\includegraphics[height=0.29\linewidth]{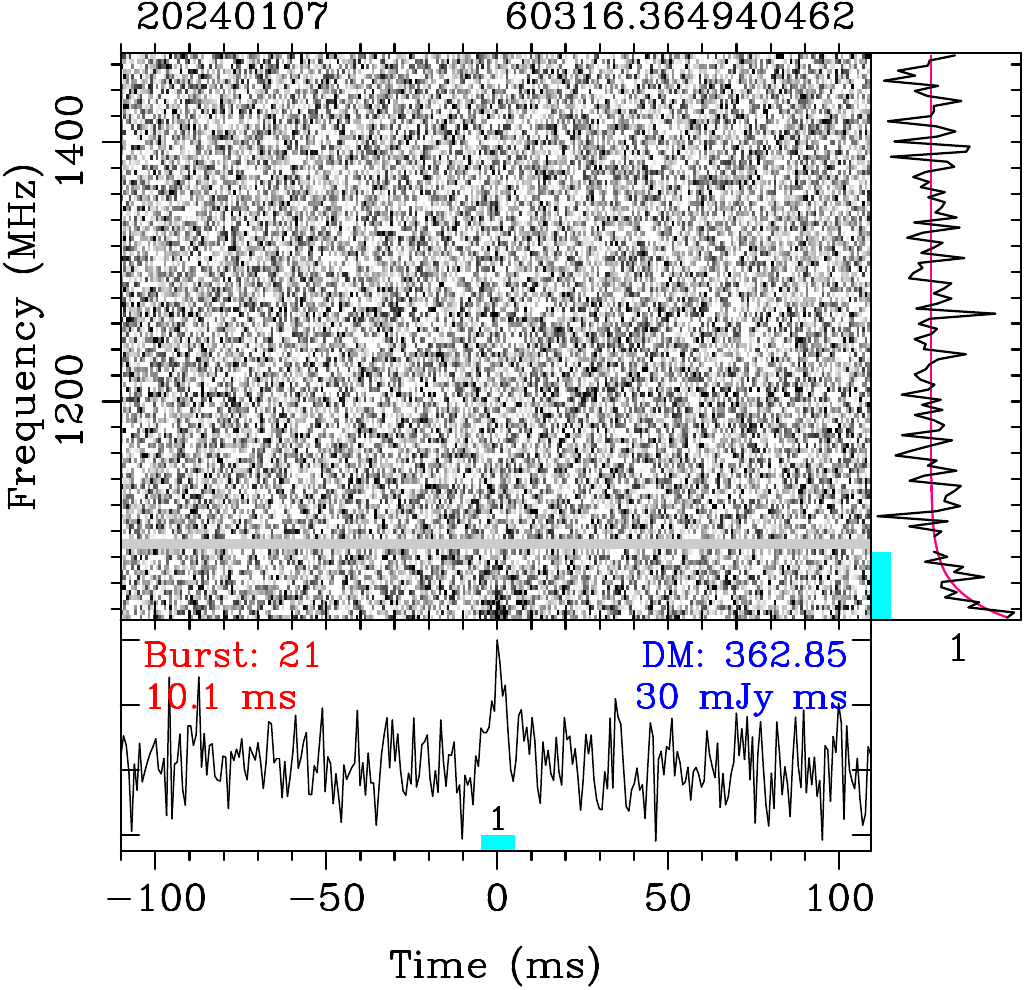}
\includegraphics[height=0.29\linewidth]{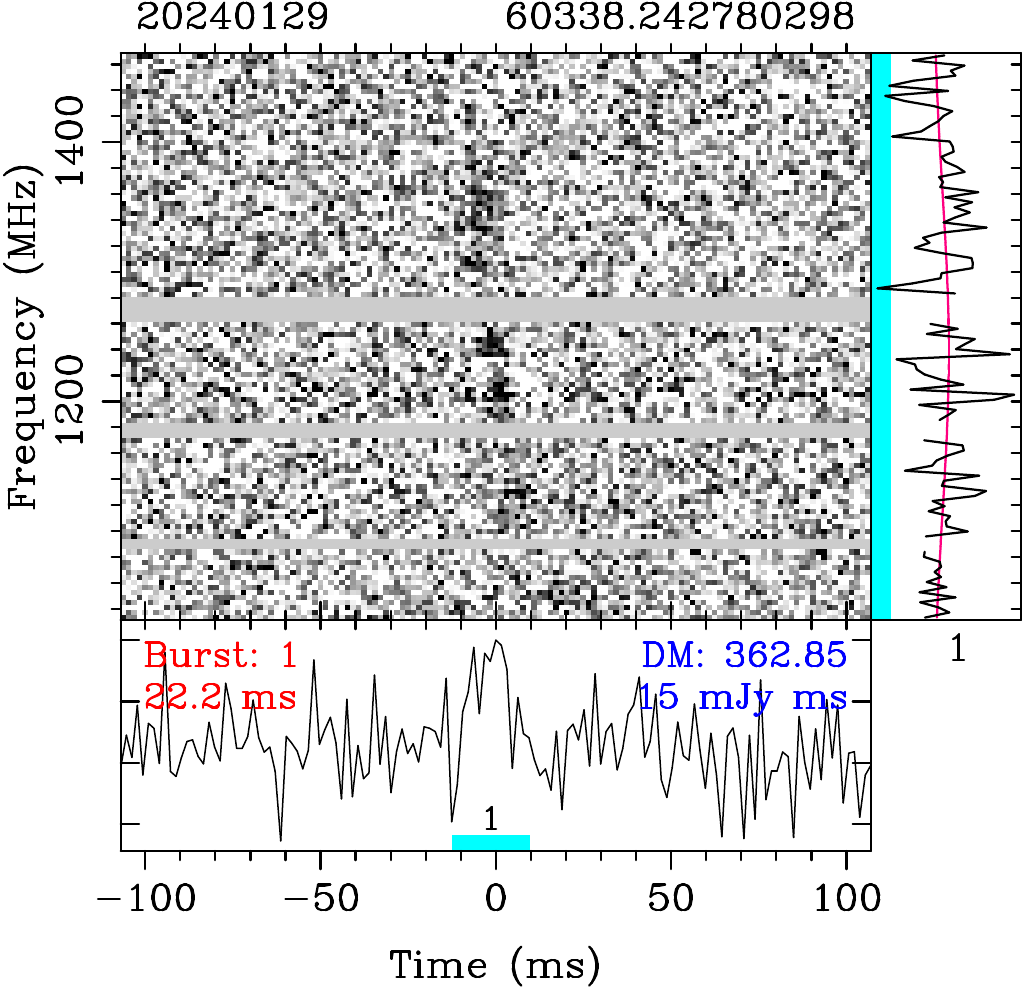}
\includegraphics[height=0.29\linewidth]{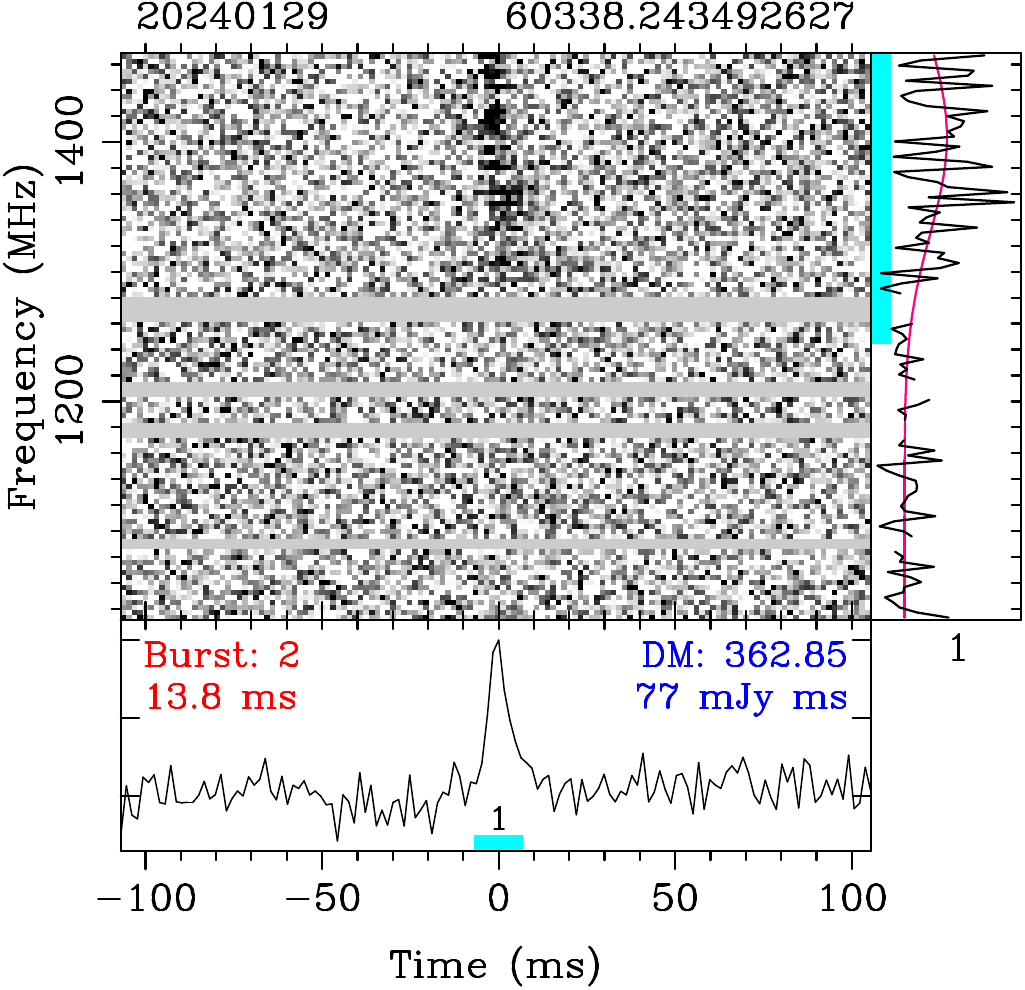}
\includegraphics[height=0.29\linewidth]{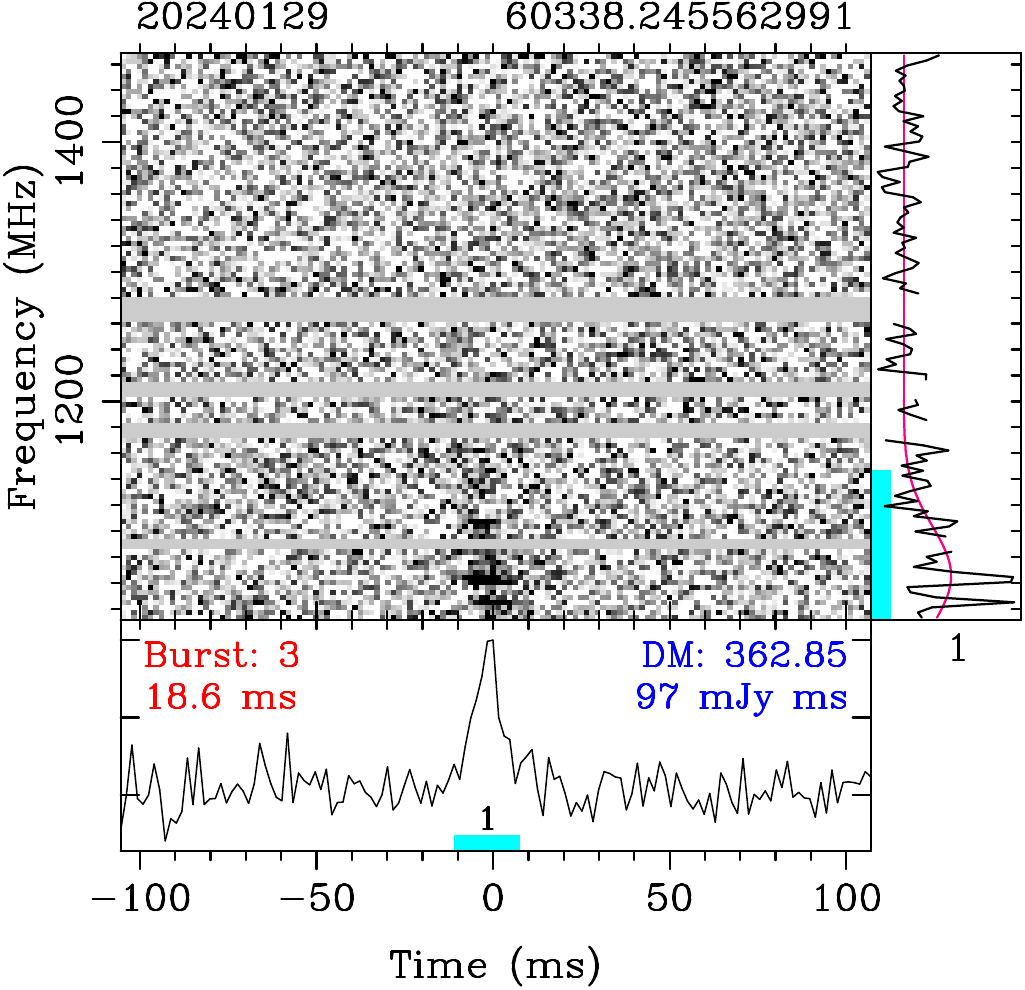}
\includegraphics[height=0.29\linewidth]{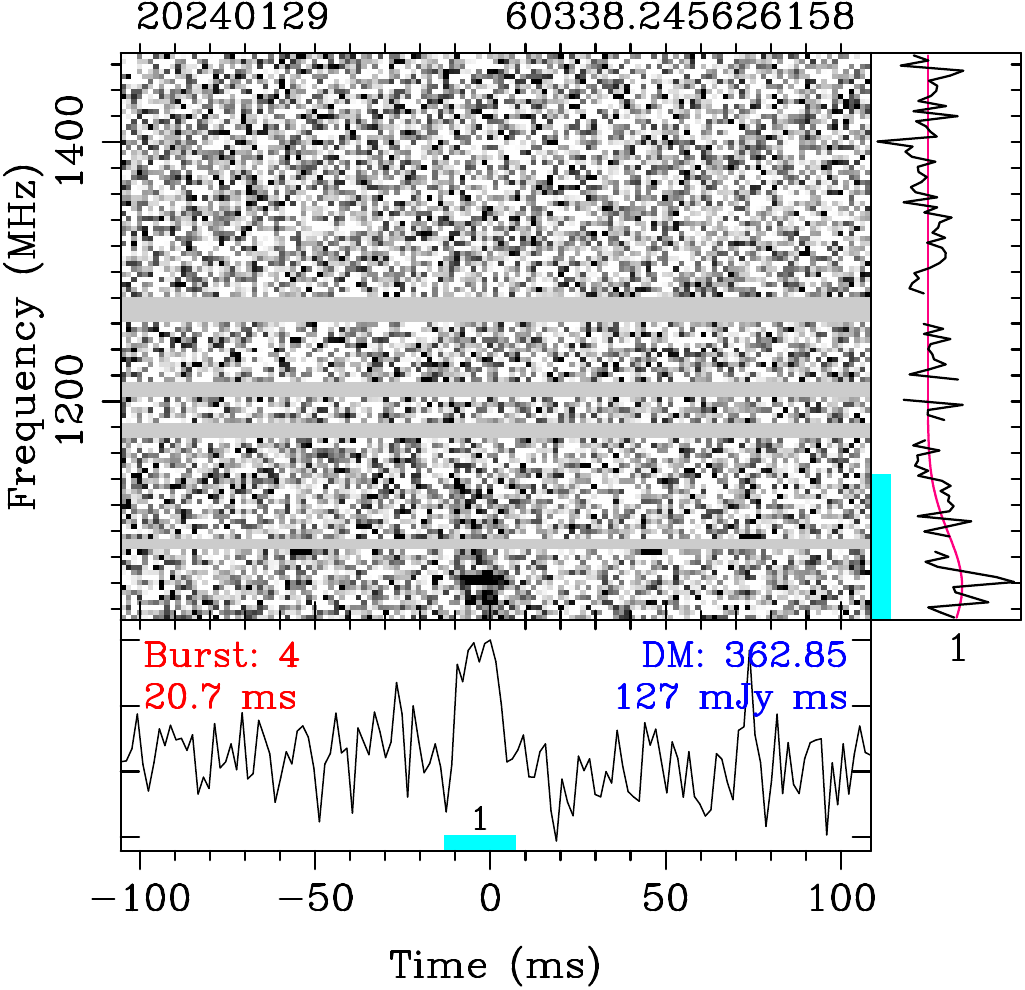}
\includegraphics[height=0.29\linewidth]{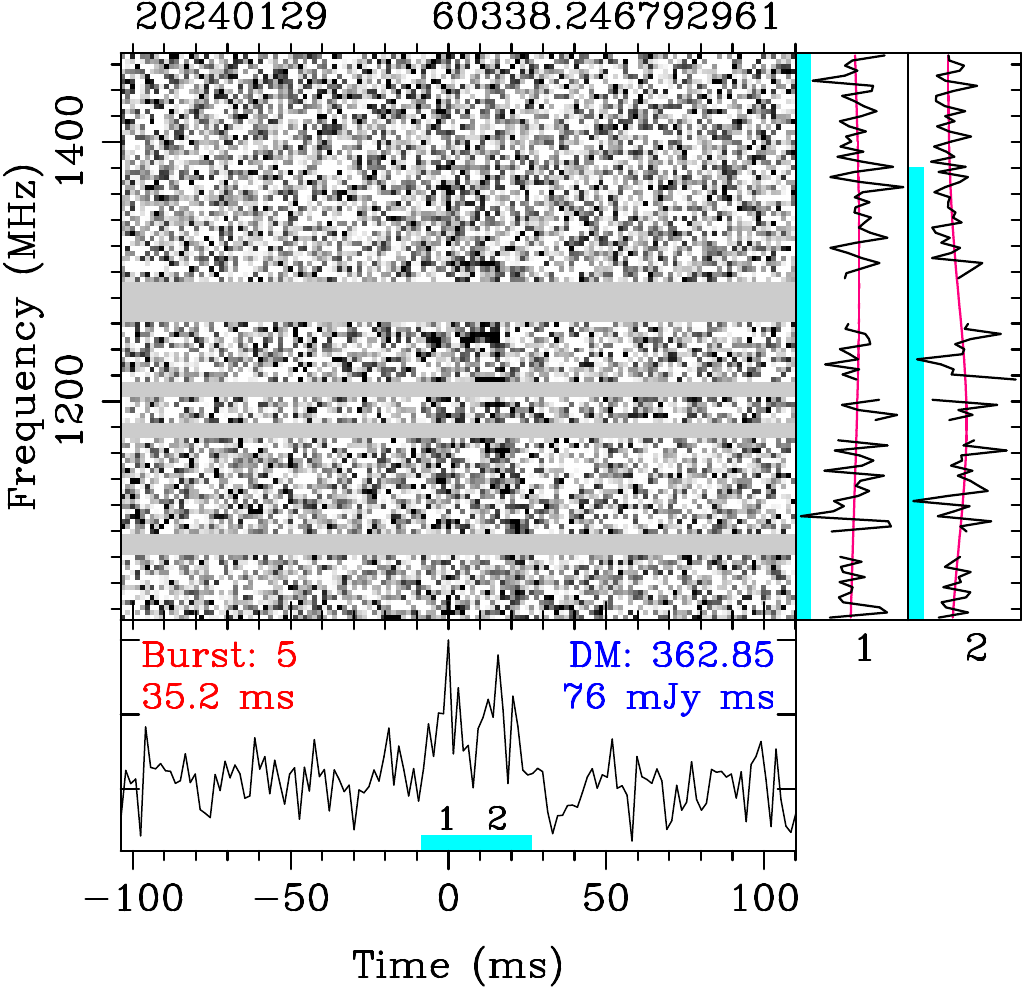}
\includegraphics[height=0.29\linewidth]{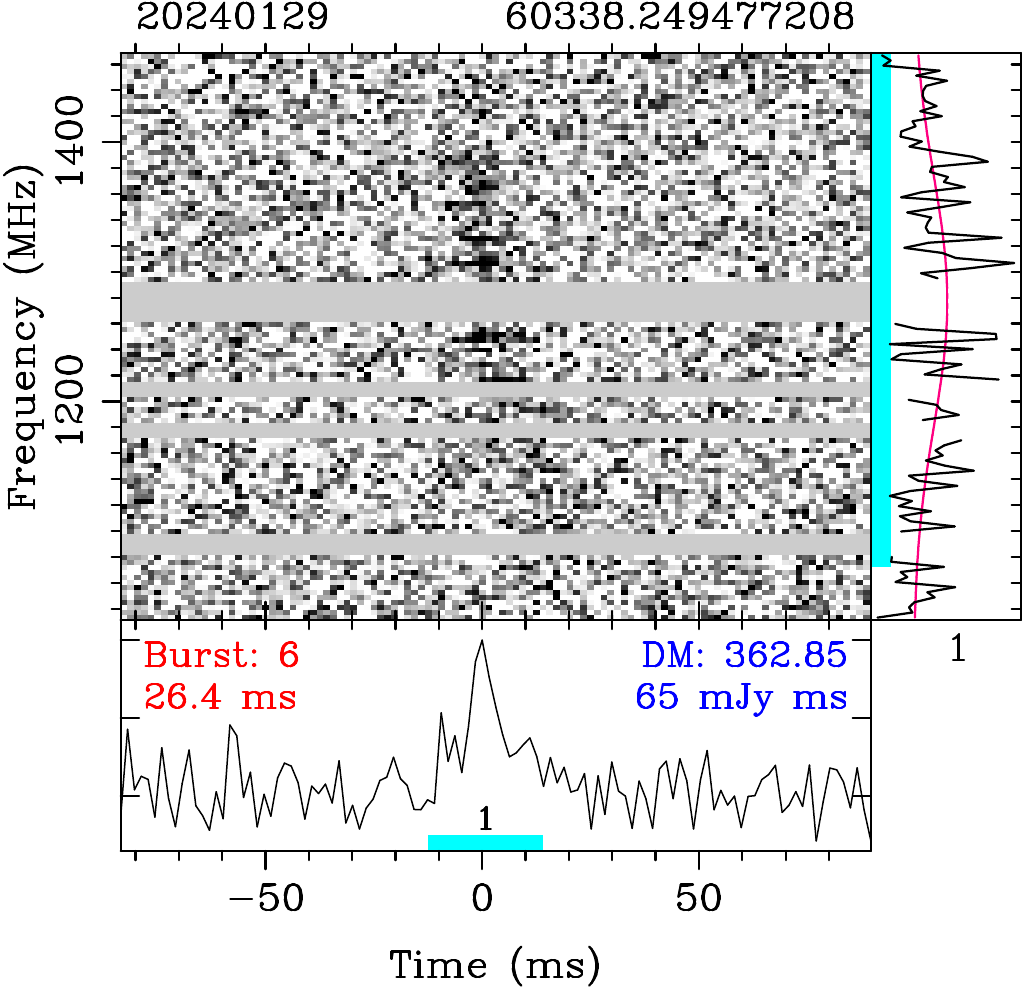}
\includegraphics[height=0.29\linewidth]{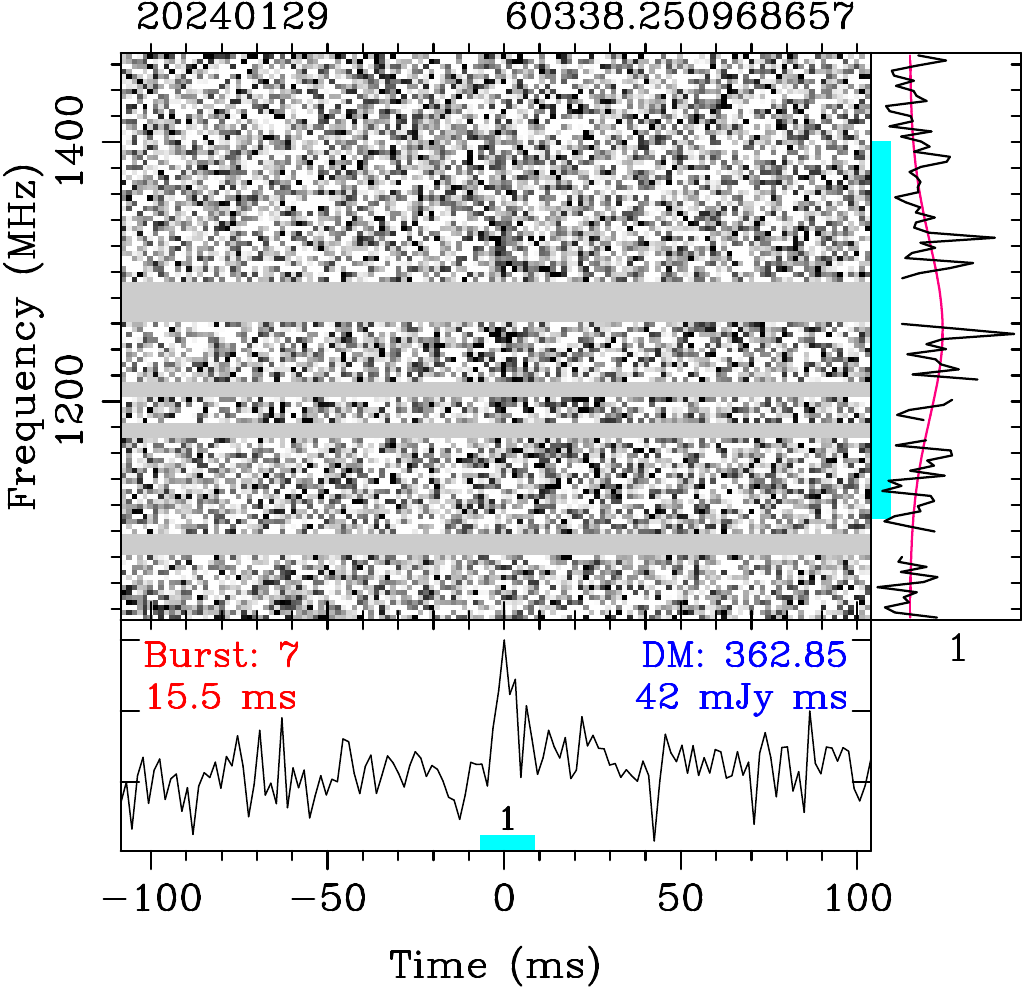}
\includegraphics[height=0.29\linewidth]{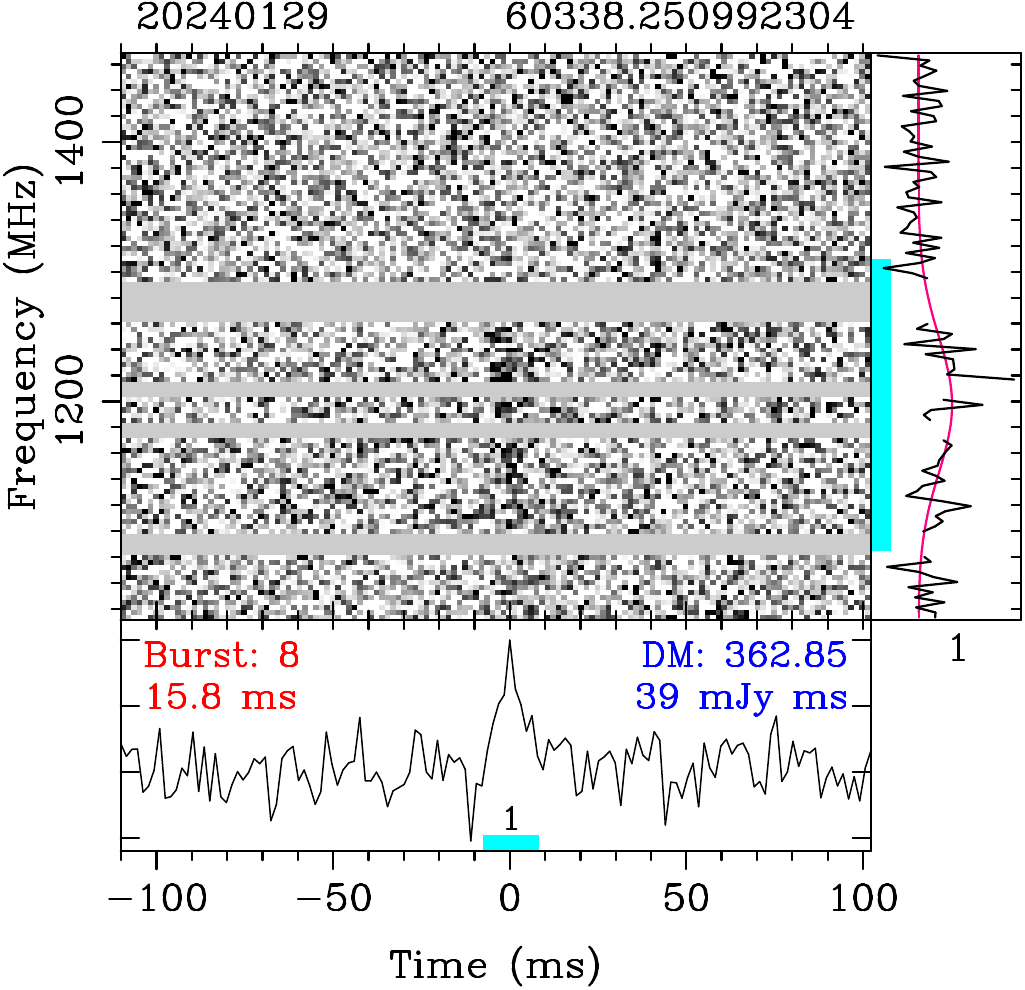}
\includegraphics[height=0.29\linewidth]{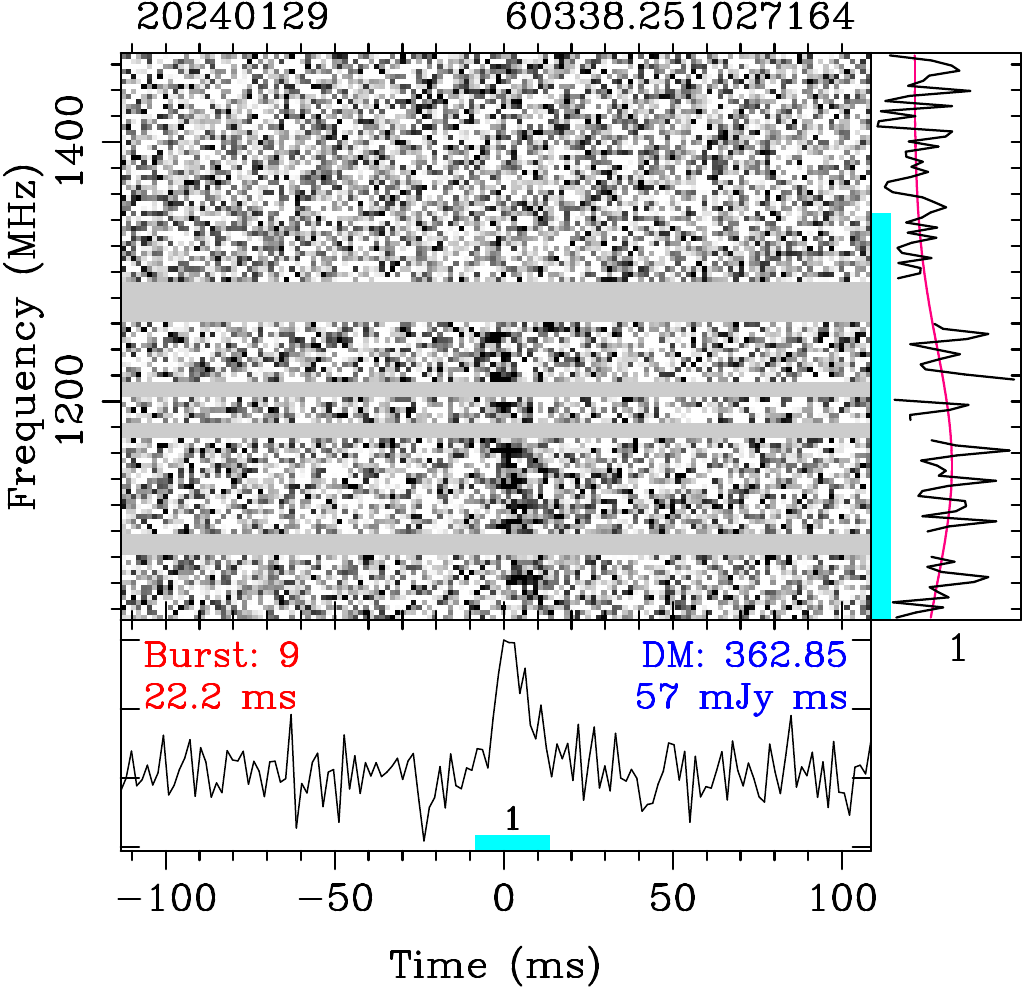}
\includegraphics[height=0.29\linewidth]{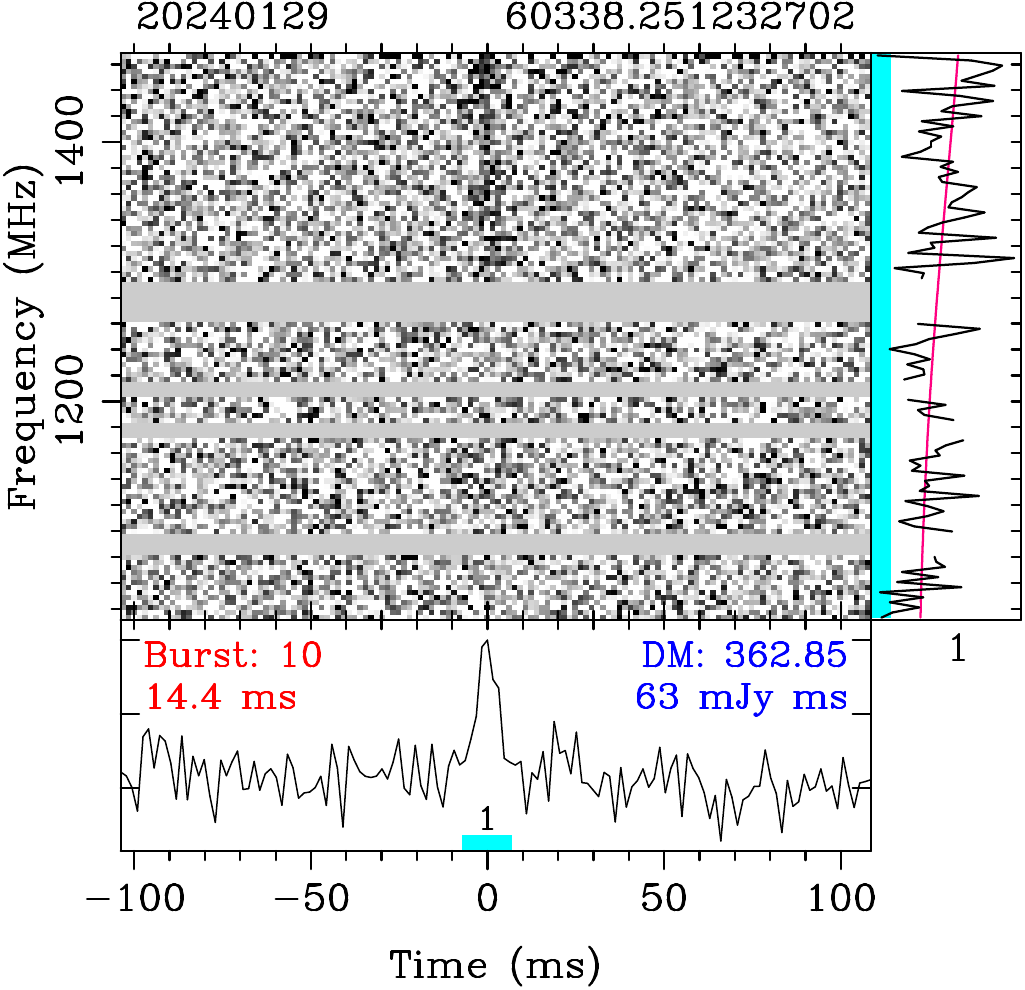}
\caption{({\textit{continued}})}
\end{figure*}
\addtocounter{figure}{-1}
\begin{figure*}
\flushleft
\includegraphics[height=0.29\linewidth]{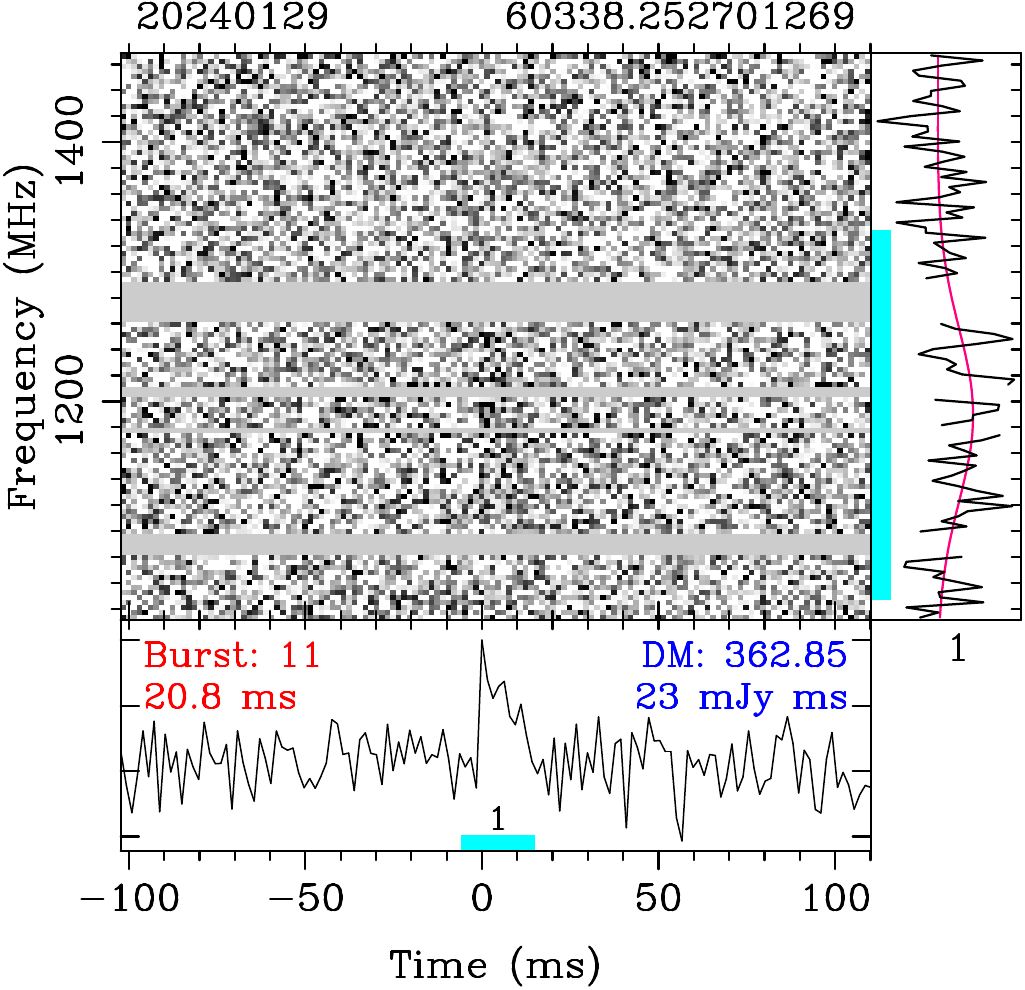}
\includegraphics[height=0.29\linewidth]{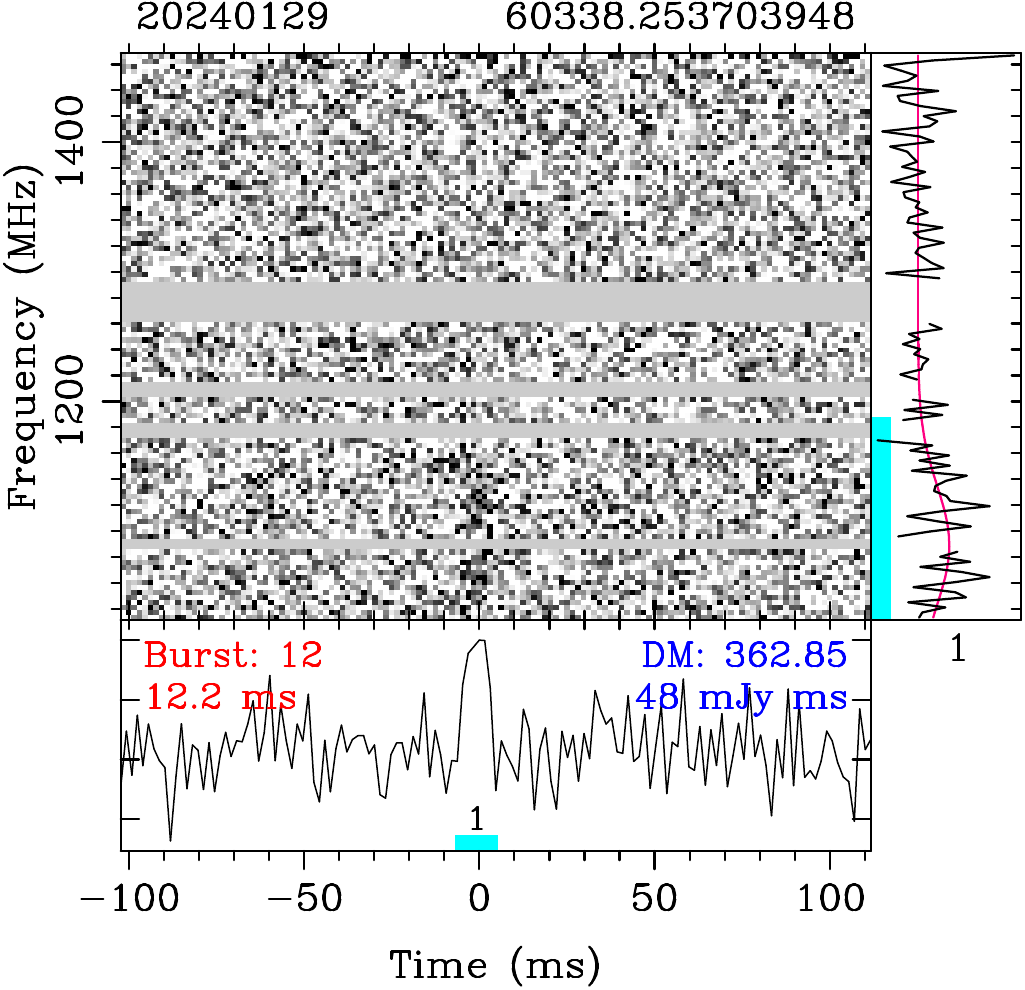}
\includegraphics[height=0.29\linewidth]{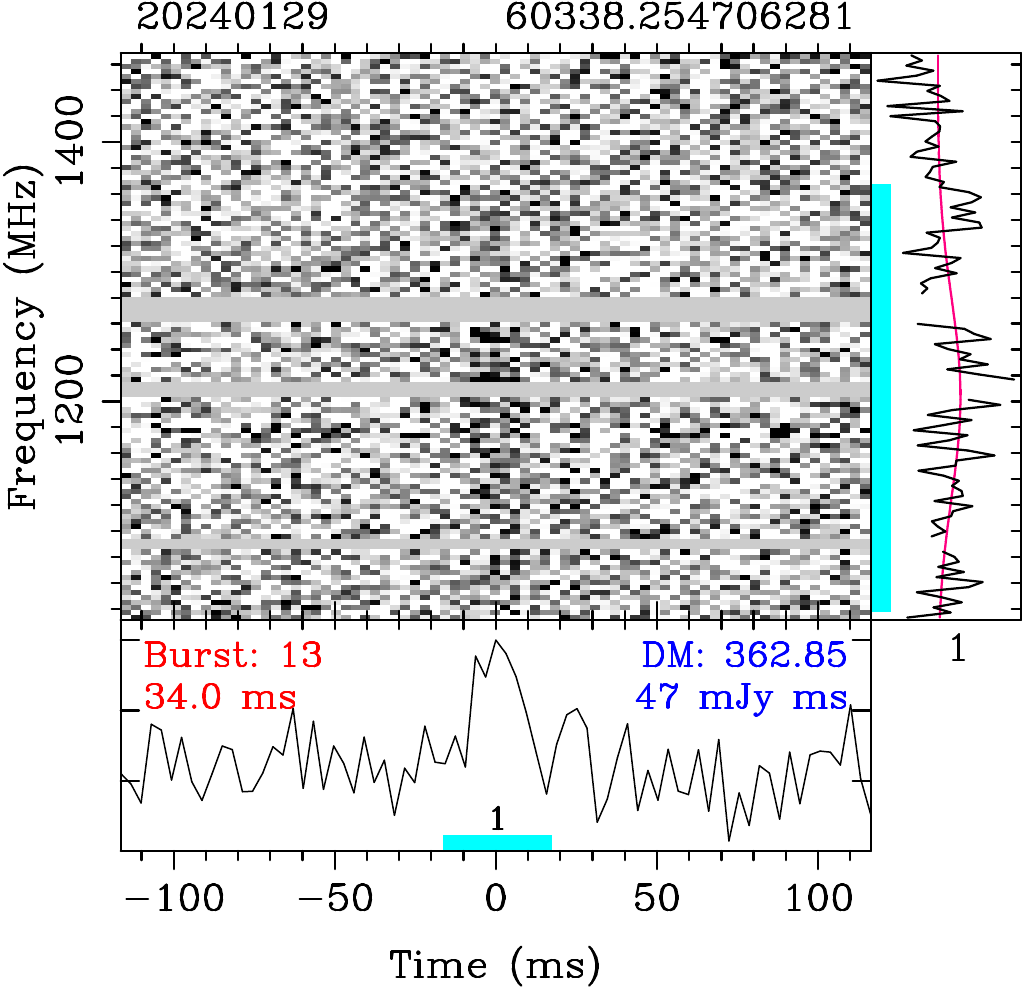}
\includegraphics[height=0.29\linewidth]{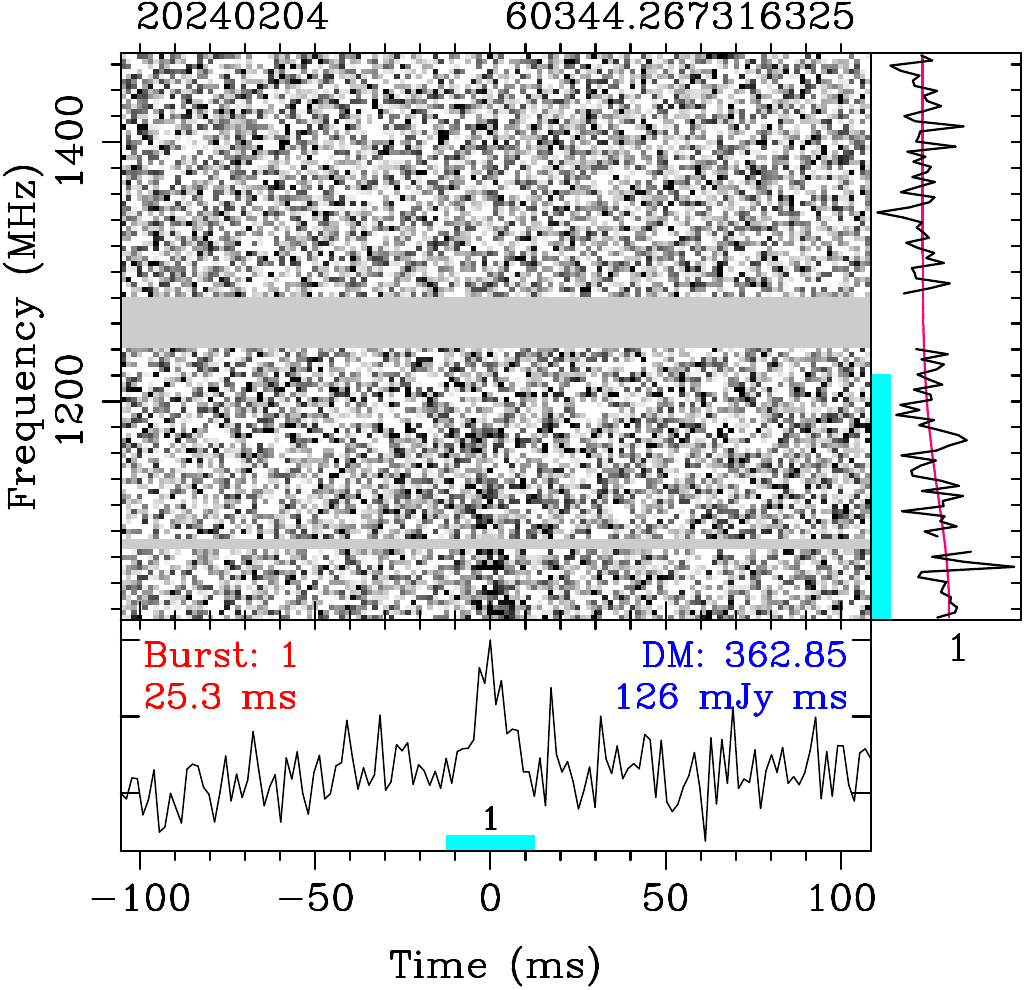}
\includegraphics[height=0.29\linewidth]{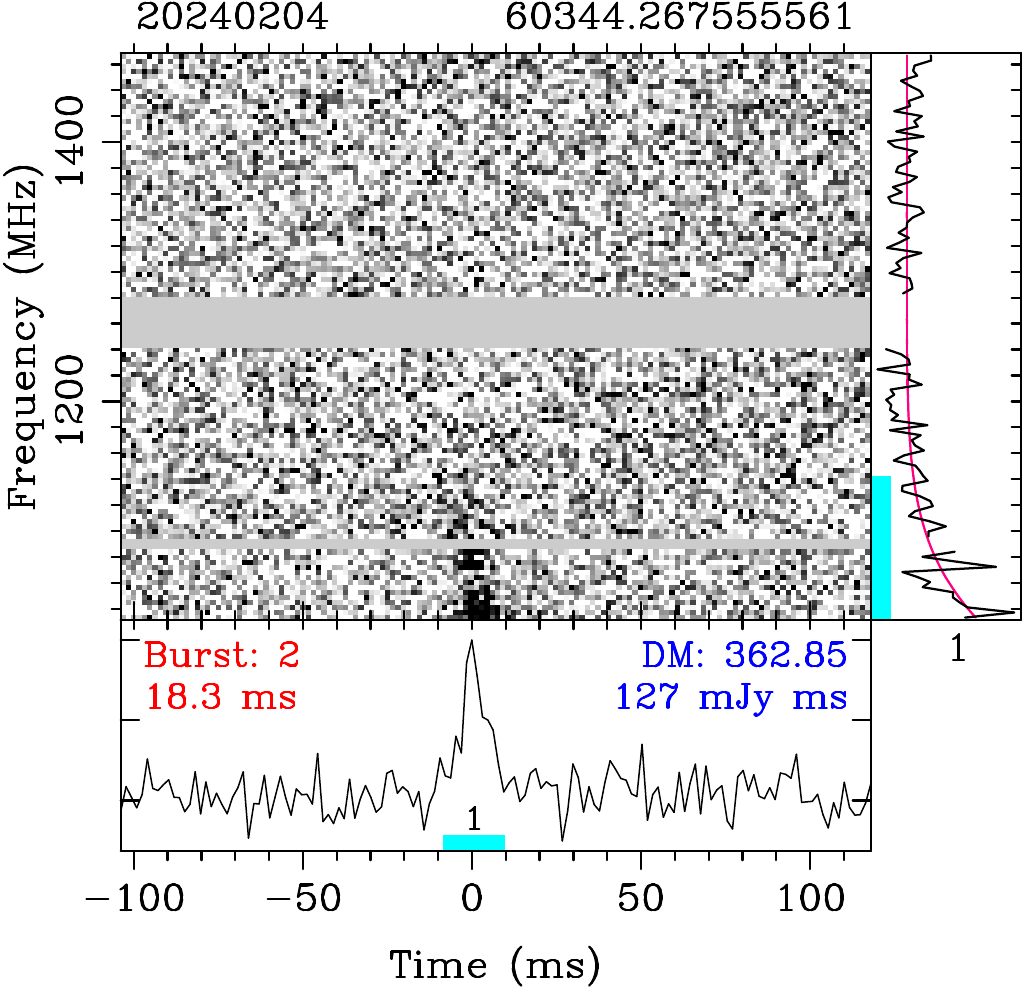}
\includegraphics[height=0.29\linewidth]{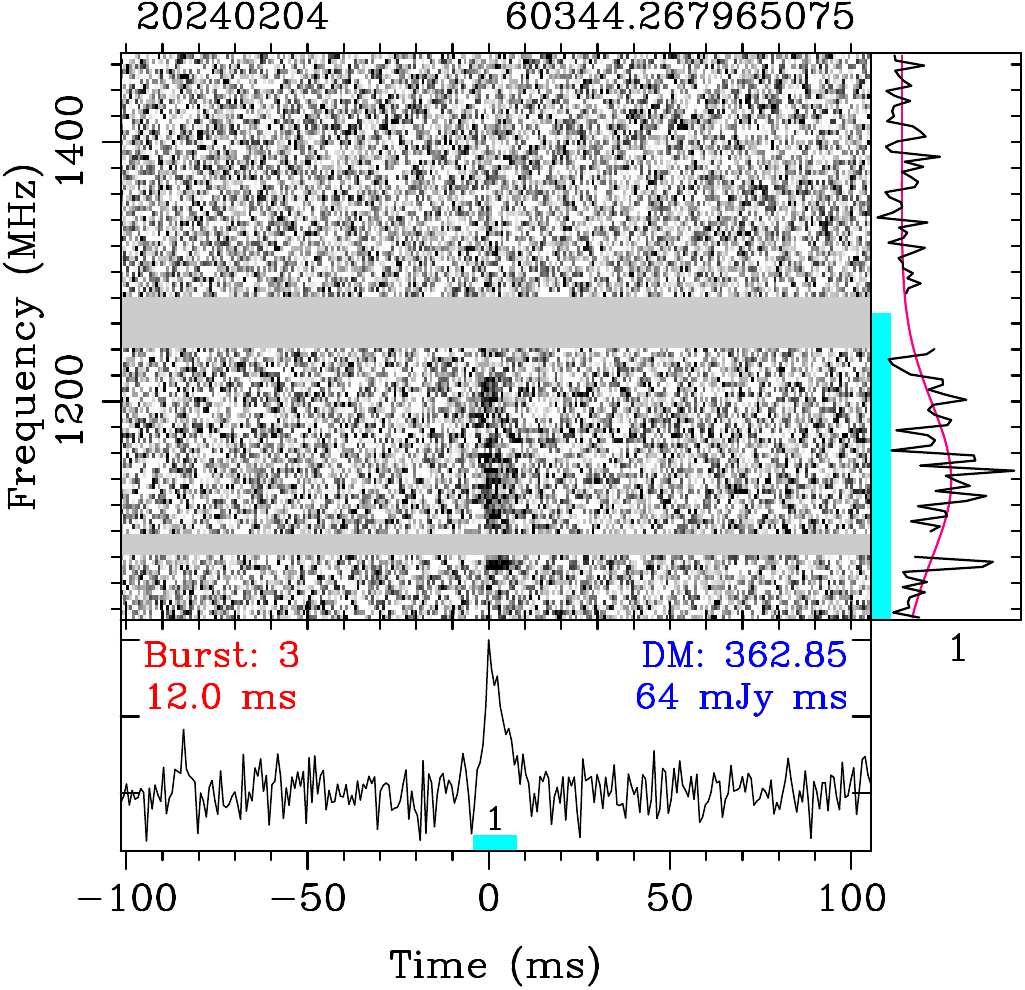}
\includegraphics[height=0.29\linewidth]{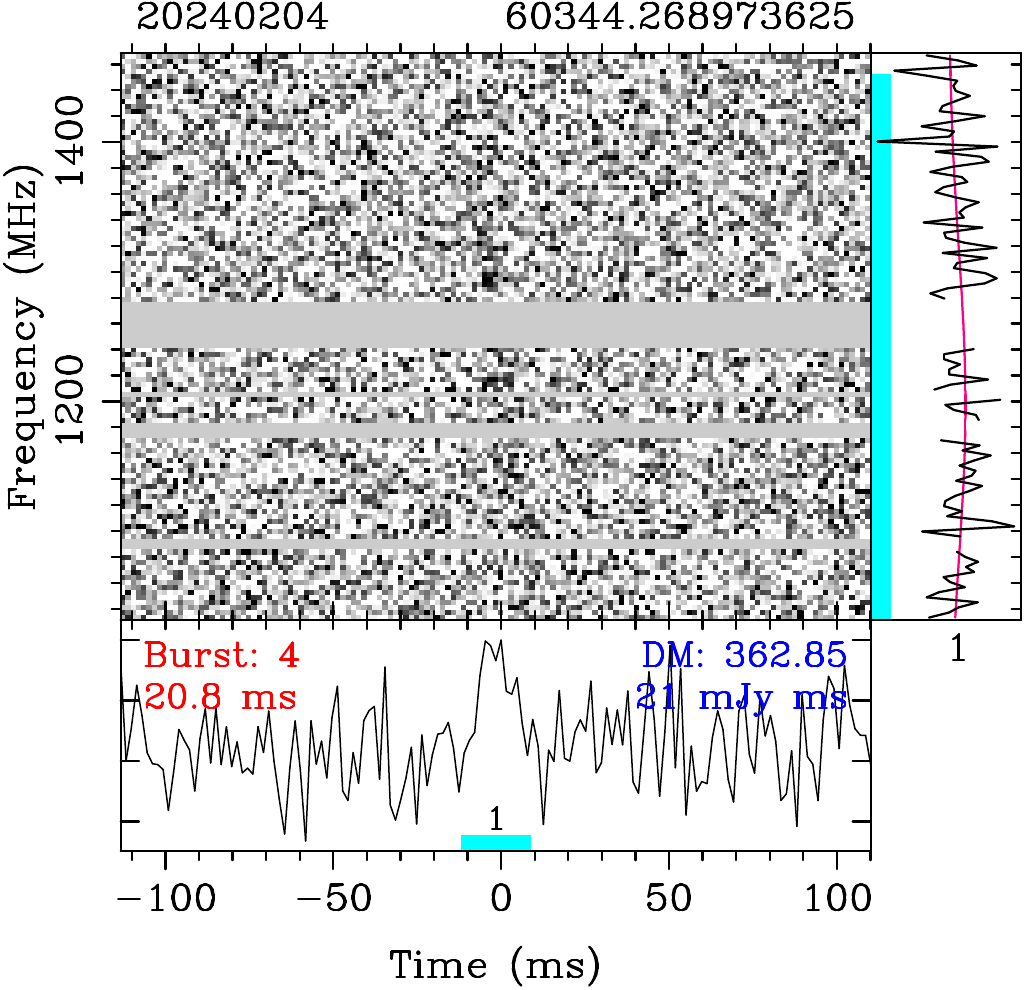}
\includegraphics[height=0.29\linewidth]{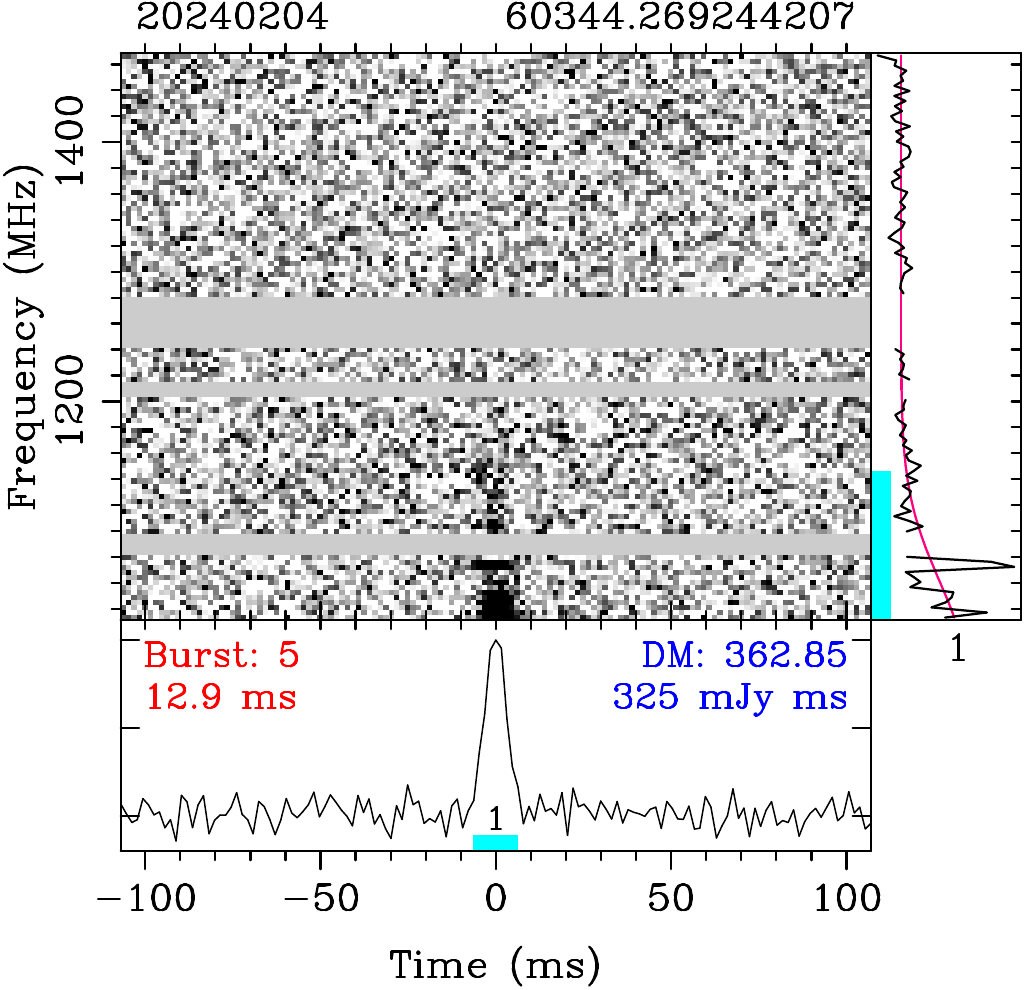}
\includegraphics[height=0.29\linewidth]{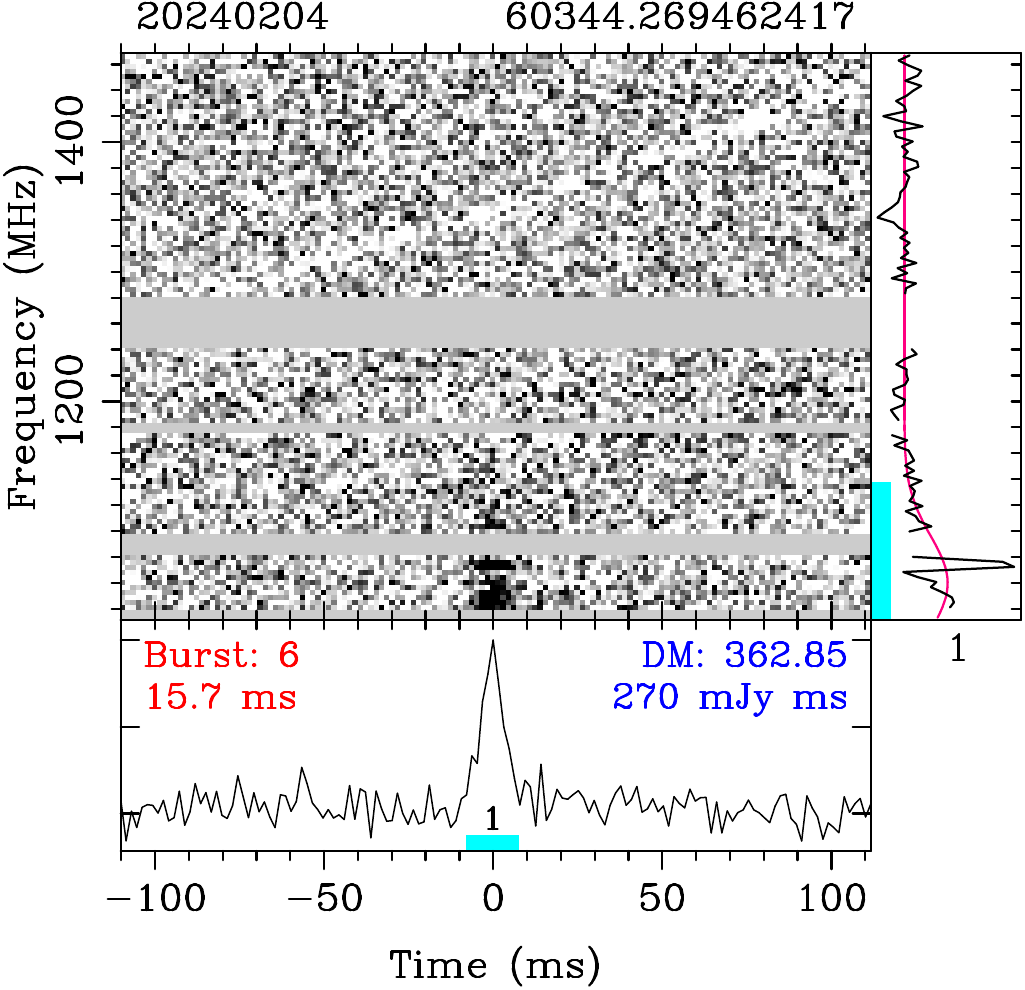}
\includegraphics[height=0.29\linewidth]{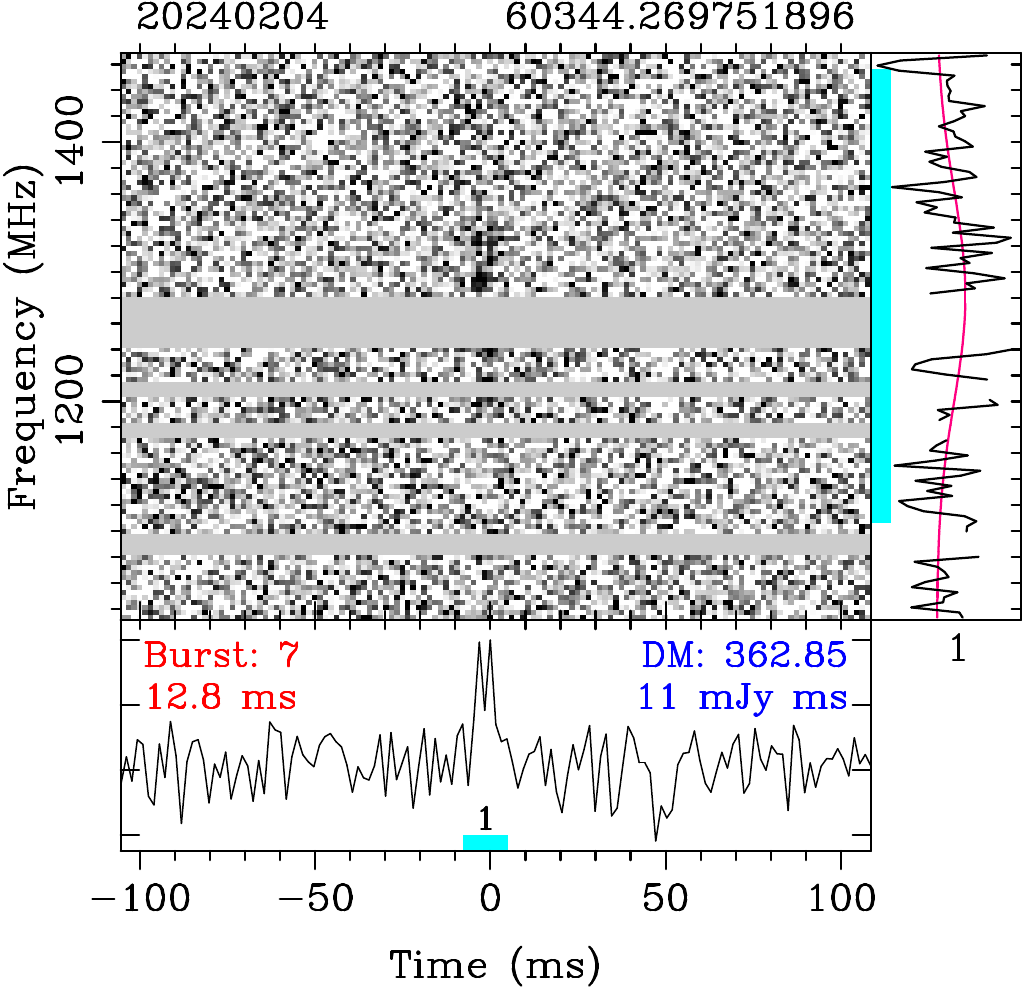}
\includegraphics[height=0.29\linewidth]{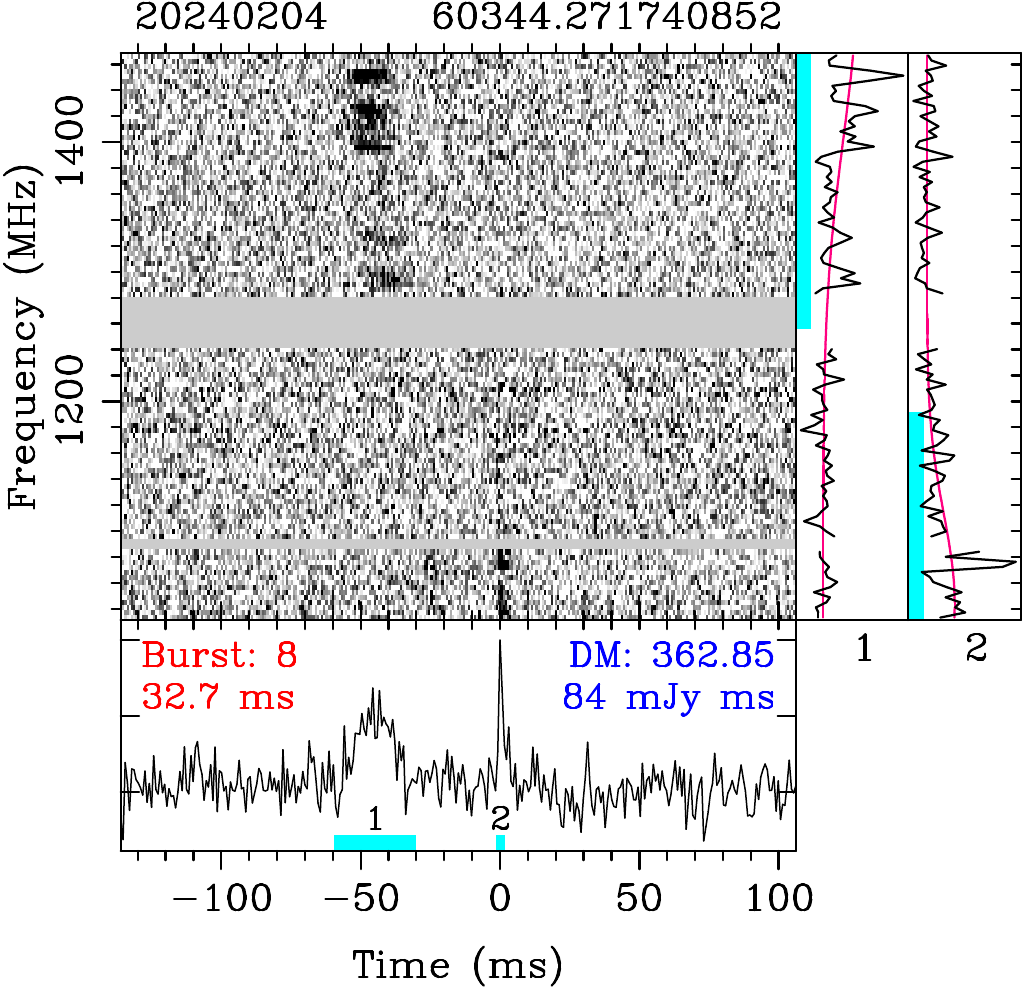}
\includegraphics[height=0.29\linewidth]{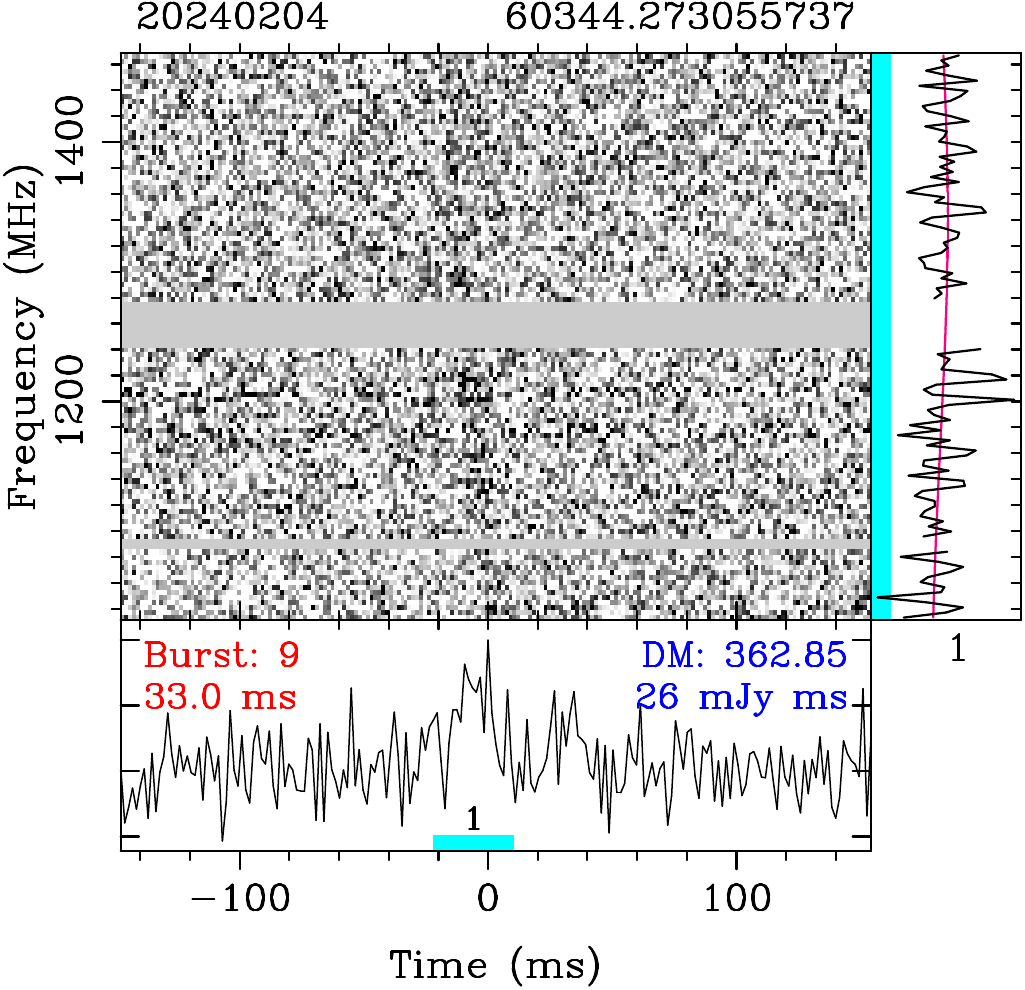}
\caption{({\textit{continued}})}
\end{figure*}
\addtocounter{figure}{-1}
\begin{figure*}
\flushleft
\includegraphics[height=0.29\linewidth]{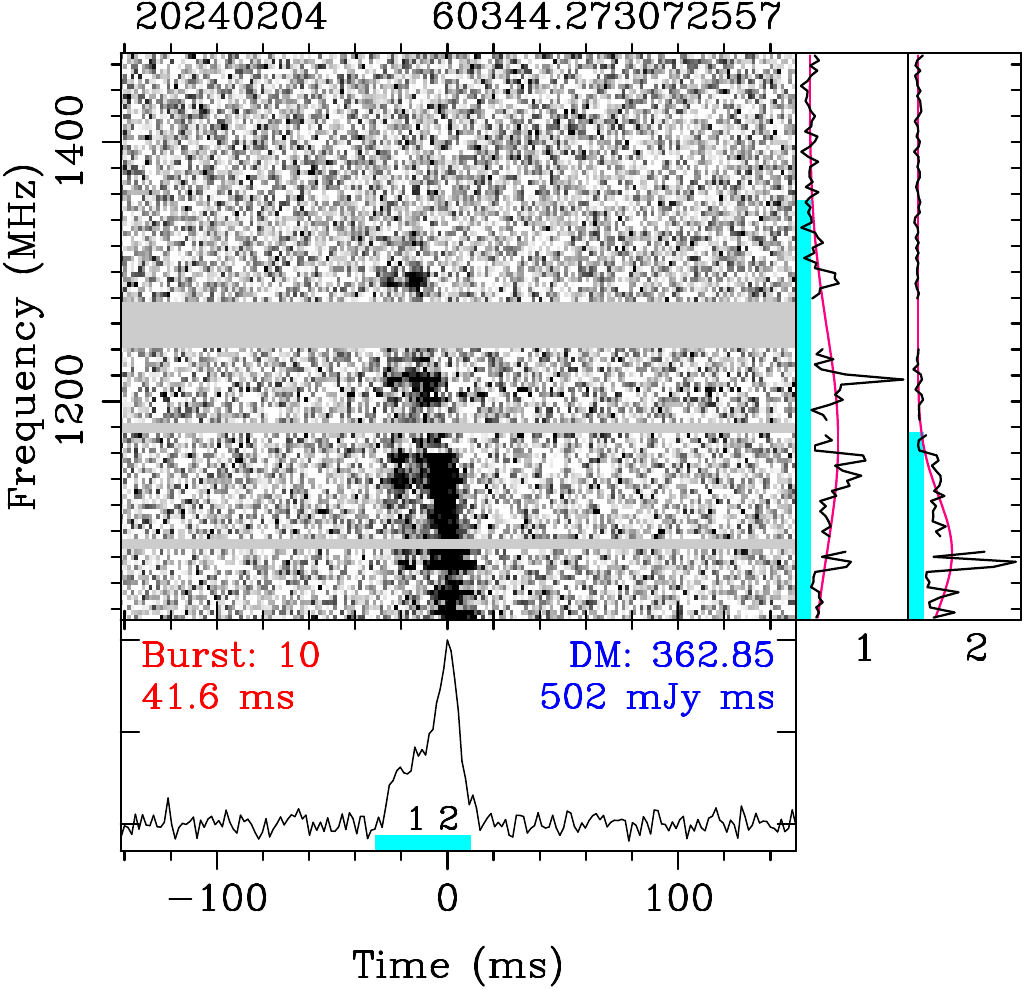}
\includegraphics[height=0.29\linewidth]{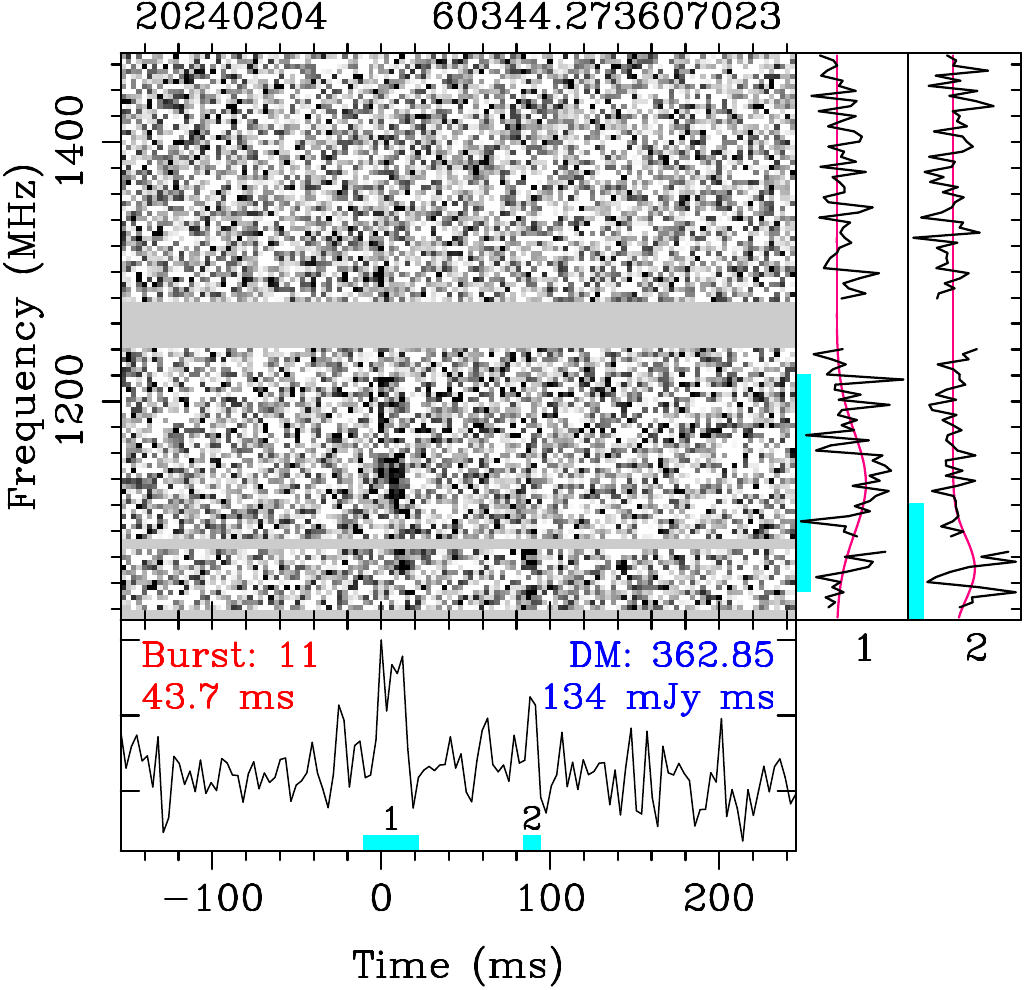}
\includegraphics[height=0.29\linewidth]{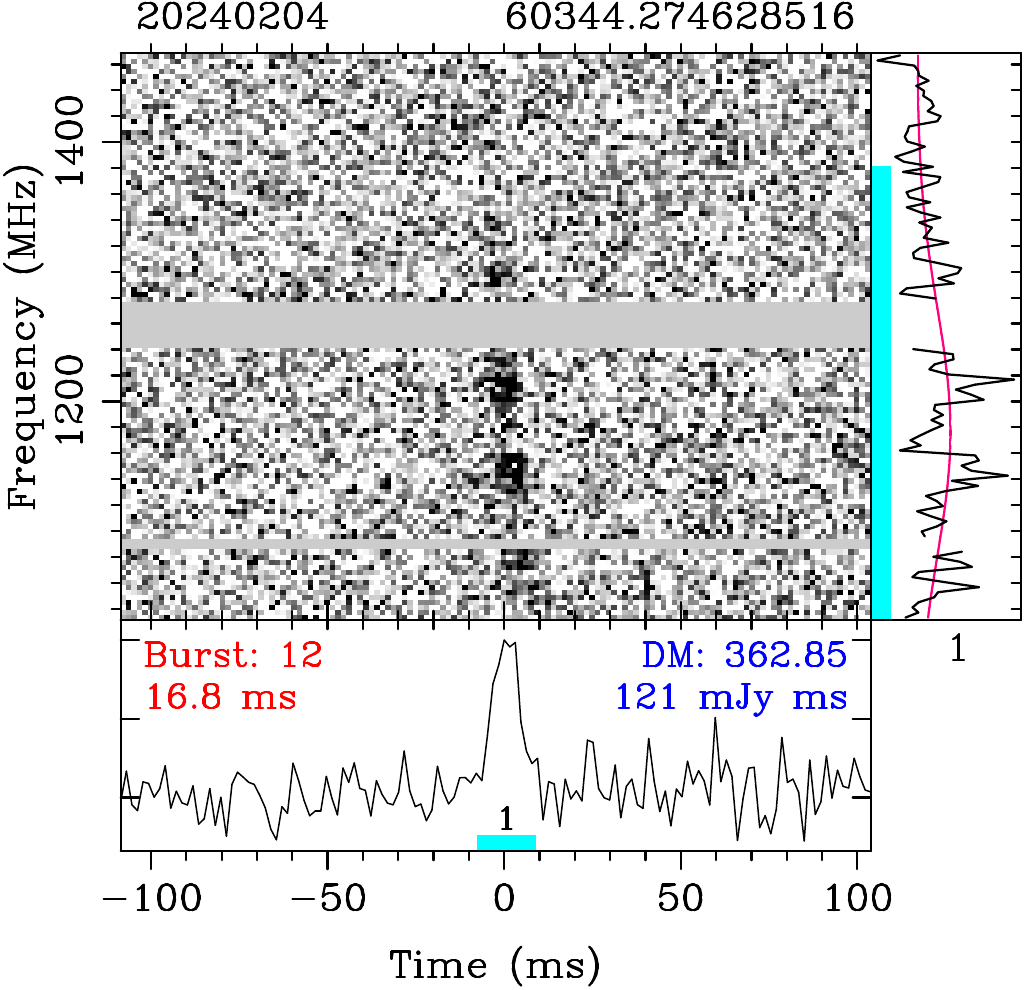}
\includegraphics[height=0.29\linewidth]{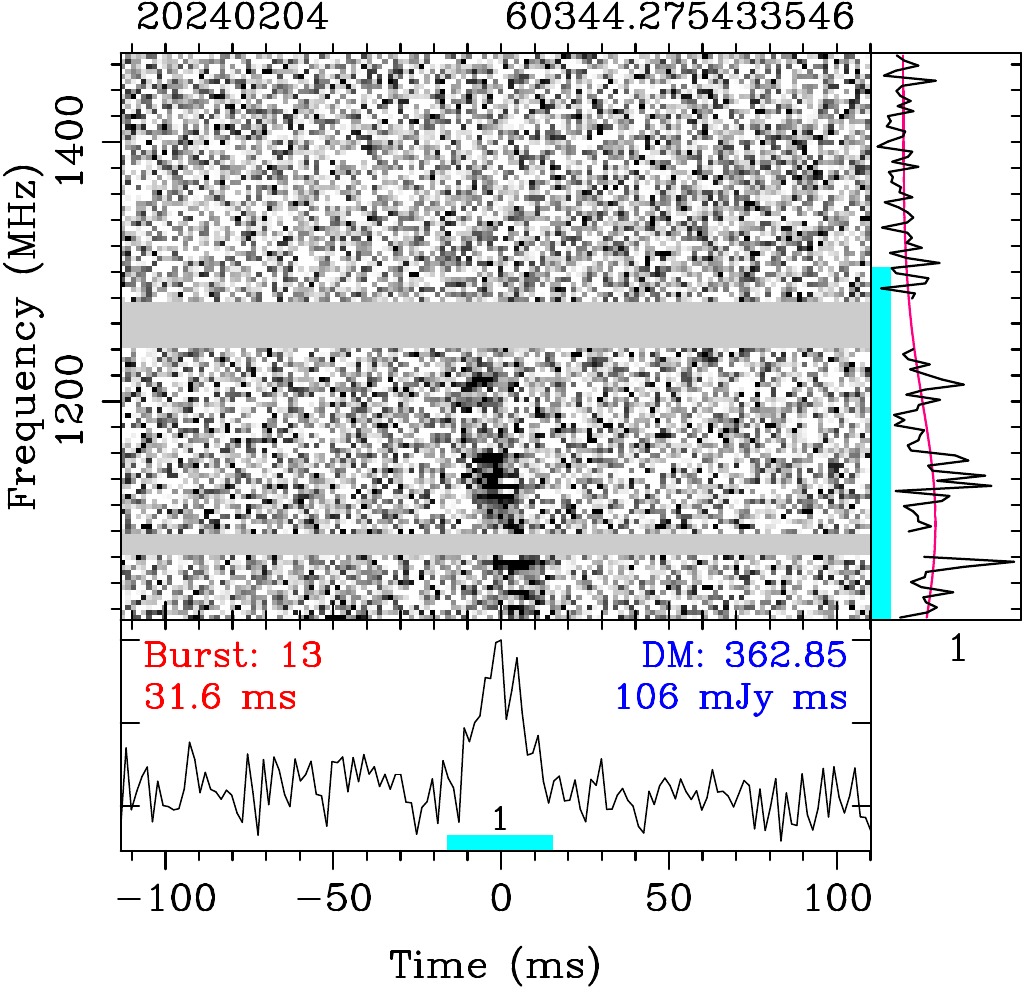}
\includegraphics[height=0.29\linewidth]{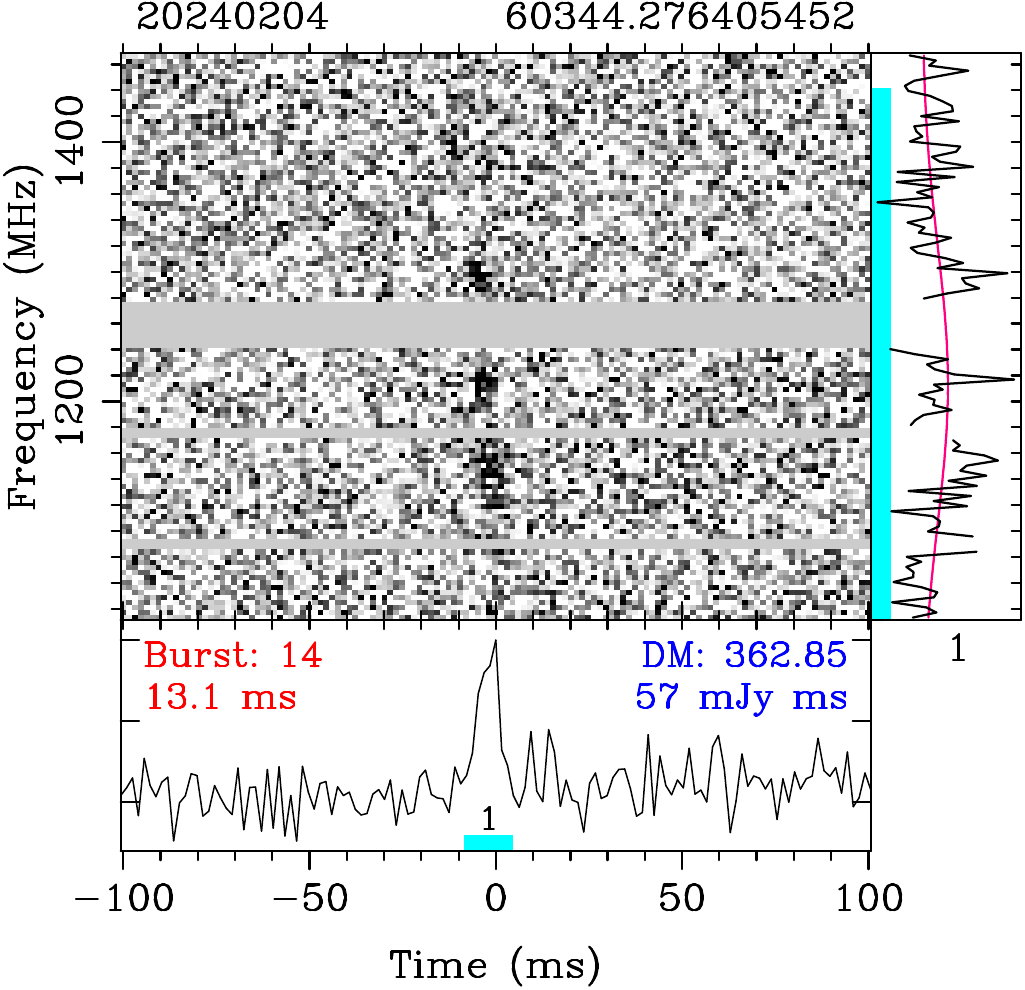}
\includegraphics[height=0.29\linewidth]{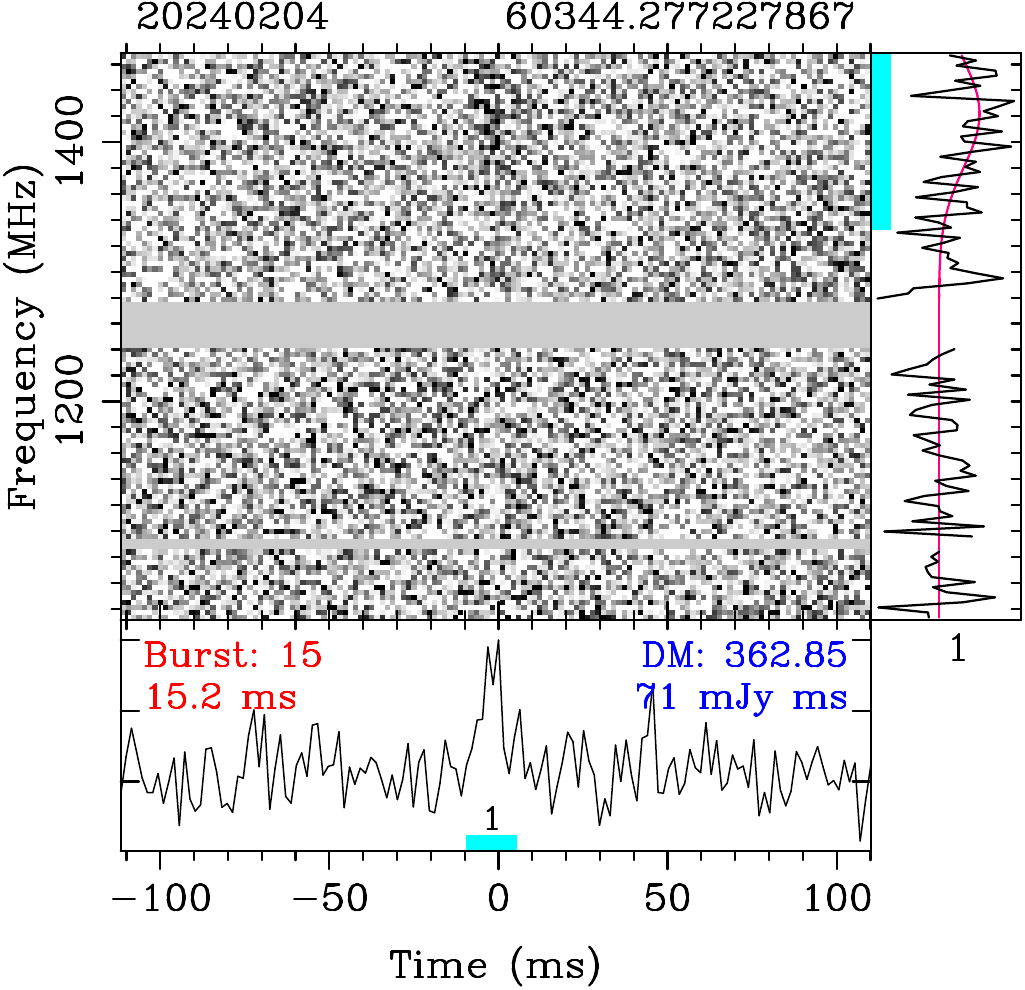}
\includegraphics[height=0.29\linewidth]{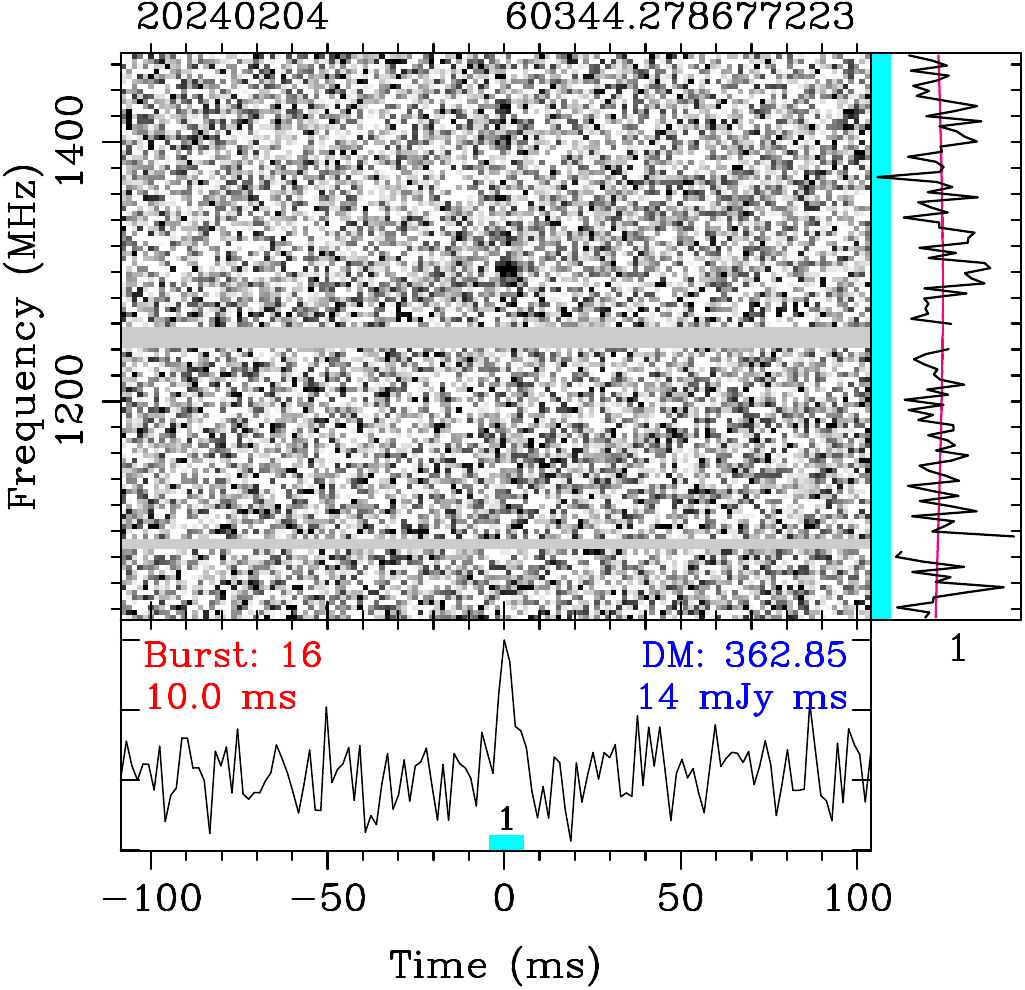}
\includegraphics[height=0.29\linewidth]{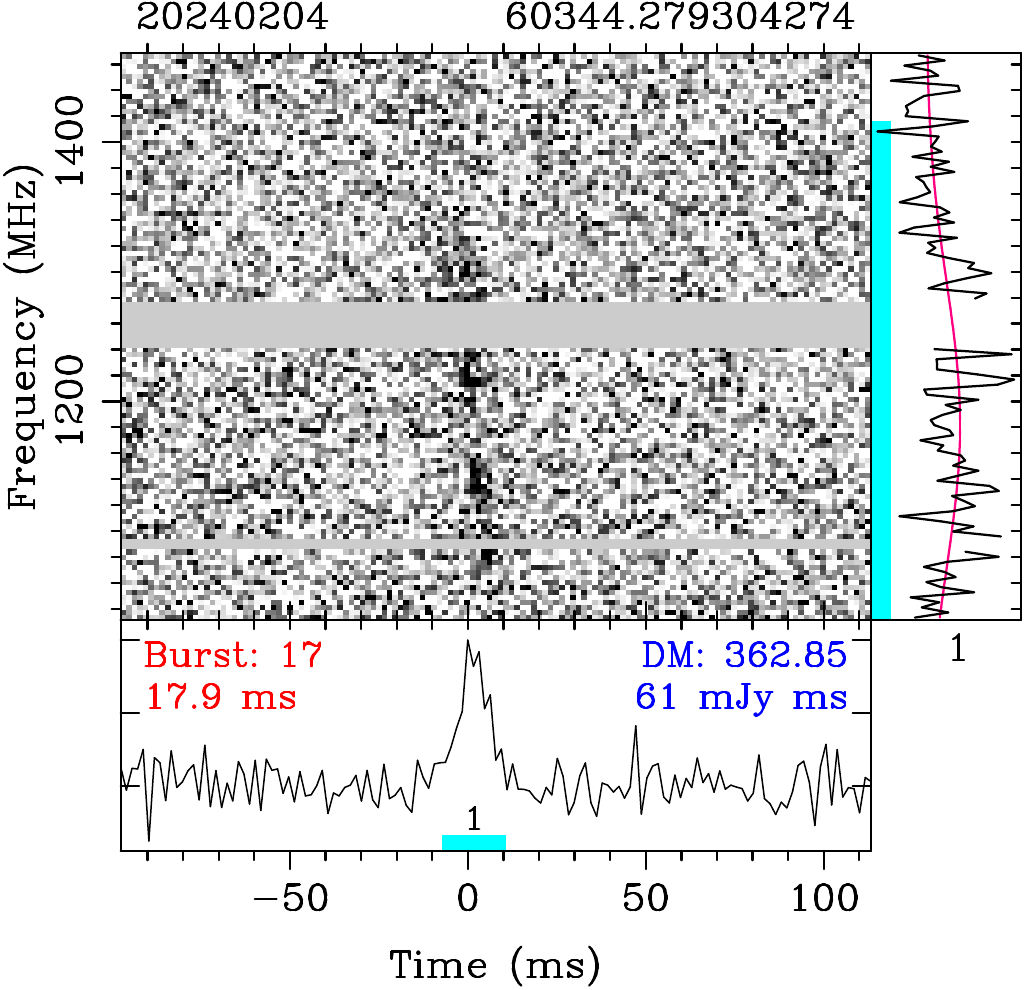}
\includegraphics[height=0.29\linewidth]{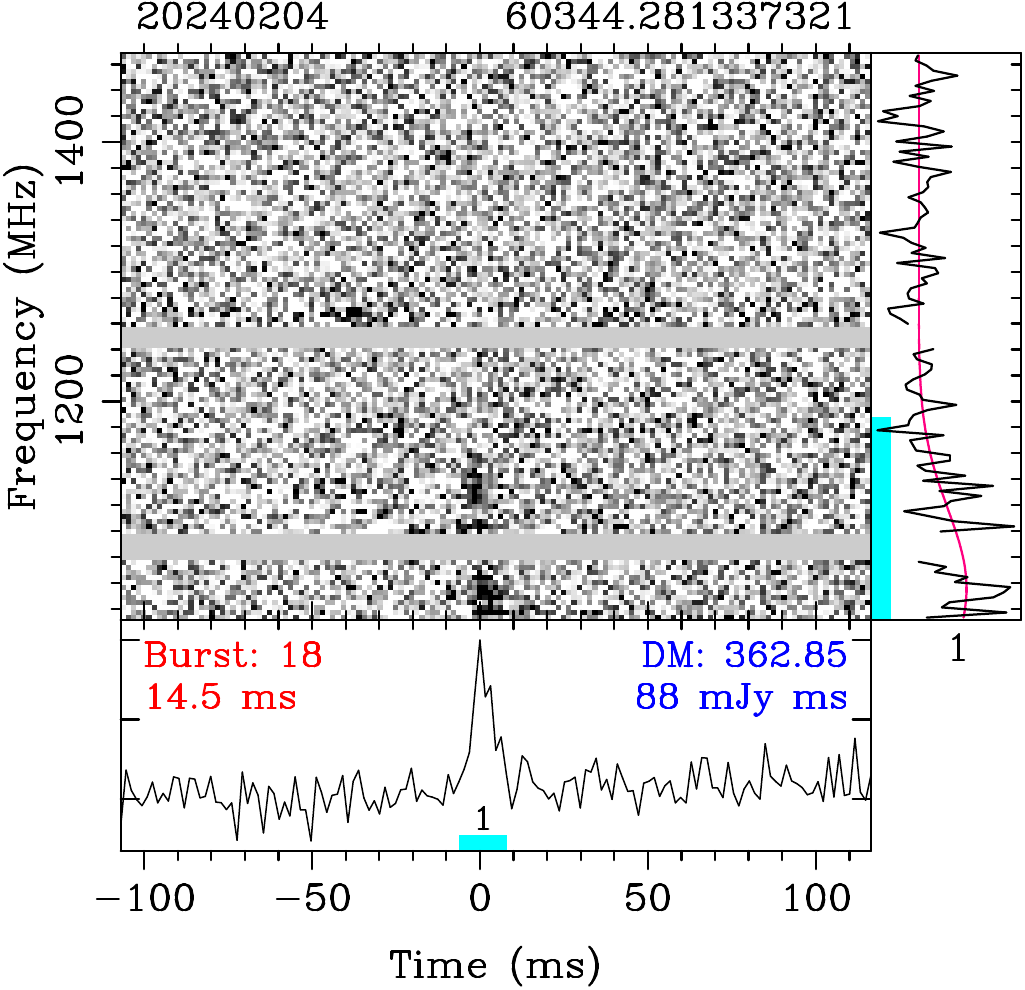}
\includegraphics[height=0.29\linewidth]{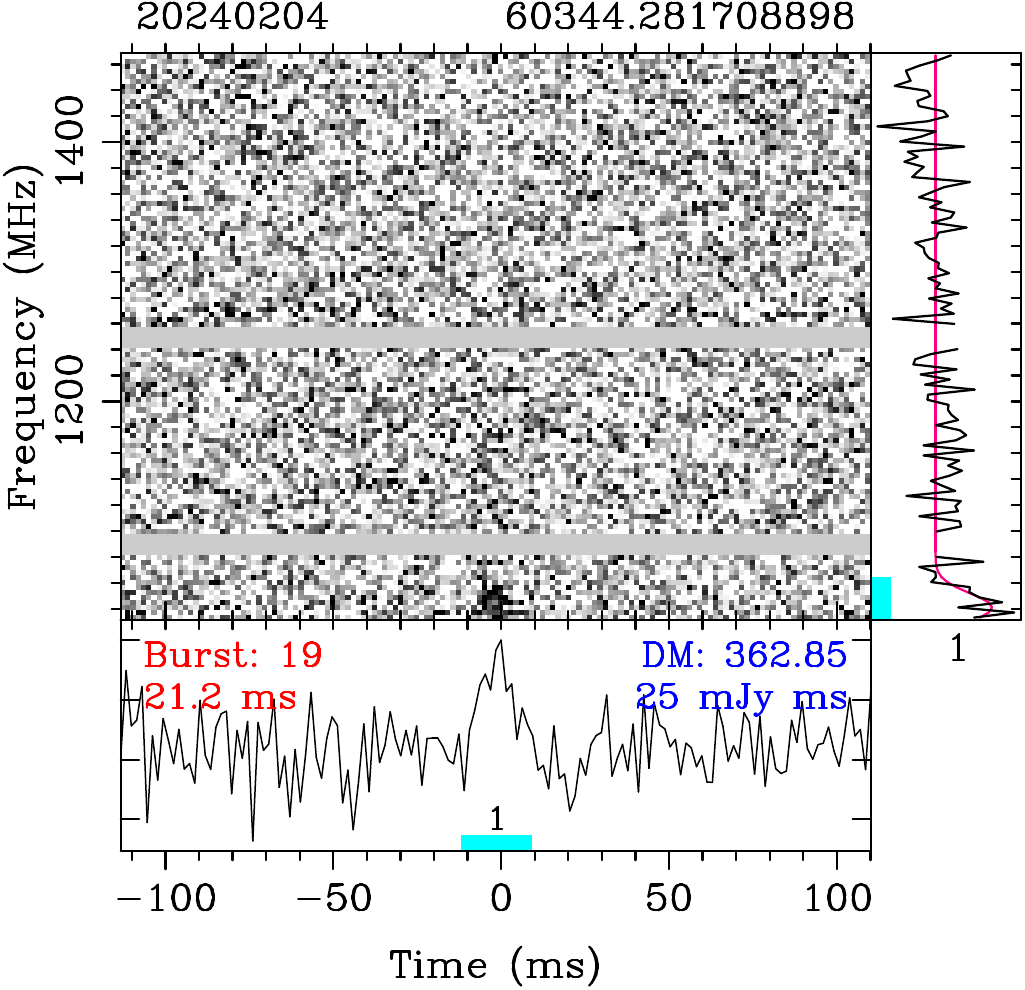}
\includegraphics[height=0.29\linewidth]{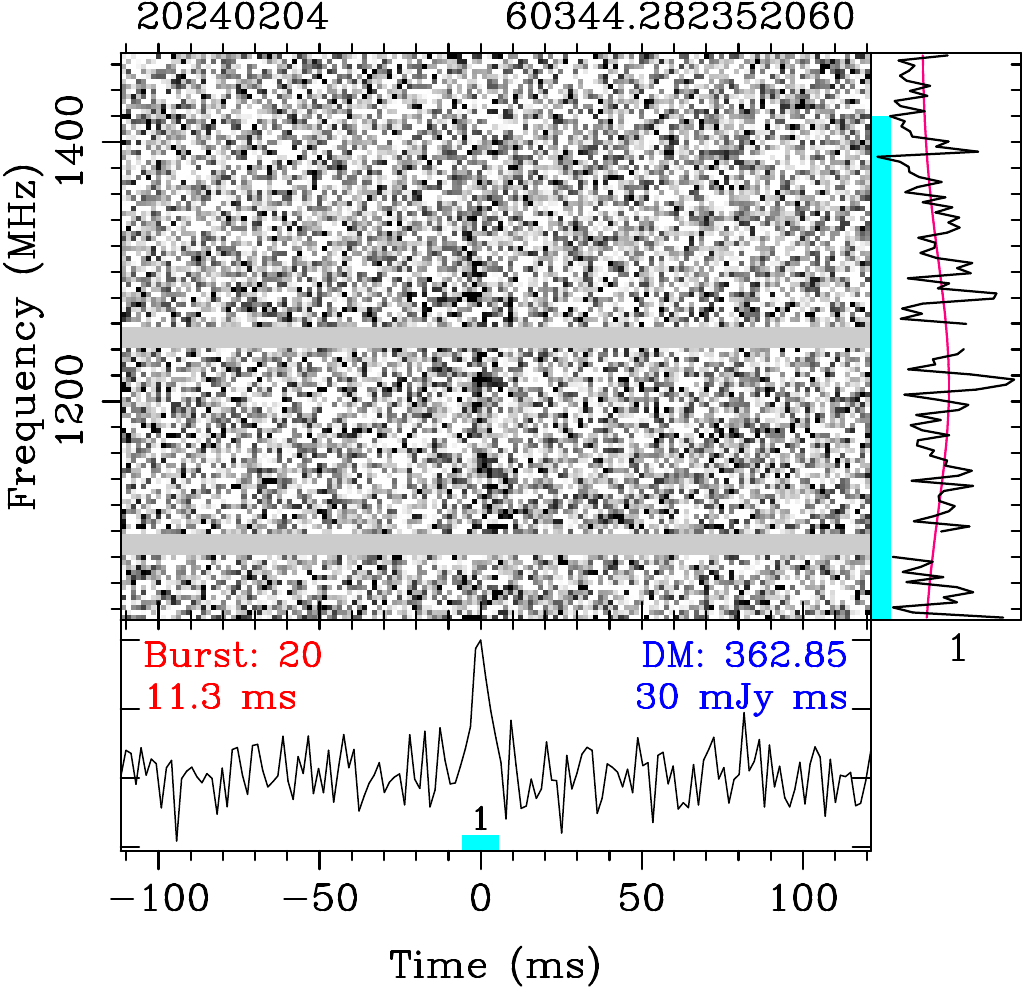}
\includegraphics[height=0.29\linewidth]{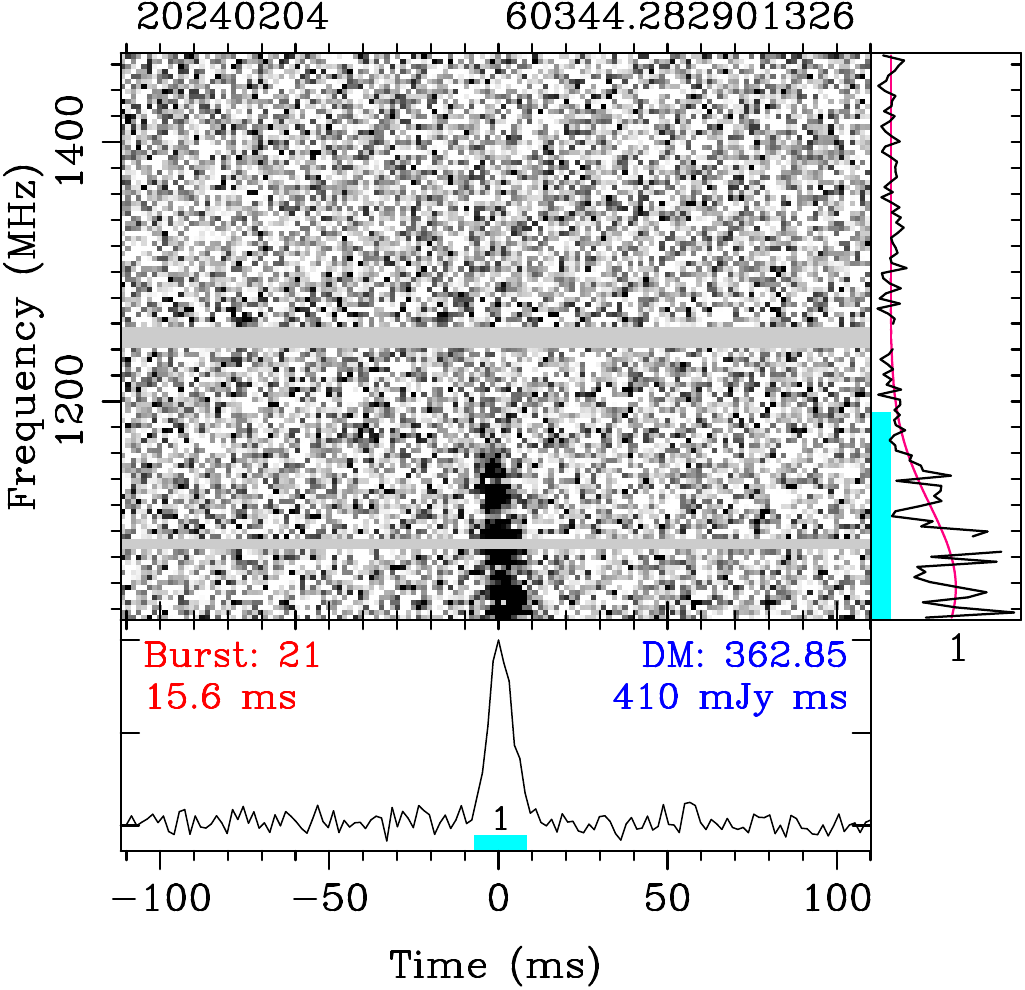}
\caption{({\textit{continued}})}
\end{figure*}
\addtocounter{figure}{-1}
\begin{figure*}
\flushleft
\includegraphics[height=0.29\linewidth]{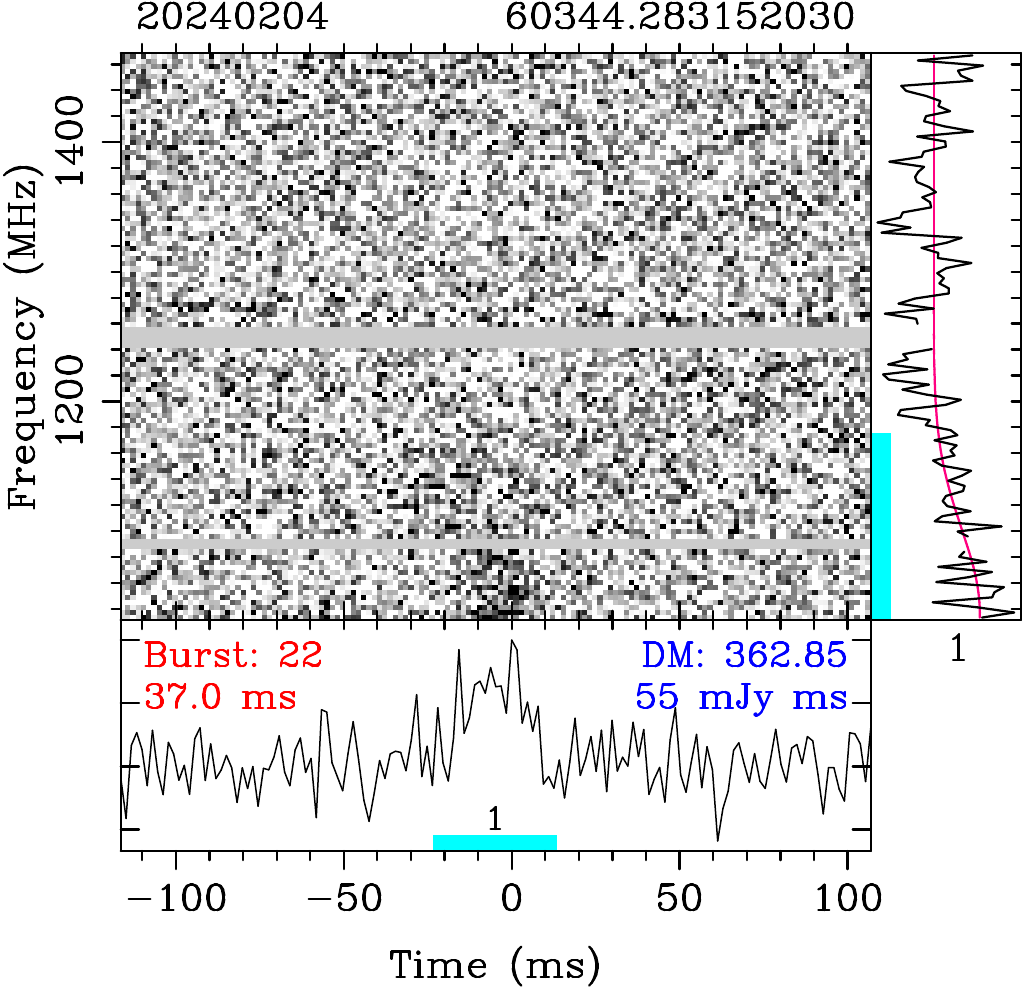}
\includegraphics[height=0.29\linewidth]{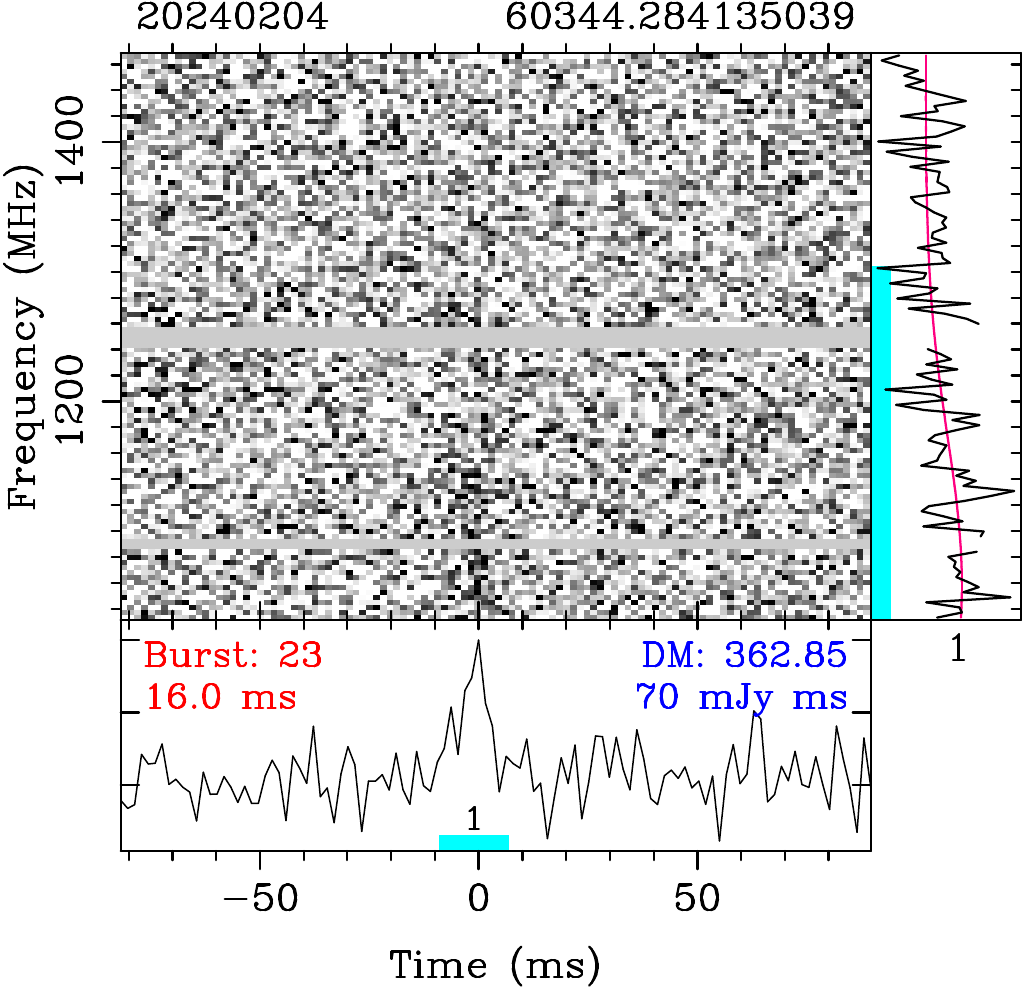}
\includegraphics[height=0.29\linewidth]{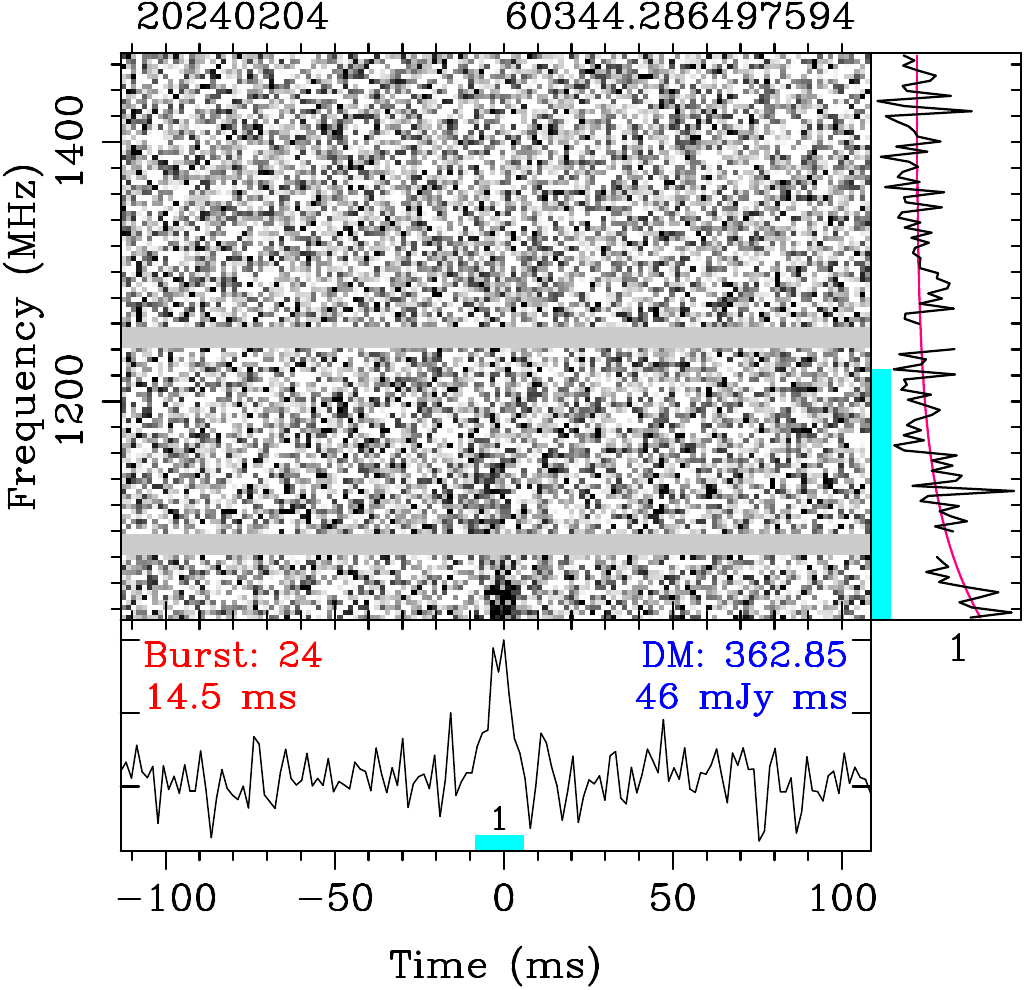}
\includegraphics[height=0.29\linewidth]{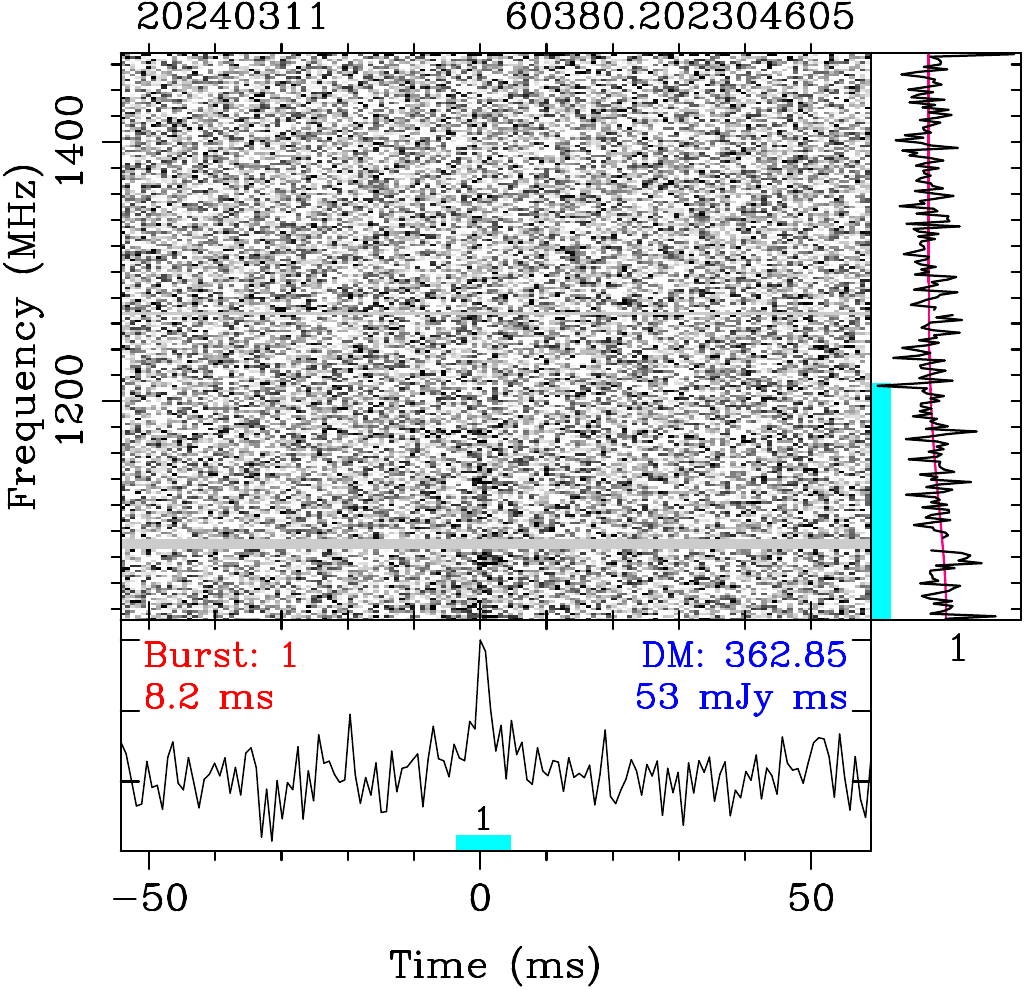}
\includegraphics[height=0.29\linewidth]{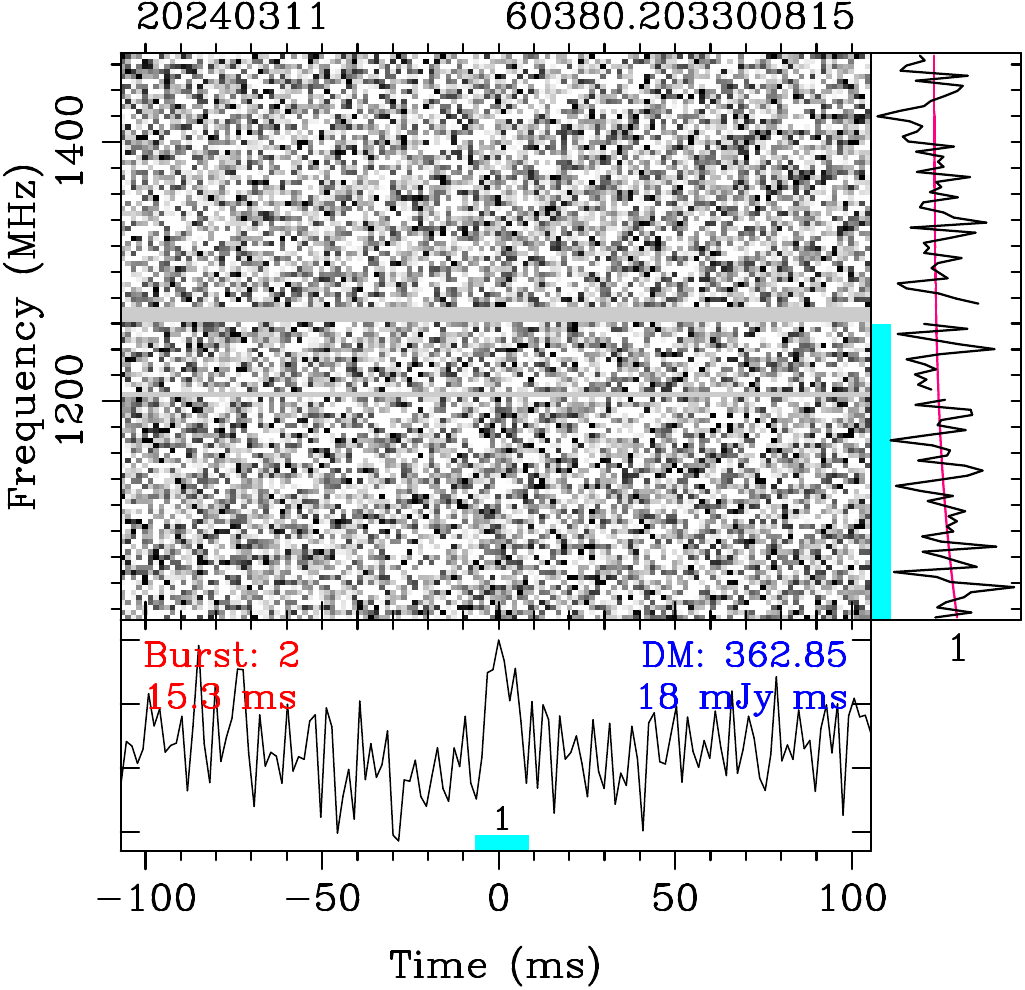}
\includegraphics[height=0.29\linewidth]{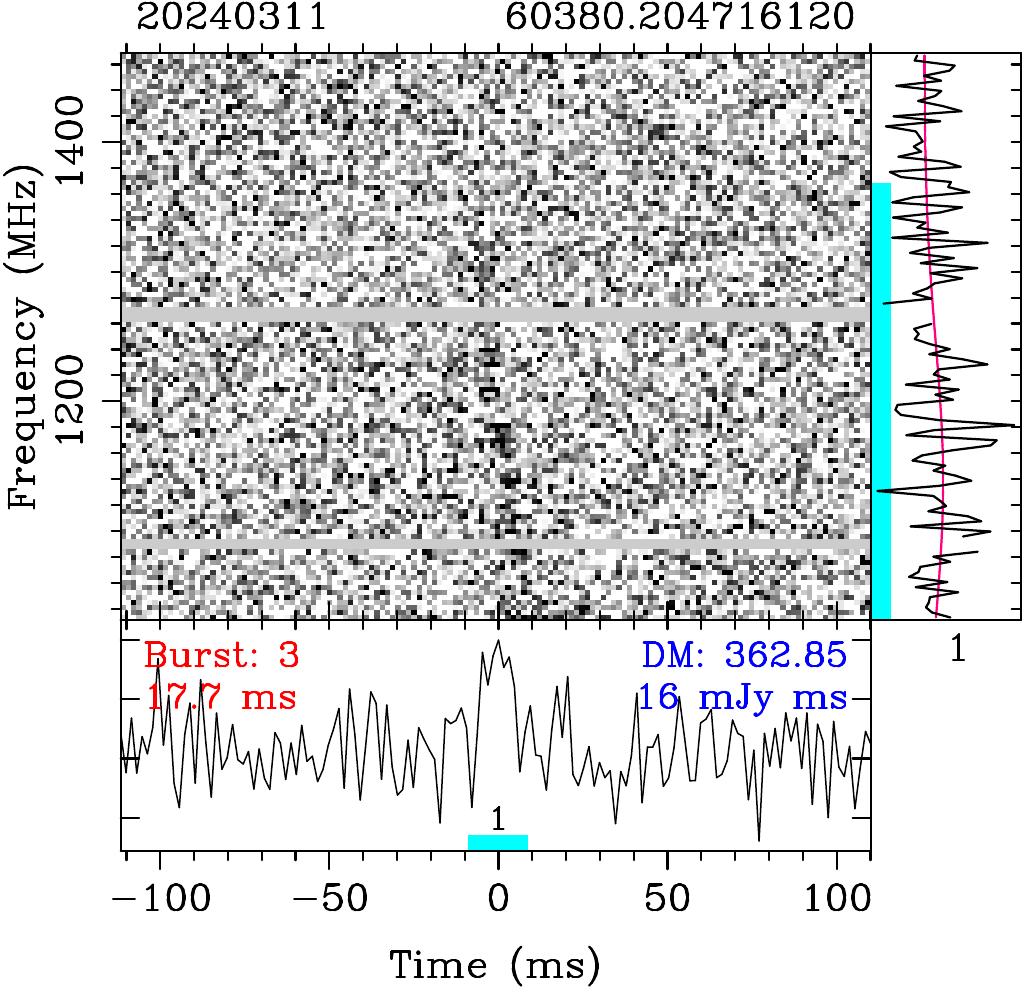}
\includegraphics[height=0.29\linewidth]{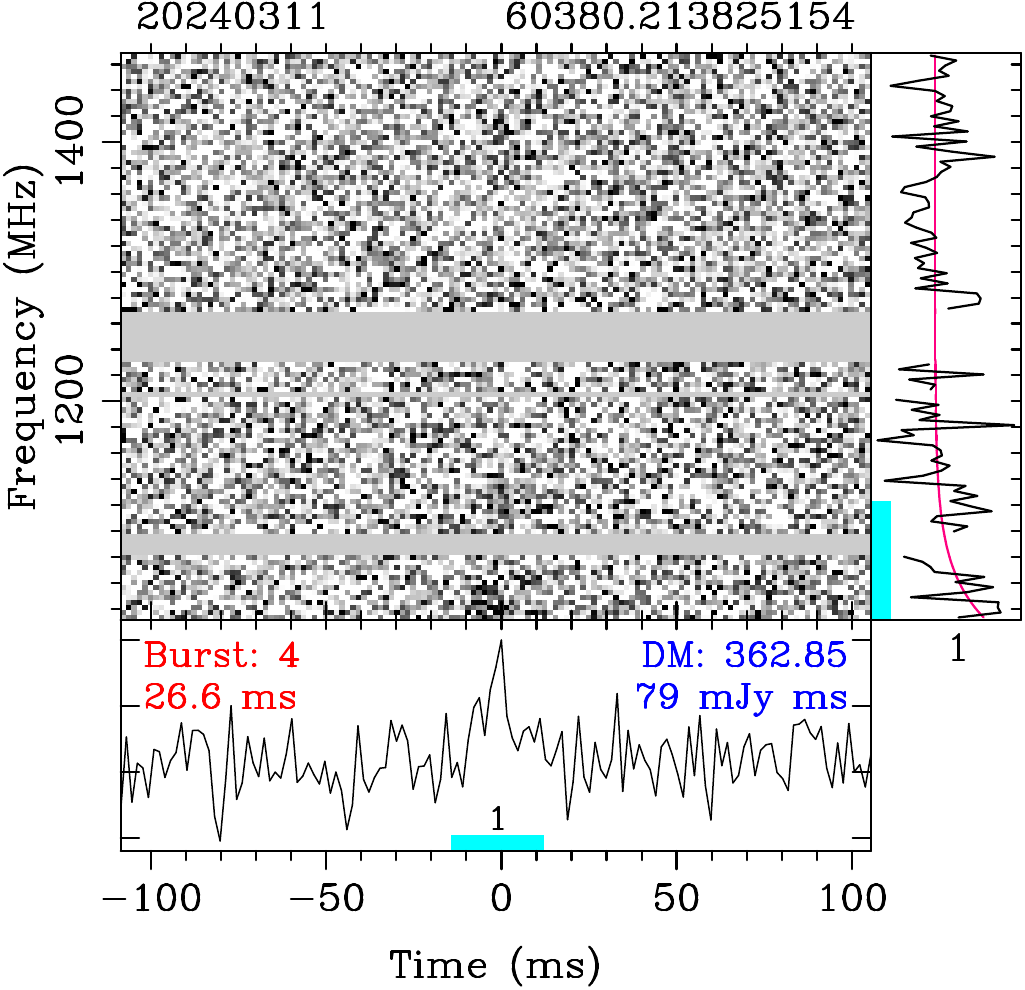}
\includegraphics[height=0.29\linewidth]{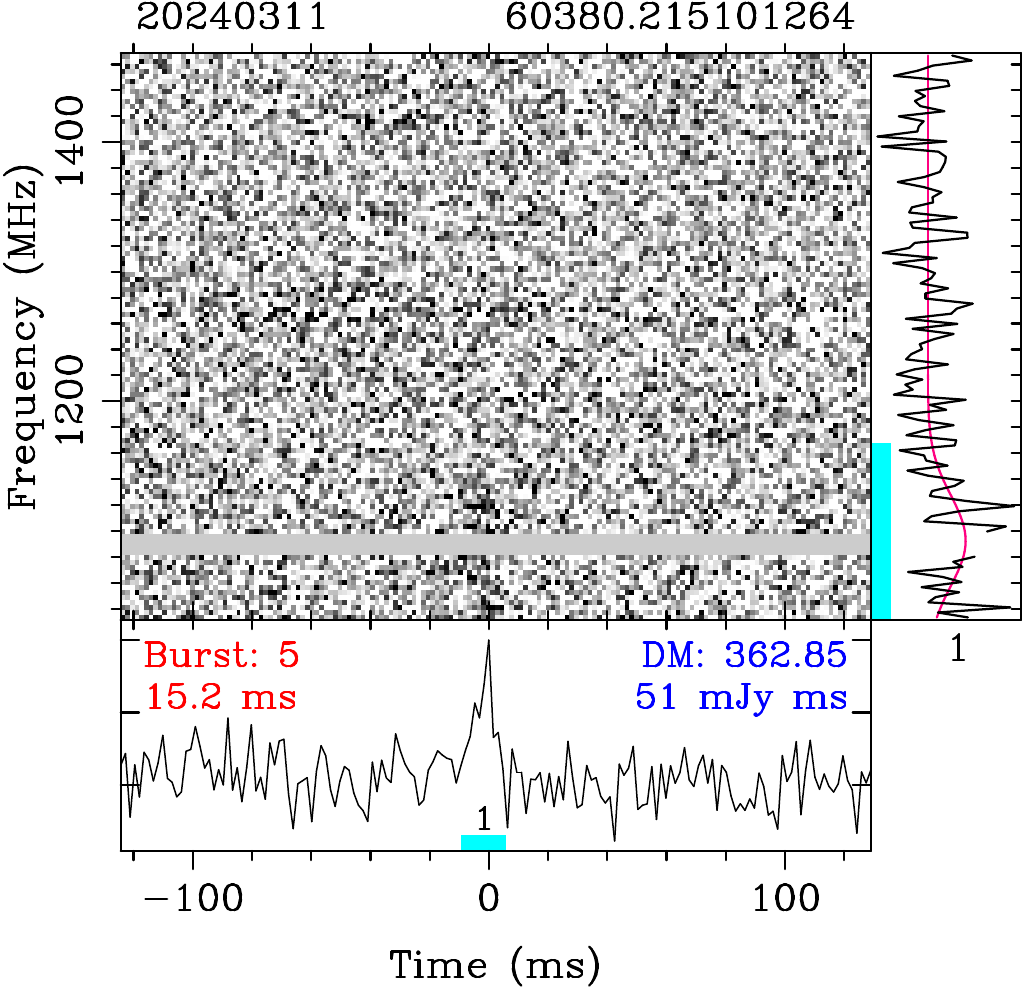}
\includegraphics[height=0.29\linewidth]{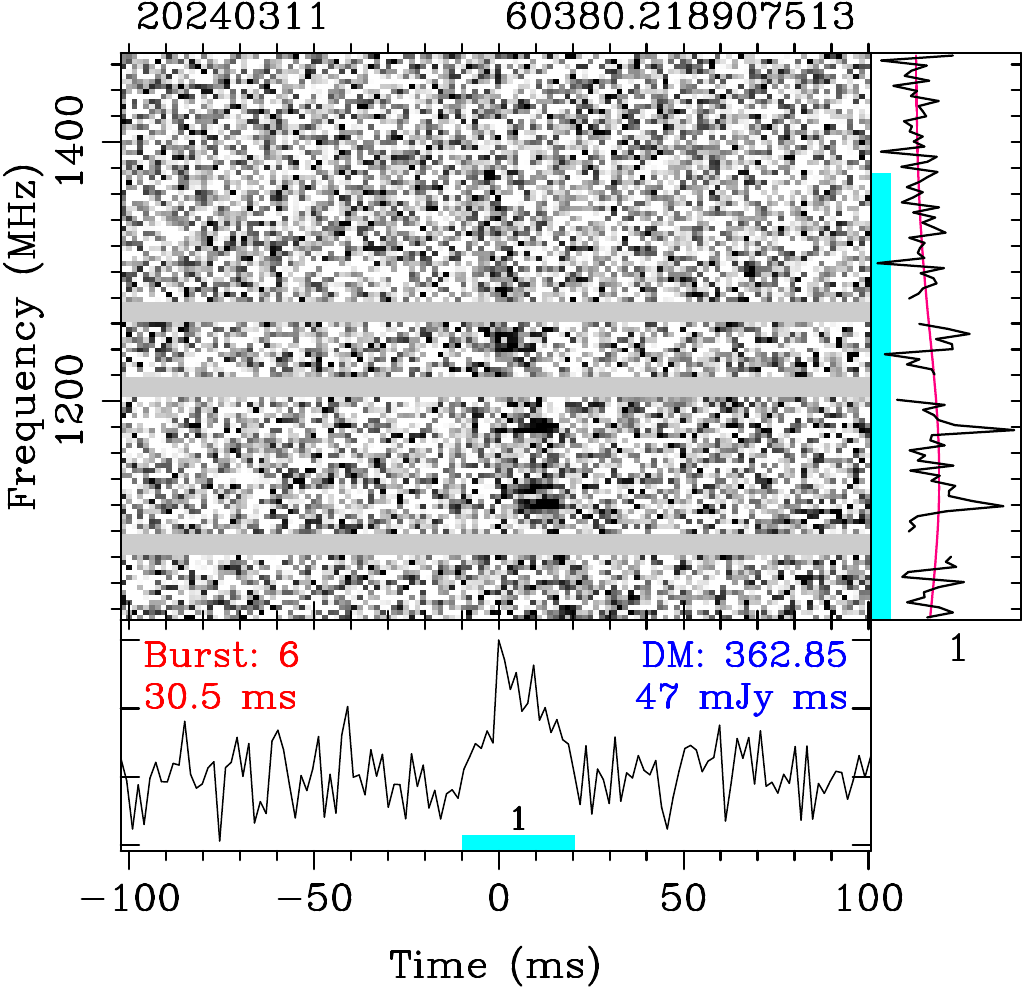}
\includegraphics[height=0.29\linewidth]{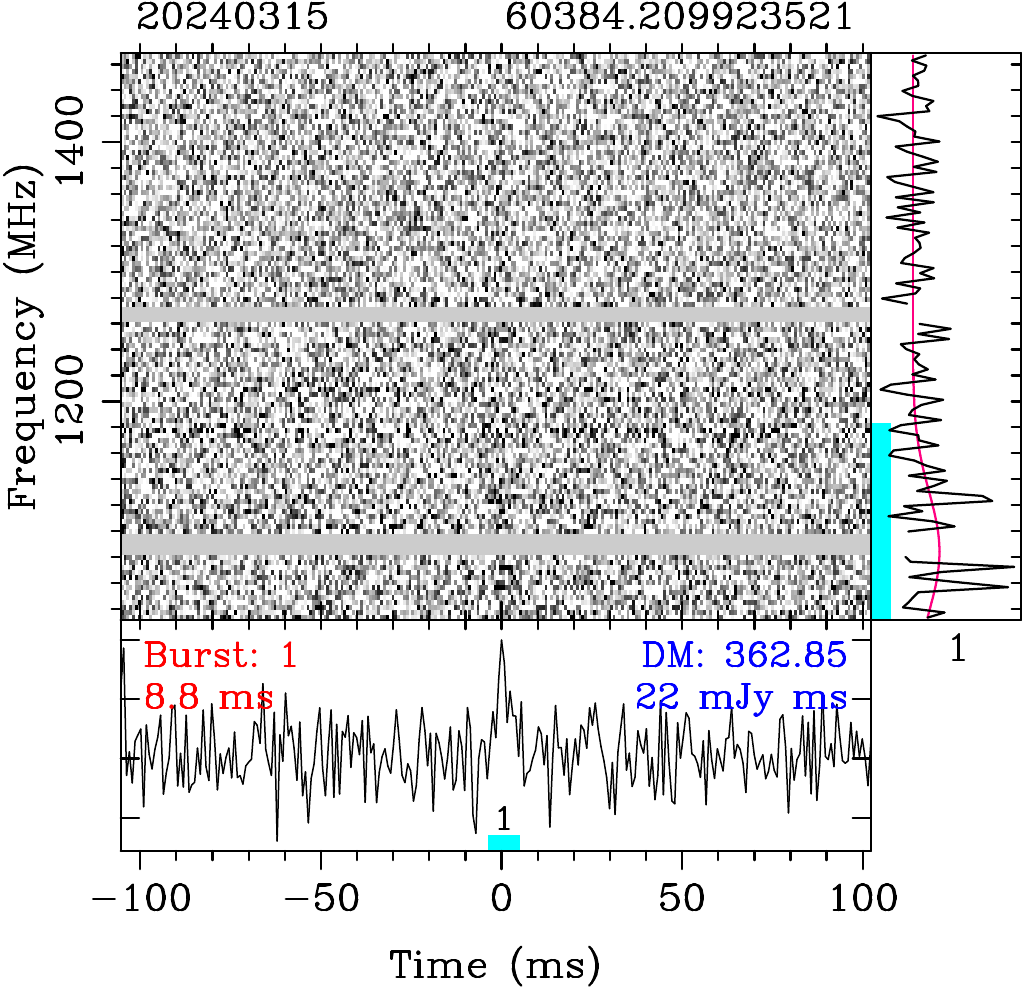}
\includegraphics[height=0.29\linewidth]{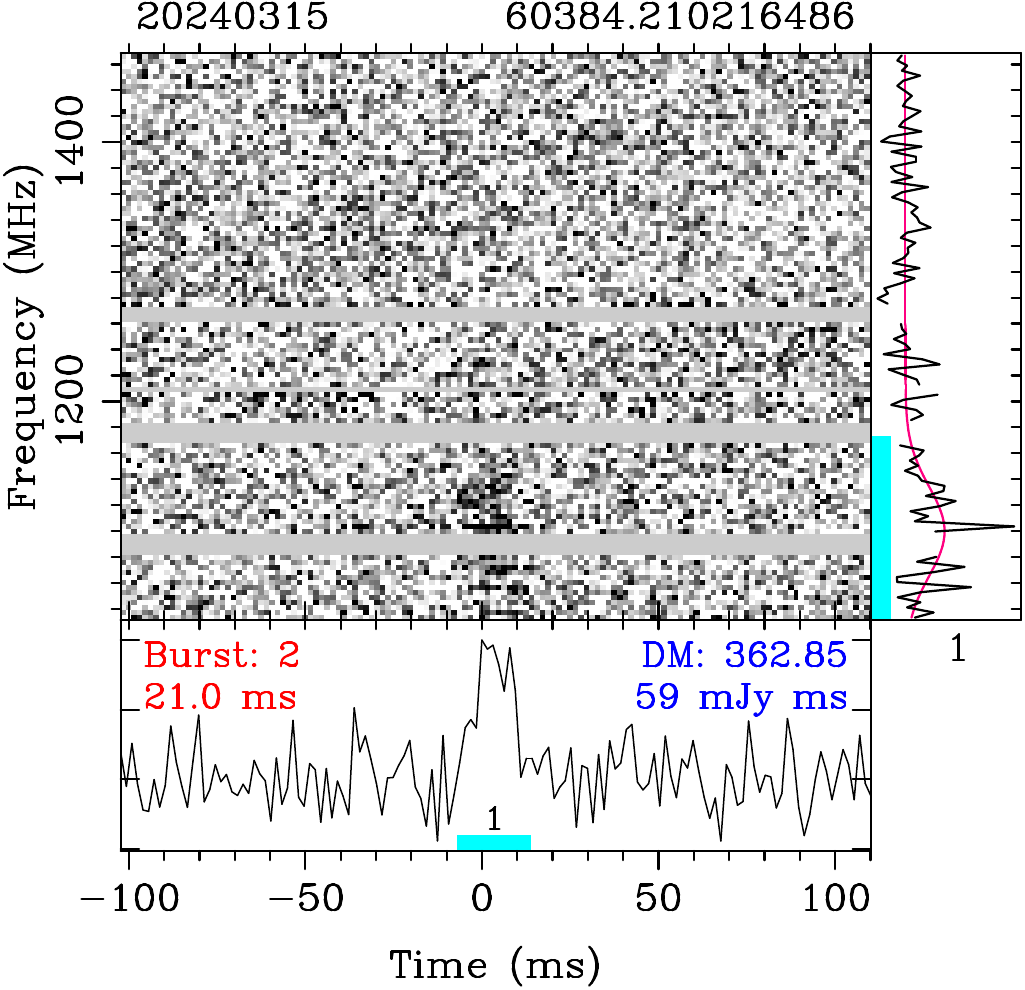}
\includegraphics[height=0.29\linewidth]{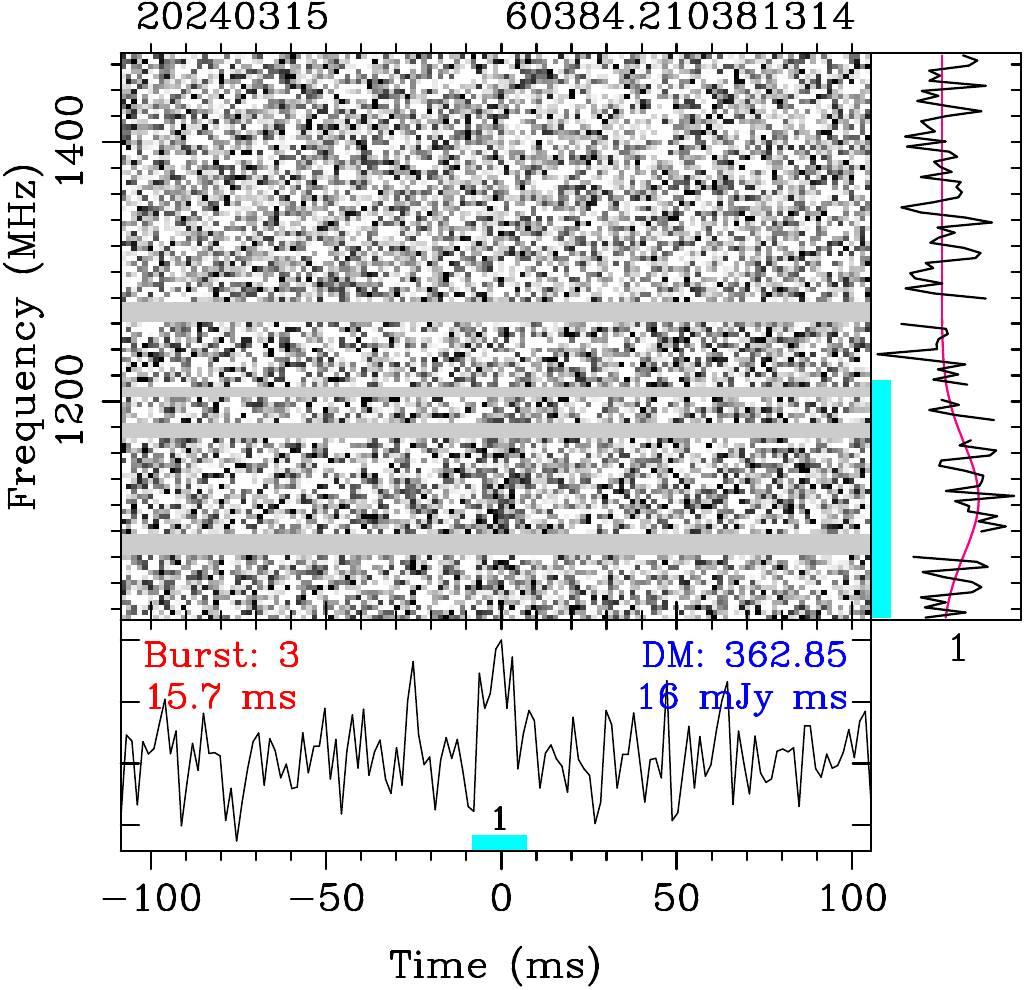}
\caption{({\textit{continued}})}
\end{figure*}
\addtocounter{figure}{-1}
\begin{figure*}
\flushleft
\includegraphics[height=0.29\linewidth]{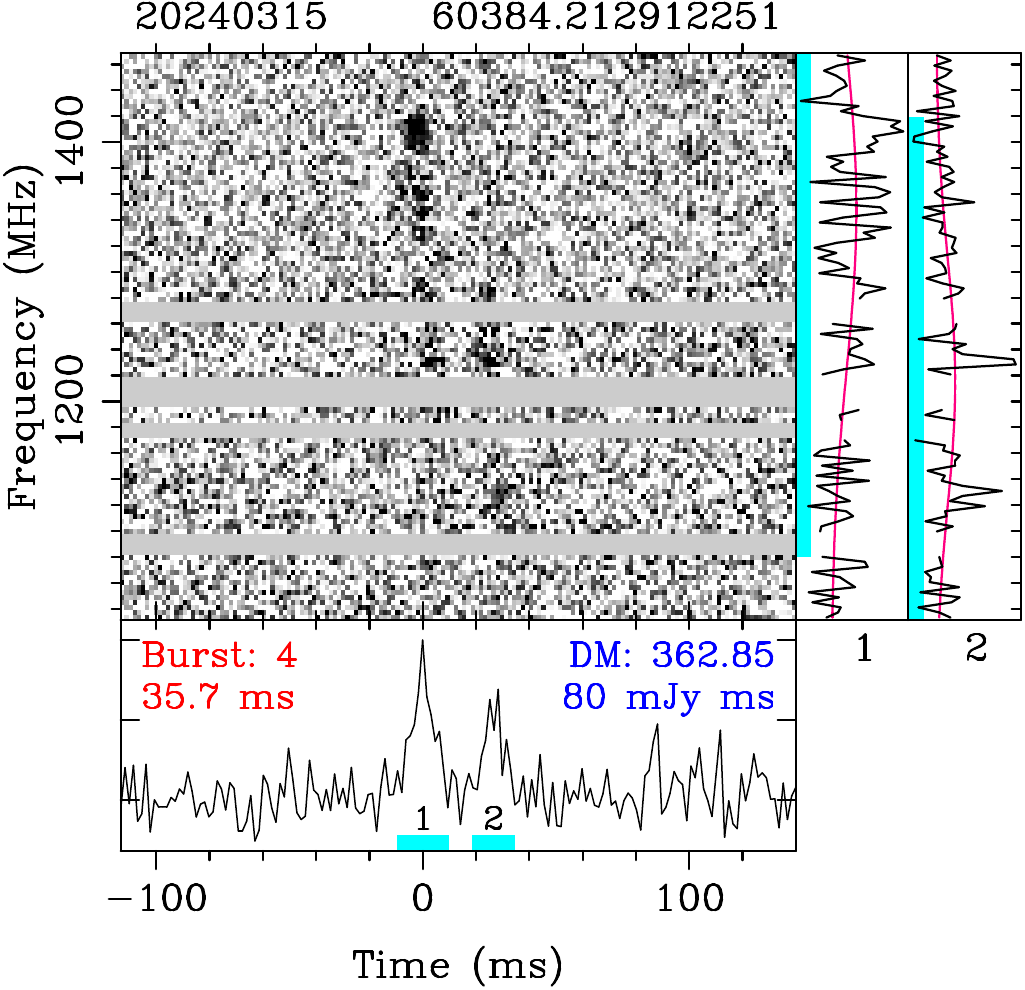}
\includegraphics[height=0.29\linewidth]{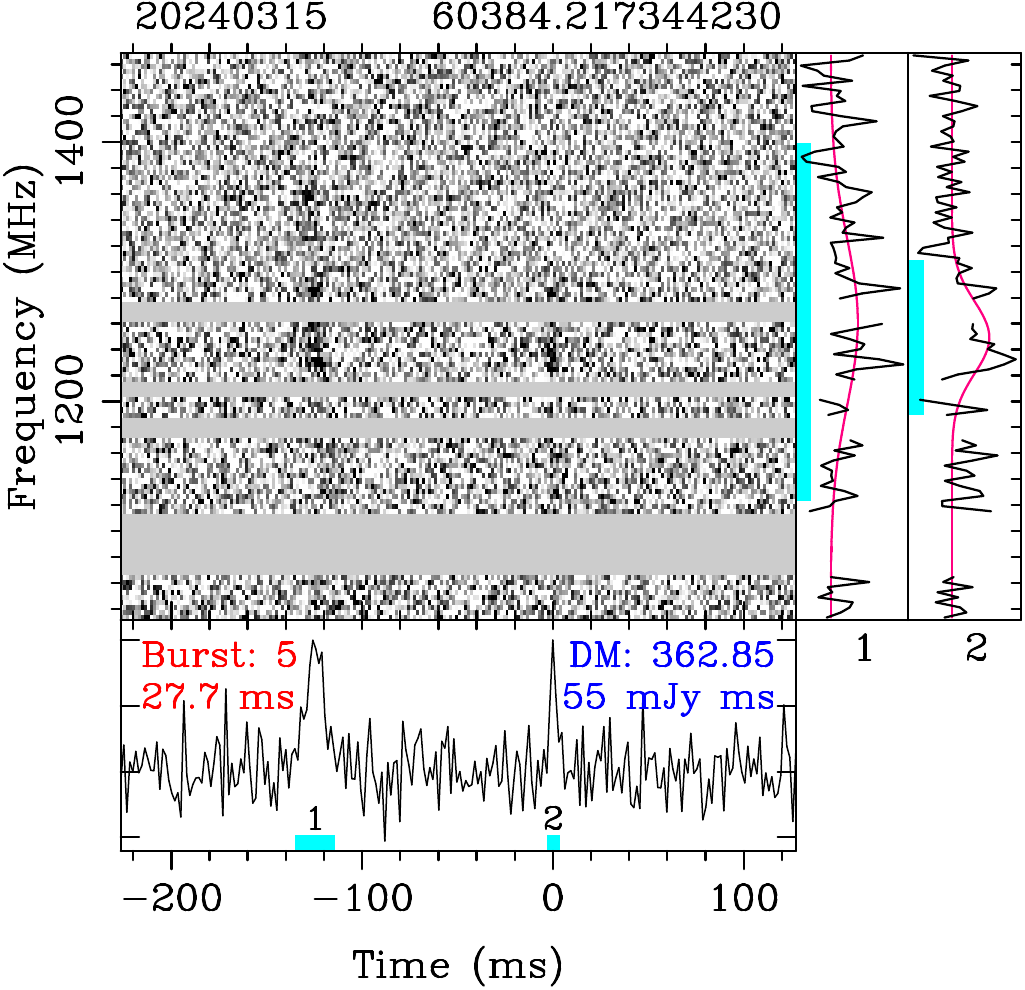}
\includegraphics[height=0.29\linewidth]{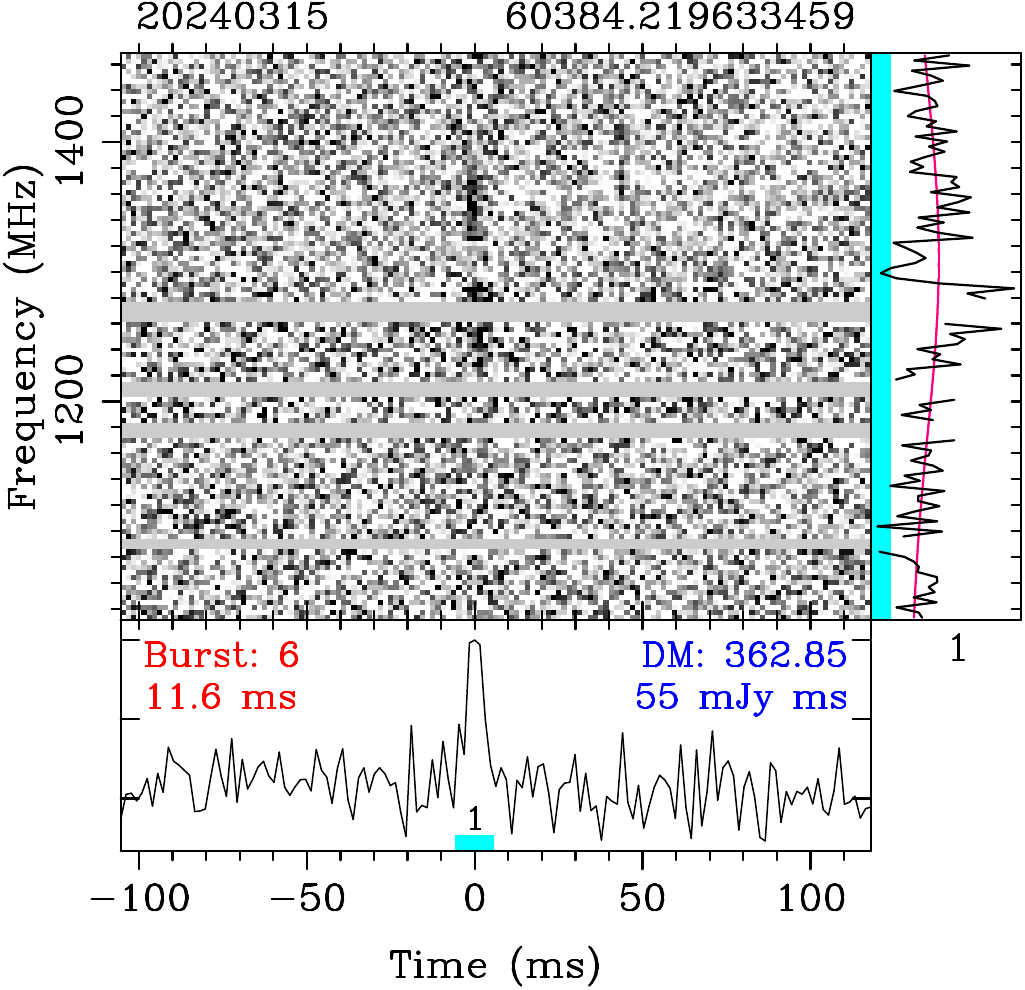}
\includegraphics[height=0.29\linewidth]{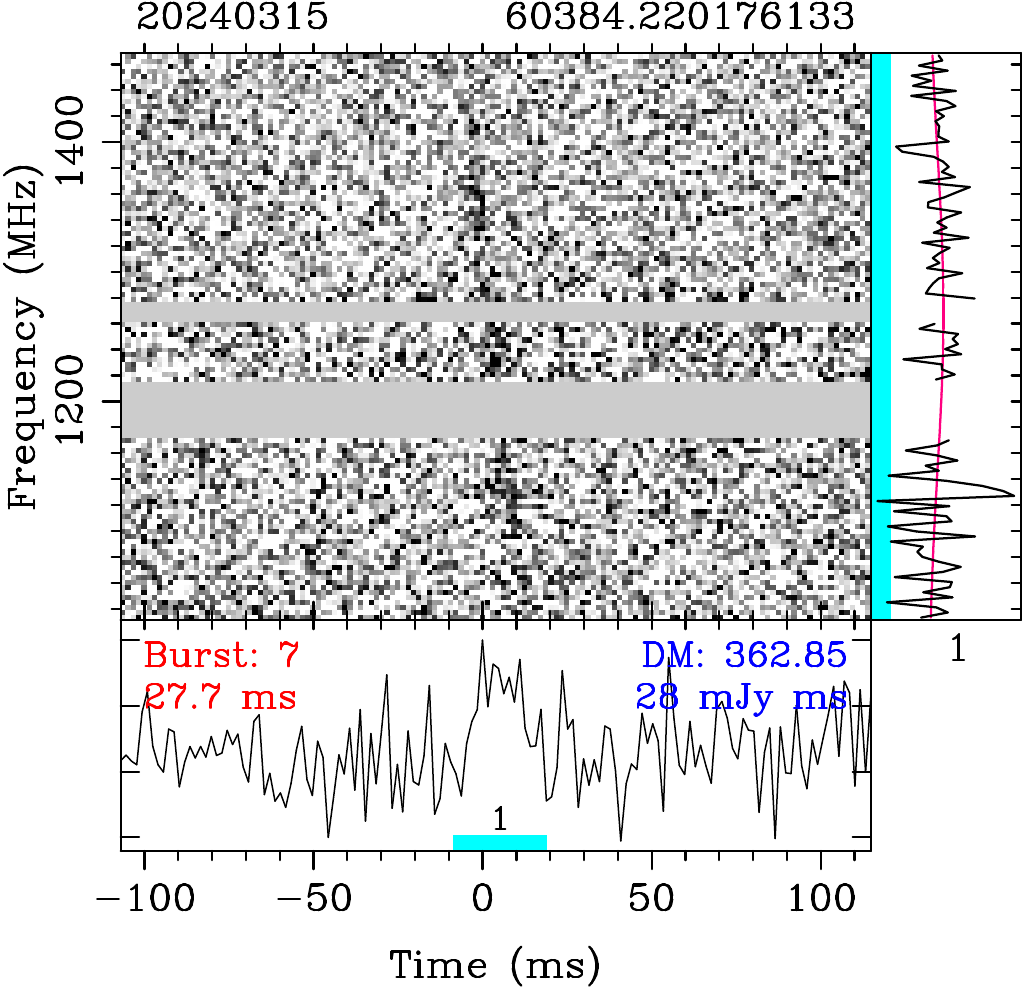}
\includegraphics[height=0.29\linewidth]{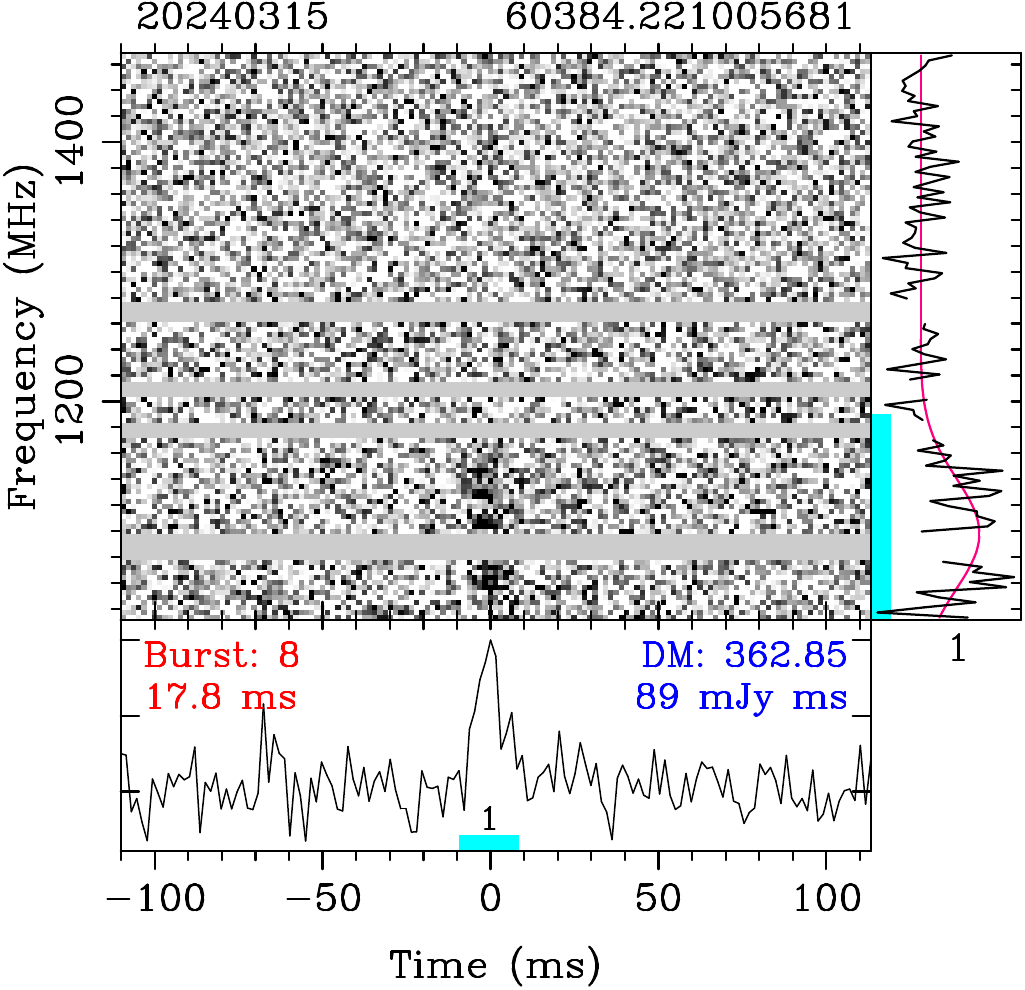}
\includegraphics[height=0.29\linewidth]{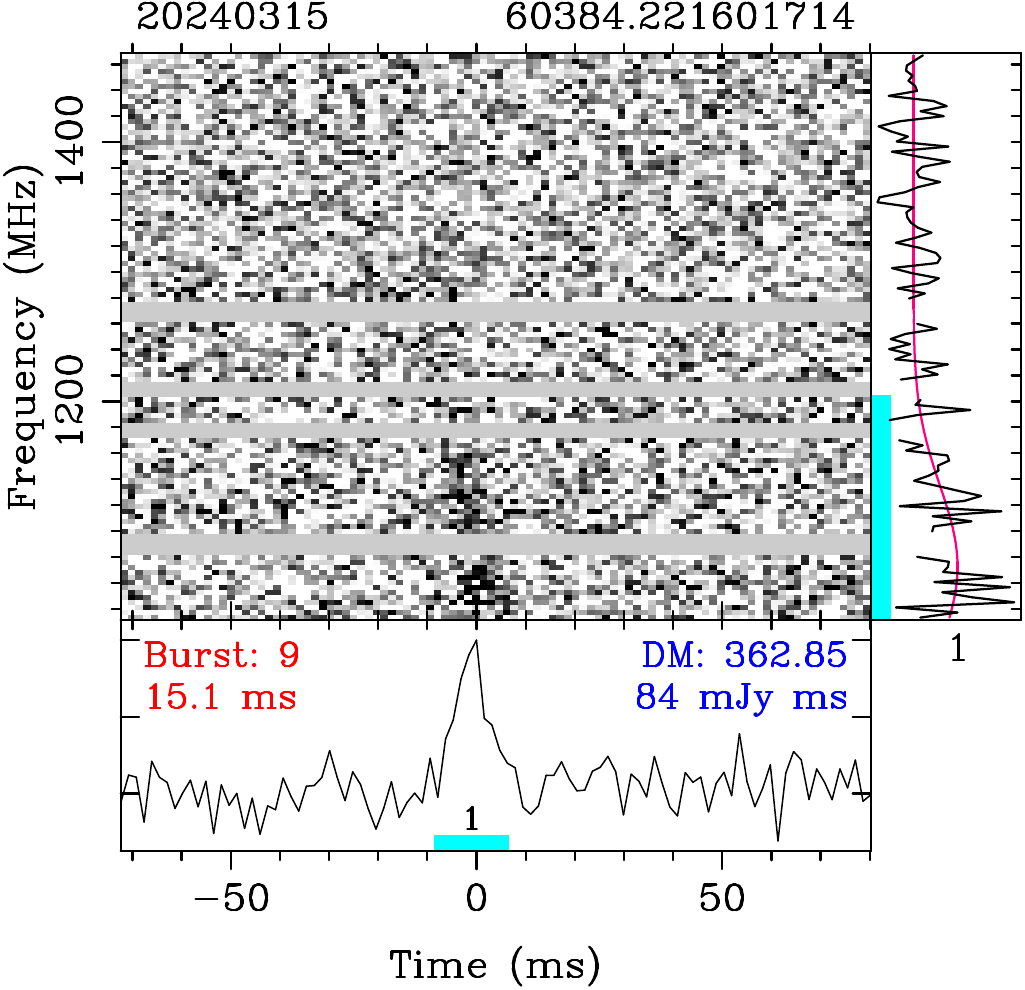}
\includegraphics[height=0.29\linewidth]{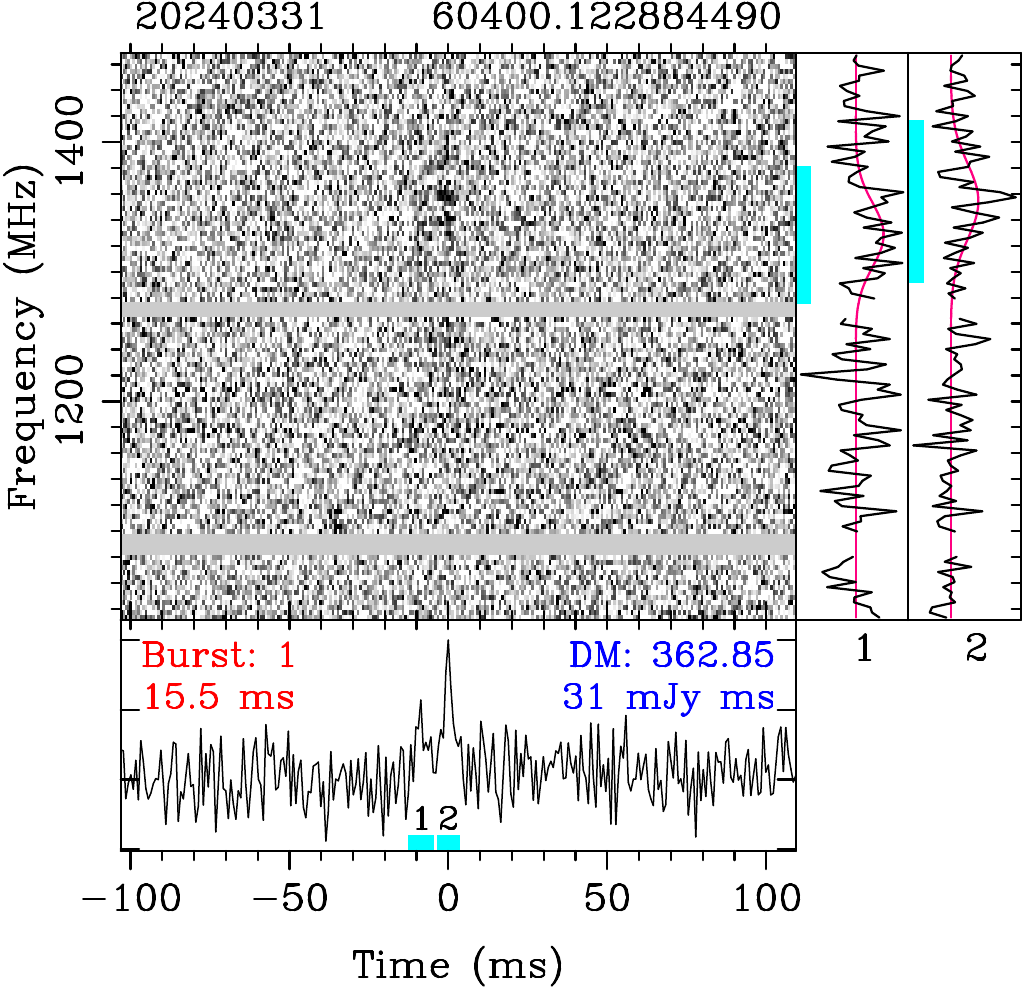}
\includegraphics[height=0.29\linewidth]{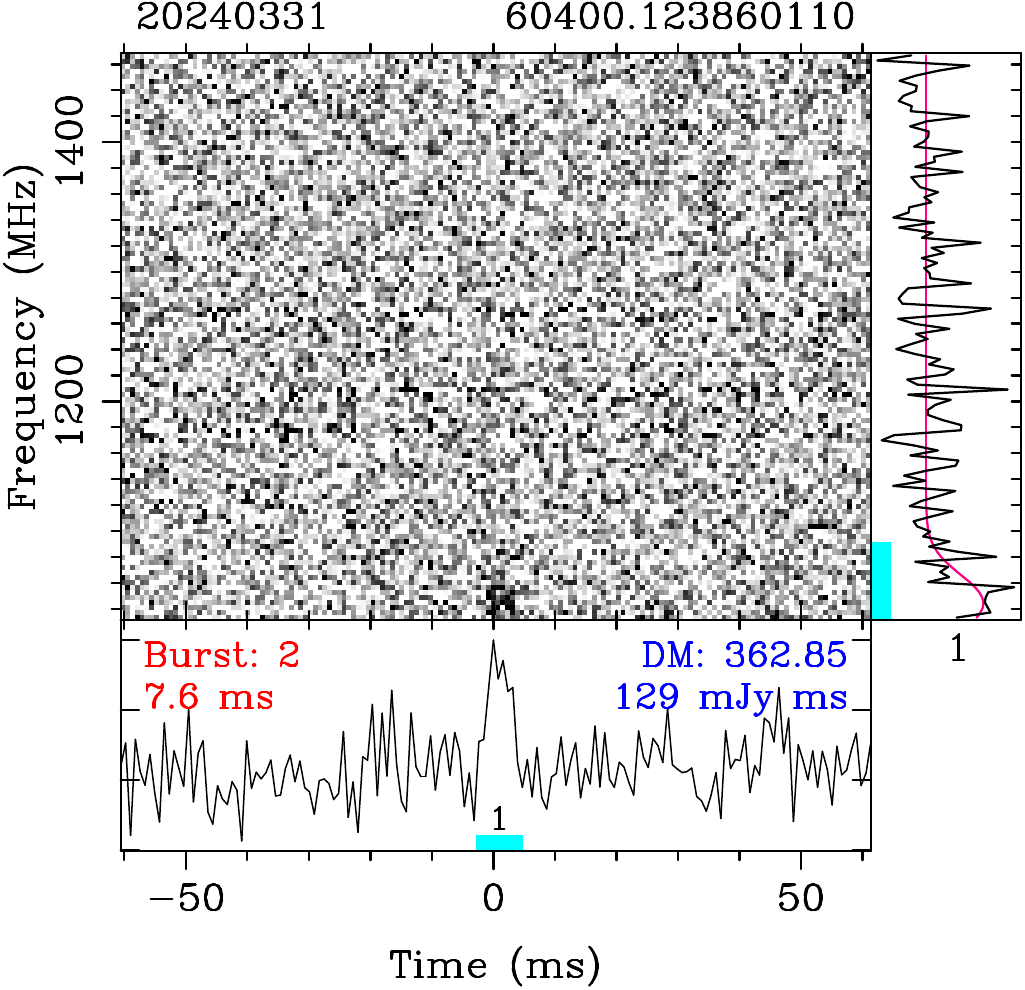}
\includegraphics[height=0.29\linewidth]{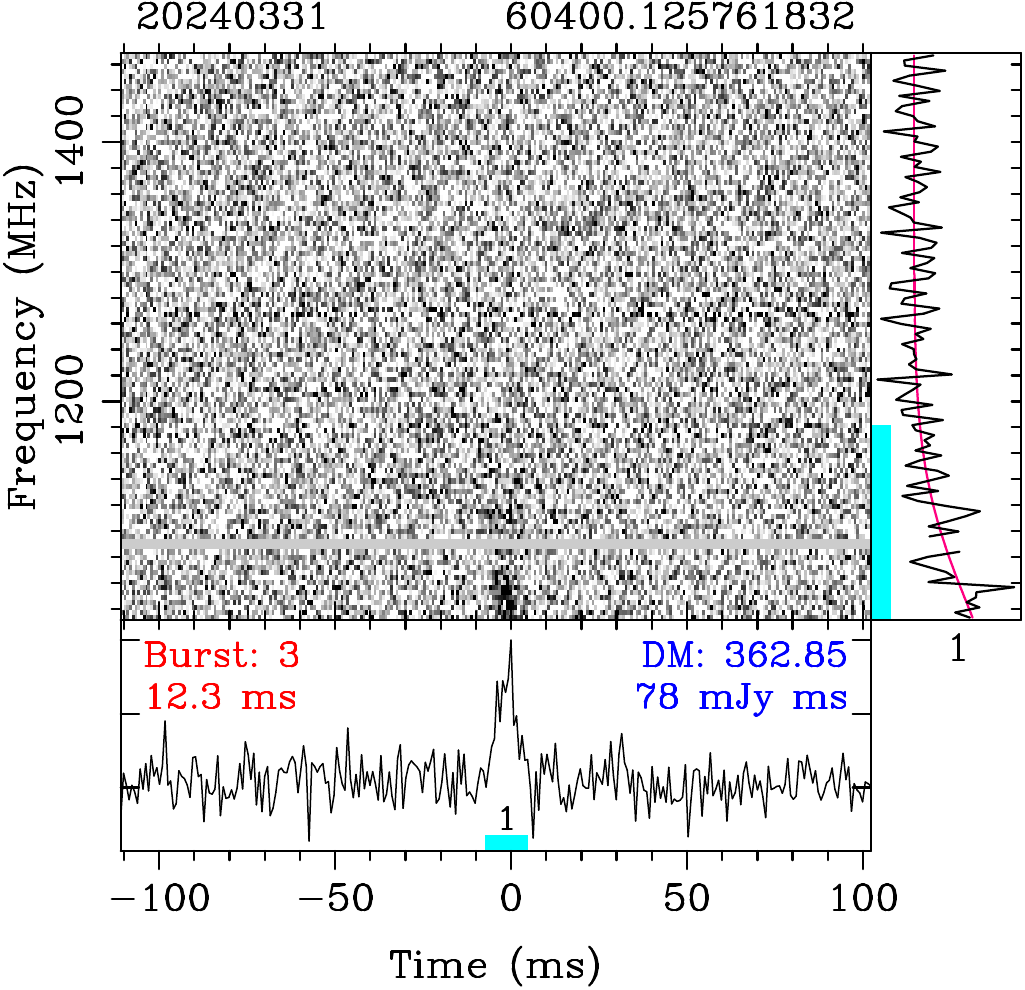}
\includegraphics[height=0.29\linewidth]{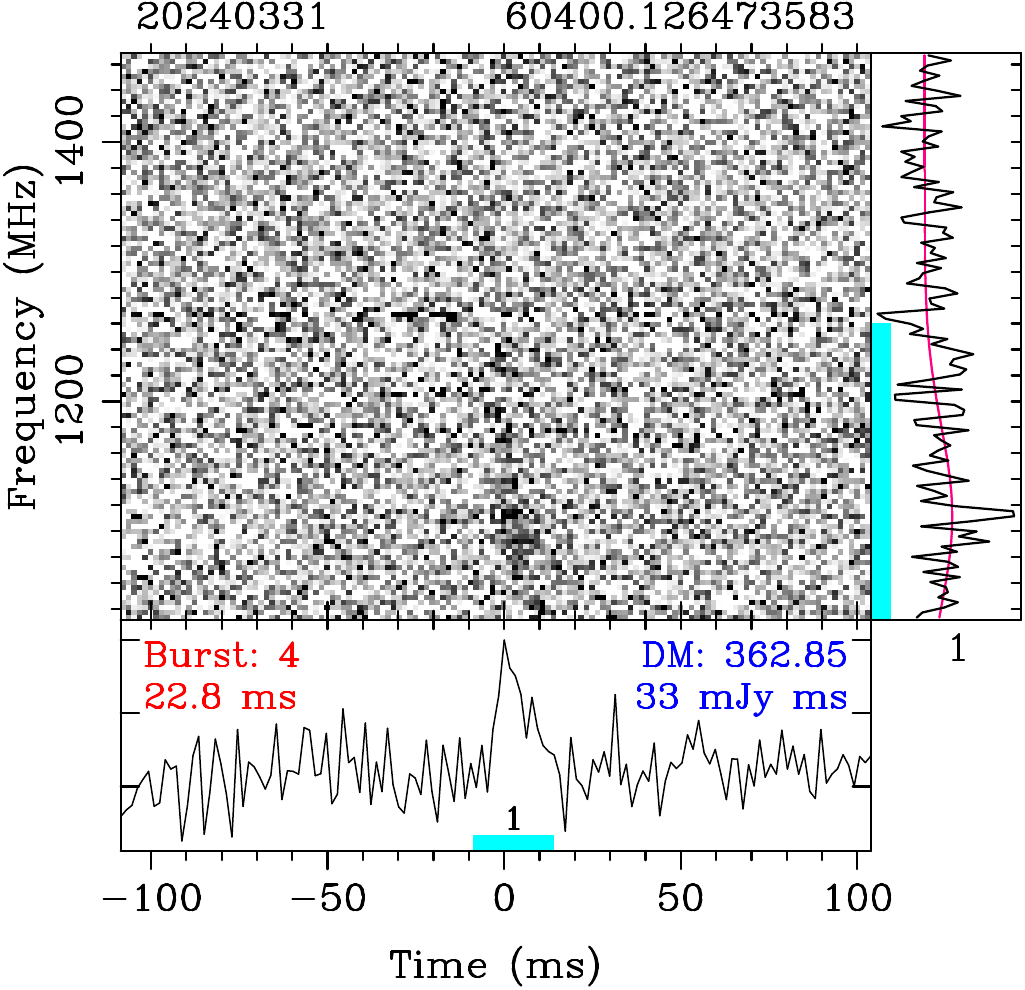}
\includegraphics[height=0.29\linewidth]{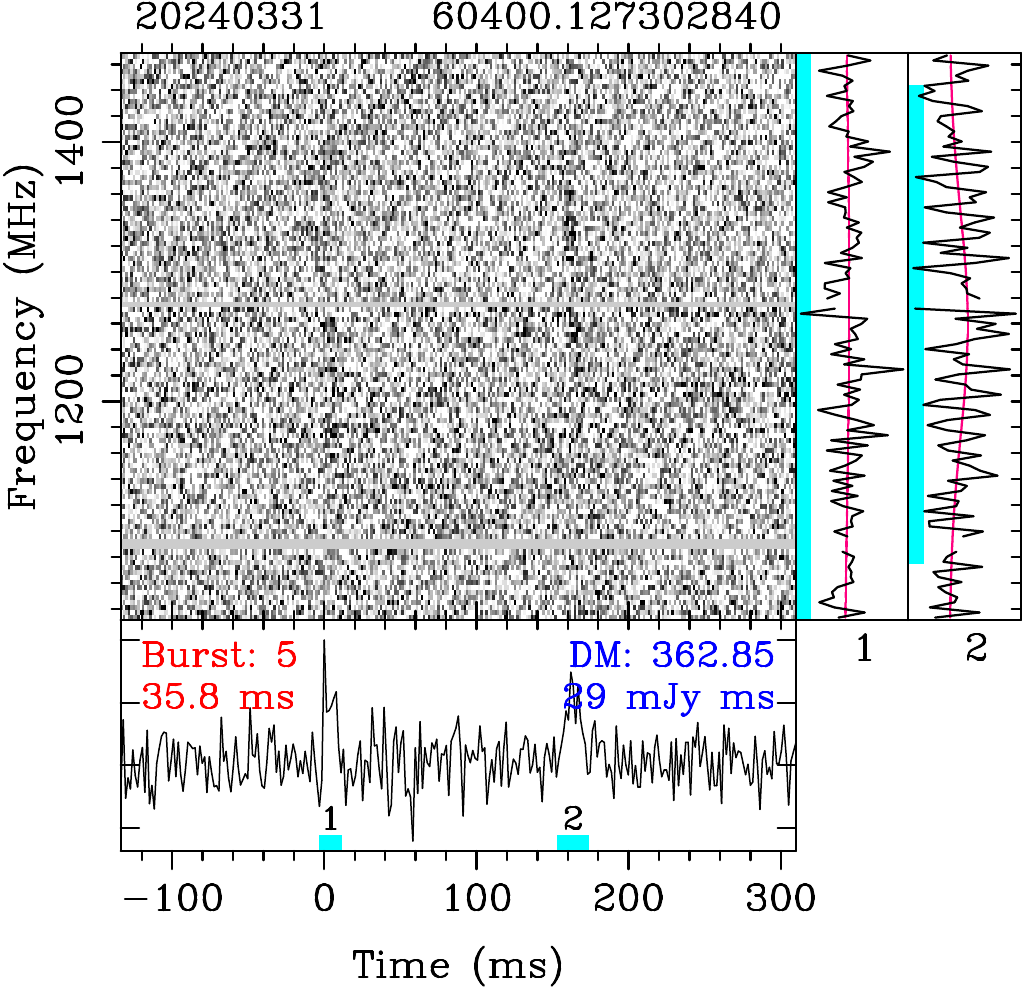}
\includegraphics[height=0.29\linewidth]{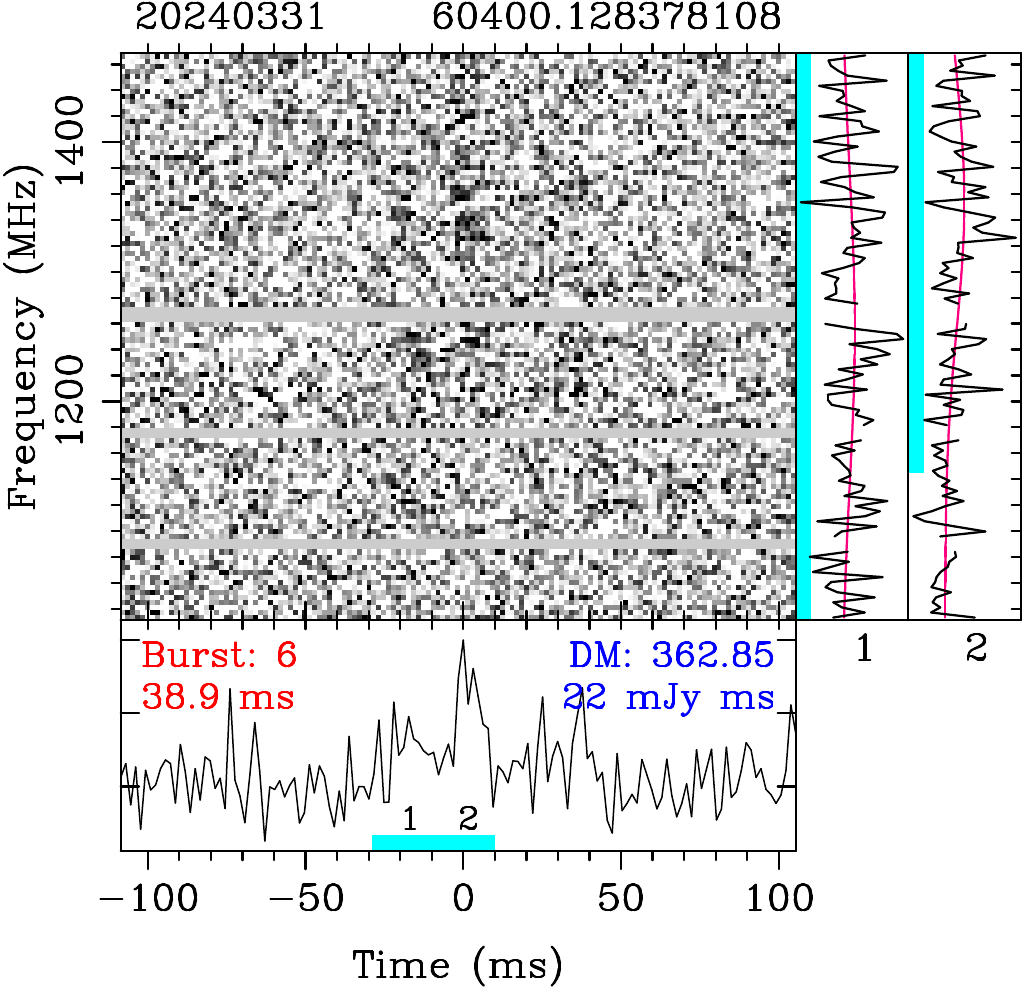}
\caption{({\textit{continued}})}
\end{figure*}
\addtocounter{figure}{-1}
\begin{figure*}
\flushleft
\includegraphics[height=0.29\linewidth]{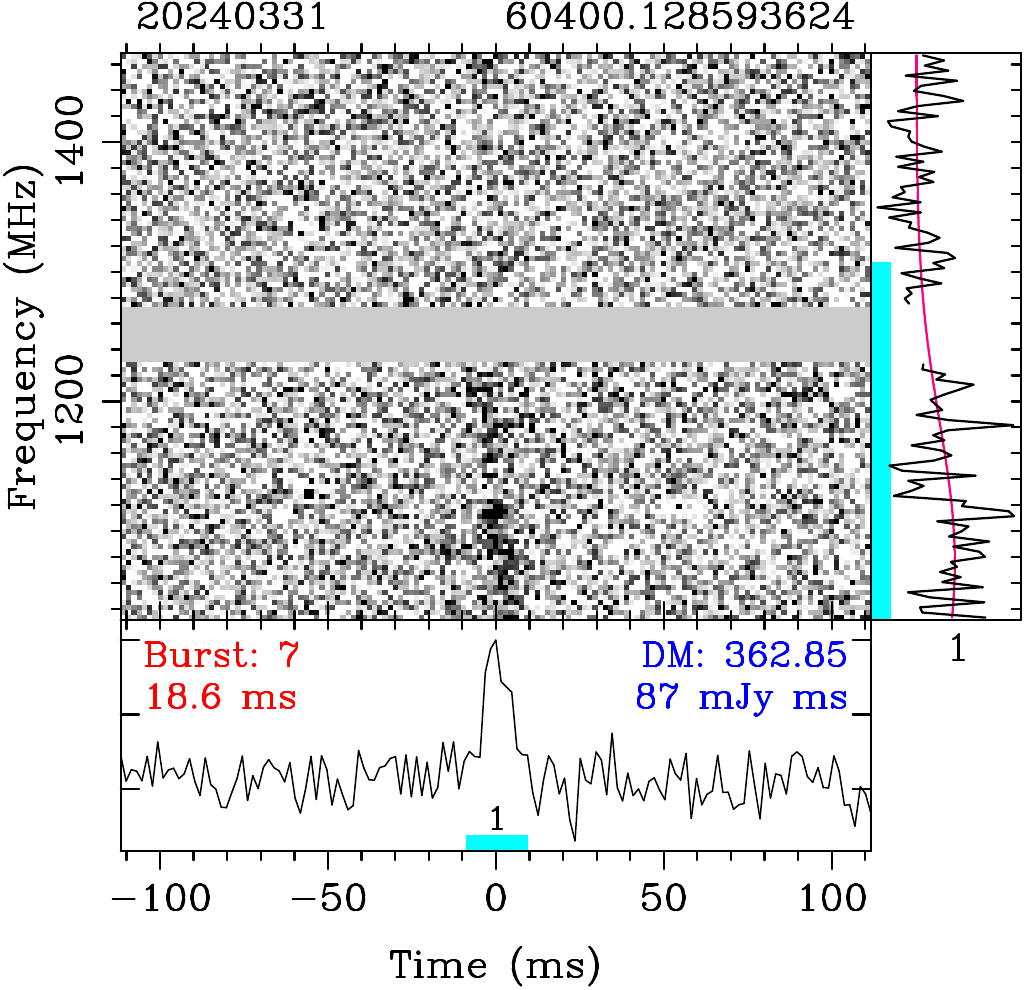}
\includegraphics[height=0.29\linewidth]{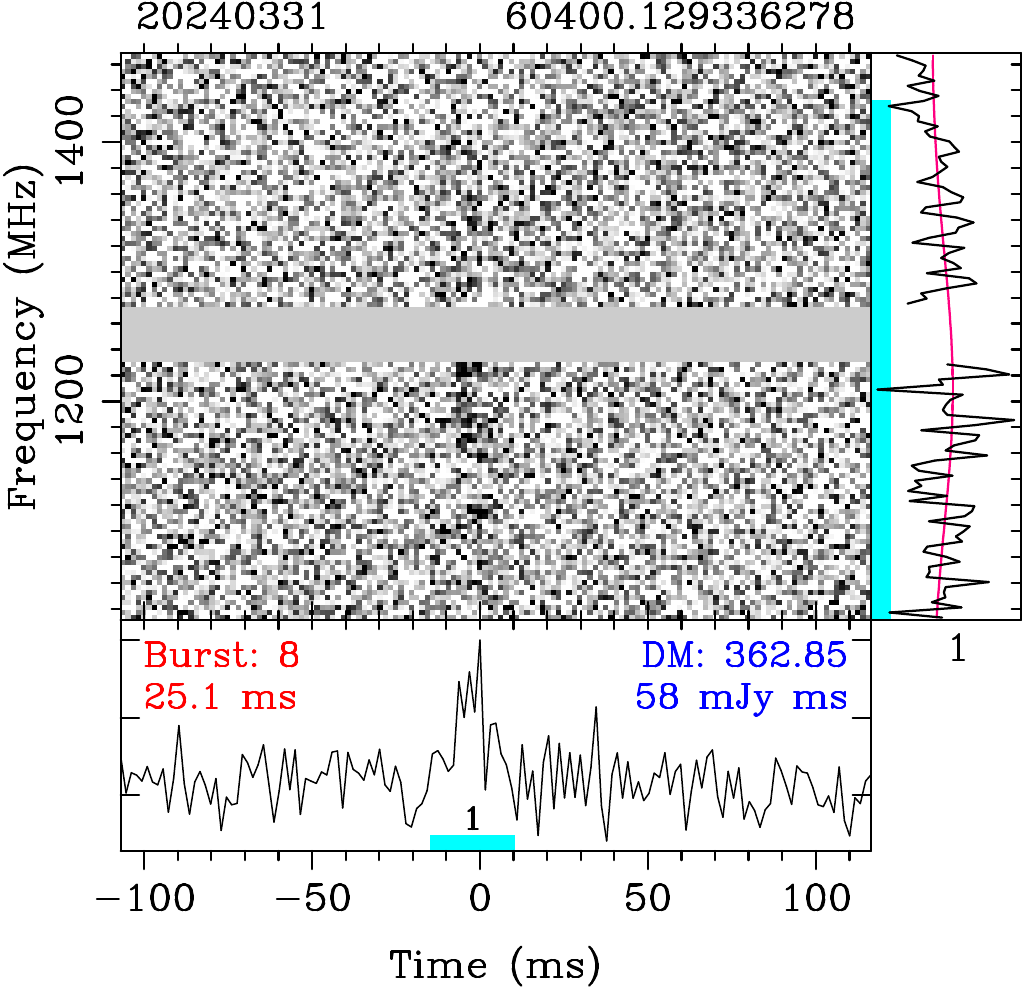}
\includegraphics[height=0.29\linewidth]{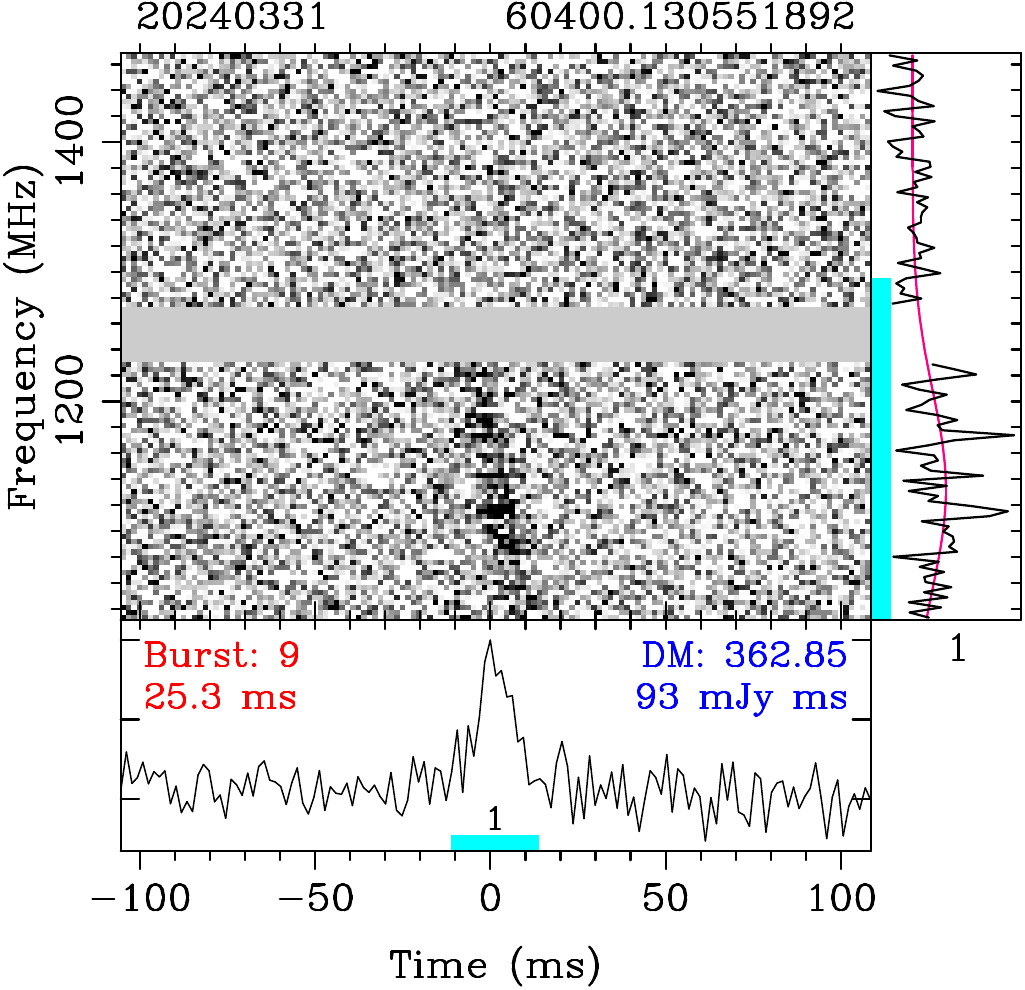}
\includegraphics[height=0.29\linewidth]{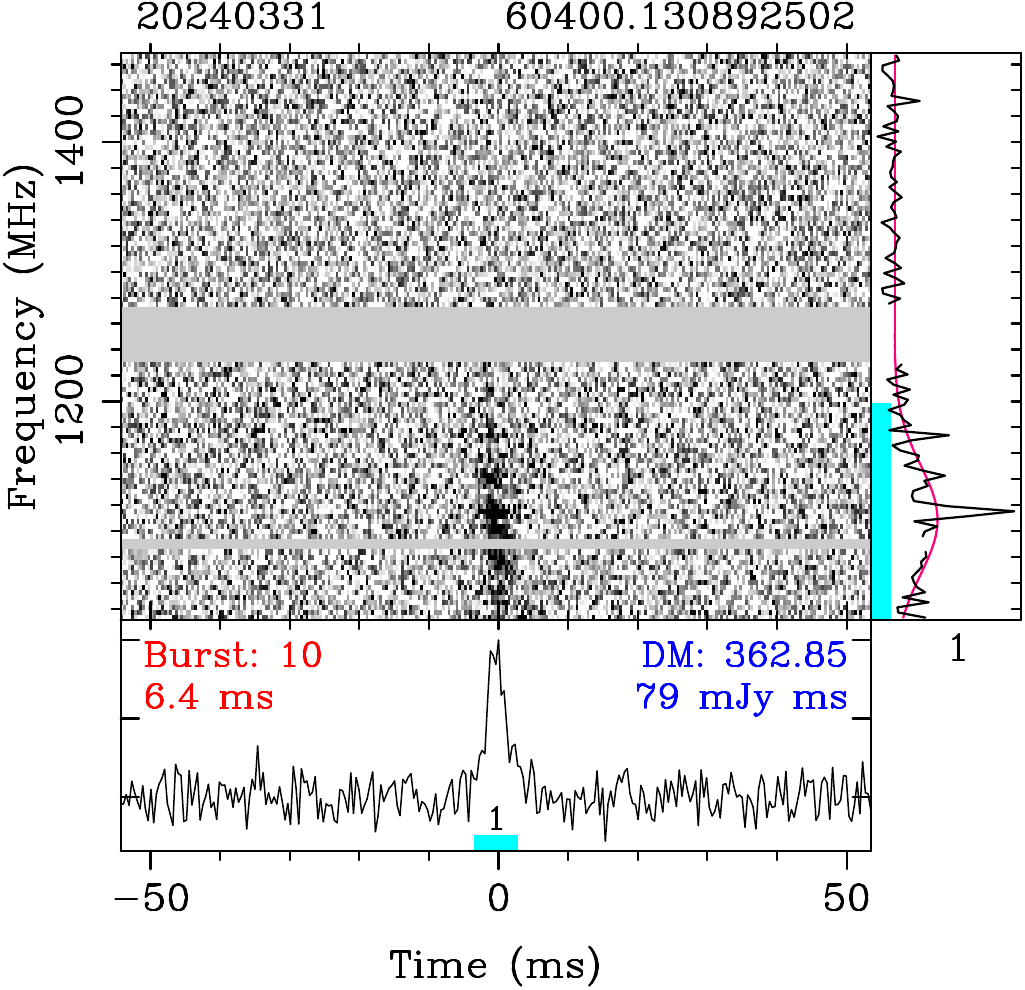}
\includegraphics[height=0.29\linewidth]{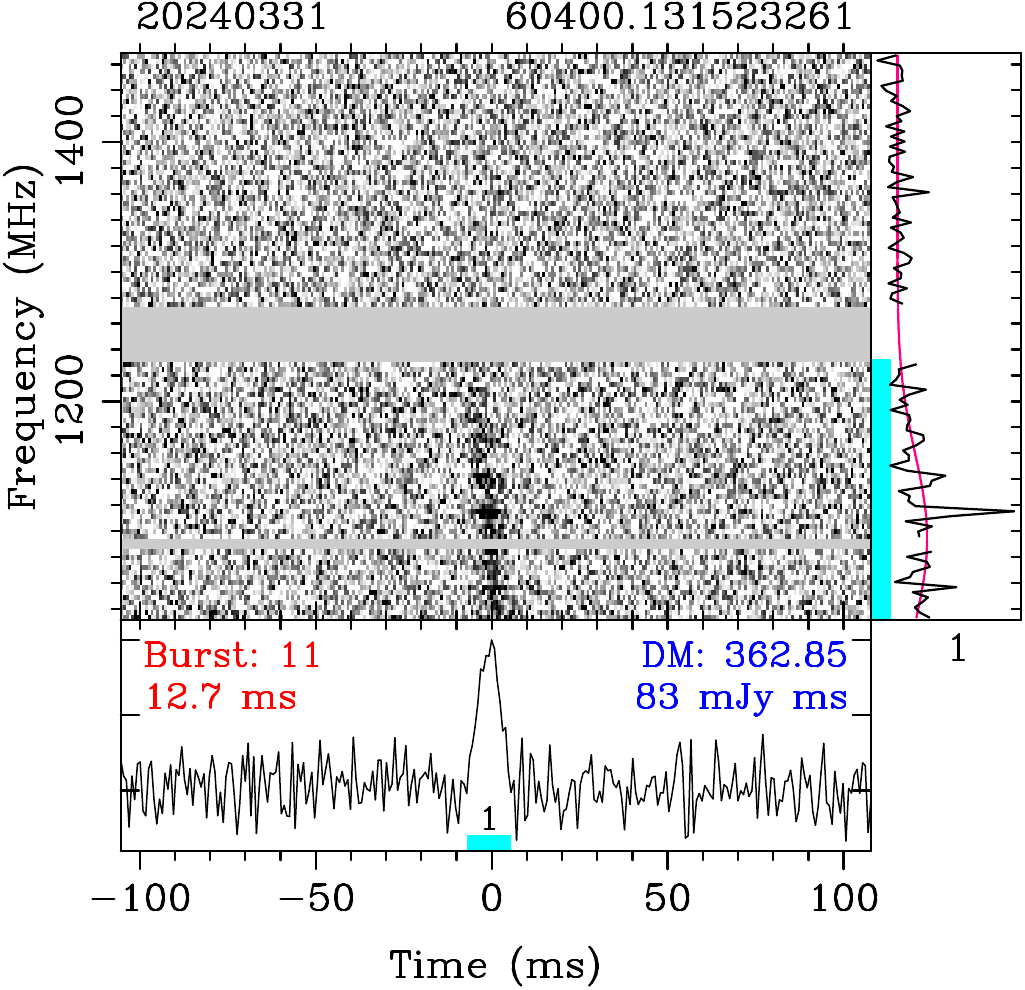}
\includegraphics[height=0.29\linewidth]{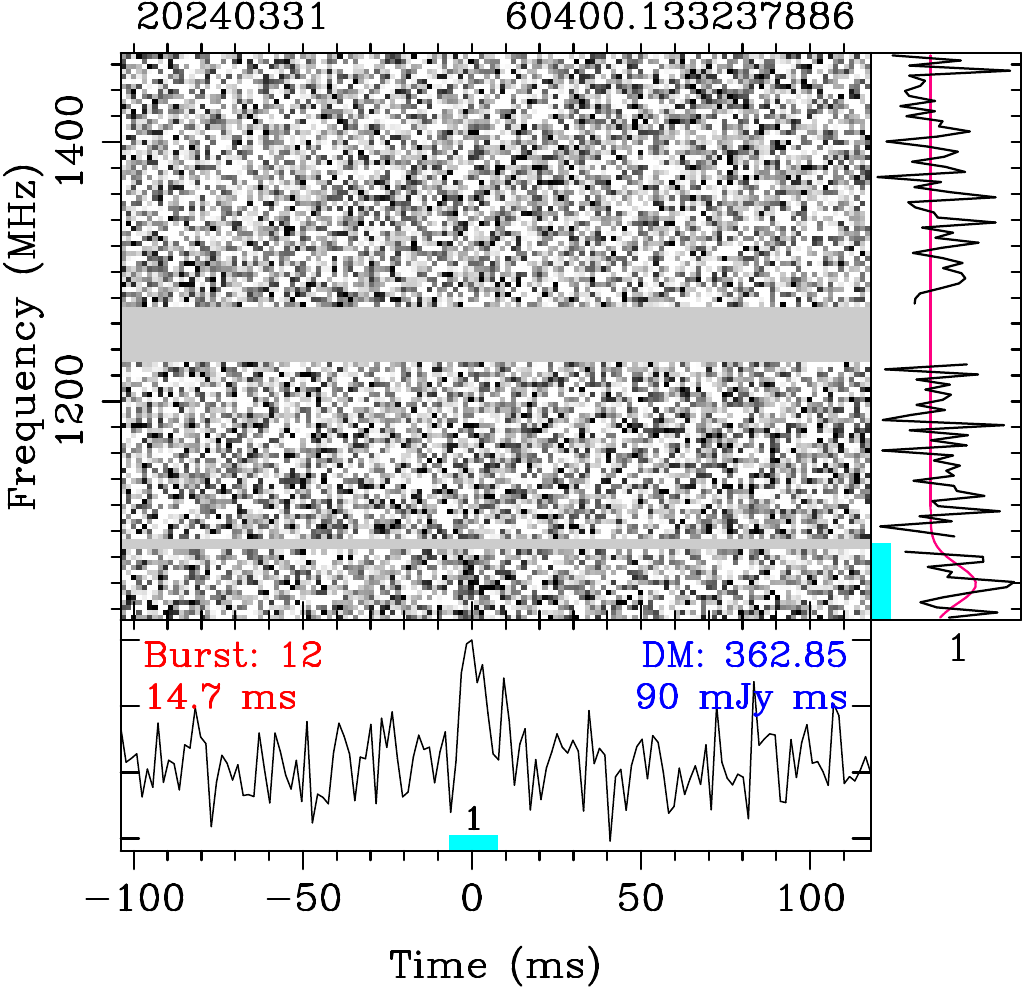}
\includegraphics[height=0.29\linewidth]{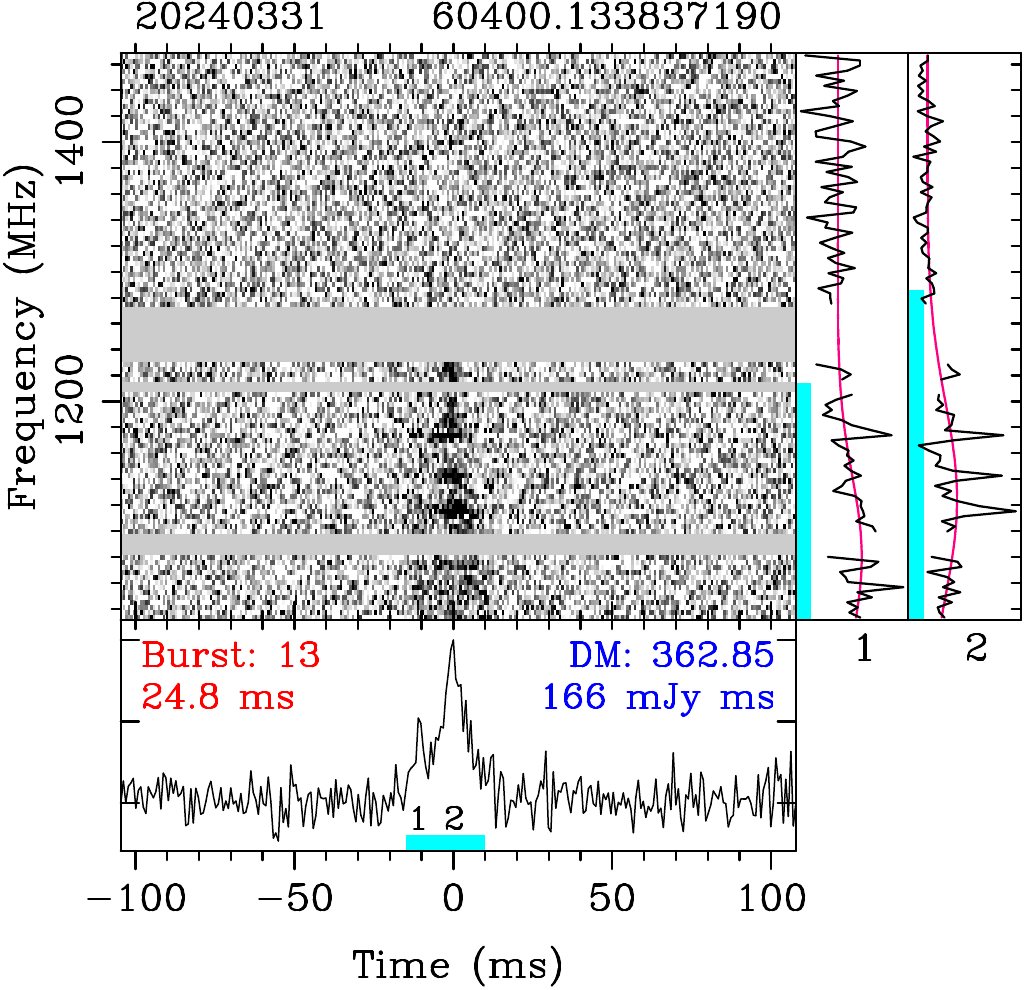}
\includegraphics[height=0.29\linewidth]{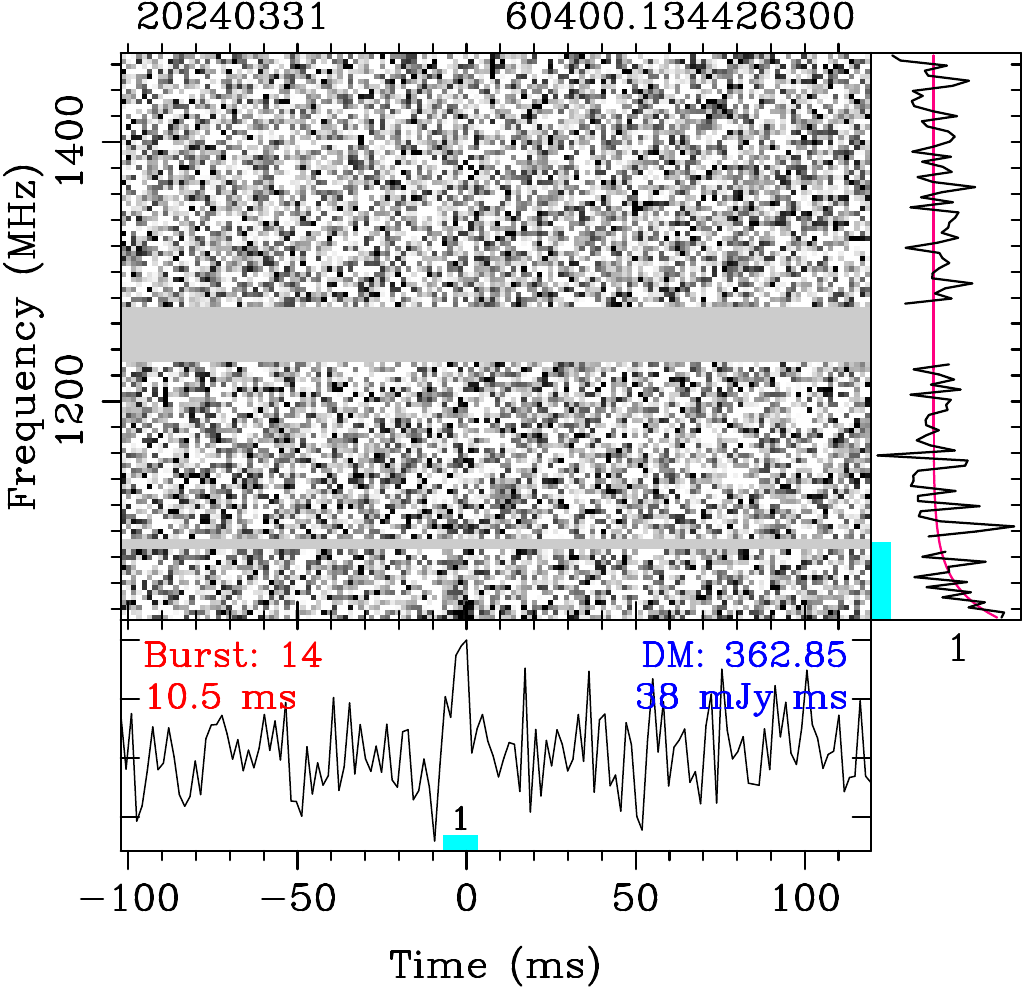}
\includegraphics[height=0.29\linewidth]{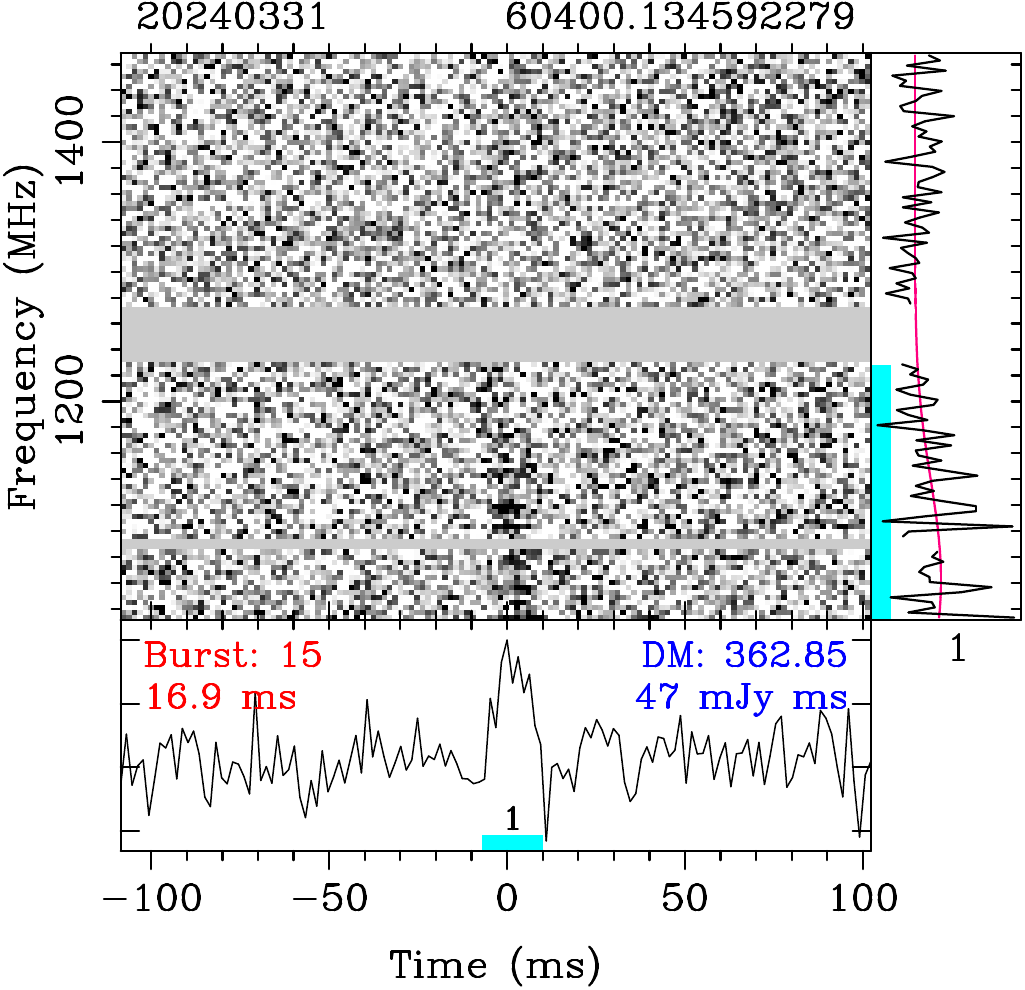}
\includegraphics[height=0.29\linewidth]{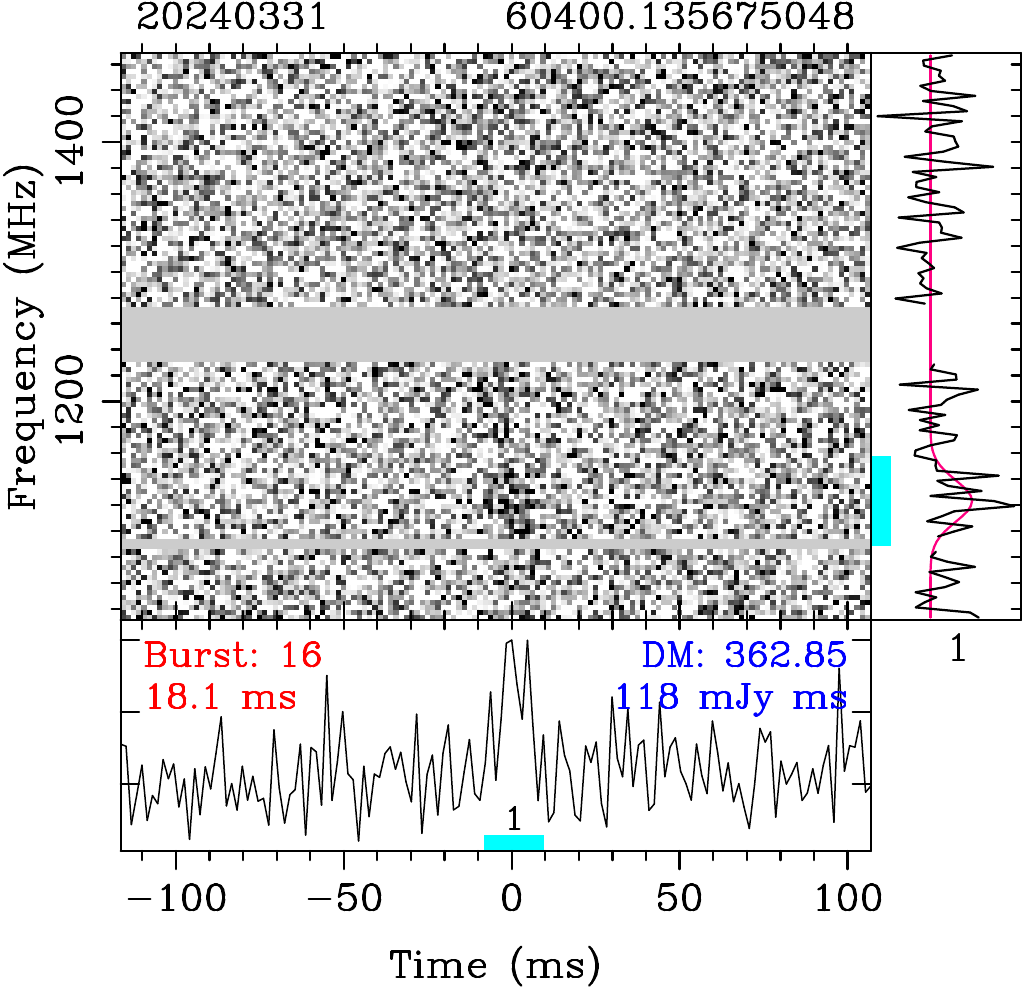}
\includegraphics[height=0.29\linewidth]{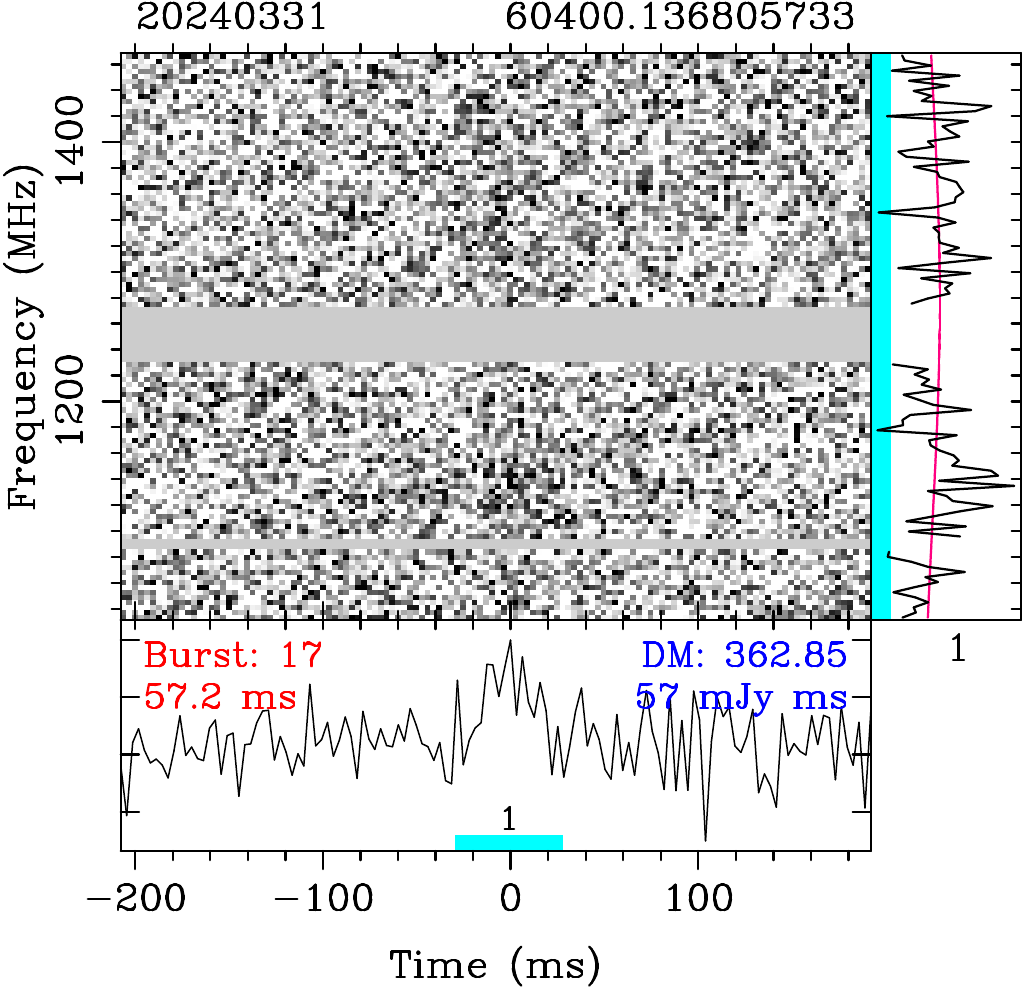}
\includegraphics[height=0.29\linewidth]{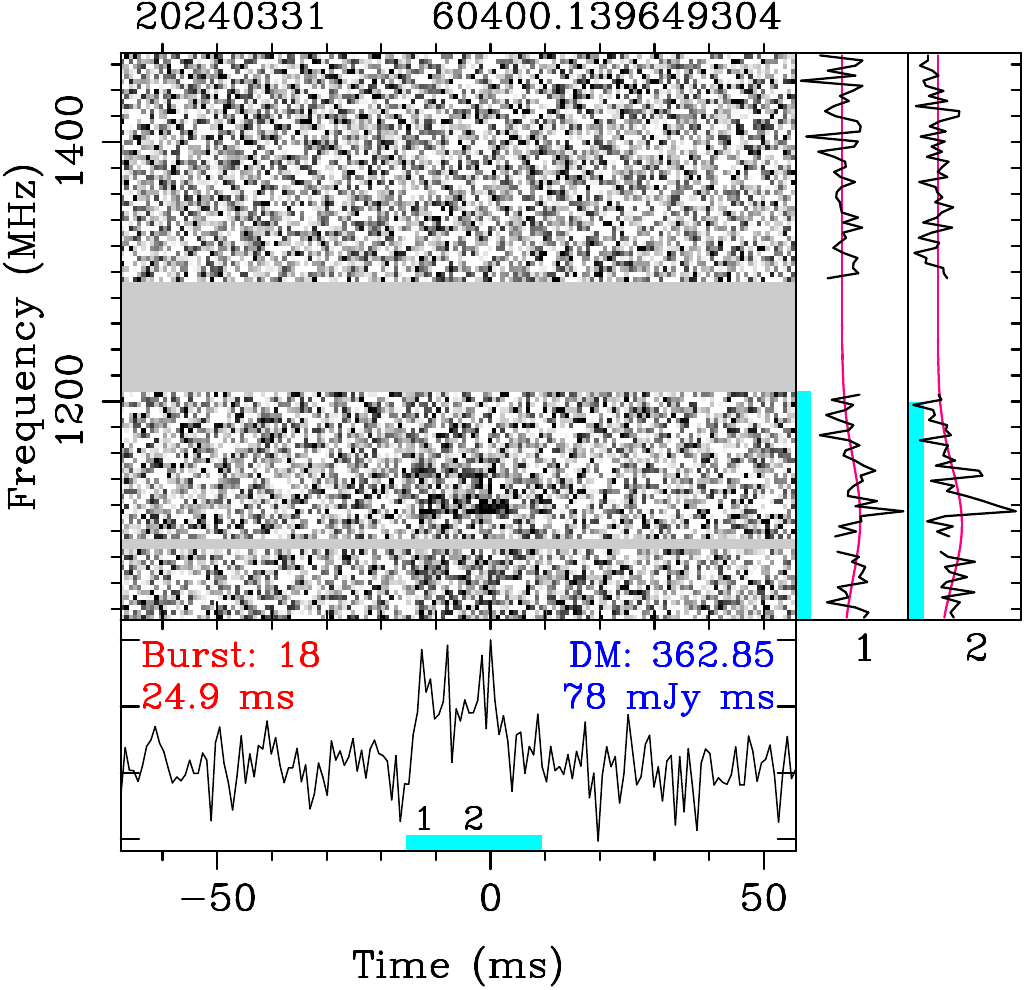}
\caption{({\textit{continued}})}
\end{figure*}
\addtocounter{figure}{-1}
\begin{figure*}
\flushleft
\includegraphics[height=0.29\linewidth]{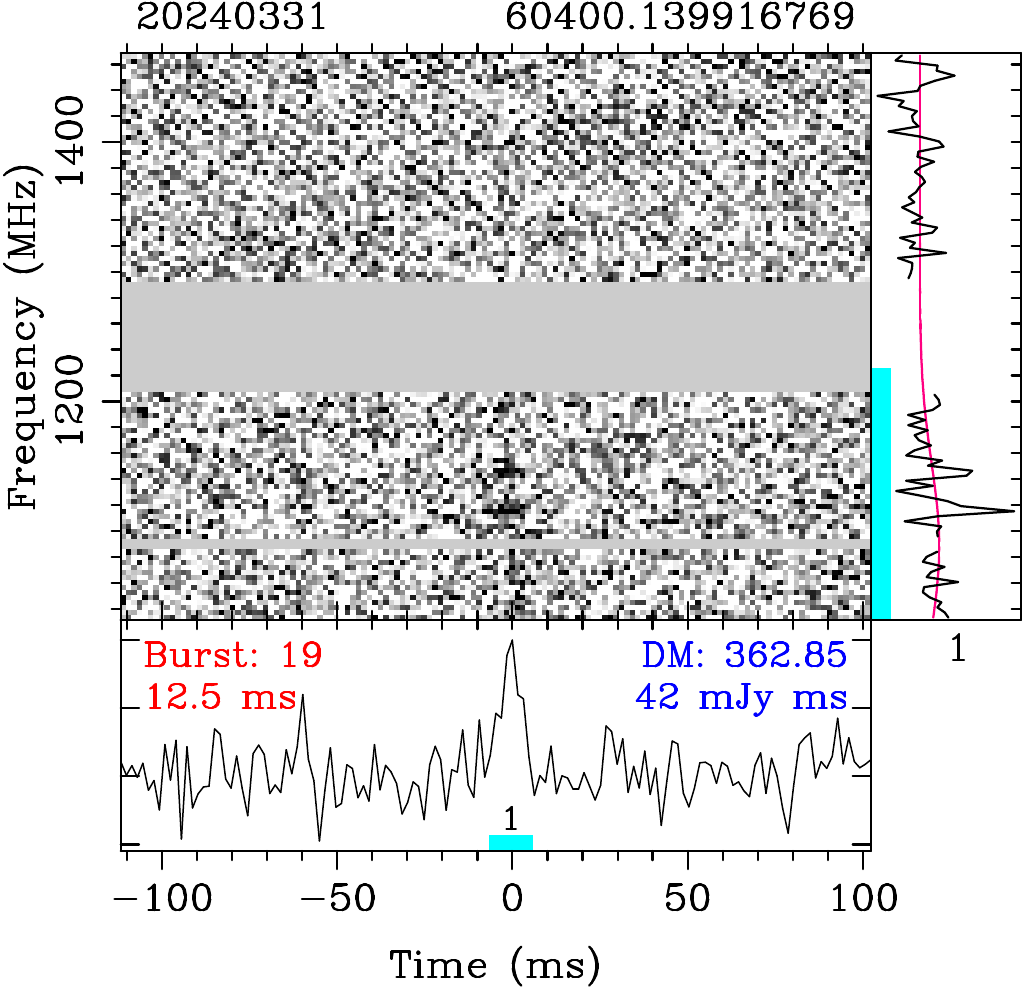}
\includegraphics[height=0.29\linewidth]{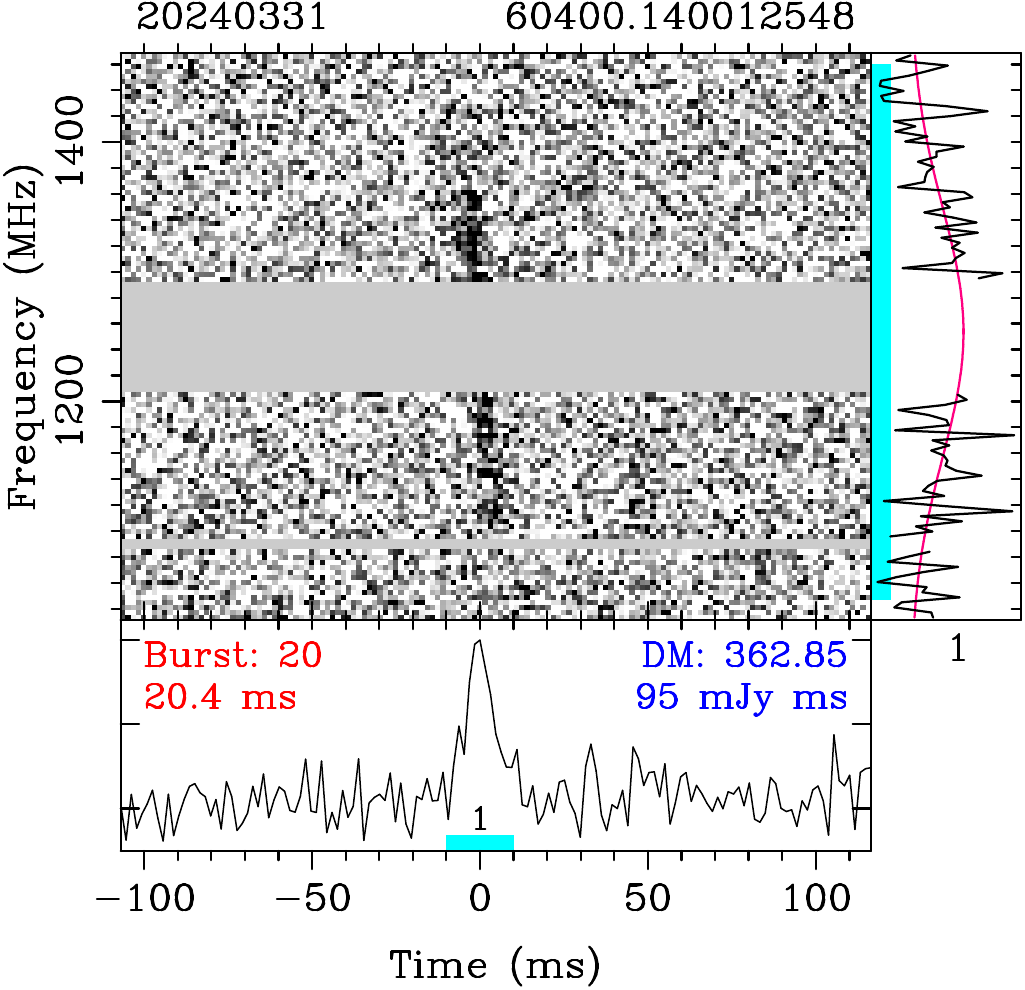}
\includegraphics[height=0.29\linewidth]{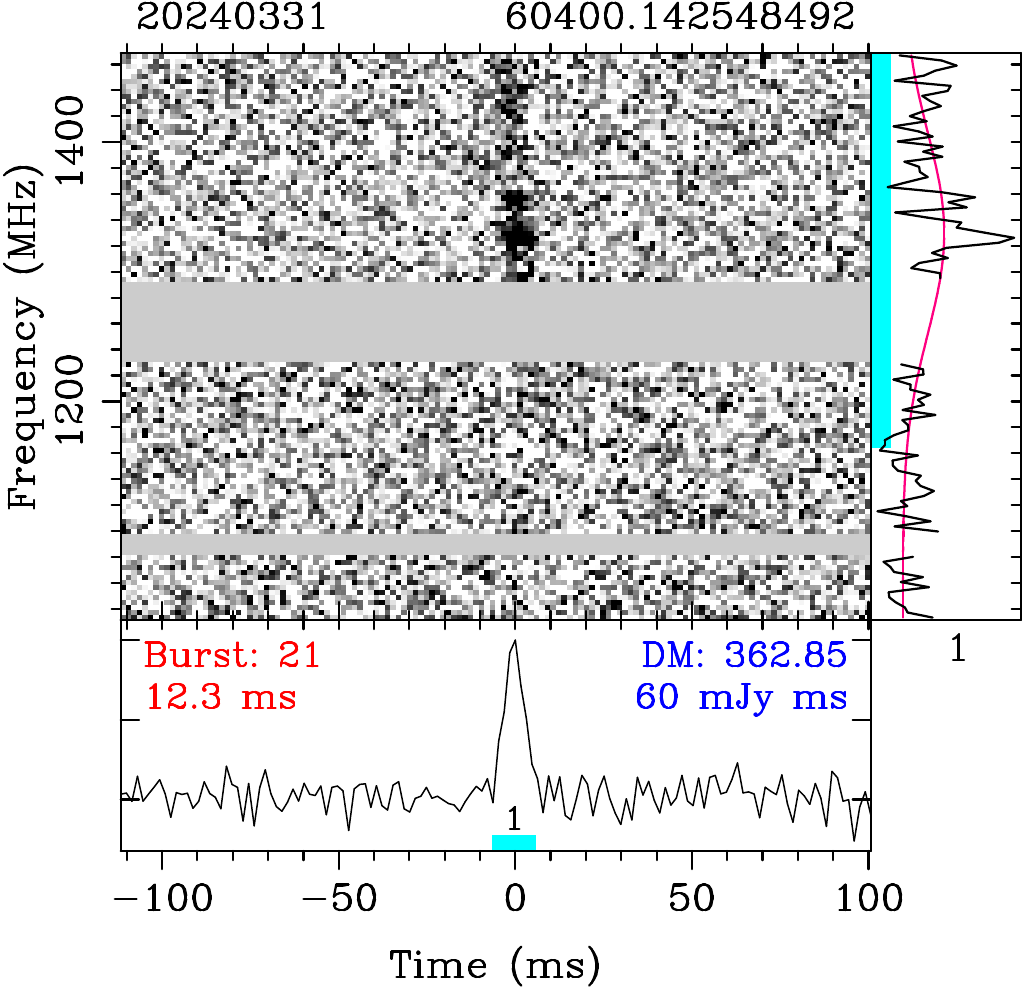}
\includegraphics[height=0.29\linewidth]{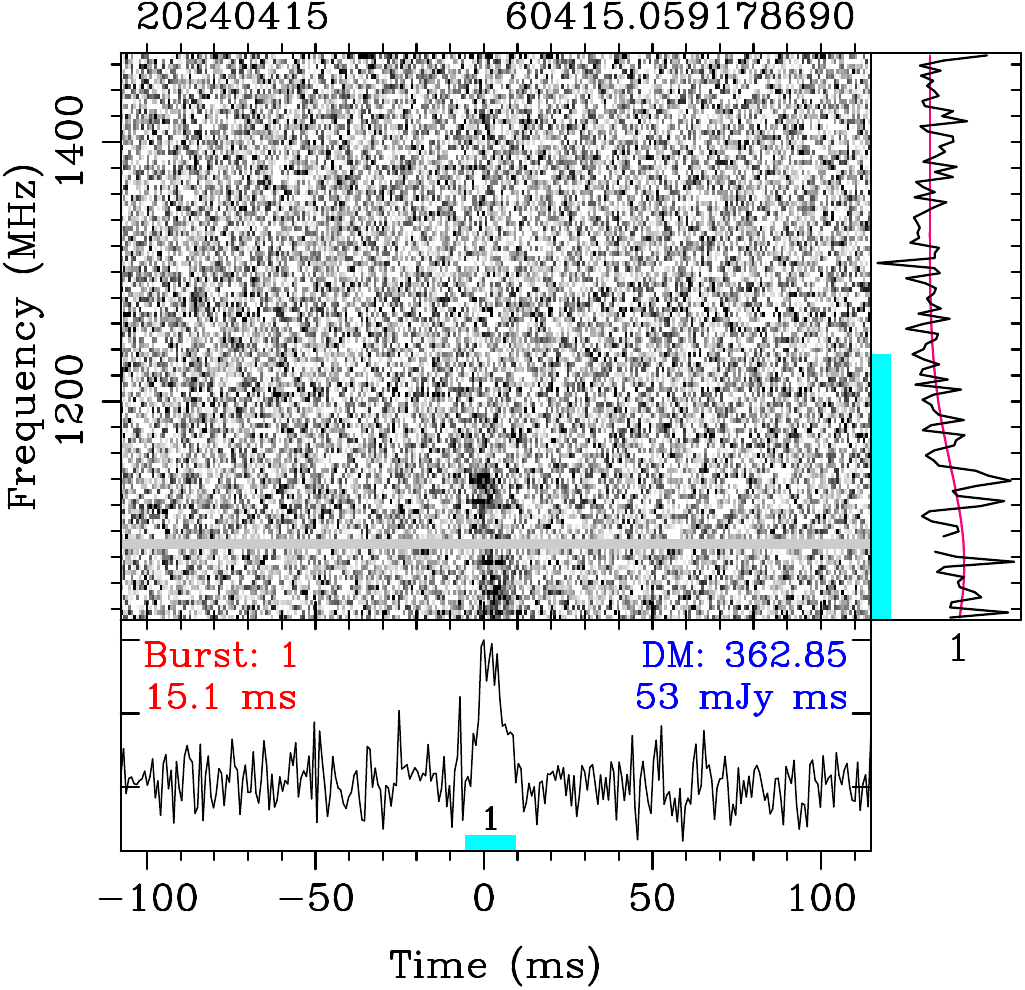}
\includegraphics[height=0.29\linewidth]{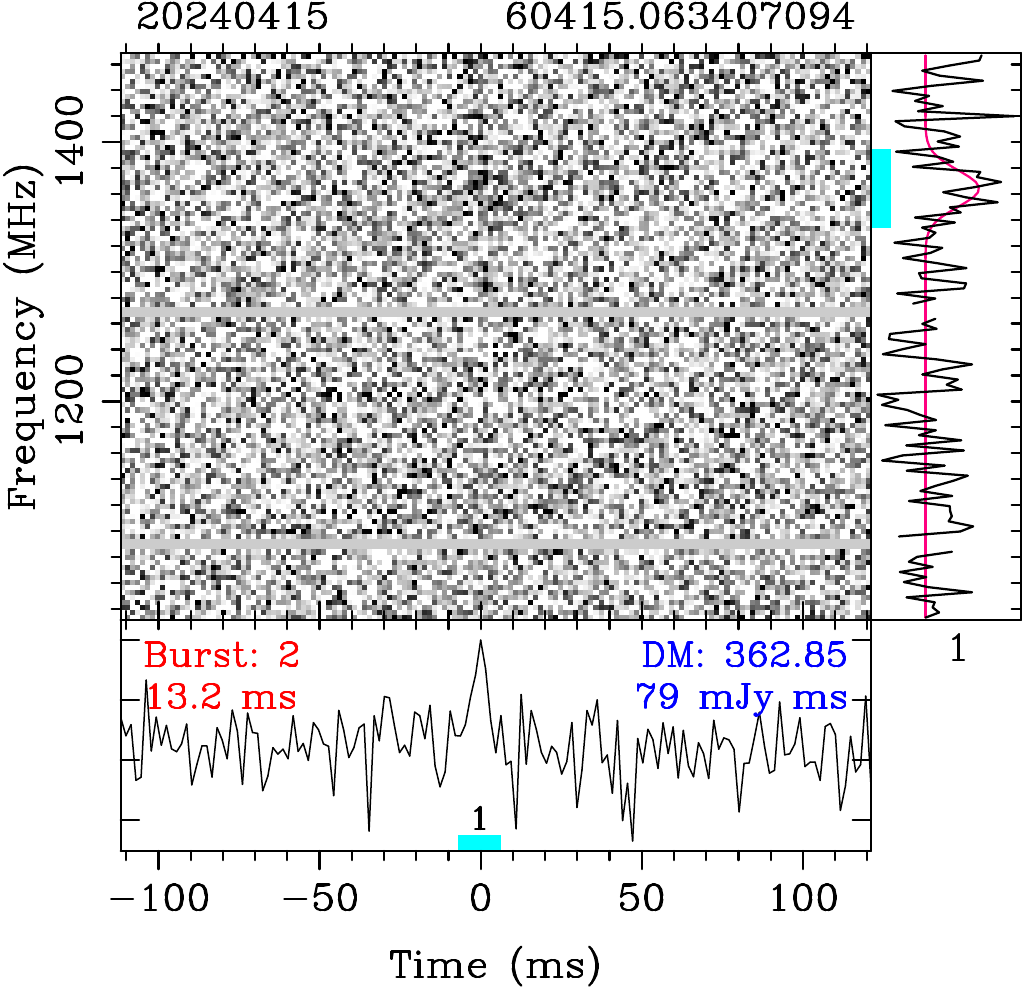}
\includegraphics[height=0.29\linewidth]{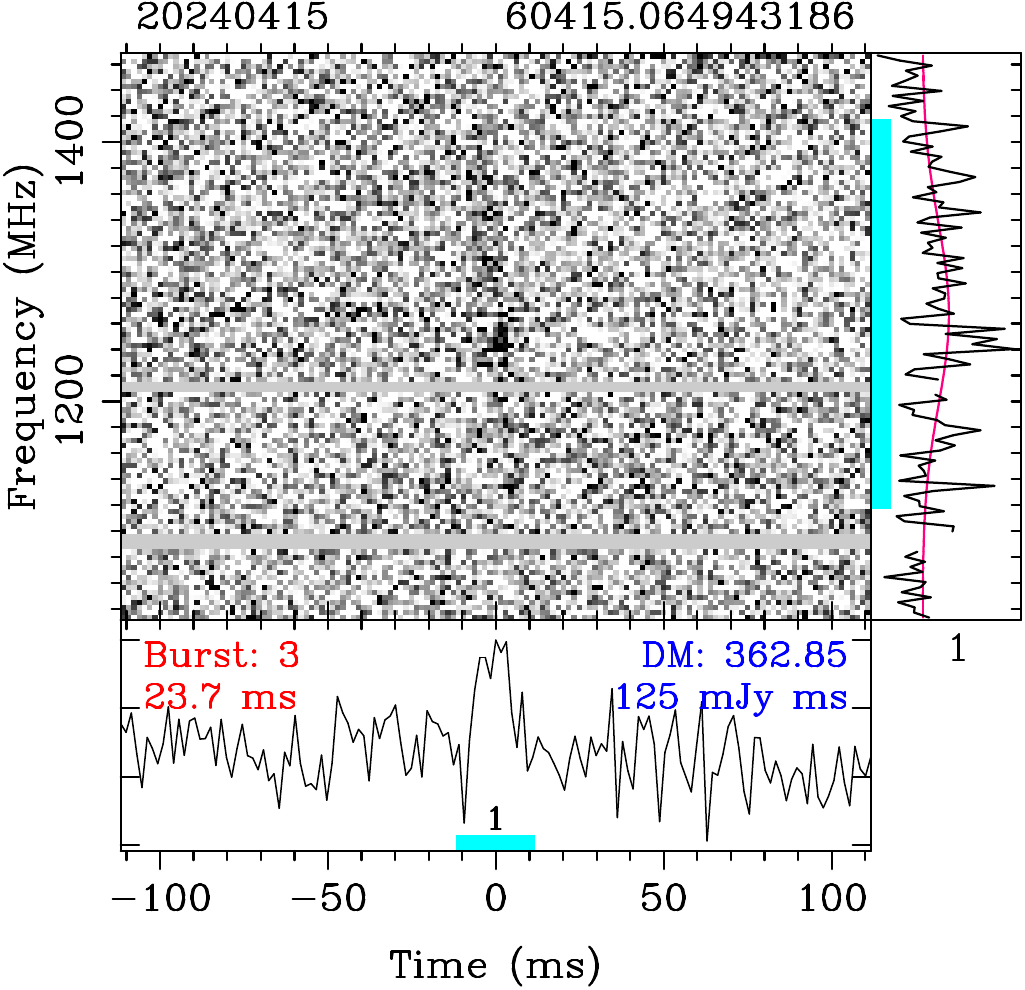}
\includegraphics[height=0.29\linewidth]{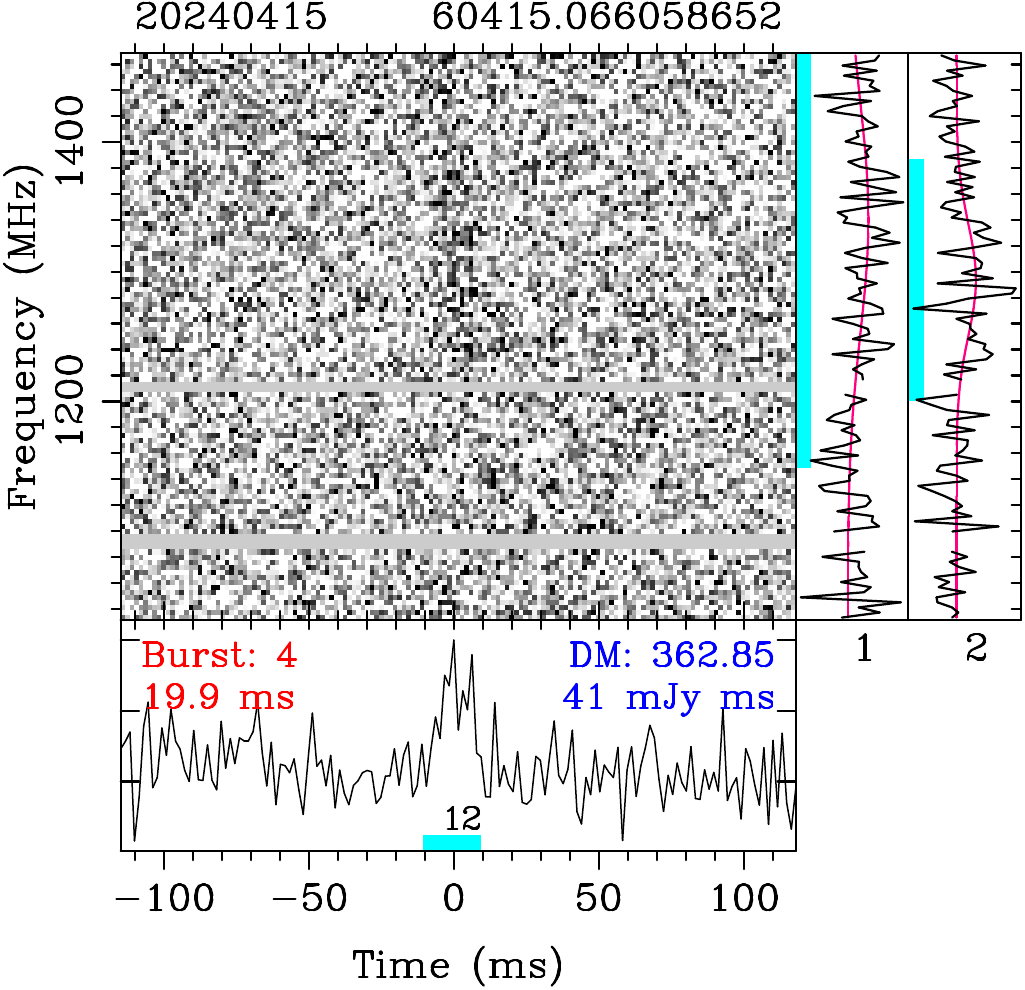}
\includegraphics[height=0.29\linewidth]{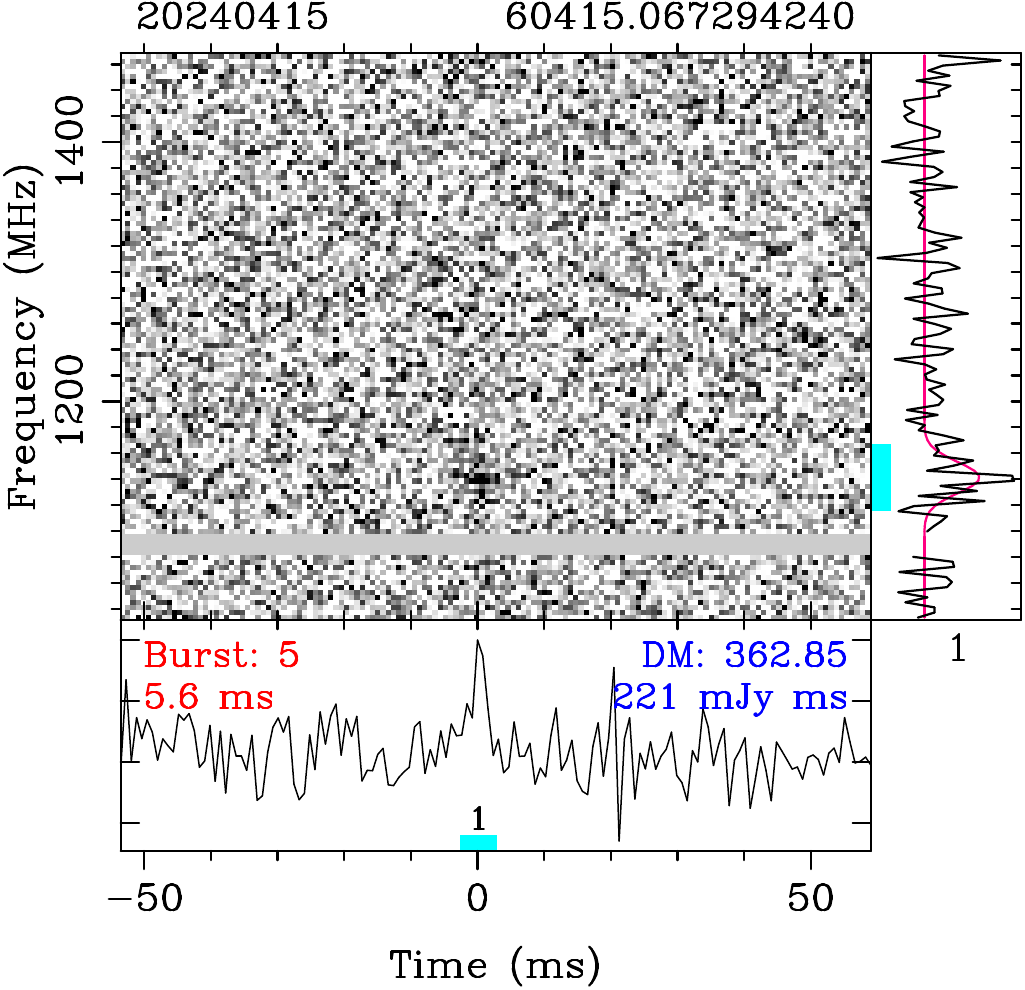}
\includegraphics[height=0.29\linewidth]{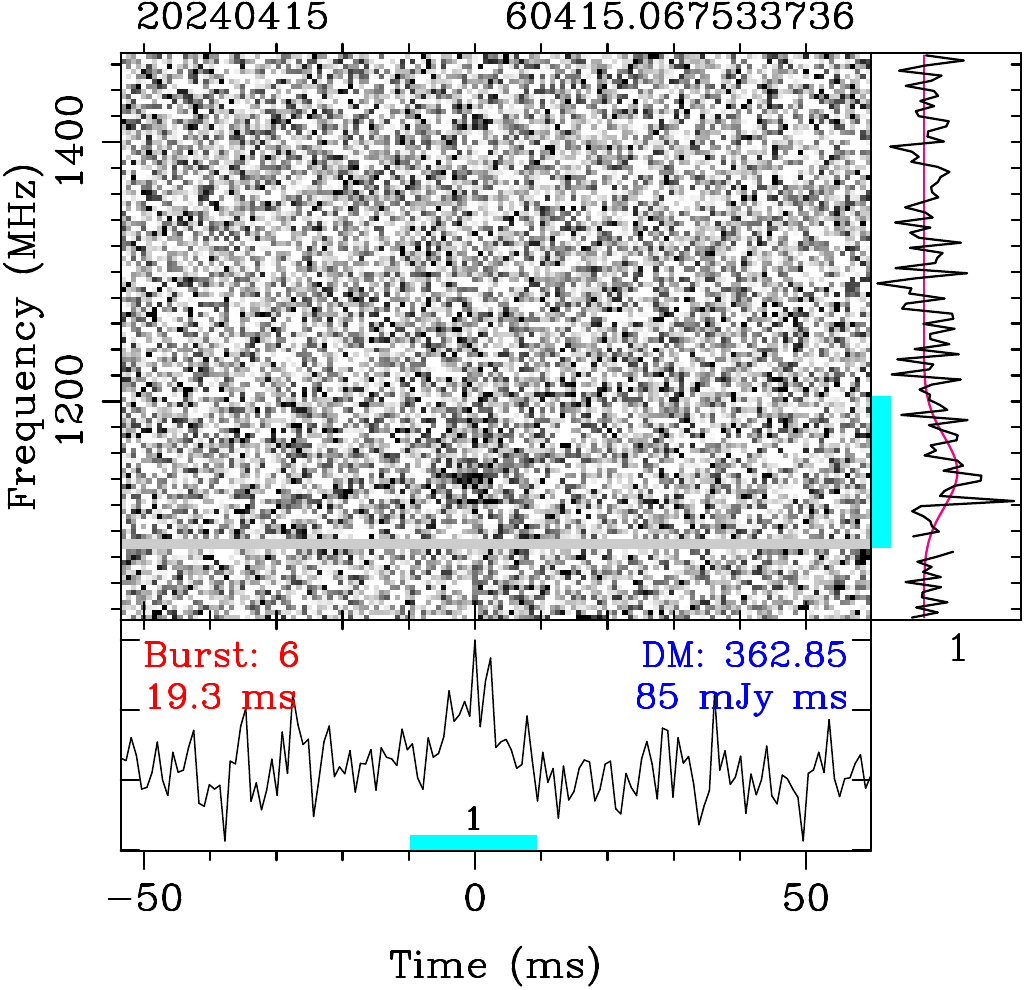}
\includegraphics[height=0.29\linewidth]{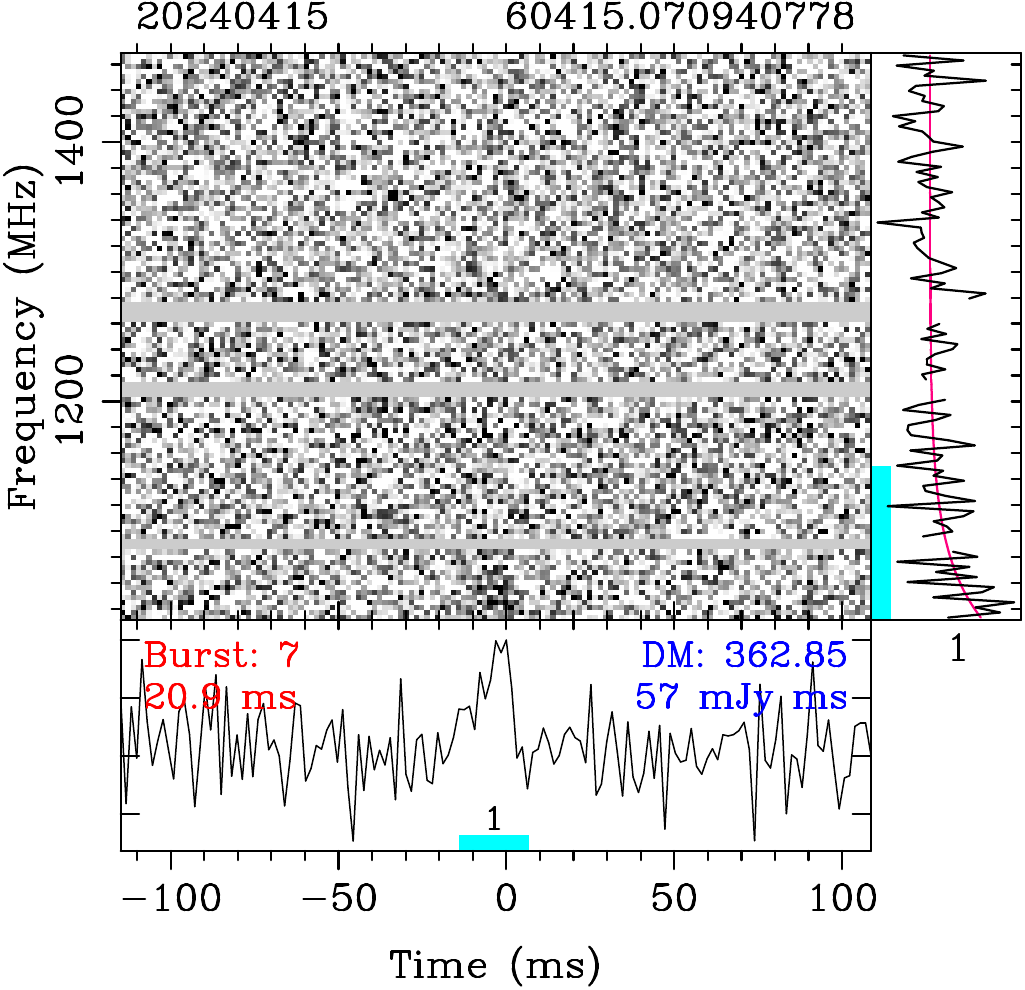}
\includegraphics[height=0.29\linewidth]{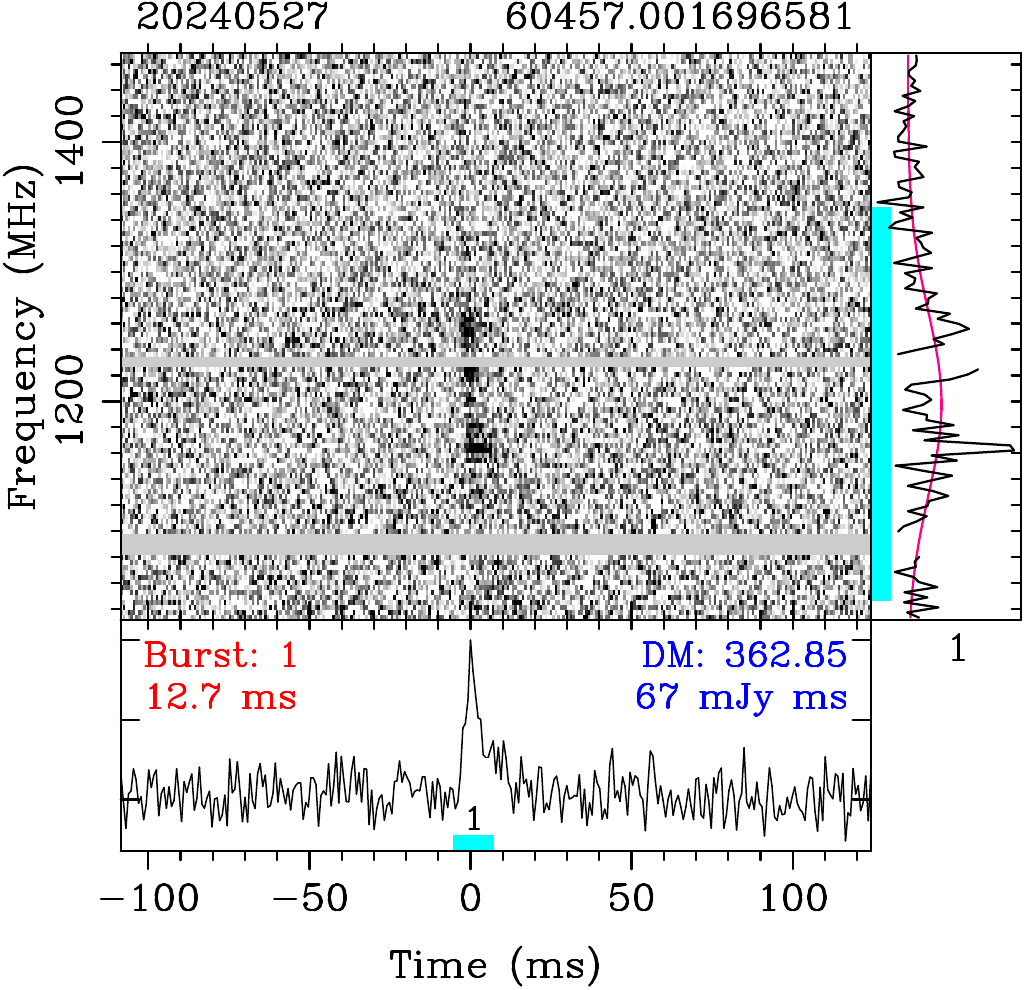}
\includegraphics[height=0.29\linewidth]{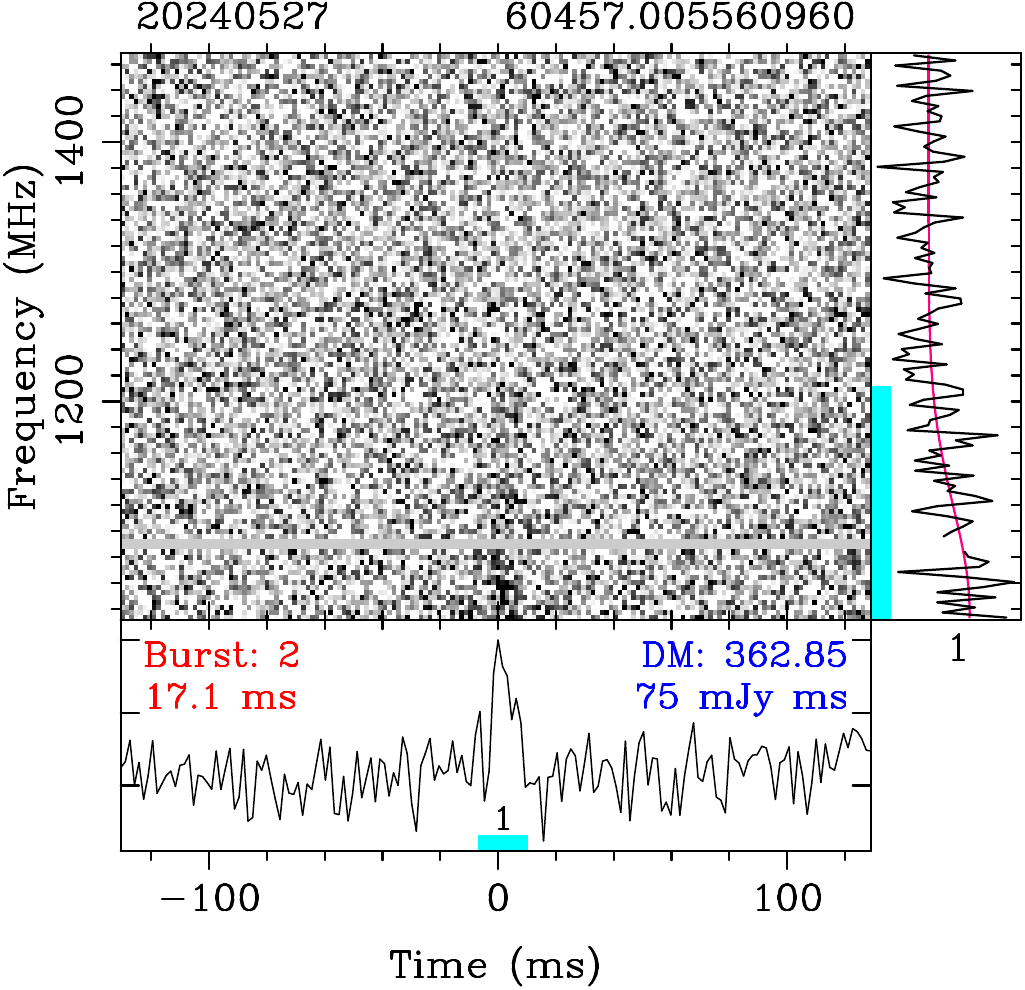}
\caption{({\textit{continued}})}
\end{figure*}
\addtocounter{figure}{-1}
\begin{figure*}
\flushleft
\includegraphics[height=0.29\linewidth]{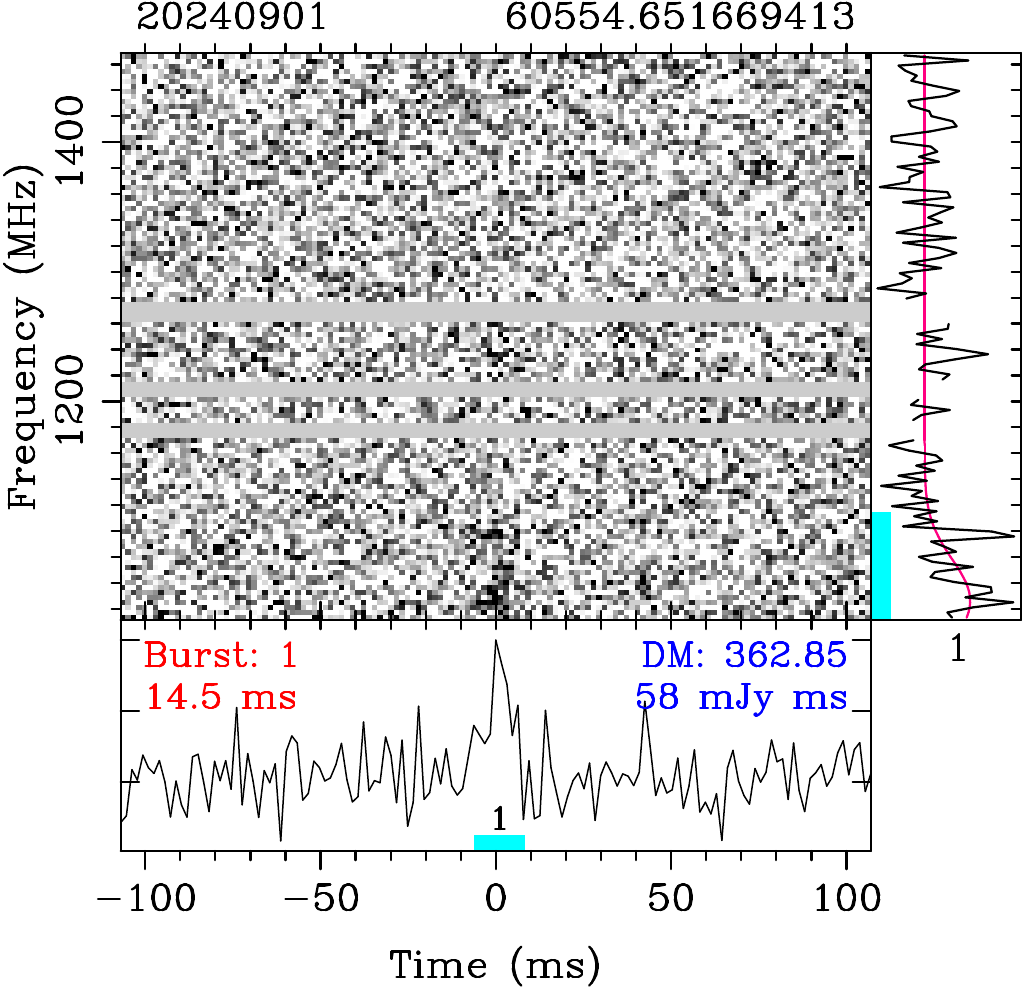}
\includegraphics[height=0.29\linewidth]{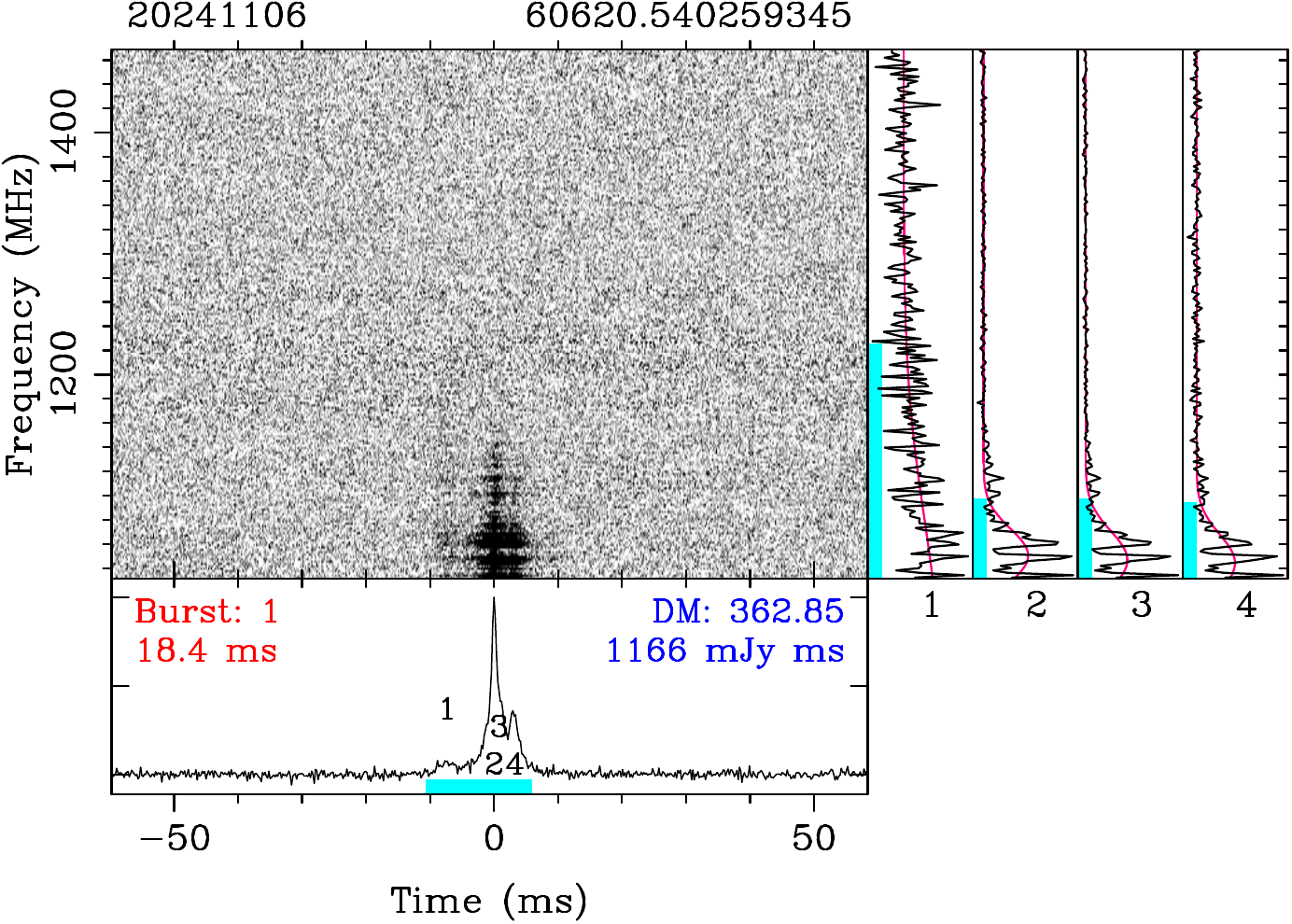}
\includegraphics[height=0.29\linewidth]{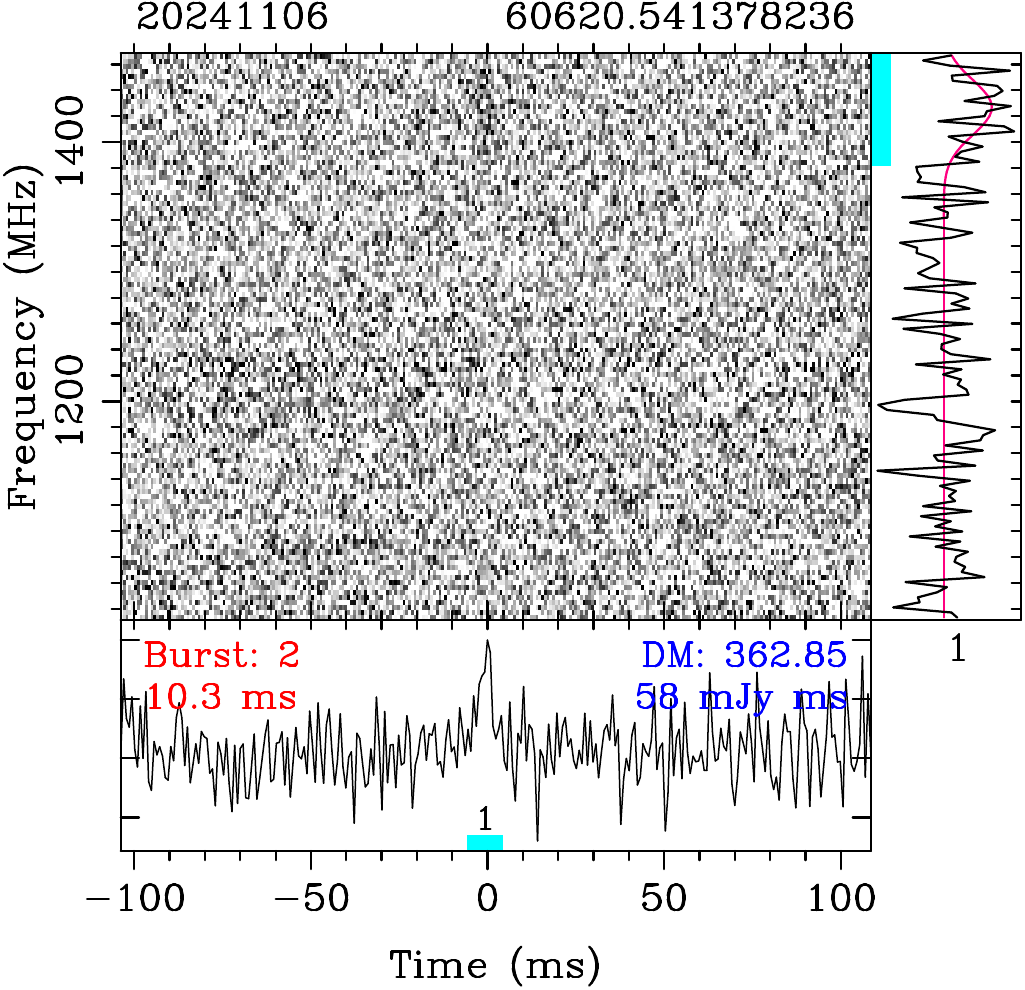}
\includegraphics[height=0.29\linewidth]{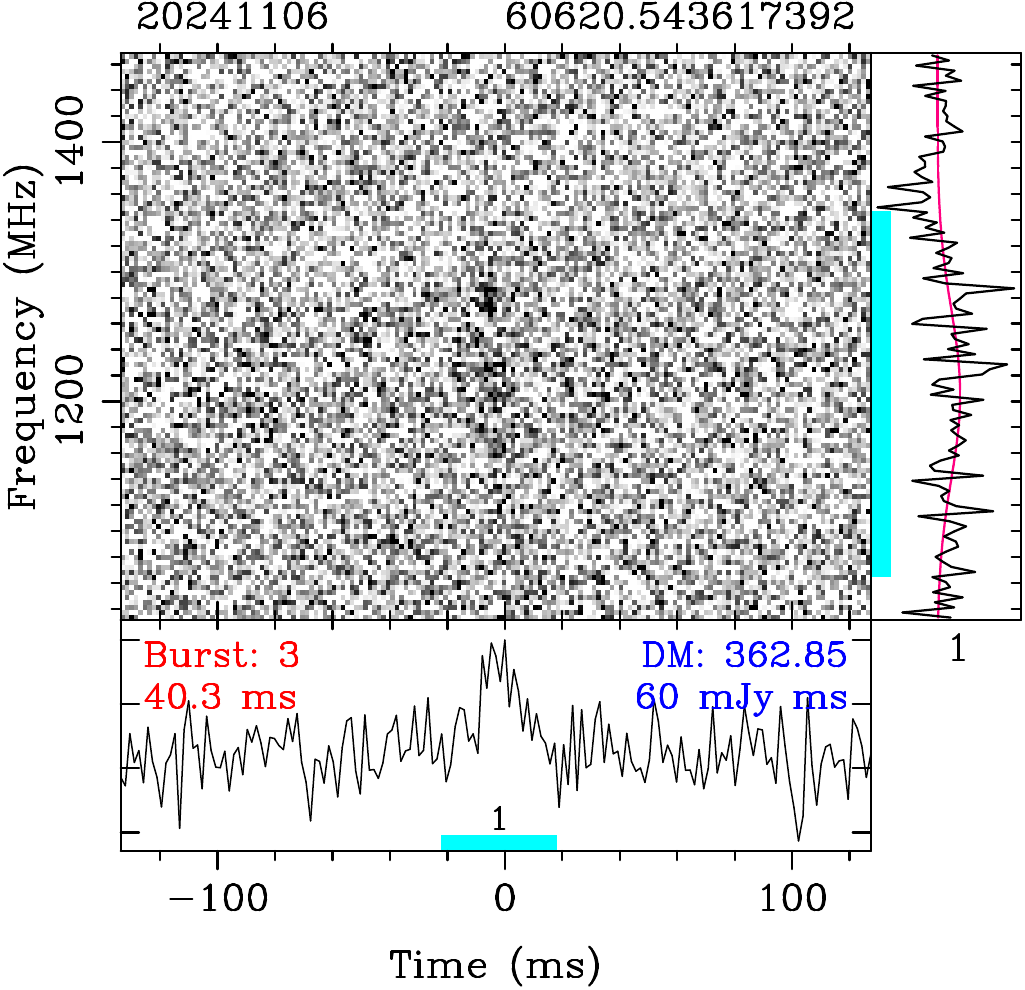}
\includegraphics[height=0.29\linewidth]{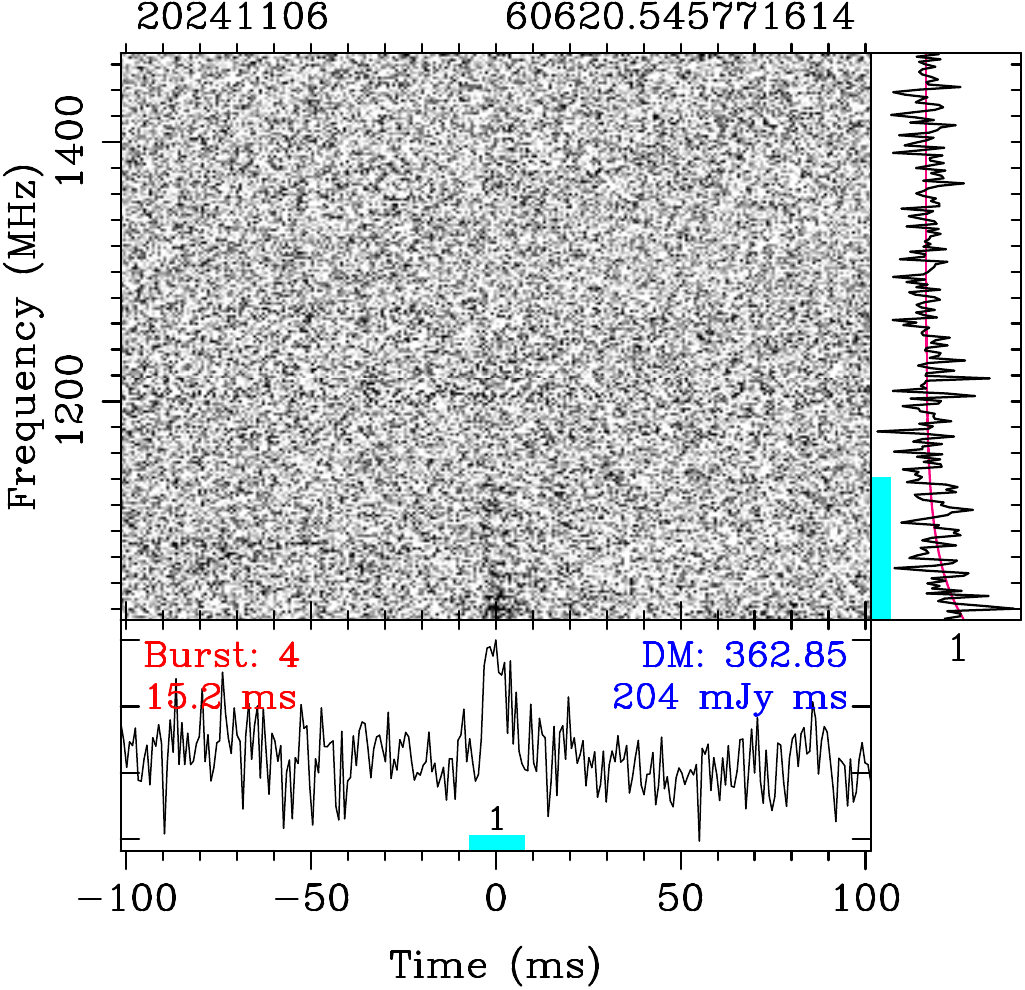}
\includegraphics[height=0.29\linewidth]{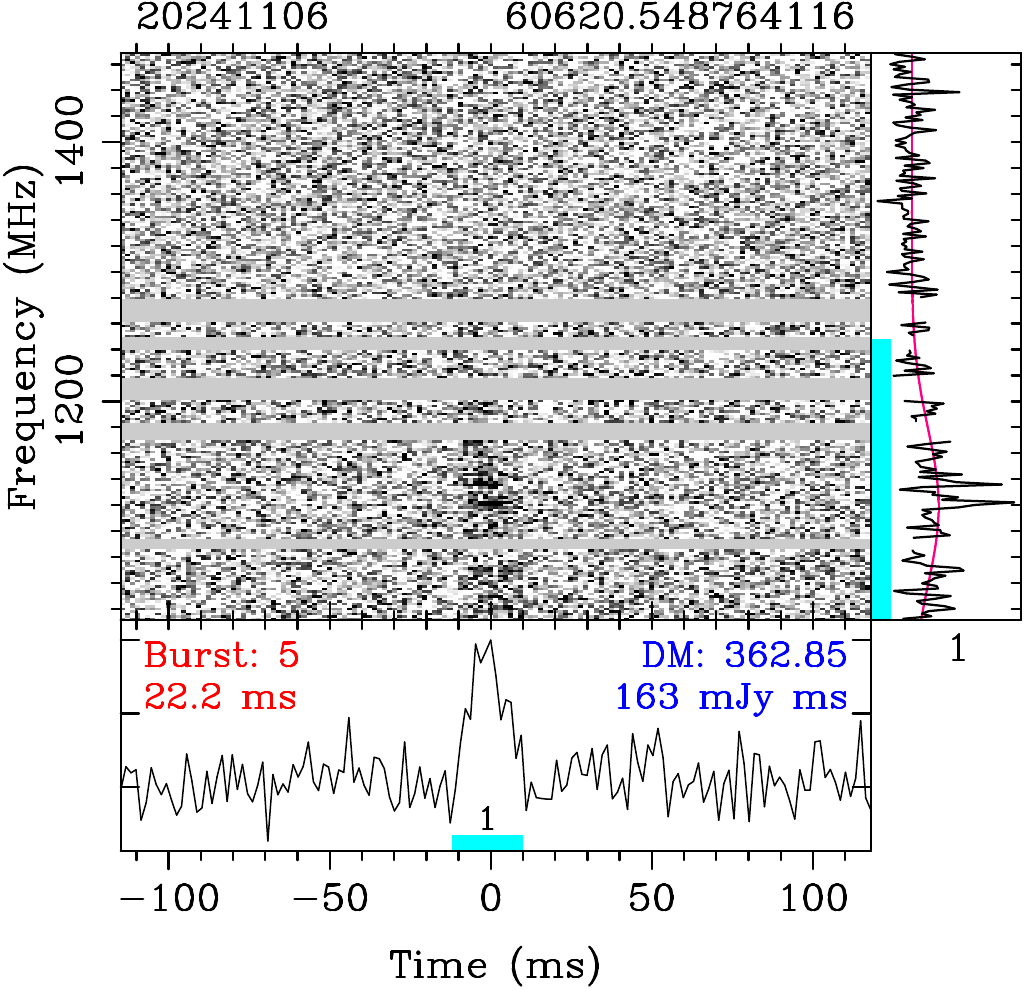}
\includegraphics[height=0.29\linewidth]{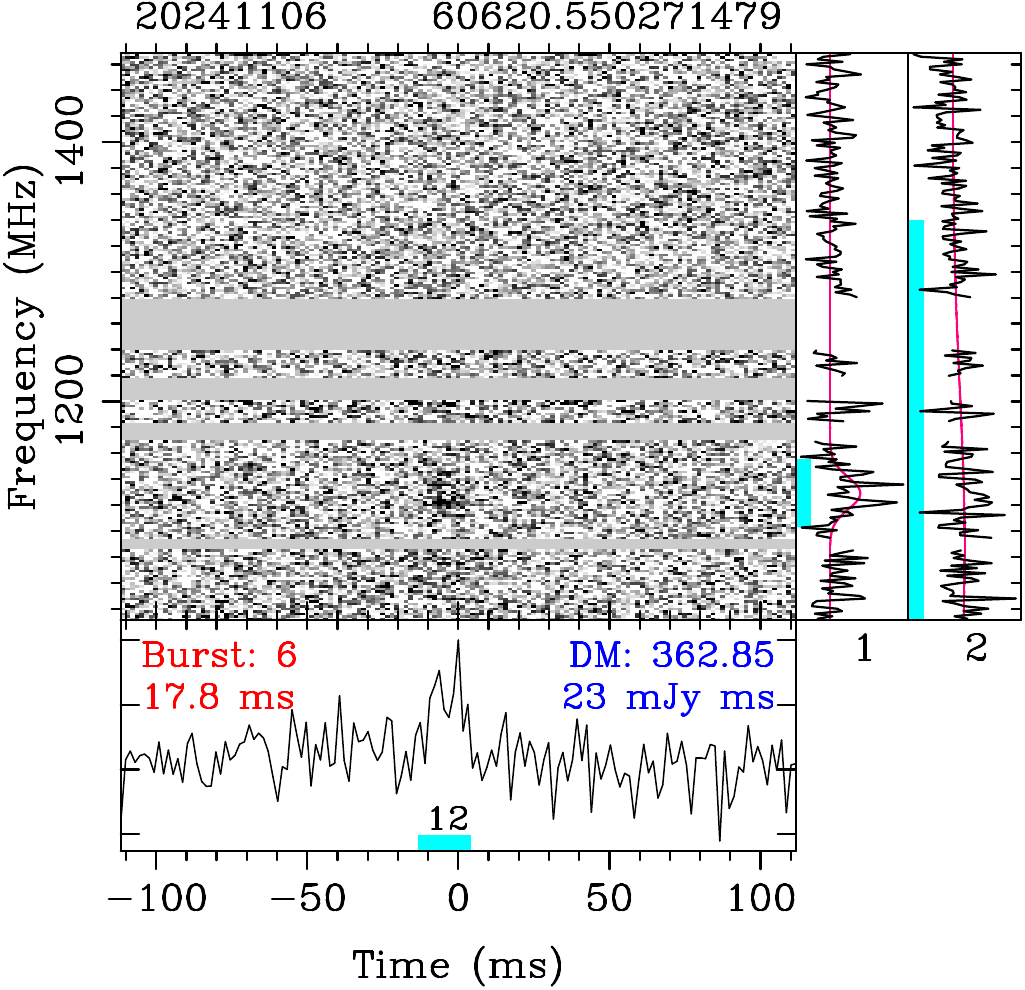}
\includegraphics[height=0.29\linewidth]{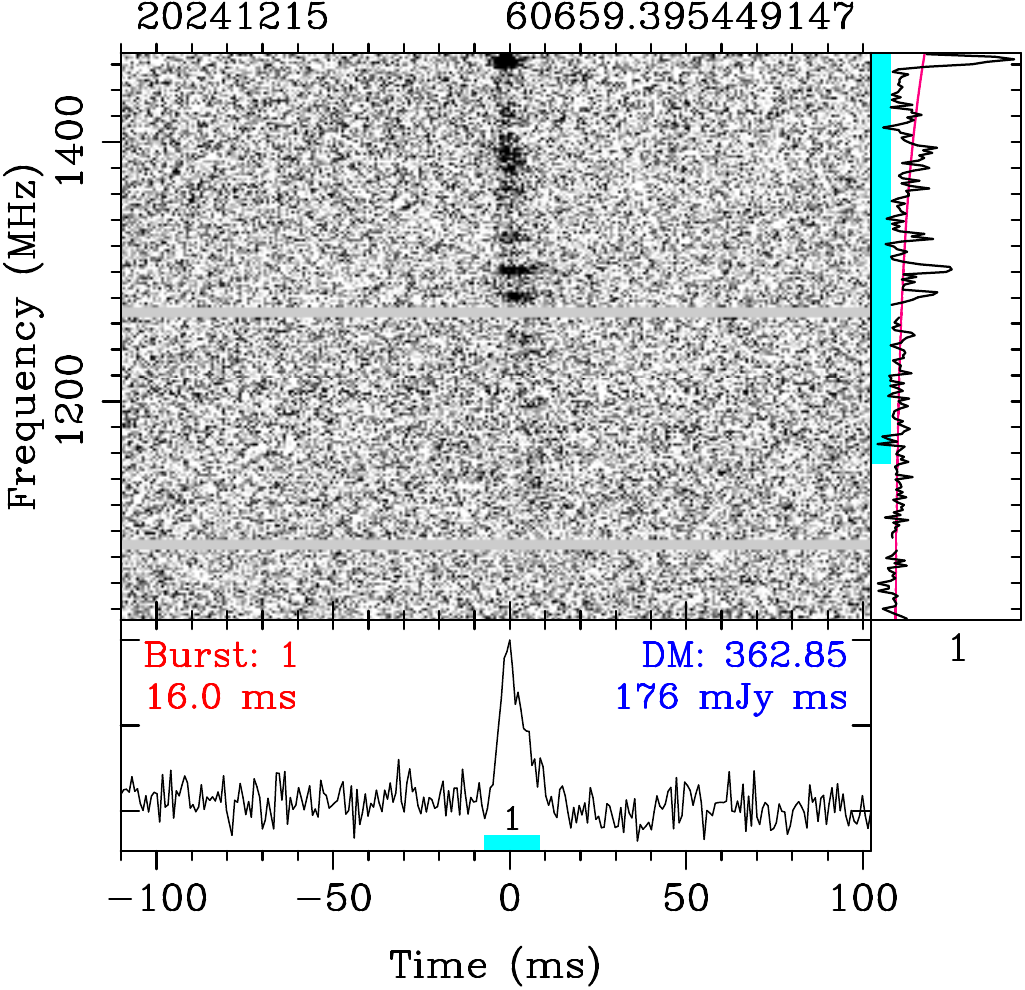}
\includegraphics[height=0.29\linewidth]{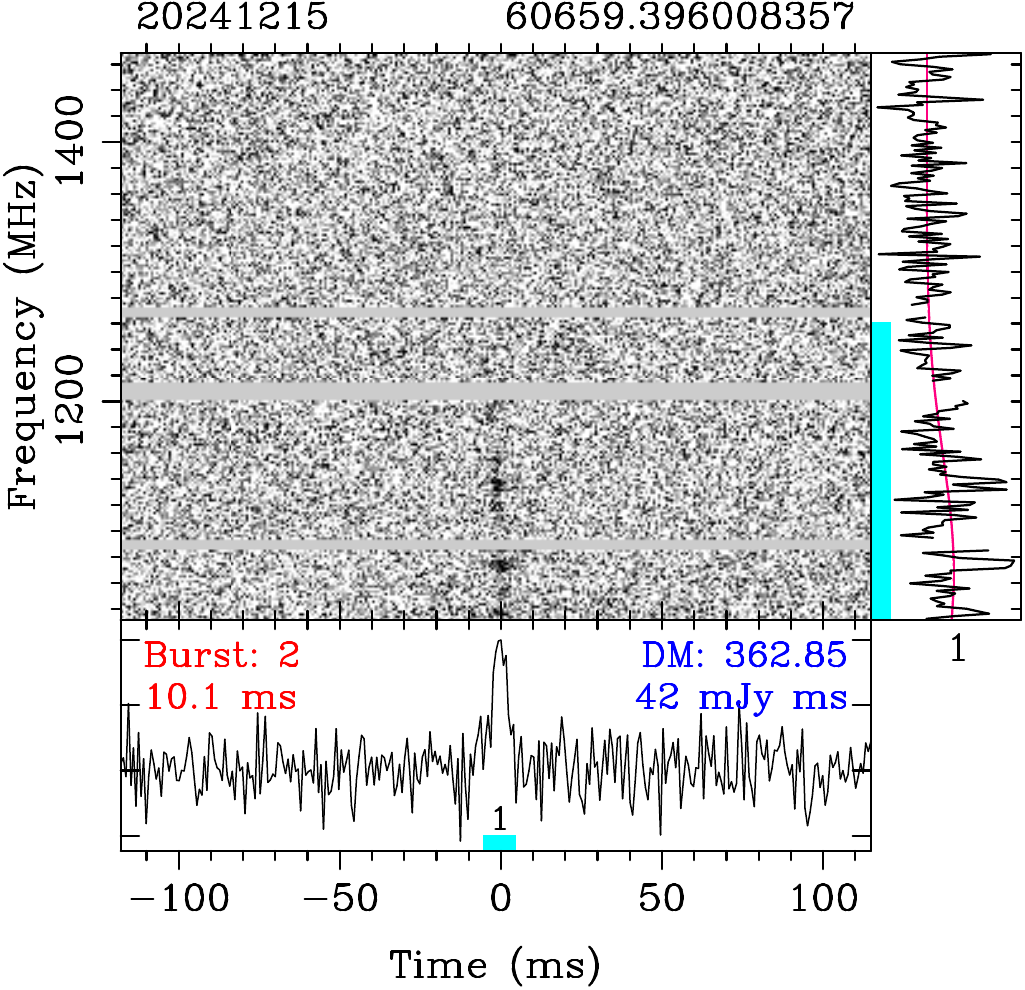}
\includegraphics[height=0.29\linewidth]{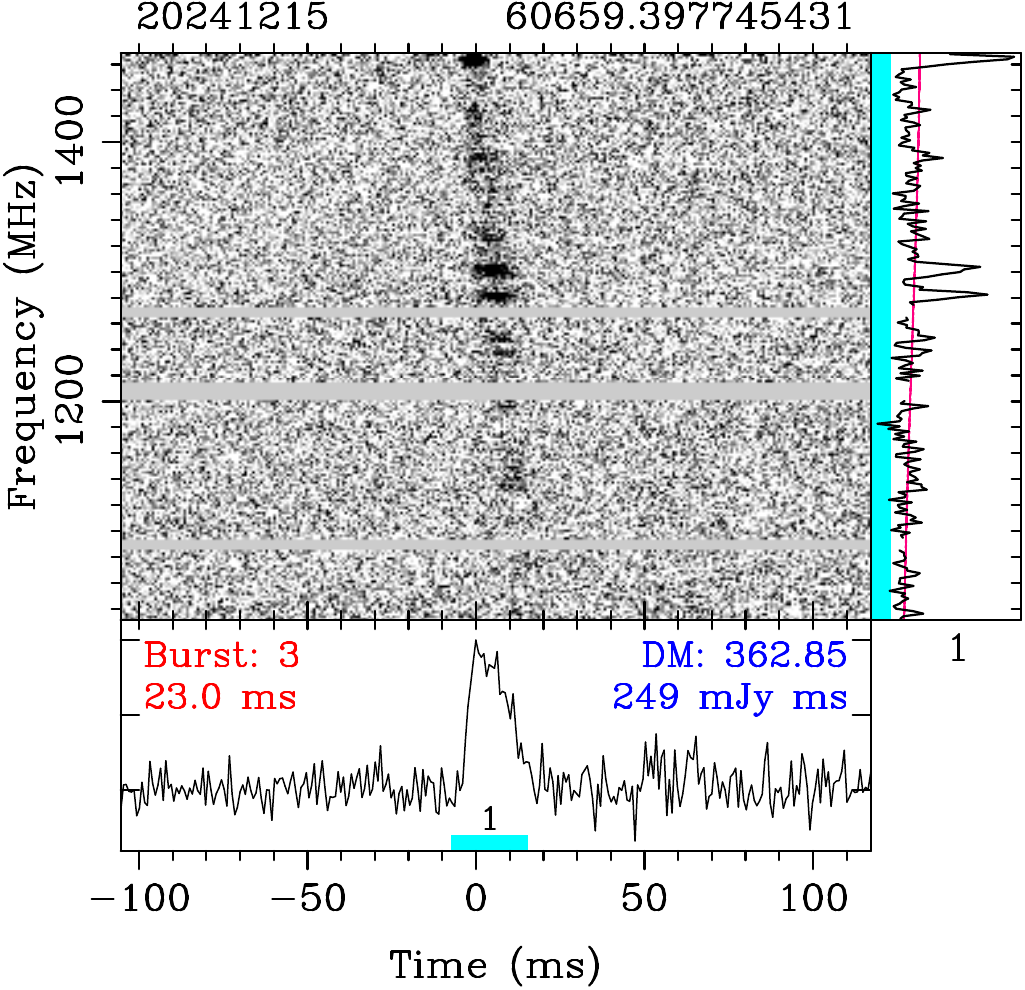}
\includegraphics[height=0.29\linewidth]{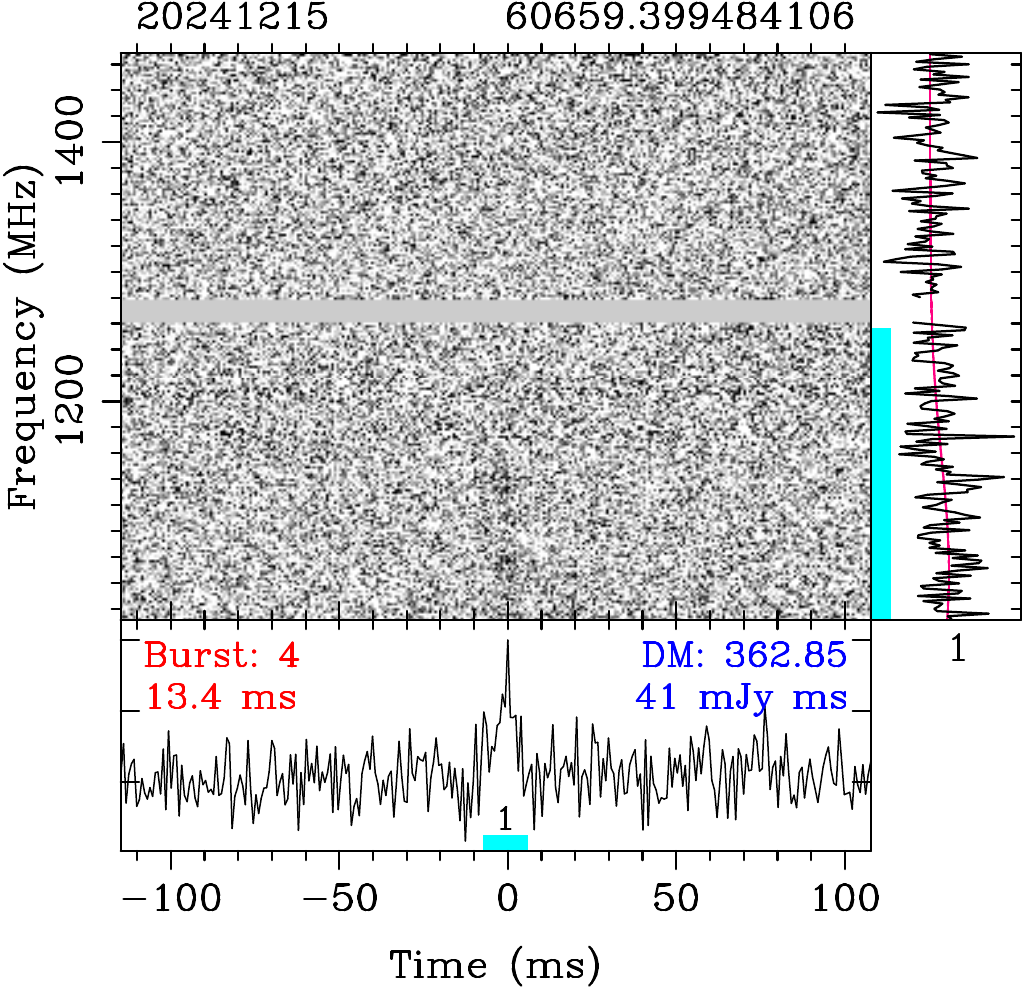}
\caption{({\textit{continued}})}
\end{figure*}
\addtocounter{figure}{-1}
\begin{figure*}
\flushleft
\includegraphics[height=0.29\linewidth]{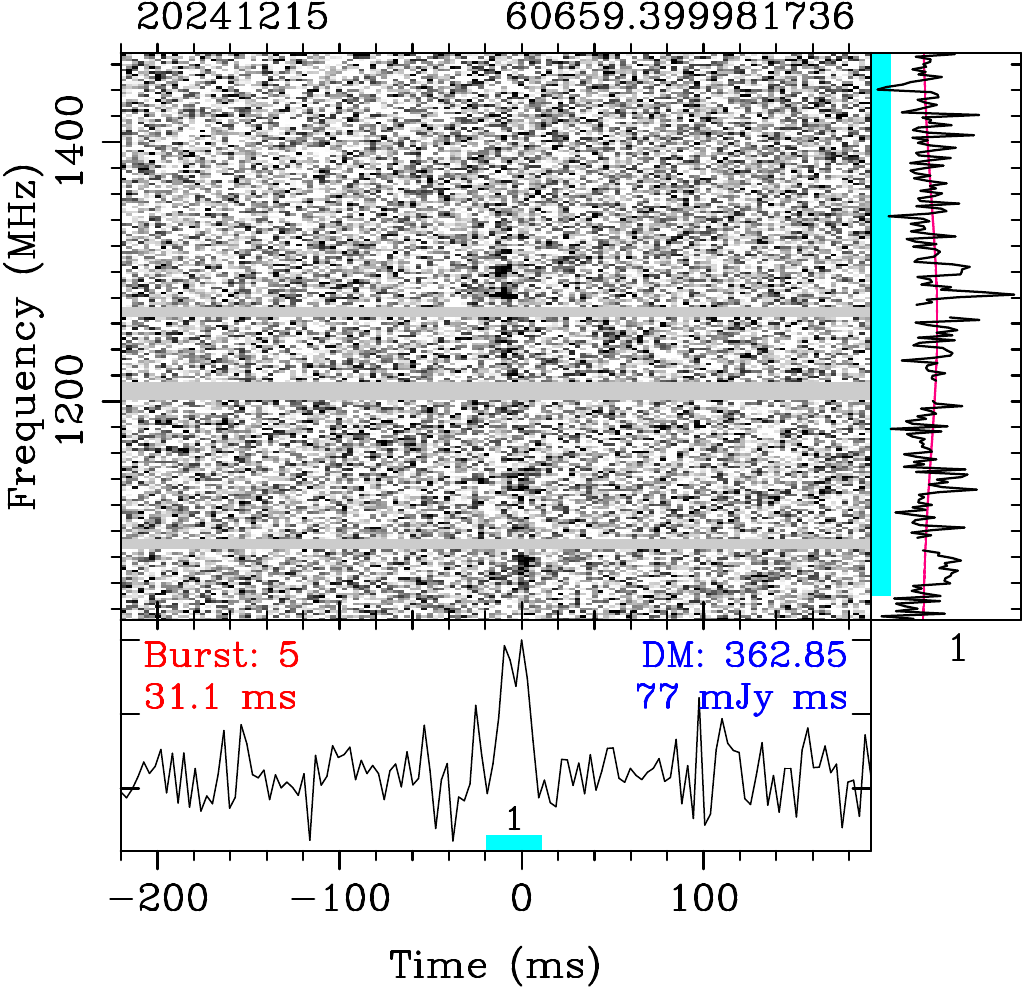}
\includegraphics[height=0.29\linewidth]{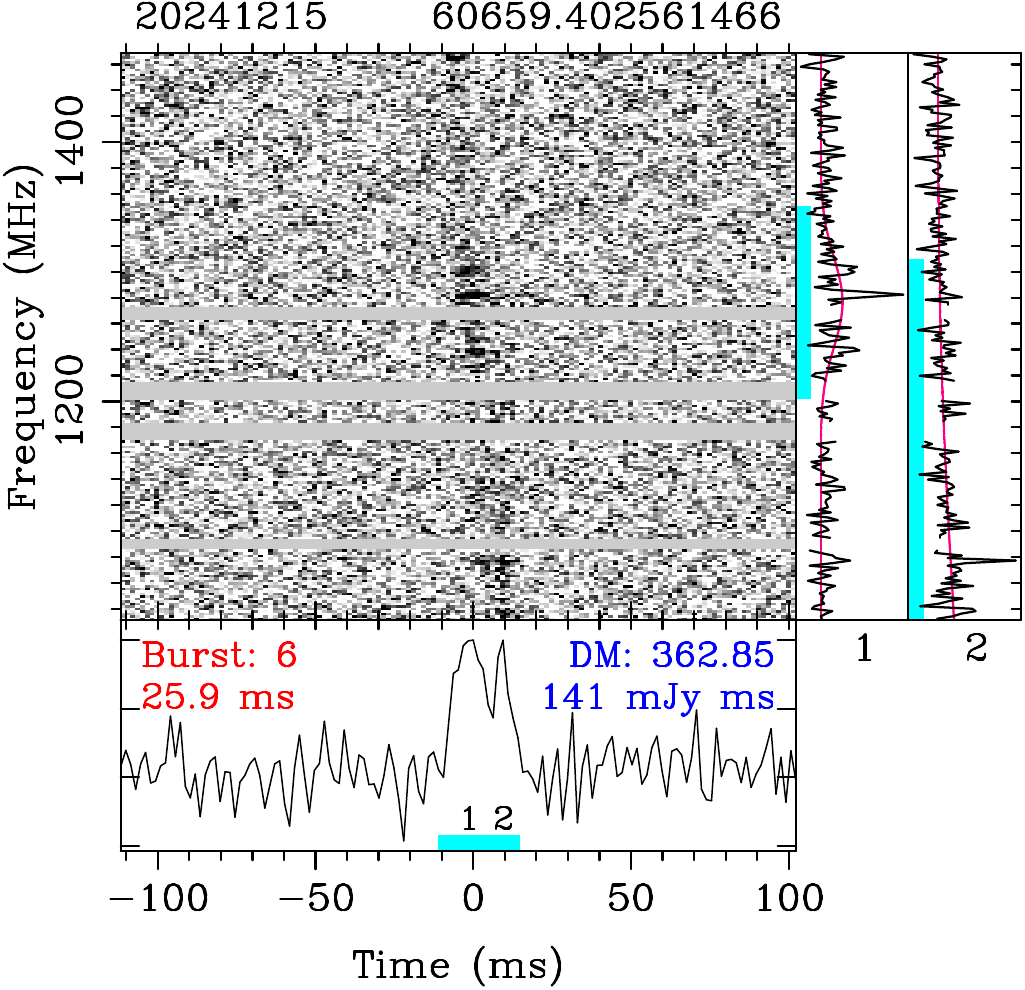}
\includegraphics[height=0.29\linewidth]{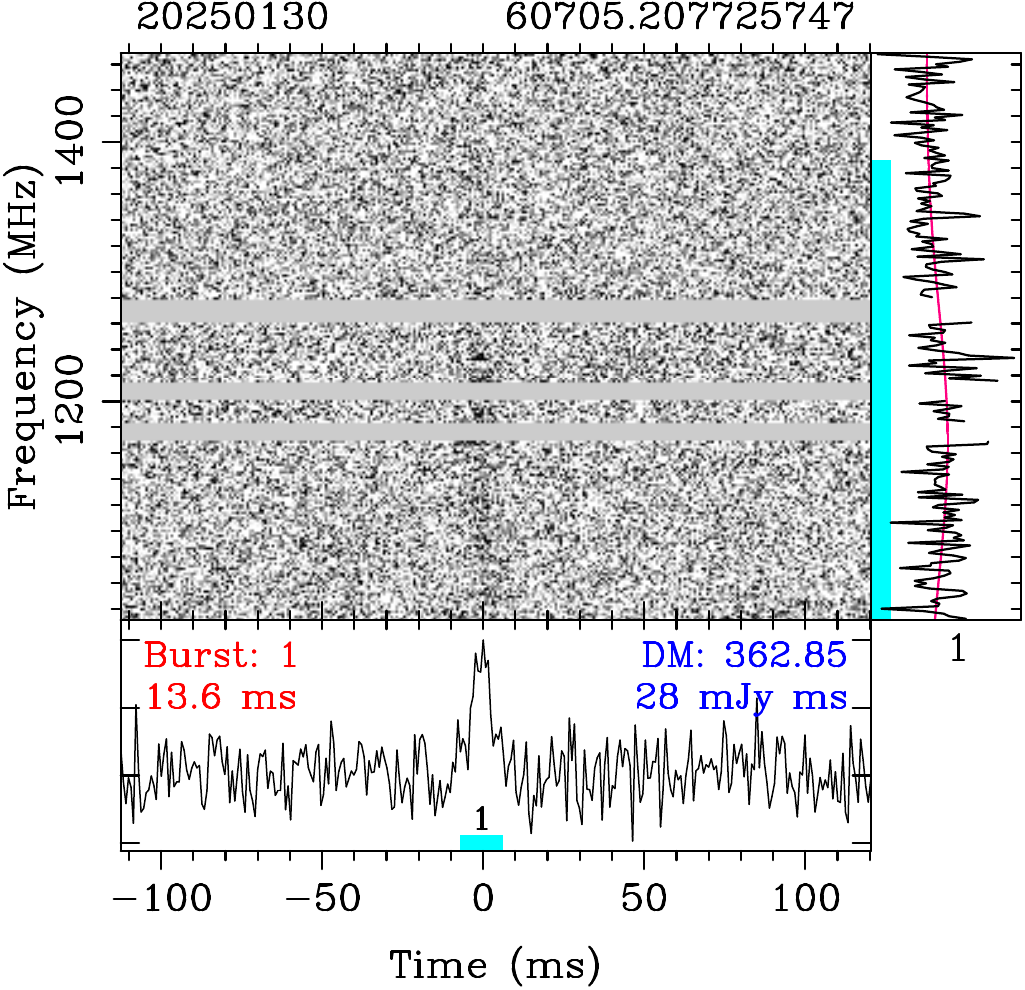}
\includegraphics[height=0.29\linewidth]{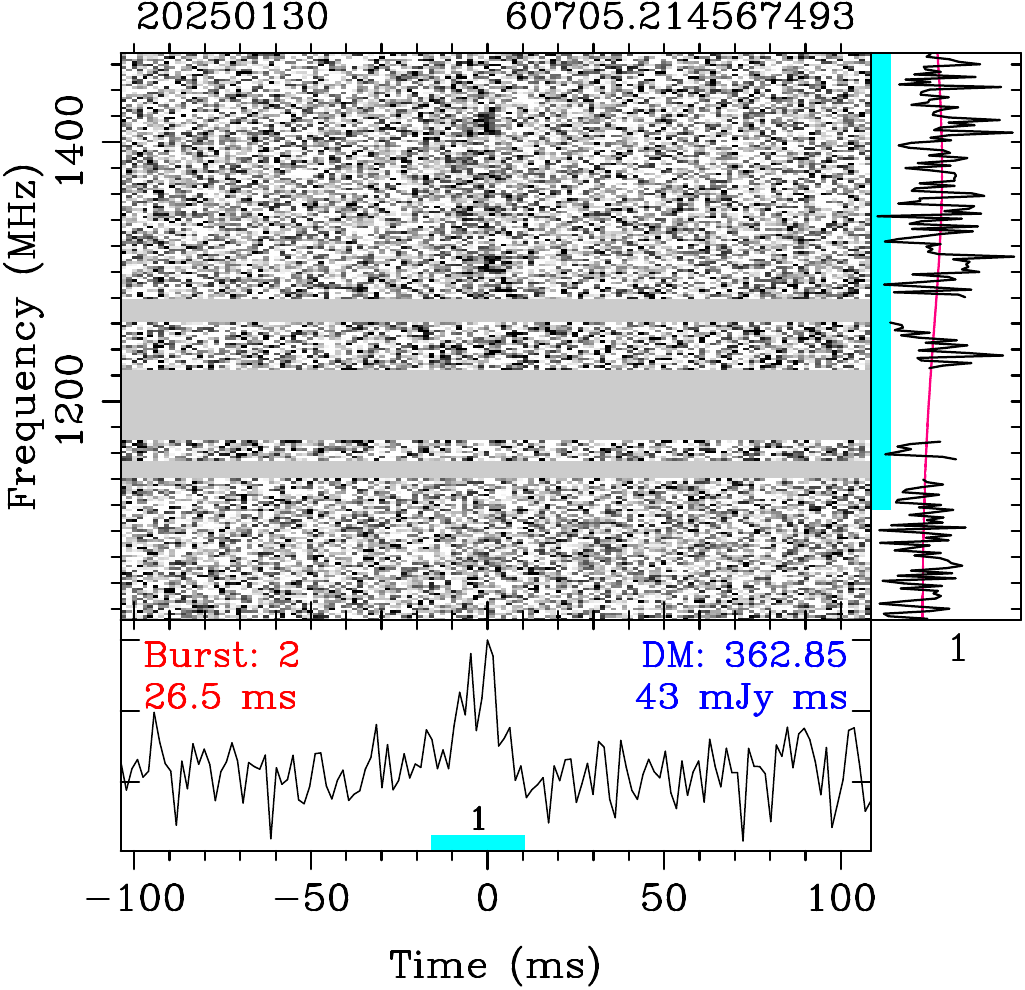}
\caption{({\textit{continued}})}
\end{figure*}

\begin{table*}[!t]
\begin{threeparttable}
\caption{FRB Sample with RM Measurements (or Upper Limits) and Flux Densities of PRS.}\label{tab:APPtab2}%
\small
\begin{tabular*}{\textwidth}{@{\extracolsep\fill}lrrr}
\hline
FRB Name    & RM                  & $L_{\nu}$                               & References  \\
            & ($\rm rad\,m^{-2}$) & ($\rm 10^{29}\,erg\,s^{-1}\,Hz^{-1}$)   &             \\
            & (1)                 & (2)                                     &             \\
\hline
FRB 20230607A & -1.2$\times 10^4$ & (predicted: 1.0)                                     & This work   \\
\hline
FRB 20121102A & 1.4$\times 10^5$  & 2.1                                     & [1,2,3,4]   \\
FRB 20190520B & -3.6$\times 10^4$ & 3.8                                     & [5,6]       \\
FRB 20201124A & -889.5            & 0.053                                   & [7,8]       \\
FRB 20181030A & 36.6              & 0.0021                                  & [9,10]    \\
FRB 20190417A & 4681              & 0.797                                   & [10,11]    \\
\hline
FRB 20110523A & -186.1            & $< 5.8_{-3.7}^{+6.6}$           & [12]         \\
FRB 20150215A & 1.5               & $< 1.4_{-0.9}^{+1.4}$           & [13]        \\
FRB 20150418A & 36                & $< 11_{-7}^{+12}$               & [14]        \\
FRB 20150807A & 12                & $< 2.1_{-1.8}^{+3.7}$           & [15]        \\
FRB 20160102A & -220.6            & $< 249_{-84}^{+112}$            & [16,17]     \\
FRB 20180309A & $< 150$           & $< 0.8_{-0.7}^{+1.5}$           & [18]        \\
FRB 20180916B & -114.6            & $< 0.0048$                      & [19,20]     \\
FRB 20180924B & 14                & $< 0.72$                        & [21]        \\
FRB 20181112A & 10.9              & $< 1.91$                        & [22]        \\
FRB 20191108A & 474               & $< 24_{-16}^{+29}$              & [23]        \\
FRB 20200120E & -36.9             & $< 3.1 \times 10^{-6}$          & [24,25]     \\
FRB 20210117A & 43                & $< 0.15$                        & [26]        \\
FRB 20220912A & -0.08             & $< 0.074$                       & [27,28]     \\
\hline
\end{tabular*}
\begin{tablenotes}
\item 
    \textbf{Note:}\\ (1): The observed RM of each FRB. \\ (2): The persistent emission luminosity.\\
    \textbf{Reference:}. 
    [1] \citet{Spitler2014ApJ...790..101S}, 
    [2] \citet{Tendulkar2017ApJ...834L...7T}, 
    [3] \citet{MarcoteB2017ApJ...834L...8M}, 
    [4] \citet{Michilli2018Natur.553..182M}, 
    [5] \citep{NiuCH2022Natur.606..873N},
    [6] \citet{AnnaT2023Sci...380..599A},    
    [7] \citet{XuH2022Natur.609..685X},
    [8] \citet{Bruni2023arXiv231215296B},
    [9] \citet{MckinvenR2023ApJ...951...82M}, 
    [10] \citet{Ibik2024arXiv240911533I},
    [11] \citet{FengY2022Sci...375.1266F},
    [12] \citet{Masui2015Natur.528..523M},
    [13] \citet{Petroff2017MNRAS.469.4465P},
    [14] \citet{Keane2016Natur.530..453K},
    [15] \citet{Ravi2016Sci...354.1249R},
    [16] \citet{Bhandari2018MNRAS.475.1427B},
    [17] \citet{Caleb2018MNRAS.478.2046C},
    [18] \citet{Oslowski2019MNRAS.488..868O},
    [19] \citet{CHIME2019ApJ...885L..24C},
    [20] \citet{Marcote2020Natur.577..190M},
    [21] \citet{Bannister2019Sci...365..565B},
    [22] \citet{Prochaska2019Sci...366..231P},
    [23] \citet{Connor2020MNRAS.499.4716C},
    [24] \citet{Bhardwaj2021ApJ...910L..18B},
    [25] \citet{Kirsten2022Natur.602..585K},
    [26] \citet{Bhandari2023ApJ...948...67B},
    [27] \citet{ZhangYK2023ApJ...955..142Z},
    [28] \citet{Hewitt2024MNRAS.529.1814H}.
\end{tablenotes}
\end{threeparttable}
\end{table*}

\end{document}